\documentclass[twocolumn]{aastex61}
\usepackage[utf8]{inputenc}
\usepackage{float}

\newcommand{\ntothosts}{358}
\newcommand{\nhosts}{116}
\newcommand{\ngoodhosts}{114}
\newcommand{\nrvhosts}{242}

\newcommand{\fbolprecper}{2\%}
\newcommand{\fbolprecperall}{1.7\%}
\newcommand{\thetaprecper}{3\%}
\newcommand{\thetaprecwunit}{0.8~$\mu$as}
\newcommand{\thetaprecwunitall}{1.6~$\mu$as}
\newcommand{\thetaprecperall}{1.8\%}
\newcommand{\rstarprecper}{8\%}
\newcommand{\mstarprecper}{30\%}
\newcommand{\rpprecper}{9\%}
\newcommand{\rplitprecper}{5\%}
\newcommand{\rpdrtwoprecper}{$\approx$3\%}
\newcommand{\mpprecper}{22\%}
\newcommand{\mplitprecper}{6\%}
\newcommand{\mpdrtwoprecper}{$\approx$5\%}
\newcommand{\rvrstarprecper}{$\approx$2\%}

\newcommand{\tycparprec}{240~$\mu$as}
\newcommand{\drtwoparprec}{20~$\mu$as}

\newcommand{\lbol}{\ensuremath{L_{\rm bol}}}
\newcommand{\teff}{\ensuremath{T_{\rm eff}}}
\newcommand{\logg}{\ensuremath{\log g}}
\newcommand{\loggp}{\ensuremath{\log g_p}}
\newcommand{\feh}{[Fe/H]}
\newcommand{\fbol}{\ensuremath{F_{\rm bol}}}
\newcommand{\rsun}{\ensuremath{R_\odot}}
\newcommand{\msun}{\ensuremath{M_\odot}}
\newcommand{\rjup}{\ensuremath{R_J}}
\newcommand{\mjup}{\ensuremath{M_J}}
\newcommand{\rprstar}{\ensuremath{R_p / R_\star}}

\newcommand{\rhostar}{\ensuremath{\rho_\star}}
\newcommand{\rhoplanet}{\ensuremath{\rho_p}}
\newcommand{\rstar}{\ensuremath{R_\star}}
\newcommand{\mstar}{\ensuremath{M_\star}}
\newcommand{\rplanet}{\ensuremath{R_p}}
\newcommand{\mplanet}{\ensuremath{M_p}}
\newcommand{\rearth}{\ensuremath{R_\oplus}}

\newcommand{\porb}{\ensuremath{P}}
\newcommand{\depth}{\ensuremath{\delta_{\rm tr}}}
\newcommand{\ar}{\ensuremath{a/R_\star}}
\newcommand{\rvamp}{\ensuremath{K_{\rm RV}}}
\newcommand{\iorb}{\ensuremath{i}}
\newcommand{\eorb}{\ensuremath{e}}
\newcommand{\insolation}{\ensuremath{\langle F \rangle}}
\newcommand{\sini}{\ensuremath{{\rm sin}~i}}

\usepackage{color}
\usepackage[normalem]{ulem}
\newcommand{\kgsins}[1]{{#1}}
\newcommand{\kgsdel}[1]{{}}

\begin{document}

\title{Accurate, Empirical Radii and Masses of Planets \kgsins{and their Host Stars} with {\it Gaia\/} Parallaxes}

\author{Keivan G.\ Stassun}
\affiliation{Vanderbilt University, Department of Physics \& Astronomy, 6301 Stevenson Center Ln., Nashville, TN 37235, USA}
\affiliation{Fisk University, Department of Physics, 1000 17th Ave.\ N., Nashville, TN 37208, USA}

\author{Karen A.\ Collins}
\affiliation{Vanderbilt University, Department of Physics \& Astronomy, 6301 Stevenson Center Ln., Nashville, TN 37235, USA}
\affiliation{Fisk University, Department of Physics, 1000 17th Ave.\ N., Nashville, TN 37208, USA}

\author{B.\ Scott Gaudi}
\affiliation{The Ohio State University, Department of Astronomy, 140 W.\ 18th Ave., Columbus, OH 43210, USA}
\affiliation{Jet Propulsion Laboratory, California Institute of Technology, 4800 Oak Grove Dr., Pasadena, CA 91109, USA}

\begin{abstract}
We present empirical measurements of the radii of \nhosts\ stars that host transiting planets. These radii are determined using only direct observables---the bolometric flux at Earth, the effective temperature, and the parallax provided by the {\it Gaia\/} first data release---and thus are virtually model independent, extinction being the only free parameter. We also determine each star's mass using our newly determined radius and the stellar density, itself a virtually model independent quantity from previously published transit analyses. These stellar radii and masses are in turn used to redetermine the transiting planet radii and masses, again using only direct observables. 
The \kgsins{median} uncertainties on the stellar radii and masses are \rstarprecper\ and \mstarprecper, respectively, and the resulting uncertainties on the planet radii and masses are \rpprecper\ and \mpprecper, respectively. These accuracies are generally larger than previously published model-dependent precisions of \rplitprecper\ and \mplitprecper\ on the planet radii and masses, respectively, but the newly determined values are purely empirical. 
We additionally report radii for \nrvhosts\ stars hosting radial-velocity (non-transiting) planets, with \kgsins{median} achieved accuracy of \rvrstarprecper.
\kgsins{Using our empirical stellar masses we verify that the majority of putative ``retired A stars" in the sample are indeed more massive than $\sim$1.2~\msun.}
Most importantly, the bolometric fluxes and angular radii reported here \kgsins{for a total of 498 planet host stars}---with \kgsins{median} accuracies of \fbolprecperall\ and \thetaprecperall, respectively---serve as a fundamental dataset to permit the re-determination of transiting planet radii and masses with the {\it Gaia\/} second data release to \rpdrtwoprecper\ and \mpdrtwoprecper\ accuracy, better than currently published precisions, and determined in an entirely empirical fashion.
\end{abstract}

\section{Introduction\label{sec:intro}}


Precise and accurate estimates of the radii and masses of extrasolar planets are essential for a wide variety of reasons.  In the most basic sense, these parameters allow one to estimate the bulk density of an exoplanet, and thus broadly categorize its nature, in other words, determine if it is, e.g., a gas giant, ice giant, mini-Neptune, or rocky planet.  Indeed, it was the discovery of the first transiting planet HD~209458~b \citep{Charbonneau:2000,Henry:2000} that ultimately cemented the interpretation of the Doppler signals of ``Hot Jupiters" \citep[first discovered with the detection of 51~Peg~b;][]{Mayor:1995} as due to roughly Jupiter-mass objects with roughly Jupiter-like densities, and thus that these objects must be primarily composed of hydrogen and helium.  Given that the ``Hot Jupiters" have periods of only a few days, this discovery, along with the fairly robust theoretical conclusion that the majority of gas giants must form beyond the ``snow line" at several AU \citep{pollack1996,kk2008}, also cemented the paradigm-shifting idea that a significant subset of giant planets undergo large-scale migration, thus revolutionizing our ideas about the evolution of planetary systems.

Similarly, estimates of the masses and radii of planets, when coupled with information about their demographics (e.g., their periods and host star properties), can provide important insight into both the physics of planetary atmospheres and interiors, and the physics of planet formation and evolution. For example, a significant fraction of planets in the range of $\sim$0.1--2~\mjup\ have much larger radii than are predicted from standard models of the evolution of hot Jupiters, given their probable irradiation history (e.g., \citealt{Burrows:2000}).  Despite many suggested solutions to this `inflated Hot Jupiter' problem \citep{Burrows:2007,Guillot:2002,Jackson:2009,Batygin:2011,Chabrier:2007,Arras:2010}, no one explanation has emerged as the leading contender, although empirical trends with stellar insolation \citep{Demory:2011} and perhaps age \citep{Hartman:2016} may provide clues to the correct physical model. Regardless, whichever physical mechanism turns out to be dominant, measurements of their radii as a function of the other properties of the system will provide important constraints on the physics of, e.g., tides, magnetic fields, and/or winds in these planets.  

As another example, estimates of the density of `warm Jupiters', i.e., those which do not appear to be affected by the inflation effect discussed above, can provide constraints on their heavy element content, and therefore potentially on the existence of a solid core \citep{Sato:2005}.  Such cores are a `smoking gun' of the core-accretion, bottom-up formation scenario for giant planets \citep{pollack1996}, but are generally not expected in the gravitational instability scenario \citep{Boss:1997}.  Indeed, evidence for a correlation between the inferred core mass of warm Jupiters and the heavy element composition of the host star lends credence to the idea that most, if not all, close-in giant planets form via core accretion \citep{Miller:2011}.

More recently, estimates of the masses and radii of less massive planets ($M_p\la 10 M_\oplus$) detected via Kepler \citep{Borucki:2010} have uncovered an apparent dichotomy in the properties of planets with radii $\la 1.5~R_\oplus$ compared to those larger than this \citep{rogers2015}. In particular, the larger planets appear to have significant hydrogen and helium envelopes, whereas the smaller planets appear to be much more similar to the terrestrial planets in our solar system, with little to no atmospheres.  Indeed, the most precise estimates of the masses and radii of the smallest planets reveal densities that are consistent with a Mg-Si-O composition that is identical to that of the Earth \citep{Dressing:2015}.  

Thus, accurate and precise estimates of the masses and radii of exoplanets have played, and will continue to play, an essential role in understanding the physical processes at work in these planets, and their formation and evolutionary histories.  

\subsection{\kgsins{The challenge of direct and accurate measurements of host-star radii and masses}}

Essentially all exoplanets with measured masses and radii are those found in transiting systems. 
Unfortunately, as is well known, the masses and radii of transiting planets are generally not measured directly; 
\kgsins{the planets' masses and radii depend, through direct transit observables, on the assumed masses and radii of their host stars}.  
The observables are the depth of the transit, which (in the absence of limb darkening) is simply $\depth = k^2$, where $k\equiv R_p/R_\star$ and \rplanet\ and \rstar\ are the radius of the planet and star, respectively, and the velocity semi-amplitude \rvamp, which is given by 
\begin{equation}
\rvamp \equiv \left(\frac{2\pi G}{P}\right)^{1/3}\frac{\mplanet~\sini}{(\mstar +\mplanet)^{2/3}}(1-e^2)^{-1/2},
\end{equation}
where $P$, $e$, and $i$ are period, eccentricity, and inclination of the planet's orbit, respectively, \mplanet\ is the mass of the planet and \mstar\ is the mass of the star.  The eccentricity of the orbit can be determined from the precise shape of the Doppler reflex (radial velocity, or RV) motion of the star, and the inclination can be measured from the relative duration of the ingress/egress $\tau$ and full-width half-maximum $T$ of the transit \citep{Carter:2008}.  Thus, in order to estimate \rplanet\ and \mplanet, one must be able to measure \rstar\ and \mstar.

Unfortunately, it is not possible to estimate the mass or radius of the host purely from photometric follow-up of the primary transit and RV measurements of the host star, {\it regardless of how precise these measurements are}.   This is due to a well-known degeneracy, first pointed out in the case of transiting planets by \citet{Seager:2003}.  As they note, the only parameter about the star that can be directly measured from observables is the ratio of the semimajor axis of the orbit $a$ to the radius of the star \ar\ \citep{Winn:2010},
\begin{equation}
\frac{a}{R_\star} = 
\frac{\depth^{1/4}}{\pi}
\frac{P}{\sqrt{T\tau}}
\left(\frac{\sqrt{1-e^2}}{1+e\sin\omega}\right),
\end{equation}
where $\omega$ is the argument of periastron, which is also an observable from the RV curve.  However, this quantity is closely related to the density of the star \citep{Seager:2003,Winn:2010},
\begin{equation}
\rho_\star = \frac{3\pi}{GP^2}\left(\frac{a}{R_\star}\right)^3
-k^3\rho_p \simeq \frac{3\pi}{GP^2}\left(\frac{a}{R_\star}\right)^3,
\end{equation}
where $\rho_p$ is the density of the planet (and is typically $\sim \rho_*$) and the last equality follows from the fact that typically $k^3 \ll 1$. 

Thus $\rho_*$ can essentially be inferred from direct observables. Nevertheless, there remains a one-parameter degeneracy that makes it impossible to estimate the mass and radius of the star independently.  All transiting planet systems (with only photometry of the primary transit and RV observations) are subject to this degeneracy.  To break the degeneracy, one must bring in additional external constraints, such as a measurement of the surface gravity of the star, \logg\ (which is a direct observable from high-resolution spectra), astroseismological inferences of the stellar mass and radius (e.g., \citealt{Huber:2013}), or a measurement of the radius of the star (which, as we will show is a direct observable from the bolometric flux and effective temperature of the star, and the distance to the system).  

Up until now, these observables \kgsins{of the host stars}, while preferred because they are direct, have been either poorly measured, subject to systematic errors, or not constrained at all.

\subsection{\kgsins{The value of reducing reliance on, and testing, stellar models and empirical relations}}

Instead, most authors typically use theoretical and/or empirically-calibrated relations between observable properties of the star.  Stellar evolution is reasonably well understood, and it is known that a star of a given effective temperature, metallicity, and density cannot have an arbitrary mass and radius.  Indeed, to first order, these three parameters essentially fix the luminosity and age of the star, and thus its radius and mass.  Therefore, adopting these constraints, while not direct, typically lead to much more precise estimates of the parameters of the system. 

Nevertheless, they are subject to uncertainties in stellar evolution models and second-order parameters (i.e., stellar rotation), and/or inaccurate calibrations of the empirical relations.  One might therefore be concerned that these estimates, while precise, are not {\it accurate}. One clear demonstration of this is the case of KELT-6b \citep{Collins:2014}, where the parameters inferred using the Yonsei-Yale isochrones \citep{Demarque:2004} to break the degeneracy disagreed significantly (by as much as $4\sigma$) from those inferred using the \citet{Torres:2010} empirical relations.  Likely this was due to the low metallicity [Fe/H]$\approx$$-0.3$ of the KELT-6 host star, and the fact that neither the isochrones nor the empirical relations are well-calibrated at such low metallicities.


While slightly erroneous inferences about the properties of individual systems (as in the case of \citealt{Collins:2014}) are troubling, the difficulty with estimating accurate parameters of host stars and thus their transiting planets can be, and indeed has proven to be, quite deleterious in some cases, sometimes leading to markedly incorrect or inconsistent inferences about individual systems or even entire populations of planets.  
%
An early example of this is the case of the supermassive ($\sim 12~\mjup$) planet XO-3~b \citep{JohnsKrull:2008}, in which initial estimates of \rplanet\ differed by nearly a factor of two, from $\sim 1.2~\rjup$ to $\sim 2.1~\rjup$.  The latter value would have implied that the planet was highly inflated relative to standard models, which would have been particularly interesting given its relatively large inferred mass.  With improved photometry and thus an improved estimate of $\rhostar$, \citet{Winn:2008} were able to demonstrate that the true planetary radius was likely at the lower end of this range. Indeed, as we show in this work, our revised determinations with the {\it Gaia\/} distance---which places the star at a significantly shorter distance than previously assumed---reveal the planet to be $\mplanet \approx 7$~\mjup\ and $\rplanet \approx 1.4$~\rjup. 
\kgsins{We defer additional case studies of problems arising from current poor constraints on models and empirical relations to the Discussion (Sec.~\ref{sec:disc}).}

\kgsins{However,}
the difficulties with interpreting the properties of planetary populations due to uncertainties about the properties of the host stars became quite prominent and acute with the discoveries of thousands of planets via {\it Kepler} \citep{Borucki:2010}.  Here the difficulties were threefold. First, the {\it Kepler} transiting planet hosts tended to be fairly faint compared to those found via ground based transit surveys, making characterization of the host stars more difficult.  Second, the shear number of hosts made systematic assays of their properties via high-resolution spectroscopy extremely resource-intensive; this was obviously exacerbated by the faintness of the hosts.  Finally, the wide {\it Kepler} bandpass, poor cadence, and/or low signal-to-noise ratio of the majority of the transit signals made estimates of the ingress/egress time, and thus stellar densities, generally imprecise.  This has led to herculean efforts to characterize the properties of the host stars, often resulting in quite different conclusions as to the radius distribution of the {\it Kepler} target stars and thus their transiting planetary companions (see,
e.g., \citealt{Pins:2012,Mann:2012,Dressing:2013,Huber:2014,Gaidos:2016}).

\subsection{\kgsins{Aim of this paper: A path to precision exoplanetology in the era of {\it Gaia\/}}}

Three recent advances now permit the determination of accurate {\it and} empirical radii and masses for a large sample of transiting planets. 
First, there now exist all-sky, broadband photometric measurements for stars spanning a very broad range of wavelengths, from the {\it GALEX\/} far-UV at $\sim$0.15~$\mu$m to the {\it WISE} mid-IR at $\sim$22~$\mu$m. These measurements permit construction of spectral energy distributions (SEDs) that effectively sample the majority of the flux for all but the hottest stars. Consequently the bolometric fluxes (\fbol), and in turn the stars' angular radii ($\Theta$, via the stellar effective temperature, \teff), can in principle be determined in a largely empirical manner. Using a set of eclipsing binary stars as benchmarks, \citet{Stassun:2016} have shown that with such data \fbol\ can be measured with a precision that is typically $\lesssim$3\%. 
Second, the {\it Gaia\/} mission's first data release 
\citep{Gaia:2016}
has delivered trigonometric parallaxes for $\sim 2\times 10^6$ stars in common with {\it Tycho-2\/} \citep{Hog:2000}, with a precision for the best $\sim$10\% of stars of \tycparprec. These parallaxes permit $\Theta$ to be converted to \rstar. 
Third, a large sample of transiting planets orbiting stars that are sufficiently bright to have been included in {\it Tycho-2\/}, and consequently in the {\it Gaia\/} first data release, have been published with quantities that follow from direct observables such as the stellar density, \rhostar, the ratio of planet-to-star radii, and the orbital radial-velocity semi-amplitude. With \rstar, these quantities yield $R_p$ as well as the stellar mass, which in turn yields the planet mass. 

In this paper, we perform this procedure 
\kgsins{to measure \fbol\ and $\Theta$ for 498 planet host stars, which will serve as fundamental stellar parameters for use with upcoming data releases from {\it Gaia\/}. We also report empirical stellar and planet radii and masses as described above}
for \nhosts\ stars that host transiting planets, have the necessary direct observables published in the literature, and have parallaxes newly reported in the {\it Gaia\/} first data release \citep{Gaia:2016}. 
We additionally perform this procedure for \nrvhosts\ stars that host non-transiting (radial-velocity only) planets, for which the newly derived planet properties remain modulo factors of $\sin i$. 

Importantly, the stellar and planet properties that we determine are 
\kgsins{independent of stellar models and of empirically calibrated stellar relations; thus the properties that we determine for the stars and their planets do not require the assumption that individual systems behave according to theoretical expectation or within the limits of mean relations. We argue, moreover, that the stellar and planet properties that we determine are}
{\it empirical and accurate}, even if the {\it Gaia\/} parallaxes do not yet yield {\it precisions} that rival those typically achieved via model-dependent analyses reported in the literature. 
However, the $\Theta$ measurements that we determine are sufficiently accurate and precise that upcoming, improved parallax measurements from {\it Gaia\/} should enable the stellar and planet properties to be re-determined with accuracies and precisions superior to those currently available in many cases---in an entirely empirical, model-free fashion. 
\kgsins{As we enter the {\it TESS\/} era of superb precision in the transit parameters of very bright stars, such empirical and accurate stellar properties will become even more important than they have been for {\it Kepler\/} targets for which precision followup often proved challenging.}

In Section~\ref{sec:data} we describe our study sample, the data used, and our methodological approach. 
The primary results of this study are presented in Section~\ref{sec:results}, including \fbol, $\Theta$, \rstar, and \mstar, followed by the planet properties, $R_p$ and $M_p$. 
In Section~\ref{sec:disc} we 
\kgsins{present additional motivation for the importance of reducing reliance on stellar models, explore the degree to which the approach laid out in this paper is truly empirical,}
discuss our results in the context of previously published results, and briefly discuss the prospects for improving on the stellar and planet properties reported here with the anticipated advent of improved parallaxes from the {\it Gaia\/} second data release. 
Finally, Section~\ref{sec:summary} provides a summary of our results and conclusions.

\section{Data and Methods\label{sec:data}} 

\subsection{Study Sample, Data from the Literature, and {\it Gaia\/} Parallaxes} 

We began by selecting all planet hosting stars found in the {\tt \url{exoplanets.org}} database \kgsins{\citep[][accessed on 31 August 2016]{Han:2014}} and added 12 well characterized transiting planets that were present in the NASA Exoplanet Archive\footnote{\tt \url{exoplanetarchive.ipac.caltech.edu}} but missing from {\tt exoplanets.org}. We then selected systems with host stars that are also present in the {\it Tycho-2\/} catalog, resulting in 560 unique stars. 
\kgsins{Of these, 62 stars were removed because they lacked one or more of}
the minimal set of parameters required for our analysis (see Sec.~\ref{sec:method});
\kgsins{nearly all of these were {\it Kepler} planets for which radial-velocity semi-amplitudes were not reported}. 
\kgsins{The remaining 498 stars form our master study sample, for which we perform our SED fitting procedures, resulting in fundamental \fbol\ and $\Theta$ measurements, as discussed below.}
\kgsins{The {\it Gaia\/} DR1 provides parallaxes for \ntothosts\ of these stars,}
of which \nhosts\ were listed as hosting transiting planets
and \nrvhosts\ were
listed as hosting radial-velocity planets.
We updated the XO-3 stellar radius and distance from the {\tt \url{exoplanets.org}} database to the more recent values reported in \citet{Winn:2008} (see \S \ref{sec:disc}).  

For each of the transiting planet hosts, we adopted the literature values from the {\tt \url{exoplanets.org}} database for each of the following quantities: 
the orbital period, \porb, and its uncertainty;
the transit depth, \depth, and its uncertainty; 
the ratio of the orbital semi-major axis to the stellar radius, \ar, and its uncertainty; 
the orbital inclination angle to the line of sight, $i$, and its uncertainty; 
and the orbital radial-velocity semi-amplitude, \rvamp, and its uncertainty. 
In this paper we analyze only the `b' planet in each system; however, the stellar properties that we newly determine here (e.g., \fbol, $\Theta$, \rstar, \mstar) can be readily applied to other planets in the case of currently known or future discovered additional planets.

We also adopted the parallax measurements newly available from {\it Gaia\/} \citep{Gaia:2016}. We adopt the {\it Gaia\/} parallax, $\pi$, and its uncertainty as provided by the {\it Gaia\/} first data release\footnote{Accessed on 14 September 2016.} (see Table~\ref{tab:gaiadata}). 
We note that, at the time of this writing, the {\it Gaia\/} $\pi$ values potentially have systematic uncertainties that are not yet fully characterized but that could reach $\sim$300~$\mu$as\footnote{See \url{http://www.cosmos.esa.int/web/gaia/dr1}.}. 
Preliminary assessments suggest a global offset of $-0.25$~mas (where the negative sign indicates that the {\it Gaia\/} parallaxes are underestimated) for $\pi \gtrsim 1$~mas \citep{StassunGaiaError:2016}, corroborating the {\it Gaia\/} claim, based on comparison to directly-measured distances to well-studied eclipsing binaries by \citet{Stassun:2016}.  
\citet{Gould:2016} similarly claim a systematic uncertainty of 0.12~mas.
\citet{Casertano:2016} used a large sample of Cepheids to show that there is likely little to no systematic error in the {\it Gaia\/} parallaxes for $\pi \lesssim 1$~mas, but find evidence for an offset at larger $\pi$ consistent with \citet{StassunGaiaError:2016}.
Thus the available evidence suggests that any systematic error in the {\it Gaia\/} parallaxes is likely to be small. Thus, for the purposes of this work, we use and propagate the reported {\it random} uncertainties on $\pi$ only, emphasizing that (a) the fundamental \fbol\ and $\Theta$ measurements that we report are independent of $\pi$, and (b) additional (or different) choices of $\pi$ uncertainties may be applied to our \fbol\ and $\Theta$ measurements following the methodology, equations, and error propagation coefficients supplied below. 

The assembled set of literature parameters for the 
\kgsins{498 stars in our master sample---including the}
\nhosts\ transiting planet hosts and the \nrvhosts\ non-transiting planet hosts in our study
\kgsins{for which {\it Gaia\/} DR1 parallaxes are available---}are tabulated in Table~\ref{tab:trsample} and Table~\ref{tab:rvsample}, respectively.

\subsection{Basic Methodology\label{sec:method}}

For all planets in the study sample, we first collect high-quality spectroscopic values of host star \teff, \logg, and \feh\ as described in \S \ref{sec:spectroscopicvals} and broadband photometric data as described in \S \ref{sec:broadband}. Then, for each planetary system, we fit a spectroscopic parameter-based stellar atmosphere model to a broadband photometry-based SED as described in \S \ref{sec:fitting}. We directly sum the unreddened SED model over all wavelengths to obtain \fbol, and use the Stefan-Boltzmann law to calculate the host star radius, \rstar, from \fbol, \teff, and the {\it Gaia\/} parallax, as described in \S \ref{sec:stellarpars}.

For the transiting planets, we calculate the planet radius \rplanet\ from the transit depth, \depth\kgsins{$\equiv$(\rplanet/\rstar)$^2$}, and the empirically calculated stellar radius, \rstar. The stellar density, \rhostar, is calculated from the transit model parameter \ar\ and the orbital period, $P$, (see \S \ref{sec:stellarpars}) and then the stellar mass, \mstar, is calculated from \rhostar\ and \rstar. Planet mass, \mplanet, is then empirically calculated from \mstar, the radial velocity semi-amplitude, \rvamp, and the orbital period, $P$, and eccentricity, $e$ (see \S \ref{sec:planetpars}). Finally, we calculate planet surface gravity, \loggp, from \rvamp, the transit parameters \ar\ and \depth, and $P$, $e$, and orbital inclination, $i$, and we calculate the insolation at the planet, \insolation, from \fbol, the distance to the star, and the semi-major axis of the orbit, $a$ (see \S \ref{sec:planetpars}).

For the non-transiting planets, we calculate \mstar\ from our empirical \rstar\ and the spectroscopic \logg. We also calculate the minimum planet mass, \mplanet\sini, in the same way that we calculate \mplanet\ for the transiting planets, but with the spectroscopic \logg-based stellar mass rather than the generally higher accuracy transit-based \rhostar. We calculate the insolation at the planet in the same way as for the transiting planets.

We determine uncertainties for all calculated parameters by propagating the measured parameter uncertainties through the relevant equations. We also include the effects of typical parameter correlations for the transit, orbital, and RV parameters, as part of error propagation (see \S \ref{sec:planetpars}).

\subsection{Stellar Effective Temperatures, Surface Gravities, and Metallicities\label{sec:spectroscopicvals}}

For most of the stars in our study sample, values of \teff, \logg, and \feh\ are available in the exoplanet archives as obtained from the original literature. We adopted these values if no other independent, spectroscopic values were available. However, where possible, we opted to instead use \teff, \logg, and \feh\ from the recently updated PASTEL catalog \citep{Soubiran:2016}, which compiles stellar properties determined from high-quality\footnote{\kgsins{The parameters compiled in the PASTEL catalog derive from high-resolution ($R\ge 25000$), high signal-to-noise ($S/N\ge 50$) spectra via a variety of analysis methods (e.g., spectral synthesis, equivalent width measurements, etc.) from a large number of published papers.}} spectroscopic analyses in the literature. 
We did this in order to ensure that the stellar properties used were as independent as possible from the transit-based parameters, so as to avoid hidden correlations in the derived parameters (see also Sec.~\ref{sec:planetpars}). 

Where multiple values were available in the PASTEL catalog, we adopted the mean and uncertainty on the mean. Where no values were available in the PASTEL catalog, we adopted the values reported in the archive. For a small number of stars with values in neither PASTEL nor in the archive, we adopted the values reported by \citet{Santos:2013}. Finally, for a few stars our initial SED fits clearly indicated a poor choice of \teff, so we adopted an alternate \teff\ from the literature, as indicated in Tables~\ref{tab:trsample} and \ref{tab:rvsample}.

\subsection{Broadband photometric data from the literature\label{sec:broadband}} 
In order to systematize and simplify our procedures, we opted to assemble for each host star the available broadband photometry from the following large, all-sky catalogs (listed here in approximate order by wavelength coverage) via the {\tt VizieR}\footnote{\url{http://vizier.u-strasbg.fr/}} query service: 
\begin{itemize}
    \item {\it GALEX} All-sky Imaging Survey (AIS): FUV and NUV at $\approx$0.15 \micron\ and $\approx$0.22 \micron, respectively. 
    \item Catalog of Homogeneous Means in the $UBV$ System for bright stars from \citet{Mermilliod:2006}: Johnson $UBV$ bands ($\approx$0.35--0.55 \micron).
    \item {\it Tycho-2\/}: Tycho $B$ ($B_{\rm T}$) and Tycho $V$ ($V_{\rm T}$) bands ($\approx$0.42 \micron\ and $\approx$0.54 \micron, respectively). 
    \item Str\"omgren Photometric Catalog by \citet{Paunzen:2015}: Str\"omgren $uvby$ bands ($\approx$0.34--0.55 \micron).
    \item AAVSO Photometric All-Sky Survey (APASS) DR6 (obtained from the UCAC-4 catalog): Johnson $BV$ and SDSS $gri$ bands ($\approx$0.45--0.75 \micron). 
    \item Two-Micron All-Sky Survey (2MASS): $JHK_S$ bands ($\approx$1.2--2.2 \micron). 
    \item All-WISE: {\it WISE1--4} bands ($\approx$3.5--22 \micron). 
\end{itemize}

We found $BV$, $JHK_S$, and {\it WISE1--3} photometry---spanning a wavelength range $\approx$0.4--10~\micron---for nearly all of the stars in our study sample. Most of the stars also have {\it WISE4\/} photometry, and many of the stars also have Str\"omgren and/or {\it GALEX\/} photometry, thus extending the wavelength coverage to $\approx$0.15--22~\micron. We adopted the reported measurement uncertainties unless they were less than 0.01 mag, in which case we assumed an uncertainty of 0.01 mag. In addition, to account for an artifact in the Kurucz atmospheres at 10~\micron\ \kgsins{\citep[see, e.g.,][and note that the contribution to \fbol\ for $\lambda\ge 10$~\micron\ is $\lesssim 10^{-4}$]{Stassun:2016}}, we artificially inflated the {\it WISE3\/} uncertainty to 0.1 mag unless the reported uncertainty was already larger than 0.1 mag.
\kgsins{\citet{Stassun:2016} found no evidence for systematic effects in any of the other passbands, although individual outlier cases do occur; see Sec.~\ref{sec:fitting}.}
The assembled SEDs are presented in Appendix~\ref{sec:sed_appendix}.

\subsection{Spectral energy distribution fitting\label{sec:fitting}} 
To measure the \fbol\ of each 
\kgsins{of the 498}
stars in our \kgsins{master} sample, we followed the SED fitting procedures described in \citet{Stassun:2016}. Briefly, 
the observed SEDs were fitted with standard stellar atmosphere models, and \fbol\ measured by summation of the model after correction for extinction. For the stars in our sample with \teff\ $>$ 4000~K, we adopted the atmospheres of \citet{Kurucz:2013}, whereas for the stars with \teff\ $<$ 4000~K we adopted the NextGen atmospheres of \citet{Hauschildt:1999}. 
As discussed in \citet{Stassun:2016}, because the photometric data span such a large portion of the SED, and because there is only one free parameter ($A_V$) in the SED fit, the determination of \fbol\ via this procedure is virtually model independent, the model atmosphere serving essentially as an ``intelligent interpolation" between the photometric measurements.

As summarized in Tables~\ref{tab:trsample} and \ref{tab:rvsample}, for each star we have \teff, \logg, and \feh. We interpolated in the model grid to obtain the appropriate model atmosphere for each star in units of emergent flux. To redden the SED model, we adopted the interstellar extinction law of \citet{Cardelli:1989} \kgsins{with the usual ratio of total-to-selective extinction, $R_V = 3.1$}\footnote{\kgsins{As noted in \citet{Stassun:2016}, experiments with varying the value of $R_V$ showed a negligible effect on \fbol, mainly because the available broadband photometry spans such a large wavelength range that the fitted atmosphere model is essentially an interpolator, and the extrapolated flux is a very small fraction of the total \fbol\ in virtually all cases.}}. We then fitted the atmosphere model to the flux measurements to minimize $\chi^2$ by varying just two fit parameters: extinction ($A_{\rm V}$) and overall normalization (effectively the ratio of the stellar radius to the distance, $R_\star^2 / d^2$). (The uncertainty in the adopted stellar \teff\ is handled in a later step via the propagation of errors through the stellar angular radius, $\Theta$; see Section \ref{sec:stellarpars}.) 
We estimated $A_{\rm V}$ from the three-dimensional Galactic dust model of \citet[][Model A]{Amores:2005} for each star's line of sight and distance, but we allowed the fit to vary $A_{\rm V}$ from this initial guess by as much as 20\%, limited by the maximum line-of-sight extinction from the dust maps of \citet{Schlegel:1998}.
The best-fit model SED with extinction is shown for each star in Appendix~\ref{sec:sed_appendix}, and the \kgsins{best-fit $A_V$ and} reduced $\chi^2$ values ($\chi_\nu^2$) are given in Table~\ref{tab:results}. 

\kgsins{Inspection of the SEDs shows generally excellent fits (see below for discussion of $\chi^2$), with only a few exhibiting clear outlier behavior. These include HD~208527, $\alpha$~Ari, XO-4, 6~Lyn, HD~190360, HAT-P-15, HD~132563, and 14~And. Such outliers could occur for a number of possible reasons, including close neighbors that are unresolved in the available photometric catalogs and/or excesses arising from circumstellar material; in principle, multi-component SED fits could improve these in the future. Here we simply note these few cases, and remind the reader that these are readily flagged by the $\chi_\nu^2$ of the fit provided in Table~\ref{tab:results}. These are also excluded from our analyses below.}

The primary quantity of interest for each star is \fbol, which we obtained via direct summation of the fitted SED, {\it without} extinction, over all wavelengths. The formal uncertainty in \fbol\ was determined according to the standard criterion of $\Delta\chi^2 = 2.30$ for the case of two fitted parameters \citep[e.g.,][]{Press:1992}, where we first renormalized the $\chi^2$ of the fits such that $\chi_\nu^2 = 1$ for the best fit model. Because $\chi_\nu^2$ is in almost all cases greater than 1 (see Table~\ref{tab:results}), this $\chi^2$ renormalization is equivalent to inflating the photometric measurement errors by a constant factor and results in a more conservative final uncertainty in \fbol\ according to the $\Delta\chi^2$ criterion. 

\begin{figure*}[!ht]
\centering
\includegraphics[width=0.49\linewidth,trim=10 10 10 50,clip]{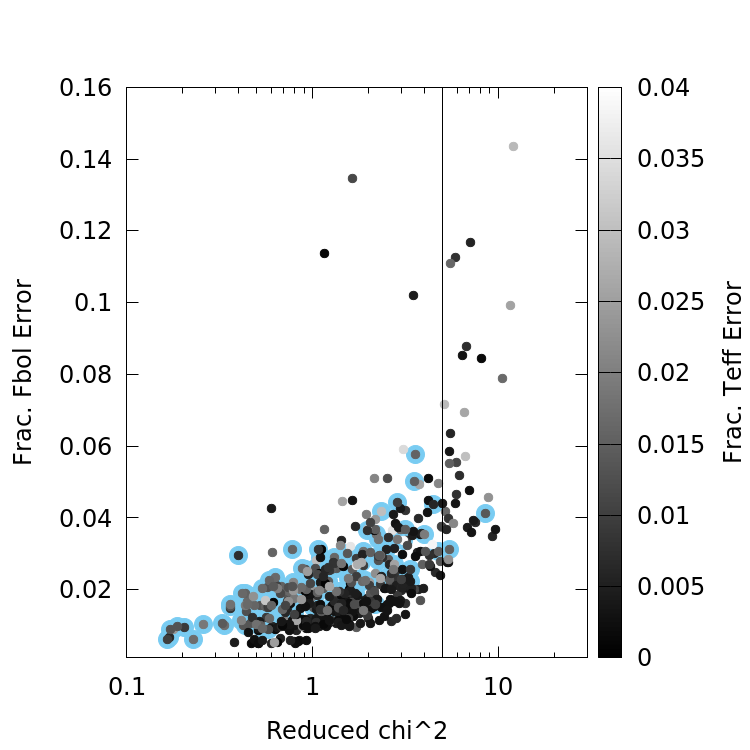}
\includegraphics[width=0.49\linewidth,trim=10 10 10 50,clip]{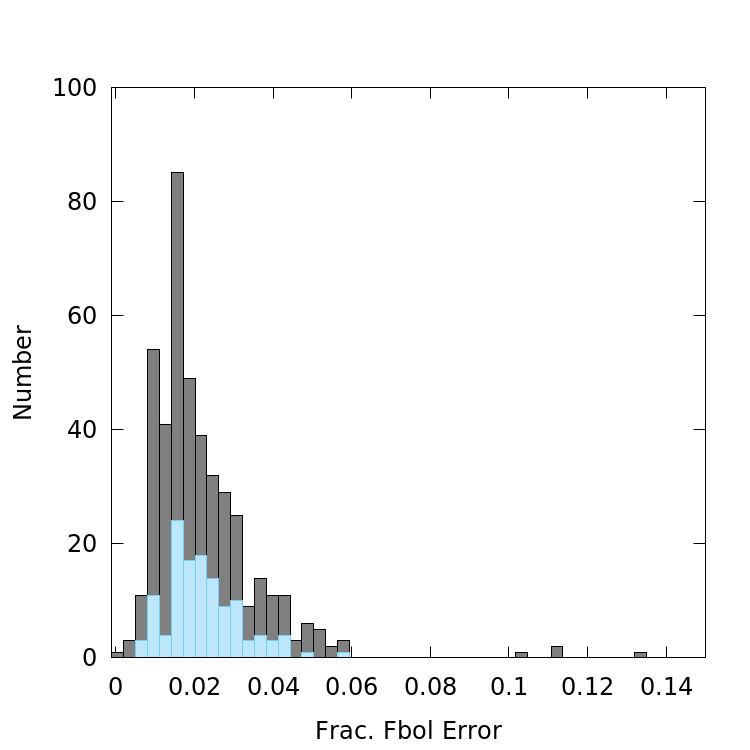}
\caption{{\it (Left:)} Fractional uncertainty on \fbol\ from the SED fitting procedure as a function of $\chi_\nu^2$ and of \teff\ uncertainty. The vertical line represents the cutoff of $\chi_\nu^2 \le 5$ for which the uncertainty on \fbol\ is at most 6\% for most stars, thus permitting a determination of \rstar\ to $\approx$3\%. Points with blue haloes represent stars with transiting planets. 
{\it (Right:)} Distribution of fractional \fbol\ uncertainties for stars with $\chi_\nu^2 \le 5$. The median uncertainty on \fbol\ for transiting-planet host stars (blue) is \fbolprecper; it is \fbolprecperall\ for all stars.
\label{fig:fbol_vs_chi2}}
\end{figure*}

Figure~\ref{fig:fbol_vs_chi2} (left) shows the dependence of the fractional \fbol\ uncertainty as functions of the goodness of the SED fit and of the uncertainty on \teff. For stars with \teff\ uncertainties of $\lesssim$1\%, the \fbol\ uncertainty is dominated by the SED goodness-of-fit. With the exception of a few outliers, we achieve an uncertainty on \fbol\ of at most 6\% for $\chi_\nu^2 \le 5$. Adopting this as a goodness-of-fit threshold thus yields a sample of \ngoodhosts\ transiting-planet host stars \kgsins{with {\it Gaia\/} DR1 parallaxes} for which we can in principle achieve an uncertainty in \rstar\ of $\approx$3\% (see Eq.~\ref{eq:frt}).

Figure~\ref{fig:fbol_vs_chi2} (right) shows the distribution of fractional \fbol\ uncertainties for the transiting-planet sample (blue) and for all stars (black). For the transiting-planet hosts with $\chi_\nu^2 \le 5$, the median uncertainty on \fbol\ is \fbolprecper\ and is at most 5.7\%. For the full sample, \kgsins{which includes the radial-velocity sample that is brighter on average than the transit sample,} the median uncertainty is \fbolprecperall, with 95\% of the sample having an uncertainty of less than 5\%.

\subsection{Host Star Parameters and Uncertainties\label{sec:stellarpars}}

Finally, we derived stellar radii \rstar\ by combining our \fbol\ values derived from SED fitting with our \teff\ values derived from photometry and spectra. These quantities are related by 
\begin{equation}\label{eq:frt}
\Theta = ( F_{\rm bol} / \sigma_{\rm SB} T_{\rm eff}^4 )^{1/2},
\end{equation}
\noindent where $\Theta$ is the stellar angular radius ($\Theta \equiv R_\star/d$), 
$d$ is the distance to the star, and $\sigma_{\rm SB}$ is the Stefan-Boltzmann constant. Errors propagating into \rstar\ come from random and systematic errors on \teff, and uncertainties on \fbol, $A_V$, and $d$. 
\kgsins{We note that the calculation of $d$ from $\pi$ can become non-trivial when $\sigma_\pi / \pi \gtrsim 20\%$ \citep[see, e.g.,][]{Bailer-Jones:2015}. The actual $\sigma_\pi / \pi$ from {\it Gaia\/} DR1 for our study sample has a median of 2.3\%, and is less than 20\% for 98\% of the sample. We therefore proceed in this paper to compute $d$ via the straightforward $1/\pi$ conversion.}

For transiting planets, we determine the stellar density, \rhostar, from the transit model parameter \ar\ and the orbital period, $P$, through the relation
\begin{equation}\label{eq:rhostar}
\rhostar = \frac{3\pi}{GP^2}(\ar)^3,
\end{equation}
for $\rhoplanet \sim \rhostar$ and $k^3 \equiv (\rprstar)^3 \ll 1$. 

Combining \rhostar\ with our determination of \rstar\ provides a direct measure of the stellar mass, \mstar. For the non-transiting planets, we calculate \mstar\ from our \rstar\ and the spectroscopic \logg.

\begin{figure*}[!ht]
    \centering
    \includegraphics[width=0.49\linewidth,trim=70 70 70 50,clip]{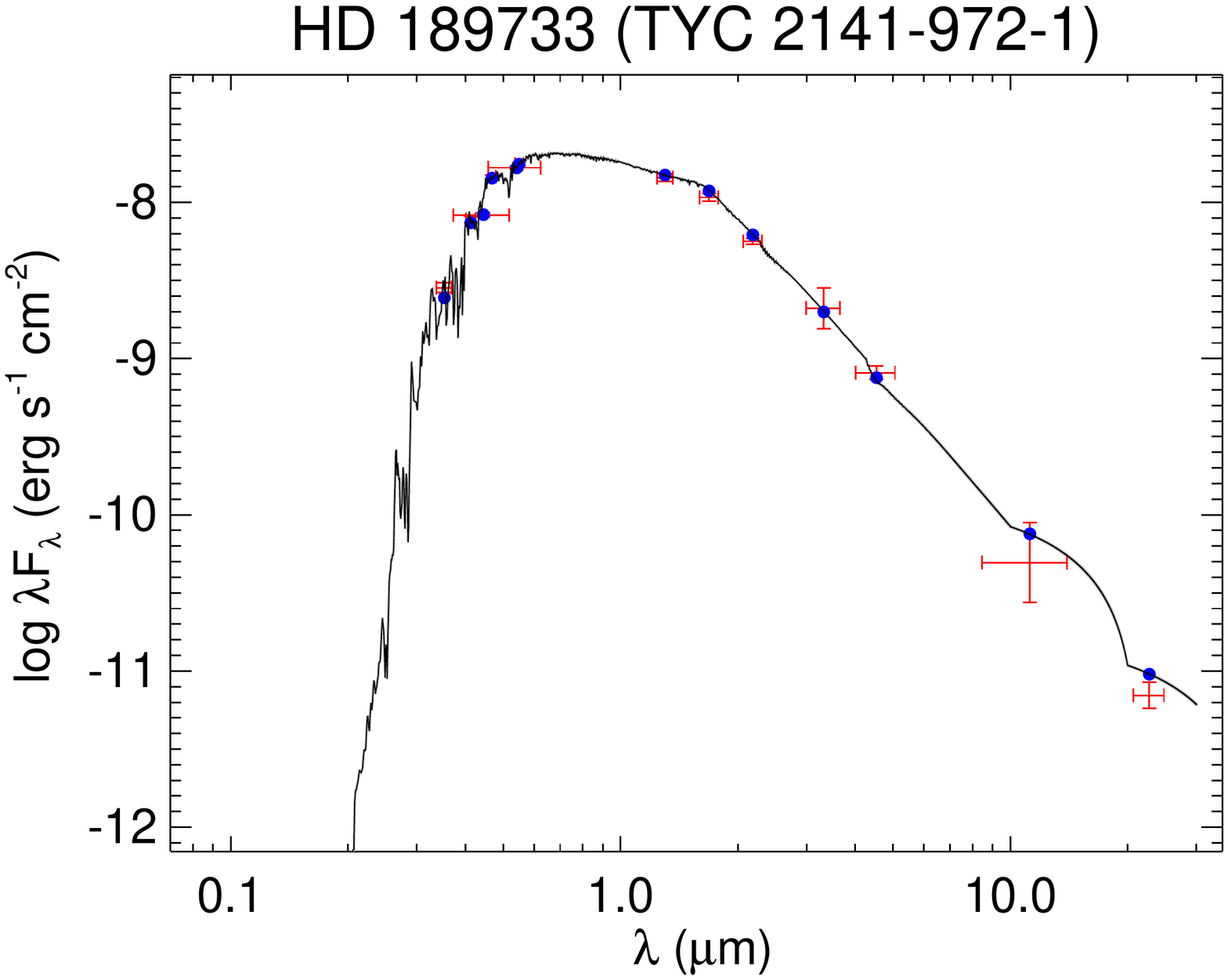}
    \includegraphics[width=0.49\linewidth,trim=70 70 70 50,clip]{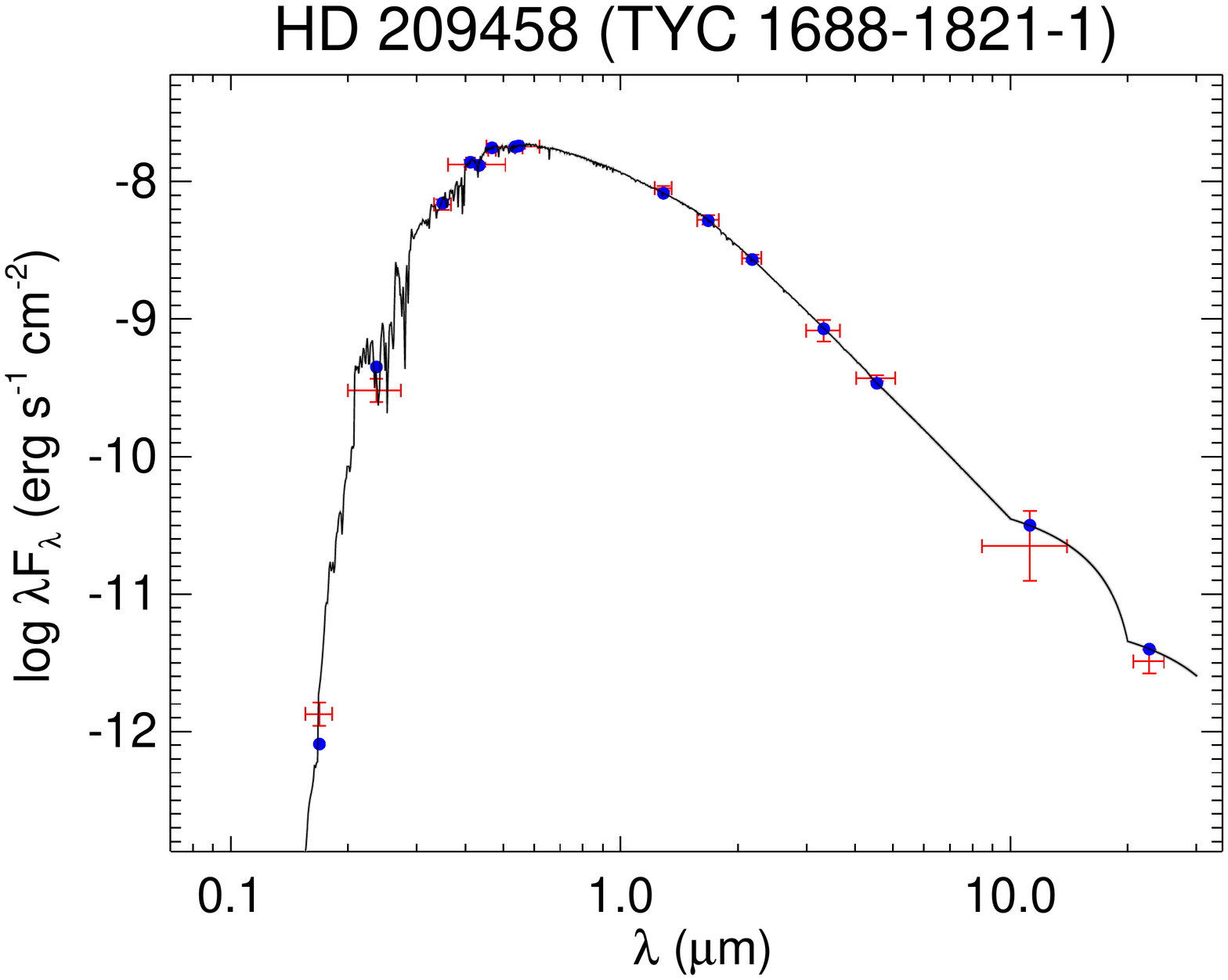}
    \caption{SED fits for the stars HD~189733 (left) and HD~209458 (right), for which interferometric angular radii have been reported \citep{Boyajian:2015} as a check on the $\Theta$ and \fbol\ values derived in this work. The two SED fits have $\chi_\nu^2$ of 1.65 and 1.67, respectively. The $\Theta$ and \fbol\ comparisons are presented in Table~\ref{tab:boyajian}. Symbols and colors are as in Figure~\ref{fig:seds}.}
    \label{fig:boyajian}
\end{figure*}

As a check that our procedures result in accurate $\Theta$ and \fbol, we applied our procedures to the interferometrically observed planet-hosting stars HD~189733 and HD~209458 reported by \citet{Boyajian:2015}. Our SED fits are shown in Figure~\ref{fig:boyajian} and the $\Theta$ and \fbol\ values directly measured by those authors versus those derived in this work are compared in Table~\ref{tab:boyajian}, where the agreement is found to be excellent and within the uncertainties. 

\begin{deluxetable}{lcc}
\tabletypesize{\scriptsize}
\tablecolumns{3}
\tablewidth{0pt}
\tablecaption{Comparison of stellar angular diameters ($2\times\Theta$) and bolometric fluxes (\fbol) for stars with interferometrically measured angular diameters from \citet{Boyajian:2015}.
\label{tab:boyajian}}
\tablehead{\colhead{} & \colhead{\citet{Boyajian:2015}} & \colhead{This work}} 
\startdata
\cutinhead{HD~189733 (\teff=4875$\pm$43~K)}
$2\times\Theta$ (mas) & 0.3848$\pm$0.0055 & 0.391$\pm$0.008 \\
\fbol\ ($10^{-8}$ erg s$^{-1}$ cm$^{-2}$) & 2.785$\pm$0.058 & 2.87$\pm$0.06 \\
\cutinhead{HD~209458 (\teff=6092$\pm$103~K)}
$2\times\Theta$ (mas) & 0.2254$\pm$0.0072 & 0.225$\pm$0.008 \\
\fbol\ ($10^{-8}$ erg s$^{-1}$ cm$^{-2}$) & 2.331$\pm$0.051 & 2.33$\pm$0.05 \\
\enddata
\end{deluxetable}

\subsection{Planet Parameters and Uncertainties\label{sec:planetpars}}
From our empirically calculated \rstar, we then directly obtain \rplanet\ via $\rprstar=\sqrt{\depth}$ for the transiting planets. From our empirically calculated \mstar, we can directly calculate \mplanet\ (\mplanet\sini\ for the RV planets) for all samples in the study via
\begin{equation}\label{eq:mplanet}
\mplanet = \frac{\rvamp\sqrt{1-e^2}}{\sini}\left(\frac{P}{2\pi G}\right)^{1/3}\mstar^{2/3},
\end{equation}
in the limit $\mplanet \ll \mstar$. 

For the transiting planets, we also directly calculate the planet surface gravity, \loggp, from the RV, orbital, and transit parameters via
\begin{equation}\label{eq:loggp}
\loggp = \frac{2\pi\rvamp\sqrt{1-e^2}(\ar)^2}{P~\depth~\sini}.
\end{equation}
Note that this is a direct observable and does {\it not} depend on the properties of the host star \citep{Winn:2010}. 

Finally, we directly calculate the insolation at the planet, \insolation, for all samples in the study from the relation 
\begin{equation}\label{eq:insolation}
\insolation = \fbol \left(\frac{d}{a}\right)^2,
\end{equation}
where $d$ is the distance from Earth to the planetary system determined from the {\it Gaia\/} parallax, and $a$ is the semi-major axis of the planetary orbit. For the transiting planets, $a$ is determined from \ar\ and our empirically determined \rstar. For the non-transiting planets, we use the values of $a$ from the literature.  

We determine uncertainties for all calculated parameters by propagating the measured parameter uncertainties through the relevant equations. 
\kgsins{For the purposes of this paper, we assume first-order linear perturbations for the error propagation, and}
we also include the effects of typical parameter correlations for the transit, orbital, and RV parameters, as part of error propagation (see below).
\kgsins{In principle our use of simple linear error propagation could underestimate the true errors for systems where the observational uncertainties are large and thus higher order terms would be needed. However, because this paper utilizes the {\it Gaia\/} DR1 parallaxes which currently are the limiting factor for most of the systems in our study sample (see Section~\ref{sec:results_planets} and Figure~\ref{fig:rstar_vs_dist}), this simplification of approach should be sufficient (see also below). The eventual arrival of the {\it Gaia\/} DR2 parallaxes may very well require more sophisticated error analysis procedures, such as through the use of full MCMC chains.}

\kgsins{It is beyond the scope of this work to calculate parameter correlations for all planets in our study, so} we use KELT-15 \citep{Rodriguez:2016} as an exemplar planetary system to estimate parameter correlations and to verify our full error propagation methodology. 
\kgsins{To be sure, the KELT-15 correlations do not exactly represent the parameter correlations for all planets in our study. Nonetheless, we expect they are reasonable representations, and since the {\it Gaia\/} DR1 parallaxes currently dominate the errors, this approximation should not compromise the accuracy of our results. To verify this expectation, we compared the results of several systems with and without the KELT-15 covariance terms included in the error propagation and found that indeed the results were identical to within $\pm1$ in the least significant digit of most reported parameter values.} 
\citet{Rodriguez:2016} used a custom version of EXOFAST \citep{Eastman:2013}, which allows multiple RV and transit datasets to be simultaneously fitted (see \citealt{,Siverd:2012} for more details), to perform a KELT-15 global system fit to spectroscopic parameters, RV data, and photometric follow-up light curves. EXOFAST uses MCMC to robustly determine system parameter uncertainties. We used the resulting MCMC chains to calculate transit parameter correlations for KELT-15b. The resulting parameter correlation values are listed in Table \ref{tab:correlations}, and are adopted for all planets for the purposes of error propagation.

\begin{deluxetable}{ccr}
\tabletypesize{\scriptsize}
\tablecolumns{3}
\tablewidth{0pt}
\tablecaption{Parameter correlation values used for error propagation as determined from the KELT-15b global system fit.
\label{tab:correlations}}
\tablehead{\colhead{Param.~1} & \colhead{Param.~2} & \colhead{Correlation}} 
\startdata
\ar     &      \depth  &   $-$0.152 \\
\ar     &      \rvamp  &   $-$0.134 \\
\ar     &      \porb   &    $-$0.217 \\
\ar     &      \iorb   &     \phs 0.517 \\
\ar     &       \eorb  &   $-$0.645 \\
\depth  &       \rvamp  &   $-$0.015 \\
\depth   &       \porb  &   \phs 0.090 \\
\depth    &      \iorb  &   $-$0.406 \\
\depth    &      \eorb  &   $-$0.007 \\
\rvamp    &     \porb   &  $-$0.049 \\
\rvamp    &     \iorb &  $-$0.051 \\
\rvamp    &     \eorb &  \phs 0.086 \\
\porb     &     \iorb & $-$0.043 \\
\porb     &     \eorb &  \phs 0.366 \\
\iorb     &     \eorb &  $-$0.187 \\
\enddata
\end{deluxetable}

We compare our calculated KELT-15b parameter values and uncertainties to those of \citet{Rodriguez:2016} in Table \ref{tab:kelt15pars}. 
We find host star and planet radii and masses that are $\sim1\sigma$ higher than the literature values. Since \rhostar\ and \loggp\ are calculated solely from the transit light curve \kgsins{and RV} parameters (i.e., they do not involve \rstar), those values are nearly identical to the literature values. Our \insolation\ is consistent with the literature value. Our propagated stellar and planet radii uncertainties are $\sim3\times$ the literature values. The stellar mass uncertainty is significantly higher and the planet mass uncertainty is $\sim2\times$ higher than the literature values.  
\kgsins{To the extent that KELT-15b is representative of the systems in our study sample, this suggests that in fact our simplified linear error propagation approach discussed above results in conservative error estimates.}

\begin{deluxetable}{llcc}
\tabletypesize{\scriptsize}
\tablecolumns{4}
\tablewidth{0pt}
\tablecaption{KELT-15b empirical parameter values and uncertainties (this work) compared to \citet{Rodriguez:2016}.
\label{tab:kelt15pars}}
\tablehead{\colhead{Parameter} & \colhead{Units} & \colhead{This Work} & \colhead{Literature}} 
\startdata
\rstar  & Stellar Radius (\rsun)   &      $1.783\pm0.205$  &   $1.481\pm0.066$ \\
\mstar  & Stellar Mass (\msun)   &      $2.065\pm0.748$   &    $1.181\pm0.051$ \\
\rhostar  & Stellar Density (cgs)   &      $0.513\pm0.056$  &   $0.514\pm0.055$ \\
\rplanet  & Planet Radius (\rjup)   &     $1.737\pm0.203$   &   $1.443\pm 0.084$\\
\mplanet  & Planet Mass (\mjup)   &       $1.311\pm0.434$  &   $0.910\pm0.220$ \\
\loggp   & Planet Surface Gravity (cgs) & $3.032\pm0.106$  &   $3.020\pm0.115$ \\
\insolation & Incident Flux ($10^9$ erg s$^{-1}$ cm$^{-2}$) & $1.642\pm0.544$ & $1.652\pm0.145$ \\
\enddata
\end{deluxetable}

\section{Results\label{sec:results}} 

\subsection{Stellar Bolometric Fluxes and Angular Radii}

Two fundamental products of this work are the newly determined \fbol\ for each of the \kgsins{498} planet host stars in our \kgsins{master} sample, and from them the newly determined $\Theta$ for each of the stars. These are presented in Table~\ref{tab:results}. 

\begin{figure*}[!ht]
\centering
\includegraphics[width=0.49\linewidth,trim=10 5 10 50,clip]{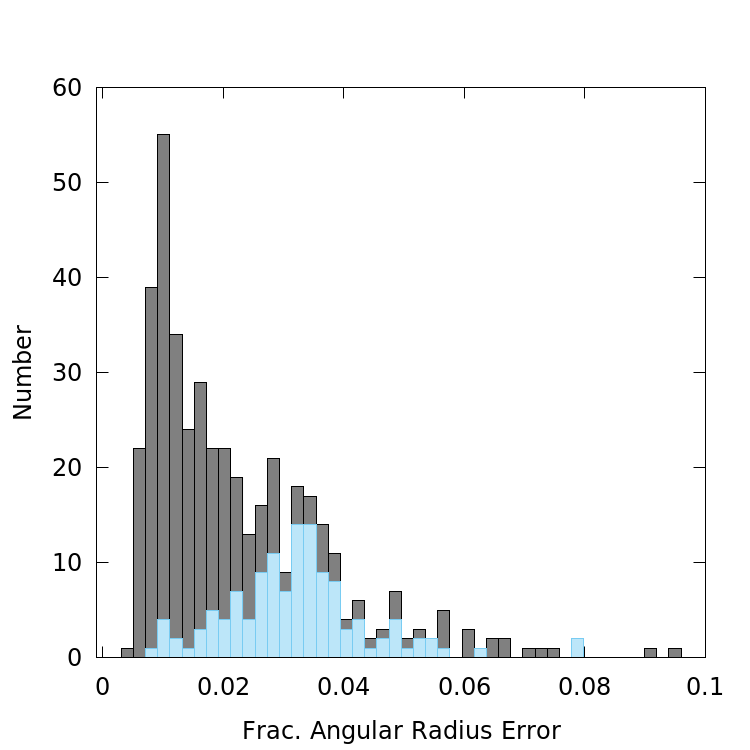}
\includegraphics[width=0.49\linewidth,trim=10 5 10 50,clip]{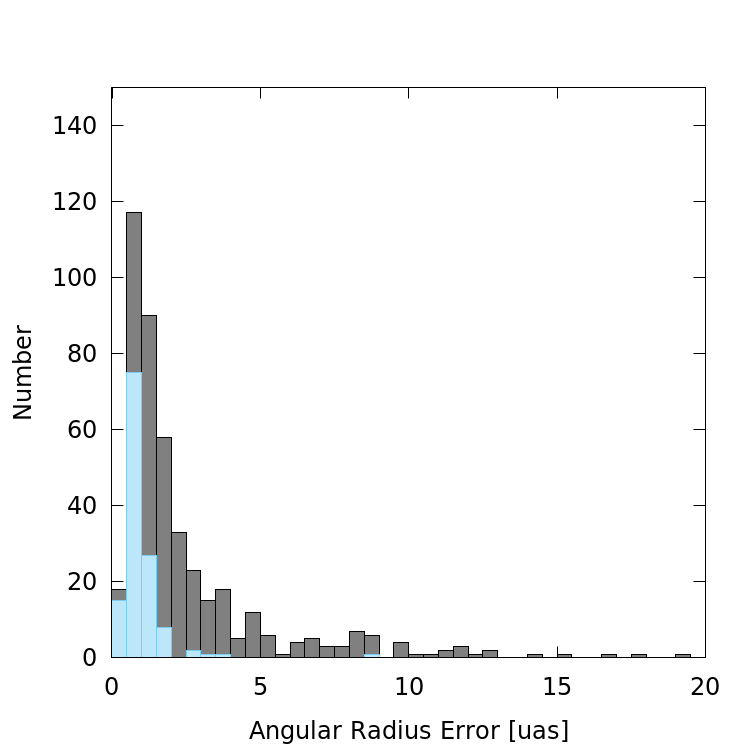}
\caption{Fractional uncertainty (left) and absolute uncertainty (right) on $\Theta$ for transiting planet host stars (blue) and for all stars (black). 
One star with a fractional uncertainty larger than 0.1 is not shown at left, and 15 stars with uncertainty larger than 20~$\mu$as are not shown at right.
\label{fig:theta_prec_dist}}
\end{figure*}

Because the precision on \fbol\ is typically \fbolprecper\ and the precision on \teff\ is typically 1--2\% (Fig.~\ref{fig:fbol_vs_chi2}), the achieved median precision on $\Theta$ is \thetaprecper\ for the transiting hosts and \thetaprecperall\ for all stars (Fig.~\ref{fig:theta_prec_dist}, left). In absolute units, the median precision is \thetaprecwunit\ for the transiting hosts and \thetaprecwunitall\ for all stars (Fig.~\ref{fig:theta_prec_dist}, right).
\kgsins{The transiting planet hosts are at greater distances, on average, than the radial-velocity planet hosts, and thus have smaller absolute $\Theta$ uncertainties despite having larger relative $\Theta$ uncertainties.}

Importantly, these two newly determined properties---\fbol\ and $\Theta$---do not depend on $d$. Thus, these results serve as a fundamental, accurate and purely empirical data set to permit the re-determination of stellar and planet radii and masses as measurements of $d$ improve.

\subsection{Host Star Radii and Masses}

\begin{figure*}[!ht]
\centering
\includegraphics[width=0.49\linewidth,trim=10 5 5 50,clip]{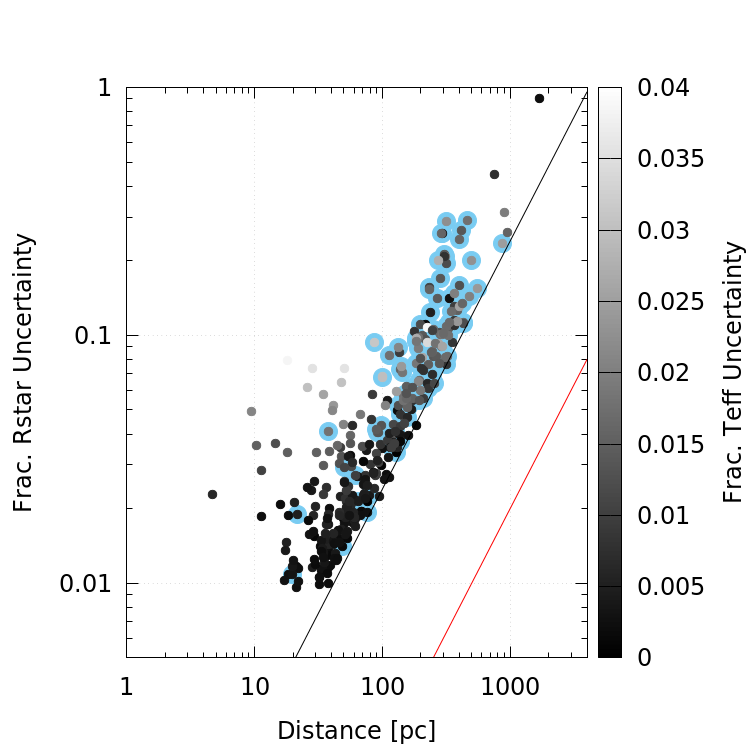}
\includegraphics[width=0.49\linewidth,trim=10 5 5 50,clip]{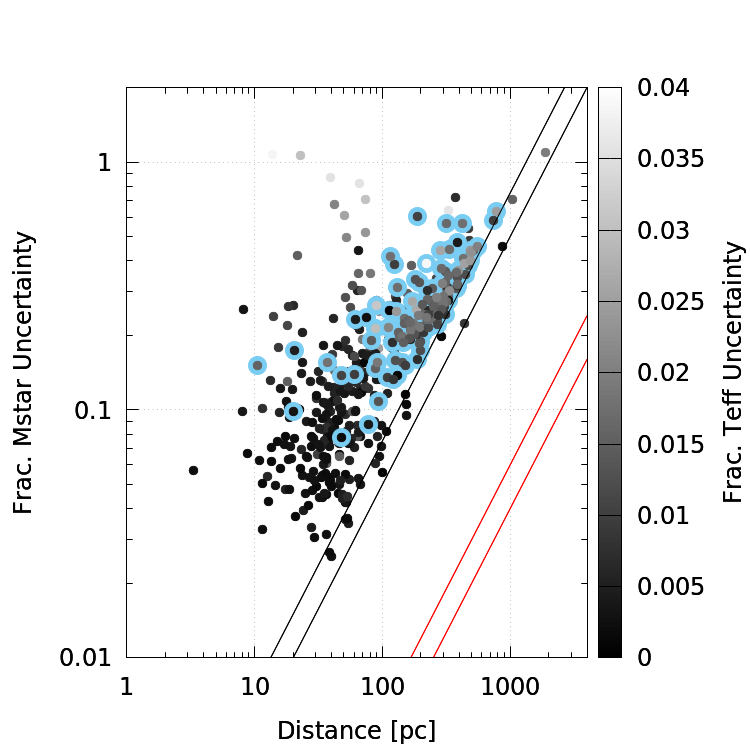}
\caption{Fractional uncertainties in stellar radii (left) and stellar mass (right) versus distance. Points with blue halos represent transiting planet host stars. The diagonal lines represents the fractional error expected for a current nominal parallax error floor of \tycparprec\ (black) and for an expected future parallax error of 20~$\mu$as (red). In the case of stellar mass (right), there are two diagonal lines representing the dependence of $\sigma_{\mstar} / \mstar$ on $\sigma_d$ for the transiting planets (where \mstar\ is determined directly from \rhostar\ and \rstar; upper line) and radial-velocity planets (where \mstar\ is determined from \logg\ and \rstar; lower line). 
\label{fig:rstar_vs_dist}}
\end{figure*}

We can use the newly available parallaxes from the {\it Gaia\/} first data release to estimate \rstar\ and \mstar\ from our newly measured \fbol\ and $\Theta$. 
The linear radii and masses newly determined here for the planet hosting stars in our sample are presented in Table~\ref{tab:results}. In Figure~\ref{fig:rstar_vs_dist} we show the achieved precision on the stellar radii and masses as a function of the {\it Gaia\/} distance and the \teff\ precision. 
The stellar radii and masses for the transiting-planet host stars are determined with a \kgsins{median} precision of \rstarprecper\ and \mstarprecper, respectively. Importantly, as shown in the figure, for most of the transiting planet host stars, the limiting factor on the radius precision is the precision on the current {\it Gaia\/} distance. 

\begin{figure}[!ht]
    \centering
    \includegraphics[width=\linewidth,trim=10 5 5 10,clip]{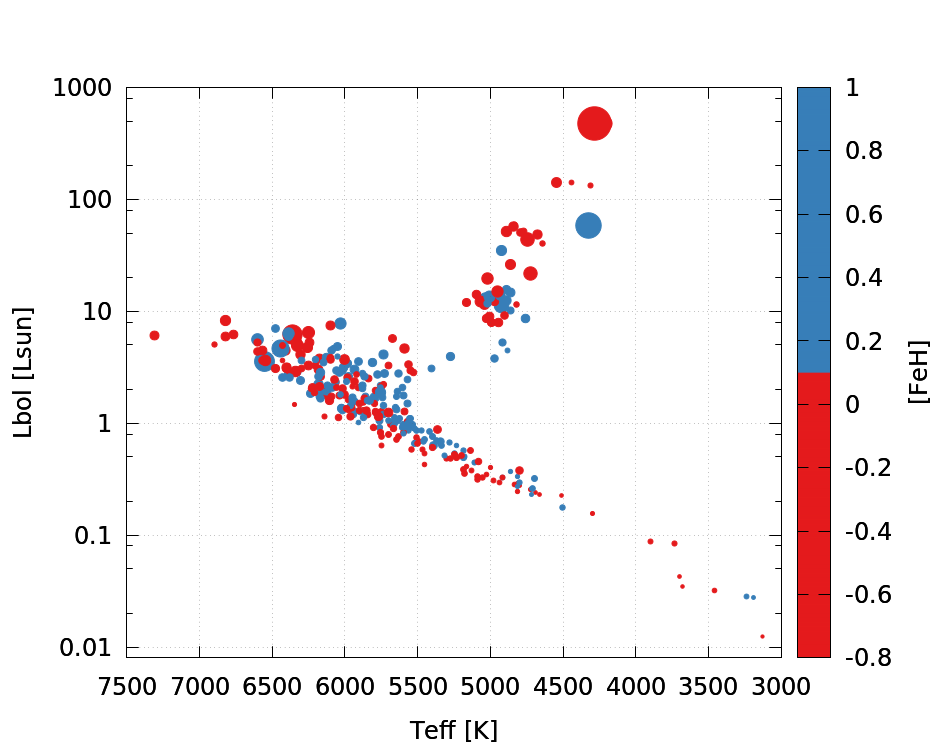}
    \caption{Hertzsprung-Russel Diagram of all planets in our study sample. Symbol size is proportional to \mstar, which for transiting planets is derived from the transit-based \rhostar\ and for RV planets is derived from \logg\ (see the text). Color represents [Fe/H]. The separation of metal-poor stars below the metal-rich stars on the main sequence is readily apparent. All quantities represented are determined empirically.}
    \label{fig:hrd}
\end{figure}

Figure~\ref{fig:hrd} presents the H-R diagram of the planet host stars, where all of the quantities represented are now measured entirely empirically. The separation of metal-poor stars below the metal-rich stars on the main sequence is readily apparent. 

\subsection{Planet Radii and Masses}\label{sec:results_planets}
\begin{figure}[!ht]
\centering
\includegraphics[width=\linewidth]{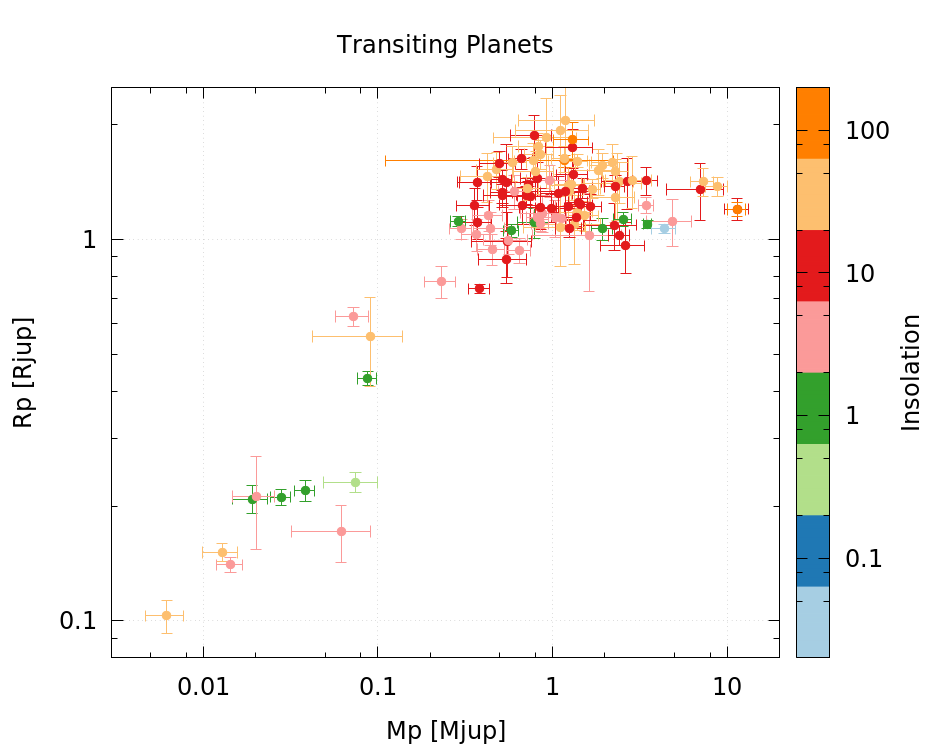}
\includegraphics[width=\linewidth]{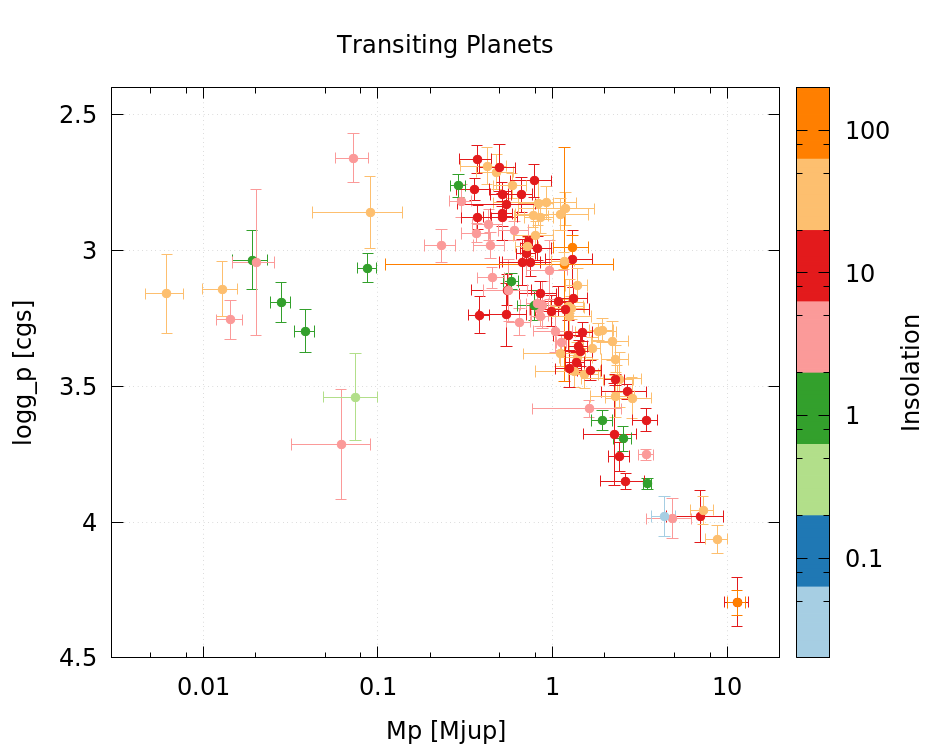}
\caption{{\it (Top)} Radius versus mass, and {\it (bottom)} $\logg_p$ versus mass, for transiting planets in our study sample. Color represents the received insolation by the planet, in units of $10^8$ erg s$^{-1}$ cm$^{-2}$, where warm colors represent insolation above the empirical threshold of $\sim 2\times10^8$ erg s$^{-1}$ cm$^{-2}$ for planet inflation \citep{Demory:2011}.}
\label{fig:planet_mass_radius}
\end{figure}

The resulting planet radii and masses, $R_p$ and $M_p$, are presented in Table~\ref{tab:results}. We also calculate directly the planet surface gravity, \loggp, and the insolation received by the planet, \insolation. These quantities are displayed together in Figure~\ref{fig:planet_mass_radius} for the transiting planets. 
The effect of insolation on the planet radii and surface gravities is evident, in the sense that more highly insolated planets of high mass have significantly larger radii and lower surface gravities at fixed mass. This effect is not clearly evident among lower mass planets below $\mplanet \lesssim 0.1 \mjup$. This may simply be due to the small sample size, but it has been suggested that the lowest mass planets at high insolation may have had their atmospheres photo-evaporated 
\kgsins{\citep{Owen:2013}}. 

\begin{figure*}[!ht]
    \centering
    \includegraphics[width=0.49\linewidth,trim=10 10 10 50,clip]{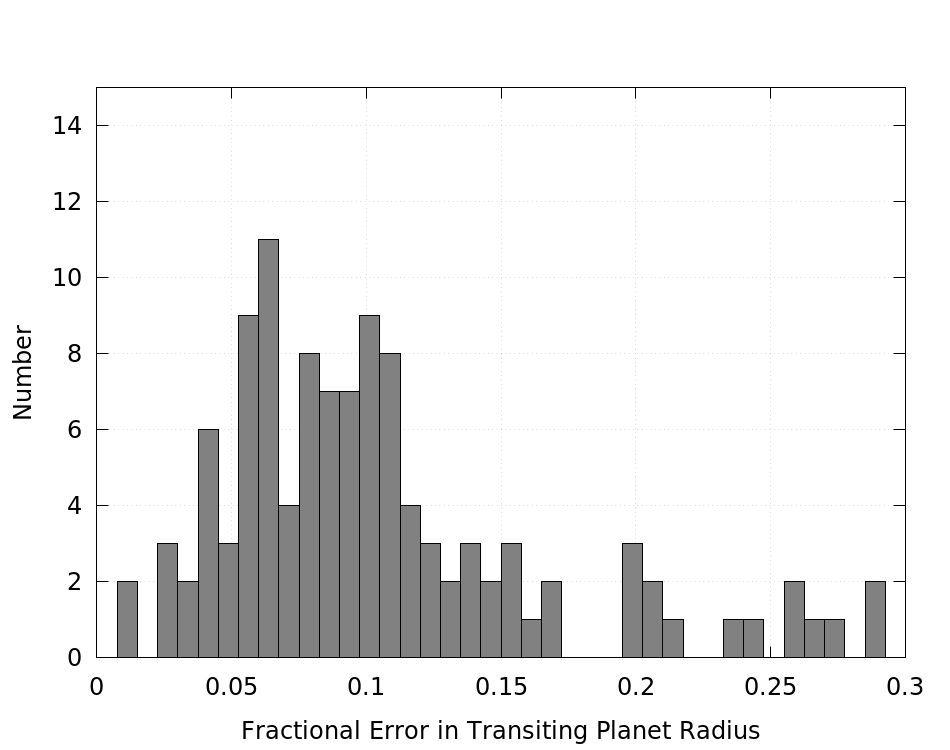}
    \includegraphics[width=0.49\linewidth,trim=10 10 10 50,clip]{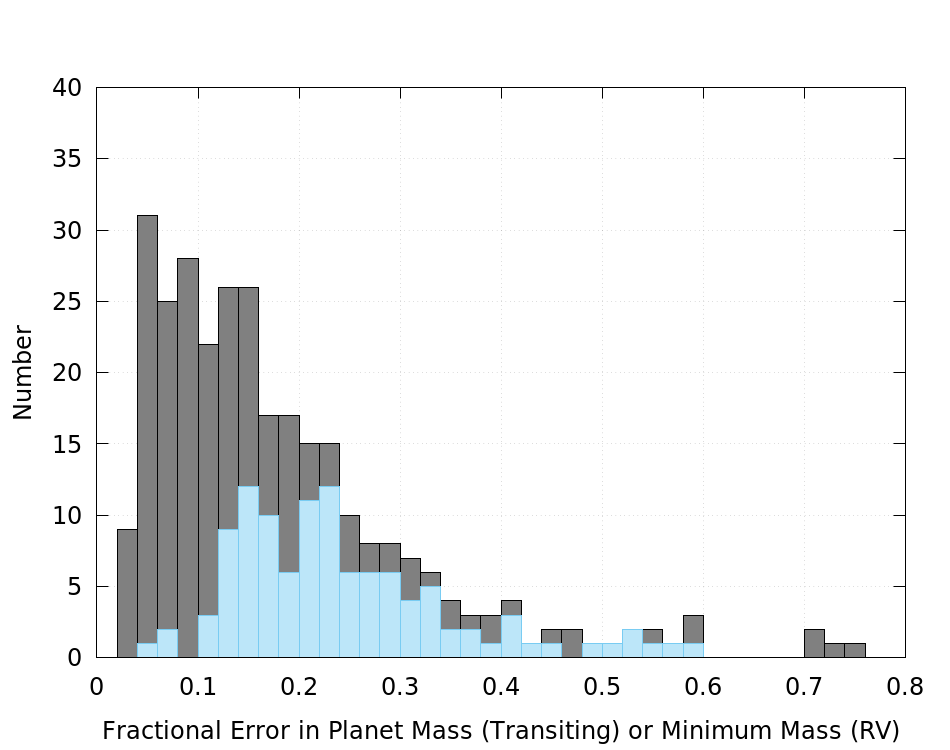}
    \caption{Distributions of fractional uncertainties on \rplanet\ (left) and \mplanet\ (\mplanet\sini\ for the RV planets) (right) determined in this work. Transiting planets are represented in blue in the right panel.}
    \label{fig:planet_err_hists}
\end{figure*}

The precisions on \rplanet\ and \mplanet\ are shown in Figure~\ref{fig:planet_err_hists}. 
The median uncertainty on \rplanet\ (we determine \rplanet\ for transiting planets only) is 9\%. The median uncertainty on \mplanet\ is 22\% for the transiting planets. The median uncertainty on \mplanet\sini\ is 10\% for the radial-velocity planets. Three planets with low signal-to-noise \rvamp\ have fractional \mplanet\ uncertainties above 0.5 and are not shown on the plot.

\section{Discussion\label{sec:disc}} 

\kgsins{In this section we compare our results to previous results, and we also consider how the results from the approach laid out in this study should improve when the second data release from {\it Gaia\/} becomes available. We begin, however, by expanding the discussion from Section~\ref{sec:intro} of the need to reduce reliance on, and to test, stellar models and empirical relations, as well as to explore the extent to which the approach laid out in this paper is truly empirical.}

\subsection{\kgsins{In Defense of the Empirical Approach Taken in this Study}}

\subsubsection{\kgsins{The importance of reducing reliance on, and testing, stellar models and empirical relations}}

\kgsins{In Section~\ref{sec:intro}, we asserted that there remain numerous areas in which theoretical stellar models, as well as empirically calibrated stellar relations, are not yet accurate and require additional testing. Here we present a number of additional examples to justify this assertion.}

One particularly relevant example, which we explore in detail in \S\ref{sec:astars}, is the controversy over the masses of subgiants and giants dubbed ``retired A stars"\footnote{We thank the referee for pointing out this particularly important application of direct stellar mass and radius measurements.}.  These are stars that were targeted by the radial velocity survey of \citet{Johnson:2007} because it was thought that their main-sequence progenitors were massive ($M\ga 1.5~M_\odot$), and thus could be used to explore the dependence of planet occurrence on host star mass.  However, because the majority of their masses and radii were not measured directly, and because of intrinsic degeneracies and uncertainties in the models of stars in the part of the H-R diagram where they reside, there has been some controversy over their true mass (e.g., \citealt{Lloyd:2013}).  Our empirically-measured radii and thus inferred masses (via \logg) speak directly to this controversy, and provide an excellent example of how such direct, empirical measurements can be important.

Another well-known discrepancy between models and empirical measurements is the ``inflated radius" problem of low-mass stars.  Specifically, stars of mass $\la M_\odot$ with empirically-measured stellar radii are often observed to be inflated by $\sim 10\%$, and their empirically-measured stellar effective temperatures to be lower by $\sim 5\%$, relative to the predictions of standard theoretical stellar
evolution models.  It is thought that this inflation problem is likely due to magnetic activity, which can suppress energy transport by increasing the spot coverage of the photosphere \citep{Stassun:2012}. Because mass-radius and mass-
\teff\ relationships are critical to our understanding of stellar evolution, resolving these discrepancies is essential for our development of accurate theoretical stellar models of low-mass stars.

We previously mentioned the case of KELT-6b \citep{Collins:2014}, where applying the empirical models of \citet{Torres:2010} to the observational data resulted in inferences for the parameters of the system that were inconsistent with those inferred by applying the YY isochrones \citep{Demarque:2004} by a much as $\sim 4\sigma$. As we suggested previously, this is likely due to the relatively low metallicity of the star (\feh$\sim -0.3$), but nevertheless clearly indicates that we need to calibrate both the empirical relations and isochrones over a much broader range of stellar parameters, thereby requiring more extensive empirical measurements.

As another example, Tayar et al. (submitted) demonstrate that theoretical models of red giants exhibit a metallicity-dependent offset in the predicted temperatures
when compared to observations, which can result in inferred ages that are incorrect by as much as a factor of two, even for modest deviations from solar metallicity.

Another final example that is both more subtle but also much more disturbing is the case of the properties of the Sun.   Standard solar models (SSMs) have existed for well over fifty years, and were used to, e.g., uncover the solar neutrino problem (e.g., \citealt{Bahcall:1992}), which eventually lead to the discovery that neutrinos have mass and thus that the standard model of particle physics was incomplete (e.g., \citealt{Fukuda:2001}).  While a seemingly impenetrable success of SSMs, a later revised estimate of the oxygen abundance of the Sun \citep{Allende:2001,Asplund:2009}, resulted in predictions for helioseismology from SSMs that were grossly inconsistent with observations \citep{Bahcall:2005}. While this ``Solar Oxygen Crisis" now appears to be largely resolved (e.g., \citealt{Caffau:2008}), this example demonstrates how even our nearest star, for which we have the most information and which we use as an anchor point for essentially all of our stellar evolutionary models, can still have uncertainties and thus can still benefit from improved direct observational constraints.   

Indeed, even the often-used \citet{Torres:2010} empirical relations between  (\teff, \logg, and \feh) and \mstar\ and \rstar\ do not reproduce the mass and radius of the Sun to better than $\sim 5\%$ and $\sim 2\%$, respectively. While such ``small" discrepancies may appear inconsequential given our current ability to measure the masses and radii of other stars, there is no reason that we cannot and should not demand better accuracy and precision in our knowledge of a larger sample of other stars.  In this paper, we demonstrate a path forward to achieving this goal. 

\subsubsection{\kgsins{Are the stellar and planet properties truly empirical?}}

\kgsins{Few measured quantities in astronomy are truly, unequivocally empirical. Arguably, the only stellar properties that one can measure in a fully empirical manner---that is, without invoking models or even basic laws of physics---are parallax, color, brightness at a given wavelength, and in some cases an interferometric angular diameter at a given wavelength. 
All other measurable stellar quantities rely on basic physical laws together with the above direct observables. For example, stellar \teff\ are determined either through fitting observed spectra to synthetic spectra, or through measurement of the equivalent widths of absorption lines and the subsequent use of models or empirical relations to convert them to \teff. Similarly, the determination of the stellar \fbol\ makes use of an adopted extinction law.}
\kgsins{In fact, even the stellar parallaxes involve analysis pipelines from the {\it Gaia\/} team that, based on the experience of the {\it Hipparcos\/} distance to the Pleiades, might be considered in doubt.}

\kgsins{The planet properties in turn depend on the stellar properties, as well as on observables that themselves involve some basic assumptions. For example, transit light curve parameters depend on adopted stellar limb darkening coefficients, algorithms for the removal of systematics in the transit light curves and for the extraction of radial velocities from spectra, and indeed the assumption of Keplerian orbits.}

\kgsins{For the reasons described in the previous section, we have attempted in this paper to lay out an approach that is empirical to the extent possible, making use only of {\it observational measures} that are either direct (e.g., apparent brightness) or that follow simply from direct observables (e.g., light curves and radial velocities) via well understood, basic physics (e.g., Kepler's laws). This then leaves only a very small number of observational quantities that depend mildly on standard assumptions (e.g., reddening) but that affect the stellar and planet properties of interest only negligibly, as we now discuss.} 


\kgsins{First, and most fundamentally for the purposes of this work, is the empirical measurement of \fbol, which is the most basic quantity that we measure for all 498 stars in our study sample. It could be argued that our procedure is dependent on the model atmospheres used, which of course it is to some extent. This model dependence is mitigated, however, by the very large wavelength range covered by the actual broadband flux measurements, which for most of the stars includes a very large fraction of the emergent stellar flux. This was examined in detail by \citet{Stassun:2016}, who calculated the fraction of the stellar \fbol\ that is from beyond the span of the flux measurements, for stars of various \teff. They found that for \teff$\lesssim$7000~K (which is the case for all but one star in our sample here), this flux fraction is in the range 1--4\%. Indeed, this is the range of \fbol\ uncertainty that we determine for our sample here (Figure~\ref{fig:fbol_vs_chi2}), which reflects this ``extrapolated" flux in the SED fitting procedure.
Other concerns may be that we have had to assume solar metallicity for some of the stars, if a spectroscopic [Fe/H] was not available. \citet{Stassun:2016} performed a check by varying the adopted [Fe/H] from $-0.5$ to $+0.3$---representing the range of metallicity for the vast majority of Milky Way stars---for several stars spanning the full range of \teff\ considered here. They found that the effect on the resulting \fbol\ is negligible for hot stars and as much as $\sim$0.5\% for cool stars, in all cases much smaller than our typical \fbol\ uncertainty of 2\%.
Arguably the most important purpose of the fitting procedure is to determine $A_V$, for determining what \fbol\ would be in the absence of extinction. We have adopted a single ratio of total-to-selective extinction, $R_V = 3.1$ in our fits. $R_V$ values in the literature span the range $\approx$2.5--4 for most Galactic sight lines, and thus in principle fitting for $R_V$ could further improve the SED fits. However, we have opted for simplicity not to introduce another free parameter to the SED fitting procedure. In any event, if any of the SED fits are poorer due to our choice of $R_V$, the resulting increased $\chi_\nu^2$ will in turn result in more conservative uncertainties on \fbol.}

\kgsins{Next, the determination of $\Theta$ from \fbol\ requires one additional datum, namely \teff\ (Equation~\ref{eq:frt}). Stellar \teff\ determinations are most reliable when measured from high-resolution spectroscopic analysis, as we have done here by drawing \teff\ from the PASTEL catalog (Sec.~\ref{sec:data}). It is true that determining the proper physical temperature of a star can be complicated by, e.g., the spectral analysis method used, temperature inhomogeneities on stellar surfaces, etc. However, \teff\ is a {\it defined} quantity, namely, defined in the context of the Stefan-Boltzmann Law. Numerous calibrations of \teff\ for stars in various color/mass/age ranges have been performed \citep[see, e.g.,][]{Casagrande:2008,Casagrande:2010,Boyajian:2012,Boyajian:2013,Mann:2015}, and thus while there remain systematic uncertainties, in general these uncertainties are well characterized and they are propagated into our $\Theta$ uncertainties \citep[see also][for extensive discussion of progress toward \teff\ and $\Theta$ at percent-level accuracy]{Casagrande:2014}.}

\kgsins{Furthermore, while spectroscopic determinations of $\log g$ and [Fe/H] can be subject to larger uncertainties and systematics \citep[e.g.,][]{Torres:2012}, our results are almost entirely independent of these quantities, at least for the transiting planet systems. Similarly, the effect of stellar limb darkening coefficients on \rhostar\ determined from the planet transit durations is exceptionally small, and in any case it is always possible to obtain transit duration measurements in the infrared where the effect is nearly nonexistent.}

\kgsins{In summary, we have attempted in this work to lay out a set of procedures that utilize the {\it Gaia\/} parallaxes to determine stellar and planet properties that are as direct and empirical as possible. For the reasons discussed here, it is our assertion that the fundamental quantities we have measured for the 498 stars in our sample---\fbol\ and $\Theta$---may be regarded as truly empirical, with effectively only one free parameter, $A_V$. Moreover, to the extent that trigonometric parallax is empirical, then so is \rstar, which follows directly from the above. And, because they follow from the above empirical stellar properties together with observables that are virtually free of assumptions, \rplanet\ and \mplanet, in particular for transit planets, are, we argue, empirically measured, at least according to the precepts and procedures laid out in this work.}

\subsection{Comparison to Previously Published Results}
\subsubsection{\kgsins{Previous compilations of planet host star radii}}

Several authors have previously published compilations of estimated stellar radii for stars with {\it Hipparcos\/} parallaxes. These include, e.g., 32 early-type stars in \citet{Jerzykiewicz:2000}, 1000 FGK stars in \citet{Valenti:2005}, 125 A--M dwarfs in \citet{Boyajian:2012,Boyajian:2013}, and 166 stars known (at the time) to host exoplanets in \citet{vanBelle:2009}. 
In general these and other works have used spectroscopic and/or photometric measures of \teff\ together with individual colors, bolometric corrections, and the {\it Hipparcos\/} distance to estimate \rstar. 
The studies of \citet[][and other papers in the series]{Boyajian:2013} used directly measured interferometric $\Theta$, including the planet-hosting stars that we have used to check our procedures \citep[see Section~\ref{sec:stellarpars};][]{Boyajian:2015}. 

The previous study most directly comparable to our work is that of \citet{vanBelle:2009}, which used full broad-band SED fitting to estimate $A_V$, spectral type templates to estimate \teff, and thereby to estimate $\Theta$ and finally \rstar\ via the {\it Hipparcos\/} distance. 
Our work complements and extends this study by including a considerably expanded set of planet-hosting stars now known possessing suitable measurements (498 here versus 166 in that work) and the newly available {\it Gaia\/} parallaxes. In addition, our work includes any and all direct observables of the planets in order to also determine \mstar\ as well as revised planet radii, masses, and surface gravities---all of these involving only direct, empirical measures. 

\subsubsection{\kgsins{Planet and host-star radii and masses: This study versus previously published values}}

\begin{figure*}[!hbt]
    \centering
    \includegraphics[width=0.49\linewidth]{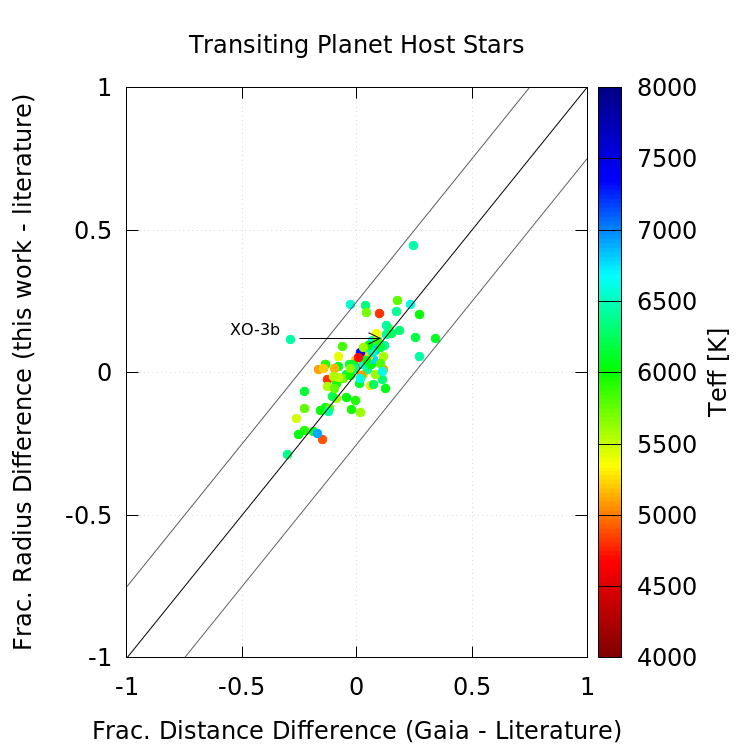}
    \includegraphics[width=0.49\linewidth]{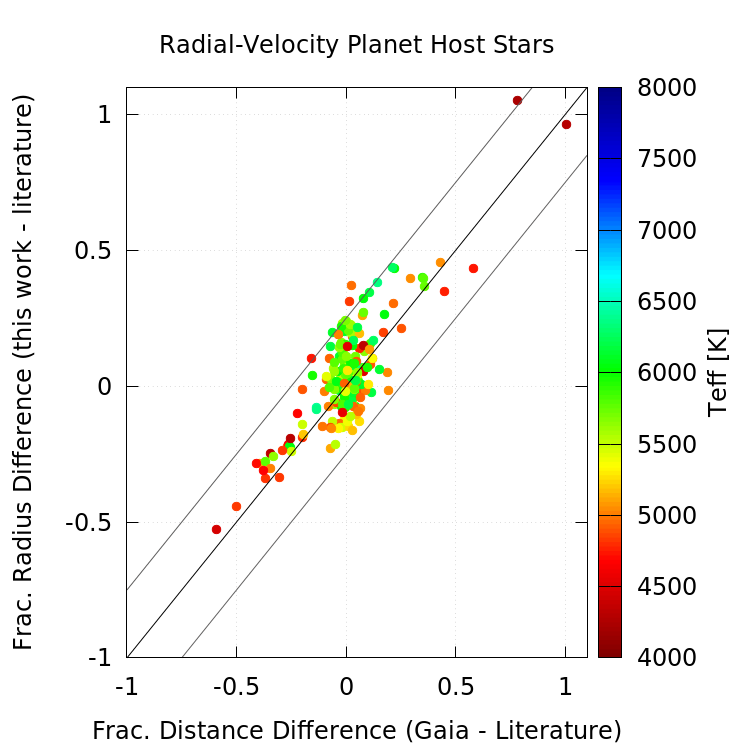}
    \caption{Fractional difference in the newly determined \rstar\ versus those previously reported in the literature,
    as a function of the fractional change in the distance, for transiting planet hosts (left) and radial-velocity planet hosts (right). Color represents \teff. Lines represent the one-to-one relationship of $\Delta$\rstar\ $\sim\Delta d$ and the range of variation arising from the \rstar\ errors determined in this work (see also Figure~\ref{fig:planet_err_hists}). The case of XO-3b is identified and discussed further in the text (see also Figure~\ref{fig:pl_radius_comp}).}
    \label{fig:st_radius_comp}
\end{figure*}

While we anticipate that the \rstar, \mstar, \rplanet, and \mplanet\ values derived from our \fbol\ and $\Theta$ determinations will be greatly improved with the {\it Gaia\/} second data release (see below), we can already assess the impact of {\it Gaia\/} parallaxes from the first data release on the inferred stellar and planet properties. 
We present a direct comparison of our newly determined \rstar\ versus those reported in {\tt exoplanets.org} in Fig.~\ref{fig:st_radius_comp}. 

\kgsins{Our revised \rstar\ values range from $\sim$50\% smaller to $\sim$100\% larger than the previous literature values, and these differences are clearly the result principally of the improved {\it Gaia\/} distances, the one exception being XO-3 for which the updated \rstar\ is small despite a change in $d$ of nearly 25\%. We have traced the XO-3 anomaly to the complicated history of this system in the literature; see below.}
For the transiting-planet host stars, \kgsins{because the fractional changes in $d$ are generally relatively small,} there is good overall agreement with previously reported \rstar; the median difference is 3\% with no strong systematic dependence on \teff, \kgsins{and for 90\% of the sample the change in \rstar\ is at most 15\%}. 
On the other hand, for the radial-velocity planet-host stars, there are significant differences, and in particular among the coolest stars there is a clear \kgsins{tendency for large changes in} \rstar. We have compared our adopted \teff\ with those previously used in the literature, and confirm that this is not the cause of these large \rstar\ differences (the difference in \teff\ being 0.0$\pm$0.3\%). 
Comparison between the new stellar distances from {\it Gaia\/} and those previously reported in the literature 
suggests that the errors in the previous distances, particularly among the cooler very nearby stars, may be the culprit; 
\kgsins{here the differences in $d$ can be as large as $\sim$100\%, most notably for the red giants in the sample (Figure~\ref{fig:st_radius_comp}, right)}. 

\begin{figure*}[!ht]
    \centering
    \includegraphics[width=0.49\linewidth]{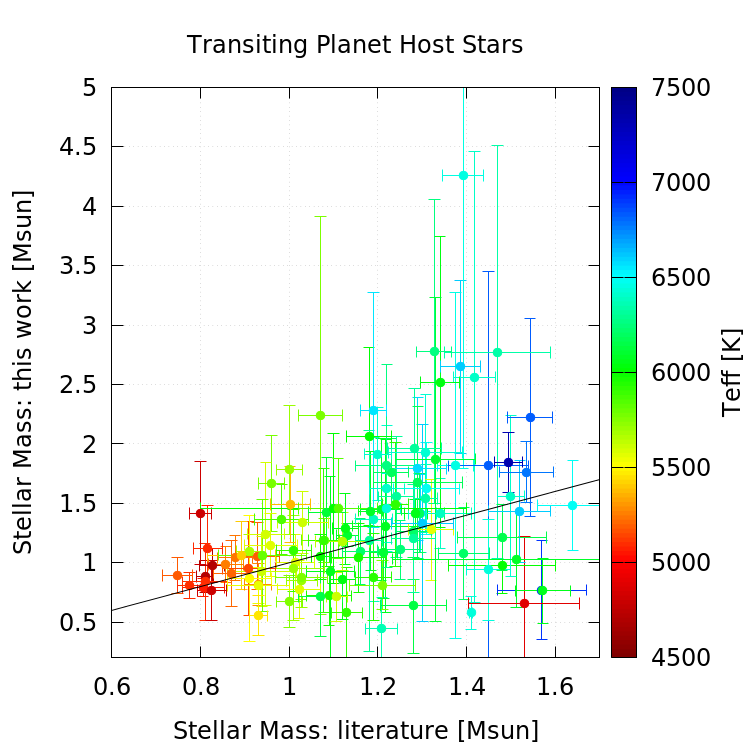}
    \includegraphics[width=0.49\linewidth]{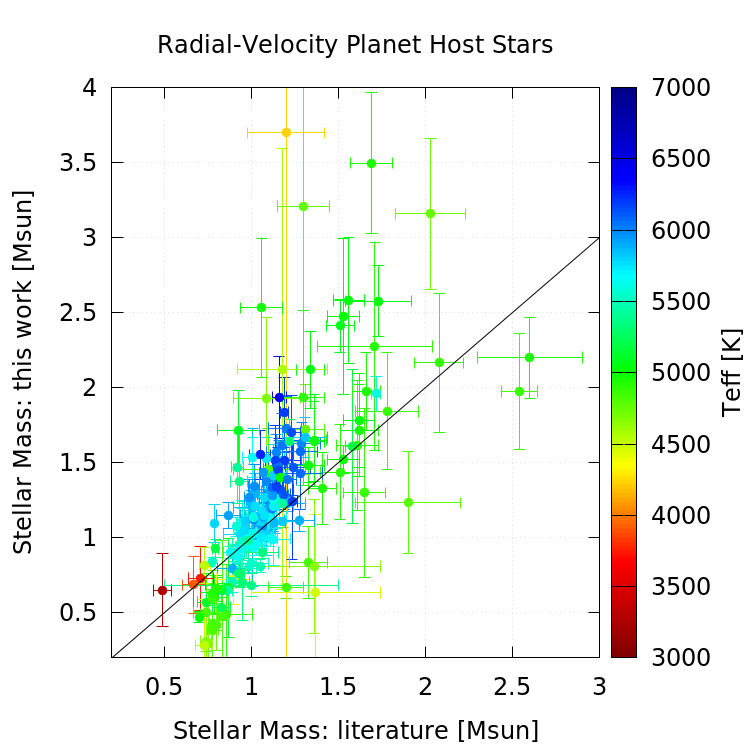}
    \caption{Comparison of newly determined \mstar\ versus those previously reported in the literature for transiting planet hosts (left) and radial-velocity planet hosts (right).}
    \label{fig:st_mass_comp}
\end{figure*}

Similarly, we compare our newly determined \mstar\ to those previously reported in Figure~\ref{fig:st_mass_comp}, where the \mstar\ for the transiting planet host stars are determined directly from transit-based \rhostar\ and \rstar, whereas for the radial-velocity planet hosts it is determined from the spectroscopic \logg\ and \rstar. 
Once again, for the transit-planet host stars, there is good overall agreement. However, again for the radial-velocity host stars, there are significant differences, and these show systematics with \teff. We suspect that these differences arise from the fact that, for the transiting systems, we have a very direct measure of \mstar\ from \rhostar\ and \rstar, whereas for the radial-velocity planet hosts we are subject to the well known challenges with spectroscopic \logg\ \citep{Torres:2012}.
Recall that we have opted to use spectroscopically determined \logg\ that are measured independent of the planet discovery papers. 

\begin{figure}[!ht]
\centering
\includegraphics[width=\linewidth]{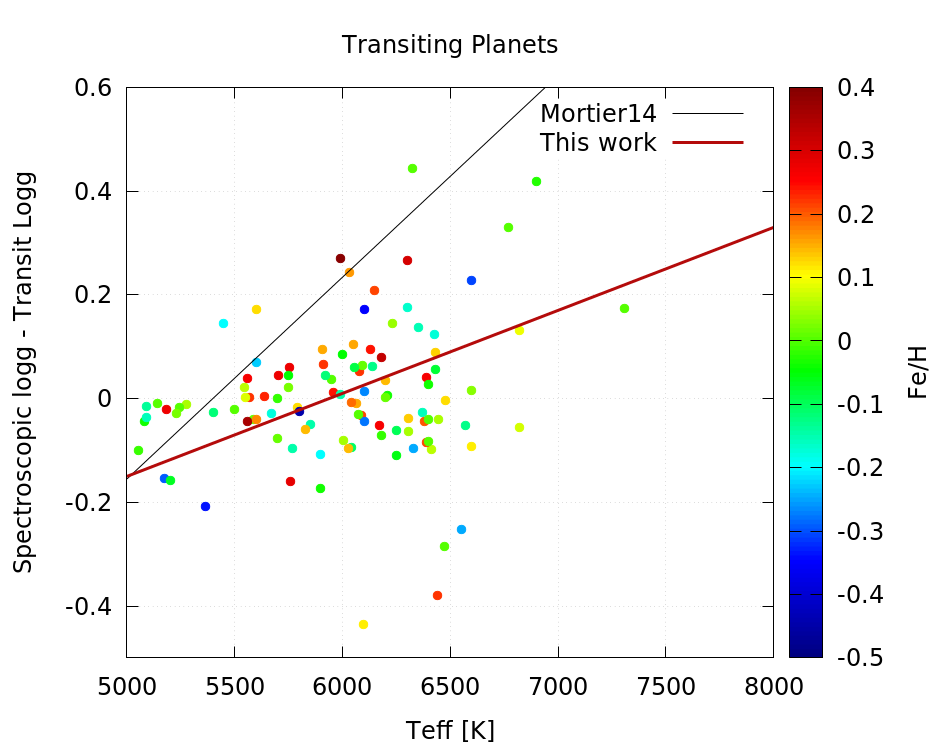}
\caption{\kgsins{Comparison of spectroscopic \logg\ versus transit-derived \logg\ for transiting planet host stars. There is a tendency (red solid line) for the spectroscopic \logg\ to be overestimated at high \teff\ and underestimated at low \teff, by $\sim$0.1~dex. The sense of this trend is in the same sense as the systematics observed in the masses derived from spectroscopic \logg\ for the radial-velocity planet hosts (Fig.~\ref{fig:st_mass_comp}, right). The correction previously suggested by \citet{Mortier:2014} (solid black line) is somewhat larger than what we find here.}}
\label{fig:logg_comp}
\end{figure}

\kgsins{\citet{Mortier:2014} proposed empirical corrections for spectroscopic based \logg\ on the basis of transits and asteroseismology. 
We can use our transit-derived \logg\ for the transiting planet host stars to test this (Figure~\ref{fig:logg_comp}). Indeed, we find a tendency for stars with \teff$\gtrsim$6000~K to have spectroscopic \logg\ that are overestimated by $\sim$0.1~dex, and for the cooler stars to have spectroscopic \logg\ that are underestimated by a similar amount. 
This trend ($\Delta\logg = 1.6\times10^{-4}\teff-0.95$, spectroscopic minus transit) is somewhat less steep than that proposed by \citet{Mortier:2014}. Nonetheless, the effect is in the correct sense to at least partially explain the systematics in \mstar\ for the radial-velocity planet hosts (Fig.~\ref{fig:st_mass_comp}, right).}

\begin{figure}[!ht]
    \centering
    \includegraphics[width=\linewidth]{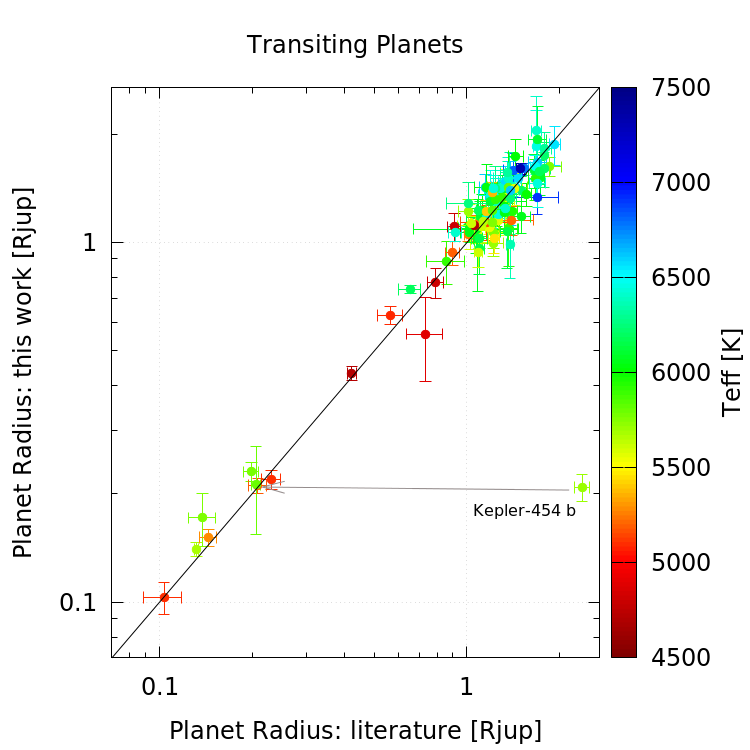}
    \caption{Comparison of newly determined \rplanet\ versus those previously reported in the literature. One notably discrepant case is labeled and discussed in the text.}
    \label{fig:pl_radius_comp}
\end{figure}

\begin{figure*}[!ht]
    \centering
    \includegraphics[width=0.49\linewidth]{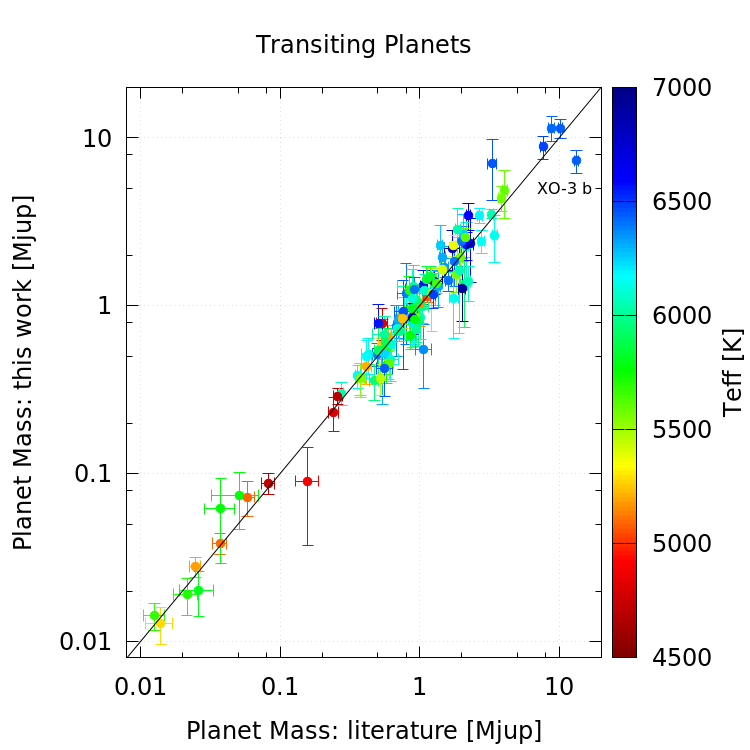}
    \includegraphics[width=0.49\linewidth]{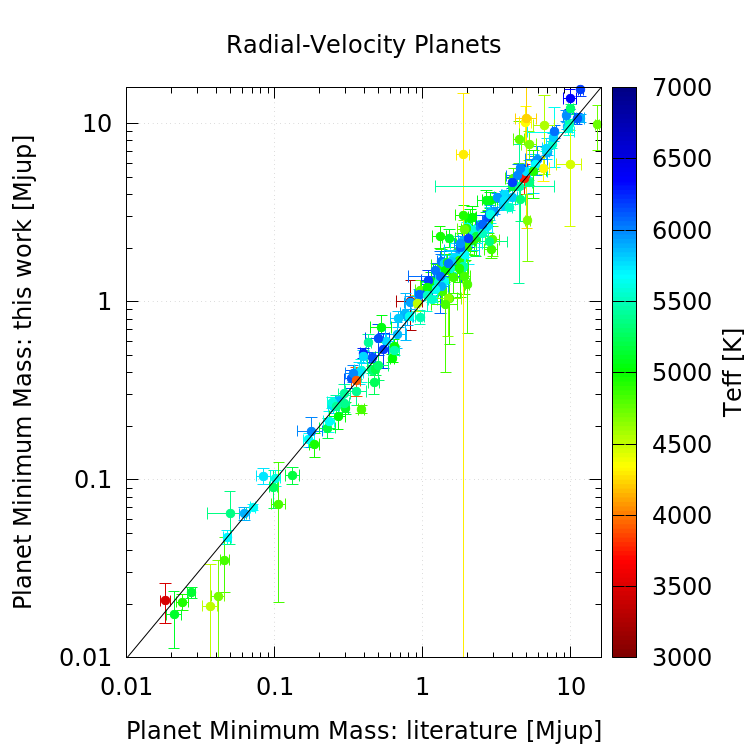}
    \caption{Comparison of newly determined \mplanet\ (\mplanet \sini\ for radial-velocity planets) versus those previously reported in the literature for transiting planets (left) and radial-velocity planets (right). One notably discrepant case is labeled and discussed in the text. 
    }
    \label{fig:pl_mass_comp}
\end{figure*}

Based on the updated stellar \rstar\ and \mstar, we can compare our newly determined \rplanet\ and \mplanet\ to those previously published. These comparisons are shown in Figures \ref{fig:pl_radius_comp} and \ref{fig:pl_mass_comp}, respectively. The outlier planet radius labeled in Figure \ref{fig:pl_radius_comp} is Kepler-454b. \citet{Gettel:2016} reported its radius as $2.37\pm0.13$ $R_\oplus$, but the value in {\tt exoplanets.org} is $2.37\pm0.13$ \rjup. The tip of the light grey arrow on the plot shows the proper position of Kepler-454b for the correct value from \citet{Gettel:2016}. The planet with mass significantly below the line in Figure~\ref{fig:pl_mass_comp} is XO-3~b. 
\kgsins{The distance to XO-3~b is reported as $d=282_{-23}^{+27}$~pc in {\tt exoplanets.org}.} 
In the XO-3~b discovery paper, \citet{JohnsKrull:2008} reported a mass of $\mplanet=13.25\pm0.64$ \mjup\ and a radius of $\rplanet=1.95\pm0.16$ \rjup\ based primarily on a host star stellar radius derived from the spectroscopic \logg. In a follow-up paper, \citet{Winn:2008} found a slightly lower mass $\mplanet=11.79\pm0.59$ \mjup\ and a significantly lower radius $\rplanet=1.217\pm0.073$ \rjup\ 
\kgsins{and distance $d=174\pm18$~pc} based the stellar density from the transit light curves\footnote{At the time of writing, the {\tt exoplanets.org} database contains the updated \rplanet, but not the updated \rstar\ \kgsins{and distance}. We have updated our literature \rstar\ 
\kgsins{and distance} 
to the values from \citet{Winn:2008}.}. 
We find a significantly lower empirical value of planet mass $\mplanet=7.290\pm1.188$ \mjup\ and a slightly higher planet radius $\rplanet=1.414\pm0.121$ \rjup\ 
\kgsins{and distance $d=201.7\pm14.2$~pc} 
compared to \citet{Winn:2008}. 

%

\subsubsection{\kgsins{Occurrence of ``retired A stars"}}\label{sec:astars}

\kgsins{Finally, one of the outstanding questions in the literature is that of the so-called ``retired A stars." 
In brief, it has been suggested that planets orbiting evolved subgiants and red giants may serve as probes of planet formation and evolution in massive stars \citep[e.g.,][]{Johnson:2007}, which is particularly useful given the paucity of known planets orbiting unevolved (i.e., main sequence) stars more massive than the Sun. The utility of the subgiant and red giant planets in this context is of course predicated on the assumption that the stars are indeed the evolved (``retired") counterparts of stars that were formerly A-type stars on the main sequence. 
However, this has been called into question \citep[e.g.,][]{Lloyd:2011,Lloyd:2013}, considering the challenge of disambiguating stars by mass in the subgiant and red giant regions of the H-R diagram, where \teff\ and \lbol\ become largely degenerate with \mstar, particularly when uncertainties and systematic errors in the determination of \feh\ and \logg\ are considered.  These degeneracies are compounded by uncertainties in the accuracy of stellar evolution models in these parts of the H-R diagram. As a result, the arguments and counter-arguments presented by, e.g.,  \citet{Lloyd:2011,Lloyd:2013}, \citet{Johnson:2013}, and \citet{Schlaufman:2013}, have largely relied on indirect lines of evidence, such as the proper application of priors, sample biases, Galactic models and stellar spins.  One exception is \citet{Johnson:2014}, who used asteroseismology to directly measure the mass of one ``retired A star" to be $\sim 2~M_\odot$.} 

\begin{figure}[!ht]
\centering
\includegraphics[width=\linewidth]{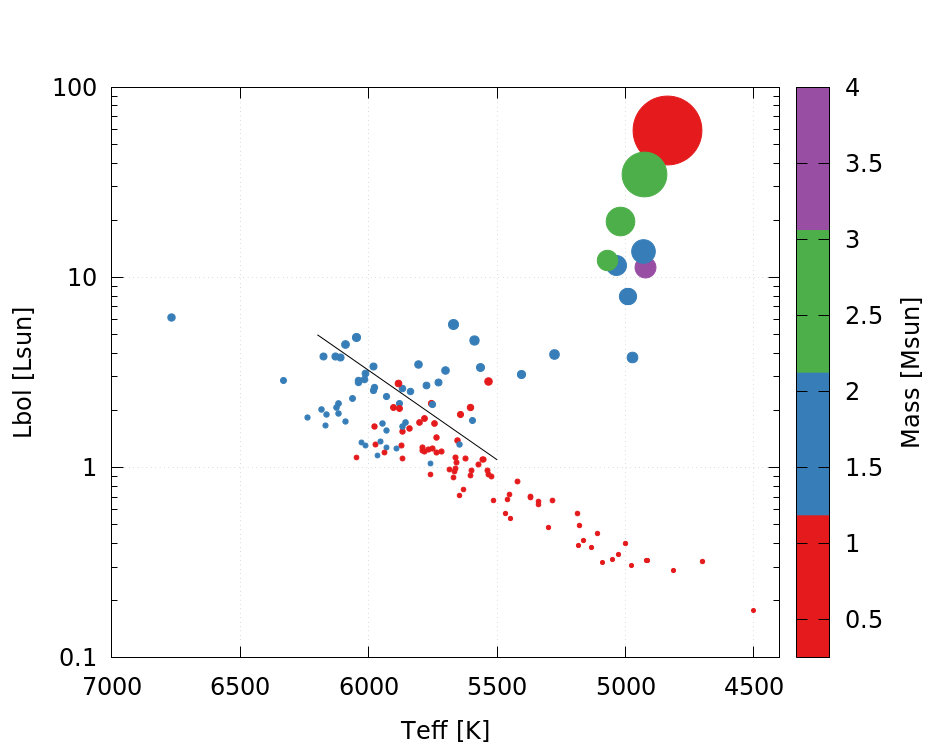}
\caption{H-R diagram of stars in our sample with the most accurate stellar mass determinations, $\sigma_{\mstar} < 15\%$. Color represents \mstar, and symbol size is proportional to \rstar. The black line denotes our separation of dwarfs from subgiants/giants. The majority of the evolved stars have \mstar$>$1.2~\msun\ (blue, green, or purple colors).}
\label{fig:ret_a_stars}
\end{figure}

\kgsins{With our new, empirically determined stellar properties---including in particular \mstar---we are in a position to assess the occurrence of stars with \mstar$\gtrsim$1.2~\msun\ in the subgiant and red giant branches of the H-R diagram, at least among the systems in our study sample. 
Figure~\ref{fig:ret_a_stars} shows the 134 stars in our study with the currently most accurately determined \mstar, for which 
$\sigma_{\mstar} < 15\%$. Thirty of these are in the region of the H-R diagram often dubbed ``retired A stars." The mean and median stellar mass of these 30 stars is \mstar~$=1.58$~\msun\ and 1.45~\msun, respectively, and the inter-quartile range is 1.22--1.66~\msun. Only $\sim$20\% of these stars have \mstar\ $<1.2$~\msun.
Thus, we conclude that in our sample there is strong evidence for a preponderance of {\it bona fide} ``retired A stars," though there do appear to be some stars that are less massive.}

\subsection{Prospects with {\it Gaia\/} Second Data Release}

The anticipated second data release from {\it Gaia\/} should enable the fundamental stellar \fbol\ and $\Theta$ reported here to achieve their full potential. In particular, whereas the accuracy that we currently achieve on \rplanet\ and \mplanet\ is limited for the transiting planets by the current {\it Gaia\/} parallax error floor of \tycparprec\ (see Figure~\ref{fig:rstar_vs_dist}), the parallax precision for the {\it Gaia\/} second data release should be \drtwoparprec\ for the bright stars in our study sample, and consequently the final \rplanet\ and \mplanet\ will in most cases be limited only by the accuracy of $\Theta$, which in this work we already determine to be \thetaprecperall. 
Thus, we may expect that the application of our $\Theta$ to the {\it Gaia\/} second data release parallaxes should yield \rplanet\ and \mplanet\ for the transiting planets in our sample to \rpdrtwoprecper\ 
and \mpdrtwoprecper, respectively, \kgsins{assuming that the uncertainties in the observed \depth\ and \rhostar\ are negligible}.
\kgsins{At present, the median uncertainties on \depth\ and \rhostar\ for the transit planet host stars are $\approx$2\% and $\approx$14\%; thus the accuracy in \rplanet\ should indeed be close to $\approx$3\%, whereas for \mplanet\ it would be closer to $\approx$10\%. Still, improvements in the \rhostar\ from high-precision transit observations, such as expected from the bright {\it TESS\/} targets, will allow the methodology laid out here to achieve $\approx$5\% accuracy in \mplanet\ with the {\it Gaia\/} DR2 parallaxes.}

\kgsins{Finally, we note that once we have many more precise {\it and} accurate empirical measurements of these stars from {\it Gaia\/} DR2, it will then be possible to refine and extend both stellar evolutionary models and empirical relations. These can then be more confidently applied to systems where direct empirical measurements are not possible, in order to derive precise and presumably more accurate parameters for vastly larger numbers of star/planet systems.}

\section{Summary and Conclusions\label{sec:summary}} 

We have demonstrated that several new observational advances have enabled direct measurements of the fundamental properties for a much larger sample of bright ($V\la 12$) stars than has heretofore been available.  These advances include the availability of broadband photometric measurements spanning a very broad range of wavelengths, thanks to several all-sky panchromatic surveys (e.g., GALEX, APASS, 2MASS, WISE). These photometric measurements permit construction of empirical SEDs that encompass a very large fraction of the stellar SED, which in turn enable nearly direct measurements of the bolometric fluxes of all but the hottest of these bright stars, to a precision of typically $\lesssim$3\% \citep{Stassun:2016}. When combined with estimates of the stellar effective temperatures, ideally measured from high-resolution spectra, the angular diameter of the star (as well as the extinction) can in principle be determined in a largely empirical manner. Finally, the newly available {\it Gaia\/} parallaxes can then be used to estimate the radii of the stars, essentially directly.

With the current data available, we find that stellar radii can be determined to a precision of $\sim$\rstarprecper.  This precision is limited by the current {\it Gaia\/} parallaxes themselves.  However, with the final {\it Gaia\/} data release, we can expect typical (systematics-limited) parallax uncertainties of $\sim$\drtwoparprec\ for stars with $V\la 12$ \citep{Gaia:2016}.  In this regime, the uncertainty in \rstar\ will be dominated by the uncertainties in \fbol\ and \teff.  Fortunately, there are excellent prospects for improving the uncertainties (and accuracy) in \fbol\ even beyond the few-percent precision already provided in this paper.  {\it Gaia\/} DR3 will release low-resolution ($R\sim 10-20$) spectrophotometry covering wavelengths of 0.33--1.05~\micron, and the proposed Explorer mission Spectro-Photometer for the History of the Universe, Epoch of Reionization, and Ices Explorer \citep[SPHEREx;][]{Dore:2016}, should it be selected, will provide low-resolution ($R\sim 40-100$) spectrophotometry from 0.75--5.0~\micron.  Together these would {\it directly} measure $\sim 98\%$ of the bolometric flux of the majority of the $V\la 12$ stars with $\teff\la 8500$K (as well as the extinction as a function of wavelength to these stars from $\sim 0.3-5$~\micron). Ultimately, the precision with which \rstar\ can be measured for these stars will likely be limited by the uncertainty in \teff, which can optimistically be reduced to $\sim$1\% with a combination of high-resolution spectroscopy and SED fitting.  The precision with which the stellar mass, and planetary mass and radius, can then be inferred will then ultimately be limited by the precision of the follow-up photometry of the primary transit, and the radial velocity measurements of the stellar reflex motion due to the planet, which together allow one to estimate \rhostar\ and  the velocity semiamplitude, \rvamp.  These parameters, together with the measurement of \rstar, allow a complete solution of the system, and thus a measurement of \mstar, \mplanet, and \rplanet.  In principle, with sufficient perseverance and strict control systematic errors, these can be reduced to arbitrarily low precisions.  Ultimately, we can expect precisions on \rplanet\ and \mplanet\ of several percent in the best cases \citep{Stevens:2016}.  

We are therefore poised to enter the era of precision exoplanetology, whereby we will be able to accurately measure the host star and planetary masses and radii of bright transiting systems to a precision of, at best, a few percent, likely limited by the total amount of available follow-up resources.  Most importantly, these measurements will be direct, and will not rely on stellar models (e.g., \citealt{Demarque:2004}) or (externally-calibrated) empirical relations (e.g., \citealt{Torres:2010}). Given that the Transiting Exoplanet Survey Satellite (TESS, \citealt{Ricker:2014}) is expected to find $\sim 1700$ transiting planets, including $\sim 600$ with $\rplanet \la 2\rearth$, this will clearly be transformative for our understanding of the physical properties of exoplanets, as well as their formation and evolution.  For example, it will be possible to estimate the masses and radii of essentially all known transiting Hot Jupiters in a consistent, uniform, precise, and accurate manner, thereby potentially allowing for the identification of trends within this population that have been heretofore hidden by large uncertainties, systematics, and/or inhomogeneous analysis methodology. Perhaps even more exciting is the prospect of constraining the properties of low-mass terrestrial planets, given the availability of such precise estimates of their masses, radii, and densities.  This may allow for the identification of terrestrial planets that have bulk heavy element compositions that differ significantly from that of the Earth (e.g., \citealt{Rogers:2010,Unterborn:2016}), and to look for trends of these parameters with the atmospheric composition of the host stars themselves. 

Importantly, these precise estimates of the planetary properties will be anchored to the equally precise estimates of their host star properties.  For stars more massive than the sun, and older than a few Gyr, it will be possible to estimate a robust (albeit model-dependent) age of the star via its (directly-measured) luminosity.  Therefore, it will be possible to better recognize and quantify trends of planet properties with host star mass, radius, luminosity, effective temperature, and age. 

Notably, these host stars will themselves provide stringent tests of stellar evolutionary models, increasing the sample of stars with precise (few percent) and directly-measured masses and radii by nearly an order of magnitude above the largest such samples current available \citep{Torres:2010}.  Furthermore, these stars will sample a much broader range of effective temperatures (particularly at the low \teff\ range), and many will be effectively single, and therefore will not inherit any potential biases in their parameters arising from the formation and evolution of close binaries. 

\kgsins{Already, we have been able to use the empirical stellar masses newly determined here to verify that the majority of putative ``retired A stars" in the sample are indeed more massive than $\sim$1.2~\msun.}

\kgsins{Most importantly,} the fundamental stellar bolometric fluxes and angular radii supplied in this work will help to enable these future improvements for the 498 planet-hosting stars studied here. And the much deeper reach of future {\it Gaia\/} data releases should enable application of the methods laid out here to most if not all of the known planet-hosting stars in the Galaxy.

\acknowledgments 
The authors are grateful to R.~Oelkers for assistance with accessing and using the {\it Gaia\/} DR1 data. 
\kgsins{We also gratefully acknowledge the meticulous and helpful review by the anonymous referee.}
K.G.S.\ acknowledges partial support from NSF PAARE grant AST-1358862. Work by B.S.G.\ was partially supported by NSF CAREER Grant AST-1056524, and by the Jet Propulsion Laboratory and NASA's Exoplanet Exploration Program.  This work has made extensive use of the Filtergraph data visualization service \citep{Burger:2013} at {\tt \url{filtergraph.vanderbilt.edu}}. 
This research has made use of the VizieR catalogue access tool, CDS, Strasbourg, France.
The Exoplanet Archive is funded through NASA's Exoplanet Exploration Program,
administered by the Jet Propulsion Laboratory, California Institute of Technology. 
This work has made use of data from the European Space Agency (ESA)
mission {\it Gaia\/} (\url{http://www.cosmos.esa.int/gaia}), processed by
the {\it Gaia\/} Data Processing and Analysis Consortium (DPAC,
\url{http://www.cosmos.esa.int/web/gaia/dpac/consortium}). Funding
for the DPAC has been provided by national institutions, in particular
the institutions participating in the {\it Gaia\/} Multilateral Agreement.


\clearpage
\startlongtable
\begin{deluxetable}{llcc}
\tablecolumns{4}
\tablewidth{0pt}
\tablecaption{{\it Gaia\/} parallax data used in this study.
\label{tab:gaiadata}}
\tablehead{\colhead{Tycho} & \colhead{Name} & \colhead{$d$ (pc)} & \colhead{$\sigma_d$ (pc)\tablenotemark{a}} }
\startdata
2-1155-1 & WASP-32 b & 232.42 & 34.72 \\
12-104-1 & HD 5319 b & 119.64 & 3.83 \\
30-116-1 & WASP-71 b & 299.15 & 28.71 \\
32-383-1 & HD 10442 b & 129.65 & 4.83 \\
88-57-1 & WASP-82 b & 277.72 & 19.74 \\
90-645-1 & HD 28678 b & 191.09 & 8.98 \\
96-602-1 & GJ 179 b & \nodata & \nodata \\
107-1139-1 & HD 33636 b & 29.44 & 0.71 \\
\enddata
\tablenotetext{a}{Random uncertainty only; does not include systematic uncertainty that is potentially as large as $\sim$300~$\mu$as (see the text).}
\end{deluxetable}
\begin{longrotatetable}
\begin{deluxetable*}{llccccclcccccccccccc}
\tablecolumns{20}
\tablewidth{0pt}
\tabletypesize{\scriptsize}
\tablecaption{Sample of transiting planets used in this study.
\label{tab:trsample}}
\tablehead{
\colhead{Tycho} & \colhead{Name} & \colhead{\teff} & \colhead{$\sigma_{\teff}$} & \colhead{\logg} & \colhead{$\sigma_{\logg}$} & \colhead{[Fe/H]} & \colhead{Source\tablenotemark{a}} & \colhead{$P_{\rm orb}$} & \colhead{$\sigma_{P_{\rm orb}}$} & \colhead{$e$} & \colhead{$\sigma_e$} & \colhead{\rvamp} & \colhead{$\sigma_{\rvamp}$} & \colhead{$i$} & \colhead{$\sigma_i$} & \colhead{\depth} & \colhead{$\sigma_{\depth}$} & \colhead{\ar} & \colhead{$\sigma_{\ar}$} \\
\colhead{}      & \colhead{}     & \colhead{K} & \colhead{K} & \colhead{[cgs]} & \colhead{[cgs]} & \colhead{} & \colhead{} & \colhead{d} & \colhead{d} & \colhead{} & \colhead{} & \colhead{m~s$^{-1}$} & \colhead{m~s$^{-1}$} & \colhead{deg.} & \colhead{deg.} & \colhead{} & \colhead{} & \colhead{AU/\rsun} & \colhead{AU/\rsun} } 
\startdata
2-1155-1 & WASP-32 b & 6140 & 95 & 4.40 & 0.02 & -0.13 & EXO & 2.71866 & 2.0e-06 & 0.00 & \nodata & 478.0 & 11.0 & 85.10 & 0.20 & 1.1e-02 & 1.0e-04 & 7.72 & 0.25 \\ 
30-116-1 & WASP-71 b & 6050 & 100 & 3.90 & 0.05 & 0.15 & EXO & 2.90367 & 6.8e-06 & 0.00 & \nodata & 236.4 & 3.7 & 84.20 & 1.80 & 4.4e-03 & 1.5e-04 & 4.30 & 0.27 \\ 
88-57-1 & WASP-82 b & 6480 & 90 & 3.96 & 0.03 & 0.12 & EXO & 2.70579 & 2.0e-06 & 0.00 & \nodata & 130.7 & 1.9 & 87.90 & 1.65 & 6.3e-03 & 1.0e-04 & 4.43 & 0.15 \\ 
203-1079-1 & HAT-P-35 b & 6096 & 88 & 4.21 & 0.04 & 0.11 & EXO & 3.64671 & 2.1e-05 & 0.03 & 0.02 & 120.7 & 2.2 & 87.30 & 1.00 & 9.1e-03 & 5.2e-04 & 7.48 & 0.45 \\ 
208-722-1 & HAT-P-30 b & 6304 & 88 & 4.36 & 0.30 & 0.13 & EXO & 2.81060 & 5.0e-06 & 0.04 & 0.02 & 88.1 & 3.3 & 83.60 & 0.04 & 1.3e-02 & 4.5e-04 & 7.43 & 0.34 \\ 
211-706-1 & WASP-84 b & 5300 & 100 & 4.40 & 0.10 & 0.00 & EXO & 8.52349 & 7.0e-06 & 0.00 & \nodata & 77.4 & 2.0 & 88.37 & 0.05 & 1.7e-02 & 1.5e-04 & 22.22 & 0.58 \\ 
320-1027-1 & HAT-P-26 b & 5079 & 88 & 4.56 & 0.06 & -0.04 & EXO & 4.23452 & 1.5e-05 & 0.12 & 0.06 & 8.5 & 1.0 & 88.60 & 0.70 & 5.4e-03 & 1.3e-05 & 13.09 & 1.19 \\ 
339-329-1 & WASP-24 b & 6075 & 100 & 4.26 & 0.17 & 0.00 & EXO & 2.34121 & 2.0e-06 & 0.00 & \nodata & 148.2 & 2.5 & 83.64 & 0.29 & 1.0e-02 & 1.2e-04 & 5.91 & 0.17 \\
\enddata
\tablenotetext{a}{Source for spectroscopic stellar parameters: EXO = {\tt exoplanets.org}; PASTEL = \citet{Paunzen:2015}; SWEET = \citet{Santos:2013}; Ammons = \citet{Ammons:2006}; McDonald:2012 = \citet{McDonald:2012}; Wright:2003 = \citet{Wright:2003}.}
\end{deluxetable*}

\end{longrotatetable}
\begin{longrotatetable}
\begin{deluxetable*}{llccccclcccccccc}
\tablecolumns{16}
\tablewidth{0pt}
\tabletypesize{\scriptsize}
\tablecaption{Sample of radial-velocity (non-transiting) planets used in this study.
\label{tab:rvsample}}
\tablehead{
\colhead{Tycho} & \colhead{Name} & \colhead{\teff} & \colhead{$\sigma_{\teff}$} & \colhead{\logg} & \colhead{$\sigma_{\logg}$} & \colhead{[Fe/H]} & \colhead{Source\tablenotemark{a}} & \colhead{$P_{\rm orb}$} & \colhead{$\sigma_{P_{\rm orb}}$} & \colhead{$e$} & \colhead{$\sigma_e$} & \colhead{\rvamp} & \colhead{$\sigma_{\rvamp}$} & \colhead{$a$} & \colhead{$\sigma_a$} \\
\colhead{}      & \colhead{}     & \colhead{K} & \colhead{K} & \colhead{[cgs]} & \colhead{[cgs]} & \colhead{} & \colhead{} & \colhead{d} & \colhead{d} & \colhead{} & \colhead{} & \colhead{m~s$^{-1}$} & \colhead{m~s$^{-1}$} & \colhead{AU} & \colhead{AU} } 
\startdata
12-104-1 & HD 5319 b & 4900 & 25 & 3.35 & 0.09 & 0.05 & PASTEL & 641.00000 & 2.0e+00 & 0.02 & 0.03 & 31.6 & 1.2 & 1.67 & 0.04 \\ 
32-383-1 & HD 10442 b & 5034 & 44 & 3.50 & 0.06 & 0.11 & EXO & 1043.00000 & 9.0e+00 & 0.11 & 0.06 & 31.5 & 2.2 & 2.34 & 0.05 \\ 
90-645-1 & HD 28678 b & 4722 & 100 & 3.25 & 0.04 & -0.16 & McDonald:2012 & 387.10001 & 3.4e+00 & 0.17 & 0.07 & 33.5 & 2.2 & 1.32 & 0.04 \\ 
96-602-1 & GJ 179 b & 3370 & 100 & 4.83 & \nodata & 0.30 & EXO & 2288.00000 & 5.9e+01 & 0.21 & 0.08 & 25.8 & 2.2 & 2.41 & 0.08 \\ 
107-1139-1 & HD 33636 b & 5964 & 22 & 4.56 & 0.04 & -0.08 & PASTEL & 2127.69995 & 8.2e+00 & 0.48 & 0.01 & 164.2 & 2.0 & 3.27 & 0.06 \\ 
112-182-1 & HD 34445 b & 5879 & 16 & 4.21 & 0.03 & 0.18 & PASTEL & 1049.00000 & 1.1e+01 & 0.27 & 0.07 & 15.7 & 1.4 & 2.07 & 0.04 \\ 
116-1316-1 & HD 38529 b & 5610 & 17 & 3.89 & 0.03 & 0.35 & PASTEL & 14.31019 & 8.1e-04 & 0.24 & 0.03 & 57.0 & 1.2 & 0.13 & 0.00 \\ 
127-402-1 & HD 37605 b & 5340 & 36 & 4.32 & 0.04 & 0.29 & PASTEL & 55.01307 & 6.4e-04 & 0.68 & 0.00 & 203.0 & 0.7 & 0.28 & 0.05 \\
\enddata
\tablenotetext{a}{Source for spectroscopic stellar parameters: EXO = {\tt exoplanets.org}; PASTEL = \citet{Paunzen:2015}; SWEET = \citet{Santos:2013}; Ammons = \citet{Ammons:2006}; McDonald:2012 = \citet{McDonald:2012}; Wright:2003 = \citet{Wright:2003}.}
\end{deluxetable*}

\end{longrotatetable}
\begin{longrotatetable}
\movetabledown=0.5in
\begin{deluxetable*}{llccccccccccccccccc}
\tablecolumns{19}
\tablewidth{0pt}
\tabletypesize{\scriptsize}
\tablecaption{Results.\label{tab:results}}
\tablehead{
\colhead{Tycho} & \colhead{Name} & \colhead{$\chi_\nu^2$\tablenotemark{a}} & \colhead{$A_V$} & \colhead{$\sigma_{A_V}$} & \colhead{\fbol} & \colhead{$\sigma_{\fbol}$} & \colhead{$\Theta$} & \colhead{$\sigma_\Theta$} & \colhead{\rstar} & \colhead{$\sigma_{\rstar}$} & \colhead{$\rho_\star$} & \colhead{$\sigma_{\rho_\star}$} & \colhead{\mstar\tablenotemark{b}} & \colhead{$\sigma_{\mstar}$} & \colhead{$R_p$} & \colhead{$\sigma_{R_p}$} & \colhead{$M_p$\tablenotemark{c}} & \colhead{$\sigma_{M_p}$} \\
\colhead{} & \colhead{} & \colhead{} & \colhead{mag.} & \colhead{mag.} & \colhead{erg s$^{-1}$ cm$^{-2}$} & \colhead{erg s$^{-1}$ cm$^{-2}$} & \colhead{$\mu$as} & \colhead{$\mu$as} & \colhead{\rsun} & \colhead{\rsun} & \colhead{g~cm$^{-3}$} & \colhead{g~cm$^{-3}$} & \colhead{\msun} & \colhead{\msun} & \colhead{$R_{\rm Jup}$} & \colhead{$R_{\rm Jup}$} & \colhead{$M_{\rm Jup}$} & \colhead{$M_{\rm Jup}$}
}
\startdata
2-1155-1 & WASP-32 b & 0.58 & 0.06 & 0.03 & 6.85e-10 & 1.54e-11 & 19.02 & 0.63 & 0.95 & 0.15 & 1.18 & 0.11 & 0.72 & 0.34 & 0.96 & 0.15 & 2.63 & 0.82 \\
12-104-1 & HD 5319 b & 3.03 & 0.06 & 0.01 & 2.04e-08 & 4.24e-10 & 163.12 & 2.37 & 4.19 & 0.15 & \nodata & \nodata & 1.44 & 0.31 & \nodata & \nodata & 1.71 & 0.26 \\
30-116-1 & WASP-71 b & 0.45 & 0.00 & 0.00 & 1.43e-09 & 2.26e-11 & 28.28 & 0.96 & 1.82 & 0.19 & 0.18 & 0.03 & 0.76 & 0.27 & 1.18 & 0.12 & 1.39 & 0.33 \\
32-383-1 & HD 10442 b & 2.31 & 0.06 & 0.00 & 2.47e-08 & 5.31e-10 & 169.77 & 3.48 & 4.73 & 0.20 & \nodata & \nodata & 2.58 & 0.42 & \nodata & \nodata & 2.94 & 0.38 \\
88-57-1 & WASP-82 b & 0.62 & 0.13 & 0.05 & 2.92e-09 & 6.08e-11 & 35.22 & 1.05 & 2.10 & 0.16 & 0.22 & 0.02 & 1.48 & 0.37 & 1.62 & 0.13 & 1.17 & 0.20 \\
90-645-1 & HD 28678 b & 1.41 & 0.18 & 0.01 & 1.91e-08 & 6.19e-10 & 169.89 & 7.72 & 6.98 & 0.46 & \nodata & \nodata & 3.16 & 0.51 & \nodata & \nodata & 2.55 & 0.32 \\
96-602-1 & GJ 179 b & 1.04 & 0.00 & 0.00 & 3.44e-09 & 9.64e-11 & 141.42 & 8.62 & \nodata & \nodata & \nodata & \nodata & \nodata & \nodata & \nodata & \nodata & \nodata & \nodata \\
107-1139-1 & HD 33636 b & 0.65 & 0.00 & 0.00 & 4.28e-08 & 4.02e-10 & 159.28 & 1.41 & 1.01 & 0.03 & \nodata & \nodata & 1.34 & 0.14 & \nodata & \nodata & 11.09 & 0.81 \\
\enddata
\tablenotetext{a}{Reduced $\chi^2$ of the SED fit.}
\tablenotetext{b}{For transiting planet hosts, \mstar\ is calculated from the direct observables $\rho_\star$ and \rstar. For RV planet hosts, \mstar\ is calculated from the direct observables \logg\ and \rstar.}
\tablenotetext{c}{For RV planets, $M_p$ is modulo $\sin i$.}
\end{deluxetable*}

\end{longrotatetable}

\clearpage


\appendix 

\section{Spectral Energy Distribution Measurements and Fits for Planet Host Sample\label{sec:sed_appendix}}
\begin{figure}[!ht]
\centering
\includegraphics[trim=70 70 70 50,clip,width=\linewidth]{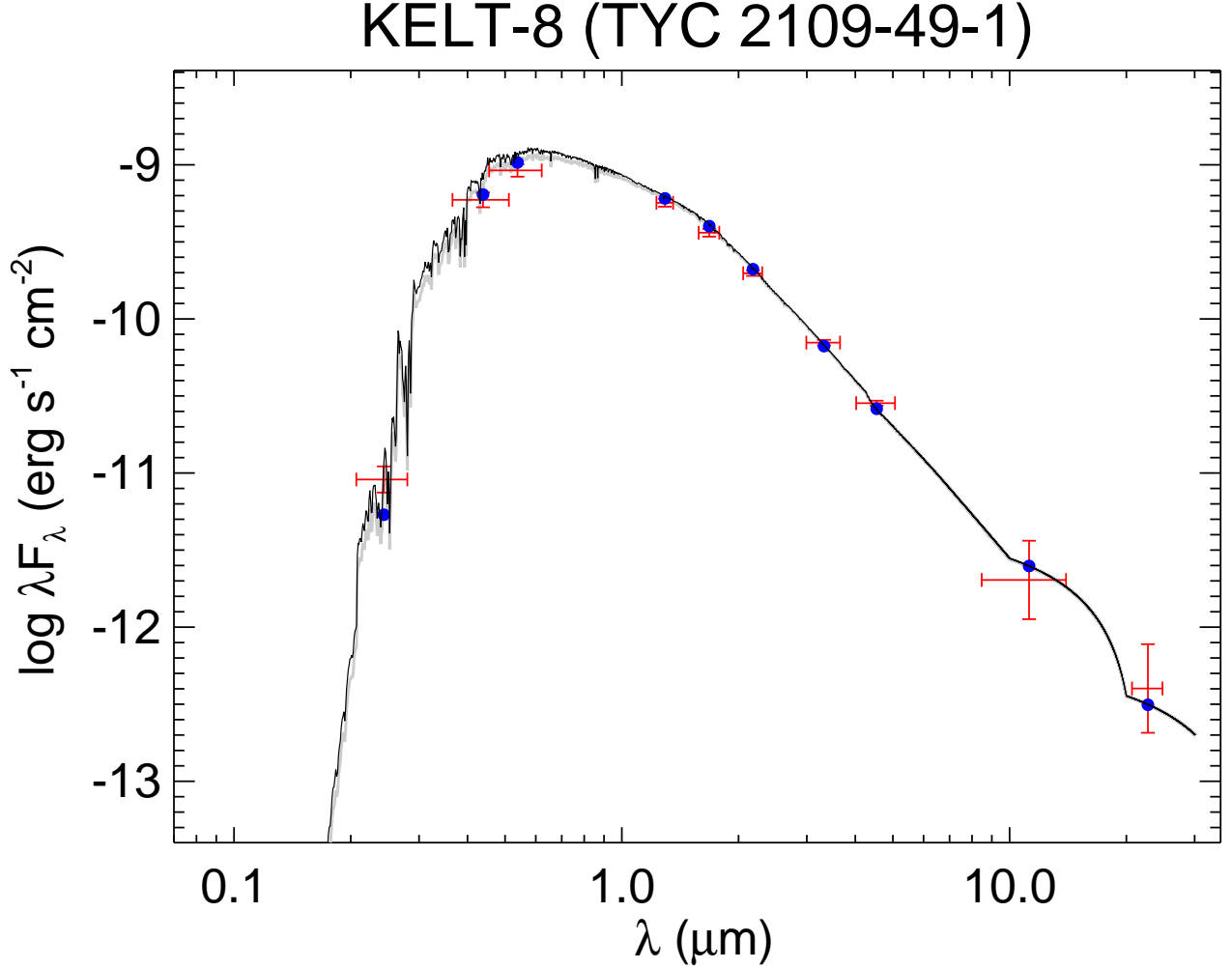}
\caption{KELT-8 is shown as an example of the figure set. Each panel in the Figure Set is labeled at top by the {\it Tycho-2} ID and common name of the star, and shows the observed fluxes (in units of erg cm$^{-2}$ s$^{-1}$) versus wavelength (in \micron) as red error bars, where the vertical error bar represents the uncertainty in the measurement and the horizontal ``error" bar represents the effective width of the passband. Also in each figure is the fitted SED model including extinction (light gray curve), on which is shown the model passband fluxes as blue dots. 
The corresponding un-extincted SED model is also shown (dark black curve); the reported \fbol\ is the sum over all wavelengths of this un-extincted model (see the text). The full figure set is displayed in Figures \ref{fig:seds_1}--\ref{fig:seds_83}.}
\label{fig:seds}
\end{figure}

In Figure Set \ref{fig:seds} we present the observed and fitted spectral energy distributions of the 
498 planet host stars for which we were able to perform our SED fitting procedures, which includes the 
\ntothosts\ planet host stars in our final study sample having {\it Gaia\/} DR1 parallaxes (Tables~\ref{tab:trsample} and \ref{tab:rvsample}). 

\clearpage
\begin{figure}[H]
  \centering
  \includegraphics[trim=60 60 60 60,clip,width=0.49\linewidth]{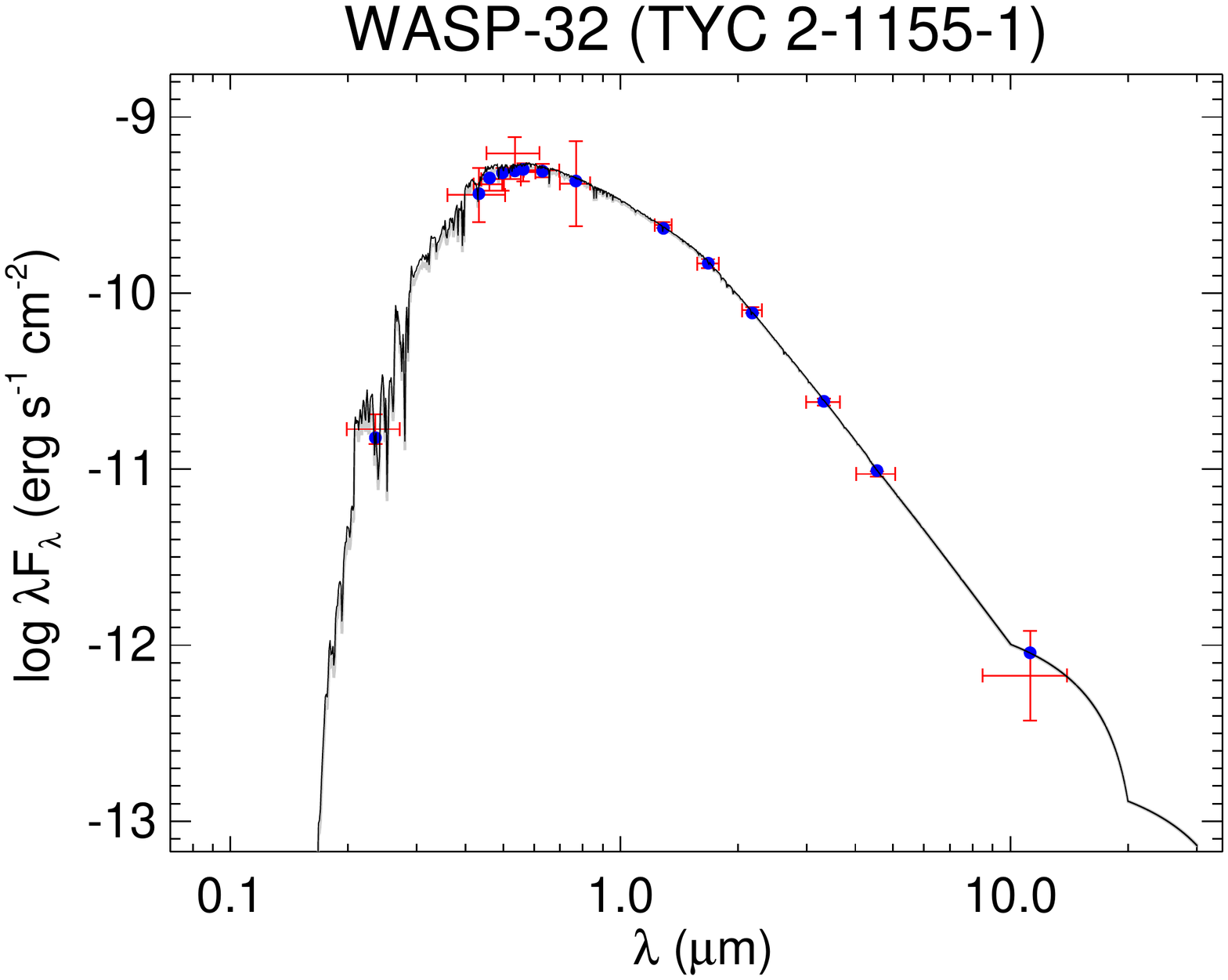}
  \includegraphics[trim=60 60 60 60,clip,width=0.49\linewidth]{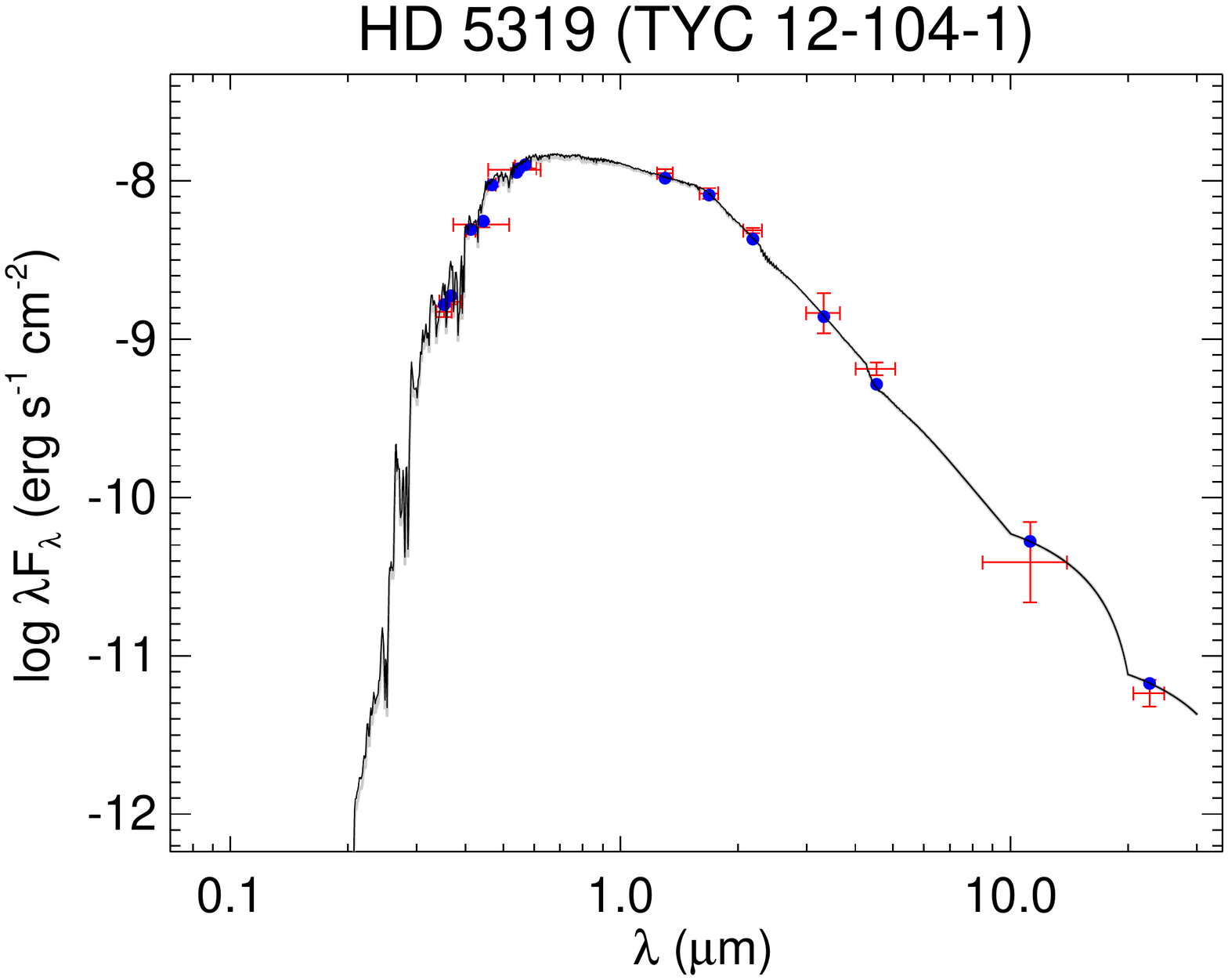}
  \includegraphics[trim=60 60 60 60,clip,width=0.49\linewidth]{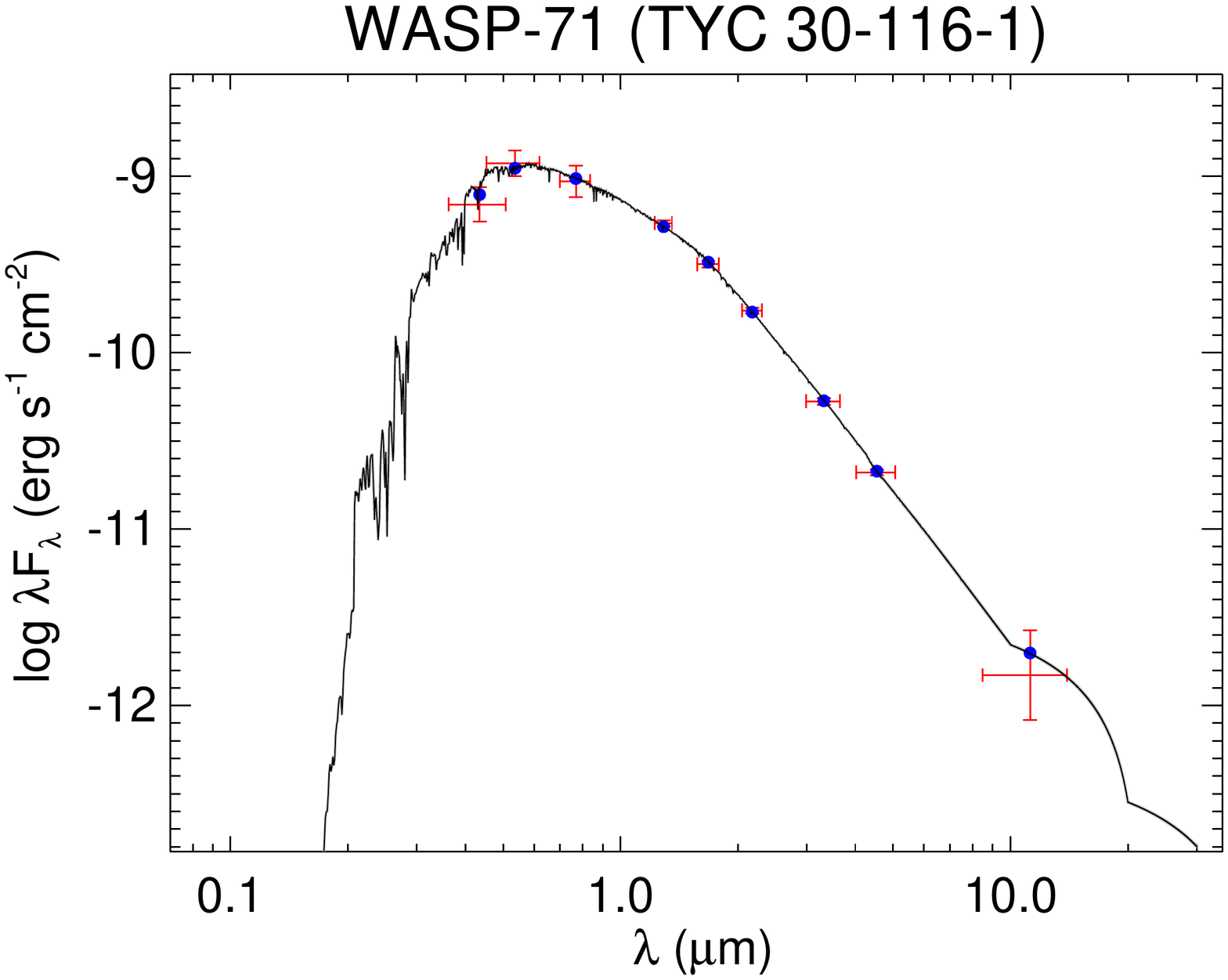}
  \includegraphics[trim=60 60 60 60,clip,width=0.49\linewidth]{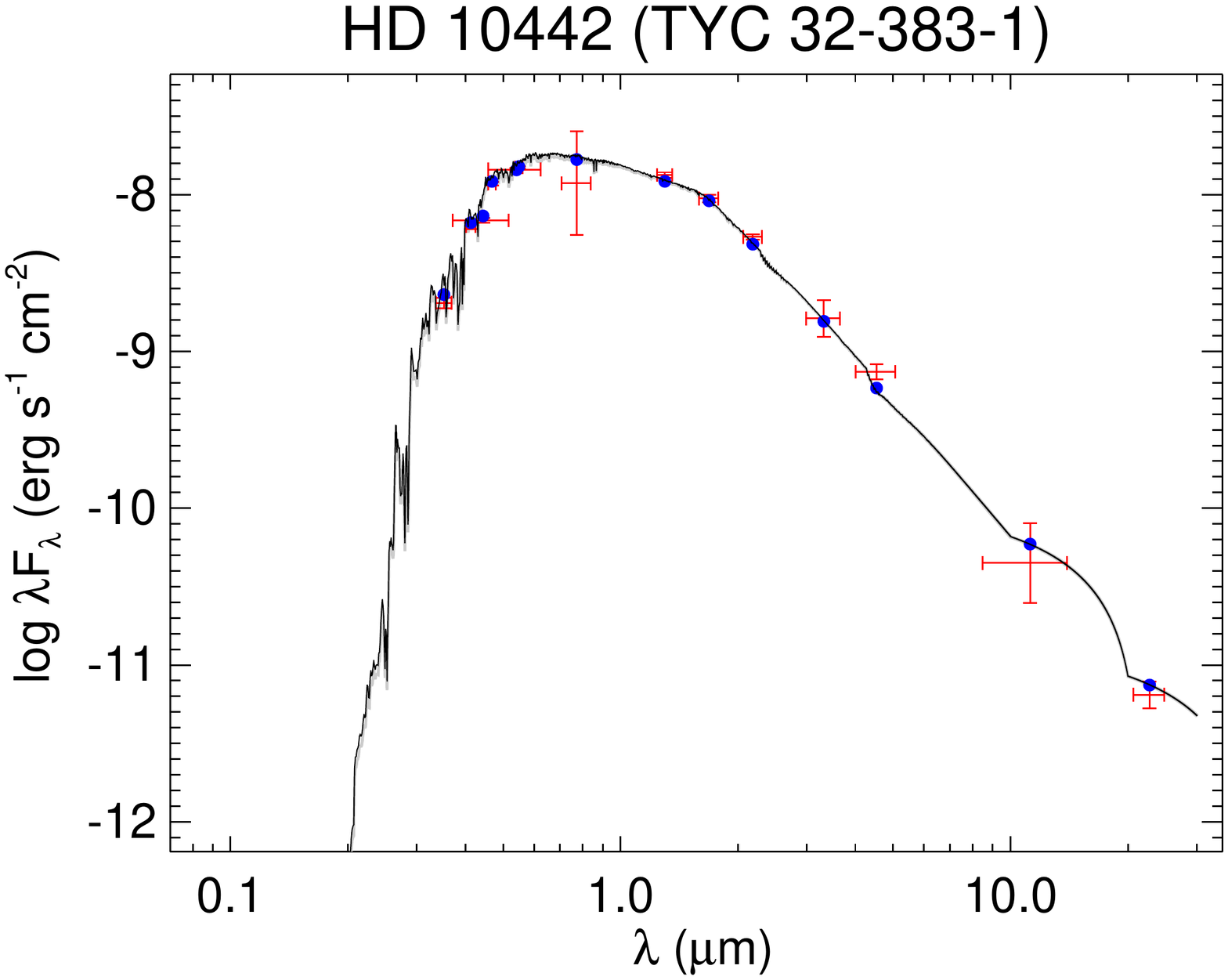}
  \includegraphics[trim=60 60 60 60,clip,width=0.49\linewidth]{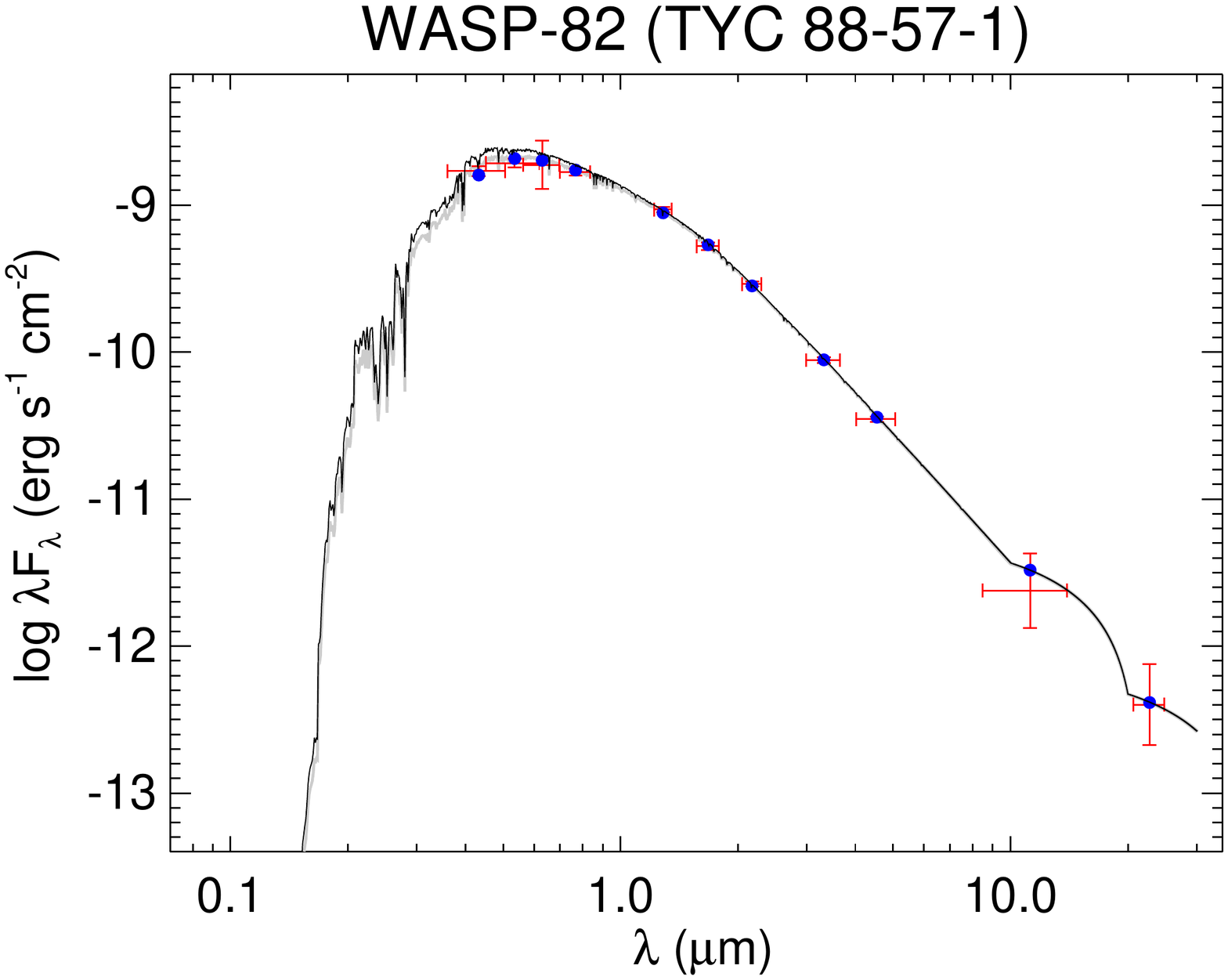}
  \includegraphics[trim=60 60 60 60,clip,width=0.49\linewidth]{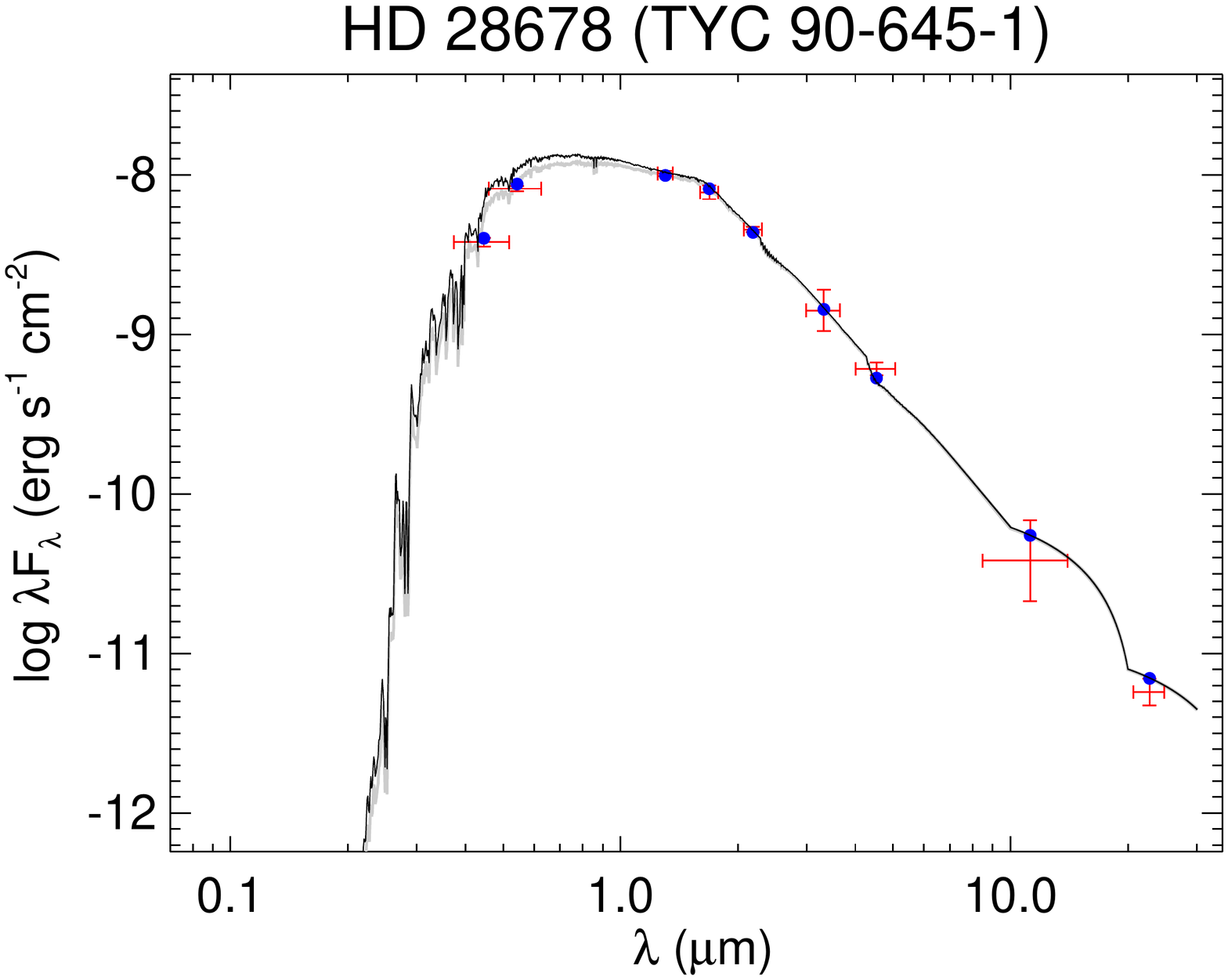}
  \caption{All labels, lines, symbols, and colors as in Figure \ref{fig:seds}.}
  \label{fig:seds_1}
\end{figure}

\begin{figure}[H]
  \centering
  \includegraphics[trim=60 60 60 60,clip,width=0.49\linewidth]{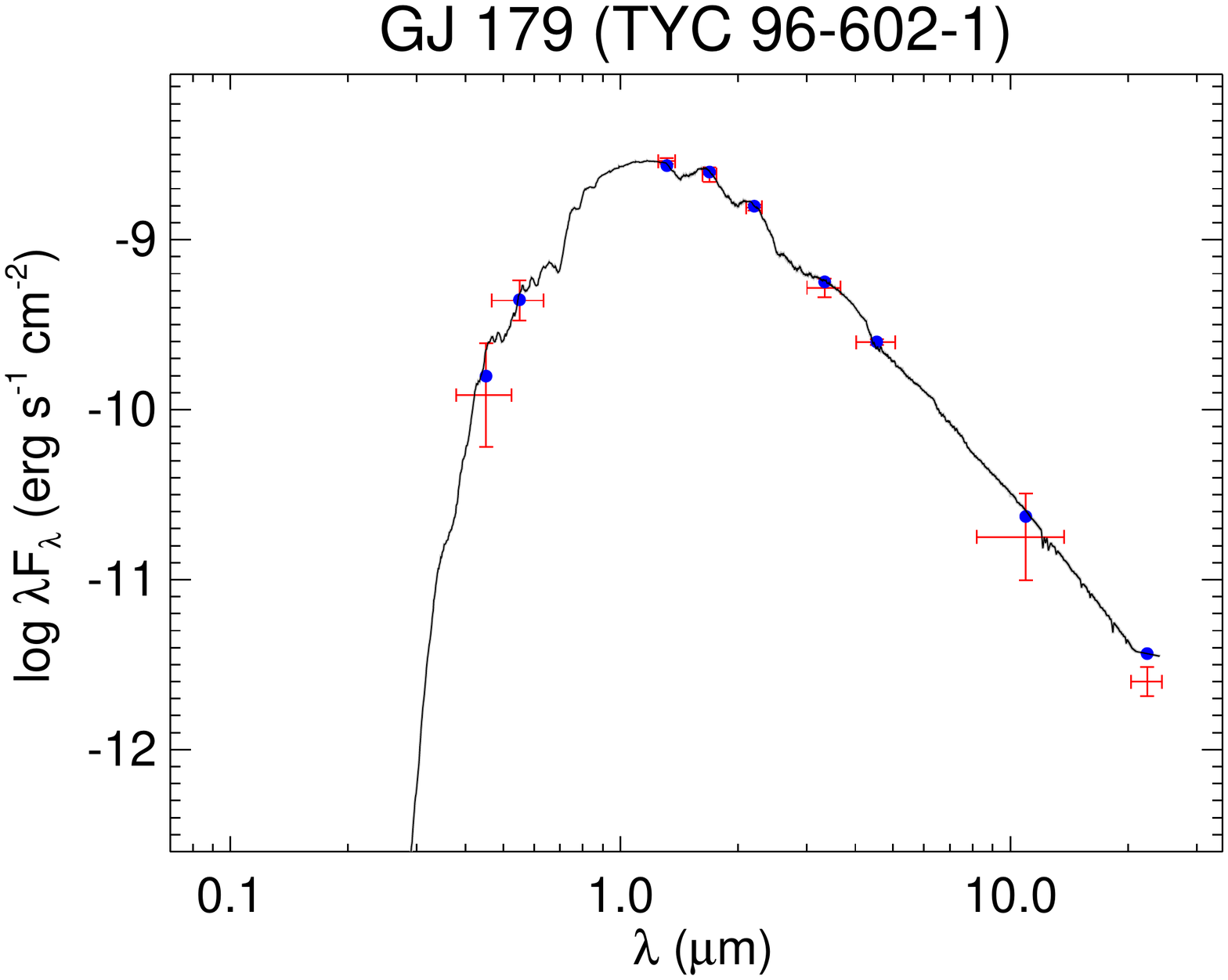}
  \includegraphics[trim=60 60 60 60,clip,width=0.49\linewidth]{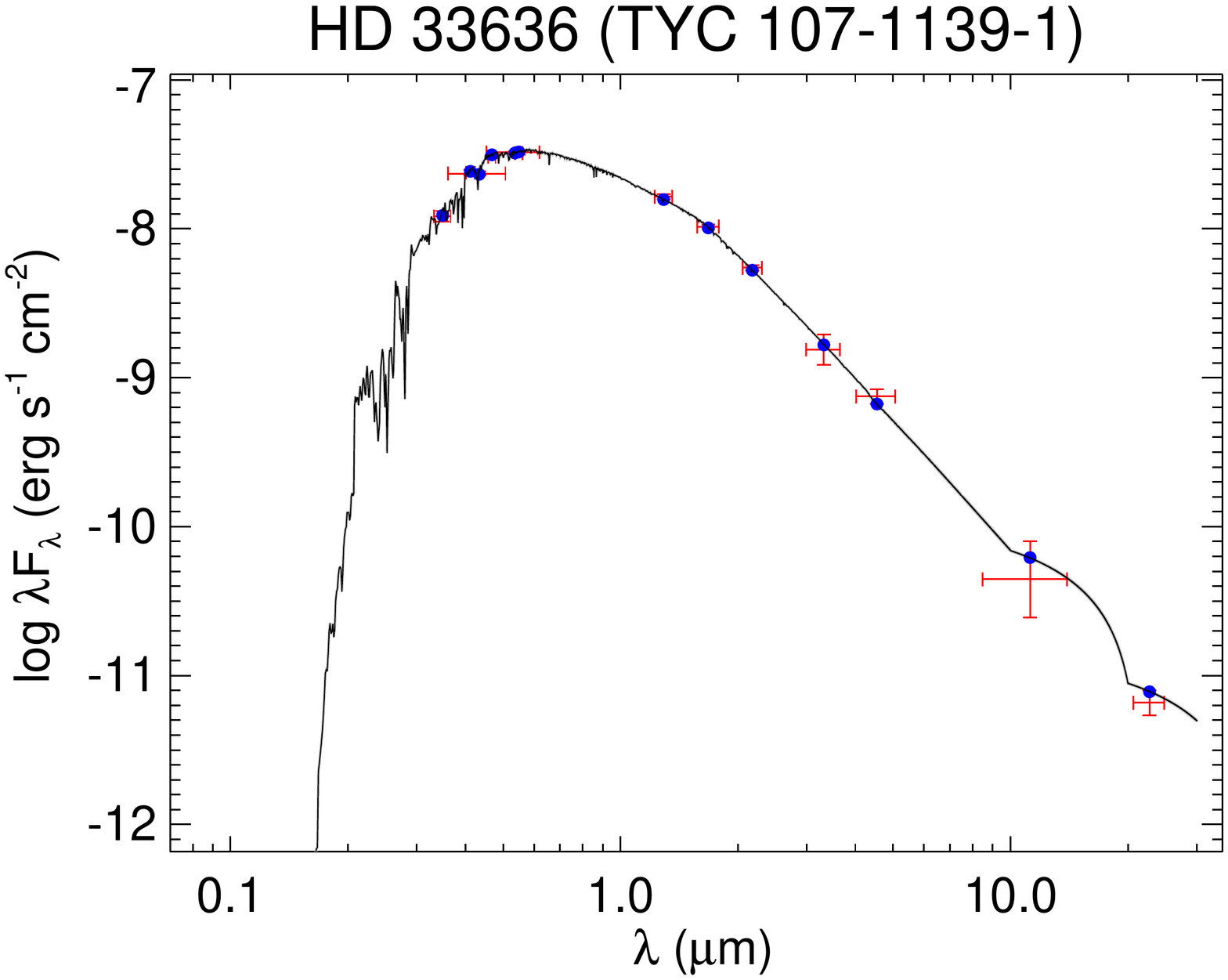}
  \includegraphics[trim=60 60 60 60,clip,width=0.49\linewidth]{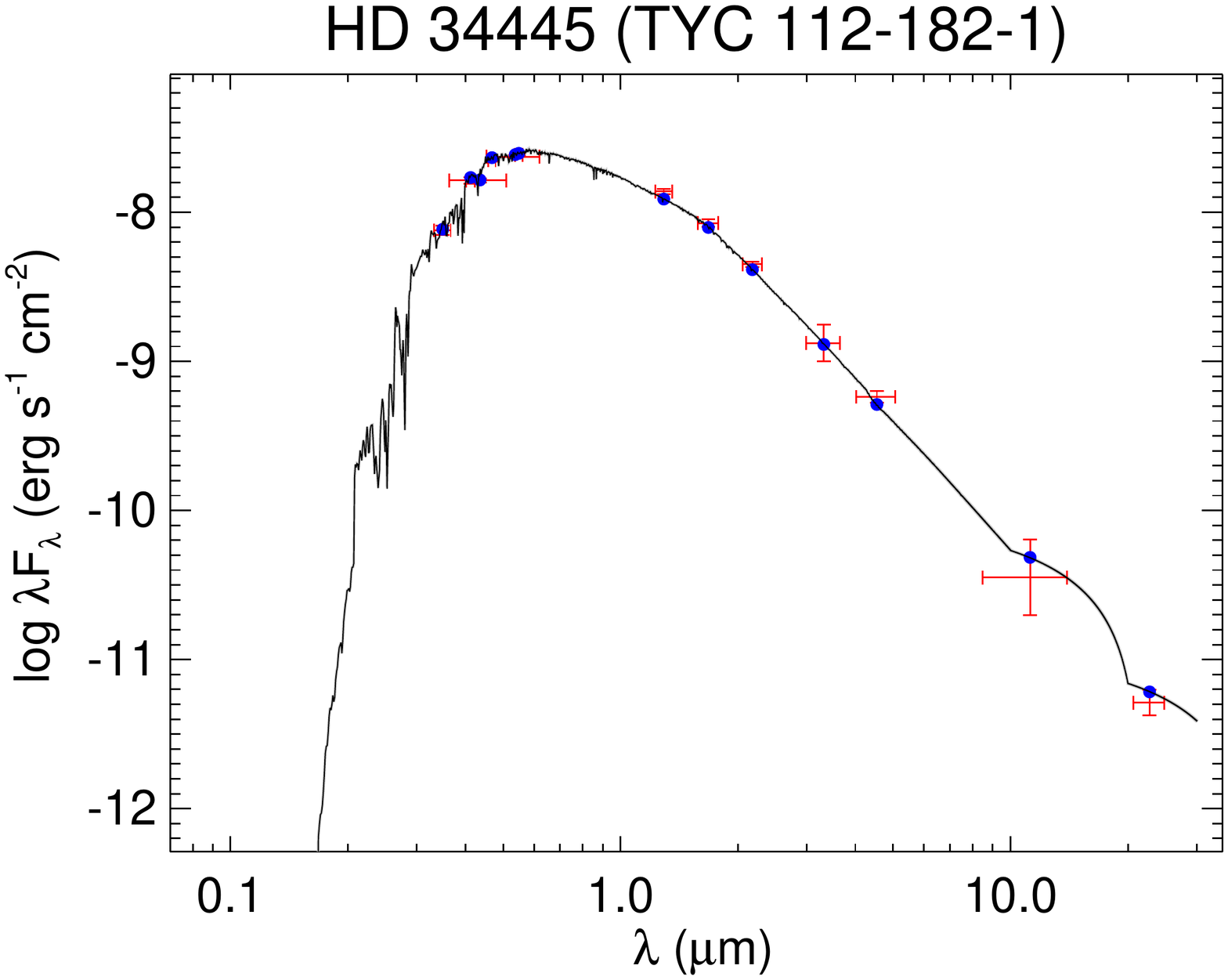}
  \includegraphics[trim=60 60 60 60,clip,width=0.49\linewidth]{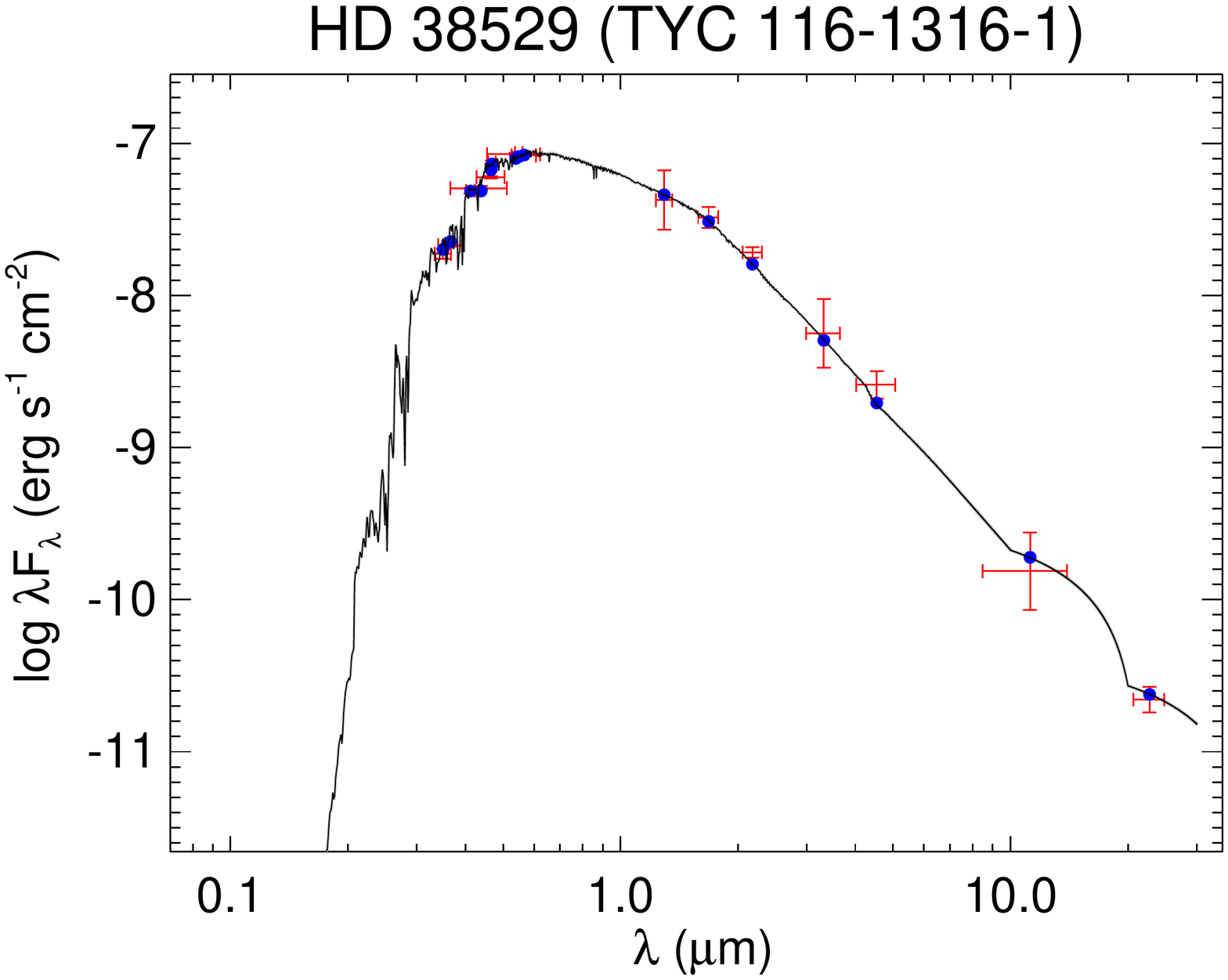}
  \includegraphics[trim=60 60 60 60,clip,width=0.49\linewidth]{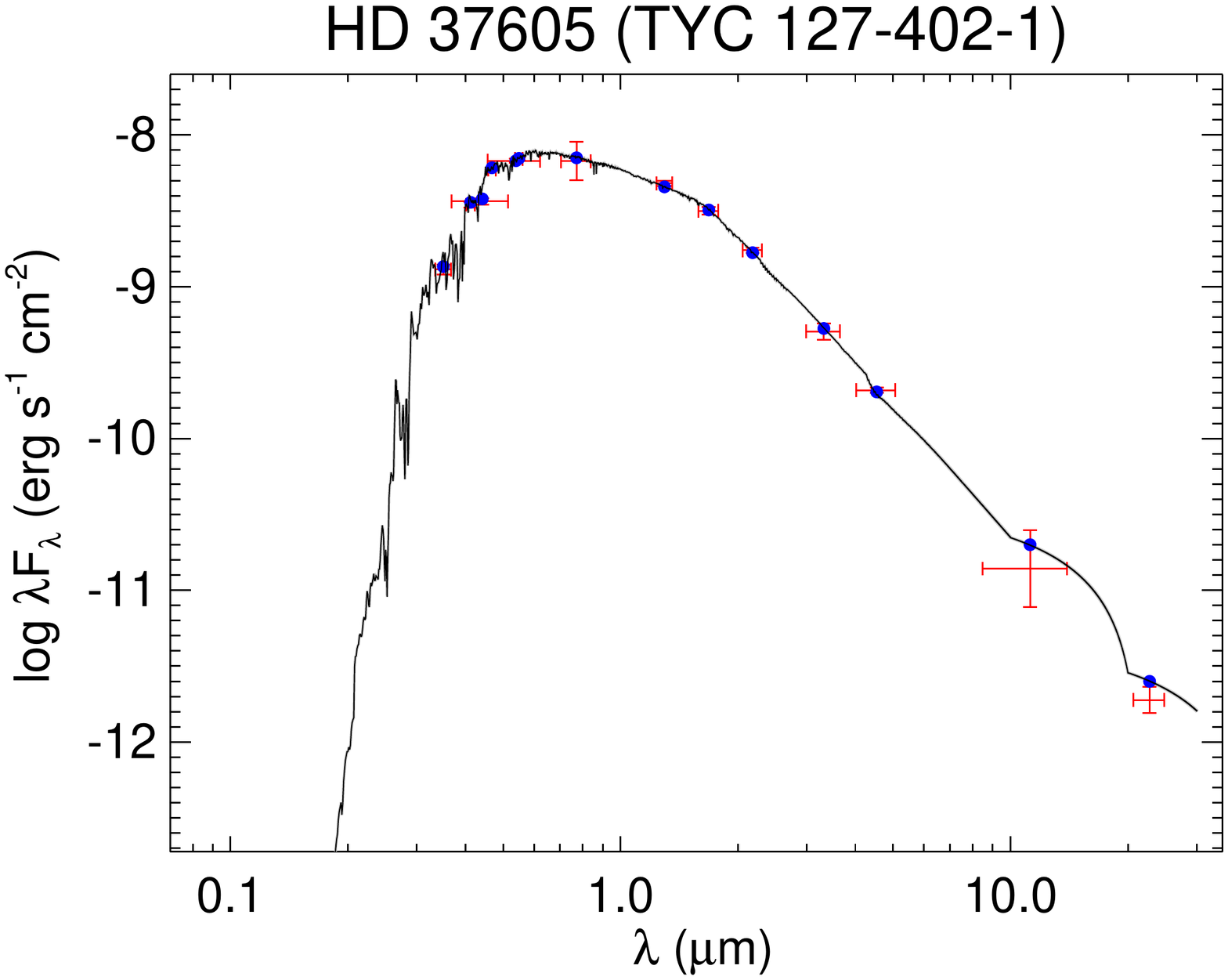}
  \includegraphics[trim=60 60 60 60,clip,width=0.49\linewidth]{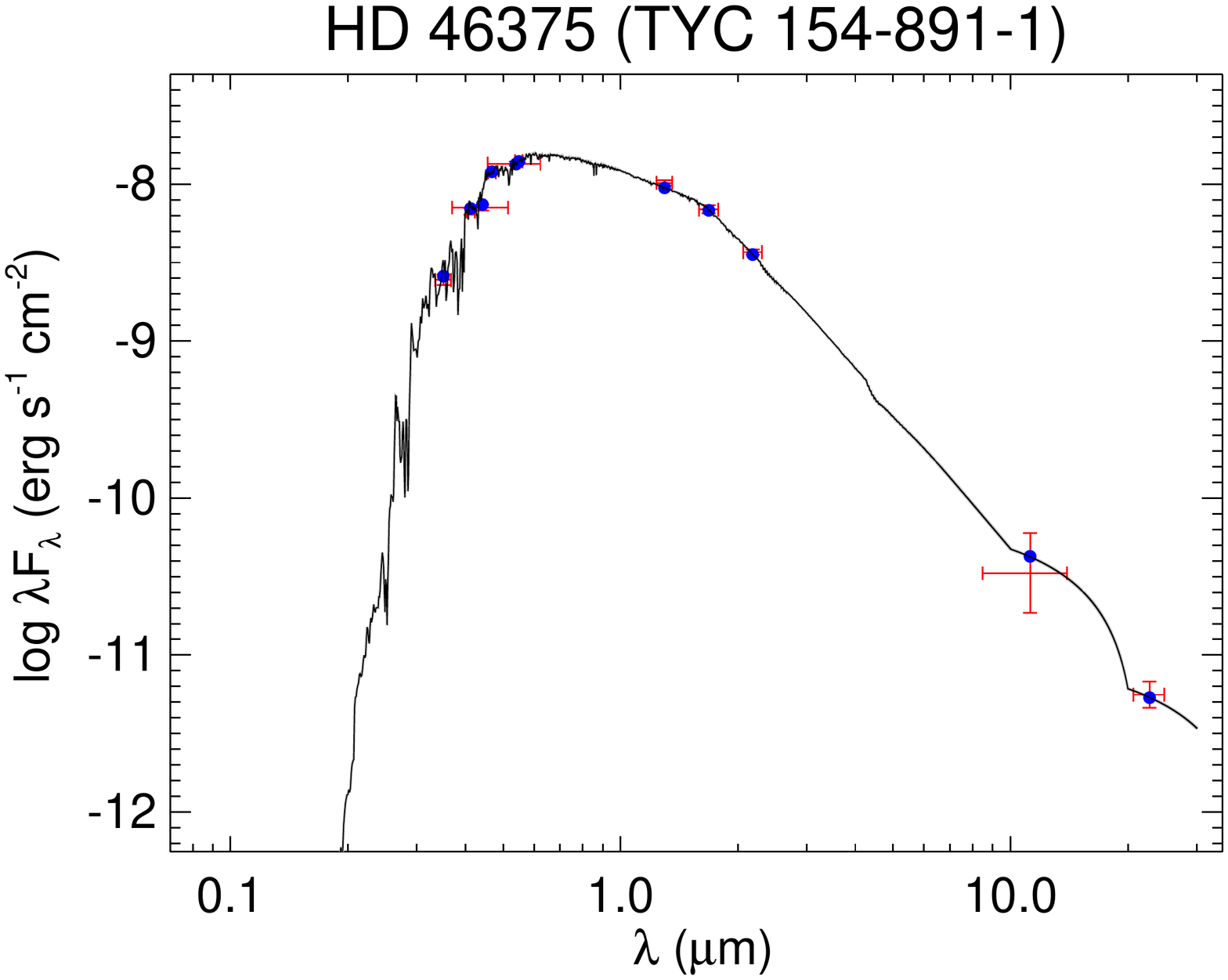}
  \caption{All labels, lines, symbols, and colors as in Figure \ref{fig:seds}.}
  \label{fig:seds_2}
\end{figure}

\begin{figure}[H]
  \centering
  \includegraphics[trim=60 60 60 60,clip,width=0.49\linewidth]{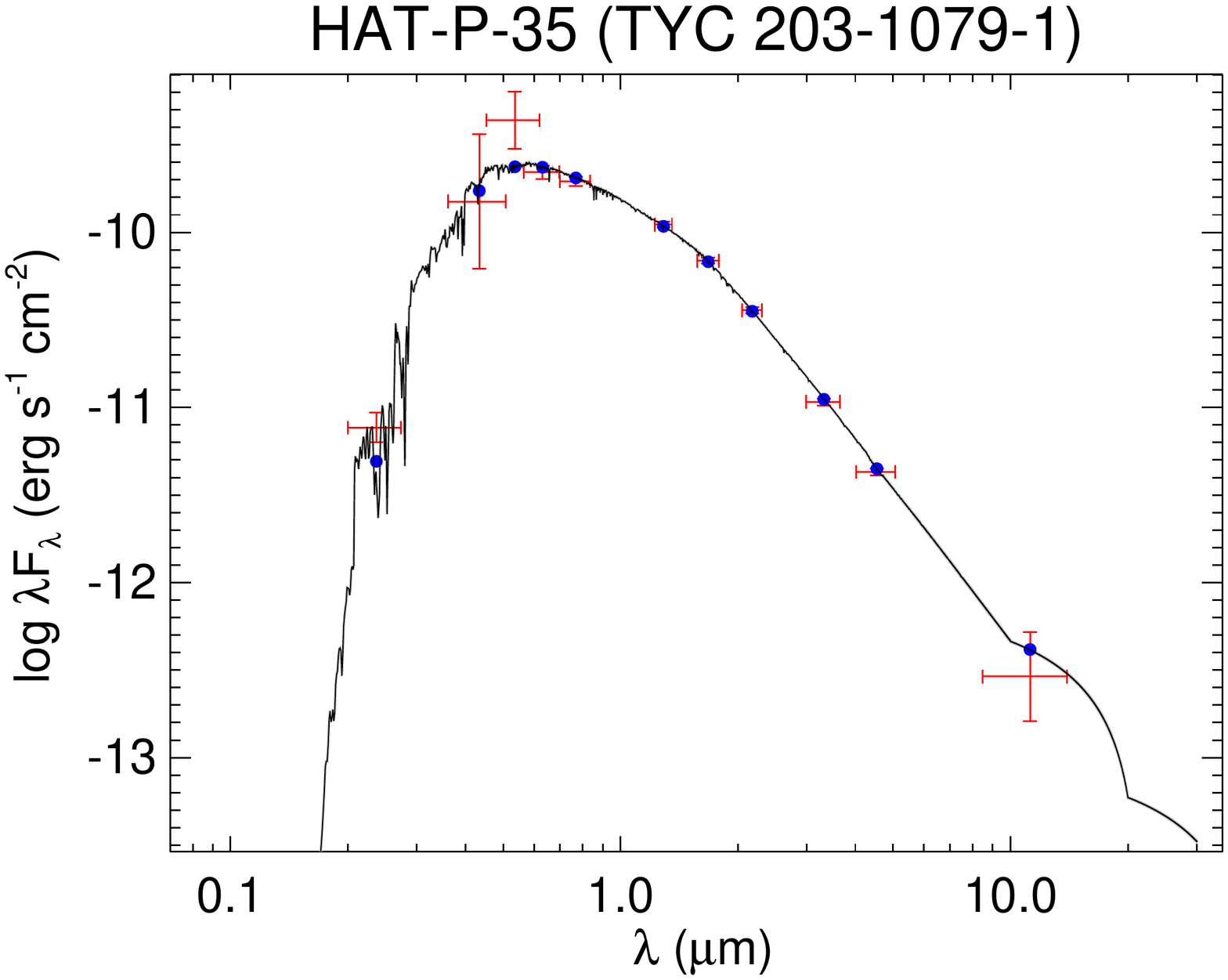}
  \includegraphics[trim=60 60 60 60,clip,width=0.49\linewidth]{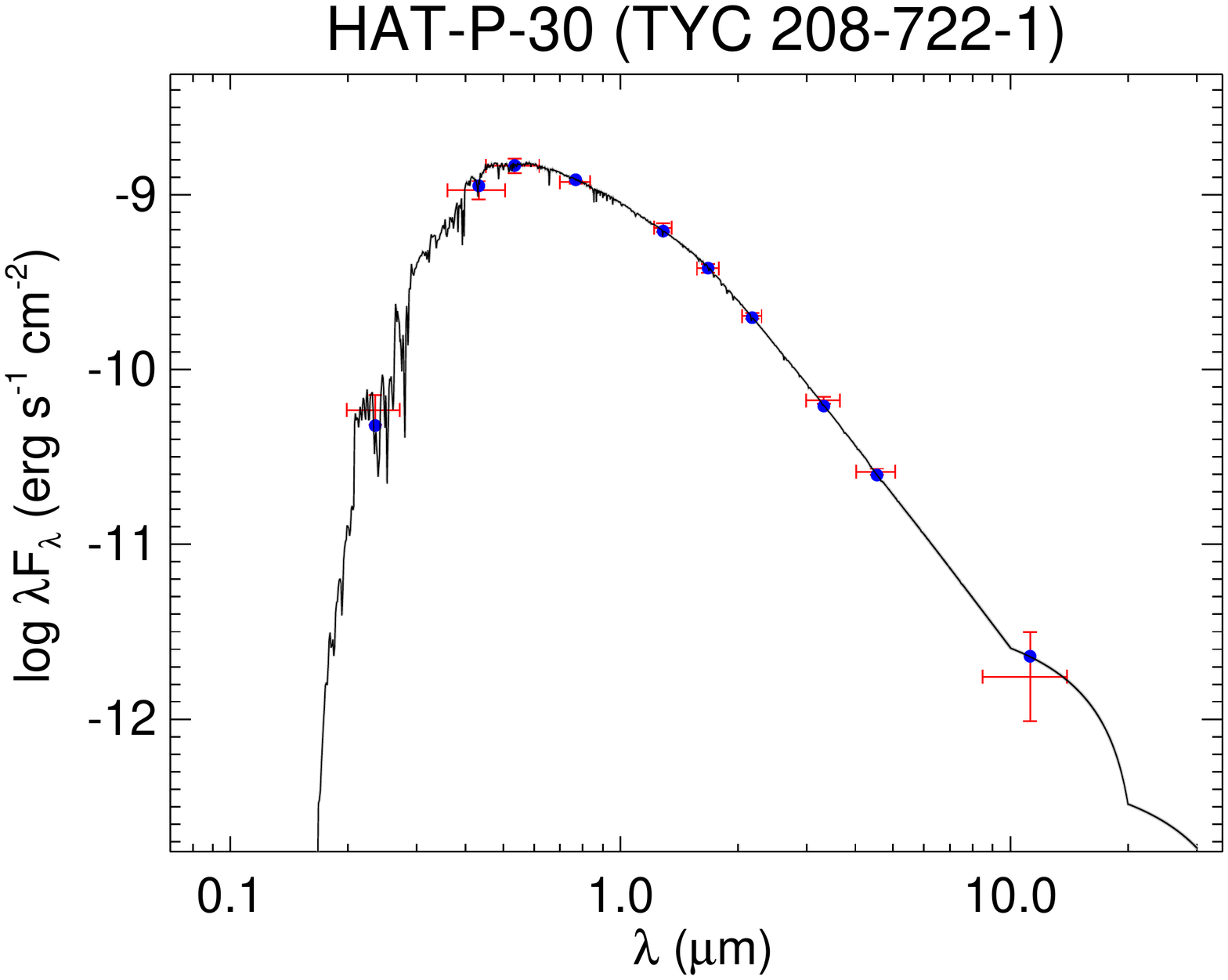}
  \includegraphics[trim=60 60 60 60,clip,width=0.49\linewidth]{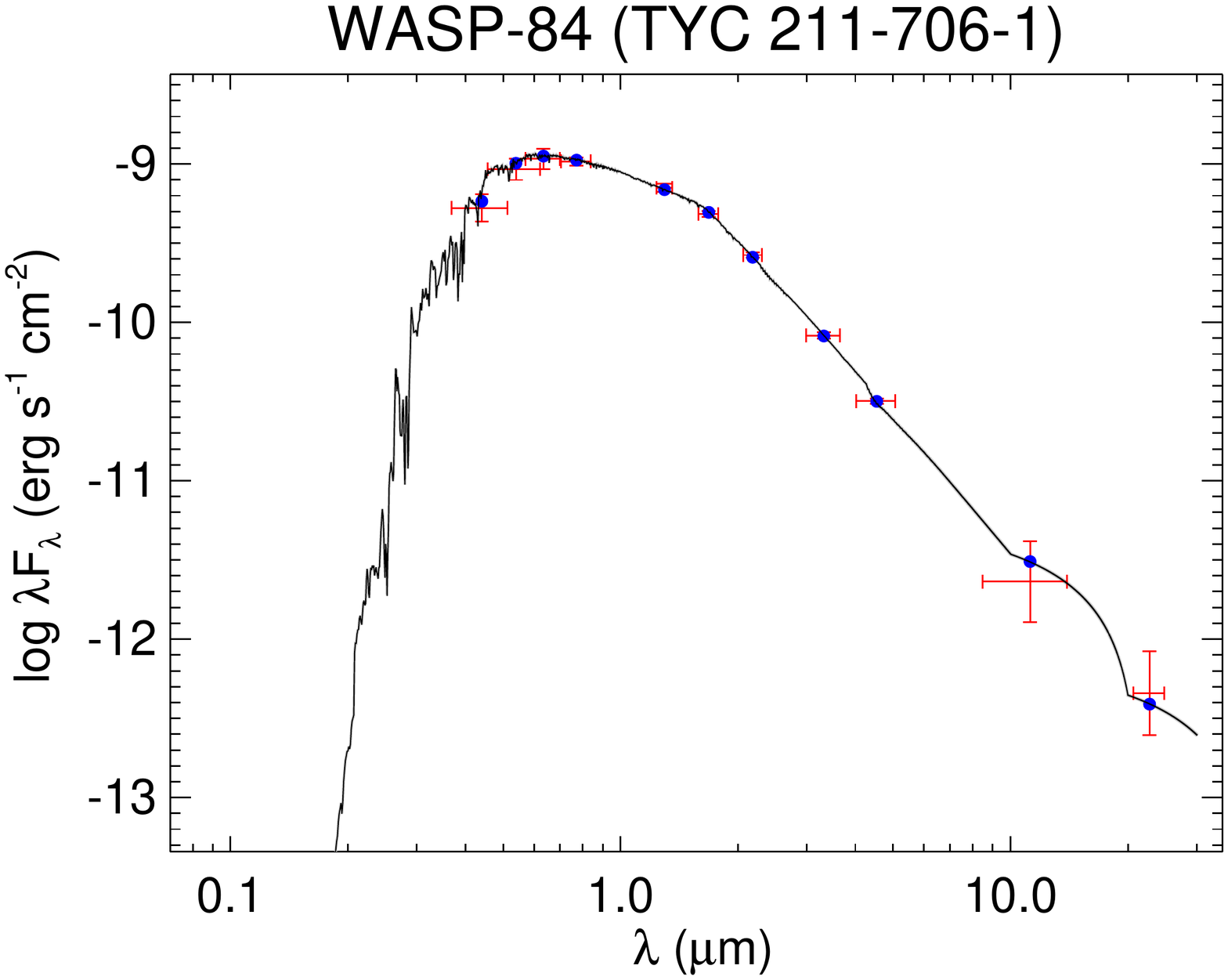}
  \includegraphics[trim=60 60 60 60,clip,width=0.49\linewidth]{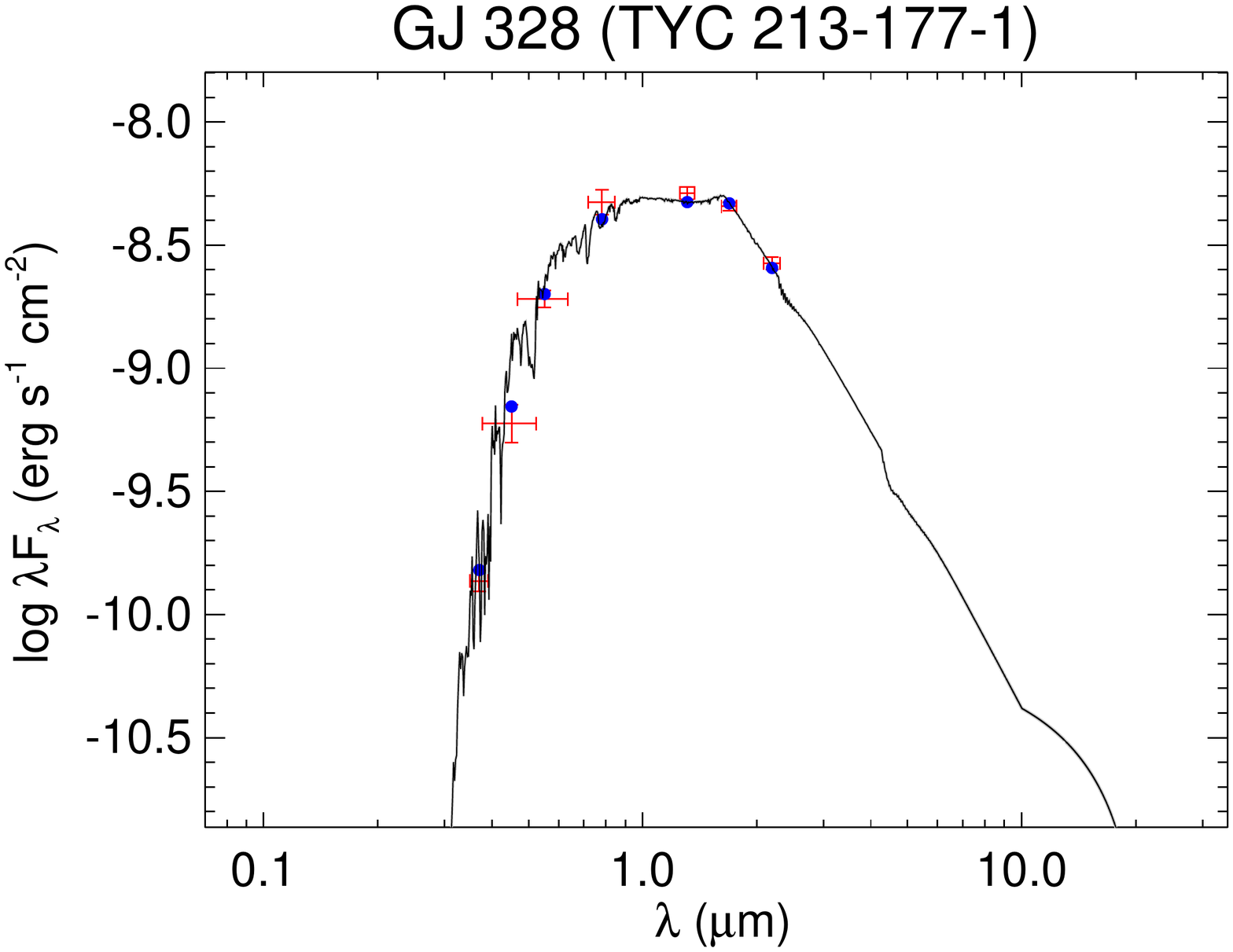}
  \includegraphics[trim=60 60 60 60,clip,width=0.49\linewidth]{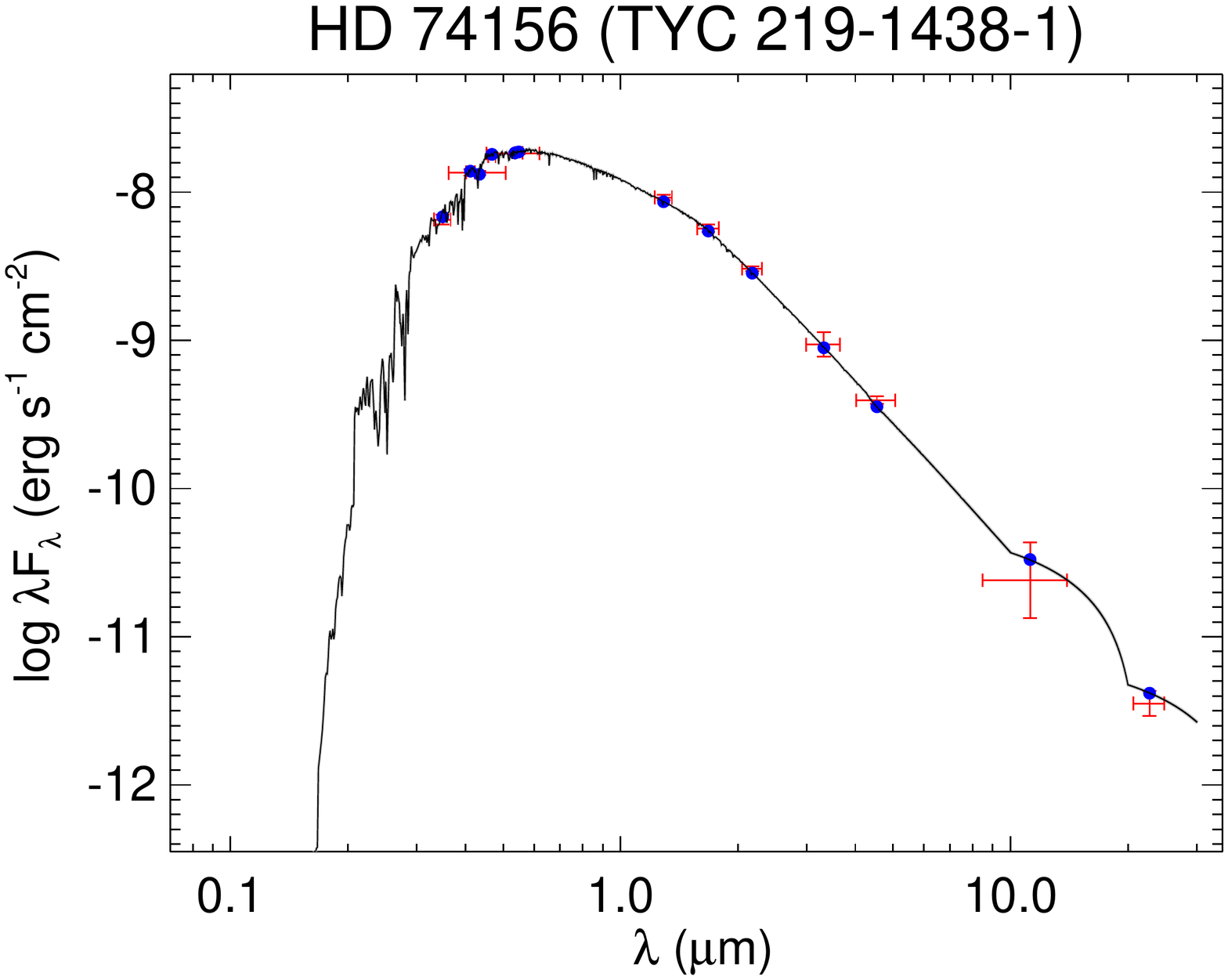}
  \includegraphics[trim=60 60 60 60,clip,width=0.49\linewidth]{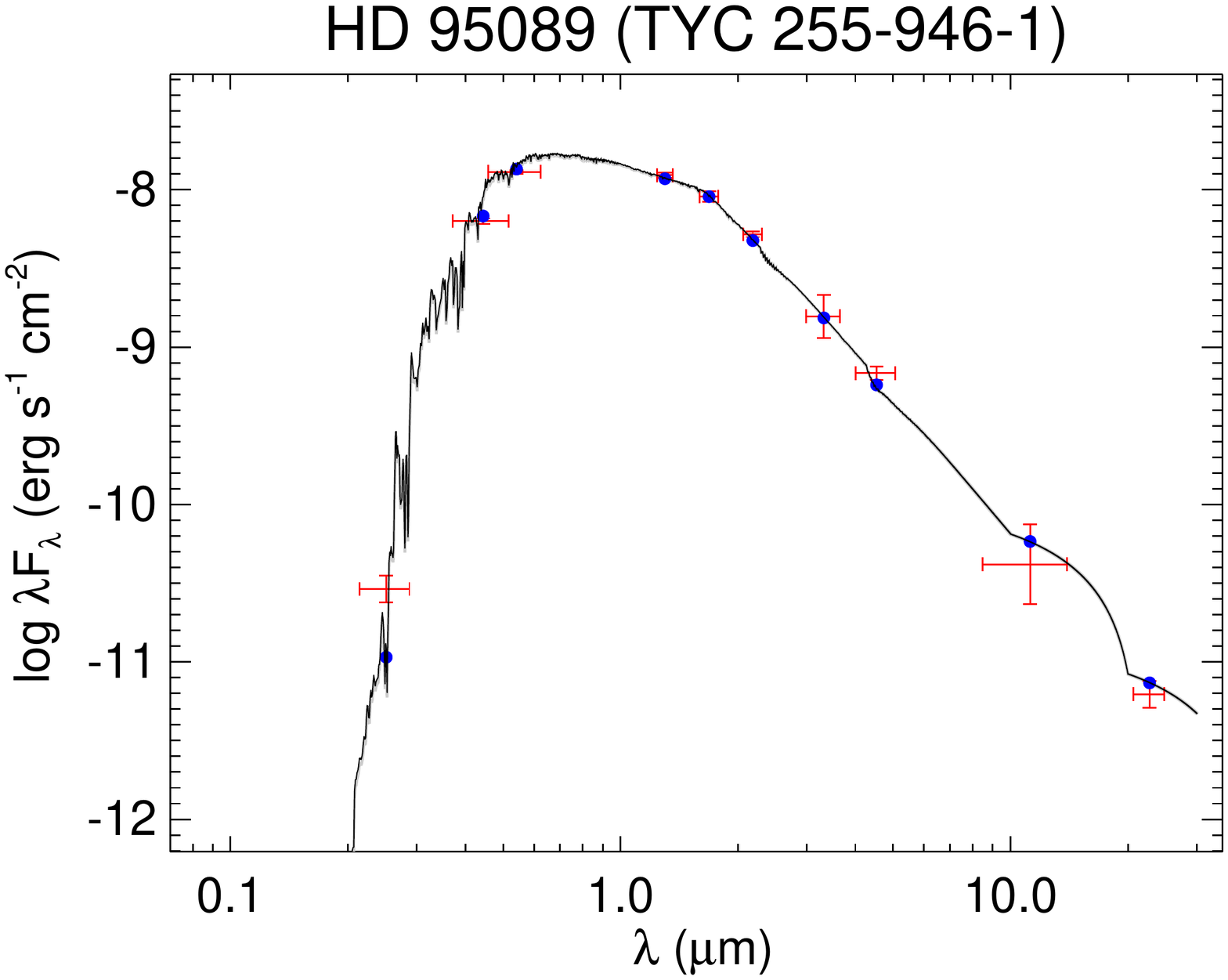}
  \caption{All labels, lines, symbols, and colors as in Figure \ref{fig:seds}.}
  \label{fig:seds_3}
\end{figure}

\begin{figure}[H]
  \centering
  \includegraphics[trim=60 60 60 60,clip,width=0.49\linewidth]{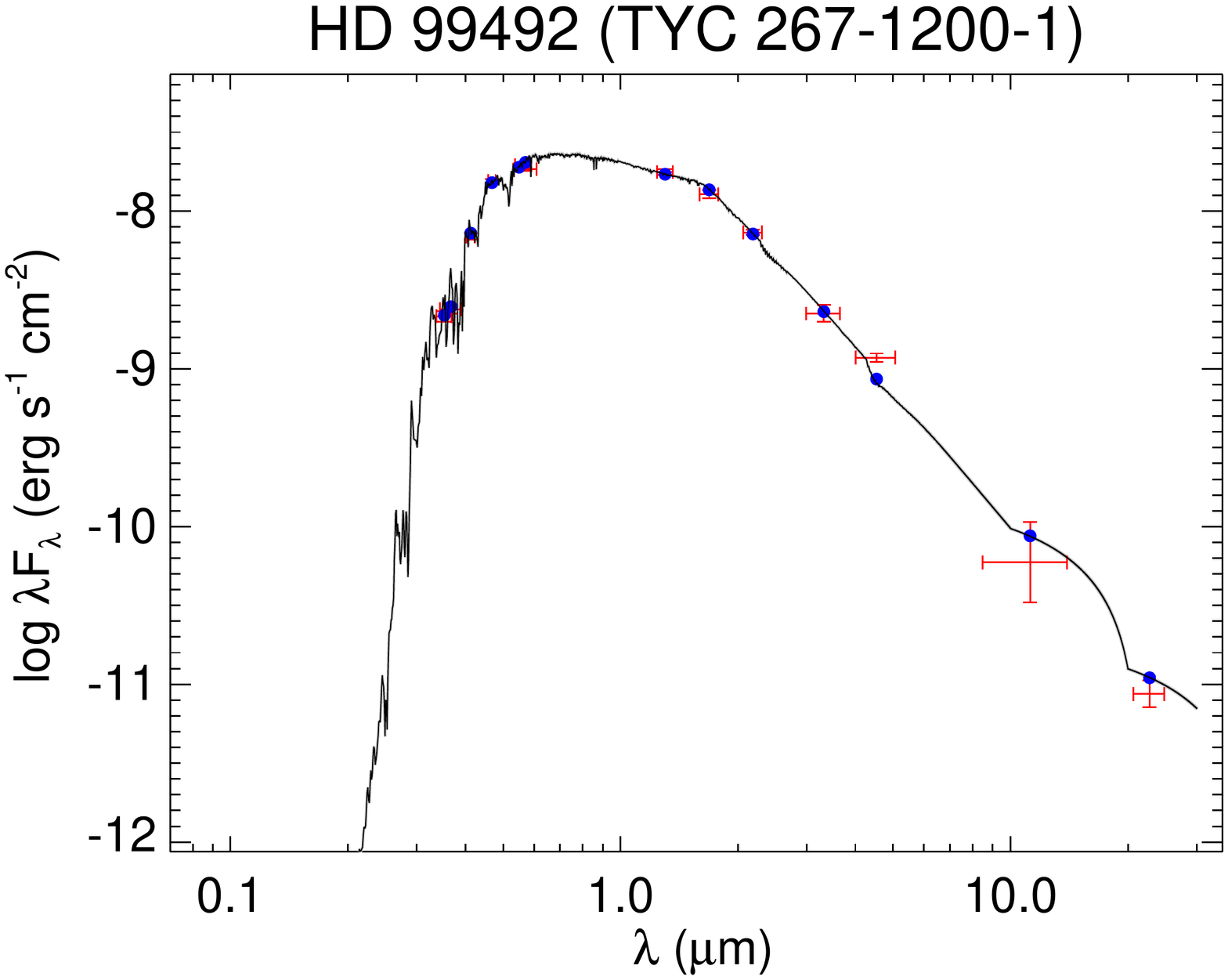}
  \includegraphics[trim=60 60 60 60,clip,width=0.49\linewidth]{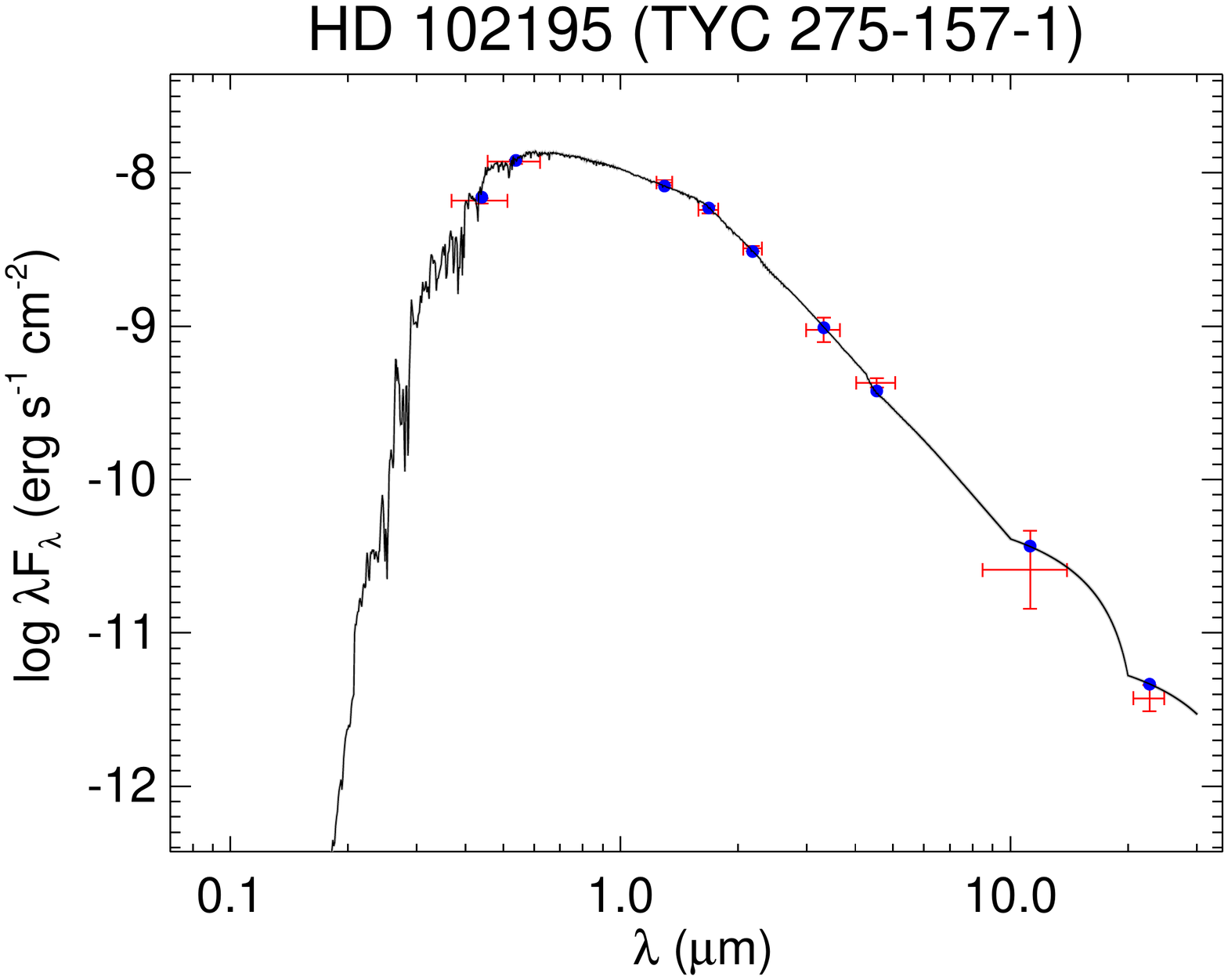}
  \includegraphics[trim=60 60 60 60,clip,width=0.49\linewidth]{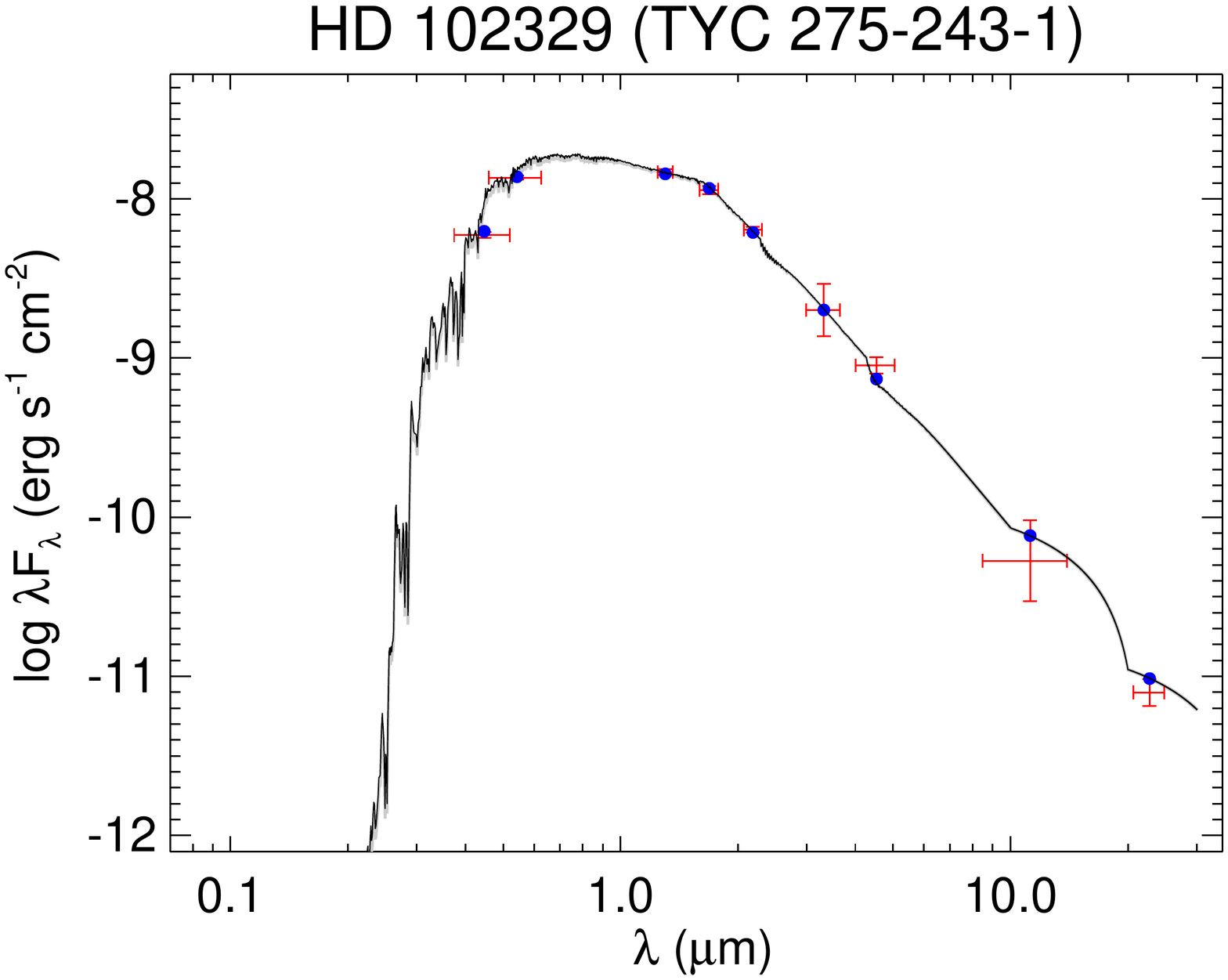}
  \includegraphics[trim=60 60 60 60,clip,width=0.49\linewidth]{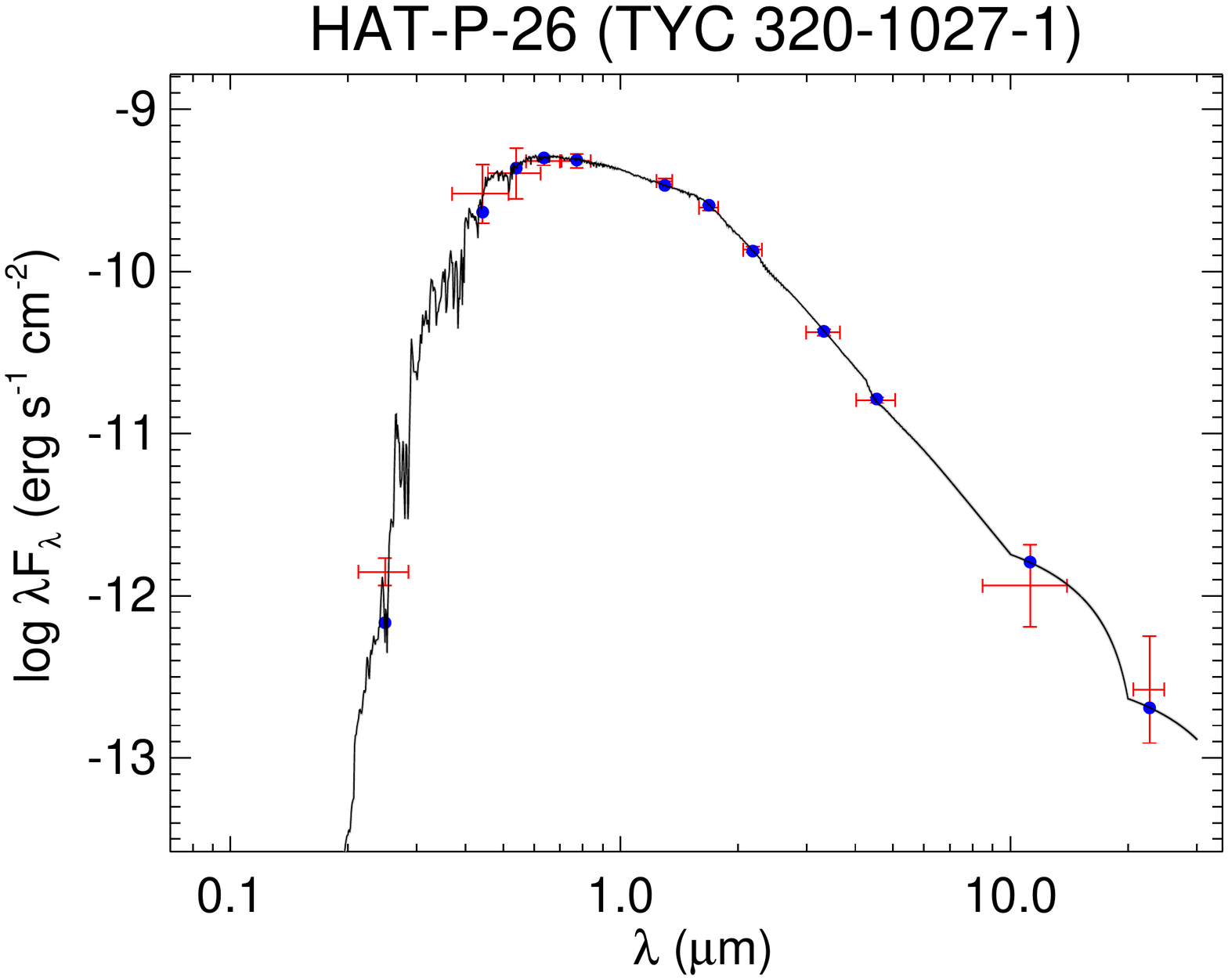}
  \includegraphics[trim=60 60 60 60,clip,width=0.49\linewidth]{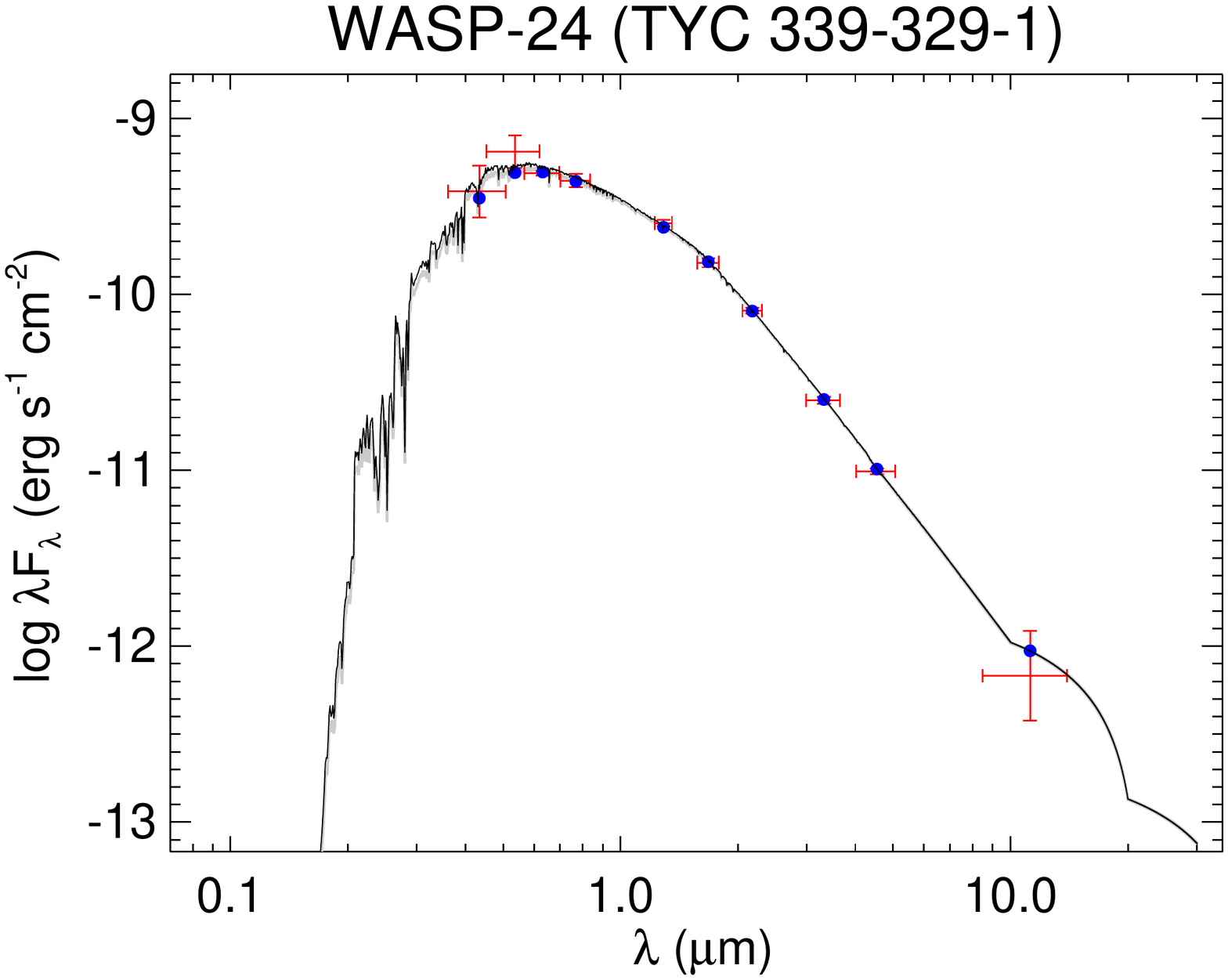}
  \includegraphics[trim=60 60 60 60,clip,width=0.49\linewidth]{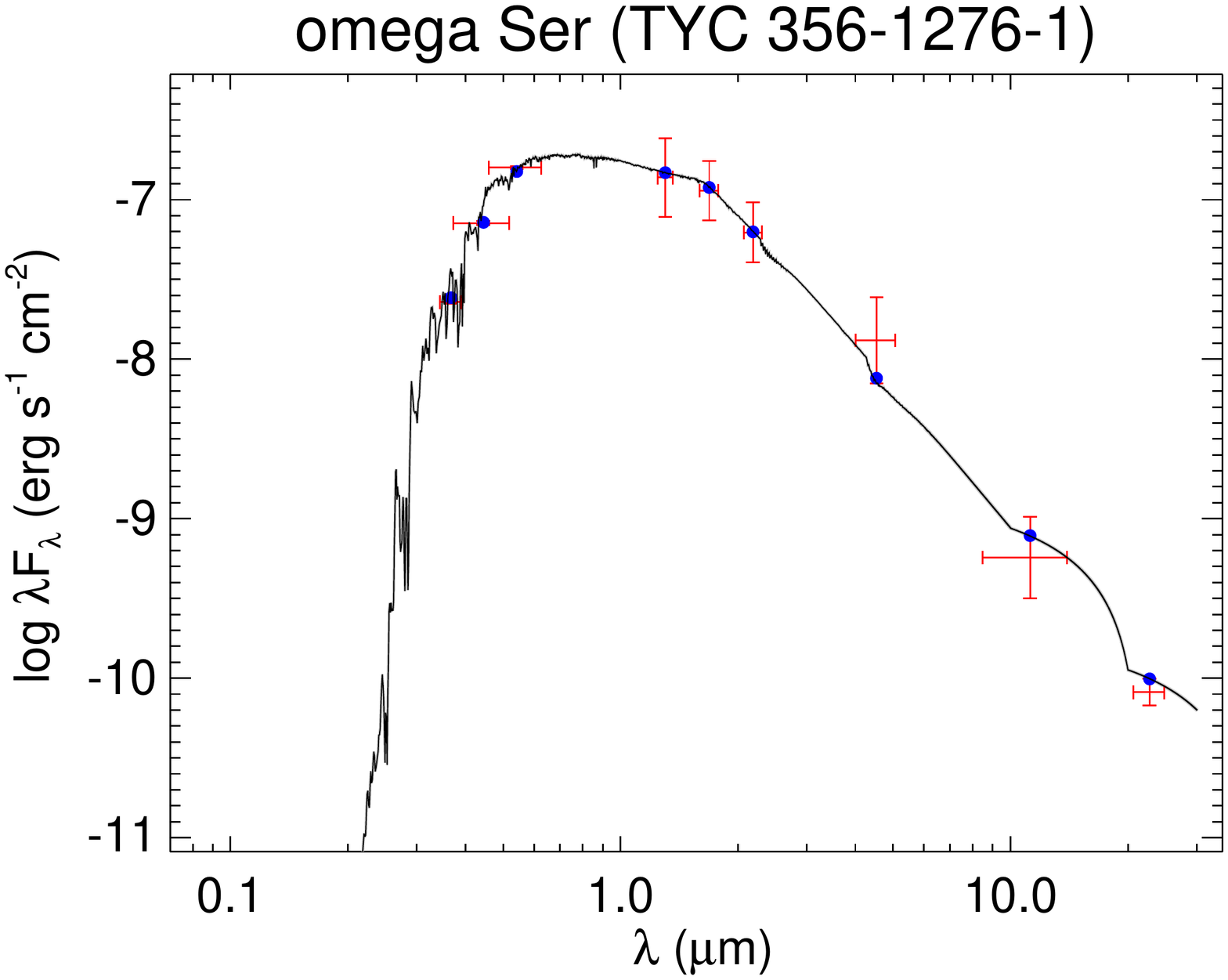}
  \caption{All labels, lines, symbols, and colors as in Figure \ref{fig:seds}.}
  \label{fig:seds_4}
\end{figure}

\begin{figure}[H]
  \centering
  \includegraphics[trim=60 60 60 60,clip,width=0.49\linewidth]{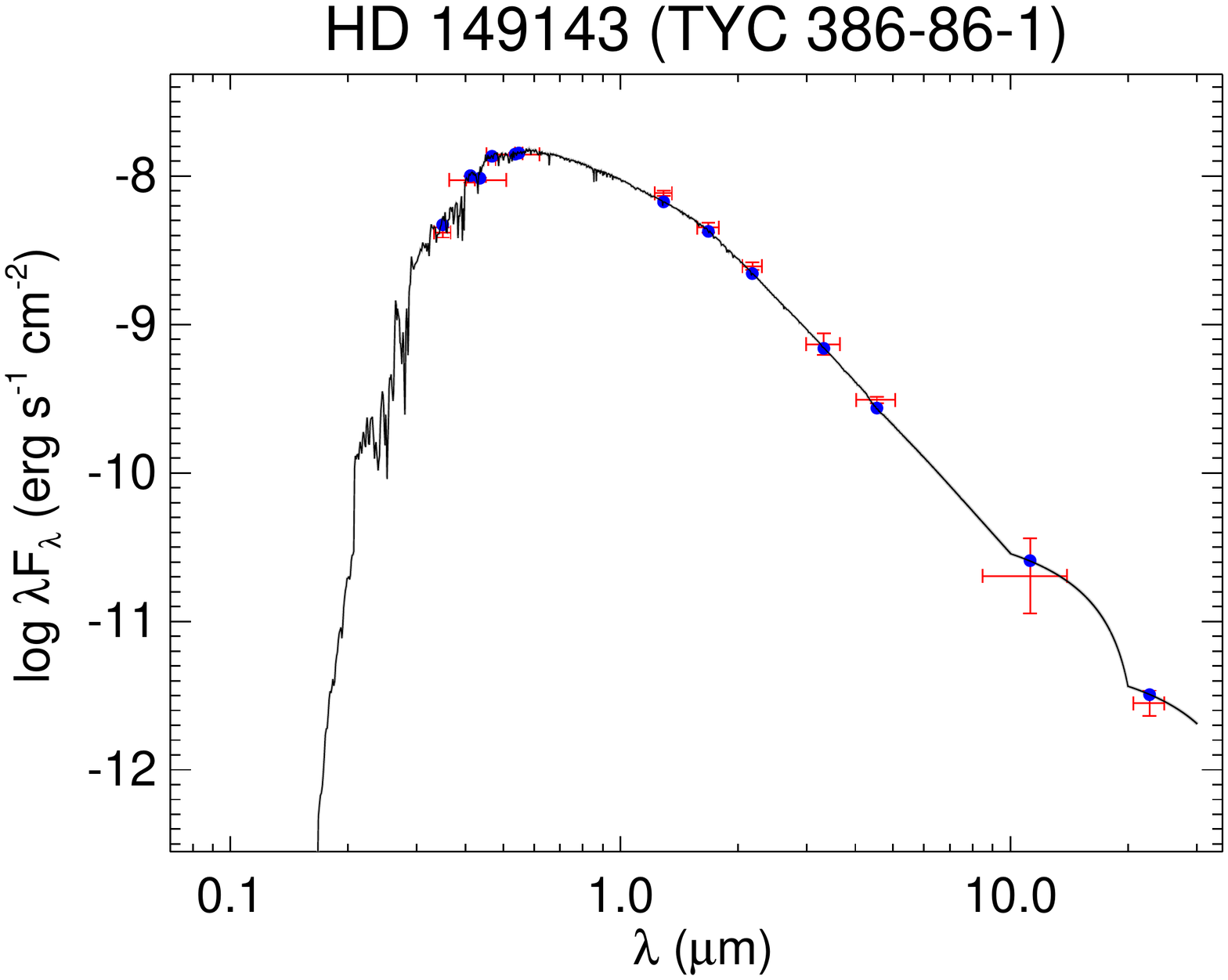}
  \includegraphics[trim=60 60 60 60,clip,width=0.49\linewidth]{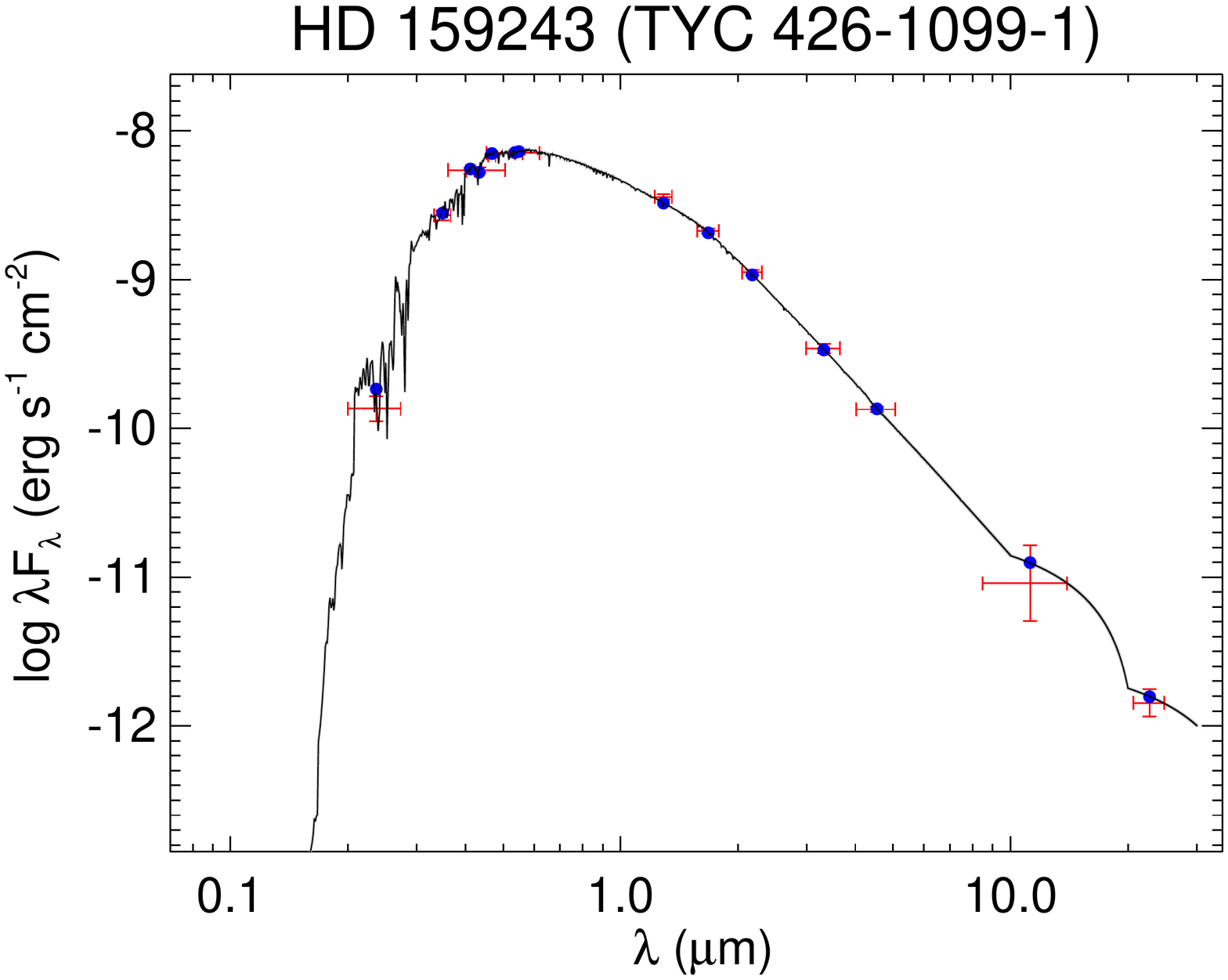}
  \includegraphics[trim=60 60 60 60,clip,width=0.49\linewidth]{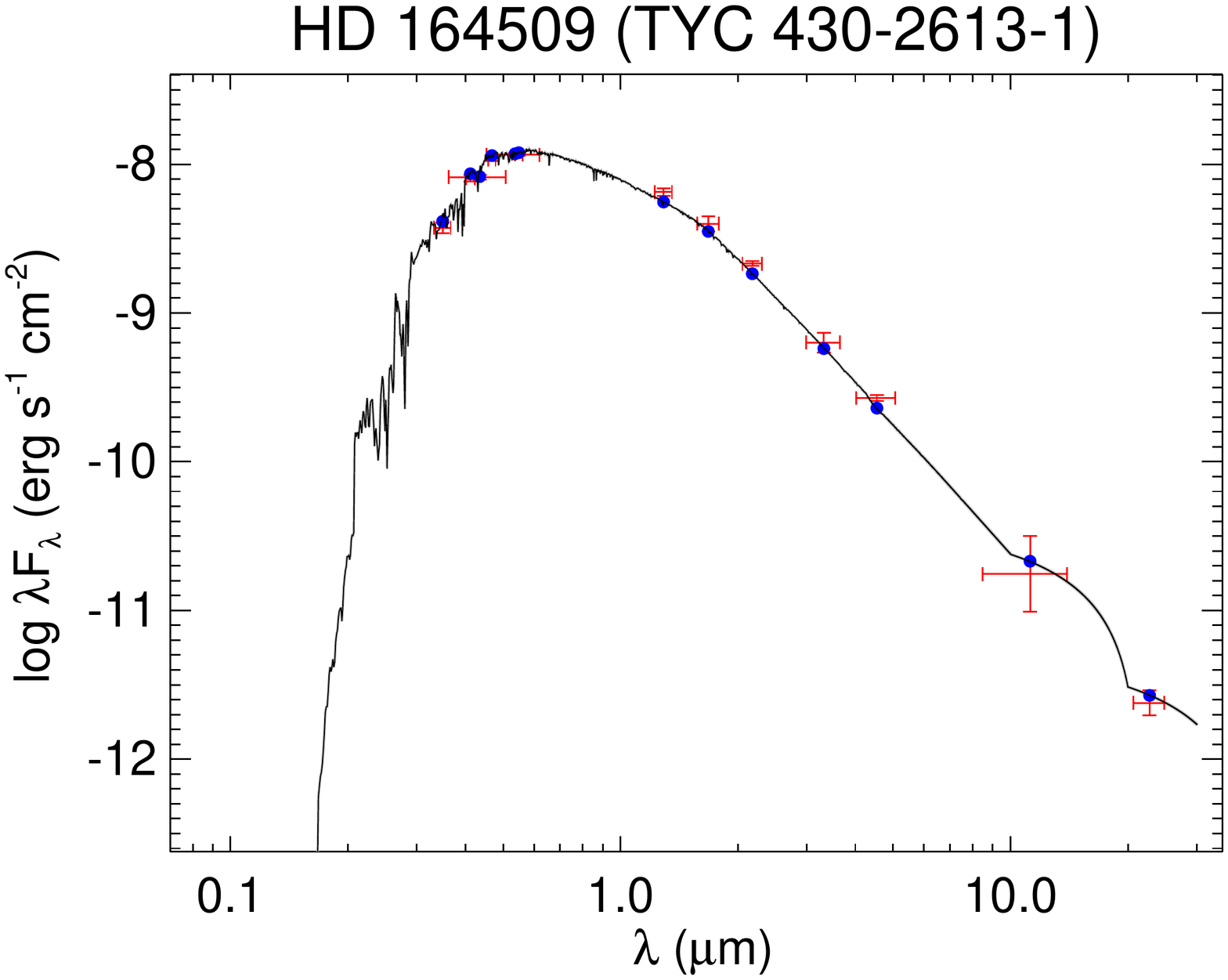}
  \includegraphics[trim=60 60 60 60,clip,width=0.49\linewidth]{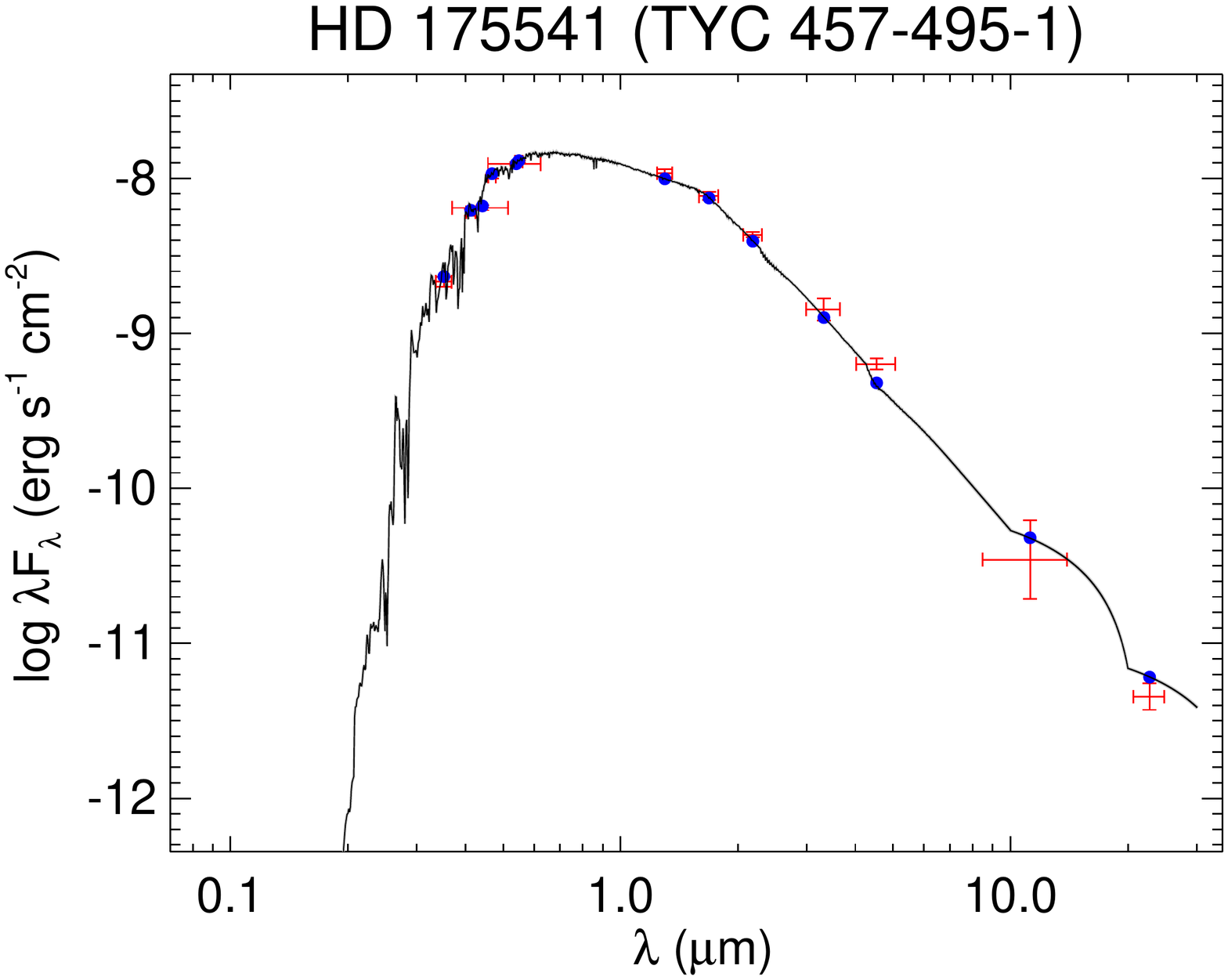}
  \includegraphics[trim=60 60 60 60,clip,width=0.49\linewidth]{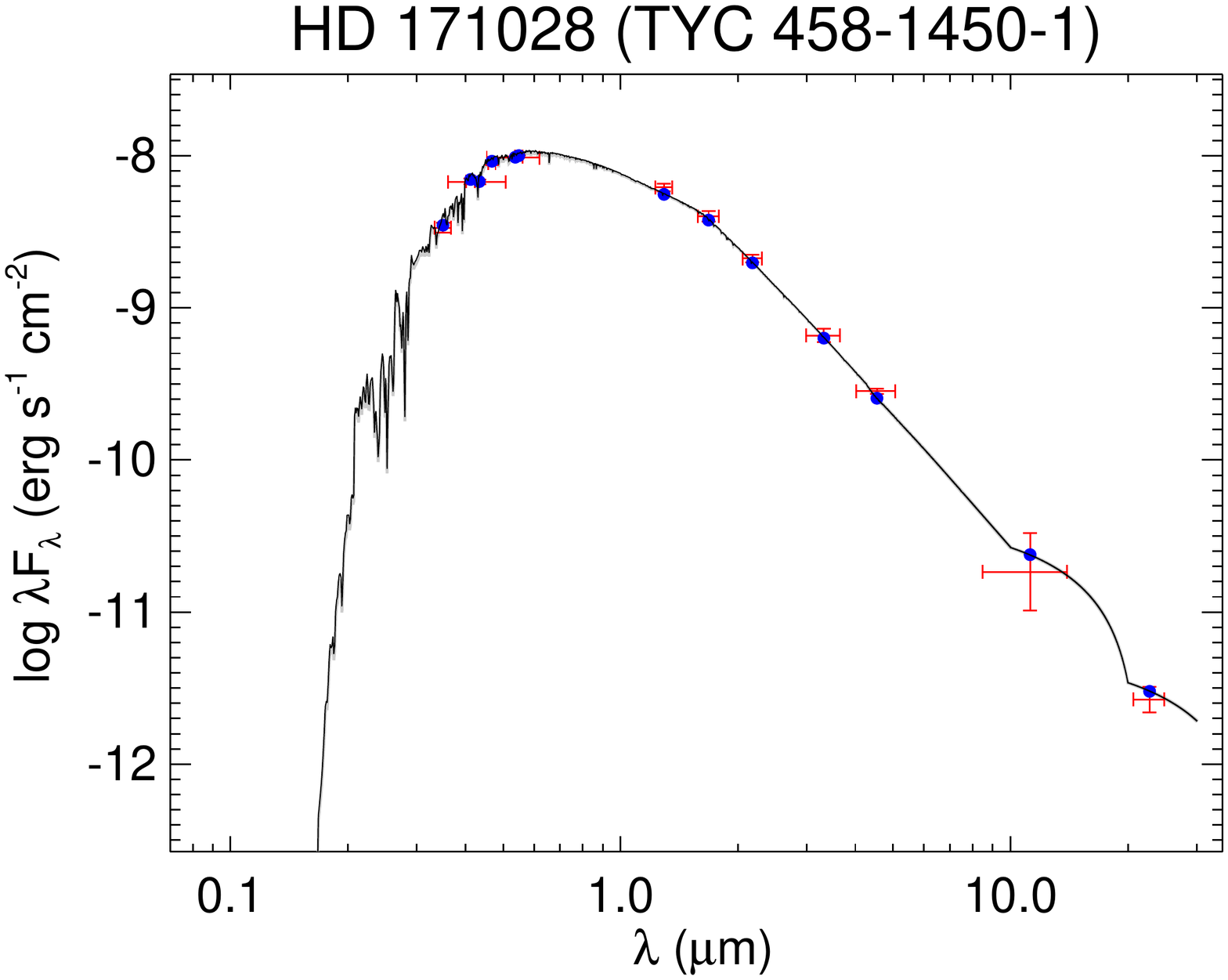}
  \includegraphics[trim=60 60 60 60,clip,width=0.49\linewidth]{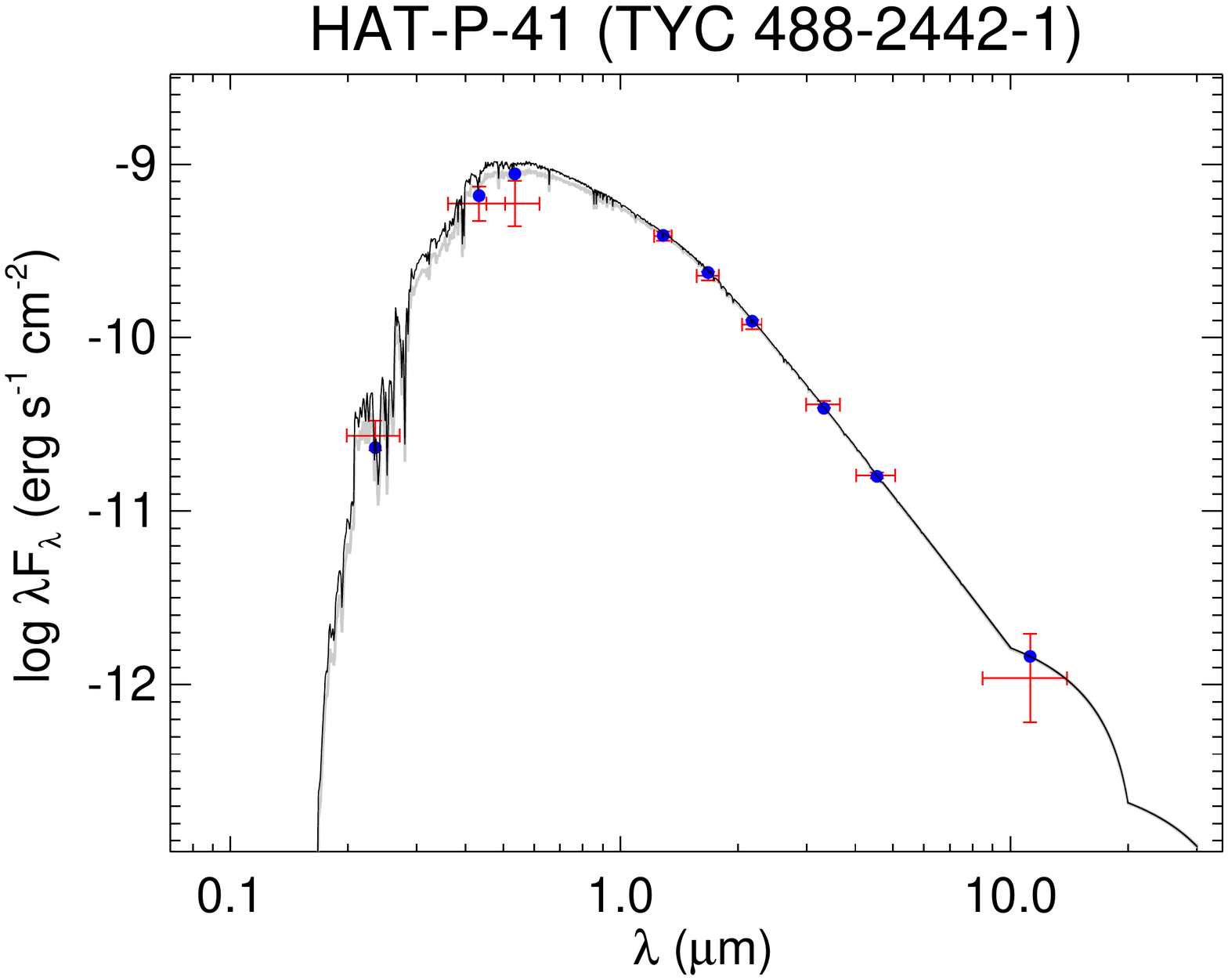}
  \caption{All labels, lines, symbols, and colors as in Figure \ref{fig:seds}.}
  \label{fig:seds_5}
\end{figure}

\begin{figure}[H]
  \centering
  \includegraphics[trim=60 60 60 60,clip,width=0.49\linewidth]{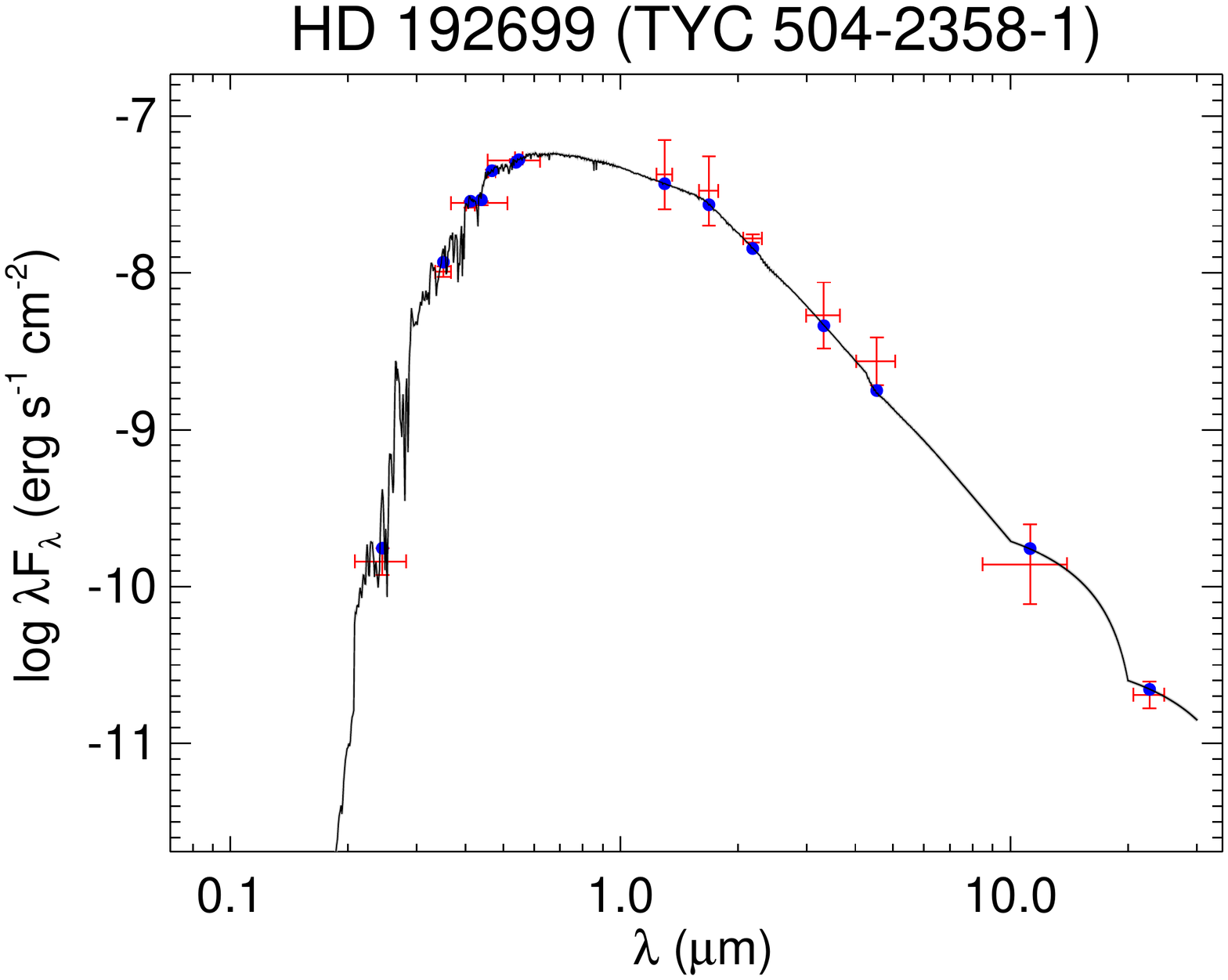}
  \includegraphics[trim=60 60 60 60,clip,width=0.49\linewidth]{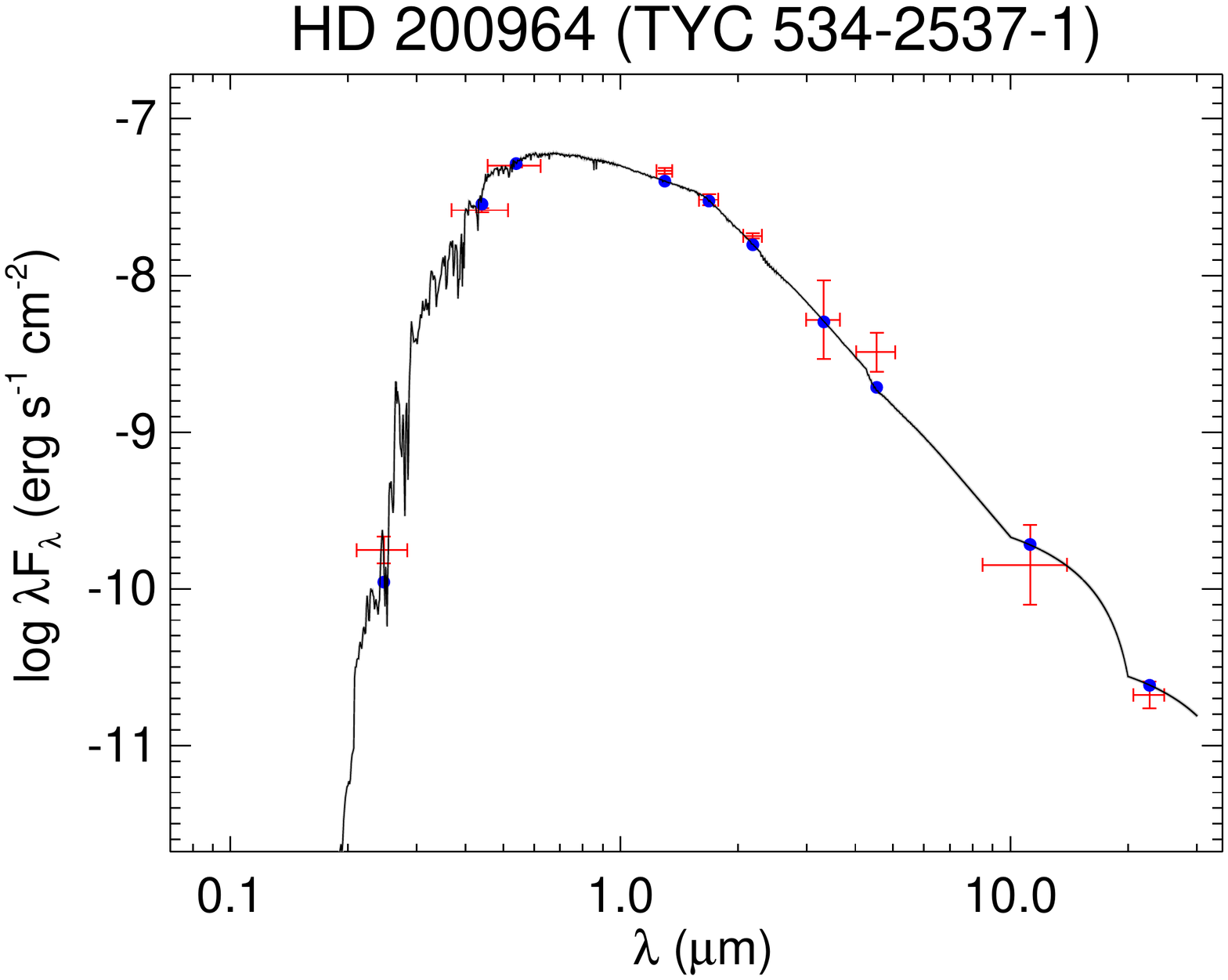}
  \includegraphics[trim=60 60 60 60,clip,width=0.49\linewidth]{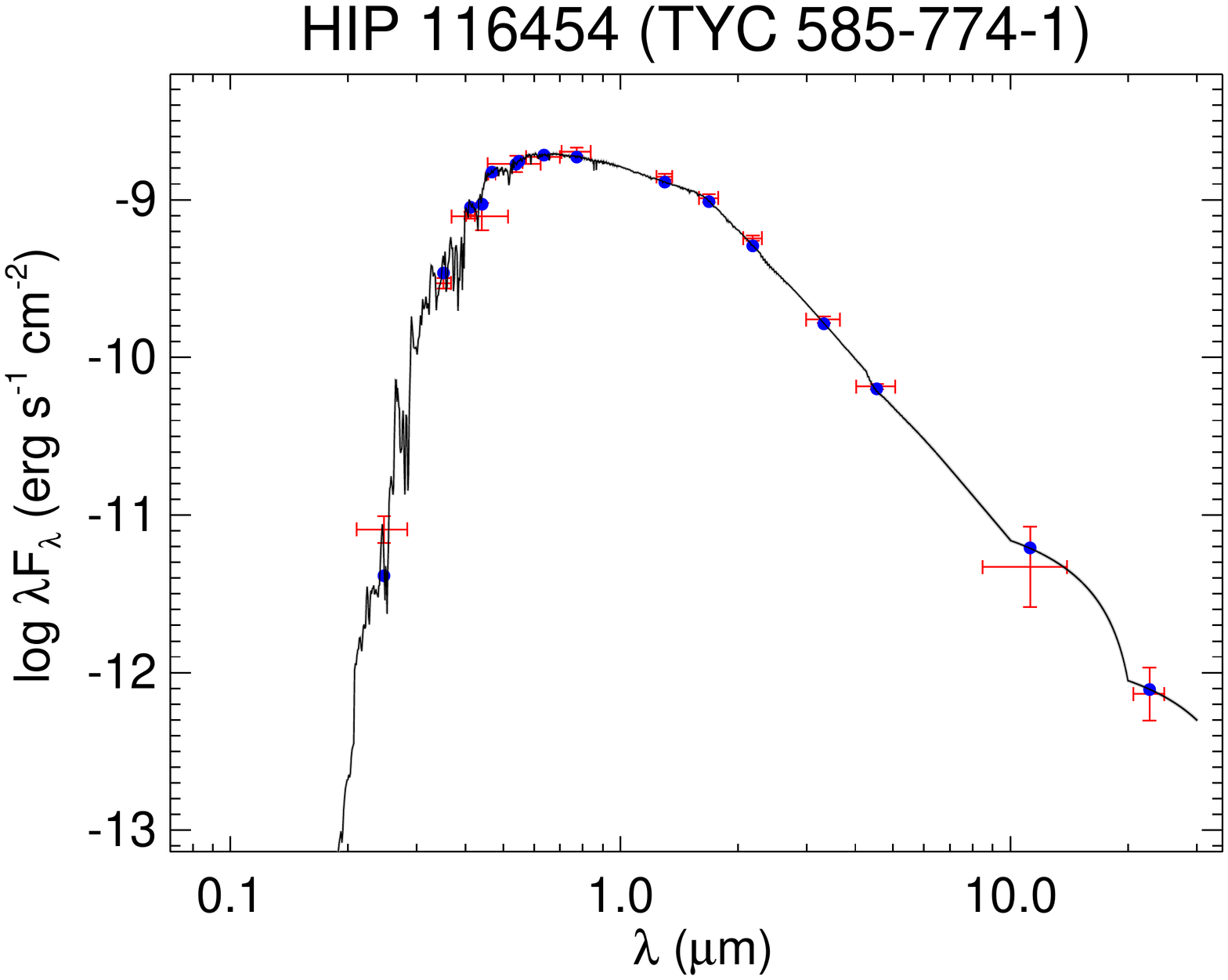}
  \includegraphics[trim=60 60 60 60,clip,width=0.49\linewidth]{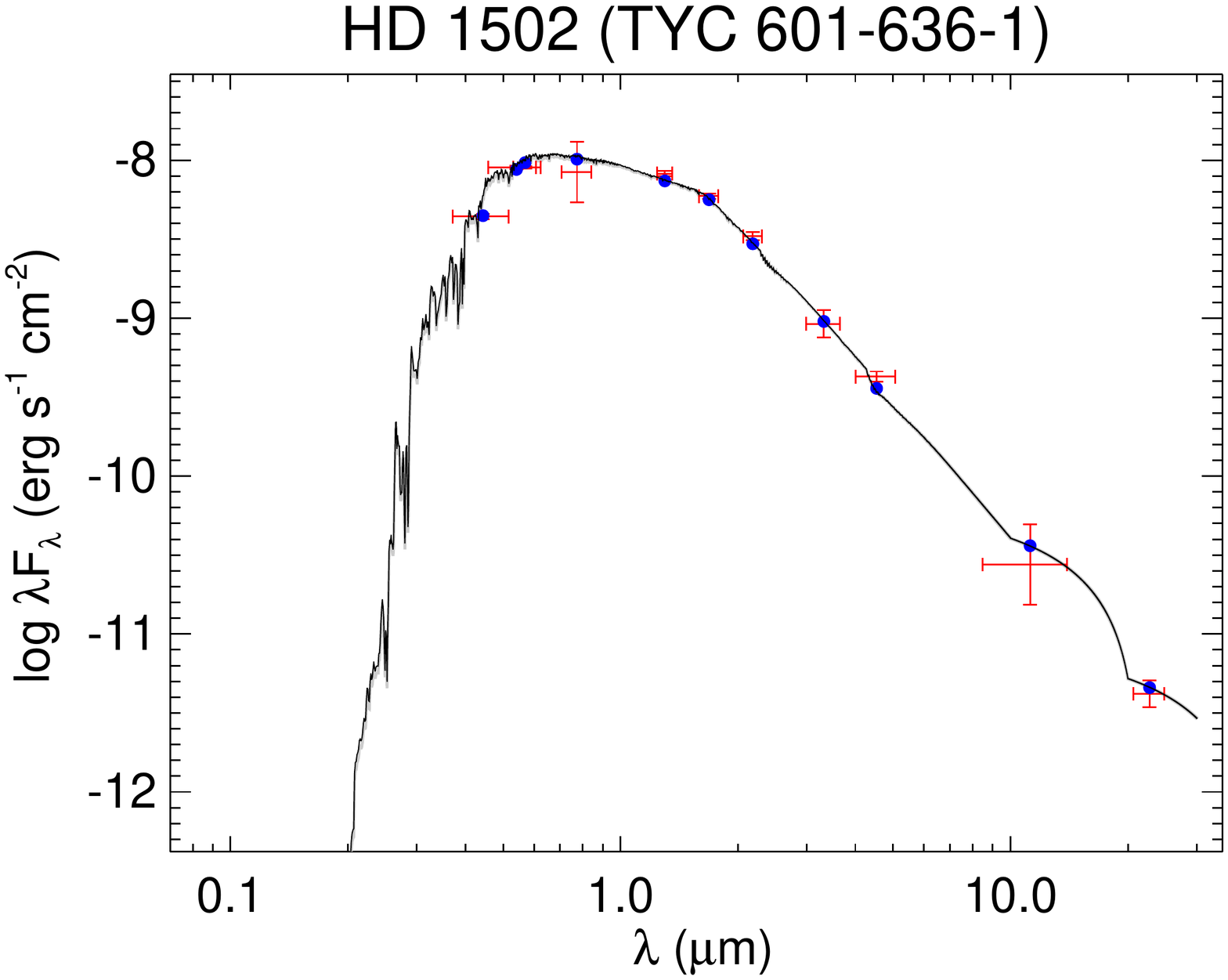}
  \includegraphics[trim=60 60 60 60,clip,width=0.49\linewidth]{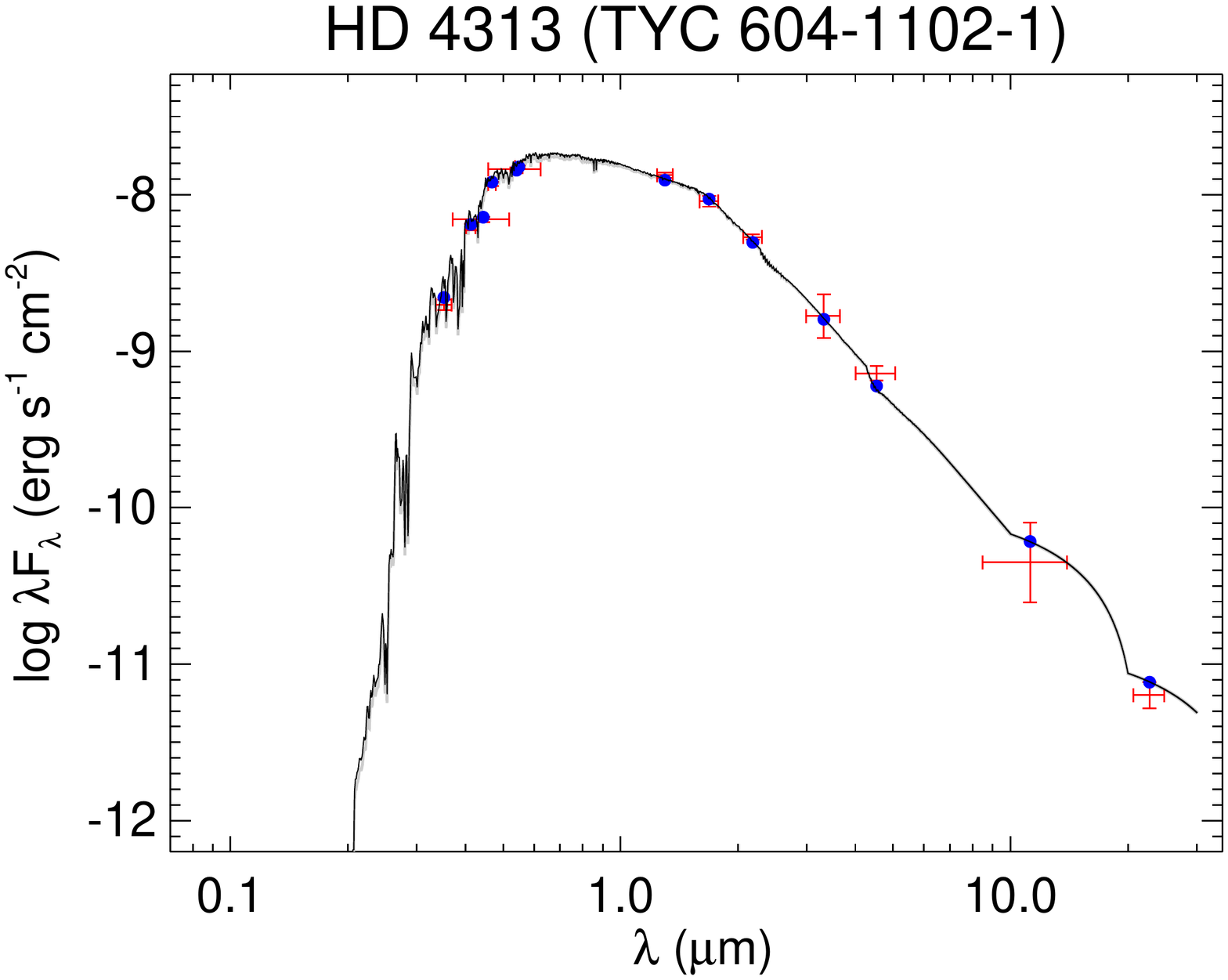}
  \includegraphics[trim=60 60 60 60,clip,width=0.49\linewidth]{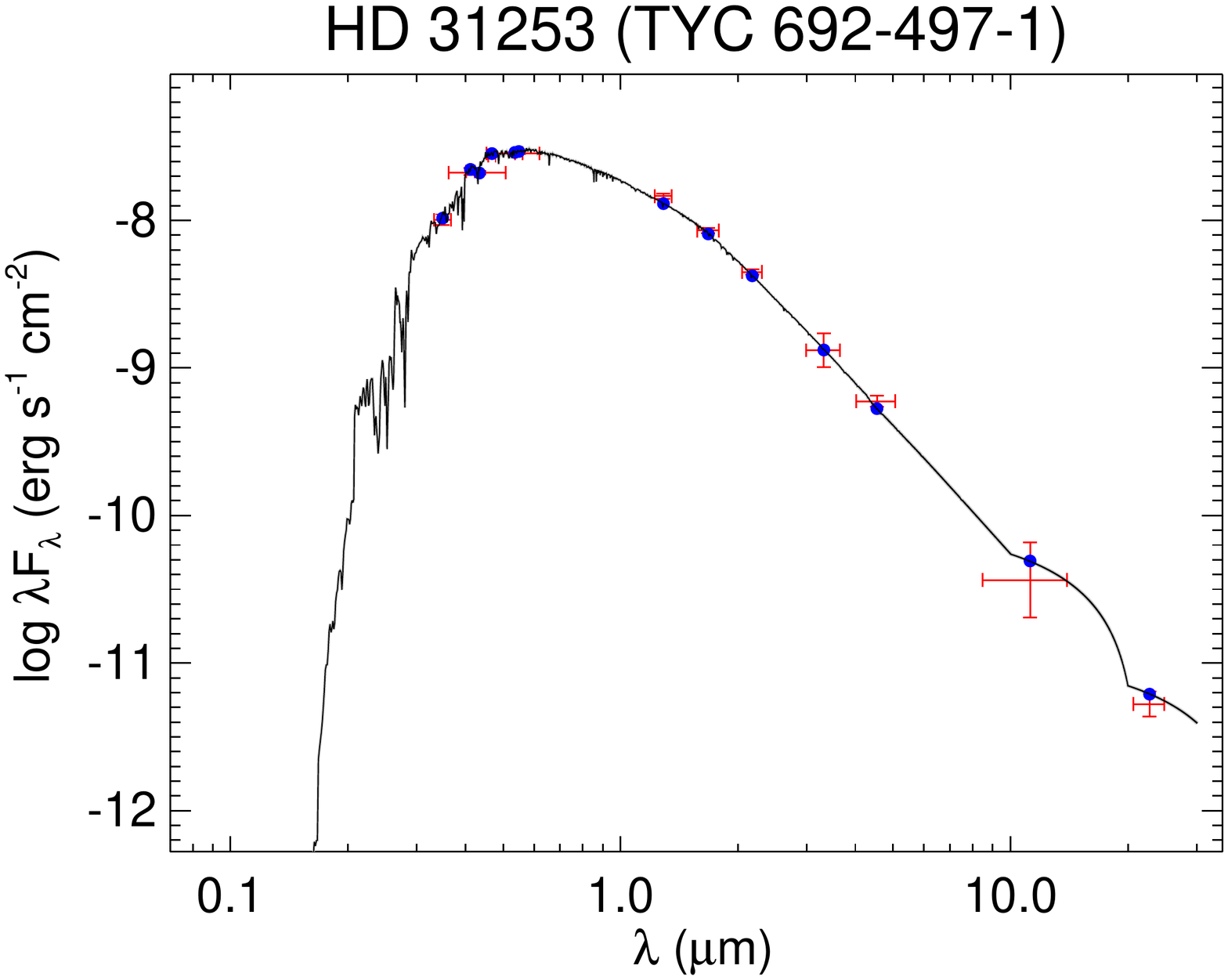}
  \caption{All labels, lines, symbols, and colors as in Figure \ref{fig:seds}.}
  \label{fig:seds_6}
\end{figure}

\begin{figure}[H]
  \centering
  \includegraphics[trim=60 60 60 60,clip,width=0.49\linewidth]{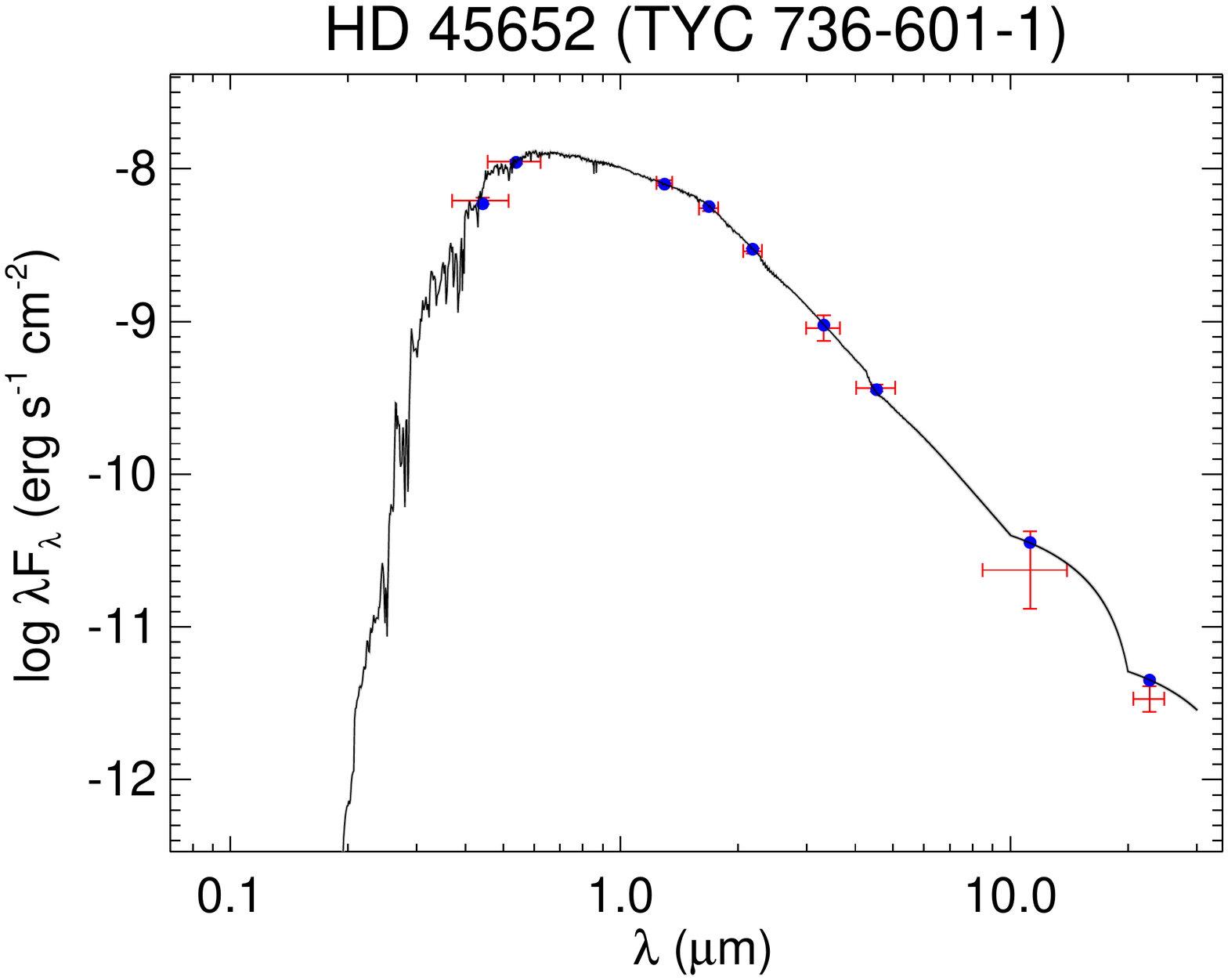}
  \includegraphics[trim=60 60 60 60,clip,width=0.49\linewidth]{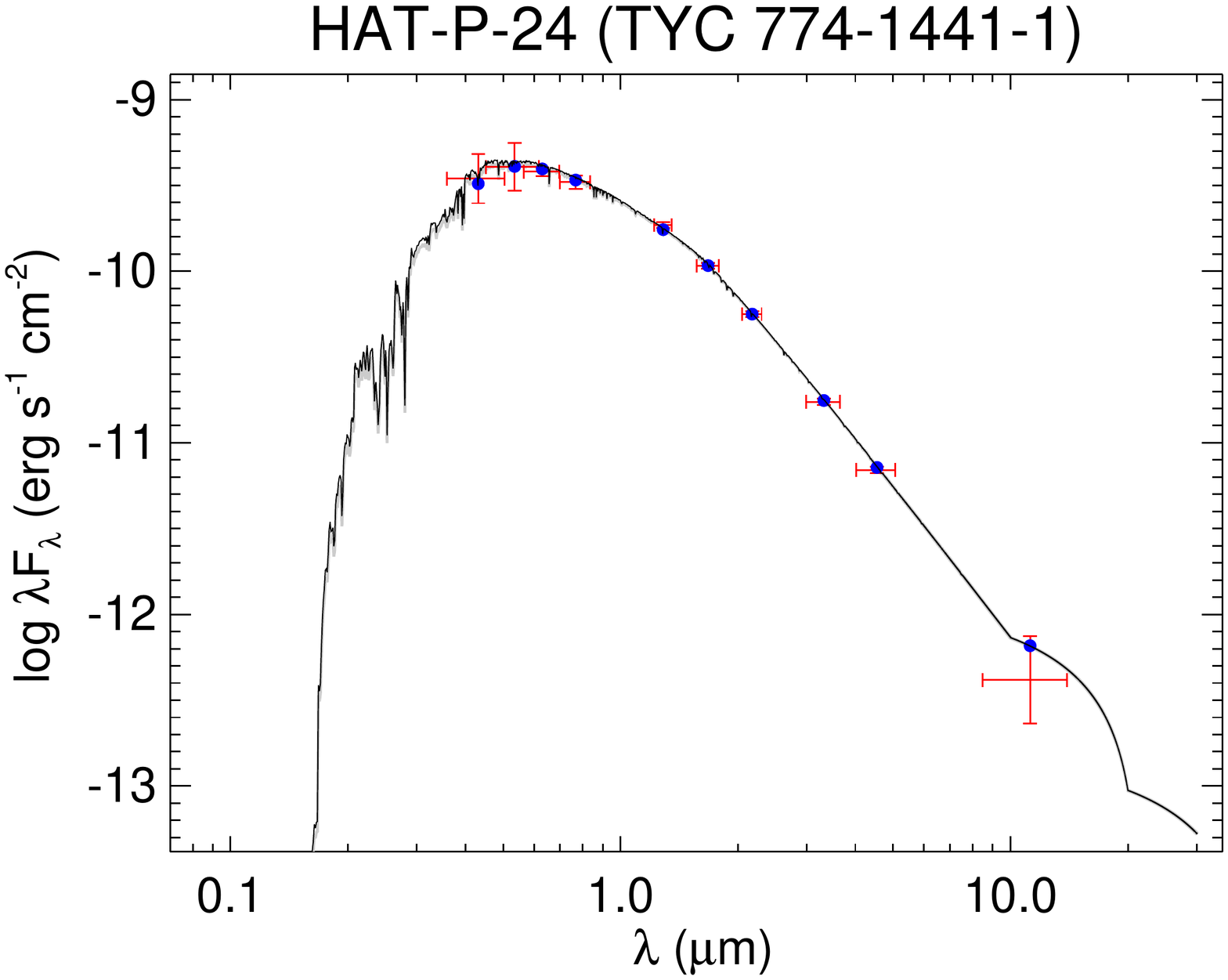}
  \includegraphics[trim=60 60 60 60,clip,width=0.49\linewidth]{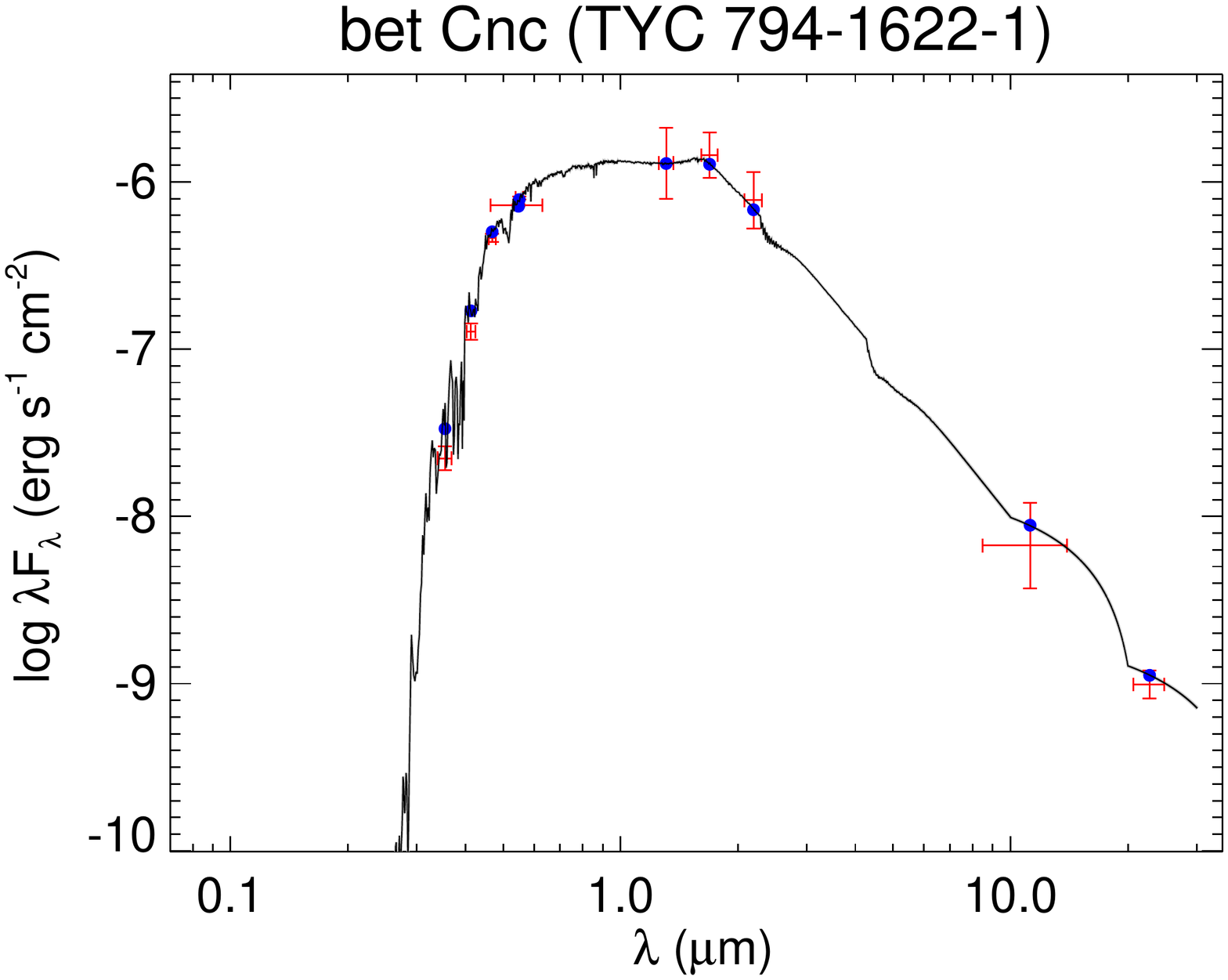}
  \includegraphics[trim=60 60 60 60,clip,width=0.49\linewidth]{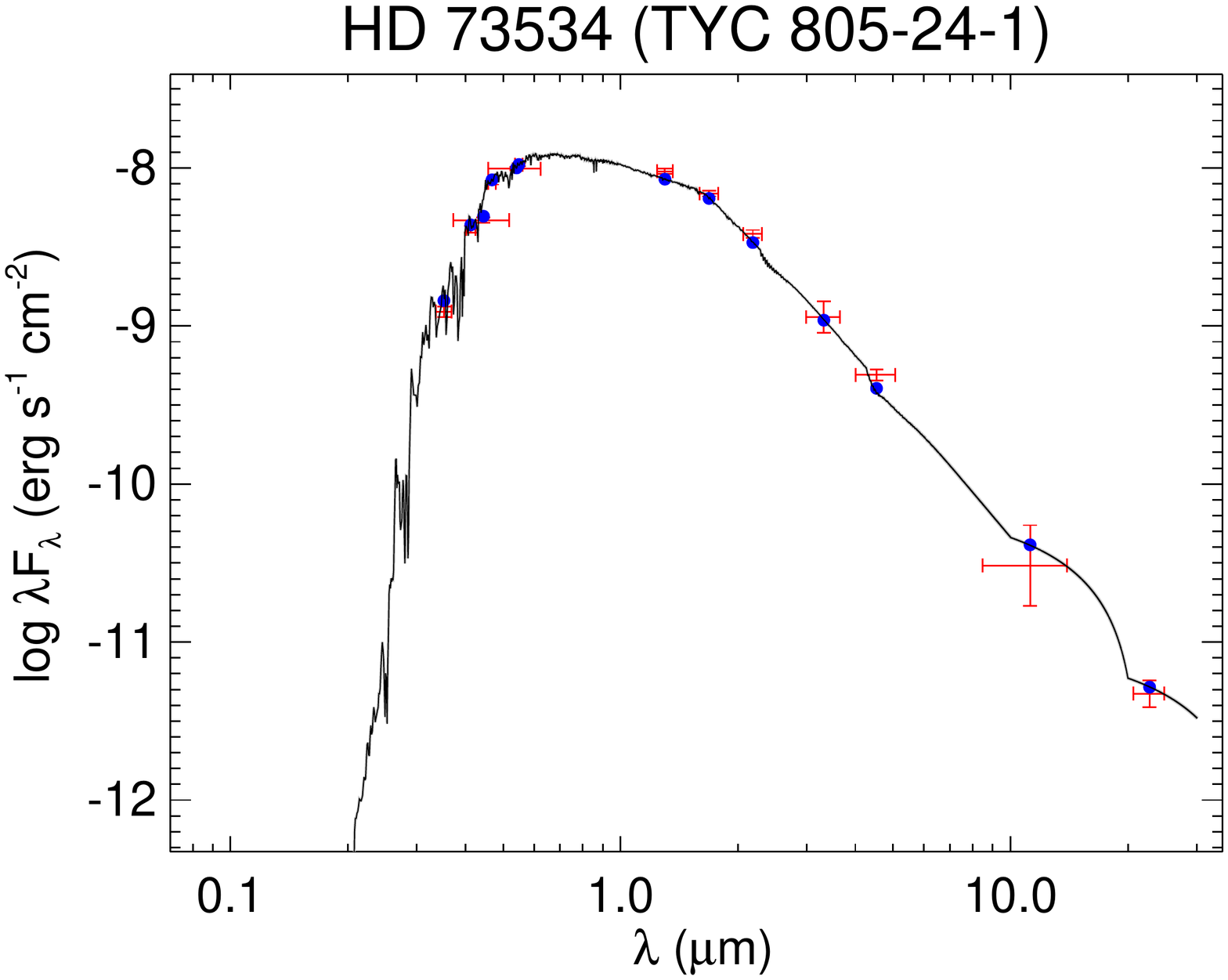}
  \includegraphics[trim=60 60 60 60,clip,width=0.49\linewidth]{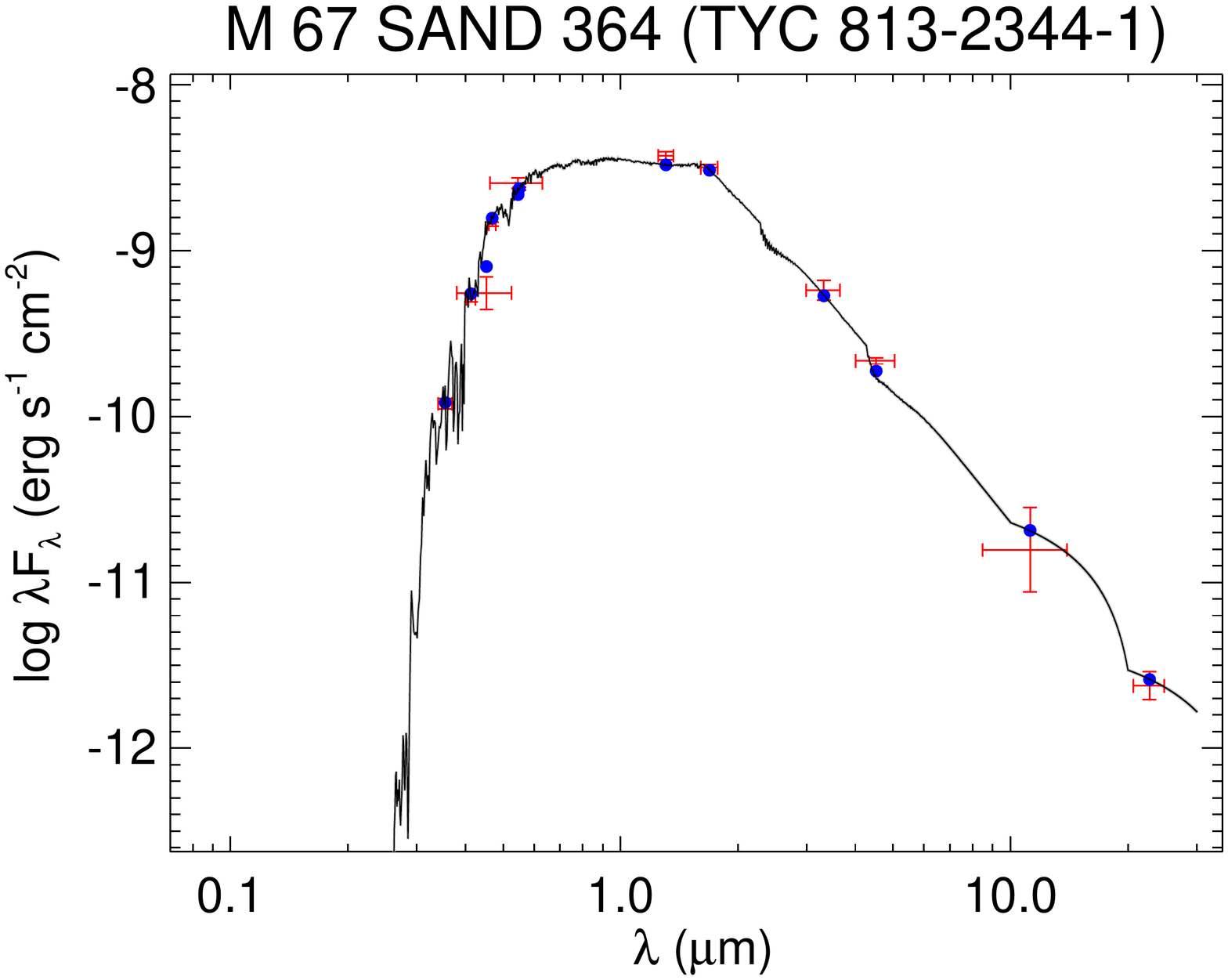}
  \includegraphics[trim=60 60 60 60,clip,width=0.49\linewidth]{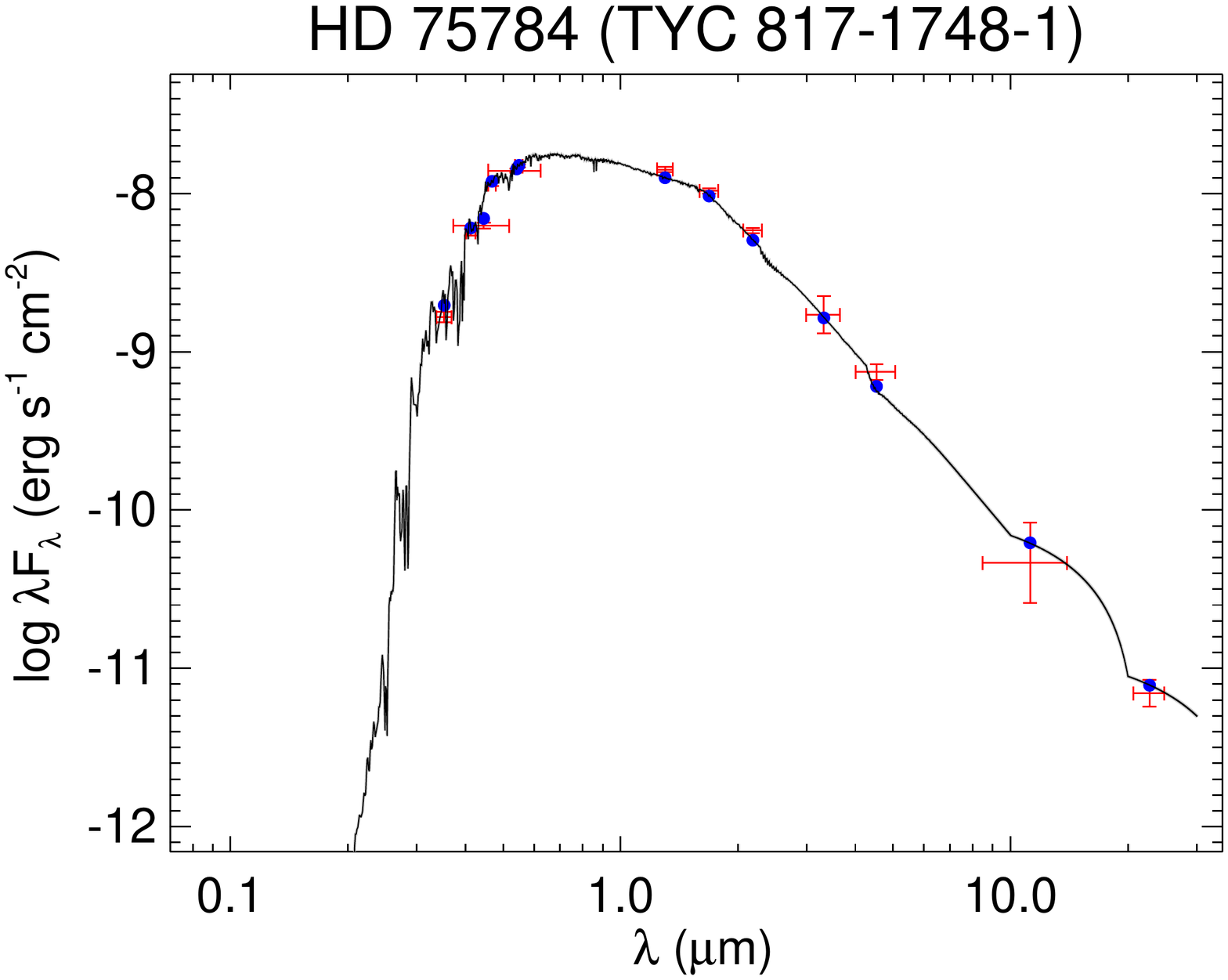}
  \caption{All labels, lines, symbols, and colors as in Figure \ref{fig:seds}.}
  \label{fig:seds_7}
\end{figure}

\begin{figure}[H]
  \centering
  \includegraphics[trim=60 60 60 60,clip,width=0.49\linewidth]{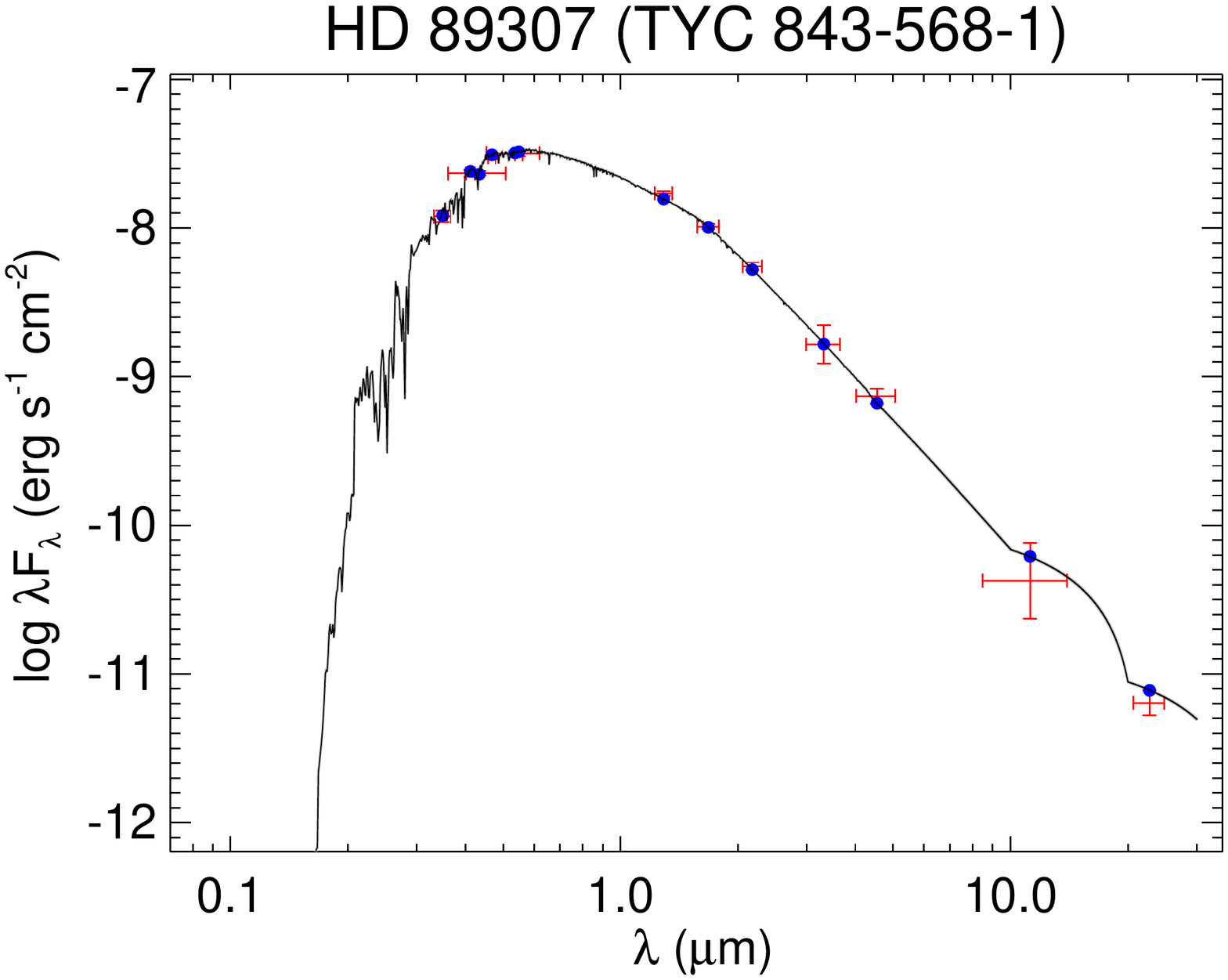}
  \includegraphics[trim=60 60 60 60,clip,width=0.49\linewidth]{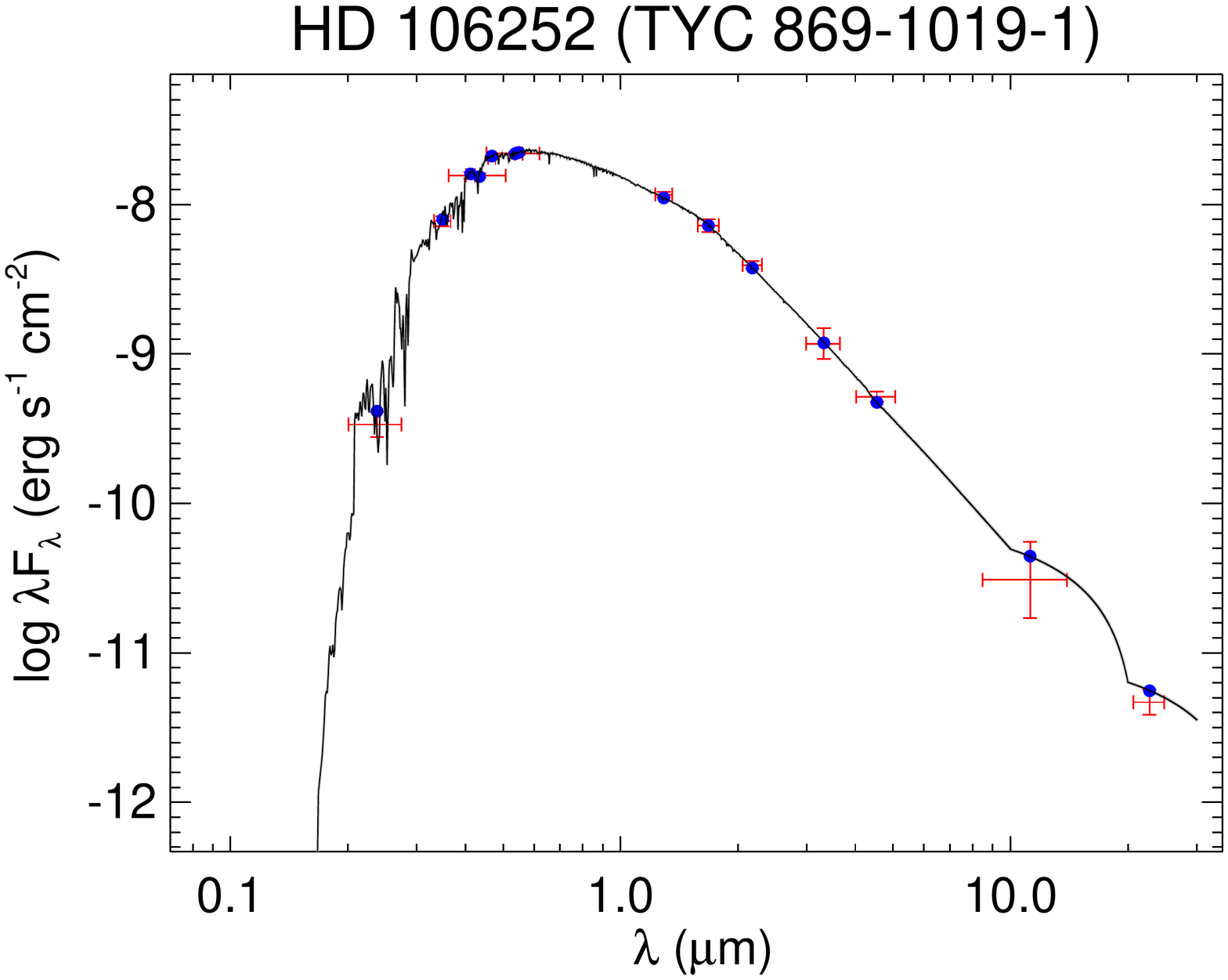}
  \includegraphics[trim=60 60 60 60,clip,width=0.49\linewidth]{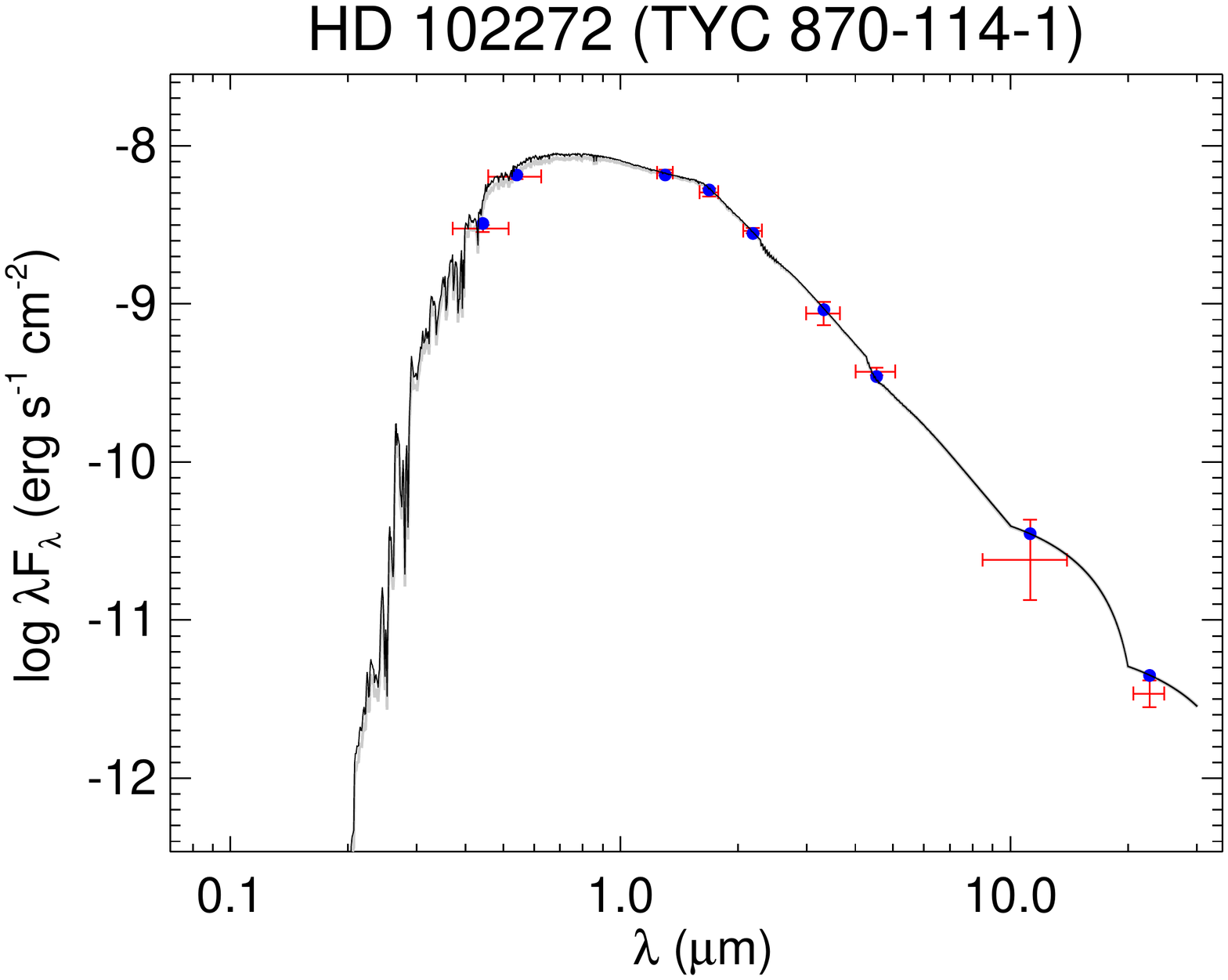}
  \includegraphics[trim=60 60 60 60,clip,width=0.49\linewidth]{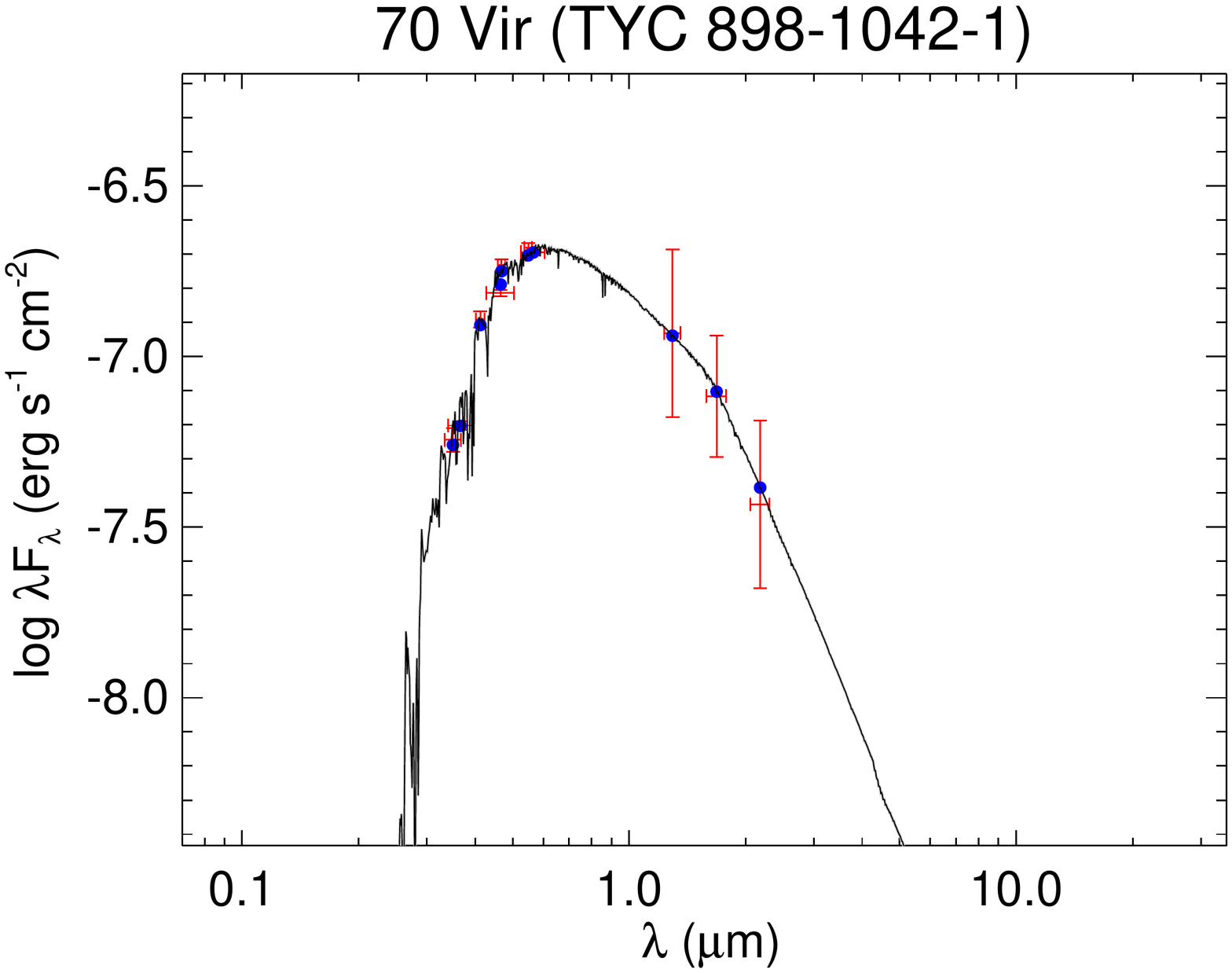}
  \includegraphics[trim=60 60 60 60,clip,width=0.49\linewidth]{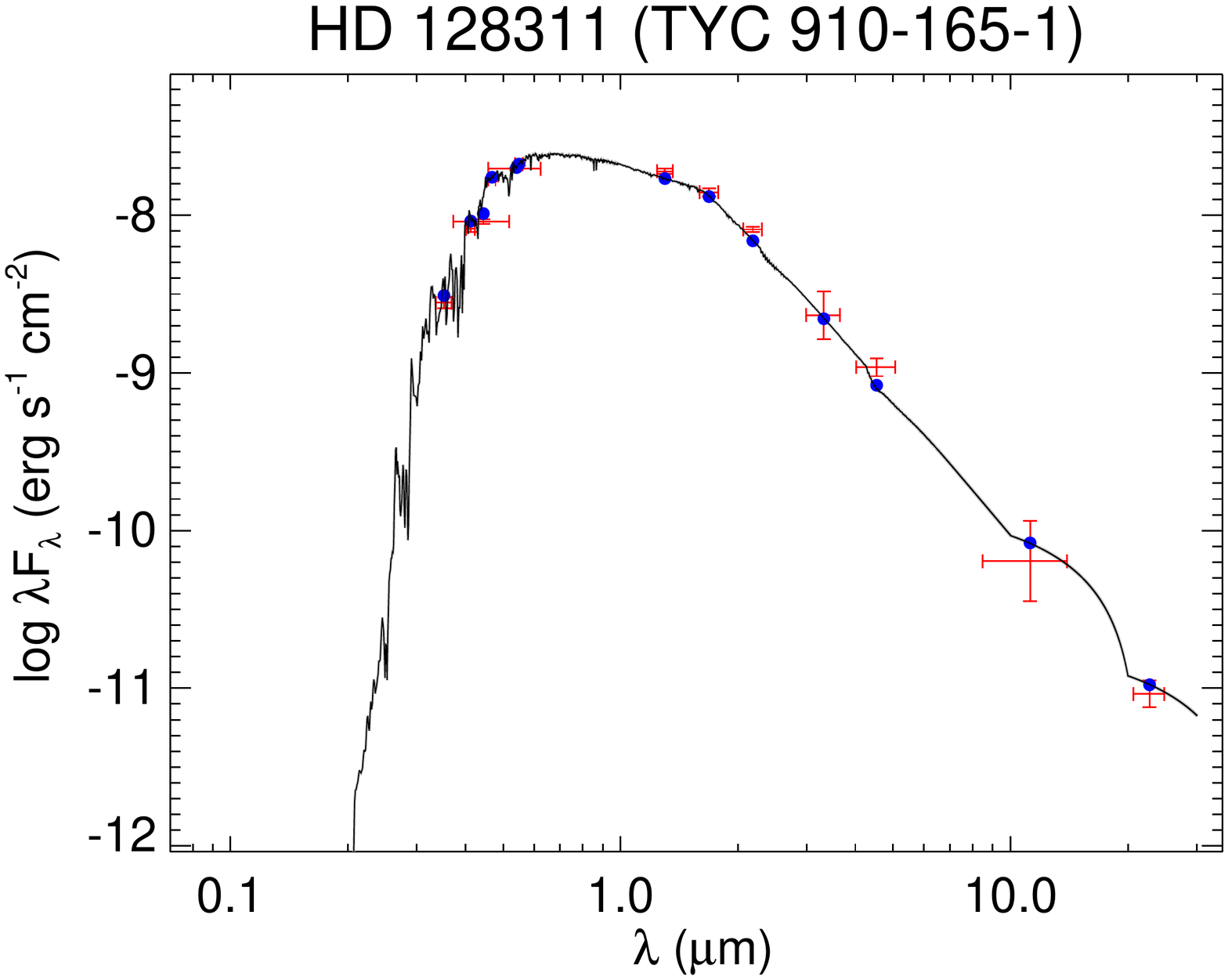}
  \includegraphics[trim=60 60 60 60,clip,width=0.49\linewidth]{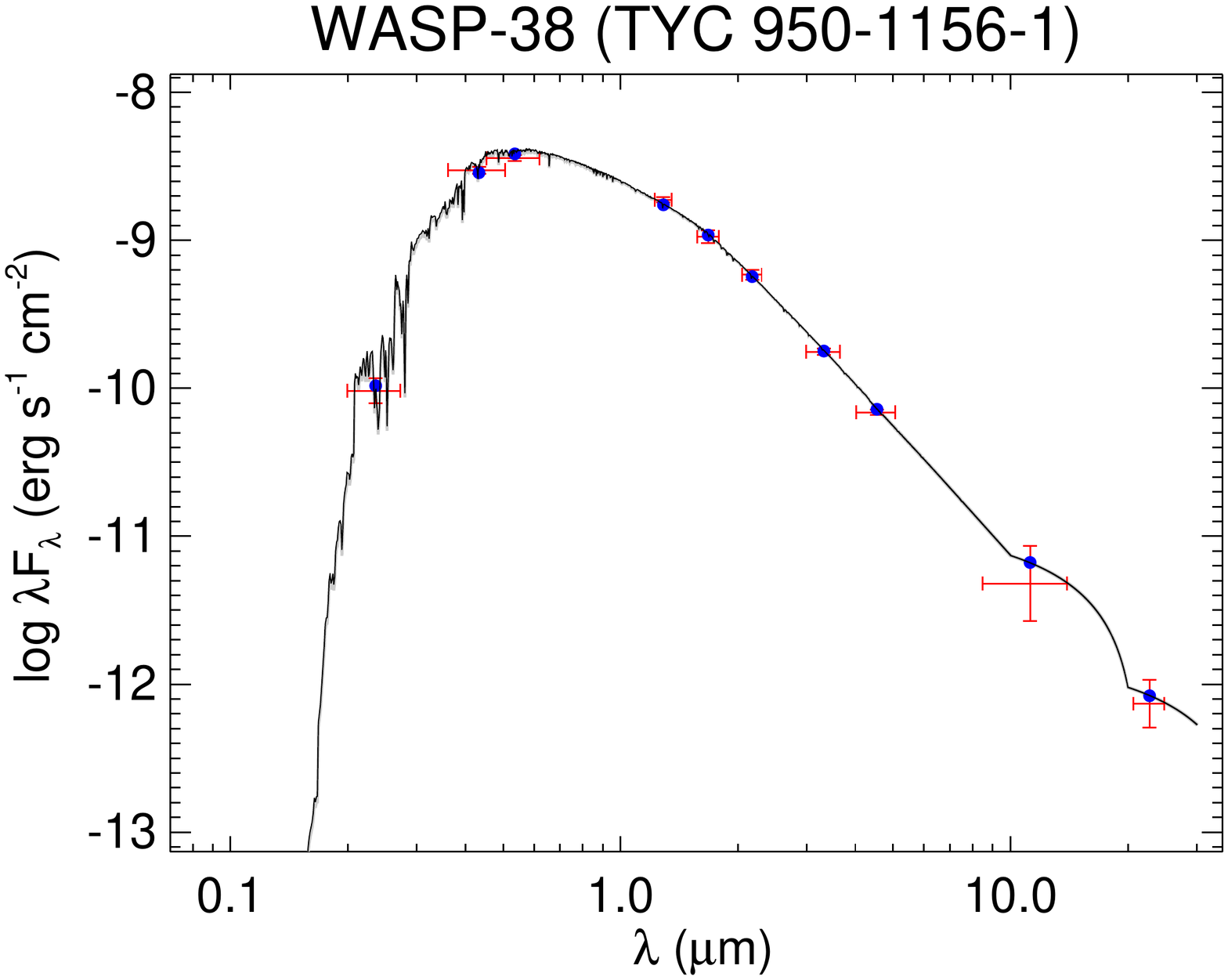}
  \caption{All labels, lines, symbols, and colors as in Figure \ref{fig:seds}.}
  \label{fig:seds_8}
\end{figure}

\begin{figure}[H]
  \centering
  \includegraphics[trim=60 60 60 60,clip,width=0.49\linewidth]{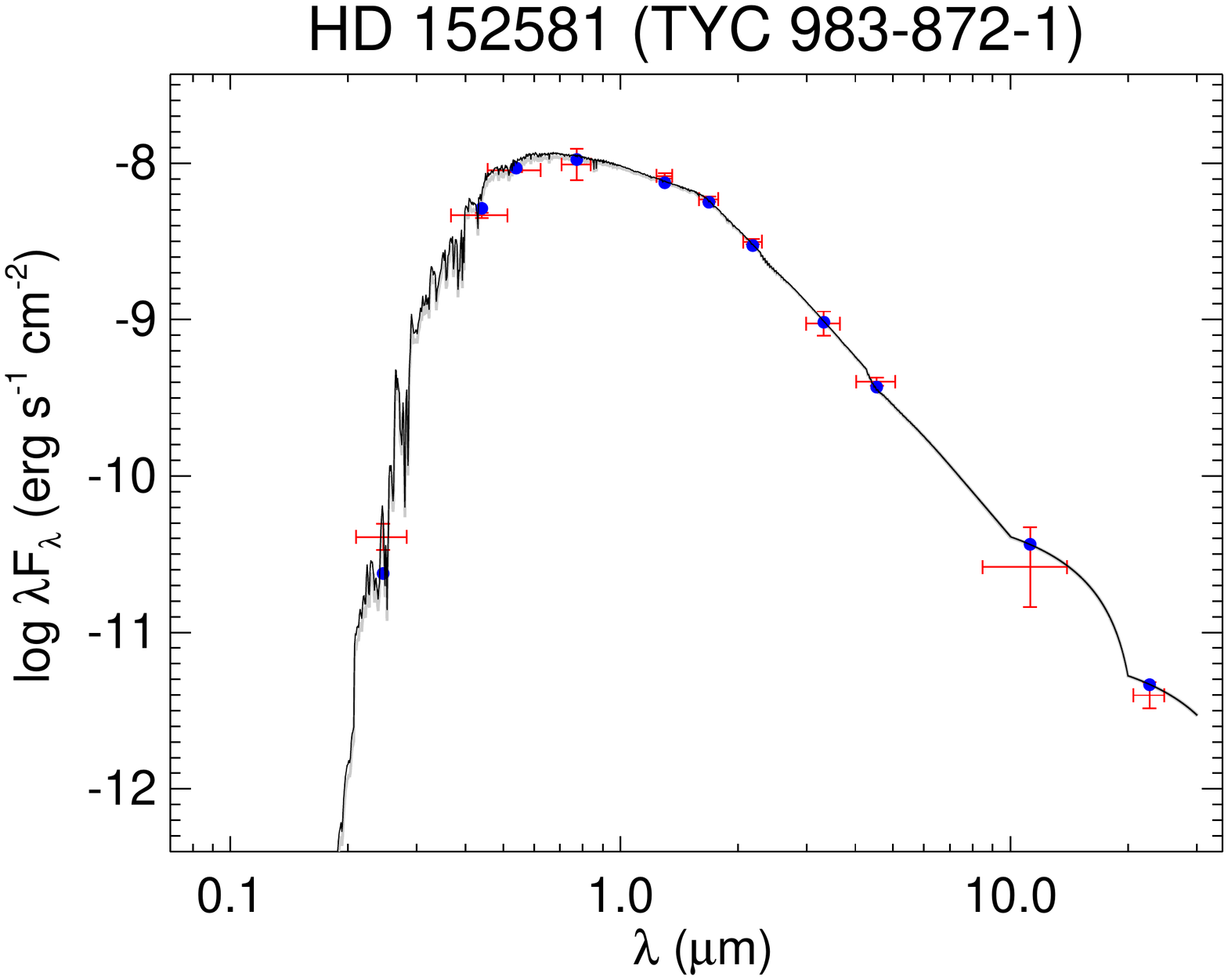}
  \includegraphics[trim=60 60 60 60,clip,width=0.49\linewidth]{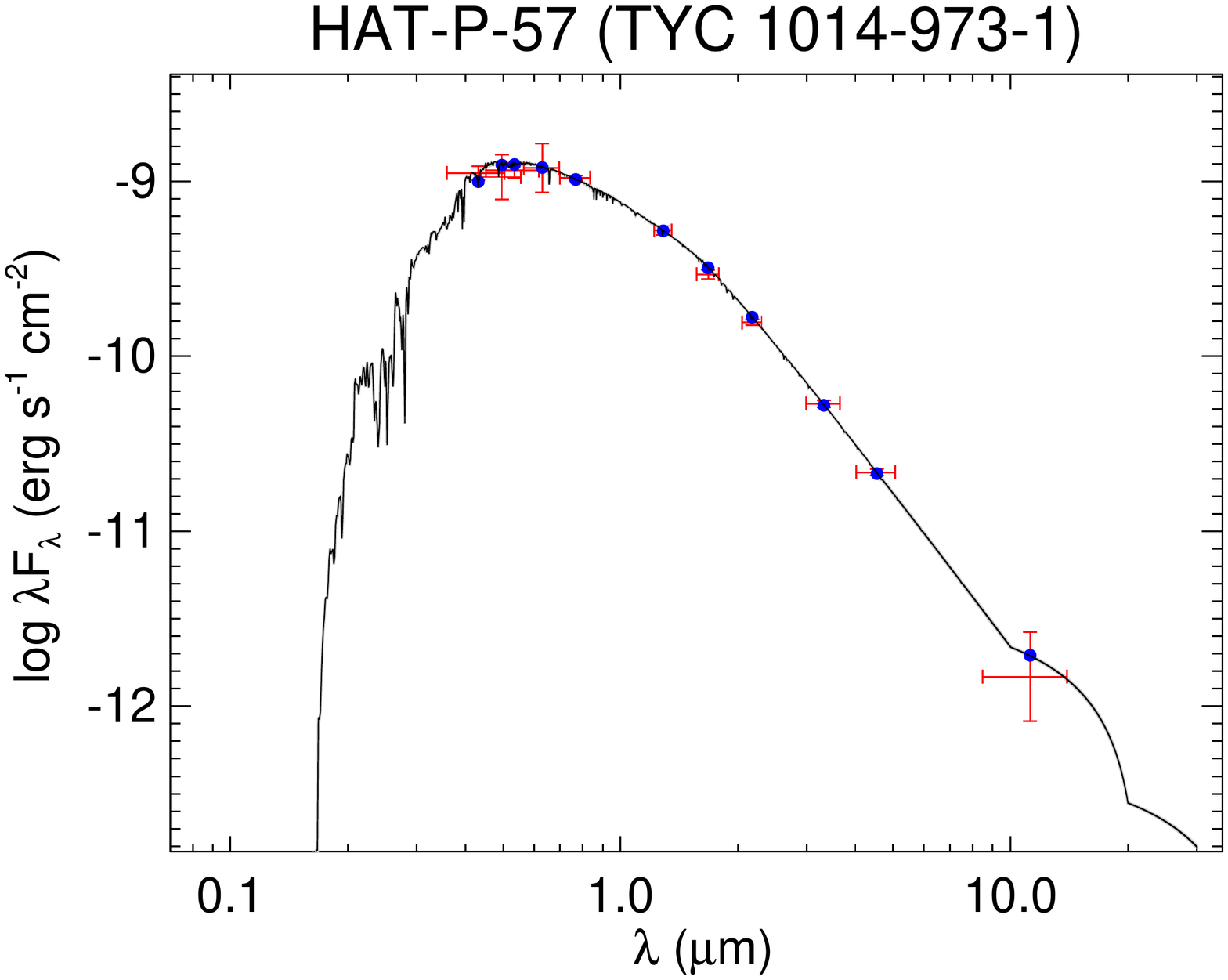}
  \includegraphics[trim=60 60 60 60,clip,width=0.49\linewidth]{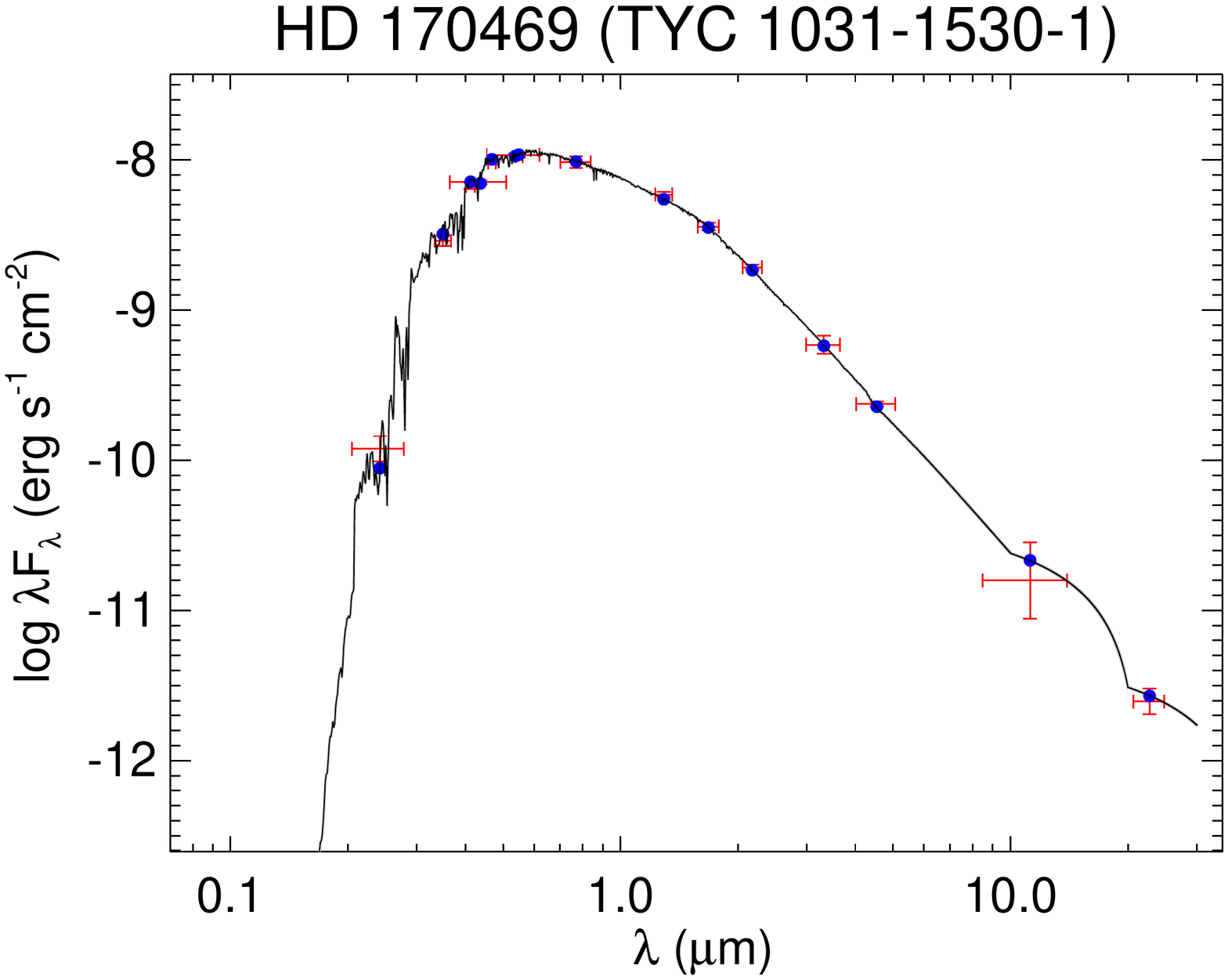}
  \includegraphics[trim=60 60 60 60,clip,width=0.49\linewidth]{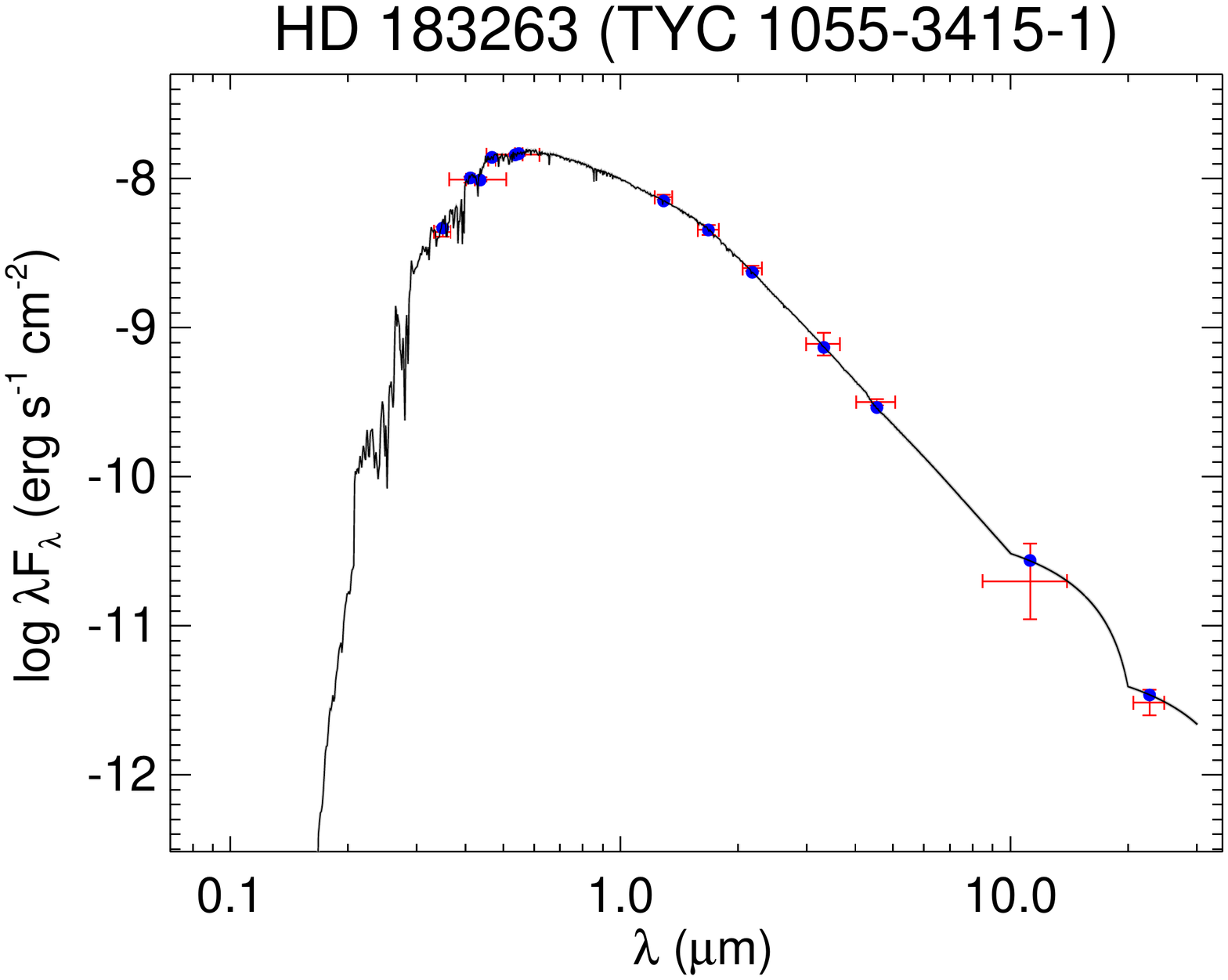}
  \includegraphics[trim=60 60 60 60,clip,width=0.49\linewidth]{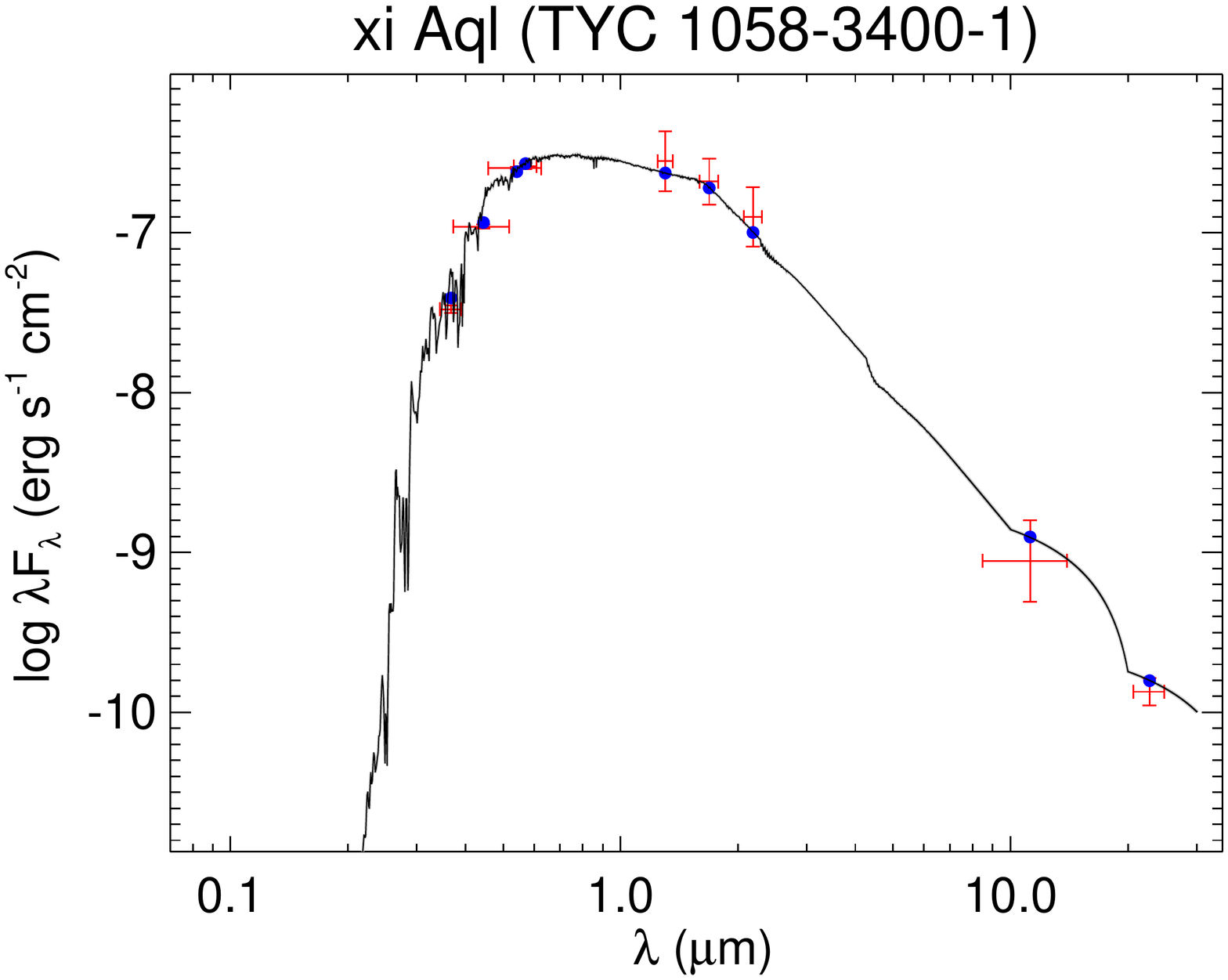}
  \includegraphics[trim=60 60 60 60,clip,width=0.49\linewidth]{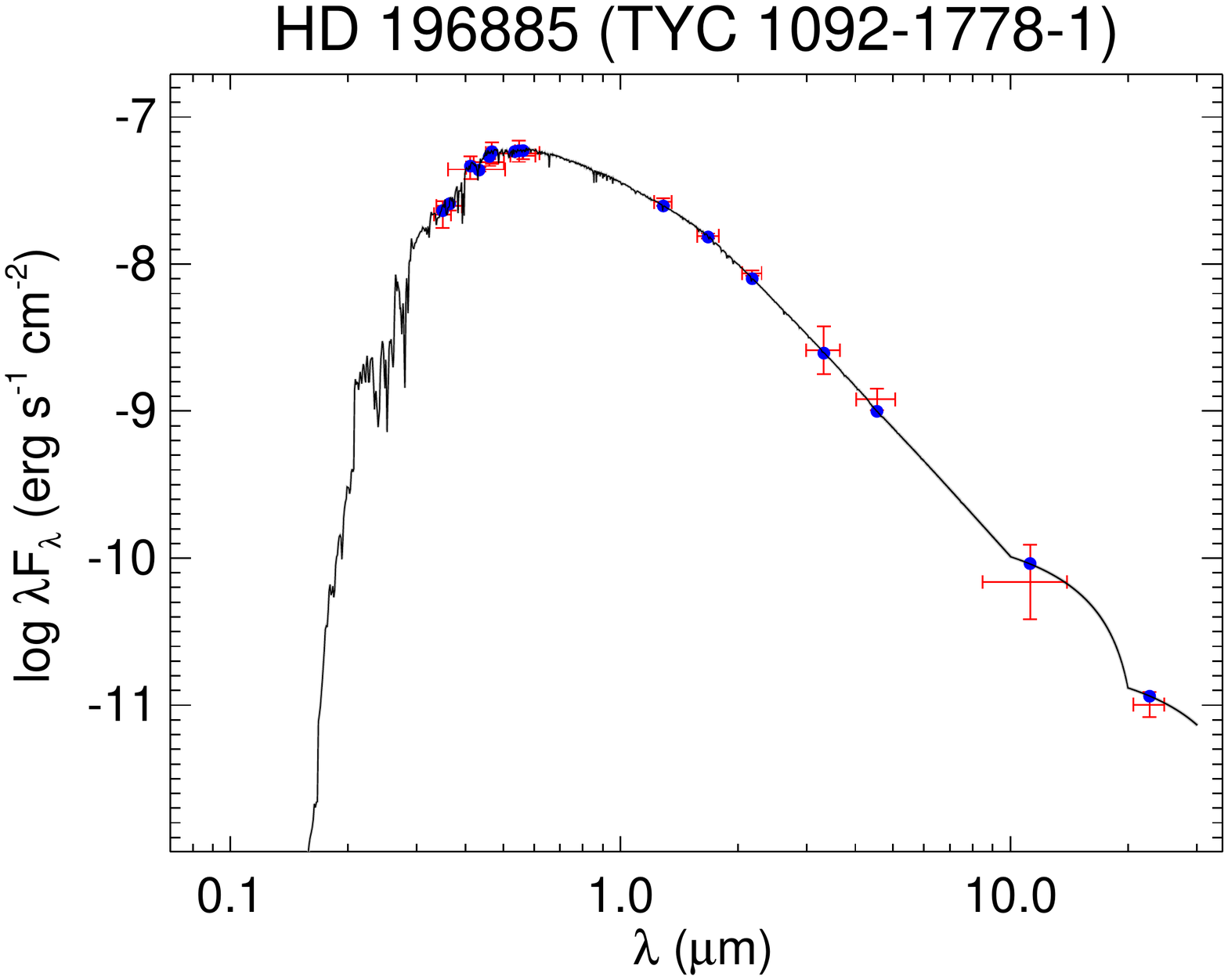}
  \caption{All labels, lines, symbols, and colors as in Figure \ref{fig:seds}.}
  \label{fig:seds_9}
\end{figure}

\begin{figure}[H]
  \centering
  \includegraphics[trim=60 60 60 60,clip,width=0.49\linewidth]{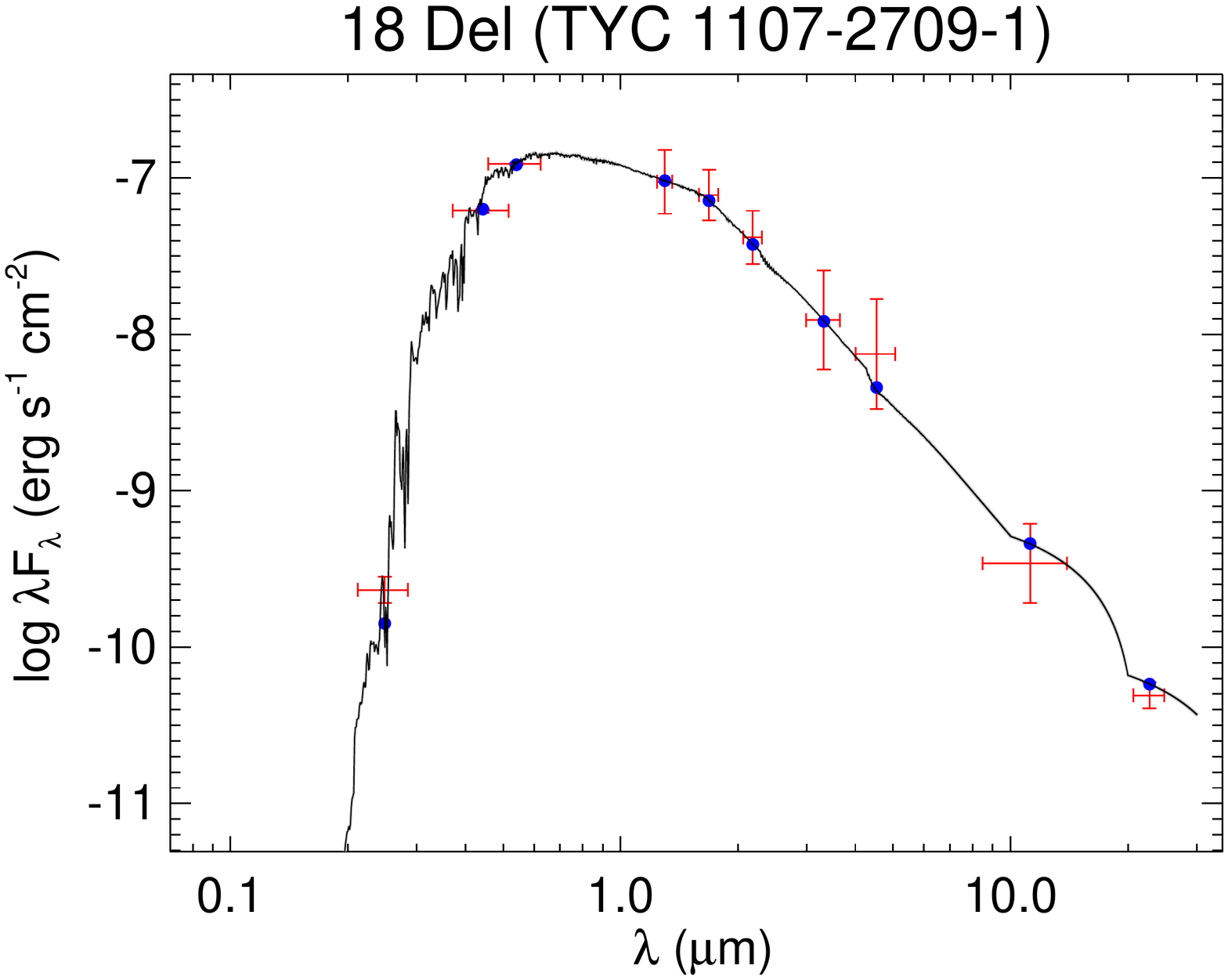}
  \includegraphics[trim=60 60 60 60,clip,width=0.49\linewidth]{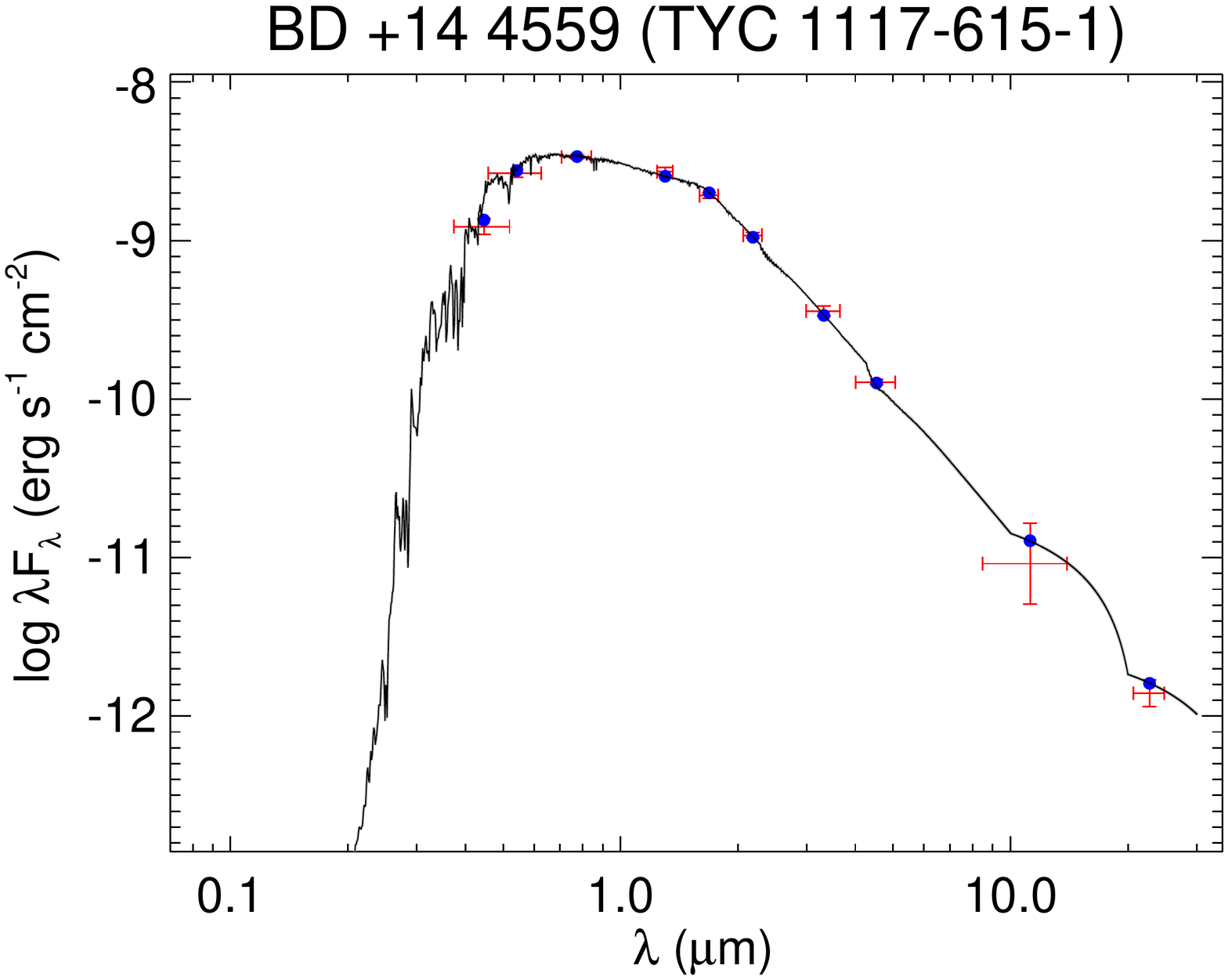}
  \includegraphics[trim=60 60 60 60,clip,width=0.49\linewidth]{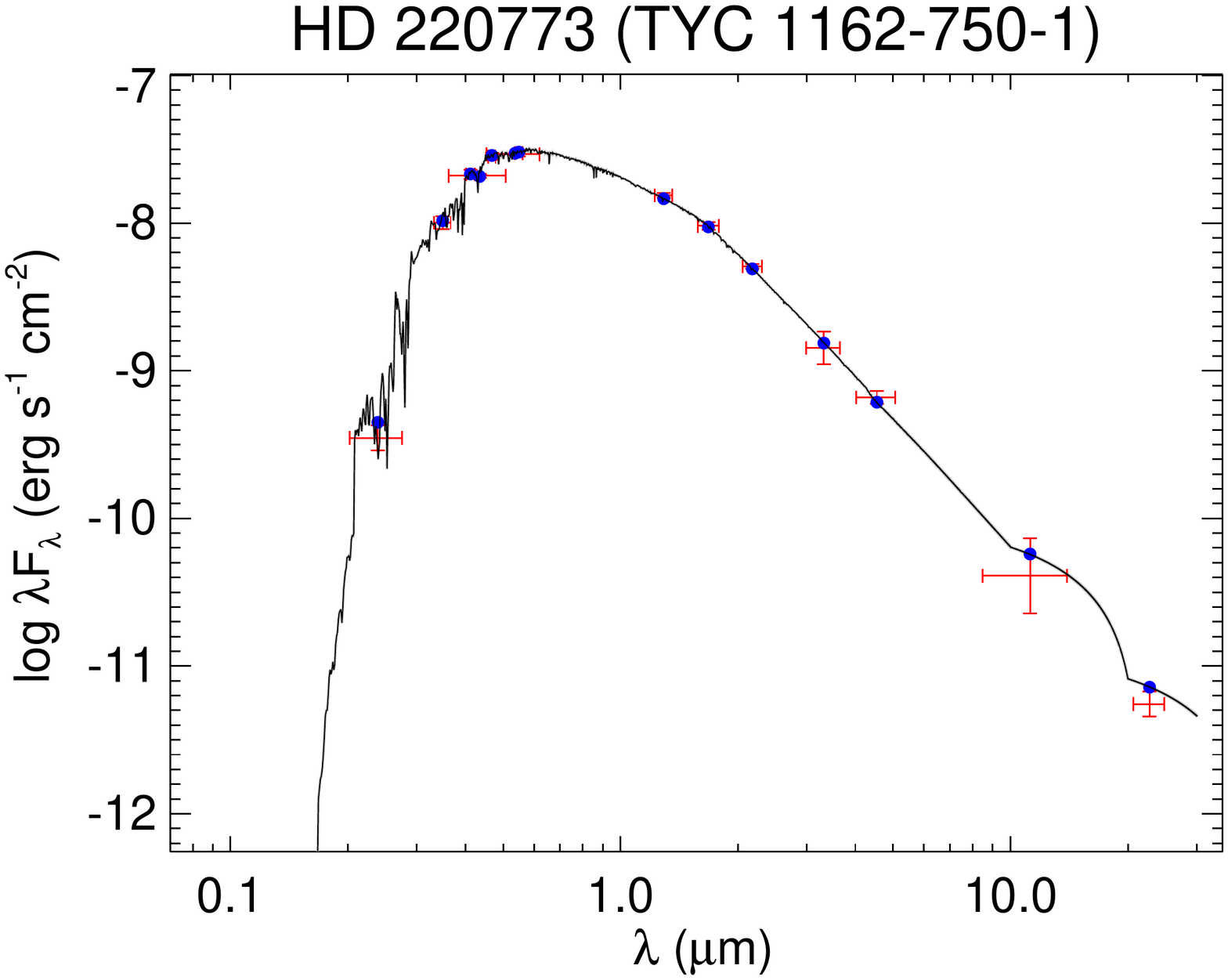}
  \includegraphics[trim=60 60 60 60,clip,width=0.49\linewidth]{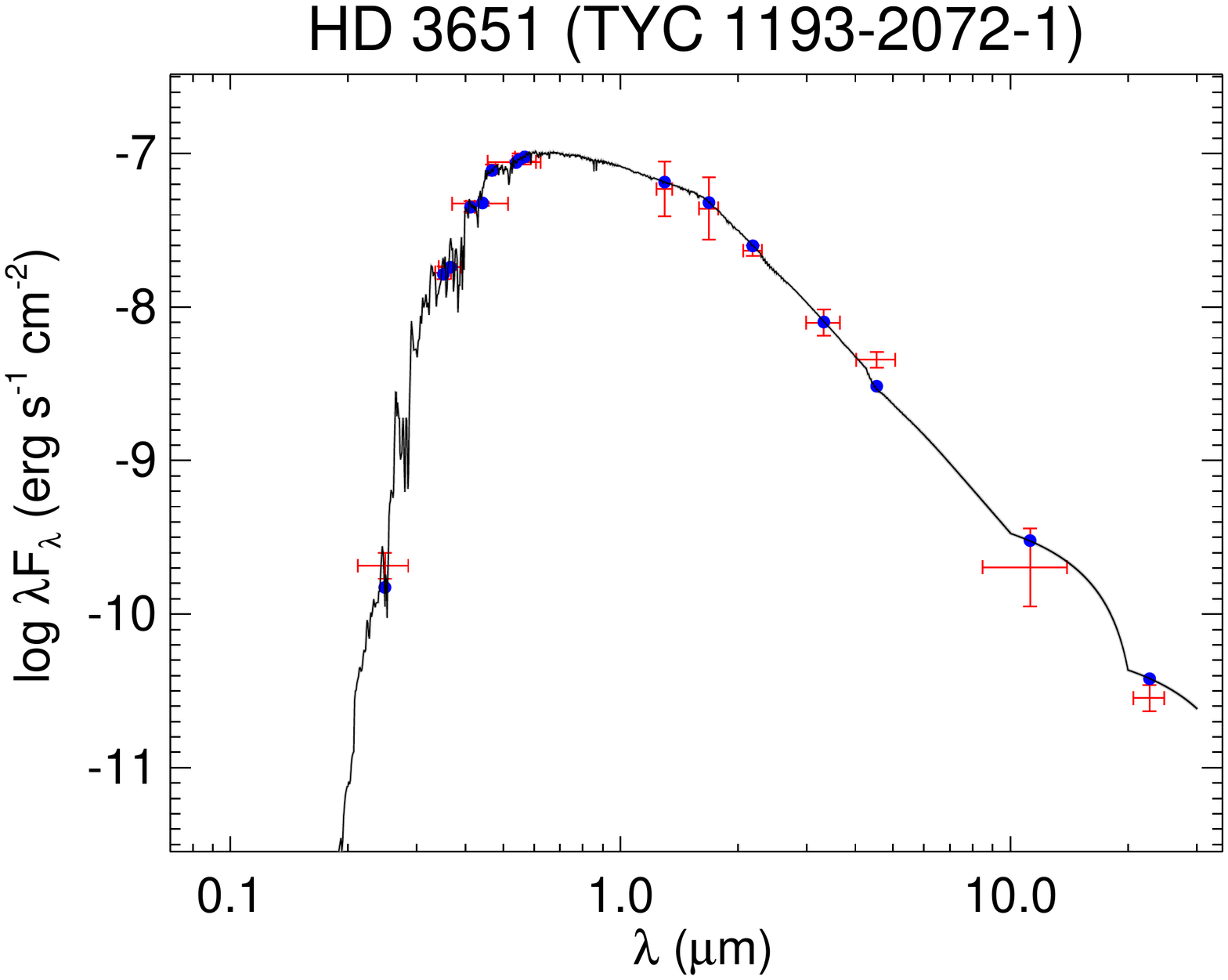}
  \includegraphics[trim=60 60 60 60,clip,width=0.49\linewidth]{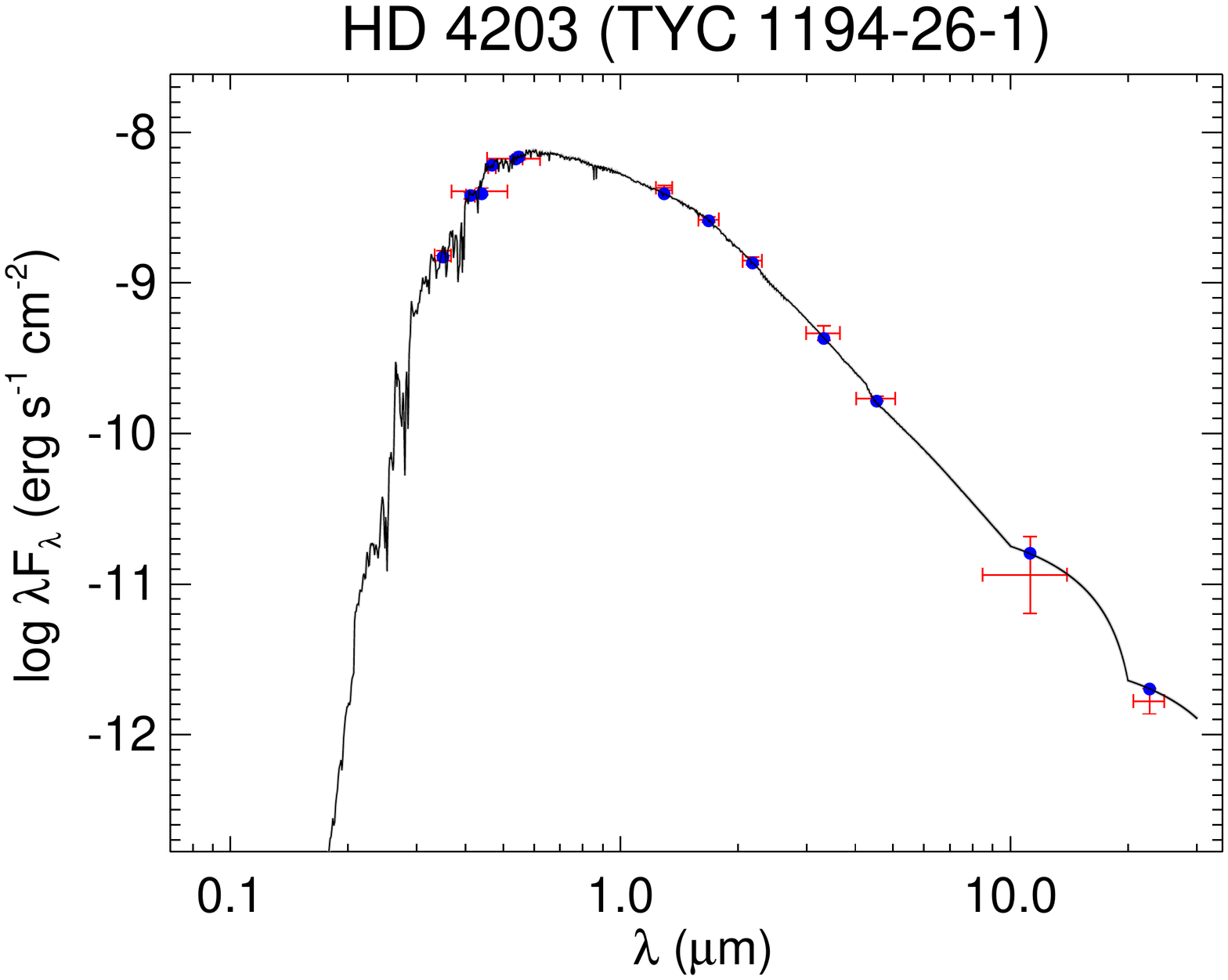}
  \includegraphics[trim=60 60 60 60,clip,width=0.49\linewidth]{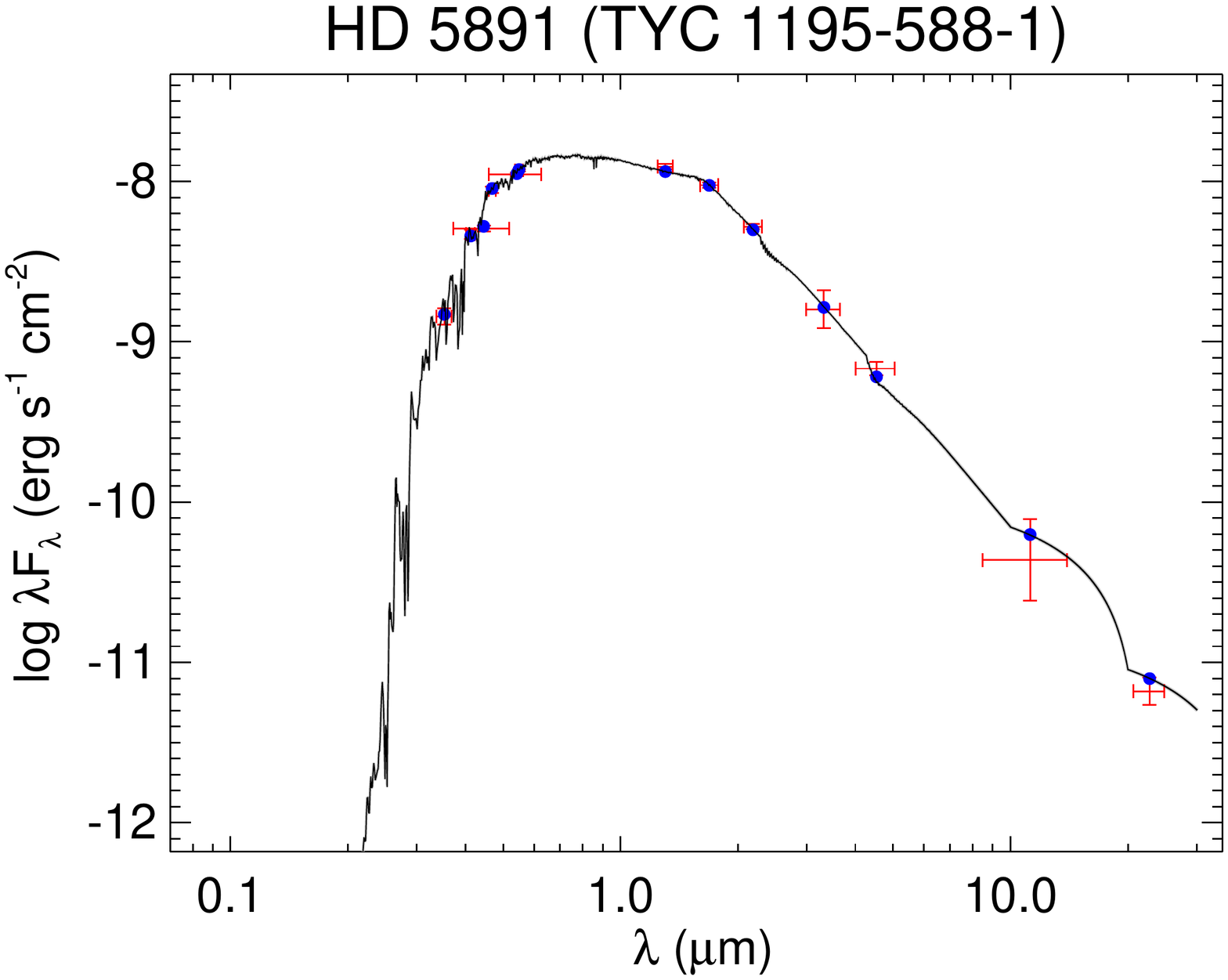}
  \caption{All labels, lines, symbols, and colors as in Figure \ref{fig:seds}.}
  \label{fig:seds_10}
\end{figure}

\begin{figure}[H]
  \centering
  \includegraphics[trim=60 60 60 60,clip,width=0.49\linewidth]{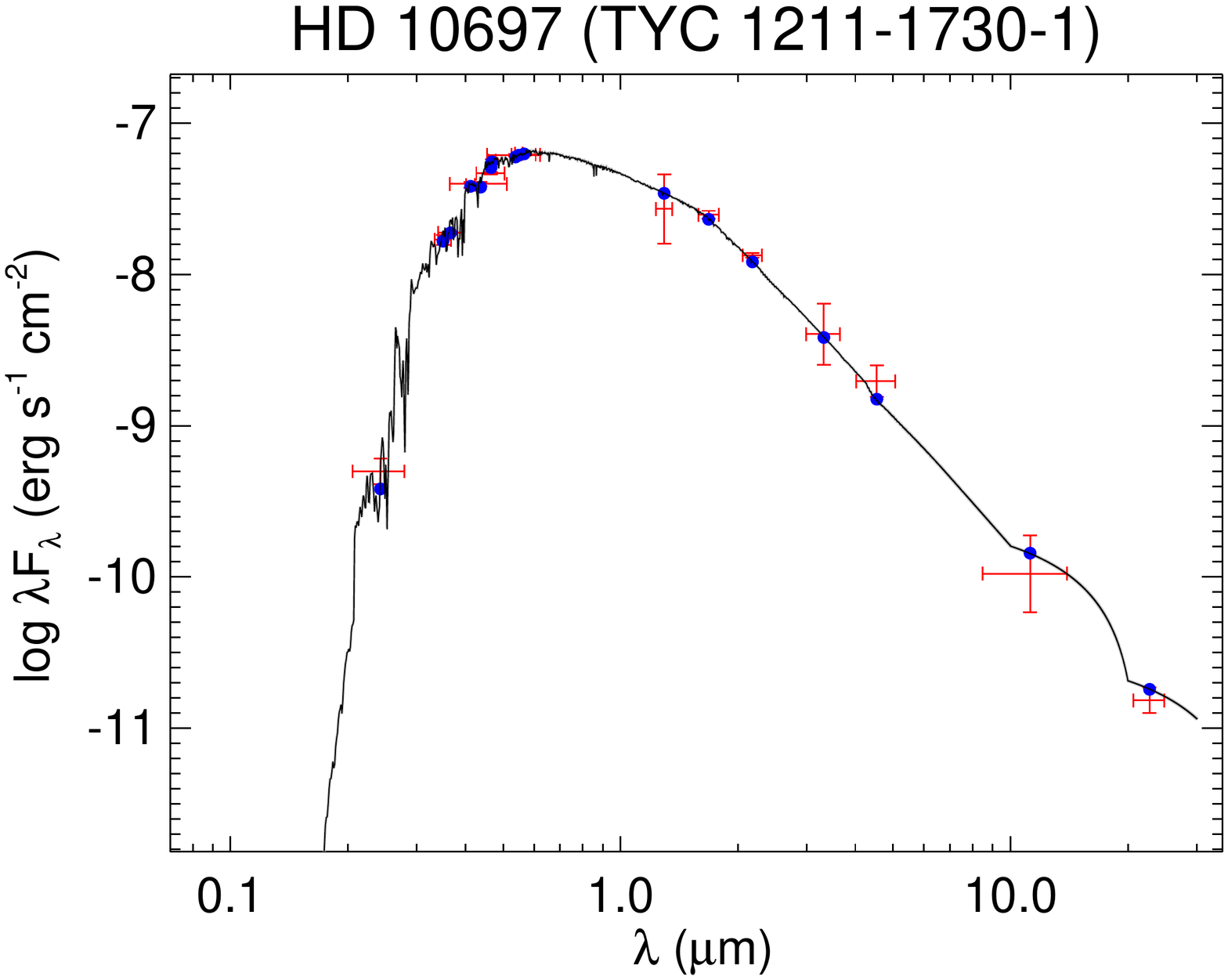}
  \includegraphics[trim=60 60 60 60,clip,width=0.49\linewidth]{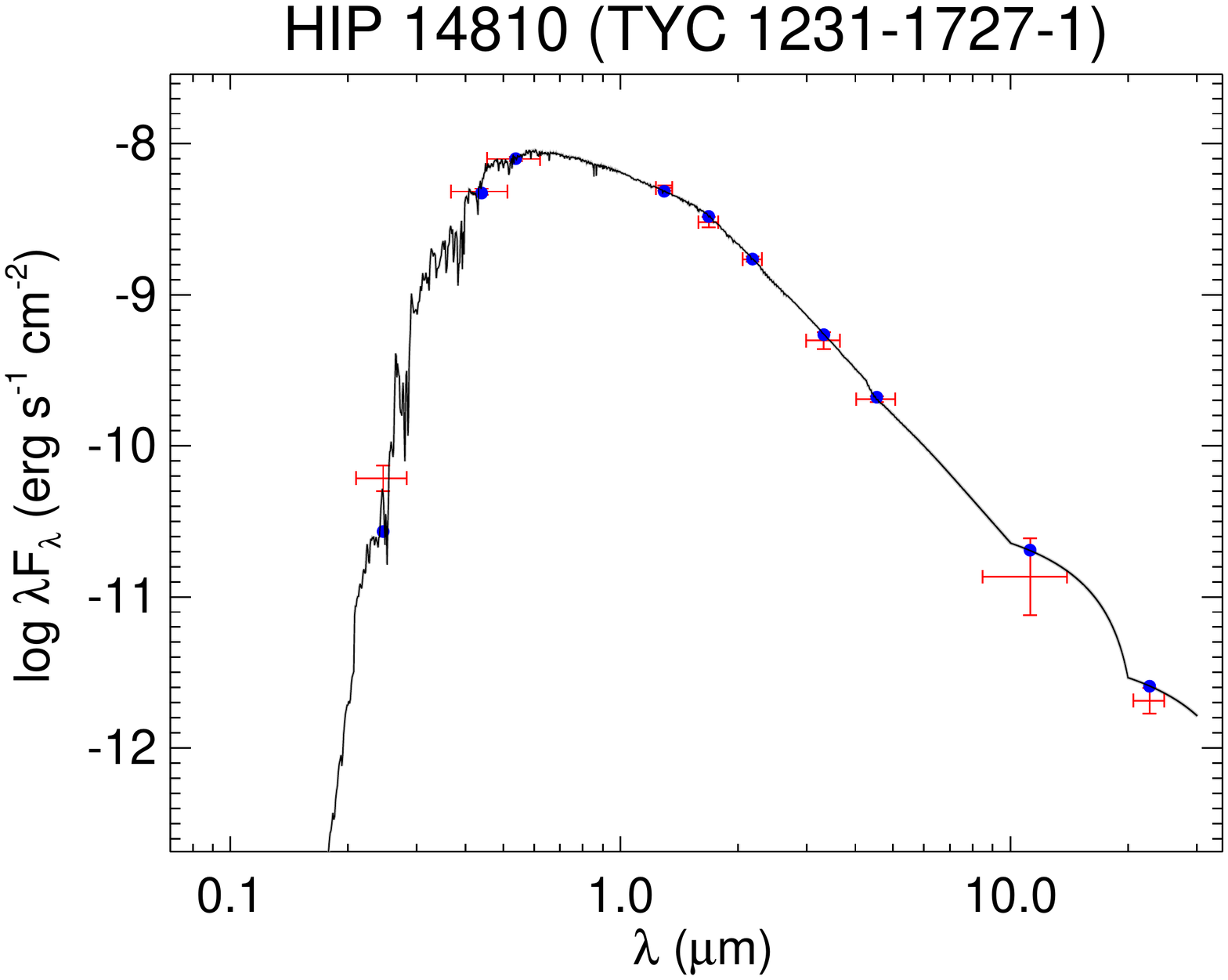}
  \includegraphics[trim=60 60 60 60,clip,width=0.49\linewidth]{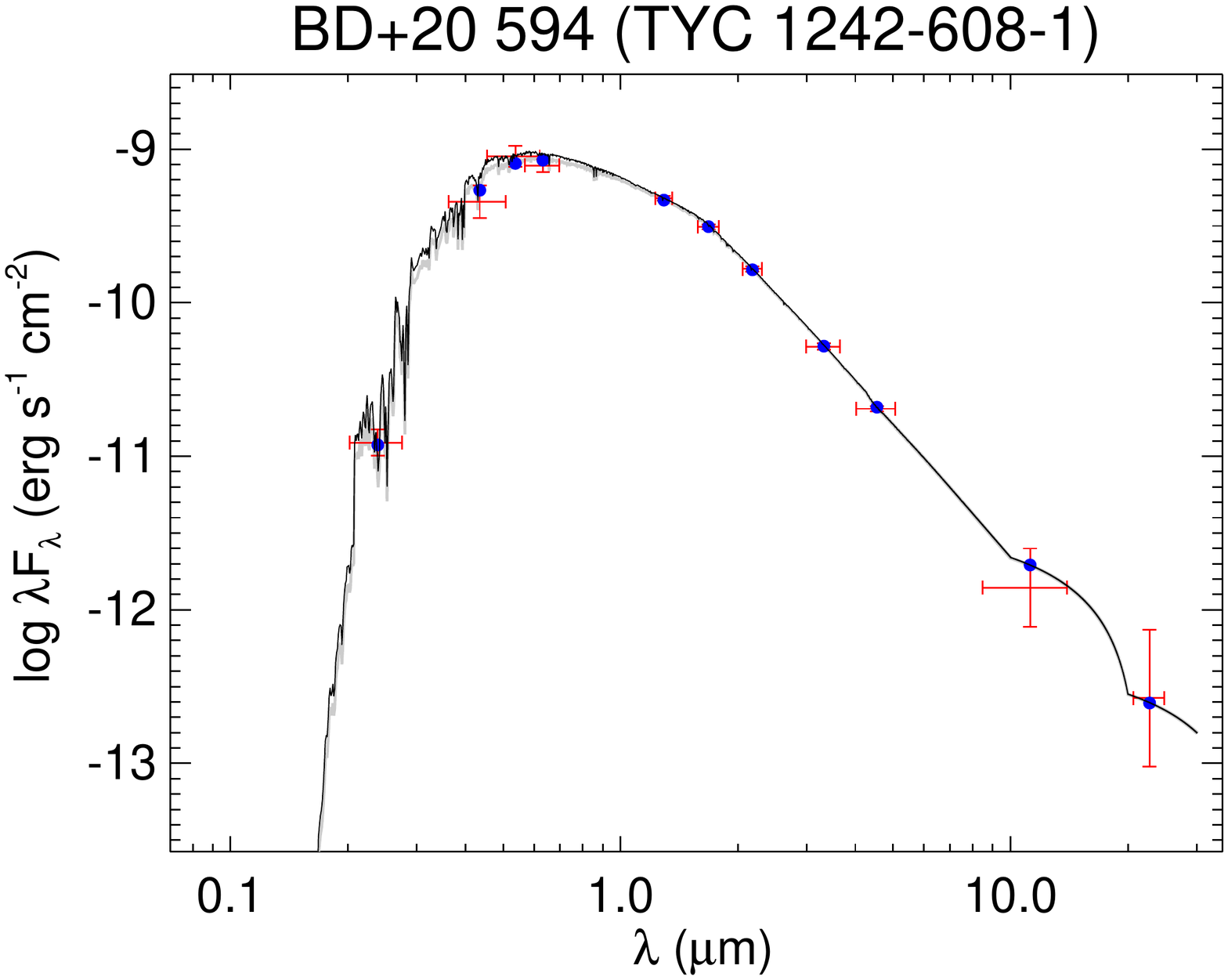}
  \includegraphics[trim=60 60 60 60,clip,width=0.49\linewidth]{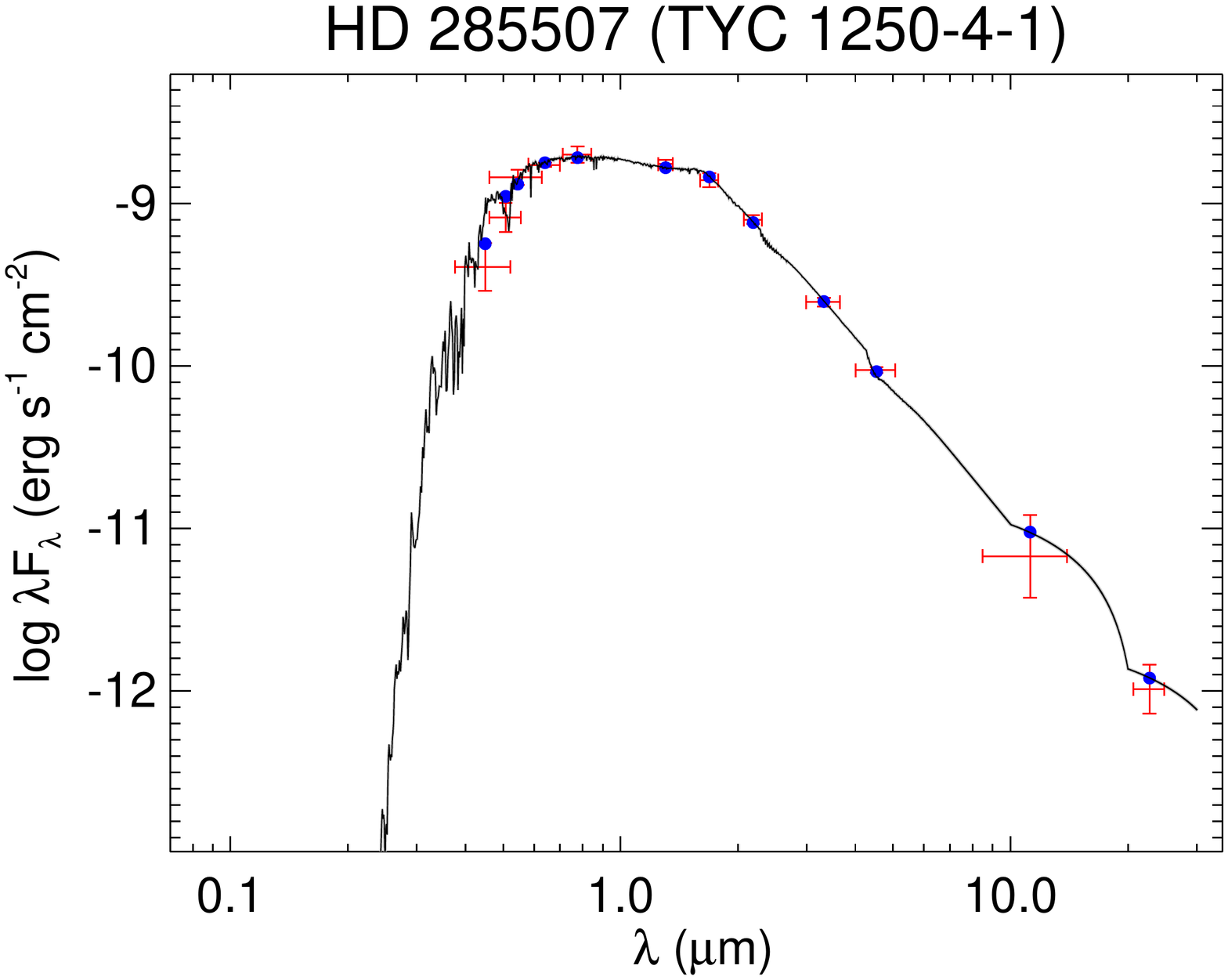}
  \includegraphics[trim=60 60 60 60,clip,width=0.49\linewidth]{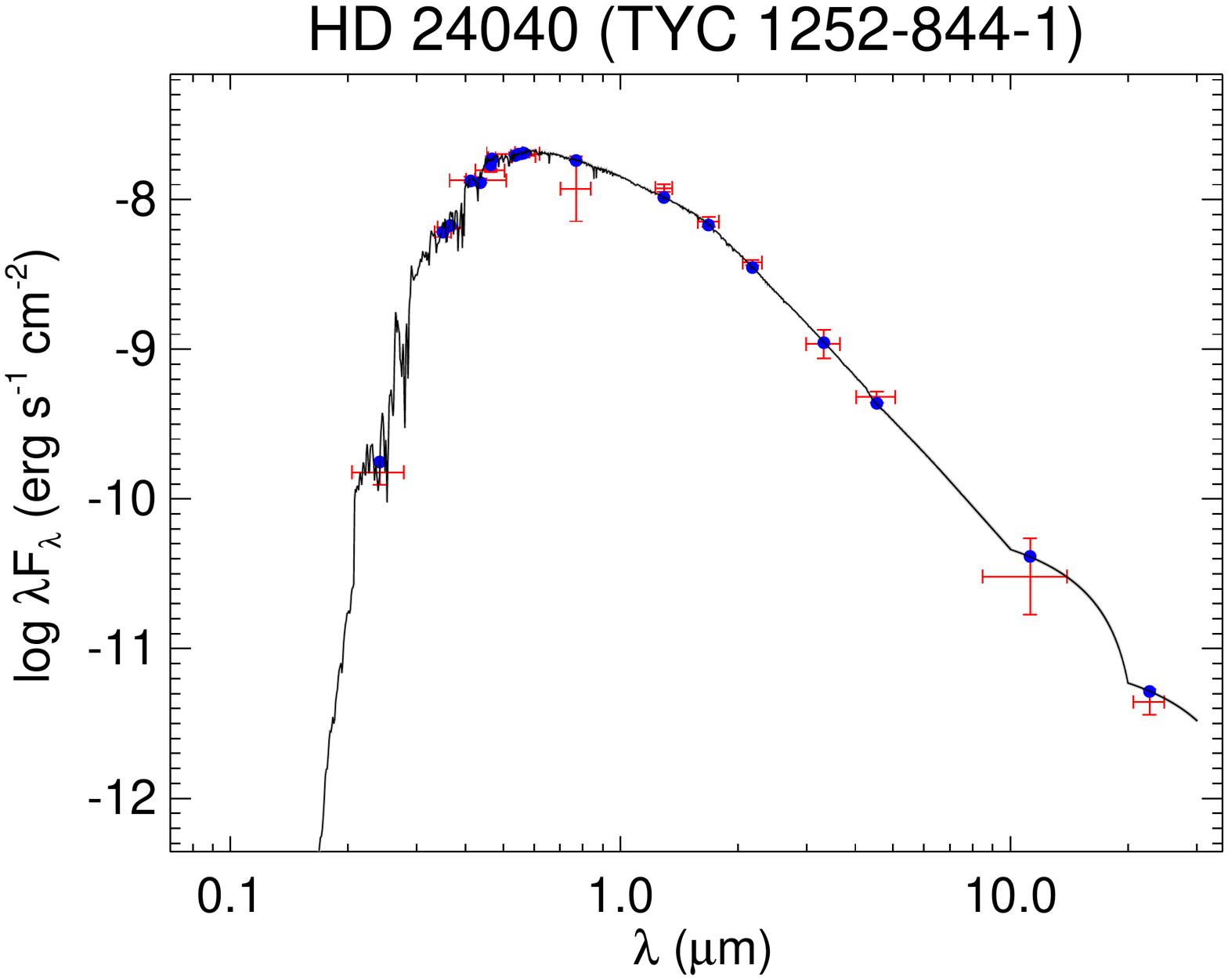}
  \includegraphics[trim=60 60 60 60,clip,width=0.49\linewidth]{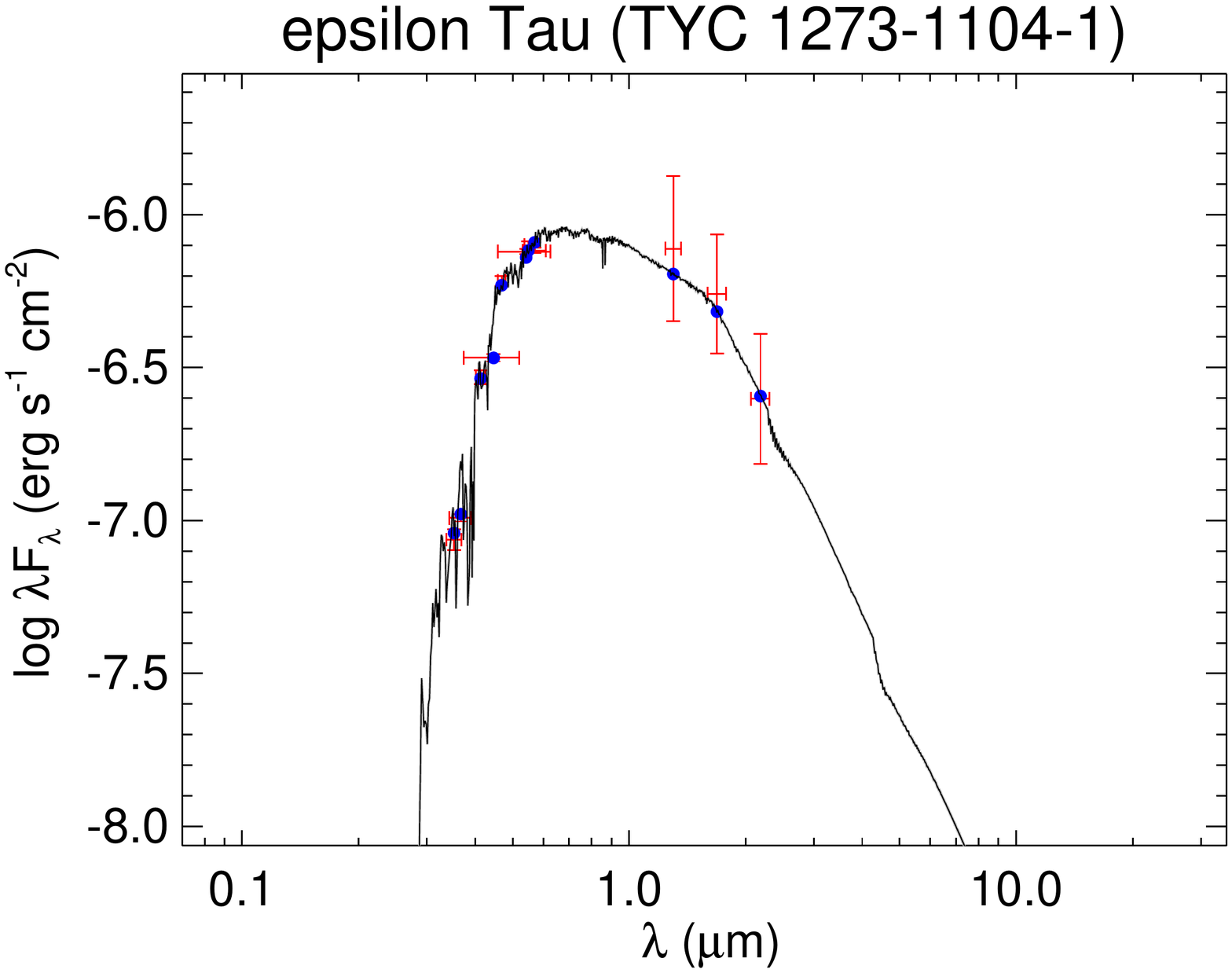}
  \caption{All labels, lines, symbols, and colors as in Figure \ref{fig:seds}.}
  \label{fig:seds_11}
\end{figure}

\begin{figure}[H]
  \centering
  \includegraphics[trim=60 60 60 60,clip,width=0.49\linewidth]{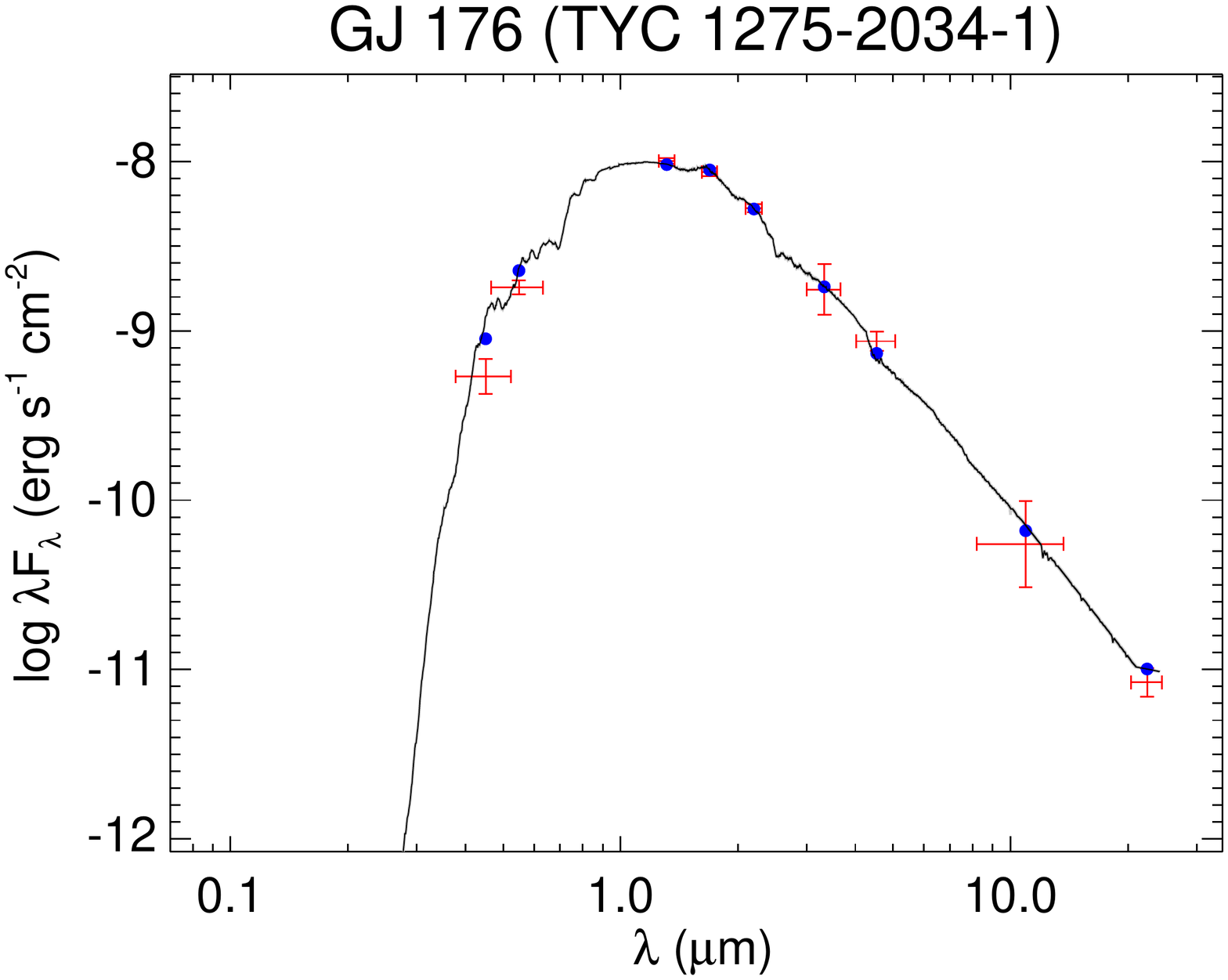}
  \includegraphics[trim=60 60 60 60,clip,width=0.49\linewidth]{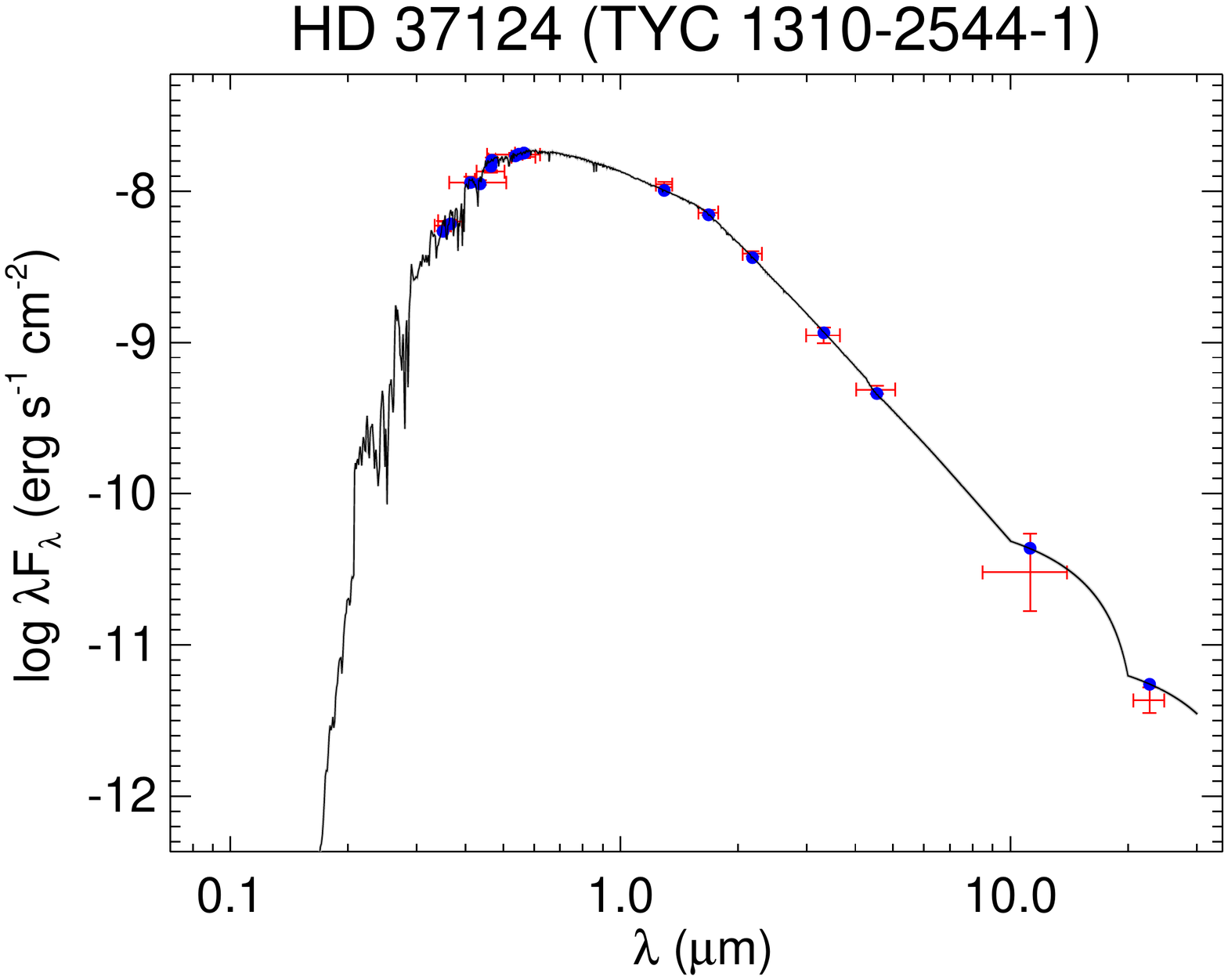}
  \includegraphics[trim=60 60 60 60,clip,width=0.49\linewidth]{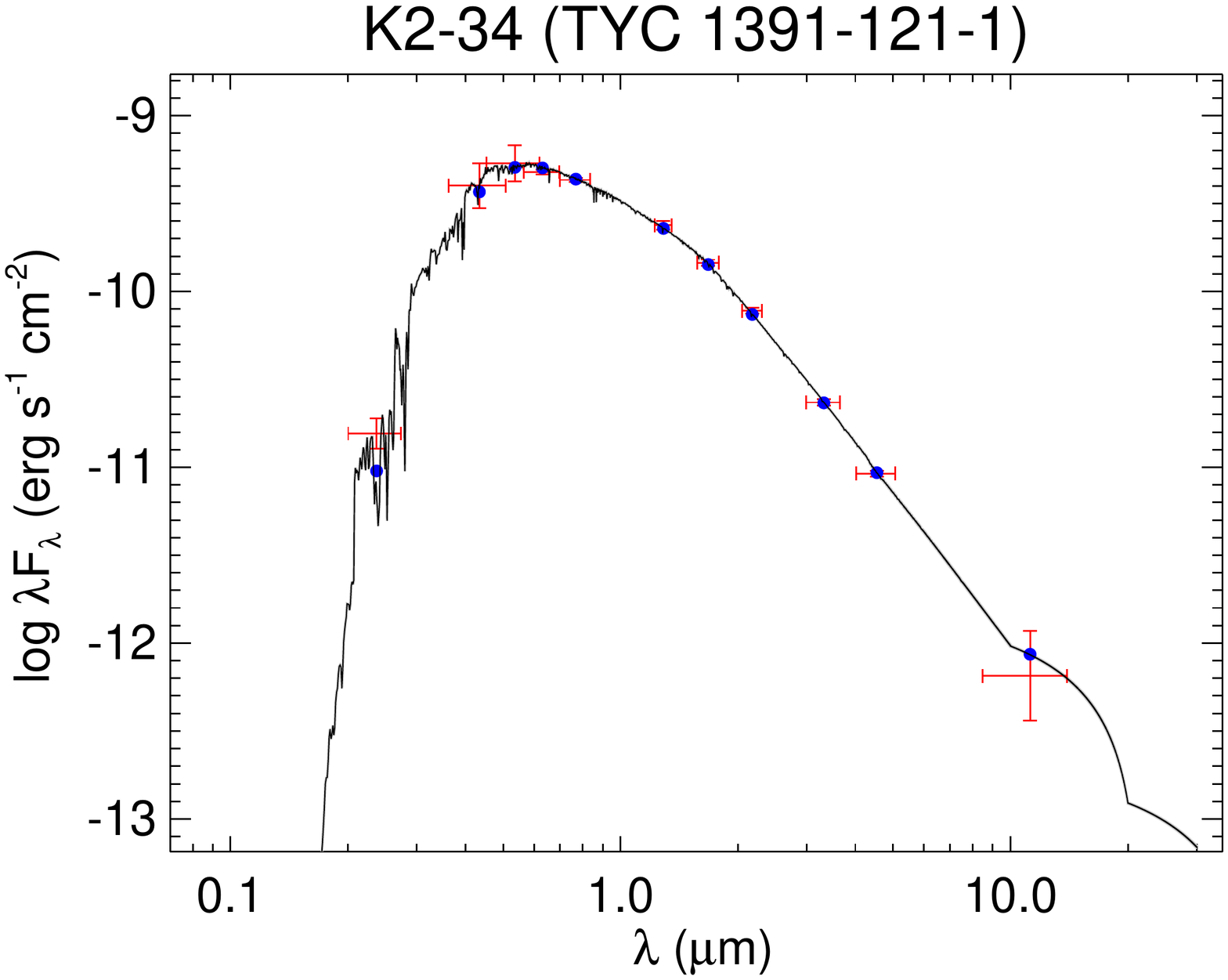}
  \includegraphics[trim=60 60 60 60,clip,width=0.49\linewidth]{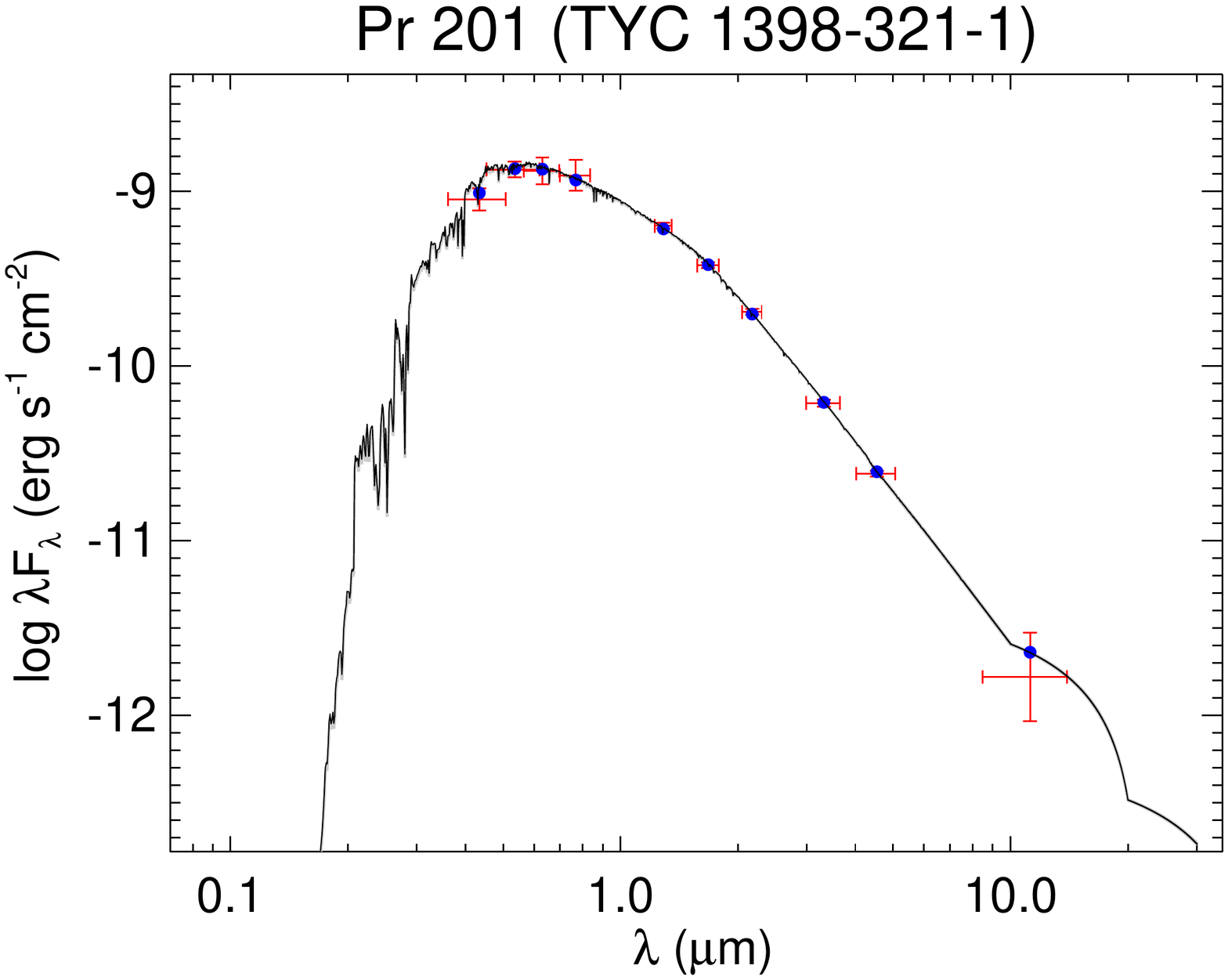}
  \includegraphics[trim=60 60 60 60,clip,width=0.49\linewidth]{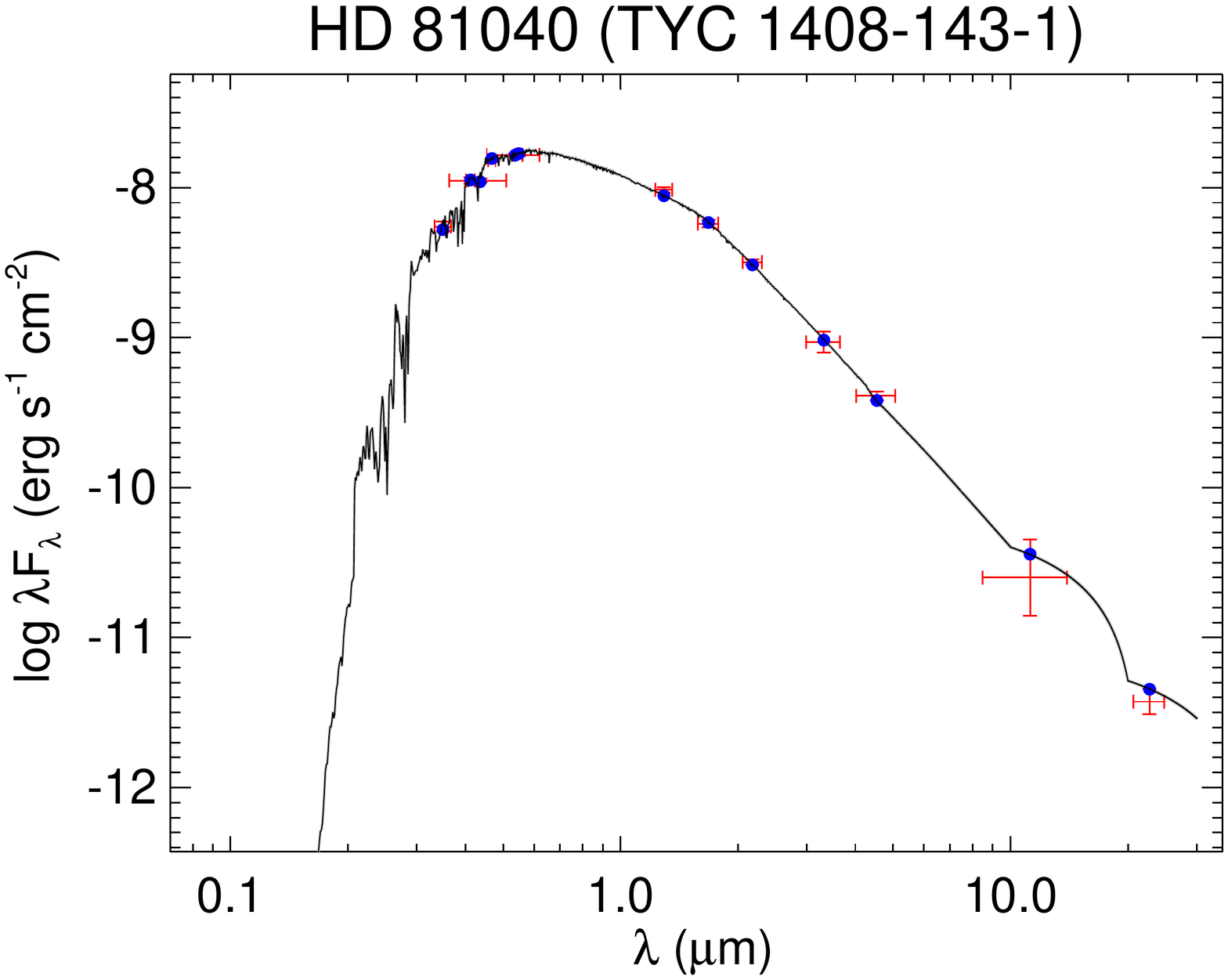}
  \includegraphics[trim=60 60 60 60,clip,width=0.49\linewidth]{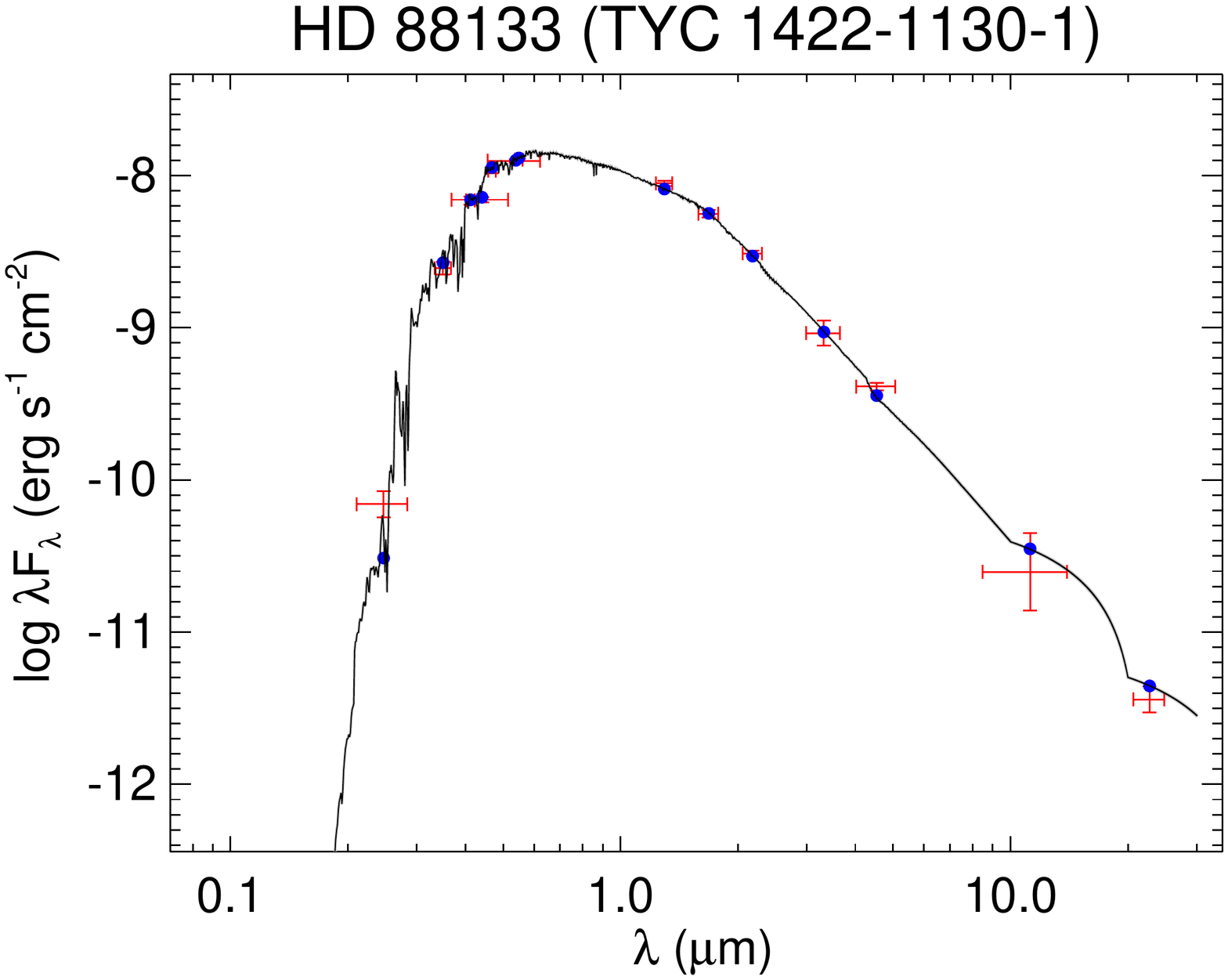}
  \caption{All labels, lines, symbols, and colors as in Figure \ref{fig:seds}.}
  \label{fig:seds_12}
\end{figure}

\begin{figure}[H]
  \centering
  \includegraphics[trim=60 60 60 60,clip,width=0.49\linewidth]{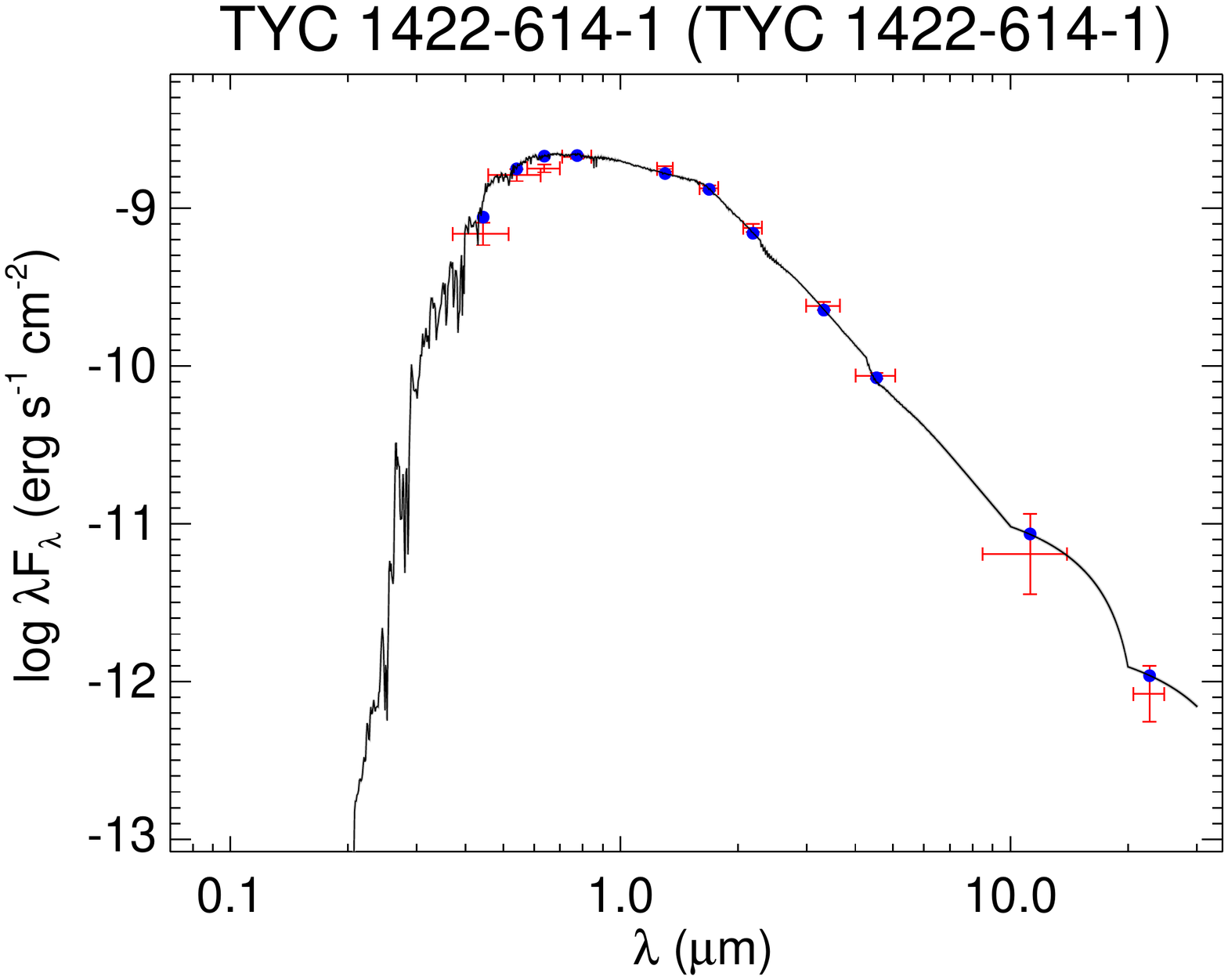}
  \includegraphics[trim=60 60 60 60,clip,width=0.49\linewidth]{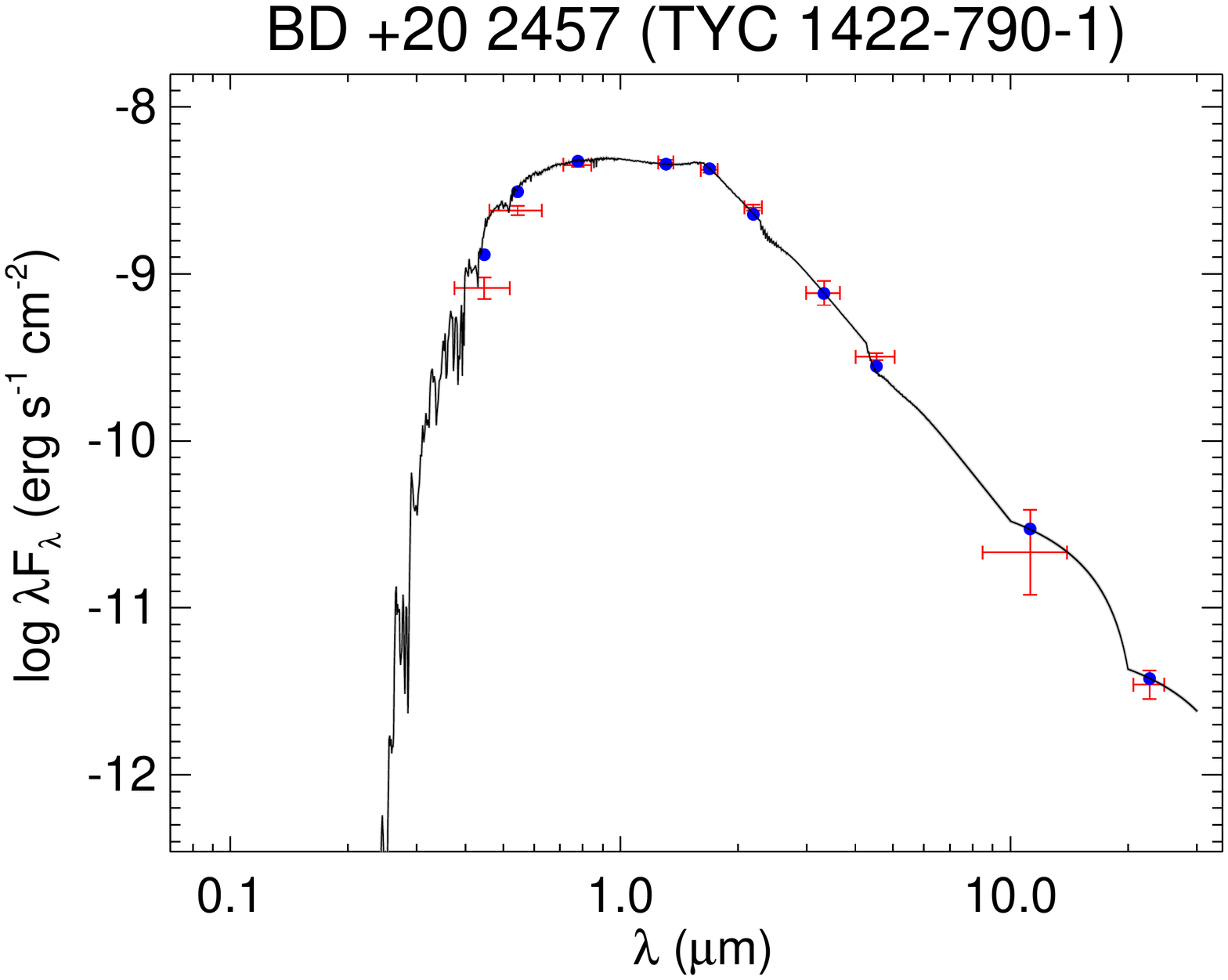}
  \includegraphics[trim=60 60 60 60,clip,width=0.49\linewidth]{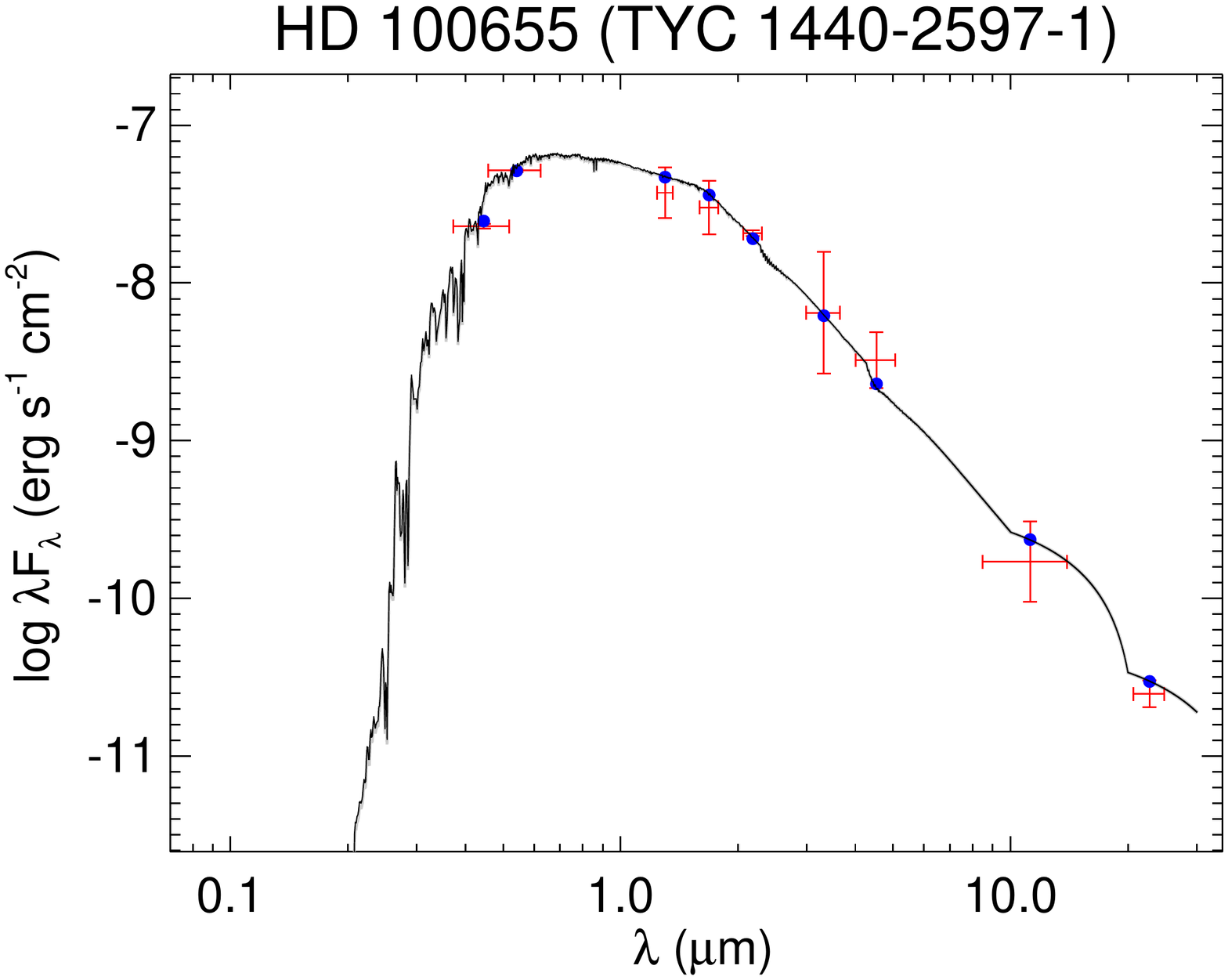}
  \includegraphics[trim=60 60 60 60,clip,width=0.49\linewidth]{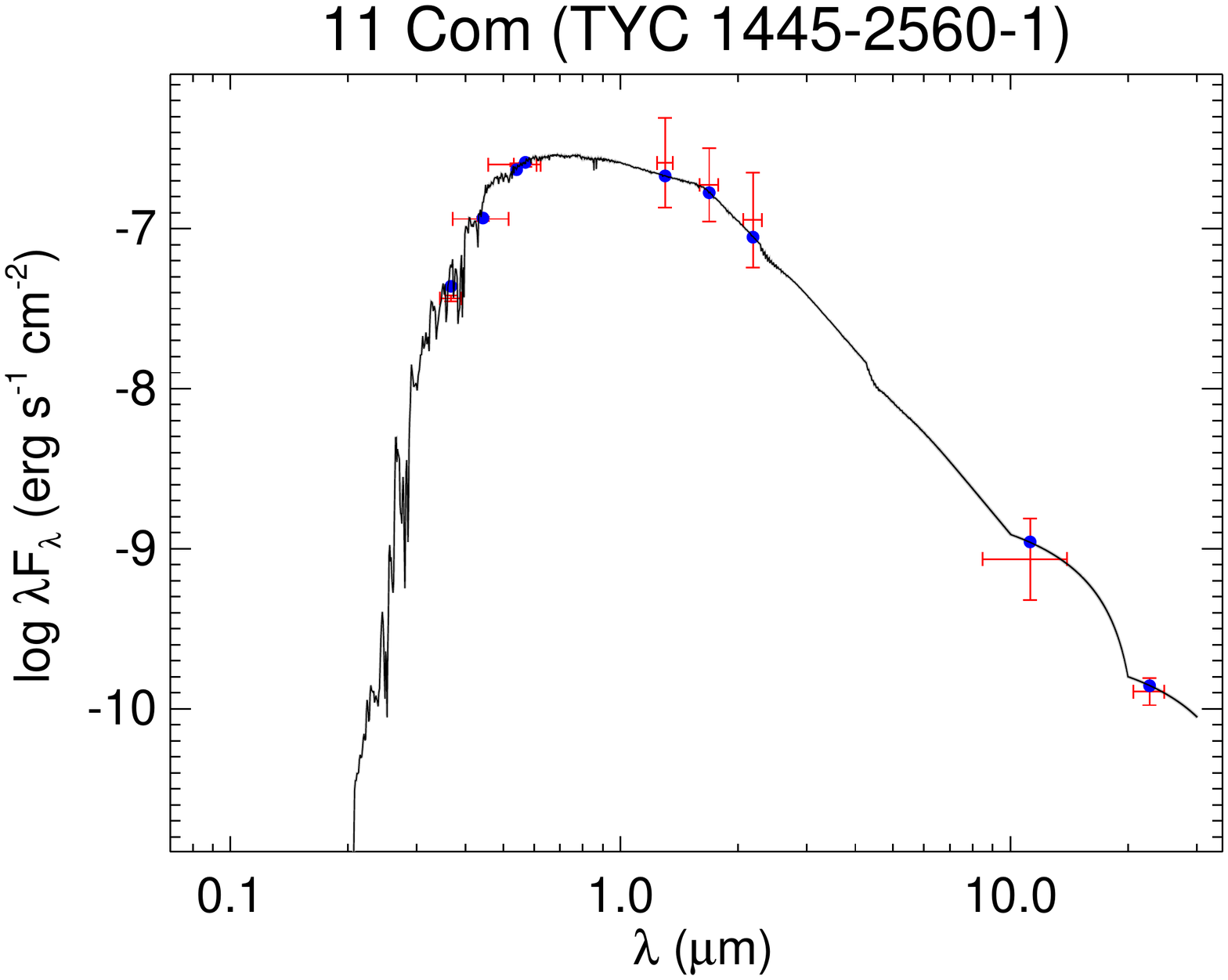}
  \includegraphics[trim=60 60 60 60,clip,width=0.49\linewidth]{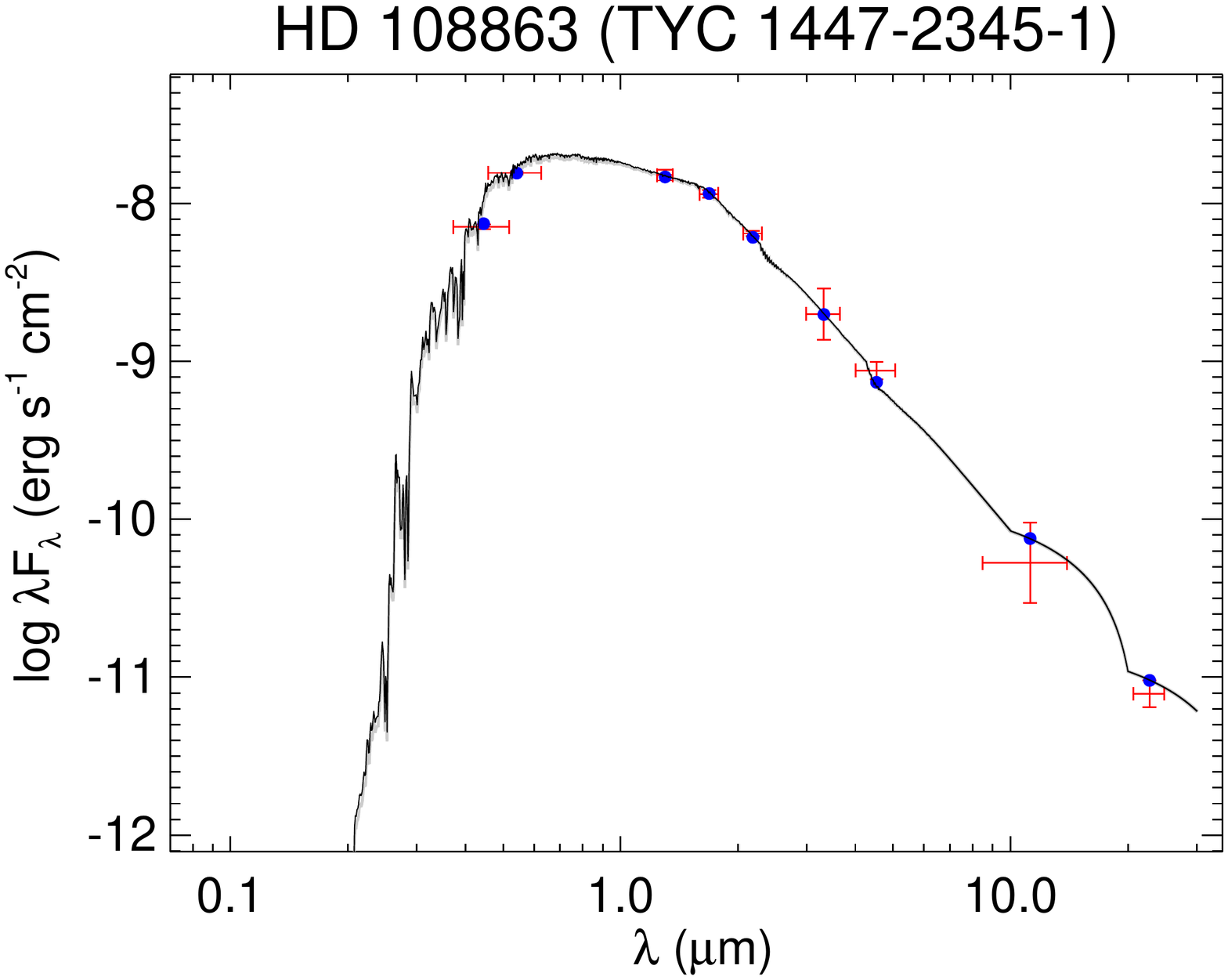}
  \includegraphics[trim=60 60 60 60,clip,width=0.49\linewidth]{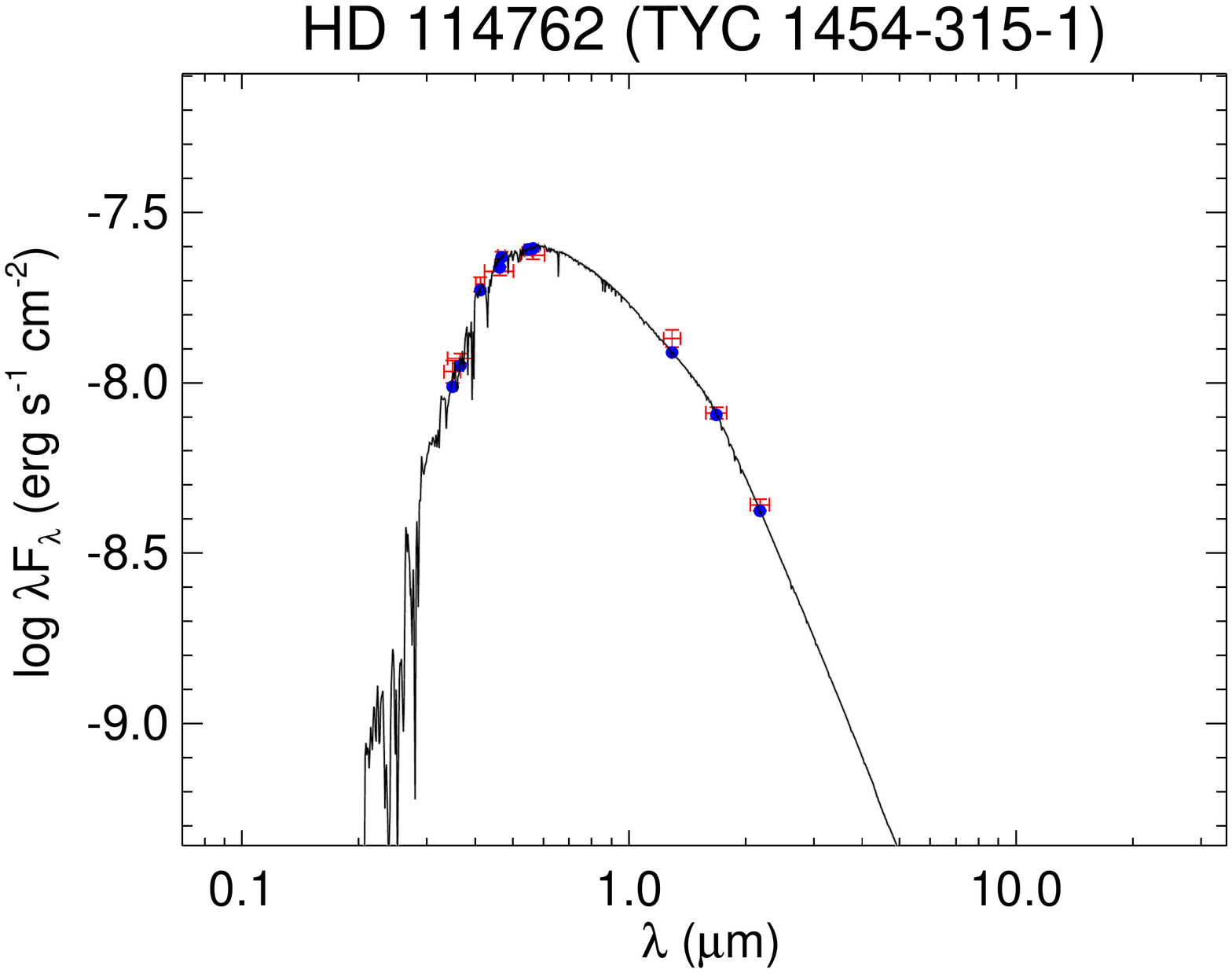}
  \caption{All labels, lines, symbols, and colors as in Figure \ref{fig:seds}.}
  \label{fig:seds_13}
\end{figure}

\begin{figure}[H]
  \centering
  \includegraphics[trim=60 60 60 60,clip,width=0.49\linewidth]{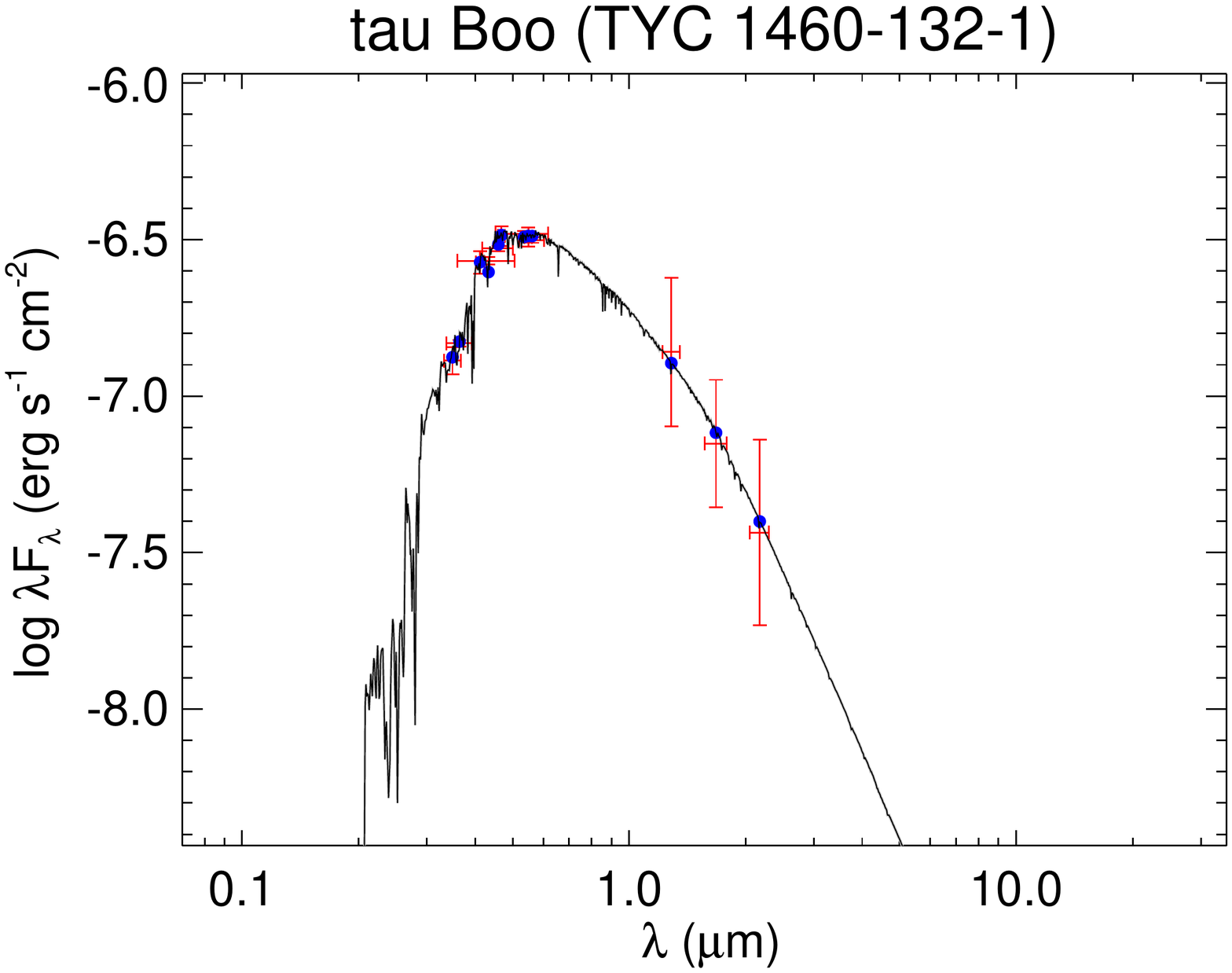}
  \includegraphics[trim=60 60 60 60,clip,width=0.49\linewidth]{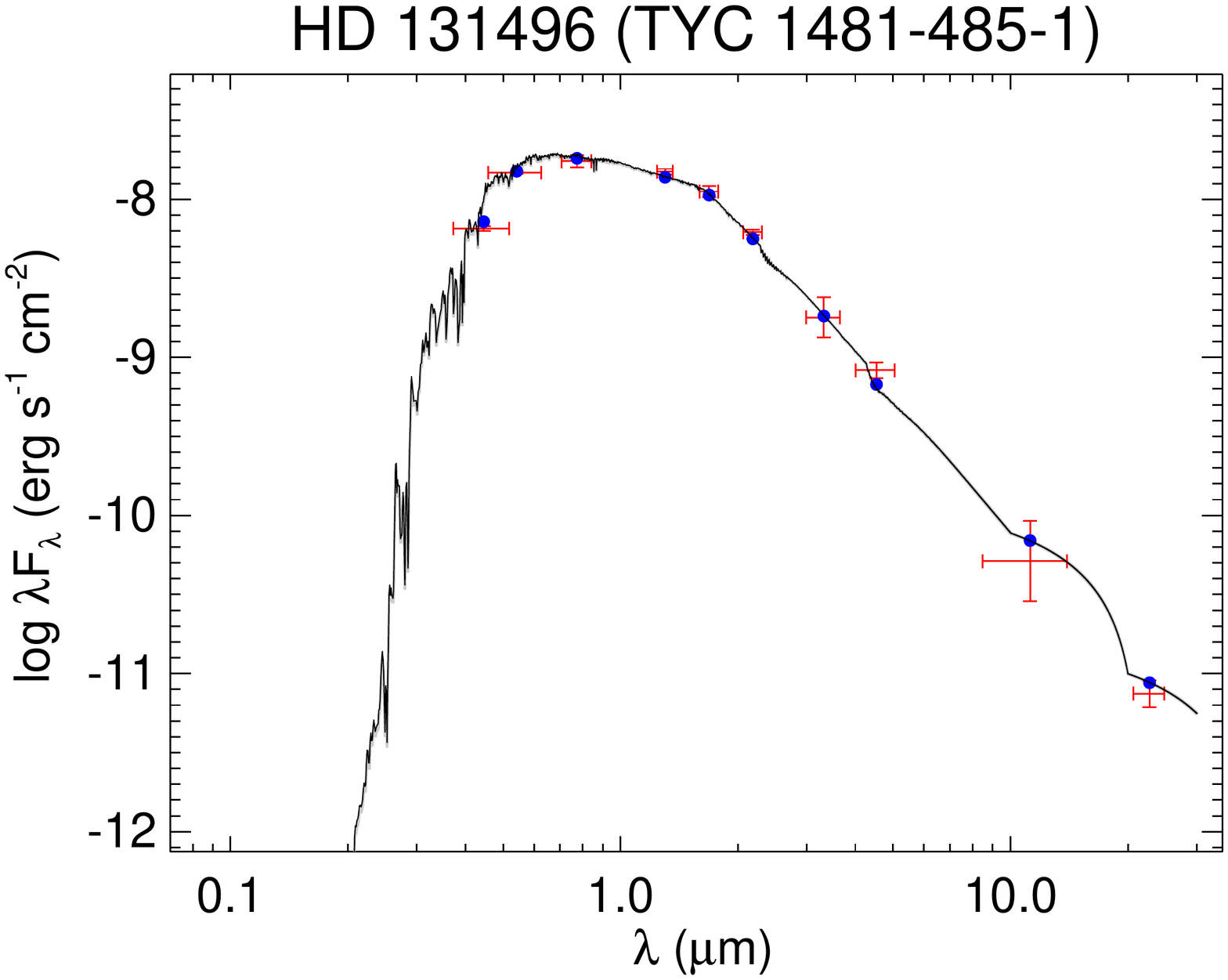}
  \includegraphics[trim=60 60 60 60,clip,width=0.49\linewidth]{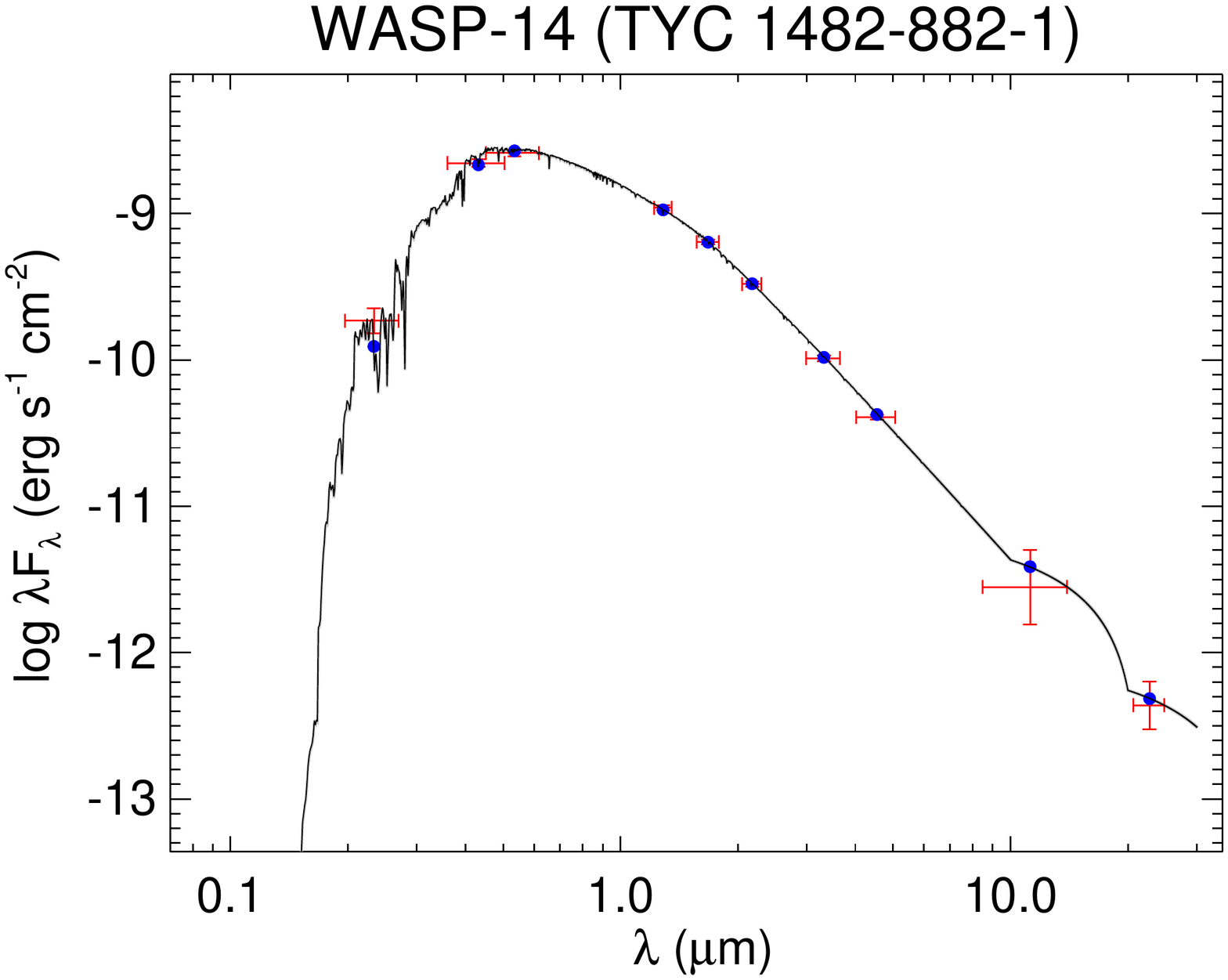}
  \includegraphics[trim=60 60 60 60,clip,width=0.49\linewidth]{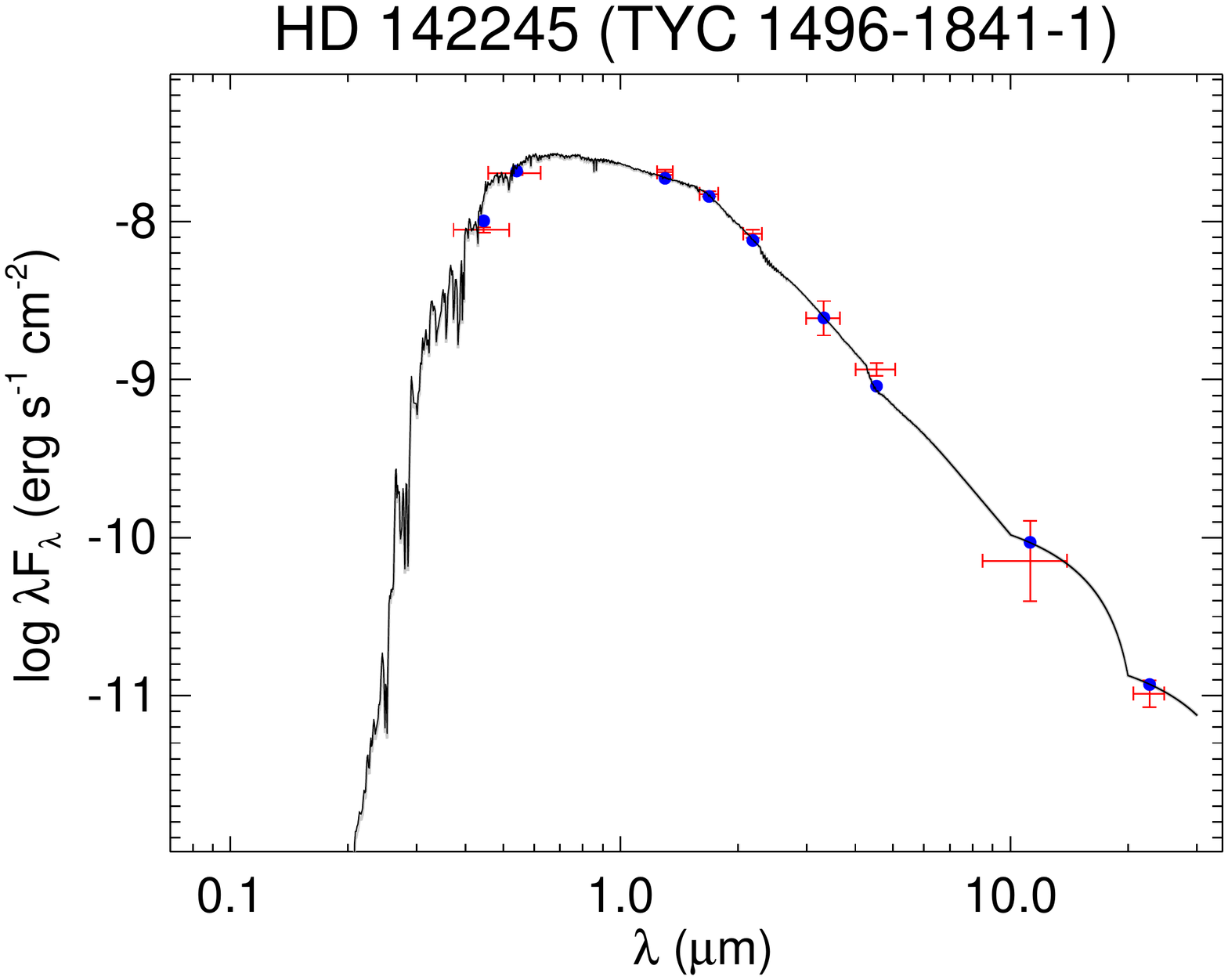}
  \includegraphics[trim=60 60 60 60,clip,width=0.49\linewidth]{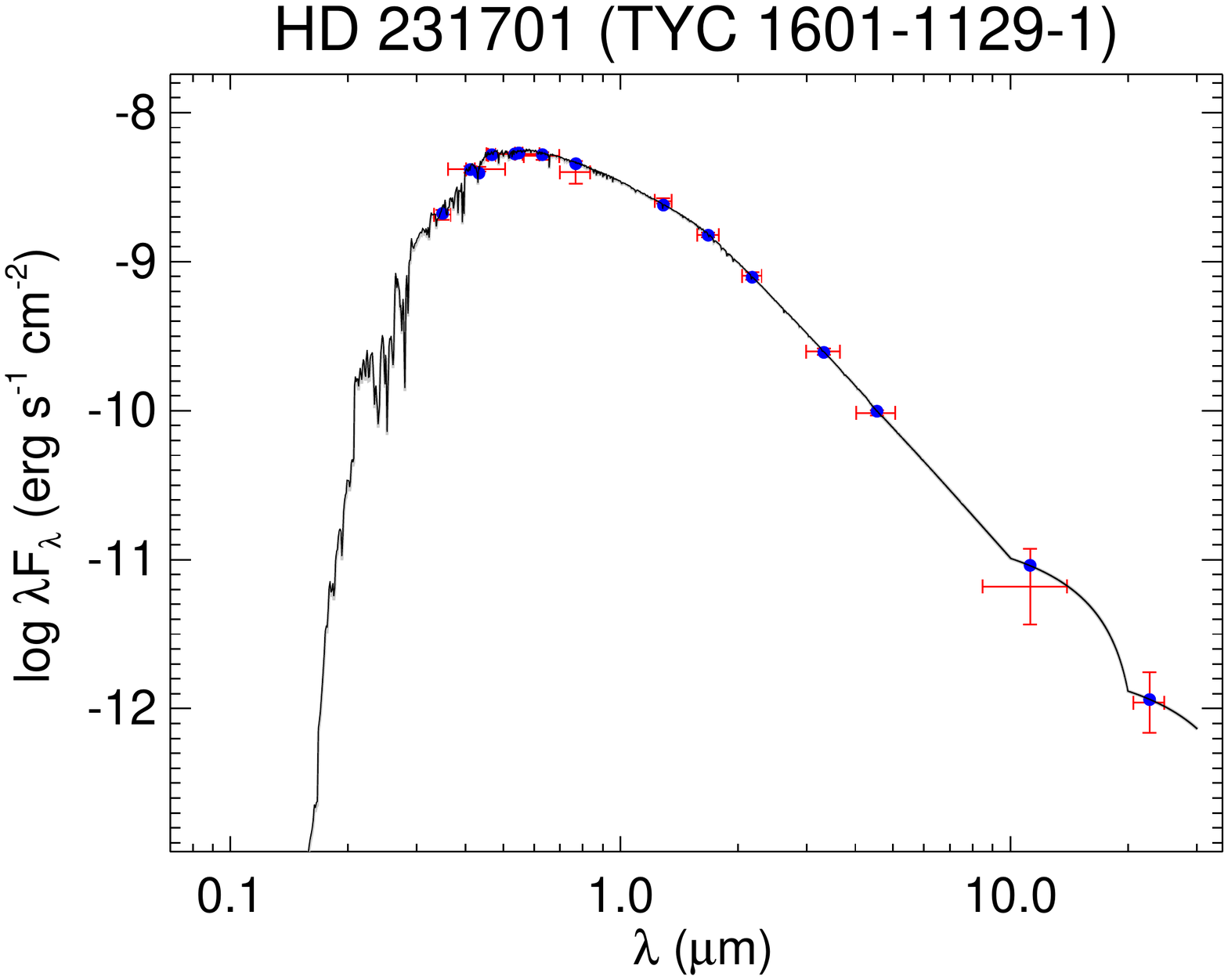}
  \includegraphics[trim=60 60 60 60,clip,width=0.49\linewidth]{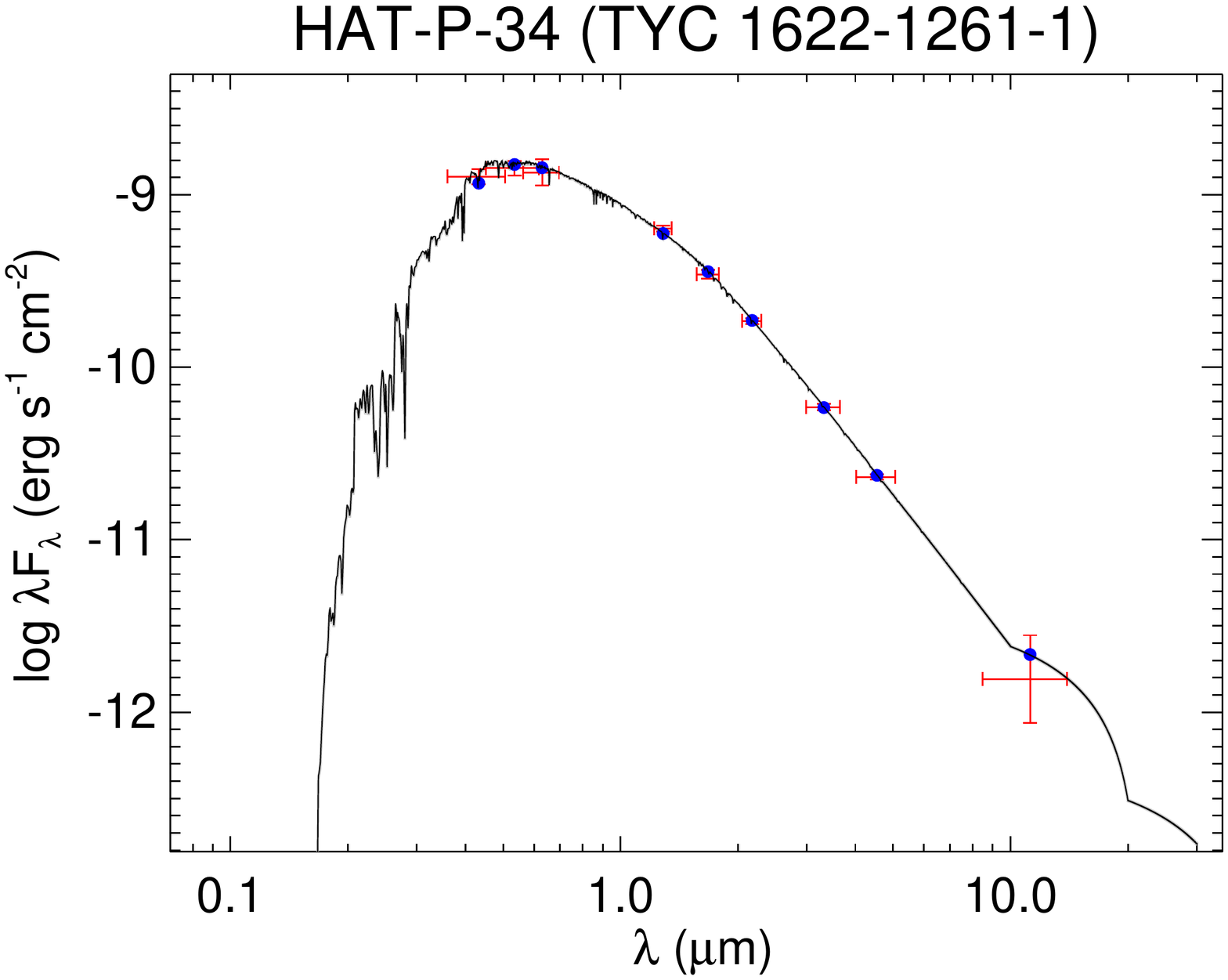}
  \caption{All labels, lines, symbols, and colors as in Figure \ref{fig:seds}.}
  \label{fig:seds_14}
\end{figure}

\begin{figure}[H]
  \centering
  \includegraphics[trim=60 60 60 60,clip,width=0.49\linewidth]{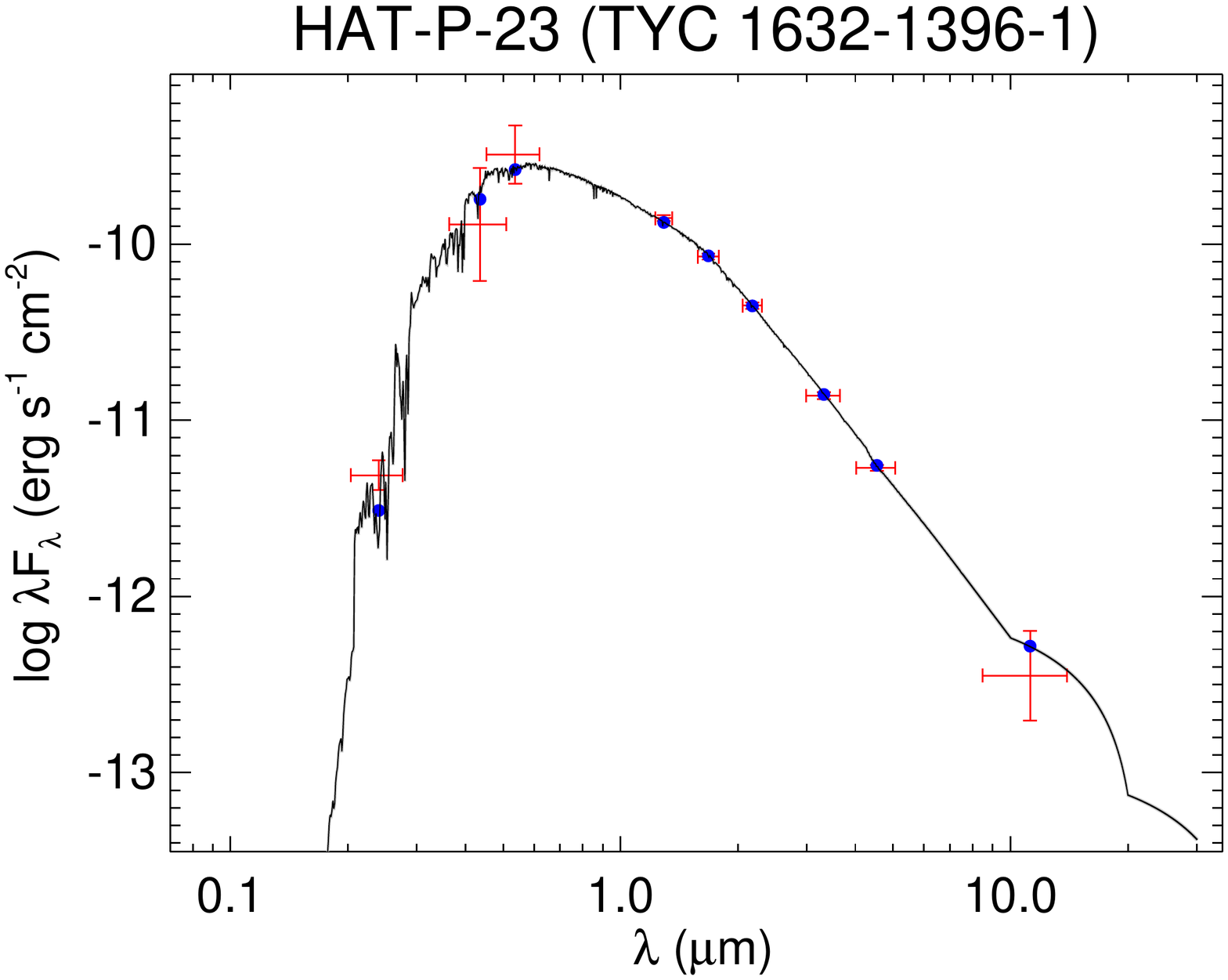}
  \includegraphics[trim=60 60 60 60,clip,width=0.49\linewidth]{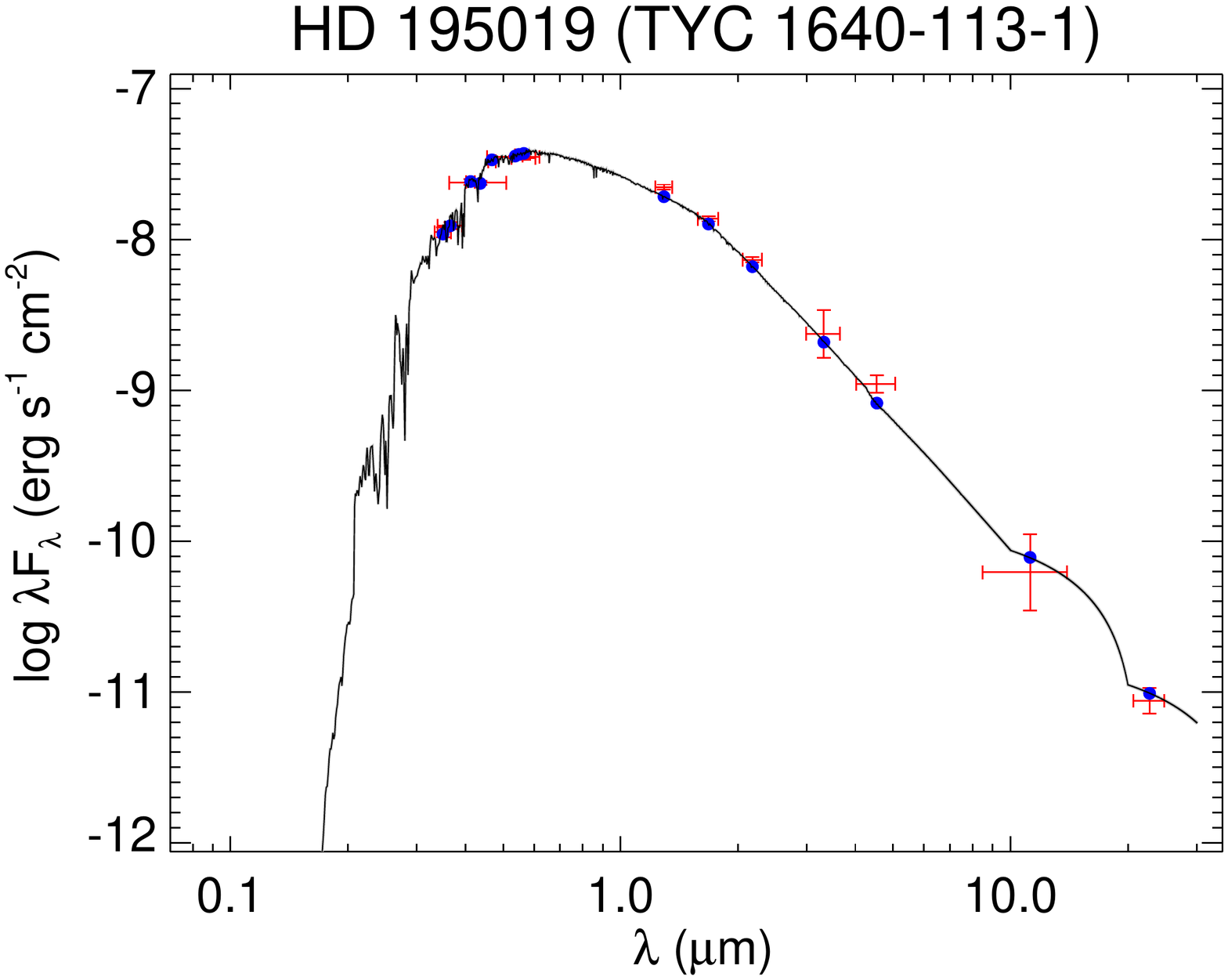}
  \includegraphics[trim=60 60 60 60,clip,width=0.49\linewidth]{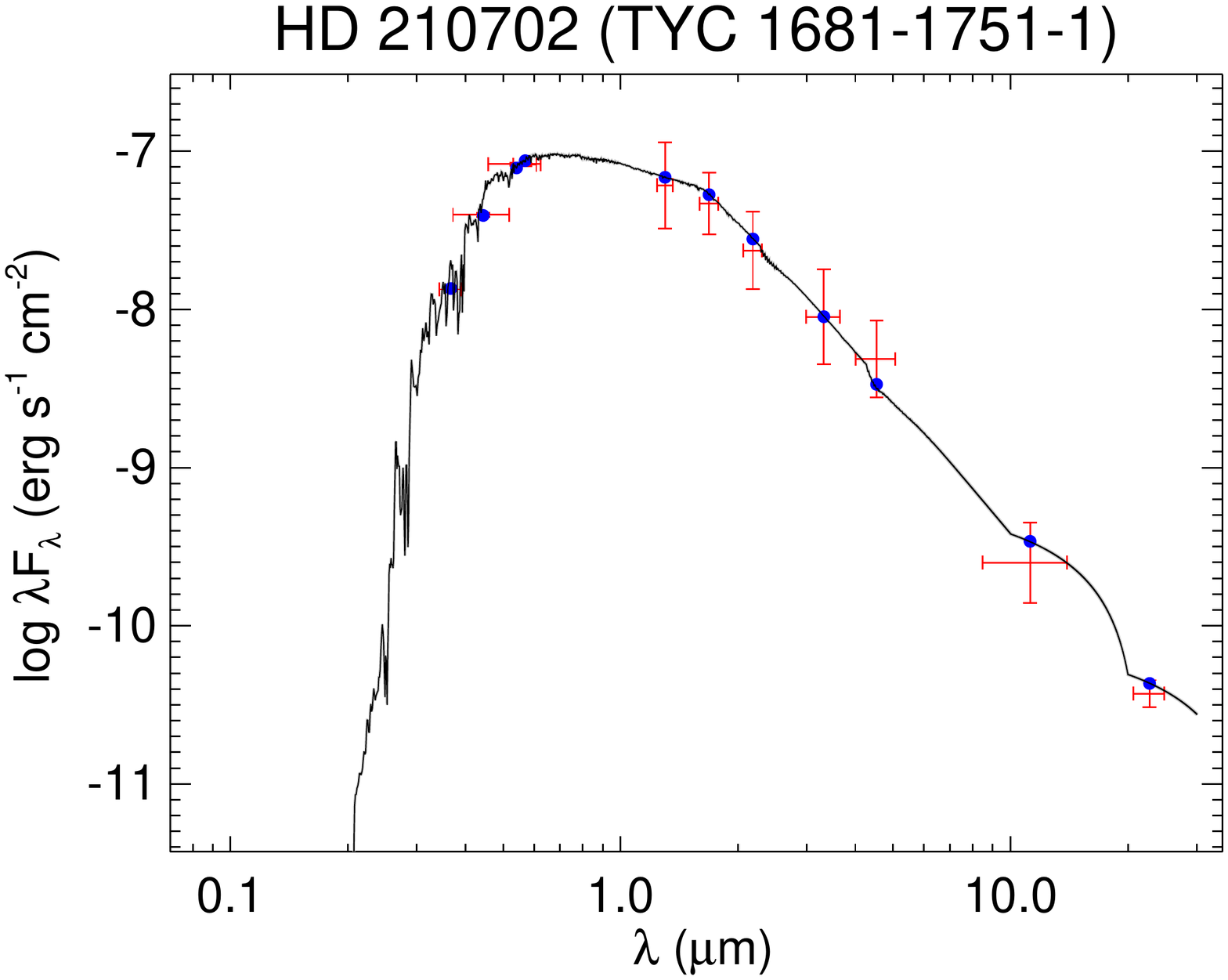}
  \includegraphics[trim=60 60 60 60,clip,width=0.49\linewidth]{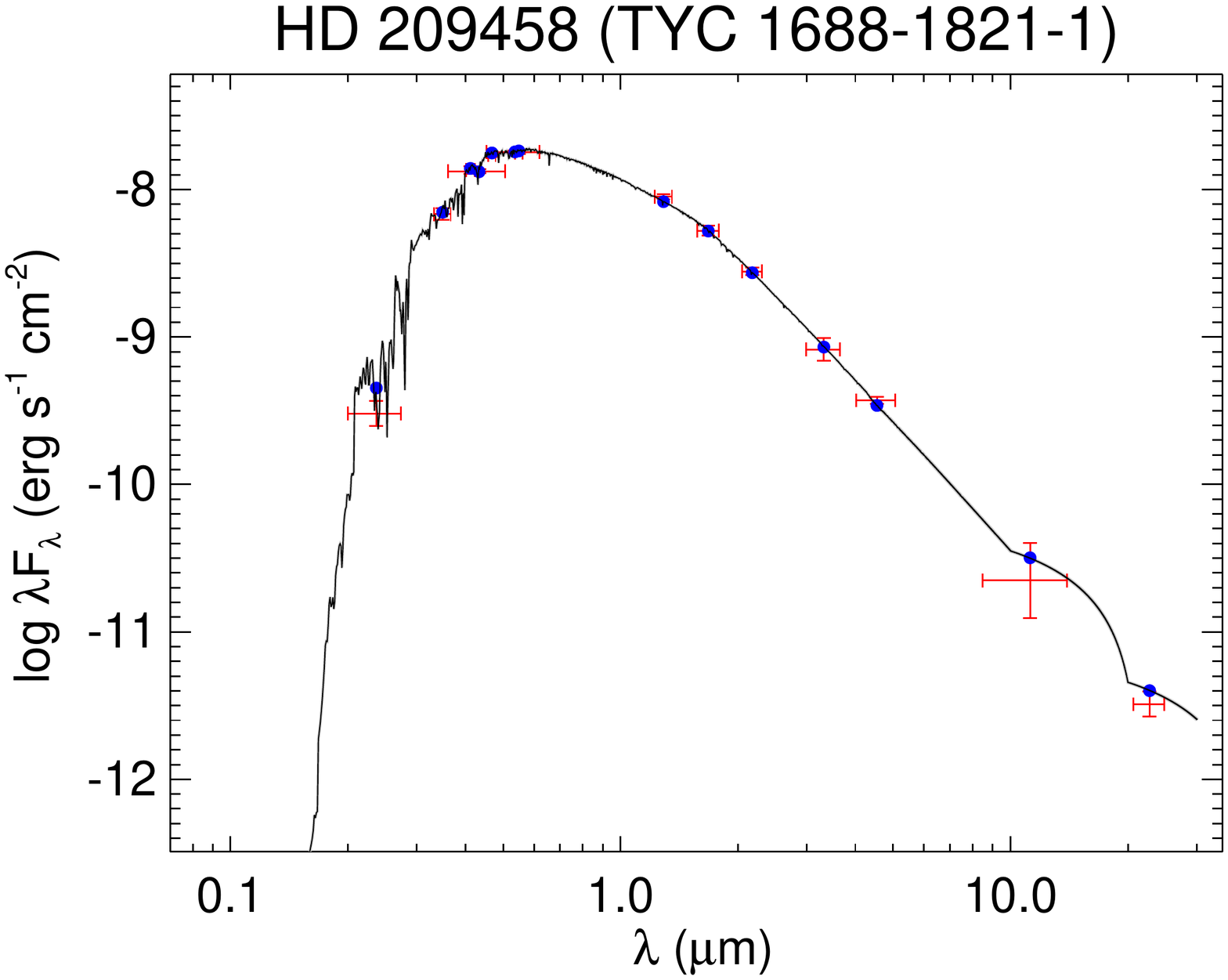}
  \includegraphics[trim=60 60 60 60,clip,width=0.49\linewidth]{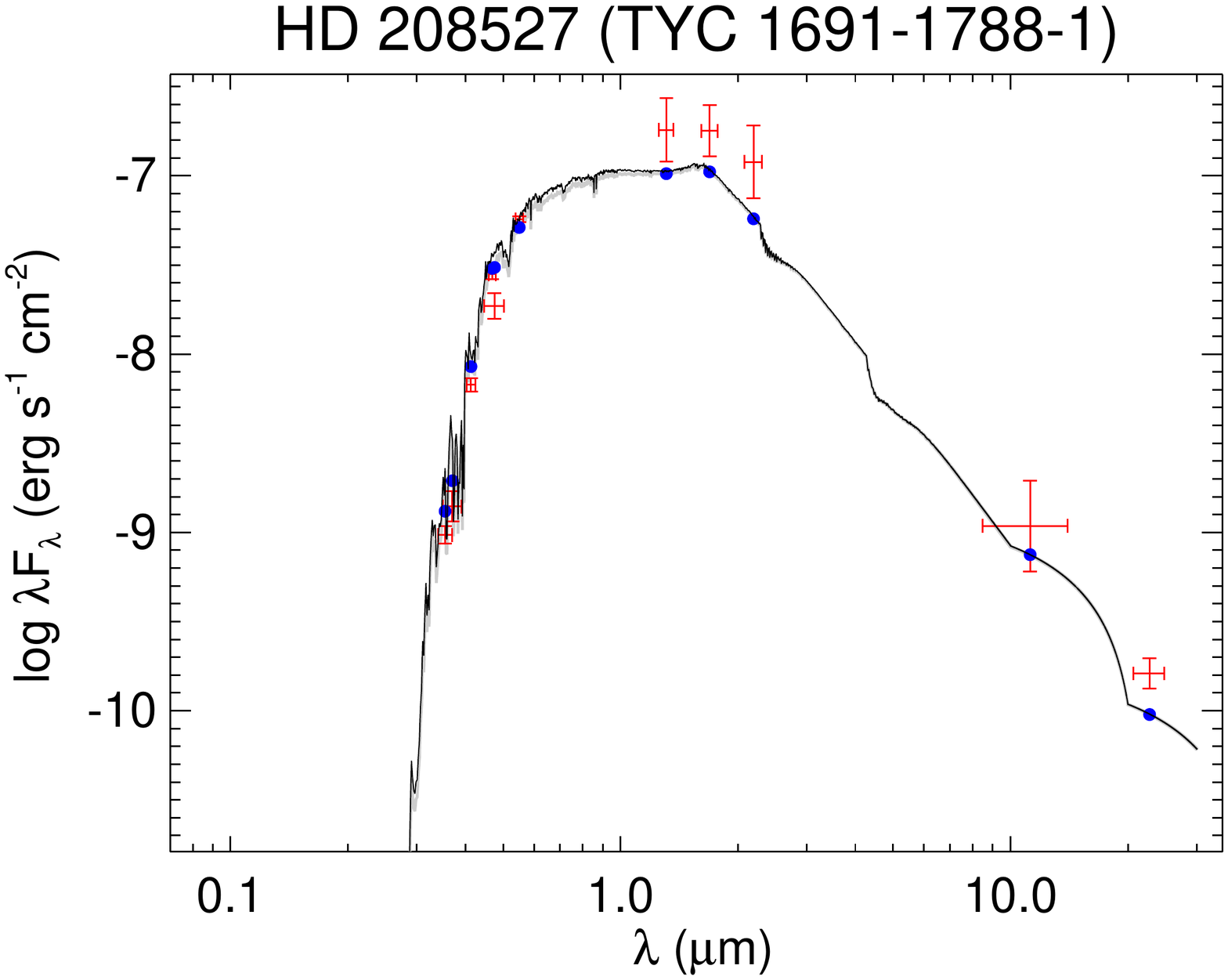}
  \includegraphics[trim=60 60 60 60,clip,width=0.49\linewidth]{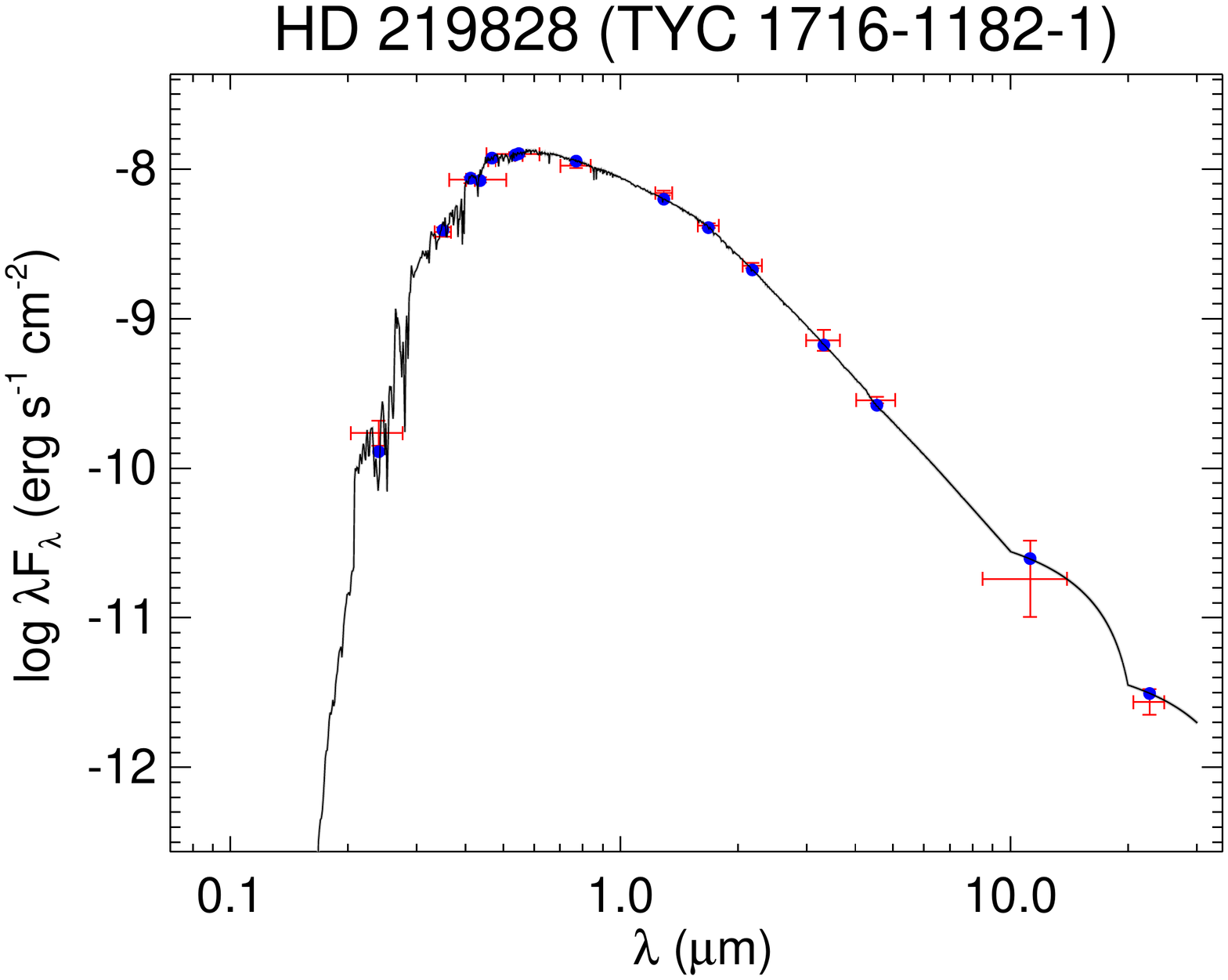}
  \caption{All labels, lines, symbols, and colors as in Figure \ref{fig:seds}.}
  \label{fig:seds_15}
\end{figure}

\begin{figure}[H]
  \centering
  \includegraphics[trim=60 60 60 60,clip,width=0.49\linewidth]{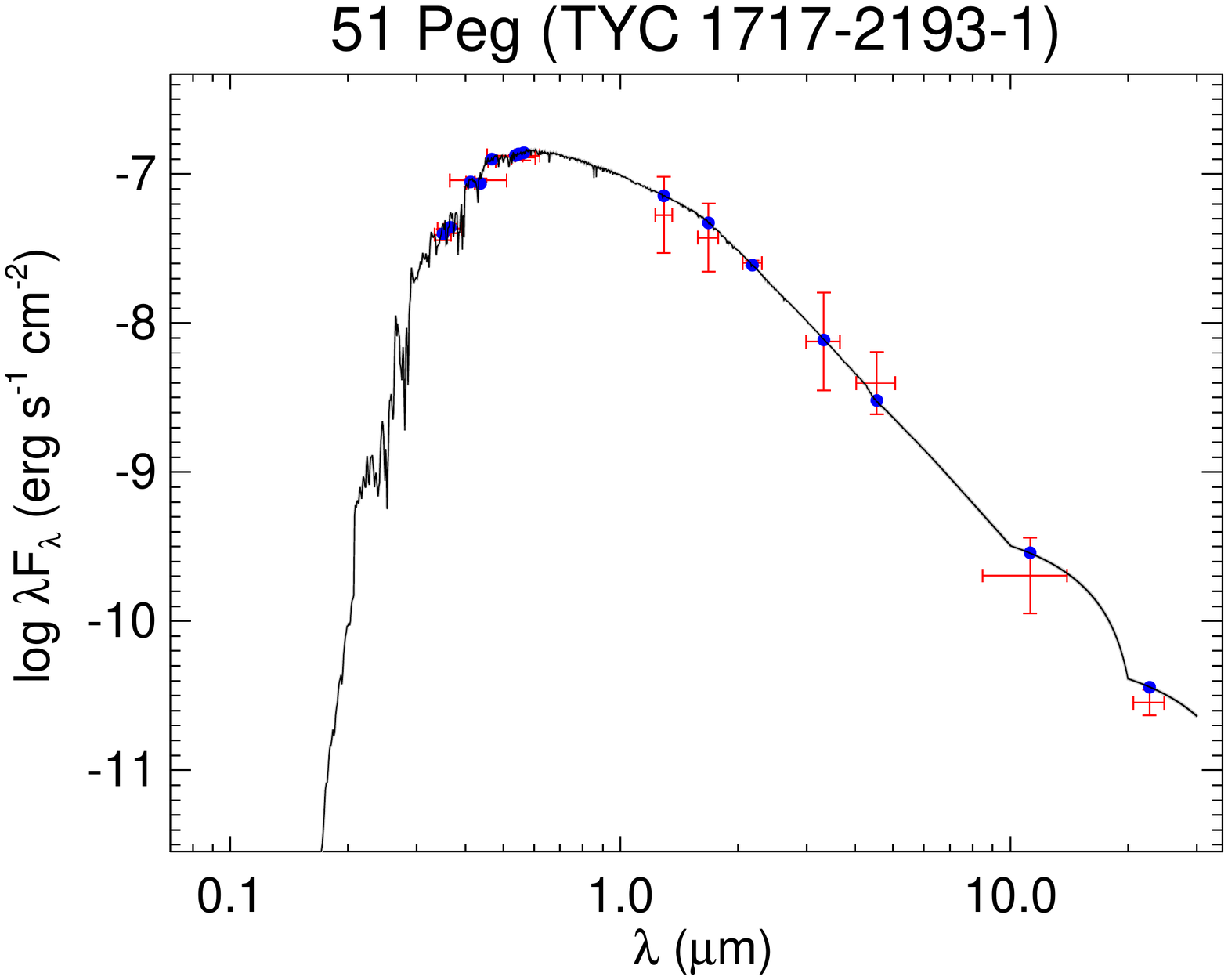}
  \includegraphics[trim=60 60 60 60,clip,width=0.49\linewidth]{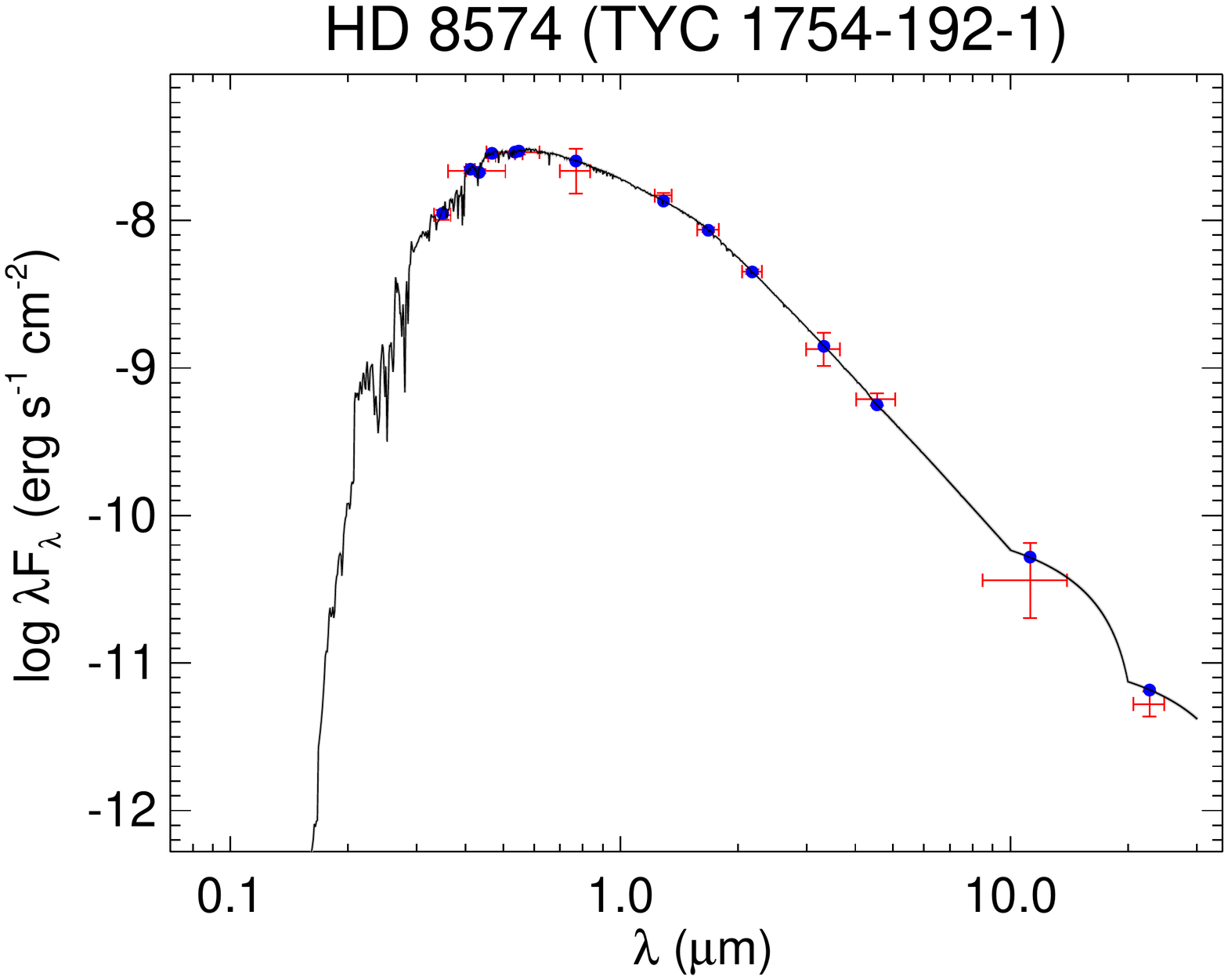}
  \includegraphics[trim=60 60 60 60,clip,width=0.49\linewidth]{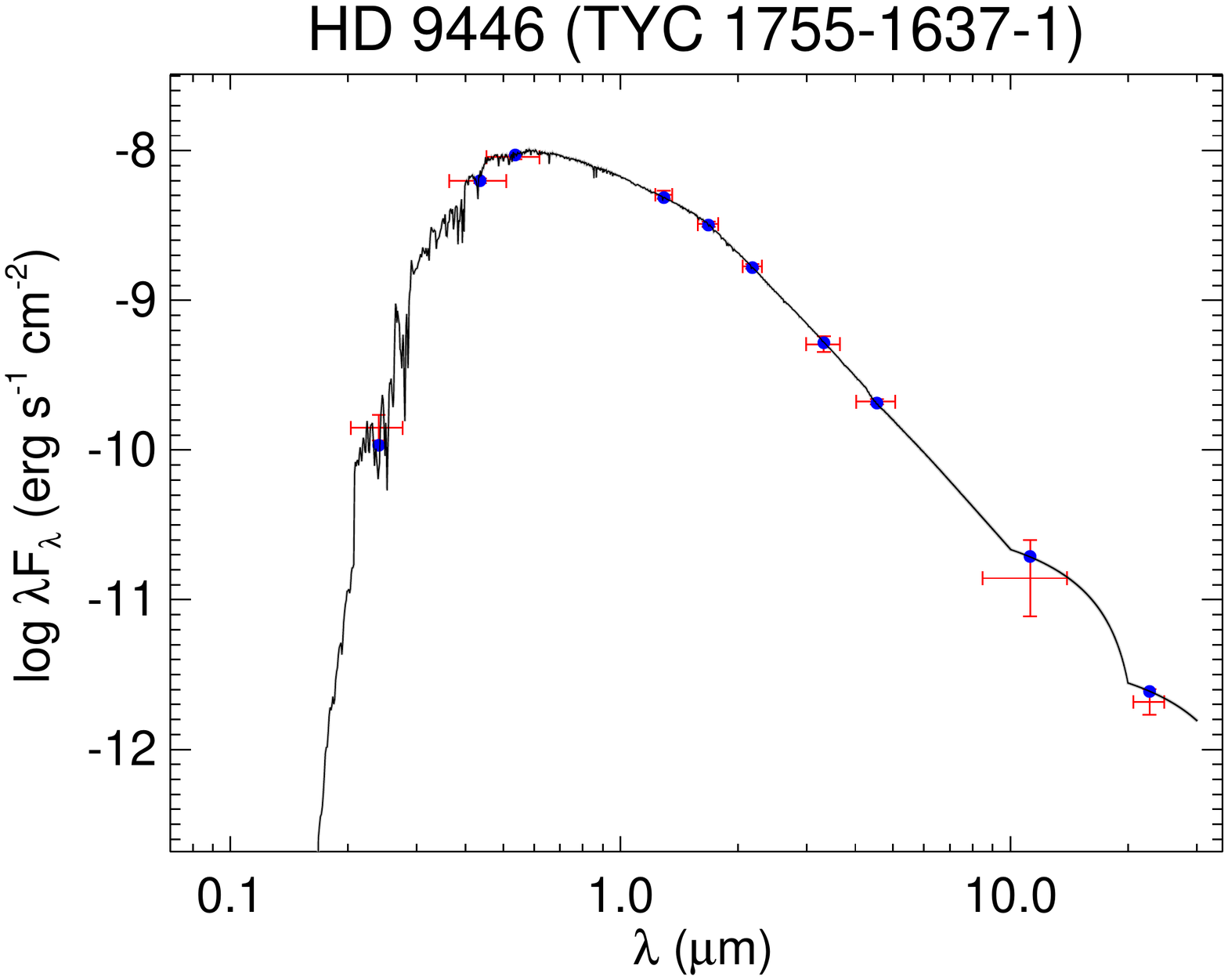}
  \includegraphics[trim=60 60 60 60,clip,width=0.49\linewidth]{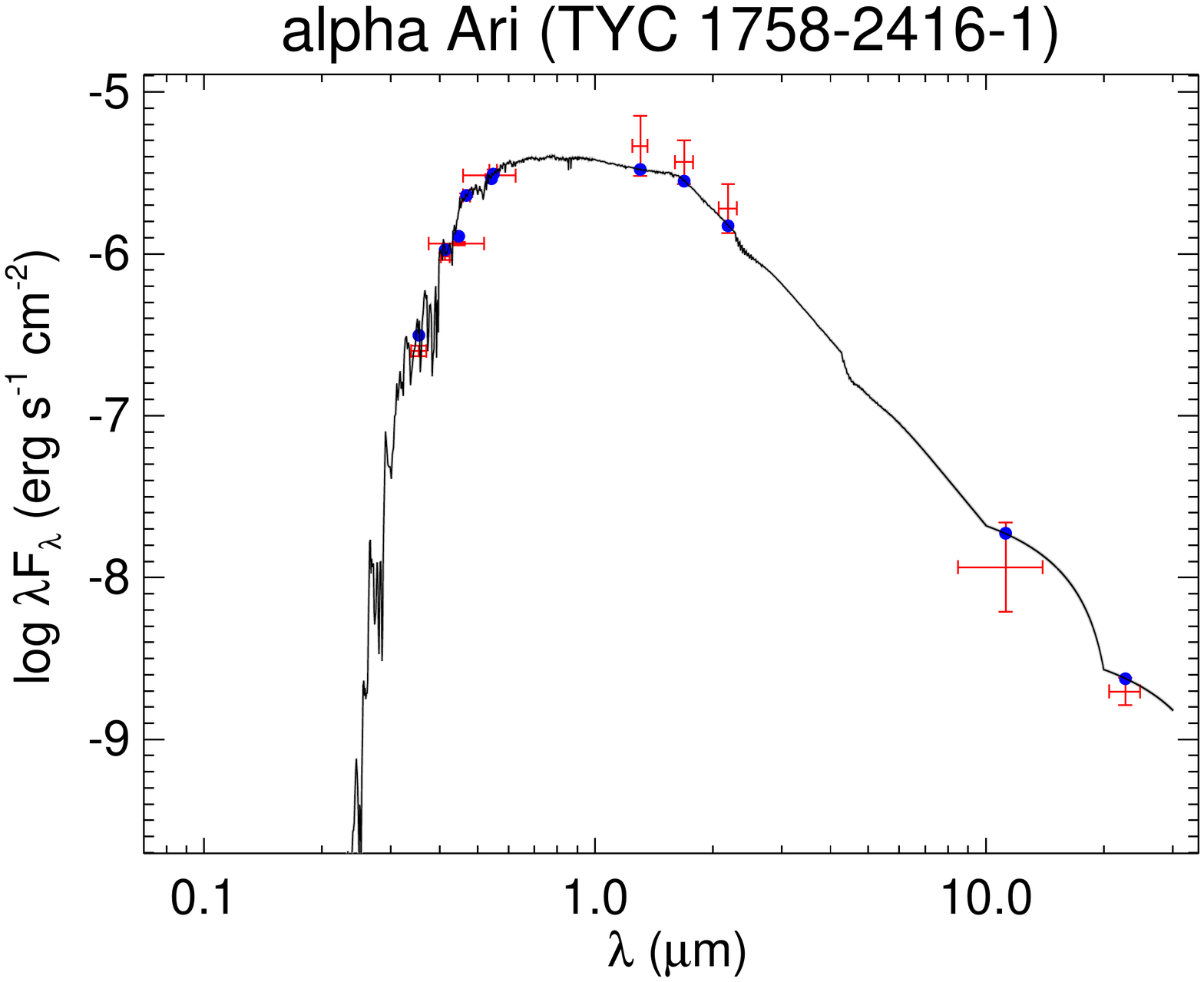}
  \includegraphics[trim=60 60 60 60,clip,width=0.49\linewidth]{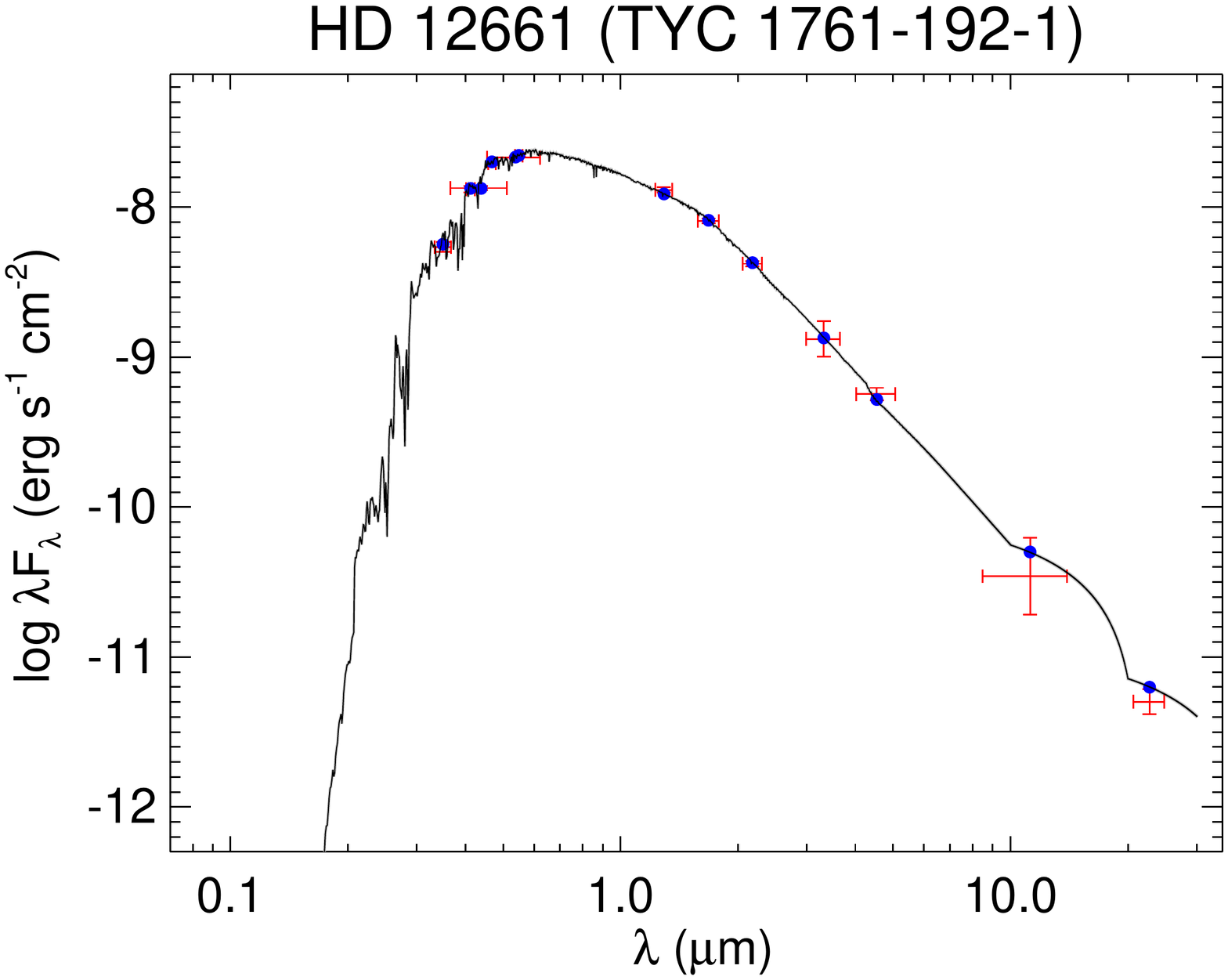}
  \includegraphics[trim=60 60 60 60,clip,width=0.49\linewidth]{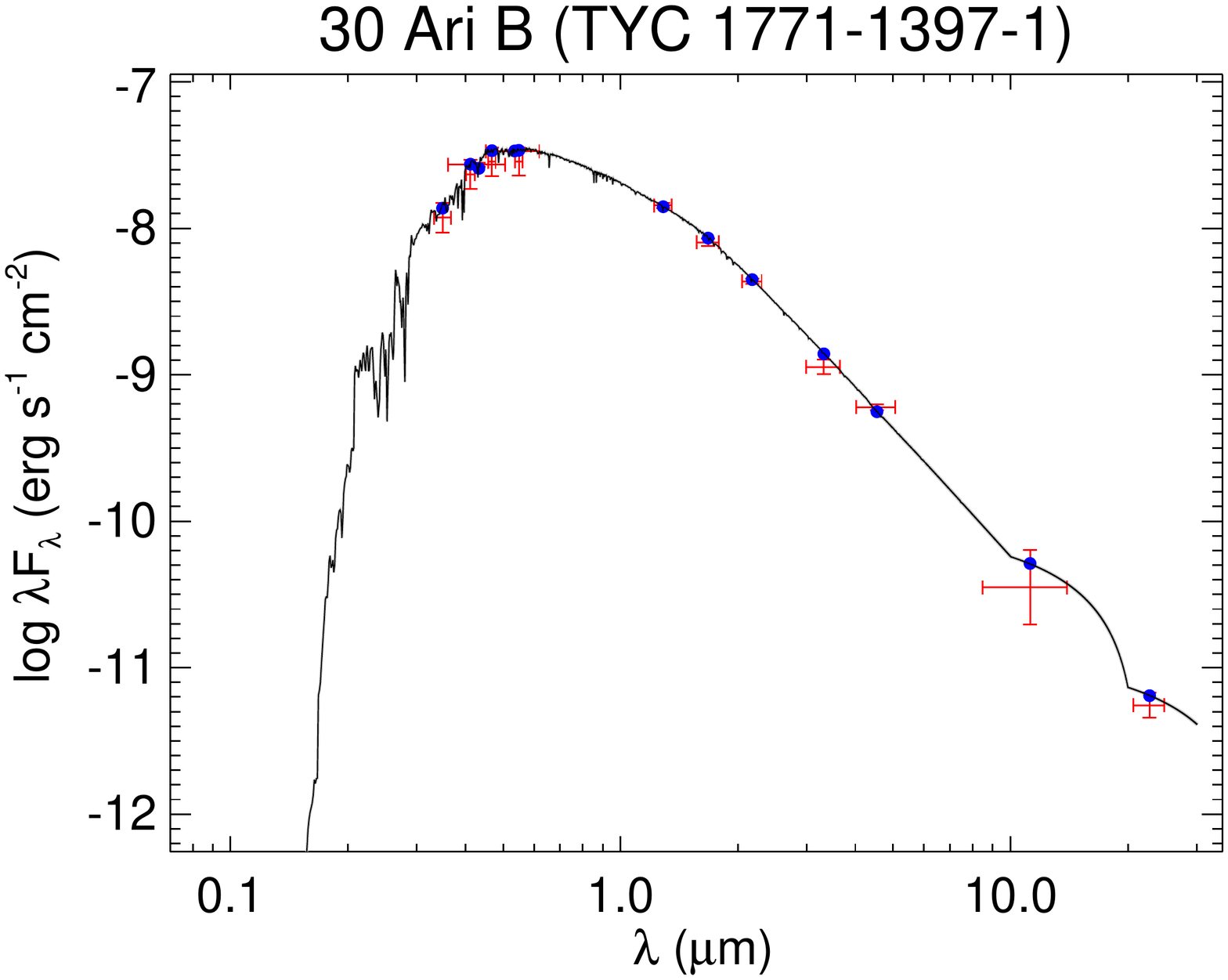}
  \caption{All labels, lines, symbols, and colors as in Figure \ref{fig:seds}.}
  \label{fig:seds_16}
\end{figure}

\begin{figure}[H]
  \centering
  \includegraphics[trim=60 60 60 60,clip,width=0.49\linewidth]{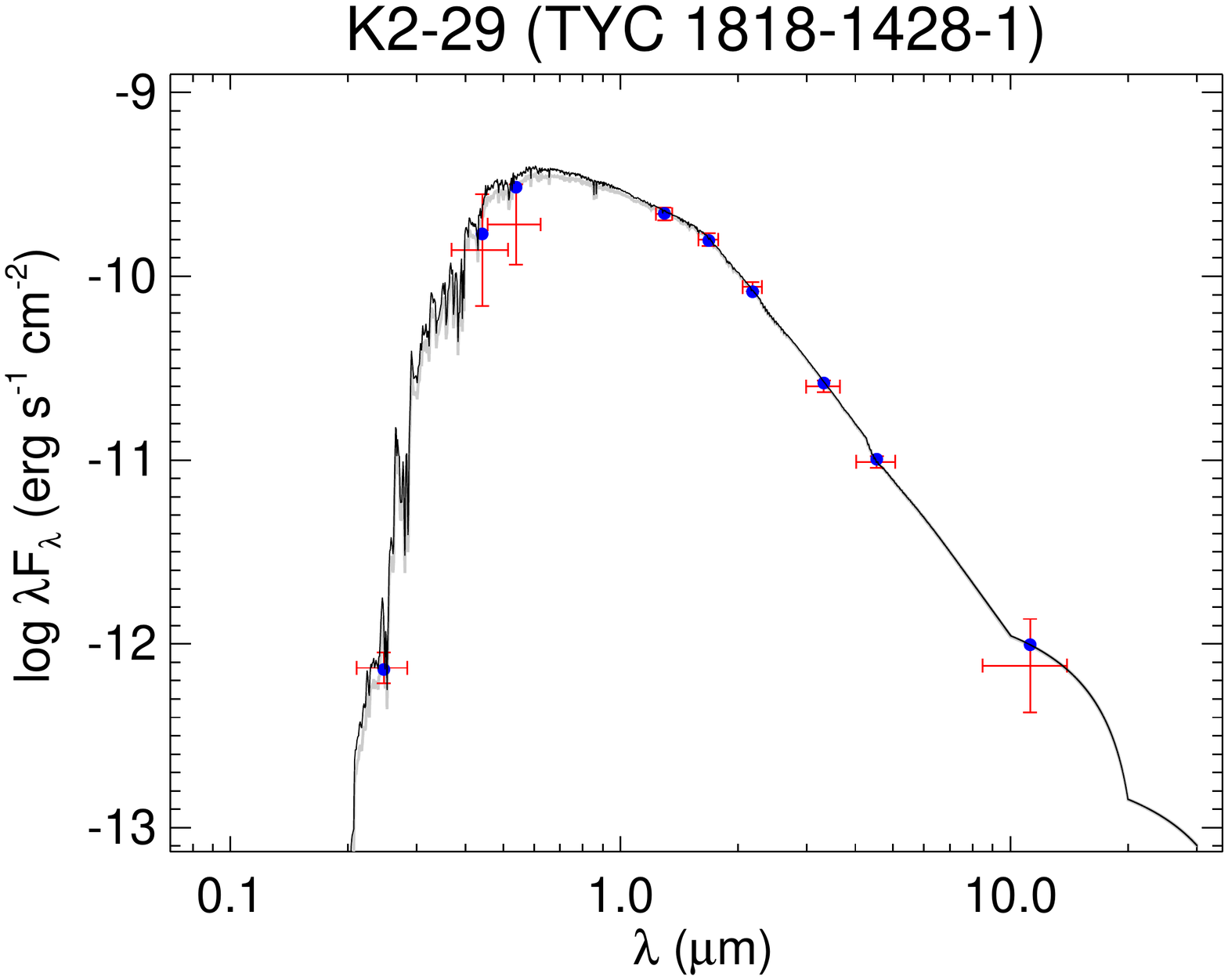}
  \includegraphics[trim=60 60 60 60,clip,width=0.49\linewidth]{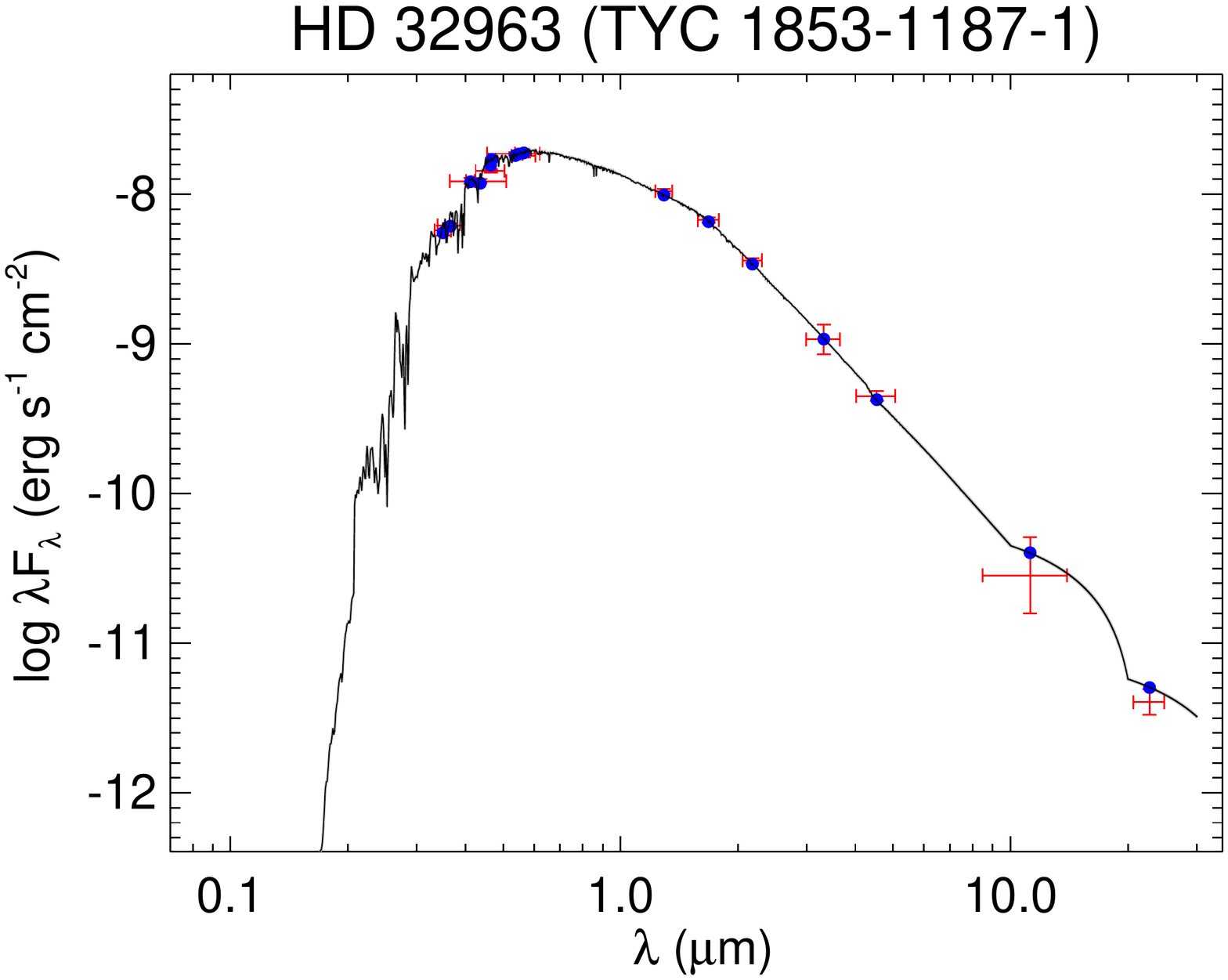}
  \includegraphics[trim=60 60 60 60,clip,width=0.49\linewidth]{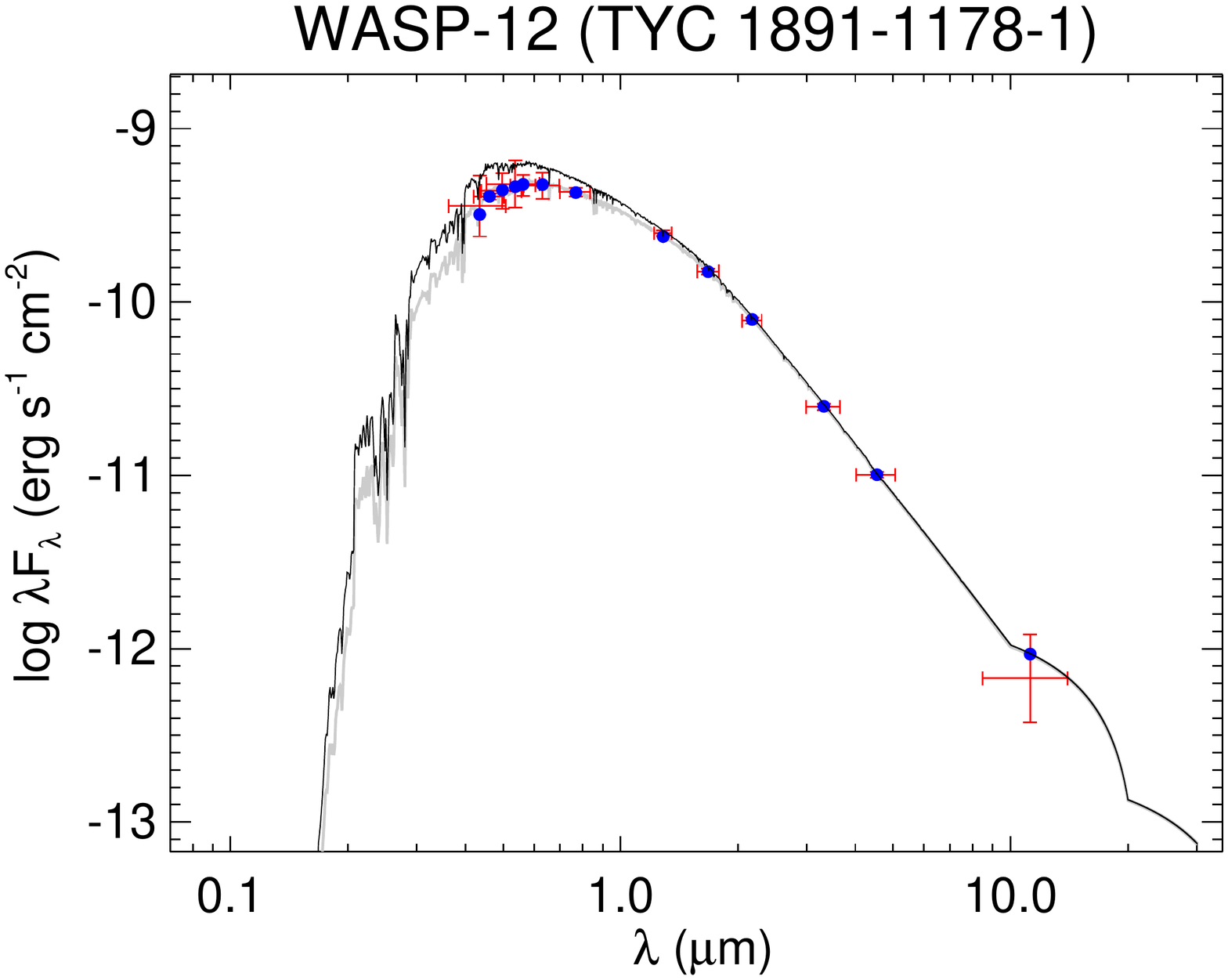}
  \includegraphics[trim=60 60 60 60,clip,width=0.49\linewidth]{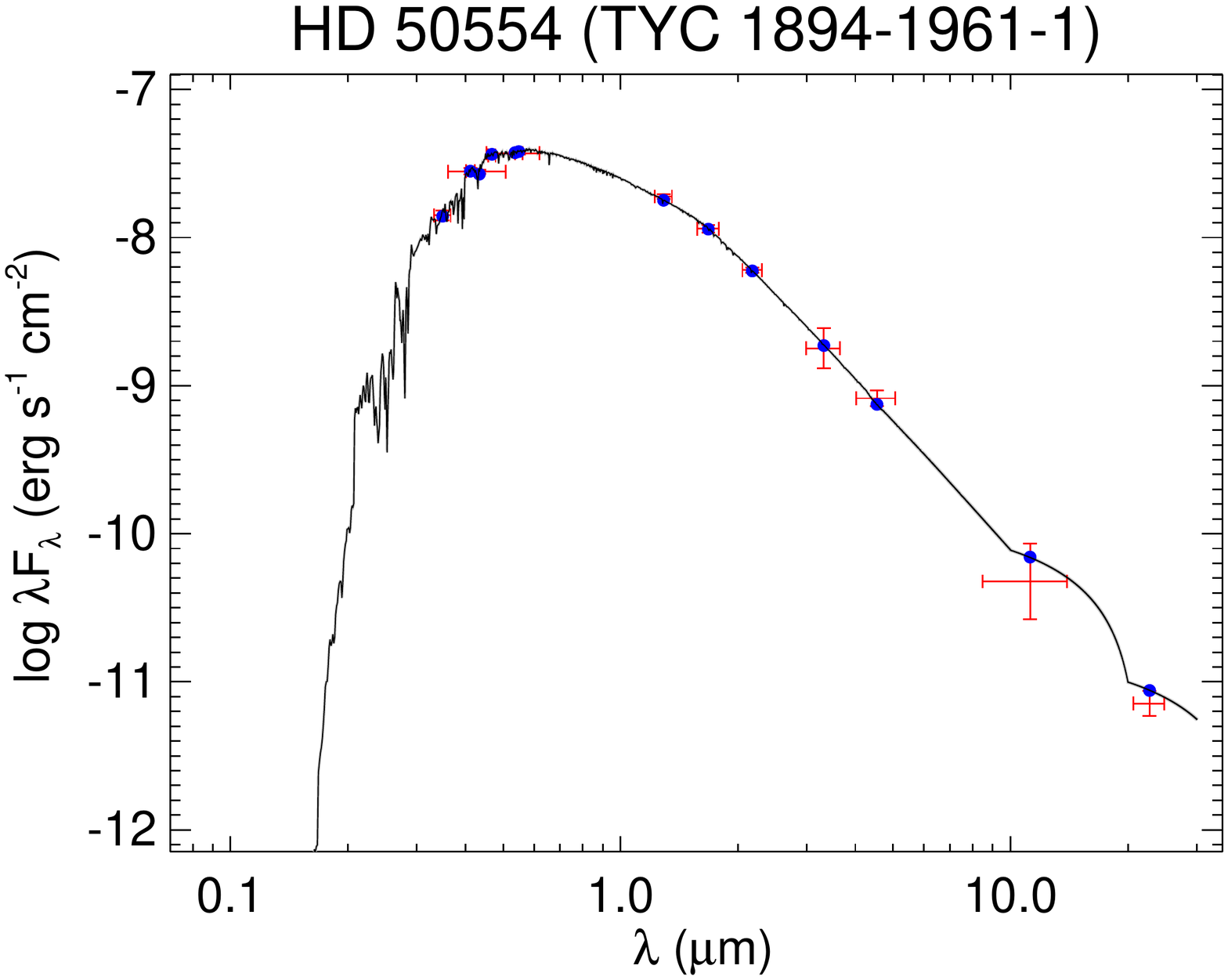}
  \includegraphics[trim=60 60 60 60,clip,width=0.49\linewidth]{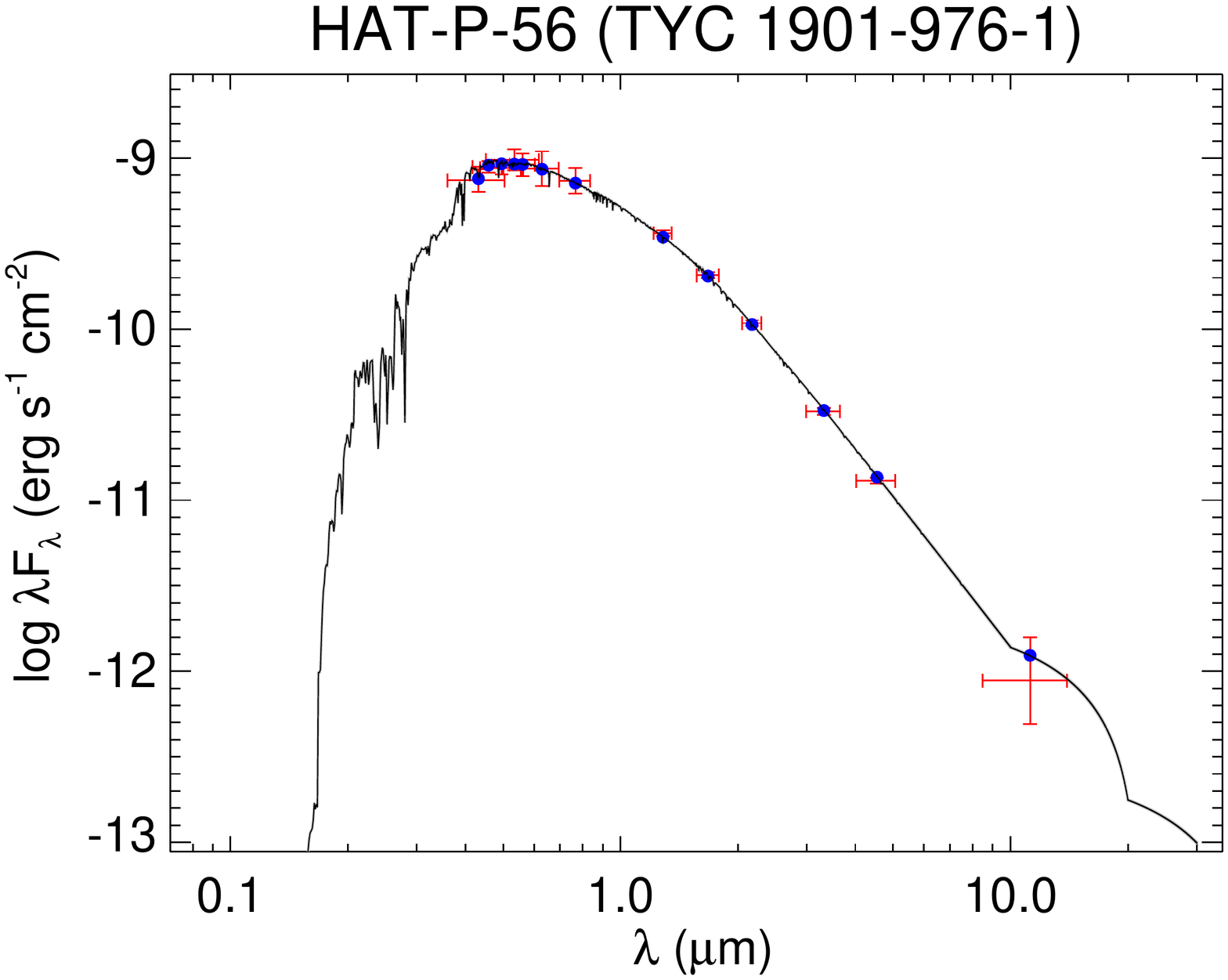}
  \includegraphics[trim=60 60 60 60,clip,width=0.49\linewidth]{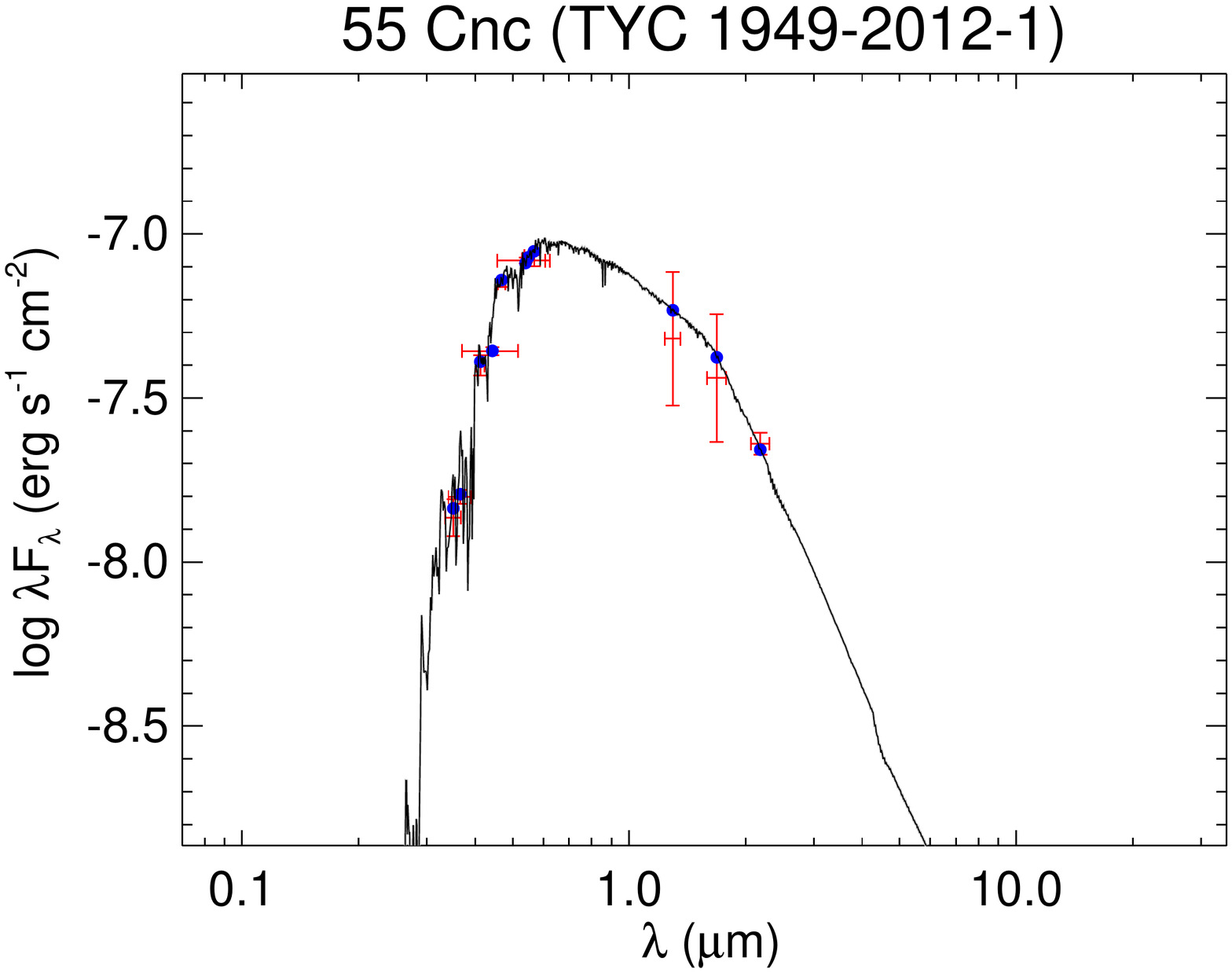}
  \caption{All labels, lines, symbols, and colors as in Figure \ref{fig:seds}.}
  \label{fig:seds_17}
\end{figure}

\begin{figure}[H]
  \centering
  \includegraphics[trim=60 60 60 60,clip,width=0.49\linewidth]{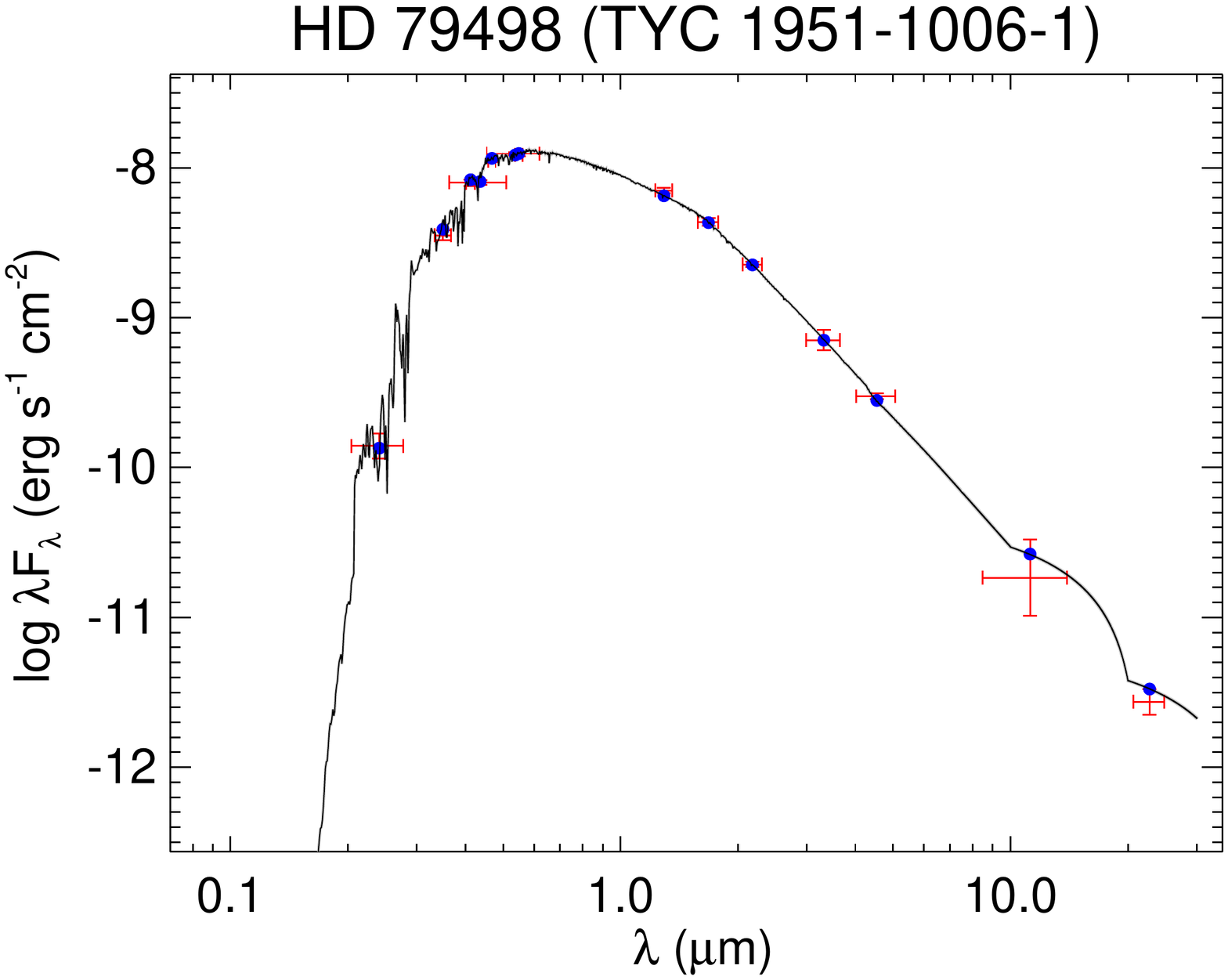}
  \includegraphics[trim=60 60 60 60,clip,width=0.49\linewidth]{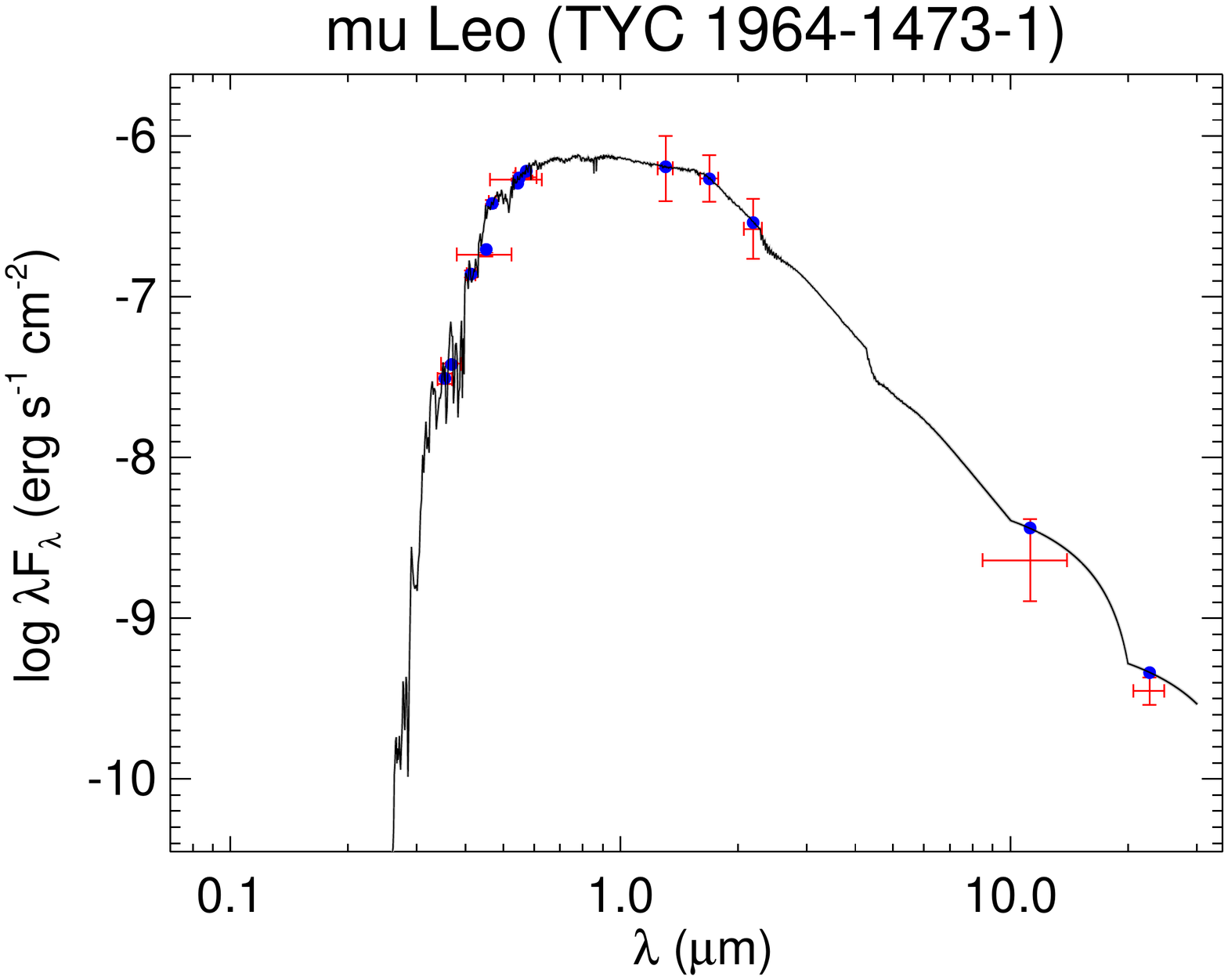}
  \includegraphics[trim=60 60 60 60,clip,width=0.49\linewidth]{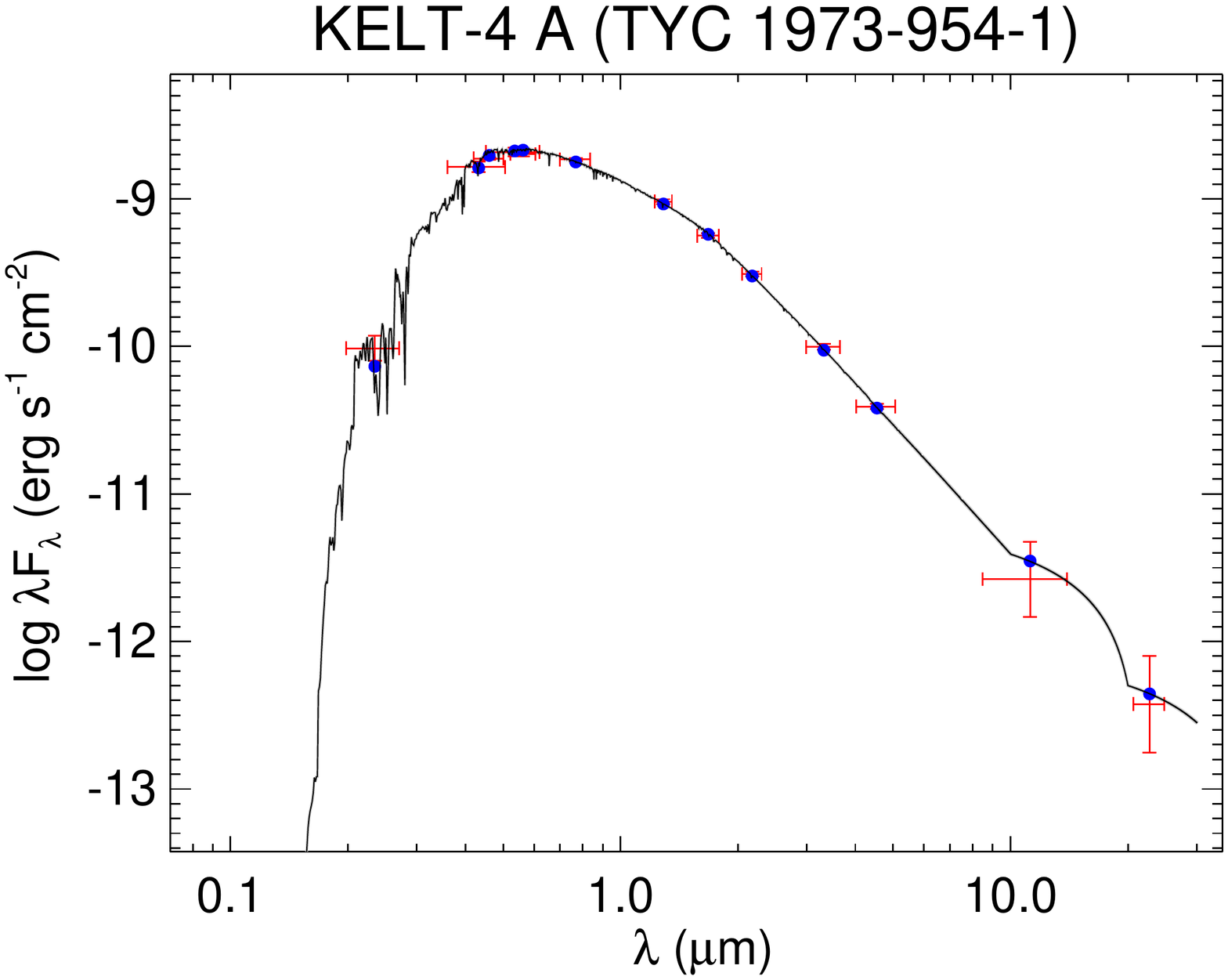}
  \includegraphics[trim=60 60 60 60,clip,width=0.49\linewidth]{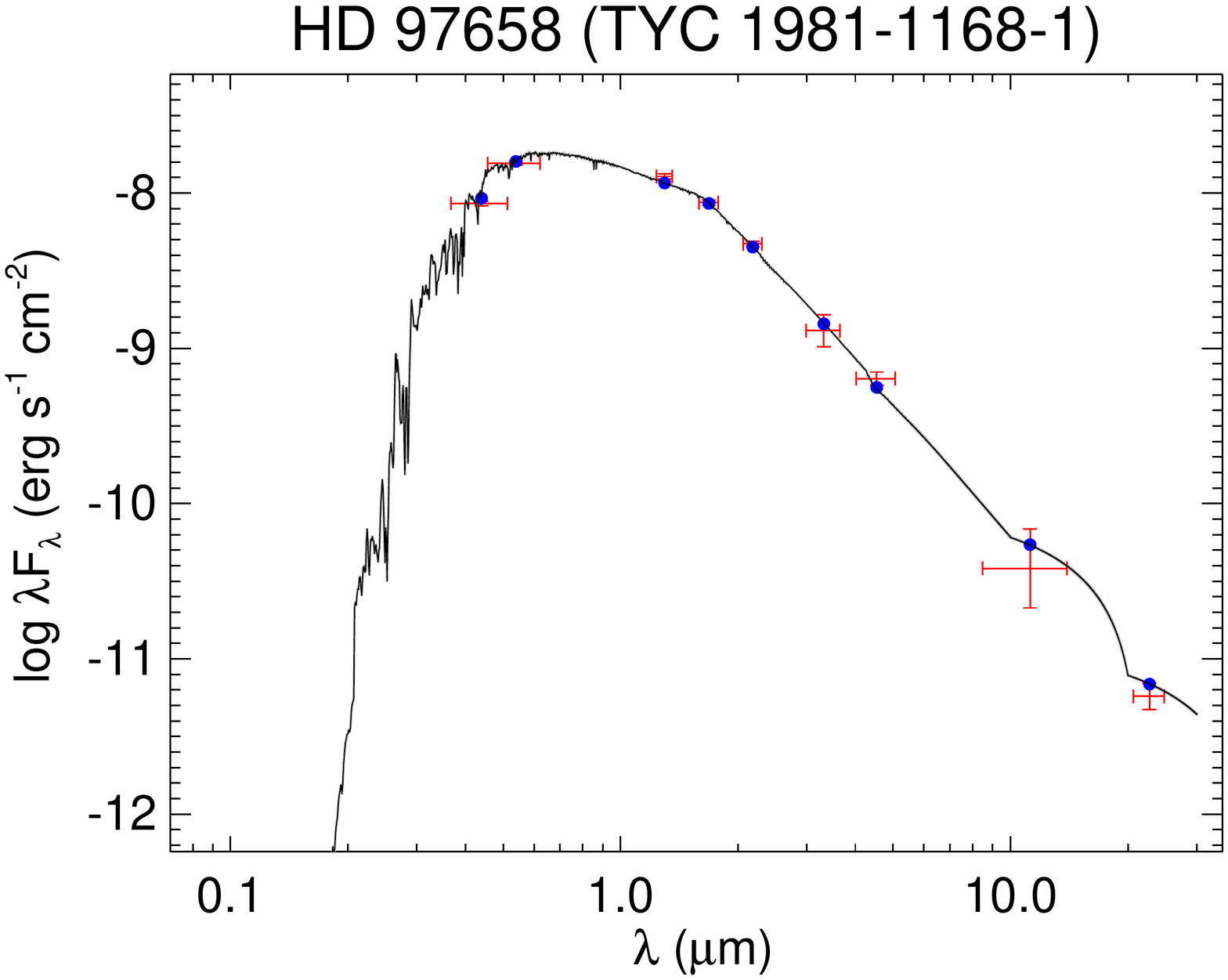}
  \includegraphics[trim=60 60 60 60,clip,width=0.49\linewidth]{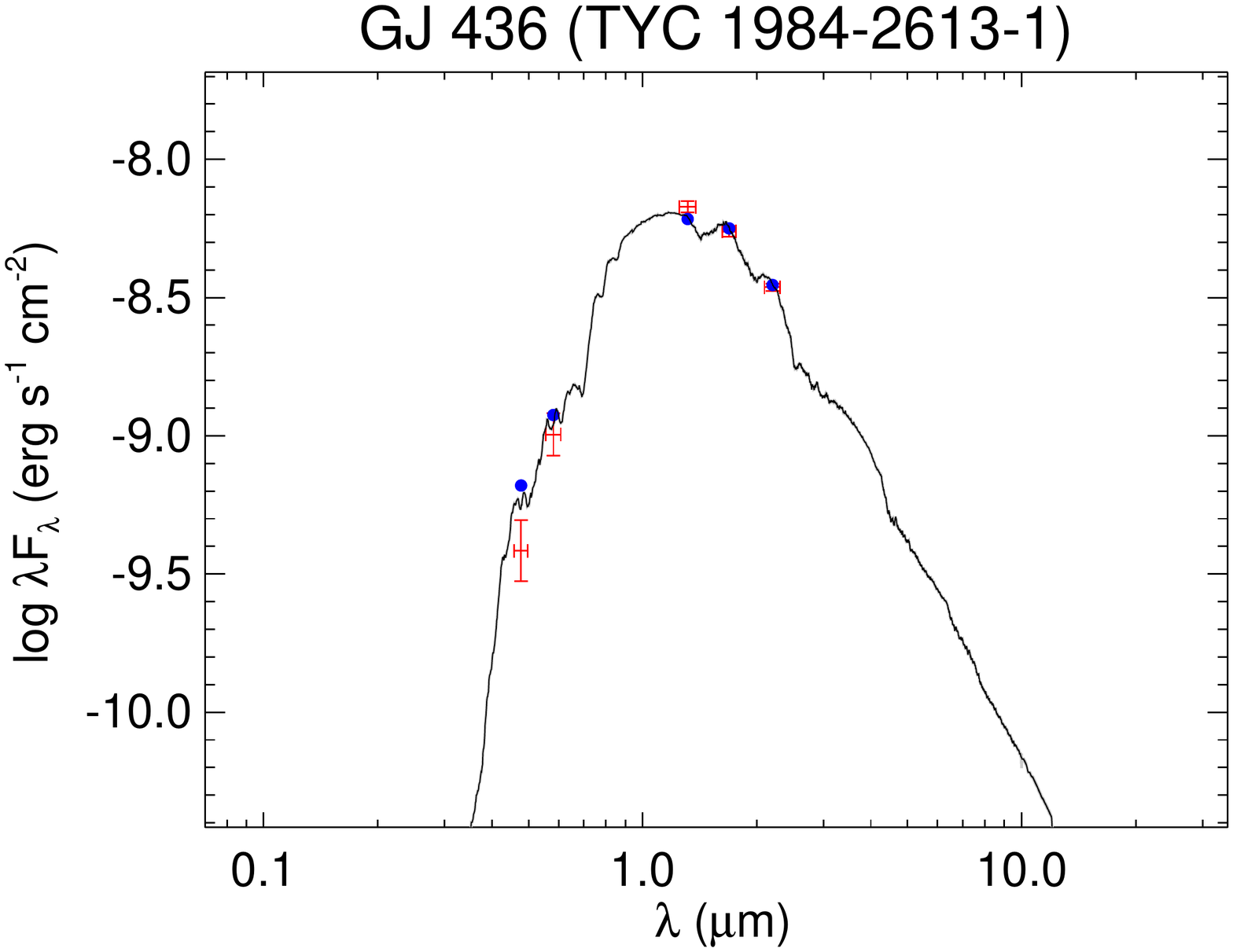}
  \includegraphics[trim=60 60 60 60,clip,width=0.49\linewidth]{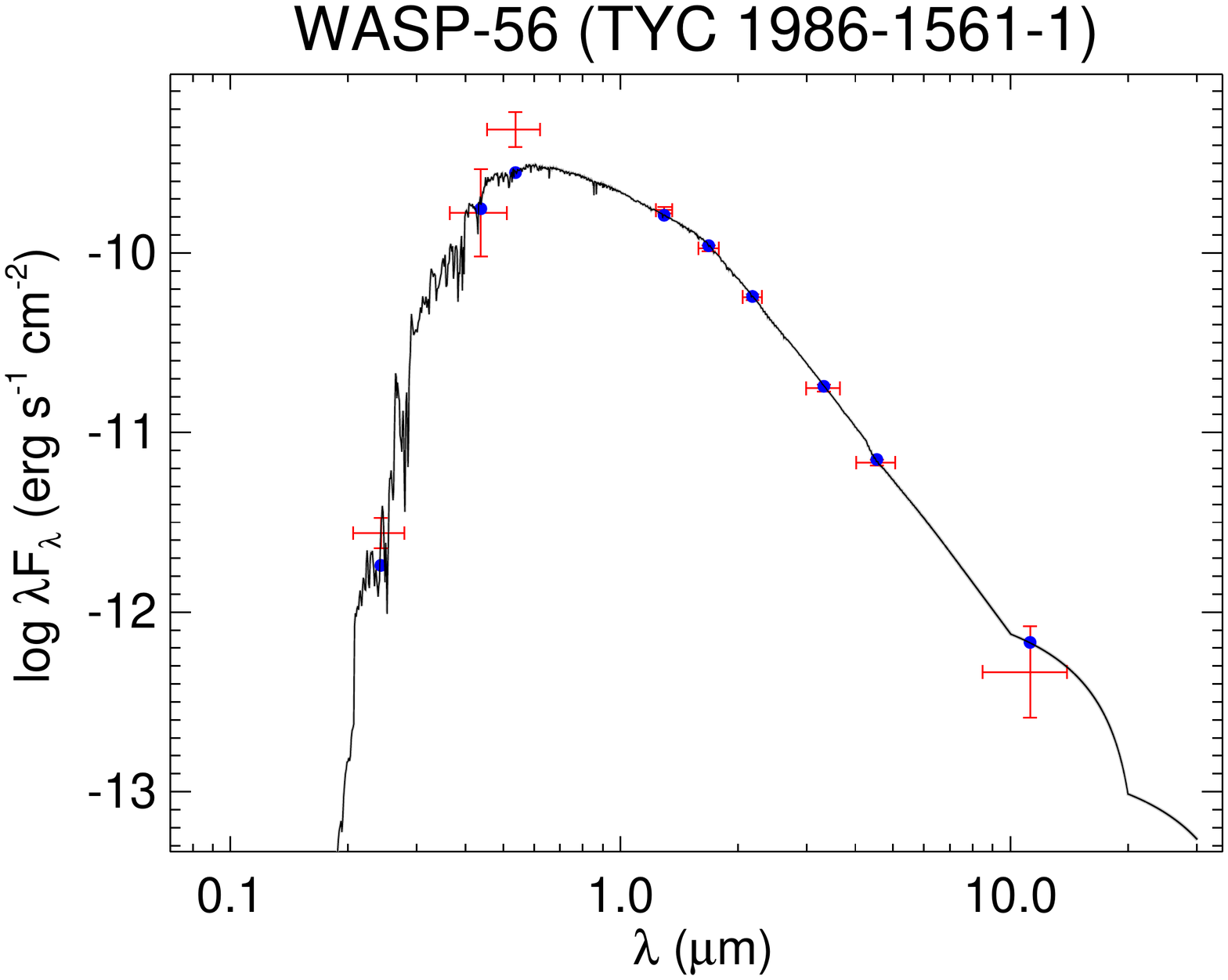}
  \caption{All labels, lines, symbols, and colors as in Figure \ref{fig:seds}.}
  \label{fig:seds_18}
\end{figure}

\begin{figure}[H]
  \centering
  \includegraphics[trim=60 60 60 60,clip,width=0.49\linewidth]{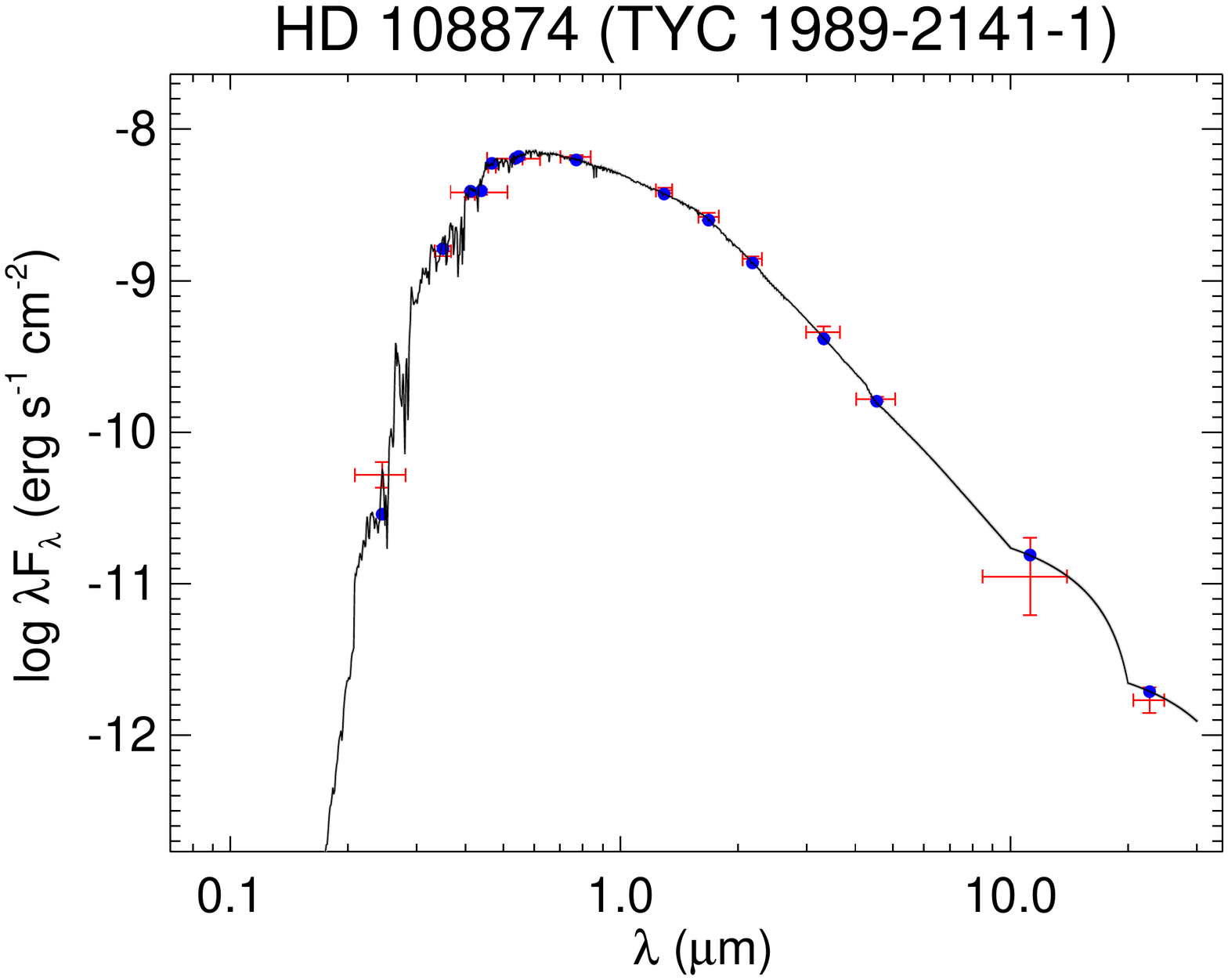}
  \includegraphics[trim=60 60 60 60,clip,width=0.49\linewidth]{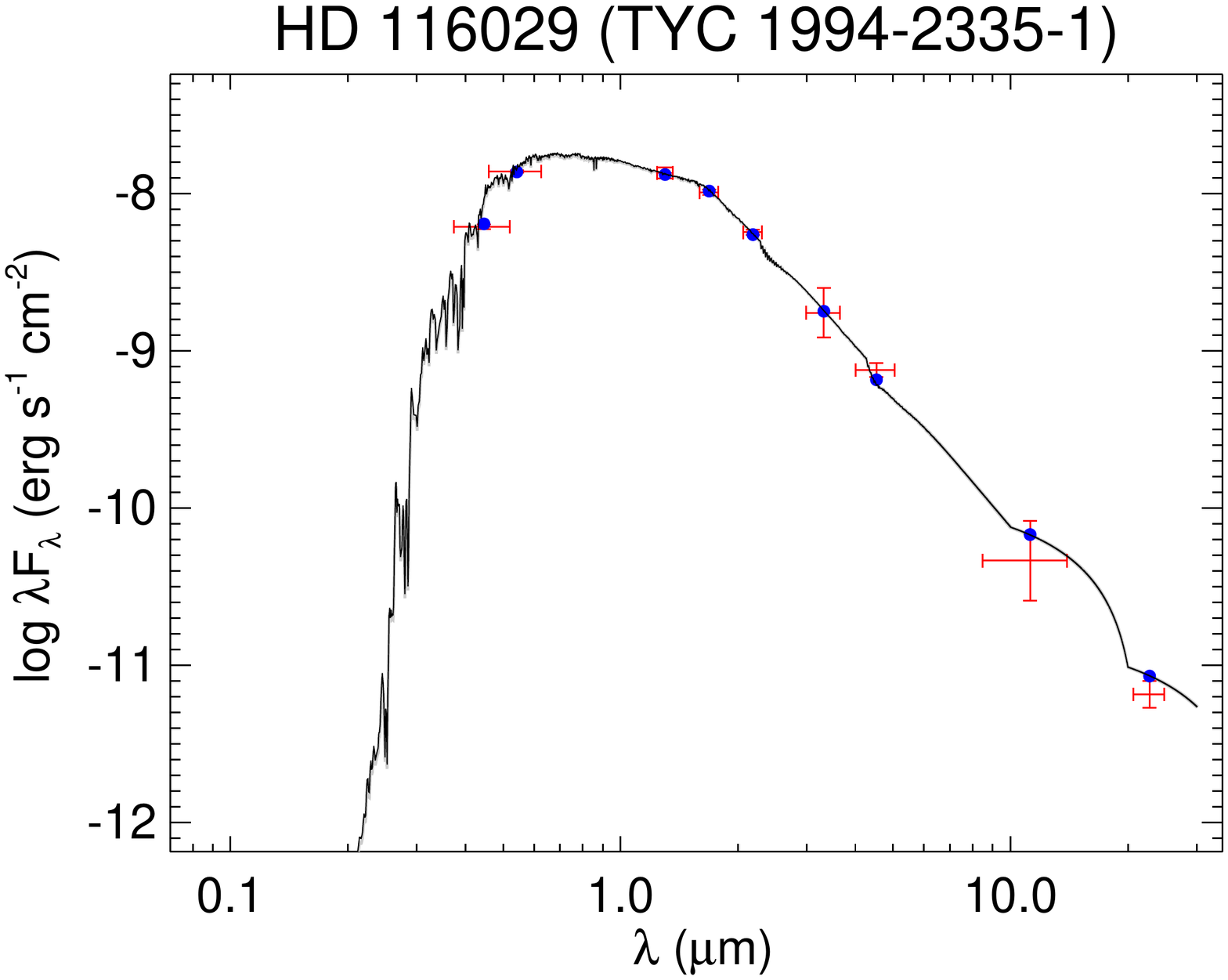}
  \includegraphics[trim=60 60 60 60,clip,width=0.49\linewidth]{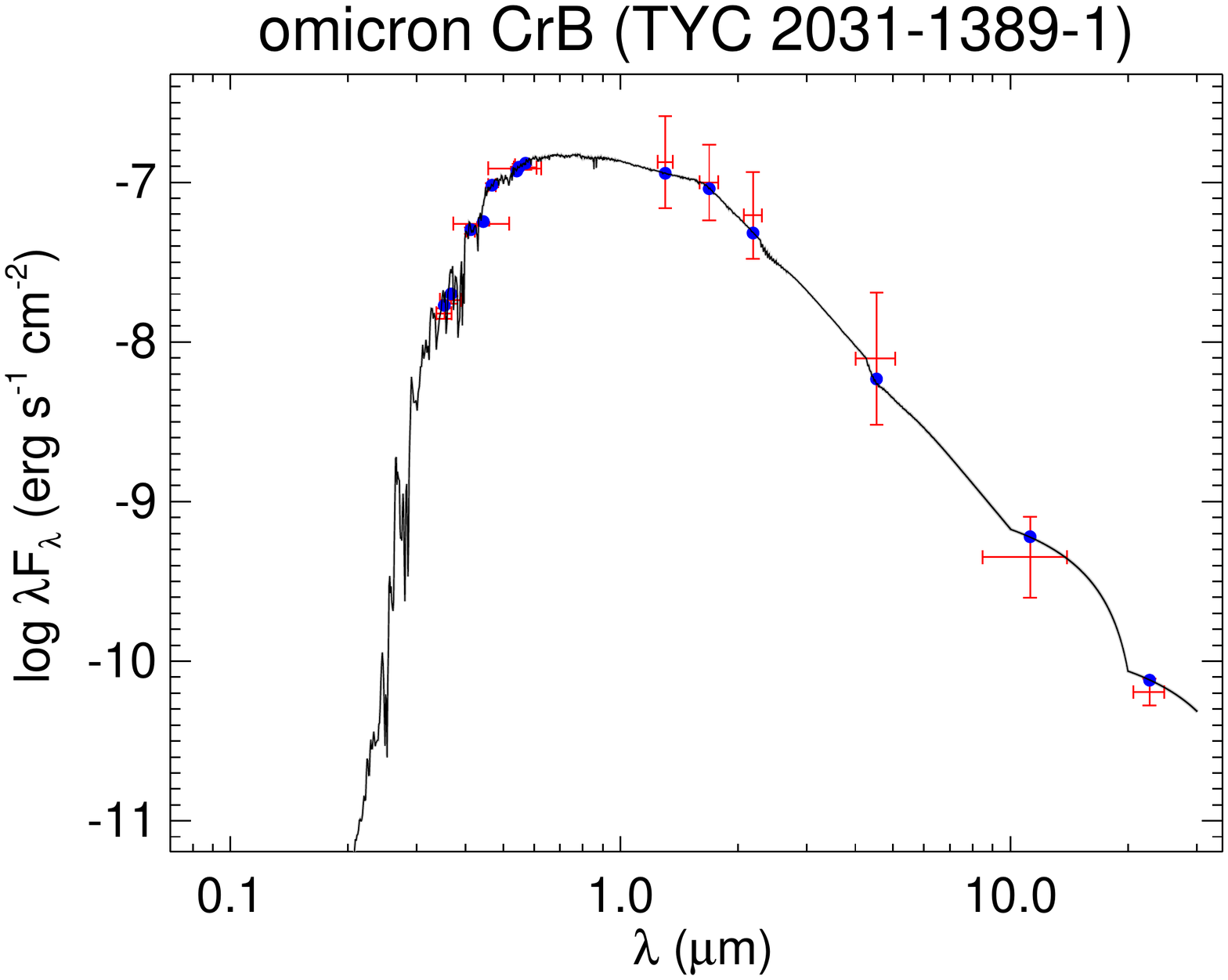}
  \includegraphics[trim=60 60 60 60,clip,width=0.49\linewidth]{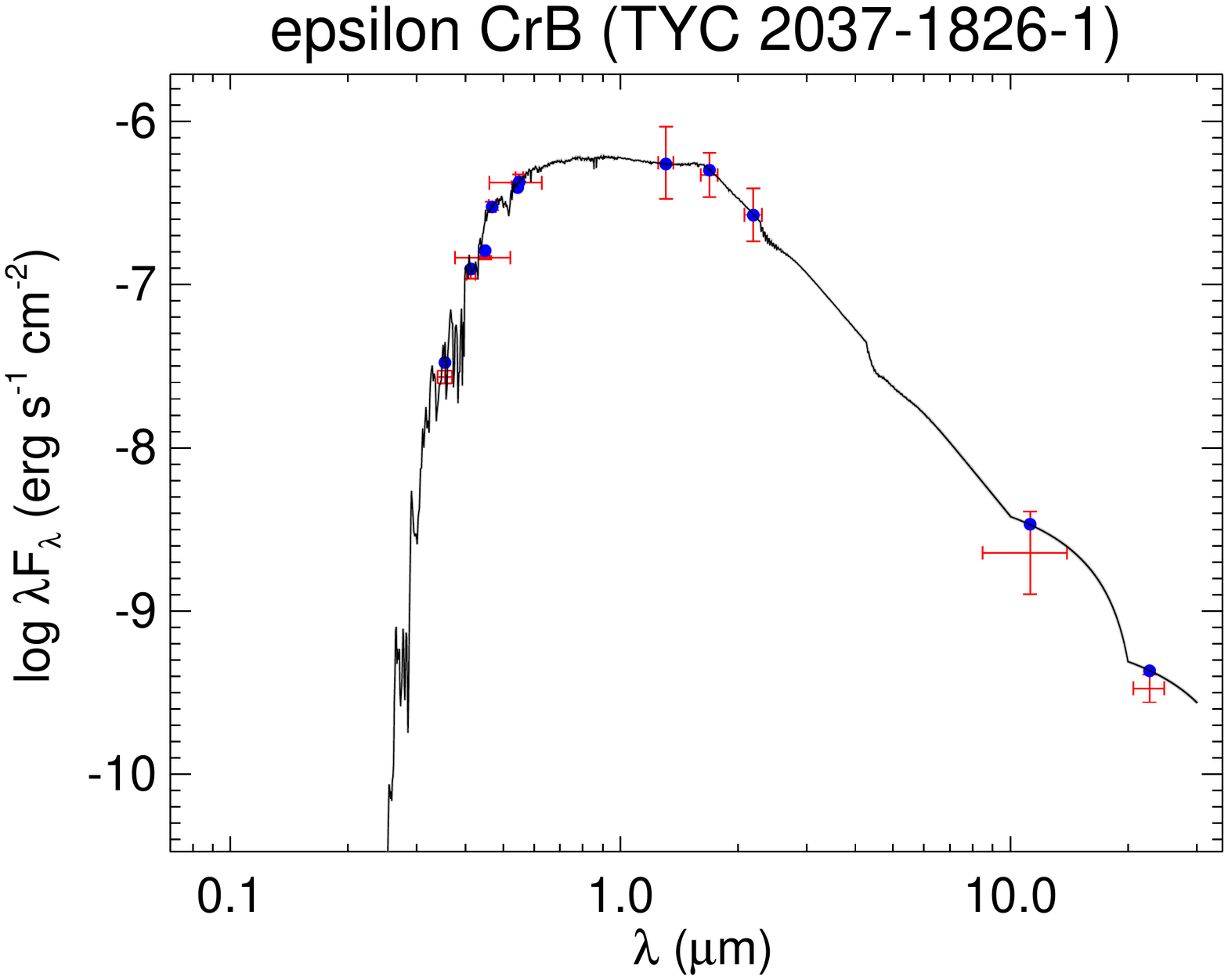}
  \includegraphics[trim=60 60 60 60,clip,width=0.49\linewidth]{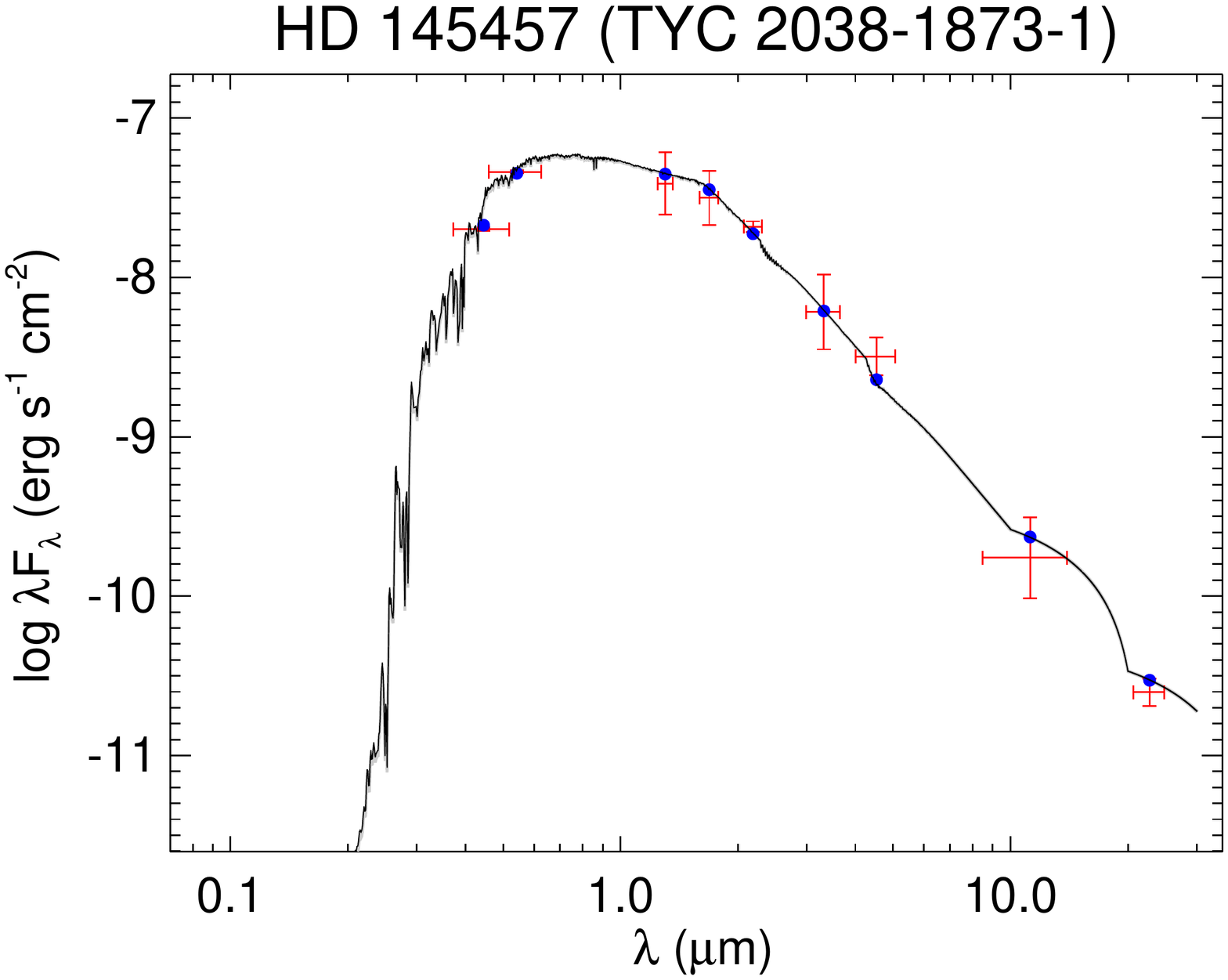}
  \includegraphics[trim=60 60 60 60,clip,width=0.49\linewidth]{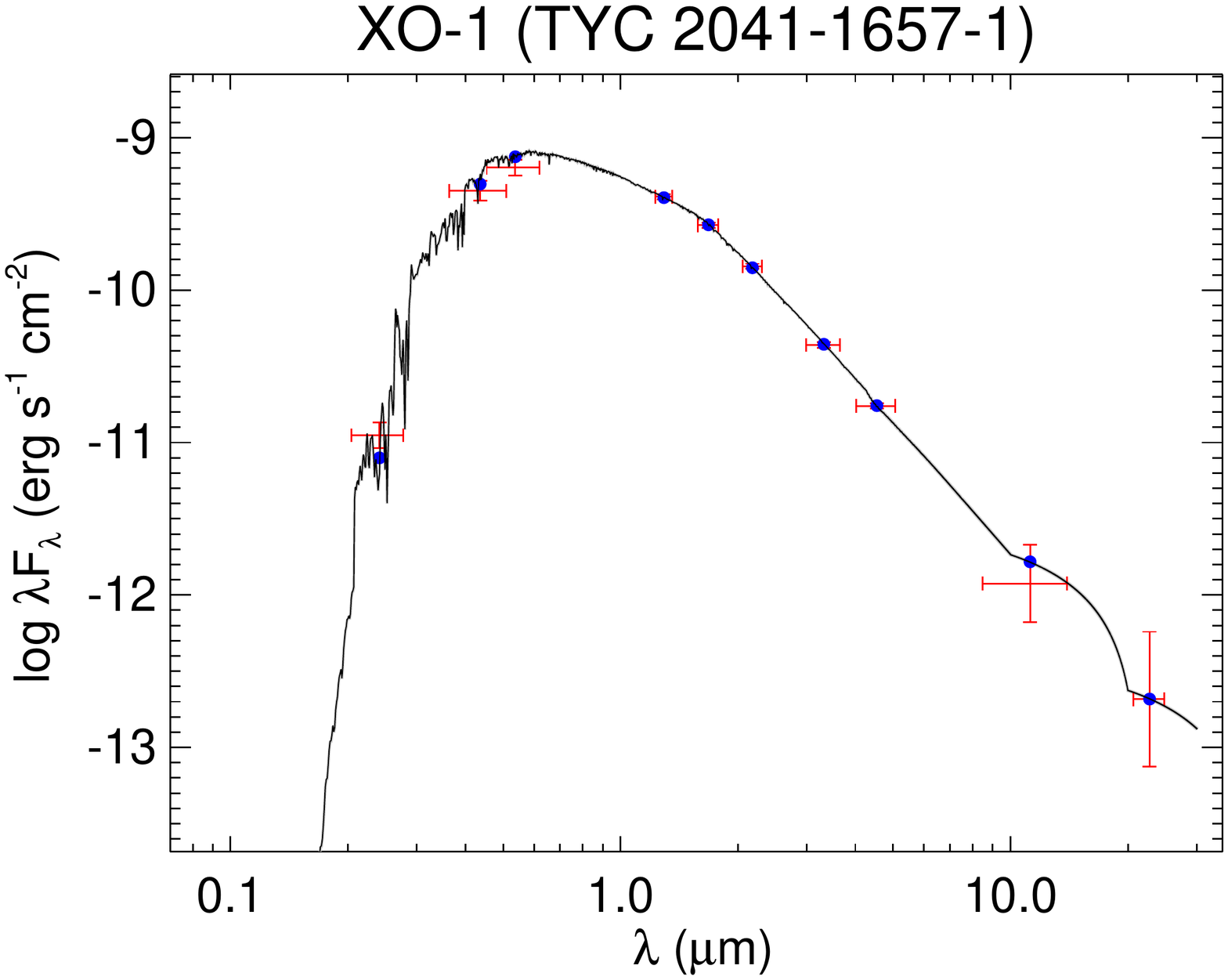}
  \caption{All labels, lines, symbols, and colors as in Figure \ref{fig:seds}.}
  \label{fig:seds_19}
\end{figure}

\begin{figure}[H]
  \centering
  \includegraphics[trim=60 60 60 60,clip,width=0.49\linewidth]{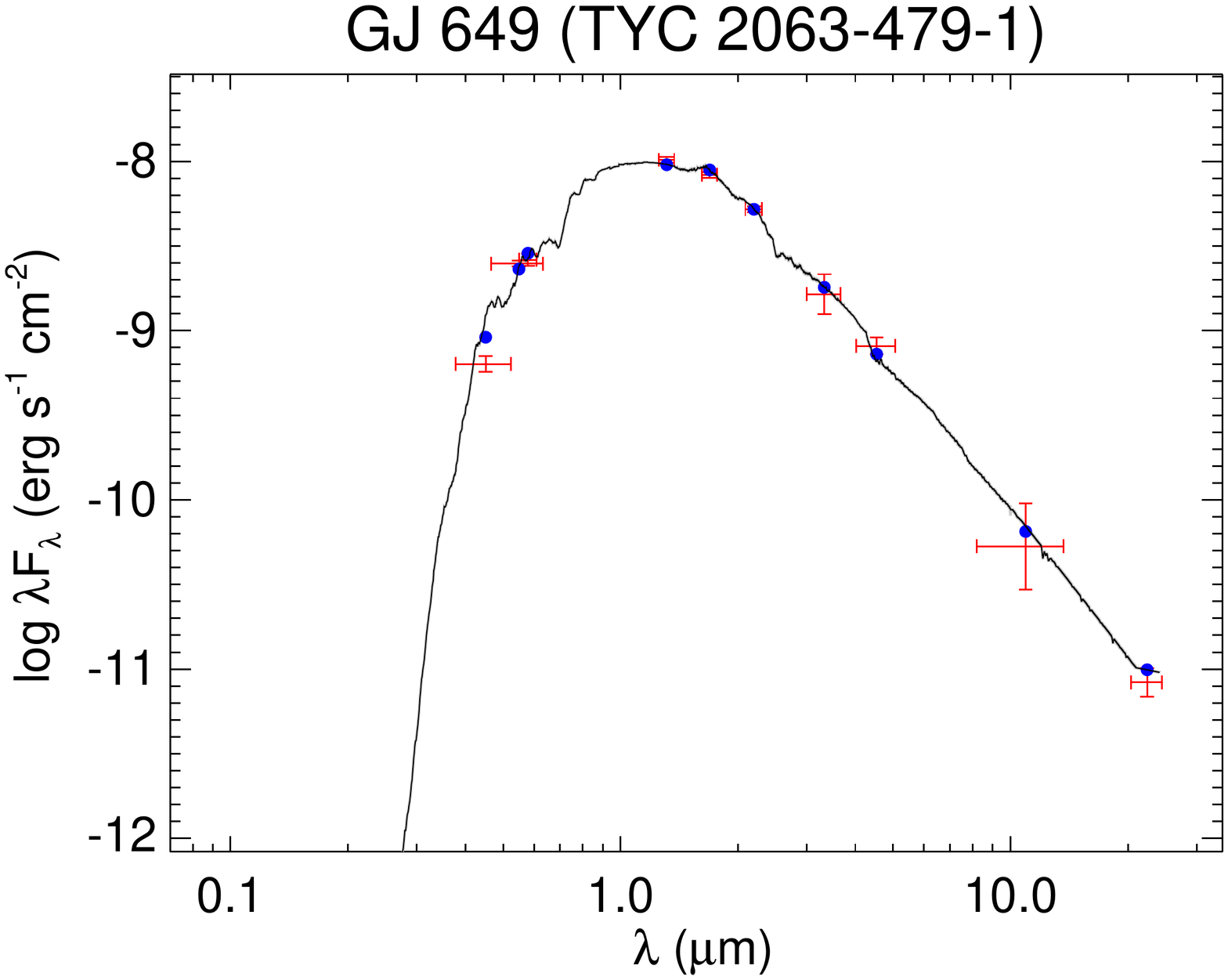}
  \includegraphics[trim=60 60 60 60,clip,width=0.49\linewidth]{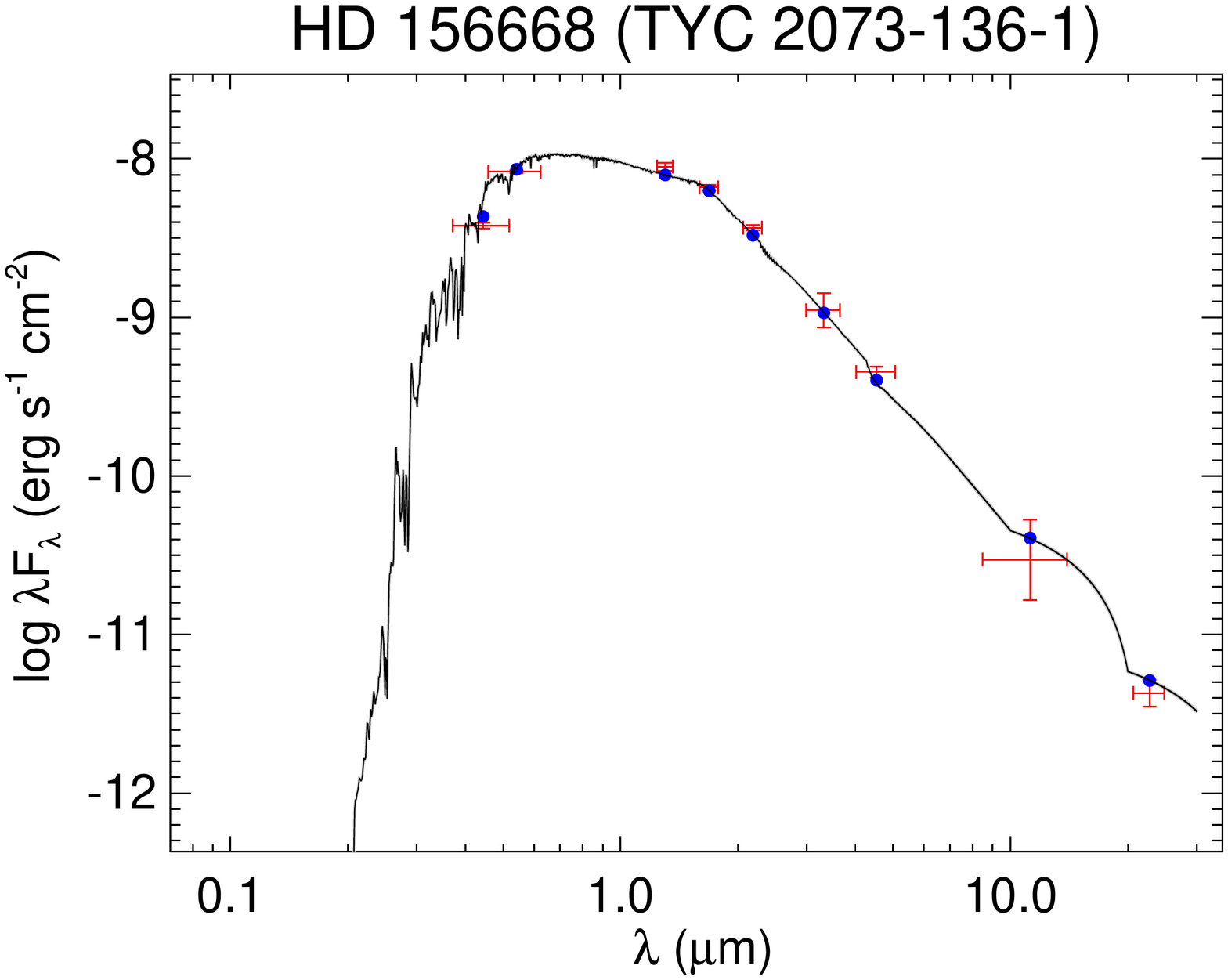}
  \includegraphics[trim=60 60 60 60,clip,width=0.49\linewidth]{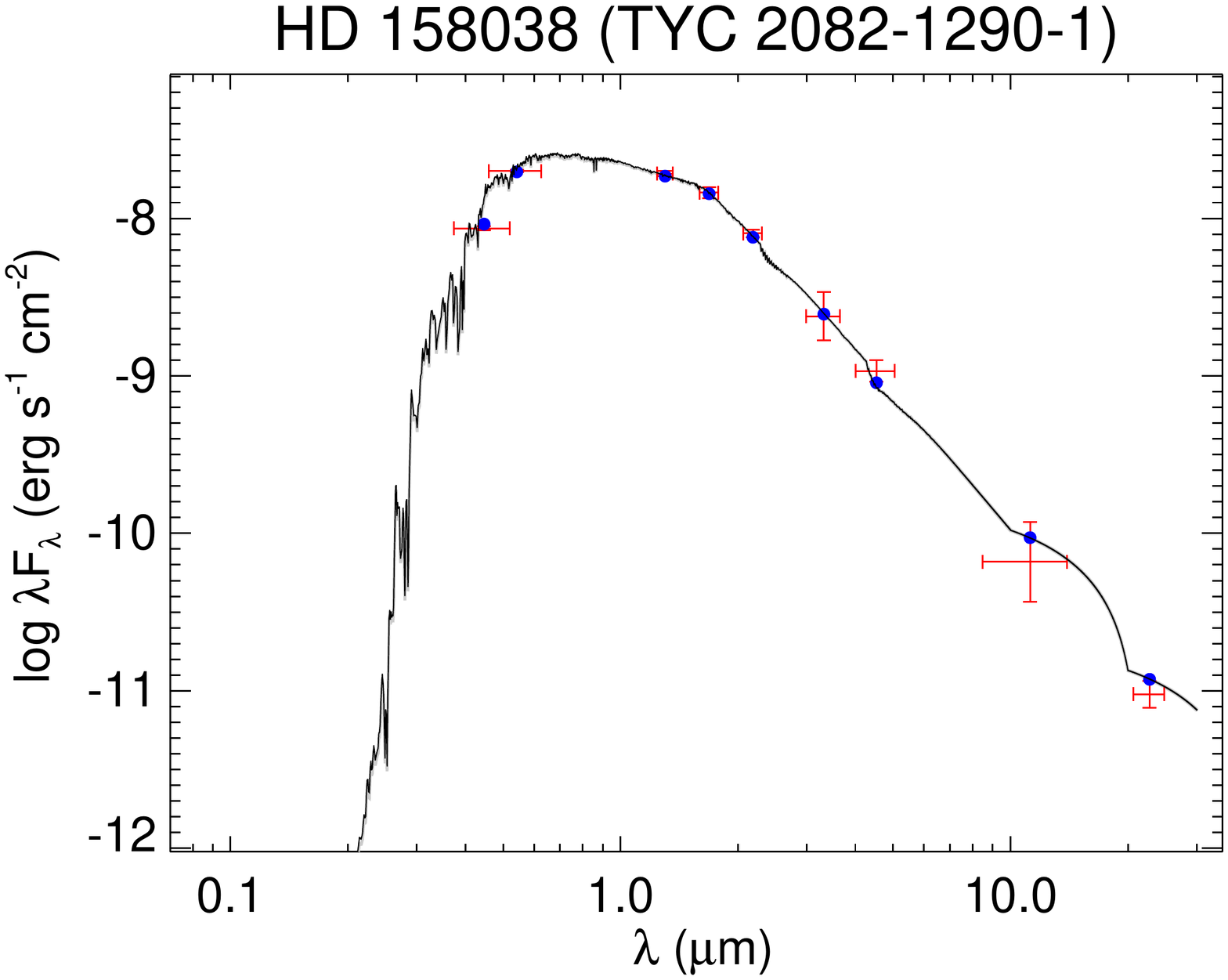}
  \includegraphics[trim=60 60 60 60,clip,width=0.49\linewidth]{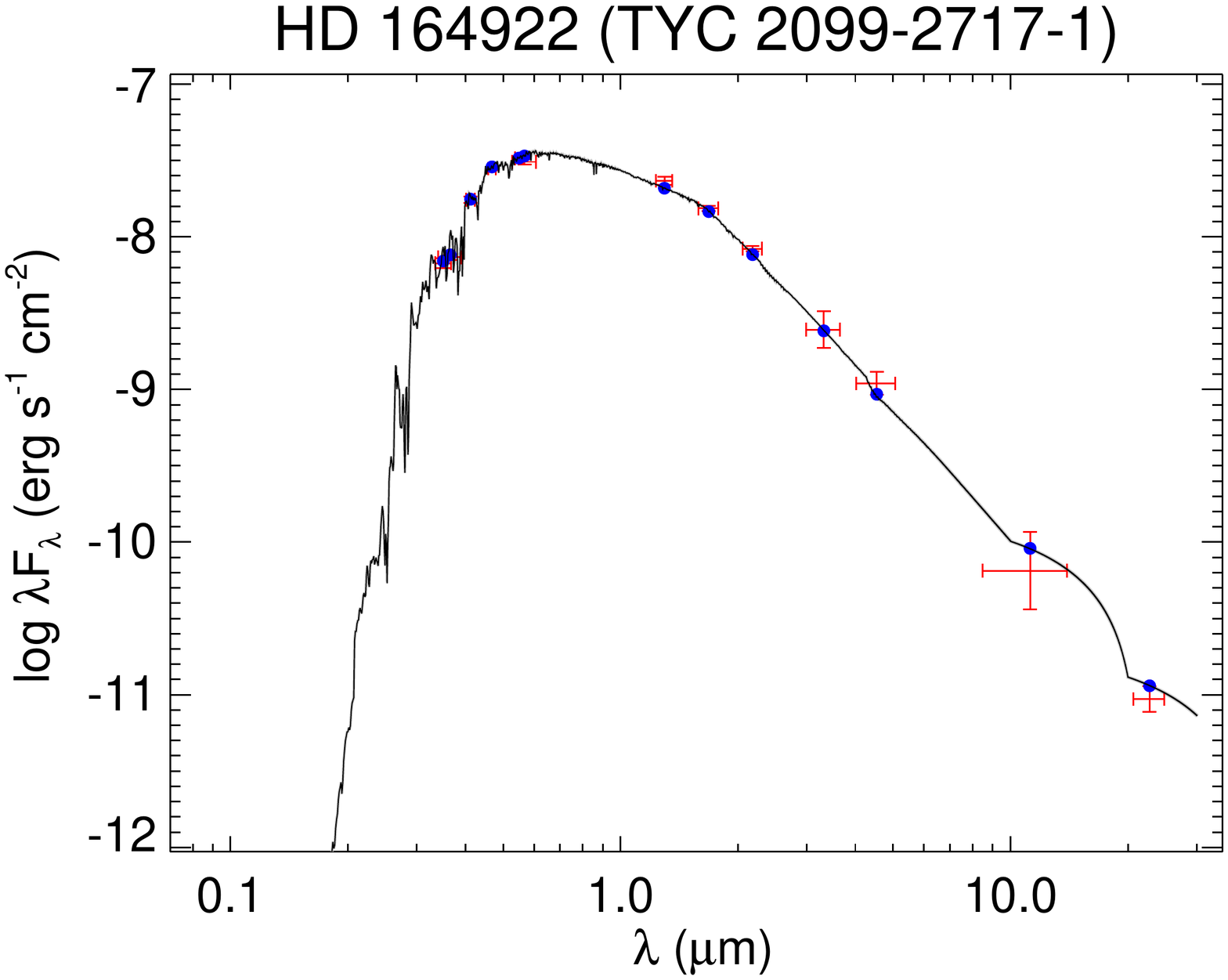}
  \includegraphics[trim=60 60 60 60,clip,width=0.49\linewidth]{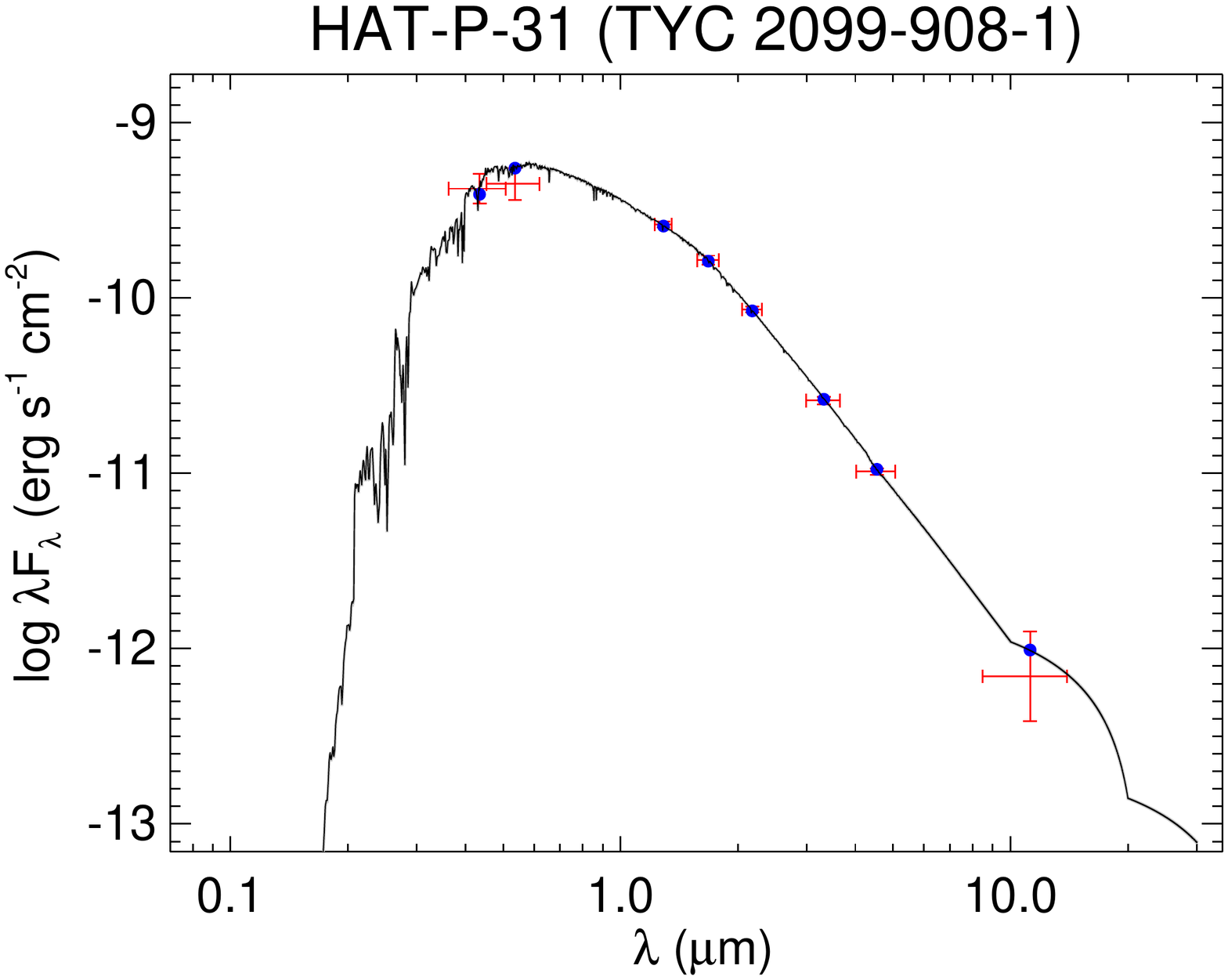}
  \includegraphics[trim=60 60 60 60,clip,width=0.49\linewidth]{sedfigs/tyc_2109-49-1.pdf}
  \caption{All labels, lines, symbols, and colors as in Figure \ref{fig:seds}.}
  \label{fig:seds_20}
\end{figure}

\begin{figure}[H]
  \centering
  \includegraphics[trim=60 60 60 60,clip,width=0.49\linewidth]{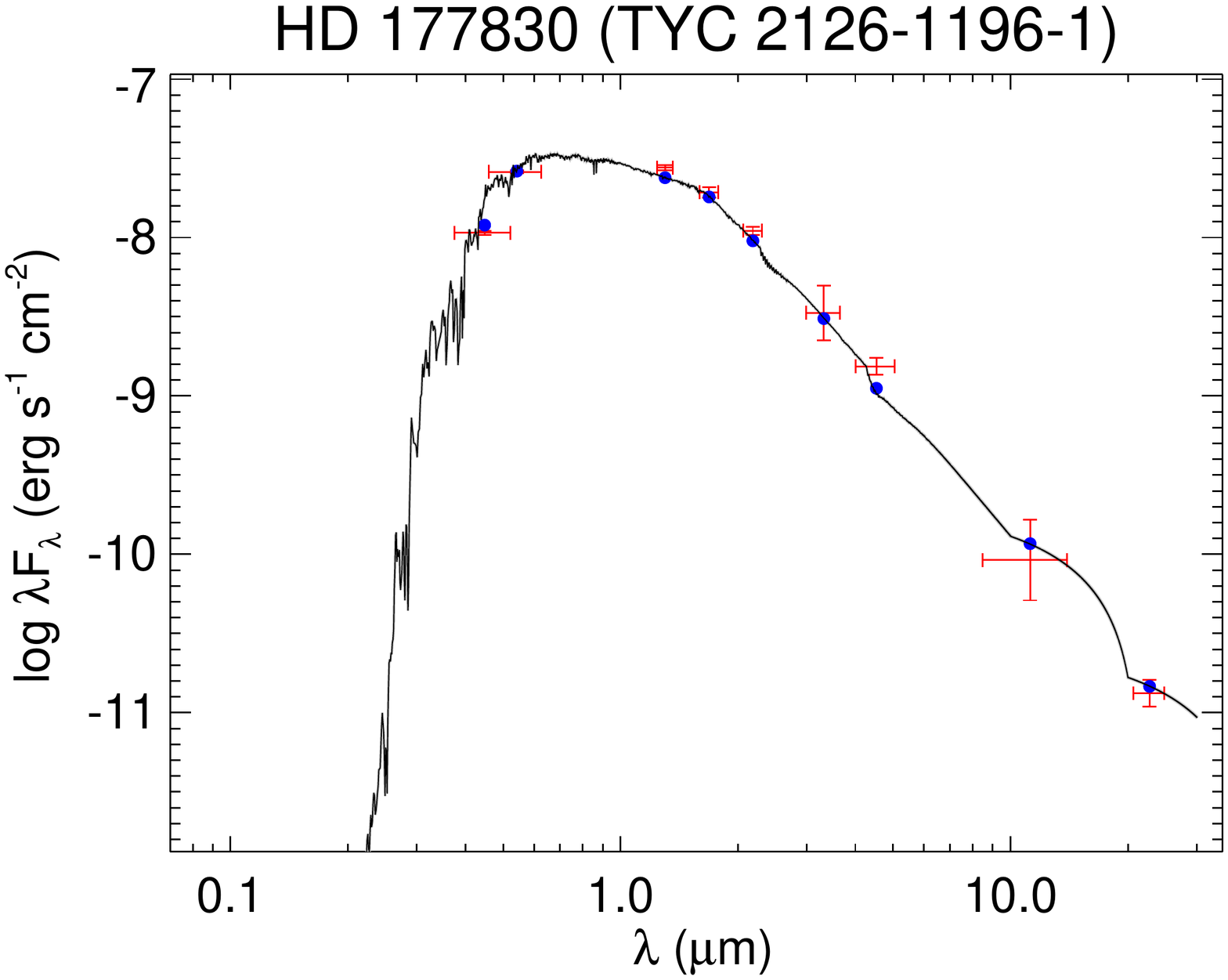}
  \includegraphics[trim=60 60 60 60,clip,width=0.49\linewidth]{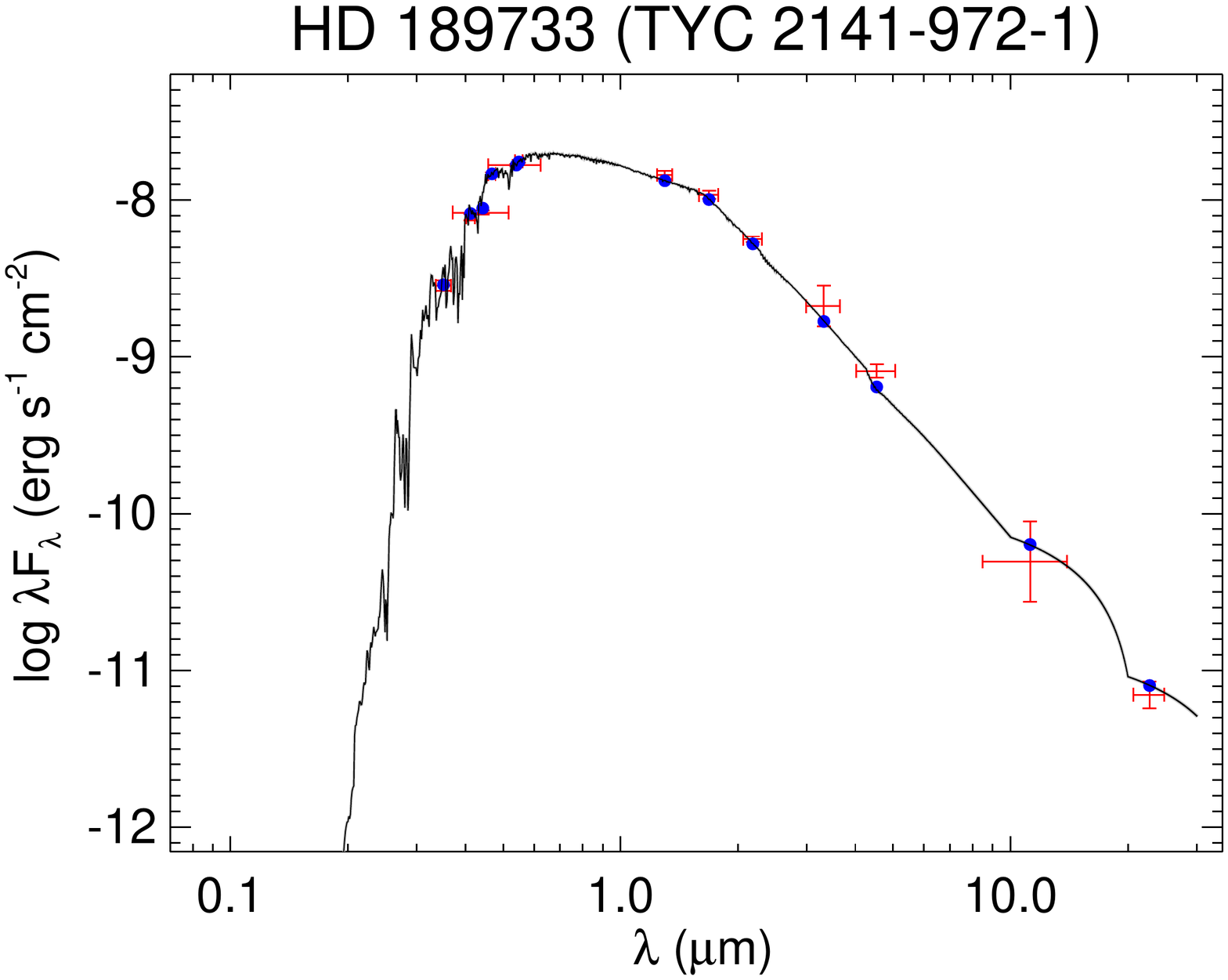}
  \includegraphics[trim=60 60 60 60,clip,width=0.49\linewidth]{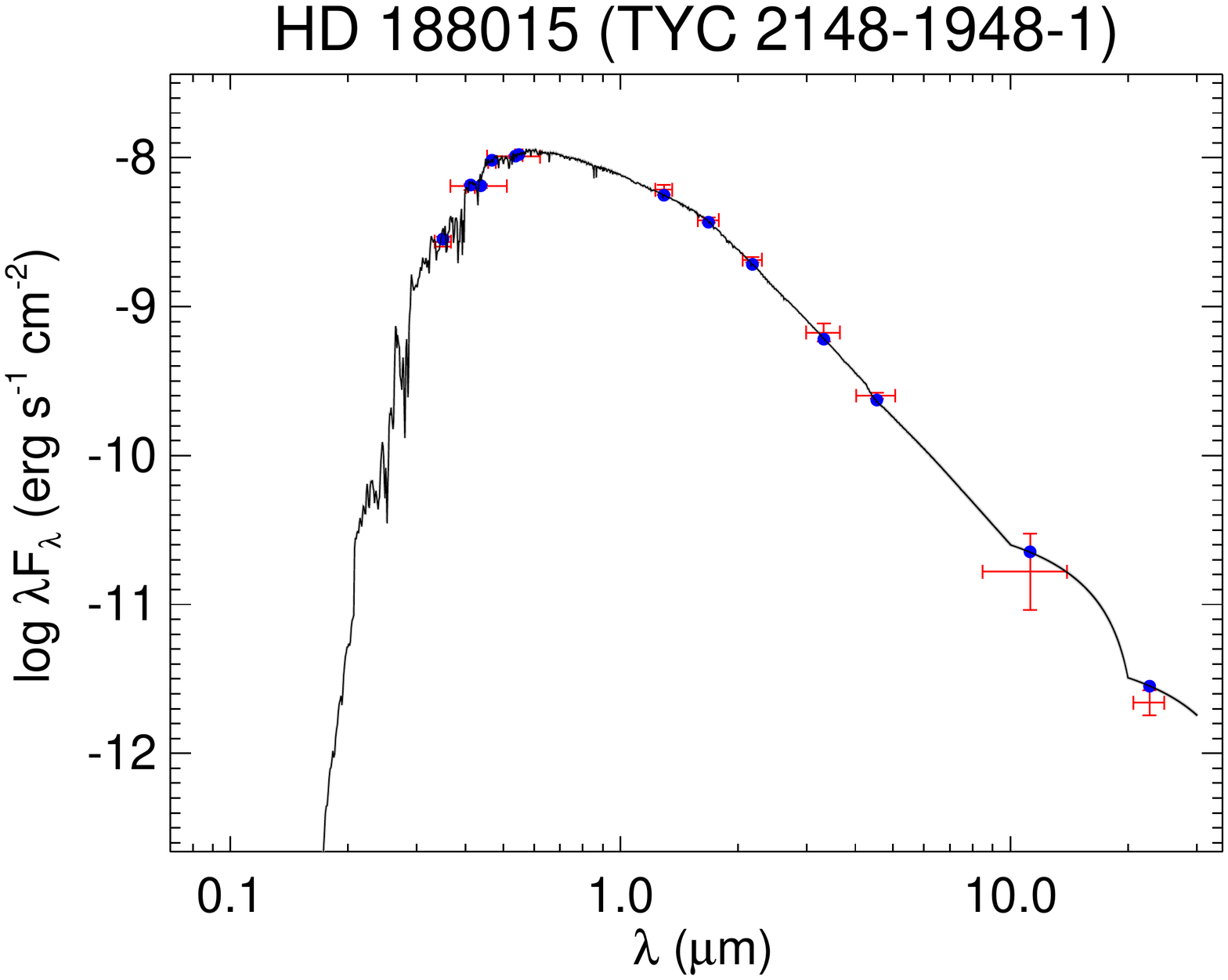}
  \includegraphics[trim=60 60 60 60,clip,width=0.49\linewidth]{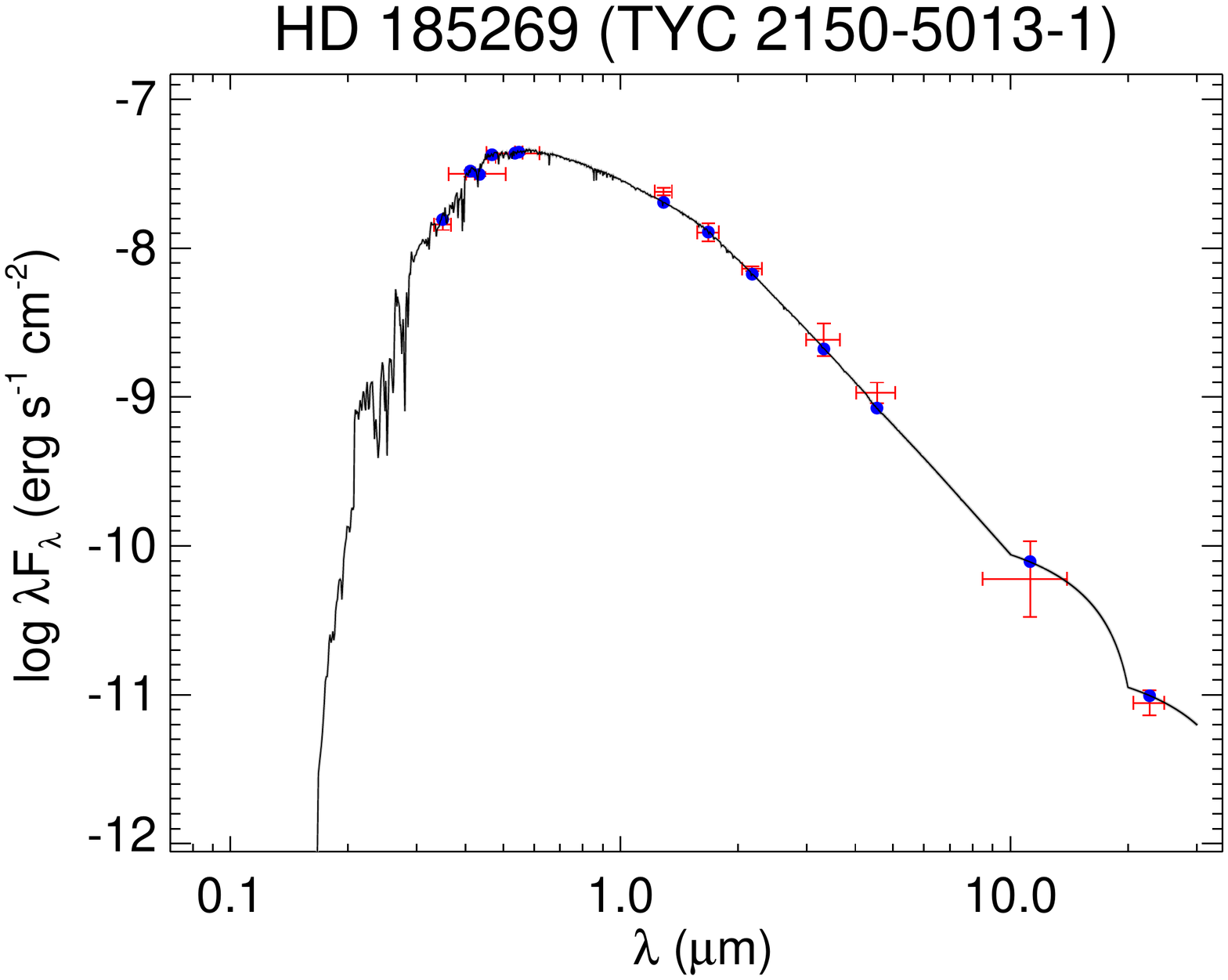}
  \includegraphics[trim=60 60 60 60,clip,width=0.49\linewidth]{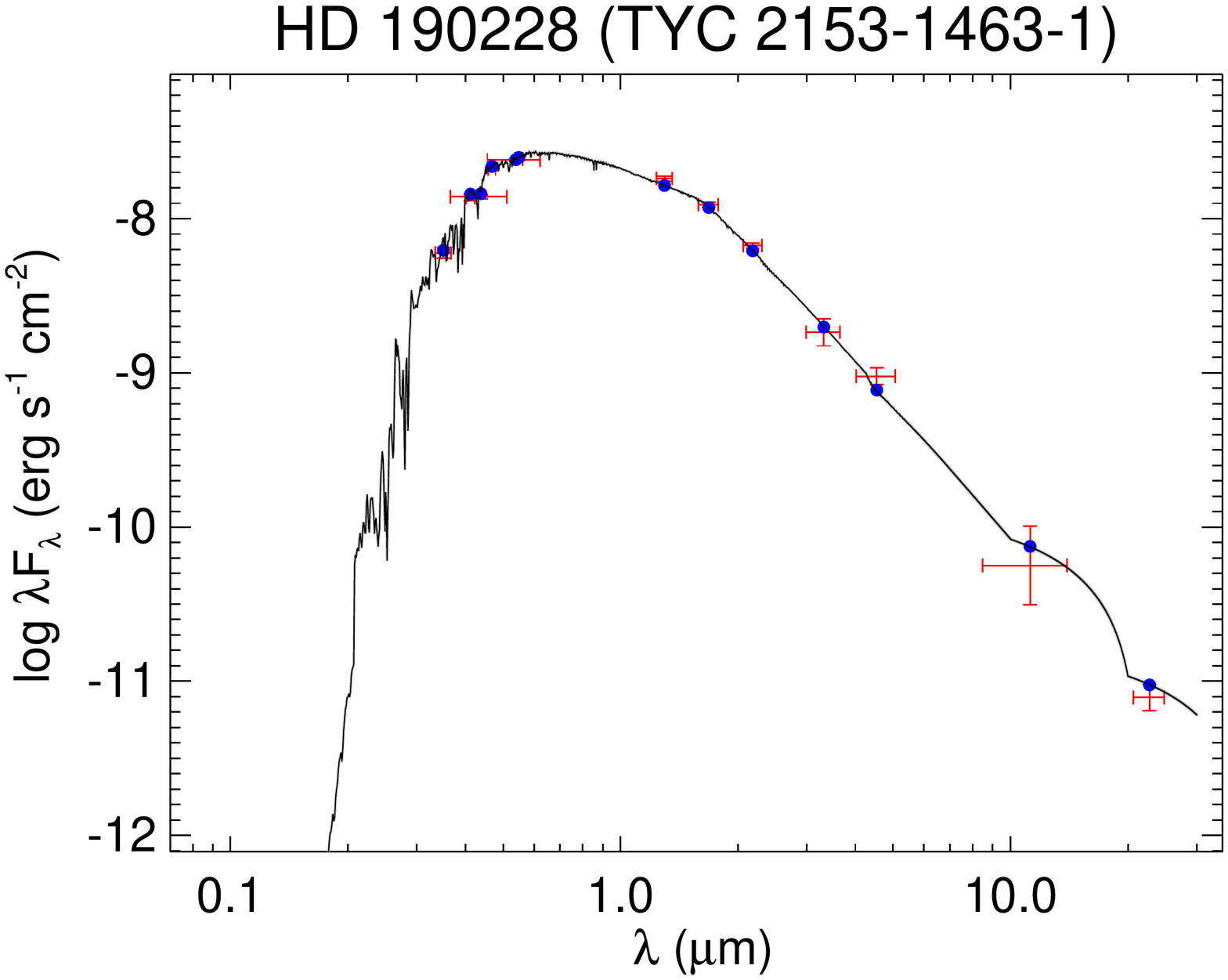}
  \includegraphics[trim=60 60 60 60,clip,width=0.49\linewidth]{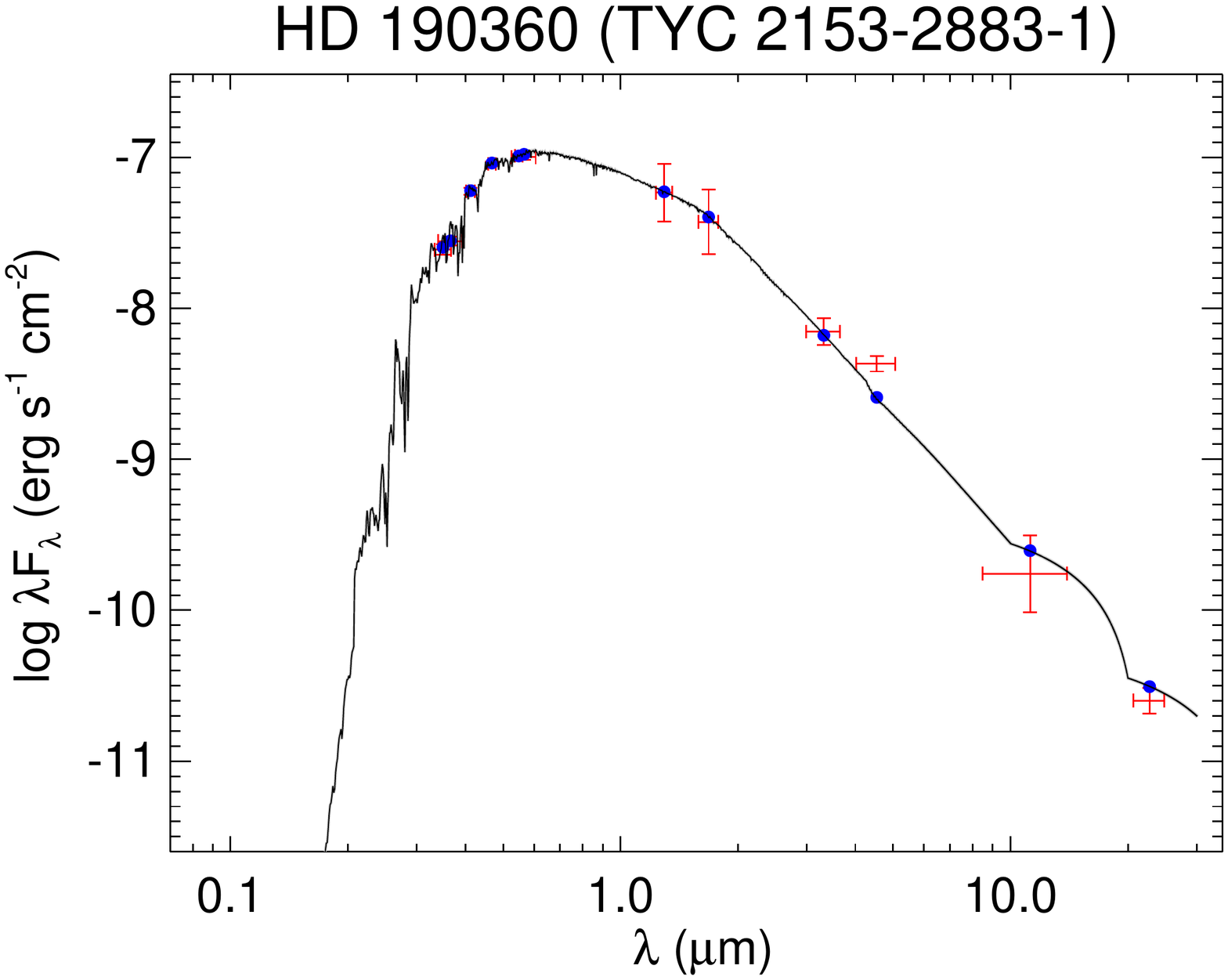}
  \caption{All labels, lines, symbols, and colors as in Figure \ref{fig:seds}.}
  \label{fig:seds_21}
\end{figure}

\begin{figure}[H]
  \centering
  \includegraphics[trim=60 60 60 60,clip,width=0.49\linewidth]{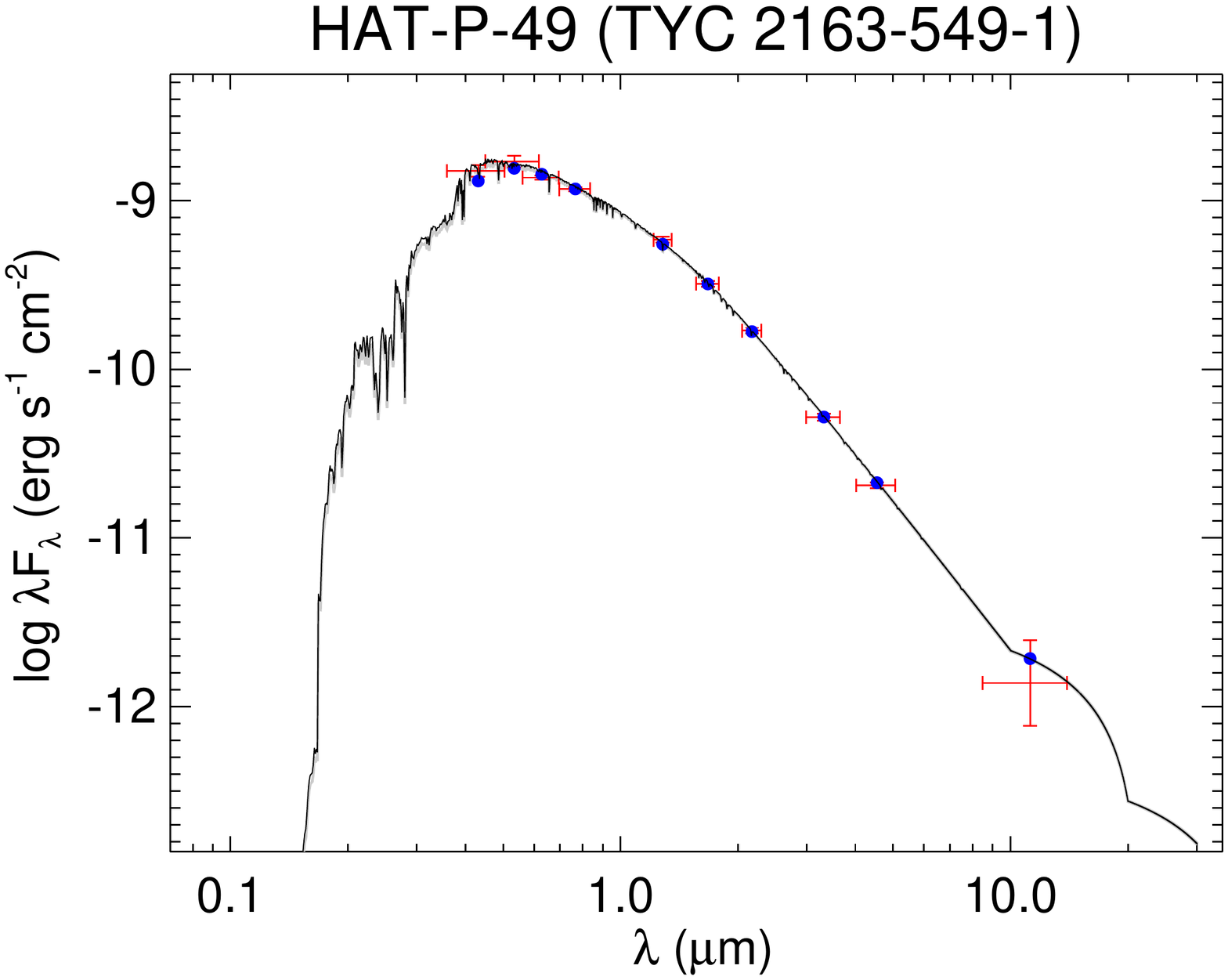}
  \includegraphics[trim=60 60 60 60,clip,width=0.49\linewidth]{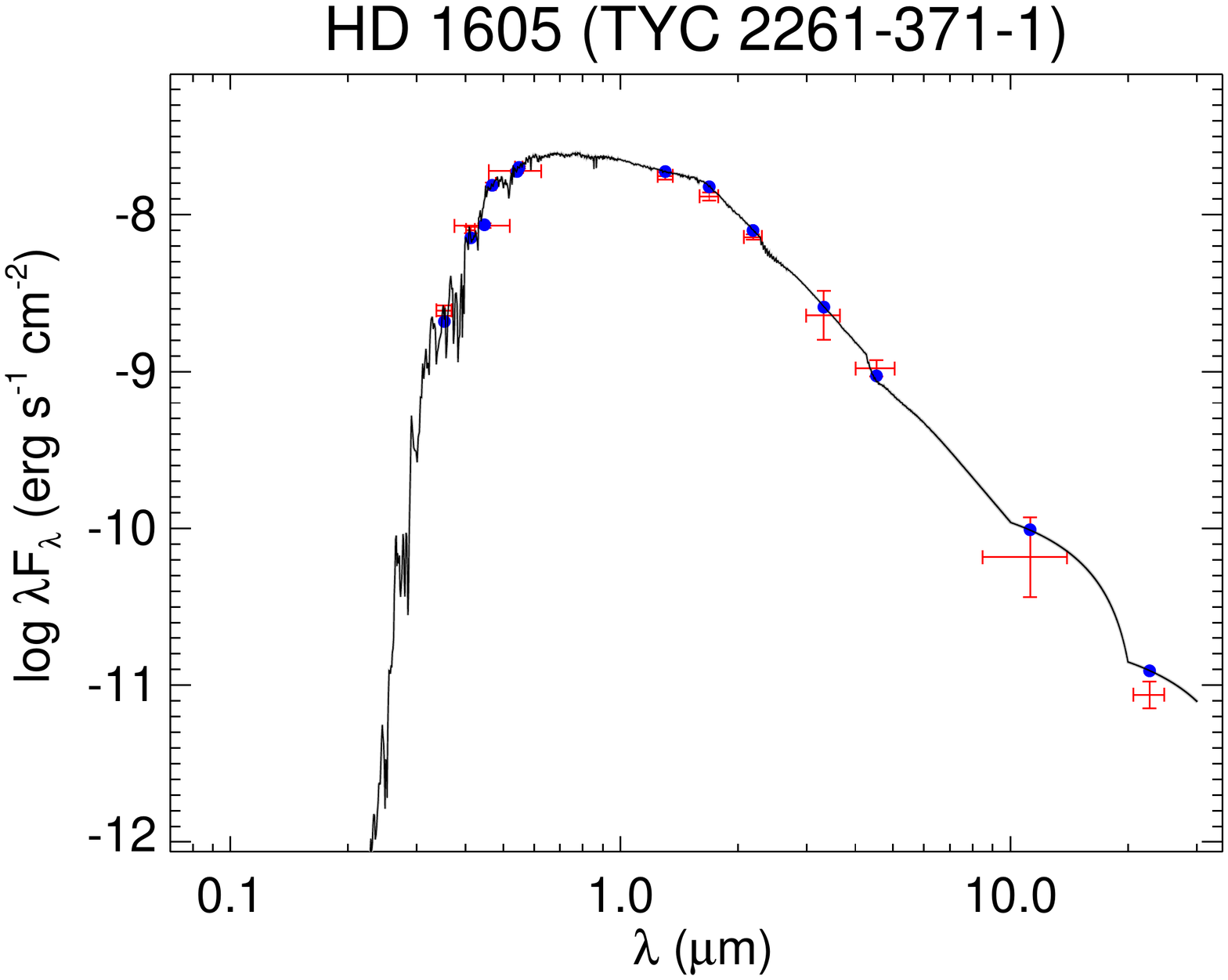}
  \includegraphics[trim=60 60 60 60,clip,width=0.49\linewidth]{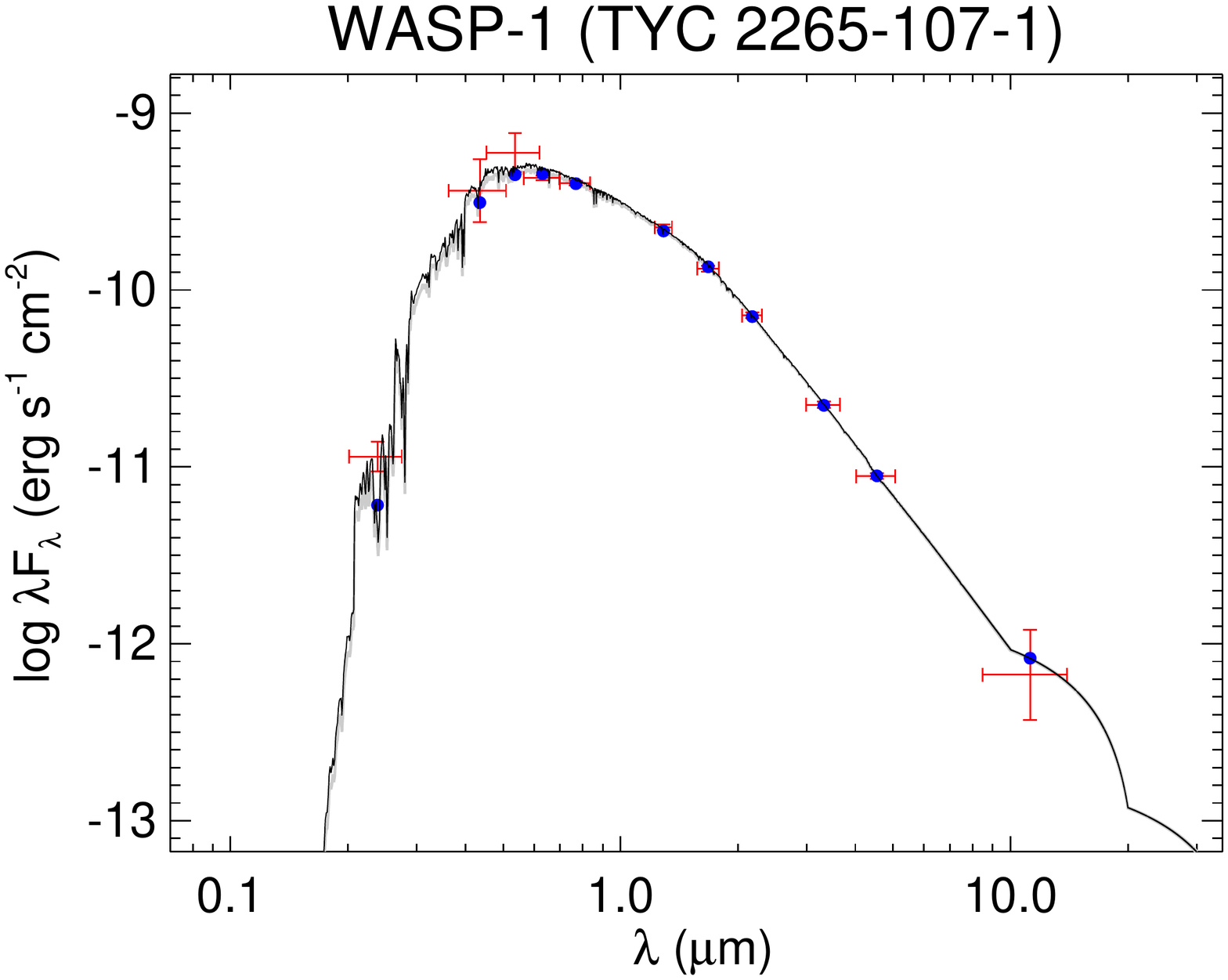}
  \includegraphics[trim=60 60 60 60,clip,width=0.49\linewidth]{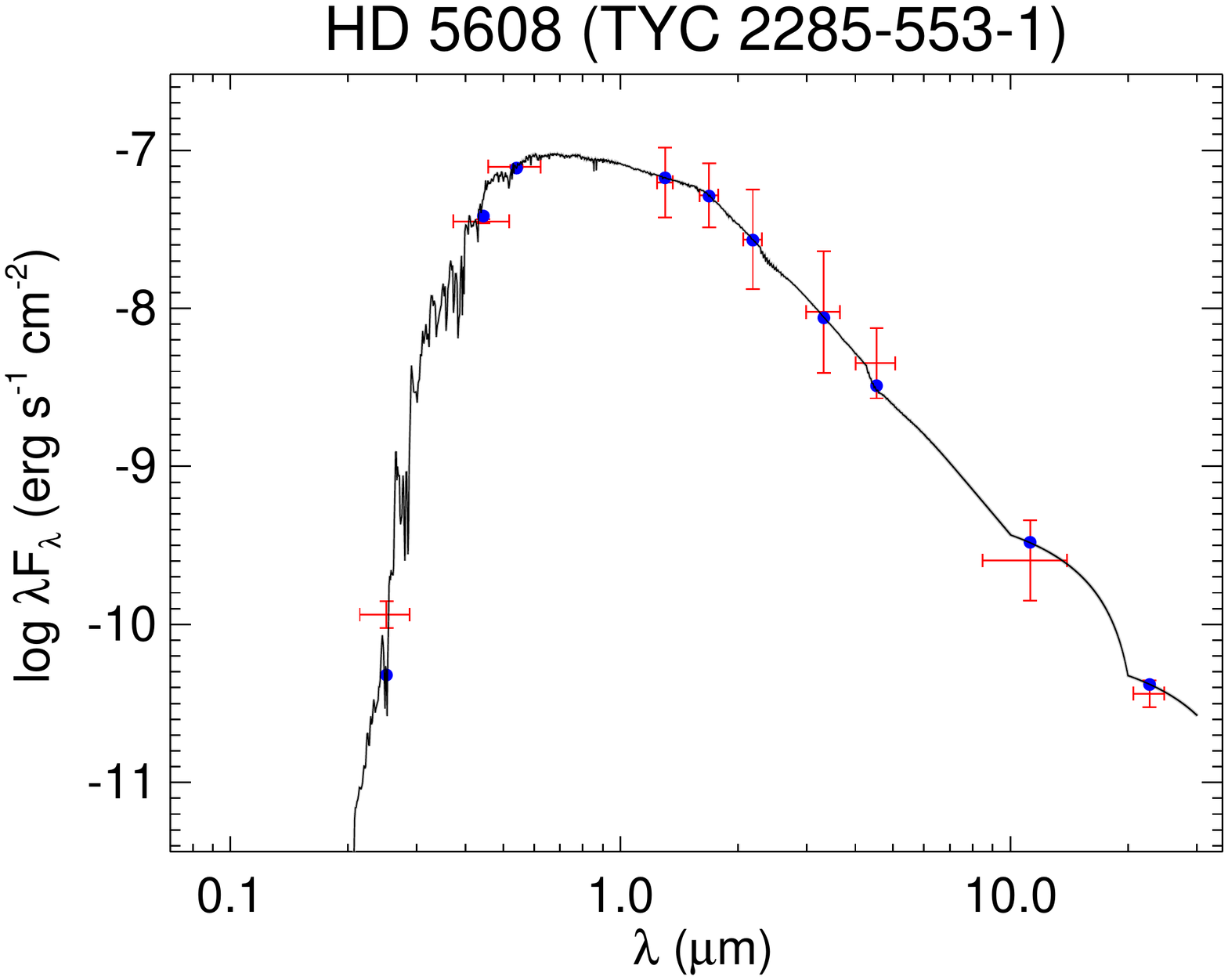}
  \includegraphics[trim=60 60 60 60,clip,width=0.49\linewidth]{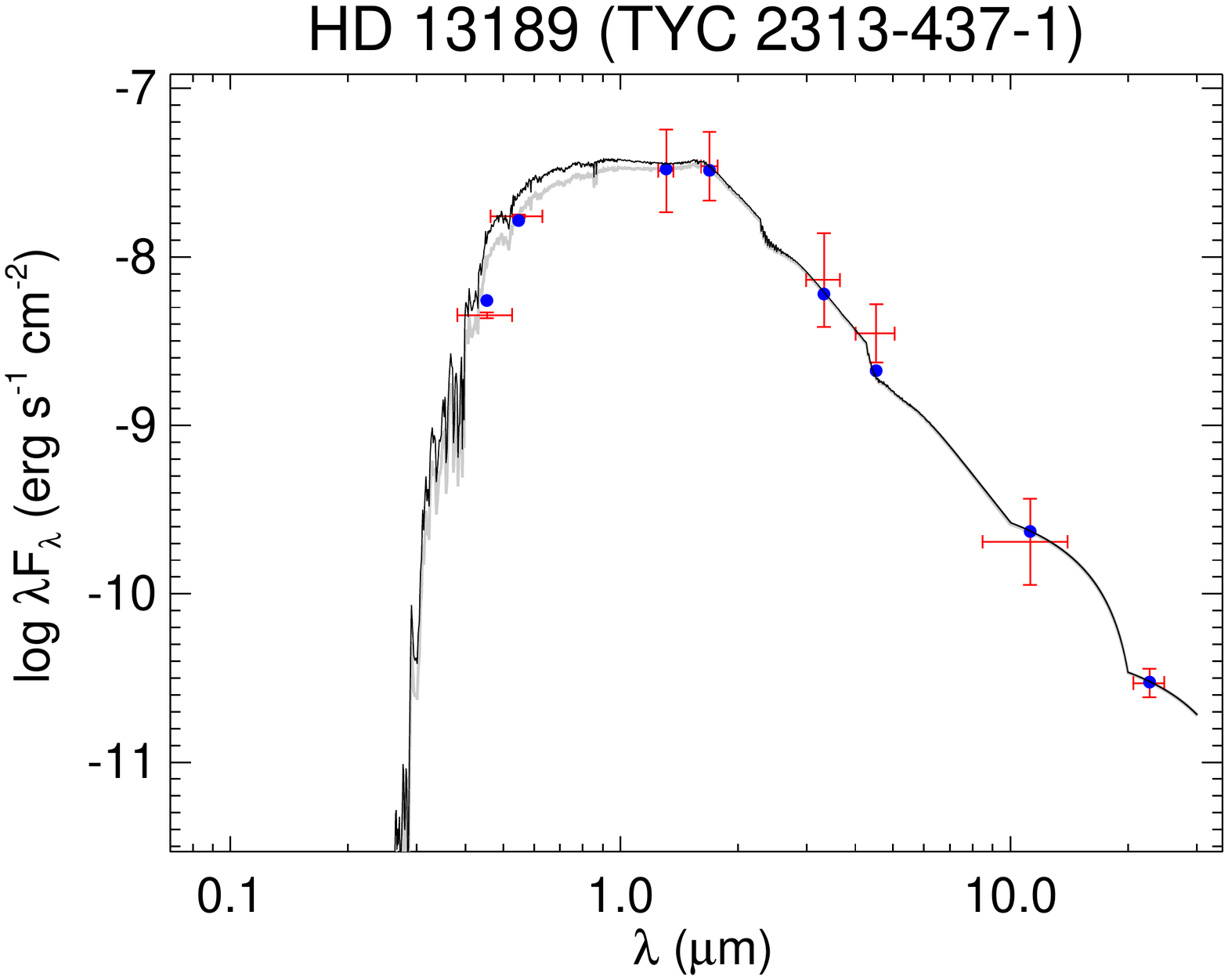}
  \includegraphics[trim=60 60 60 60,clip,width=0.49\linewidth]{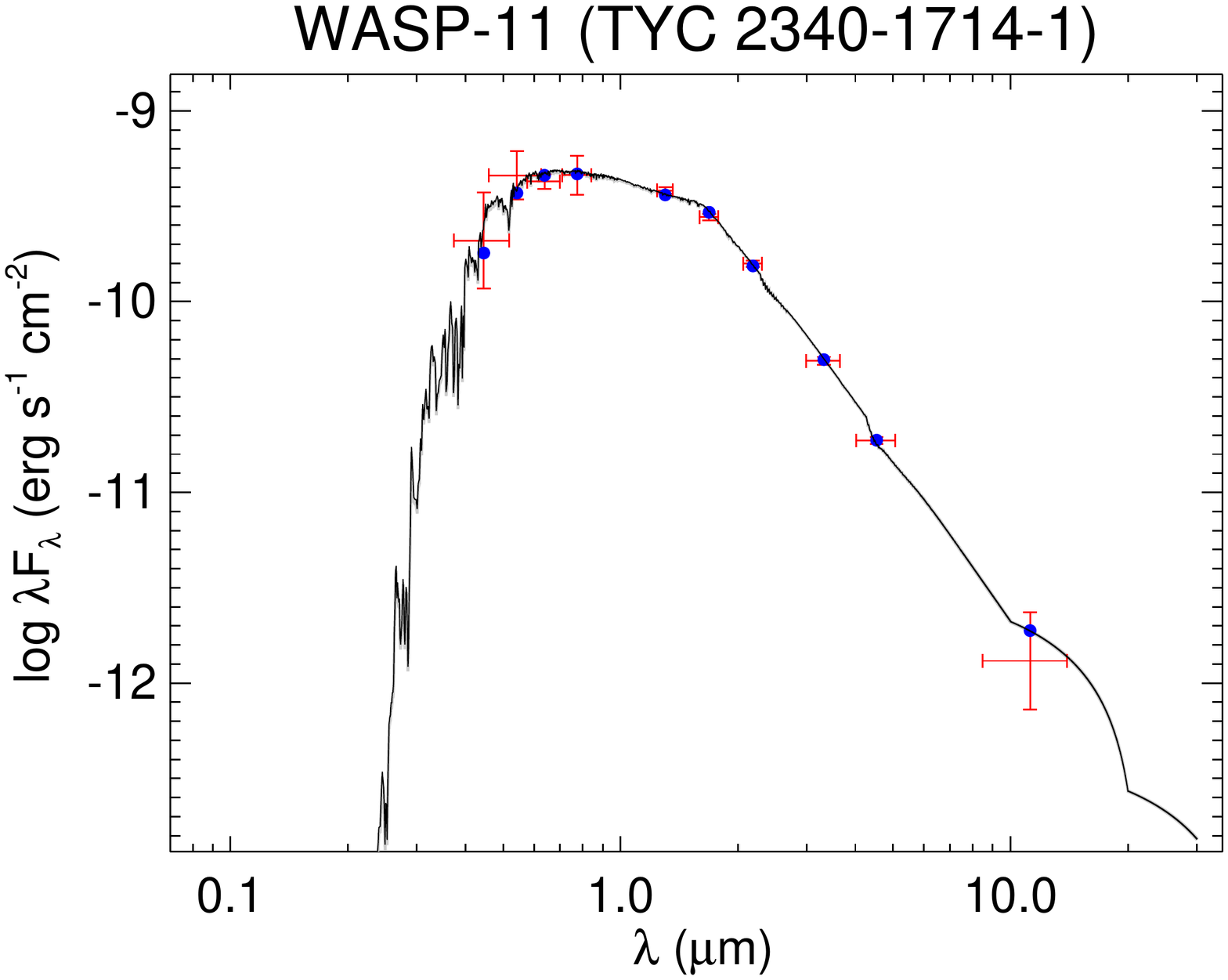}
  \caption{All labels, lines, symbols, and colors as in Figure \ref{fig:seds}.}
  \label{fig:seds_22}
\end{figure}

\begin{figure}[H]
  \centering
  \includegraphics[trim=60 60 60 60,clip,width=0.49\linewidth]{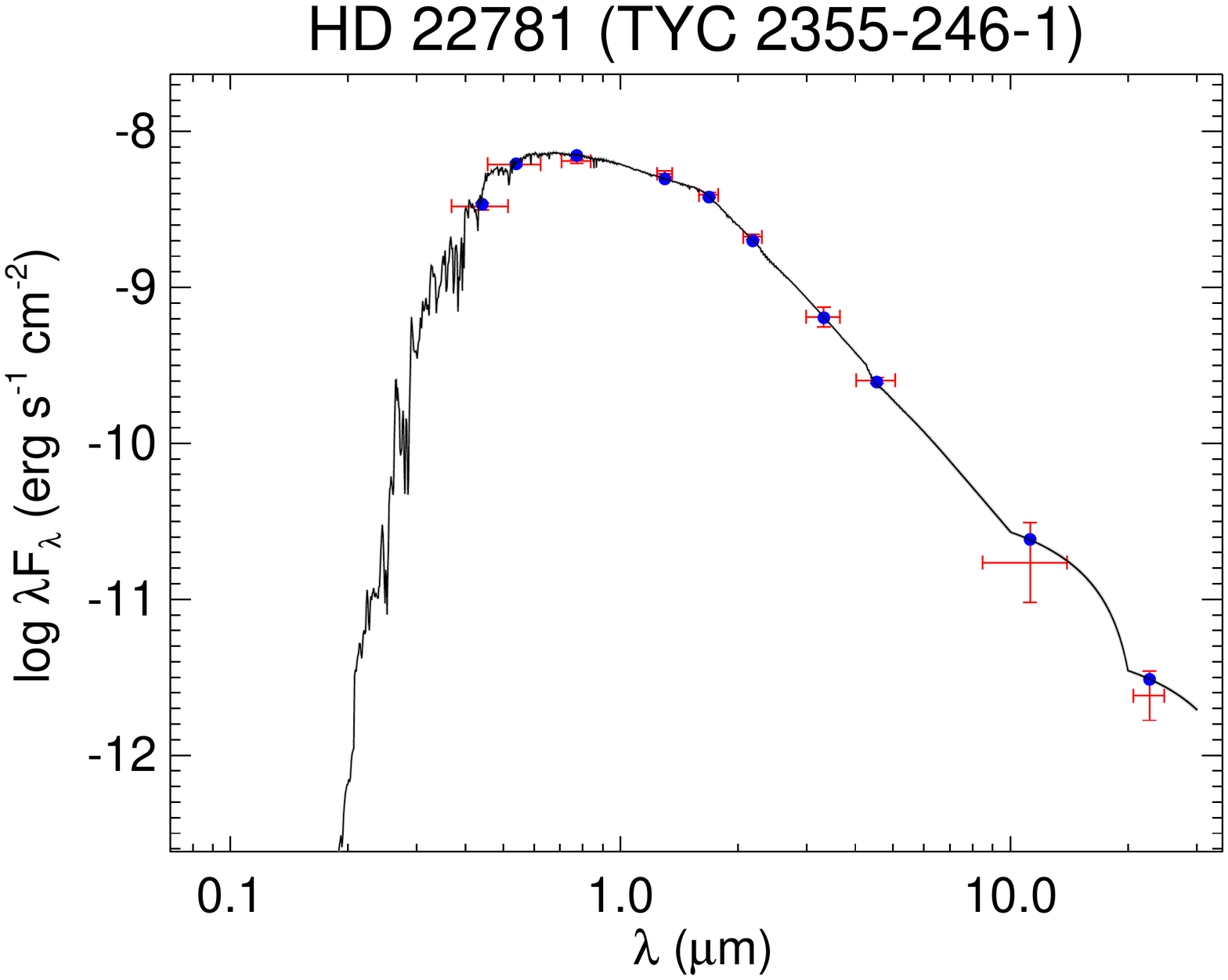}
  \includegraphics[trim=60 60 60 60,clip,width=0.49\linewidth]{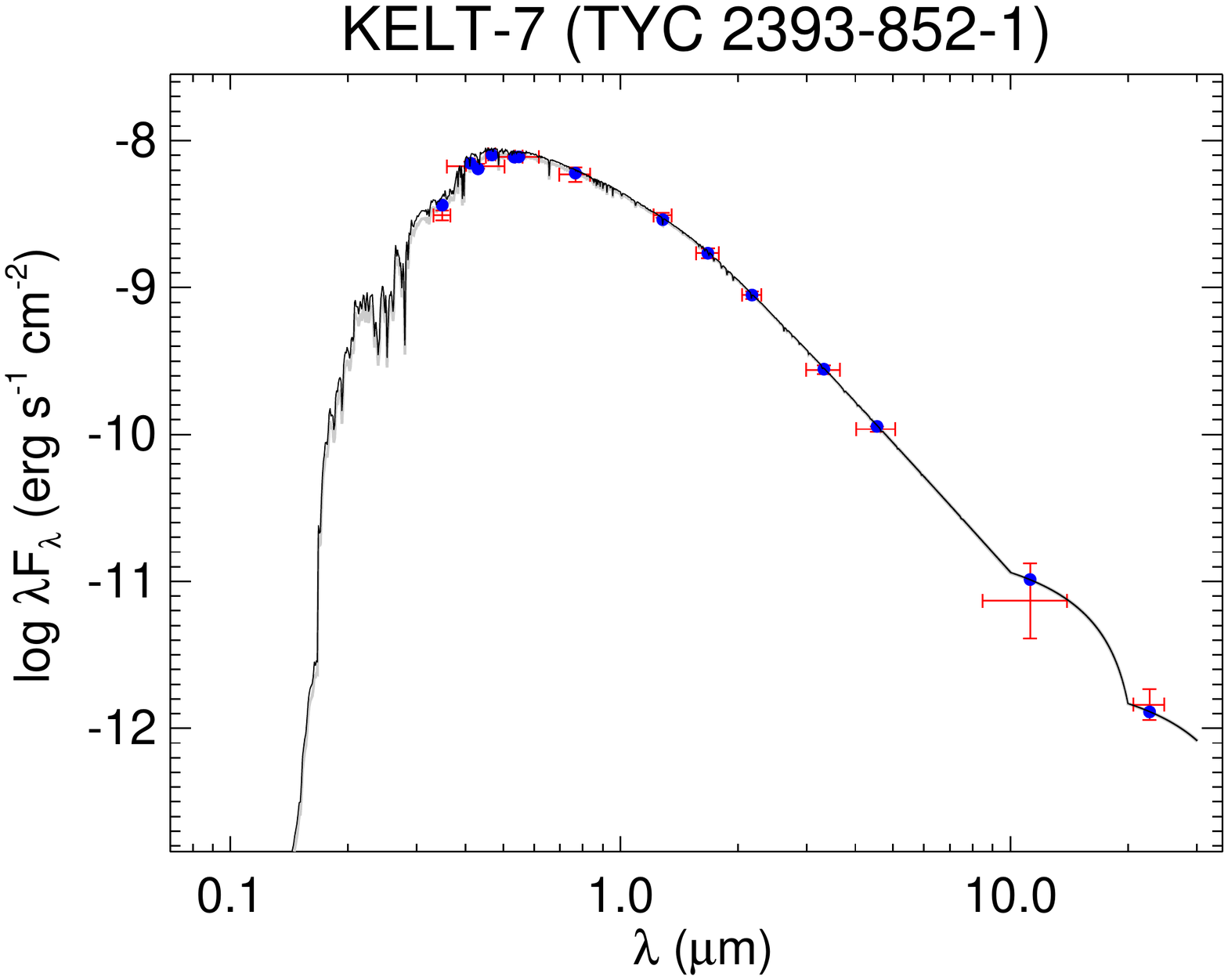}
  \includegraphics[trim=60 60 60 60,clip,width=0.49\linewidth]{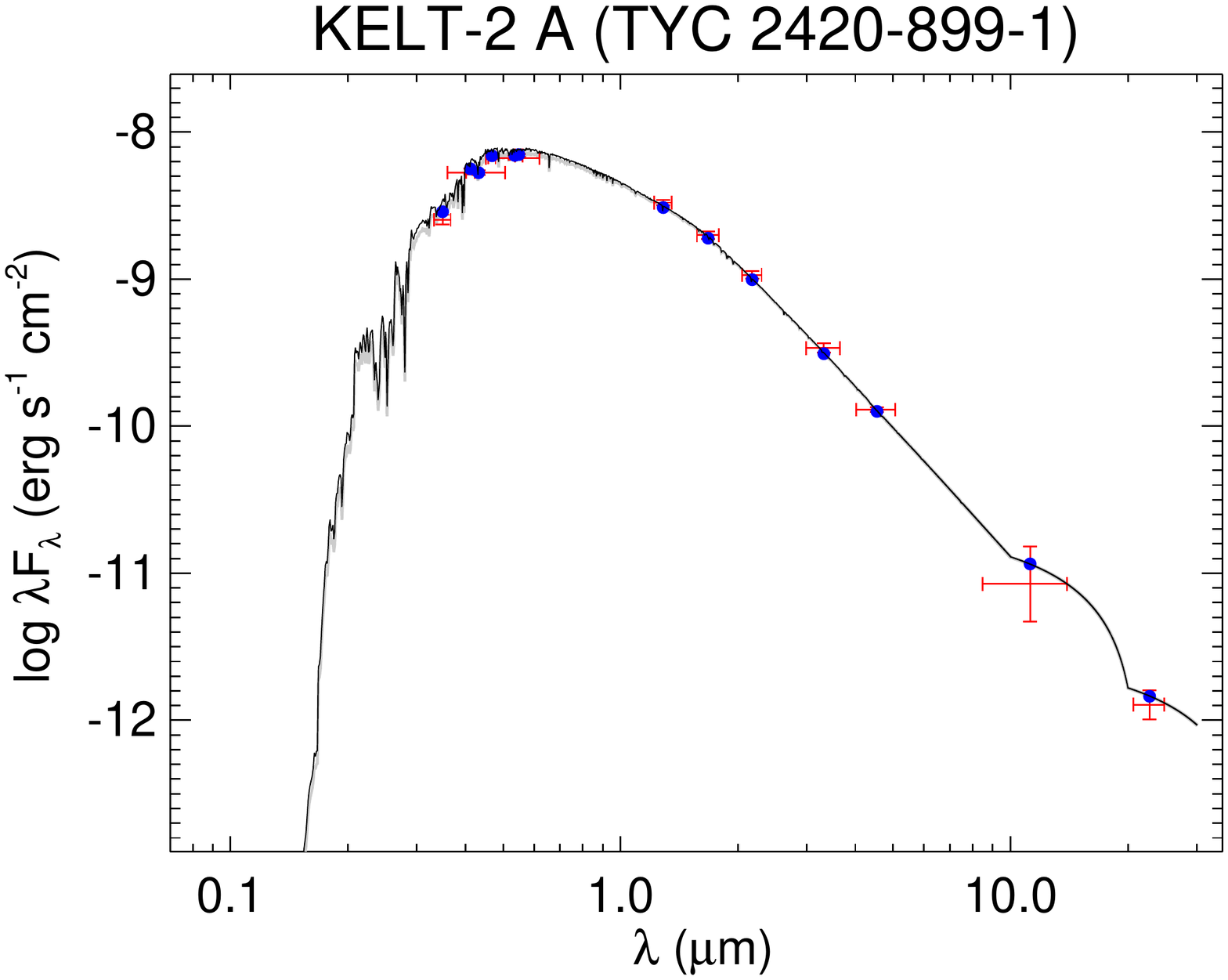}
  \includegraphics[trim=60 60 60 60,clip,width=0.49\linewidth]{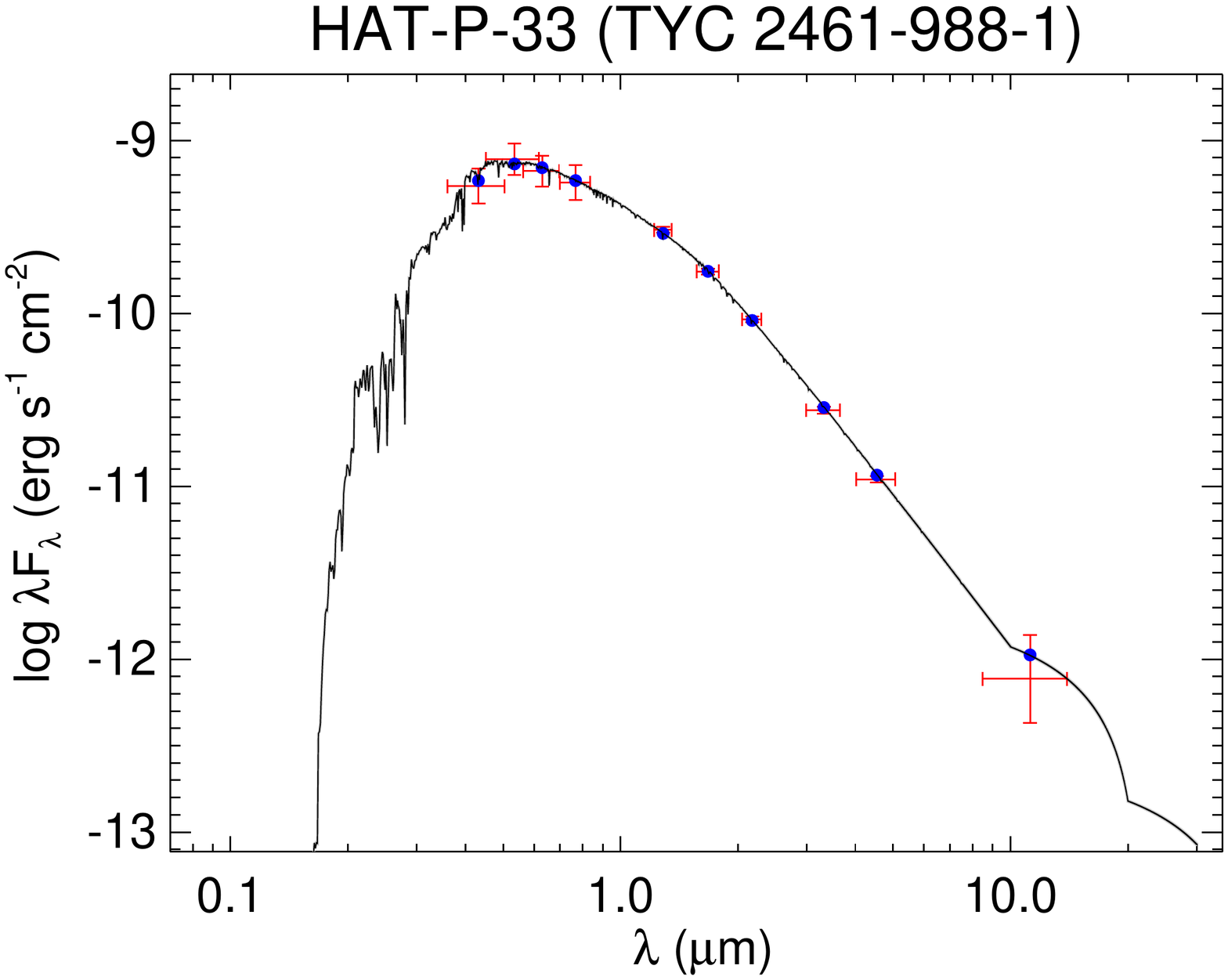}
  \includegraphics[trim=60 60 60 60,clip,width=0.49\linewidth]{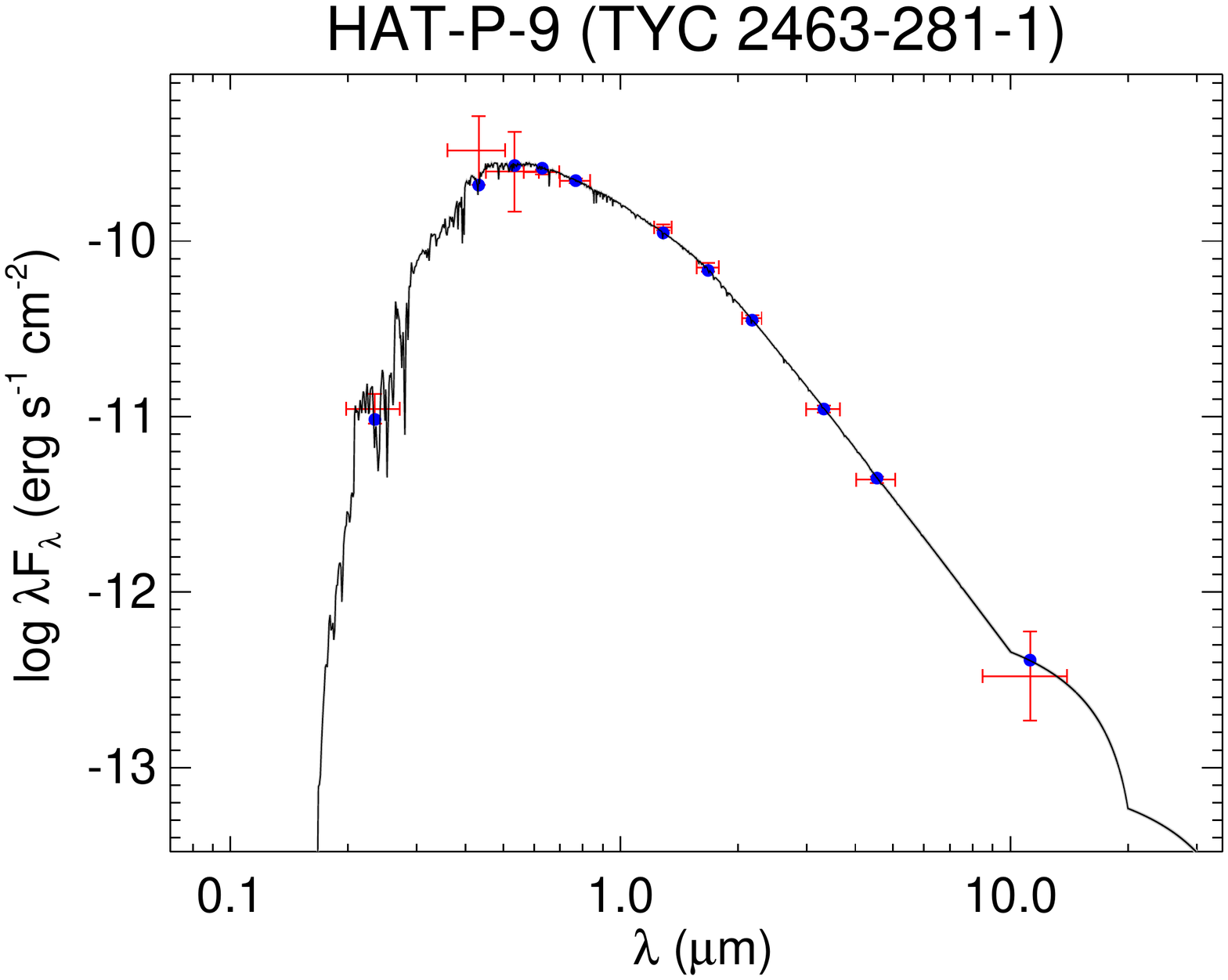}
  \includegraphics[trim=60 60 60 60,clip,width=0.49\linewidth]{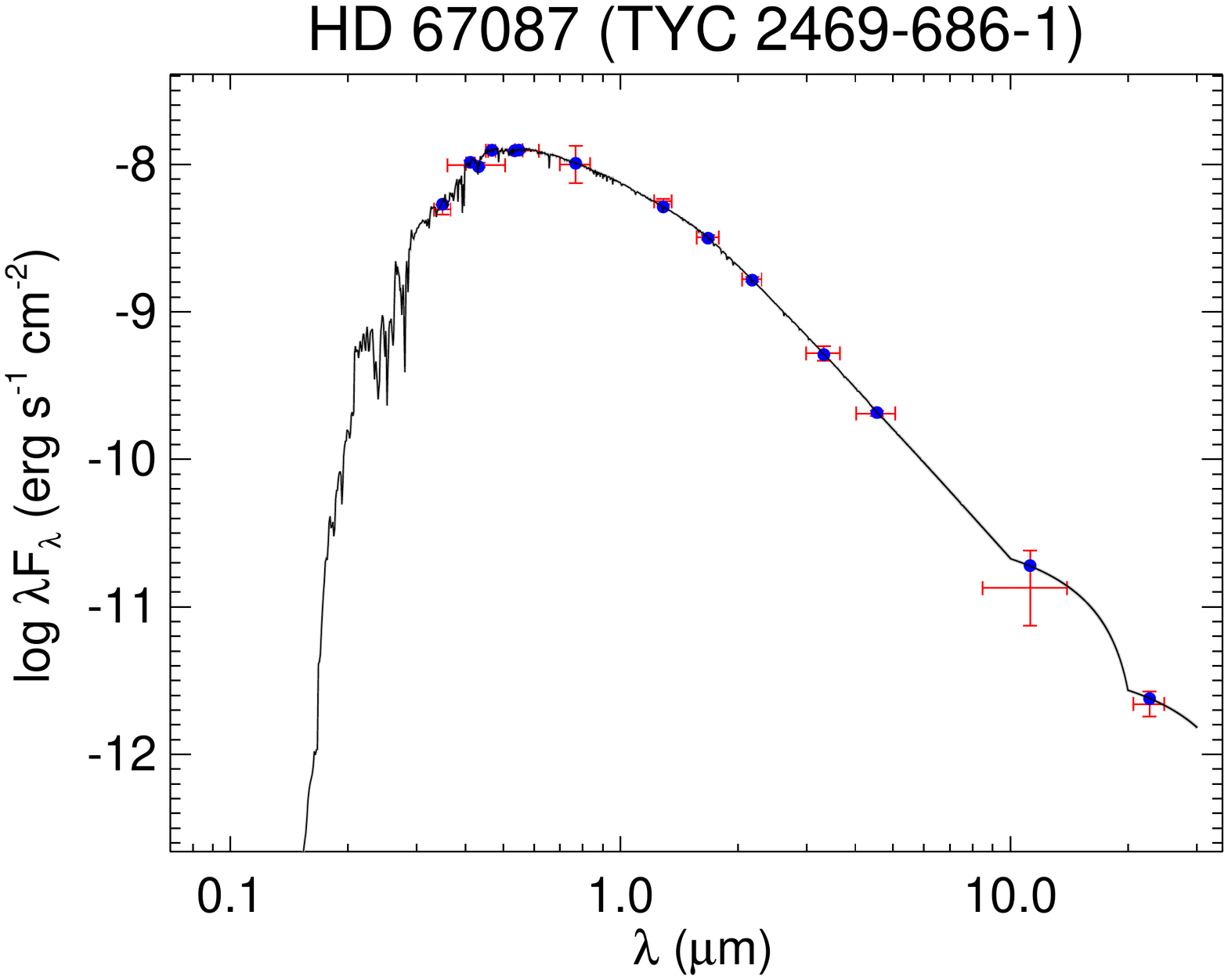}
  \caption{All labels, lines, symbols, and colors as in Figure \ref{fig:seds}.}
  \label{fig:seds_23}
\end{figure}

\begin{figure}[H]
  \centering
  \includegraphics[trim=60 60 60 60,clip,width=0.49\linewidth]{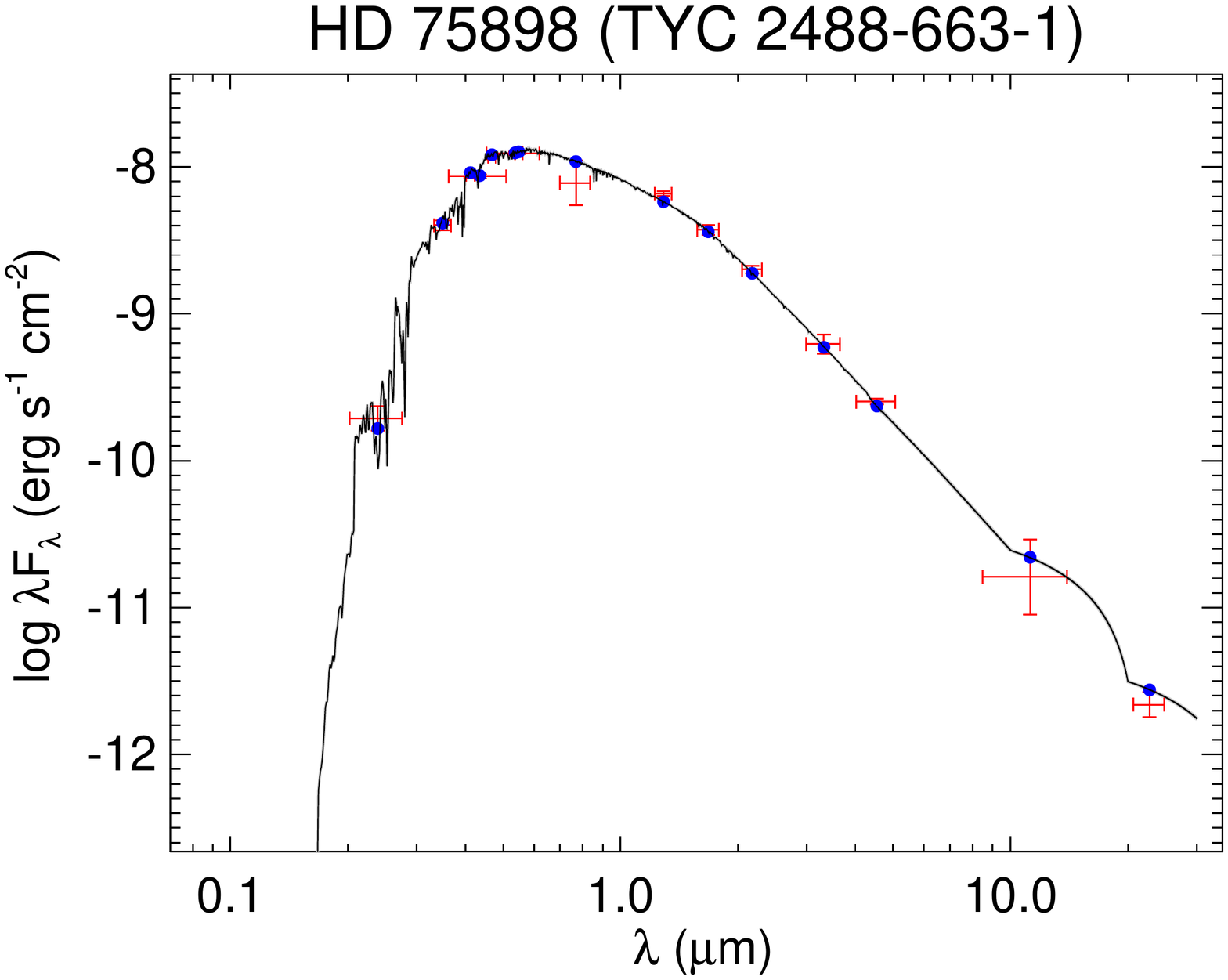}
  \includegraphics[trim=60 60 60 60,clip,width=0.49\linewidth]{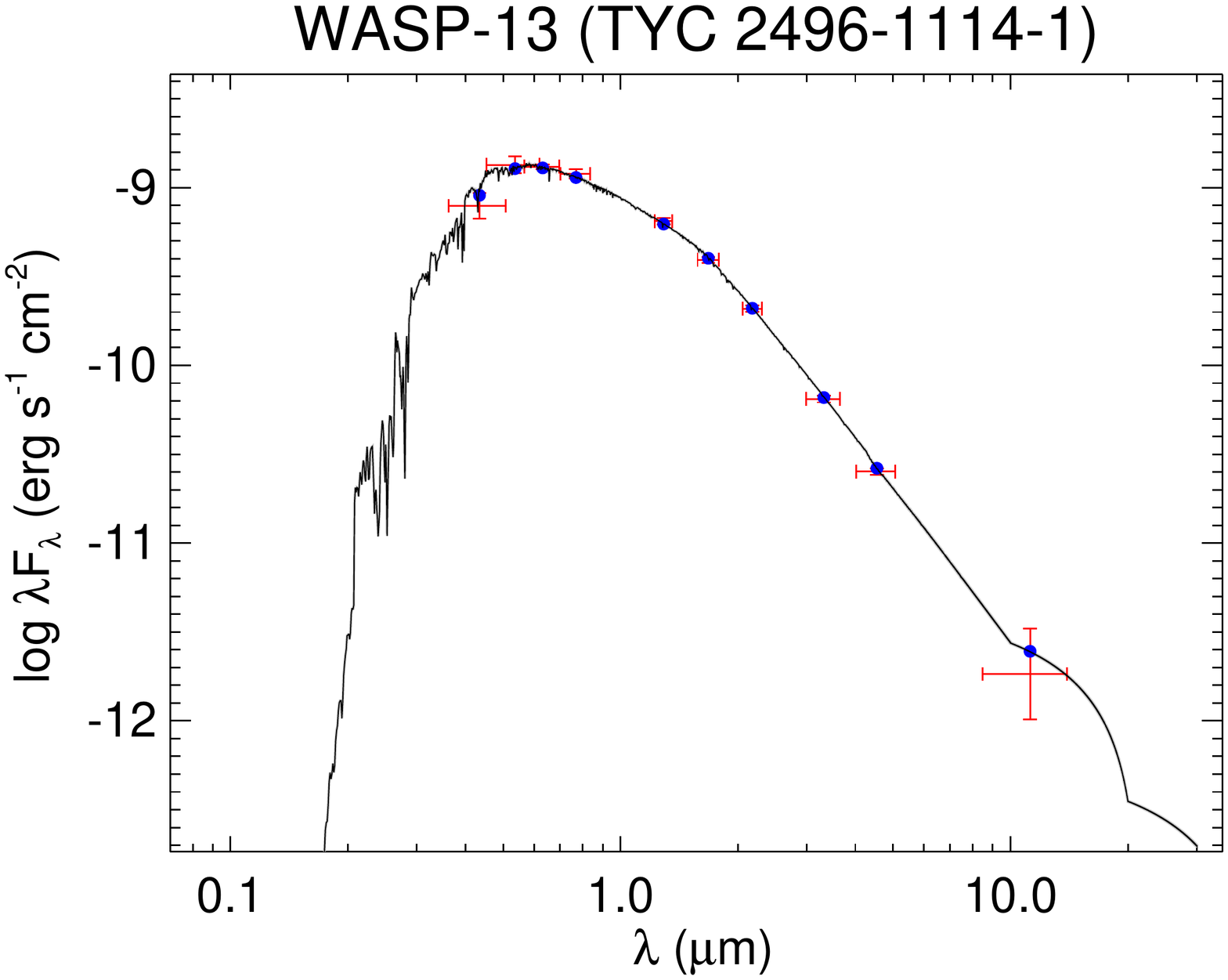}
  \includegraphics[trim=60 60 60 60,clip,width=0.49\linewidth]{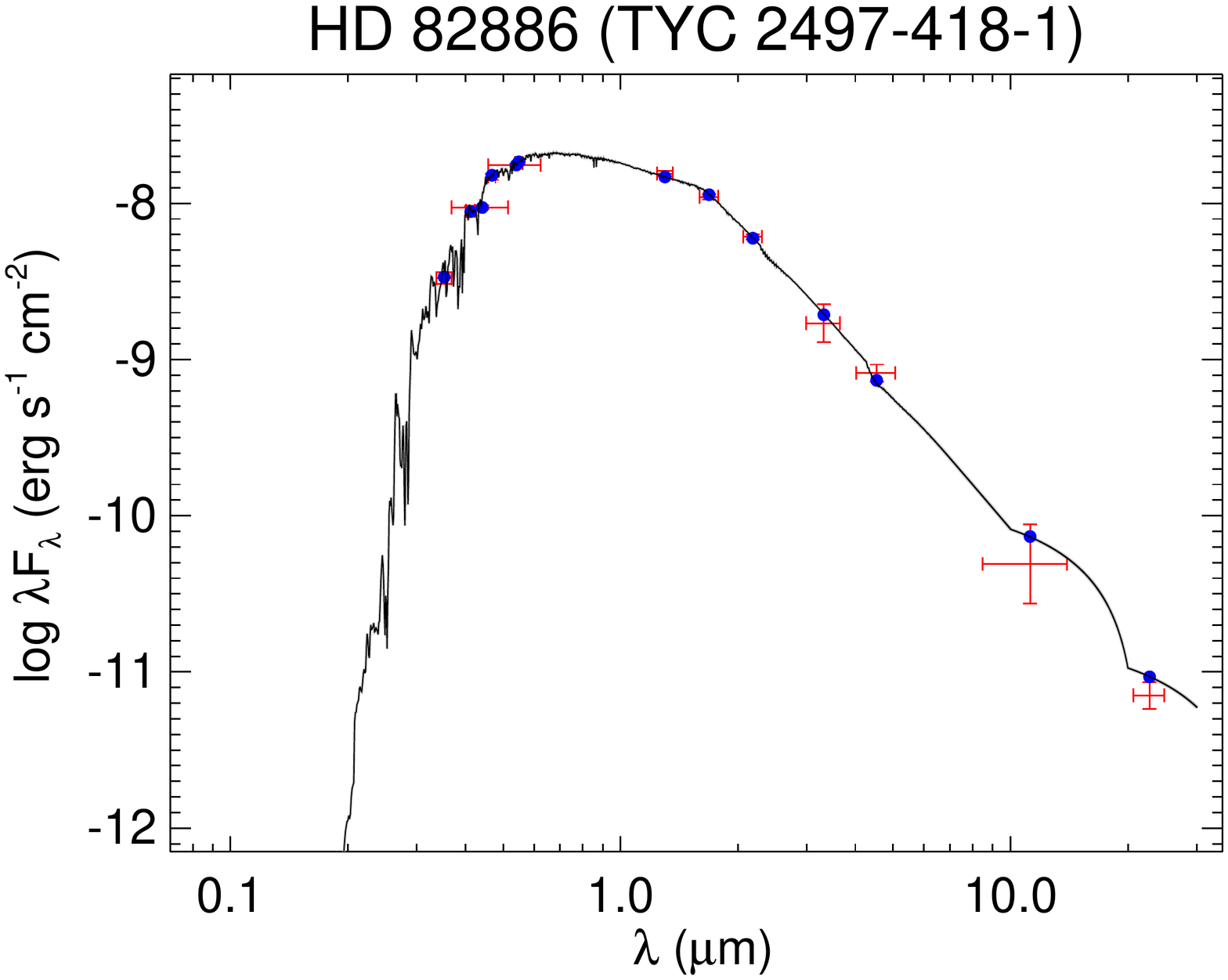}
  \includegraphics[trim=60 60 60 60,clip,width=0.49\linewidth]{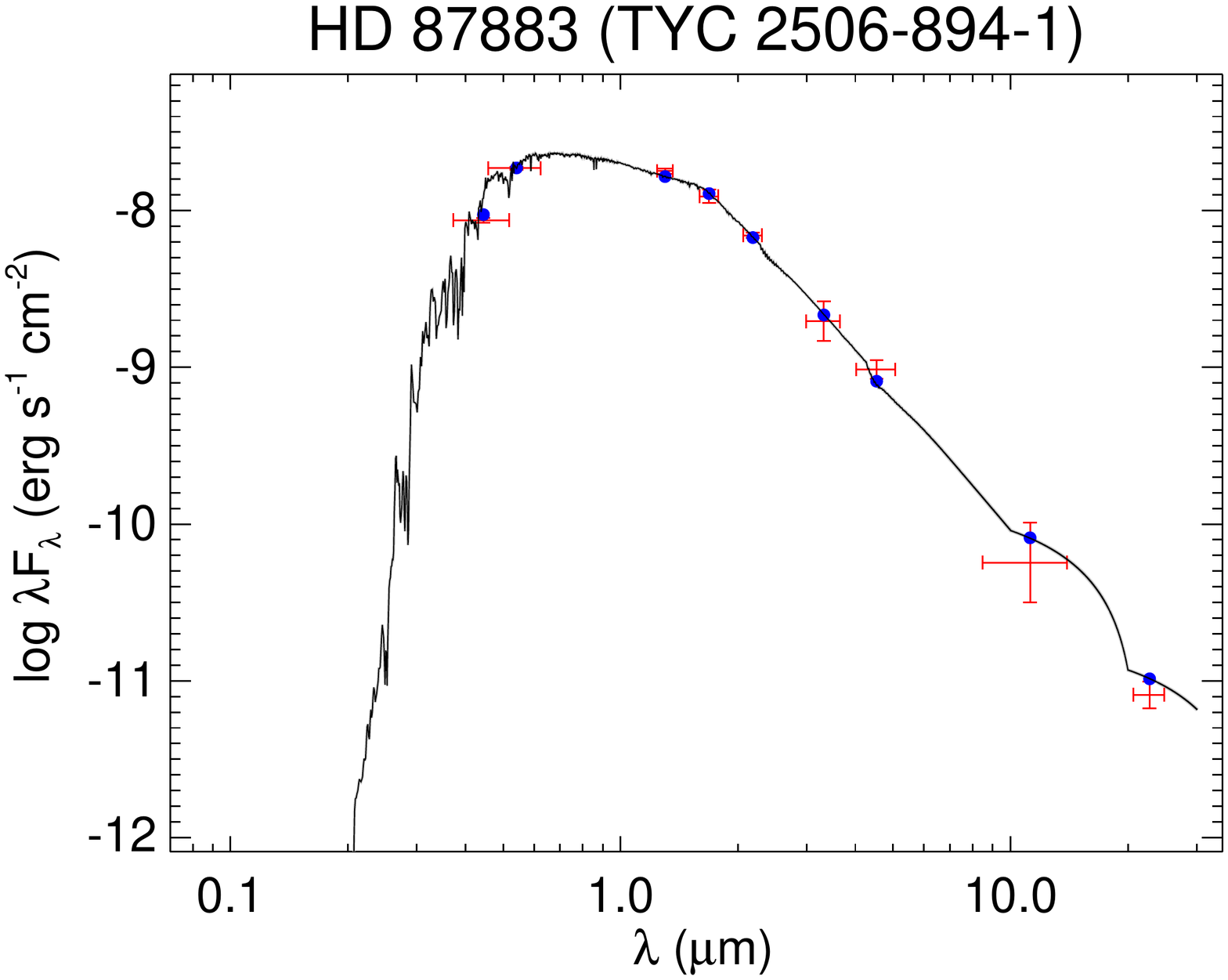}
  \includegraphics[trim=60 60 60 60,clip,width=0.49\linewidth]{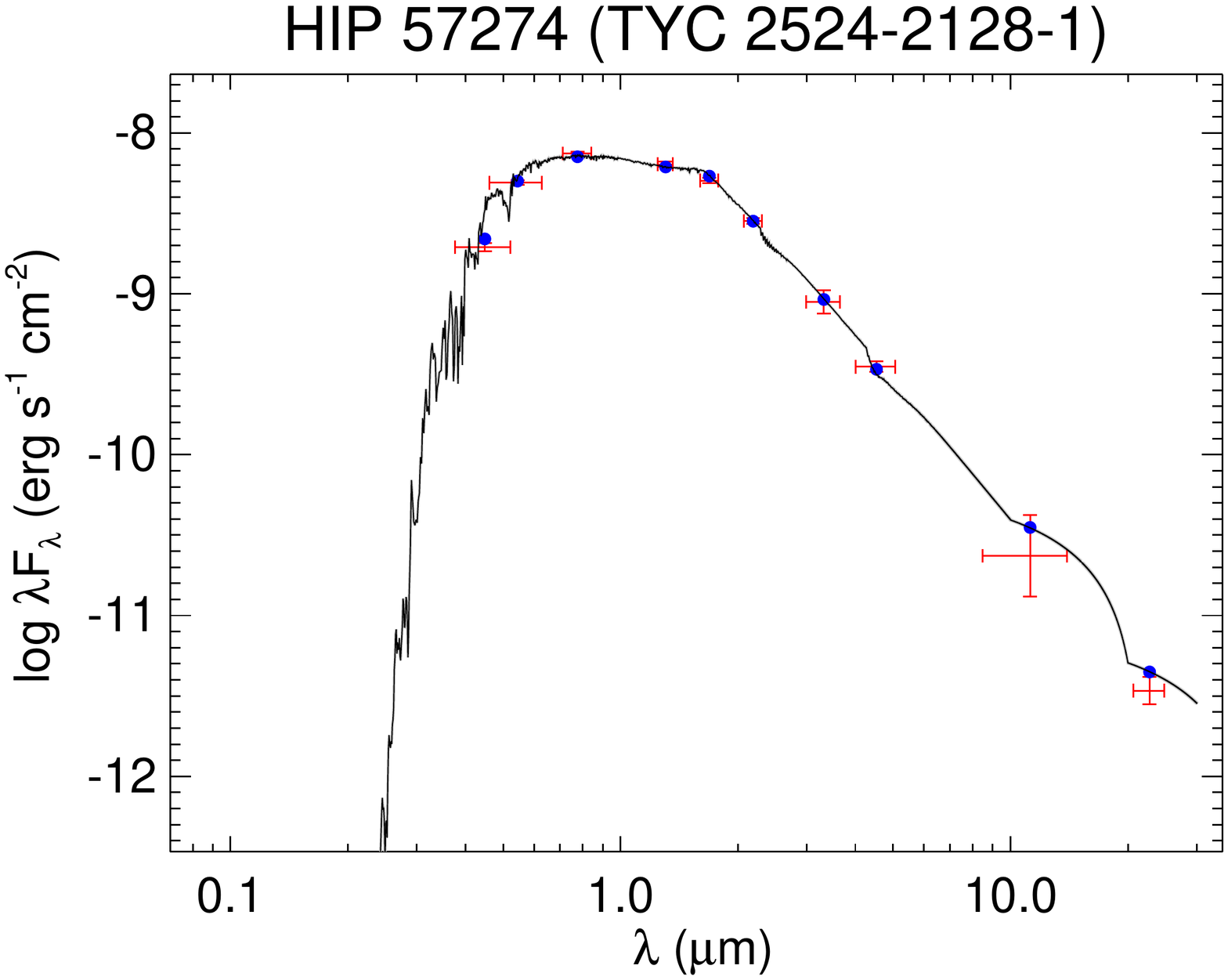}
  \includegraphics[trim=60 60 60 60,clip,width=0.49\linewidth]{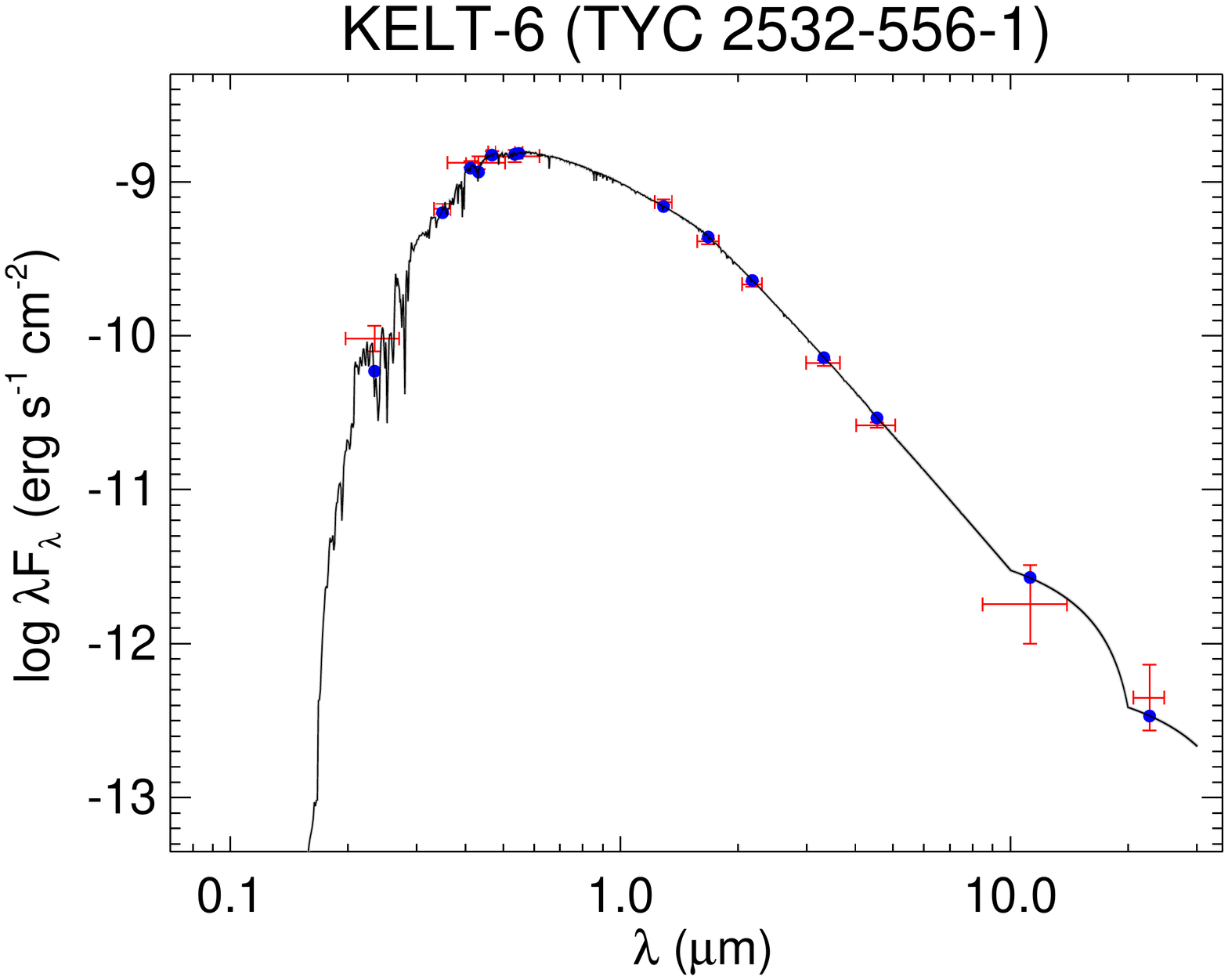}
  \caption{All labels, lines, symbols, and colors as in Figure \ref{fig:seds}.}
  \label{fig:seds_24}
\end{figure}

\begin{figure}[H]
  \centering
  \includegraphics[trim=60 60 60 60,clip,width=0.49\linewidth]{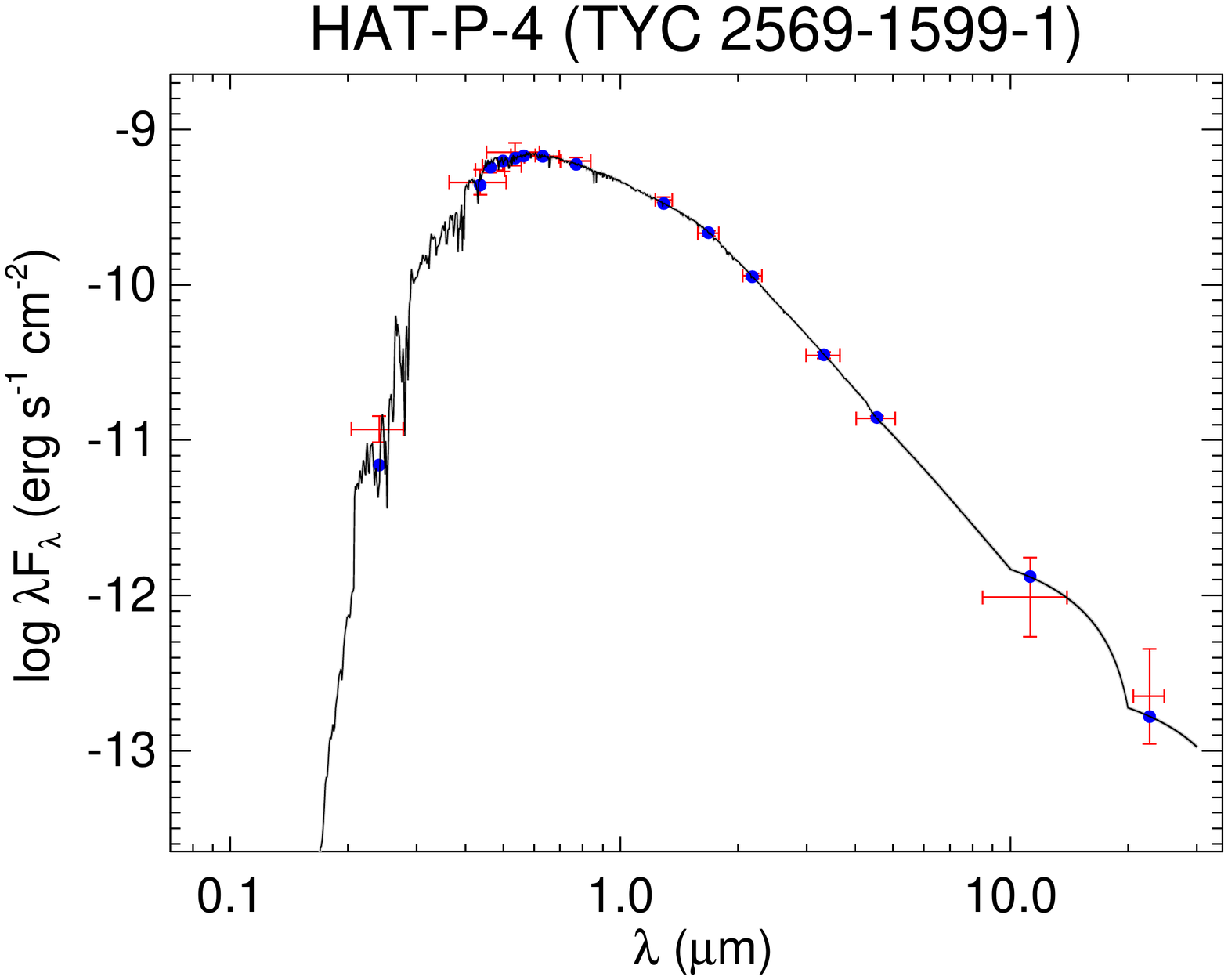}
  \includegraphics[trim=60 60 60 60,clip,width=0.49\linewidth]{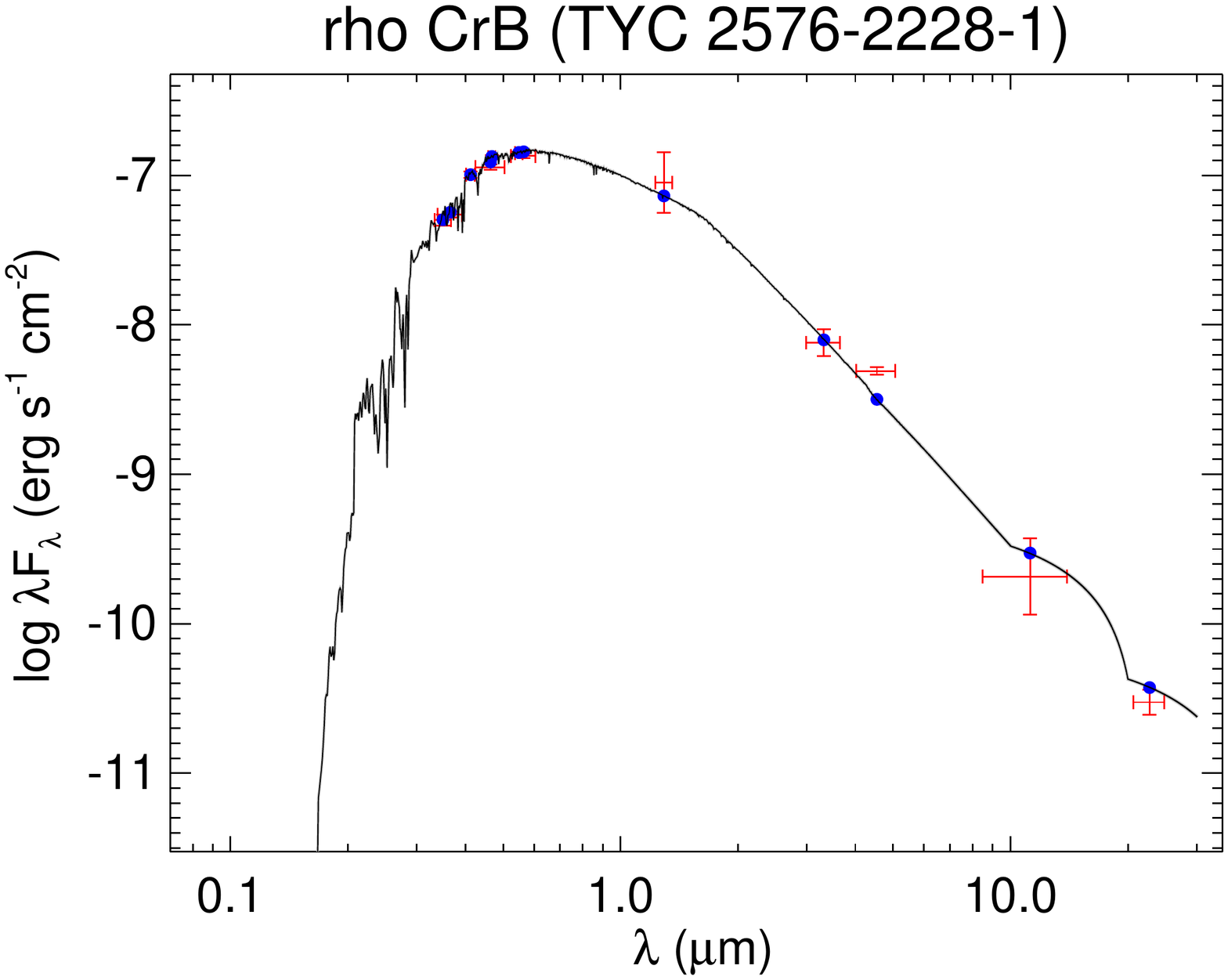}
  \includegraphics[trim=60 60 60 60,clip,width=0.49\linewidth]{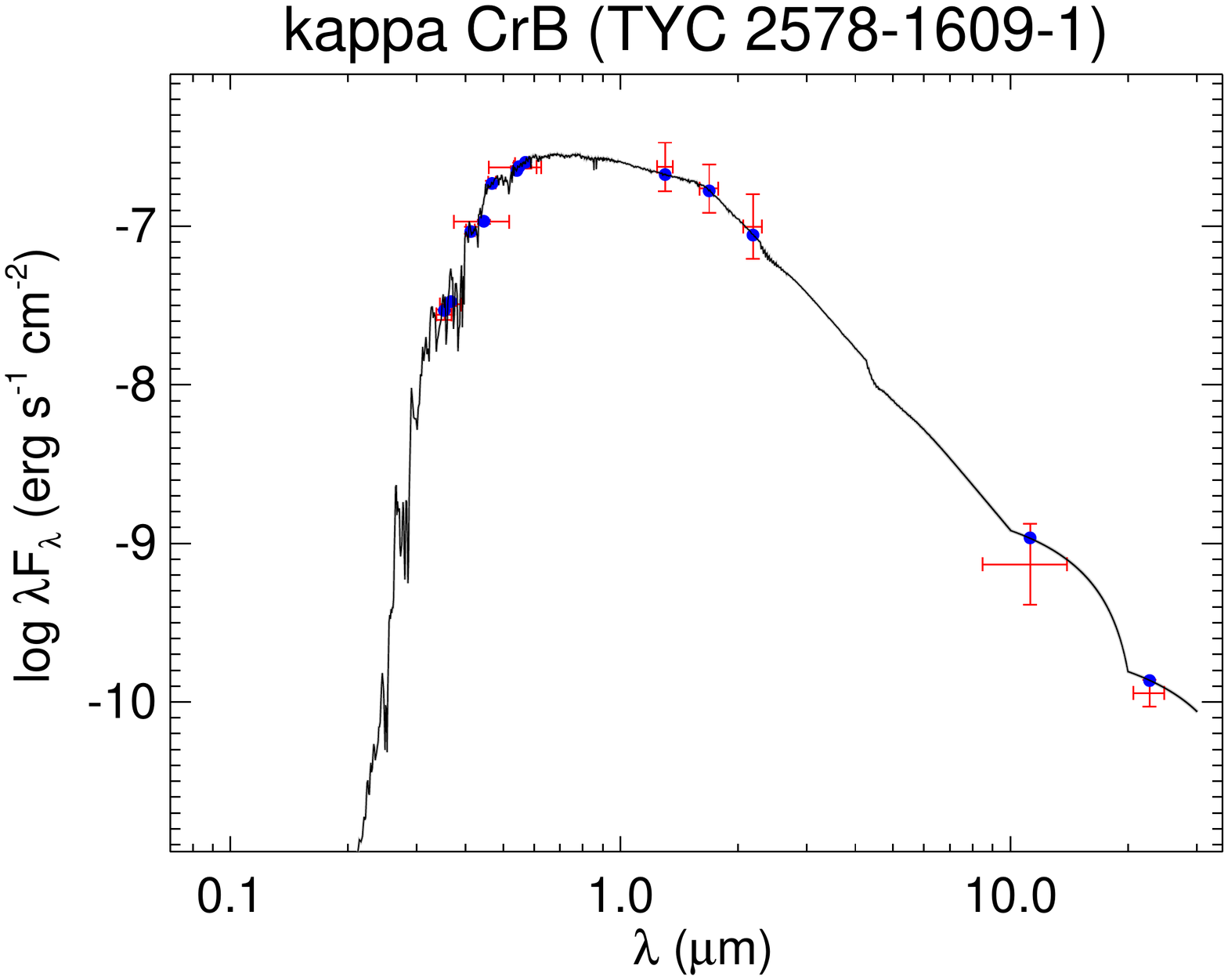}
  \includegraphics[trim=60 60 60 60,clip,width=0.49\linewidth]{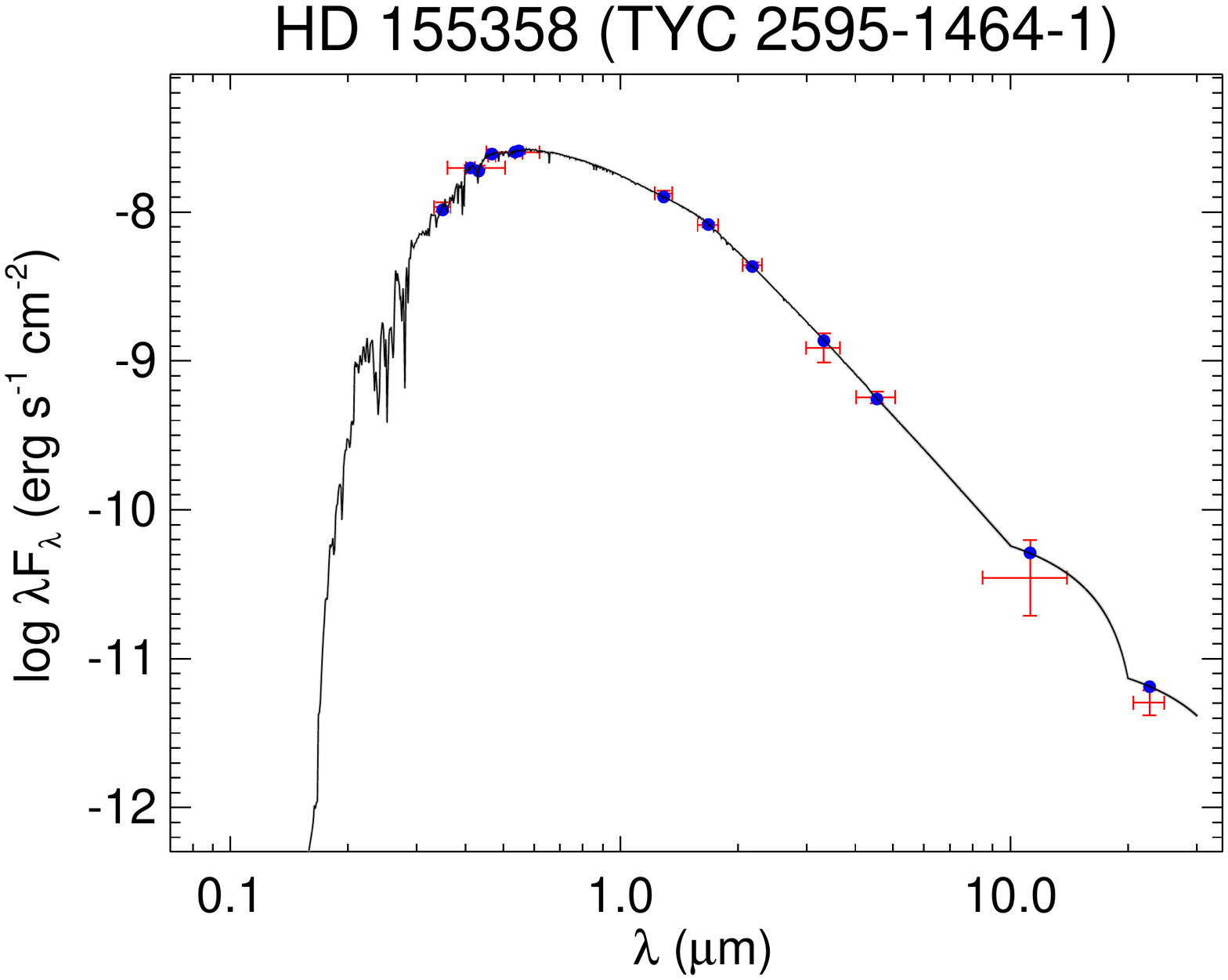}
  \includegraphics[trim=60 60 60 60,clip,width=0.49\linewidth]{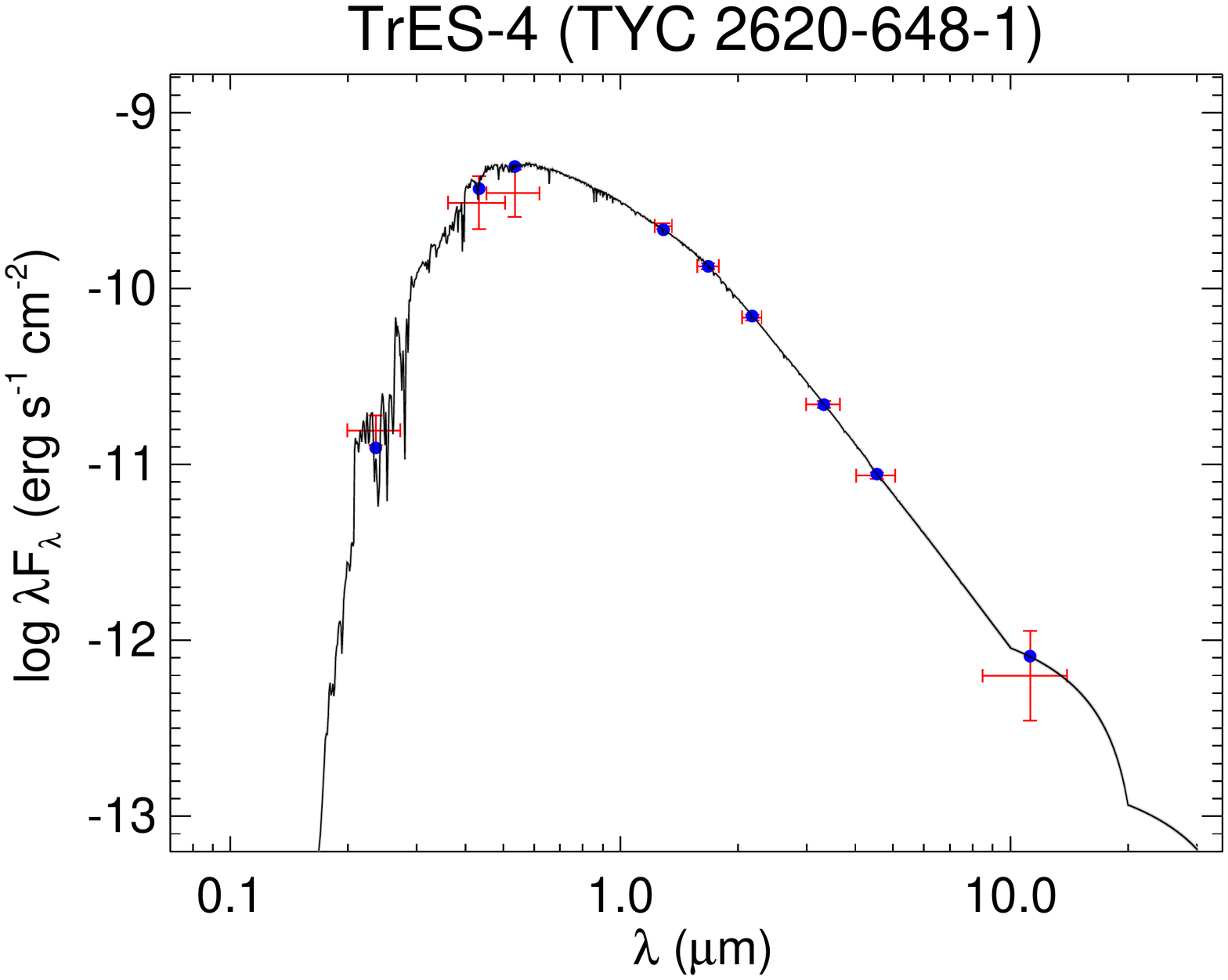}
  \includegraphics[trim=60 60 60 60,clip,width=0.49\linewidth]{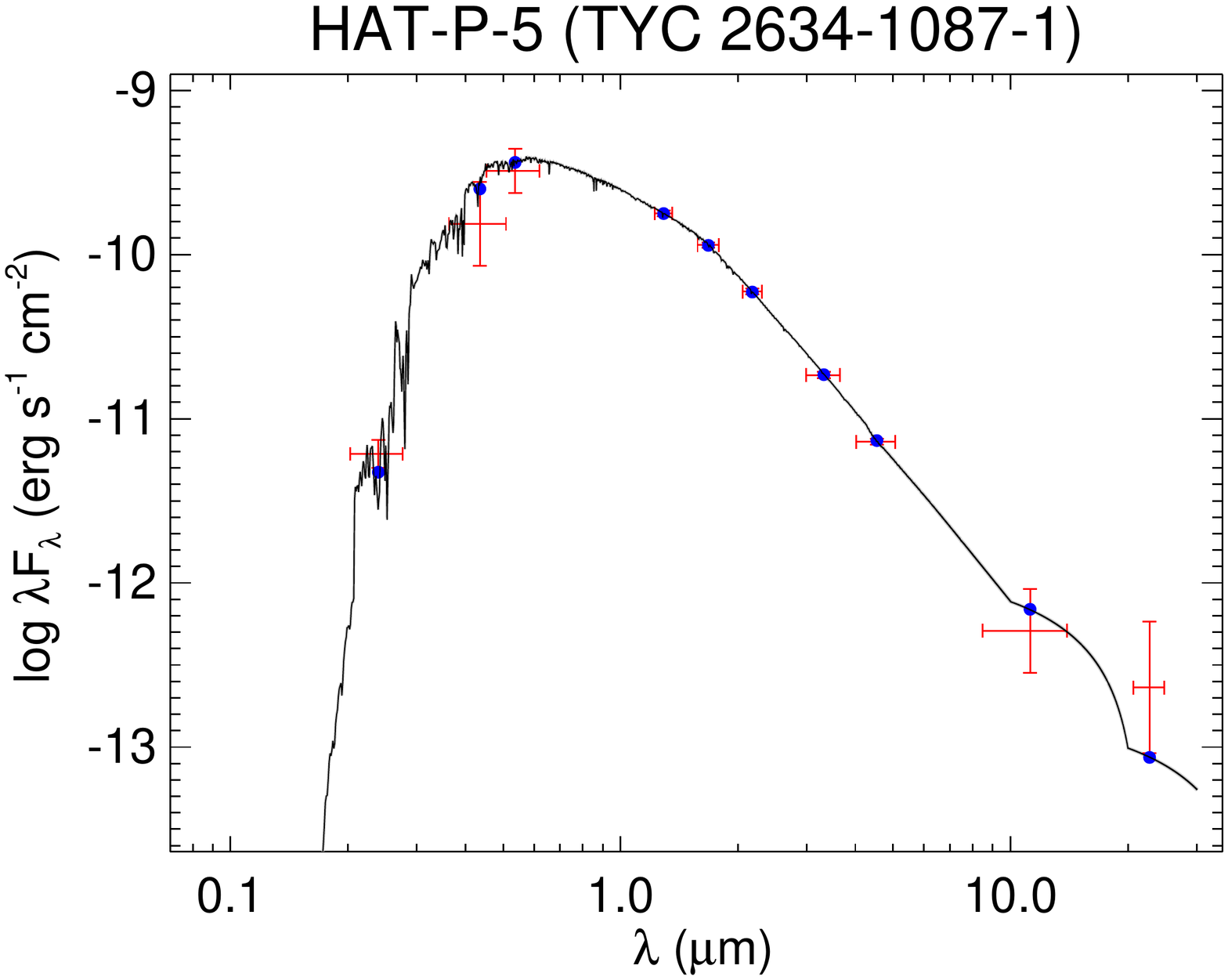}
  \caption{All labels, lines, symbols, and colors as in Figure \ref{fig:seds}.}
  \label{fig:seds_25}
\end{figure}

\begin{figure}[H]
  \centering
  \includegraphics[trim=60 60 60 60,clip,width=0.49\linewidth]{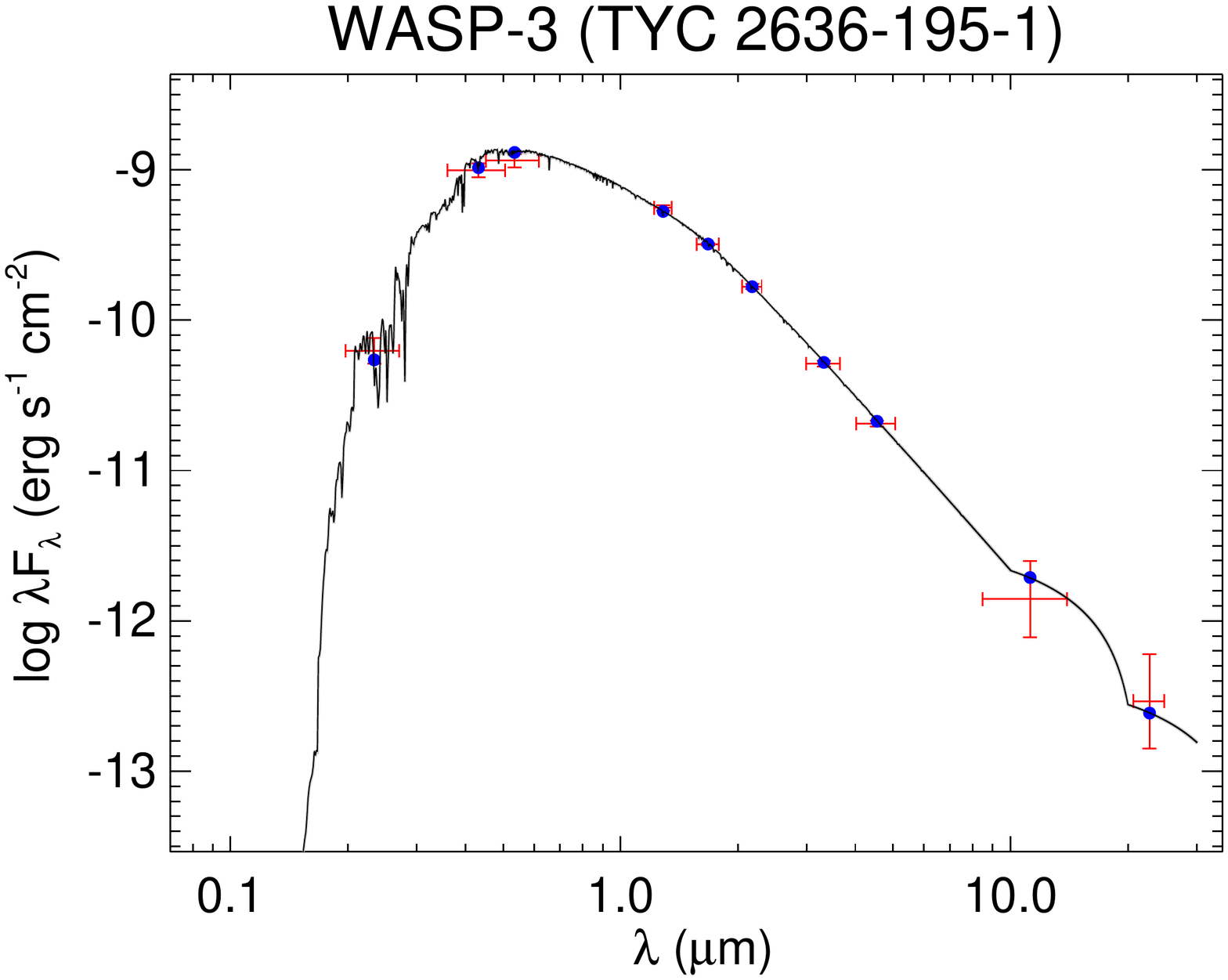}
  \includegraphics[trim=60 60 60 60,clip,width=0.49\linewidth]{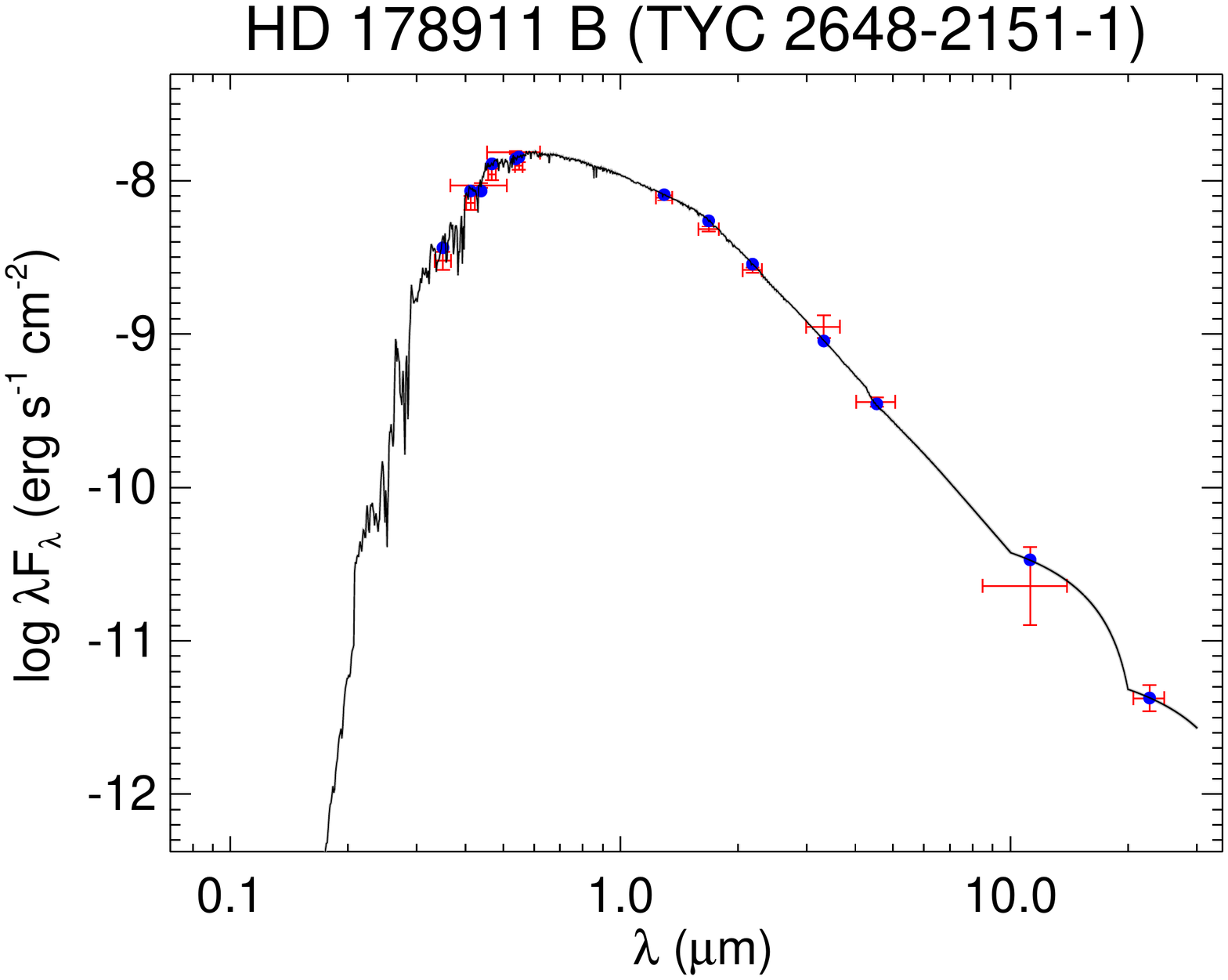}
  \includegraphics[trim=60 60 60 60,clip,width=0.49\linewidth]{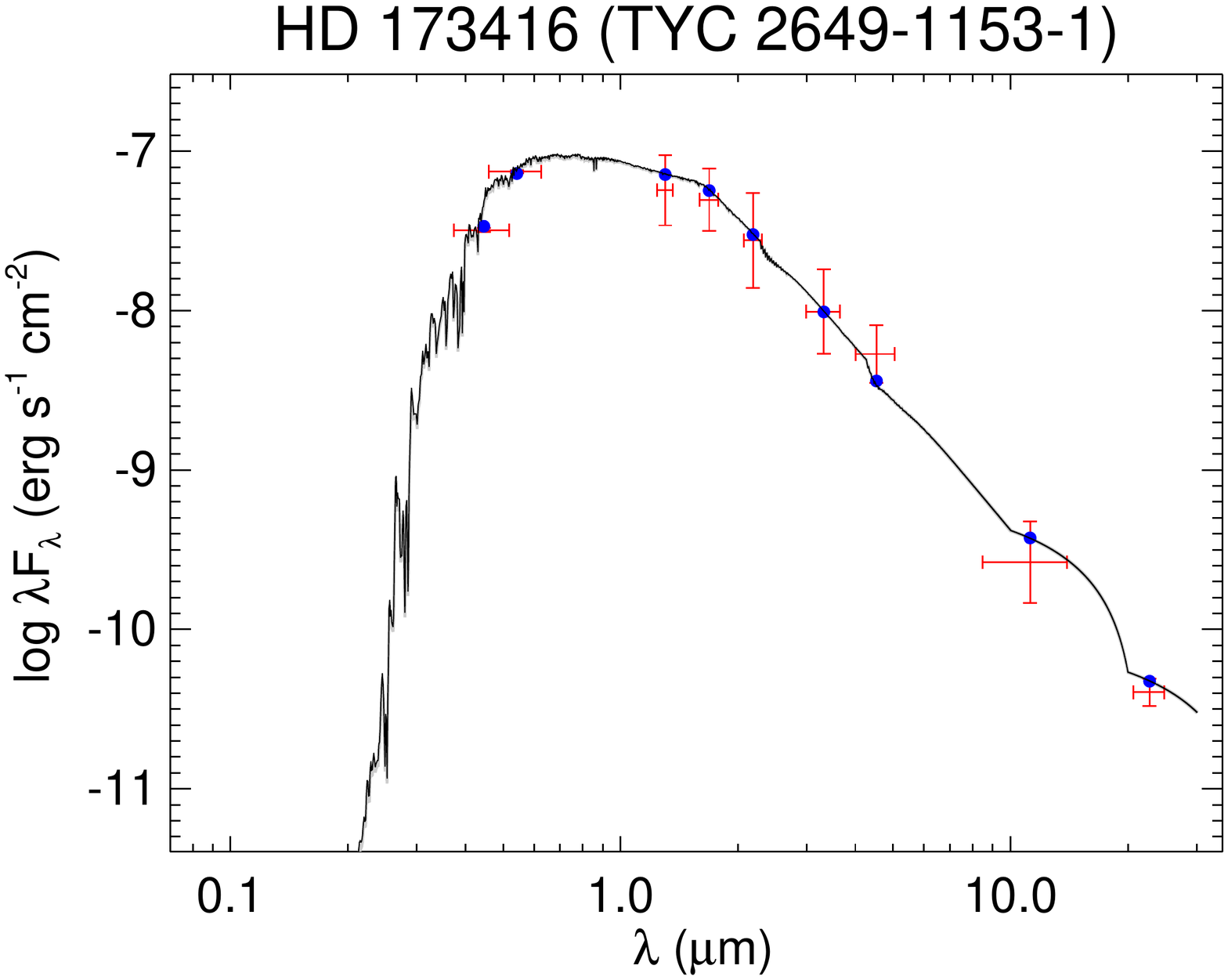}
  \includegraphics[trim=60 60 60 60,clip,width=0.49\linewidth]{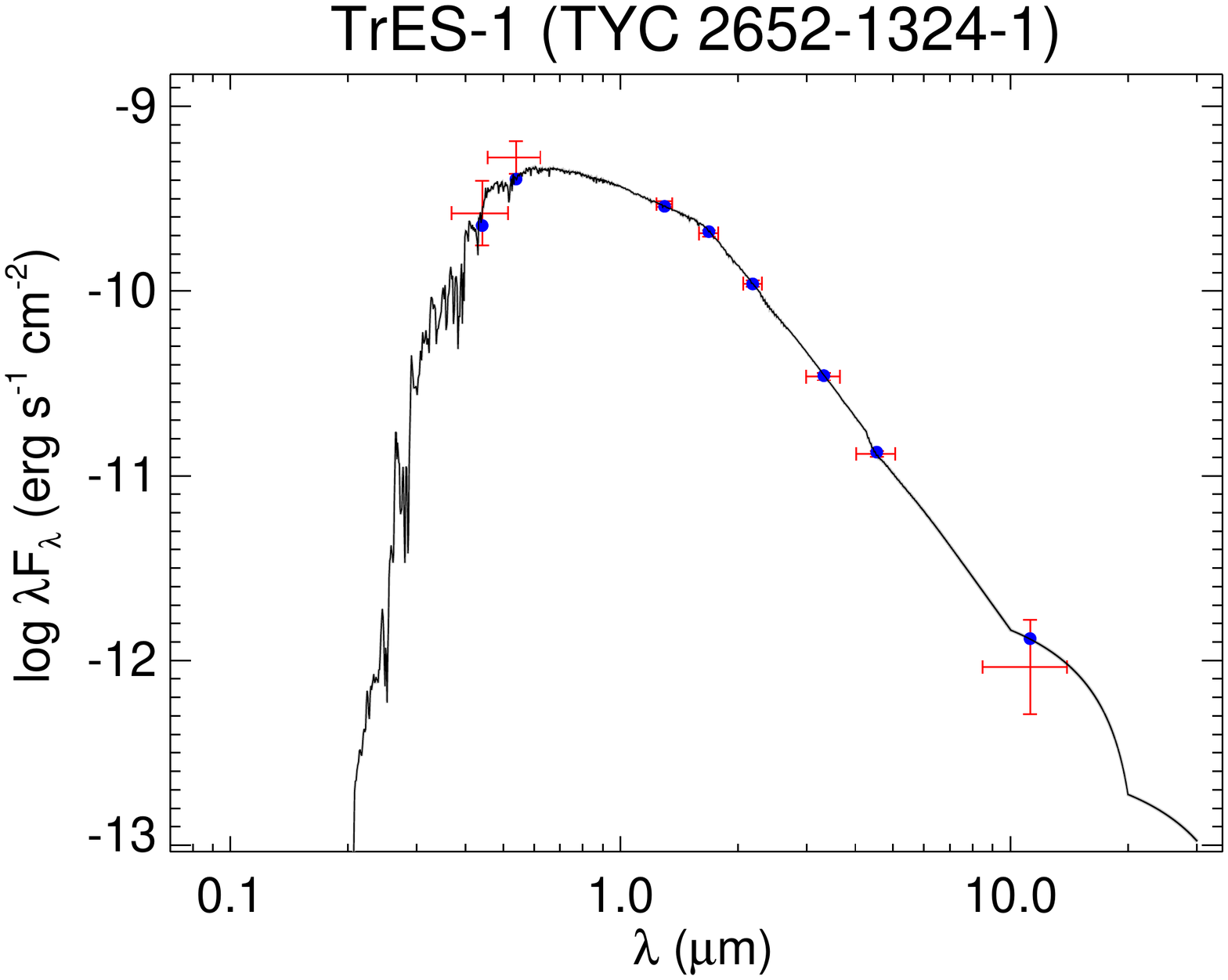}
  \includegraphics[trim=60 60 60 60,clip,width=0.49\linewidth]{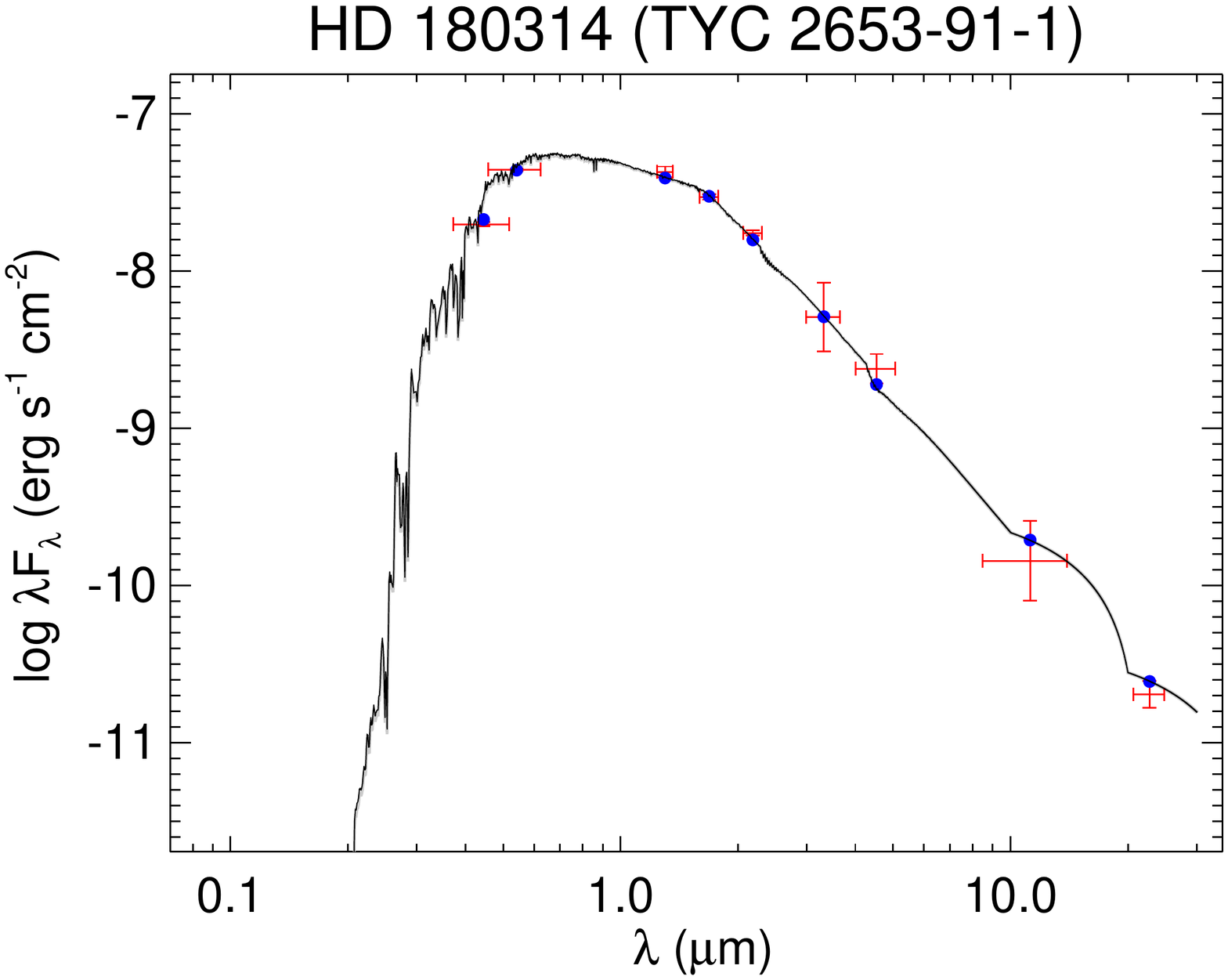}
  \includegraphics[trim=60 60 60 60,clip,width=0.49\linewidth]{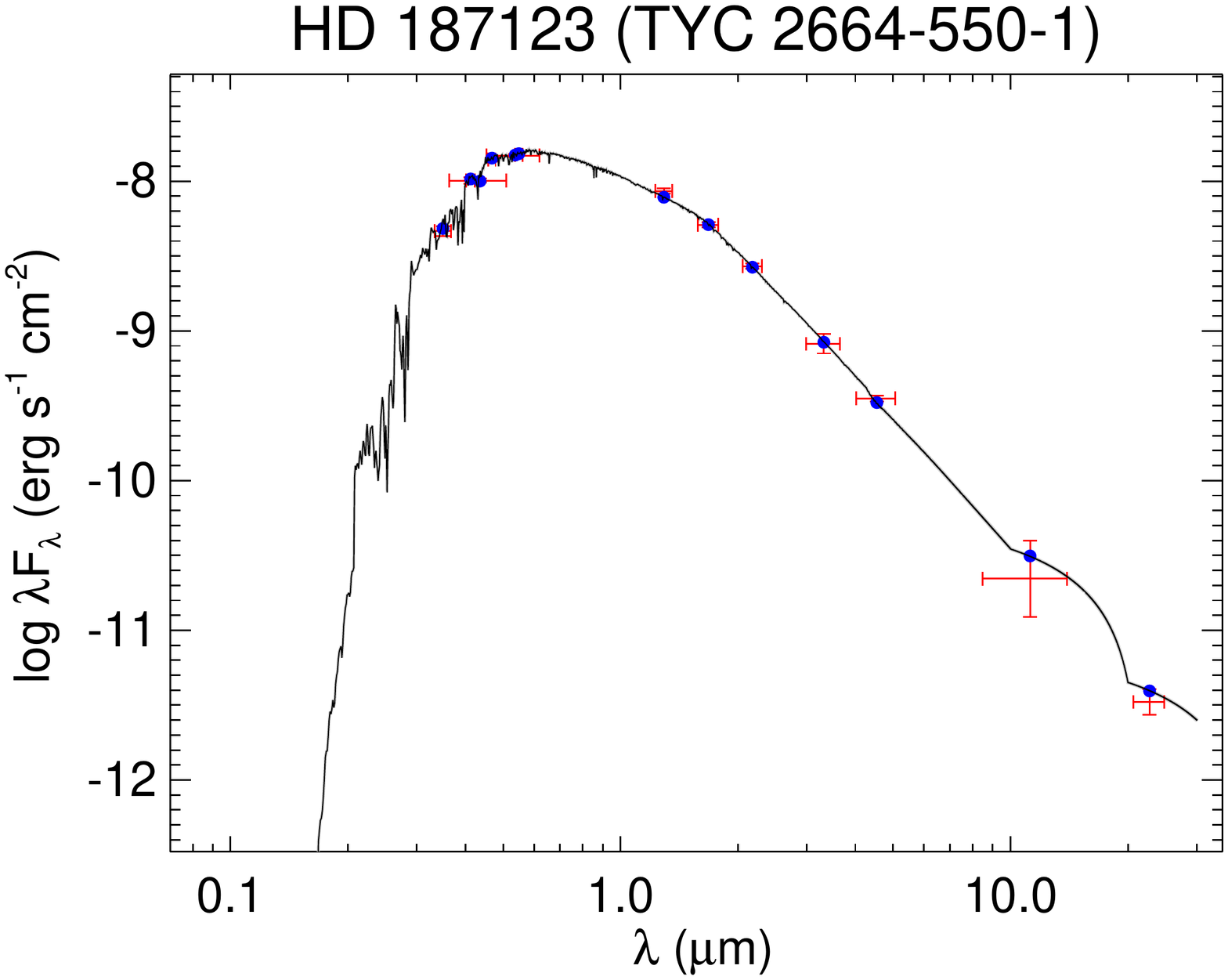}
  \caption{All labels, lines, symbols, and colors as in Figure \ref{fig:seds}.}
  \label{fig:seds_26}
\end{figure}

\begin{figure}[H]
  \centering
  \includegraphics[trim=60 60 60 60,clip,width=0.49\linewidth]{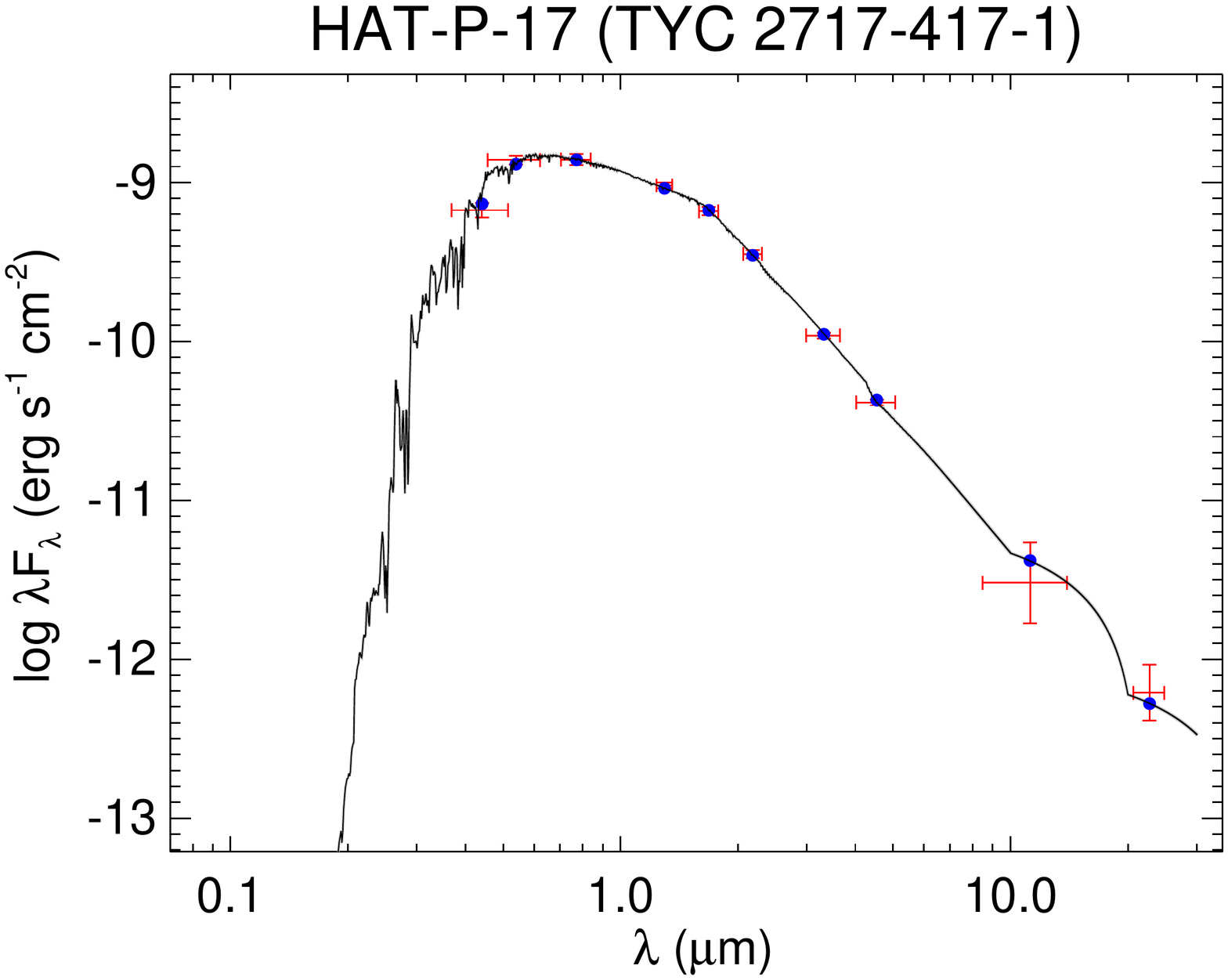}
  \includegraphics[trim=60 60 60 60,clip,width=0.49\linewidth]{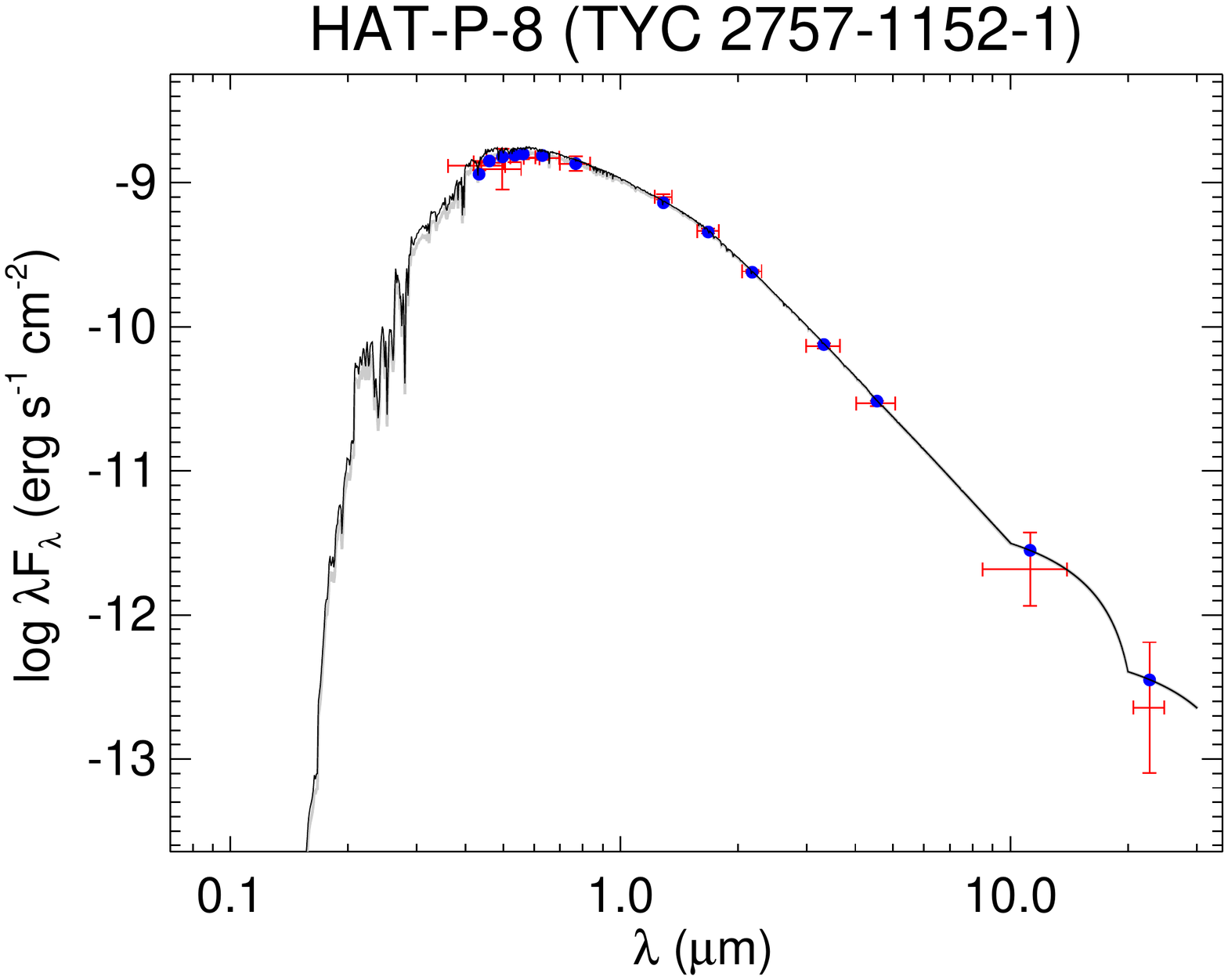}
  \includegraphics[trim=60 60 60 60,clip,width=0.49\linewidth]{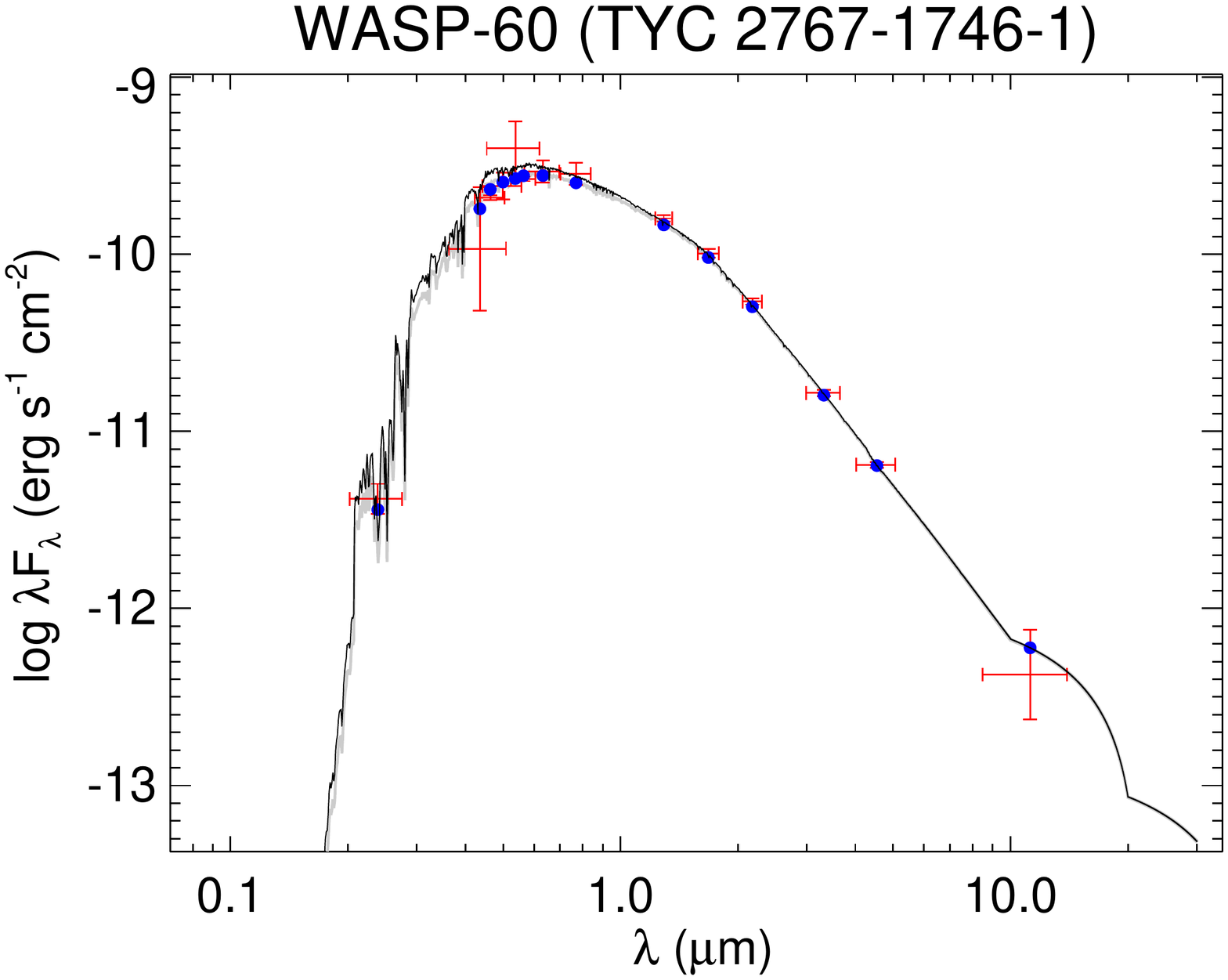}
  \includegraphics[trim=60 60 60 60,clip,width=0.49\linewidth]{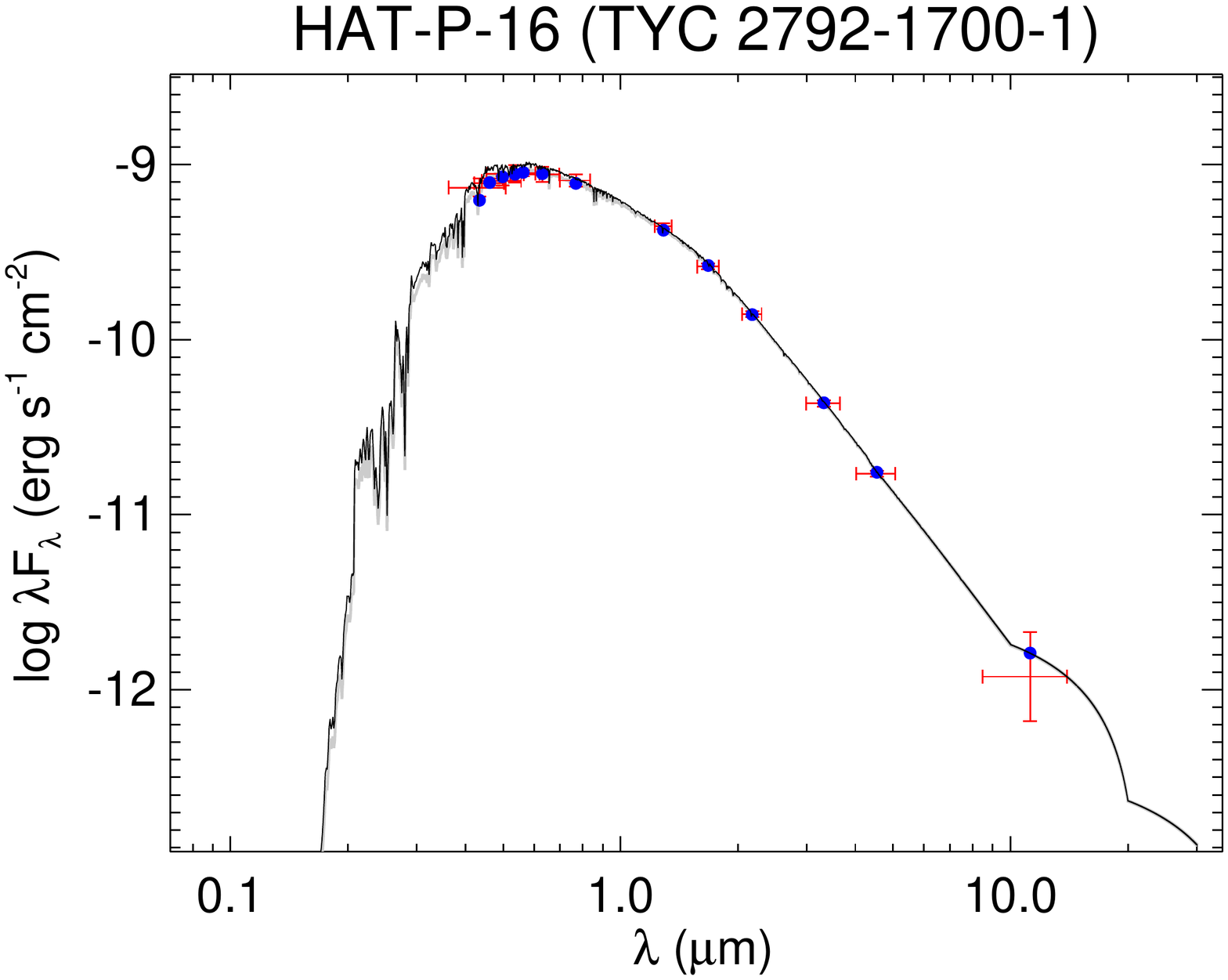}
  \includegraphics[trim=60 60 60 60,clip,width=0.49\linewidth]{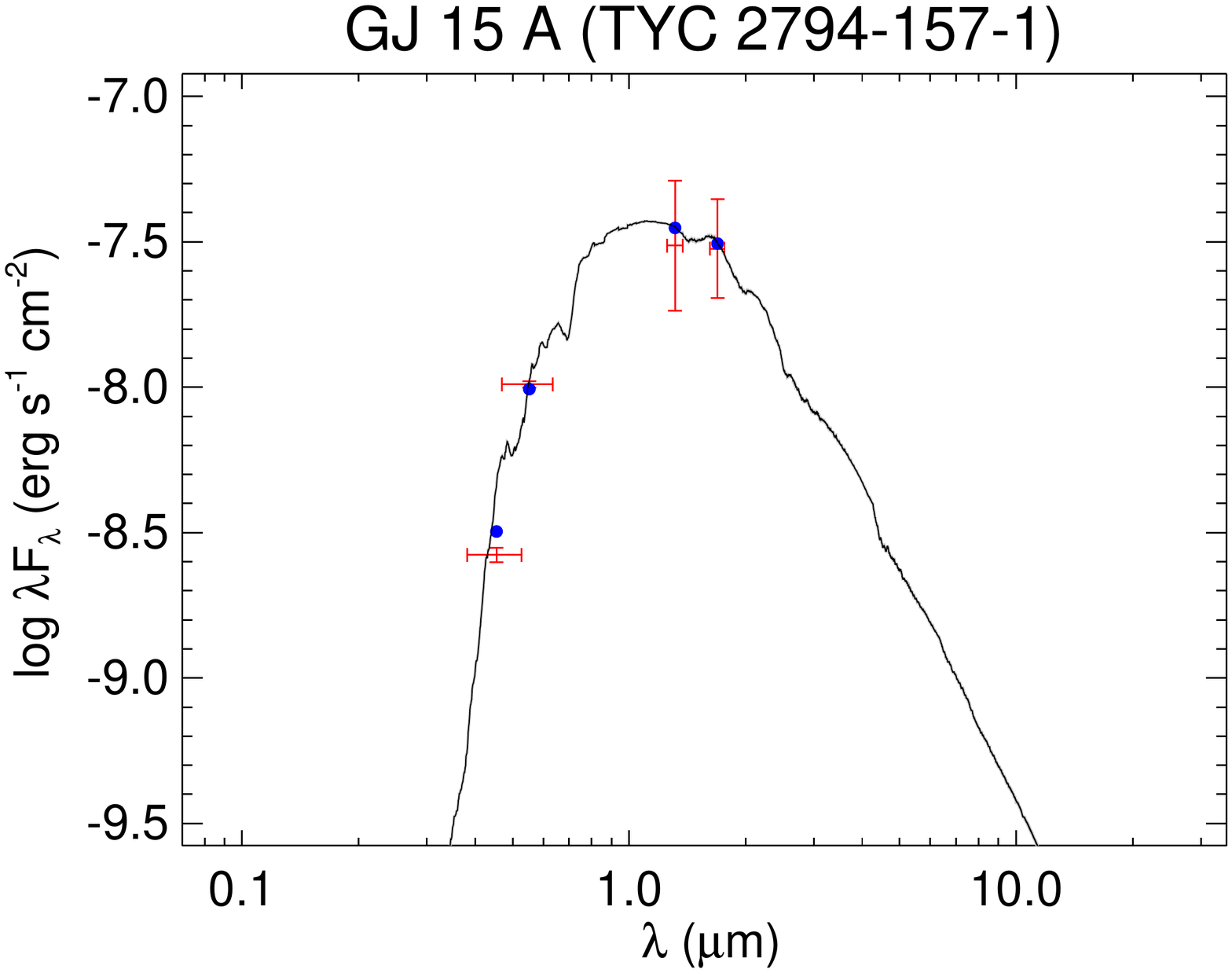}
  \includegraphics[trim=60 60 60 60,clip,width=0.49\linewidth]{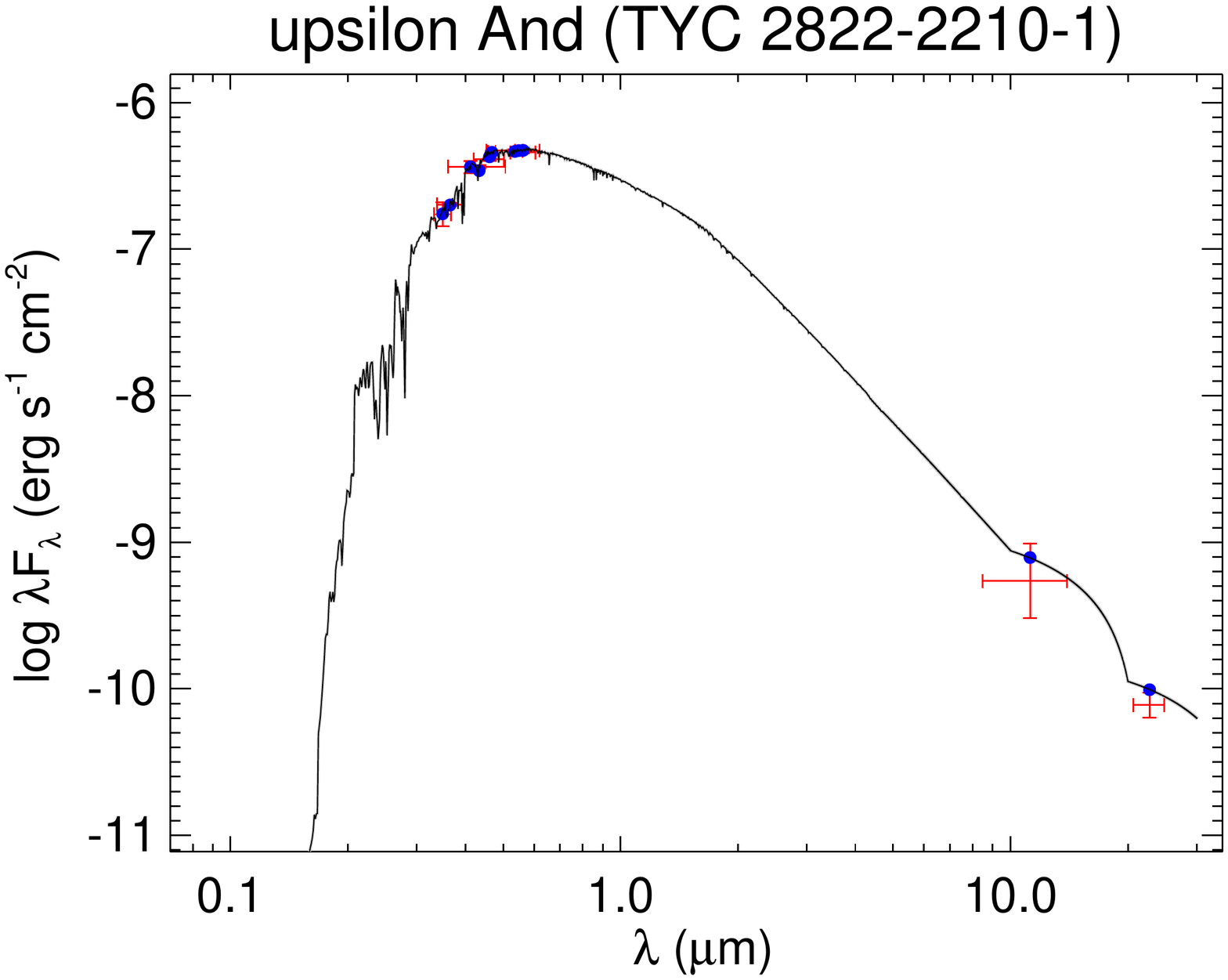}
  \caption{All labels, lines, symbols, and colors as in Figure \ref{fig:seds}.}
  \label{fig:seds_27}
\end{figure}

\begin{figure}[H]
  \centering
  \includegraphics[trim=60 60 60 60,clip,width=0.49\linewidth]{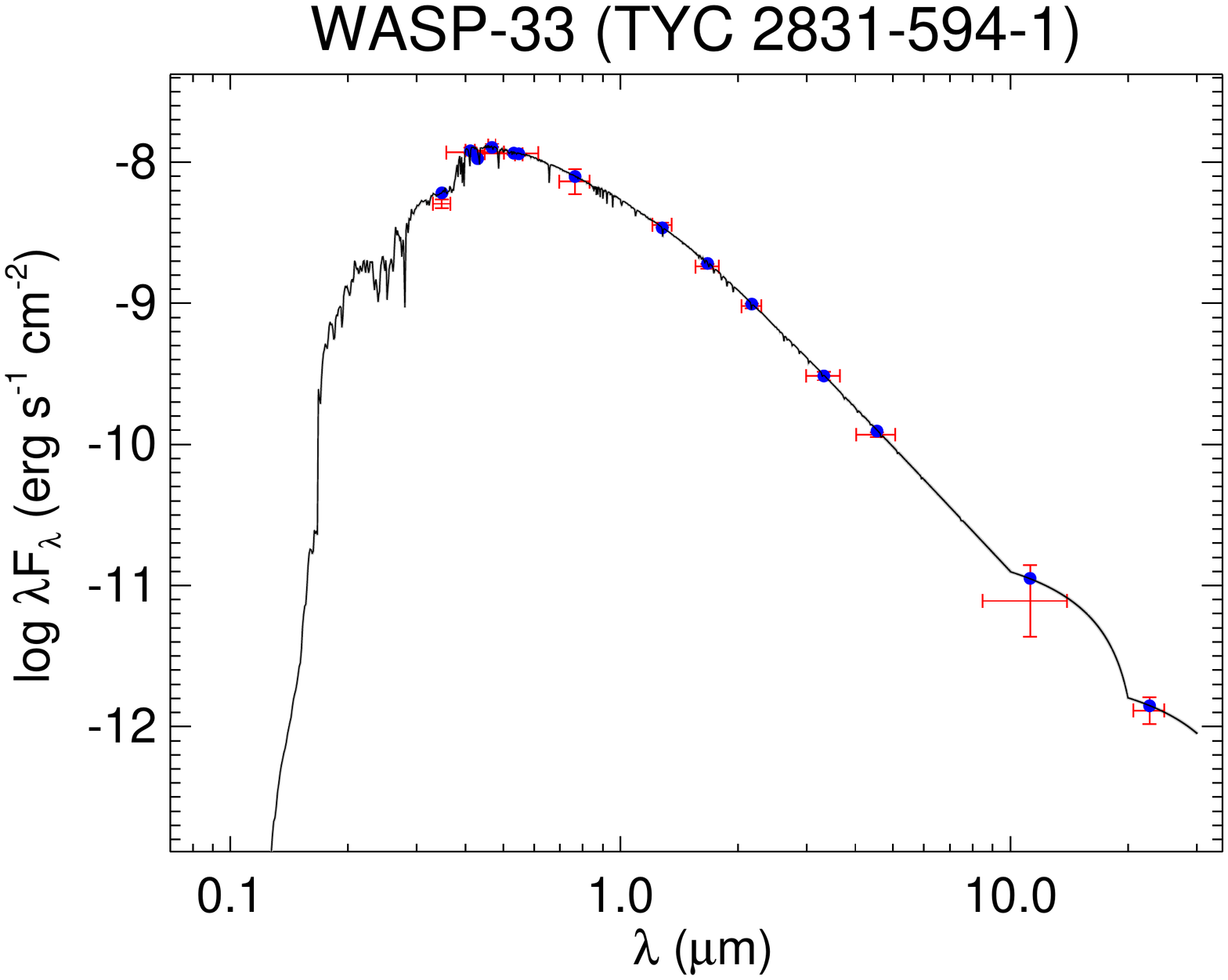}
  \includegraphics[trim=60 60 60 60,clip,width=0.49\linewidth]{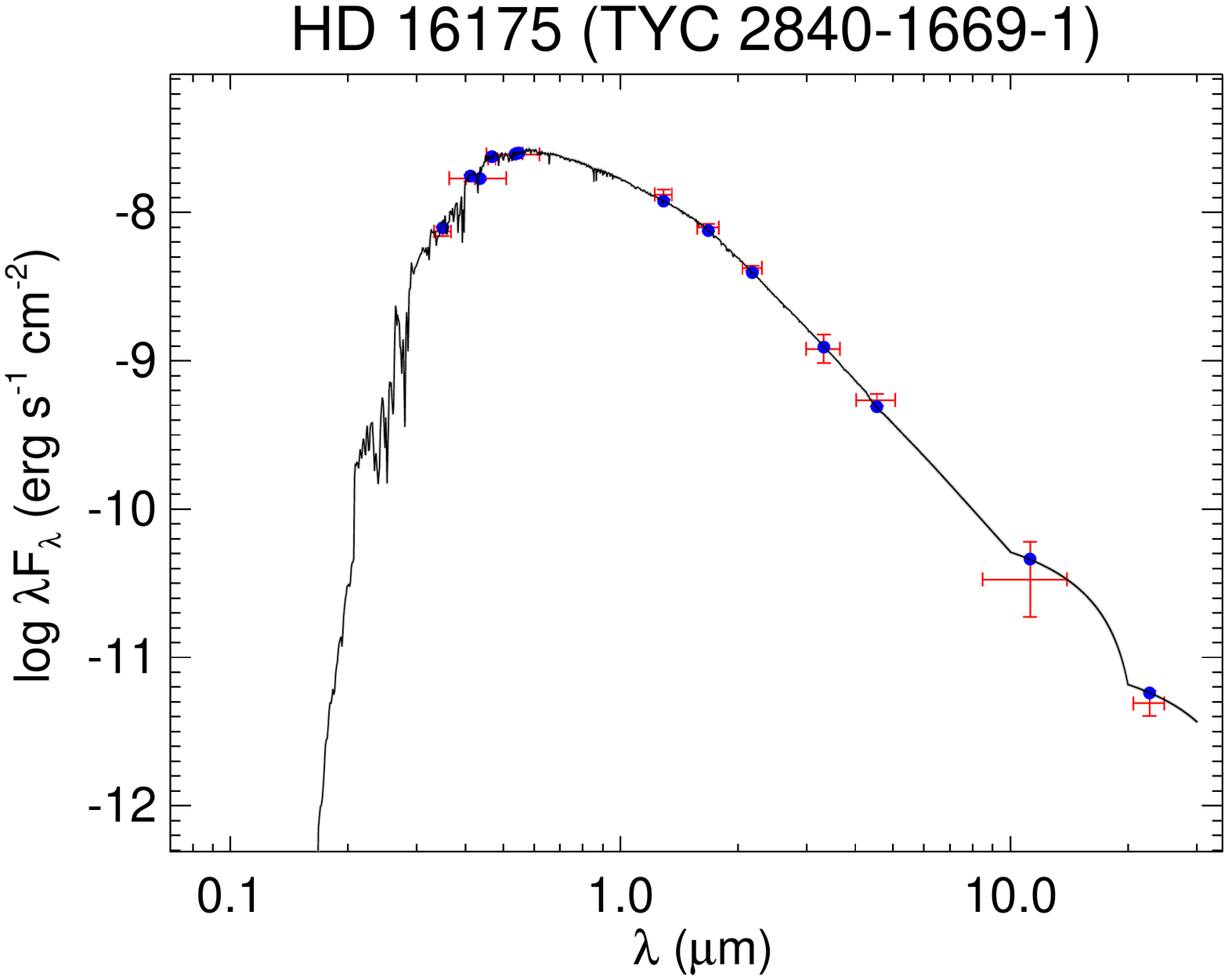}
  \includegraphics[trim=60 60 60 60,clip,width=0.49\linewidth]{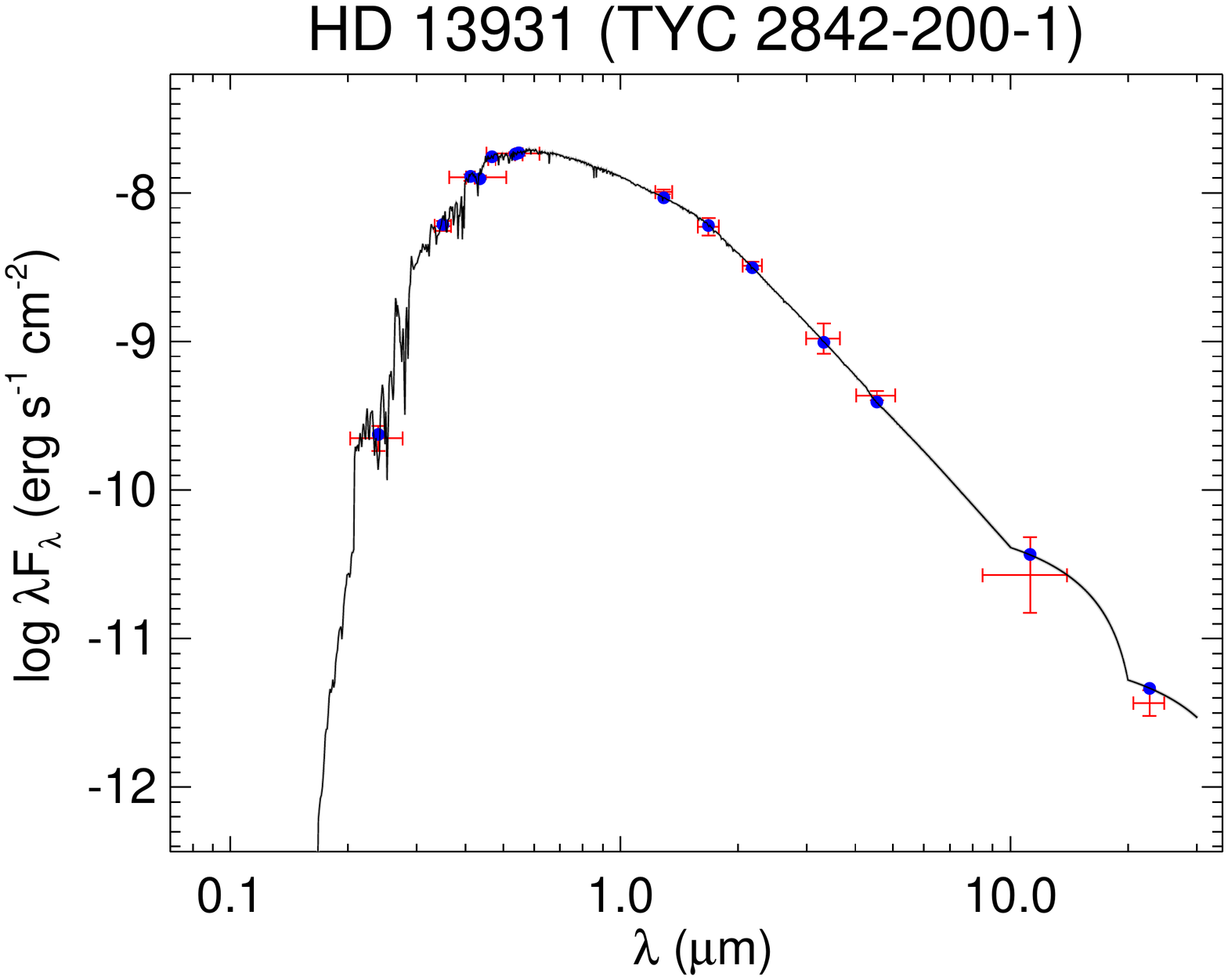}
  \includegraphics[trim=60 60 60 60,clip,width=0.49\linewidth]{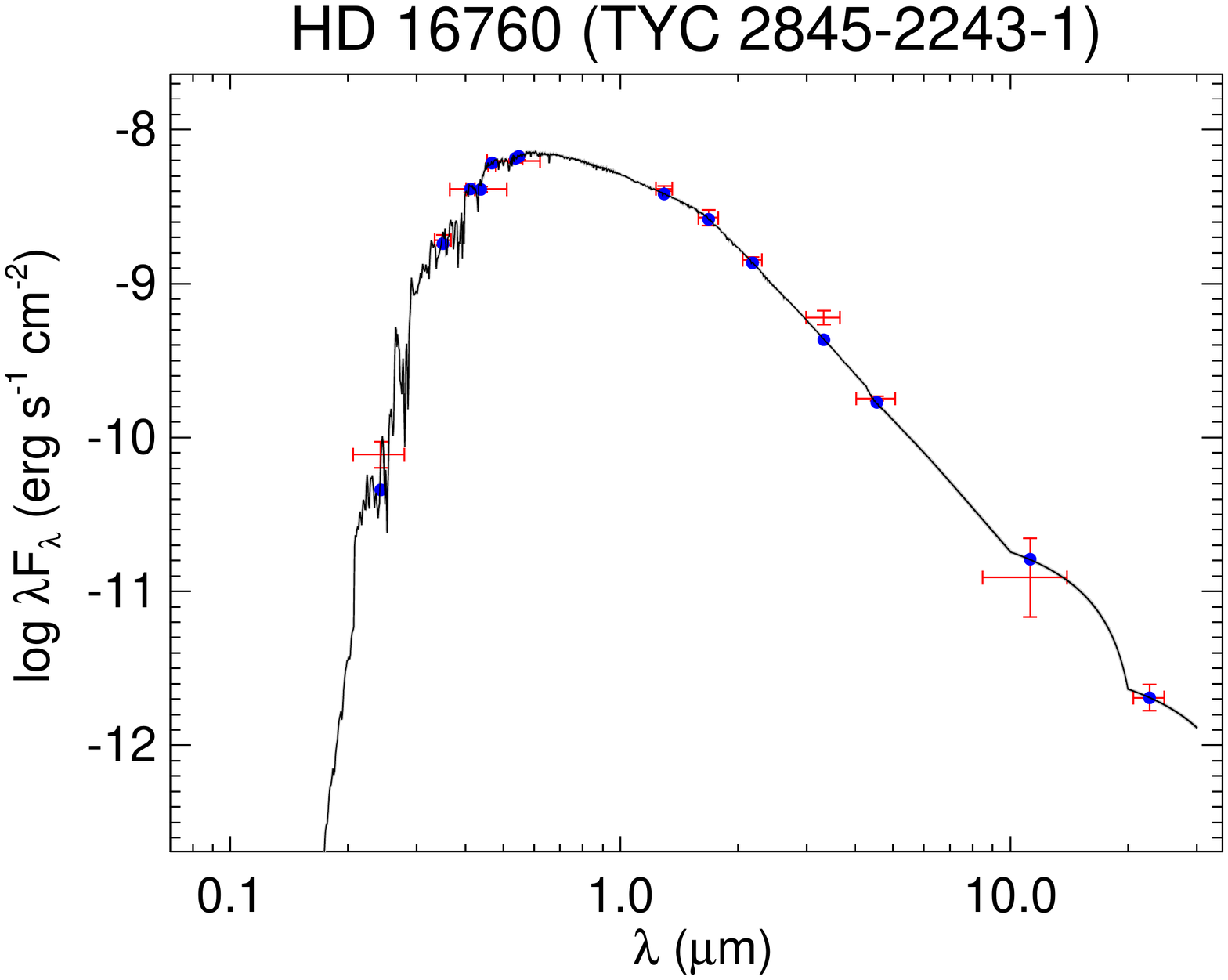}
  \includegraphics[trim=60 60 60 60,clip,width=0.49\linewidth]{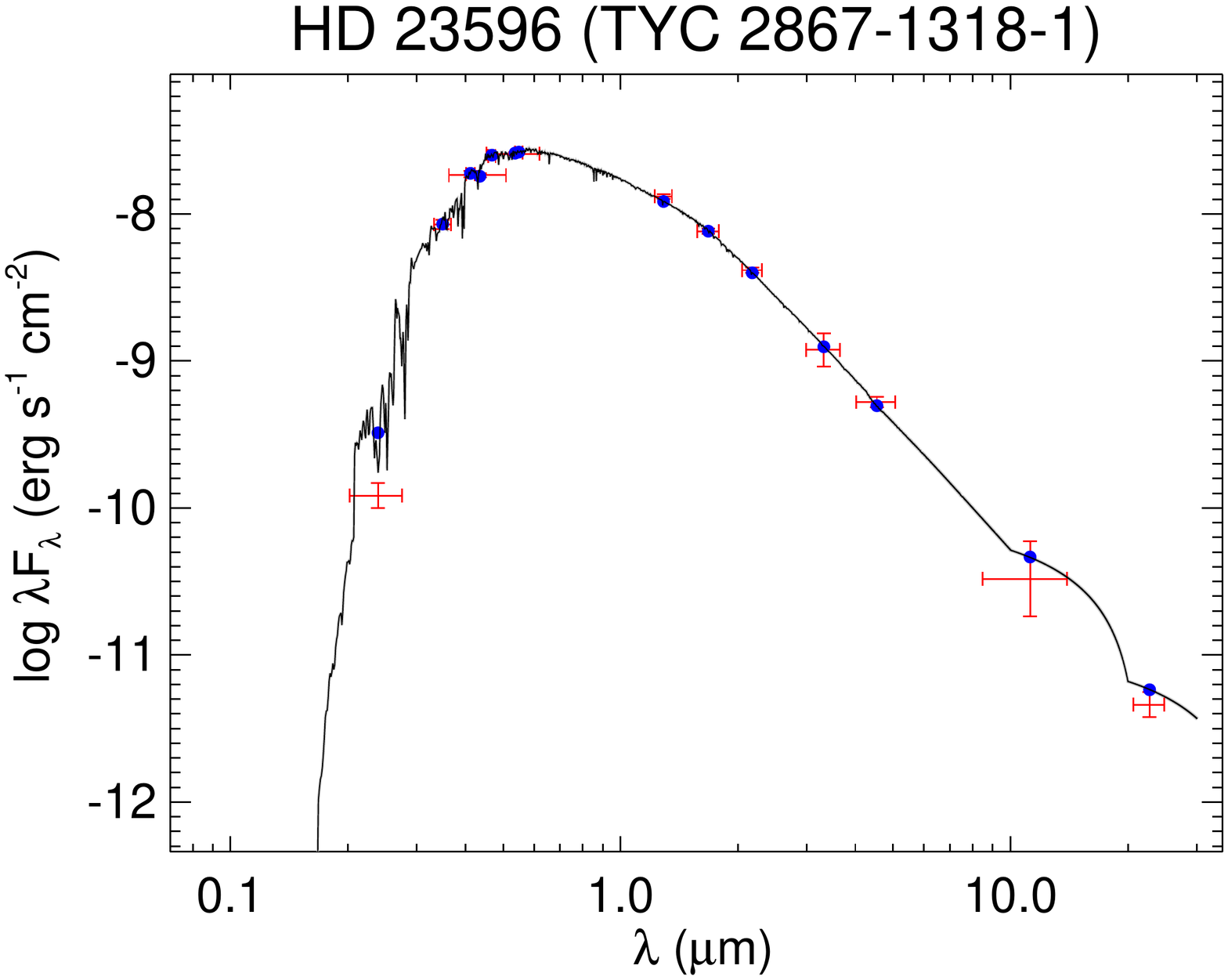}
  \includegraphics[trim=60 60 60 60,clip,width=0.49\linewidth]{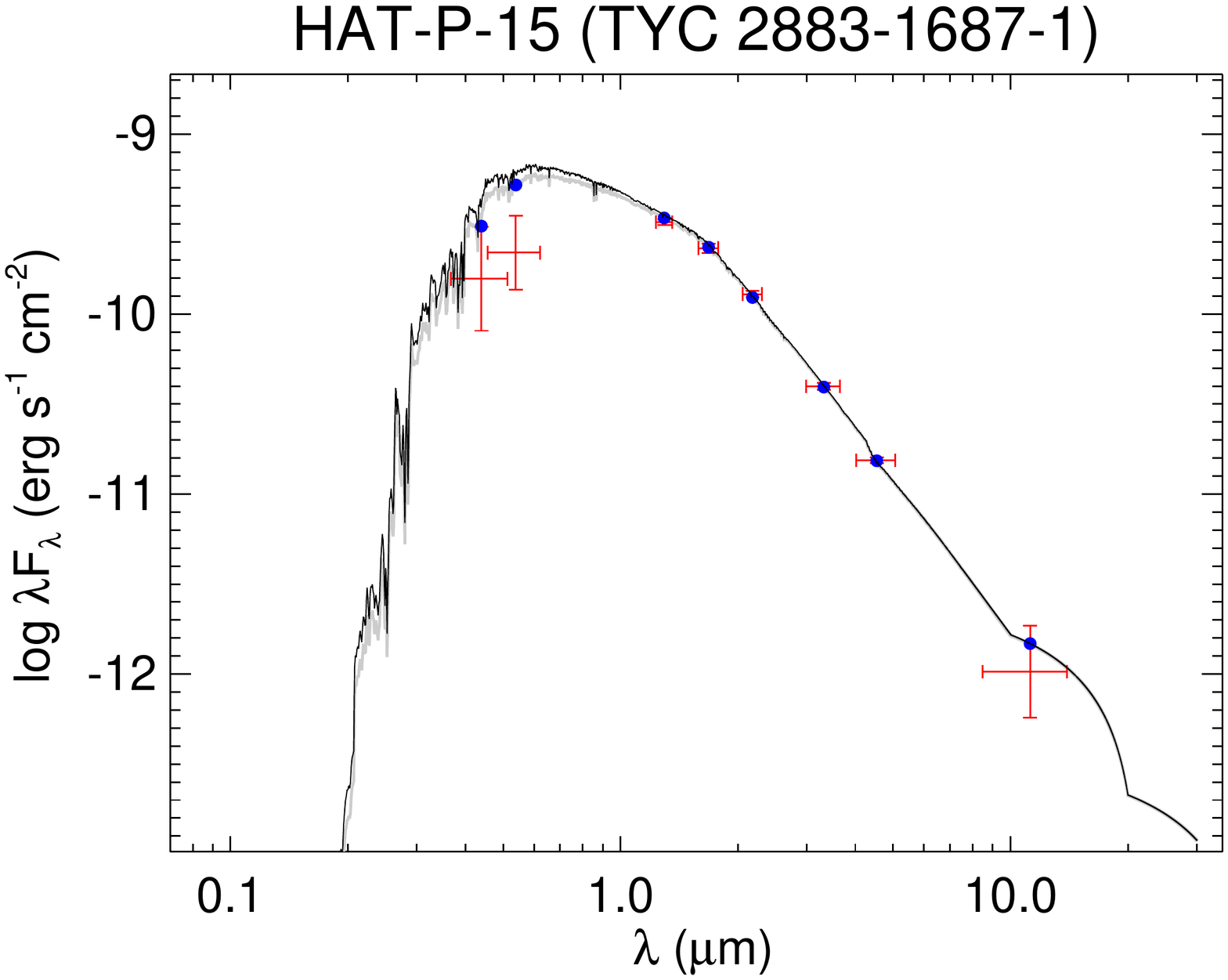}
  \caption{All labels, lines, symbols, and colors as in Figure \ref{fig:seds}.}
  \label{fig:seds_28}
\end{figure}

\begin{figure}[H]
  \centering
  \includegraphics[trim=60 60 60 60,clip,width=0.49\linewidth]{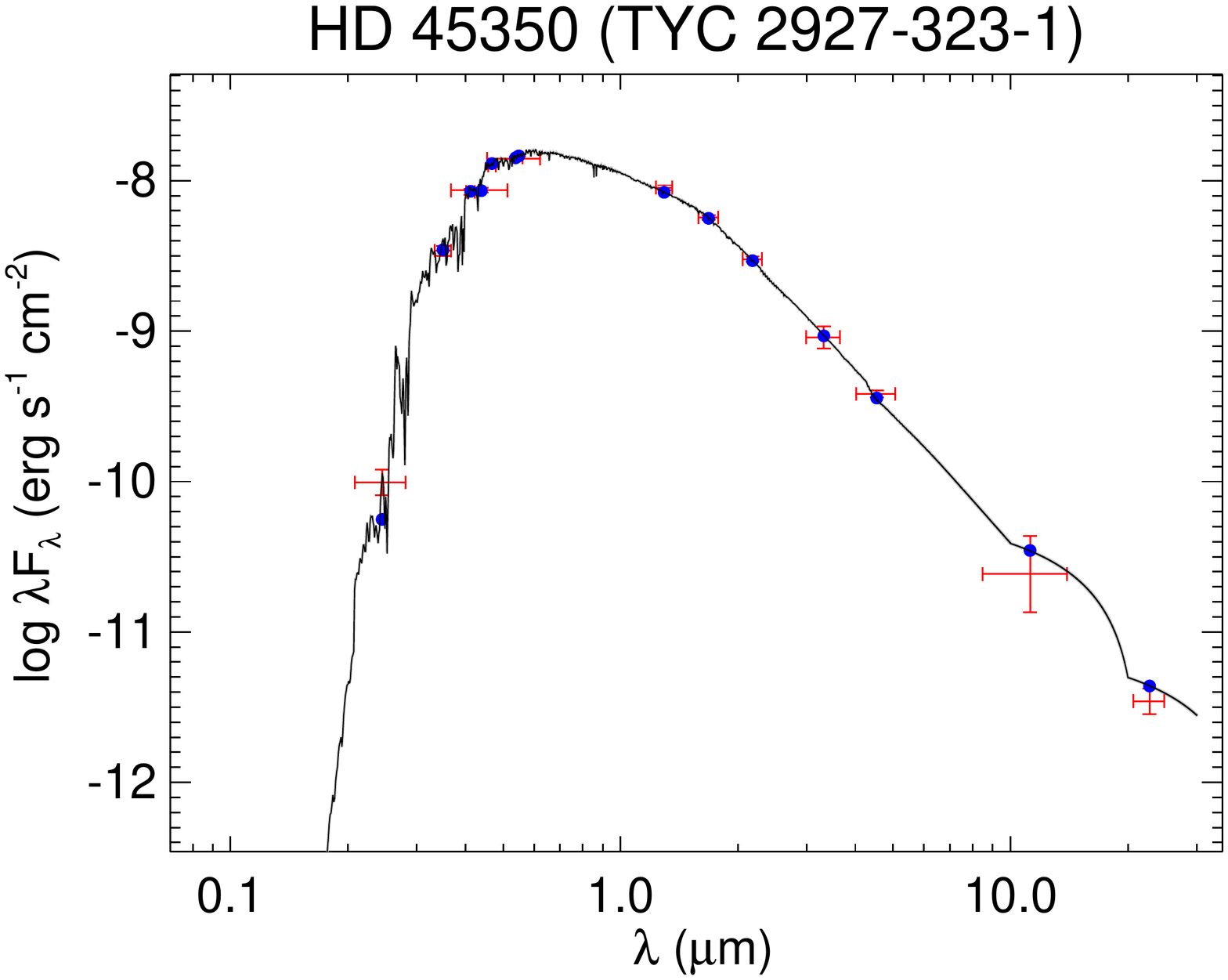}
  \includegraphics[trim=60 60 60 60,clip,width=0.49\linewidth]{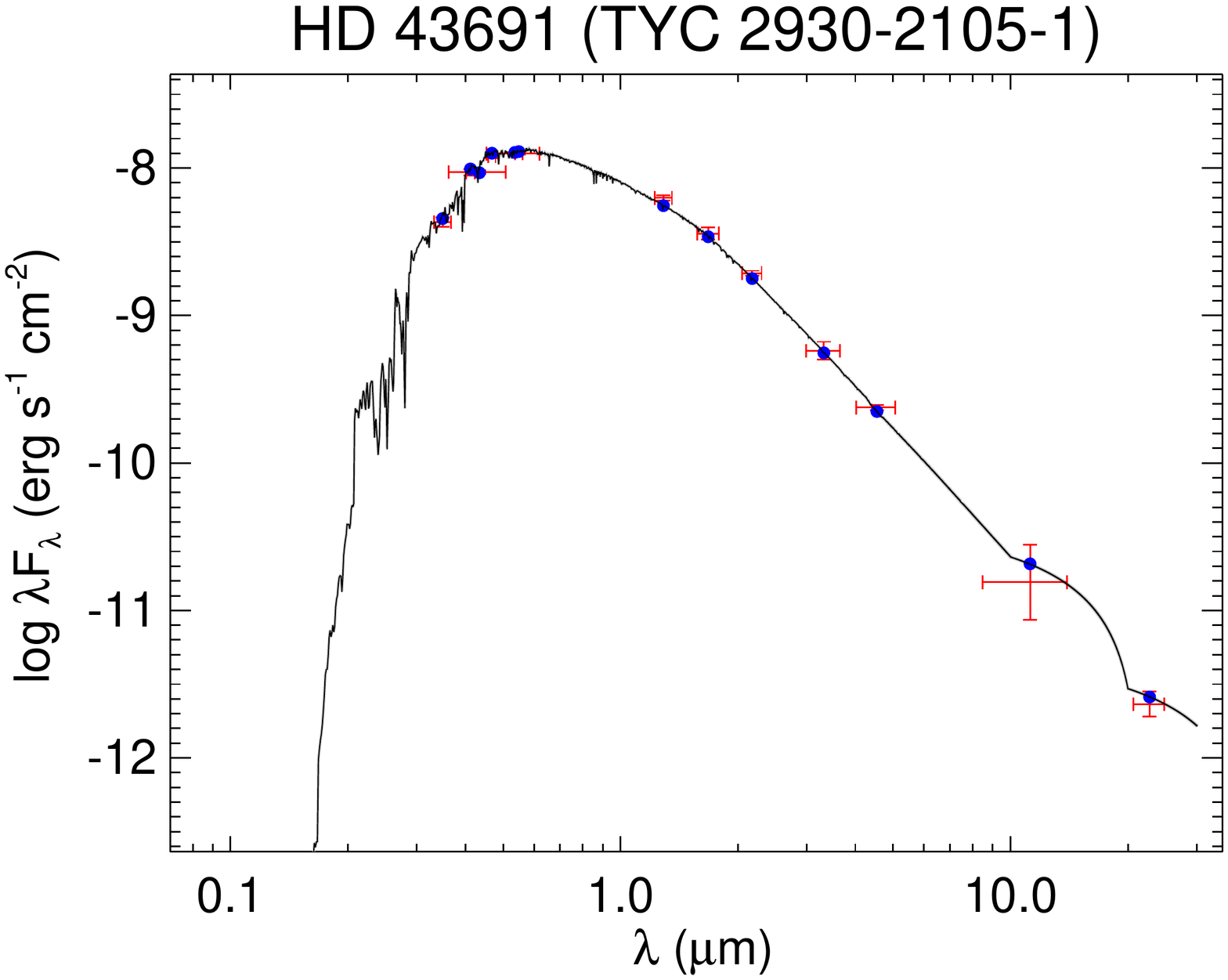}
  \includegraphics[trim=60 60 60 60,clip,width=0.49\linewidth]{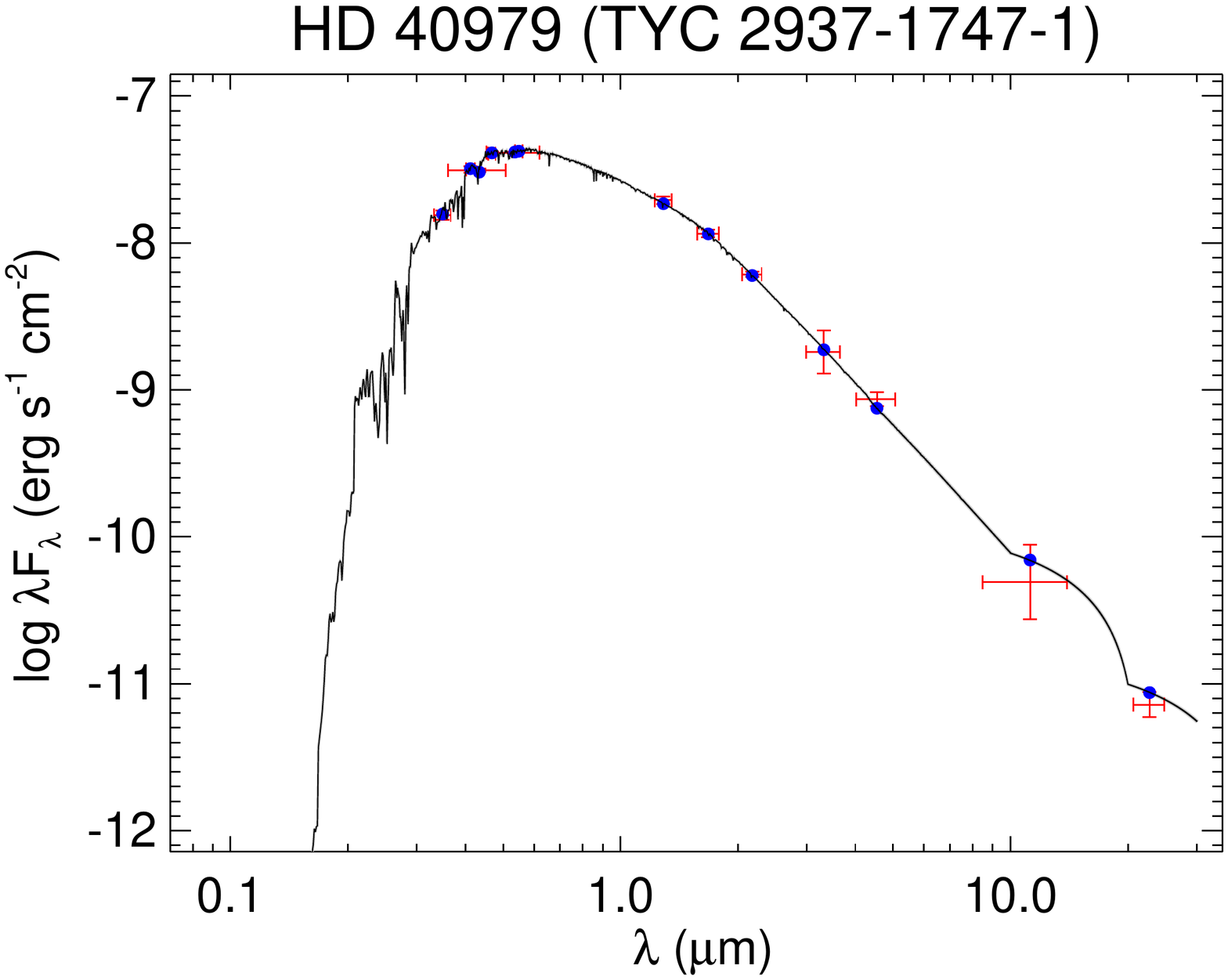}
  \includegraphics[trim=60 60 60 60,clip,width=0.49\linewidth]{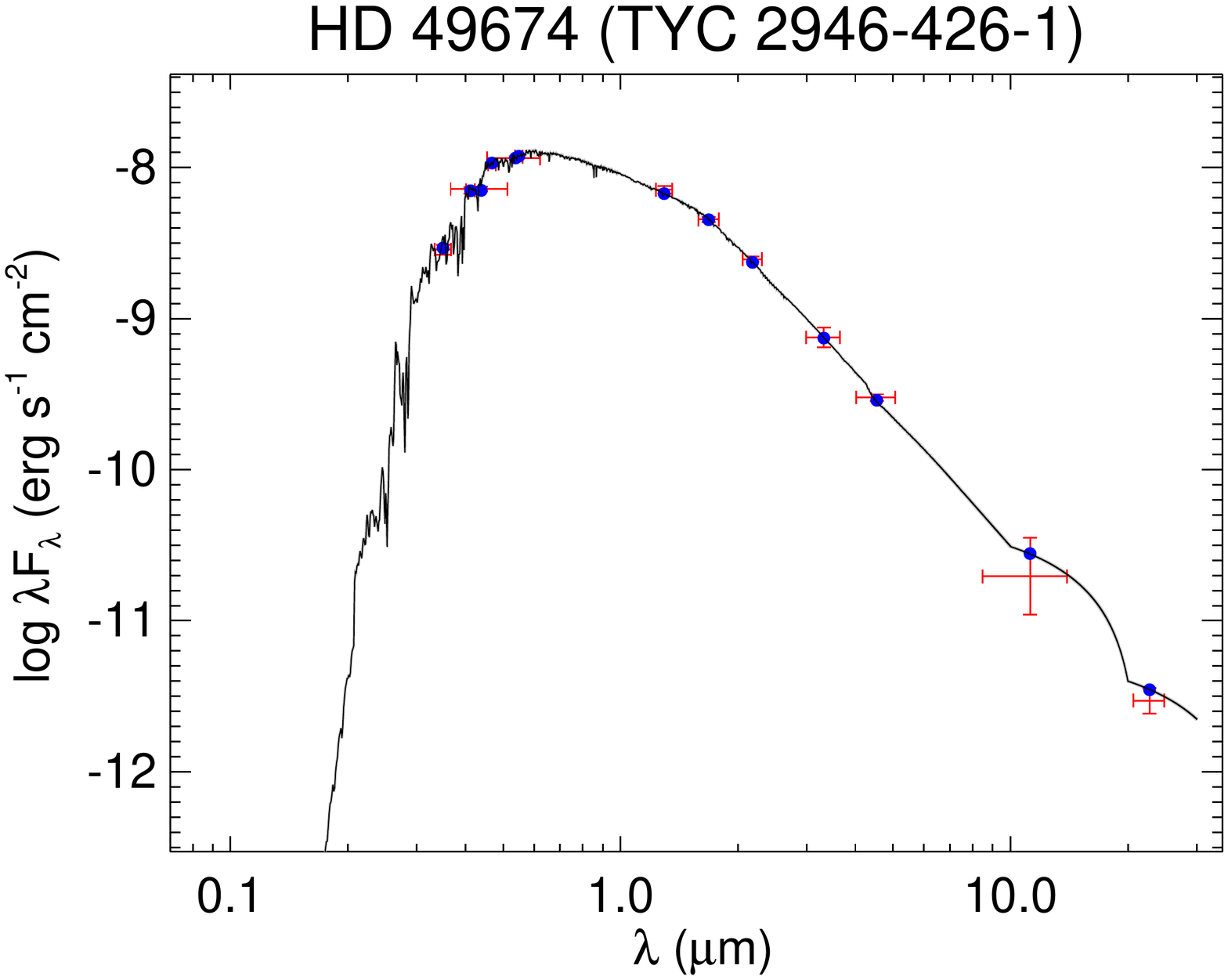}
  \includegraphics[trim=60 60 60 60,clip,width=0.49\linewidth]{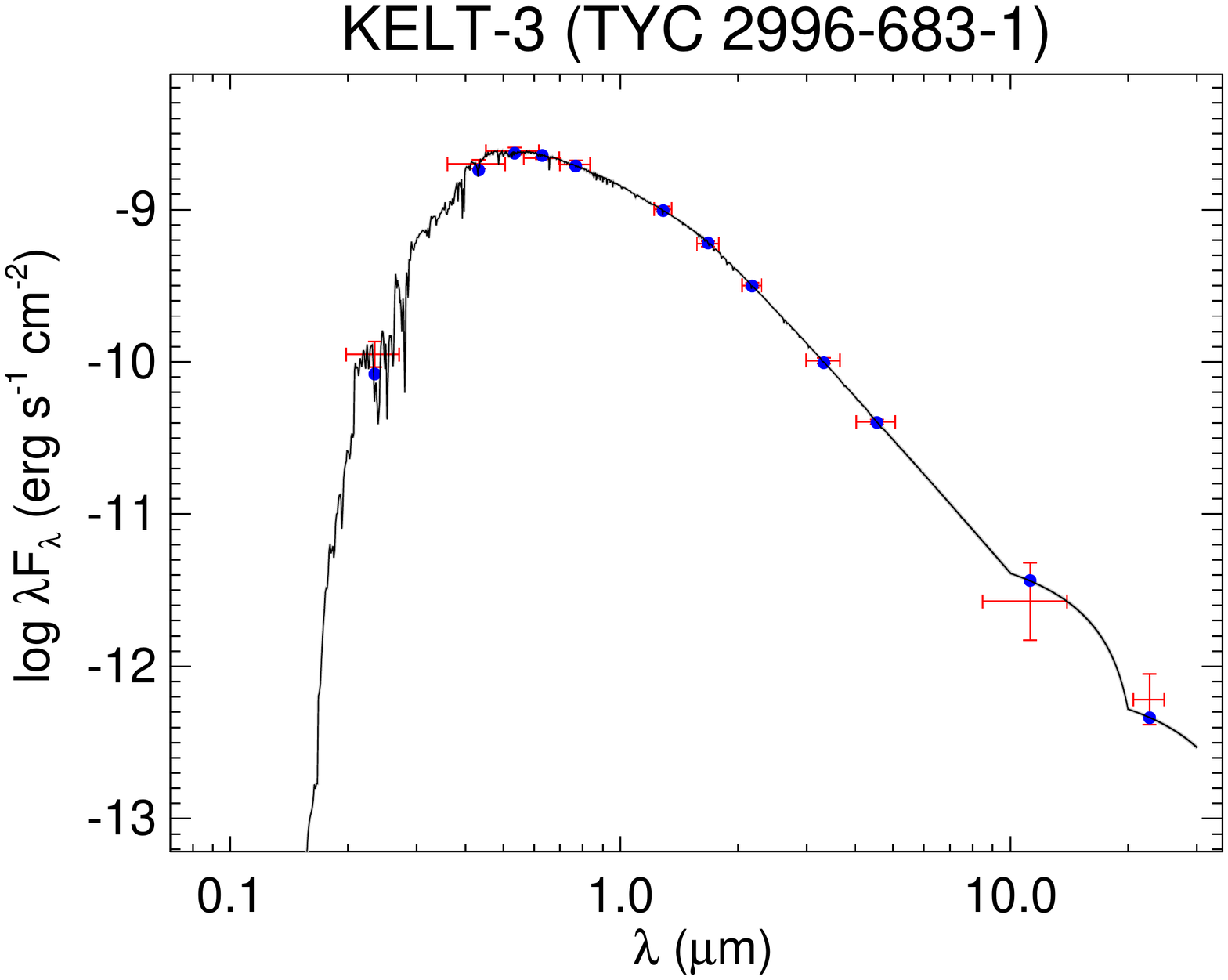}
  \includegraphics[trim=60 60 60 60,clip,width=0.49\linewidth]{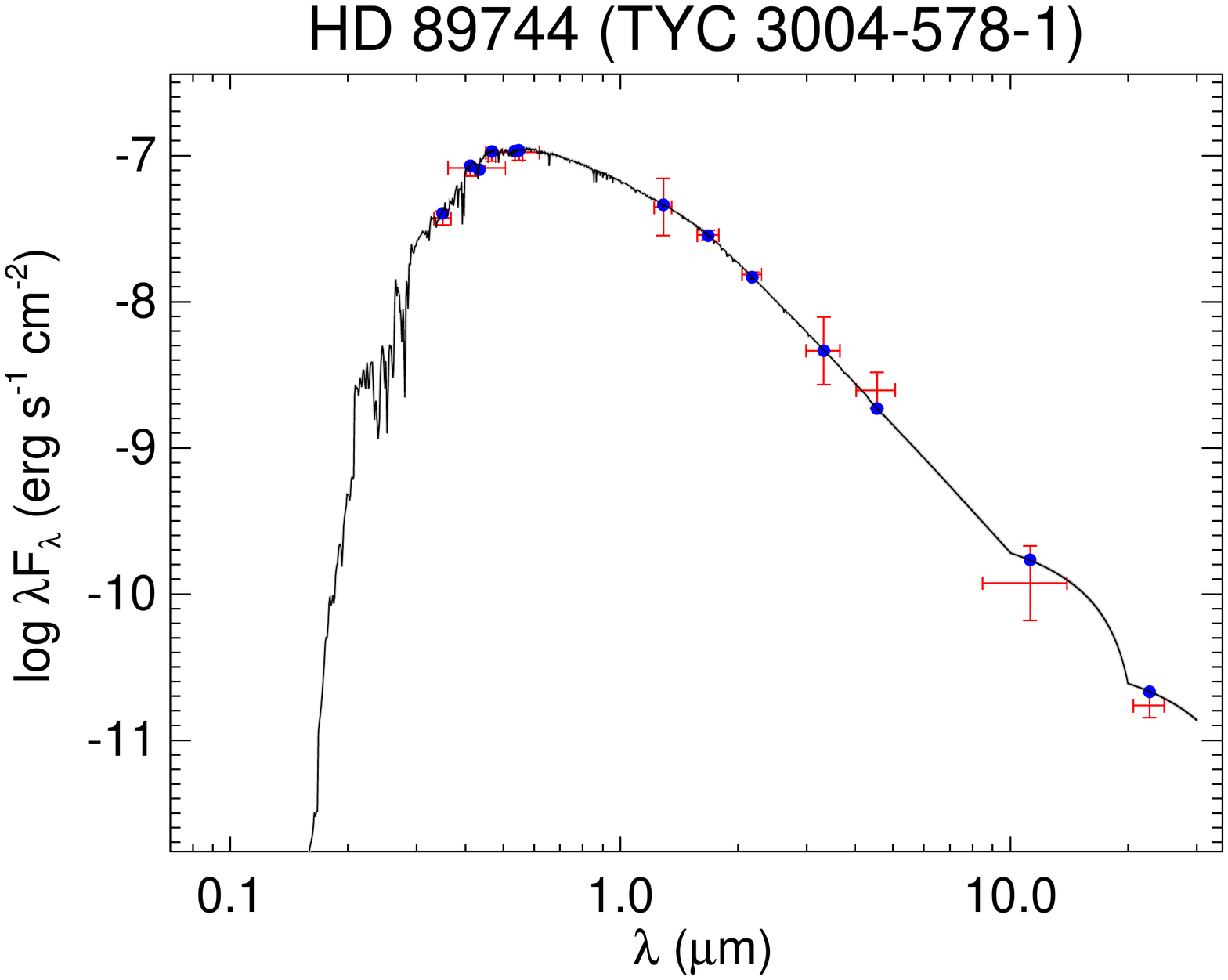}
  \caption{All labels, lines, symbols, and colors as in Figure \ref{fig:seds}.}
  \label{fig:seds_29}
\end{figure}

\begin{figure}[H]
  \centering
  \includegraphics[trim=60 60 60 60,clip,width=0.49\linewidth]{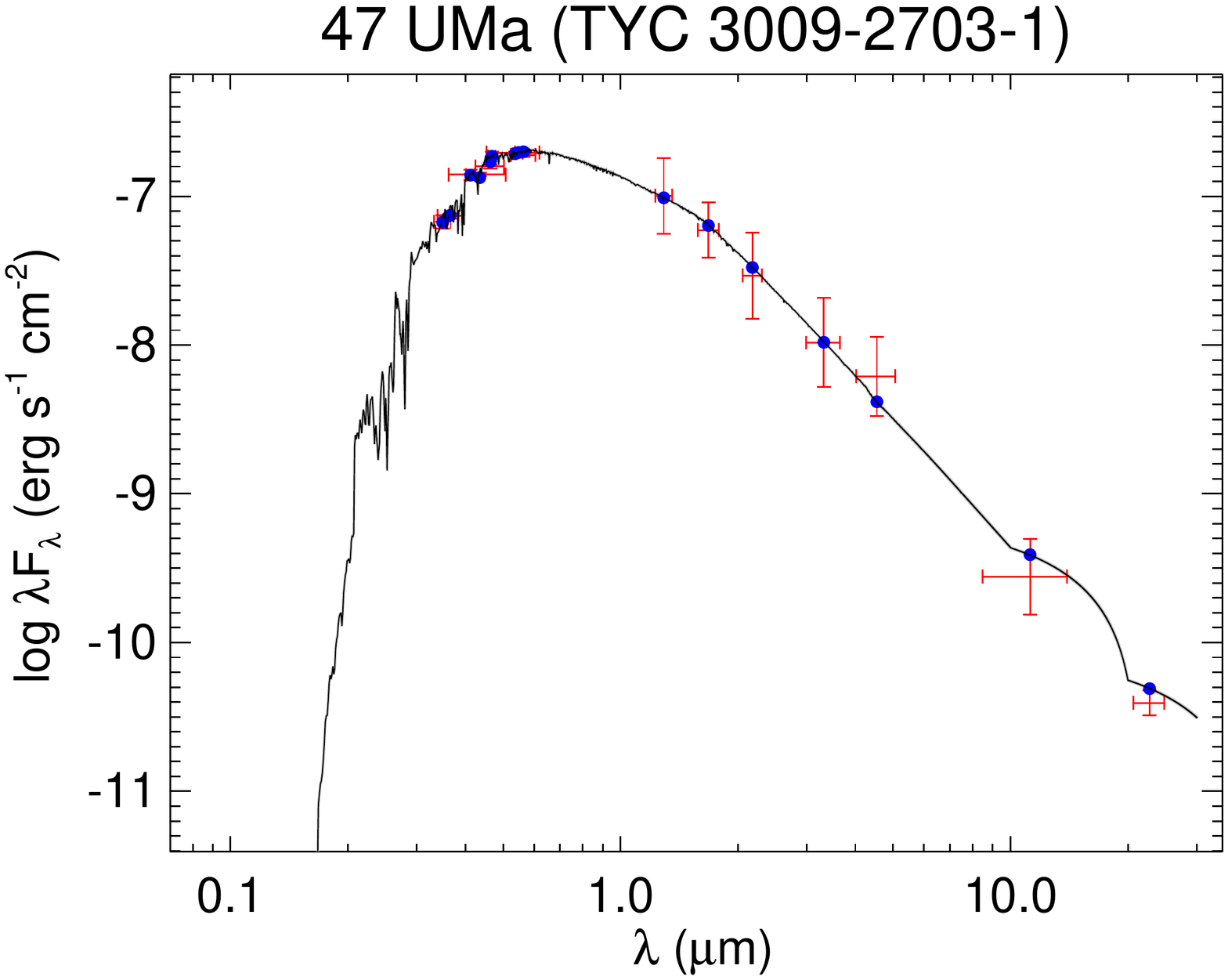}
  \includegraphics[trim=60 60 60 60,clip,width=0.49\linewidth]{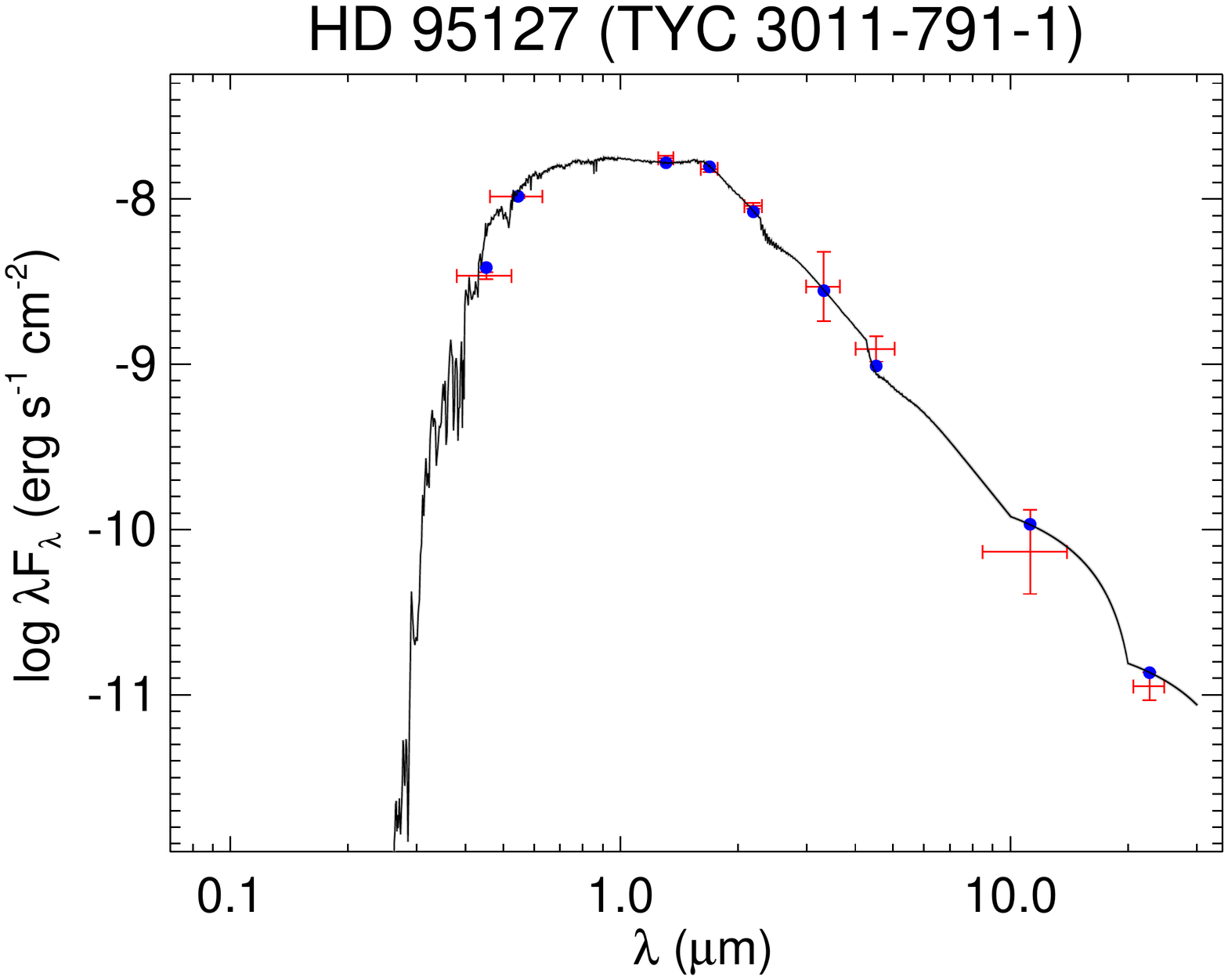}
  \includegraphics[trim=60 60 60 60,clip,width=0.49\linewidth]{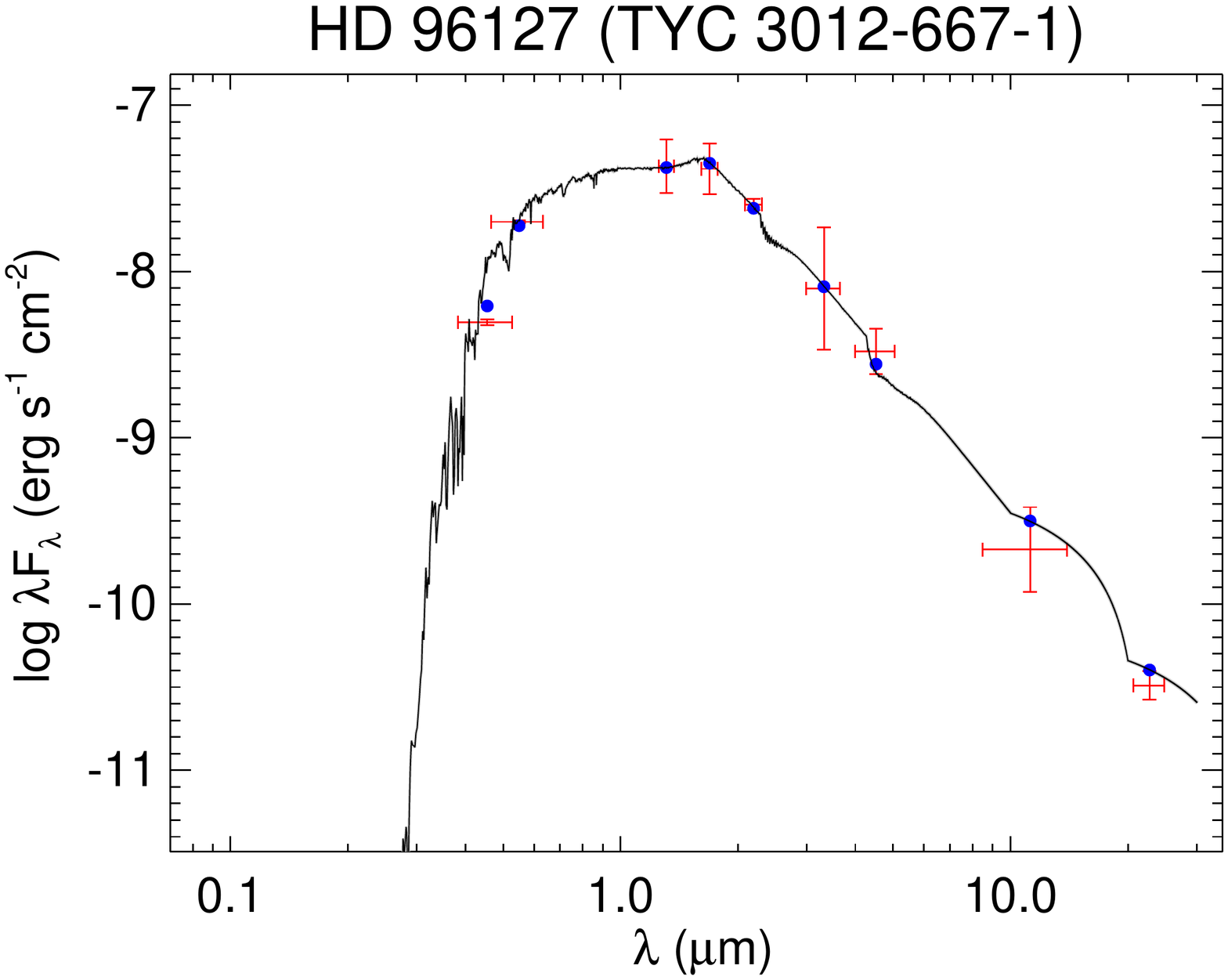}
  \includegraphics[trim=60 60 60 60,clip,width=0.49\linewidth]{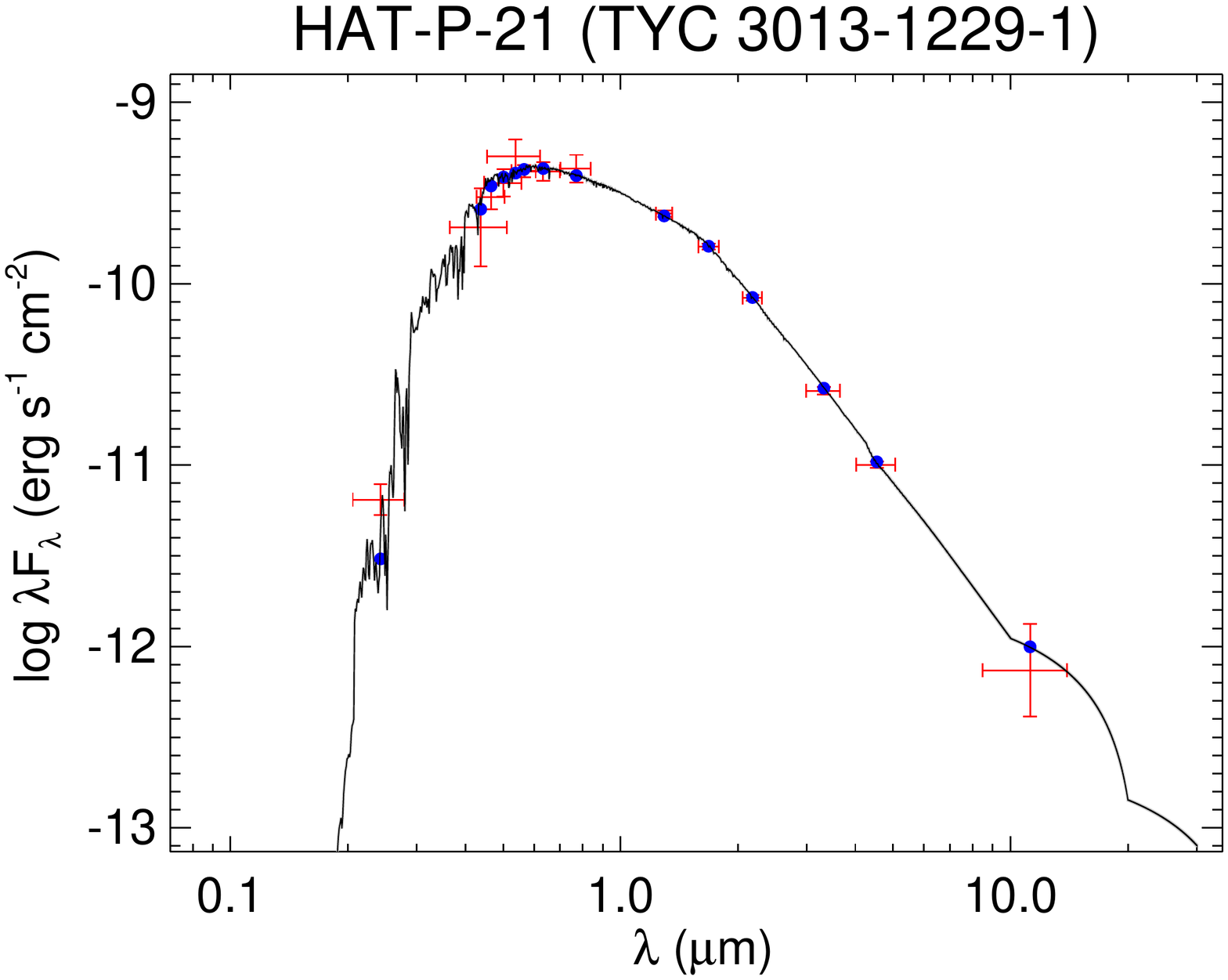}
  \includegraphics[trim=60 60 60 60,clip,width=0.49\linewidth]{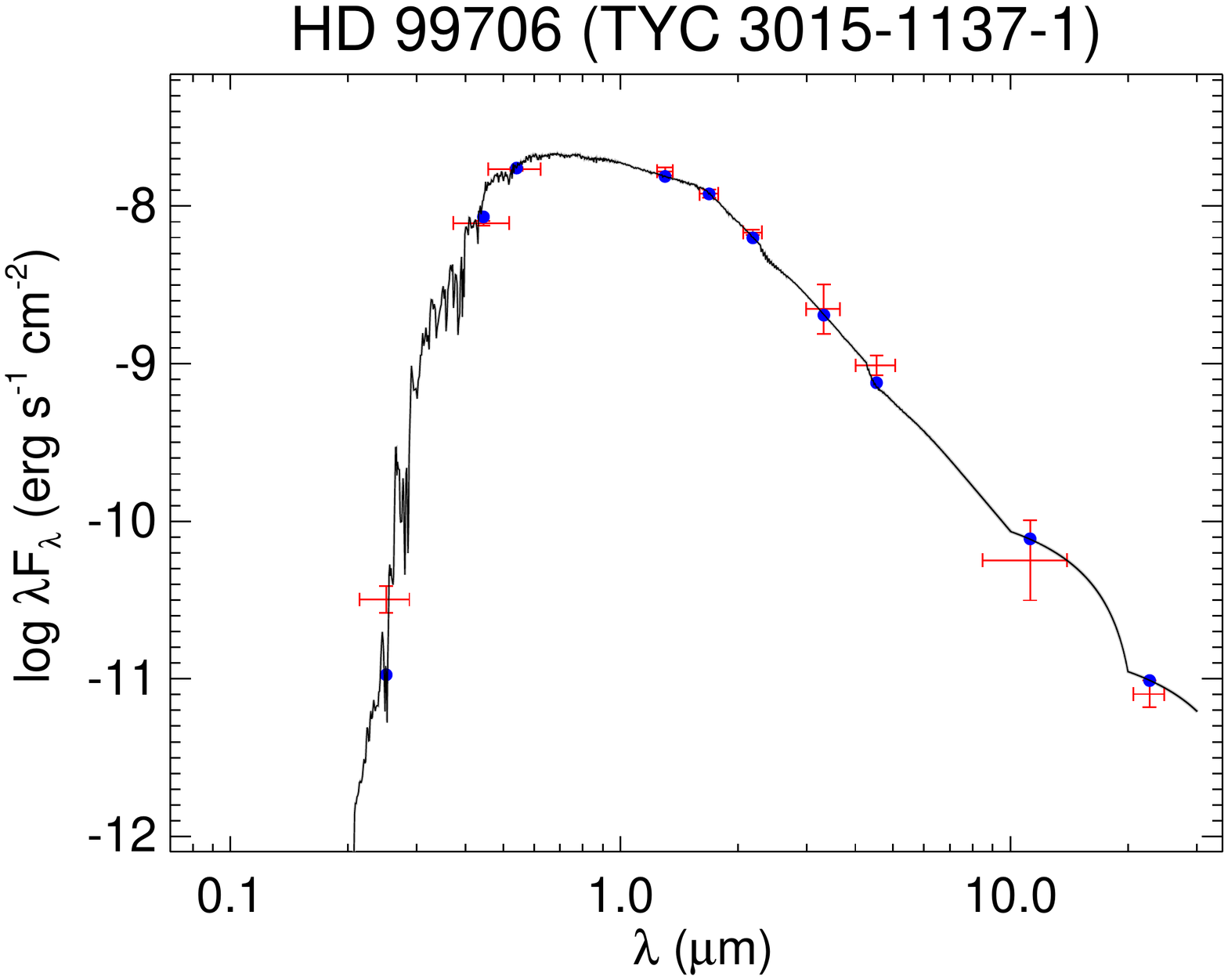}
  \includegraphics[trim=60 60 60 60,clip,width=0.49\linewidth]{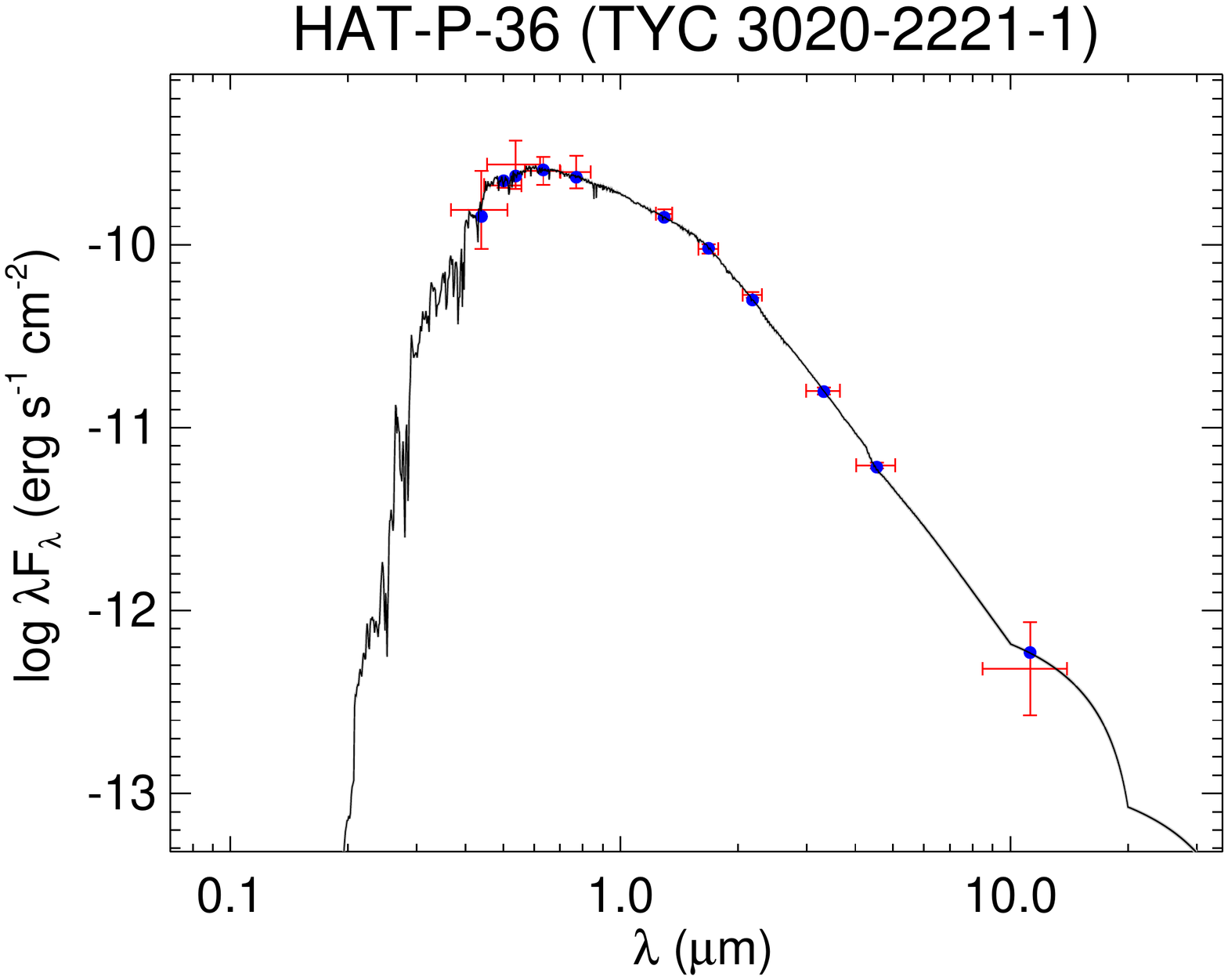}
  \caption{All labels, lines, symbols, and colors as in Figure \ref{fig:seds}.}
  \label{fig:seds_30}
\end{figure}

\begin{figure}[H]
  \centering
  \includegraphics[trim=60 60 60 60,clip,width=0.49\linewidth]{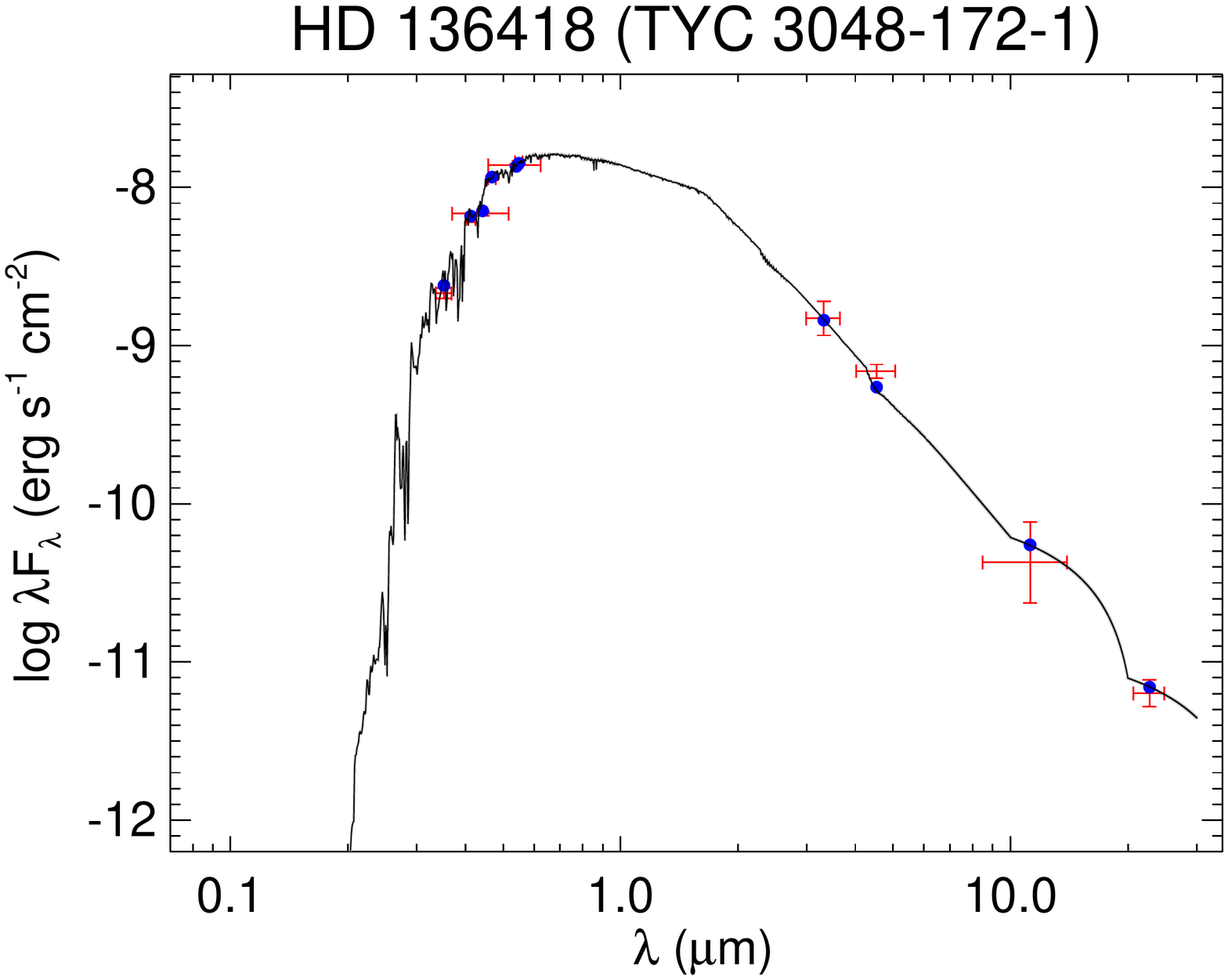}
  \includegraphics[trim=60 60 60 60,clip,width=0.49\linewidth]{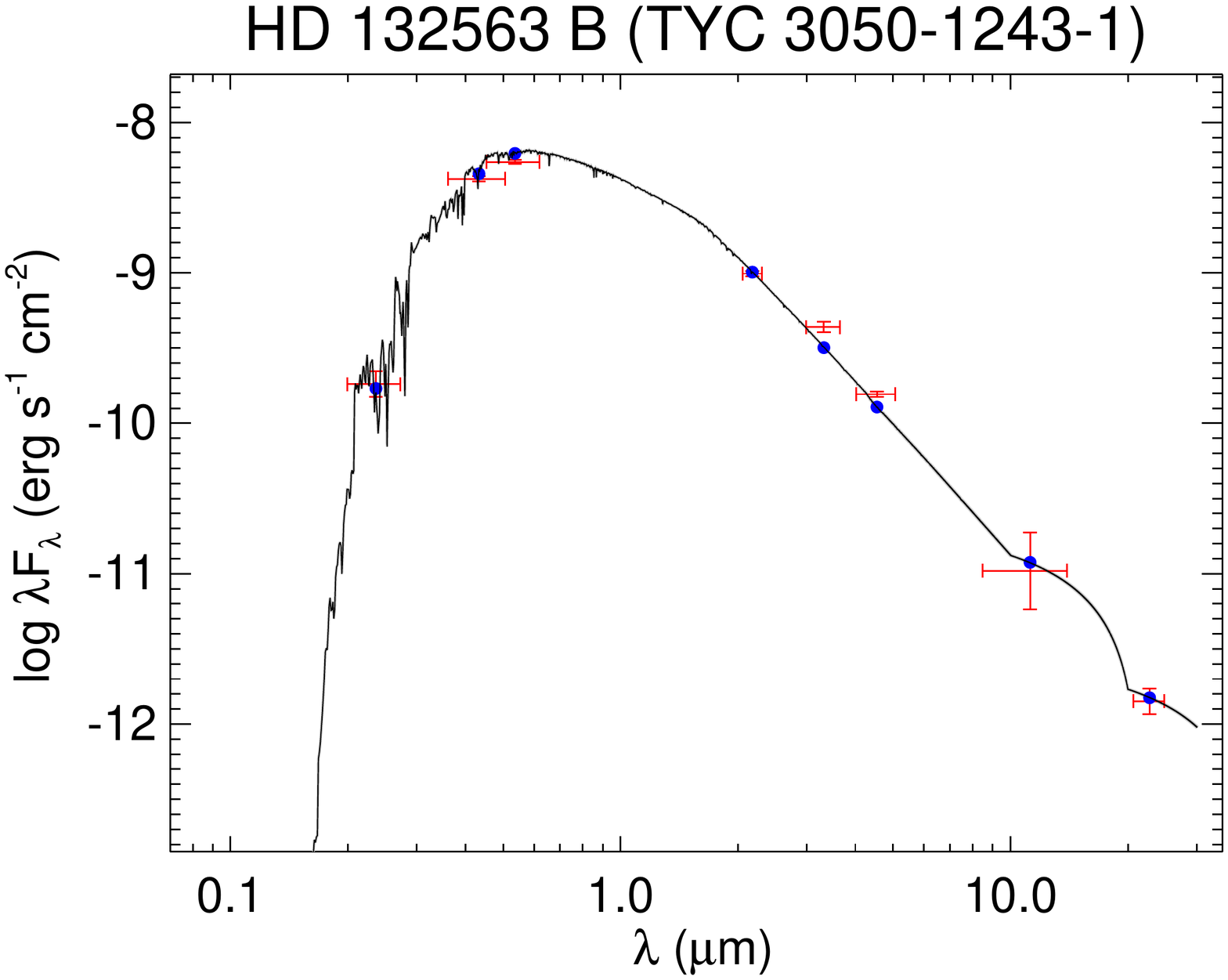}
  \includegraphics[trim=60 60 60 60,clip,width=0.49\linewidth]{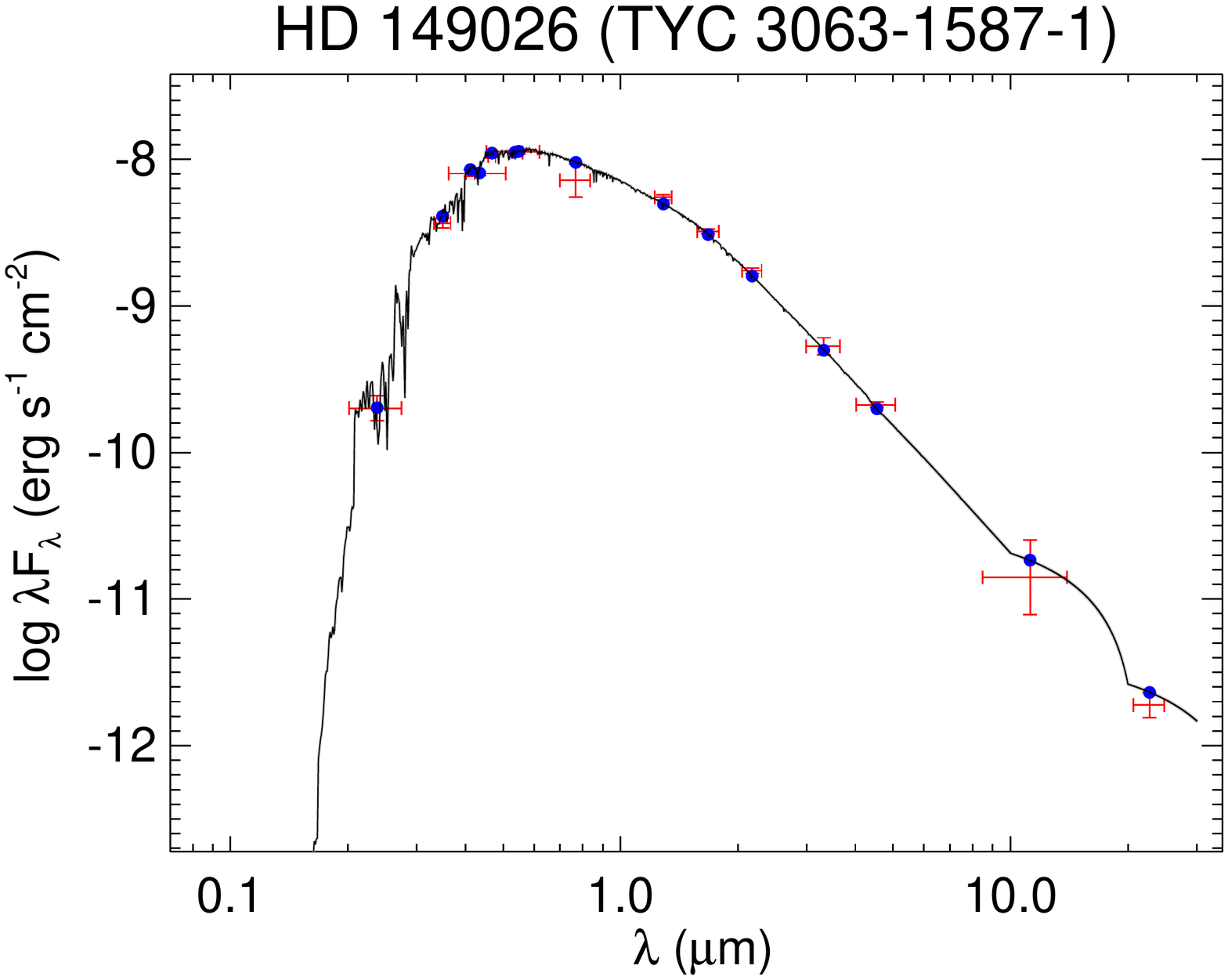}
  \includegraphics[trim=60 60 60 60,clip,width=0.49\linewidth]{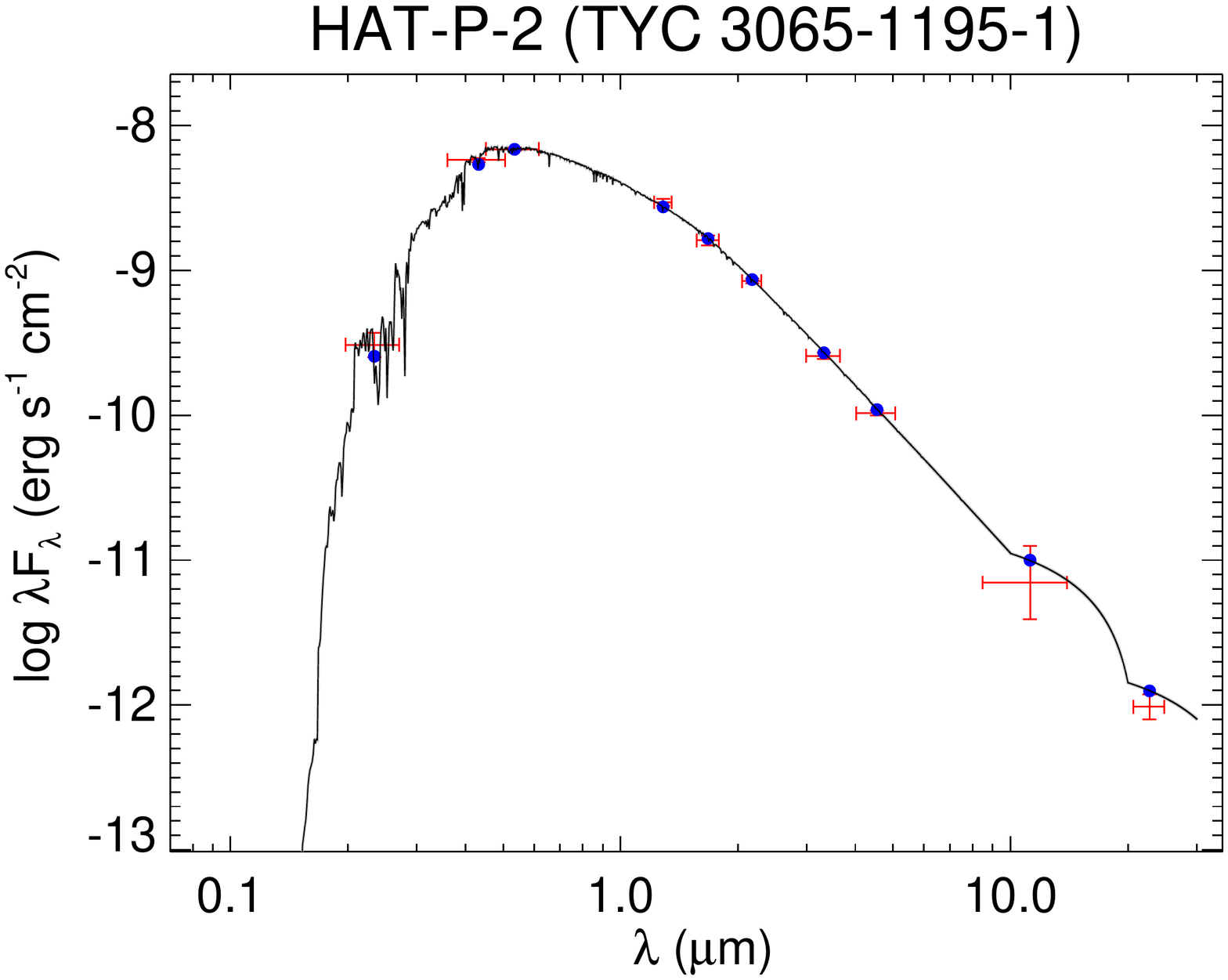}
  \includegraphics[trim=60 60 60 60,clip,width=0.49\linewidth]{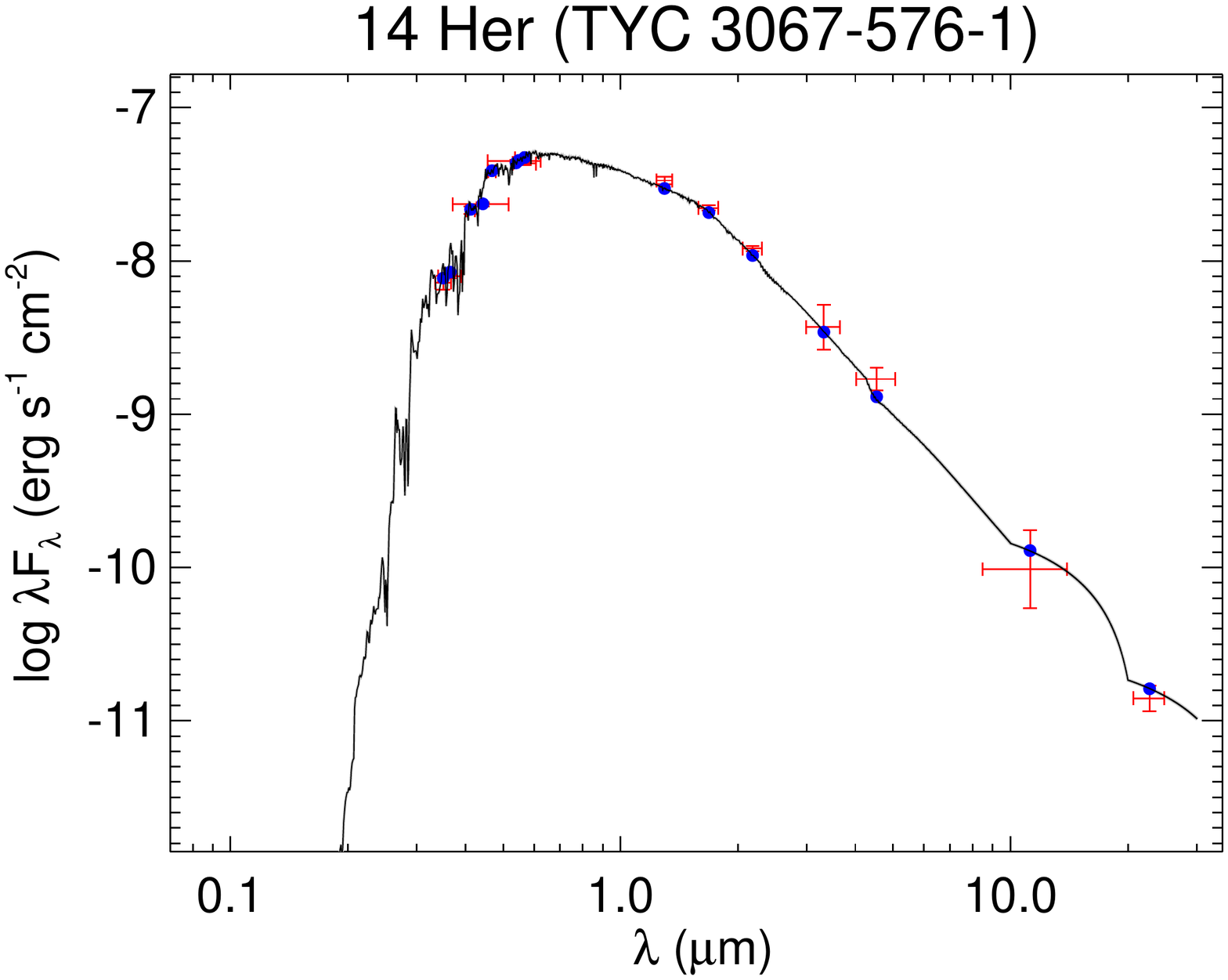}
  \includegraphics[trim=60 60 60 60,clip,width=0.49\linewidth]{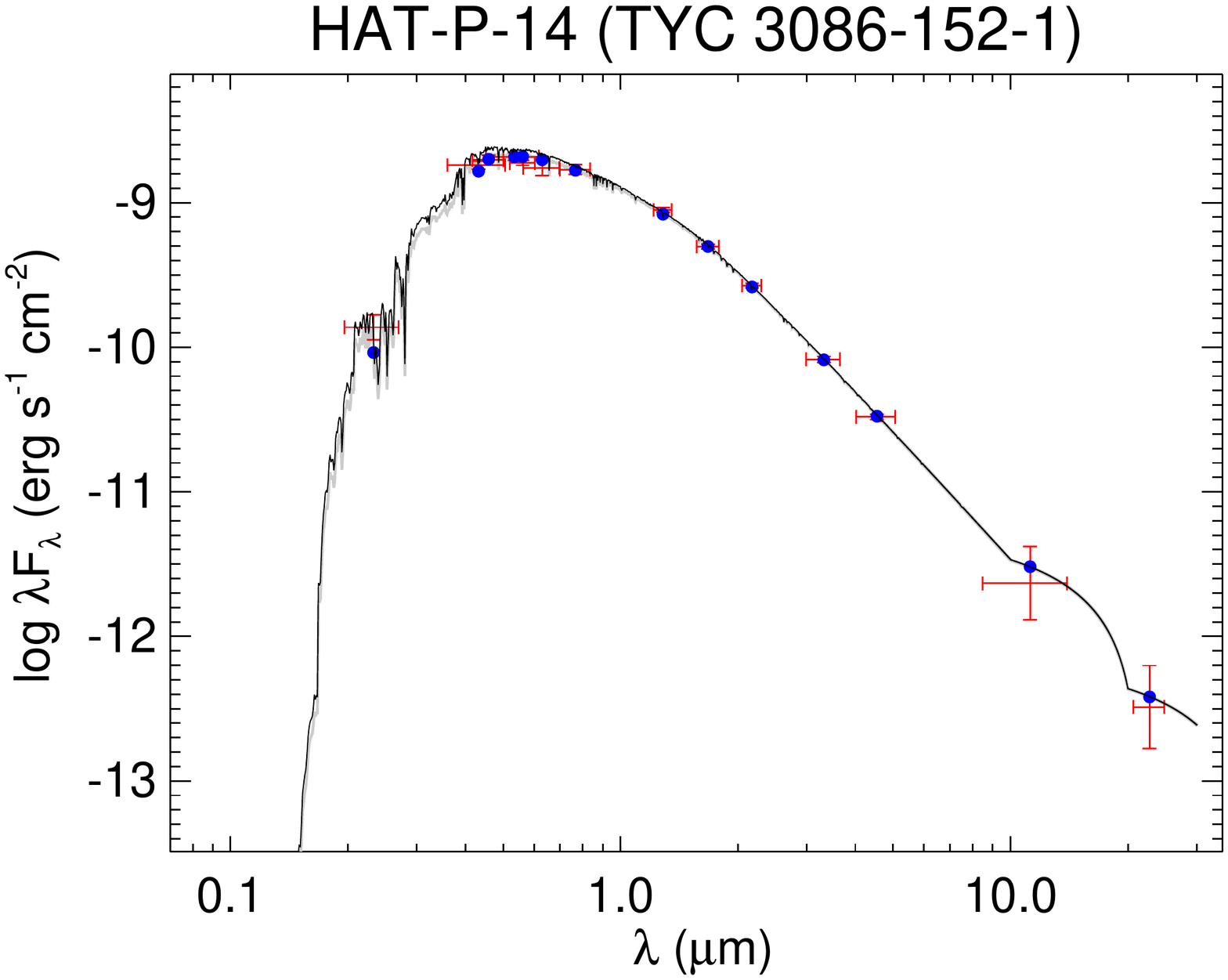}
  \caption{All labels, lines, symbols, and colors as in Figure \ref{fig:seds}.}
  \label{fig:seds_31}
\end{figure}

\begin{figure}[H]
  \centering
  \includegraphics[trim=60 60 60 60,clip,width=0.49\linewidth]{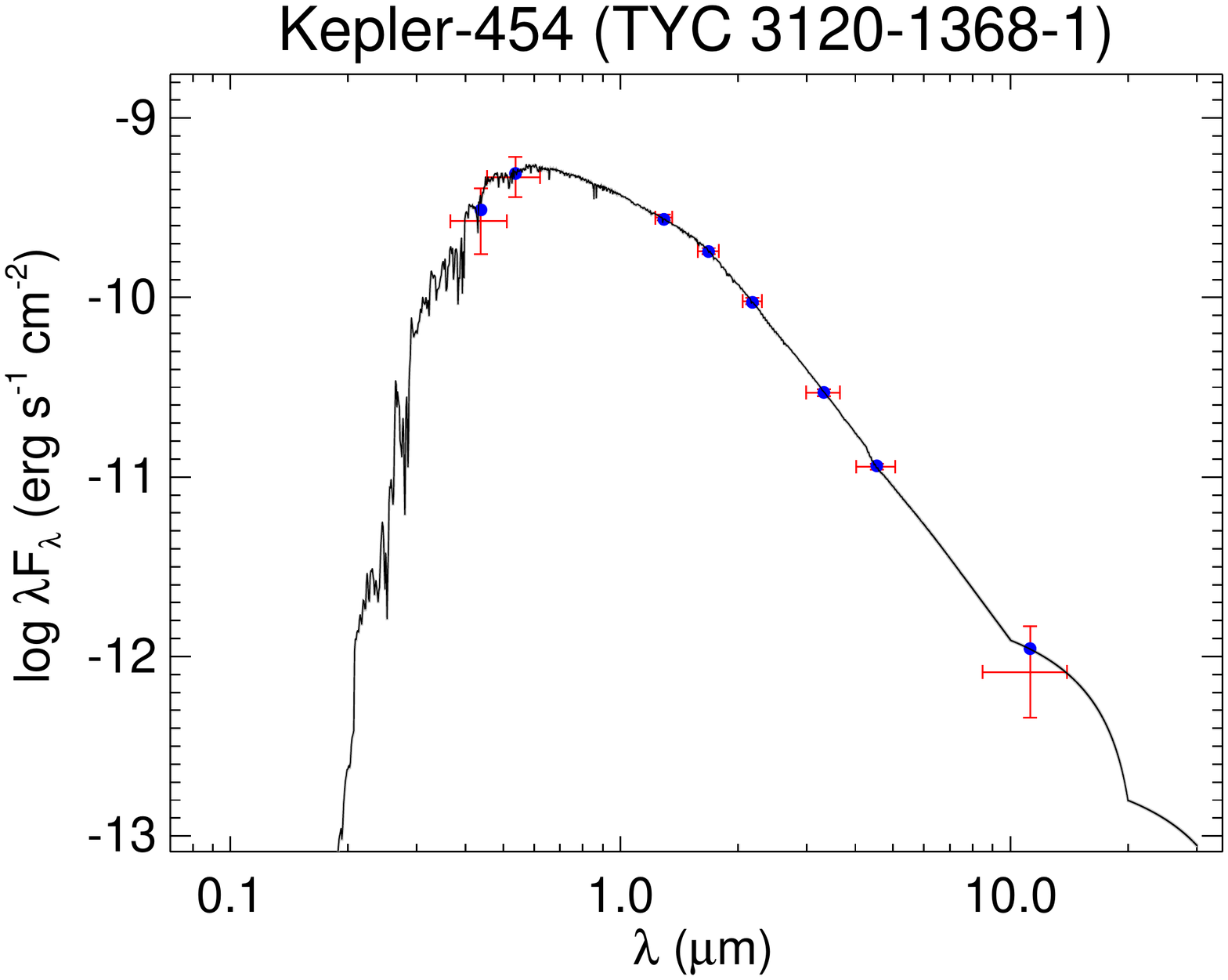}
  \includegraphics[trim=60 60 60 60,clip,width=0.49\linewidth]{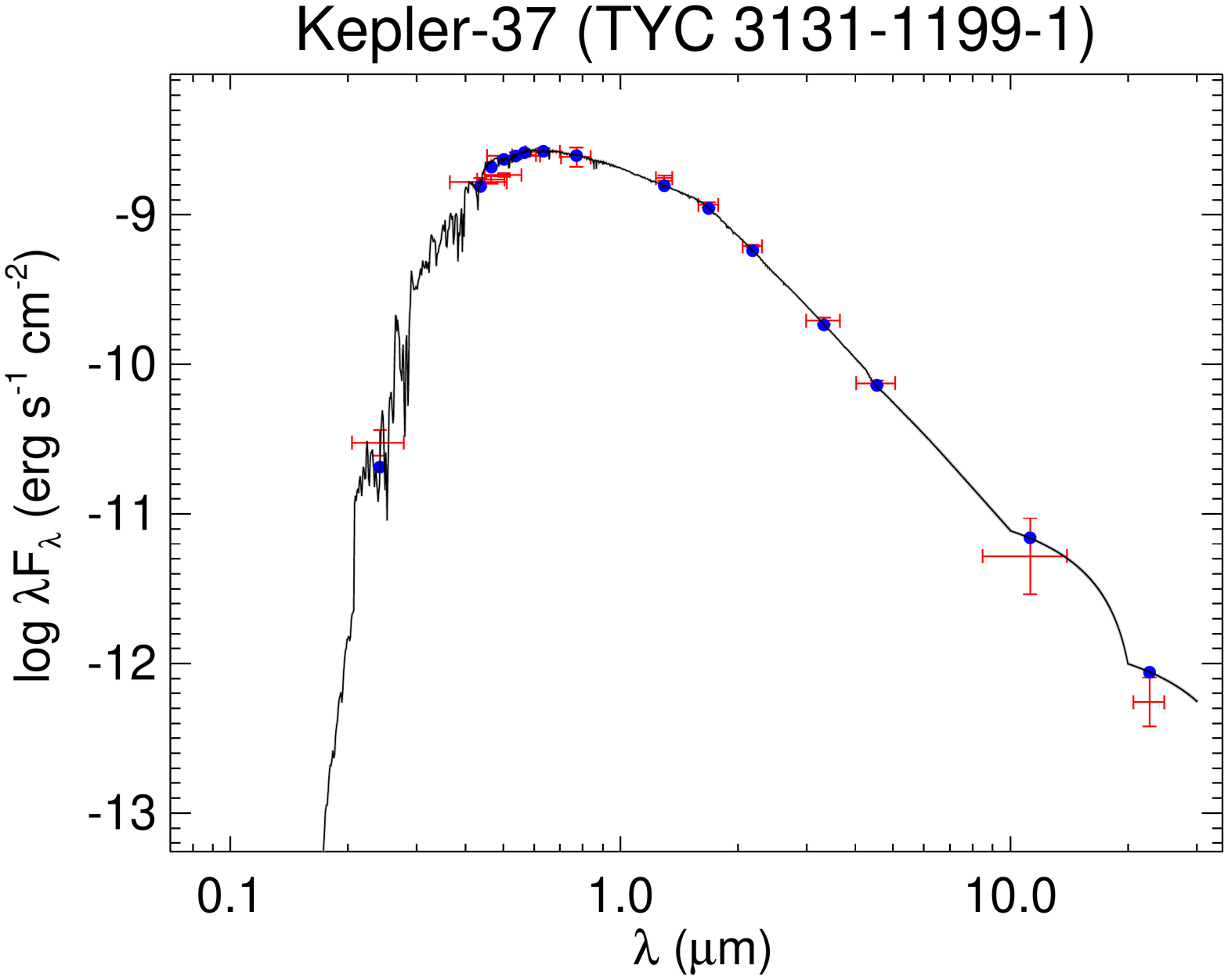}
  \includegraphics[trim=60 60 60 60,clip,width=0.49\linewidth]{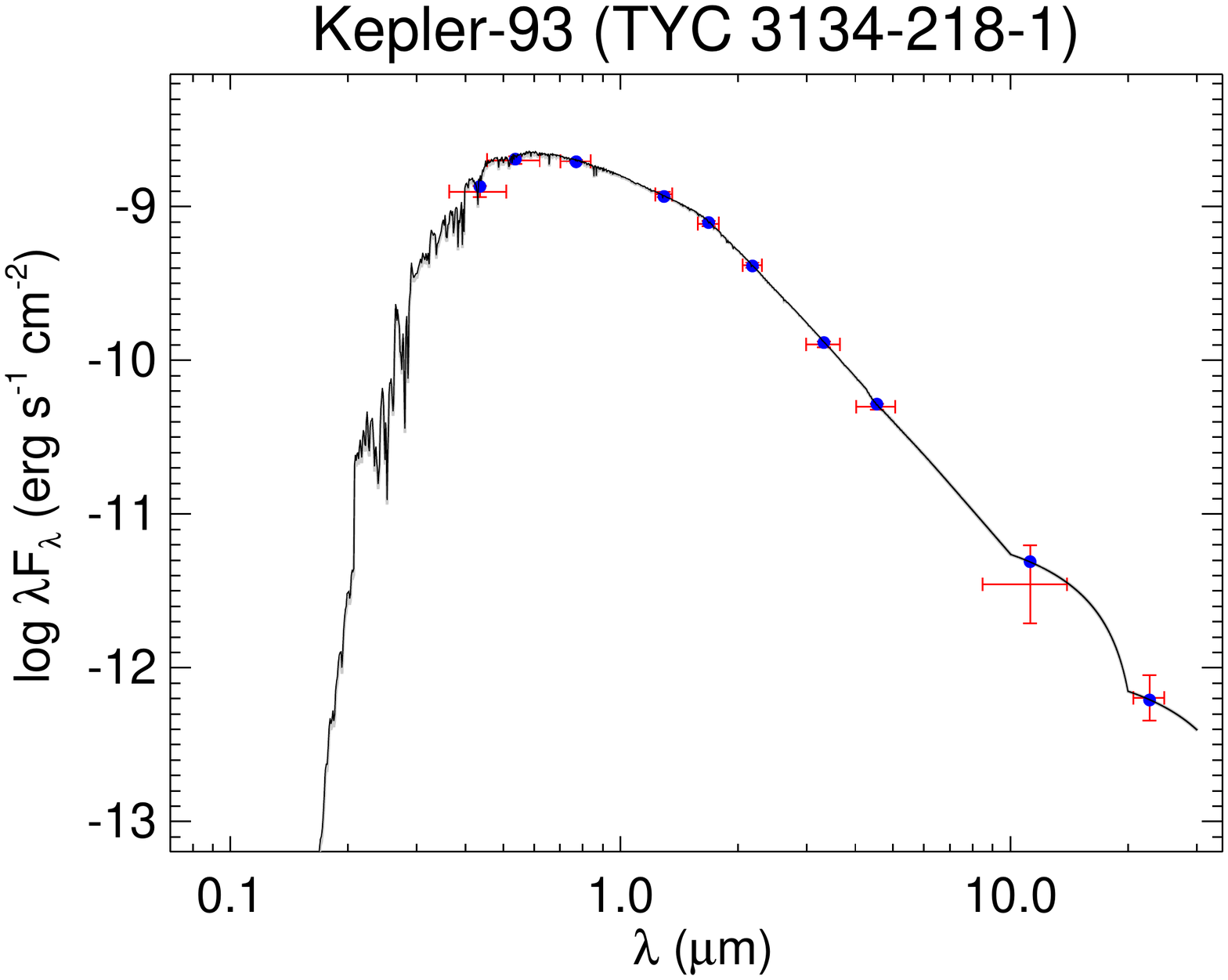}
  \includegraphics[trim=60 60 60 60,clip,width=0.49\linewidth]{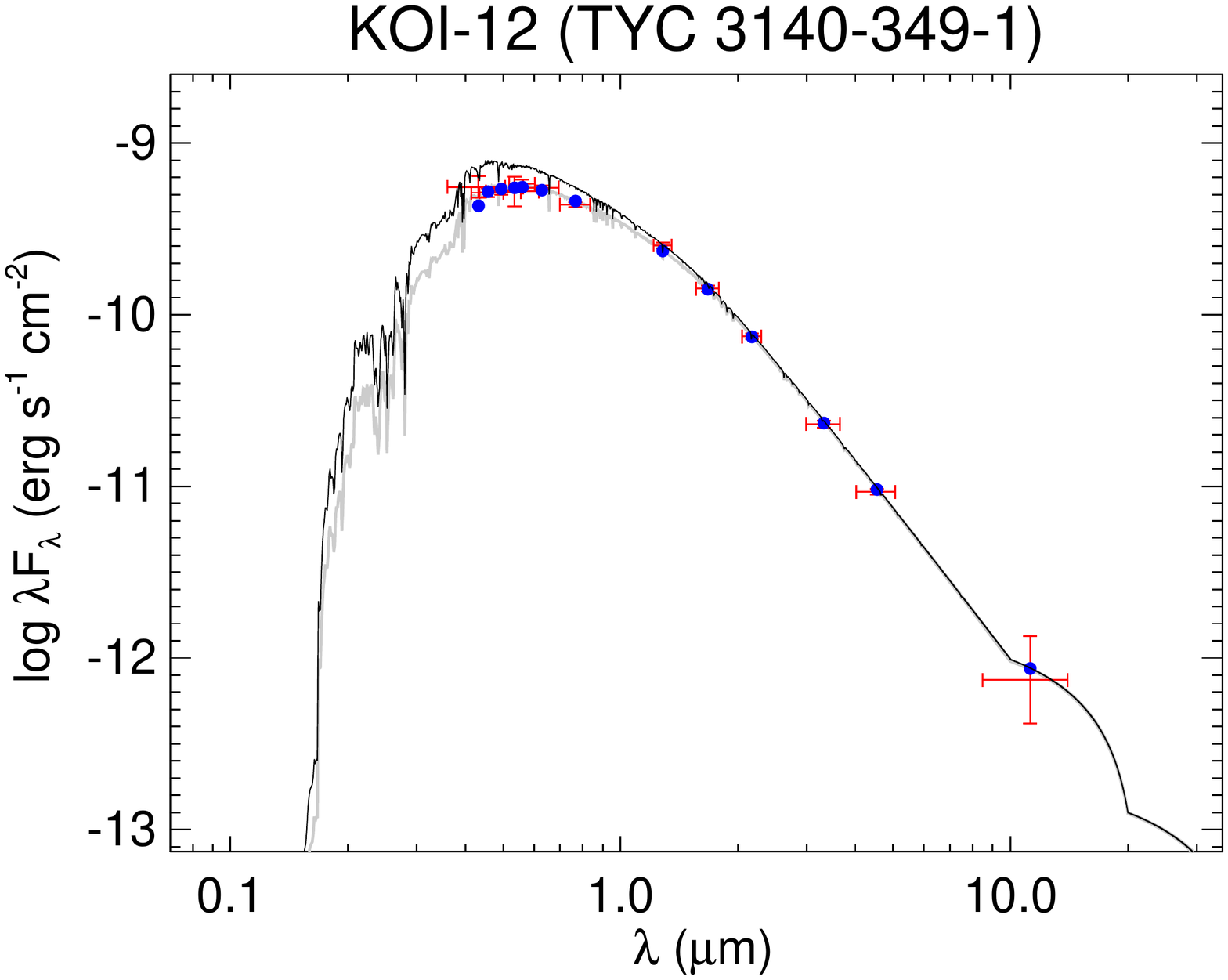}
  \includegraphics[trim=60 60 60 60,clip,width=0.49\linewidth]{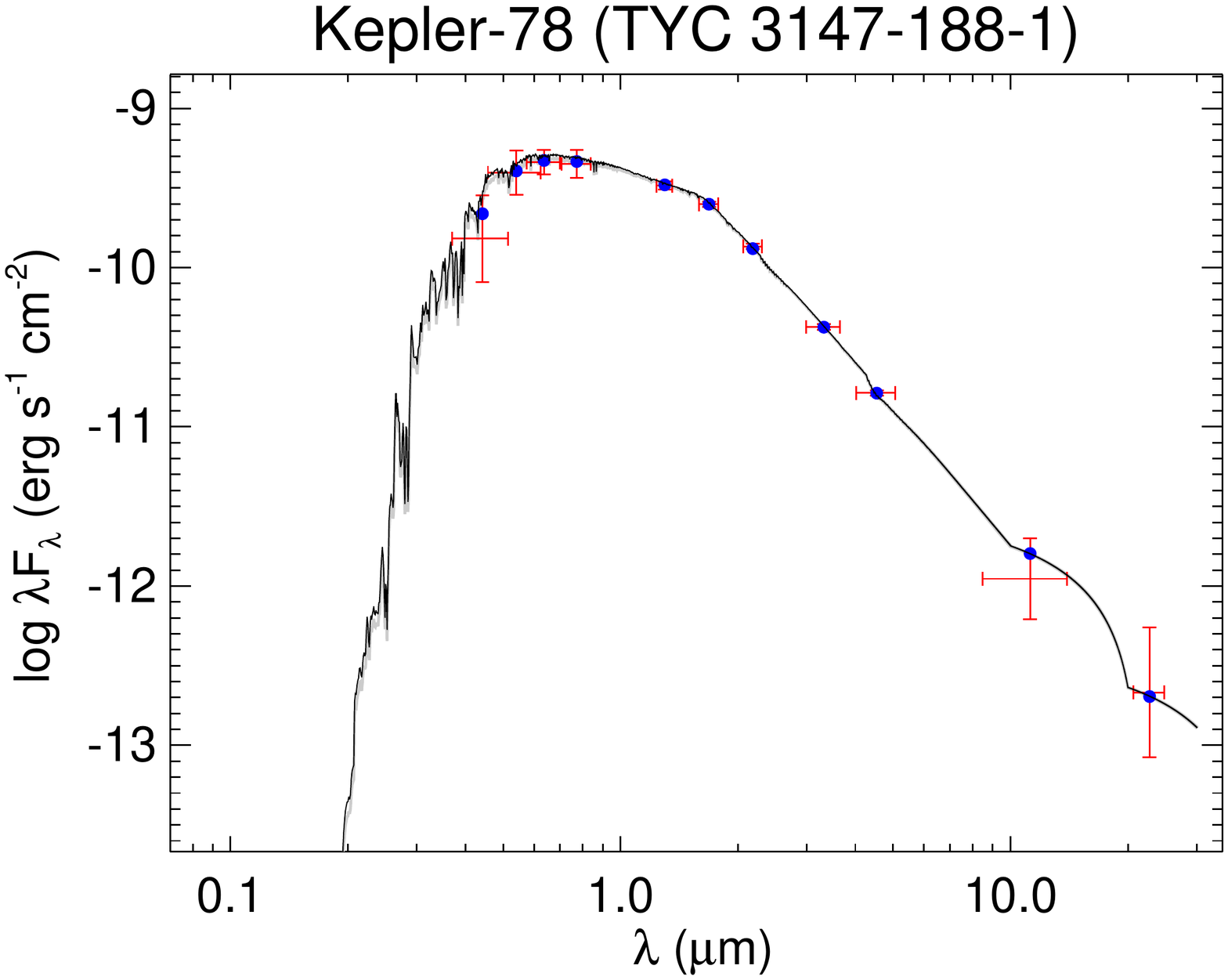}
  \includegraphics[trim=60 60 60 60,clip,width=0.49\linewidth]{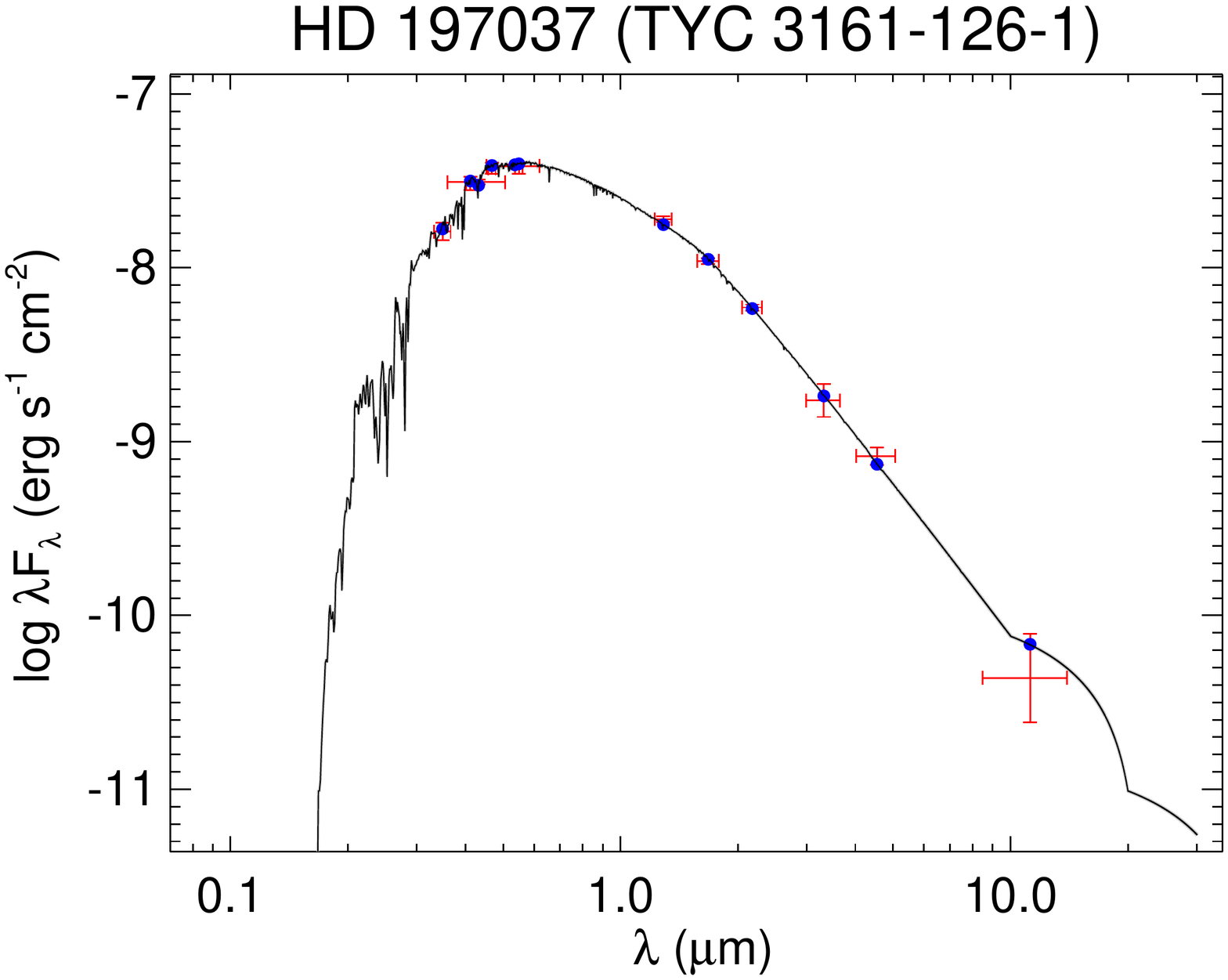}
  \caption{All labels, lines, symbols, and colors as in Figure \ref{fig:seds}.}
  \label{fig:seds_32}
\end{figure}

\begin{figure}[H]
  \centering
  \includegraphics[trim=60 60 60 60,clip,width=0.49\linewidth]{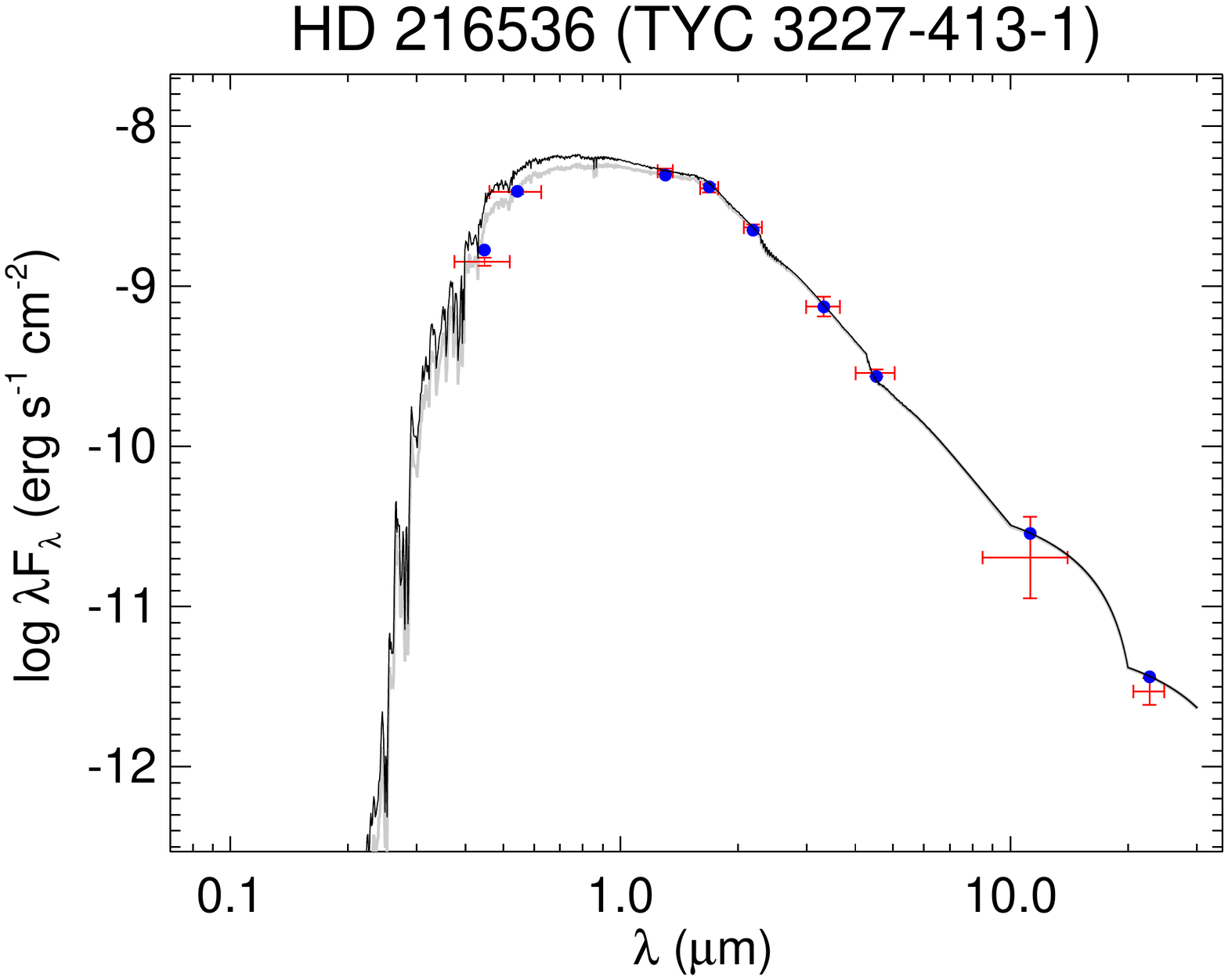}
  \includegraphics[trim=60 60 60 60,clip,width=0.49\linewidth]{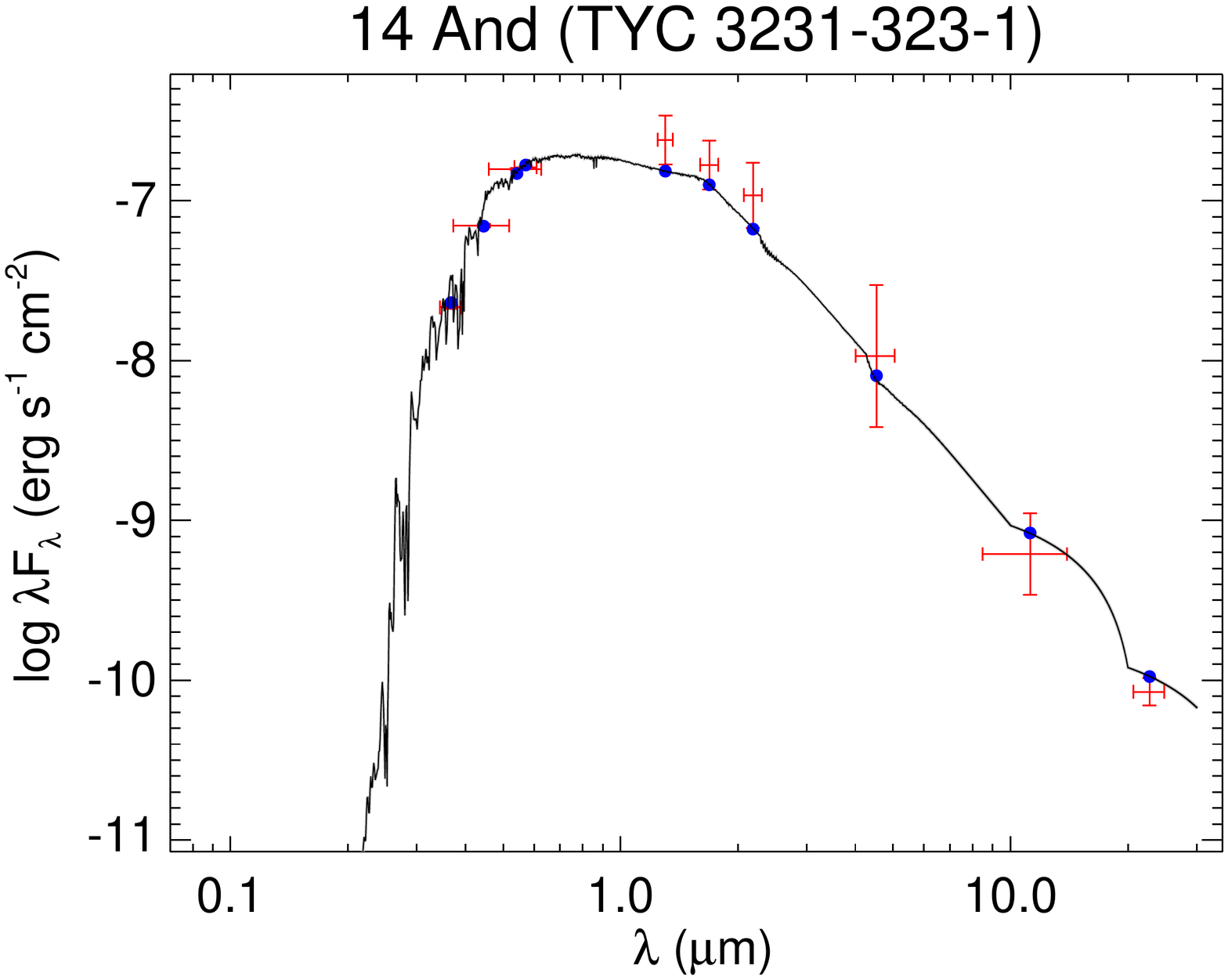}
  \includegraphics[trim=60 60 60 60,clip,width=0.49\linewidth]{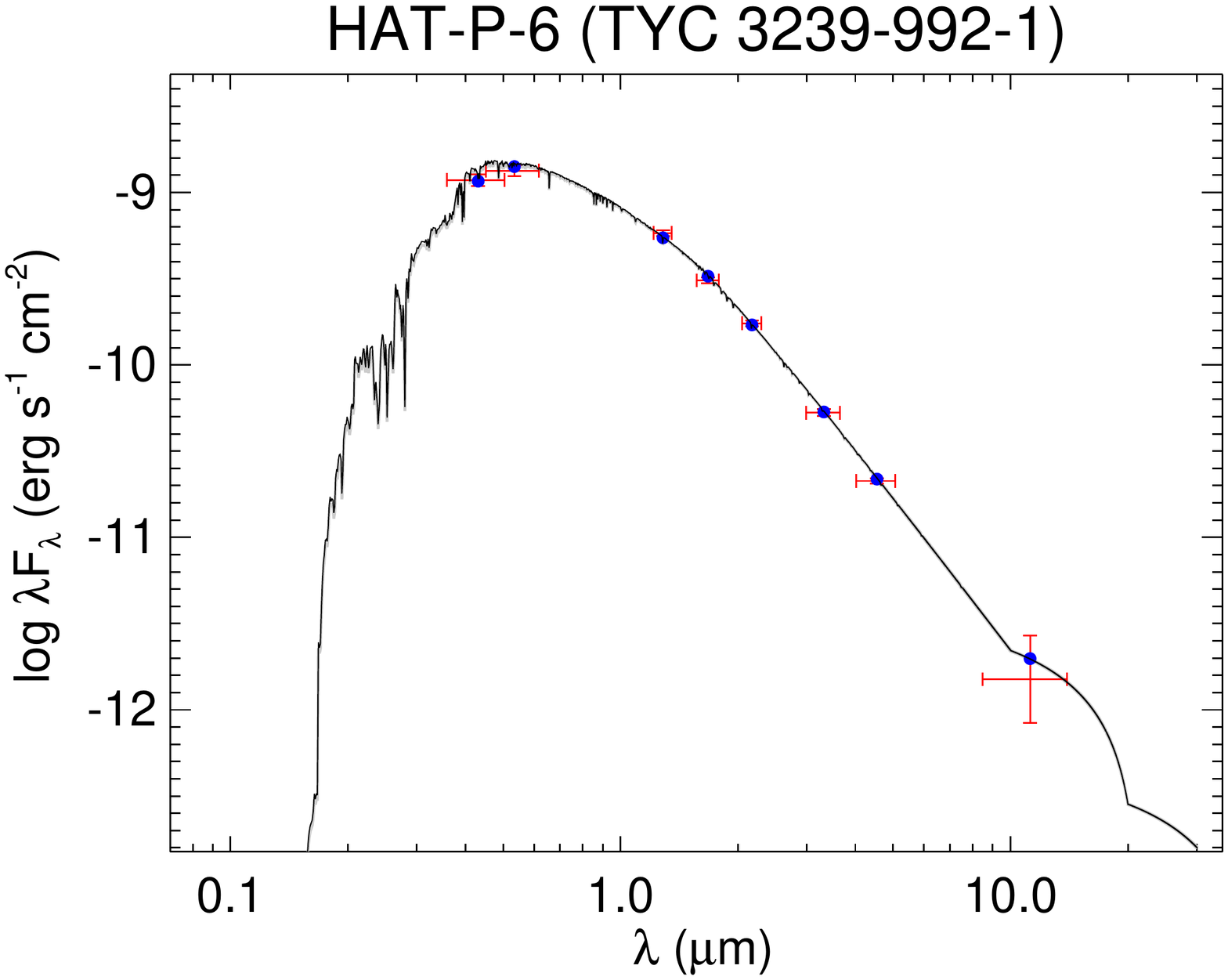}
  \includegraphics[trim=60 60 60 60,clip,width=0.49\linewidth]{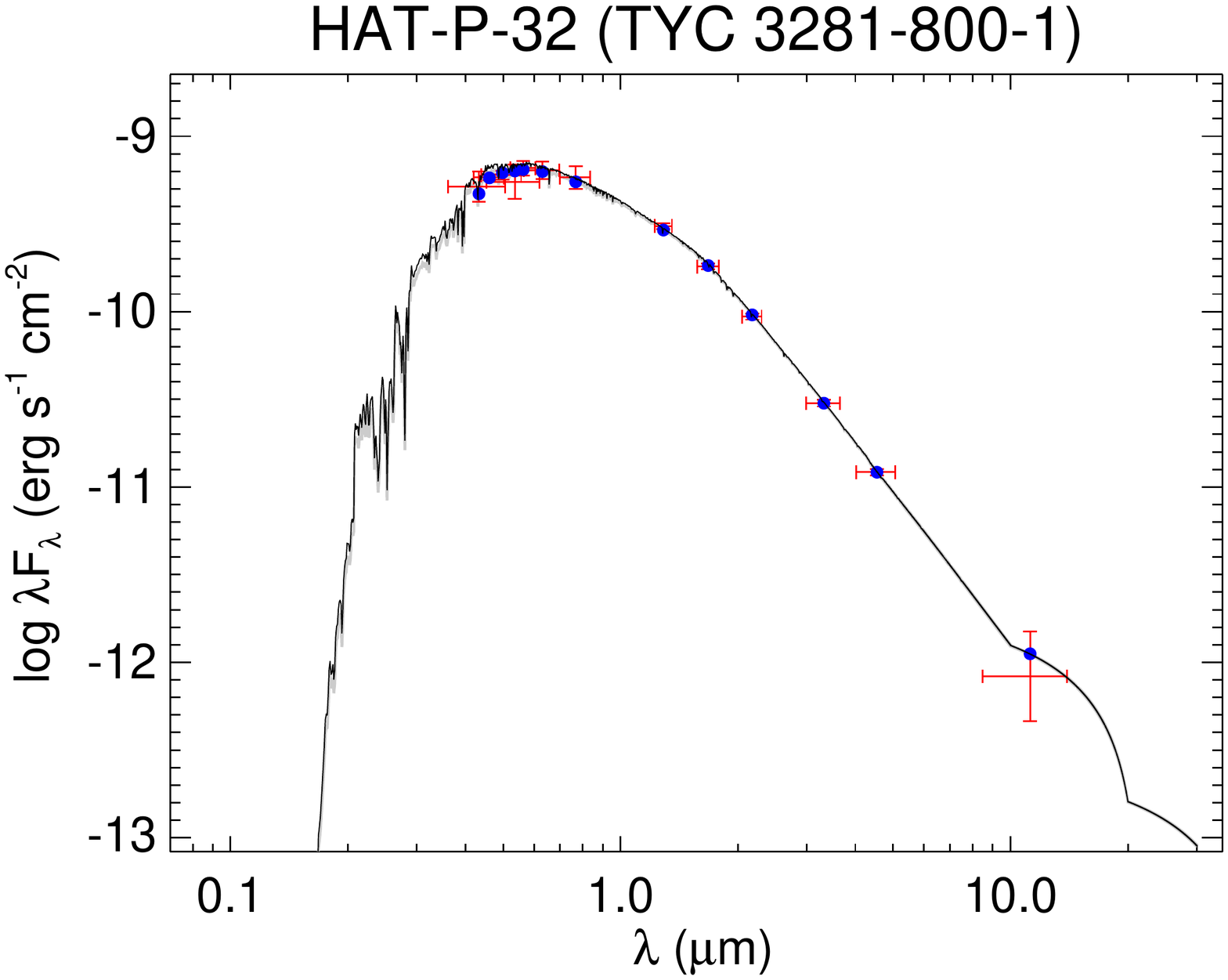}
  \includegraphics[trim=60 60 60 60,clip,width=0.49\linewidth]{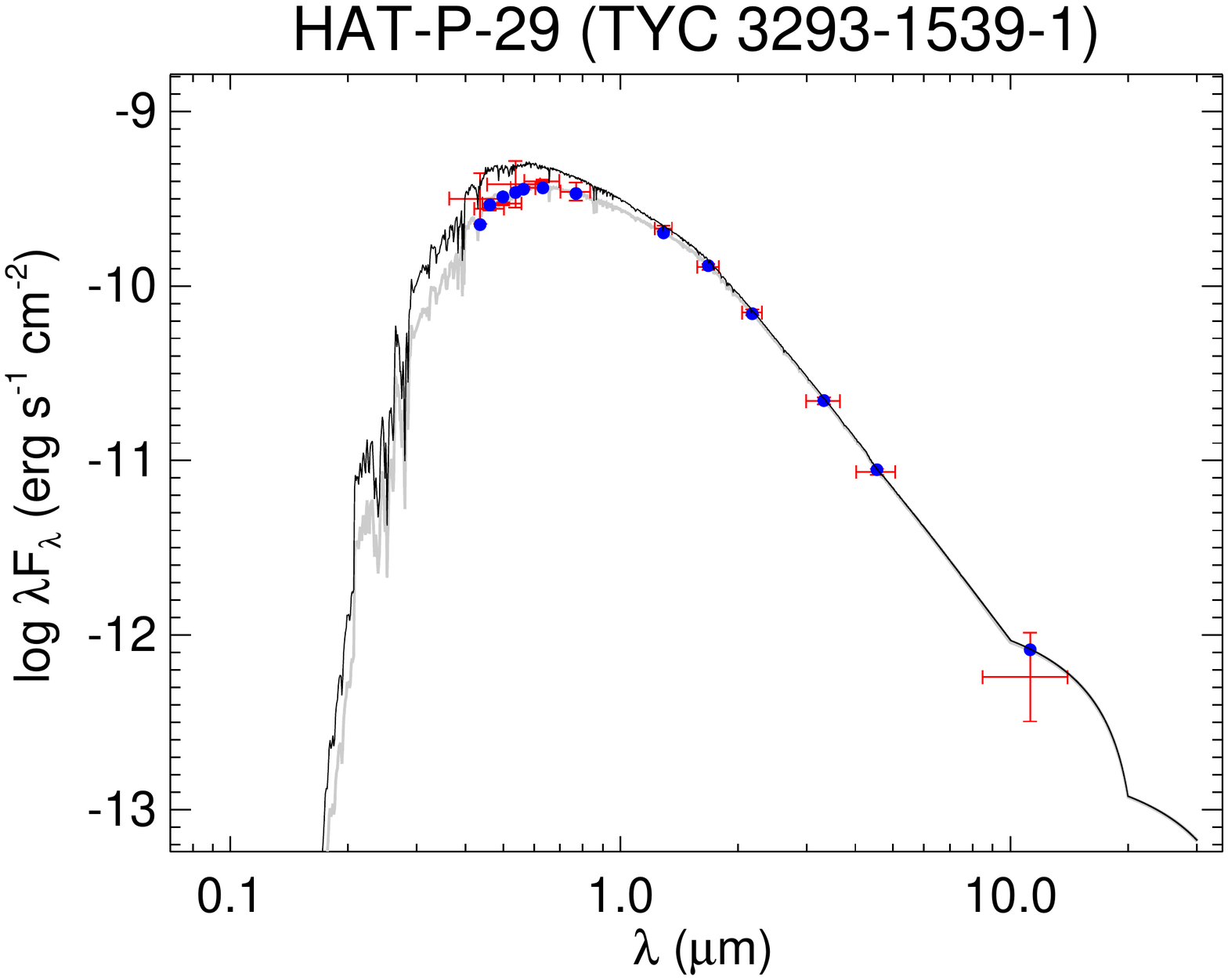}
  \includegraphics[trim=60 60 60 60,clip,width=0.49\linewidth]{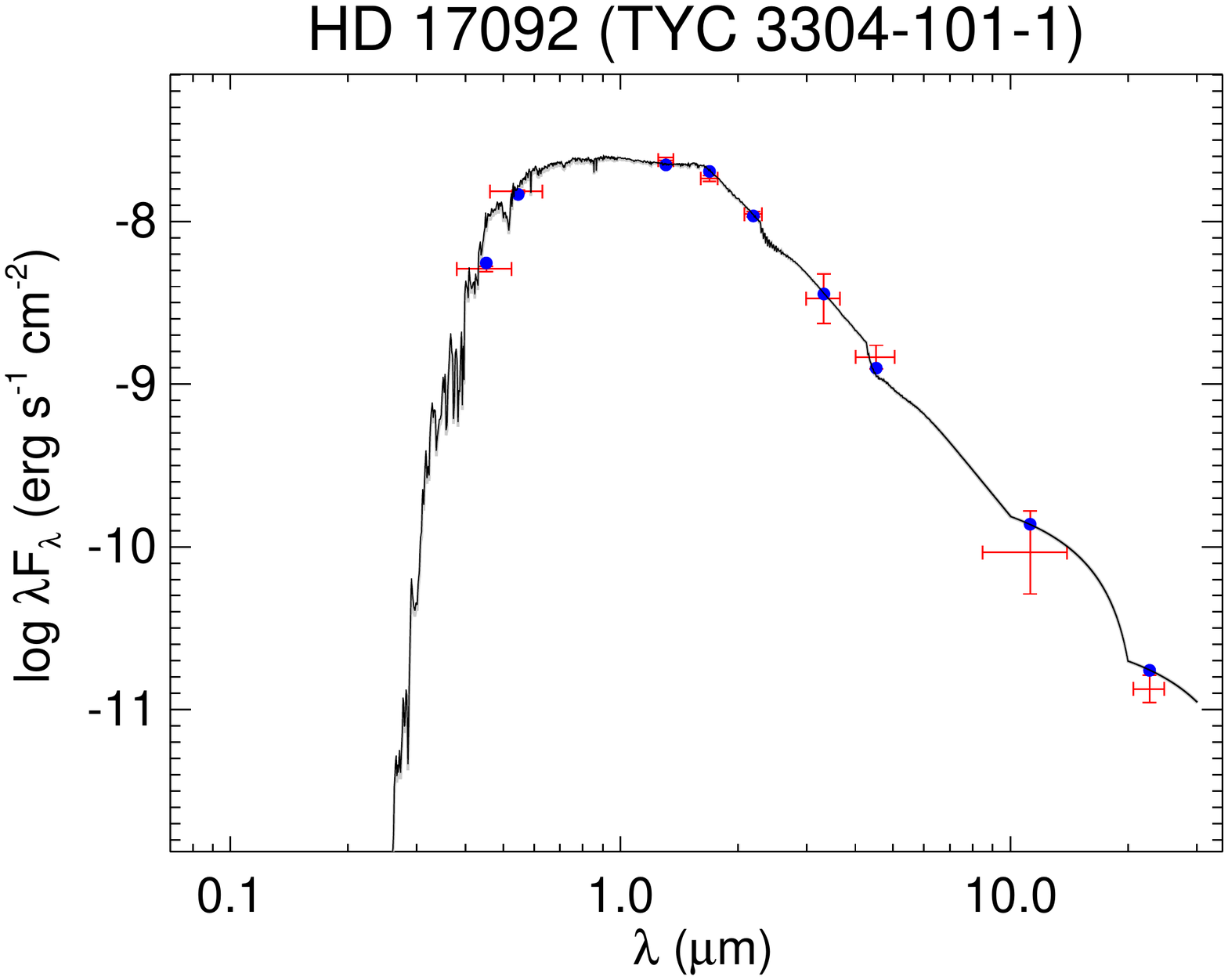}
  \caption{All labels, lines, symbols, and colors as in Figure \ref{fig:seds}.}
  \label{fig:seds_33}
\end{figure}

\begin{figure}[H]
  \centering
  \includegraphics[trim=60 60 60 60,clip,width=0.49\linewidth]{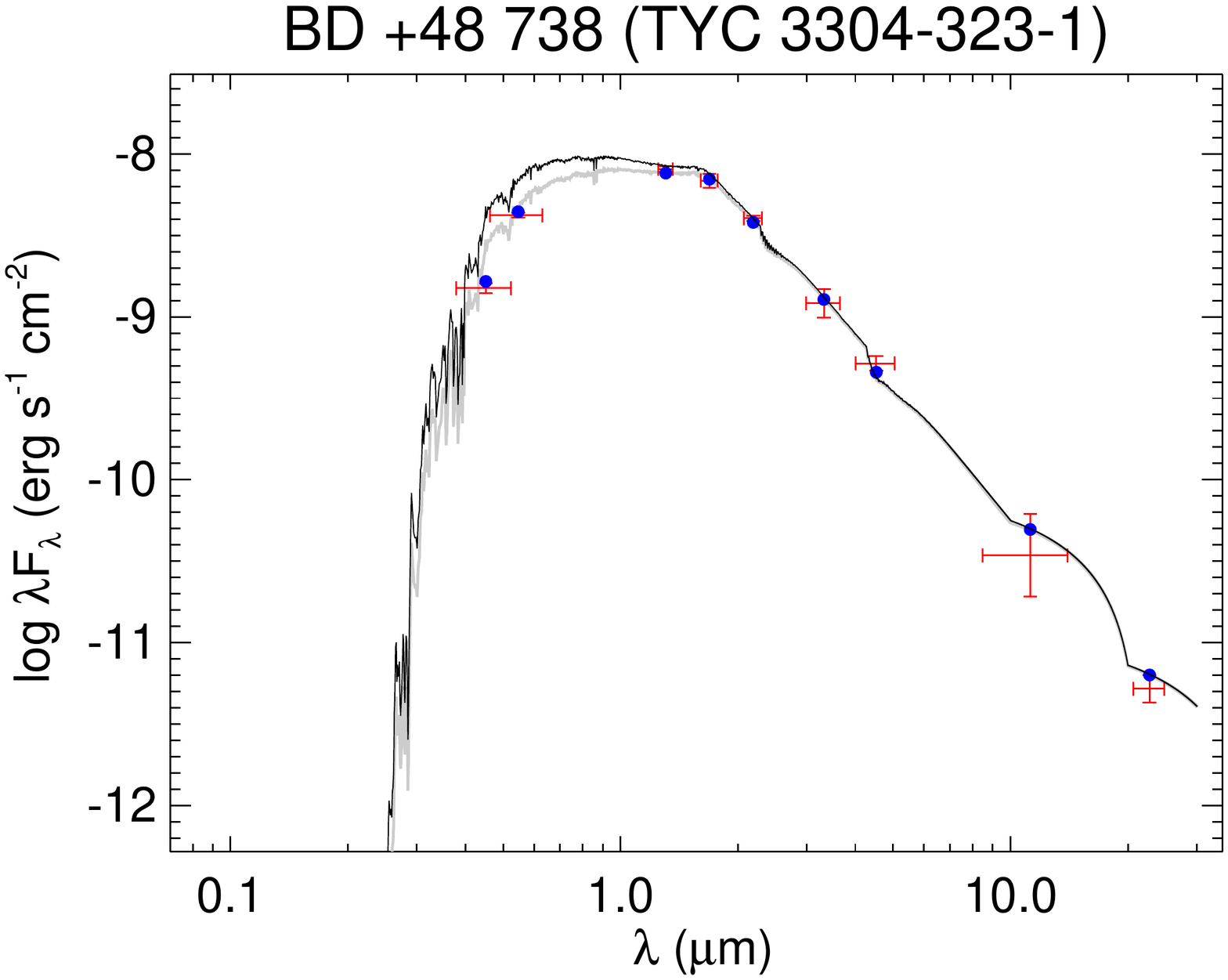}
  \includegraphics[trim=60 60 60 60,clip,width=0.49\linewidth]{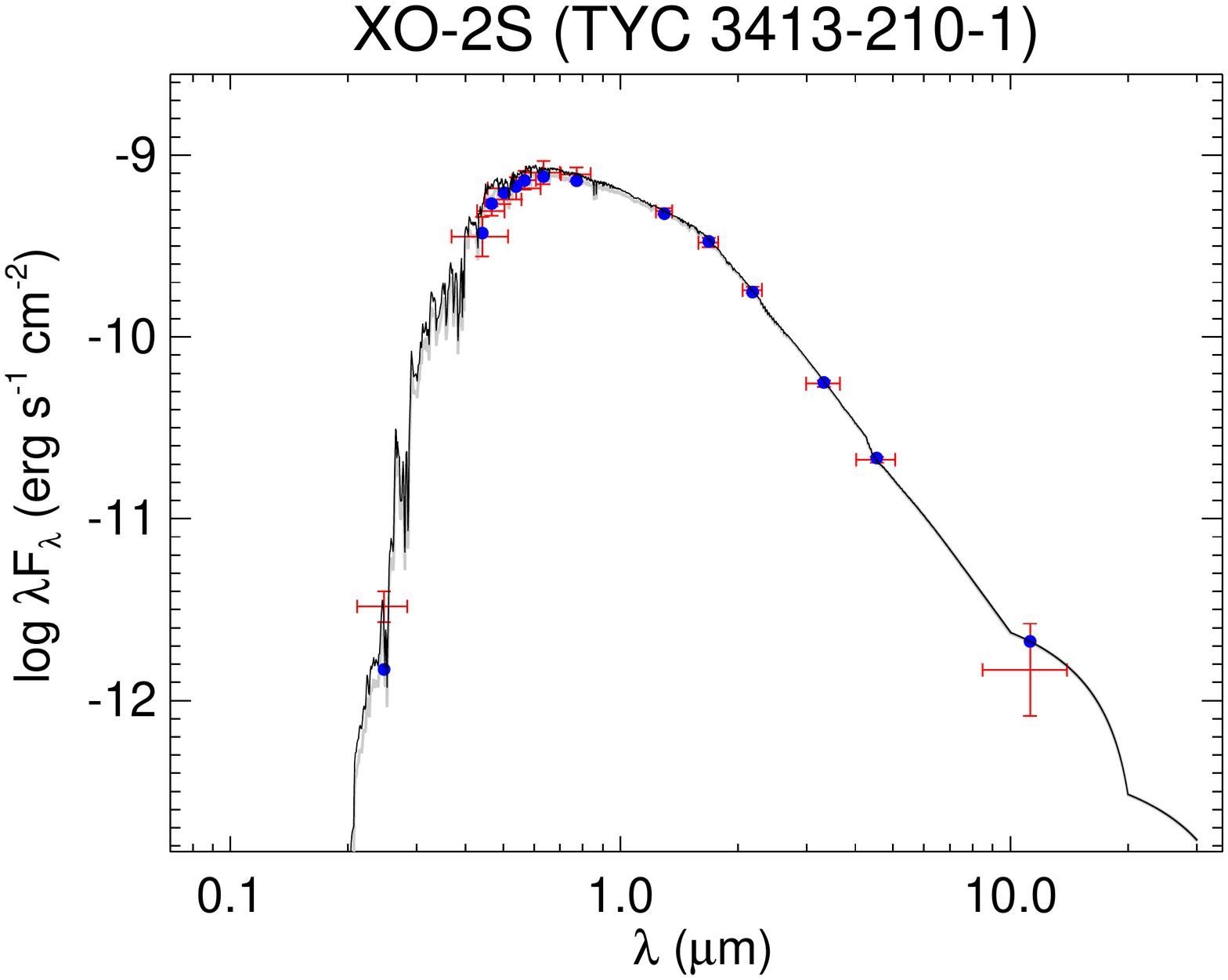}
  \includegraphics[trim=60 60 60 60,clip,width=0.49\linewidth]{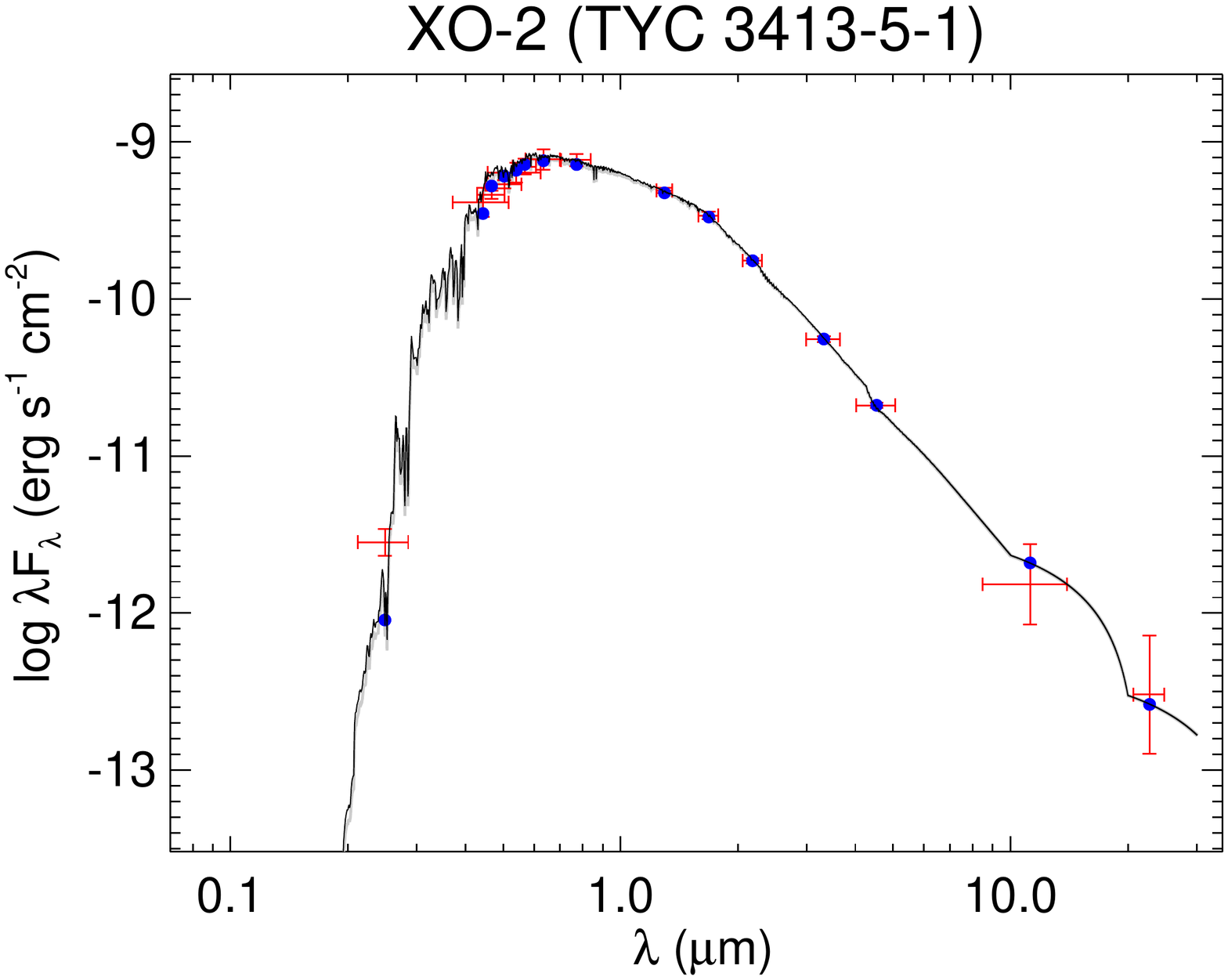}
  \includegraphics[trim=60 60 60 60,clip,width=0.49\linewidth]{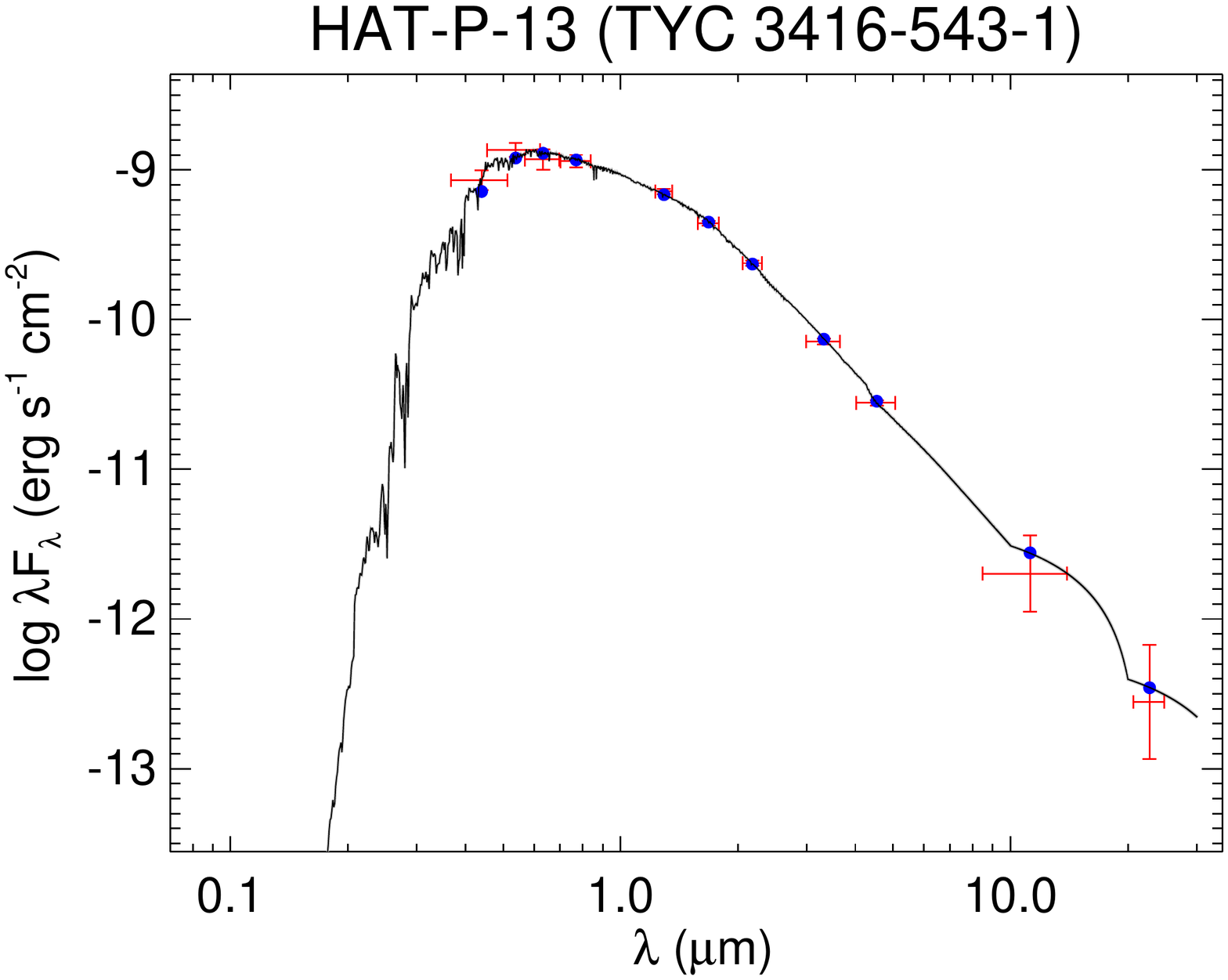}
  \includegraphics[trim=60 60 60 60,clip,width=0.49\linewidth]{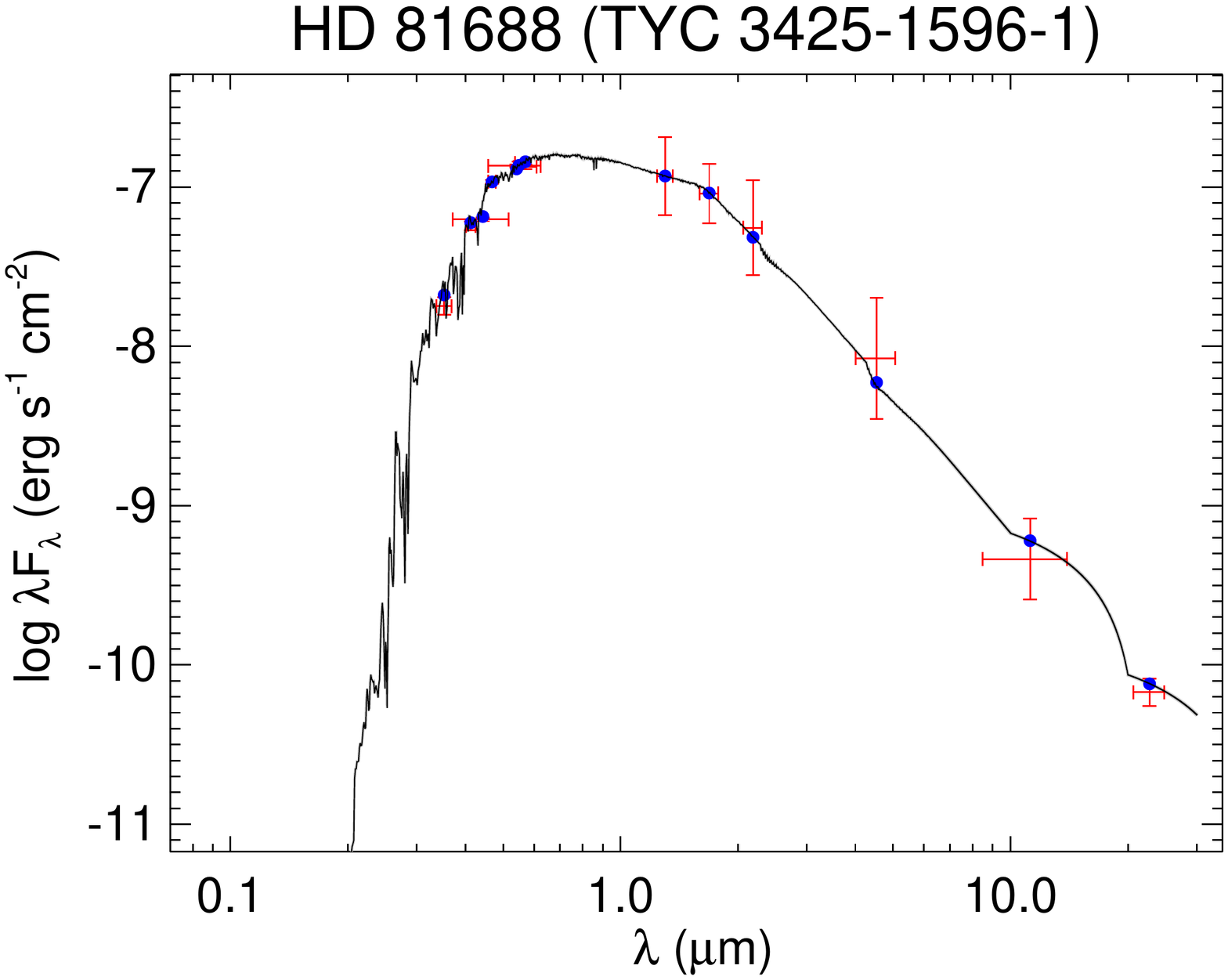}
  \includegraphics[trim=60 60 60 60,clip,width=0.49\linewidth]{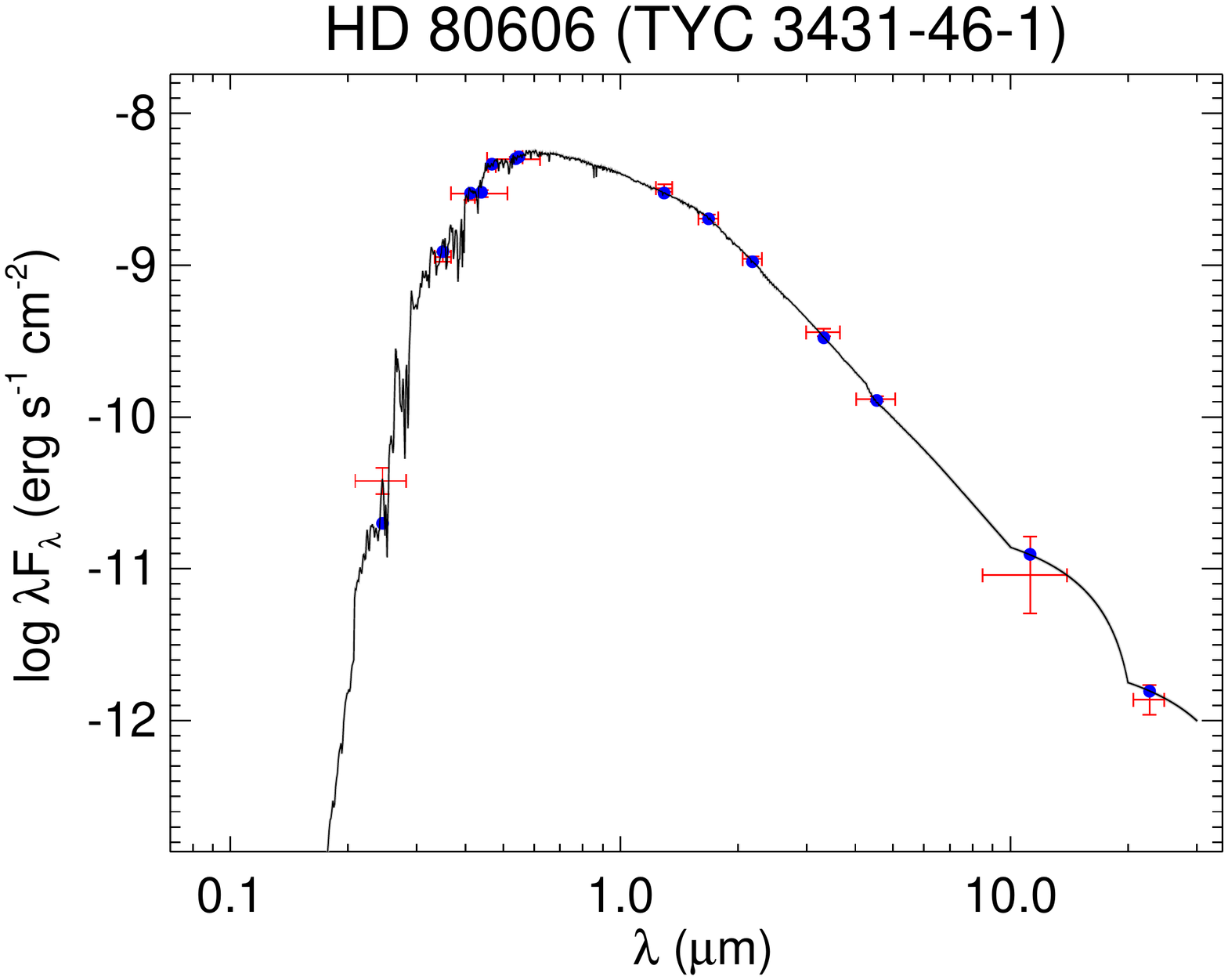}
  \caption{All labels, lines, symbols, and colors as in Figure \ref{fig:seds}.}
  \label{fig:seds_34}
\end{figure}

\begin{figure}[H]
  \centering
  \includegraphics[trim=60 60 60 60,clip,width=0.49\linewidth]{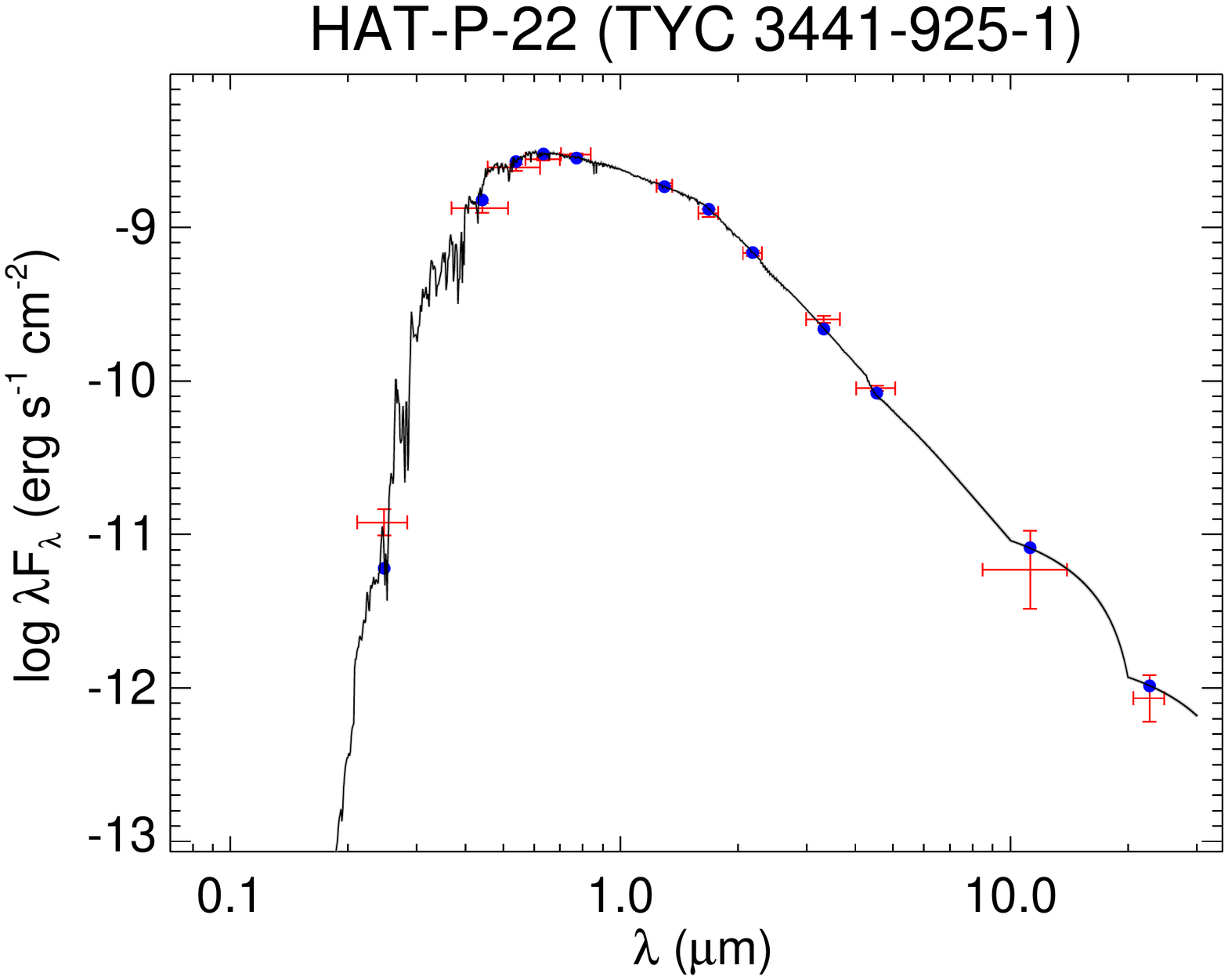}
  \includegraphics[trim=60 60 60 60,clip,width=0.49\linewidth]{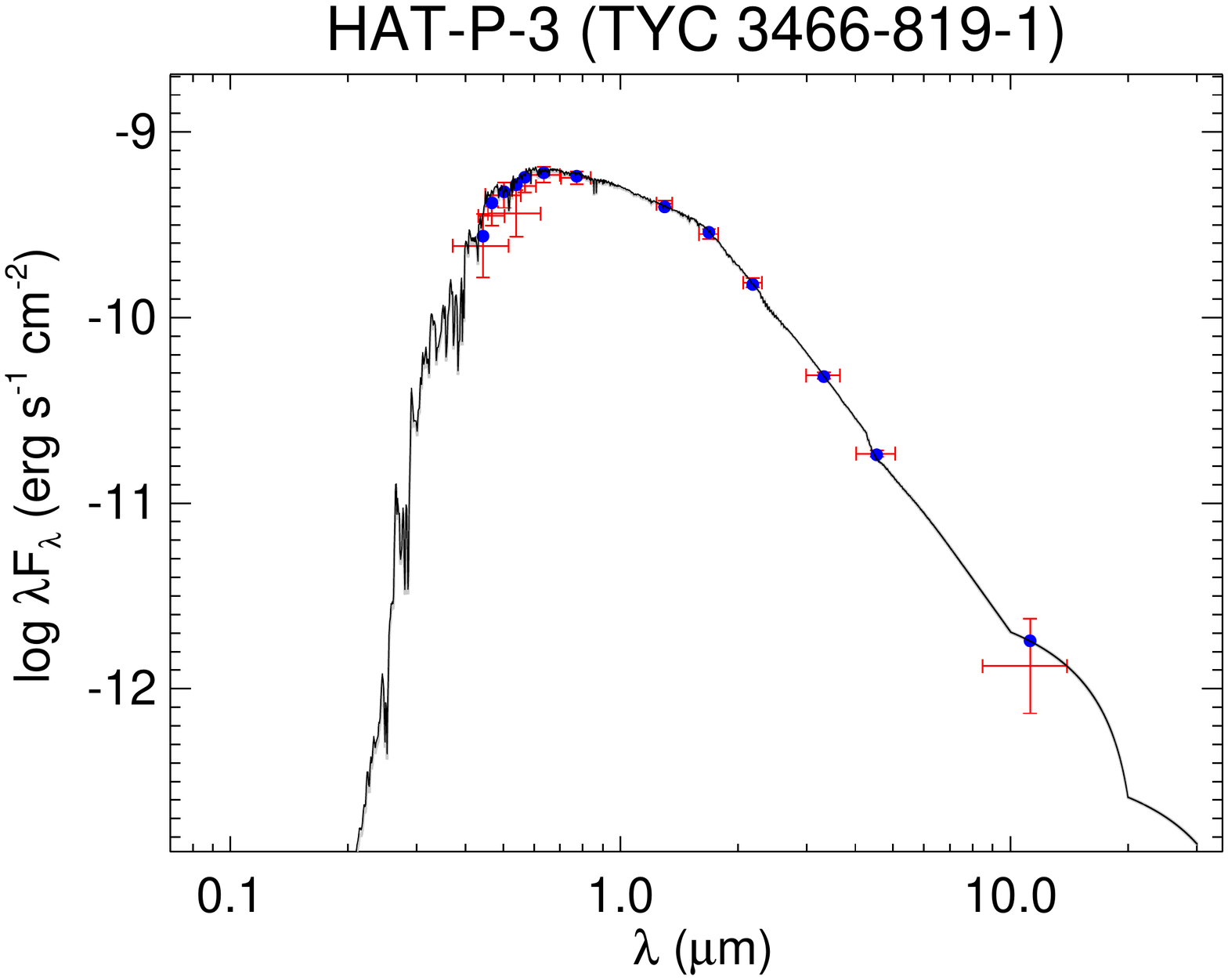}
  \includegraphics[trim=60 60 60 60,clip,width=0.49\linewidth]{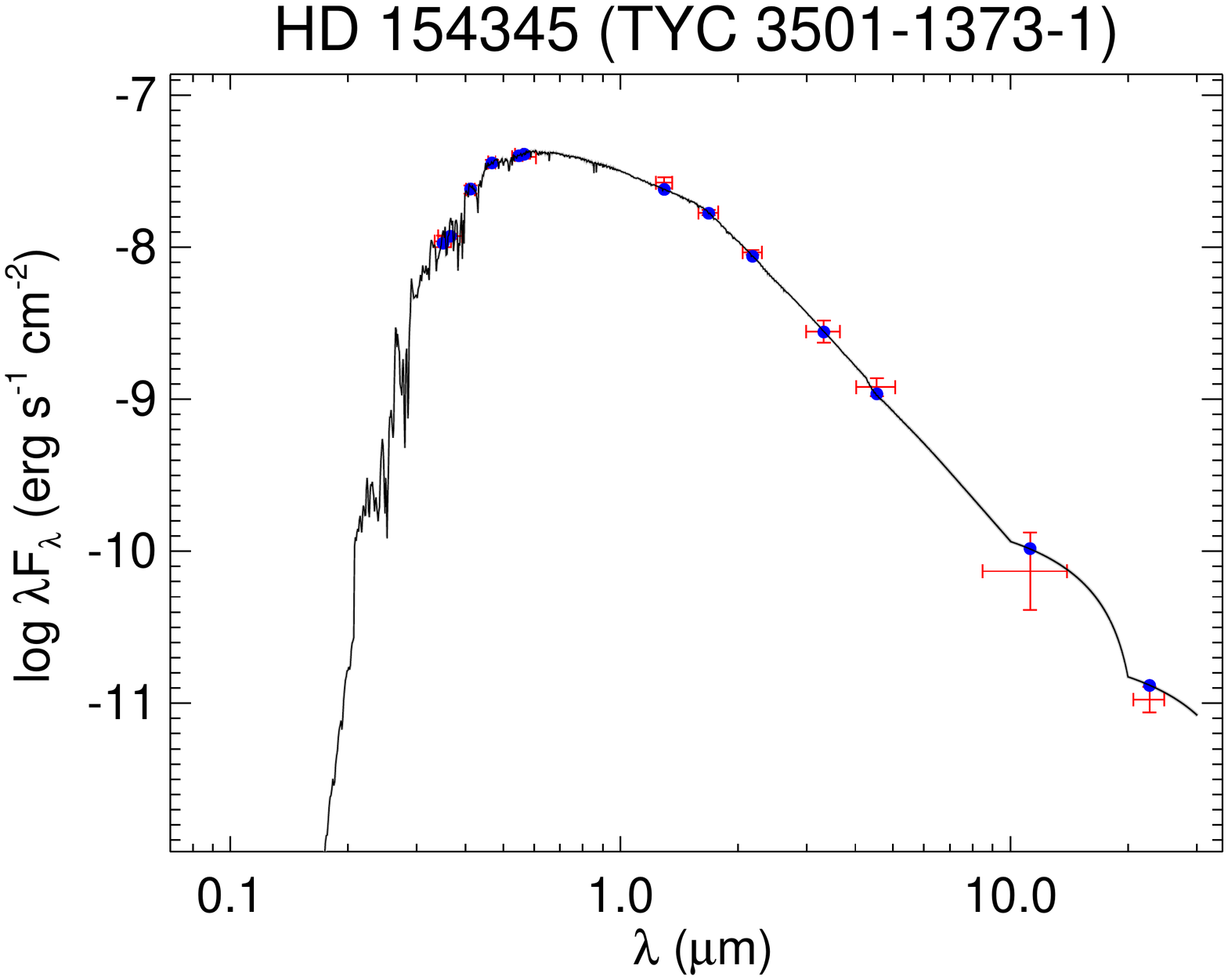}
  \includegraphics[trim=60 60 60 60,clip,width=0.49\linewidth]{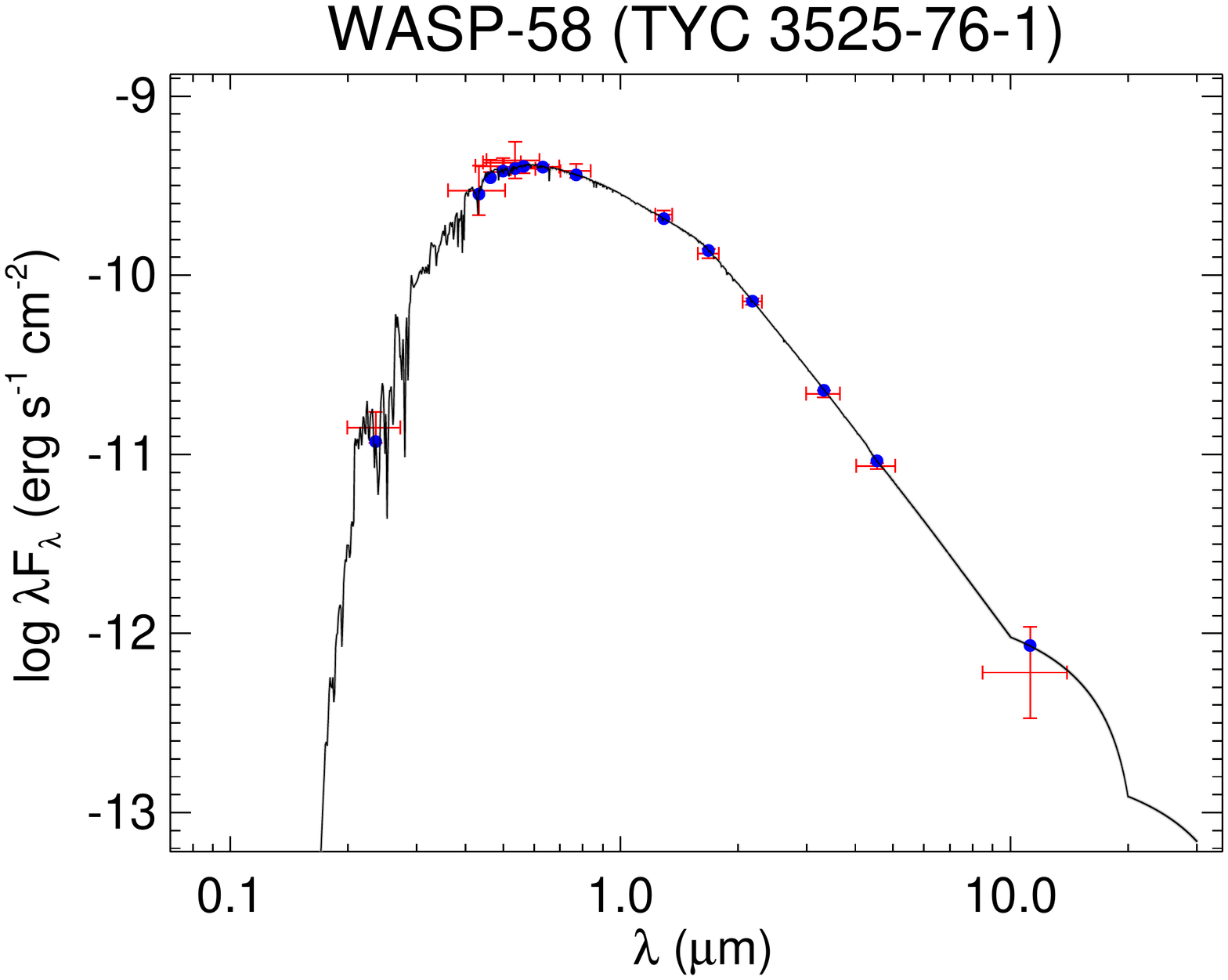}
  \includegraphics[trim=60 60 60 60,clip,width=0.49\linewidth]{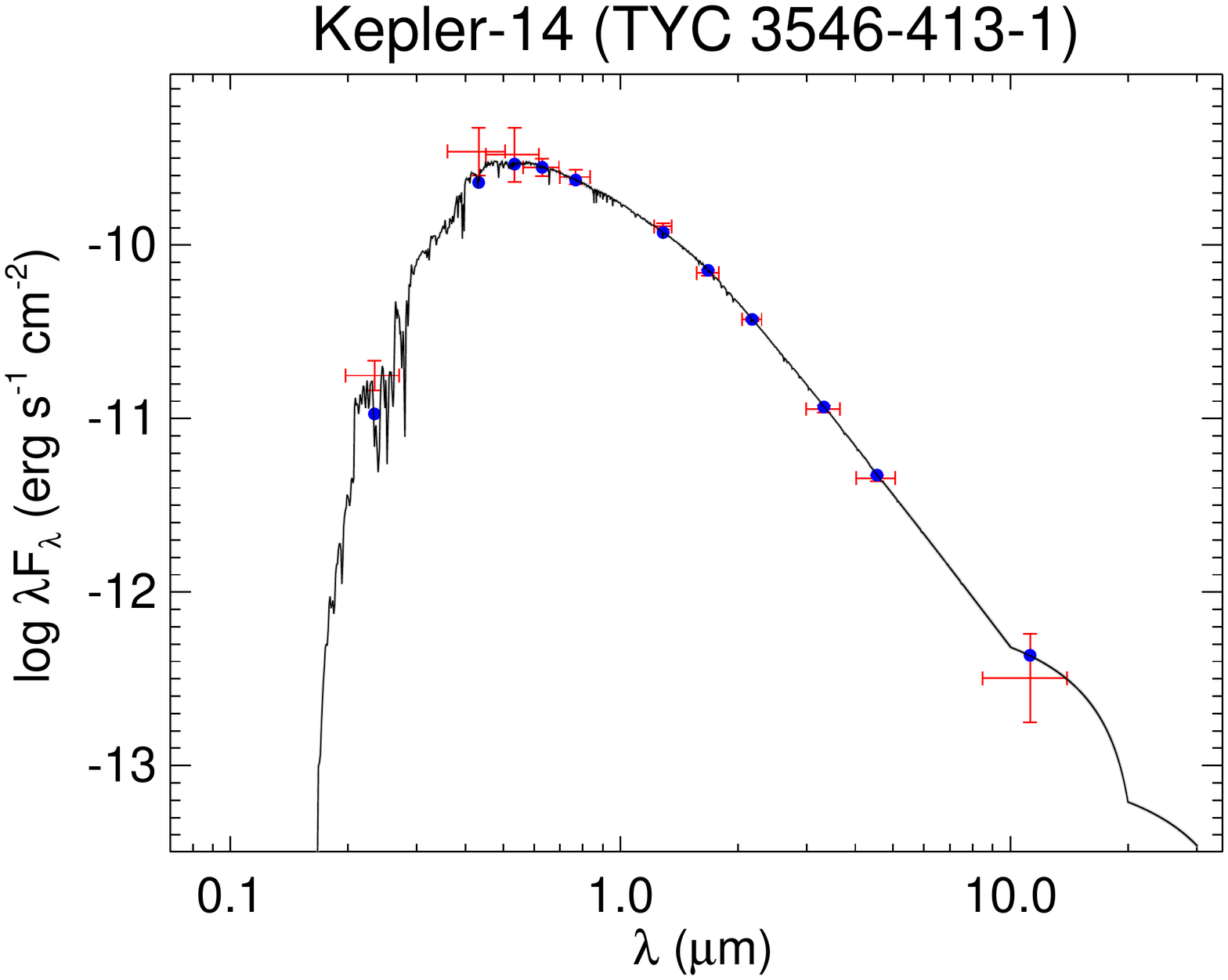}
  \includegraphics[trim=60 60 60 60,clip,width=0.49\linewidth]{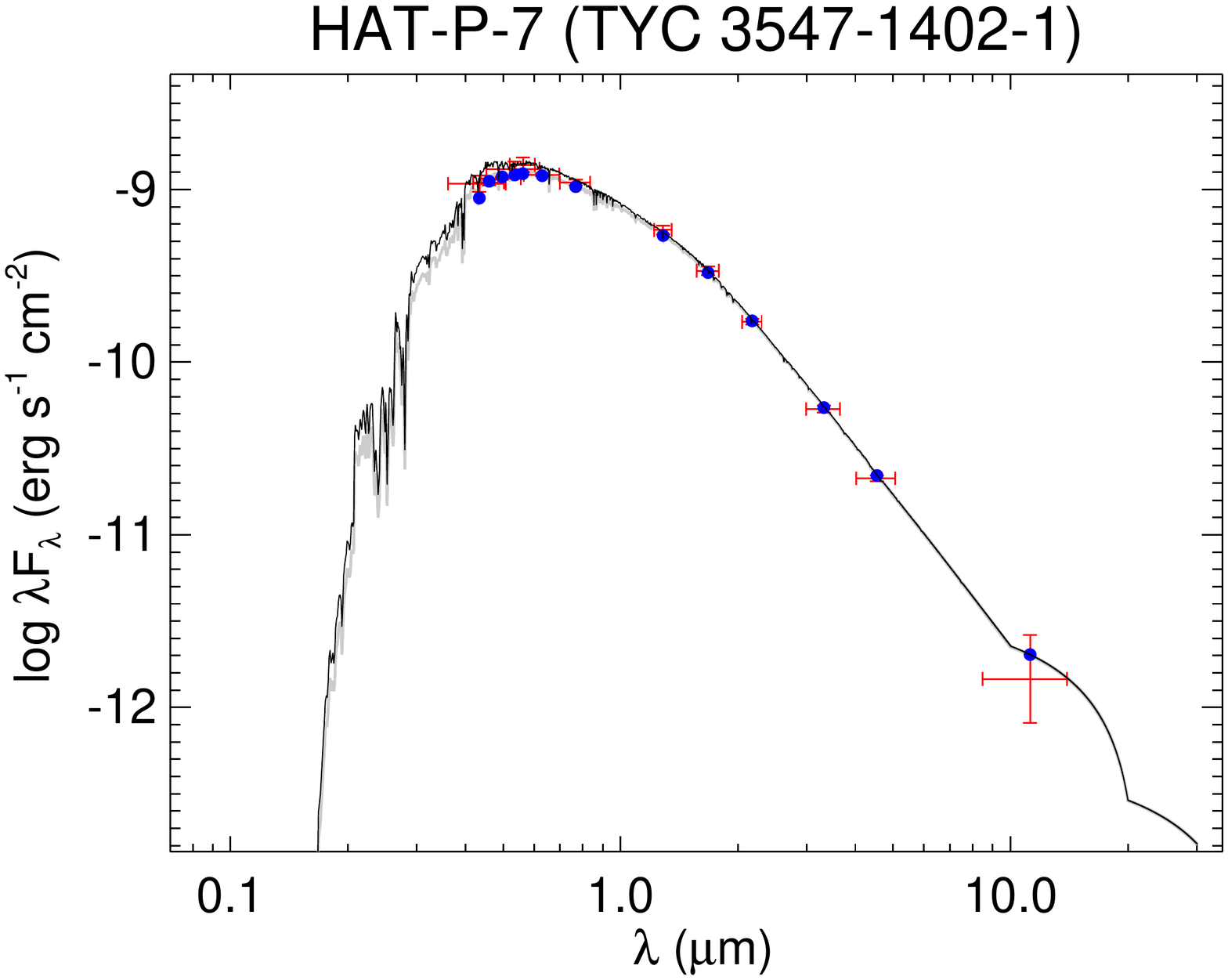}
  \caption{All labels, lines, symbols, and colors as in Figure \ref{fig:seds}.}
  \label{fig:seds_35}
\end{figure}

\begin{figure}[H]
  \centering
  \includegraphics[trim=60 60 60 60,clip,width=0.49\linewidth]{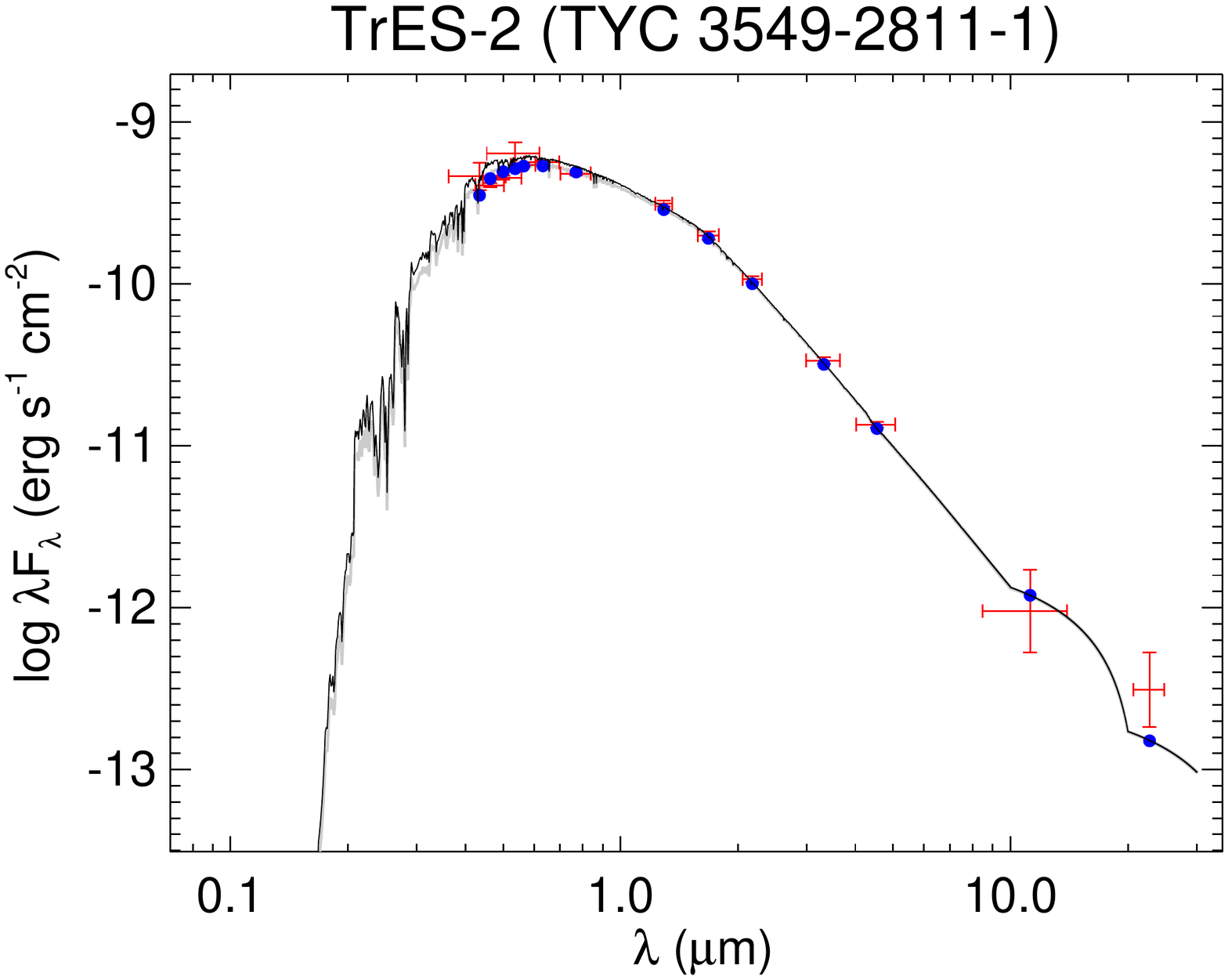}
  \includegraphics[trim=60 60 60 60,clip,width=0.49\linewidth]{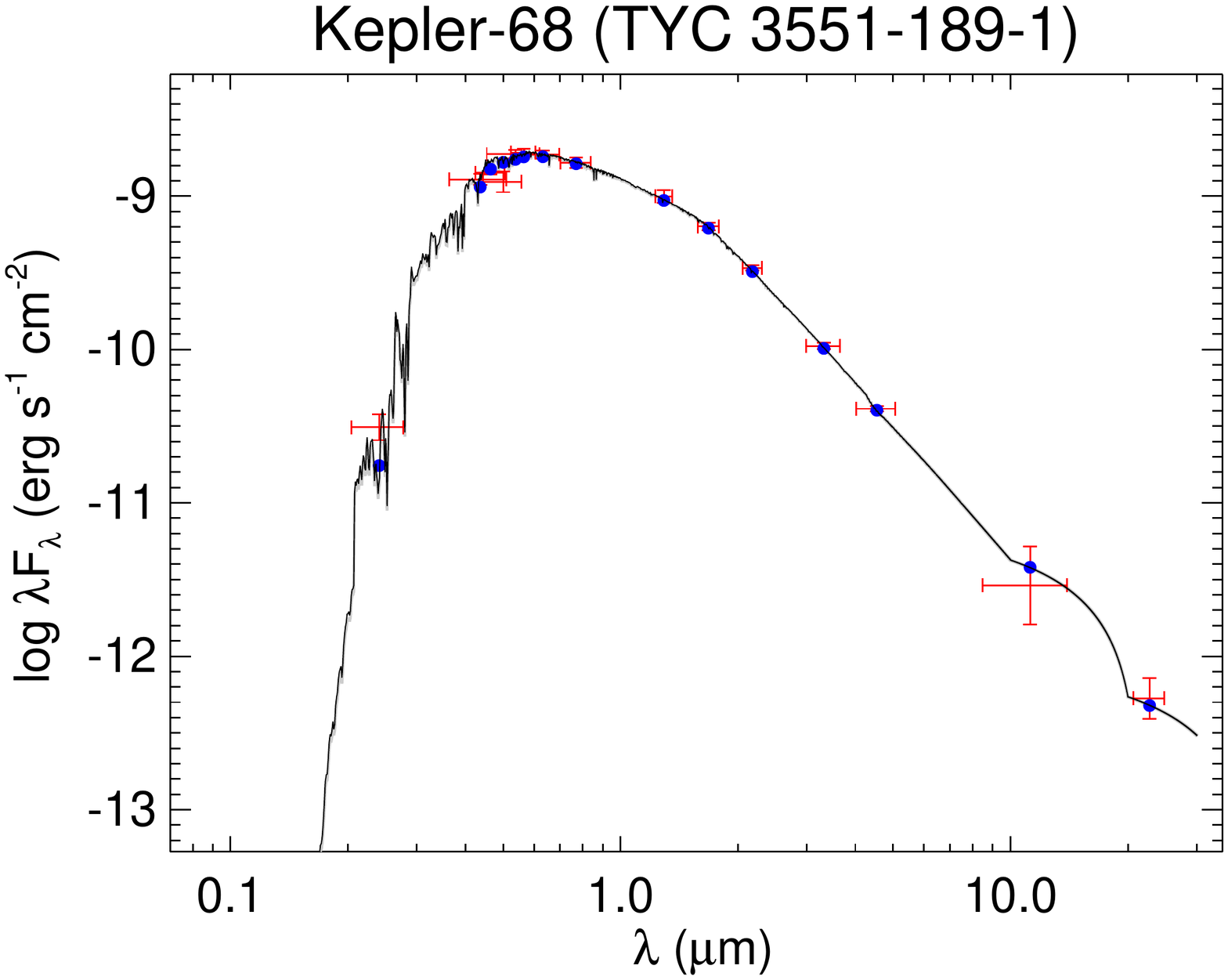}
  \includegraphics[trim=60 60 60 60,clip,width=0.49\linewidth]{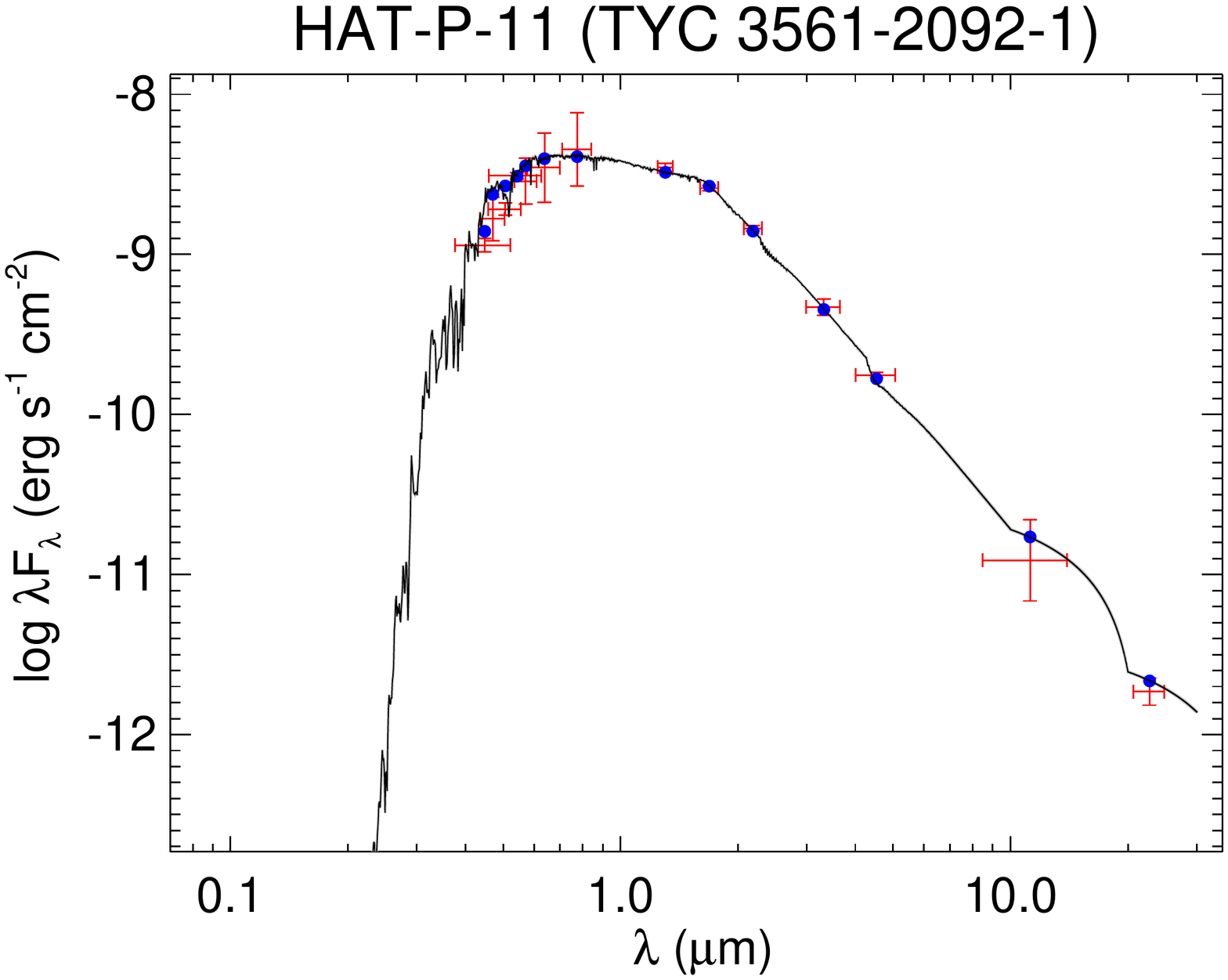}
  \includegraphics[trim=60 60 60 60,clip,width=0.49\linewidth]{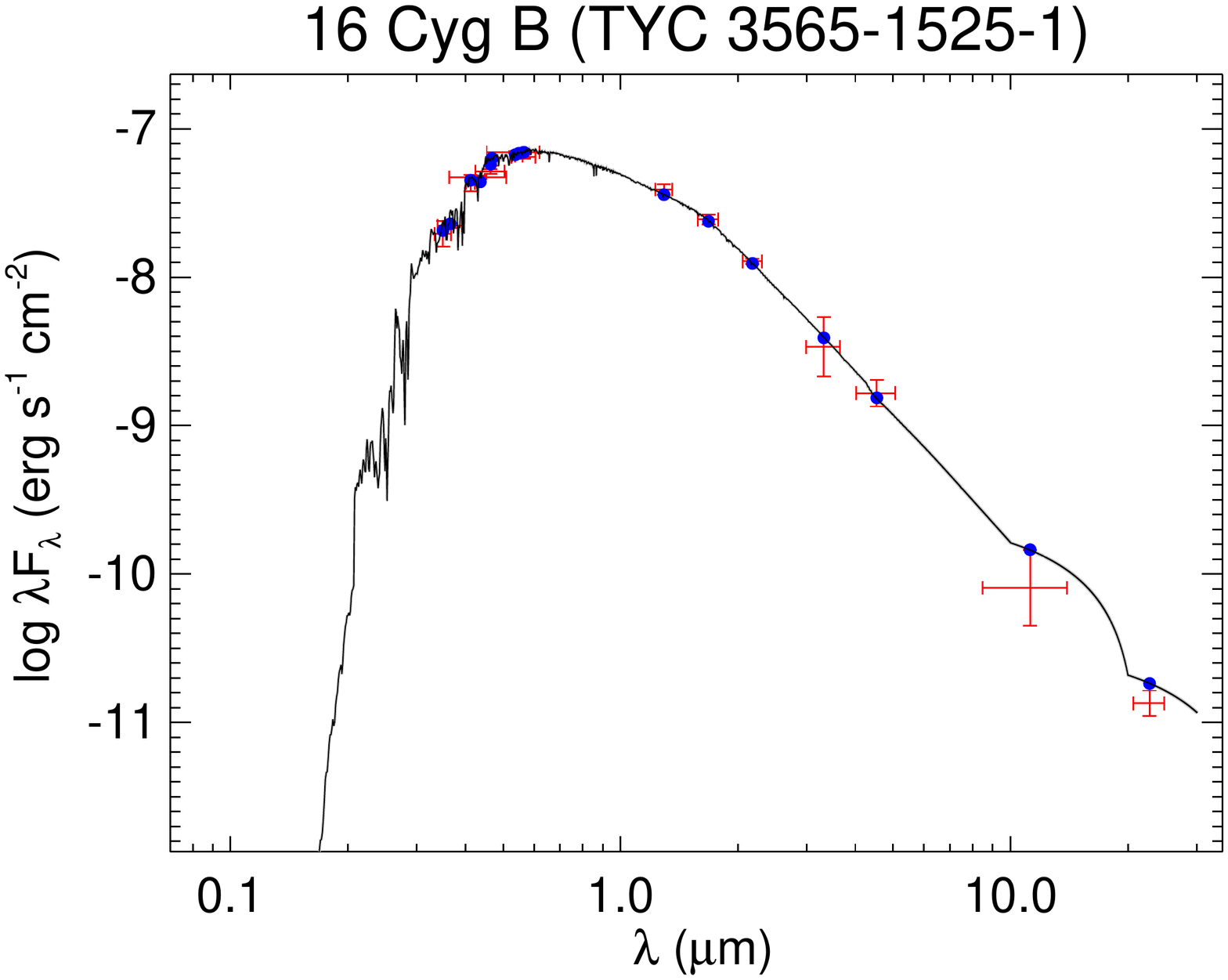}
  \includegraphics[trim=60 60 60 60,clip,width=0.49\linewidth]{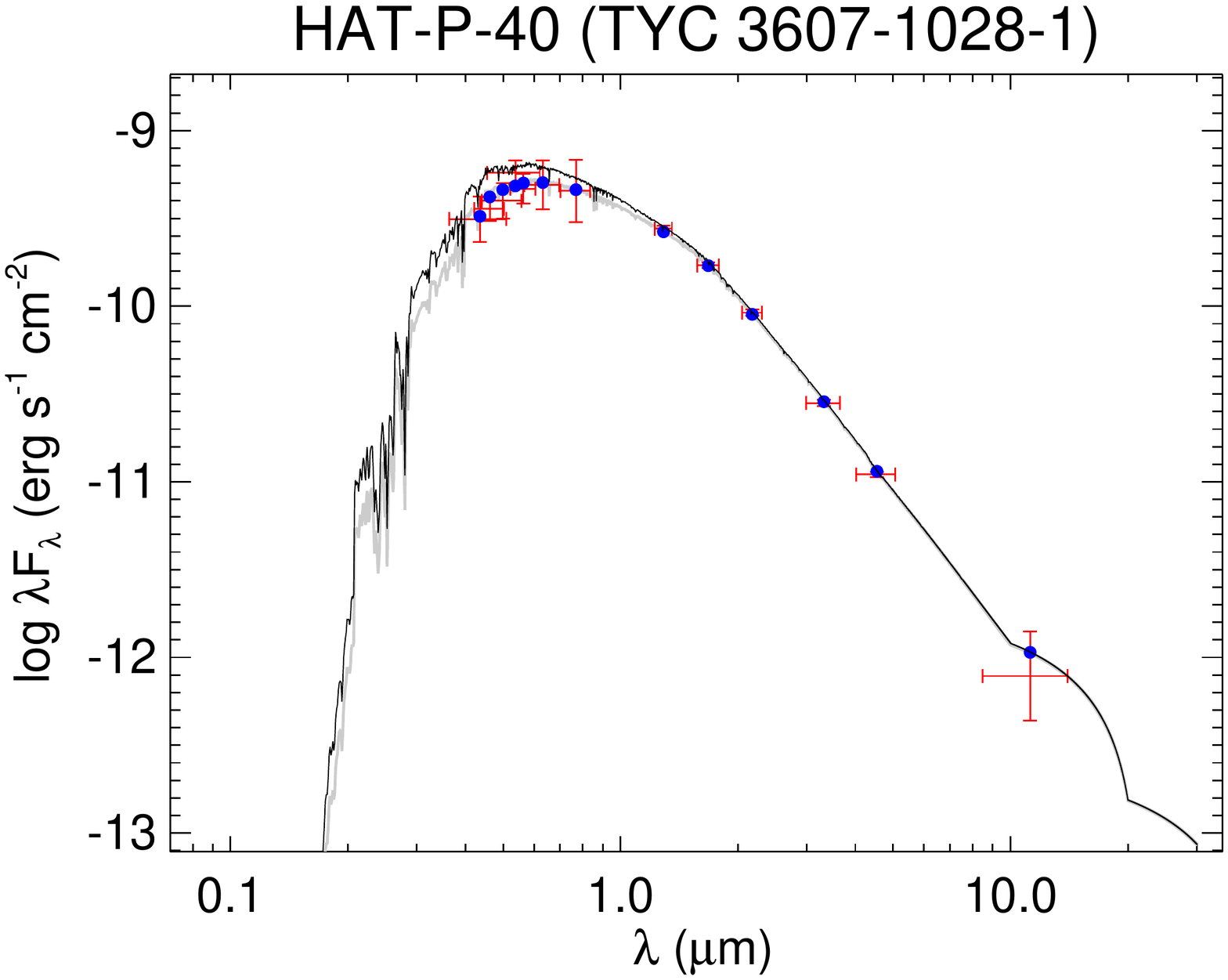}
  \includegraphics[trim=60 60 60 60,clip,width=0.49\linewidth]{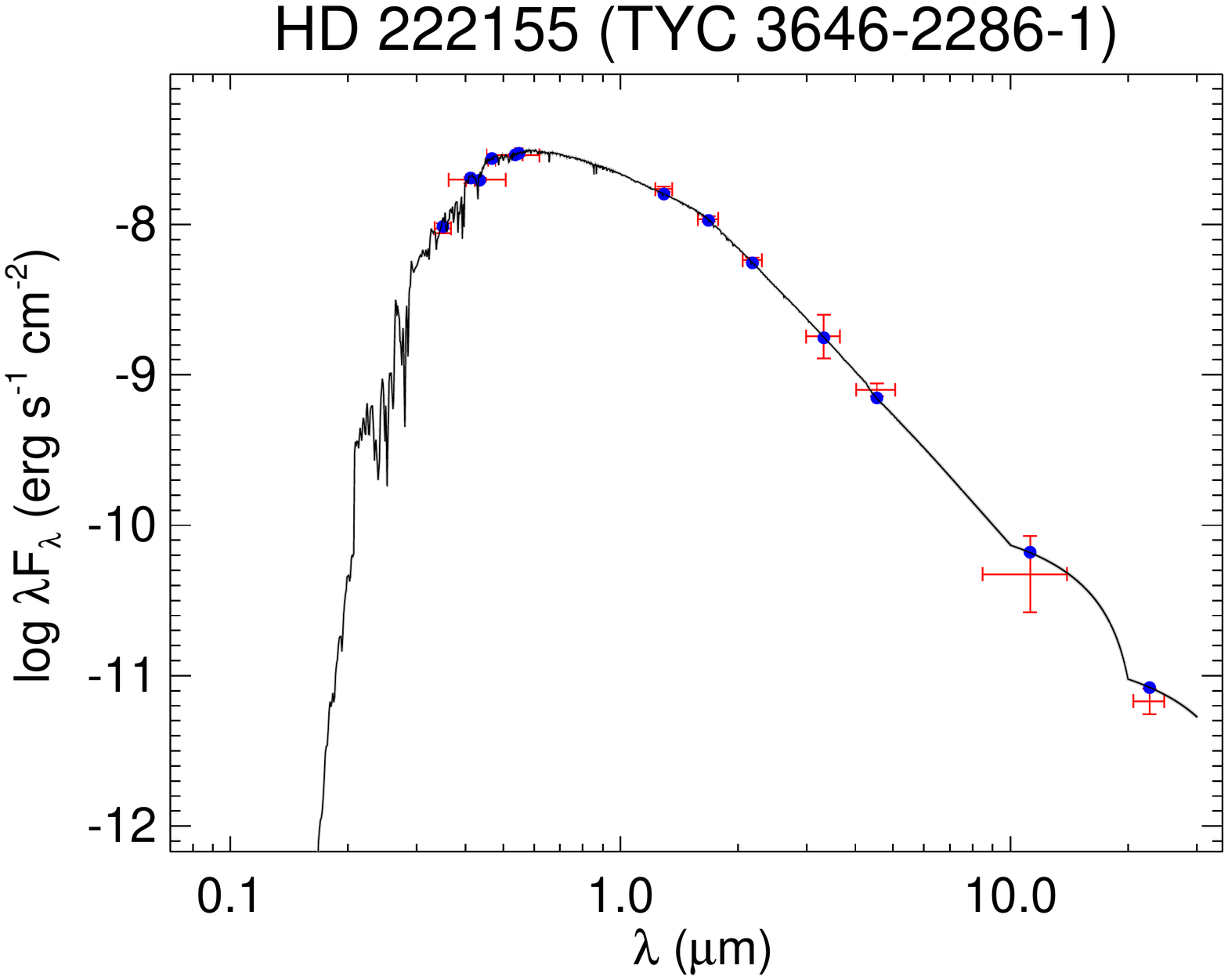}
  \caption{All labels, lines, symbols, and colors as in Figure \ref{fig:seds}.}
  \label{fig:seds_36}
\end{figure}

\begin{figure}[H]
  \centering
  \includegraphics[trim=60 60 60 60,clip,width=0.49\linewidth]{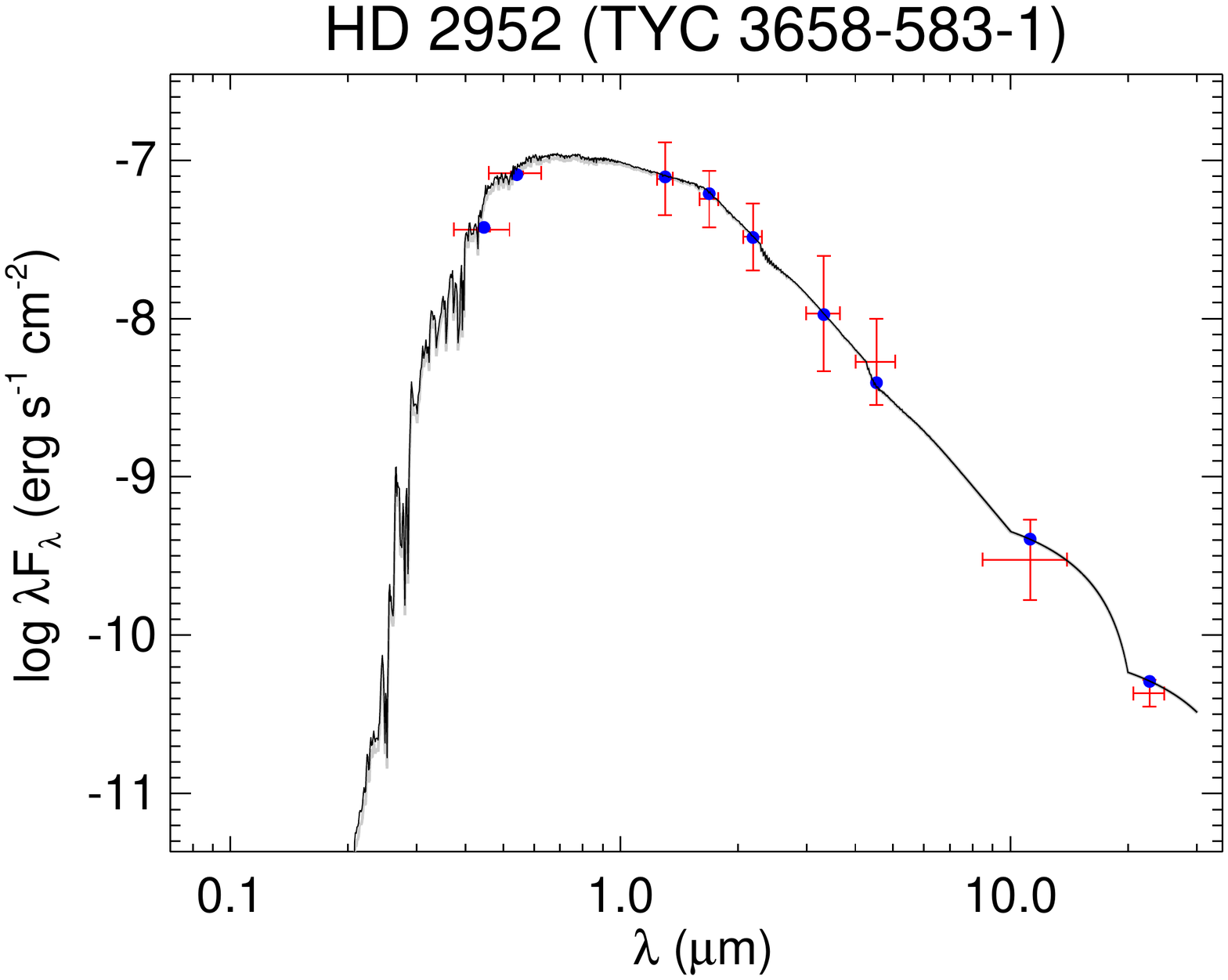}
  \includegraphics[trim=60 60 60 60,clip,width=0.49\linewidth]{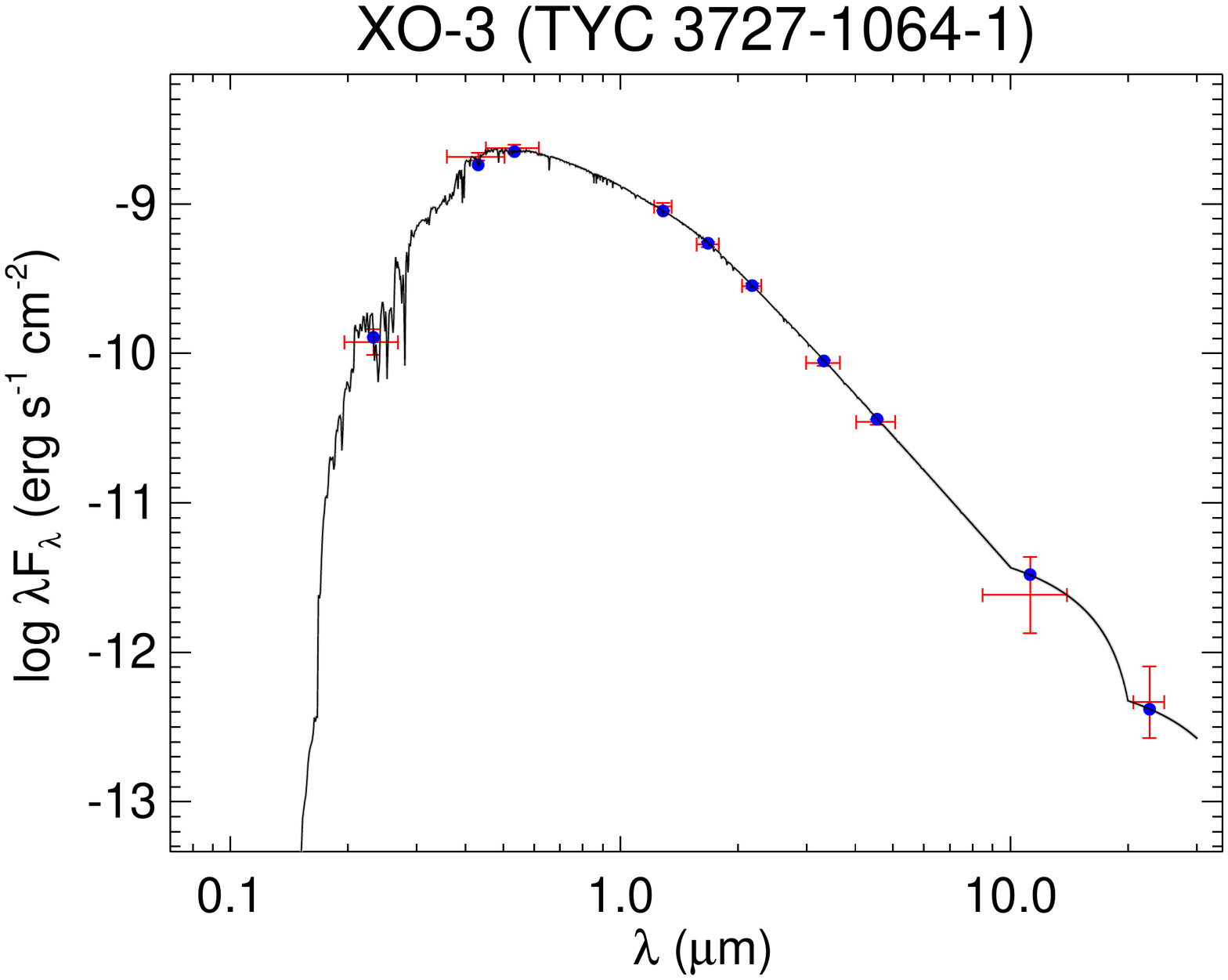}
  \includegraphics[trim=60 60 60 60,clip,width=0.49\linewidth]{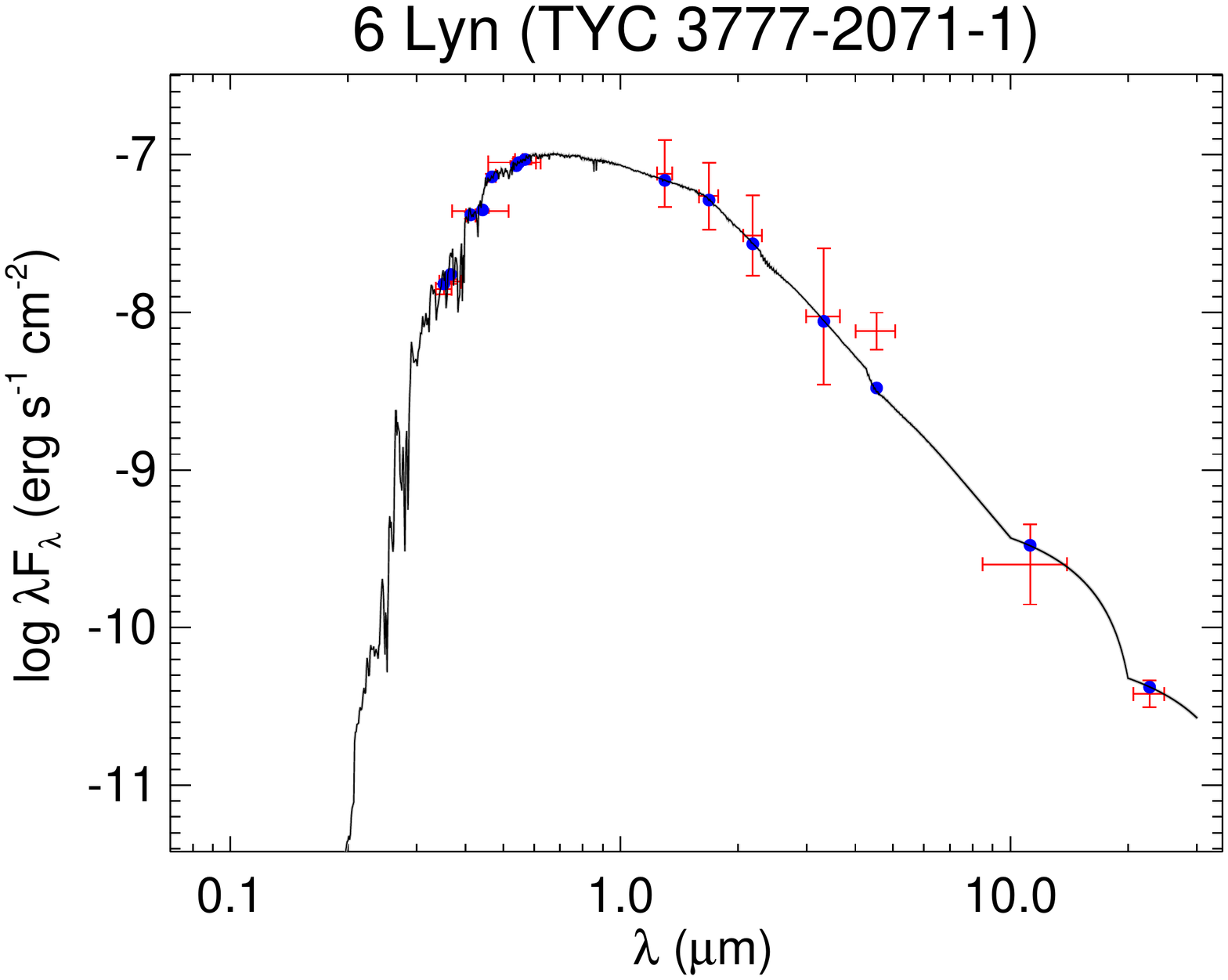}
  \includegraphics[trim=60 60 60 60,clip,width=0.49\linewidth]{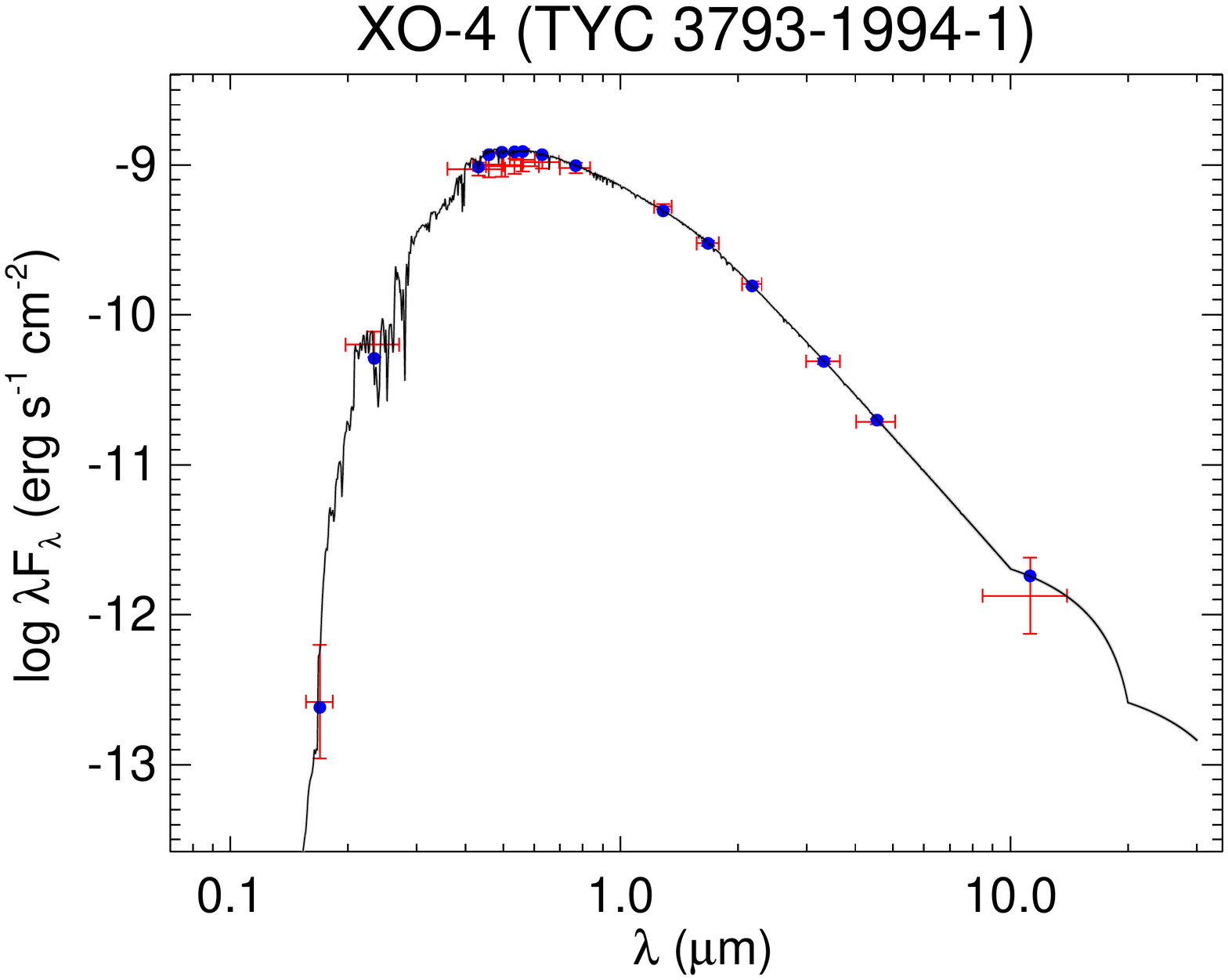}
  \includegraphics[trim=60 60 60 60,clip,width=0.49\linewidth]{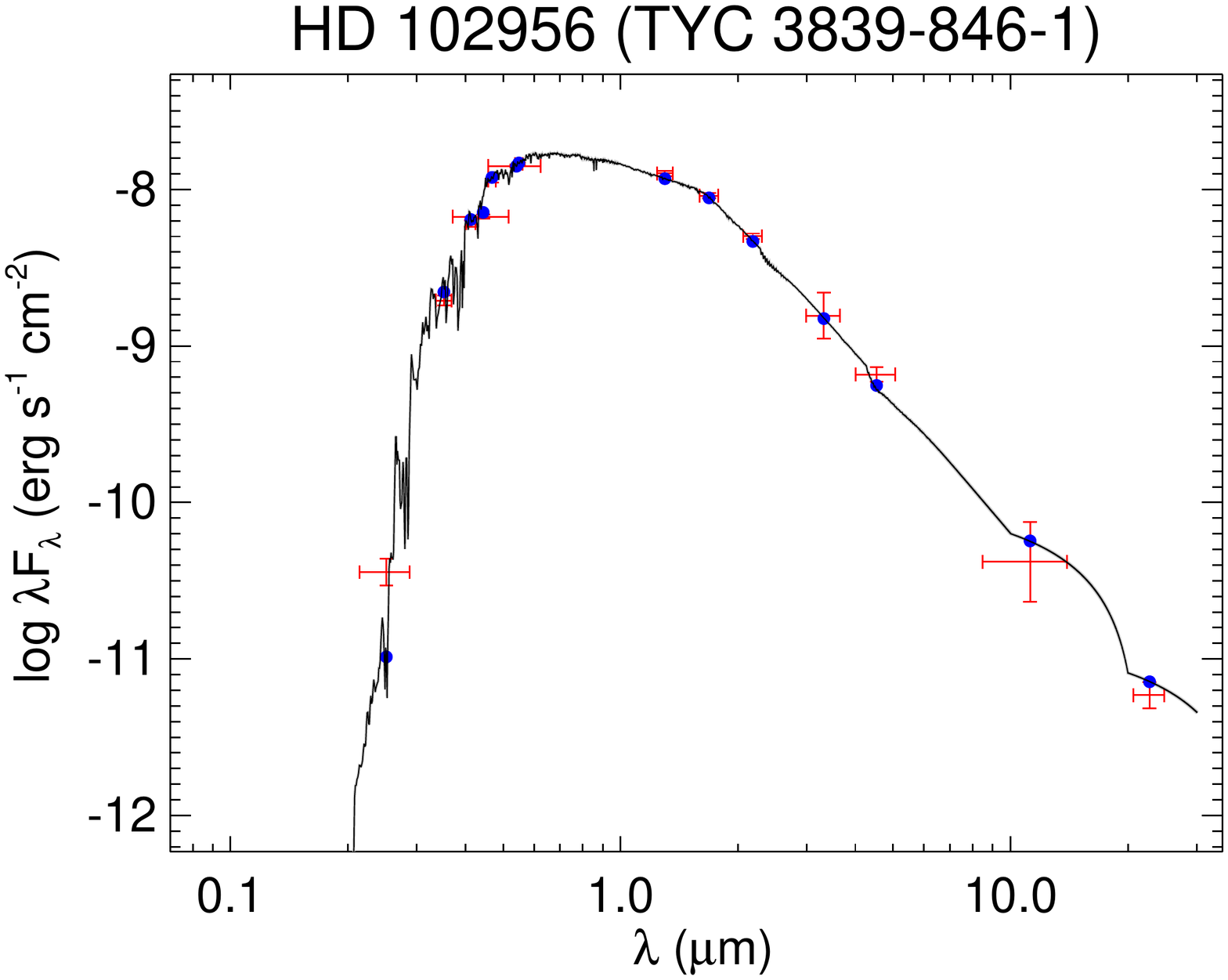}
  \includegraphics[trim=60 60 60 60,clip,width=0.49\linewidth]{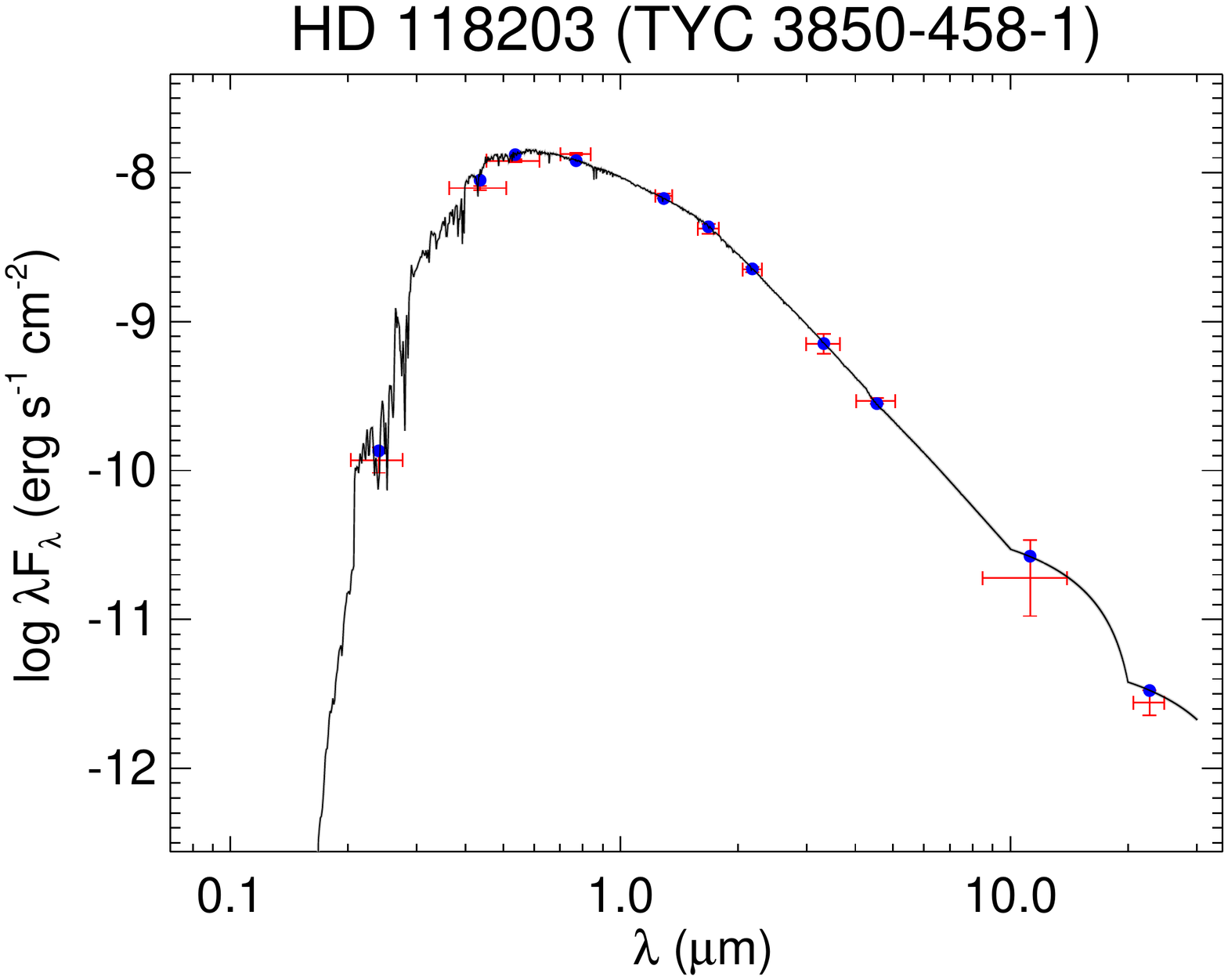}
  \caption{All labels, lines, symbols, and colors as in Figure \ref{fig:seds}.}
  \label{fig:seds_37}
\end{figure}

\begin{figure}[H]
  \centering
  \includegraphics[trim=60 60 60 60,clip,width=0.49\linewidth]{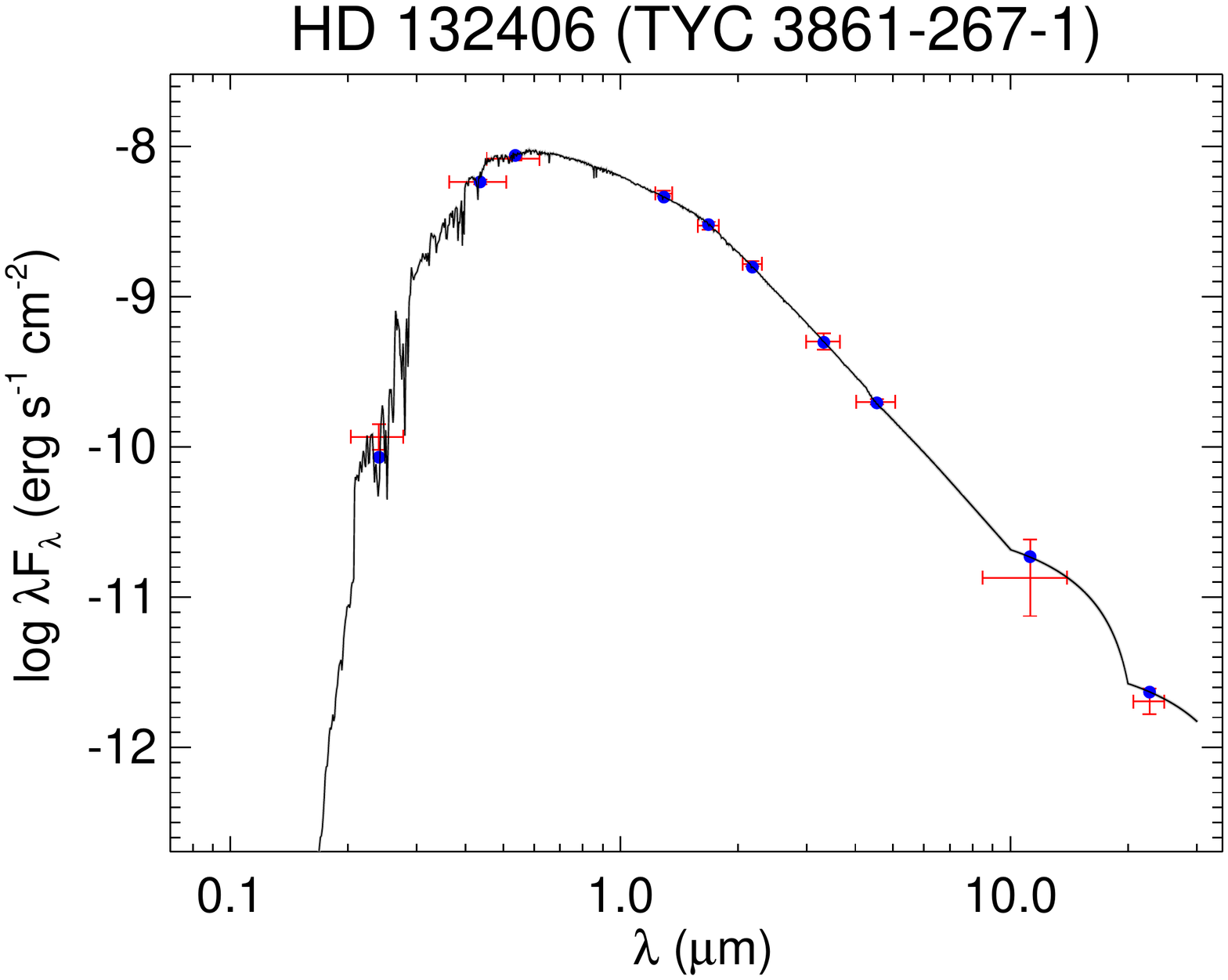}
  \includegraphics[trim=60 60 60 60,clip,width=0.49\linewidth]{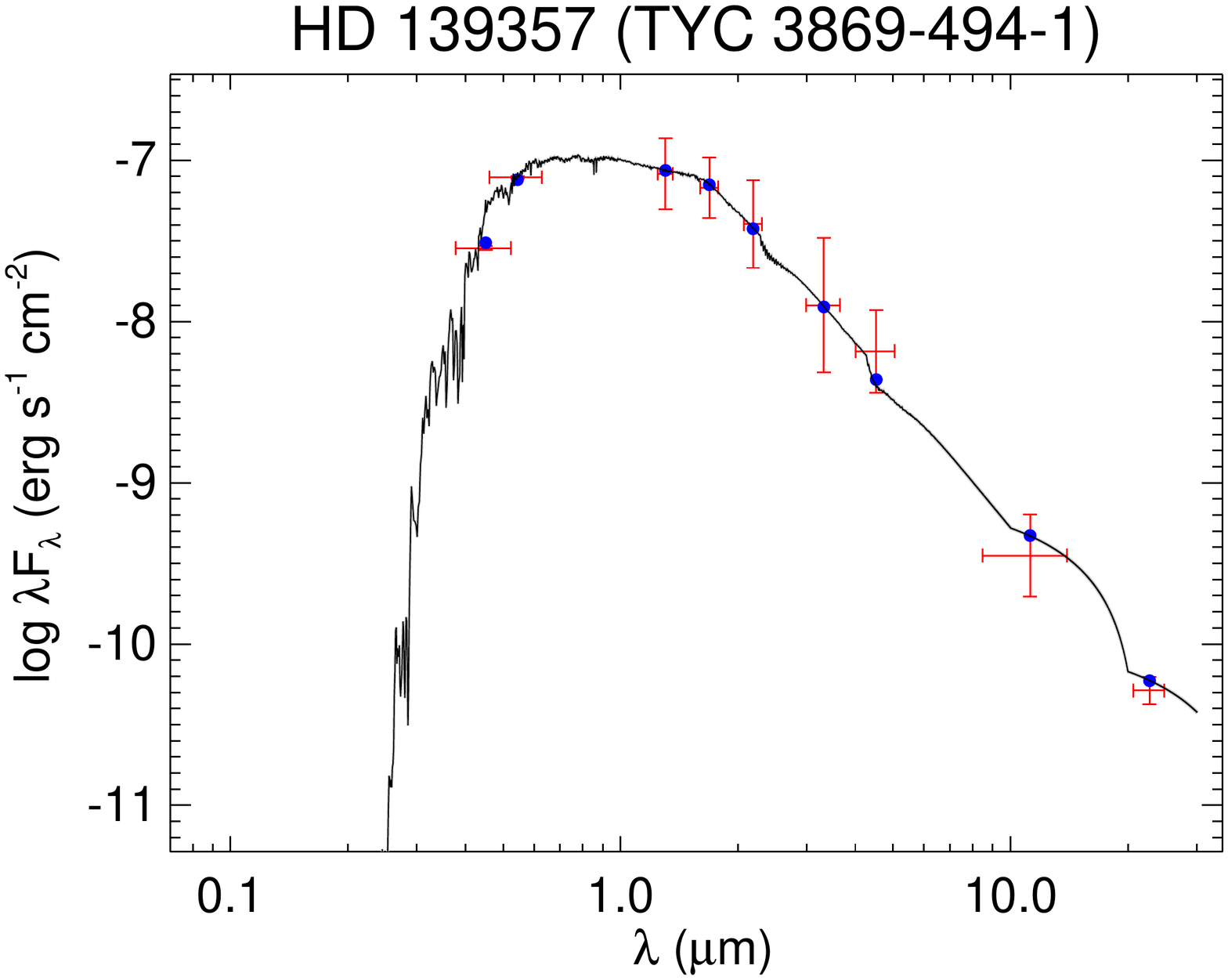}
  \includegraphics[trim=60 60 60 60,clip,width=0.49\linewidth]{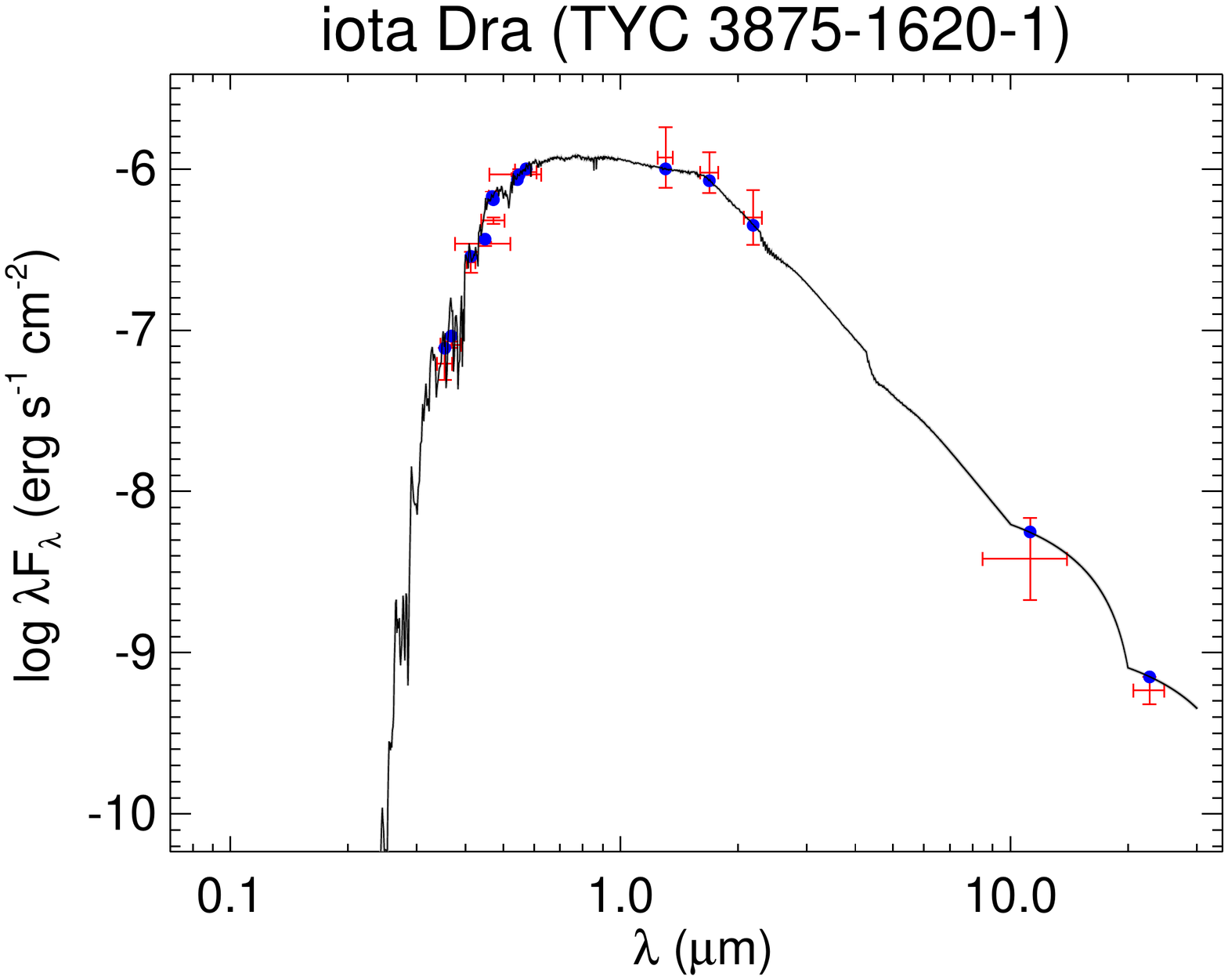}
  \includegraphics[trim=60 60 60 60,clip,width=0.49\linewidth]{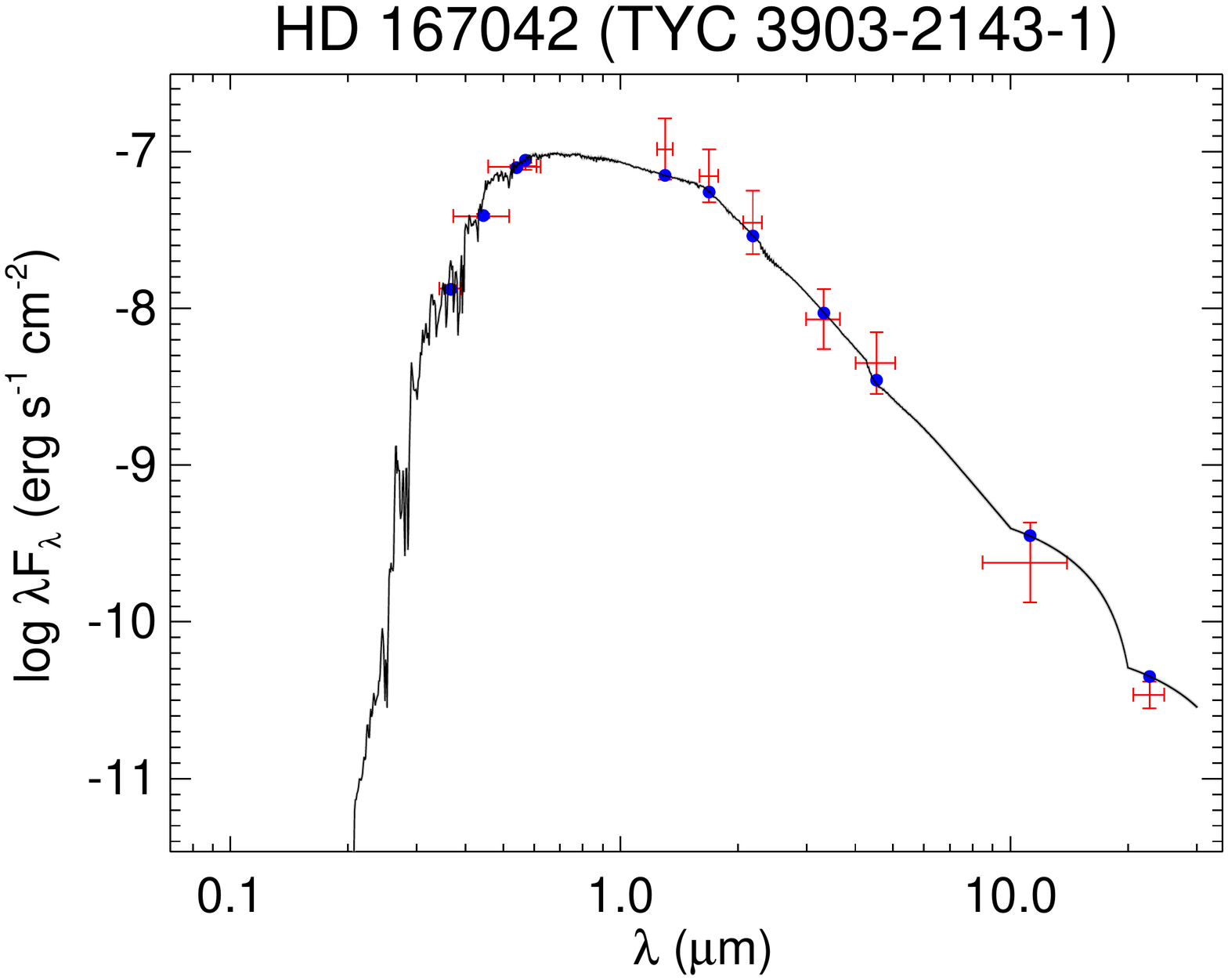}
  \includegraphics[trim=60 60 60 60,clip,width=0.49\linewidth]{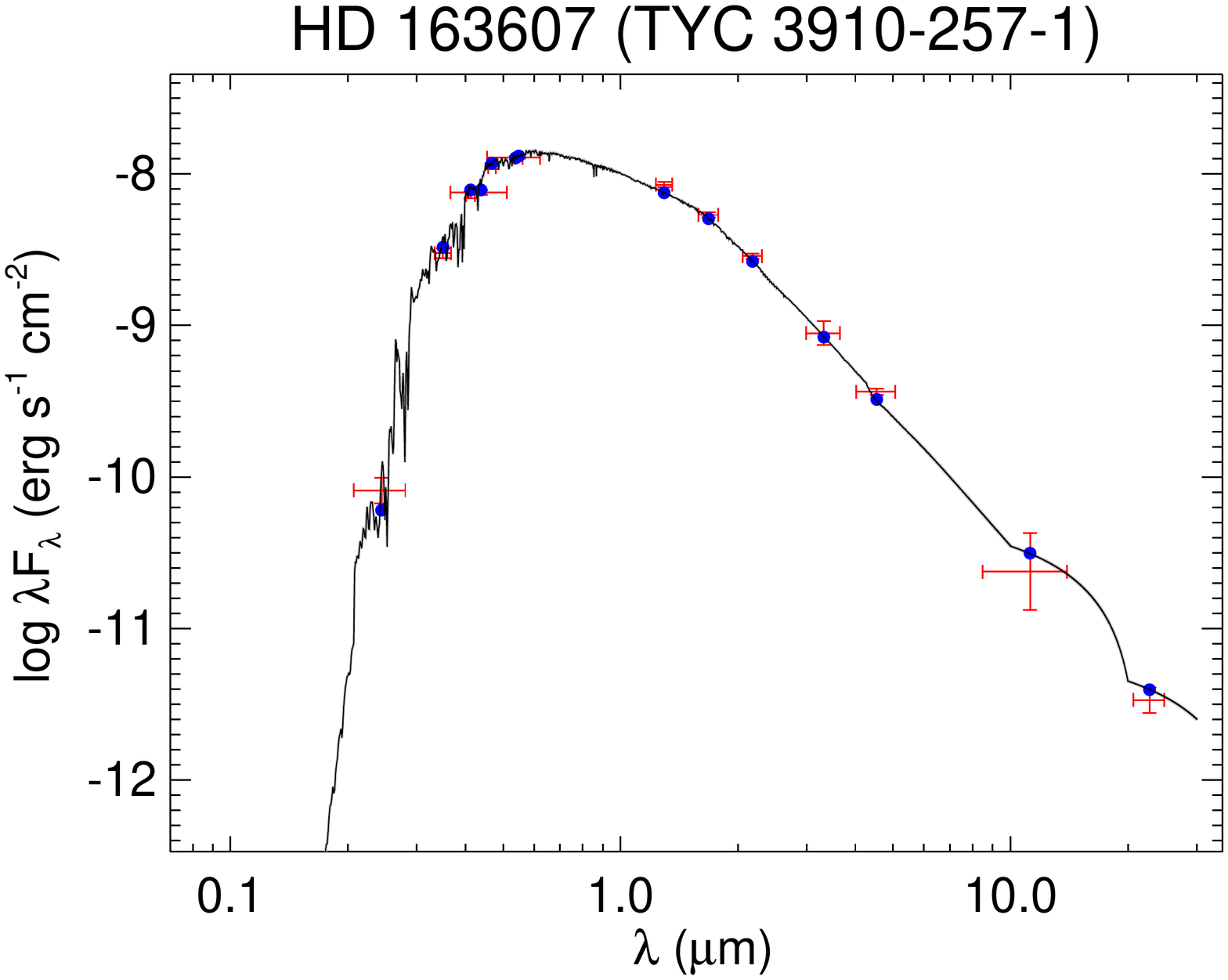}
  \includegraphics[trim=60 60 60 60,clip,width=0.49\linewidth]{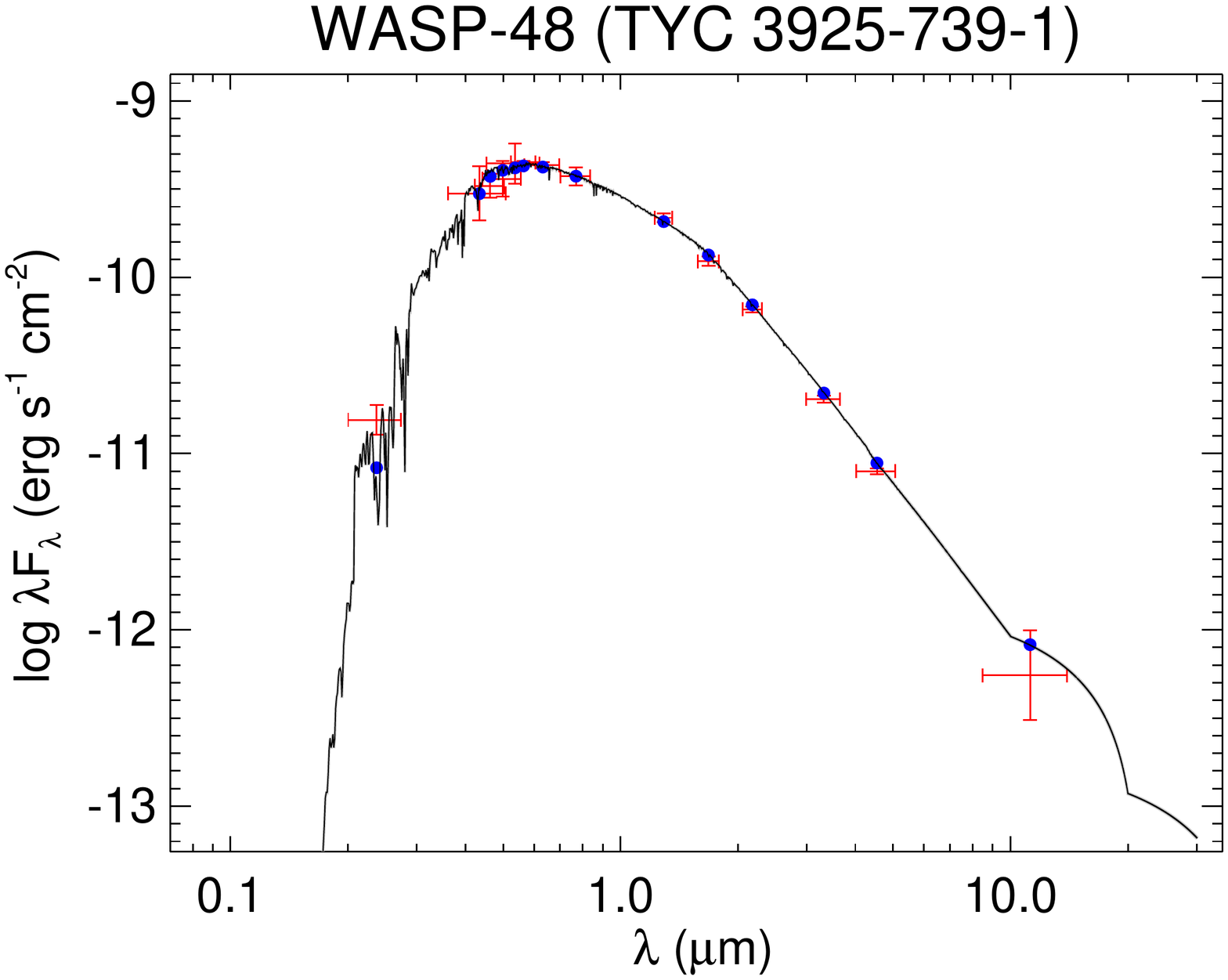}
  \caption{All labels, lines, symbols, and colors as in Figure \ref{fig:seds}.}
  \label{fig:seds_38}
\end{figure}

\begin{figure}[H]
  \centering
  \includegraphics[trim=60 60 60 60,clip,width=0.49\linewidth]{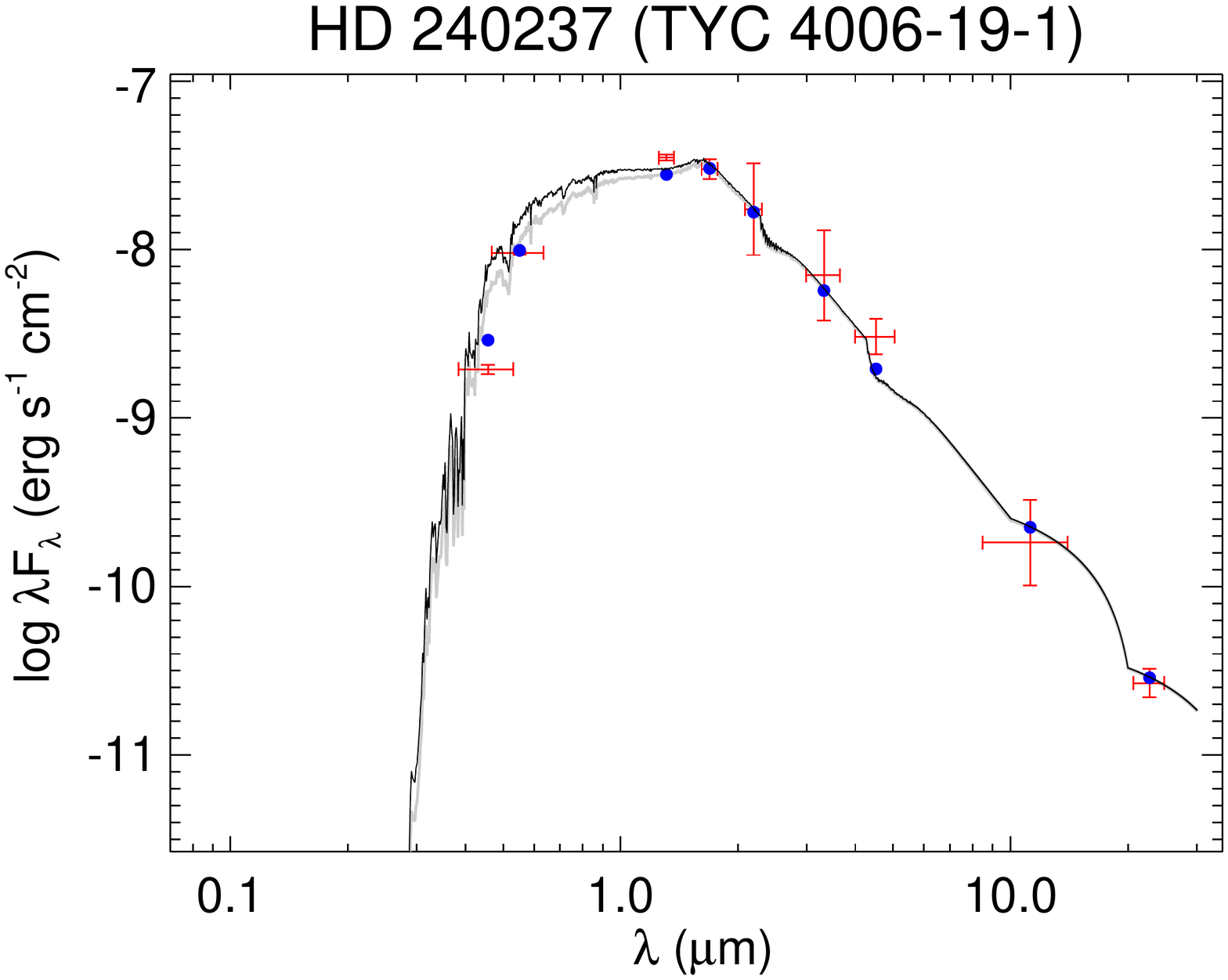}
  \includegraphics[trim=60 60 60 60,clip,width=0.49\linewidth]{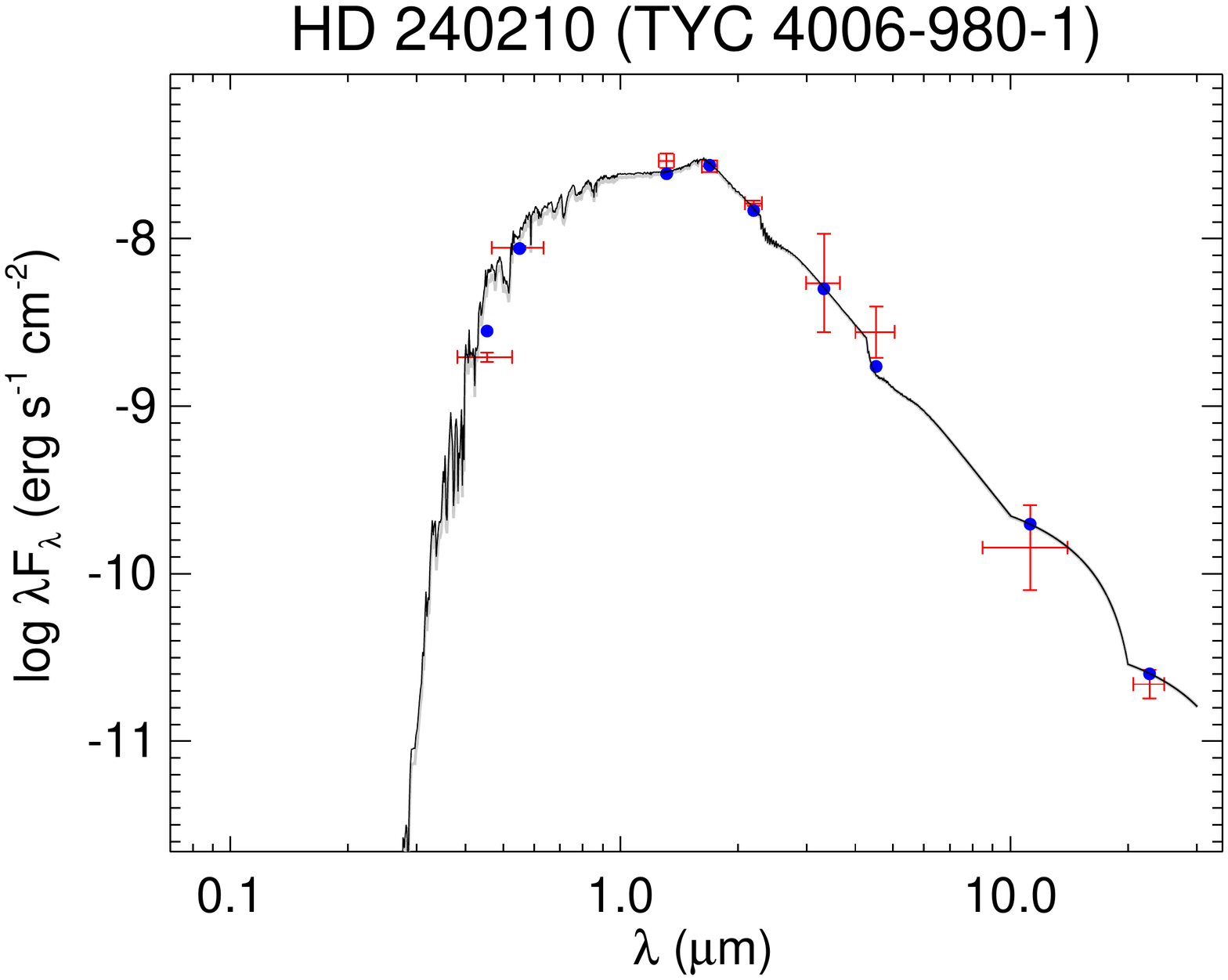}
  \includegraphics[trim=60 60 60 60,clip,width=0.49\linewidth]{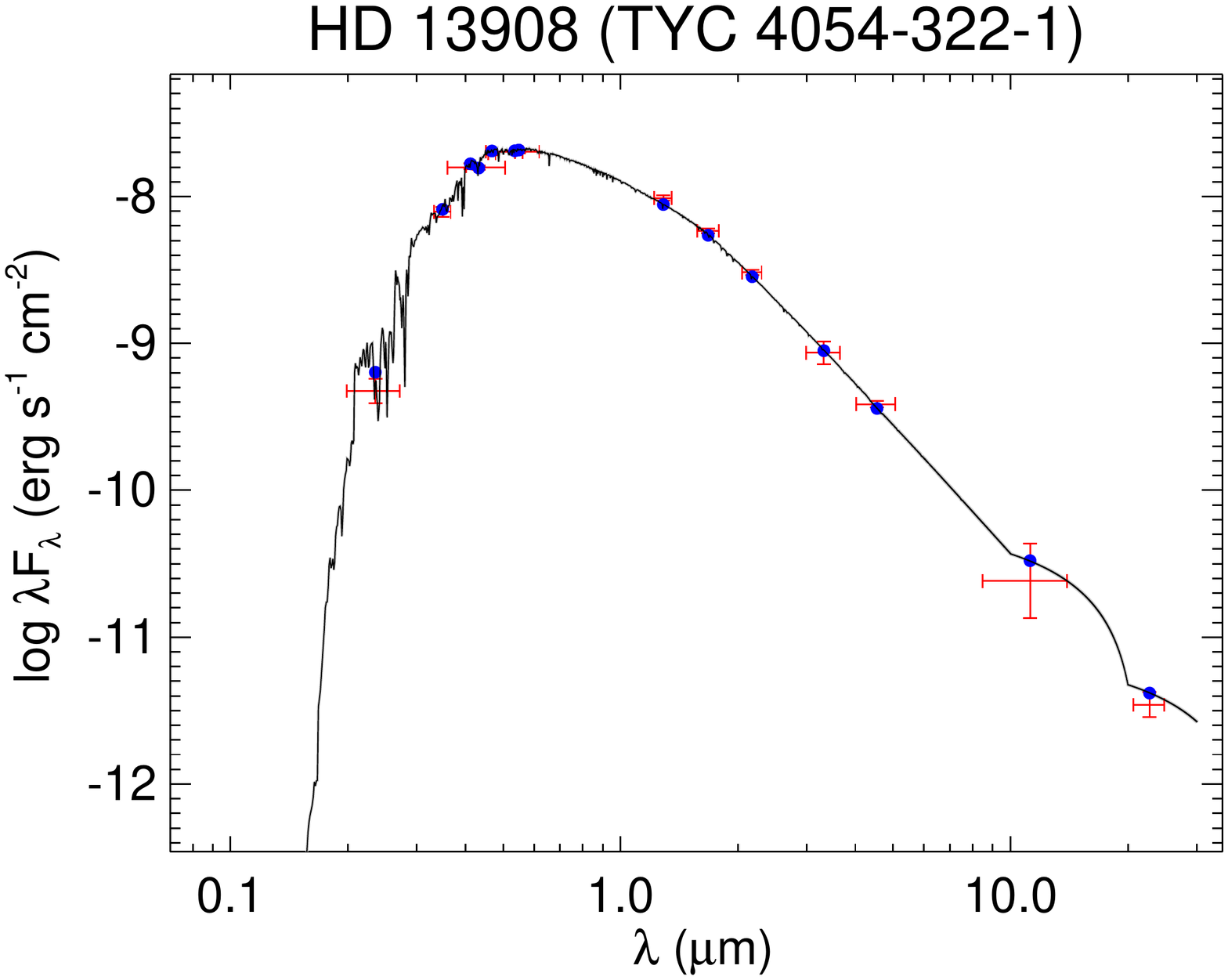}
  \includegraphics[trim=60 60 60 60,clip,width=0.49\linewidth]{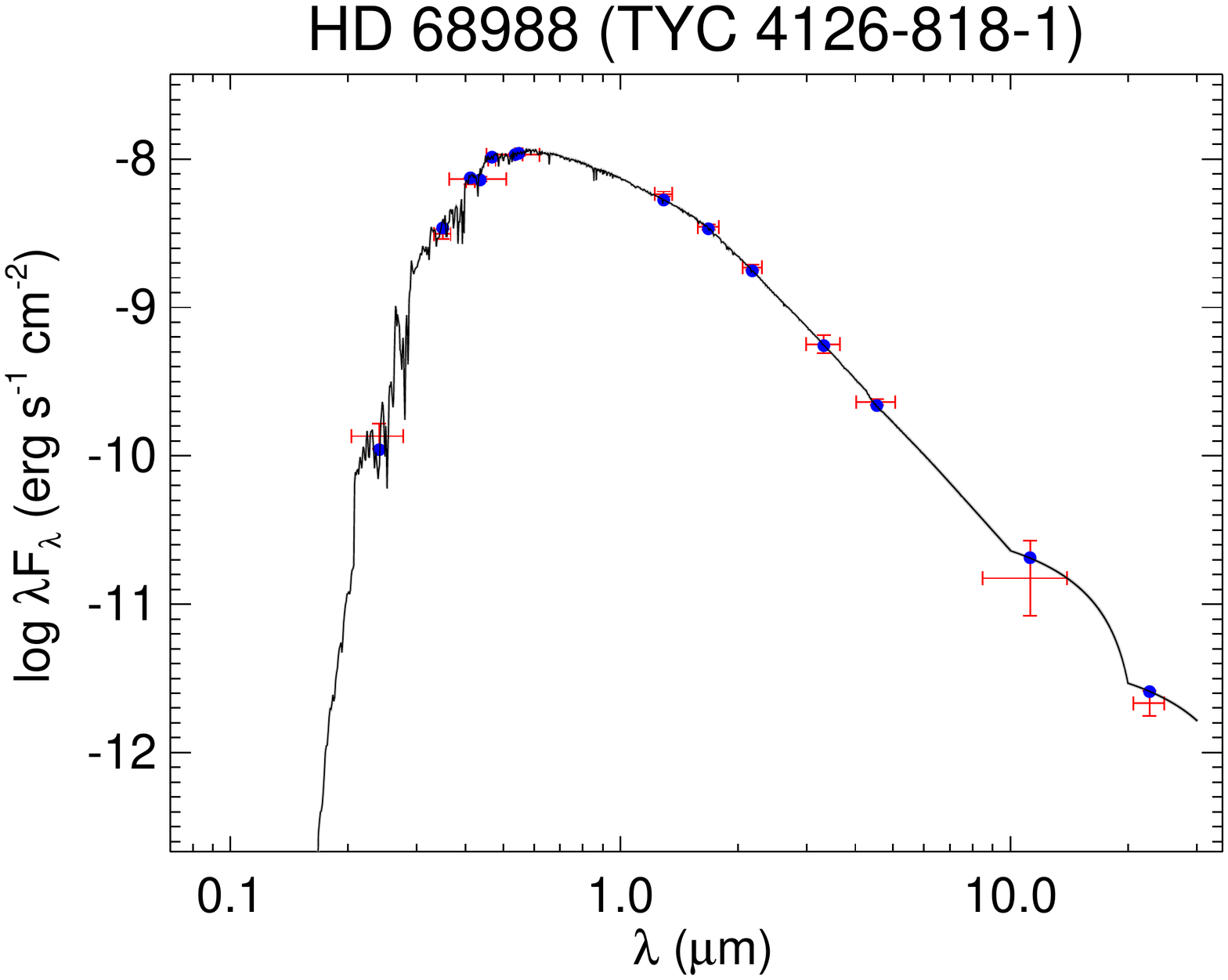}
  \includegraphics[trim=60 60 60 60,clip,width=0.49\linewidth]{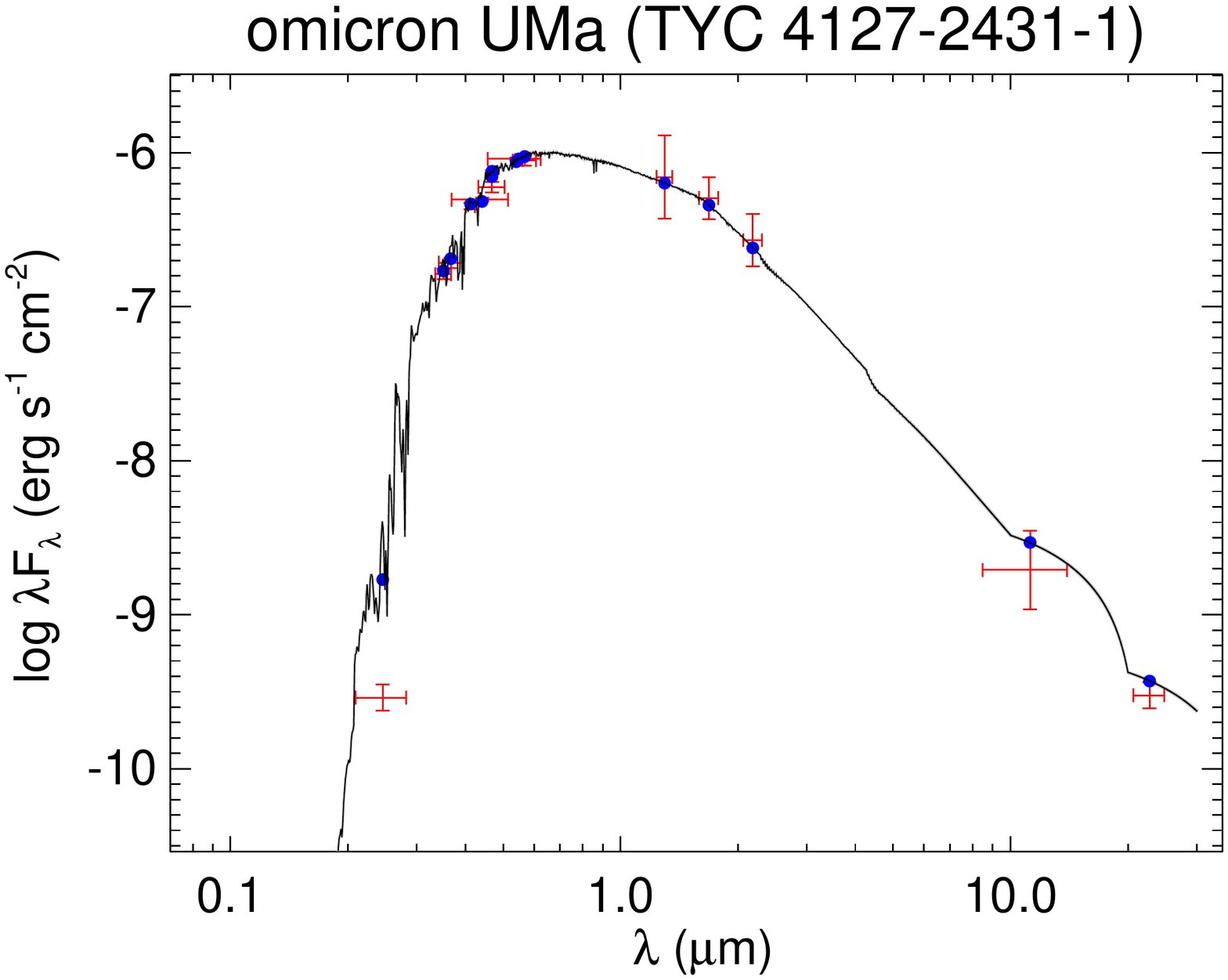}
  \includegraphics[trim=60 60 60 60,clip,width=0.49\linewidth]{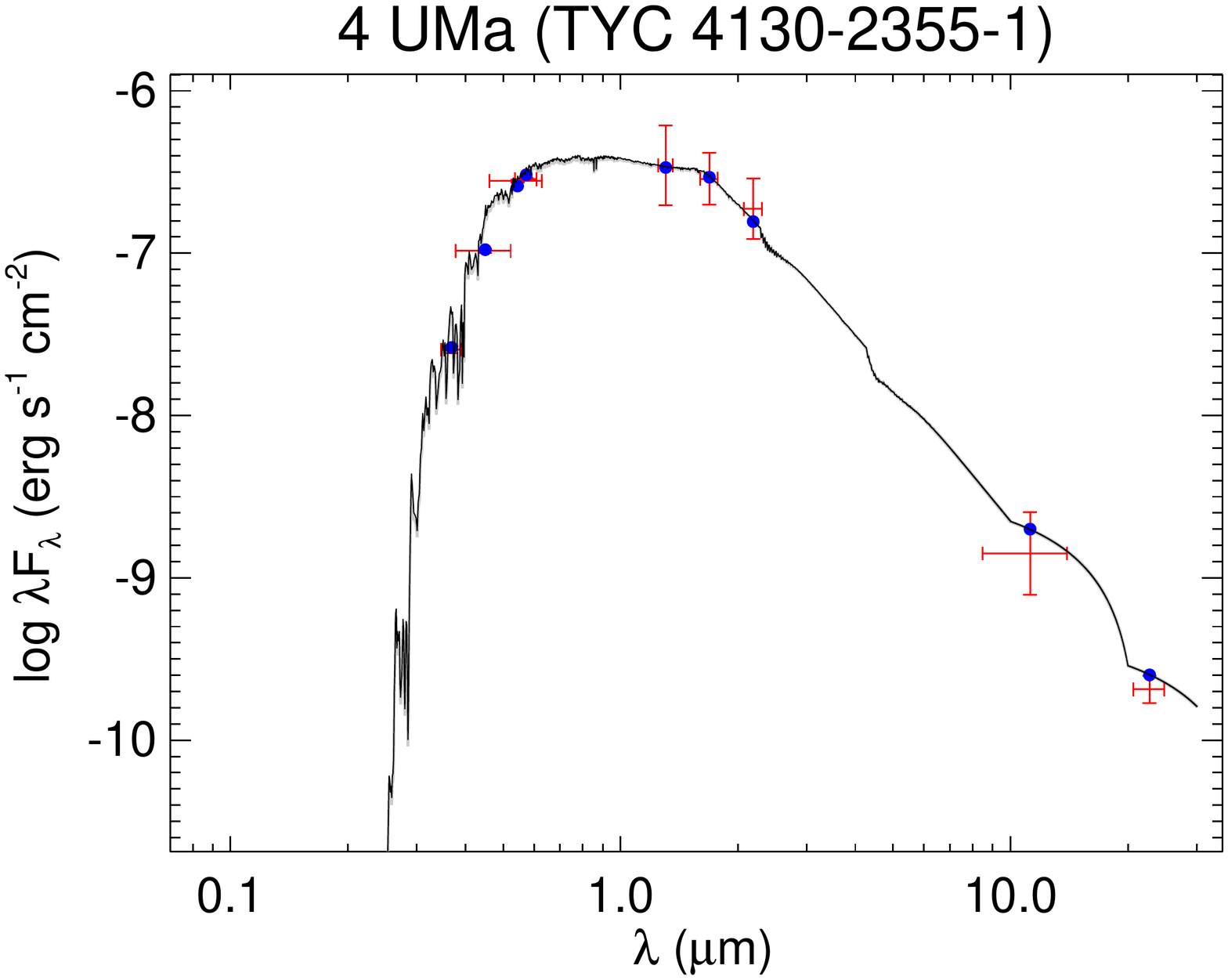}
  \caption{All labels, lines, symbols, and colors as in Figure \ref{fig:seds}.}
  \label{fig:seds_39}
\end{figure}

\begin{figure}[H]
  \centering
  \includegraphics[trim=60 60 60 60,clip,width=0.49\linewidth]{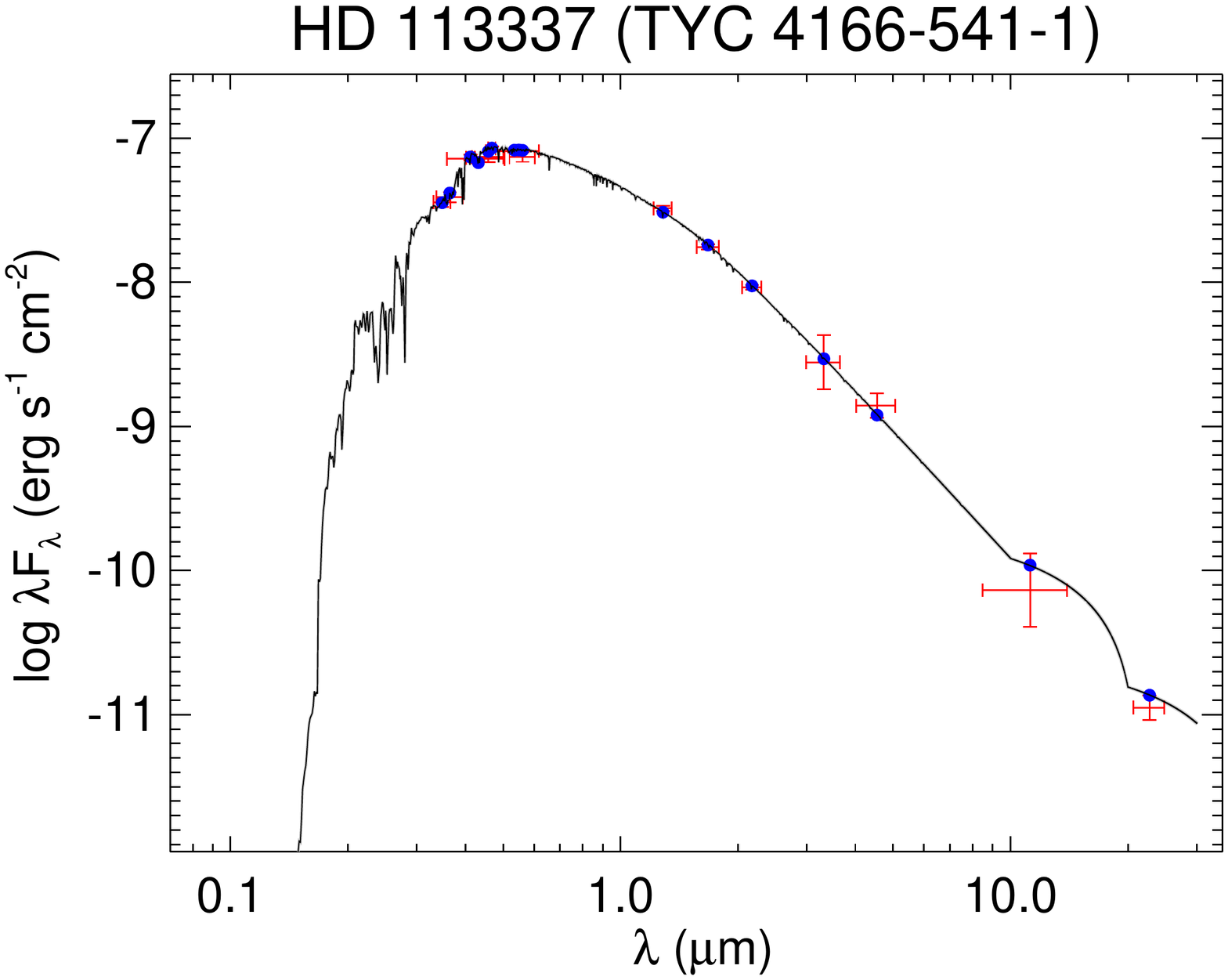}
  \includegraphics[trim=60 60 60 60,clip,width=0.49\linewidth]{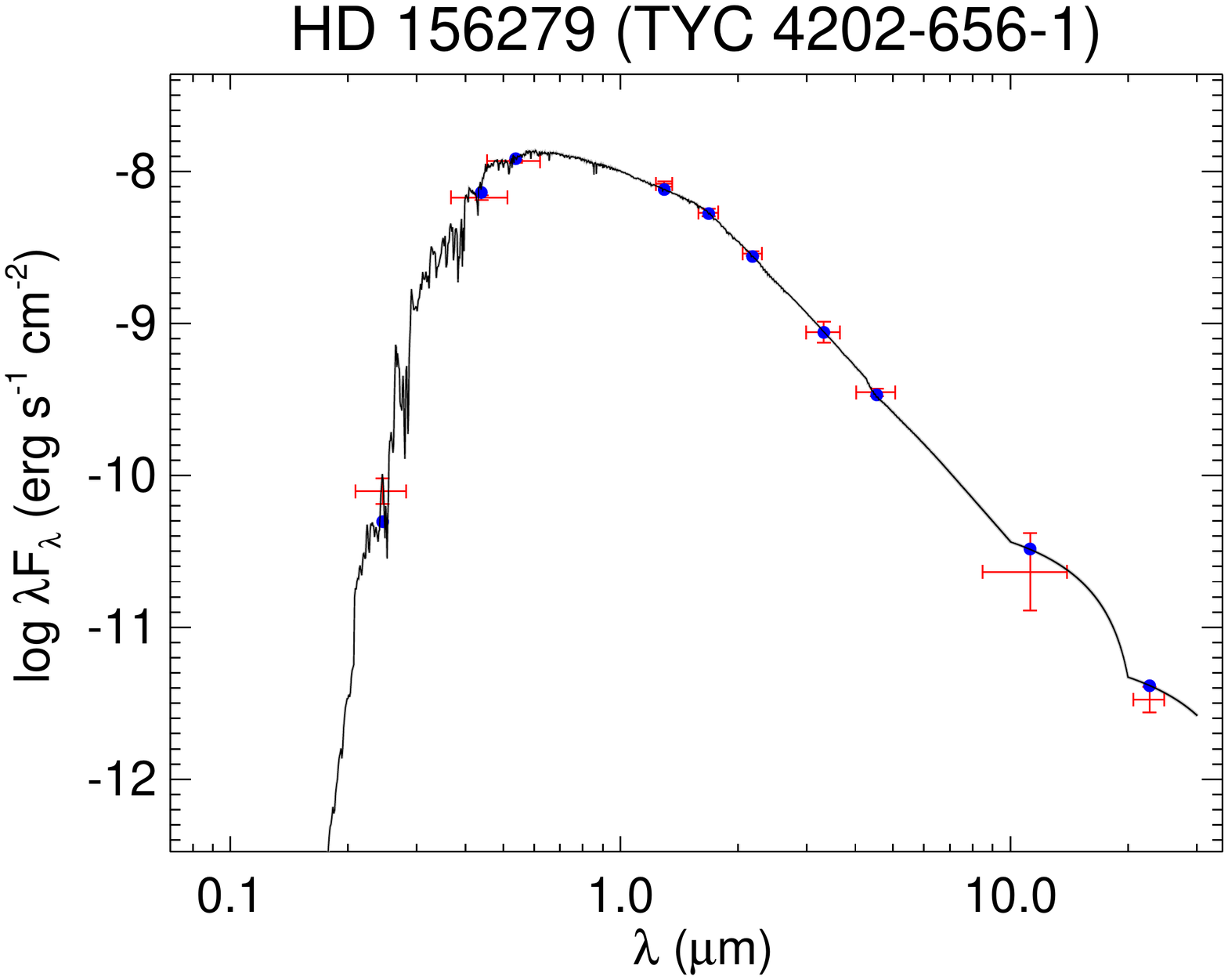}
  \includegraphics[trim=60 60 60 60,clip,width=0.49\linewidth]{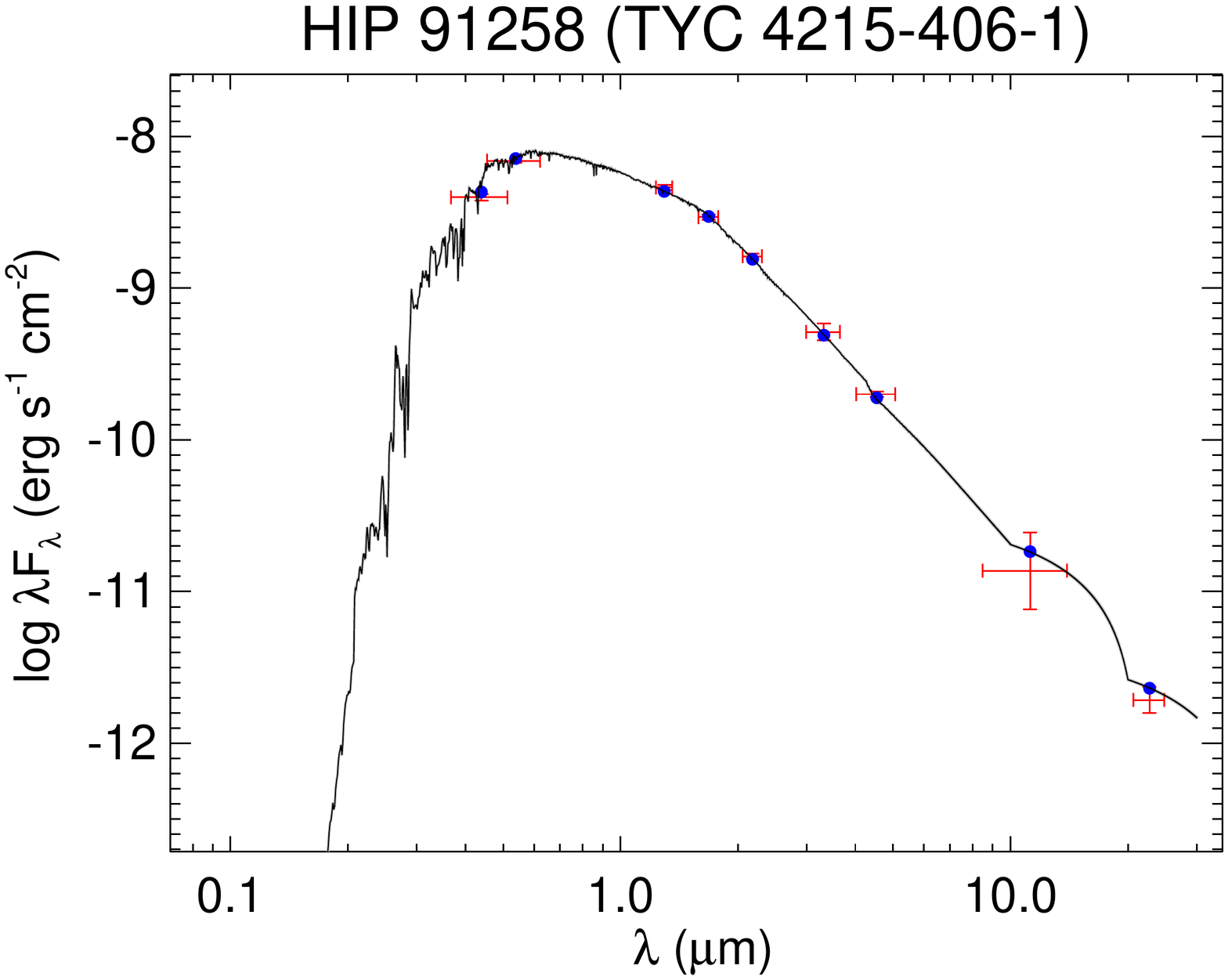}
  \includegraphics[trim=60 60 60 60,clip,width=0.49\linewidth]{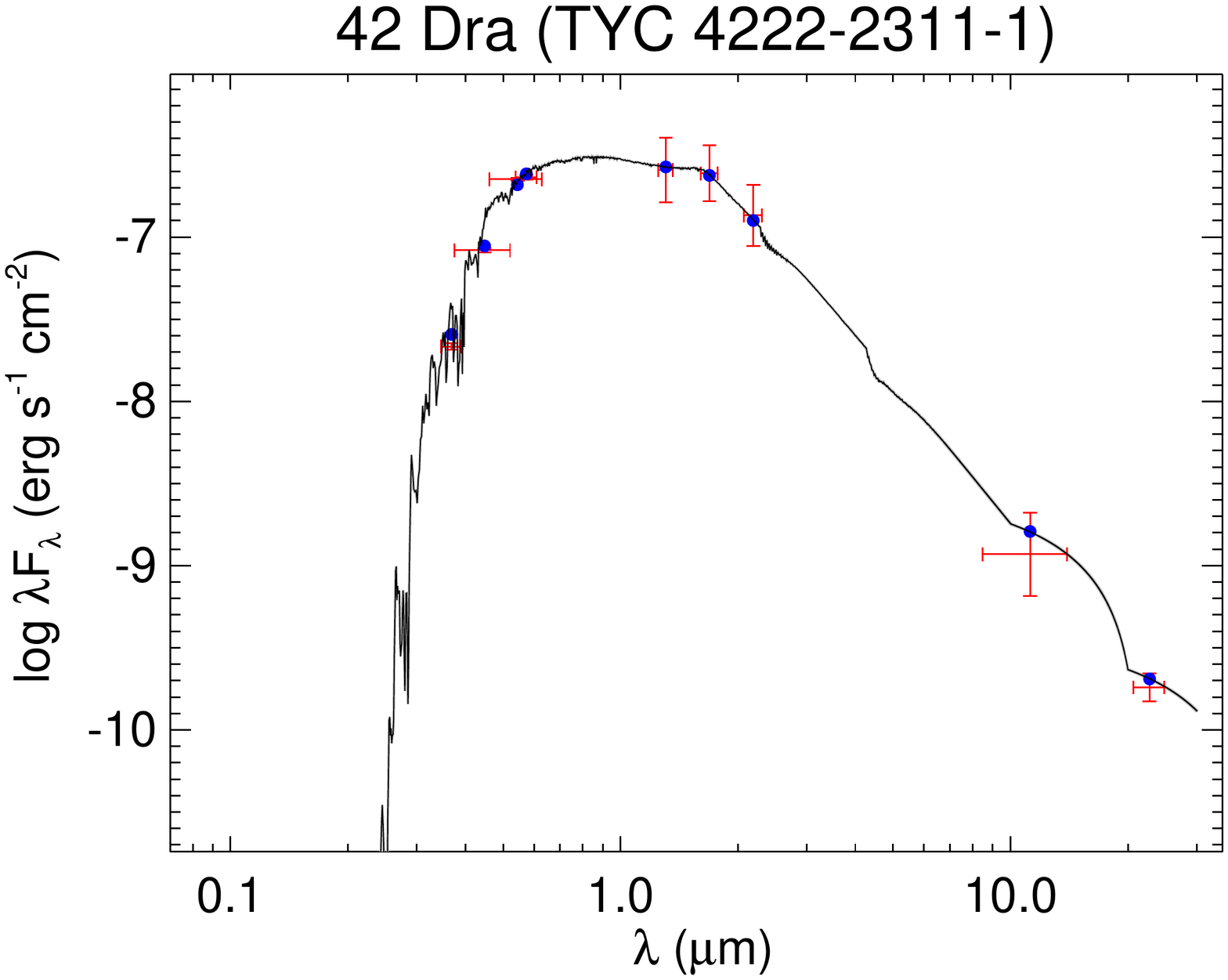}
  \includegraphics[trim=60 60 60 60,clip,width=0.49\linewidth]{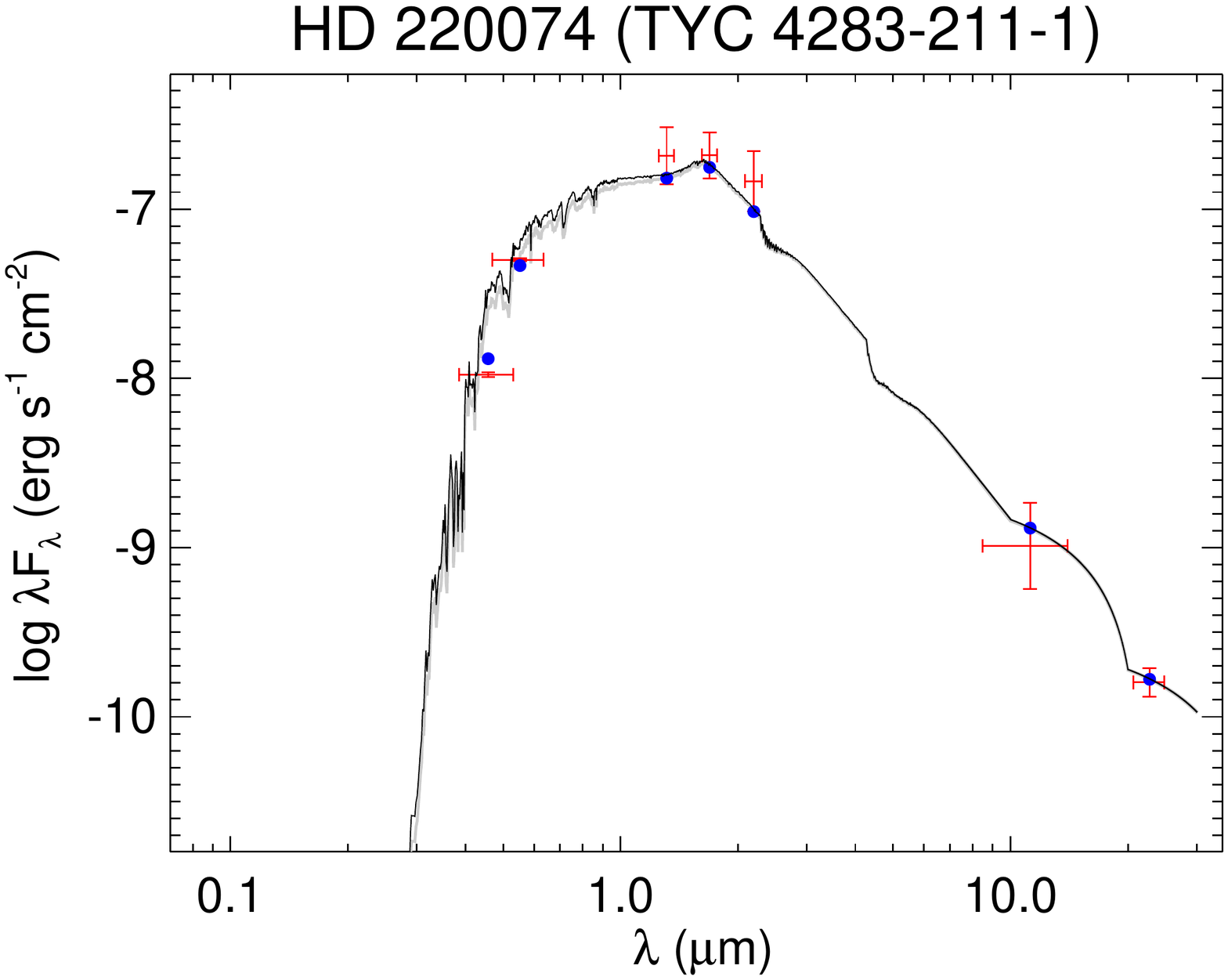}
  \includegraphics[trim=60 60 60 60,clip,width=0.49\linewidth]{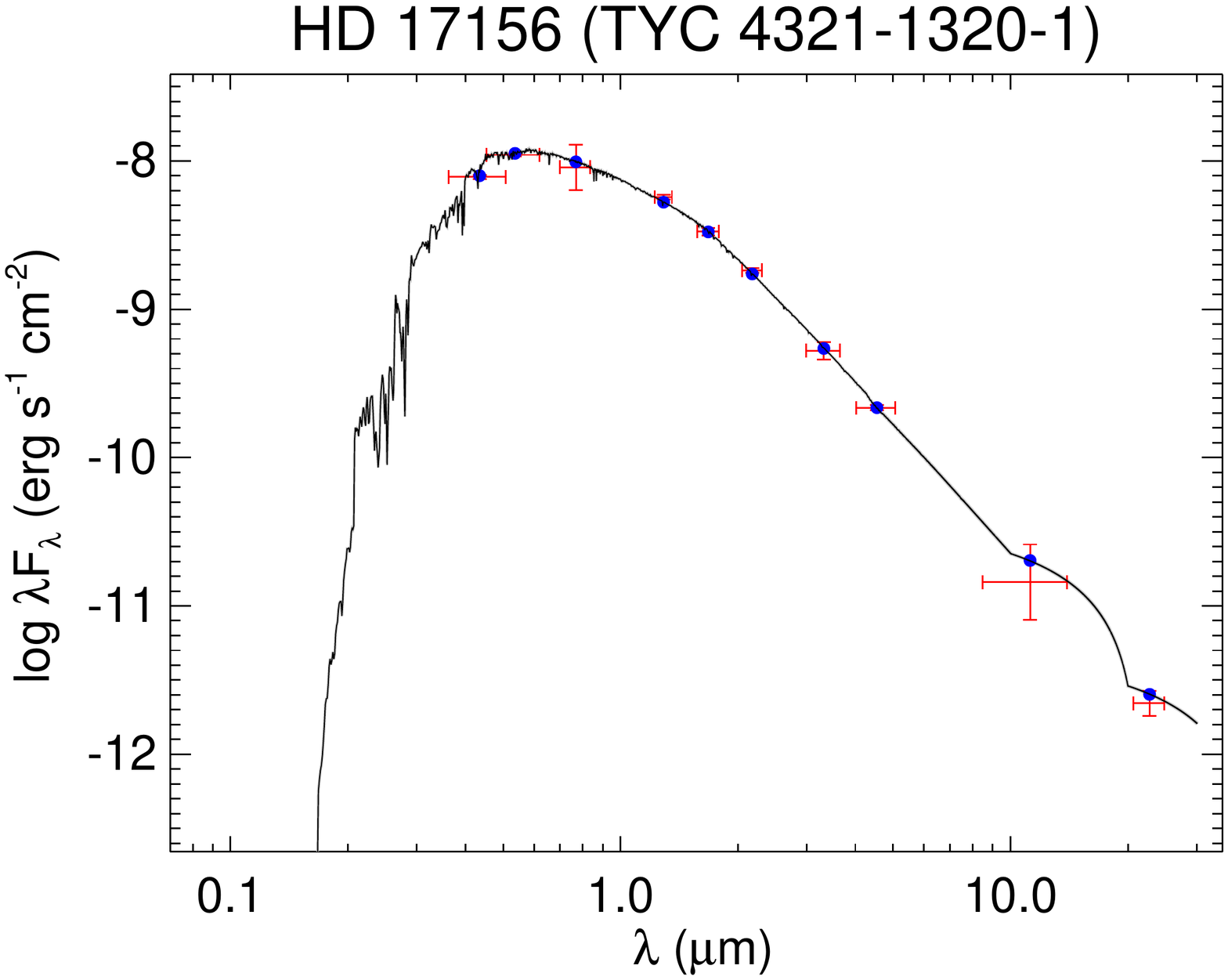}
  \caption{All labels, lines, symbols, and colors as in Figure \ref{fig:seds}.}
  \label{fig:seds_40}
\end{figure}

\begin{figure}[H]
  \centering
  \includegraphics[trim=60 60 60 60,clip,width=0.49\linewidth]{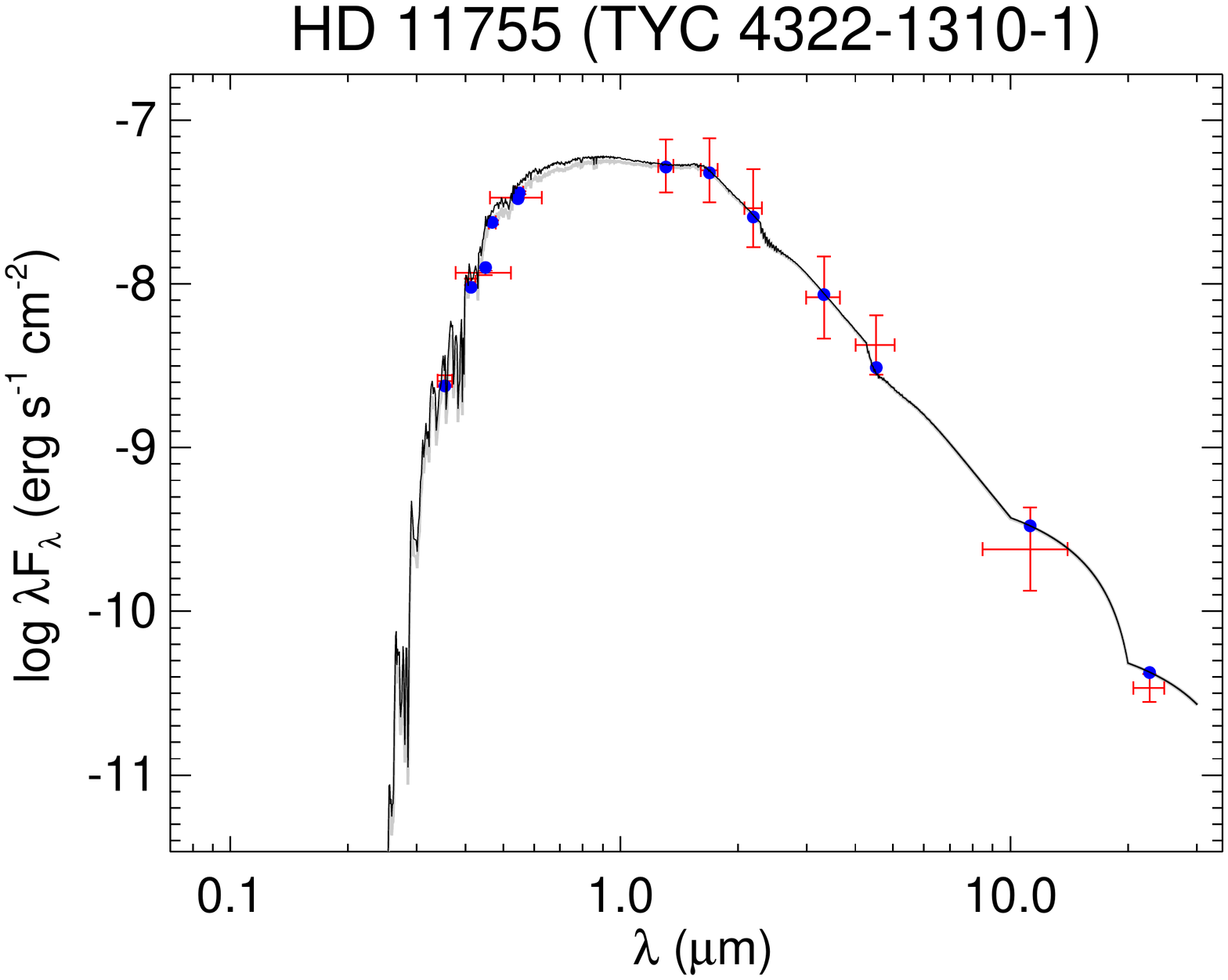}
  \includegraphics[trim=60 60 60 60,clip,width=0.49\linewidth]{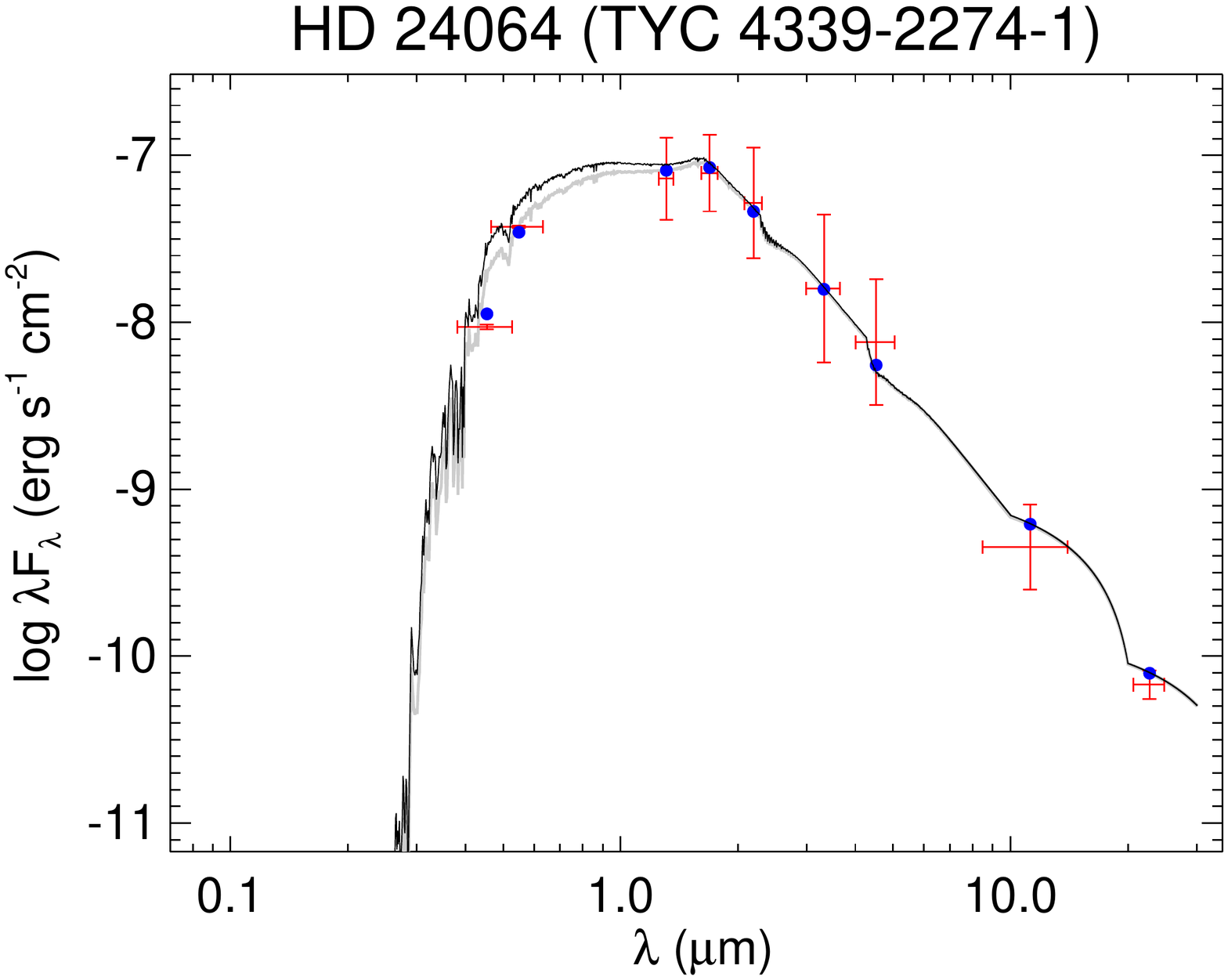}
  \includegraphics[trim=60 60 60 60,clip,width=0.49\linewidth]{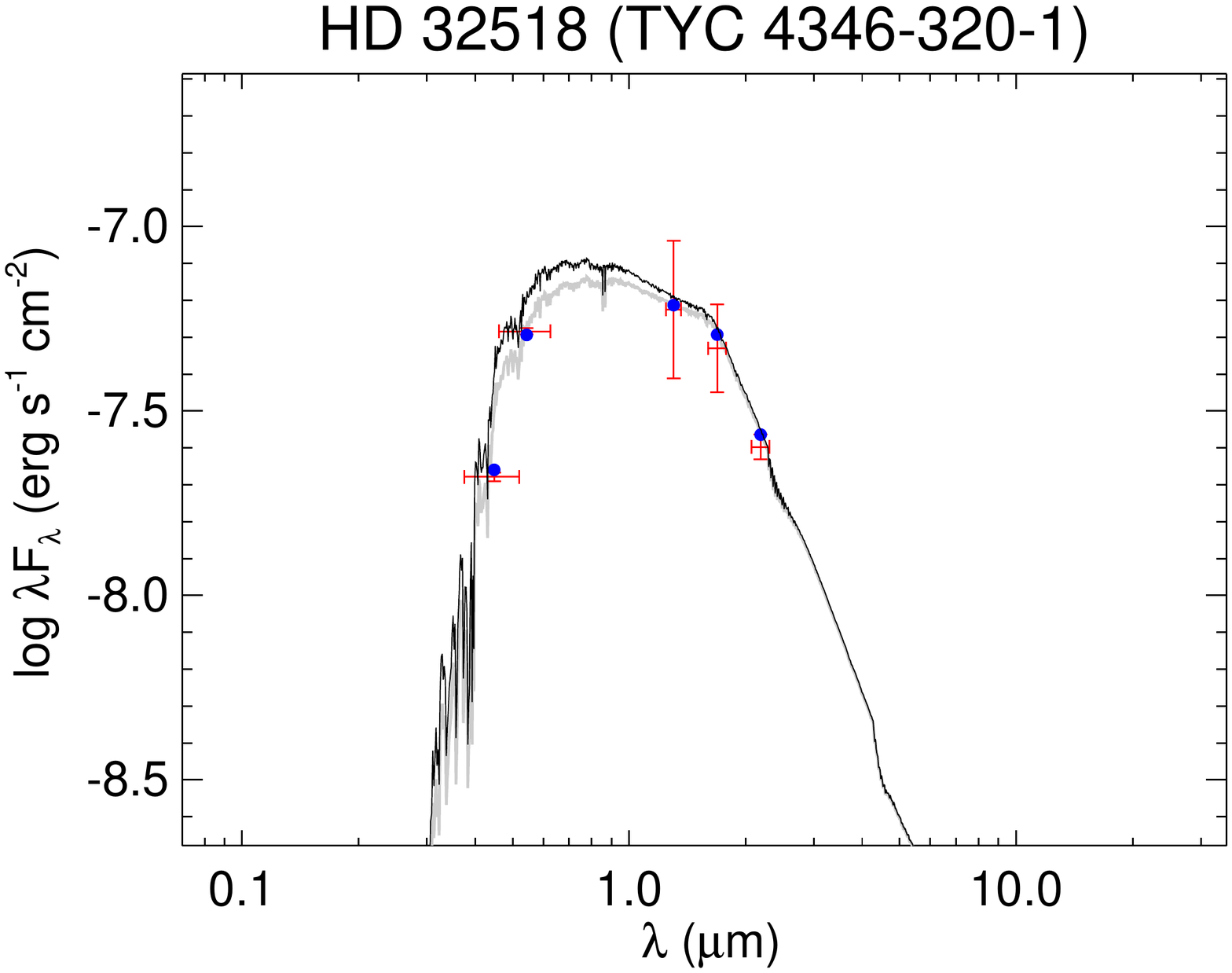}
  \includegraphics[trim=60 60 60 60,clip,width=0.49\linewidth]{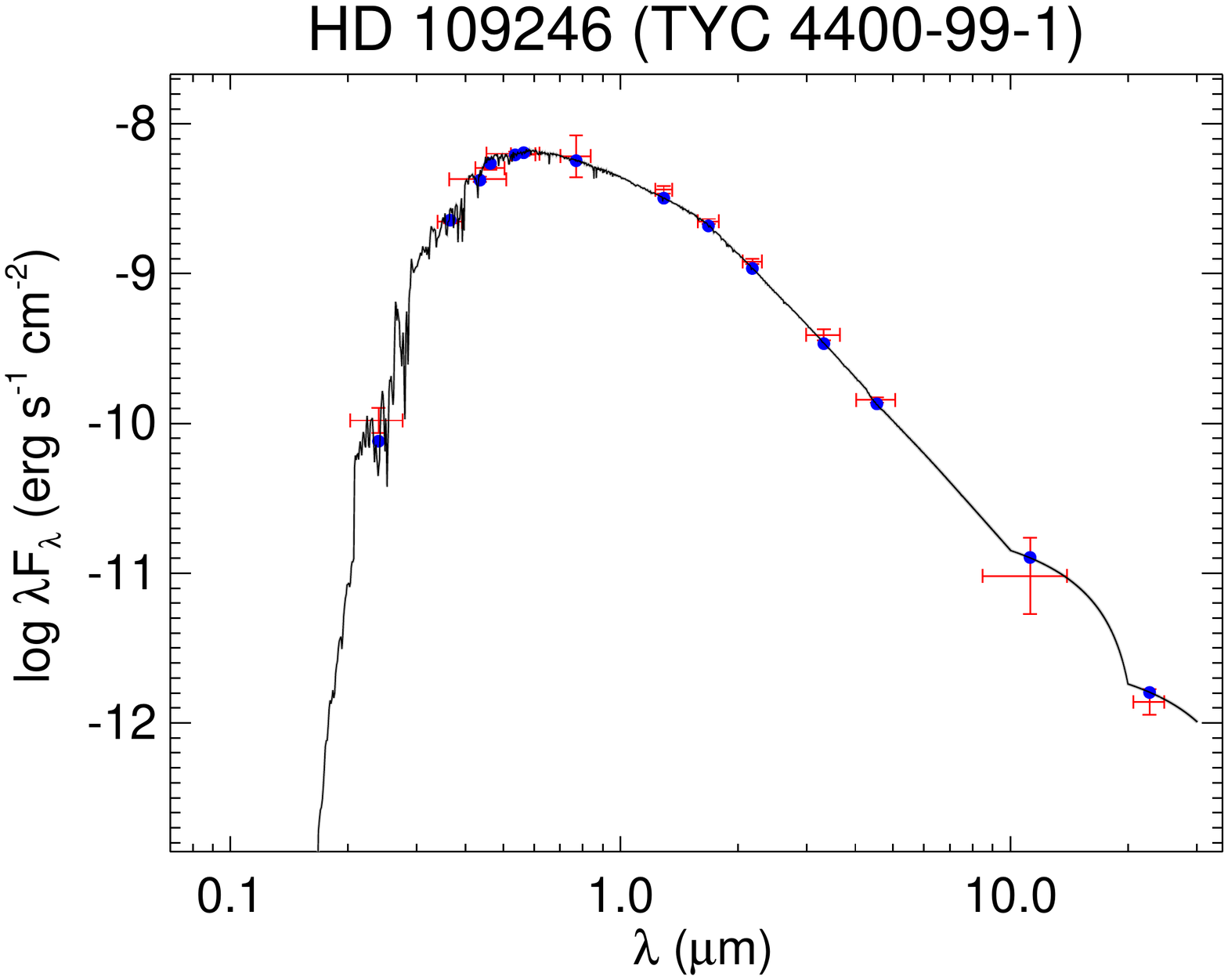}
  \includegraphics[trim=60 60 60 60,clip,width=0.49\linewidth]{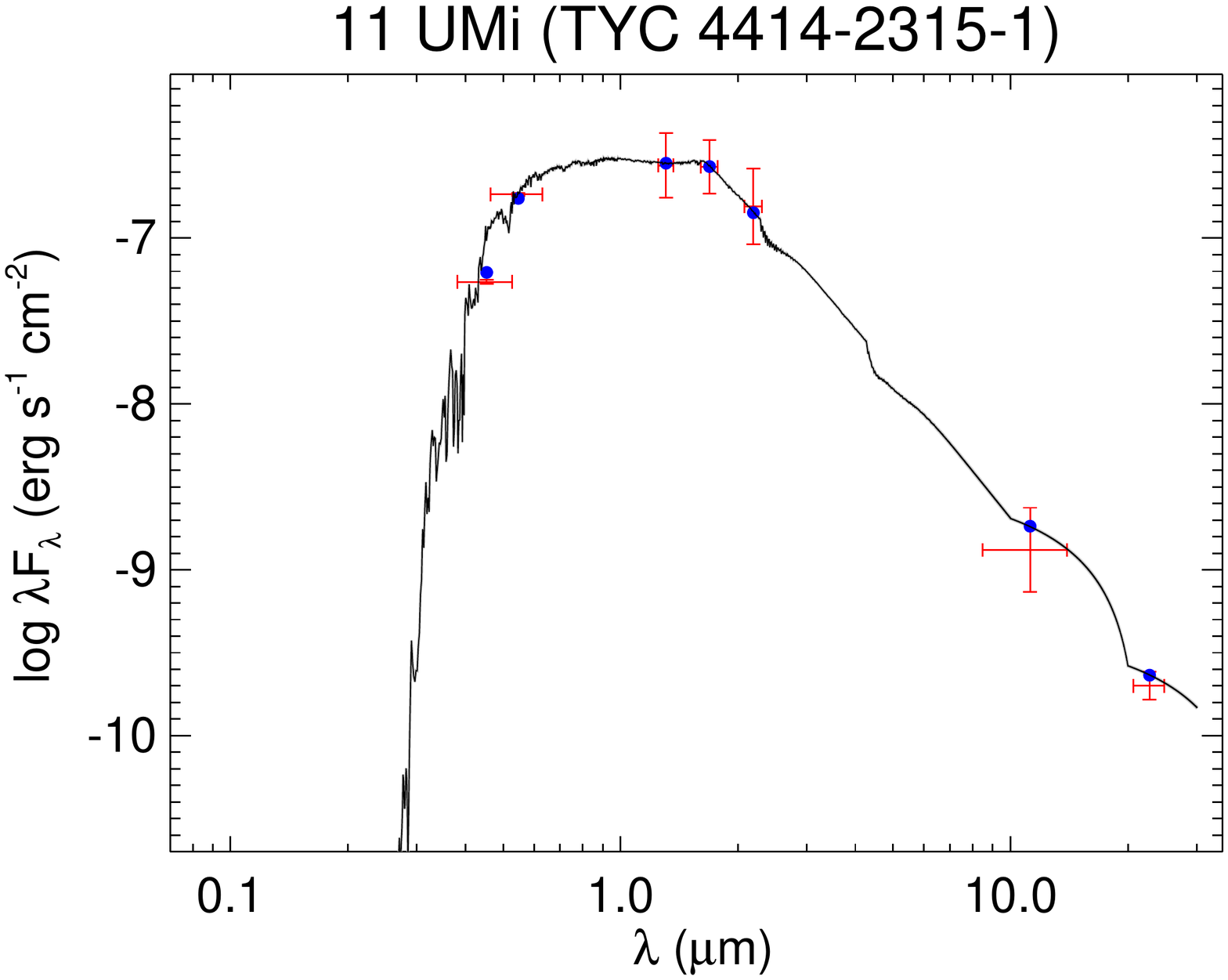}
  \includegraphics[trim=60 60 60 60,clip,width=0.49\linewidth]{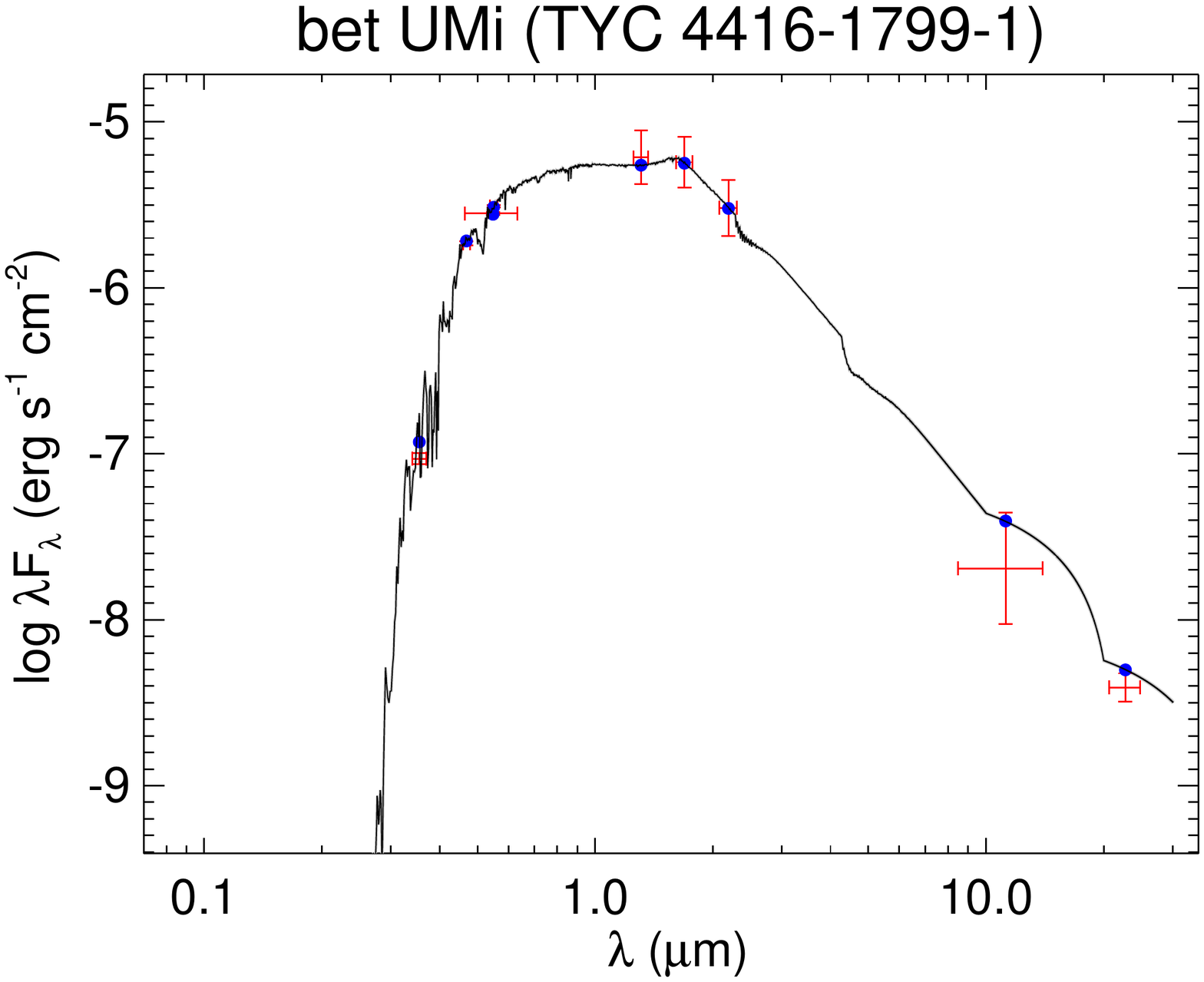}
  \caption{All labels, lines, symbols, and colors as in Figure \ref{fig:seds}.}
  \label{fig:seds_41}
\end{figure}

\begin{figure}[H]
  \centering
  \includegraphics[trim=60 60 60 60,clip,width=0.49\linewidth]{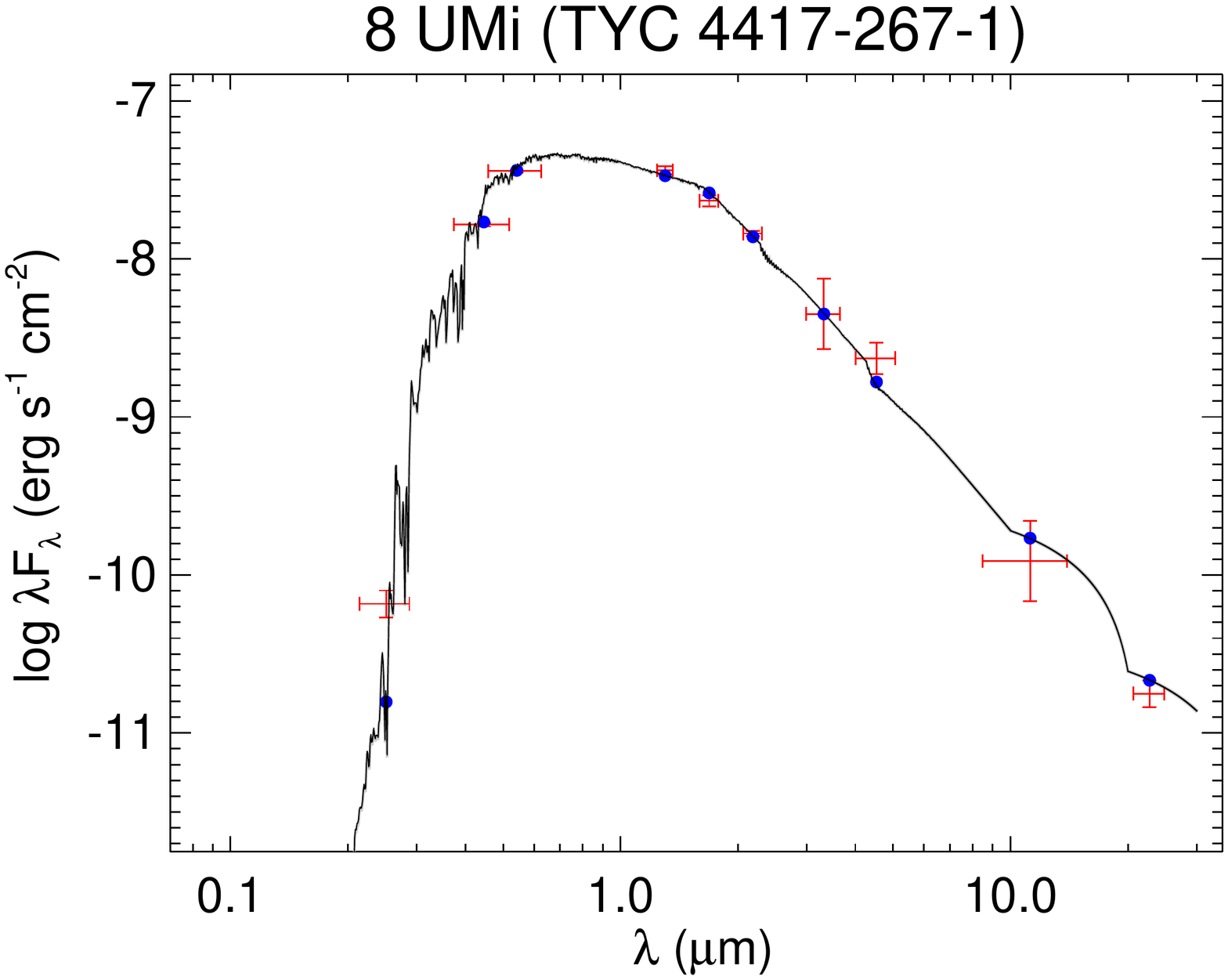}
  \includegraphics[trim=60 60 60 60,clip,width=0.49\linewidth]{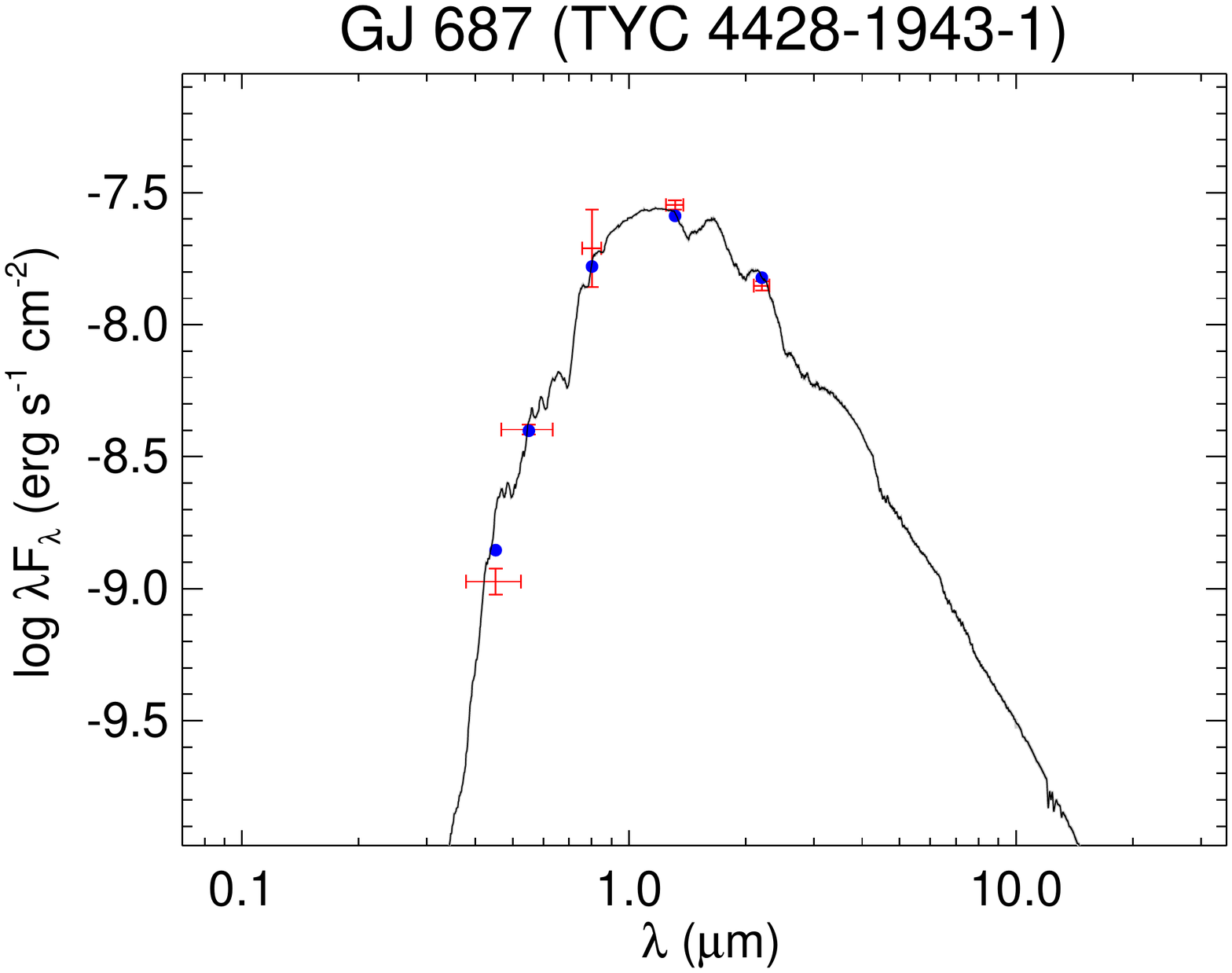}
  \includegraphics[trim=60 60 60 60,clip,width=0.49\linewidth]{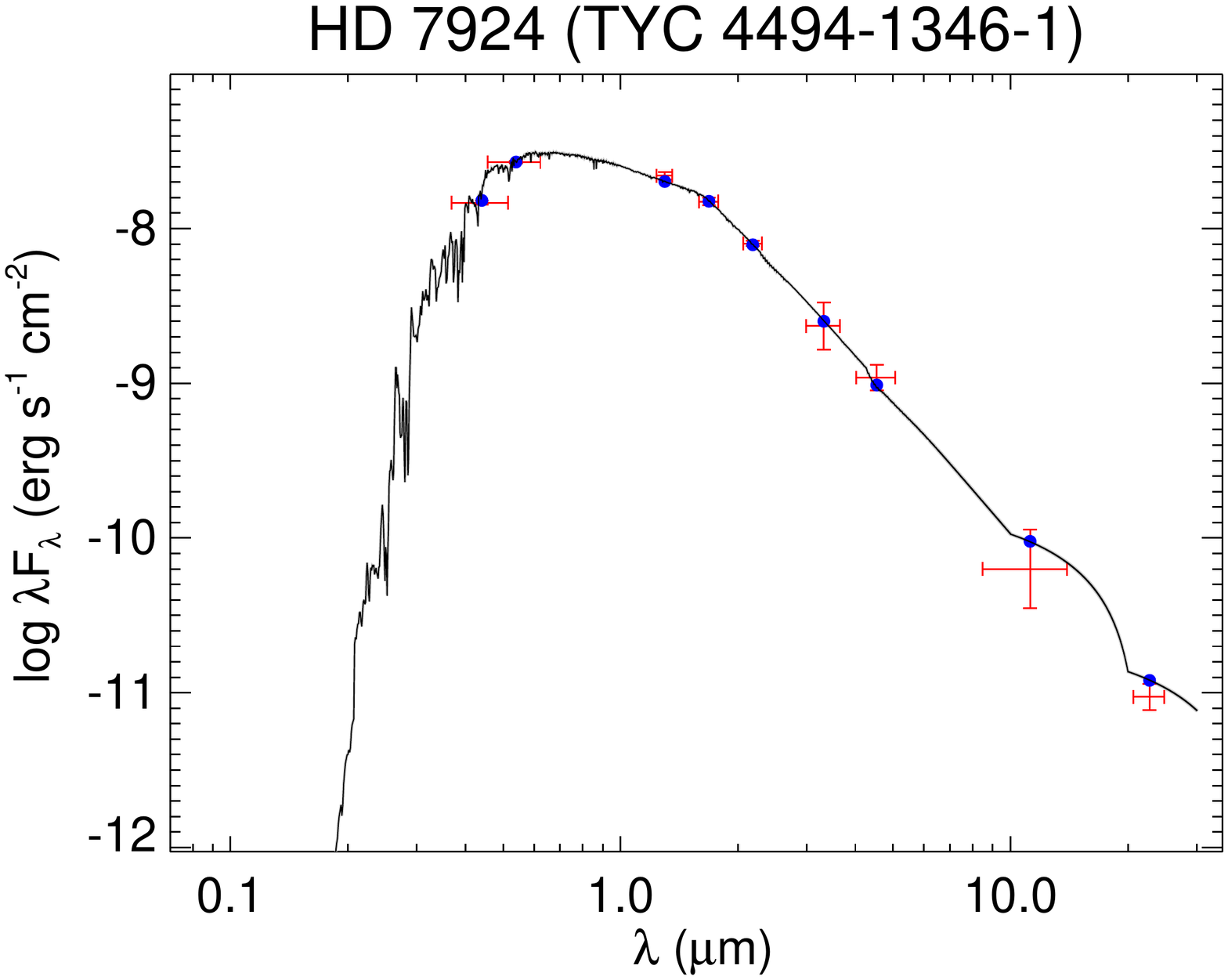}
  \includegraphics[trim=60 60 60 60,clip,width=0.49\linewidth]{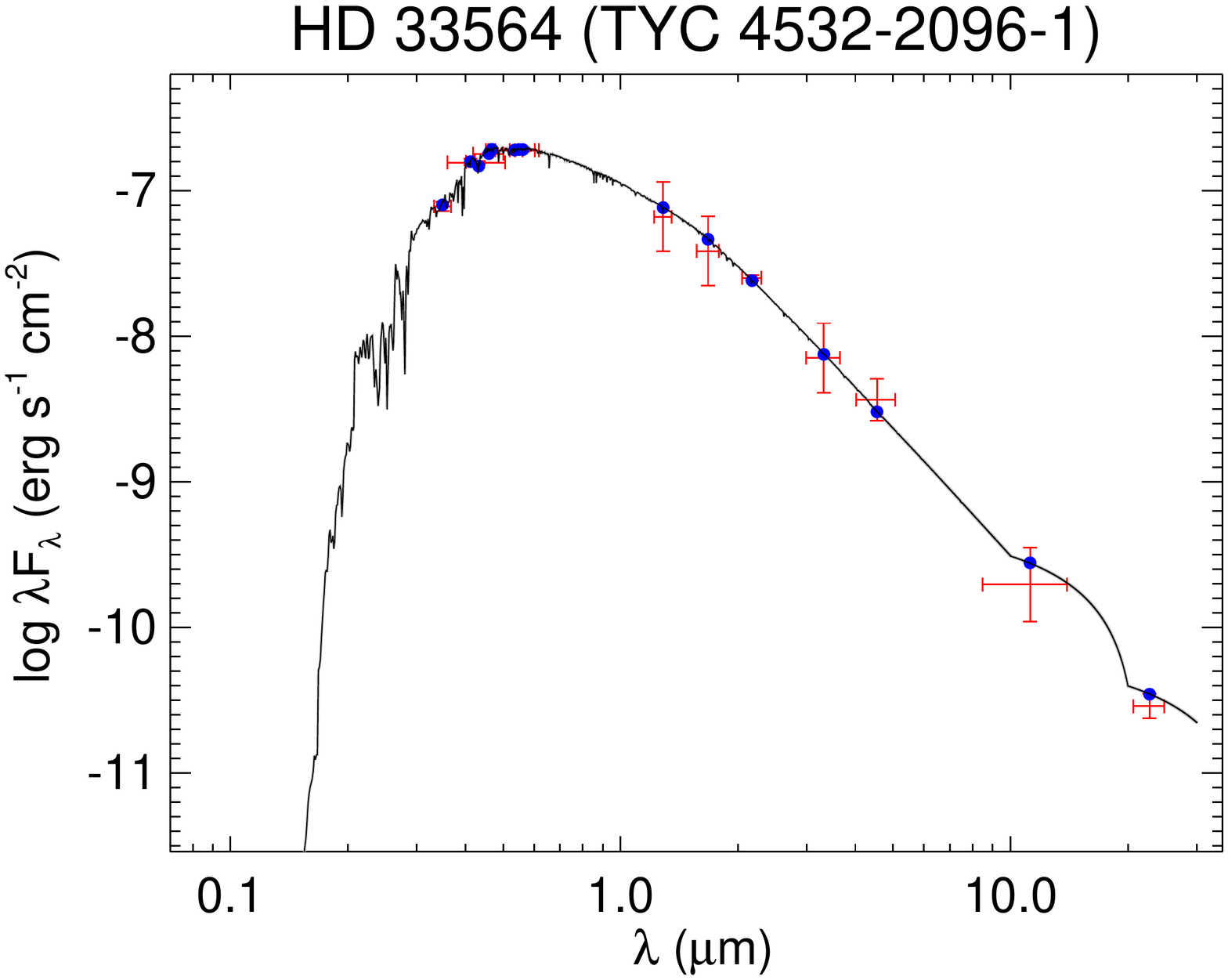}
  \includegraphics[trim=60 60 60 60,clip,width=0.49\linewidth]{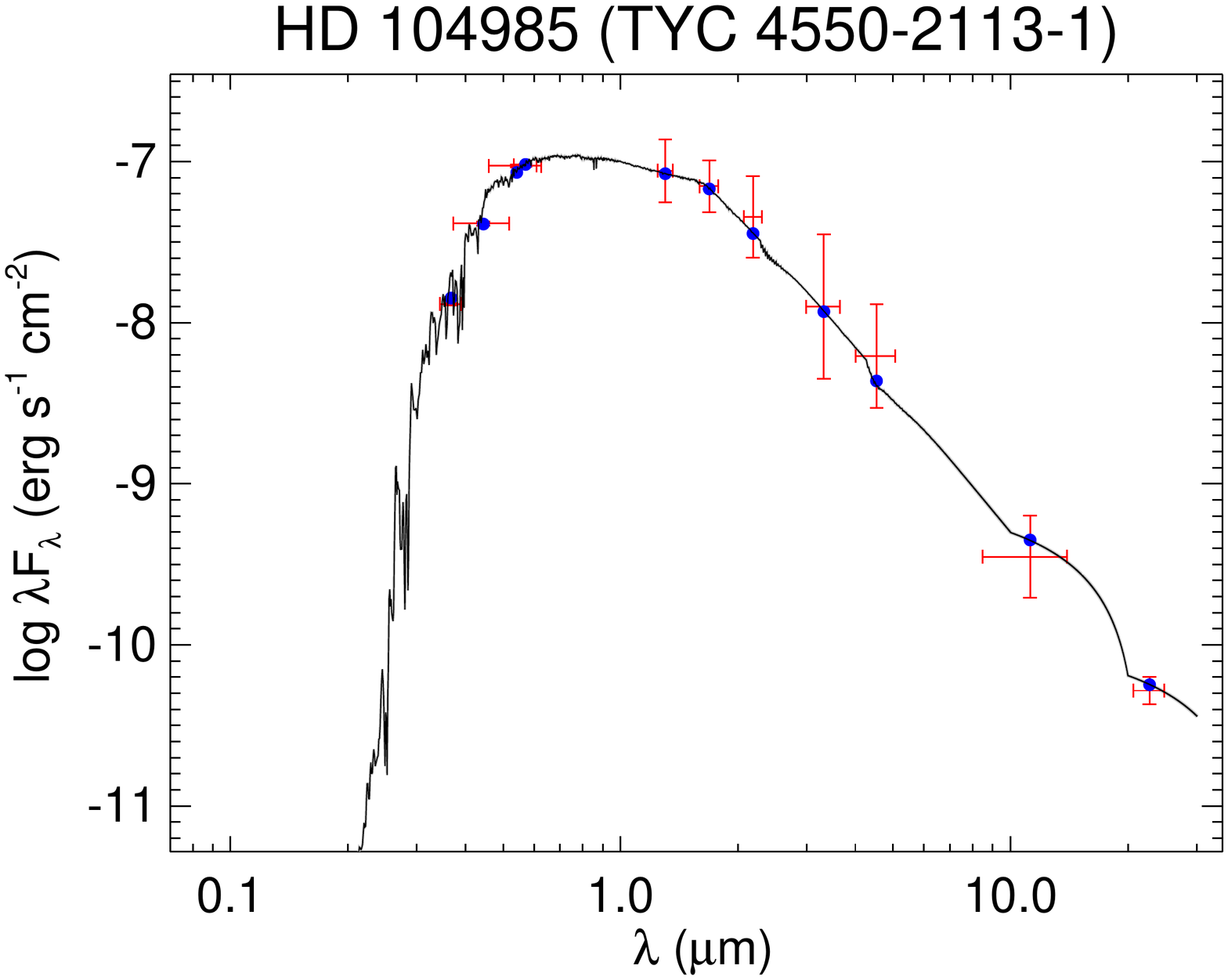}
  \includegraphics[trim=60 60 60 60,clip,width=0.49\linewidth]{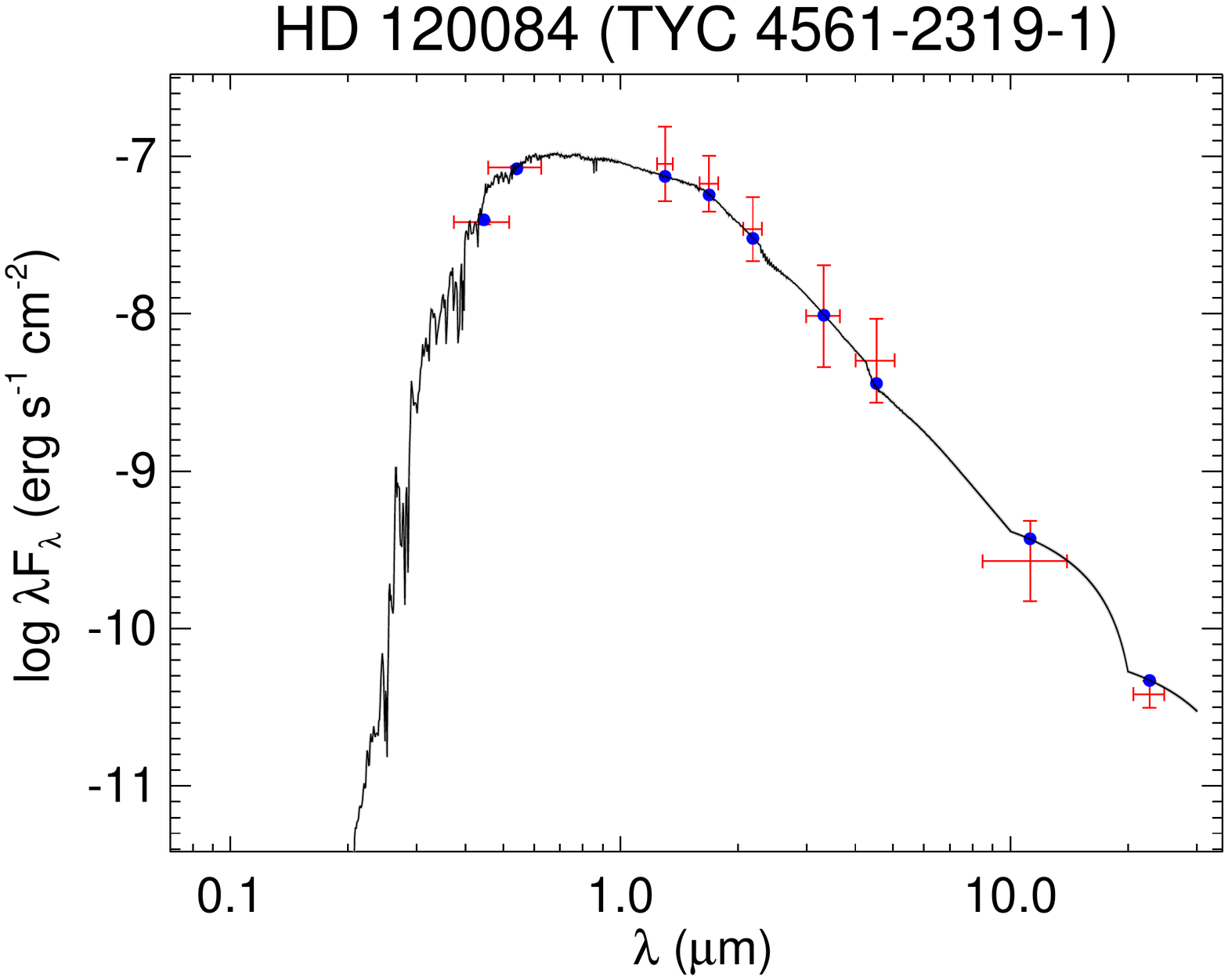}
  \caption{All labels, lines, symbols, and colors as in Figure \ref{fig:seds}.}
  \label{fig:seds_42}
\end{figure}

\begin{figure}[H]
  \centering
  \includegraphics[trim=60 60 60 60,clip,width=0.49\linewidth]{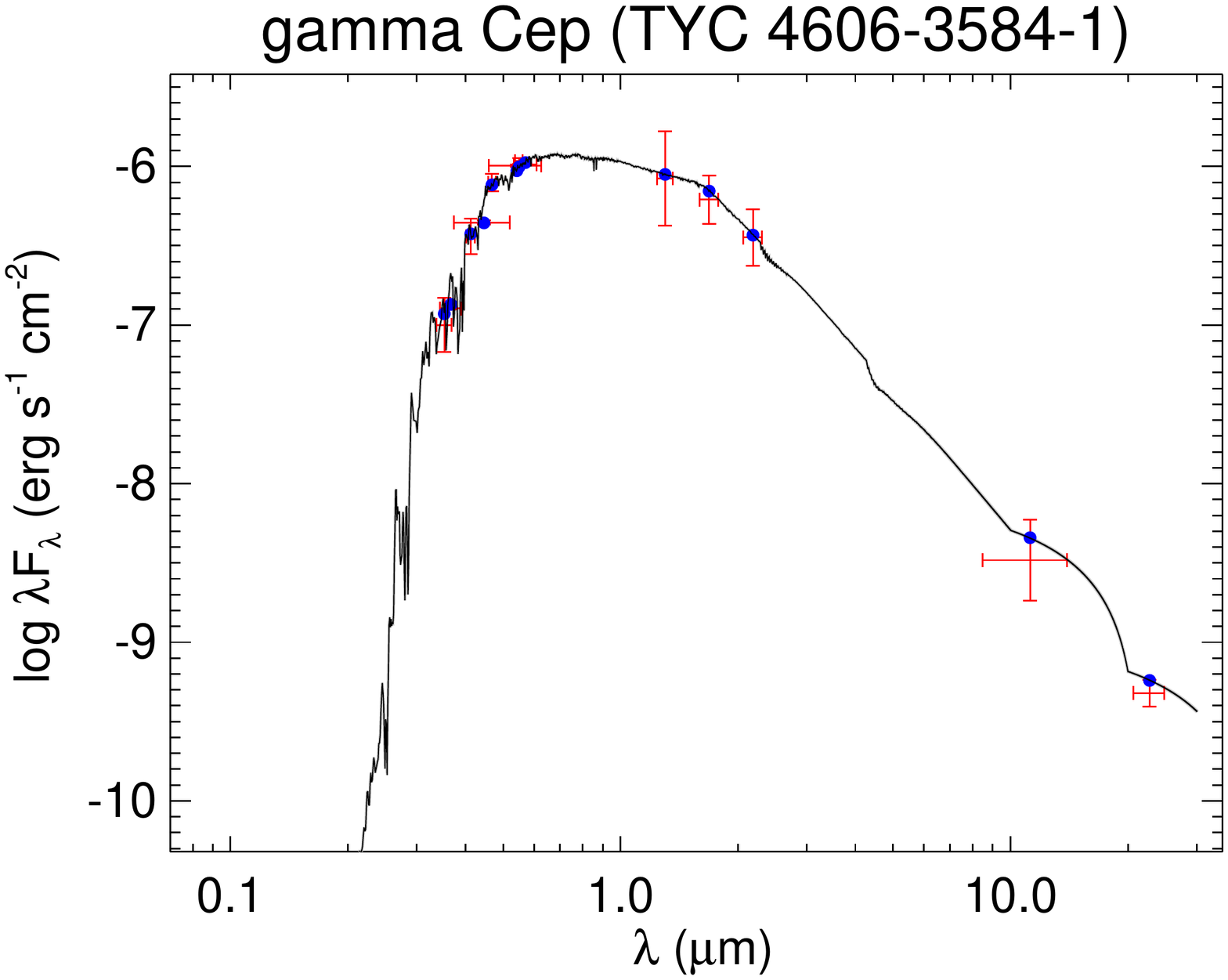}
  \includegraphics[trim=60 60 60 60,clip,width=0.49\linewidth]{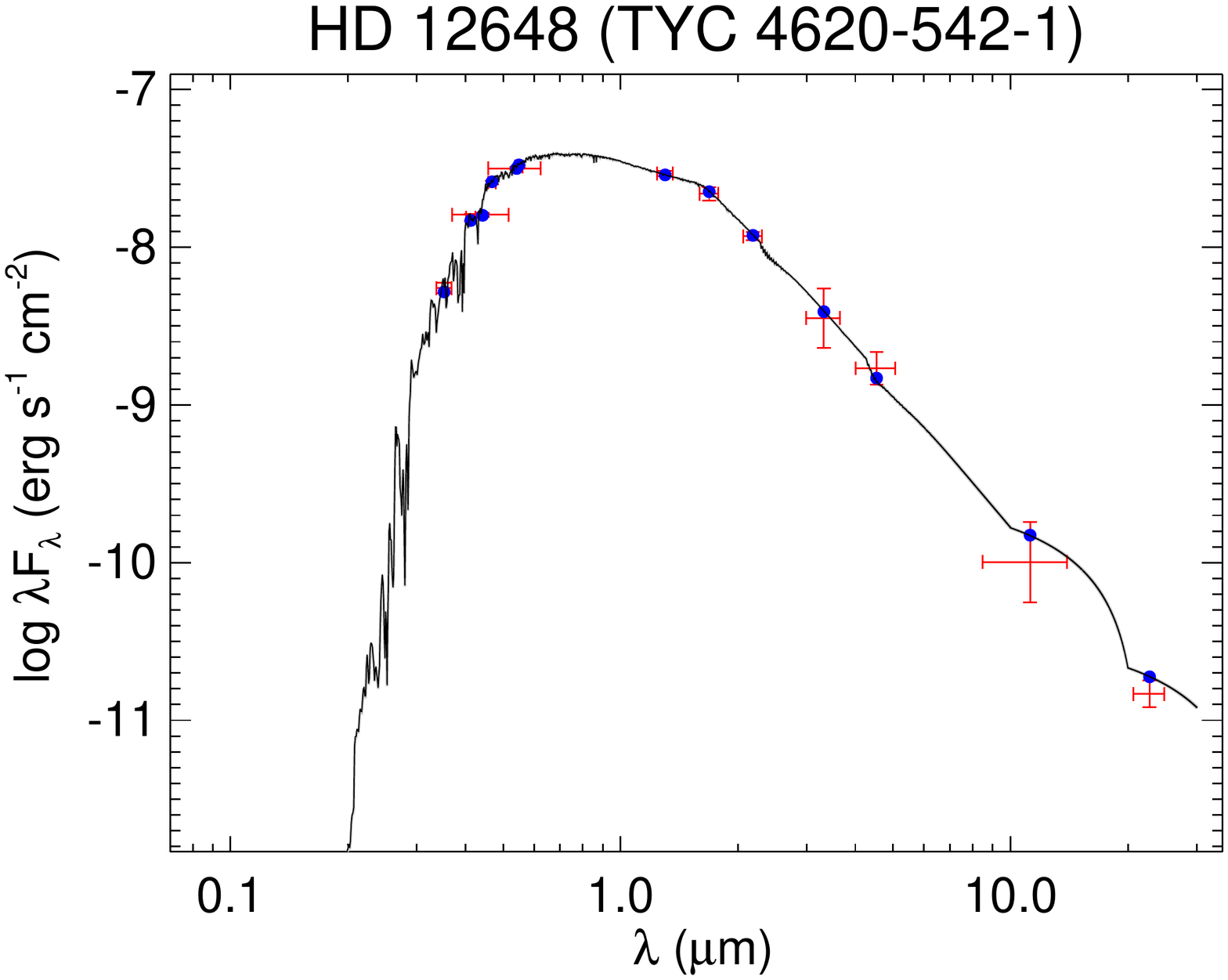}
  \includegraphics[trim=60 60 60 60,clip,width=0.49\linewidth]{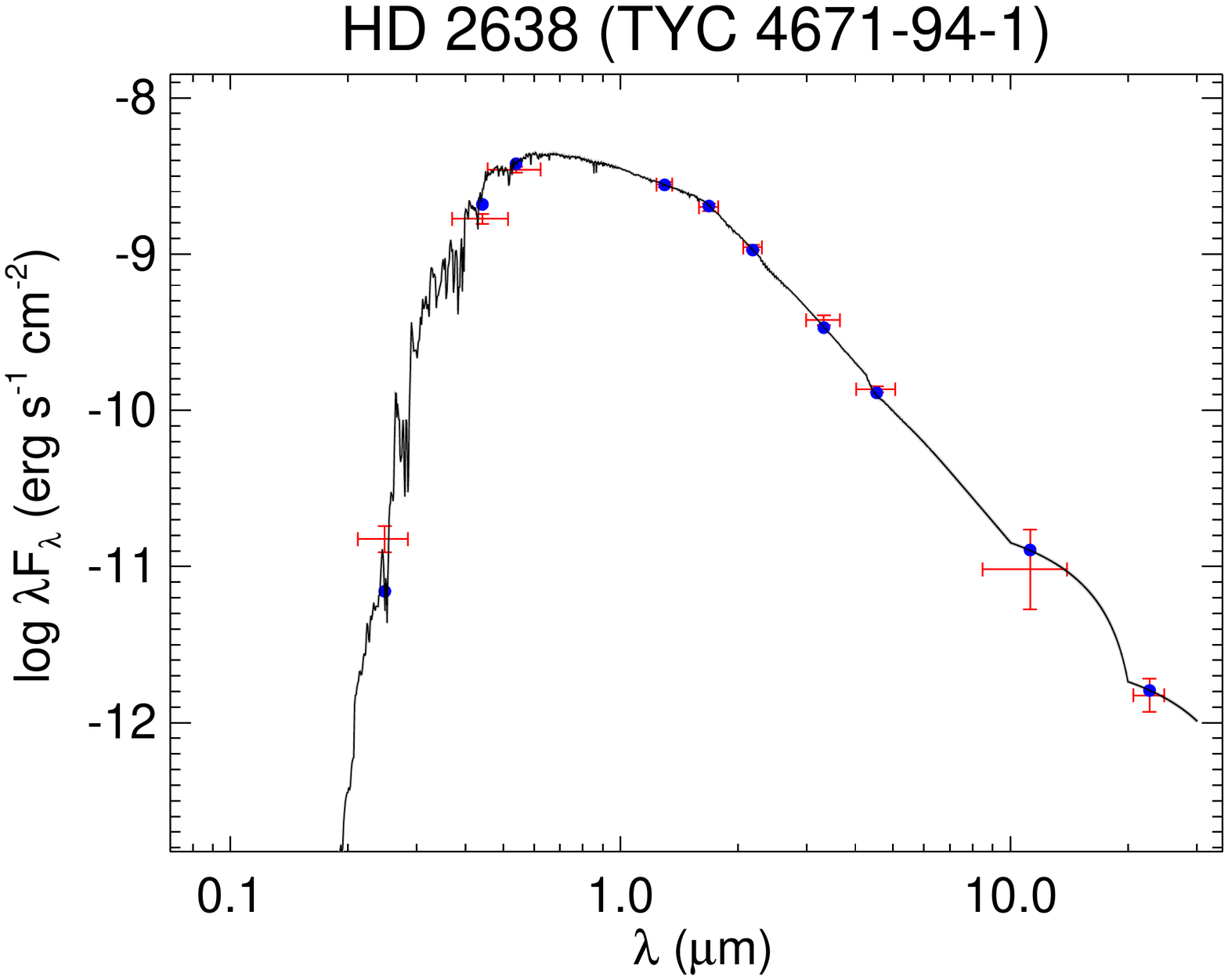}
  \includegraphics[trim=60 60 60 60,clip,width=0.49\linewidth]{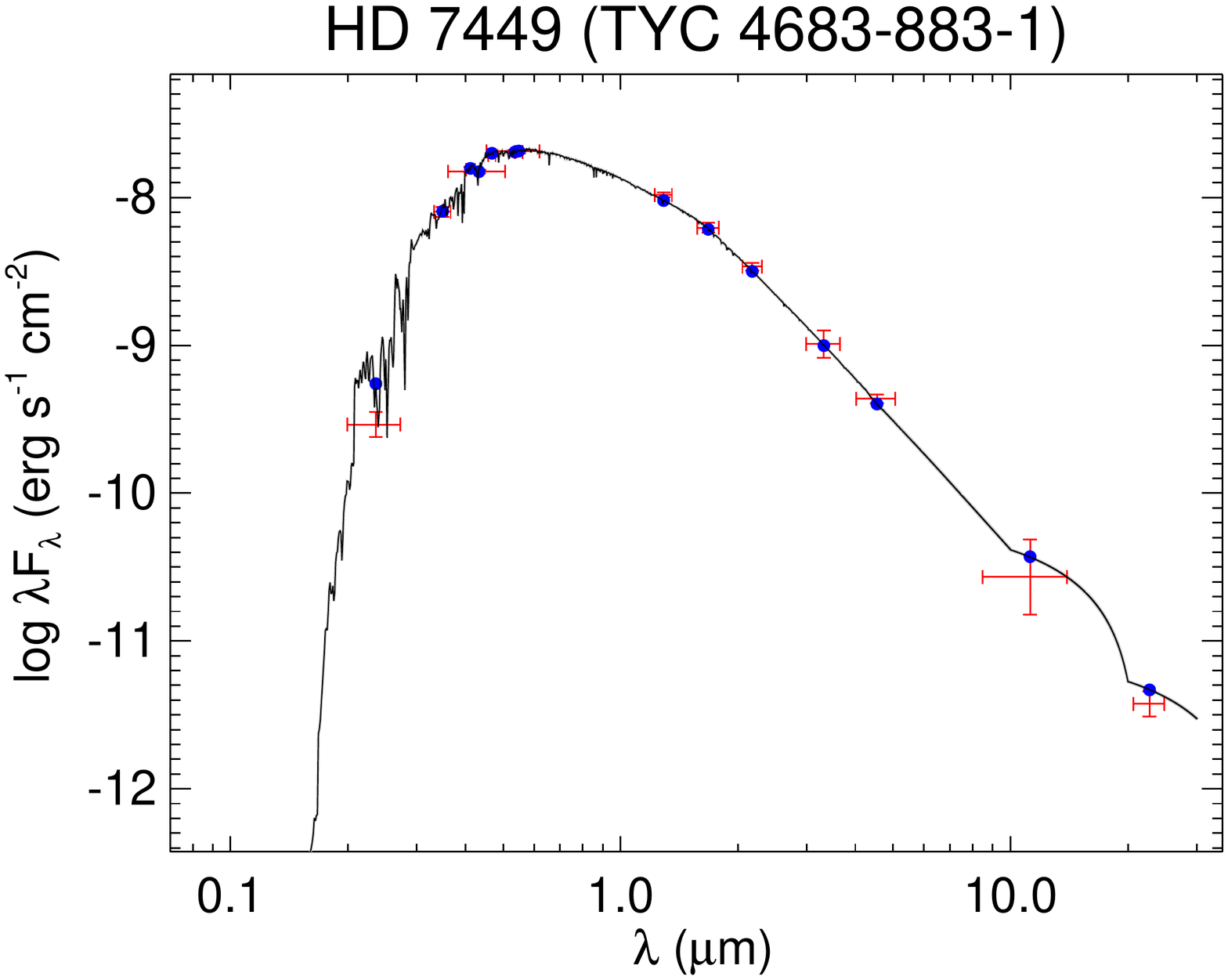}
  \includegraphics[trim=60 60 60 60,clip,width=0.49\linewidth]{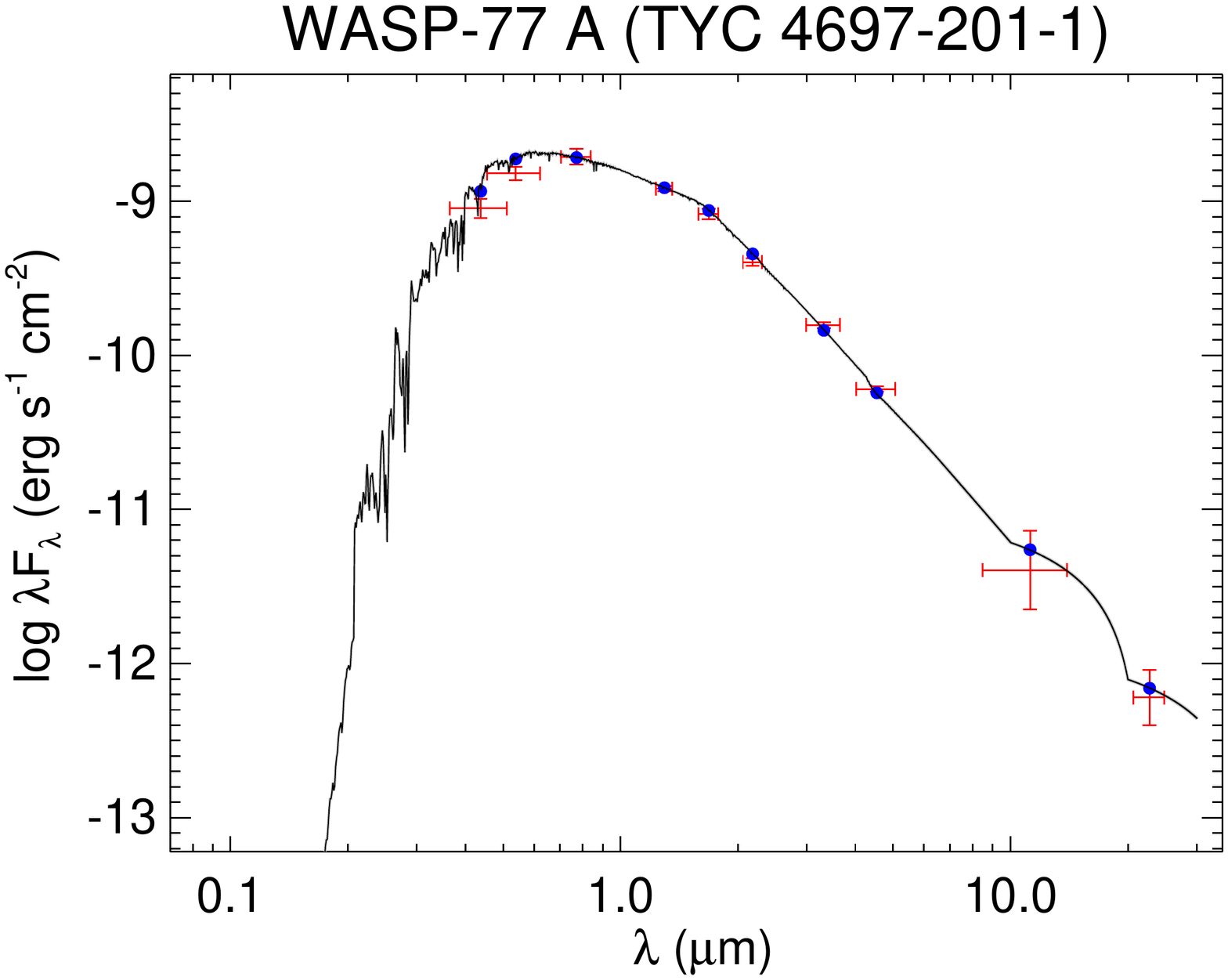}
  \includegraphics[trim=60 60 60 60,clip,width=0.49\linewidth]{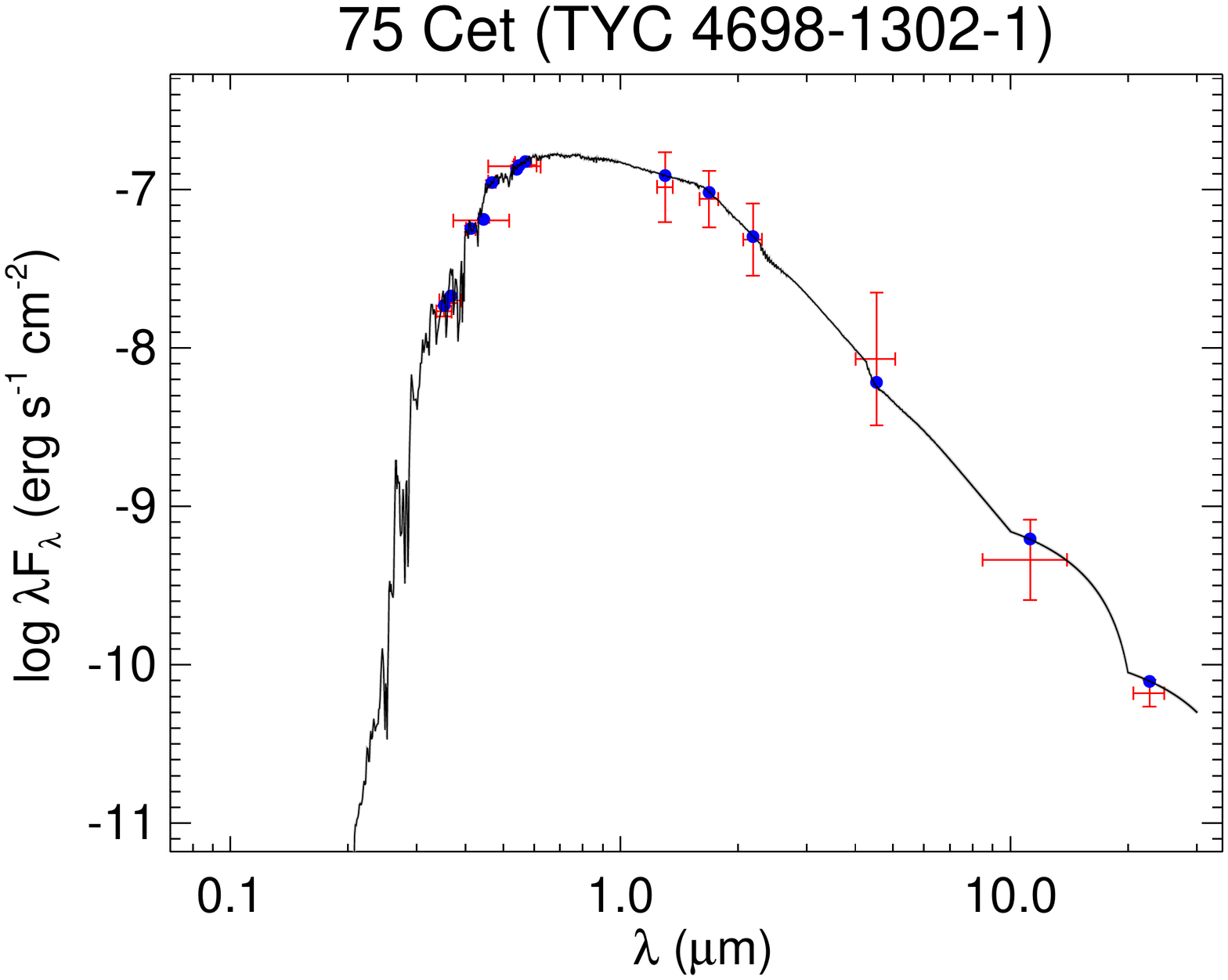}
  \caption{All labels, lines, symbols, and colors as in Figure \ref{fig:seds}.}
  \label{fig:seds_43}
\end{figure}

\begin{figure}[H]
  \centering
  \includegraphics[trim=60 60 60 60,clip,width=0.49\linewidth]{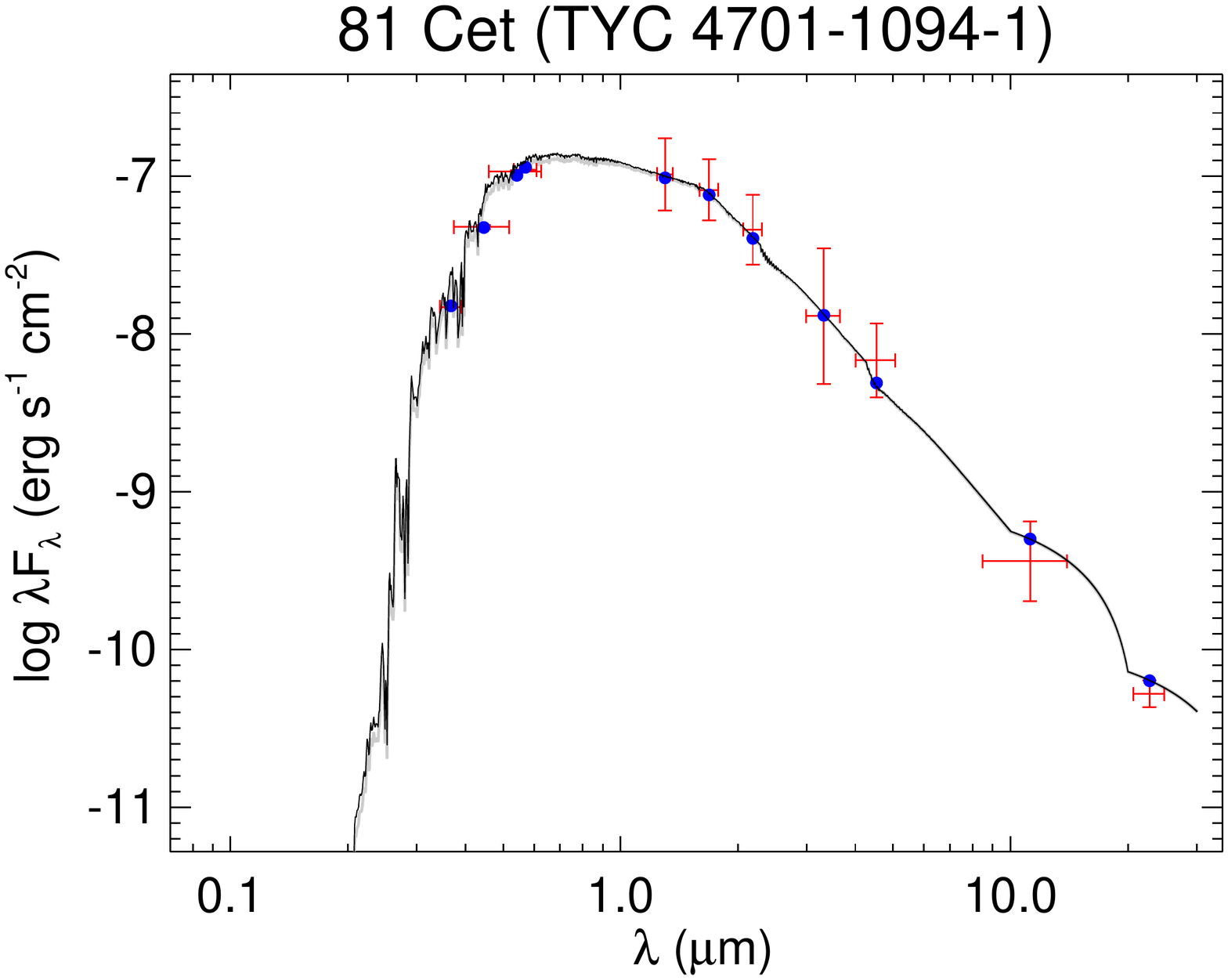}
  \includegraphics[trim=60 60 60 60,clip,width=0.49\linewidth]{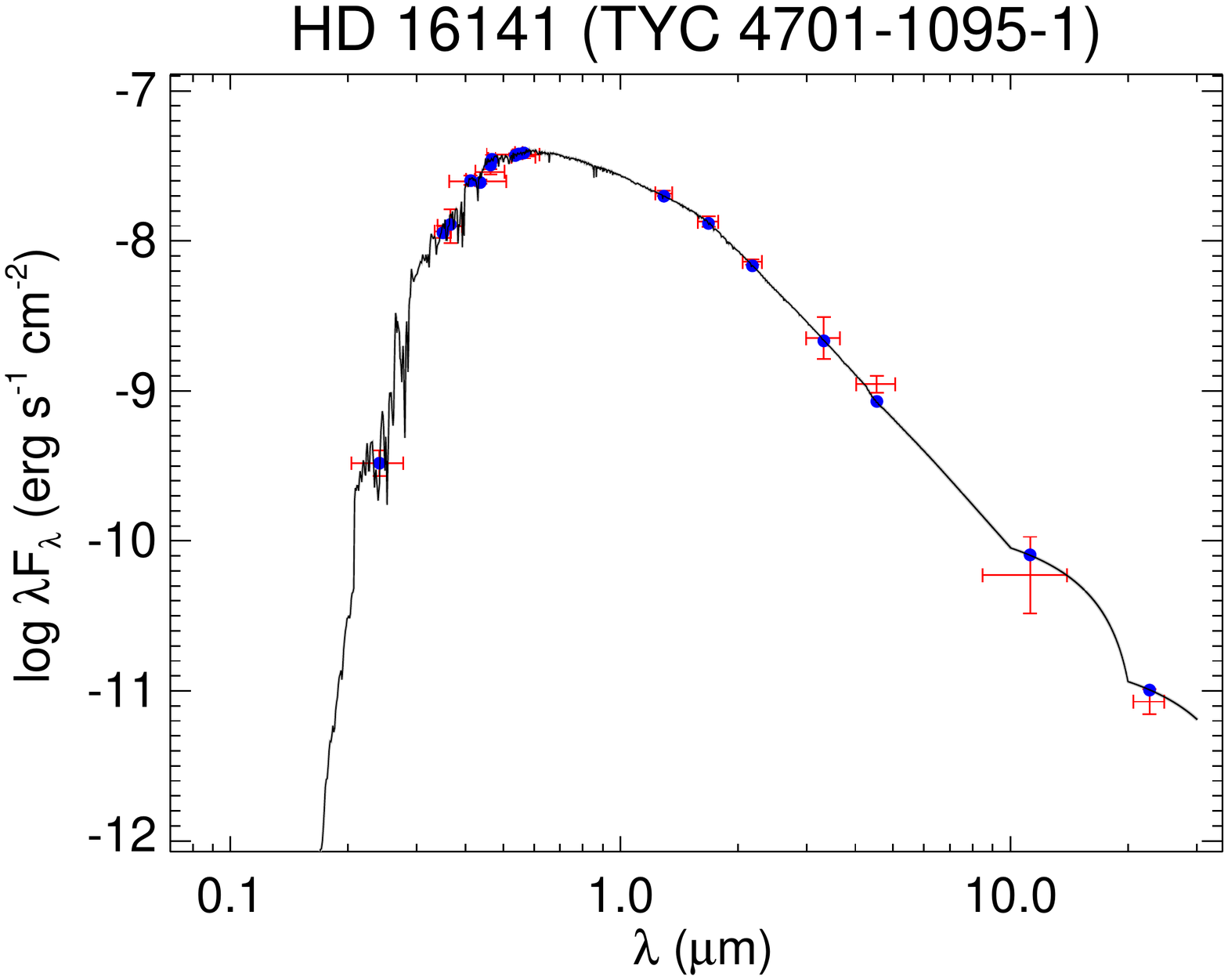}
  \includegraphics[trim=60 60 60 60,clip,width=0.49\linewidth]{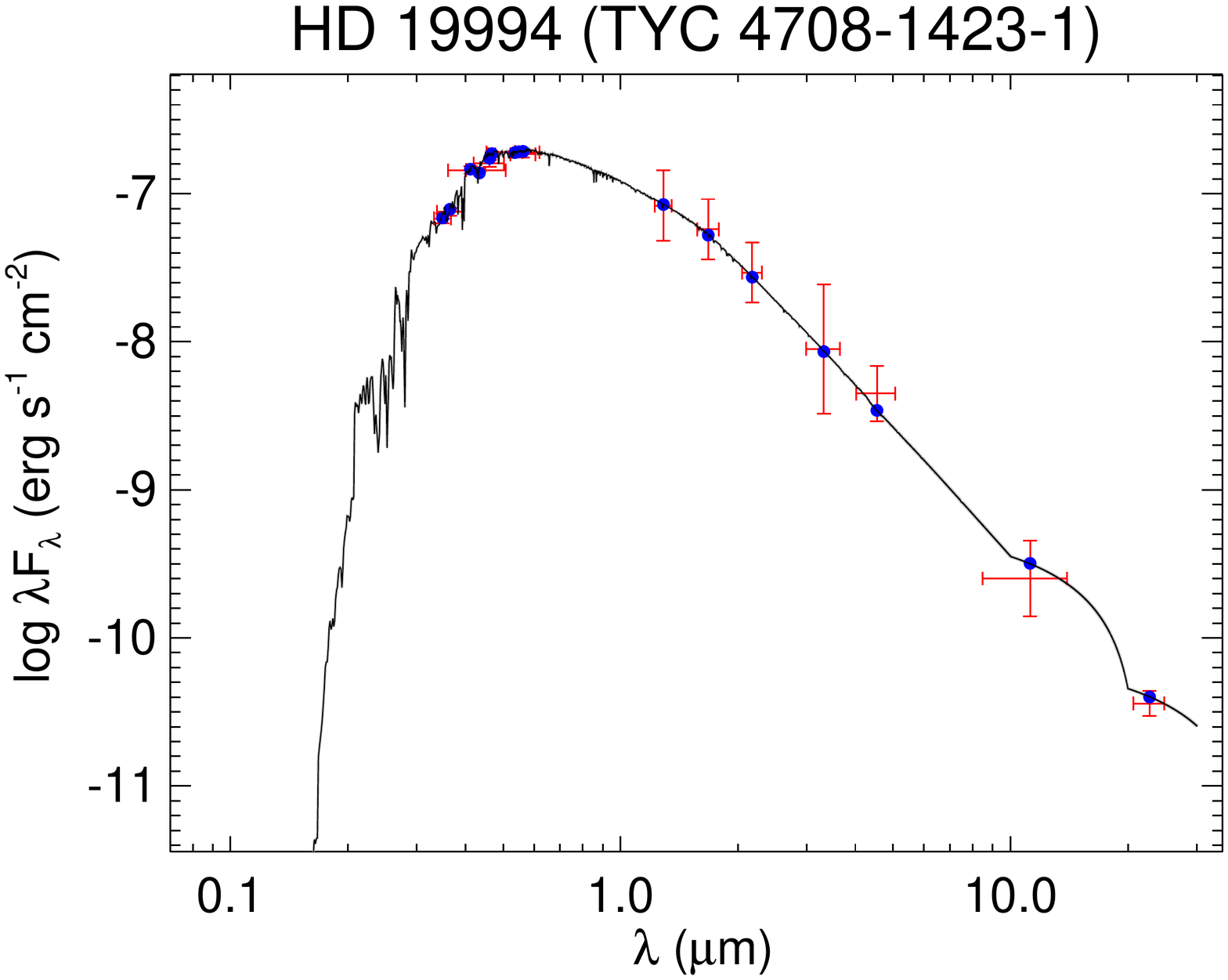}
  \includegraphics[trim=60 60 60 60,clip,width=0.49\linewidth]{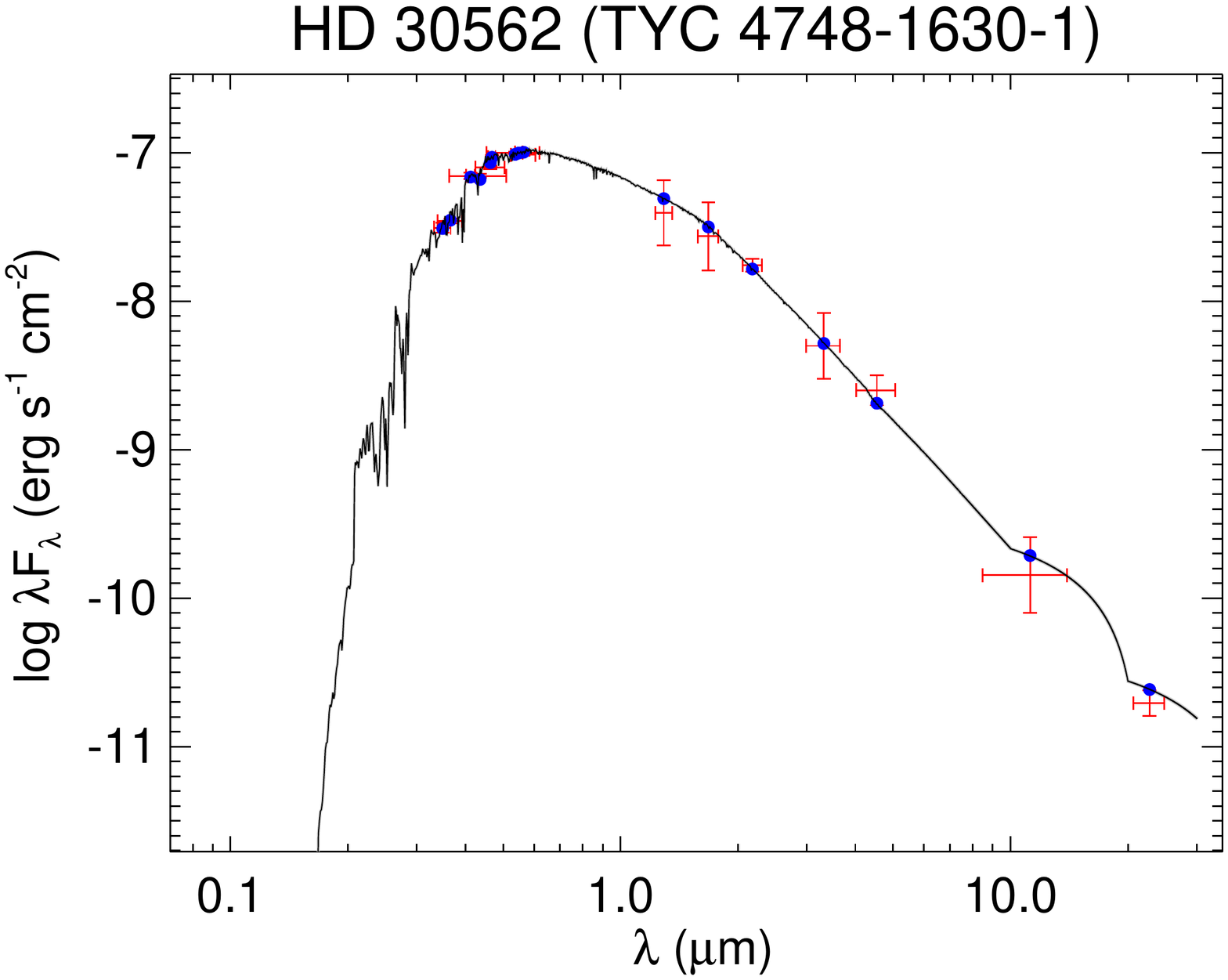}
  \includegraphics[trim=60 60 60 60,clip,width=0.49\linewidth]{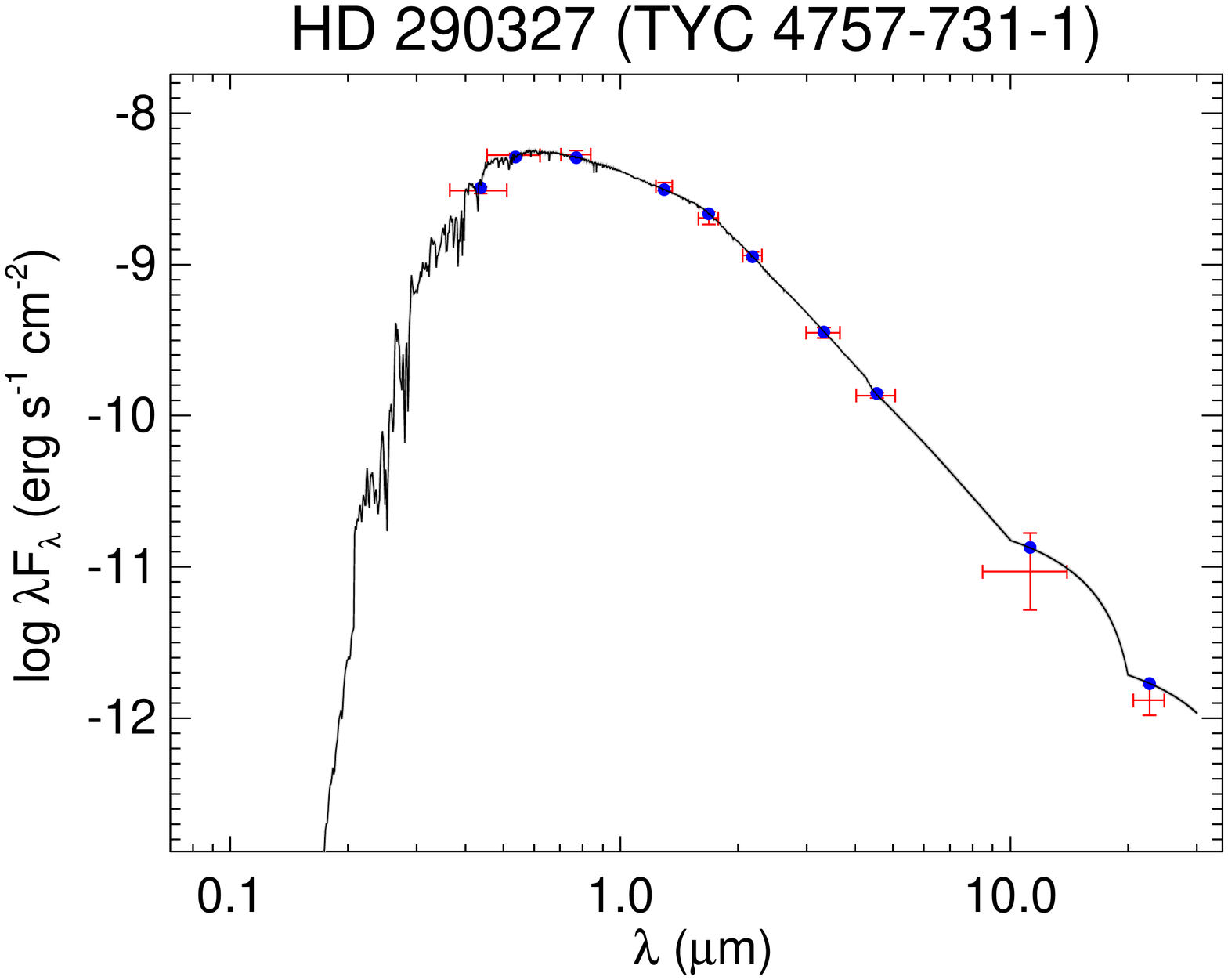}
  \includegraphics[trim=60 60 60 60,clip,width=0.49\linewidth]{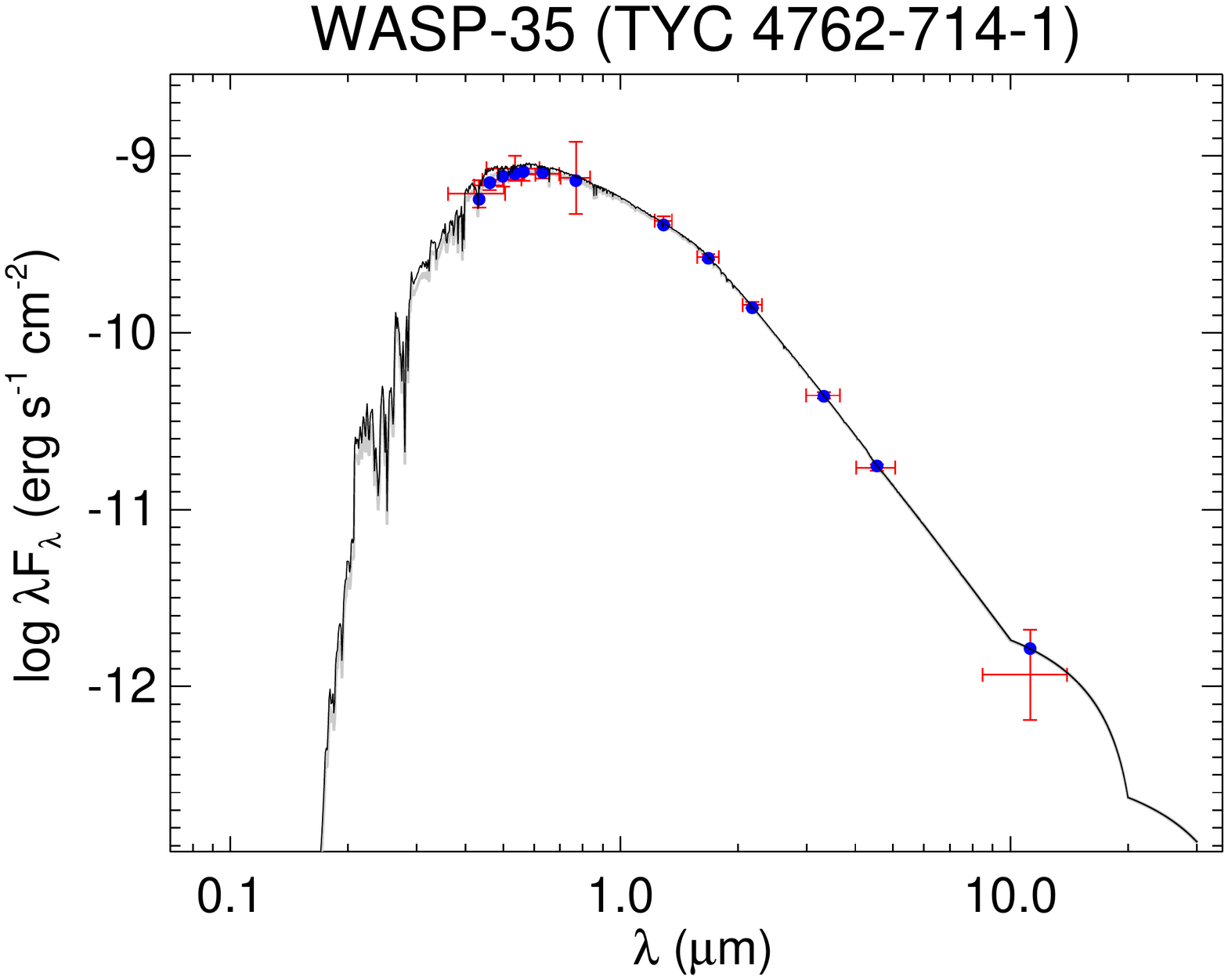}
  \caption{All labels, lines, symbols, and colors as in Figure \ref{fig:seds}.}
  \label{fig:seds_44}
\end{figure}

\begin{figure}[H]
  \centering
  \includegraphics[trim=60 60 60 60,clip,width=0.49\linewidth]{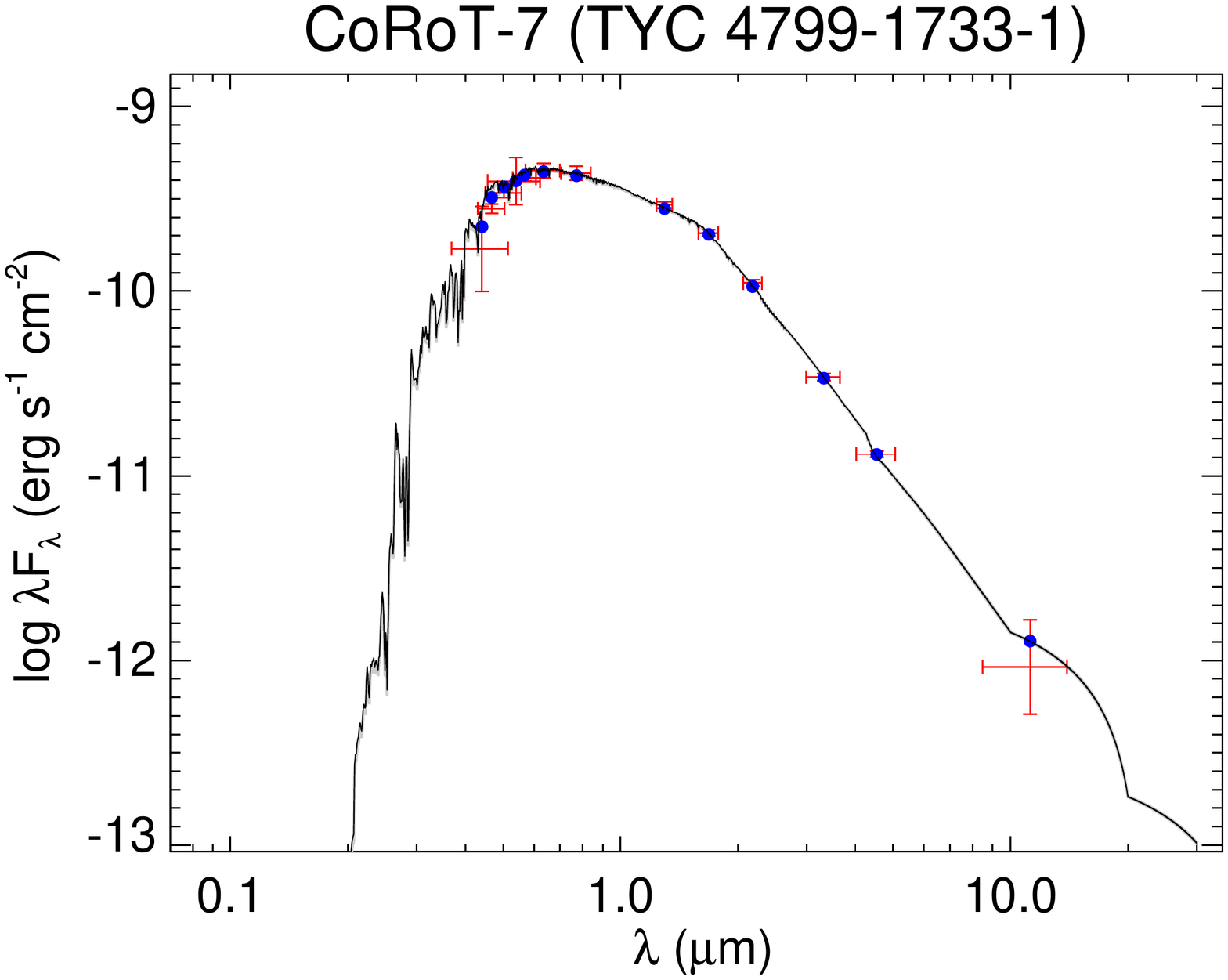}
  \includegraphics[trim=60 60 60 60,clip,width=0.49\linewidth]{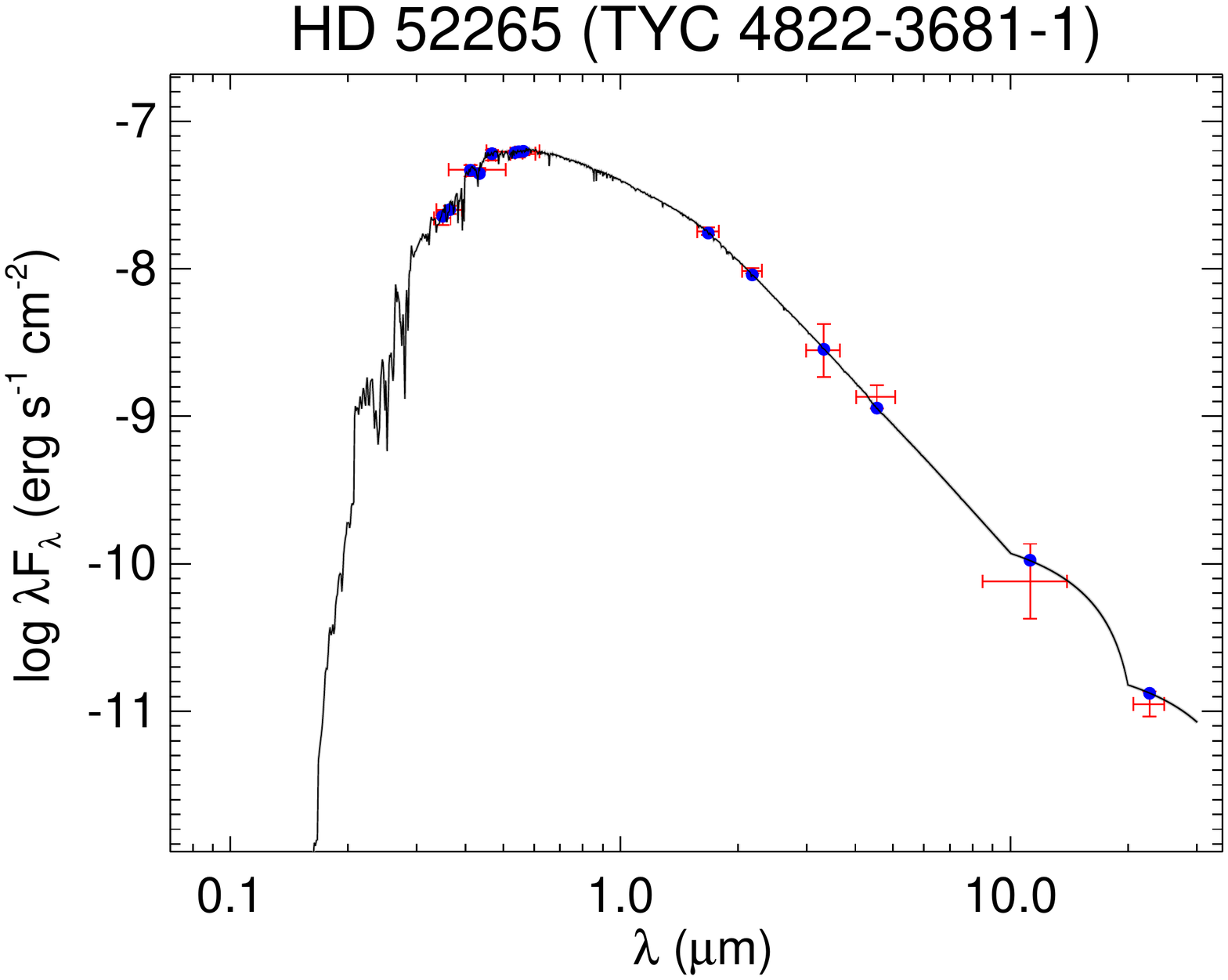}
  \includegraphics[trim=60 60 60 60,clip,width=0.49\linewidth]{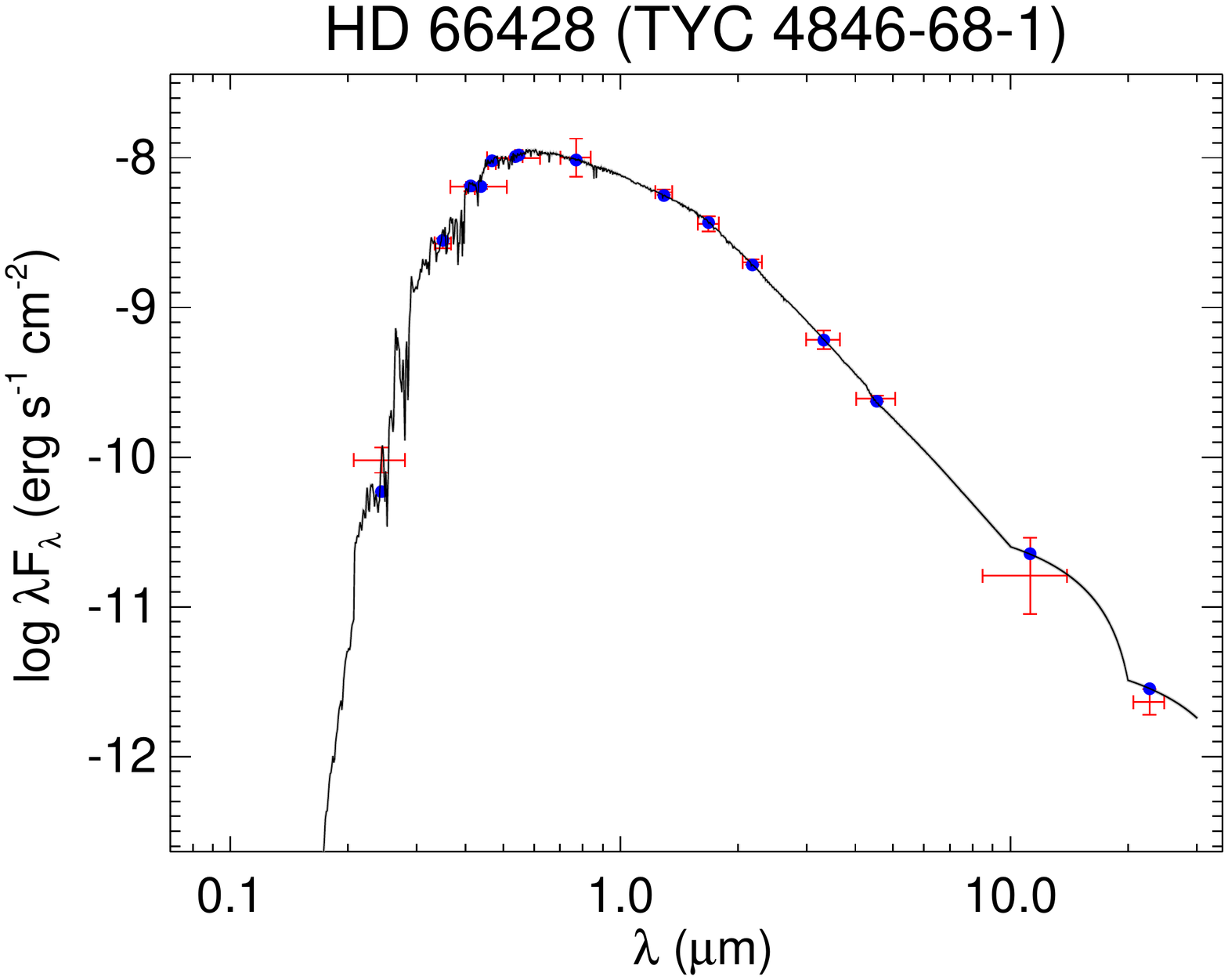}
  \includegraphics[trim=60 60 60 60,clip,width=0.49\linewidth]{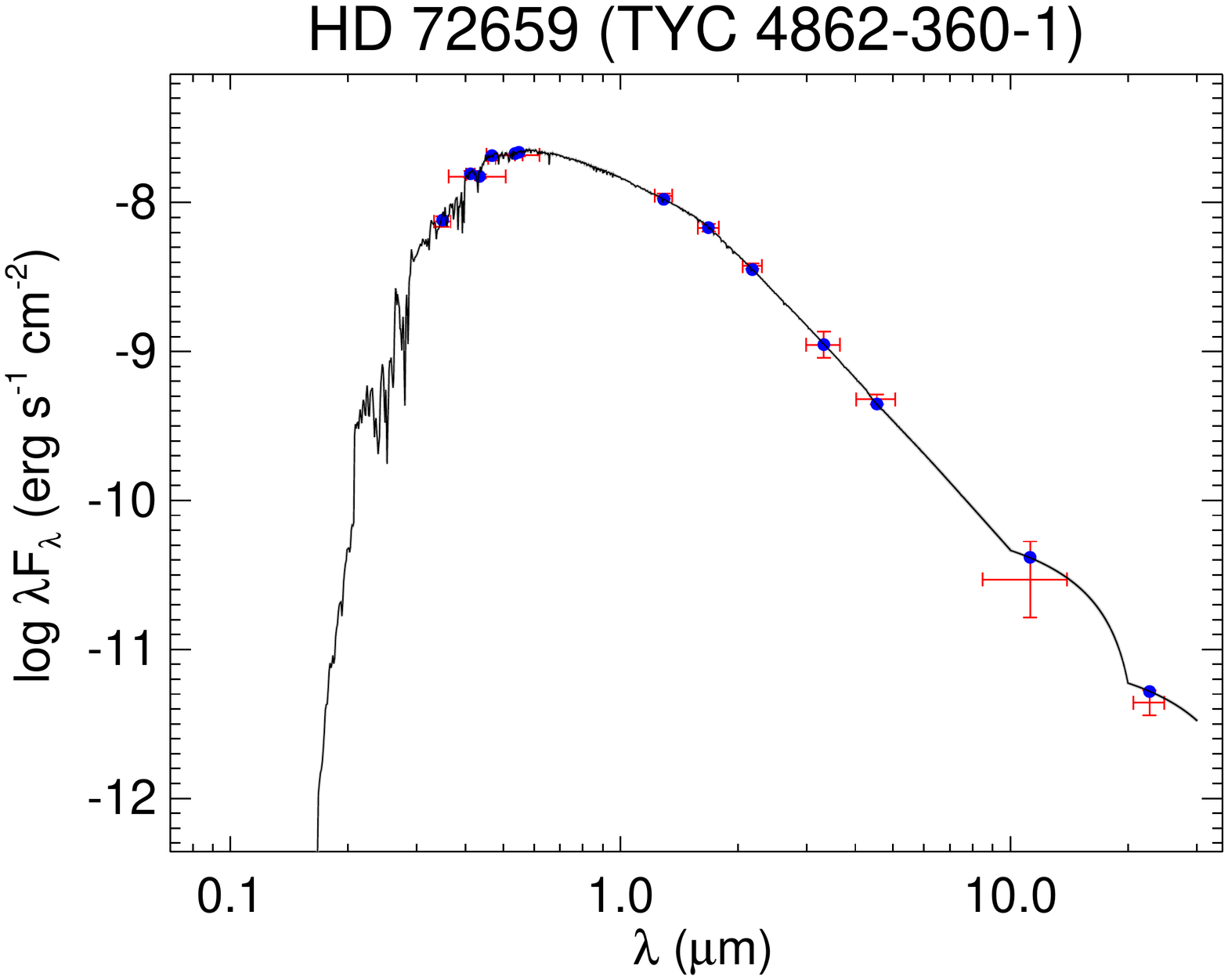}
  \includegraphics[trim=60 60 60 60,clip,width=0.49\linewidth]{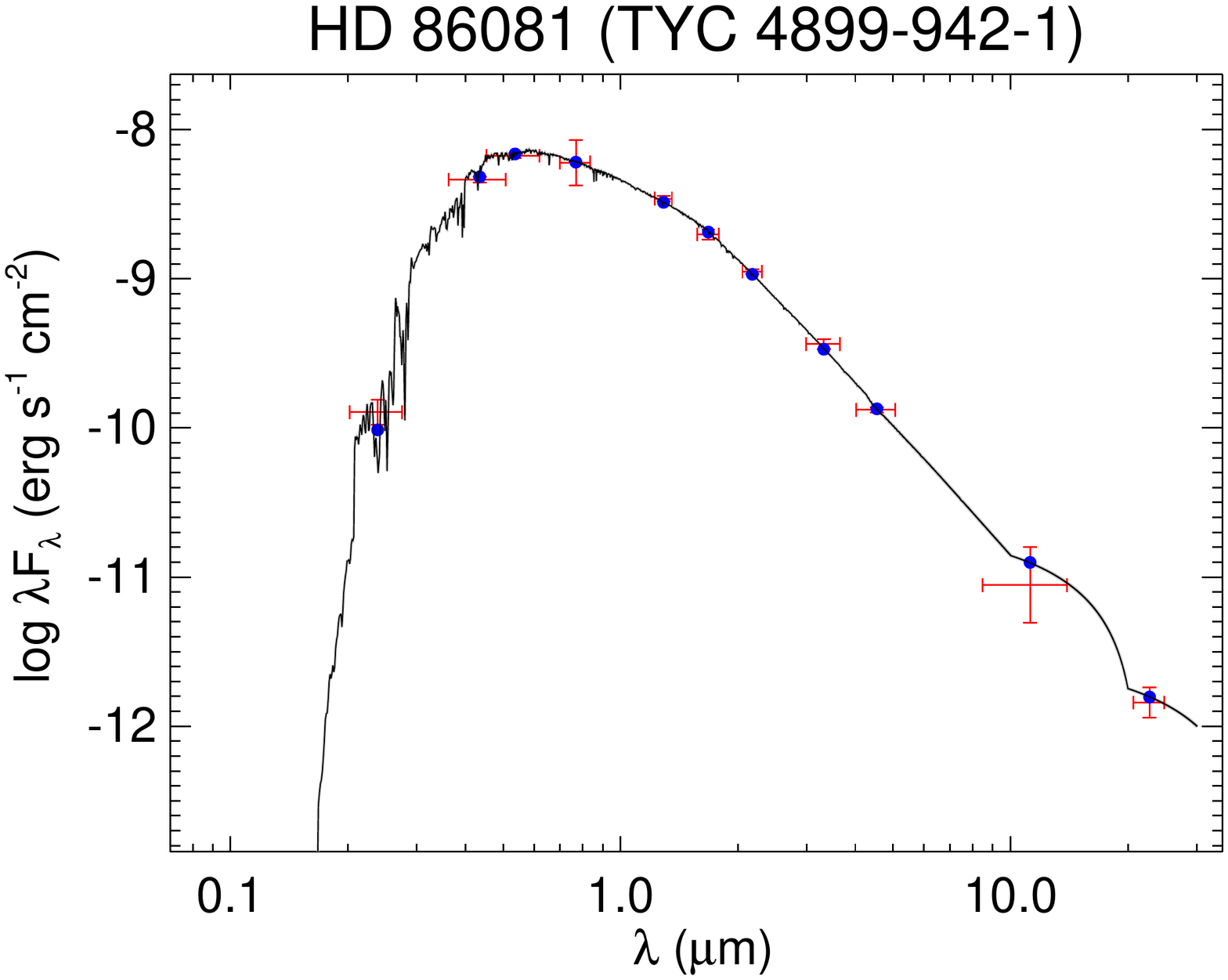}
  \includegraphics[trim=60 60 60 60,clip,width=0.49\linewidth]{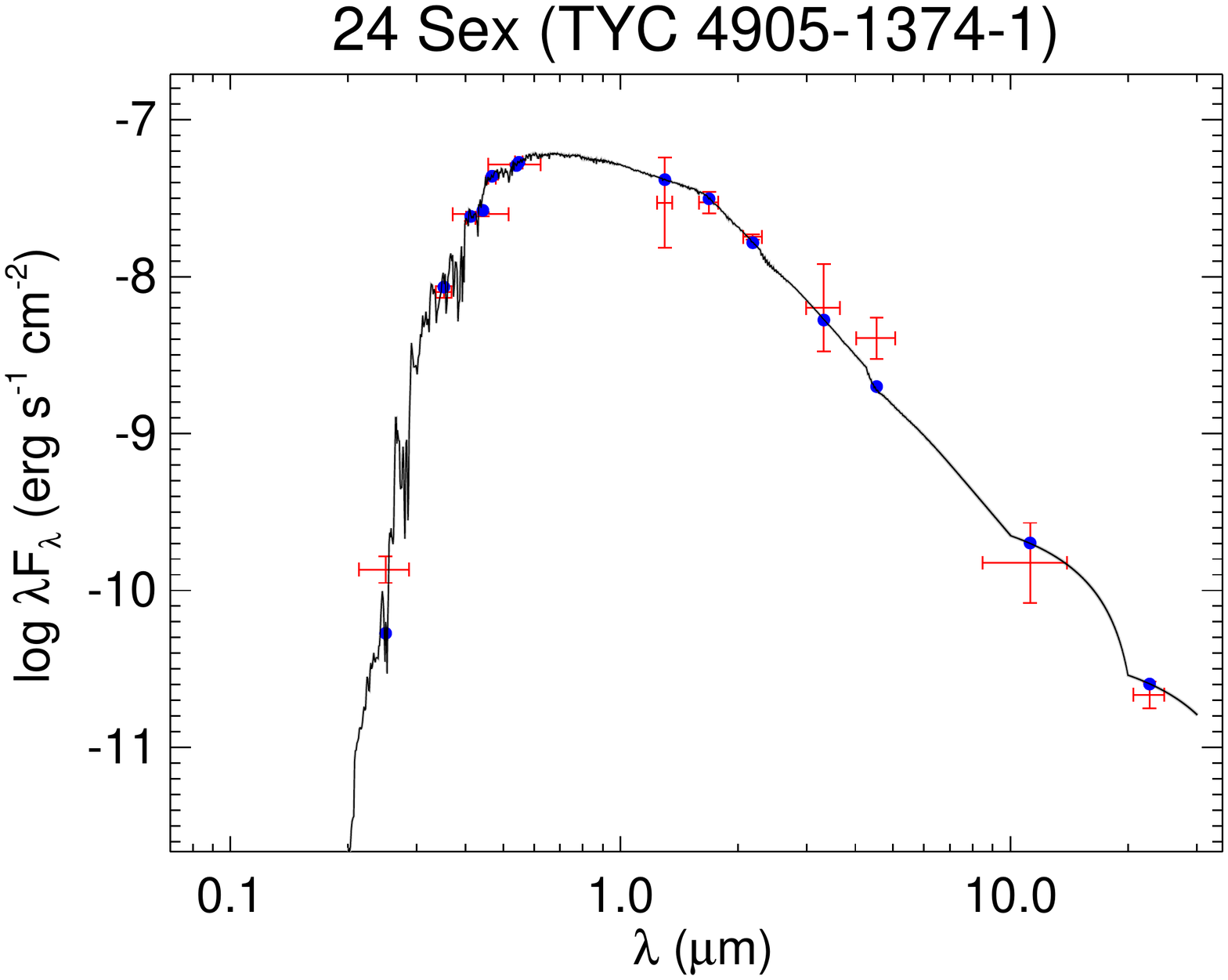}
  \caption{All labels, lines, symbols, and colors as in Figure \ref{fig:seds}.}
  \label{fig:seds_45}
\end{figure}

\begin{figure}[H]
  \centering
  \includegraphics[trim=60 60 60 60,clip,width=0.49\linewidth]{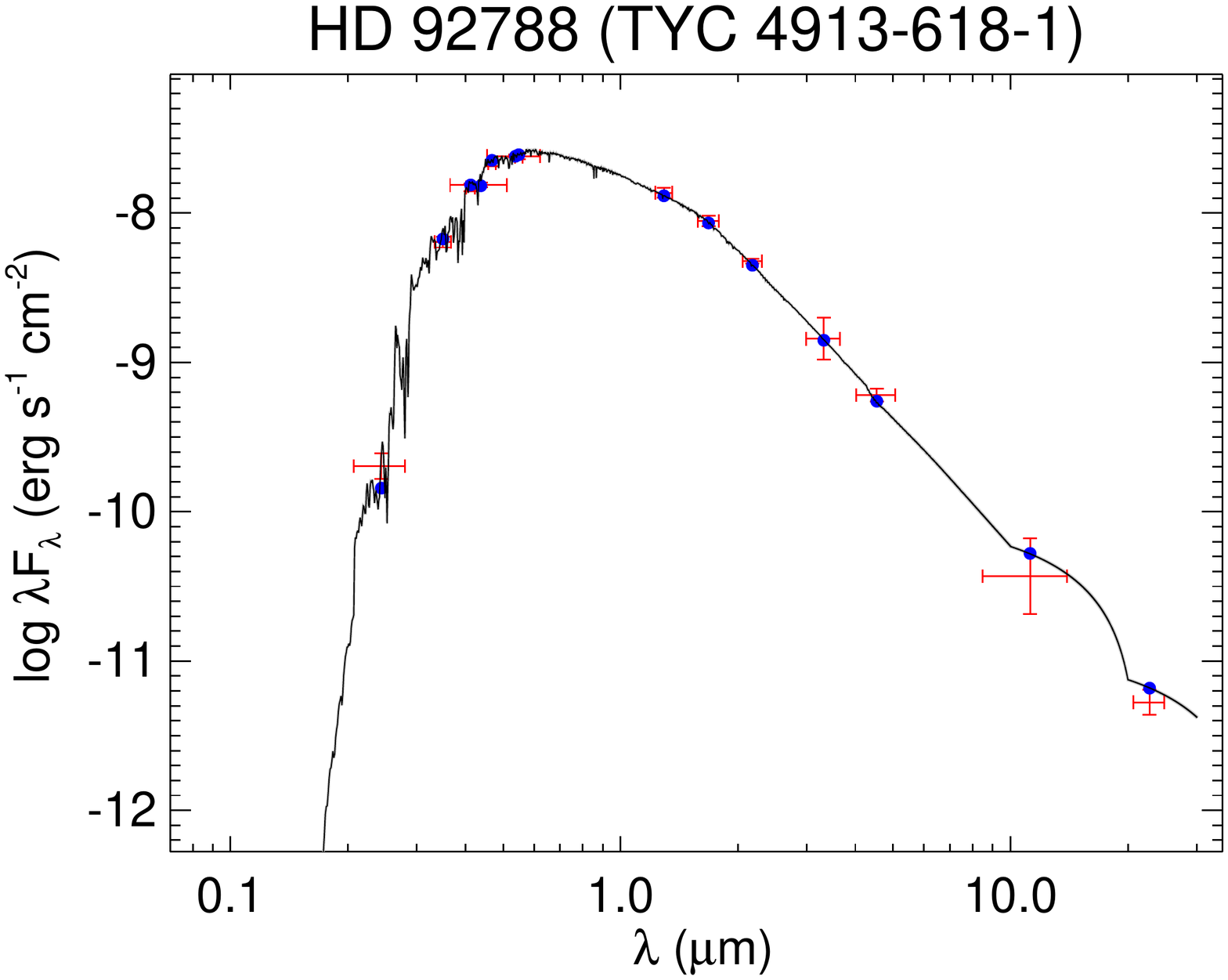}
  \includegraphics[trim=60 60 60 60,clip,width=0.49\linewidth]{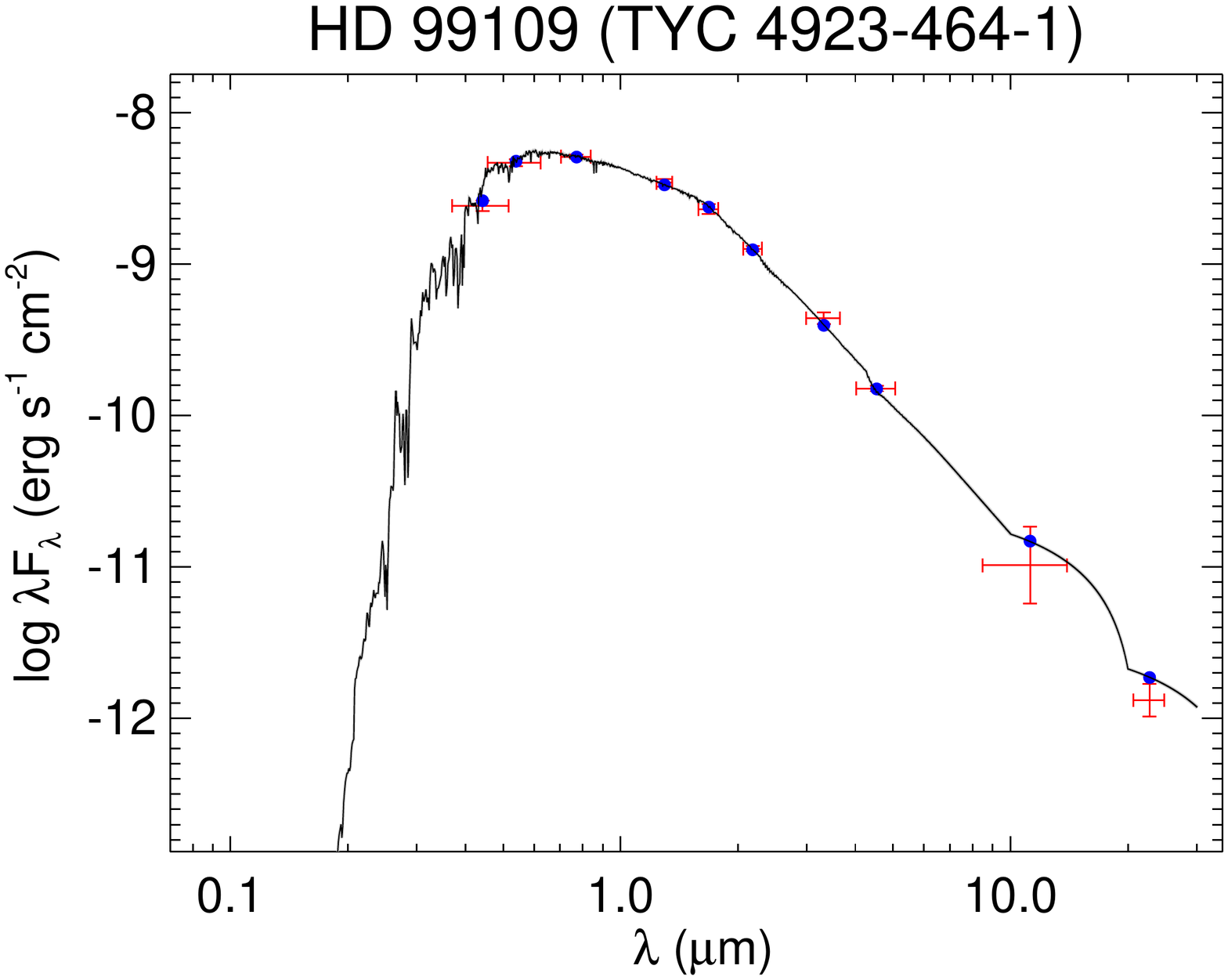}
  \includegraphics[trim=60 60 60 60,clip,width=0.49\linewidth]{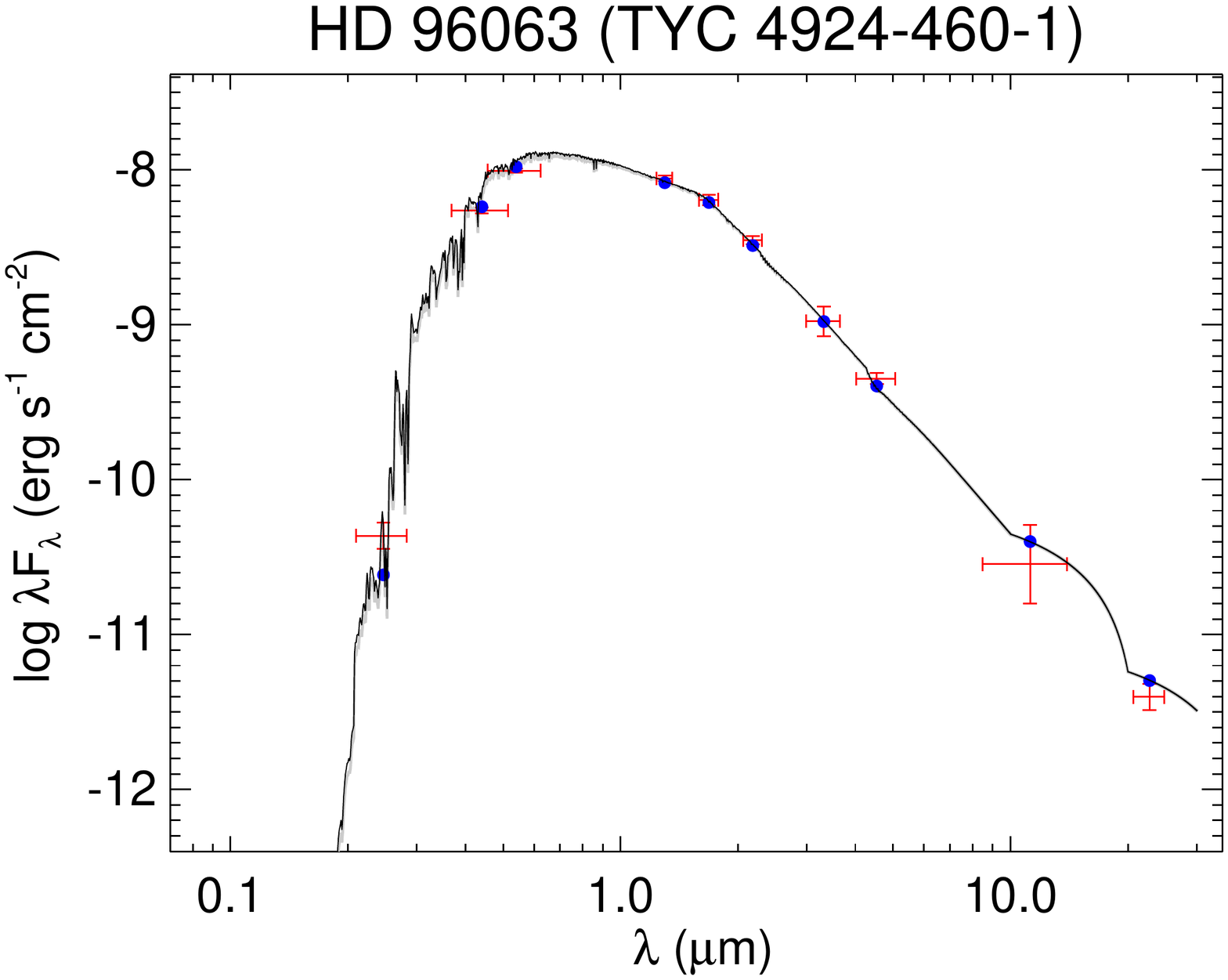}
  \includegraphics[trim=60 60 60 60,clip,width=0.49\linewidth]{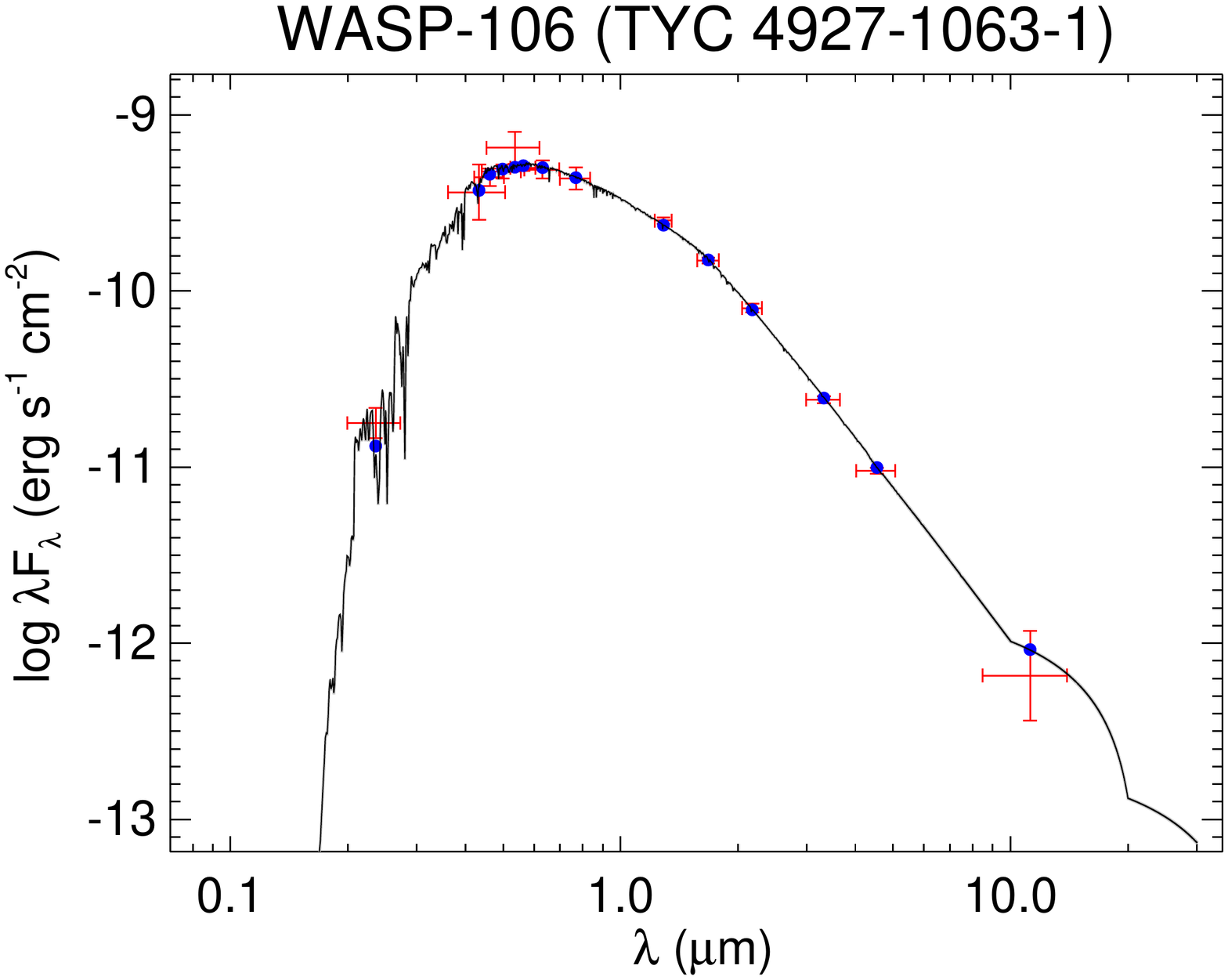}
  \includegraphics[trim=60 60 60 60,clip,width=0.49\linewidth]{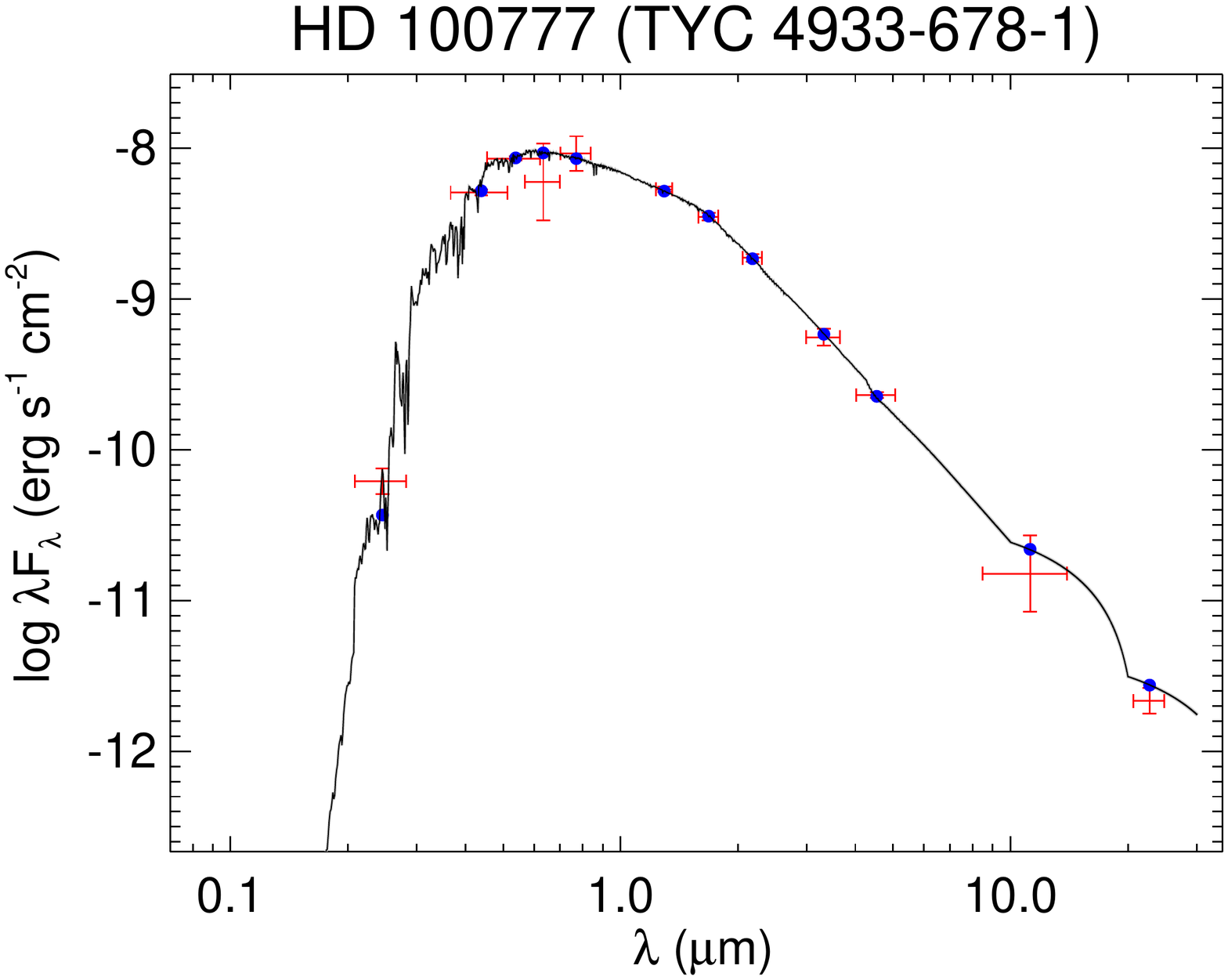}
  \includegraphics[trim=60 60 60 60,clip,width=0.49\linewidth]{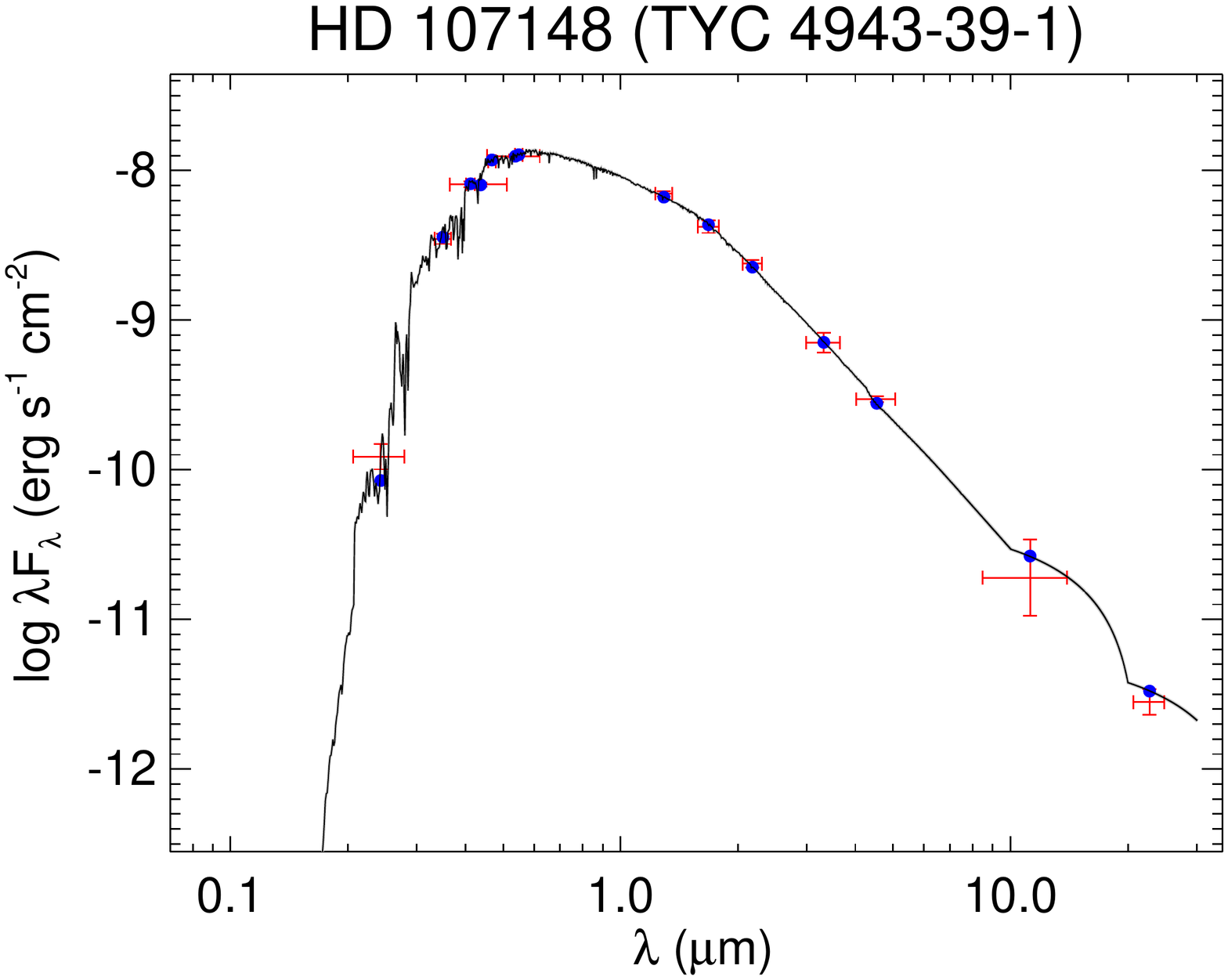}
  \caption{All labels, lines, symbols, and colors as in Figure \ref{fig:seds}.}
  \label{fig:seds_46}
\end{figure}

\begin{figure}[H]
  \centering
  \includegraphics[trim=60 60 60 60,clip,width=0.49\linewidth]{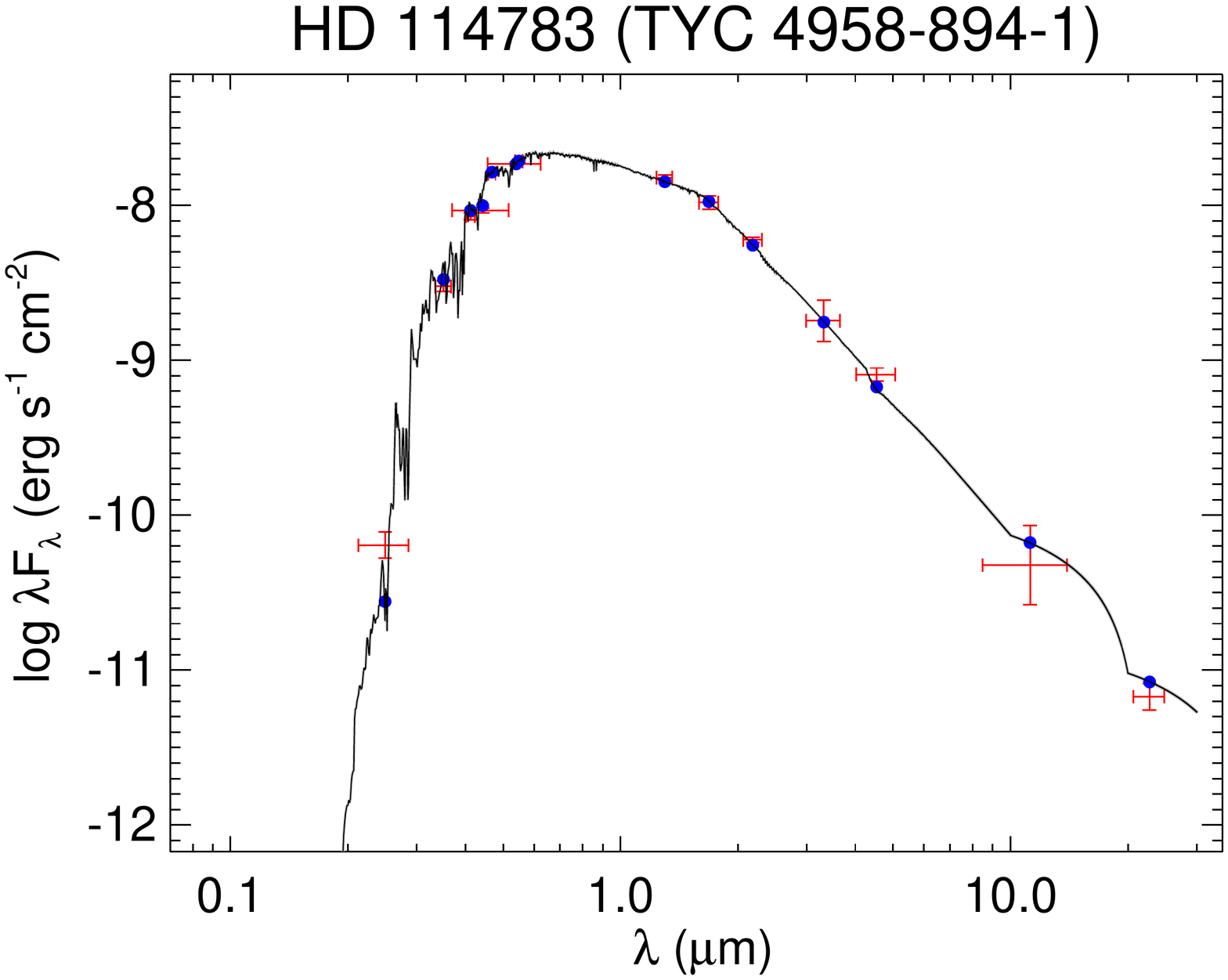}
  \includegraphics[trim=60 60 60 60,clip,width=0.49\linewidth]{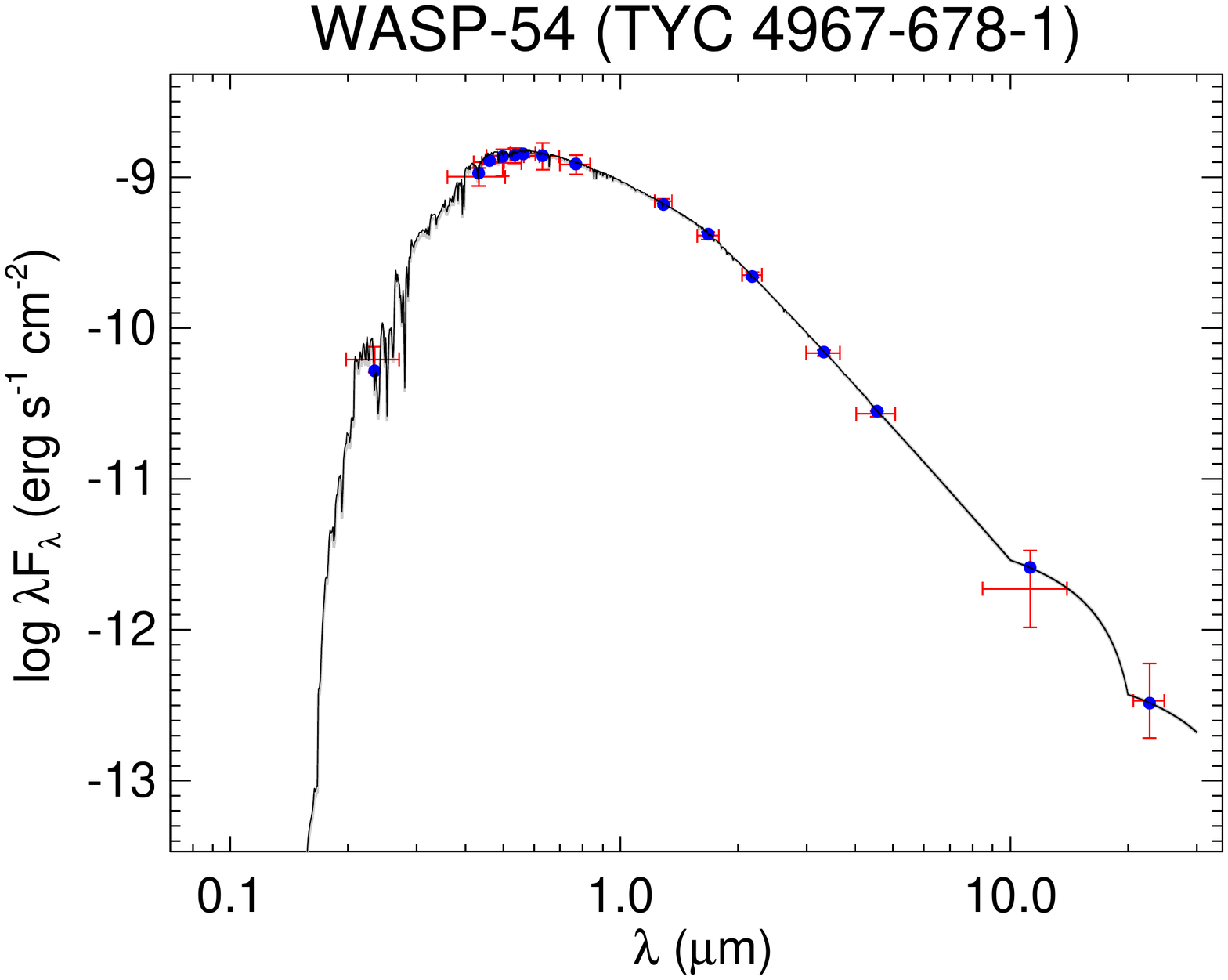}
  \includegraphics[trim=60 60 60 60,clip,width=0.49\linewidth]{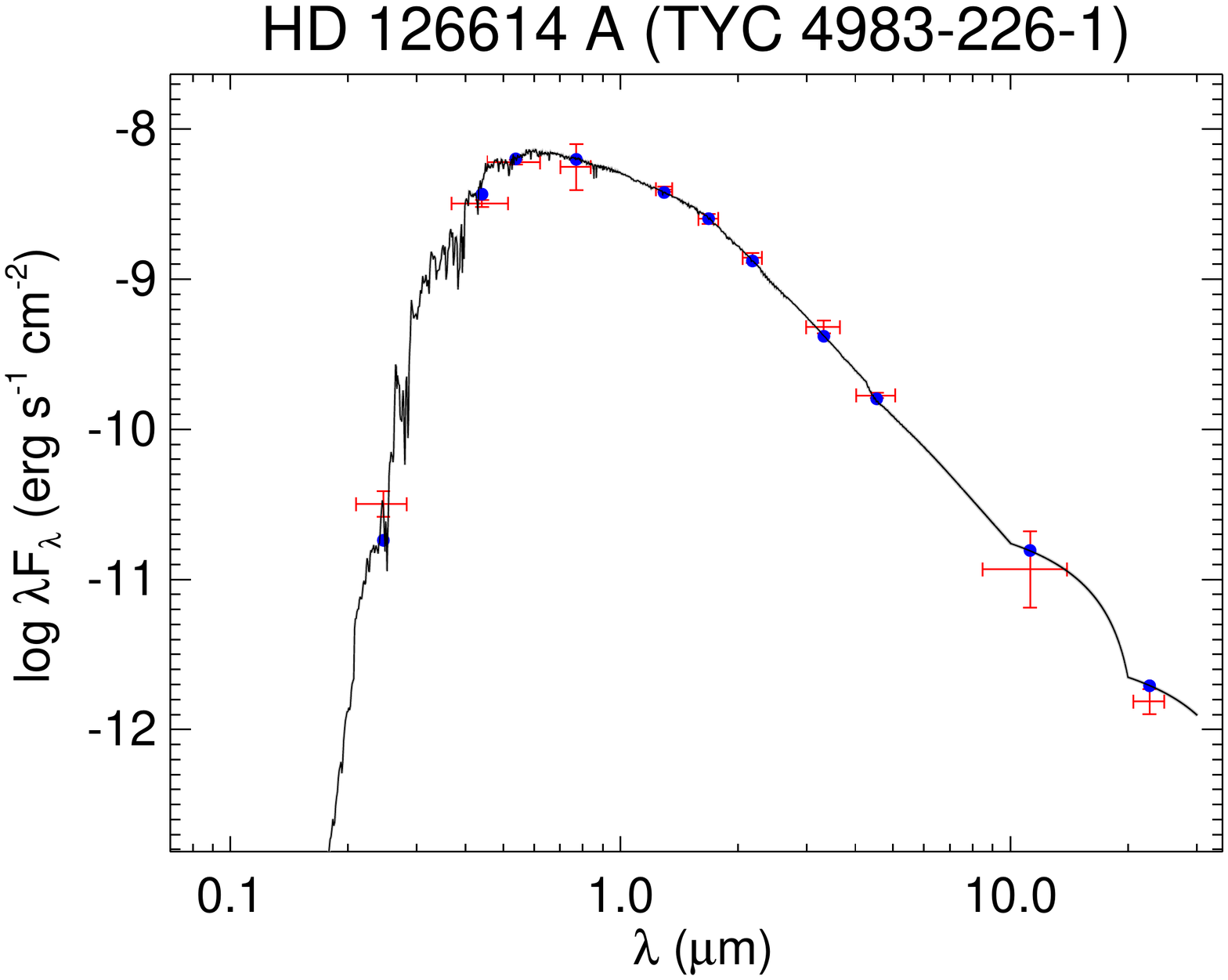}
  \includegraphics[trim=60 60 60 60,clip,width=0.49\linewidth]{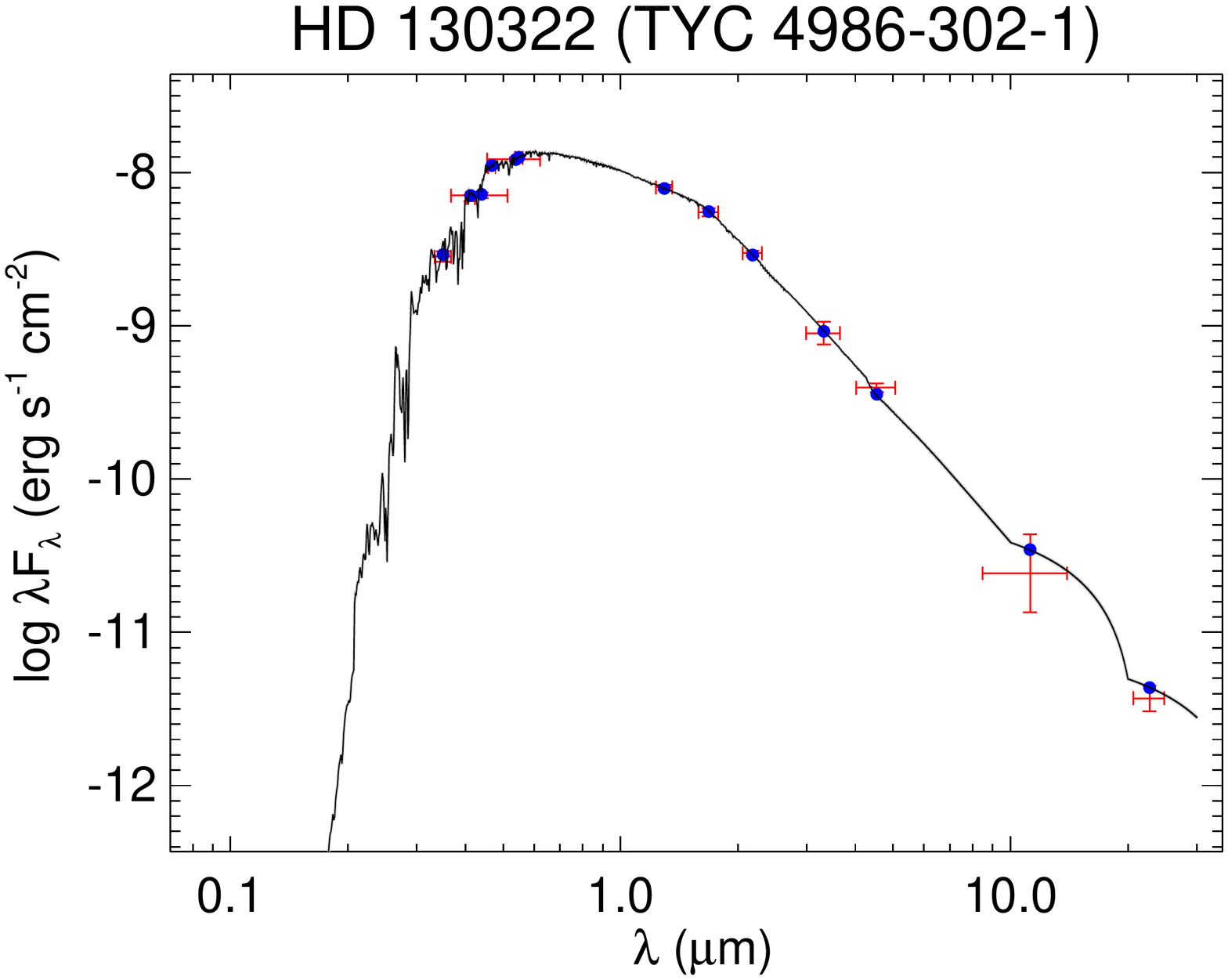}
  \includegraphics[trim=60 60 60 60,clip,width=0.49\linewidth]{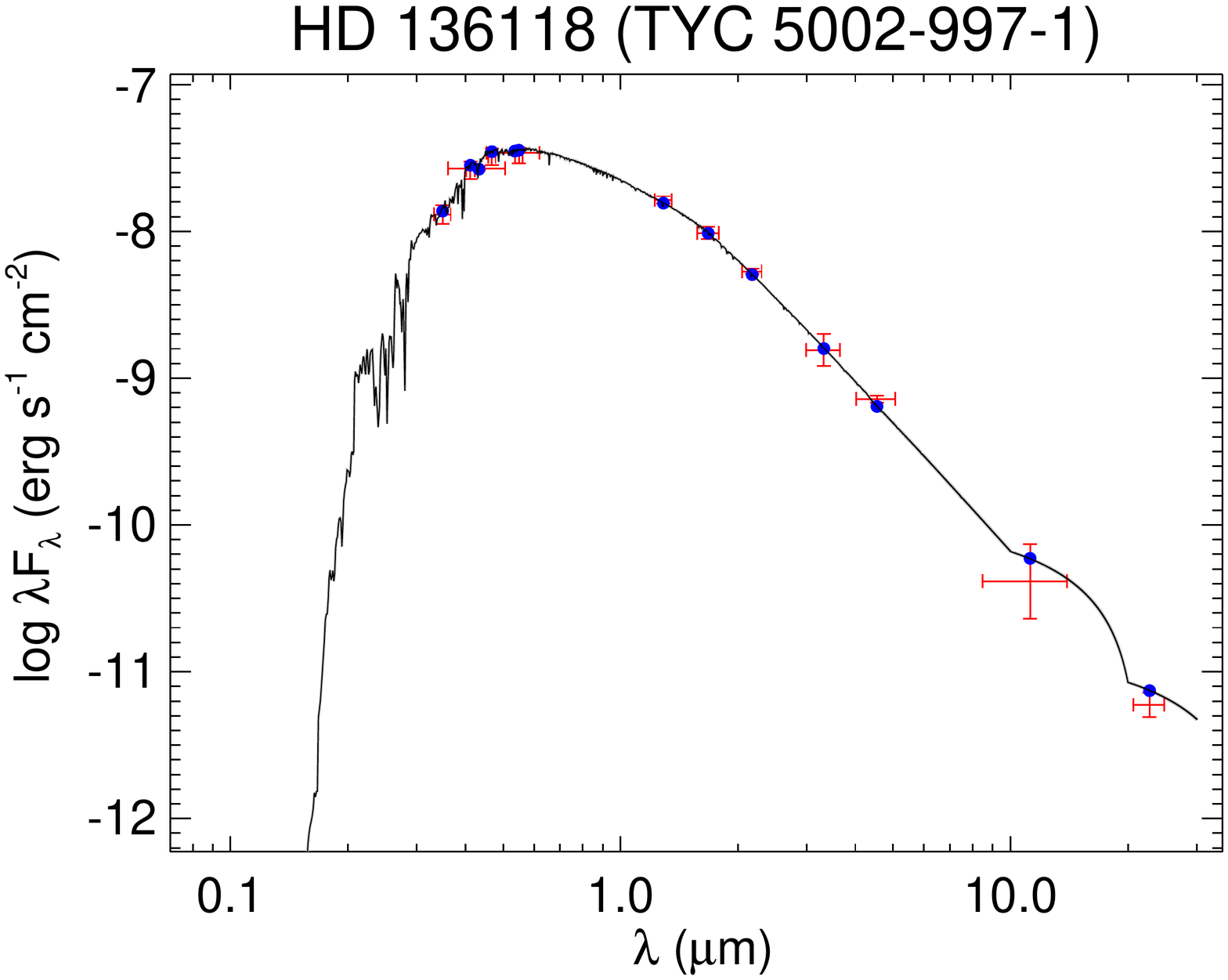}
  \includegraphics[trim=60 60 60 60,clip,width=0.49\linewidth]{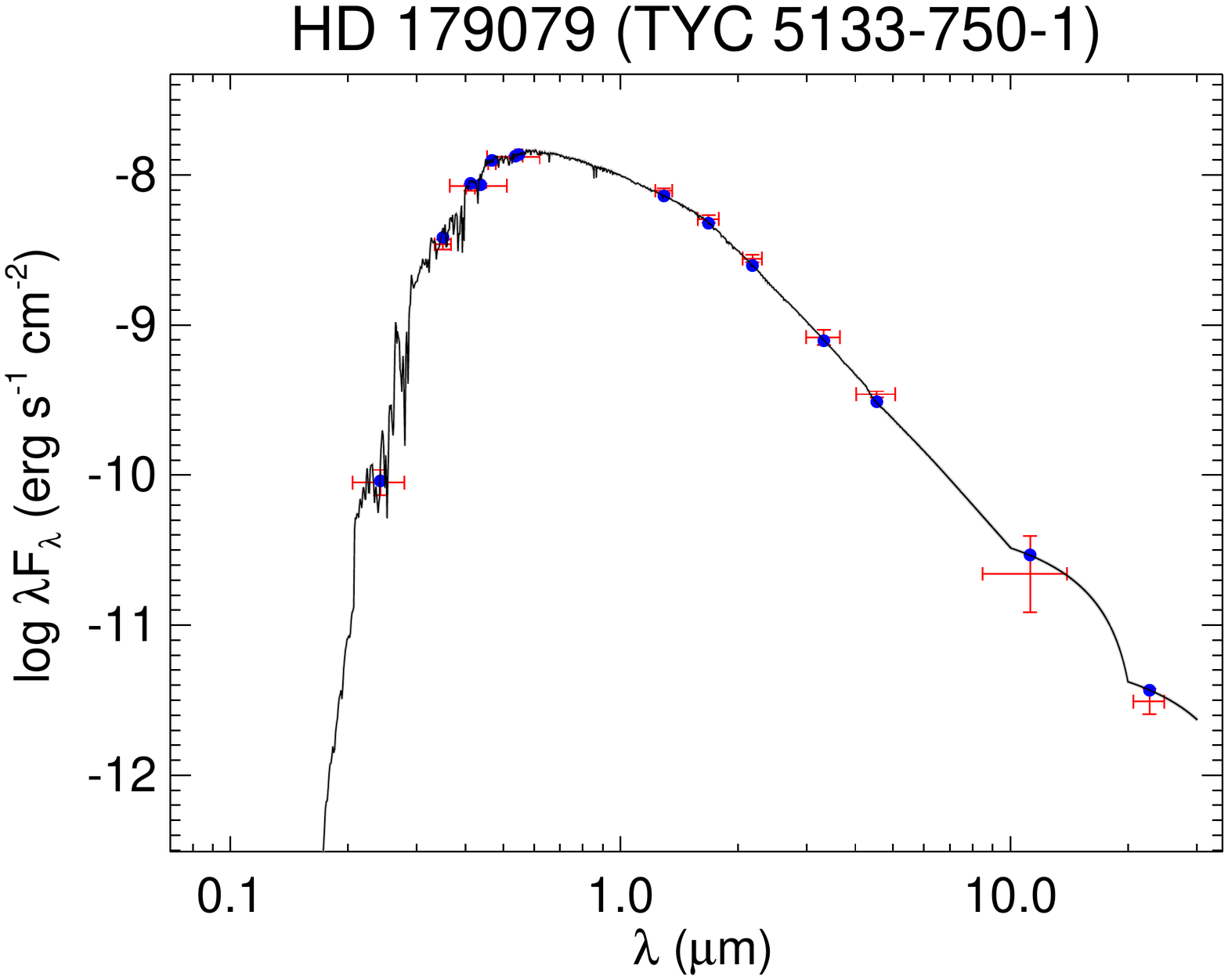}
  \caption{All labels, lines, symbols, and colors as in Figure \ref{fig:seds}.}
  \label{fig:seds_47}
\end{figure}

\begin{figure}[H]
  \centering
  \includegraphics[trim=60 60 60 60,clip,width=0.49\linewidth]{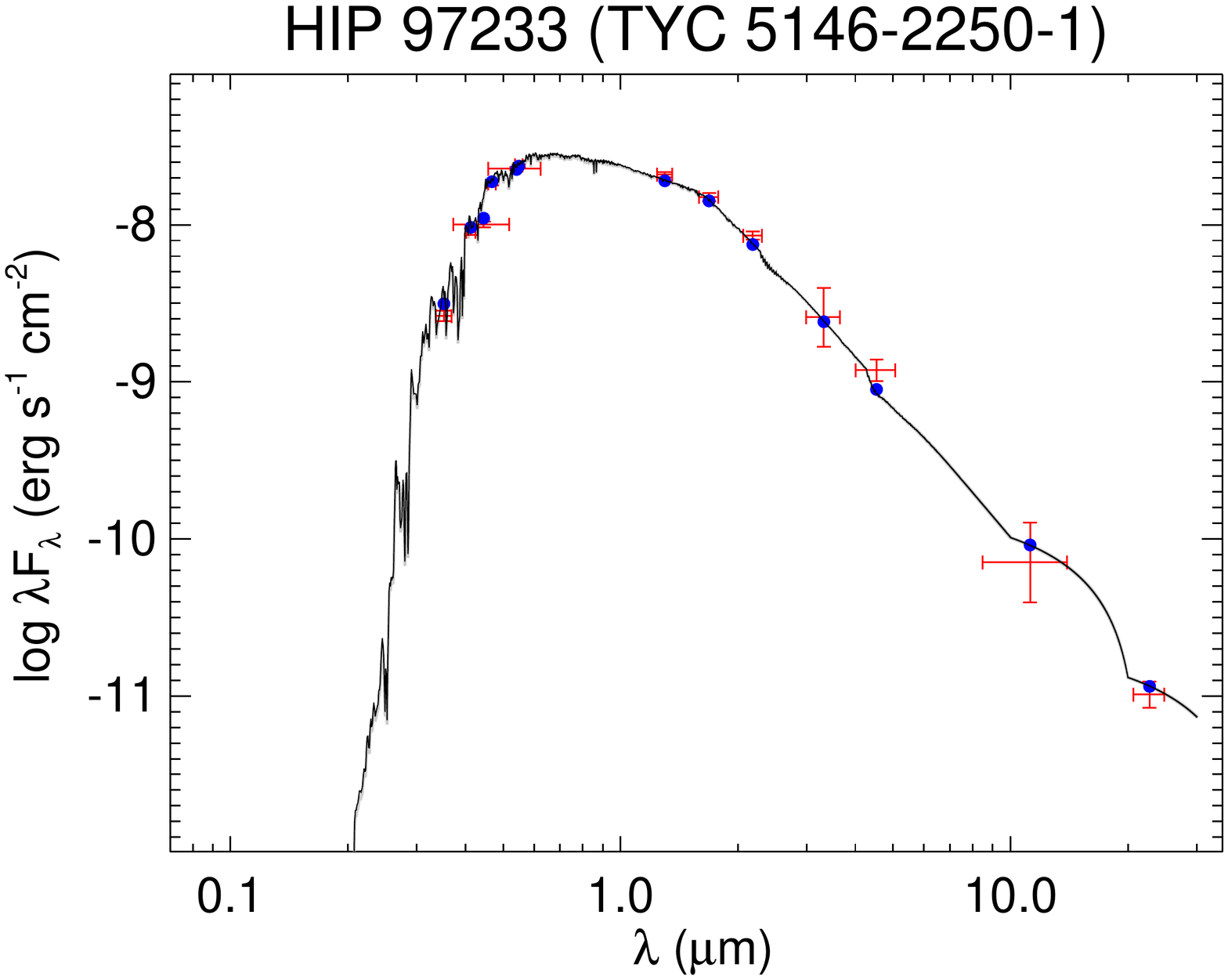}
  \includegraphics[trim=60 60 60 60,clip,width=0.49\linewidth]{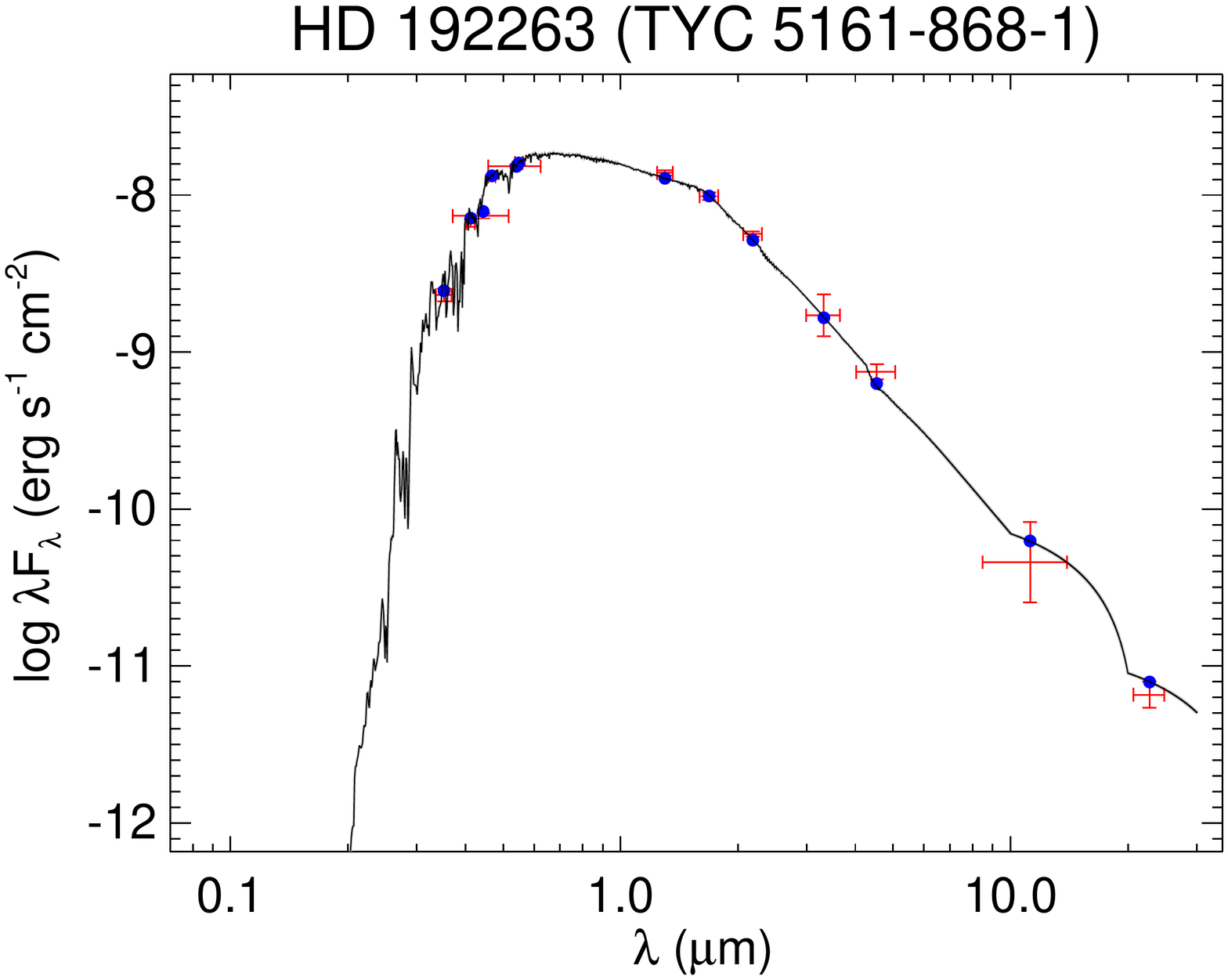}
  \includegraphics[trim=60 60 60 60,clip,width=0.49\linewidth]{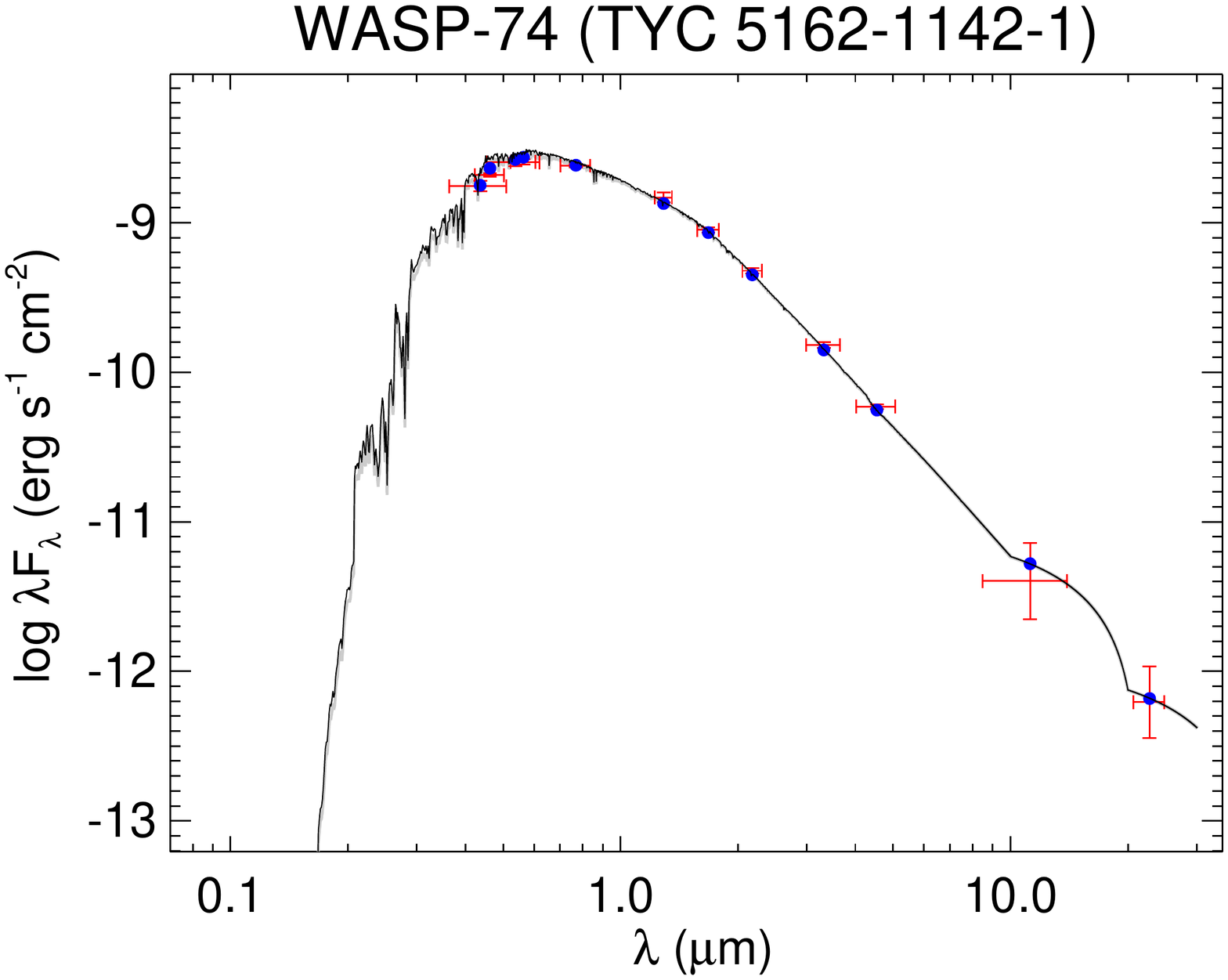}
  \includegraphics[trim=60 60 60 60,clip,width=0.49\linewidth]{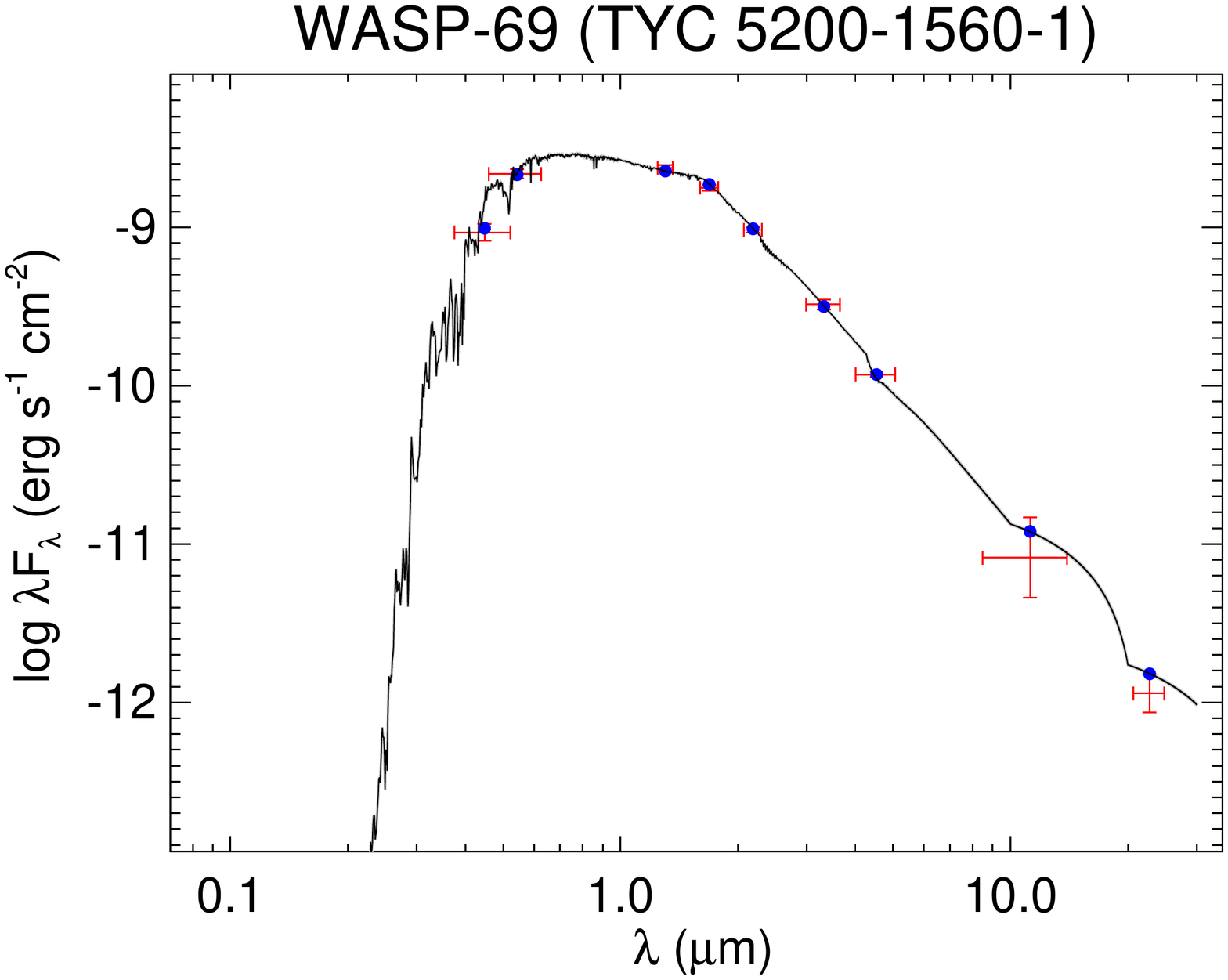}
  \includegraphics[trim=60 60 60 60,clip,width=0.49\linewidth]{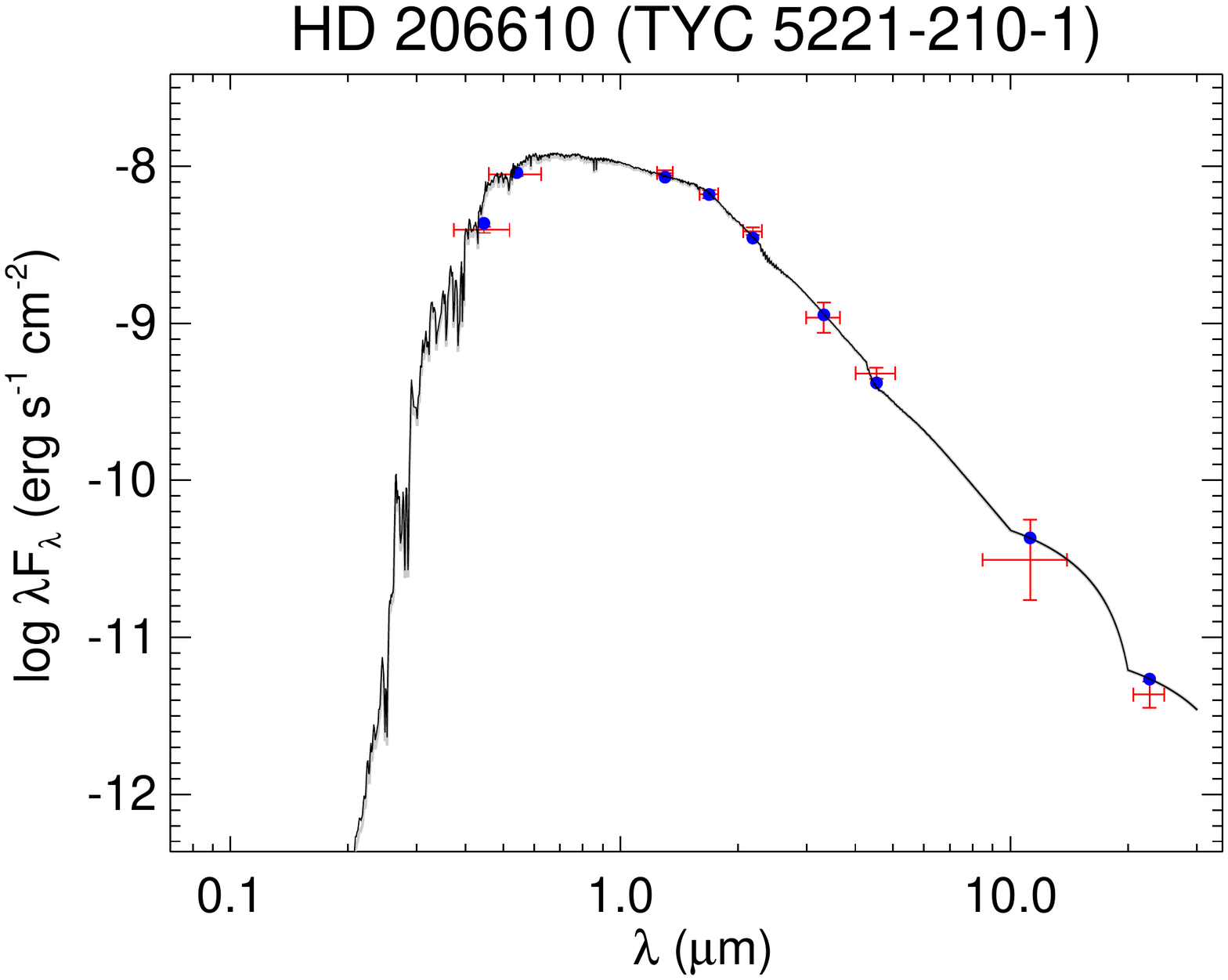}
  \includegraphics[trim=60 60 60 60,clip,width=0.49\linewidth]{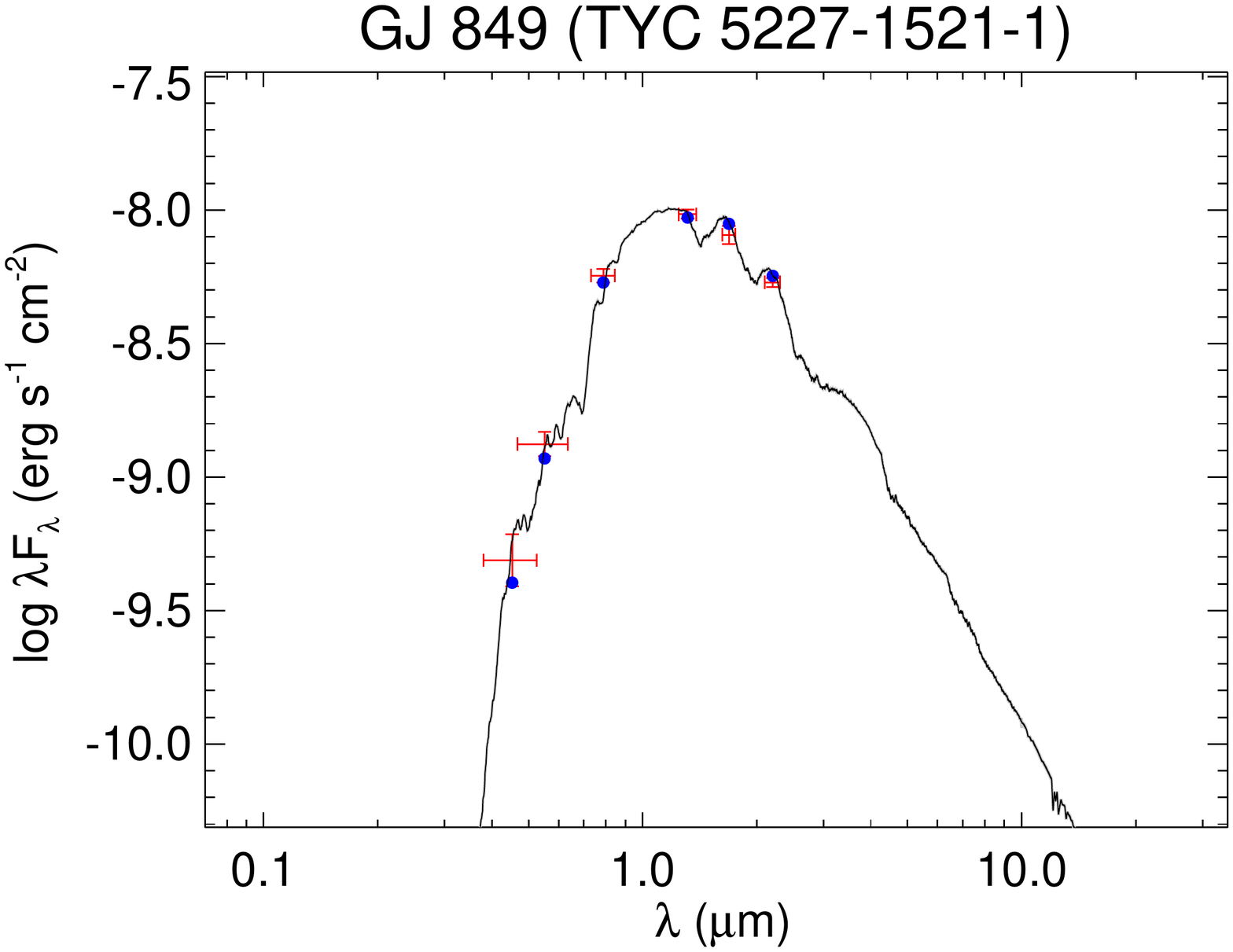}
  \caption{All labels, lines, symbols, and colors as in Figure \ref{fig:seds}.}
  \label{fig:seds_48}
\end{figure}

\begin{figure}[H]
  \centering
  \includegraphics[trim=60 60 60 60,clip,width=0.49\linewidth]{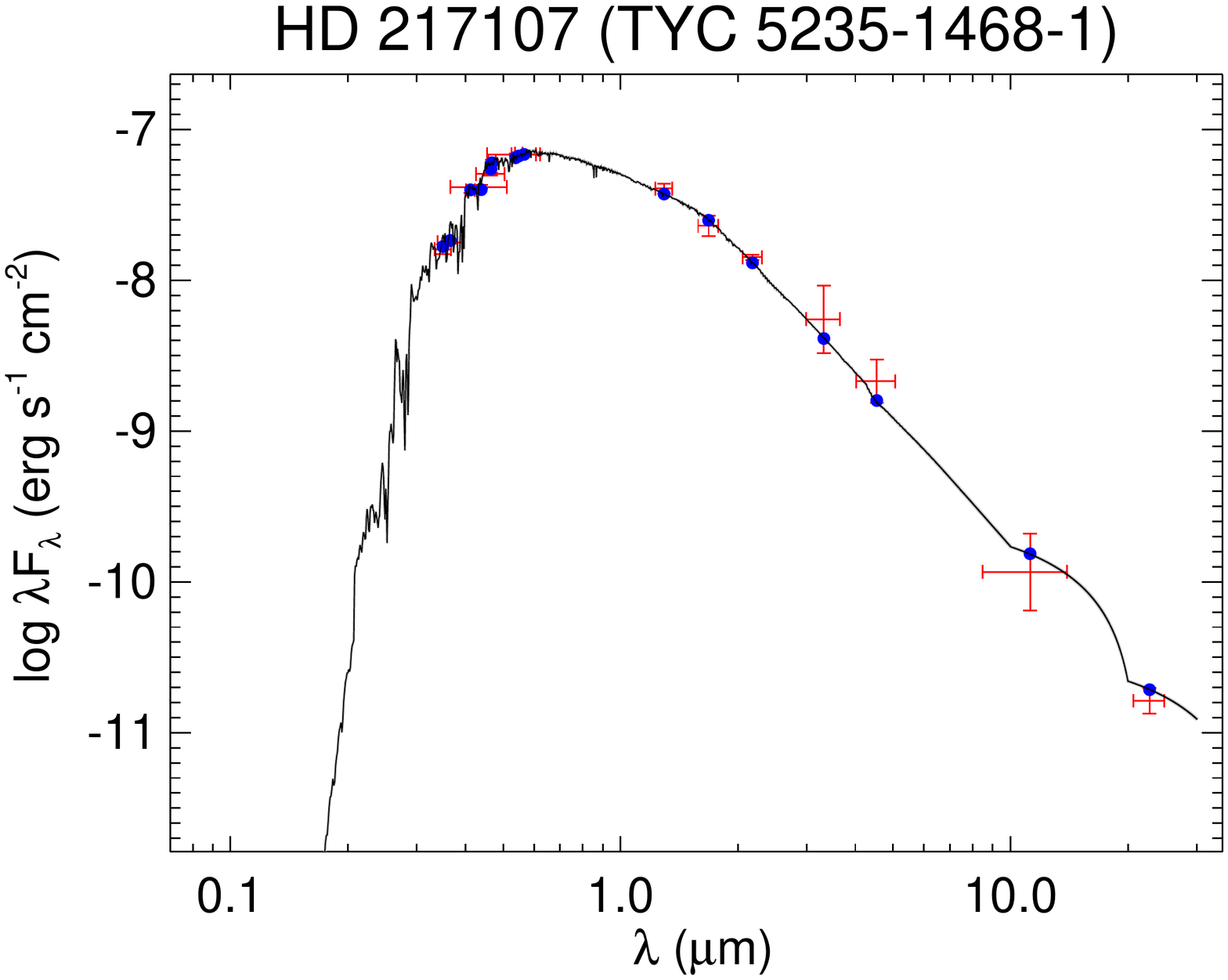}
  \includegraphics[trim=60 60 60 60,clip,width=0.49\linewidth]{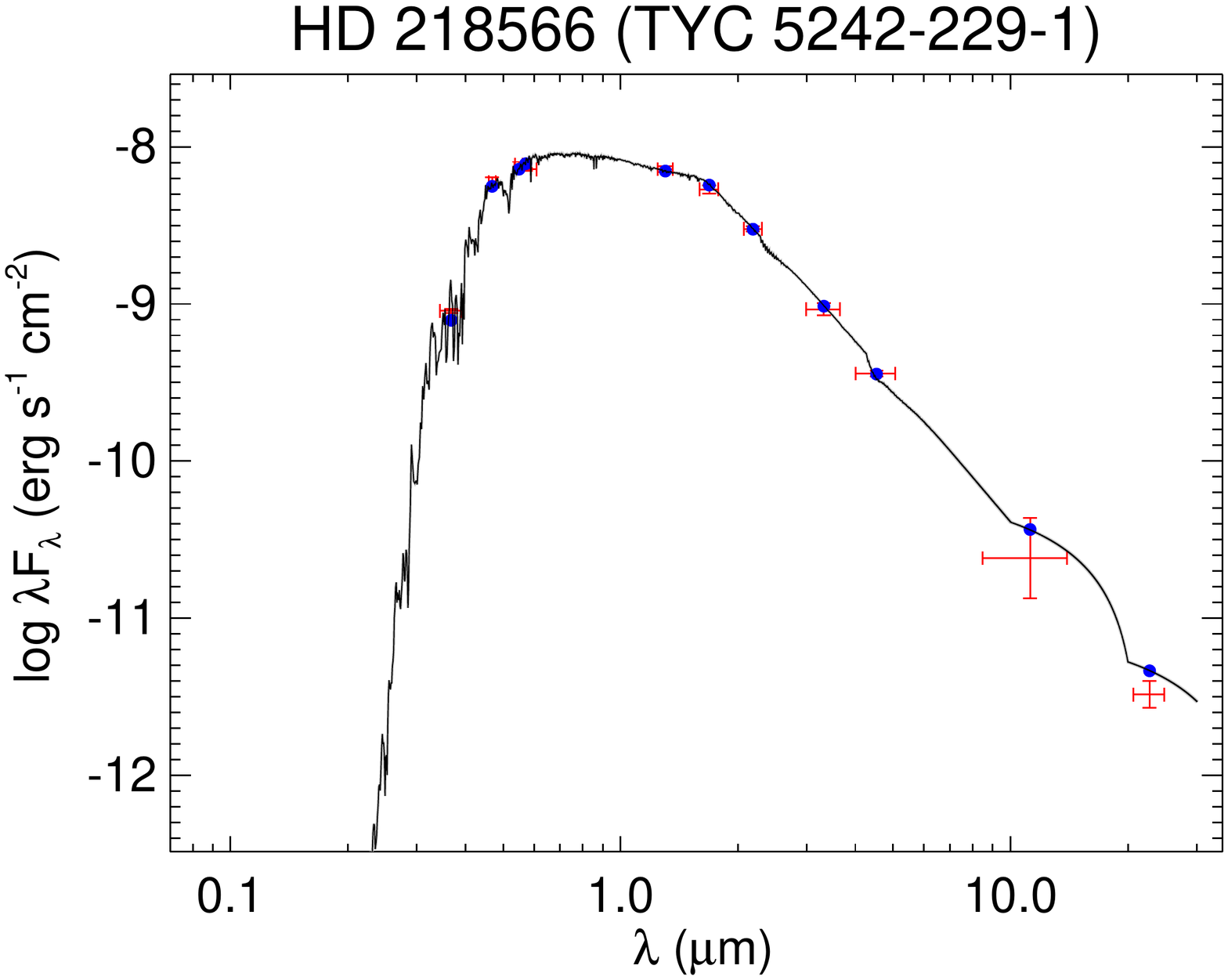}
  \includegraphics[trim=60 60 60 60,clip,width=0.49\linewidth]{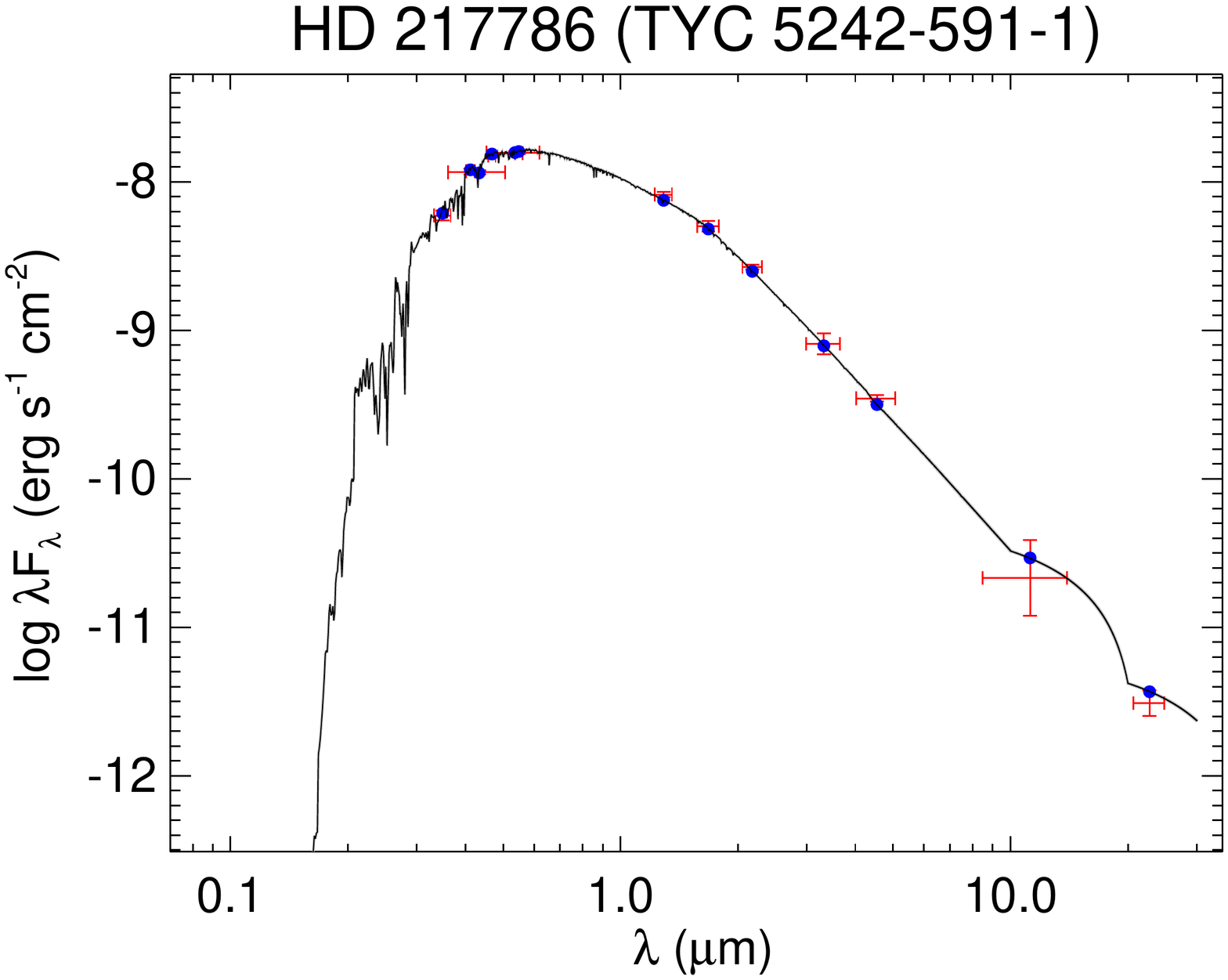}
  \includegraphics[trim=60 60 60 60,clip,width=0.49\linewidth]{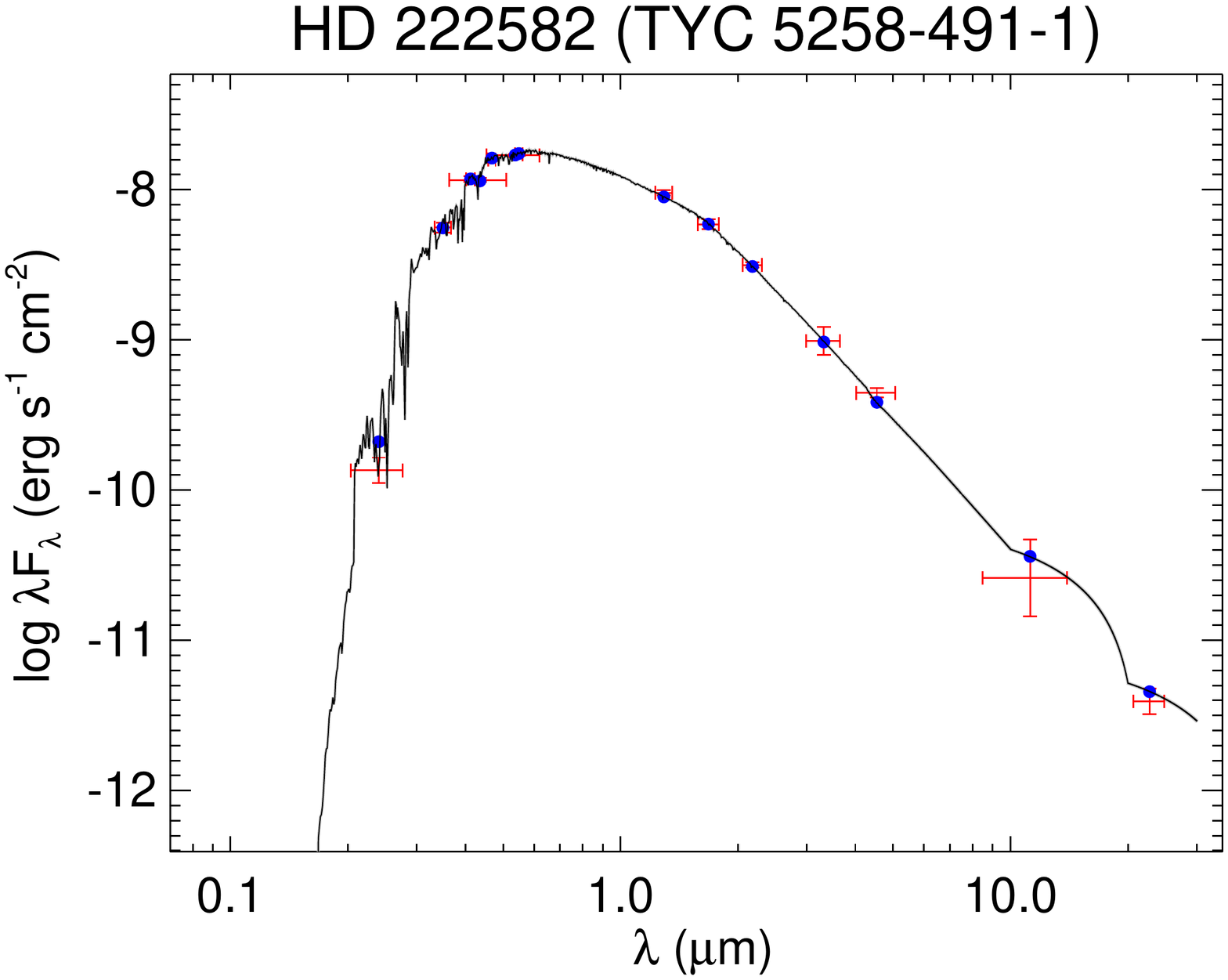}
  \includegraphics[trim=60 60 60 60,clip,width=0.49\linewidth]{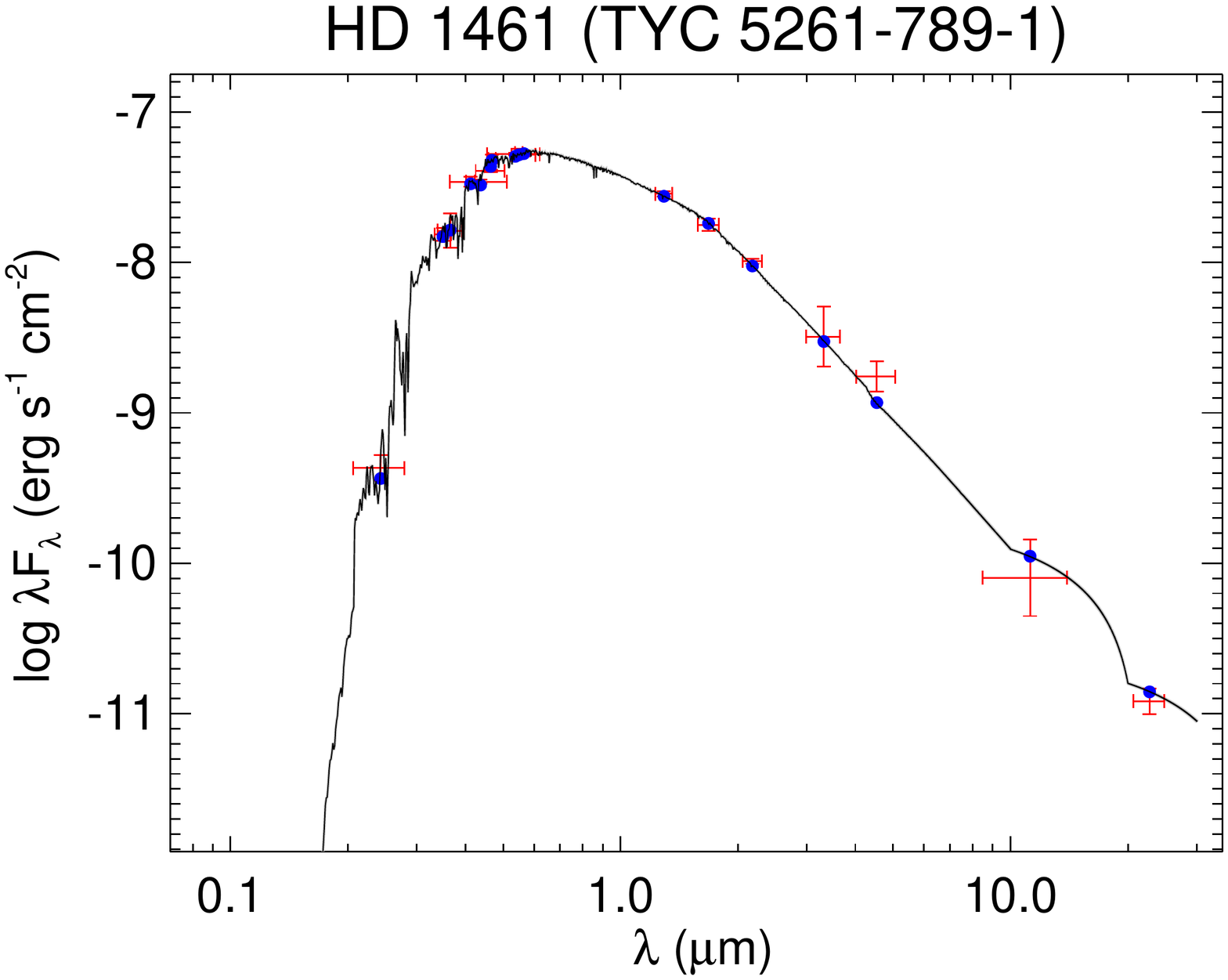}
  \includegraphics[trim=60 60 60 60,clip,width=0.49\linewidth]{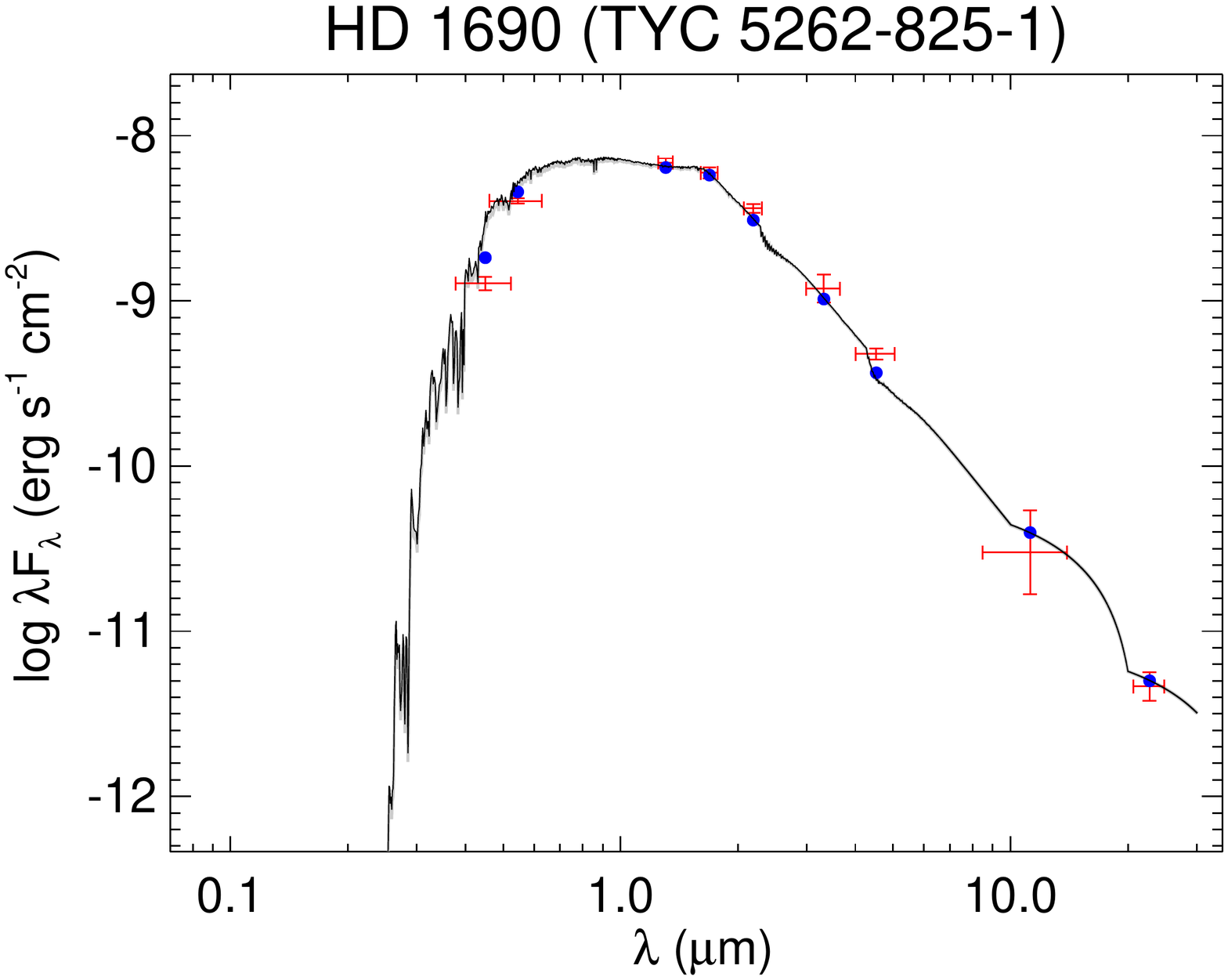}
  \caption{All labels, lines, symbols, and colors as in Figure \ref{fig:seds}.}
  \label{fig:seds_49}
\end{figure}

\begin{figure}[H]
  \centering
  \includegraphics[trim=60 60 60 60,clip,width=0.49\linewidth]{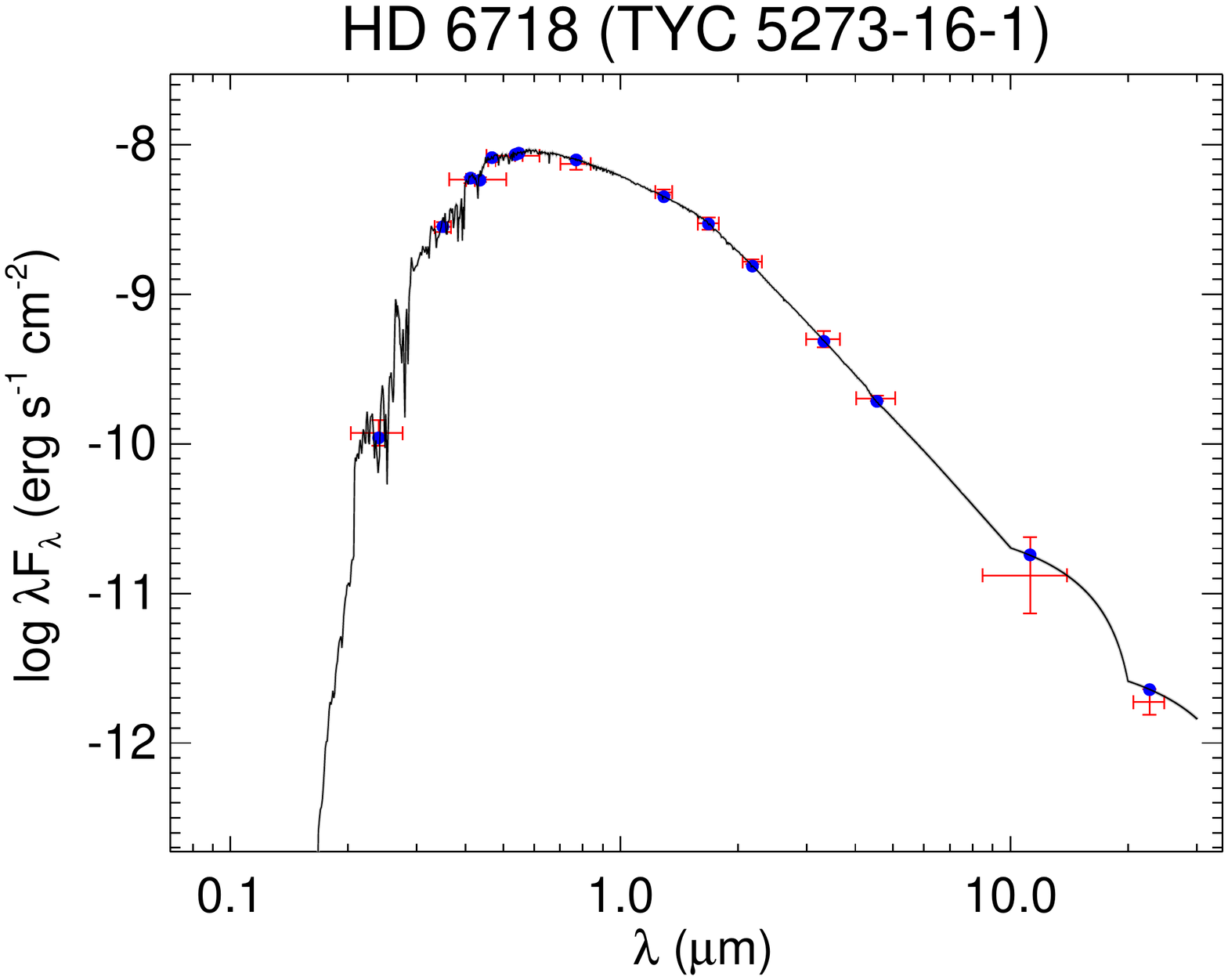}
  \includegraphics[trim=60 60 60 60,clip,width=0.49\linewidth]{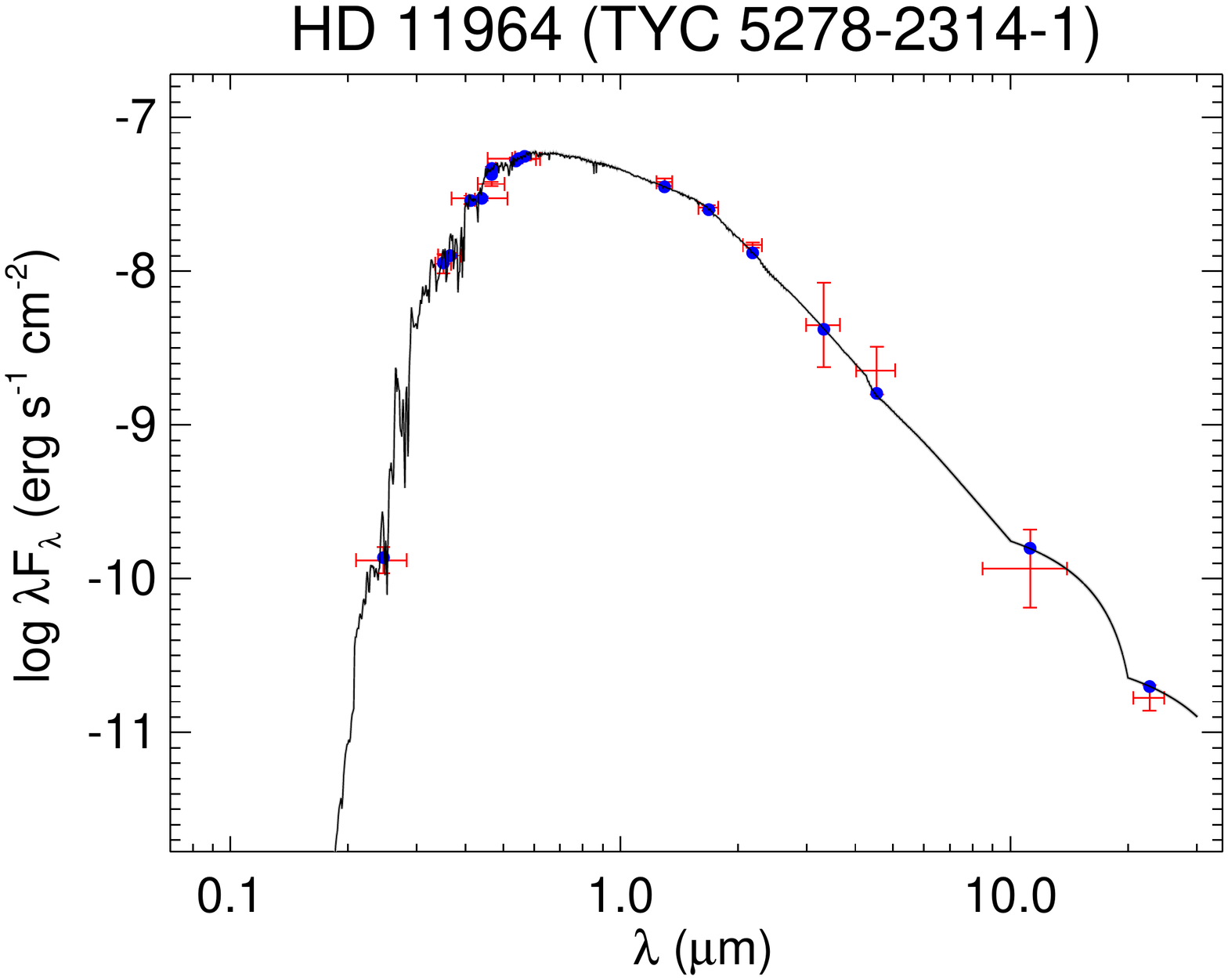}
  \includegraphics[trim=60 60 60 60,clip,width=0.49\linewidth]{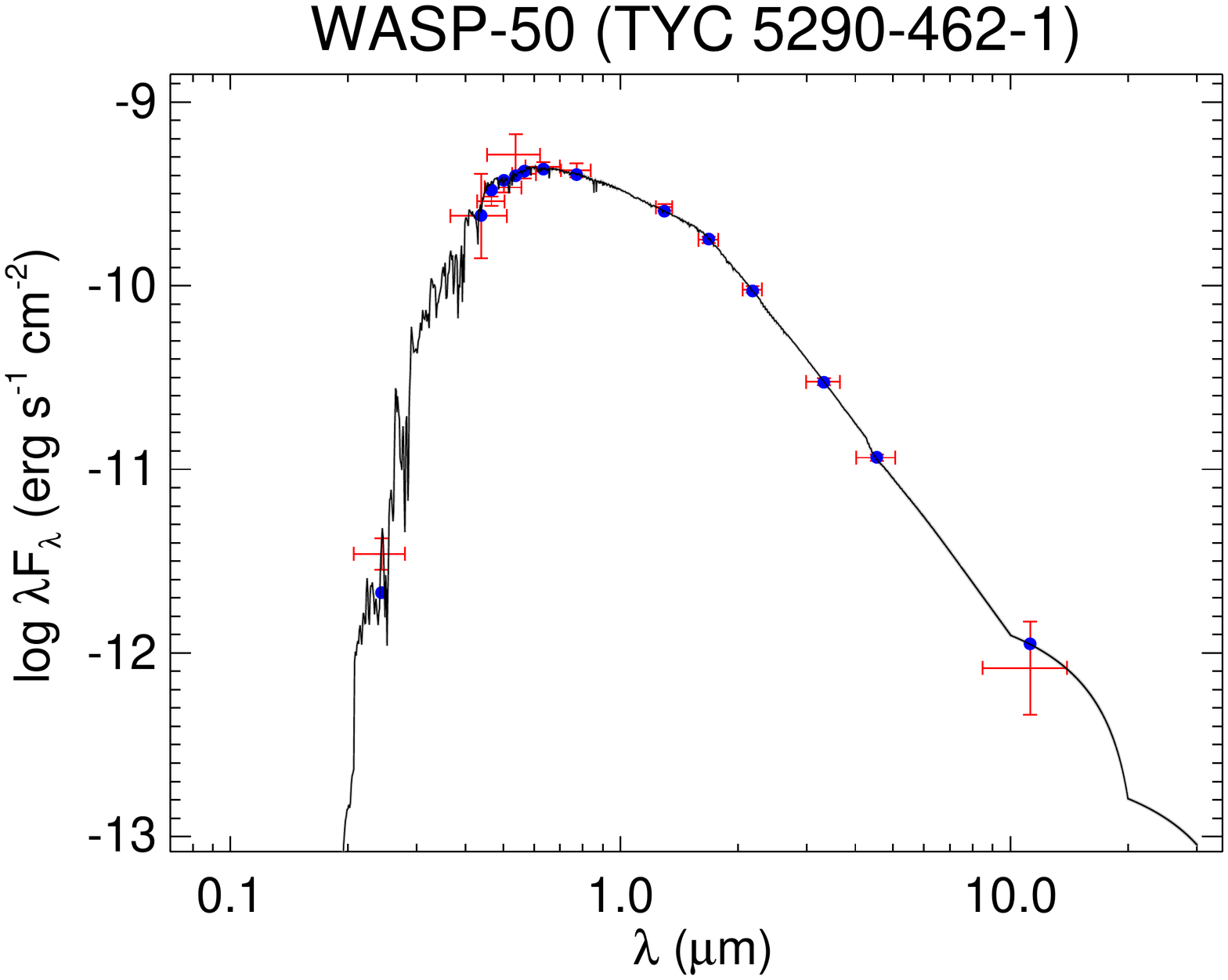}
  \includegraphics[trim=60 60 60 60,clip,width=0.49\linewidth]{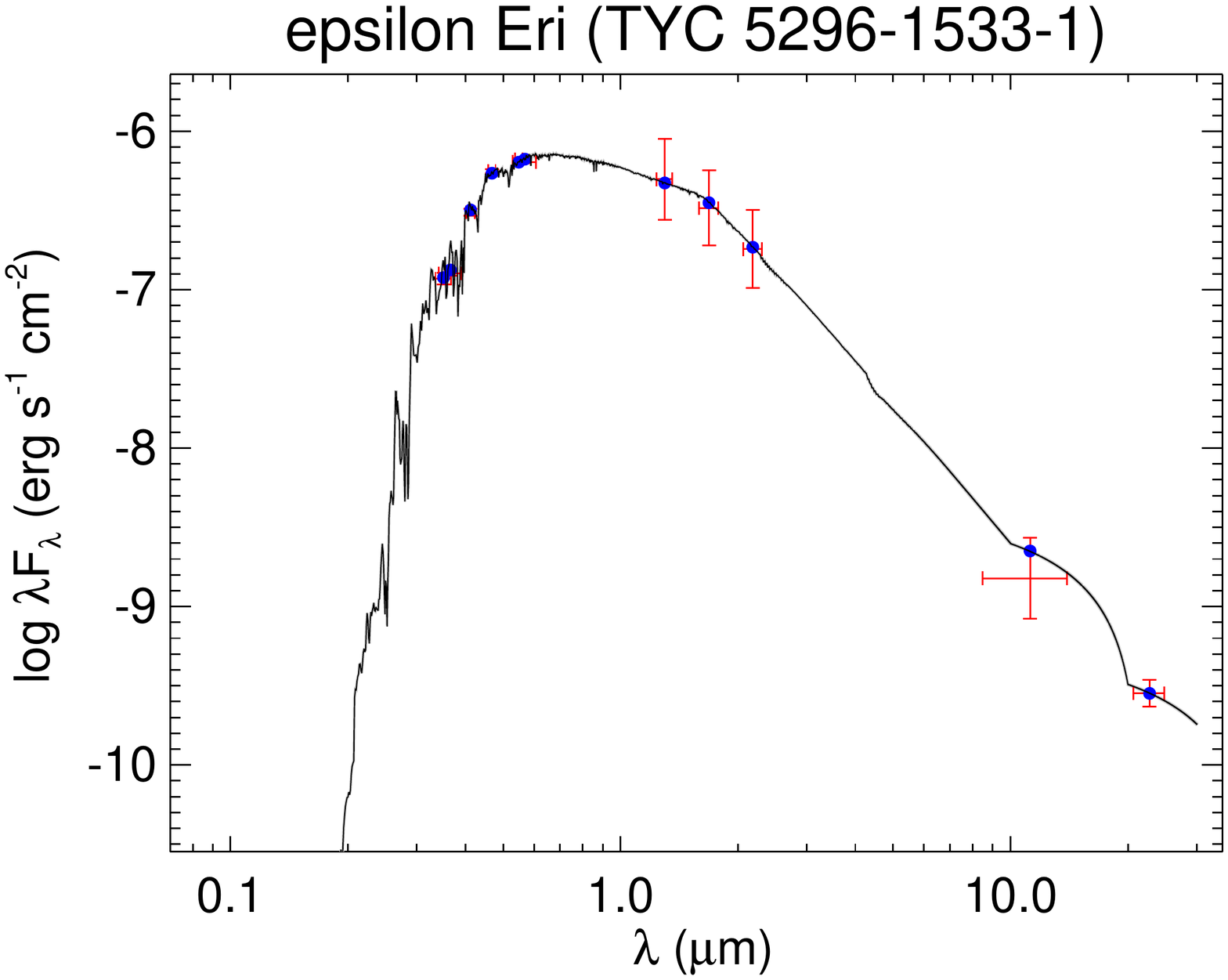}
  \includegraphics[trim=60 60 60 60,clip,width=0.49\linewidth]{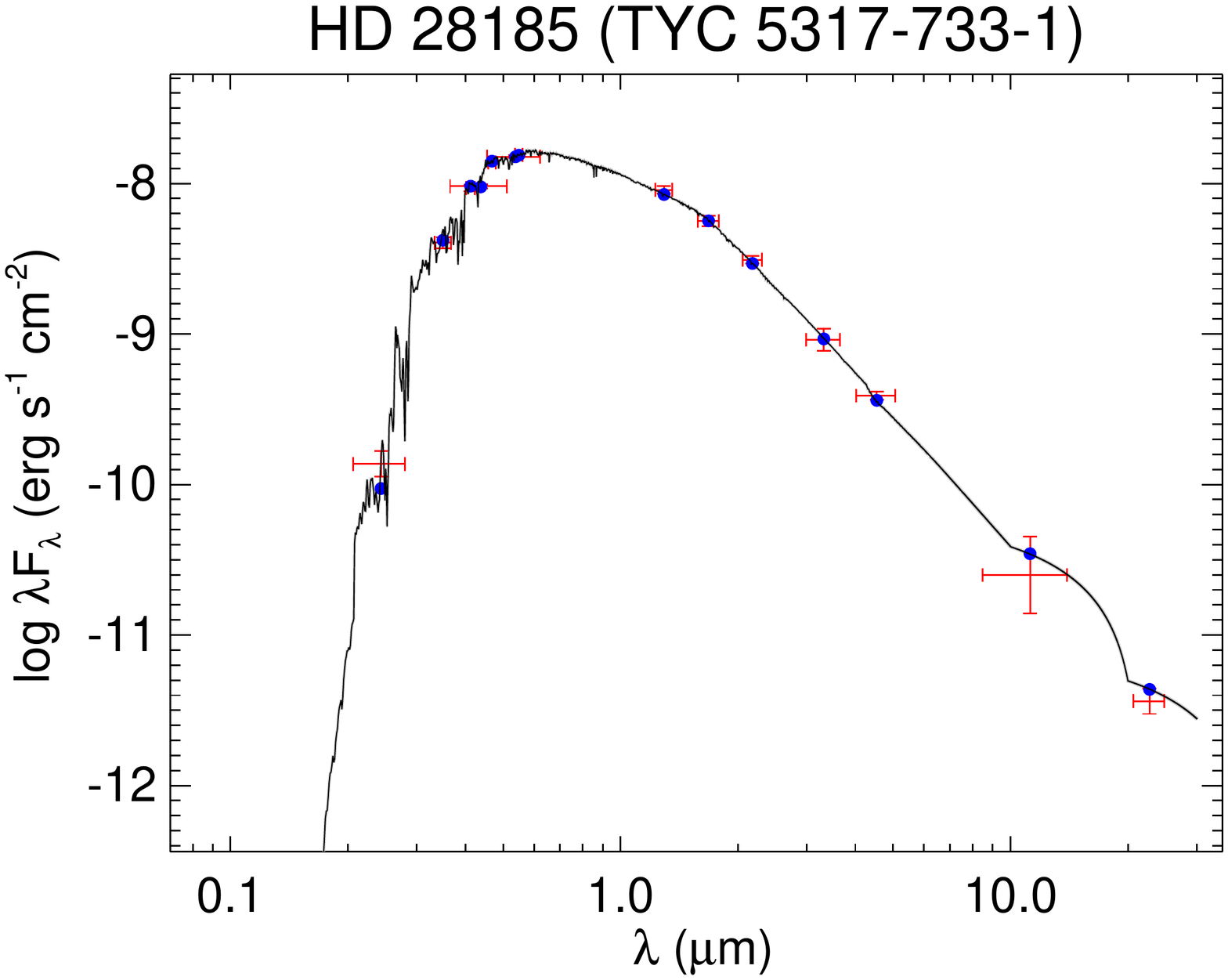}
  \includegraphics[trim=60 60 60 60,clip,width=0.49\linewidth]{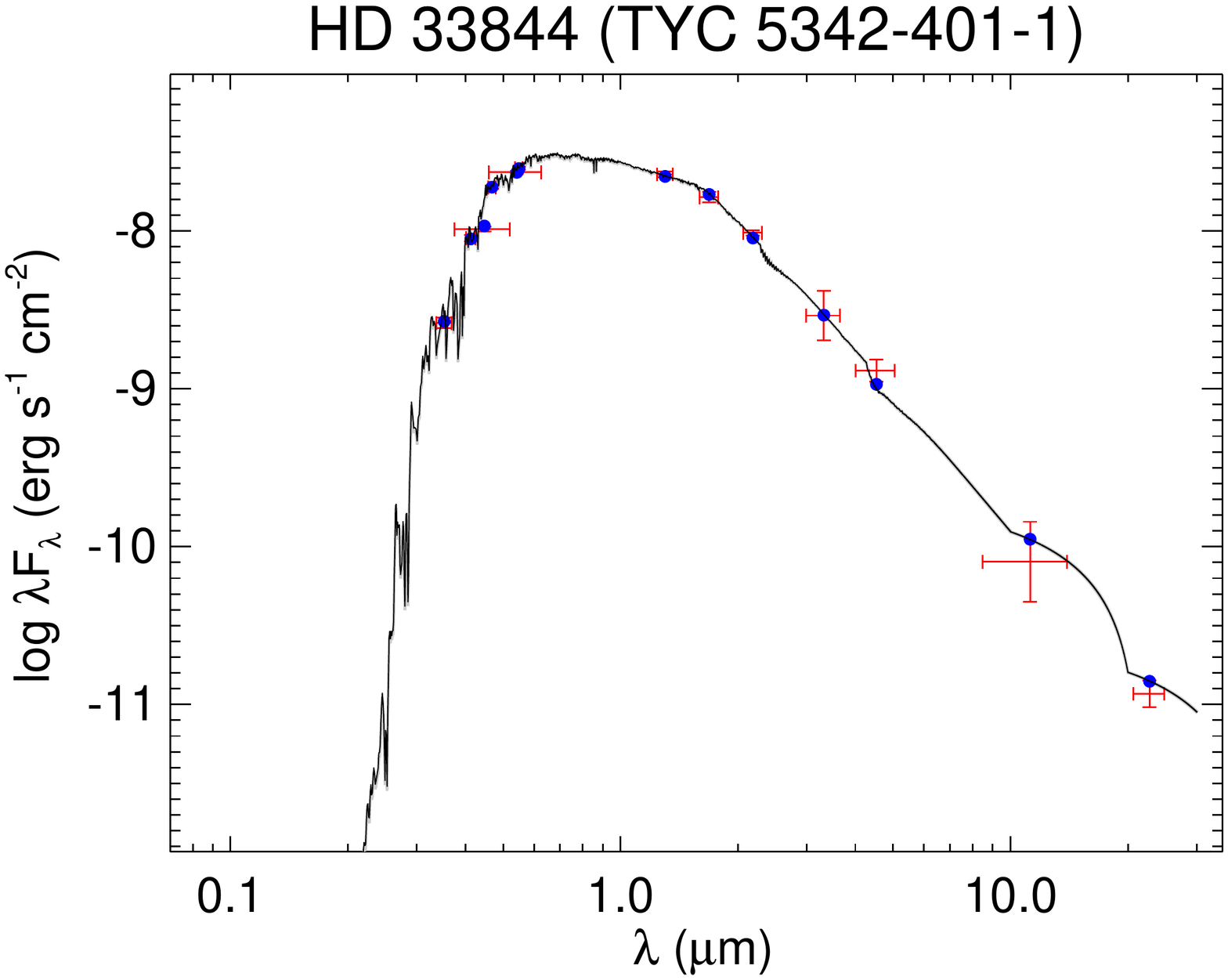}
  \caption{All labels, lines, symbols, and colors as in Figure \ref{fig:seds}.}
  \label{fig:seds_50}
\end{figure}

\begin{figure}[H]
  \centering
  \includegraphics[trim=60 60 60 60,clip,width=0.49\linewidth]{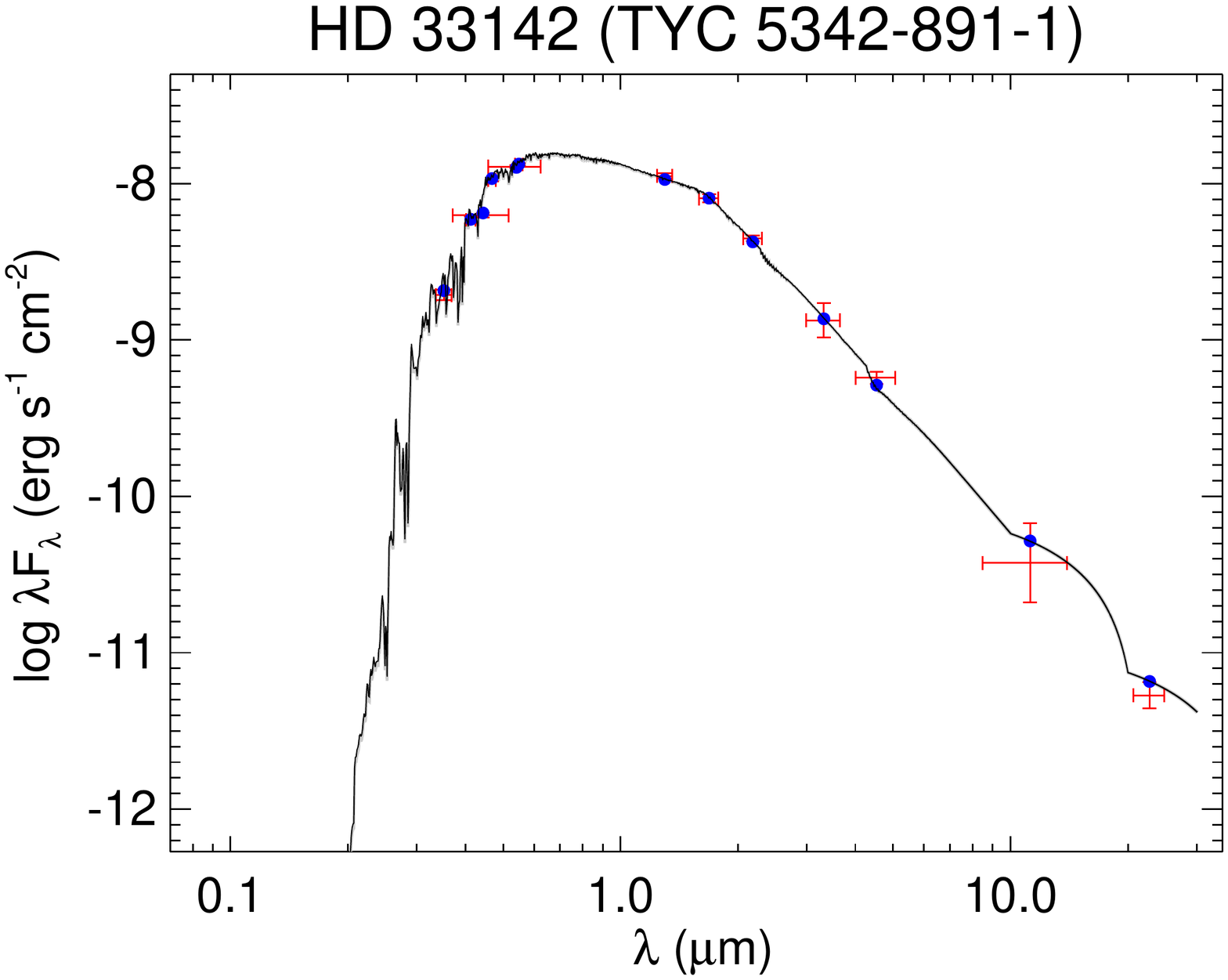}
  \includegraphics[trim=60 60 60 60,clip,width=0.49\linewidth]{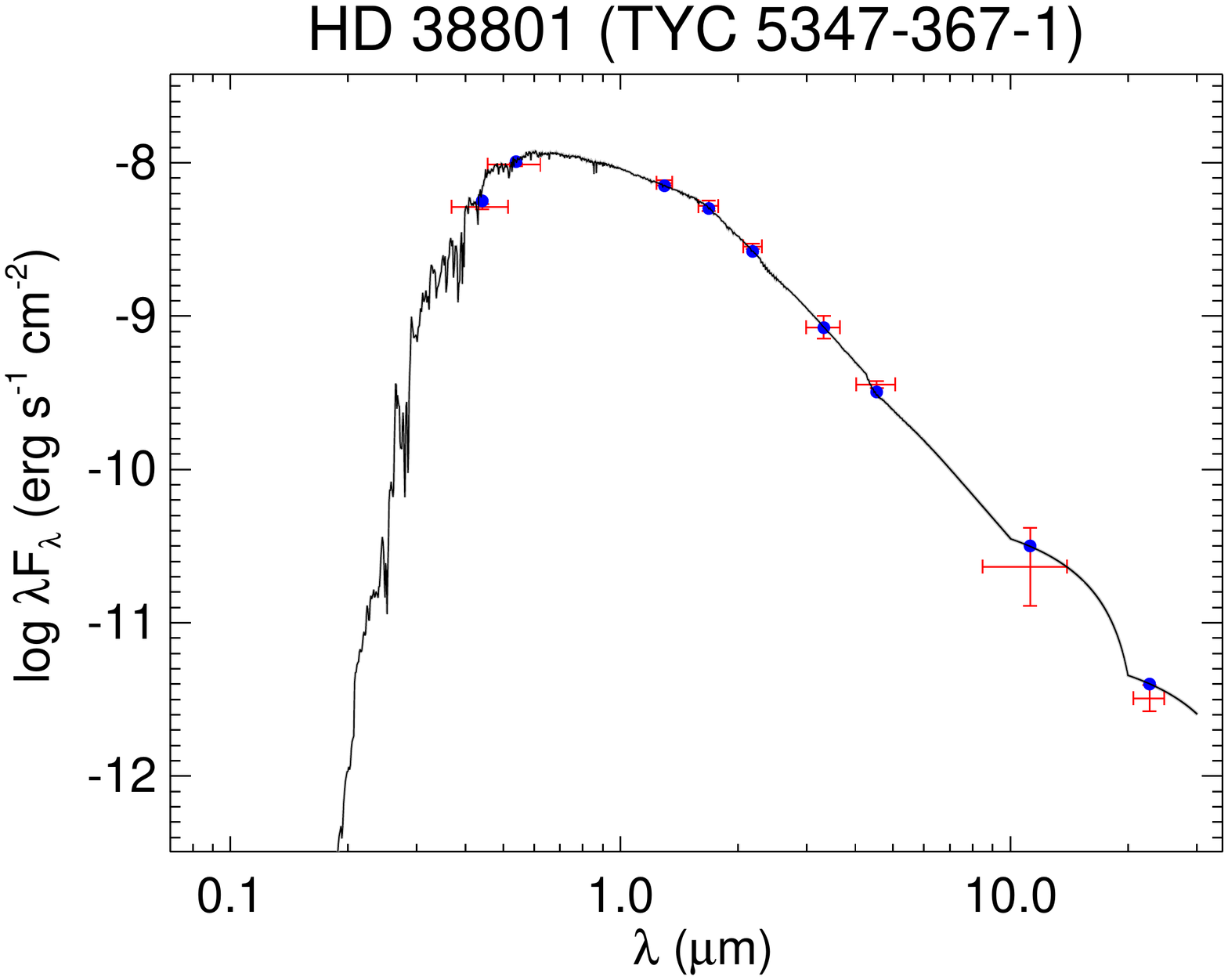}
  \includegraphics[trim=60 60 60 60,clip,width=0.49\linewidth]{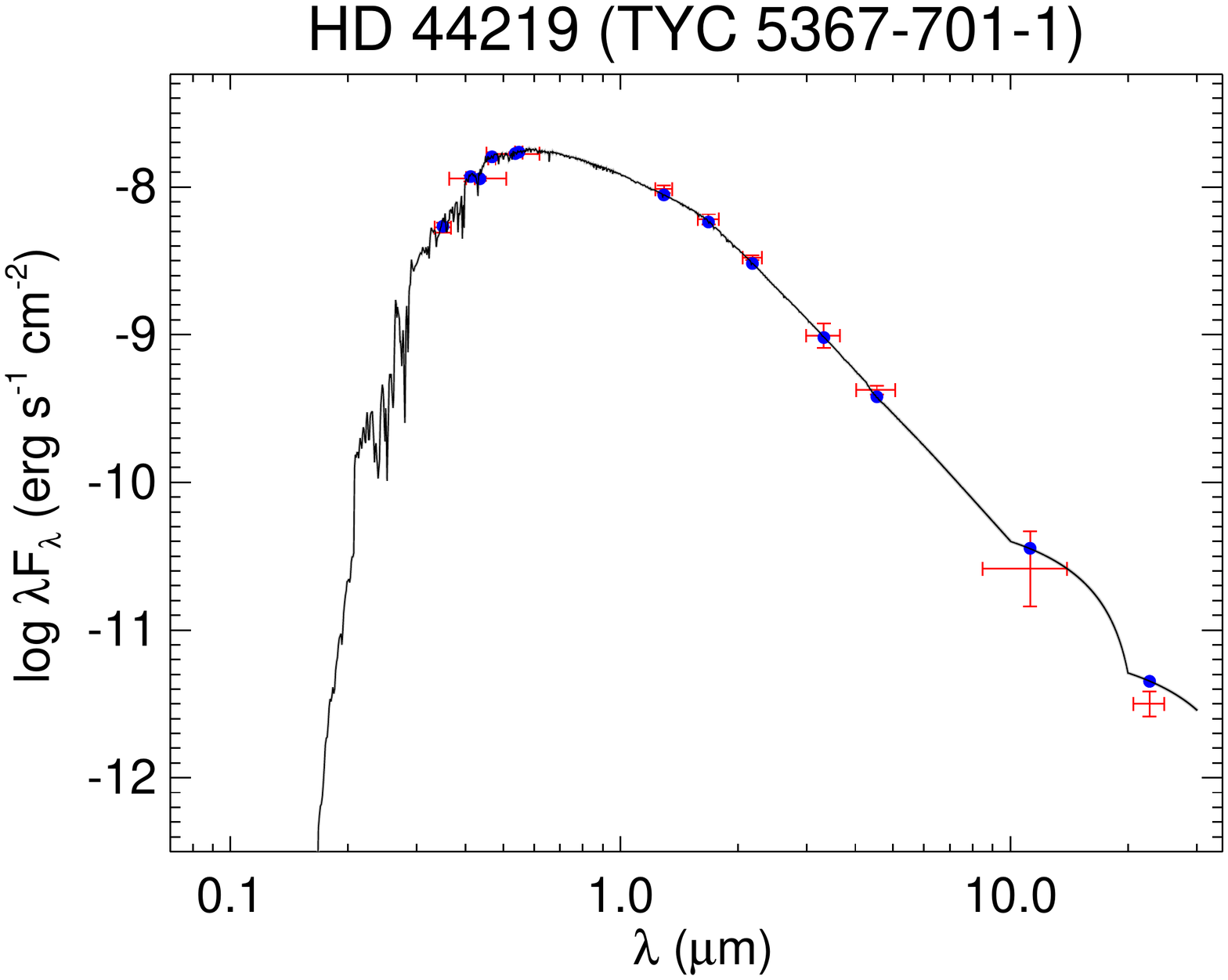}
  \includegraphics[trim=60 60 60 60,clip,width=0.49\linewidth]{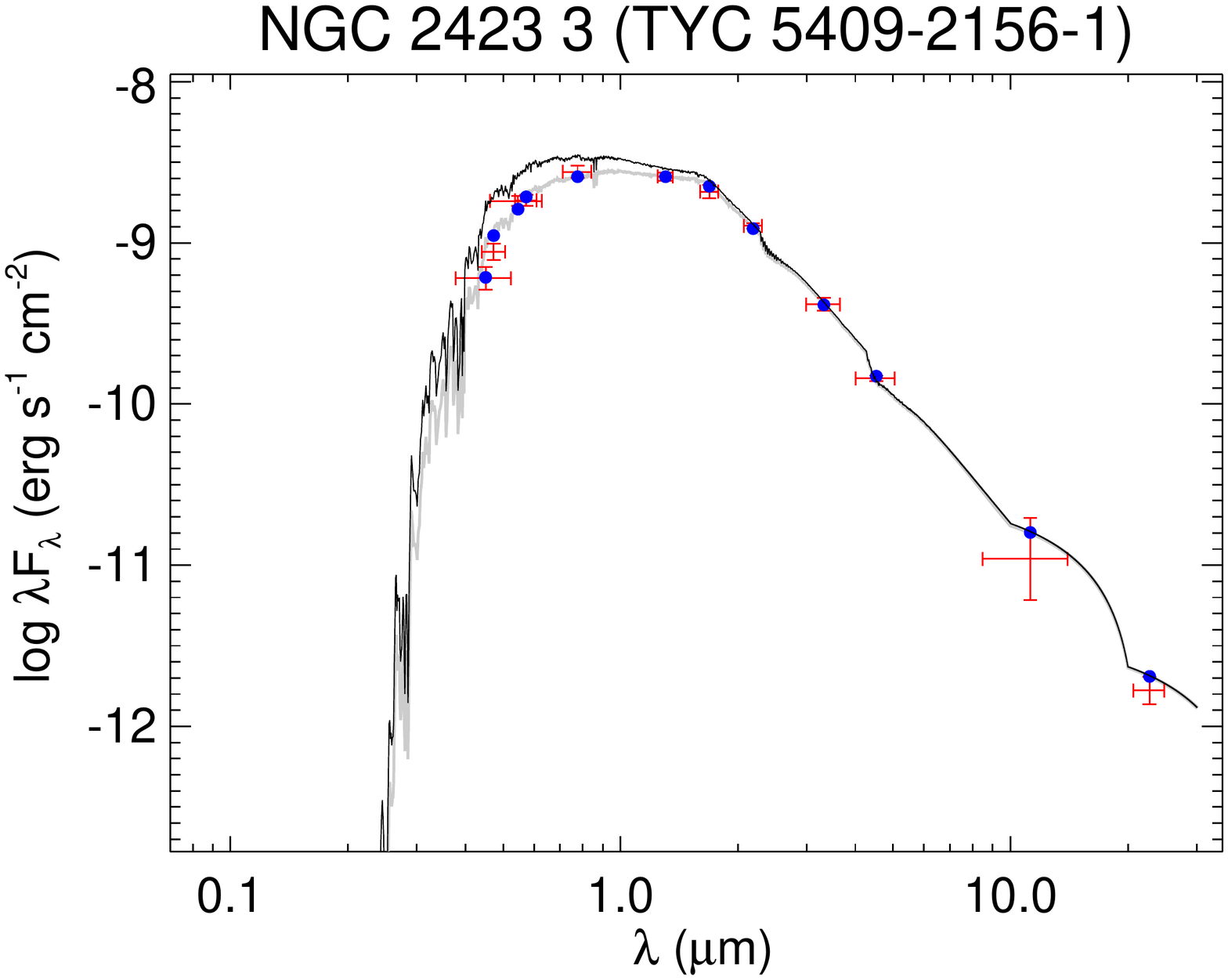}
  \includegraphics[trim=60 60 60 60,clip,width=0.49\linewidth]{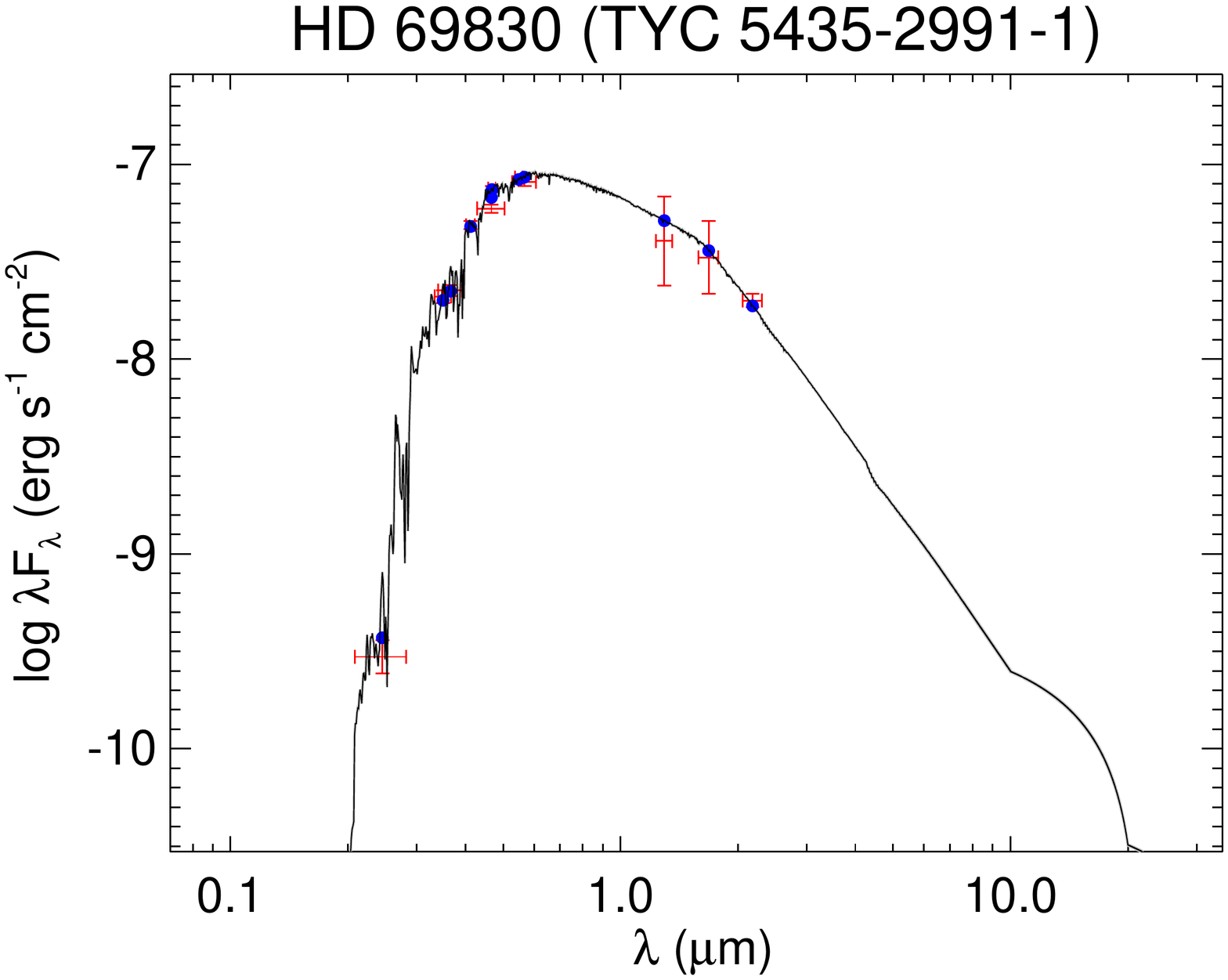}
  \includegraphics[trim=60 60 60 60,clip,width=0.49\linewidth]{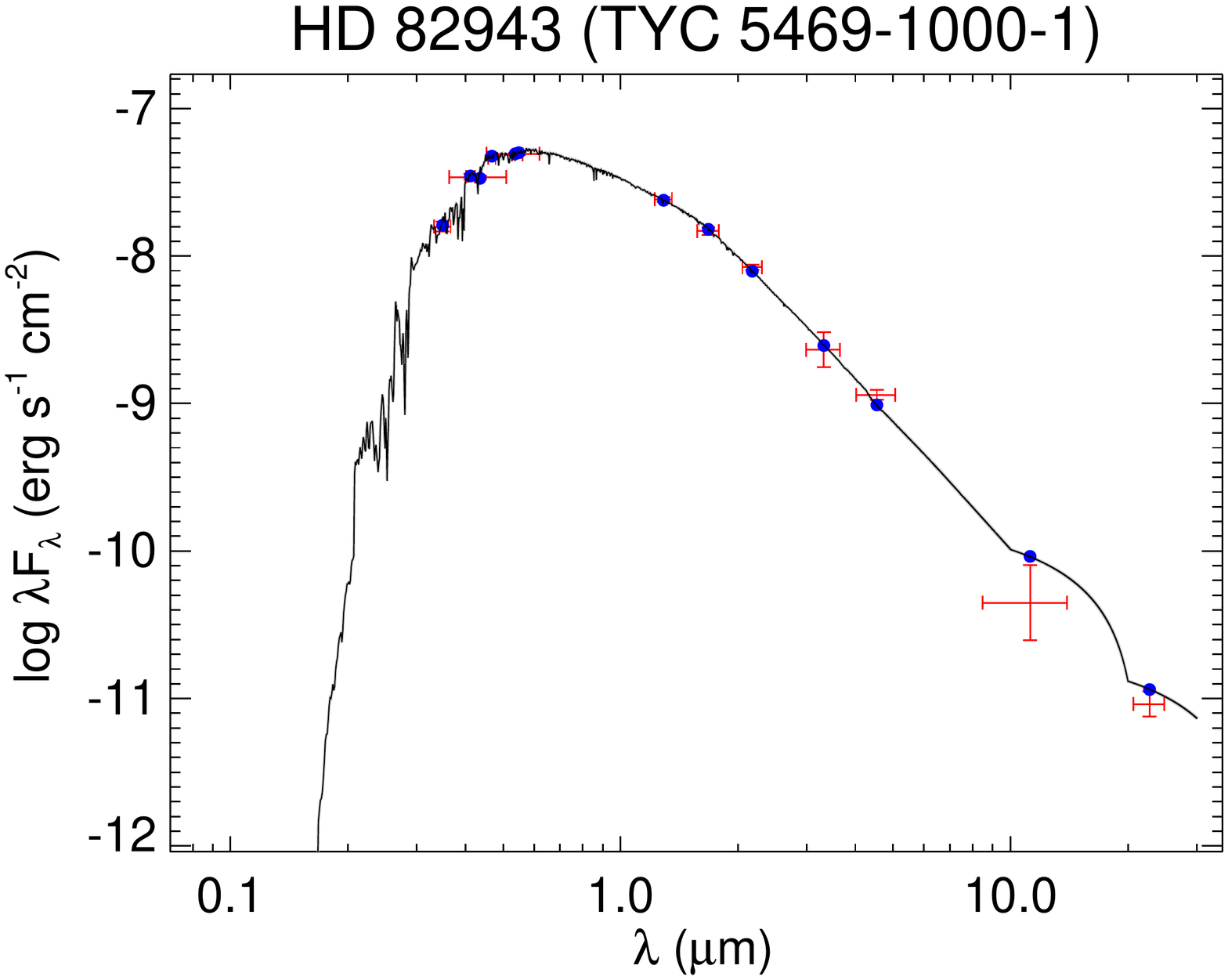}
  \caption{All labels, lines, symbols, and colors as in Figure \ref{fig:seds}.}
  \label{fig:seds_51}
\end{figure}

\begin{figure}[H]
  \centering
  \includegraphics[trim=60 60 60 60,clip,width=0.49\linewidth]{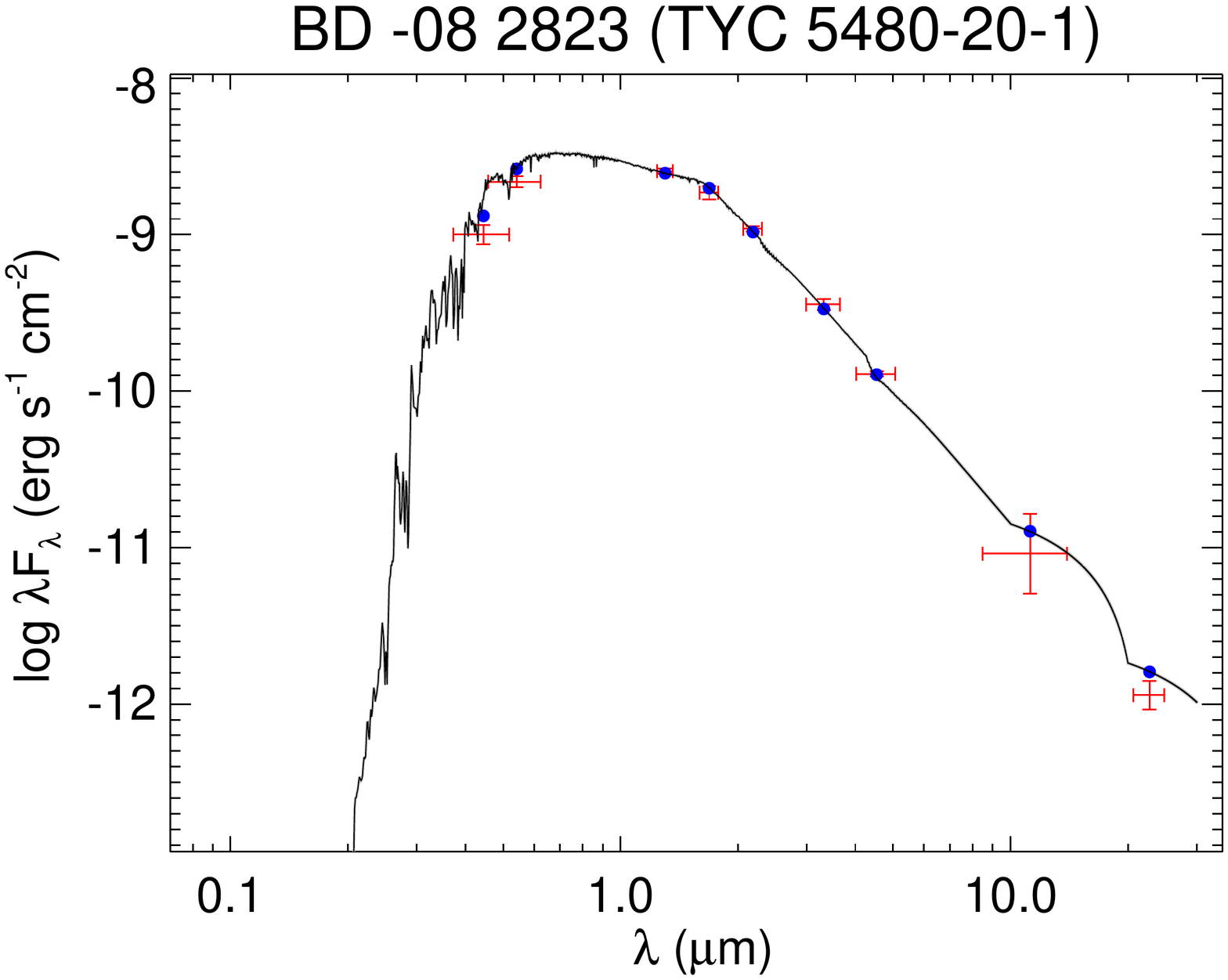}
  \includegraphics[trim=60 60 60 60,clip,width=0.49\linewidth]{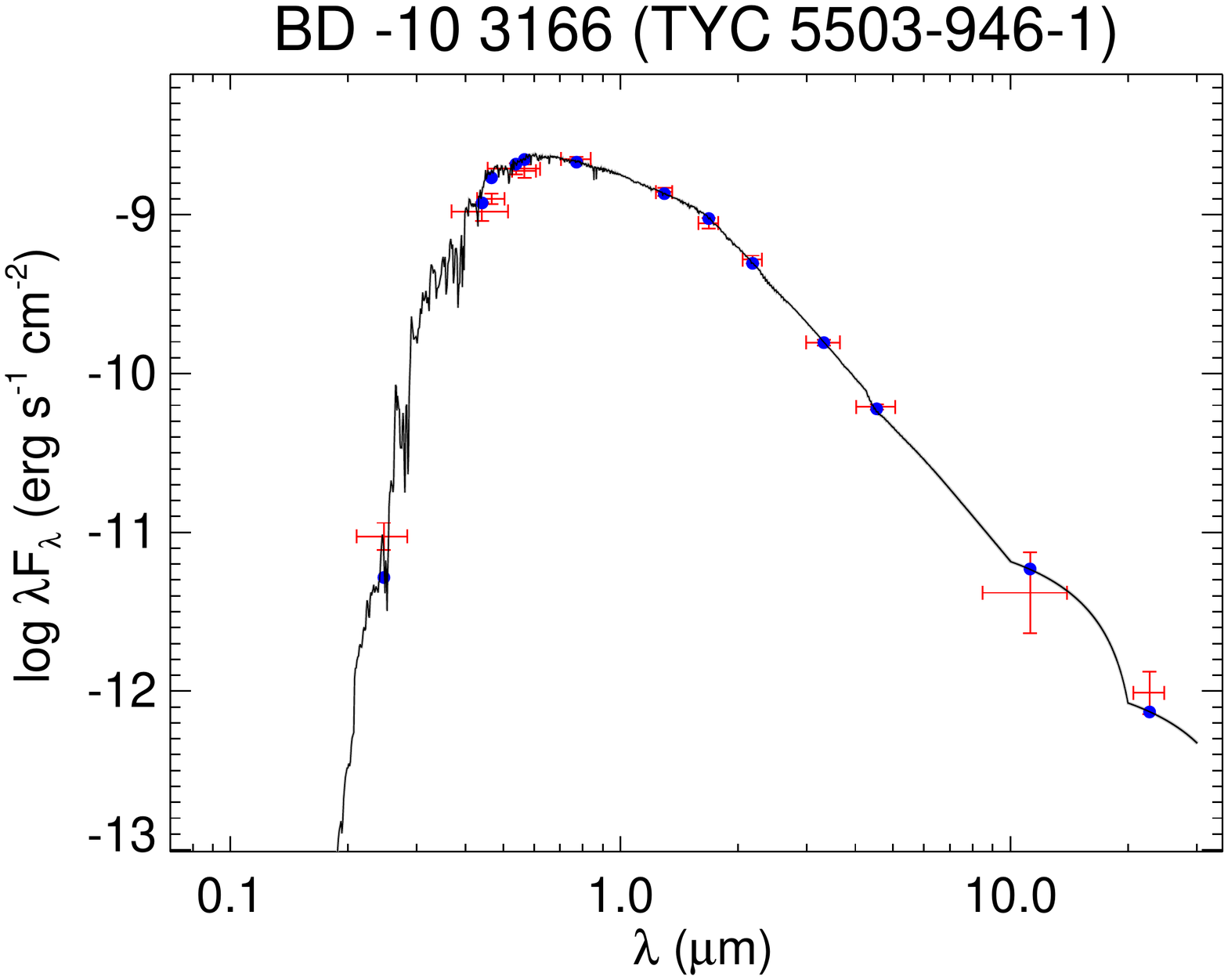}
  \includegraphics[trim=60 60 60 60,clip,width=0.49\linewidth]{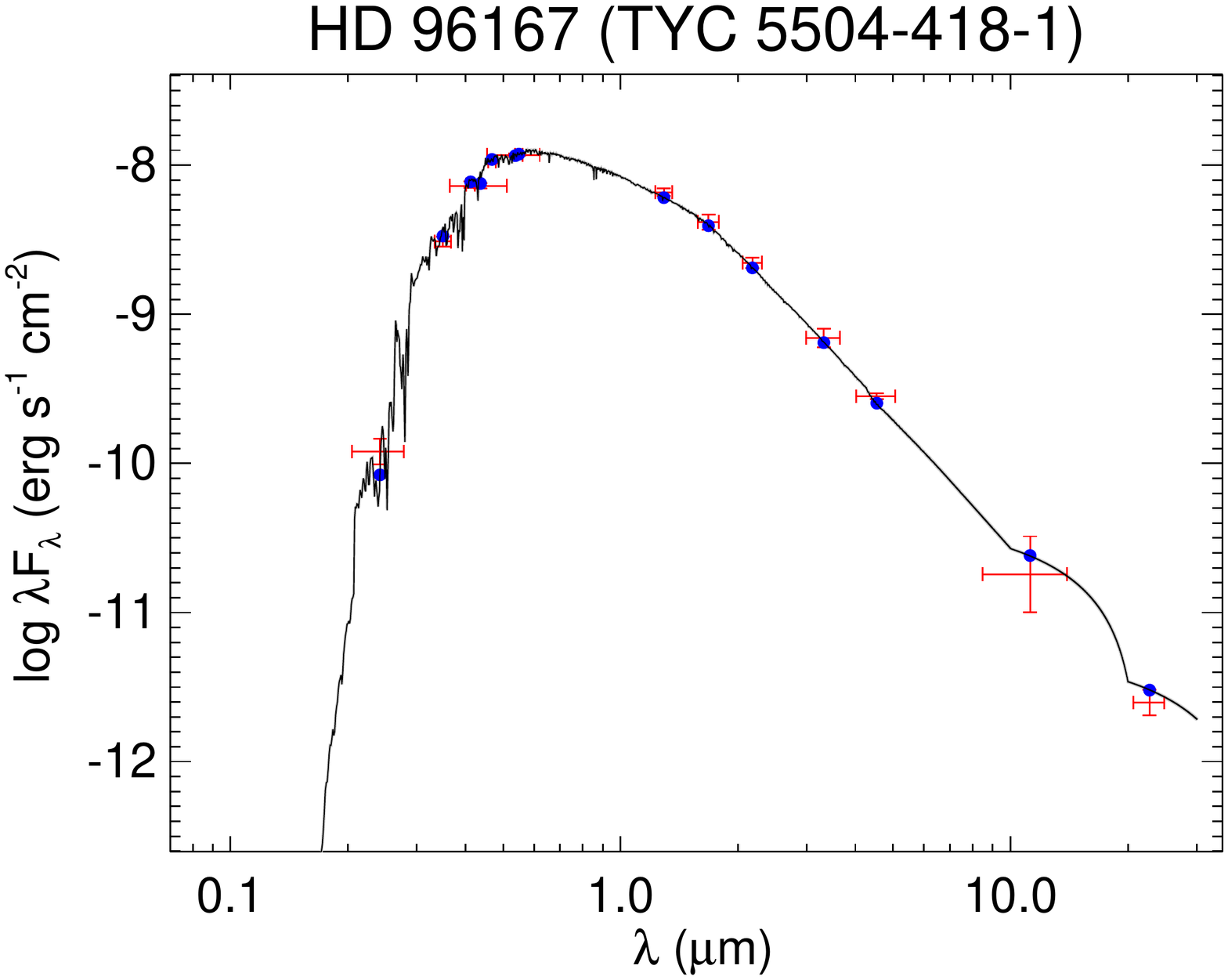}
  \includegraphics[trim=60 60 60 60,clip,width=0.49\linewidth]{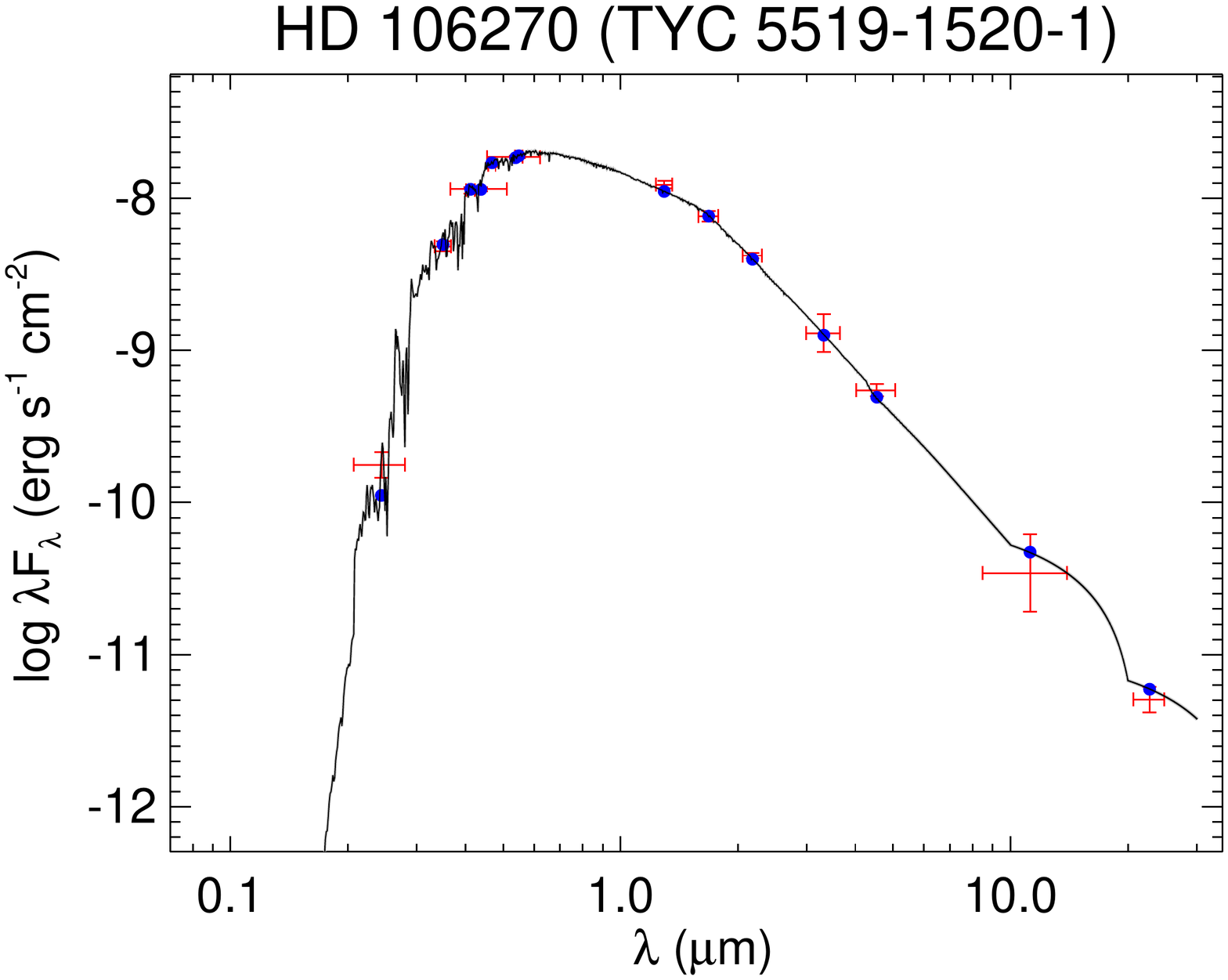}
  \includegraphics[trim=60 60 60 60,clip,width=0.49\linewidth]{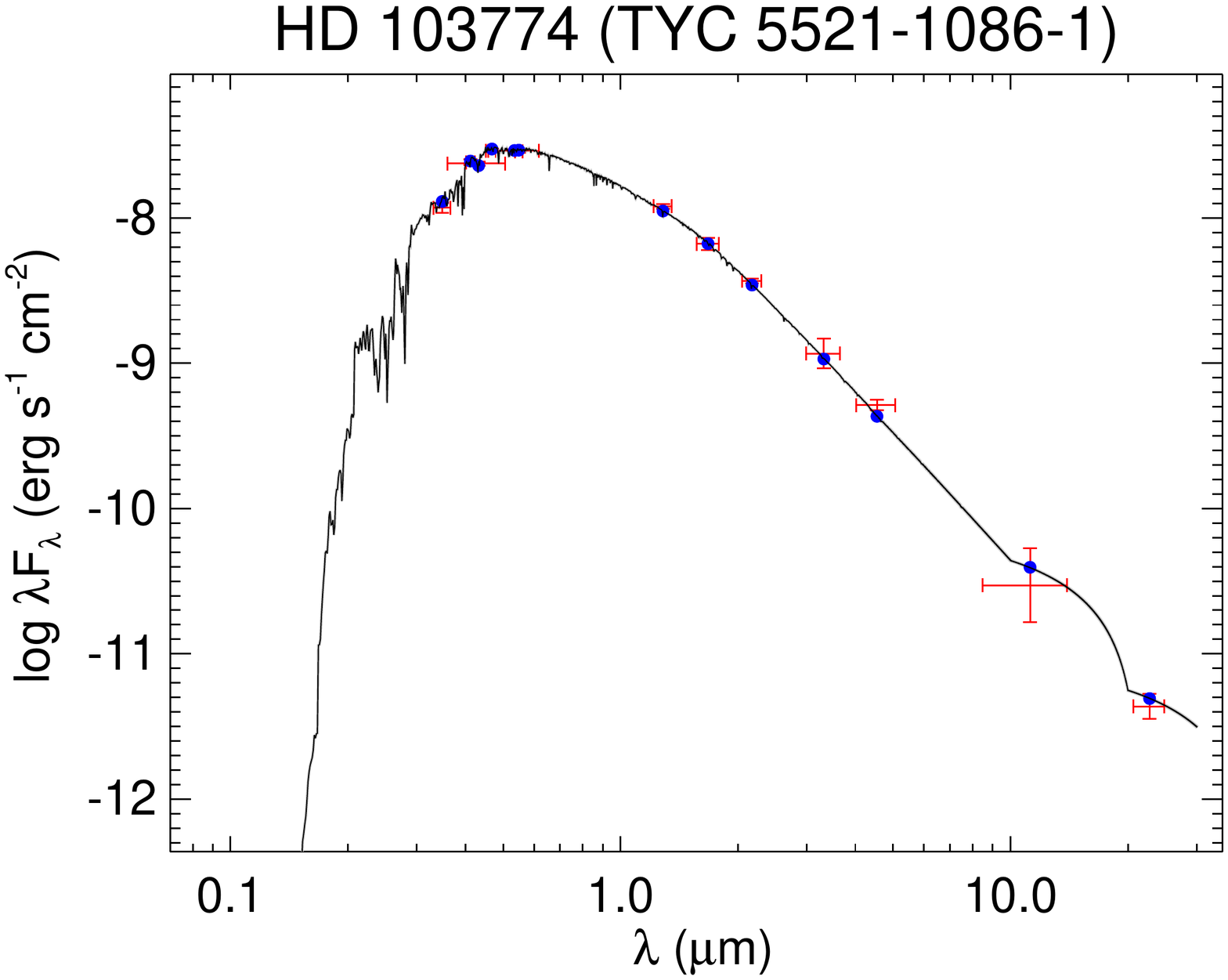}
  \includegraphics[trim=60 60 60 60,clip,width=0.49\linewidth]{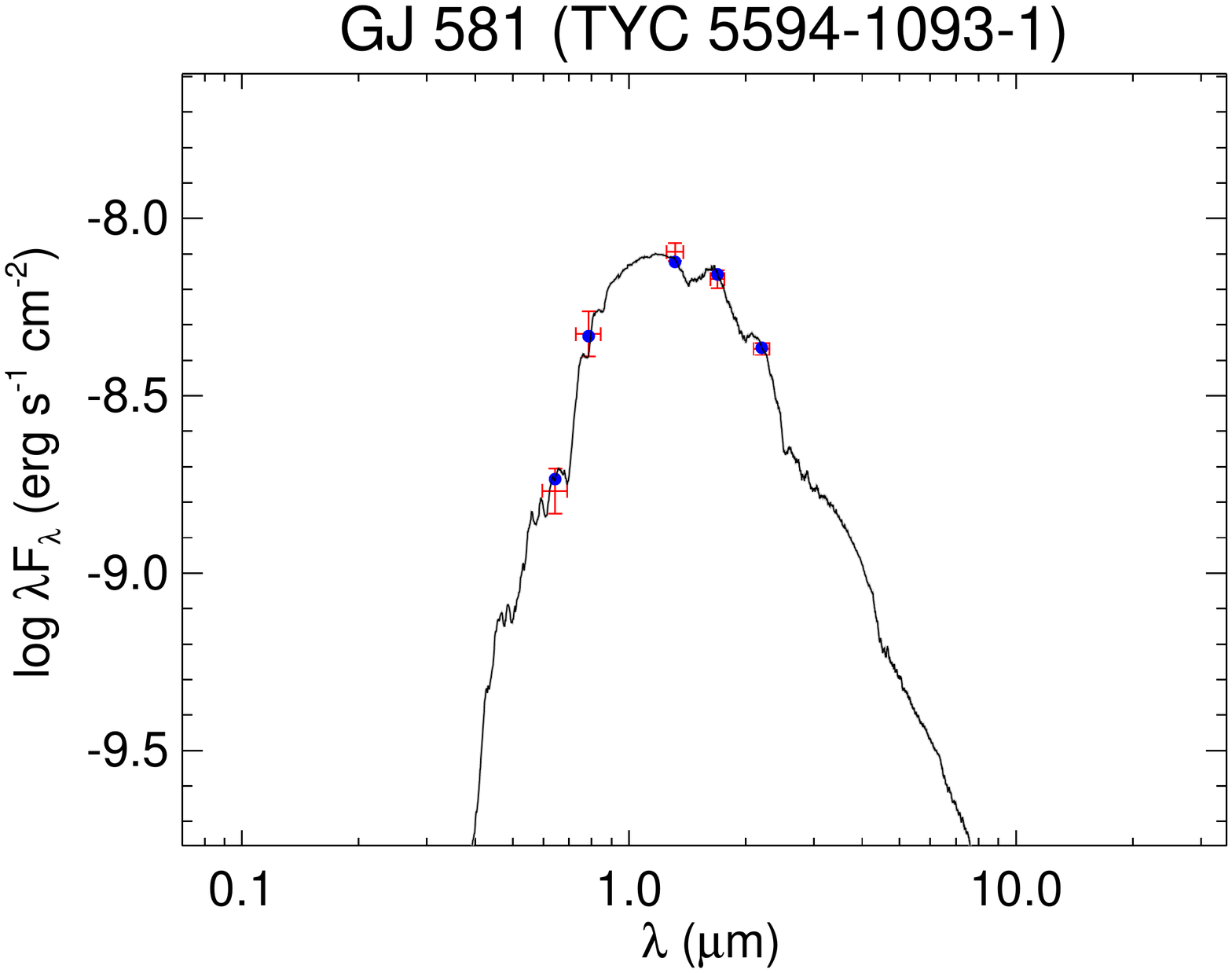}
  \caption{All labels, lines, symbols, and colors as in Figure \ref{fig:seds}.}
  \label{fig:seds_52}
\end{figure}

\begin{figure}[H]
  \centering
  \includegraphics[trim=60 60 60 60,clip,width=0.49\linewidth]{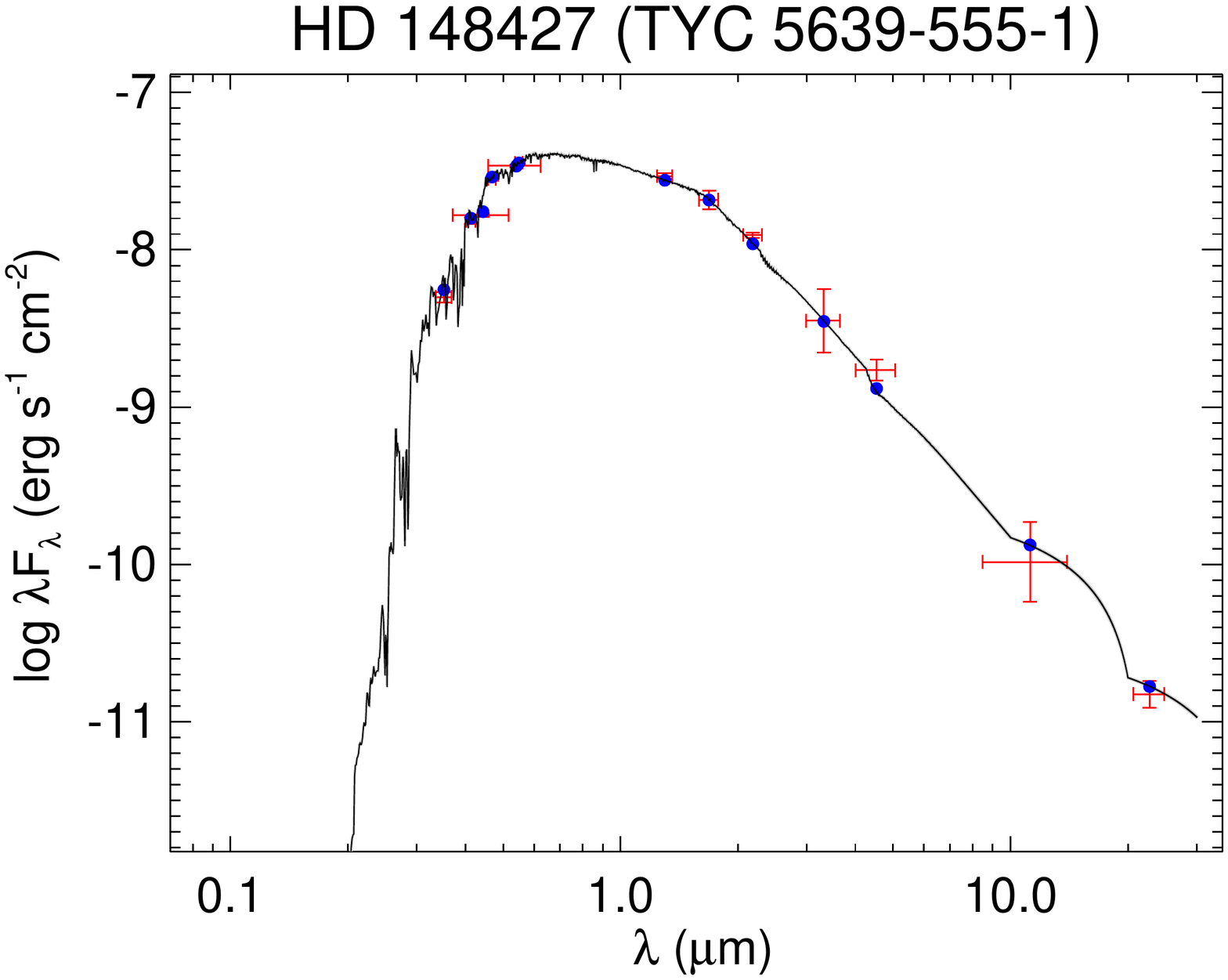}
  \includegraphics[trim=60 60 60 60,clip,width=0.49\linewidth]{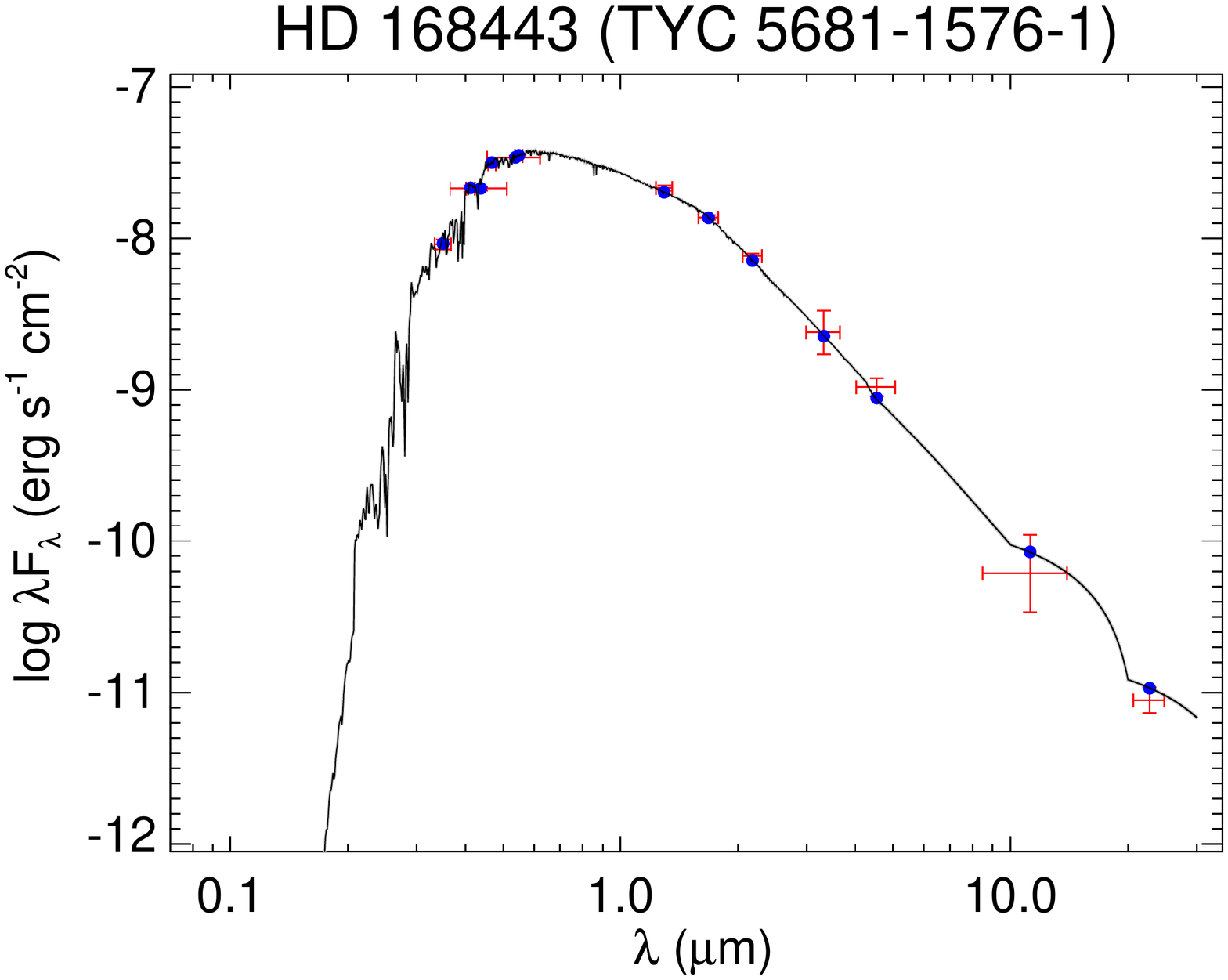}
  \includegraphics[trim=60 60 60 60,clip,width=0.49\linewidth]{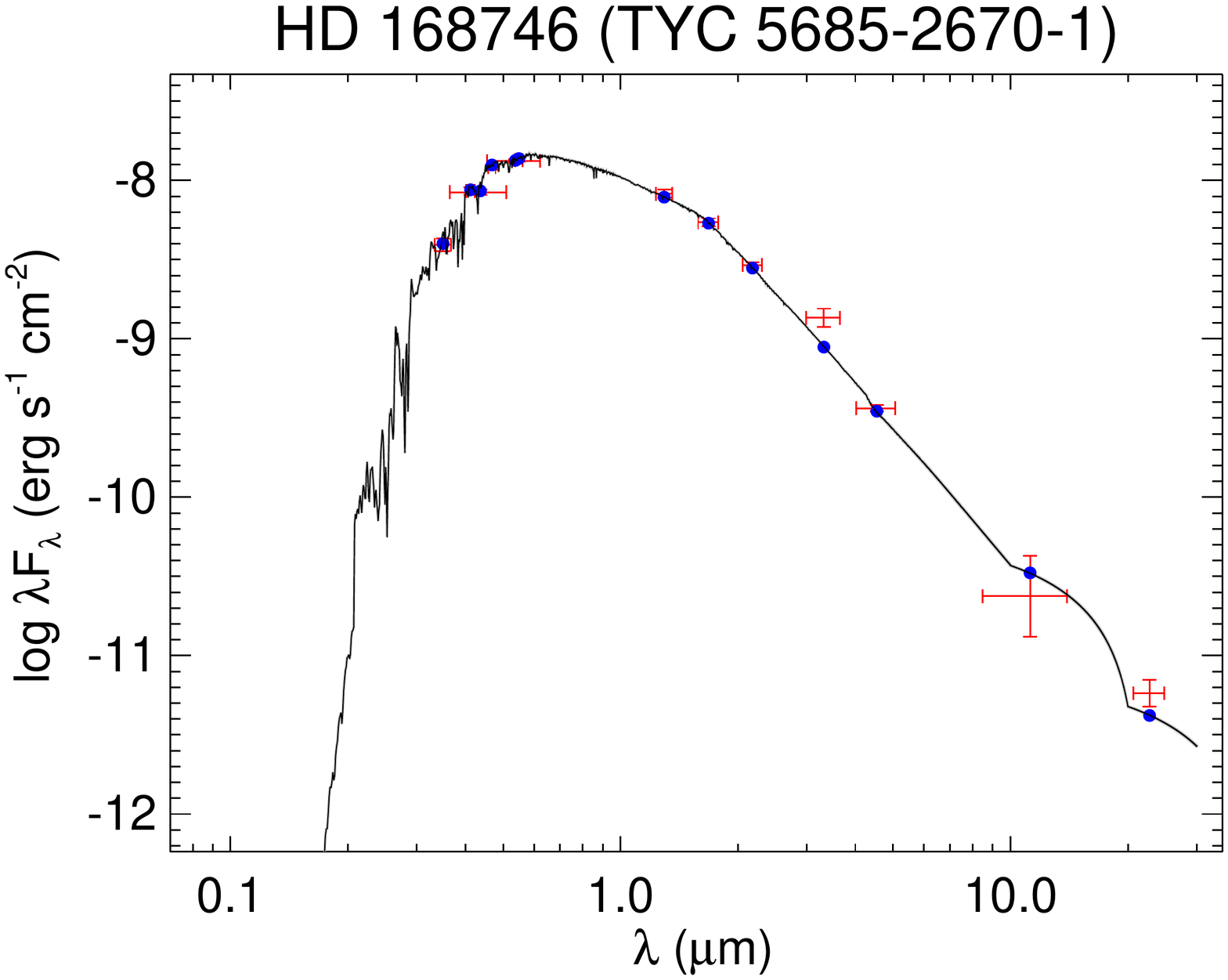}
  \includegraphics[trim=60 60 60 60,clip,width=0.49\linewidth]{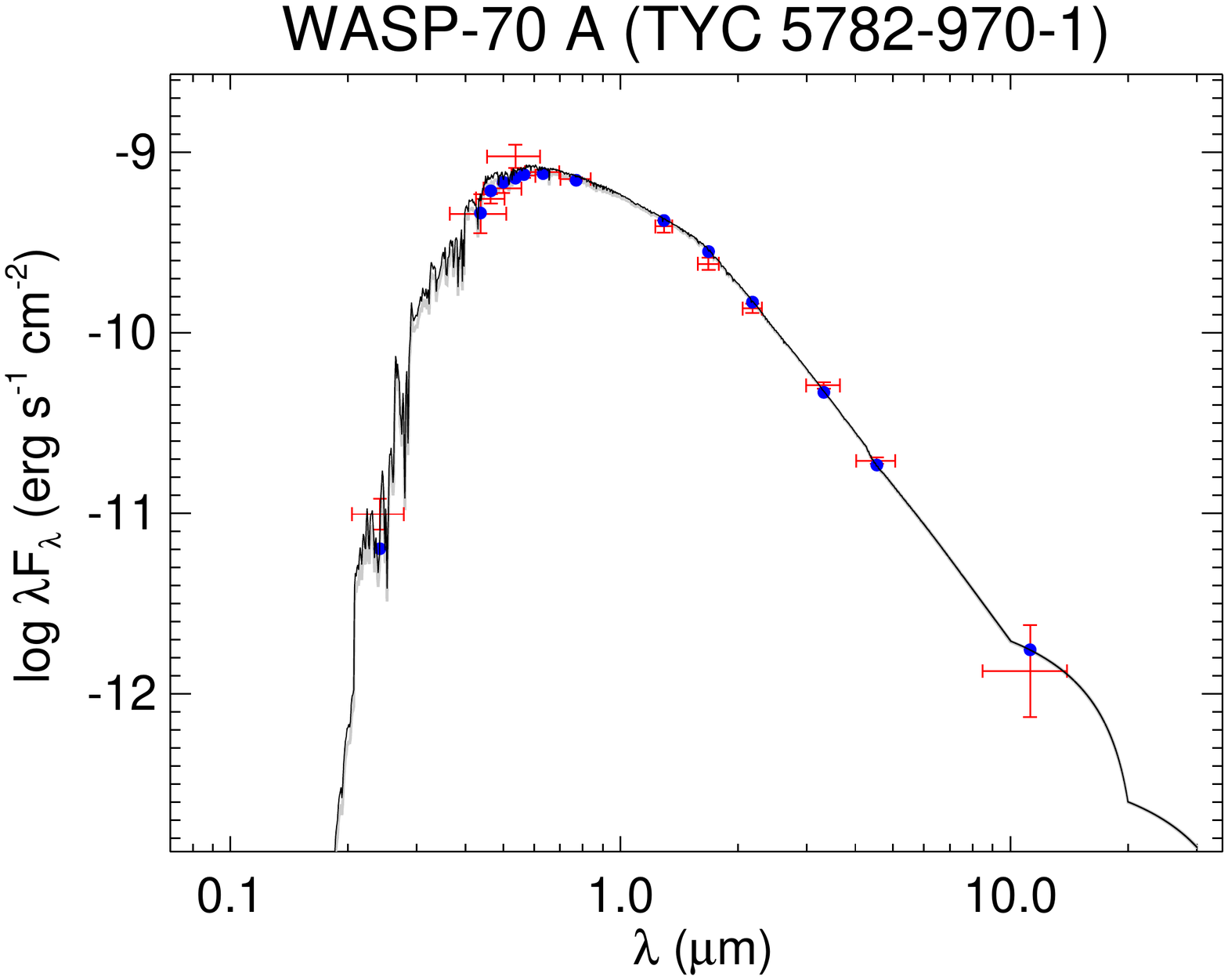}
  \includegraphics[trim=60 60 60 60,clip,width=0.49\linewidth]{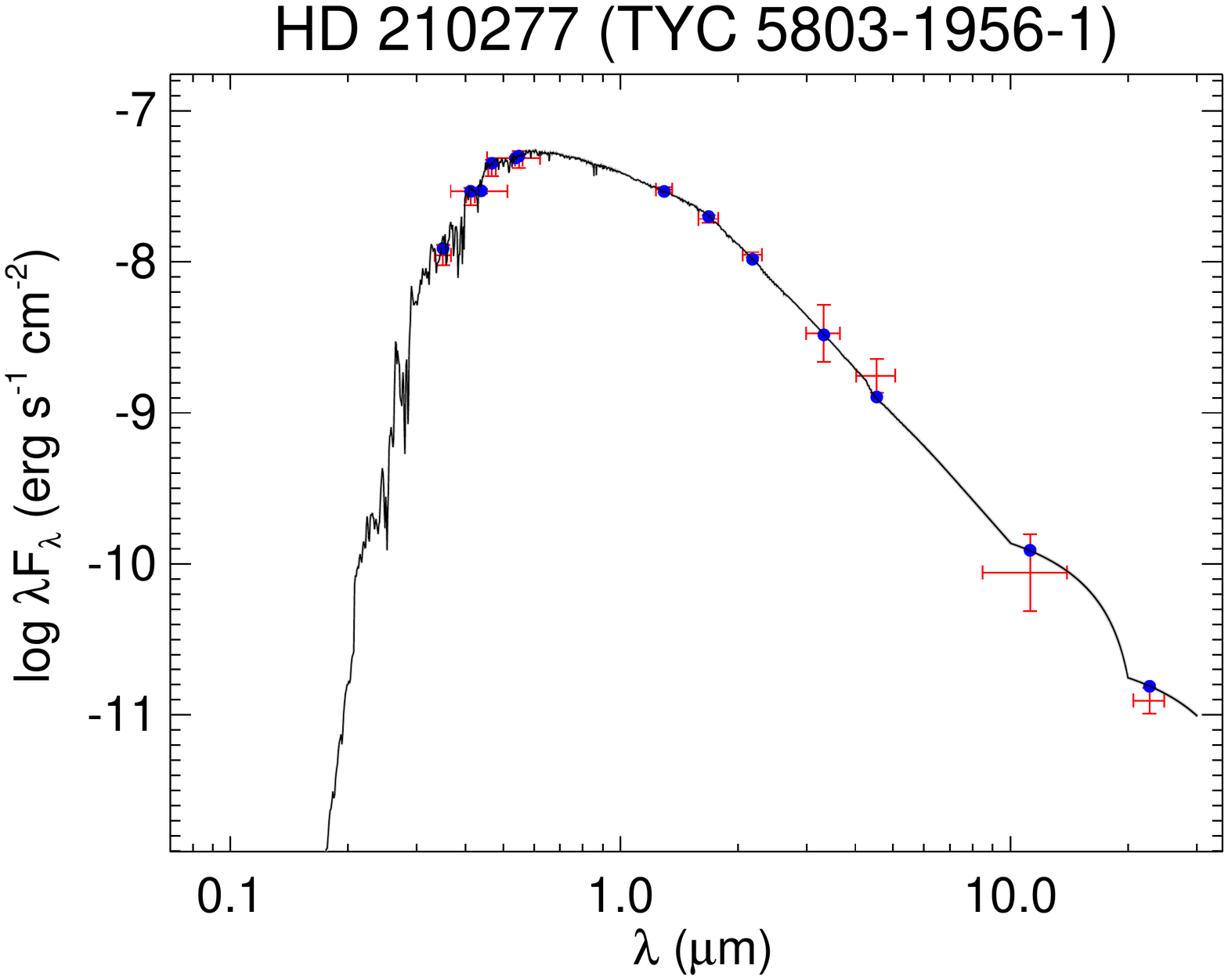}
  \includegraphics[trim=60 60 60 60,clip,width=0.49\linewidth]{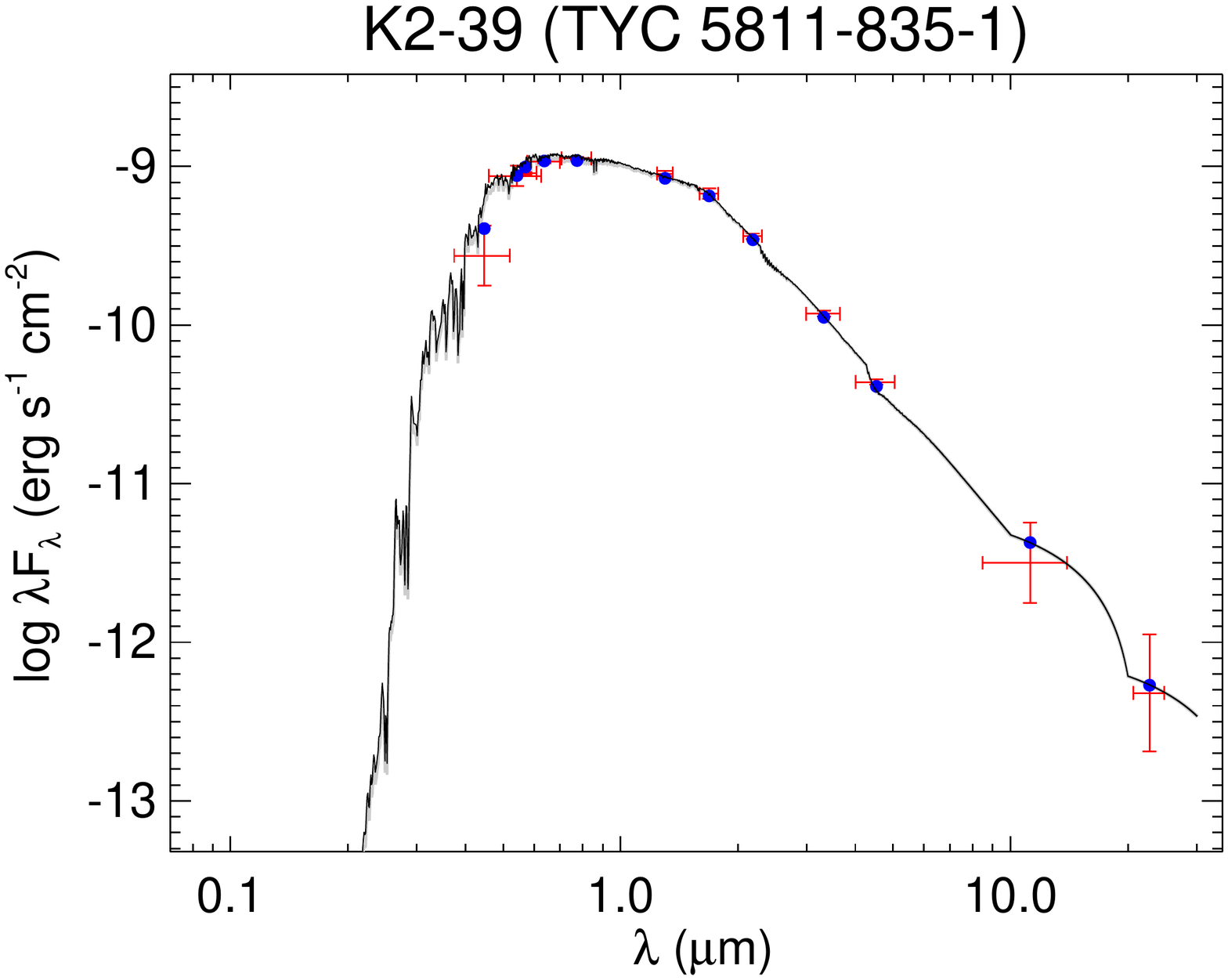}
  \caption{All labels, lines, symbols, and colors as in Figure \ref{fig:seds}.}
  \label{fig:seds_53}
\end{figure}

\begin{figure}[H]
  \centering
  \includegraphics[trim=60 60 60 60,clip,width=0.49\linewidth]{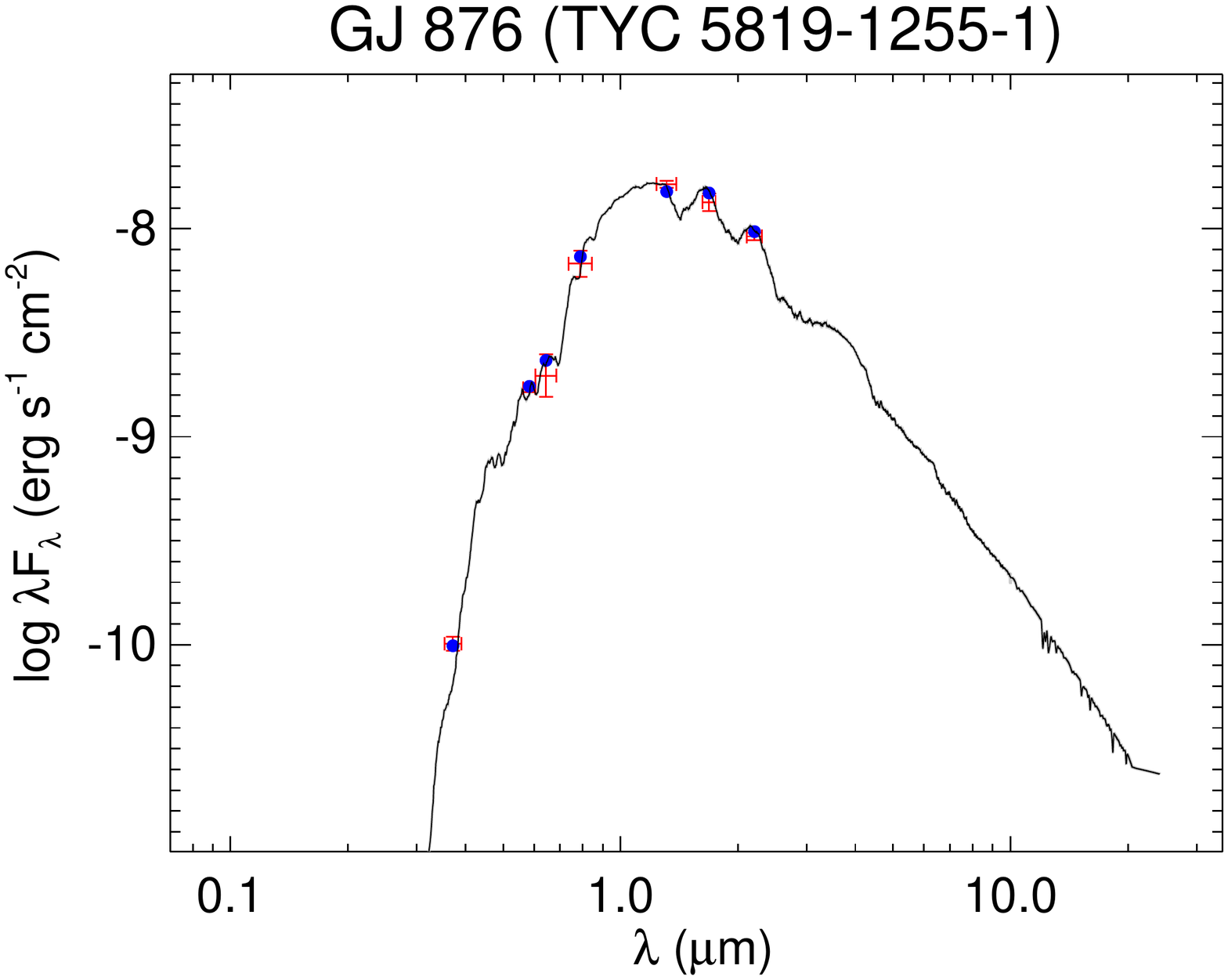}
  \includegraphics[trim=60 60 60 60,clip,width=0.49\linewidth]{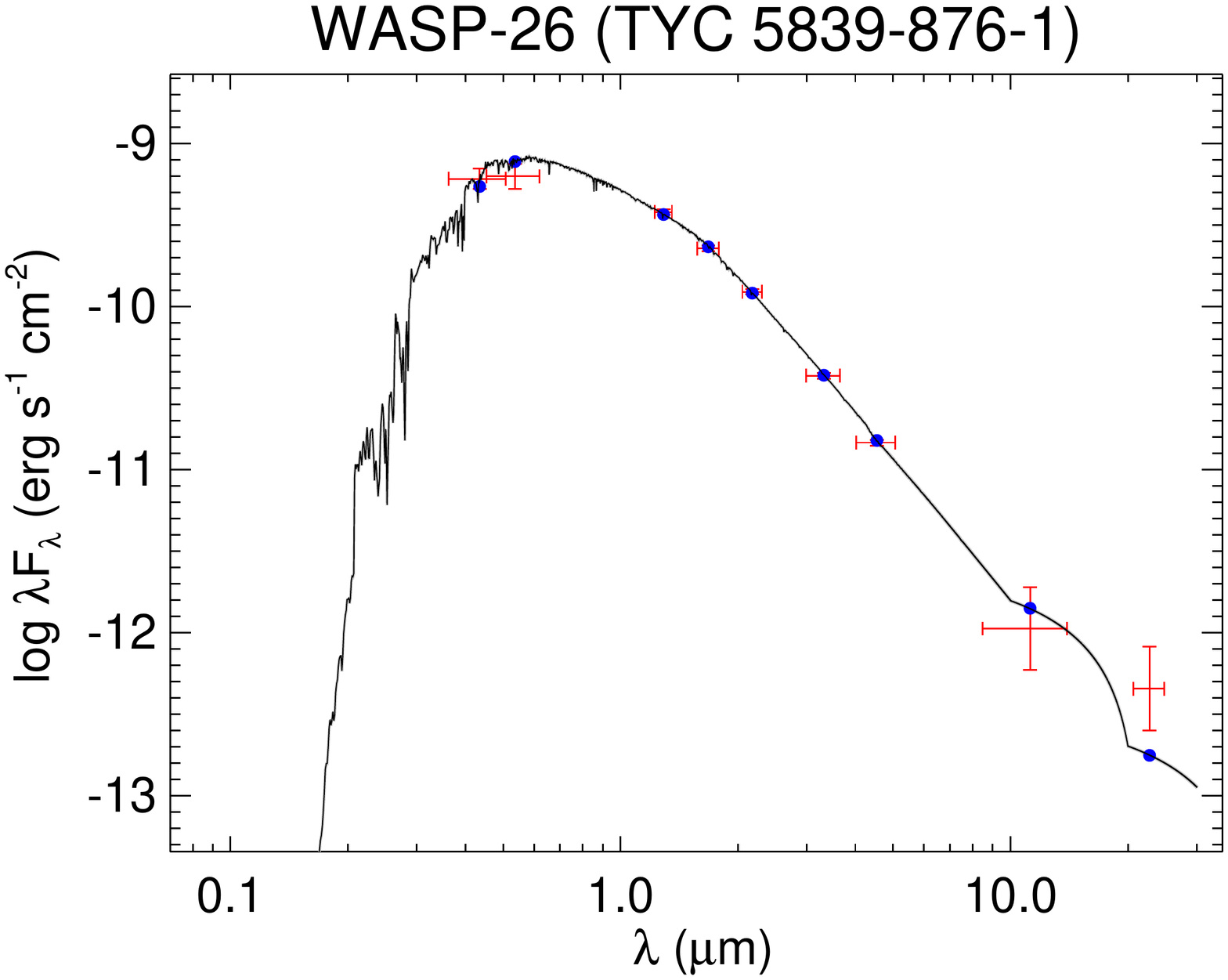}
  \includegraphics[trim=60 60 60 60,clip,width=0.49\linewidth]{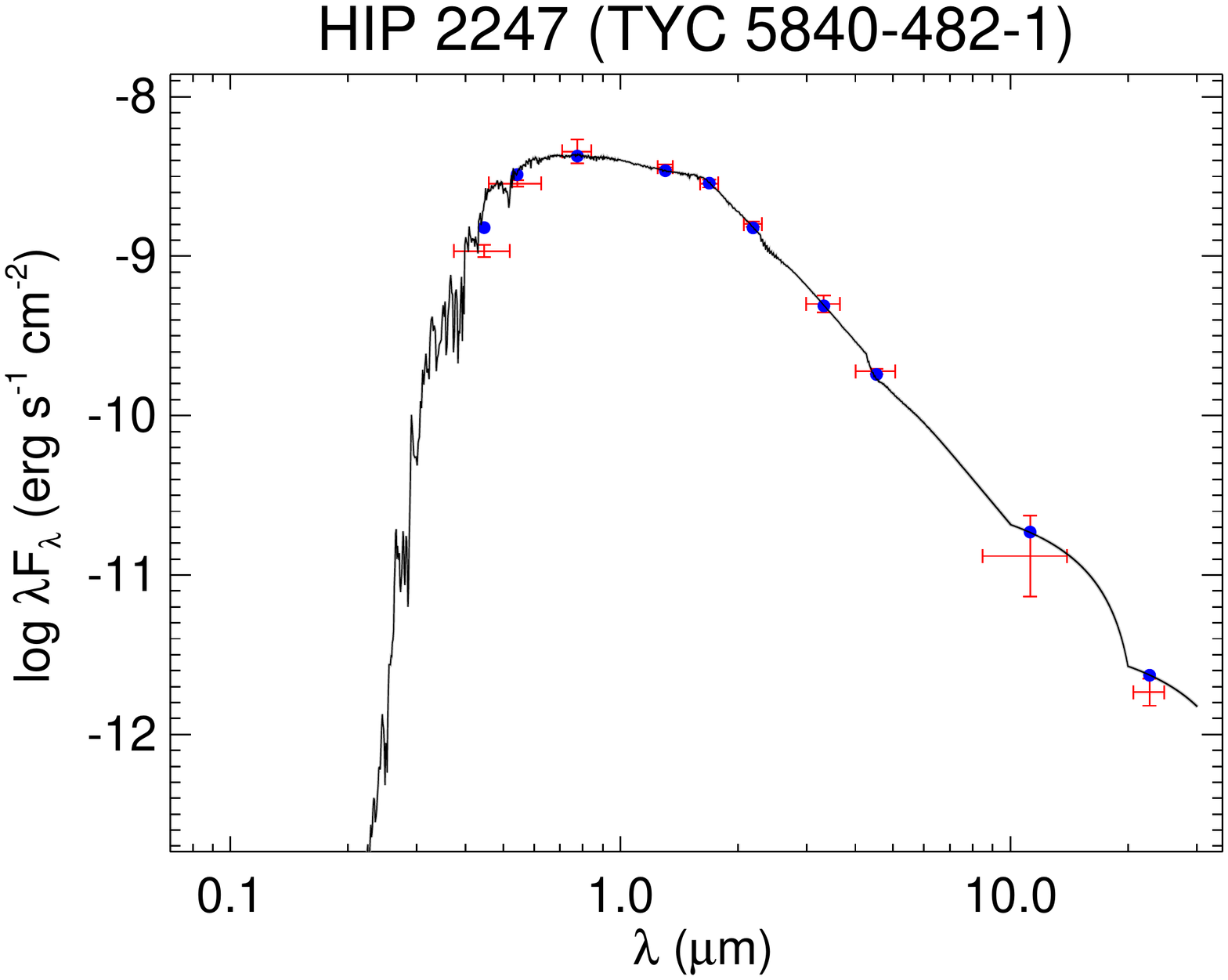}
  \includegraphics[trim=60 60 60 60,clip,width=0.49\linewidth]{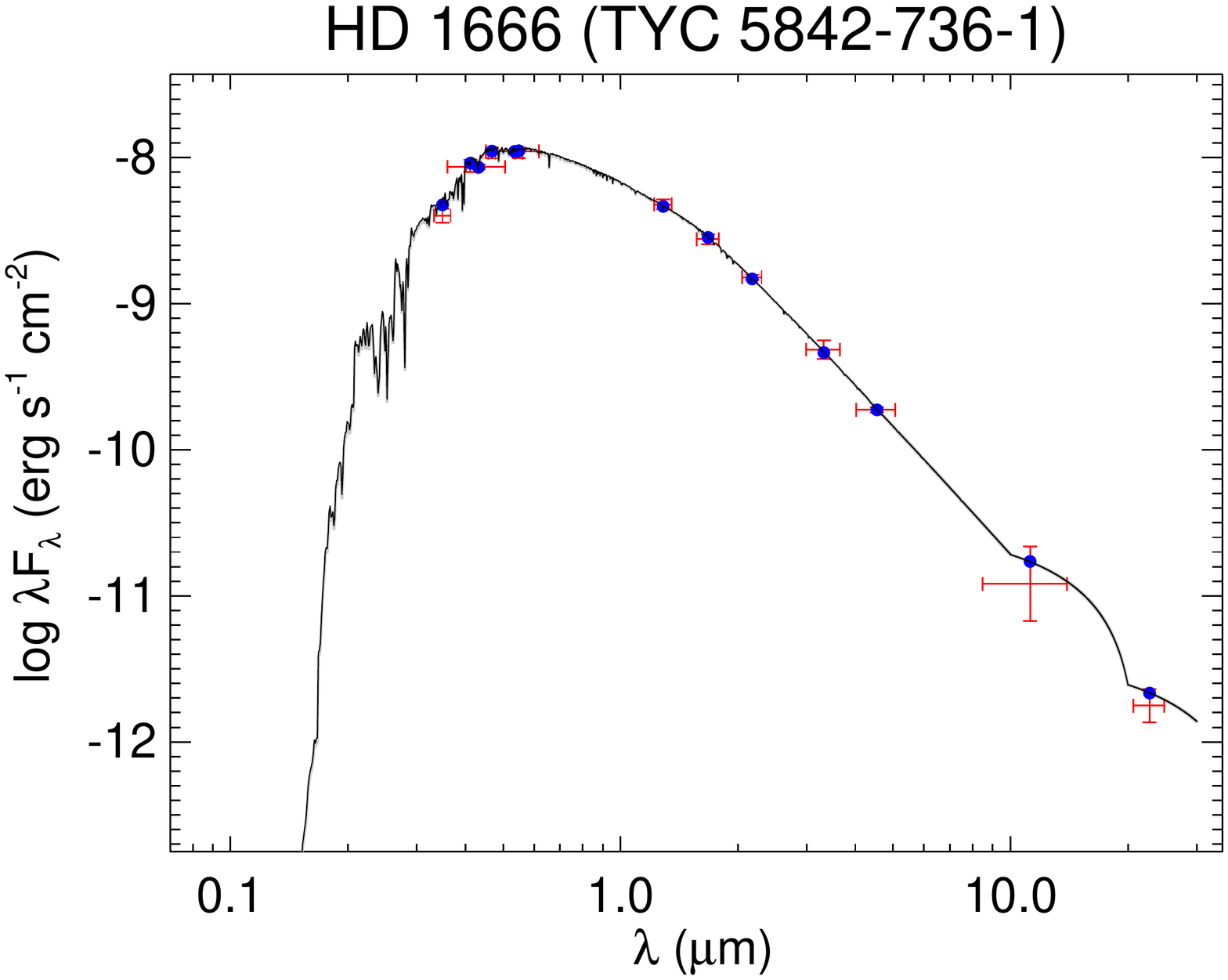}
  \includegraphics[trim=60 60 60 60,clip,width=0.49\linewidth]{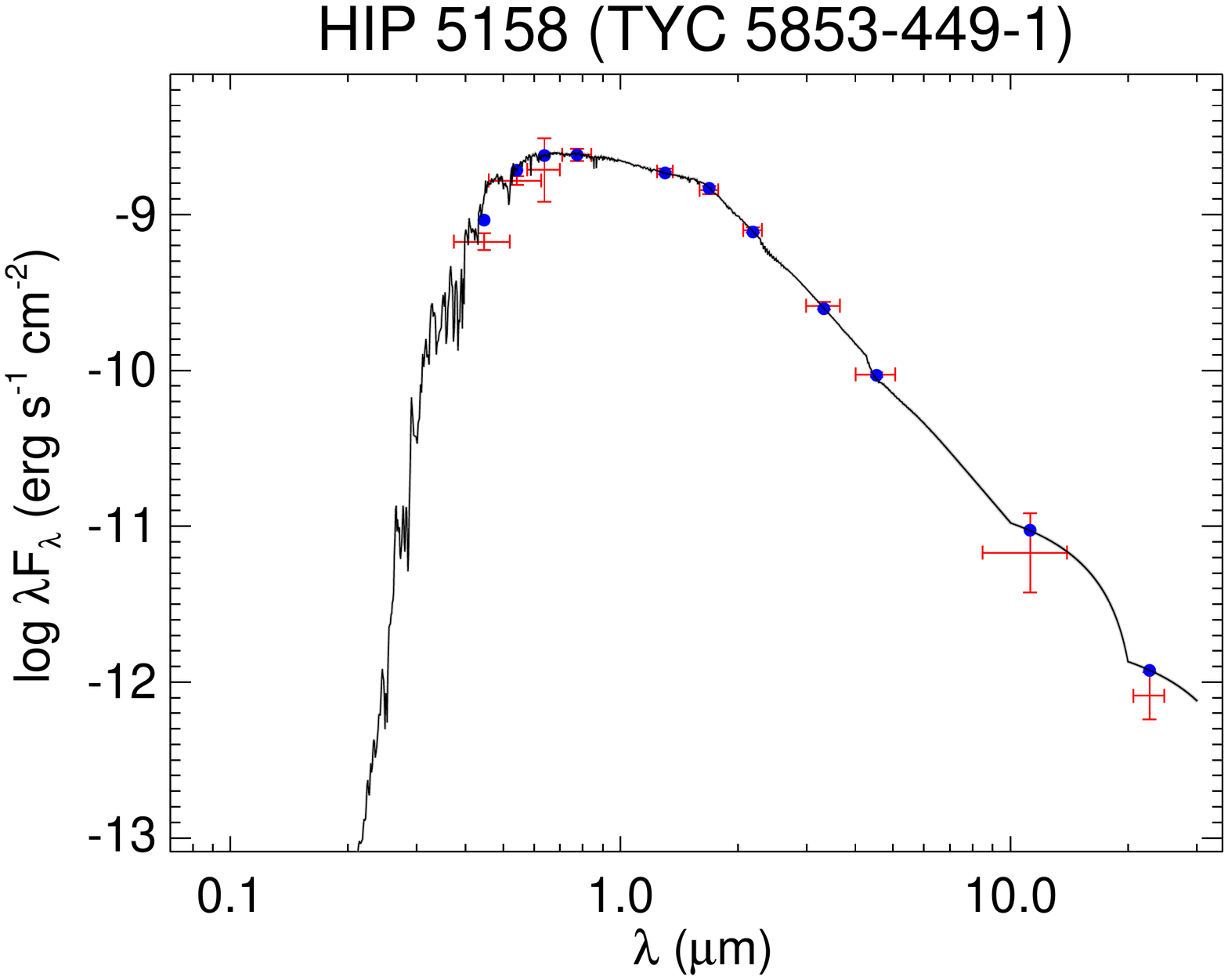}
  \includegraphics[trim=60 60 60 60,clip,width=0.49\linewidth]{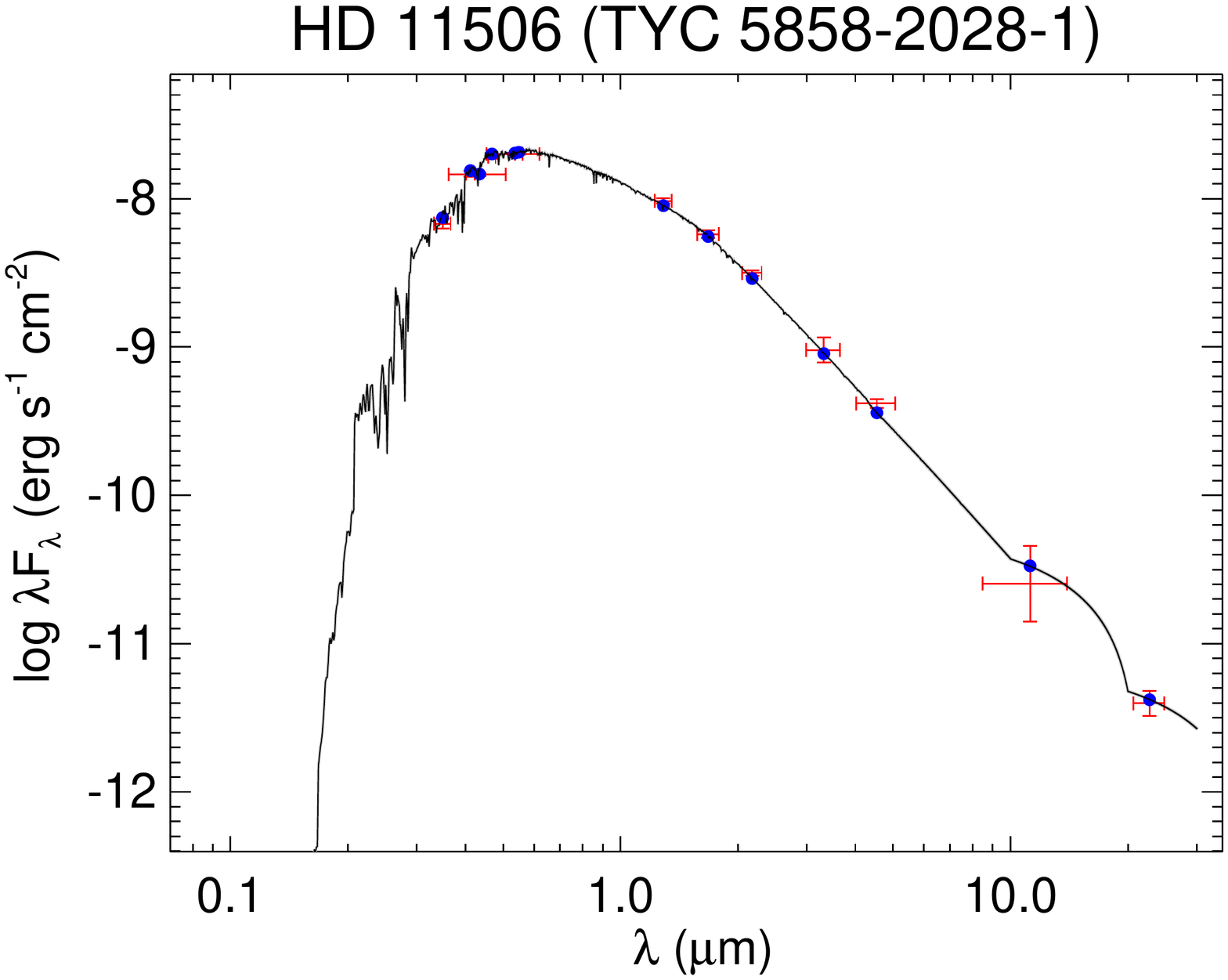}
  \caption{All labels, lines, symbols, and colors as in Figure \ref{fig:seds}.}
  \label{fig:seds_54}
\end{figure}

\begin{figure}[H]
  \centering
  \includegraphics[trim=60 60 60 60,clip,width=0.49\linewidth]{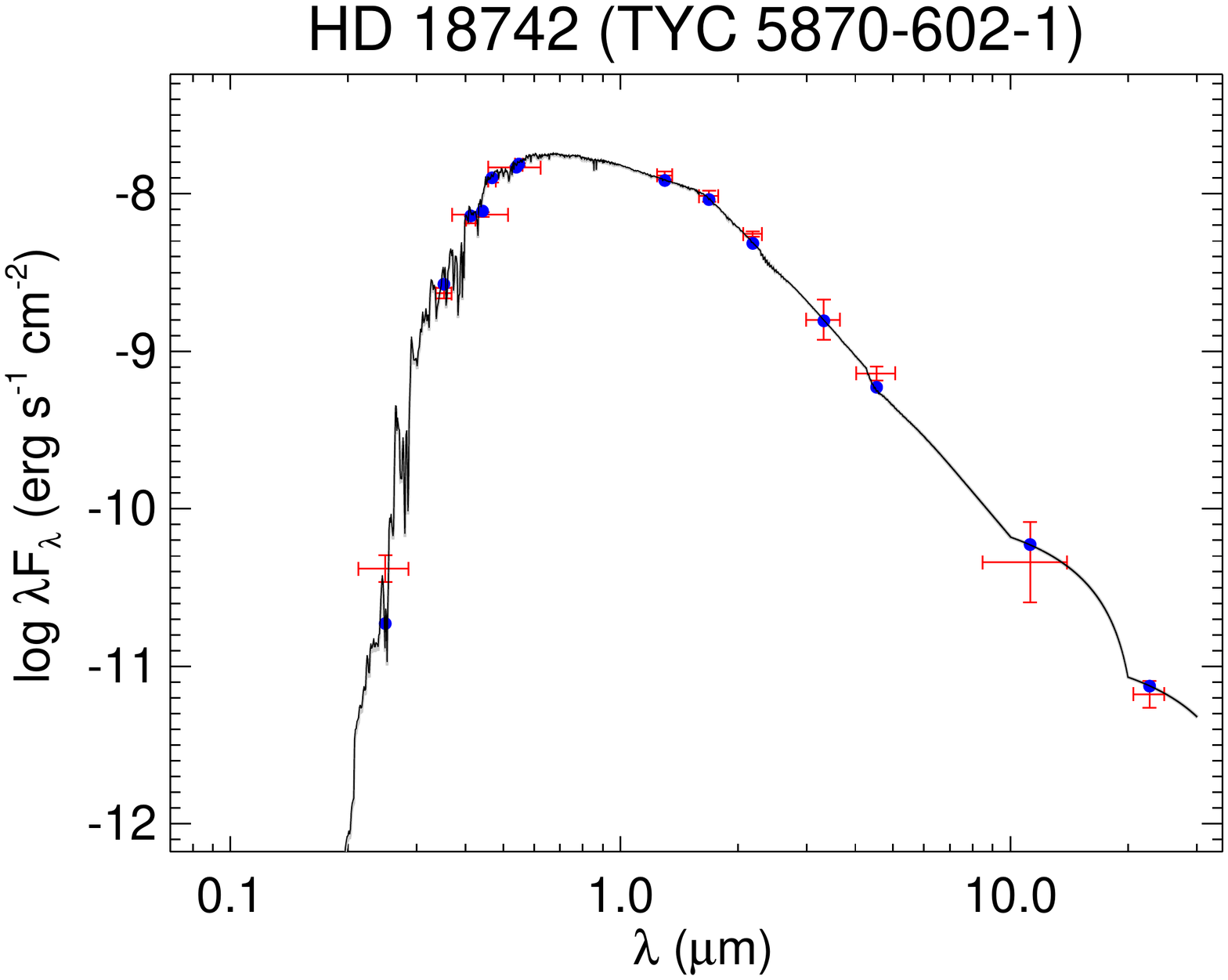}
  \includegraphics[trim=60 60 60 60,clip,width=0.49\linewidth]{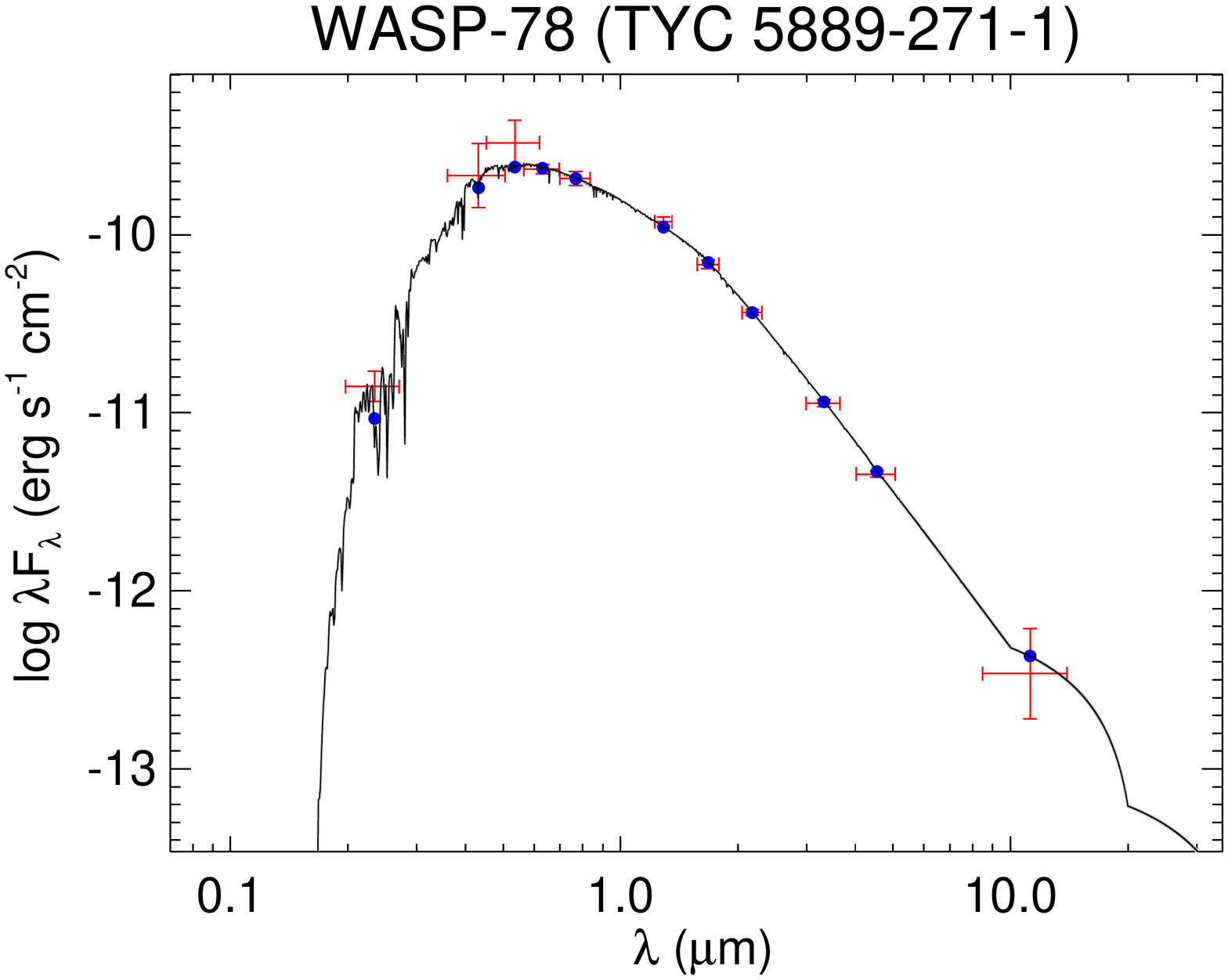}
  \includegraphics[trim=60 60 60 60,clip,width=0.49\linewidth]{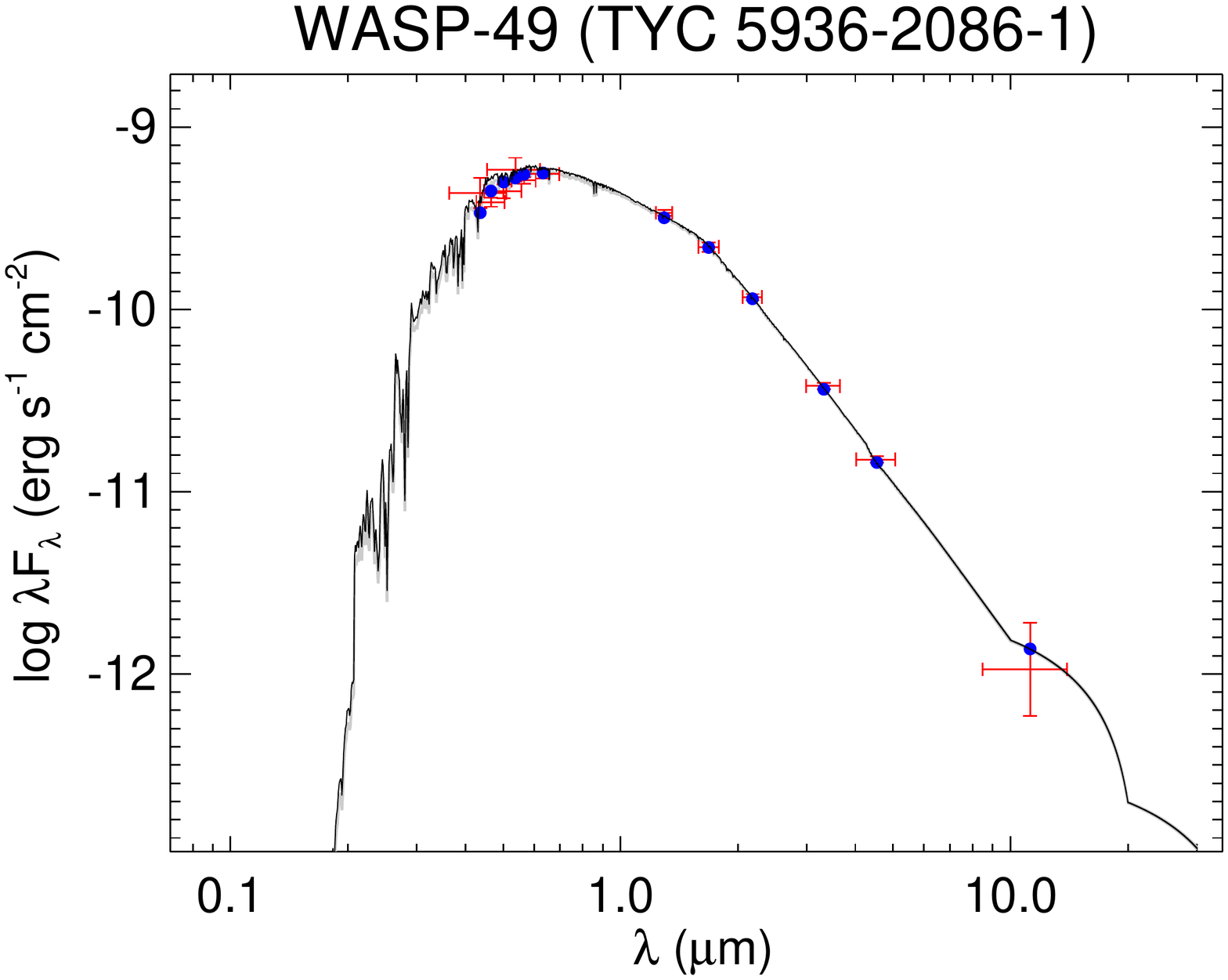}
  \includegraphics[trim=60 60 60 60,clip,width=0.49\linewidth]{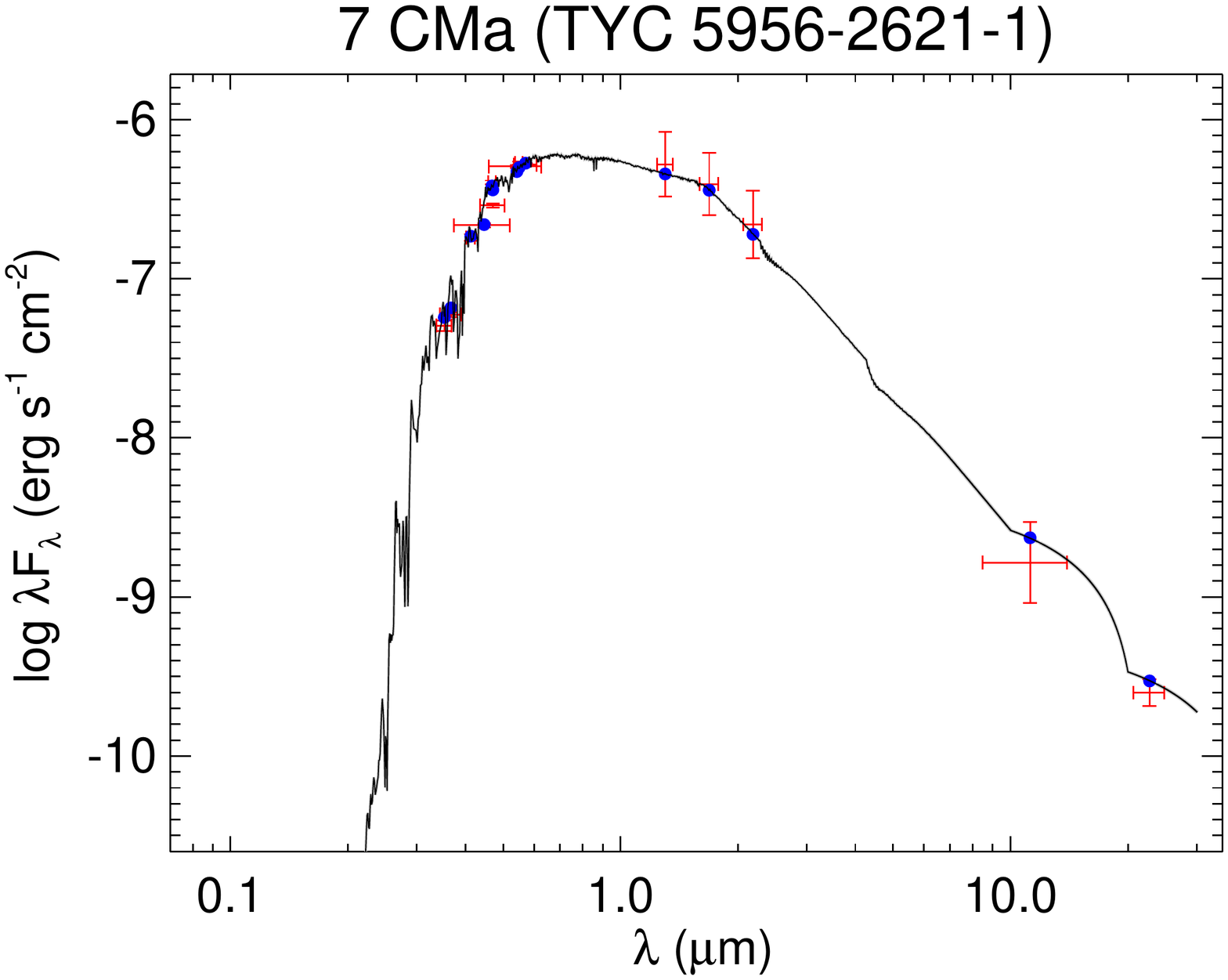}
  \includegraphics[trim=60 60 60 60,clip,width=0.49\linewidth]{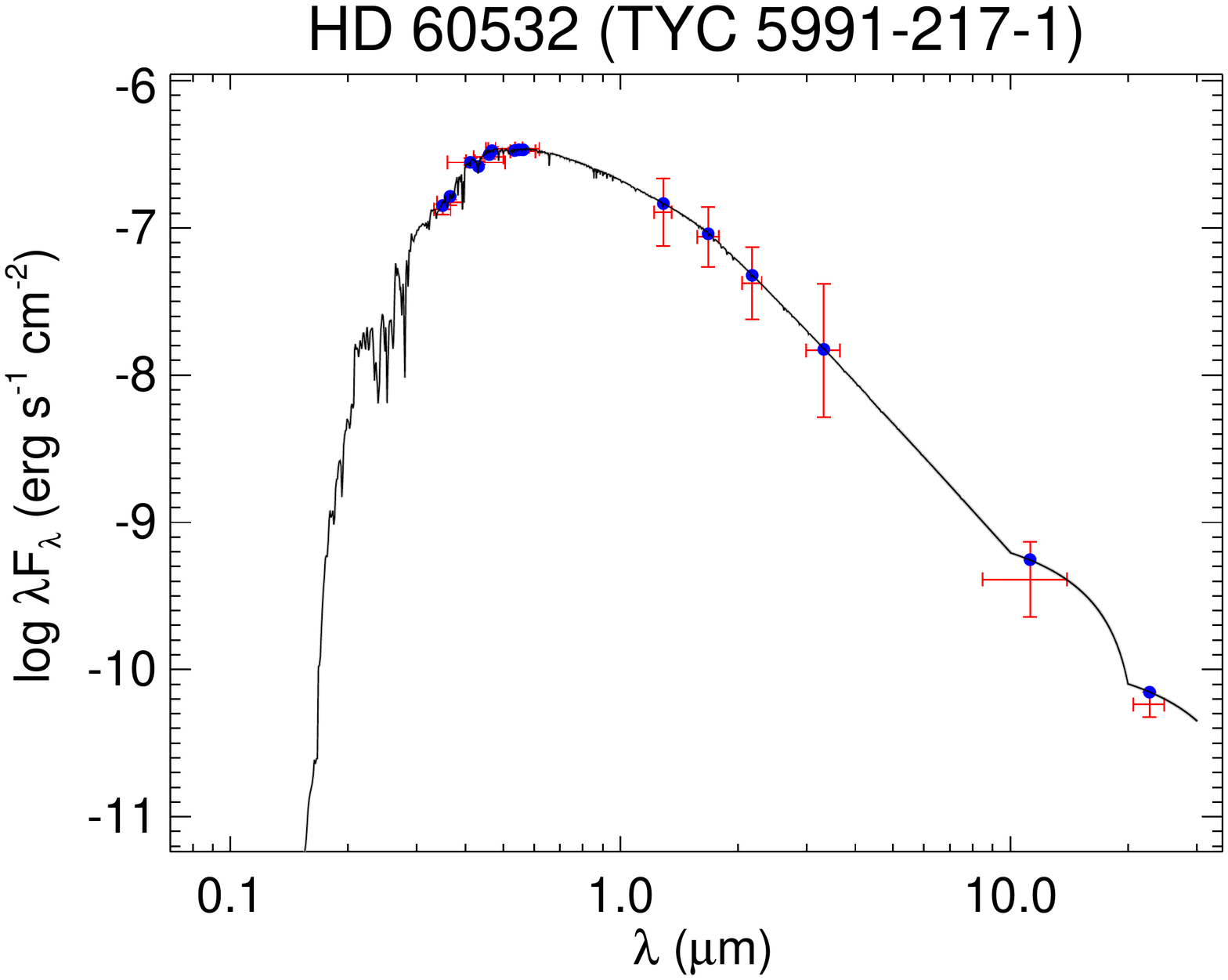}
  \includegraphics[trim=60 60 60 60,clip,width=0.49\linewidth]{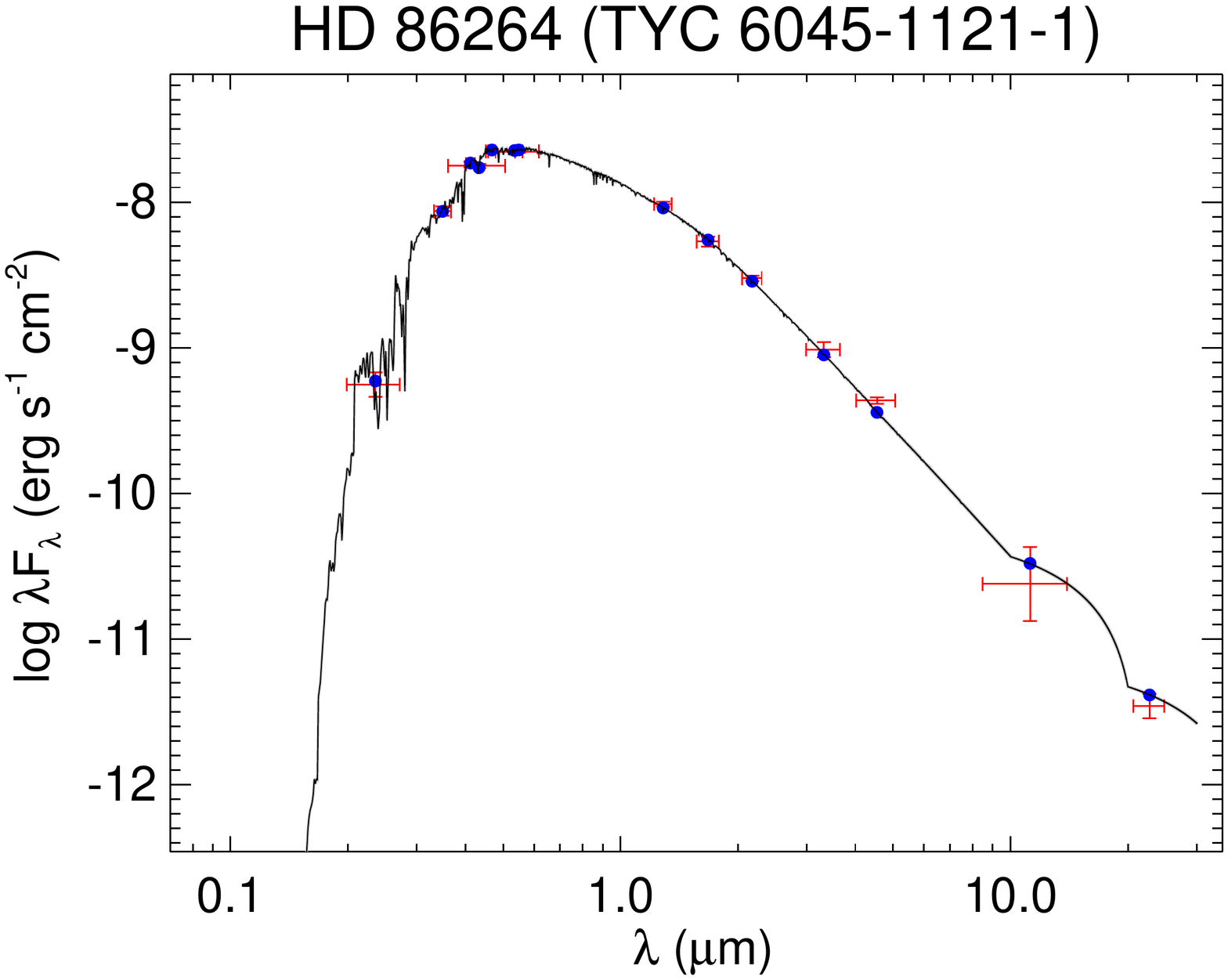}
  \caption{All labels, lines, symbols, and colors as in Figure \ref{fig:seds}.}
  \label{fig:seds_55}
\end{figure}

\begin{figure}[H]
  \centering
  \includegraphics[trim=60 60 60 60,clip,width=0.49\linewidth]{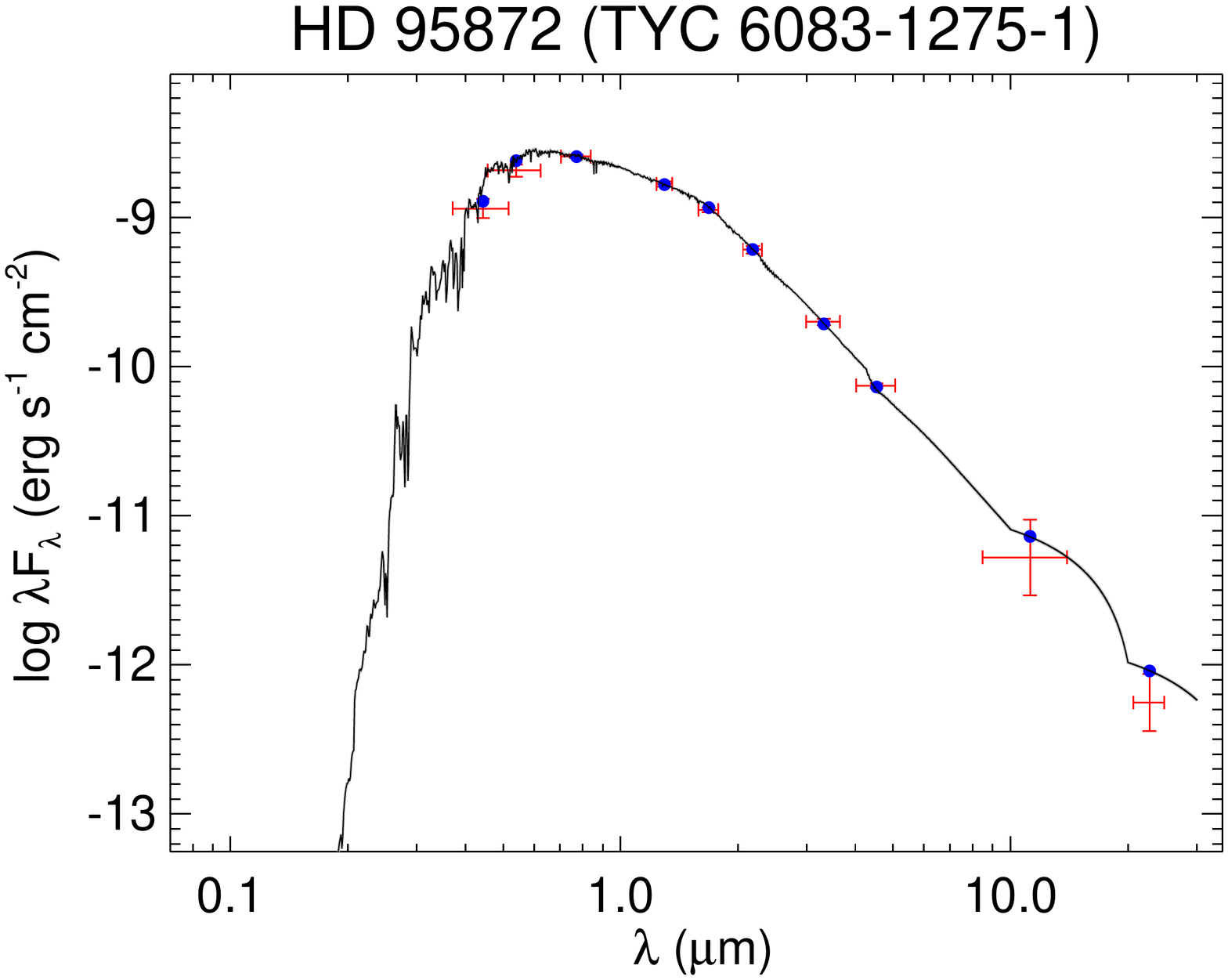}
  \includegraphics[trim=60 60 60 60,clip,width=0.49\linewidth]{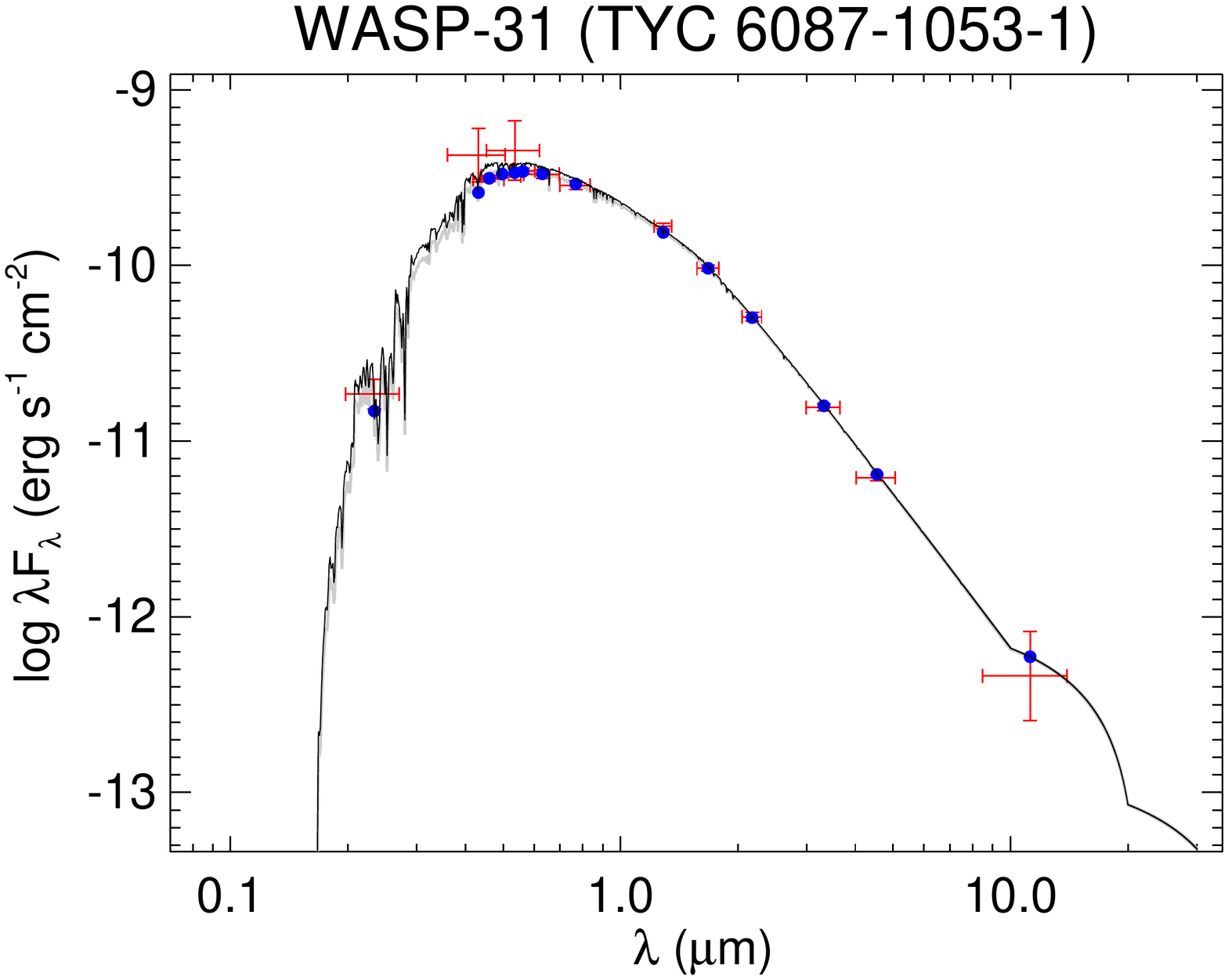}
  \includegraphics[trim=60 60 60 60,clip,width=0.49\linewidth]{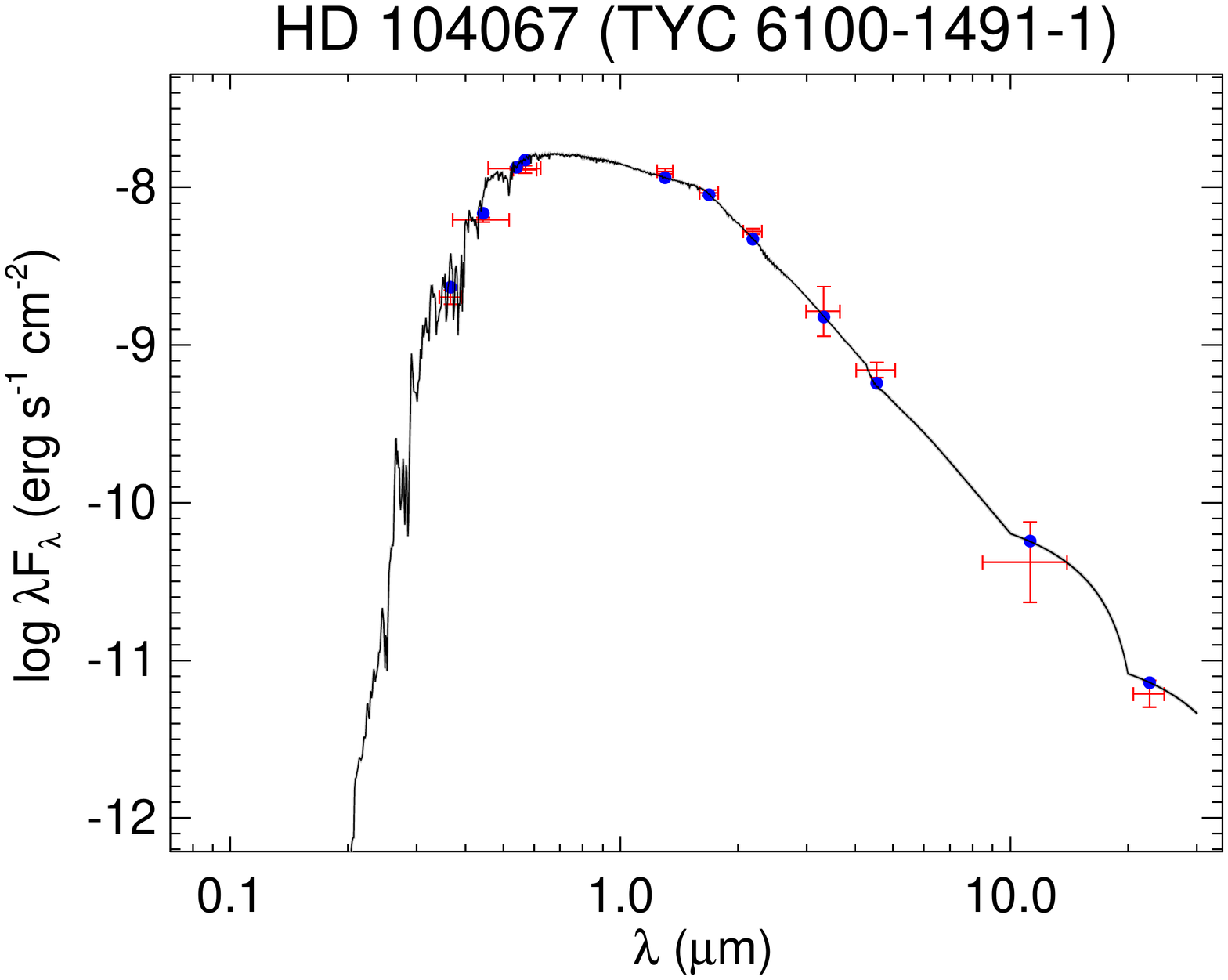}
  \includegraphics[trim=60 60 60 60,clip,width=0.49\linewidth]{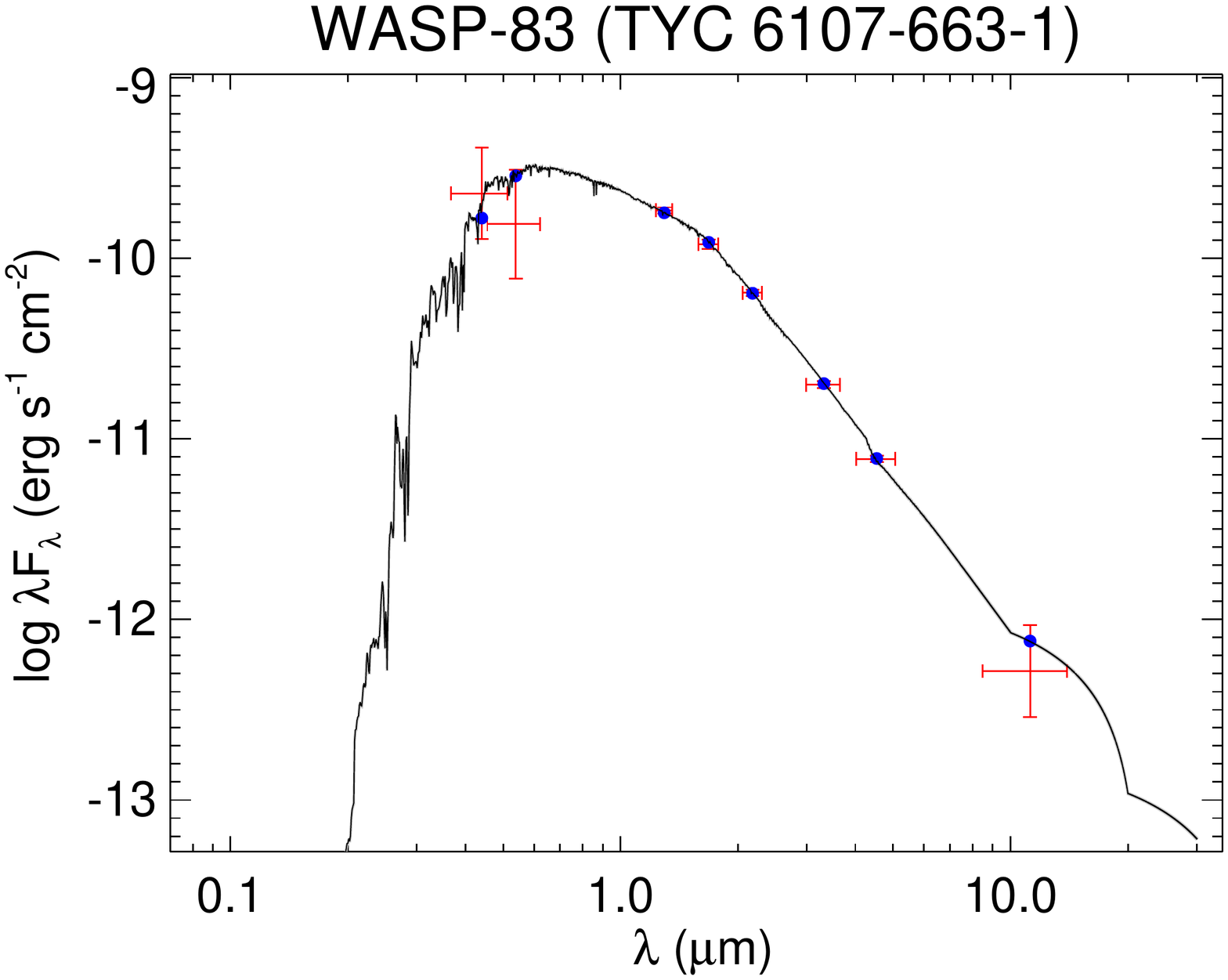}
  \includegraphics[trim=60 60 60 60,clip,width=0.49\linewidth]{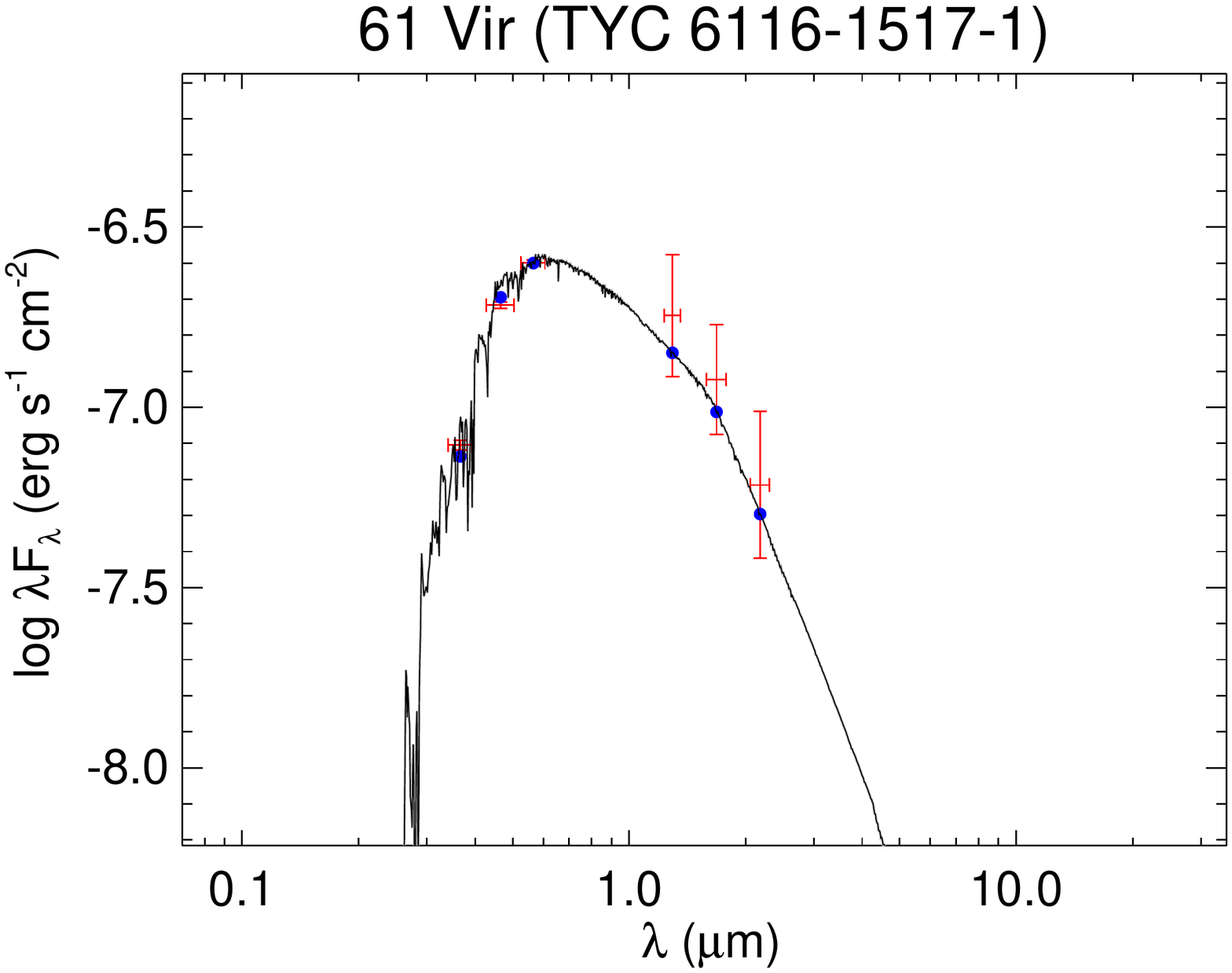}
  \includegraphics[trim=60 60 60 60,clip,width=0.49\linewidth]{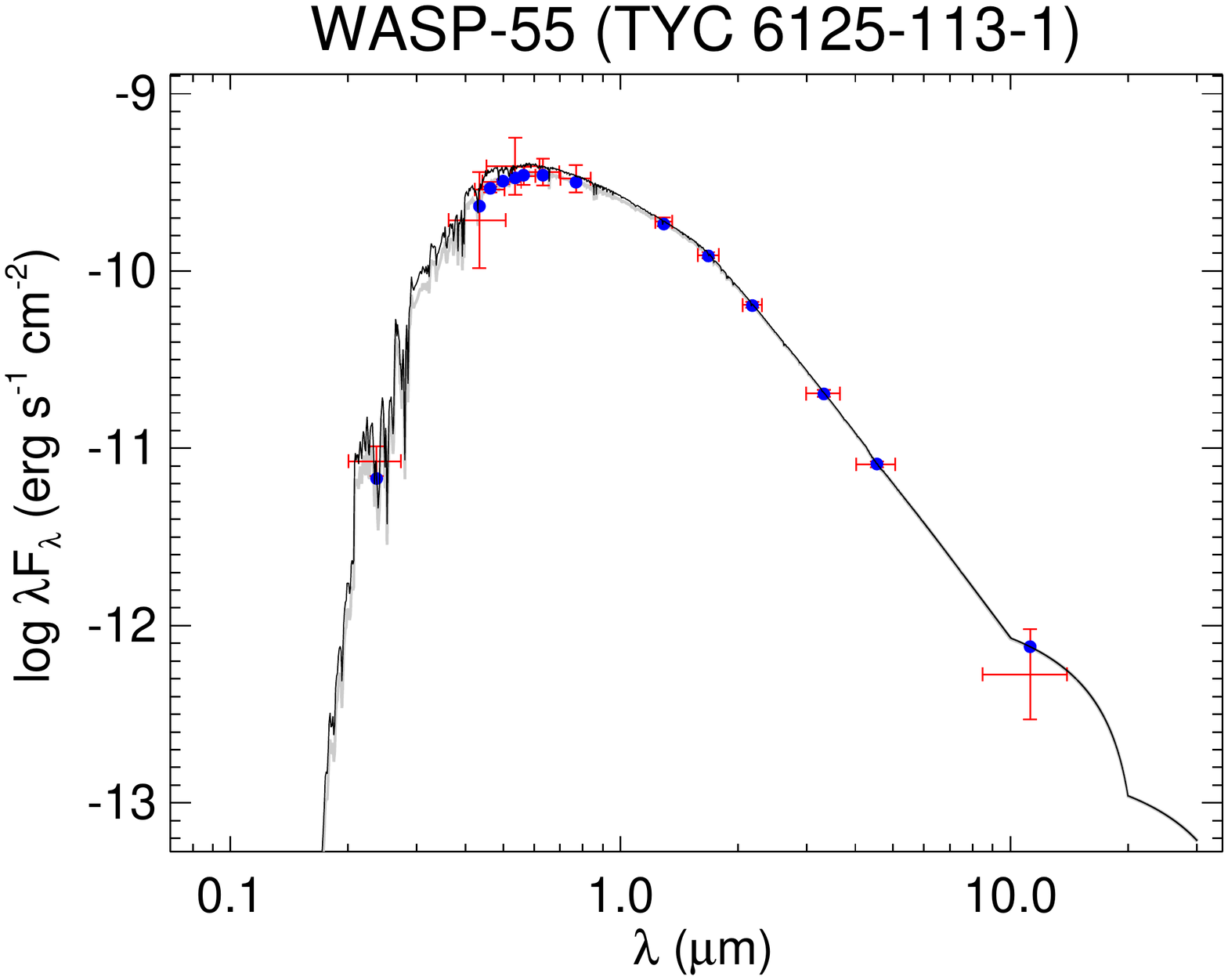}
  \caption{All labels, lines, symbols, and colors as in Figure \ref{fig:seds}.}
  \label{fig:seds_56}
\end{figure}

\begin{figure}[H]
  \centering
  \includegraphics[trim=60 60 60 60,clip,width=0.49\linewidth]{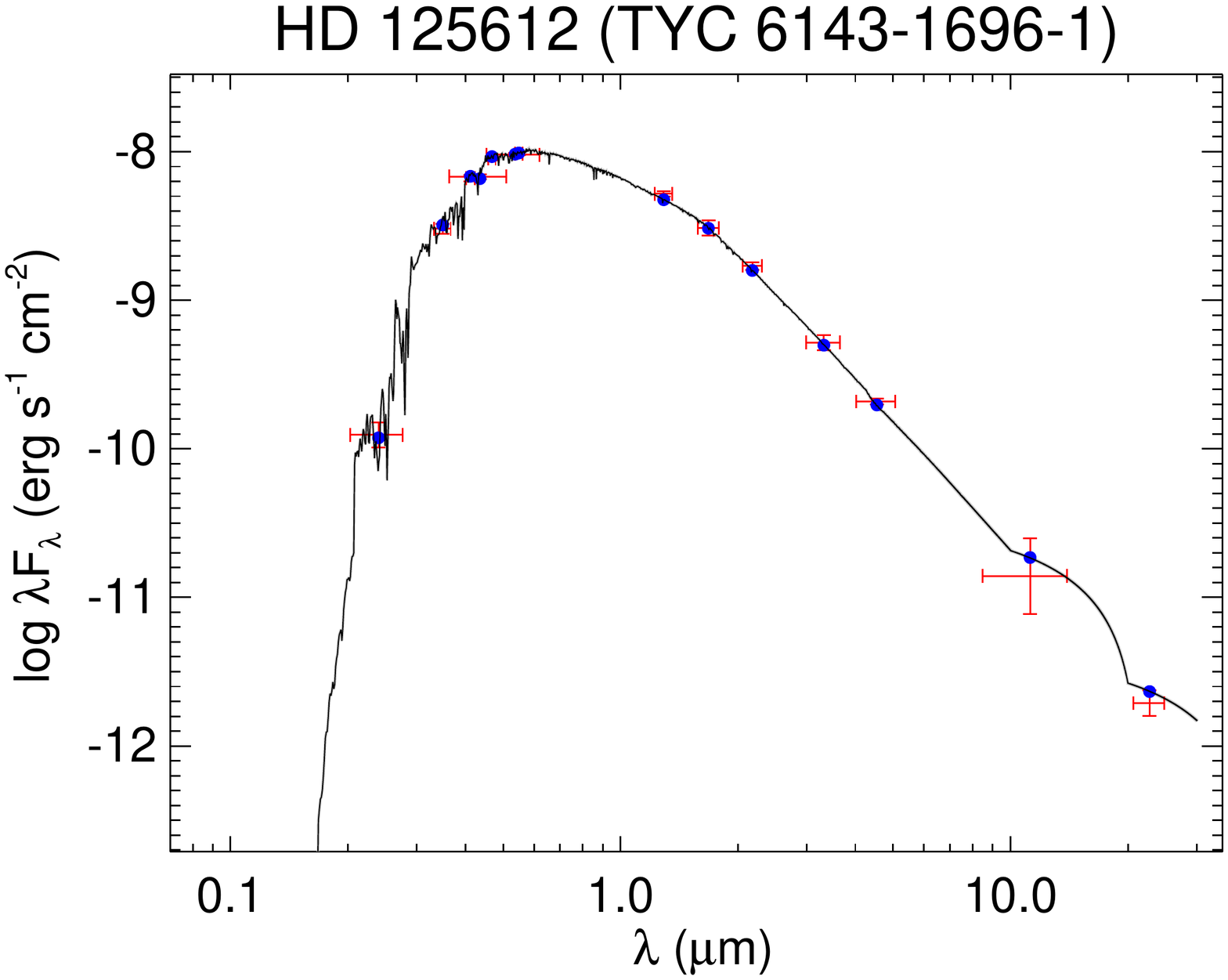}
  \includegraphics[trim=60 60 60 60,clip,width=0.49\linewidth]{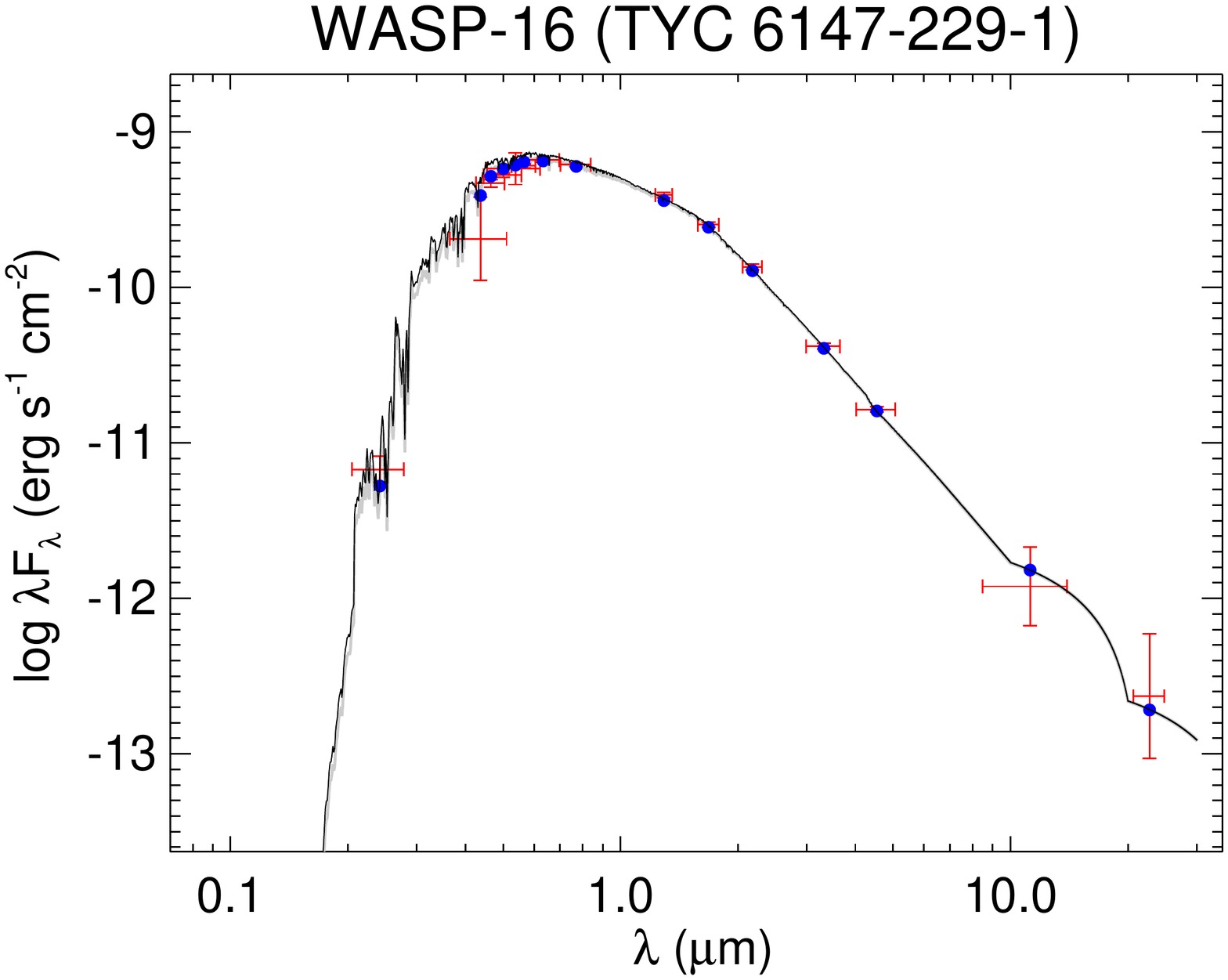}
  \includegraphics[trim=60 60 60 60,clip,width=0.49\linewidth]{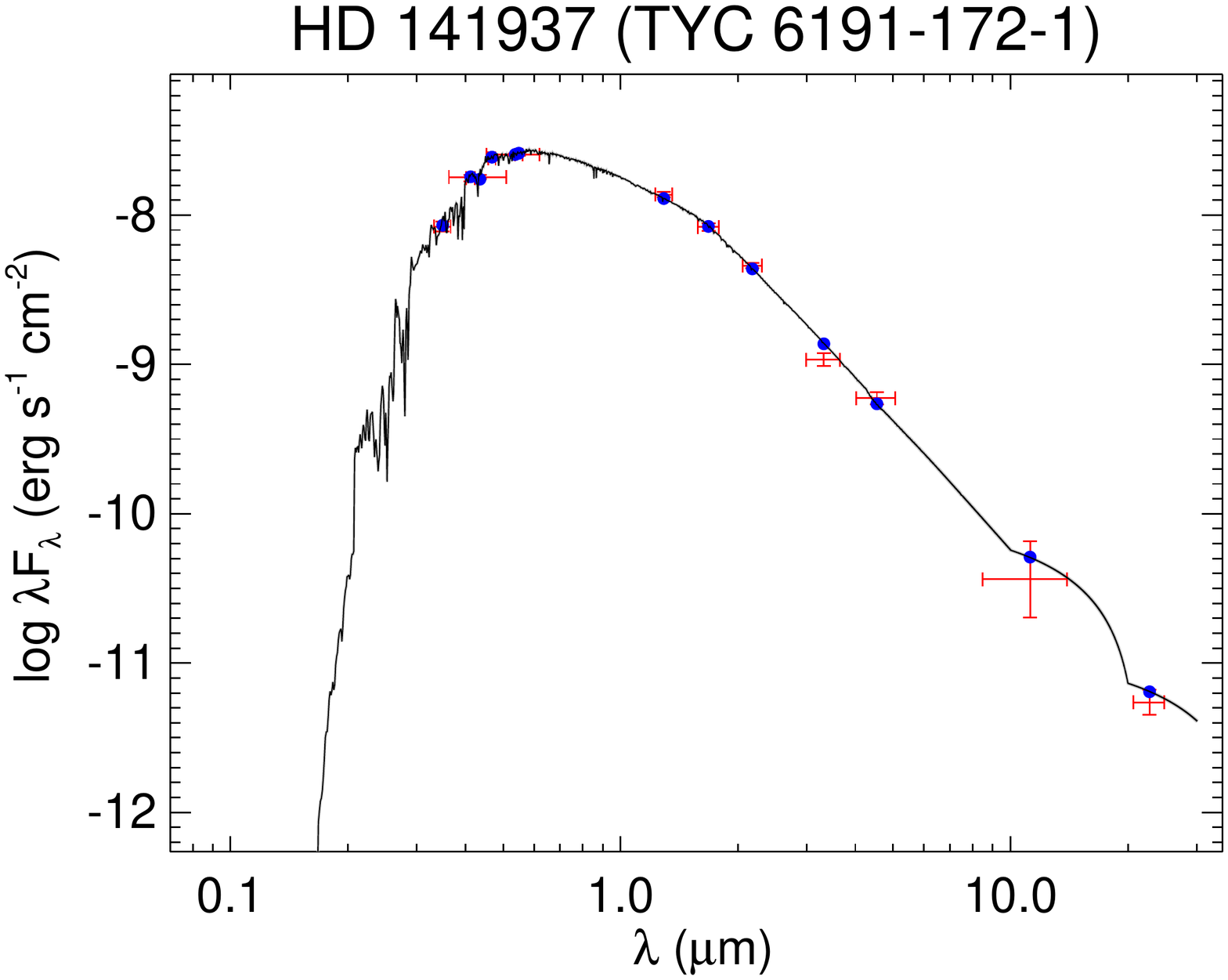}
  \includegraphics[trim=60 60 60 60,clip,width=0.49\linewidth]{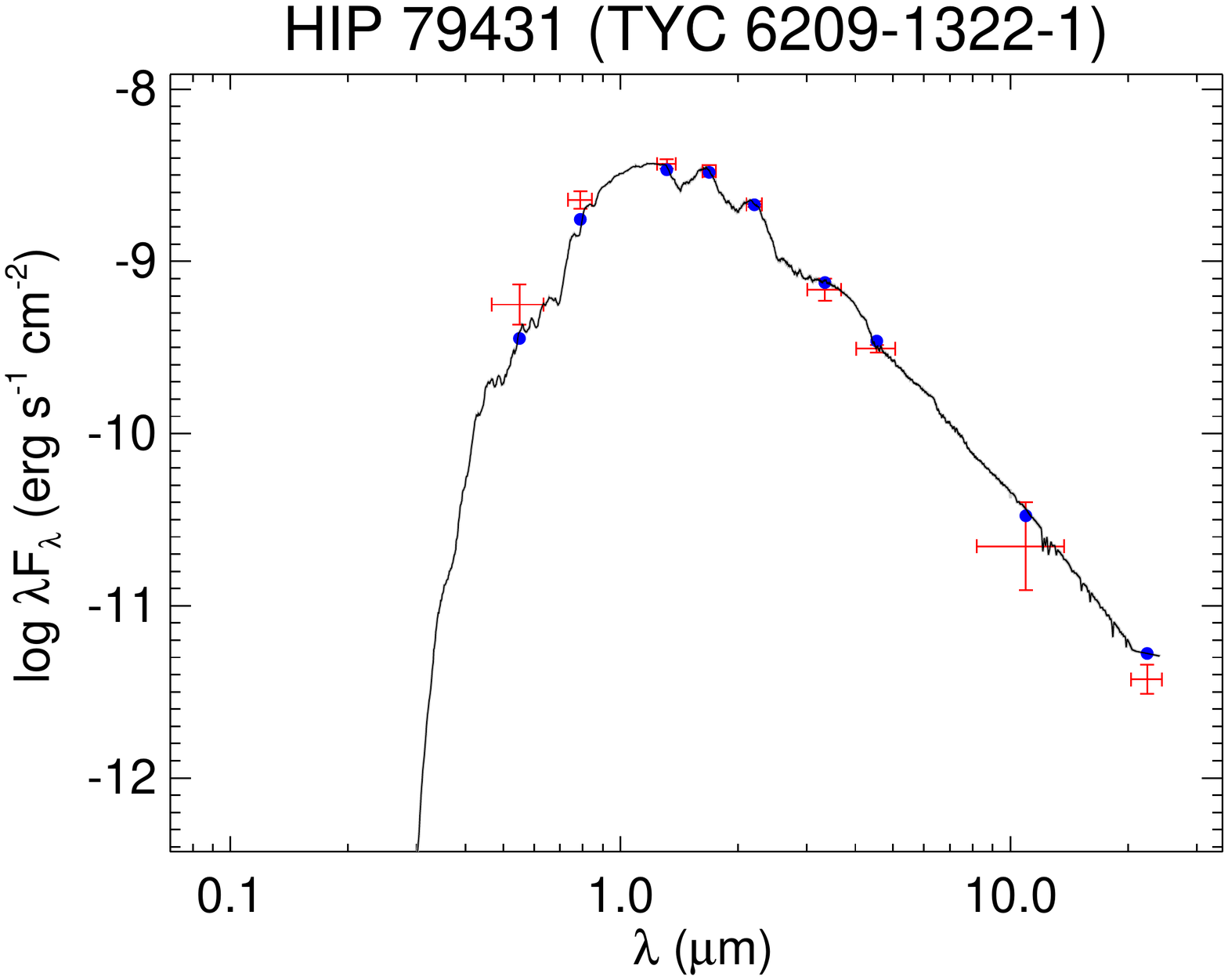}
  \includegraphics[trim=60 60 60 60,clip,width=0.49\linewidth]{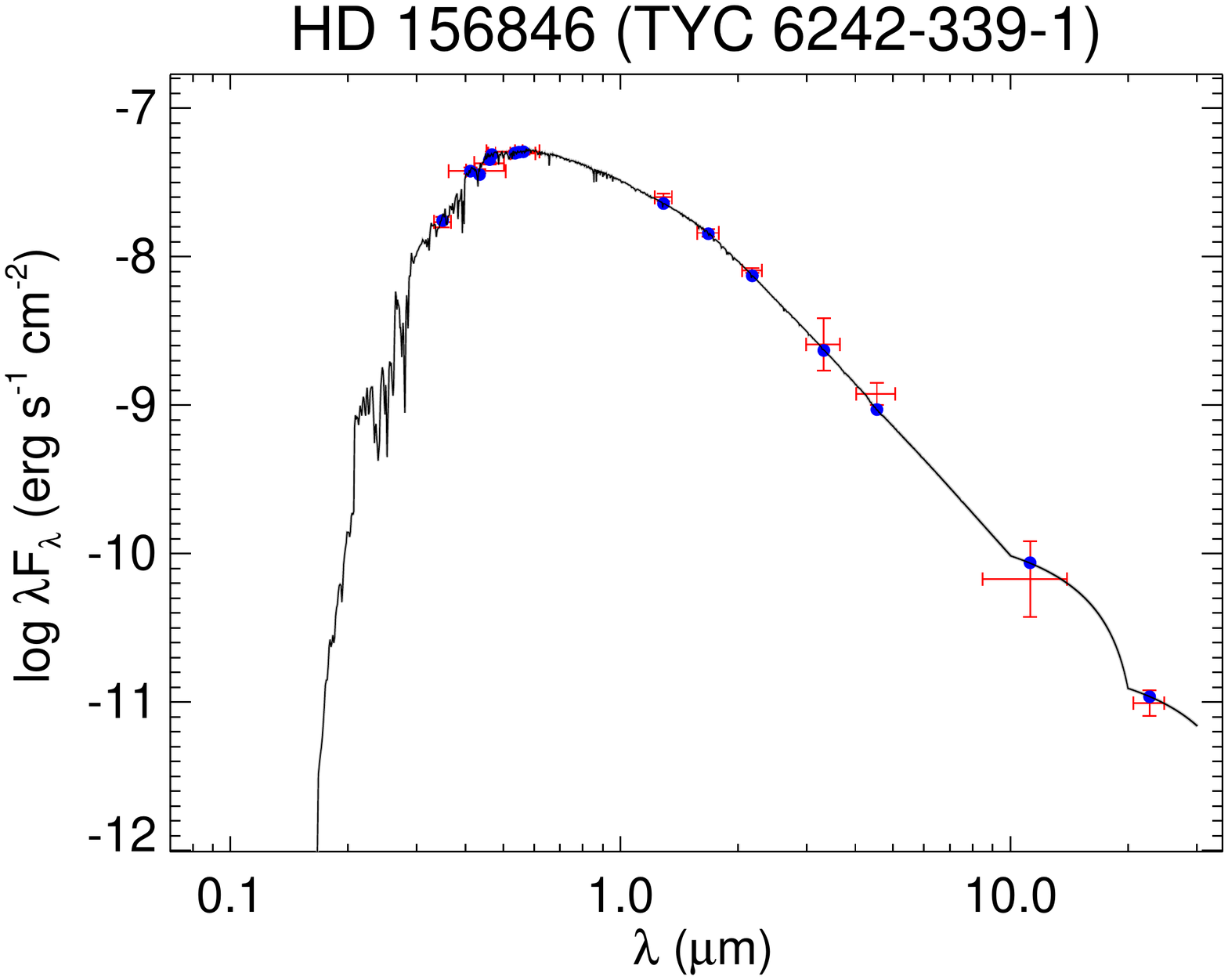}
  \includegraphics[trim=60 60 60 60,clip,width=0.49\linewidth]{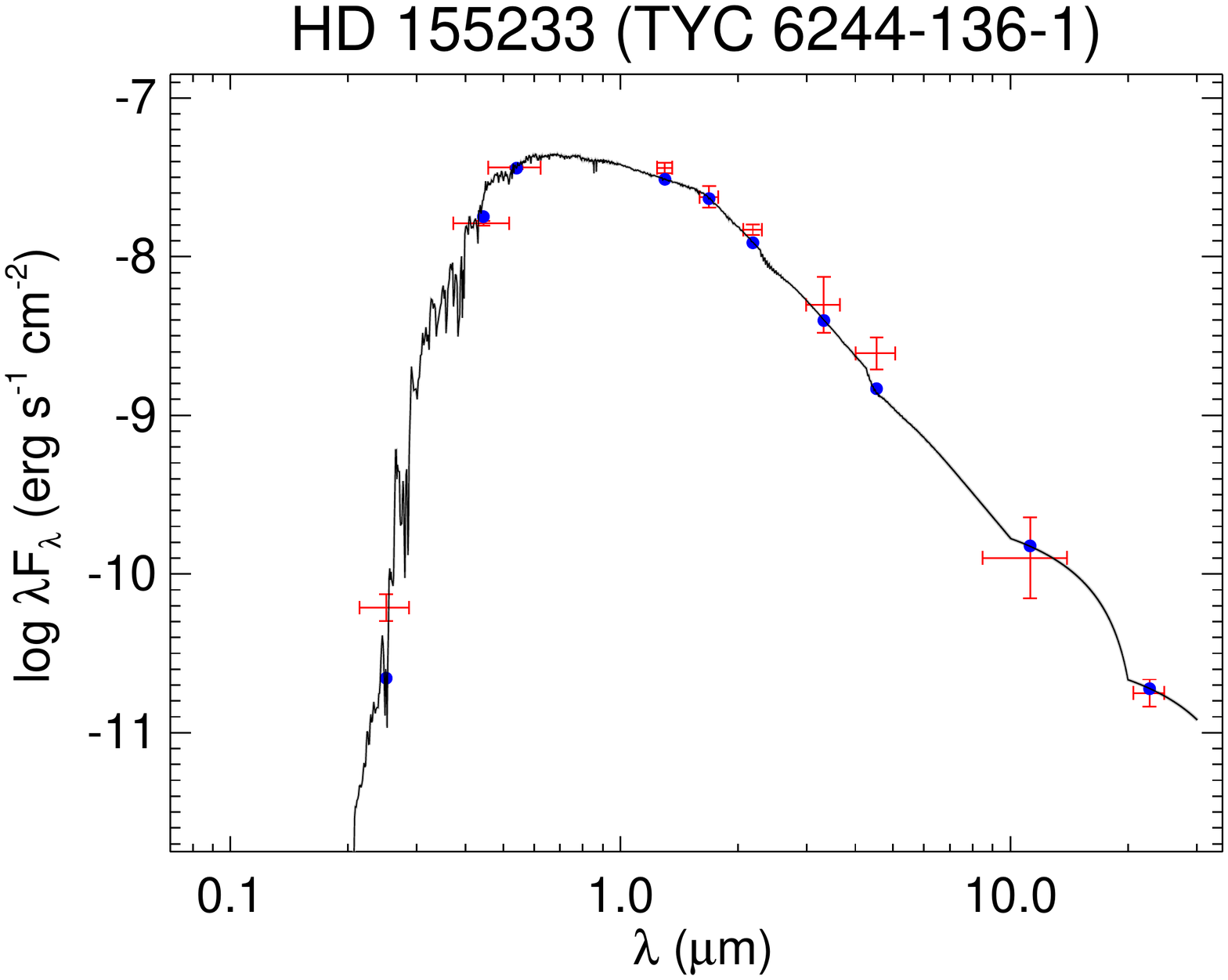}
  \caption{All labels, lines, symbols, and colors as in Figure \ref{fig:seds}.}
  \label{fig:seds_57}
\end{figure}

\begin{figure}[H]
  \centering
  \includegraphics[trim=60 60 60 60,clip,width=0.49\linewidth]{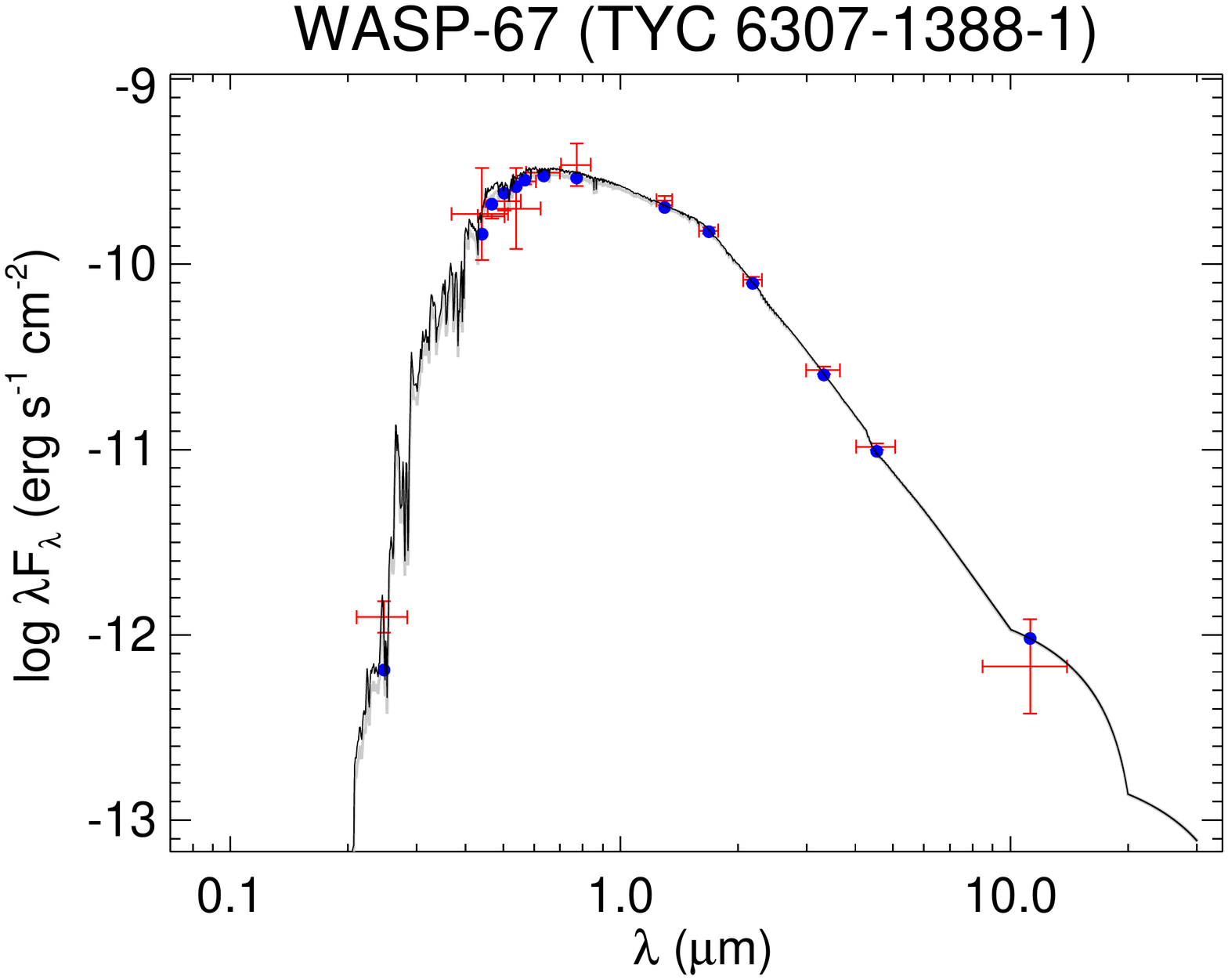}
  \includegraphics[trim=60 60 60 60,clip,width=0.49\linewidth]{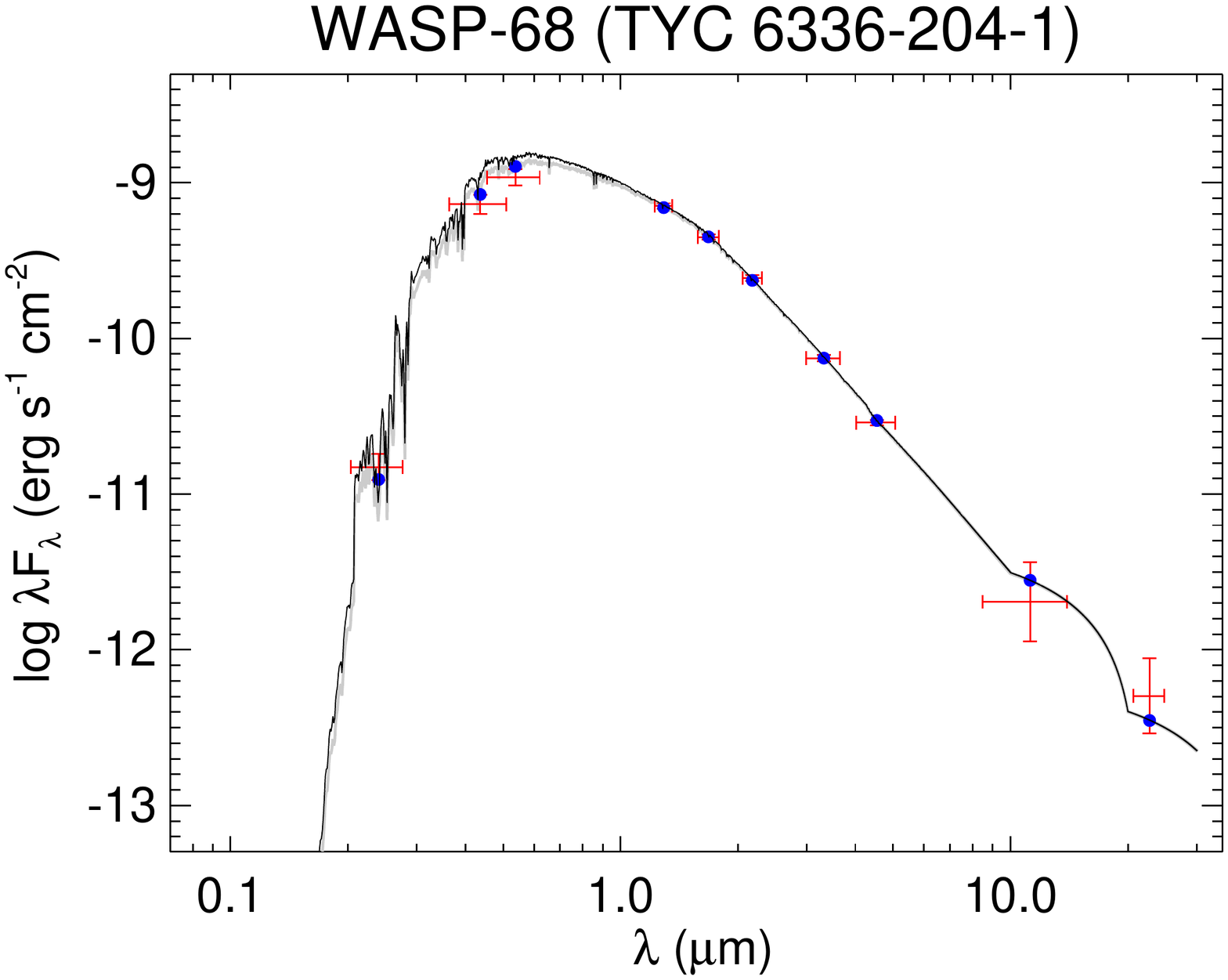}
  \includegraphics[trim=60 60 60 60,clip,width=0.49\linewidth]{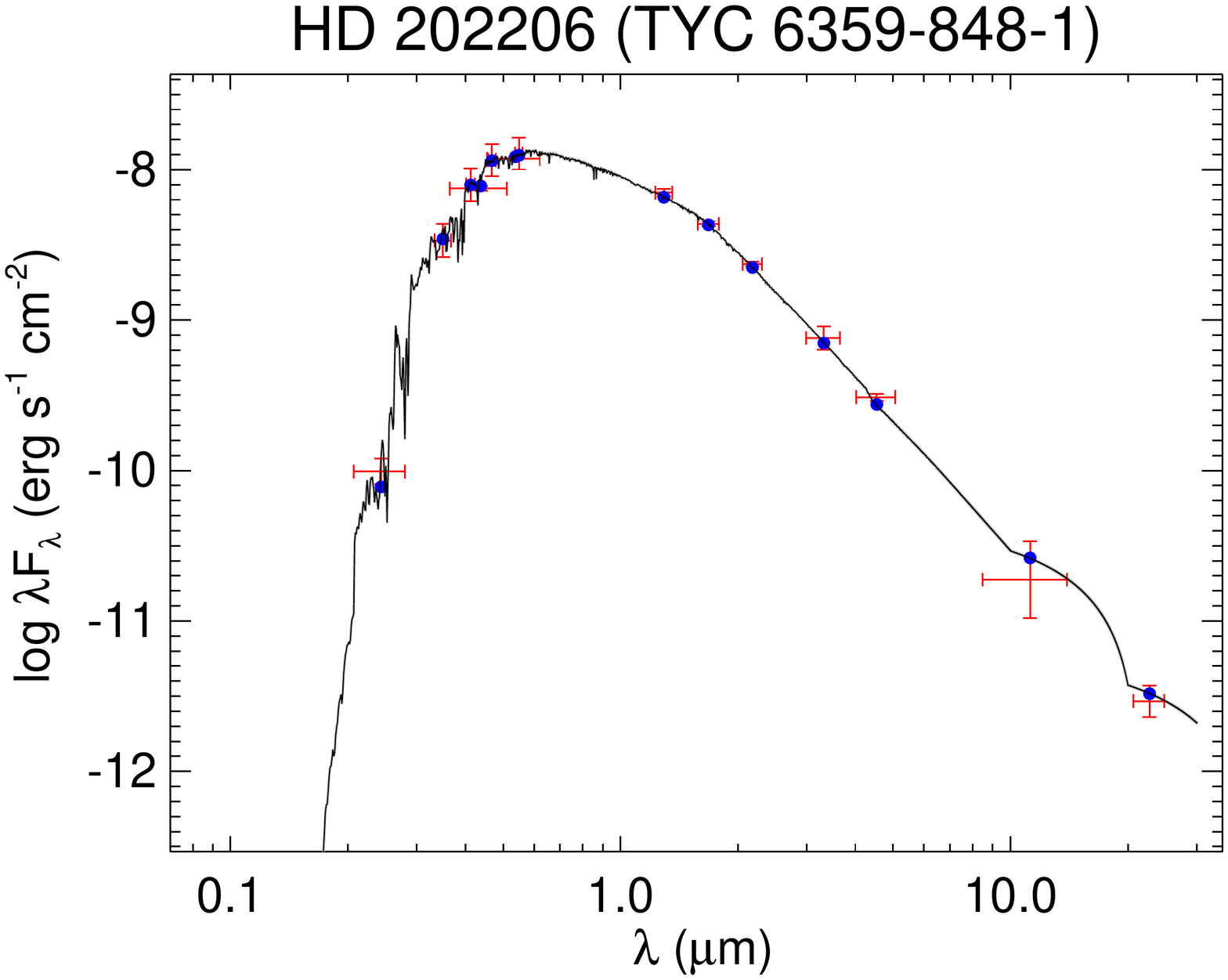}
  \includegraphics[trim=60 60 60 60,clip,width=0.49\linewidth]{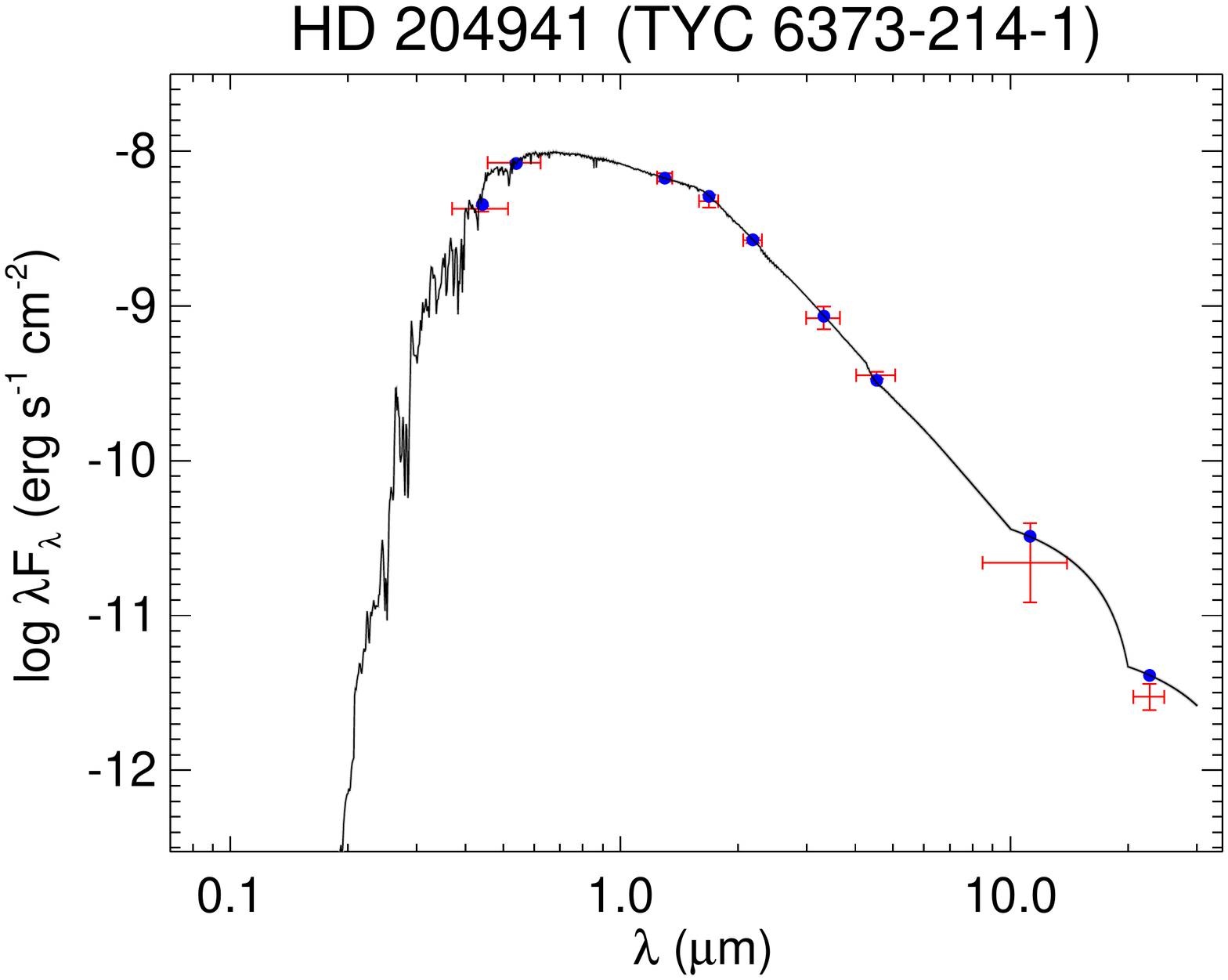}
  \includegraphics[trim=60 60 60 60,clip,width=0.49\linewidth]{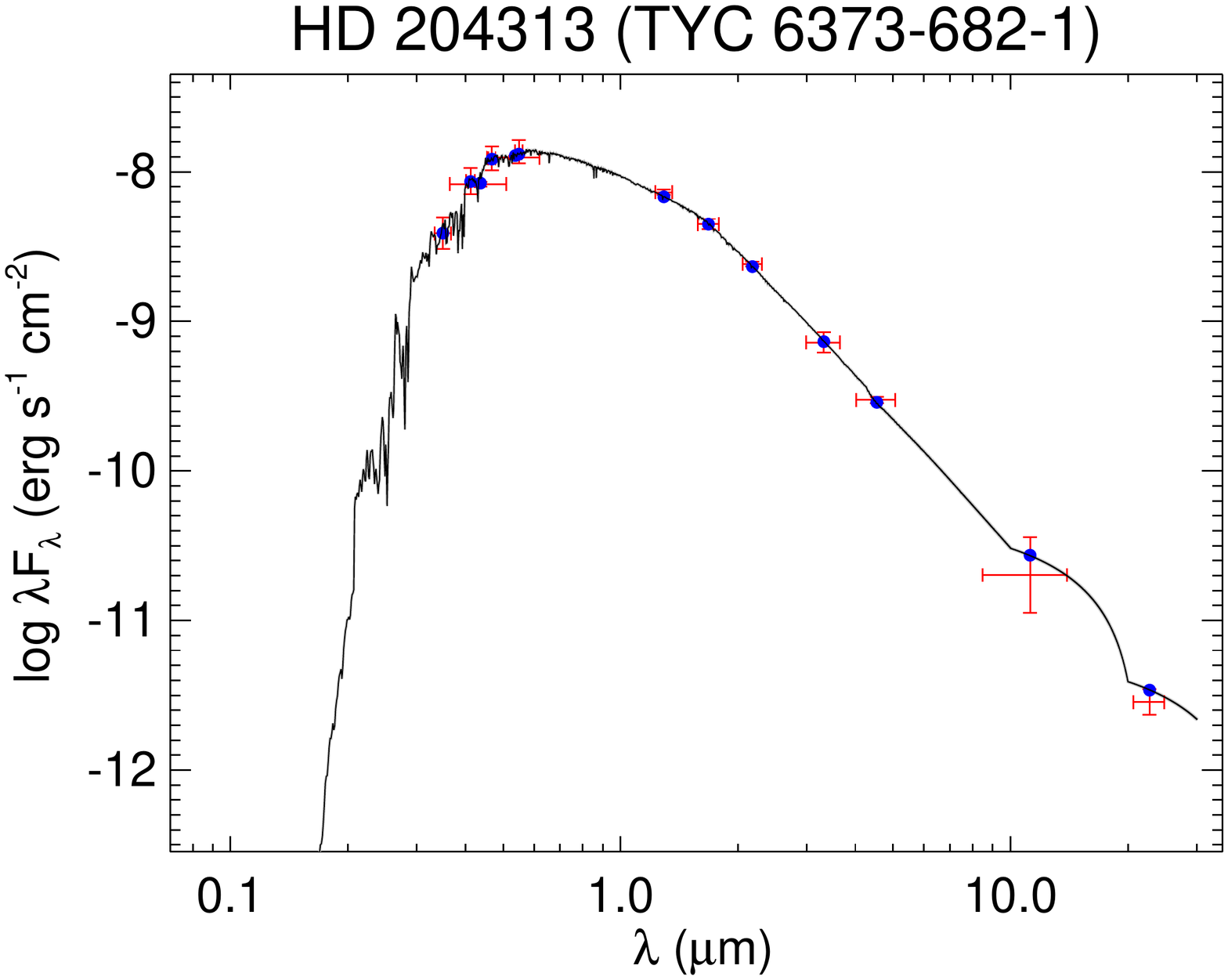}
  \includegraphics[trim=60 60 60 60,clip,width=0.49\linewidth]{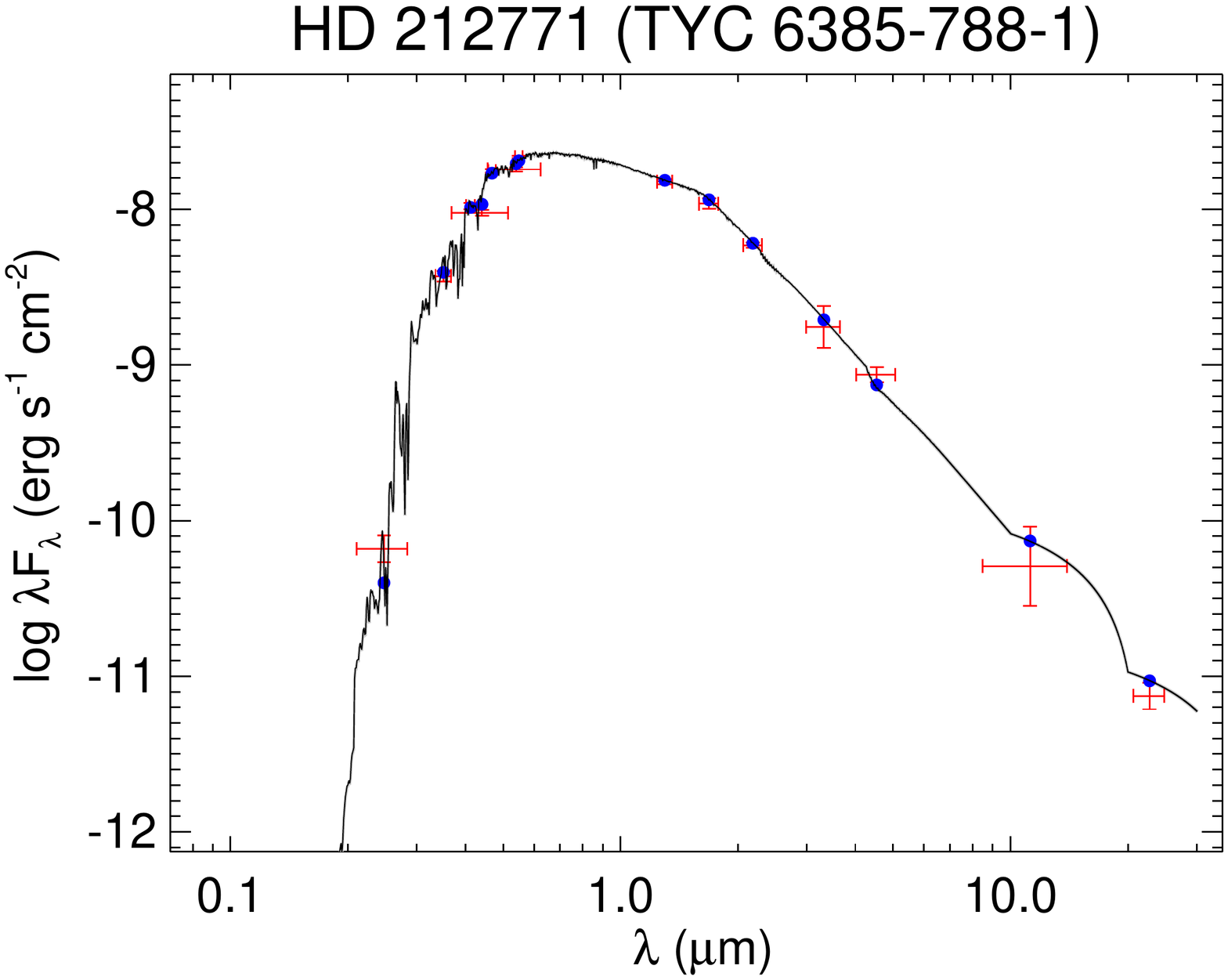}
  \caption{All labels, lines, symbols, and colors as in Figure \ref{fig:seds}.}
  \label{fig:seds_58}
\end{figure}

\begin{figure}[H]
  \centering
  \includegraphics[trim=60 60 60 60,clip,width=0.49\linewidth]{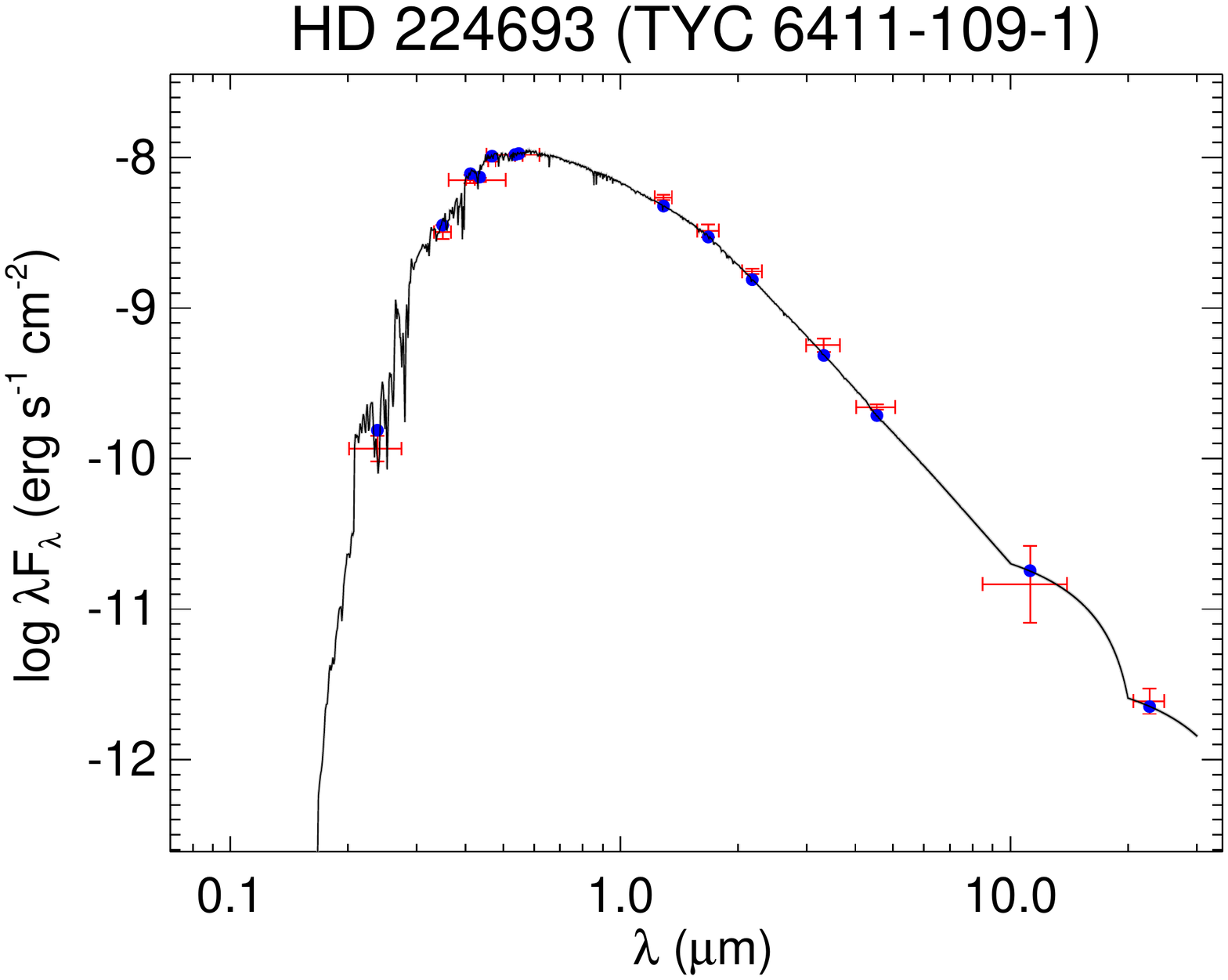}
  \includegraphics[trim=60 60 60 60,clip,width=0.49\linewidth]{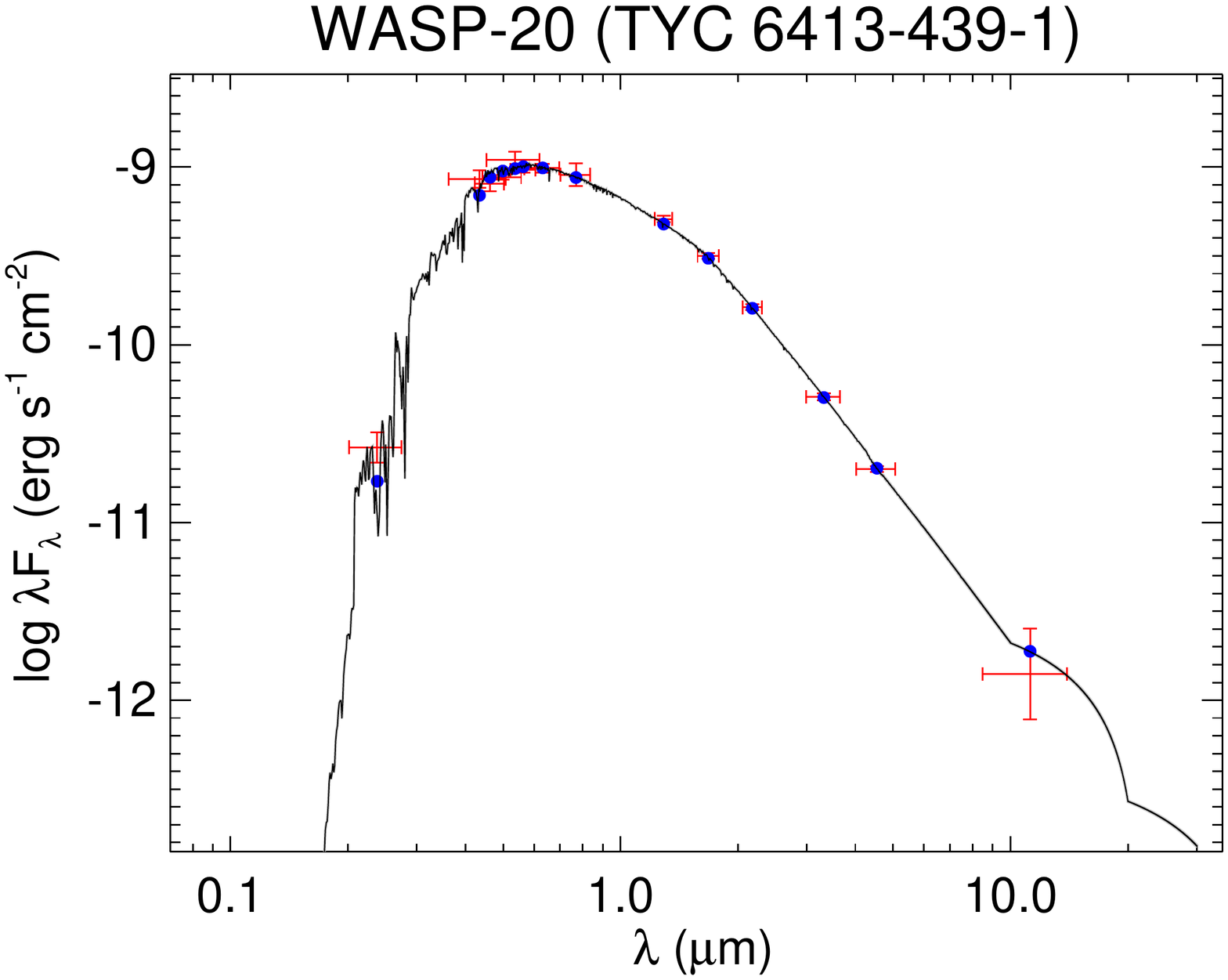}
  \includegraphics[trim=60 60 60 60,clip,width=0.49\linewidth]{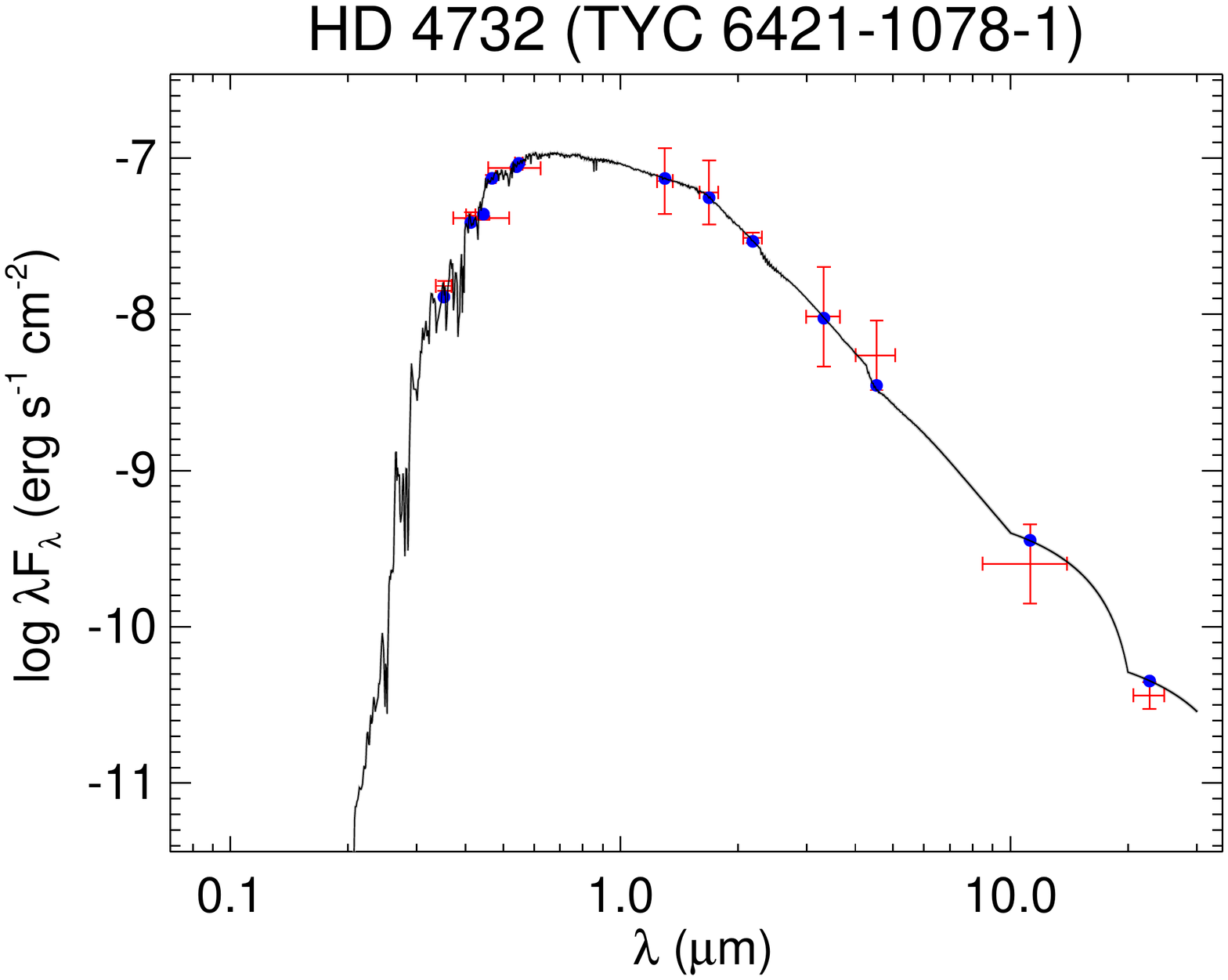}
  \includegraphics[trim=60 60 60 60,clip,width=0.49\linewidth]{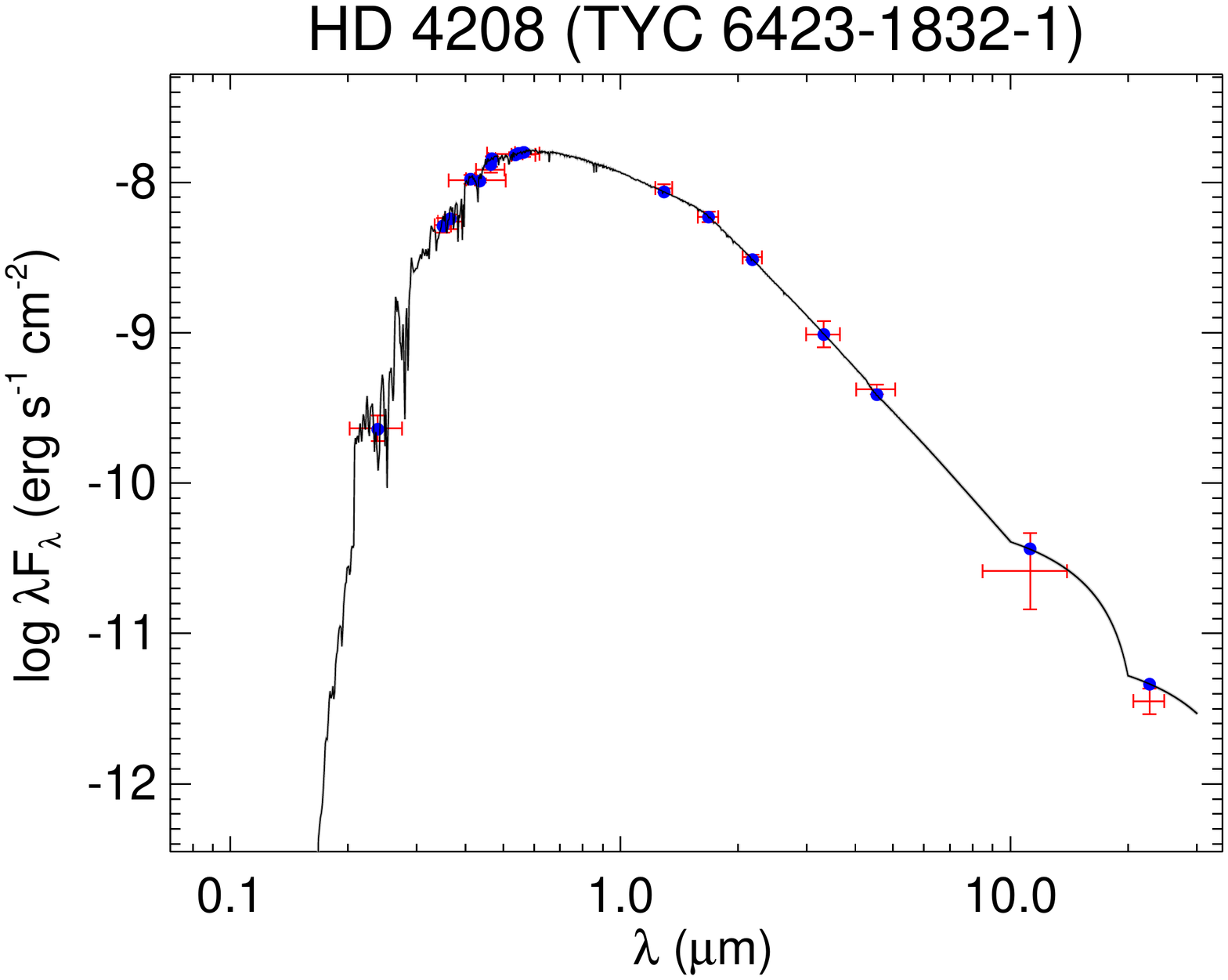}
  \includegraphics[trim=60 60 60 60,clip,width=0.49\linewidth]{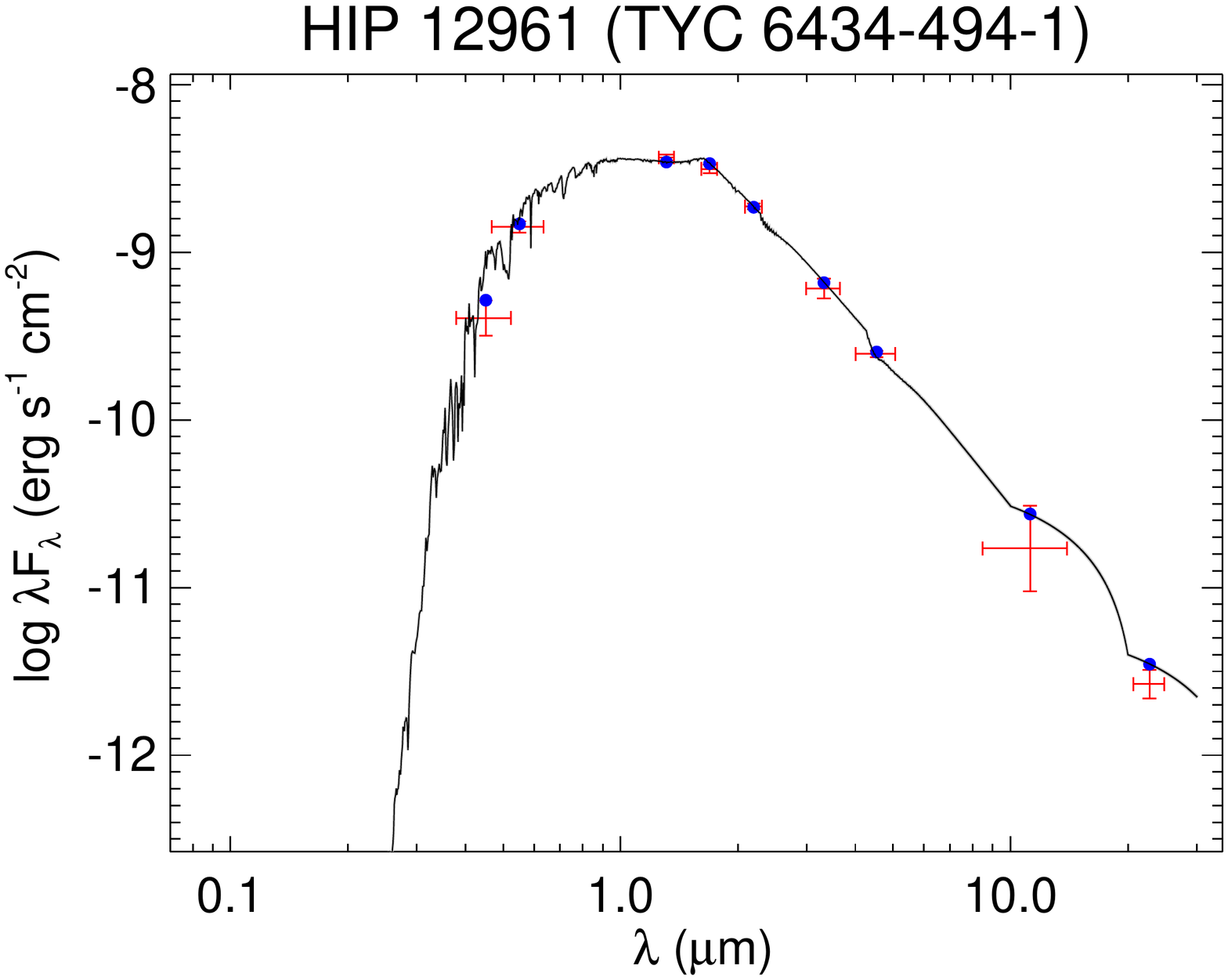}
  \includegraphics[trim=60 60 60 60,clip,width=0.49\linewidth]{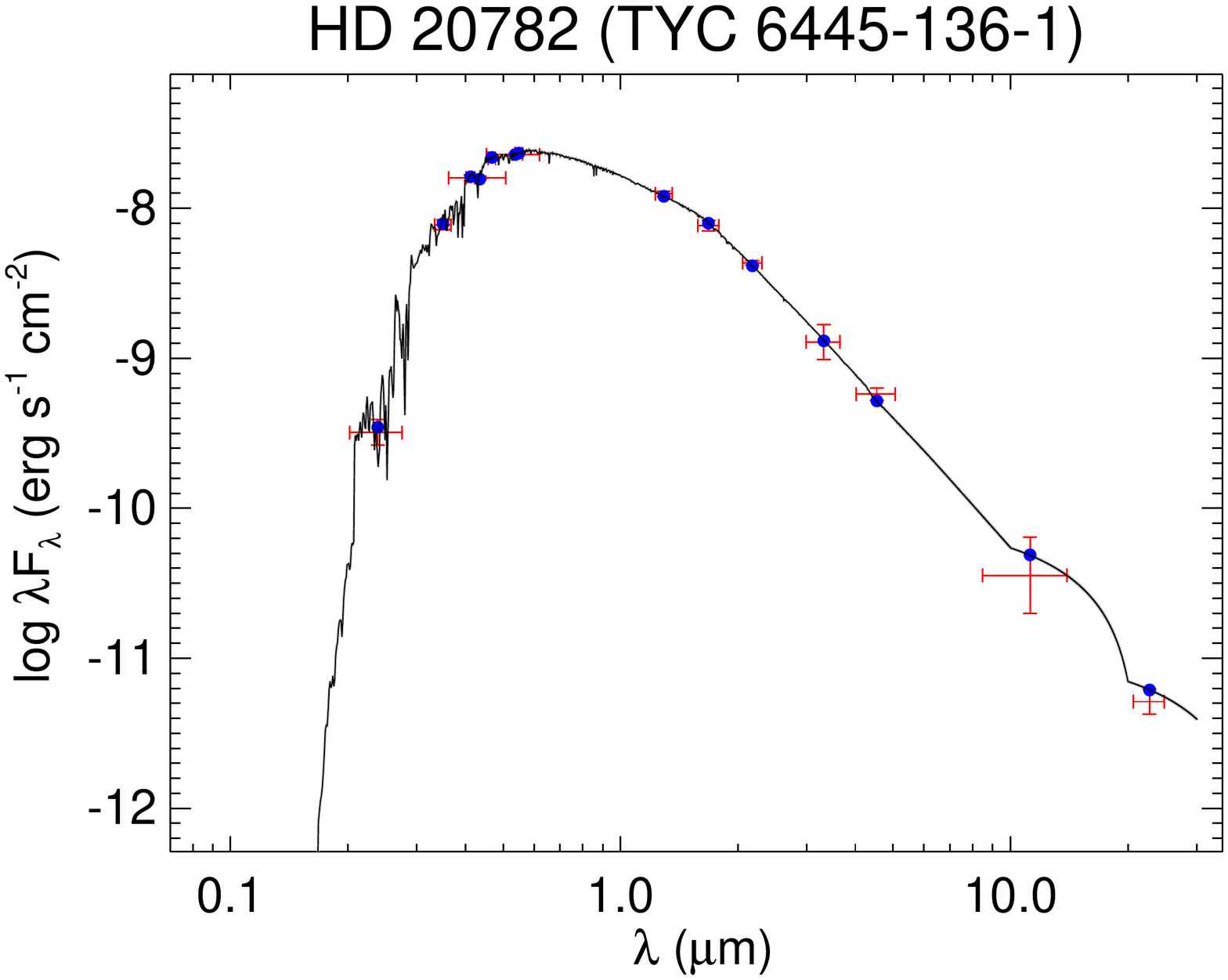}
  \caption{All labels, lines, symbols, and colors as in Figure \ref{fig:seds}.}
  \label{fig:seds_59}
\end{figure}

\begin{figure}[H]
  \centering
  \includegraphics[trim=60 60 60 60,clip,width=0.49\linewidth]{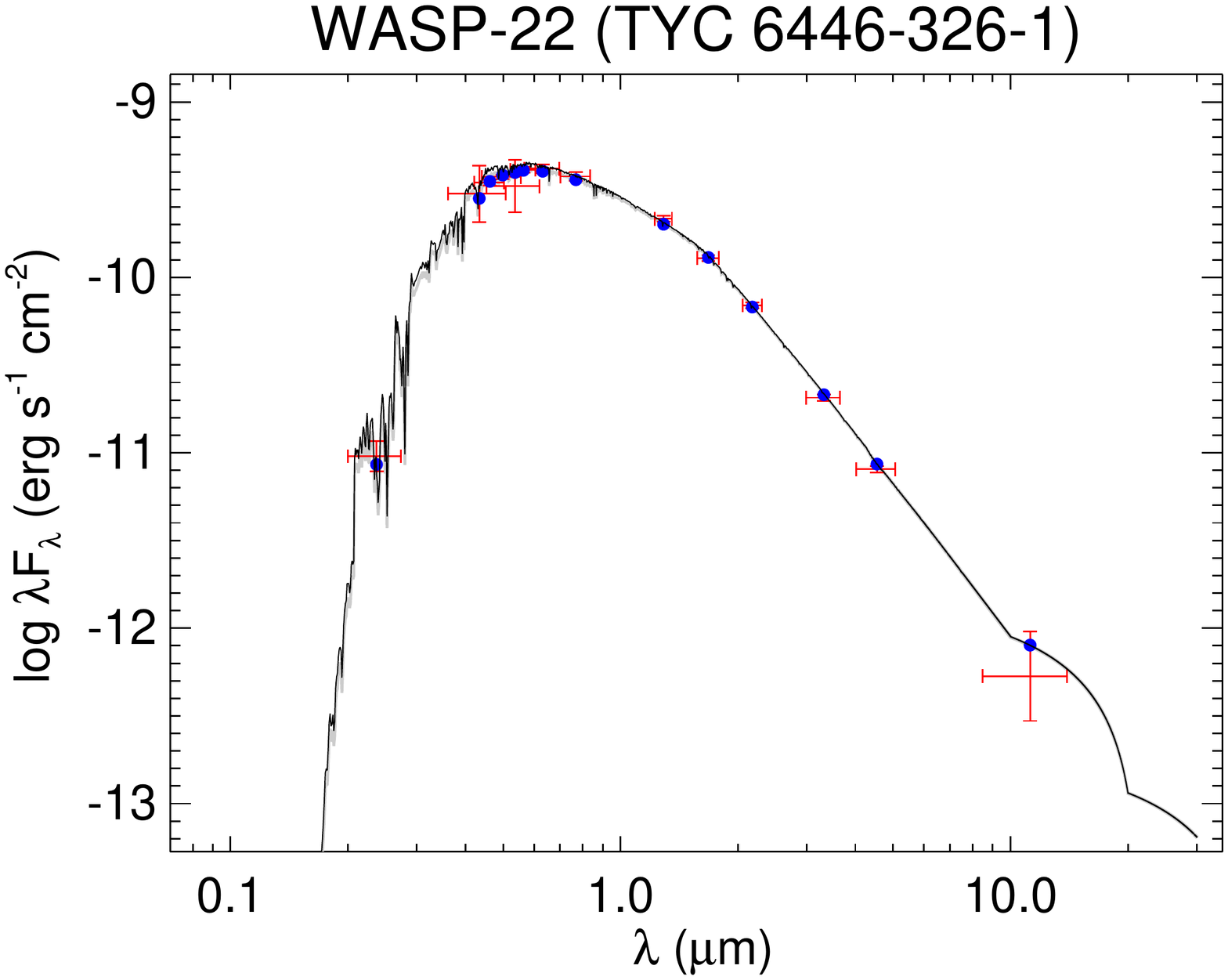}
  \includegraphics[trim=60 60 60 60,clip,width=0.49\linewidth]{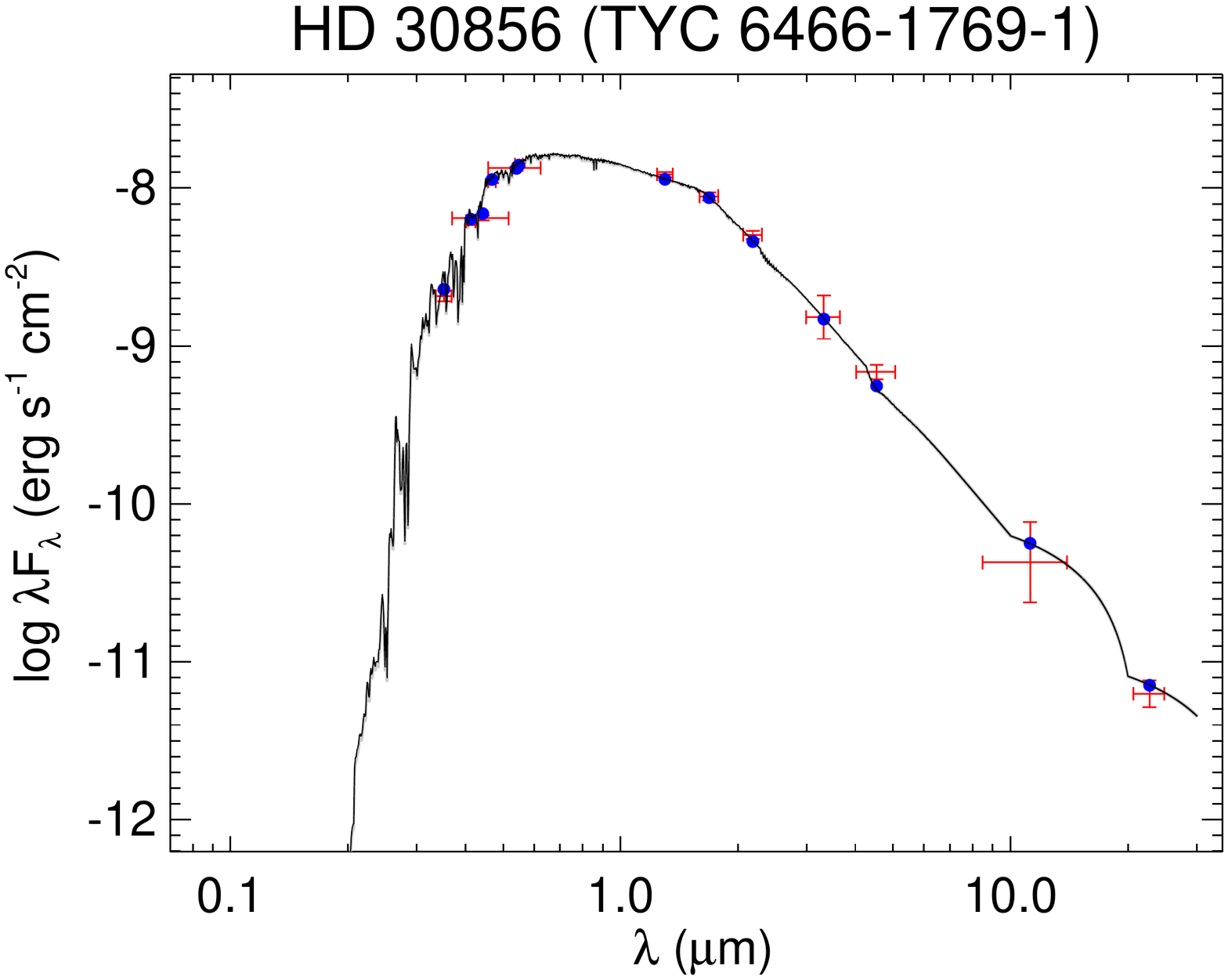}
  \includegraphics[trim=60 60 60 60,clip,width=0.49\linewidth]{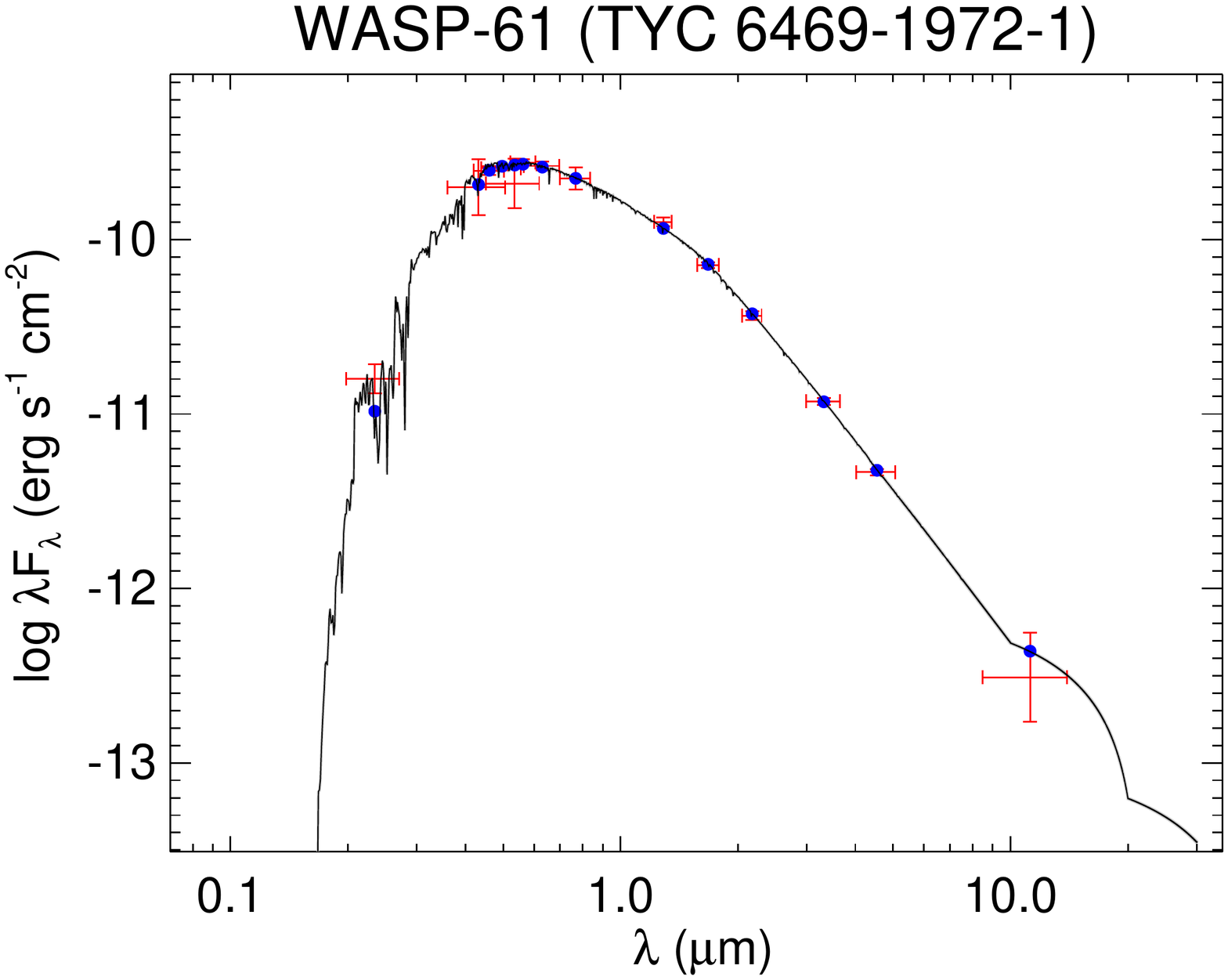}
  \includegraphics[trim=60 60 60 60,clip,width=0.49\linewidth]{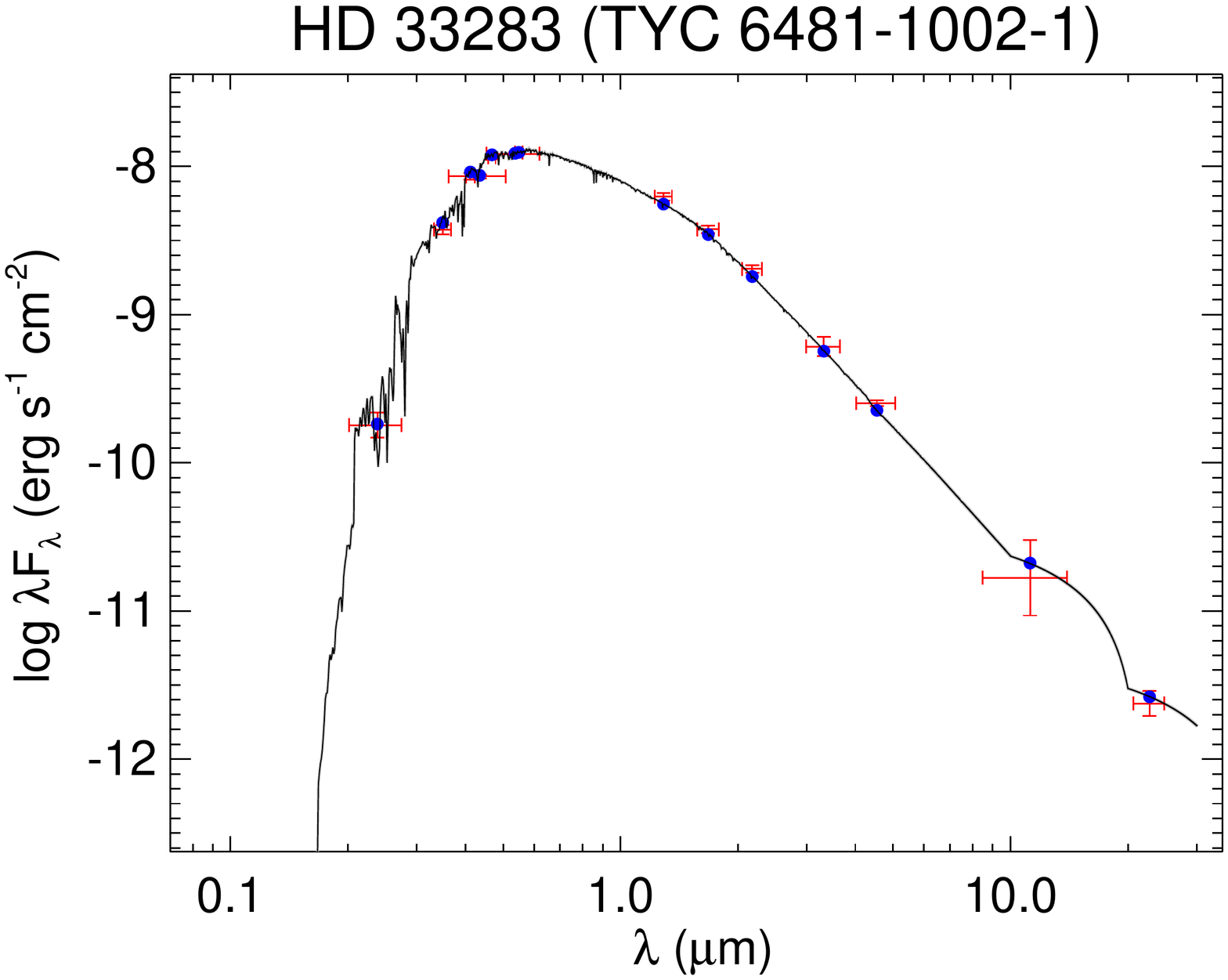}
  \includegraphics[trim=60 60 60 60,clip,width=0.49\linewidth]{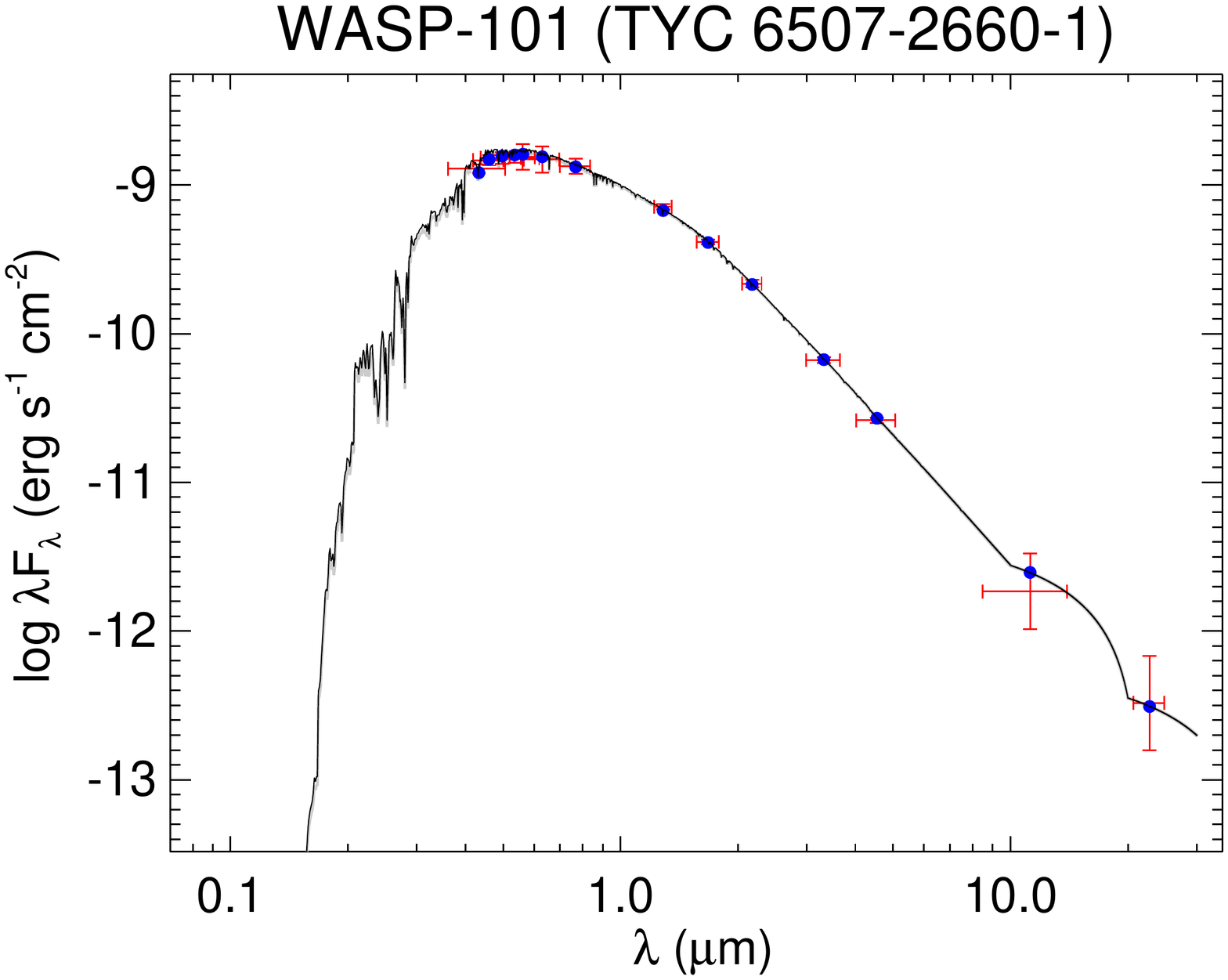}
  \includegraphics[trim=60 60 60 60,clip,width=0.49\linewidth]{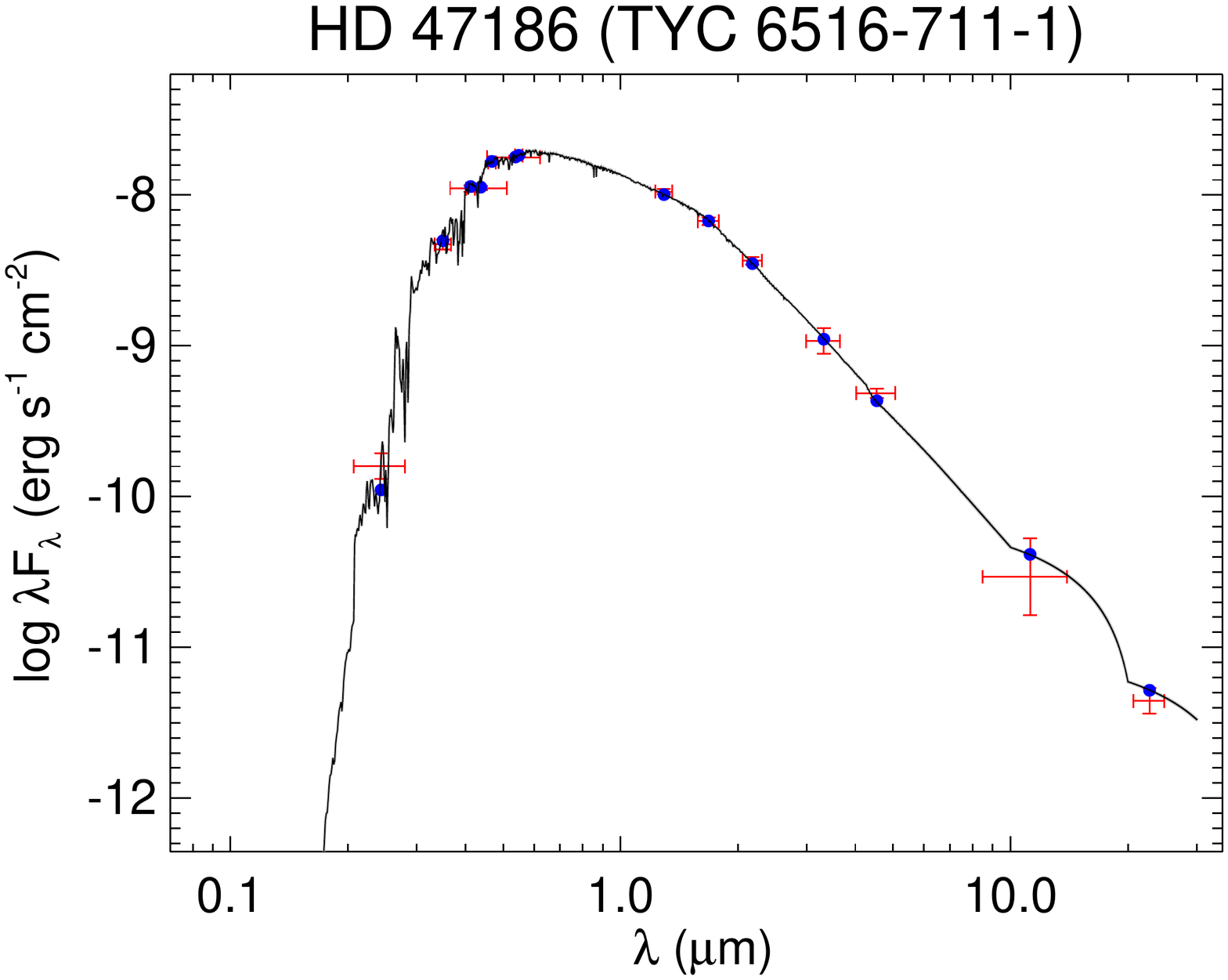}
  \caption{All labels, lines, symbols, and colors as in Figure \ref{fig:seds}.}
  \label{fig:seds_60}
\end{figure}

\begin{figure}[H]
  \centering
  \includegraphics[trim=60 60 60 60,clip,width=0.49\linewidth]{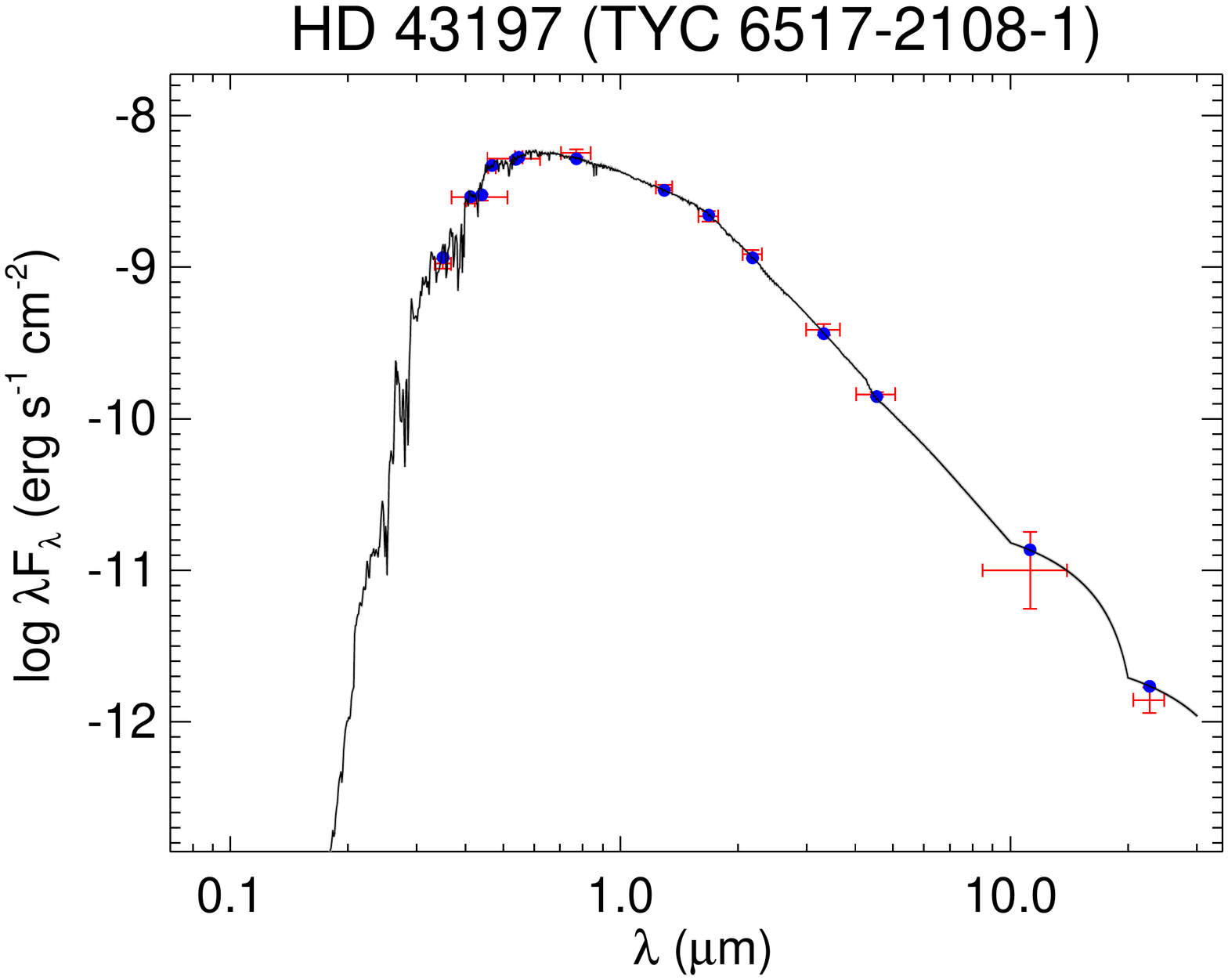}
  \includegraphics[trim=60 60 60 60,clip,width=0.49\linewidth]{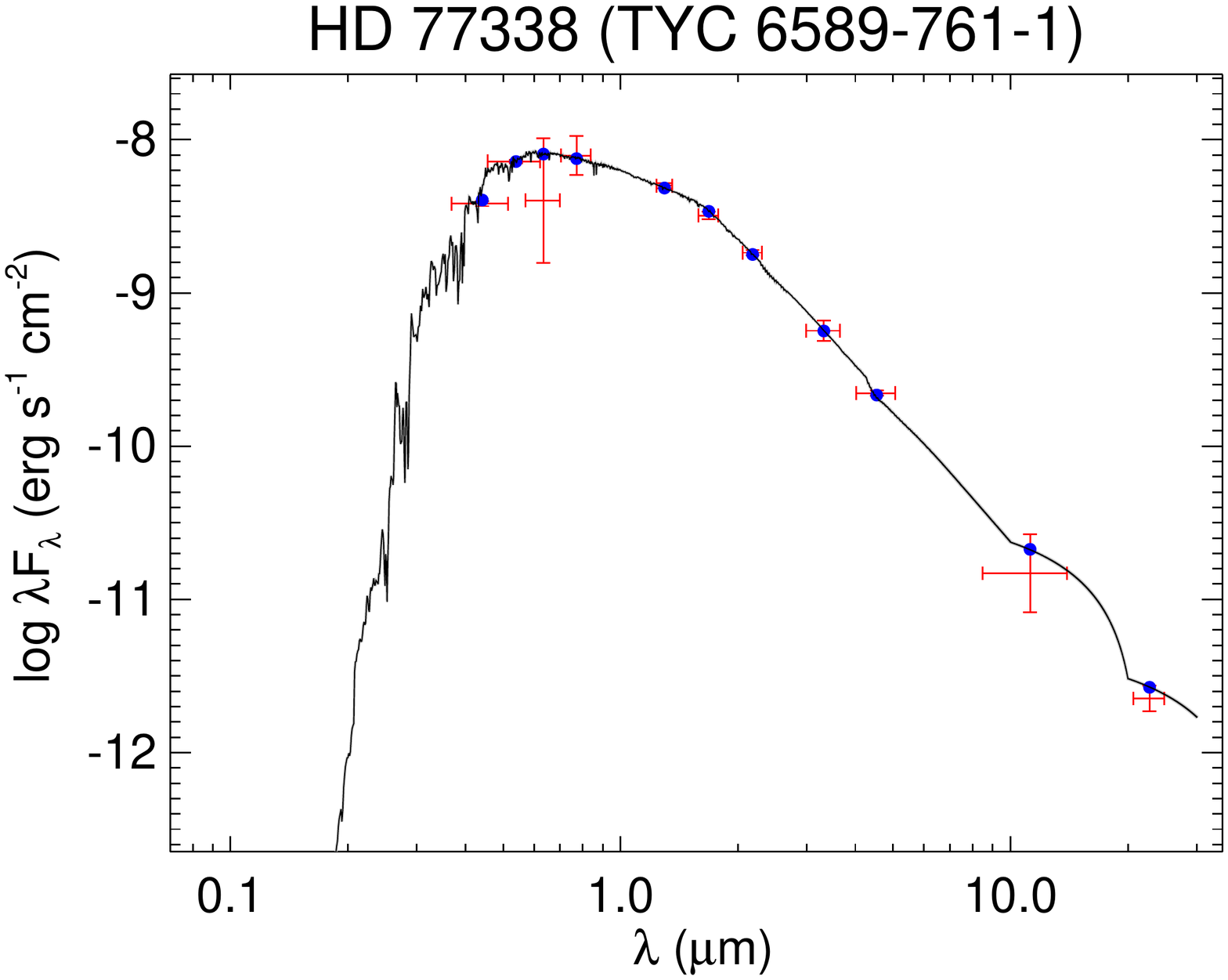}
  \includegraphics[trim=60 60 60 60,clip,width=0.49\linewidth]{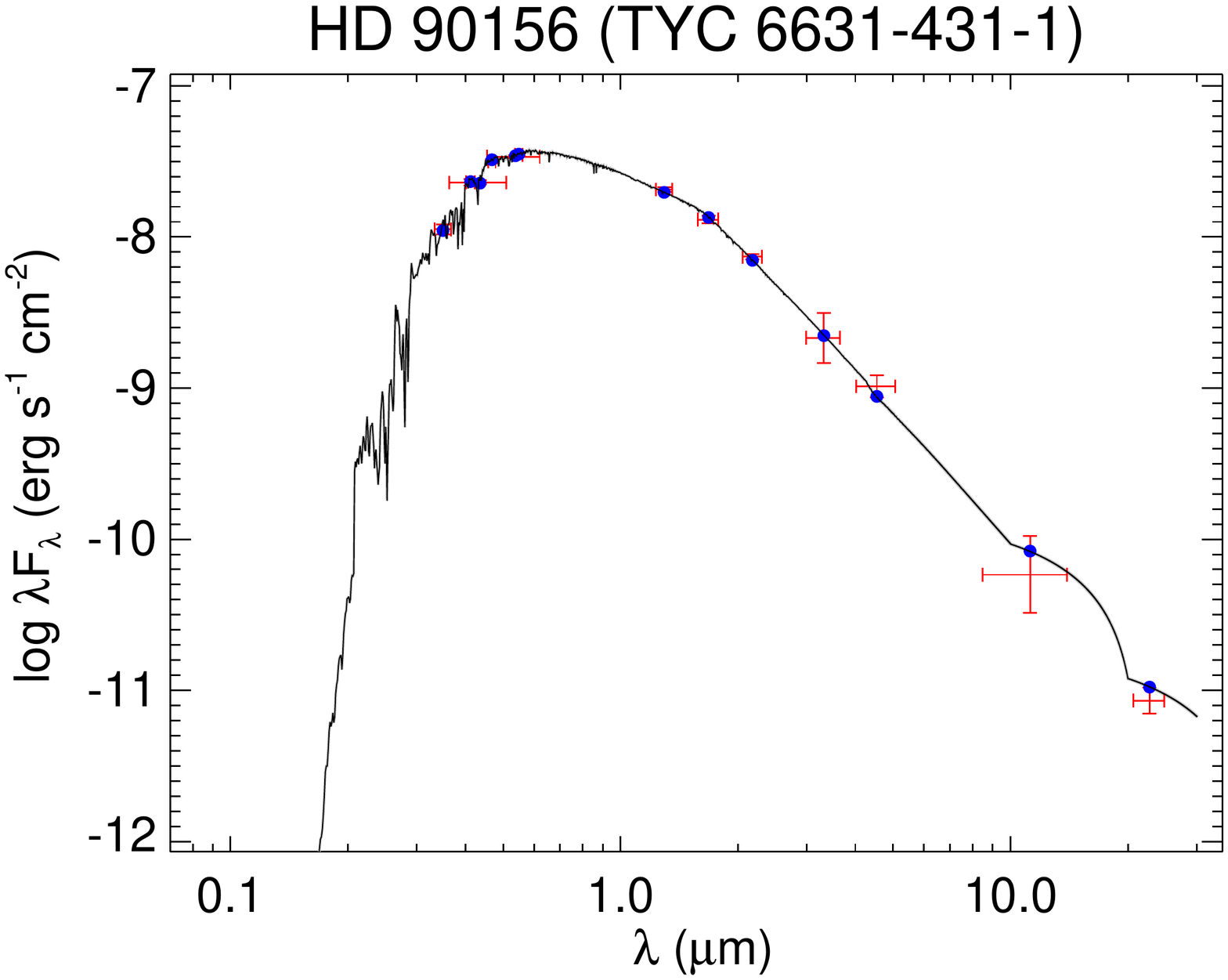}
  \includegraphics[trim=60 60 60 60,clip,width=0.49\linewidth]{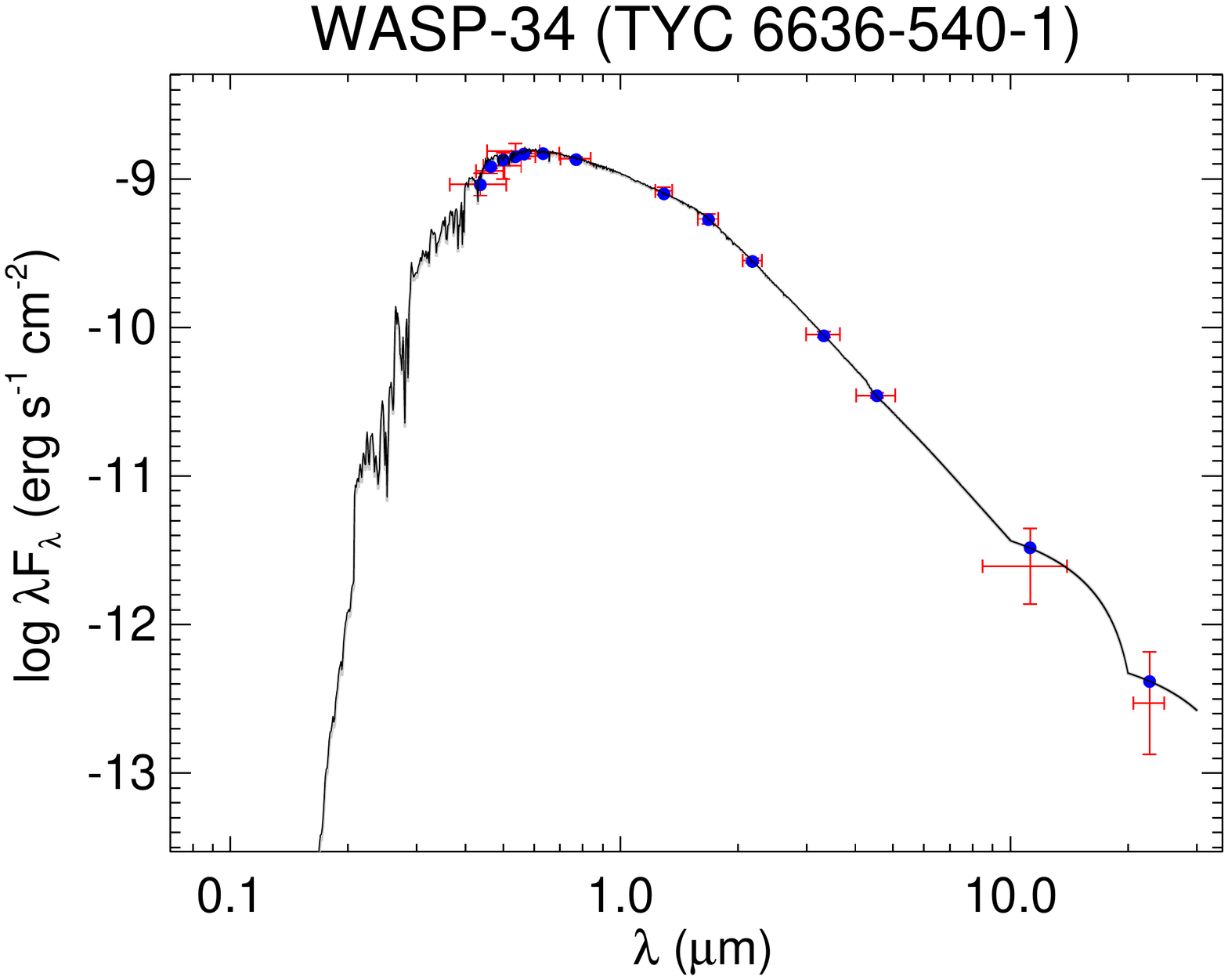}
  \includegraphics[trim=60 60 60 60,clip,width=0.49\linewidth]{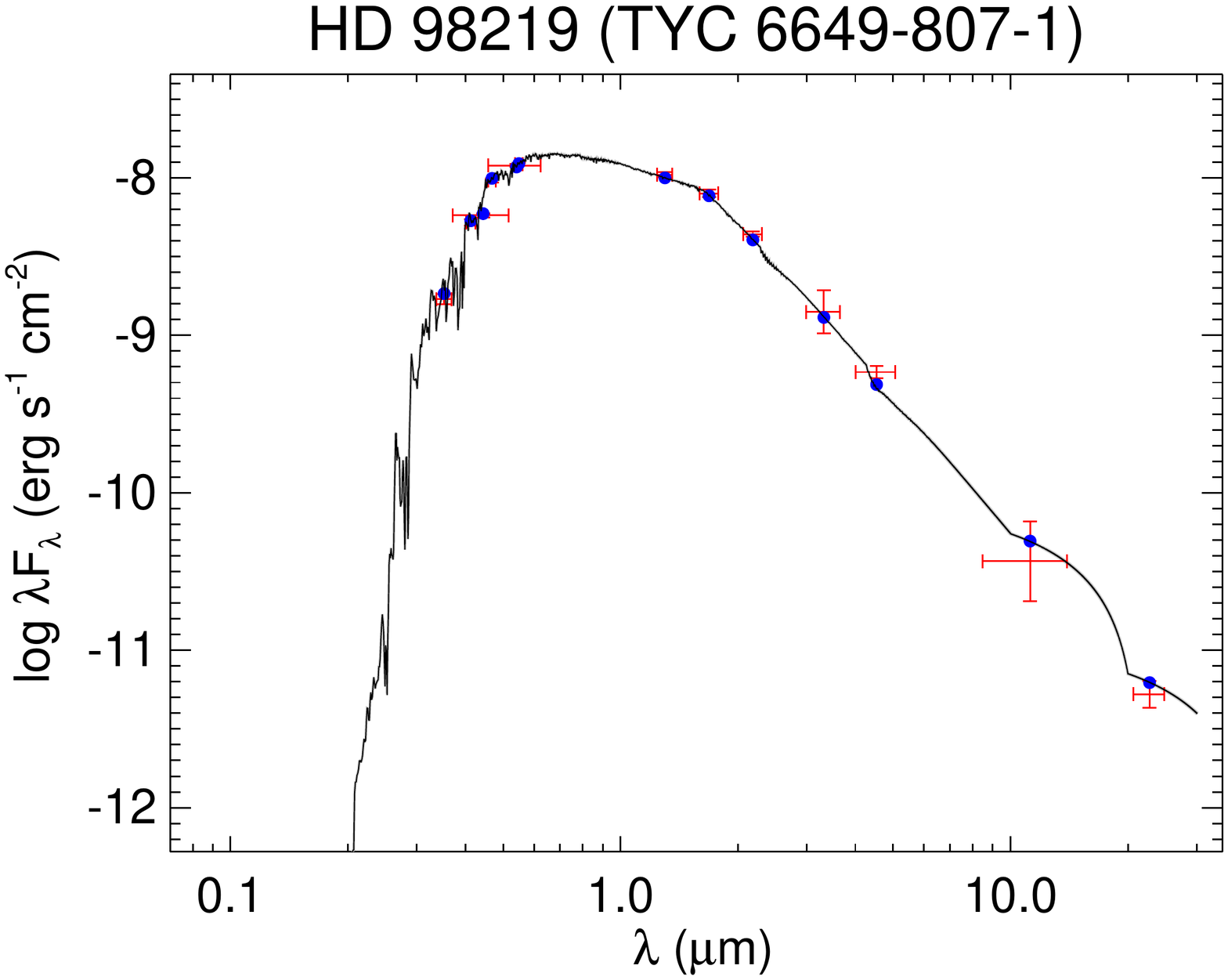}
  \includegraphics[trim=60 60 60 60,clip,width=0.49\linewidth]{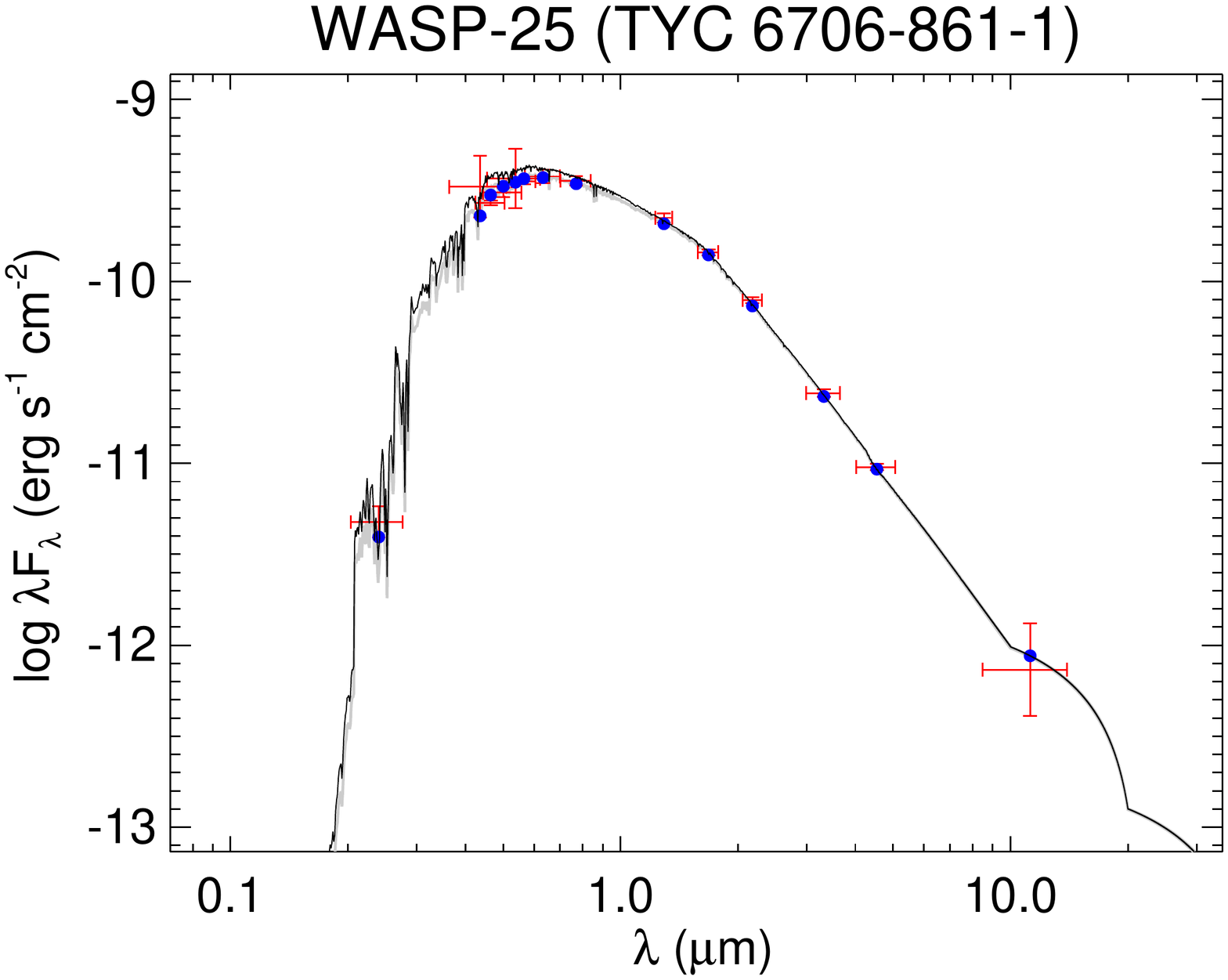}
  \caption{All labels, lines, symbols, and colors as in Figure \ref{fig:seds}.}
  \label{fig:seds_61}
\end{figure}

\begin{figure}[H]
  \centering
  \includegraphics[trim=60 60 60 60,clip,width=0.49\linewidth]{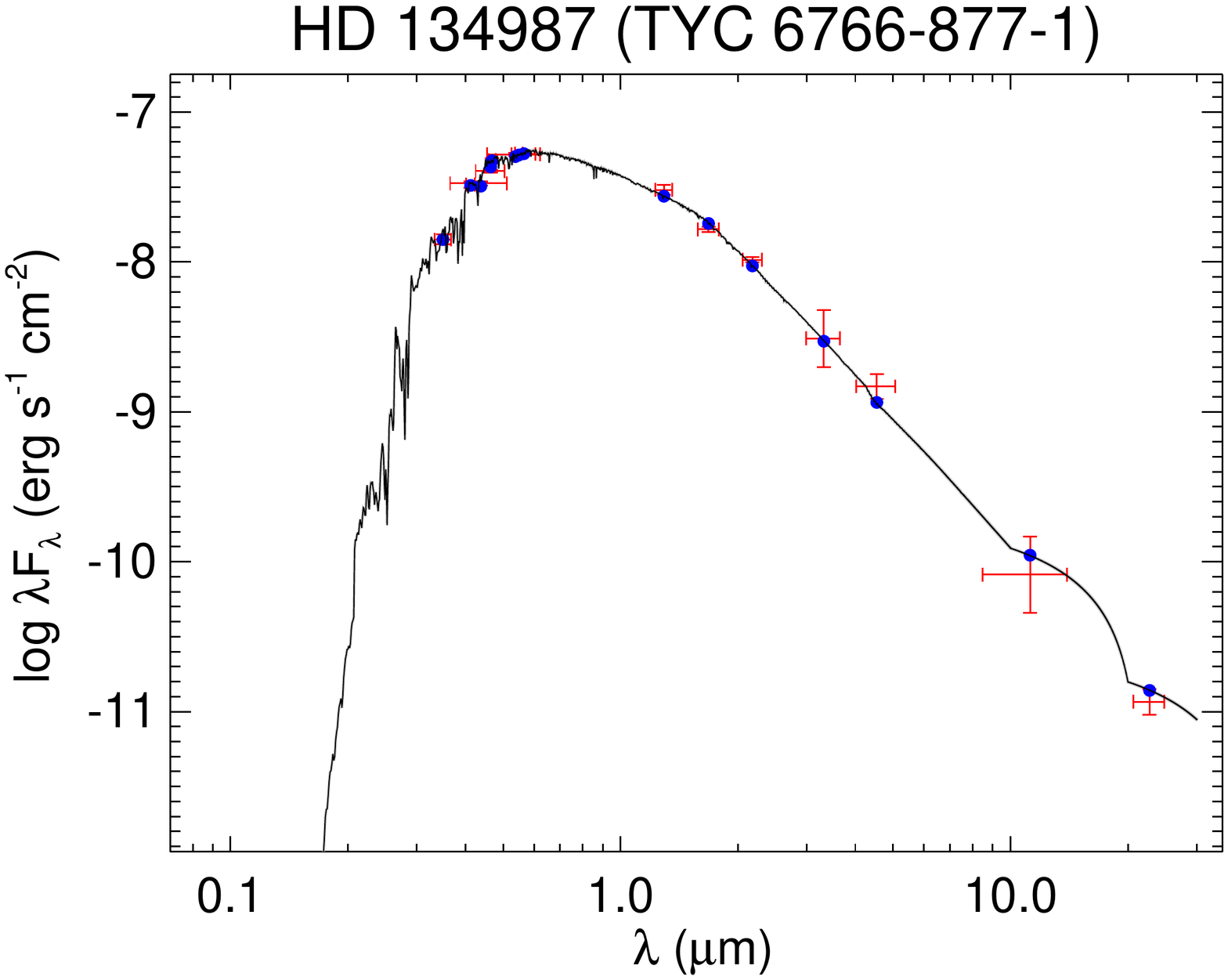}
  \includegraphics[trim=60 60 60 60,clip,width=0.49\linewidth]{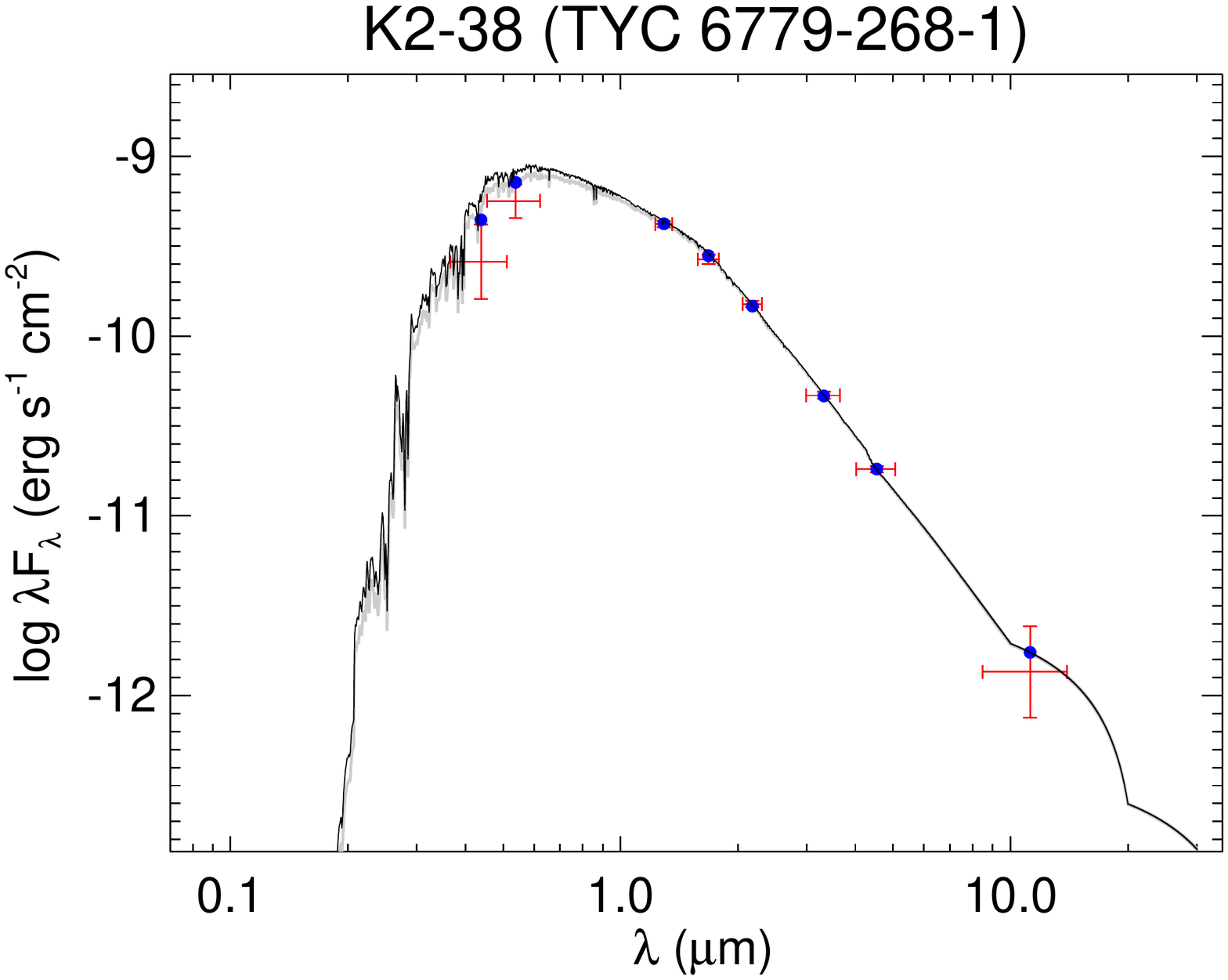}
  \includegraphics[trim=60 60 60 60,clip,width=0.49\linewidth]{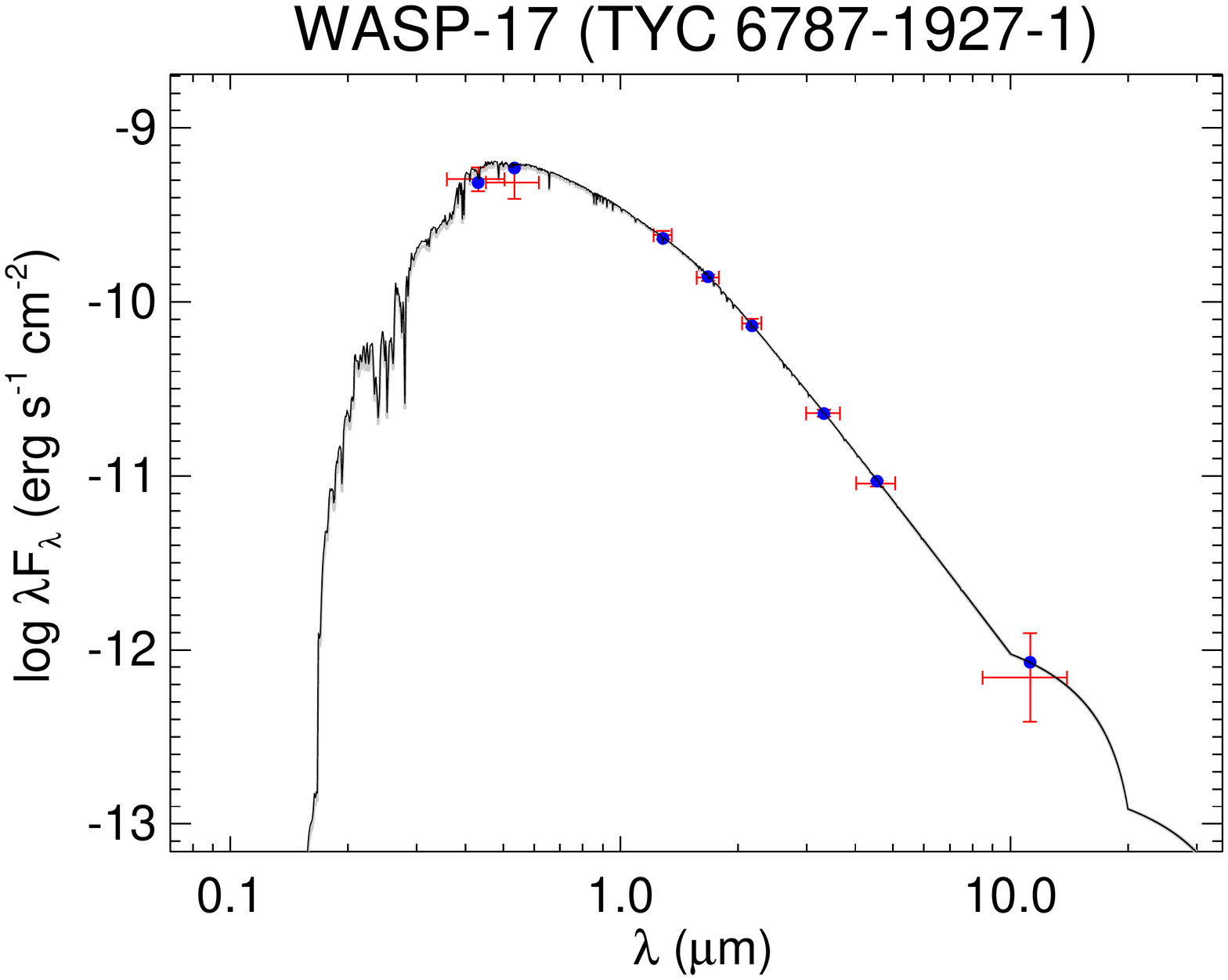}
  \includegraphics[trim=60 60 60 60,clip,width=0.49\linewidth]{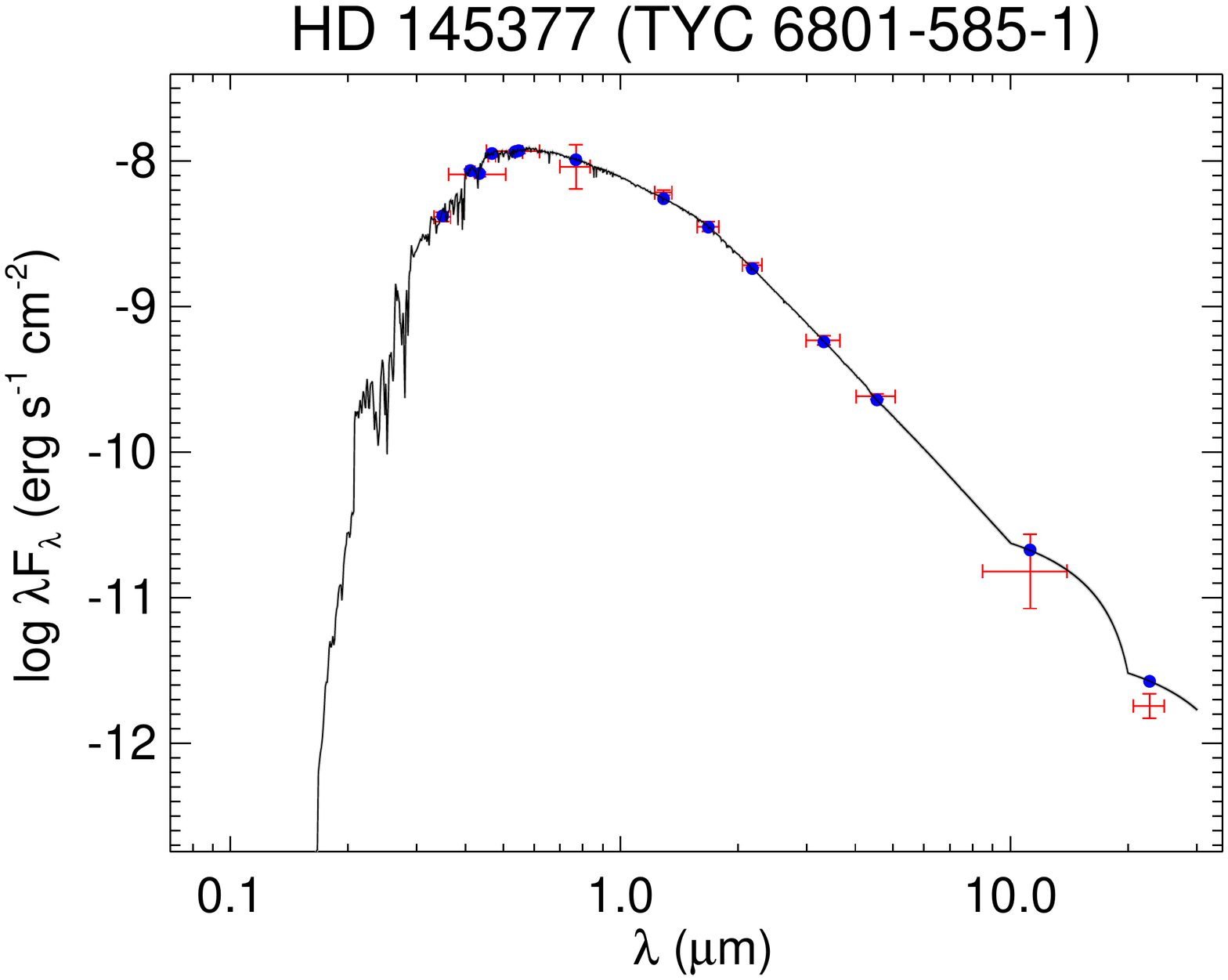}
  \includegraphics[trim=60 60 60 60,clip,width=0.49\linewidth]{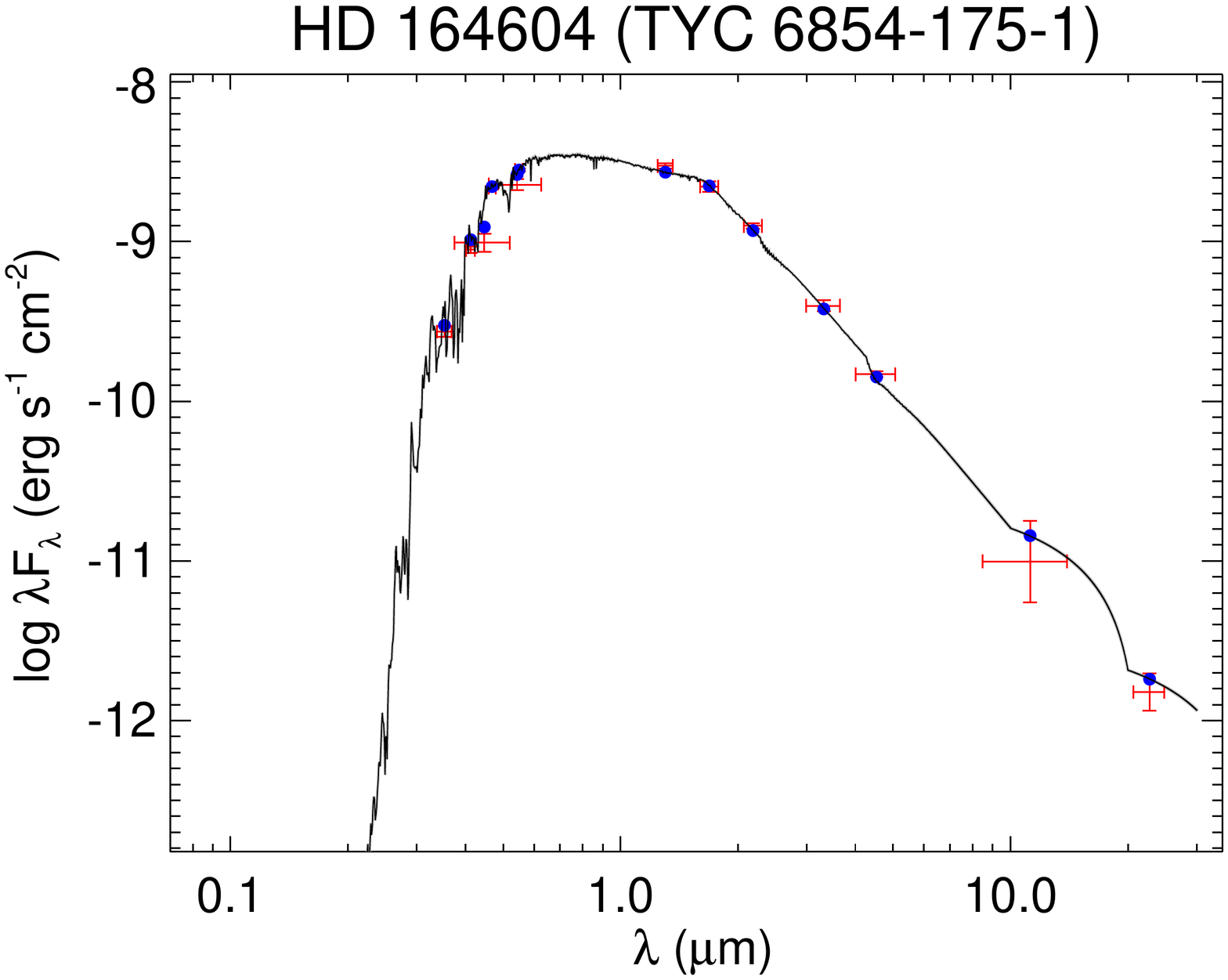}
  \includegraphics[trim=60 60 60 60,clip,width=0.49\linewidth]{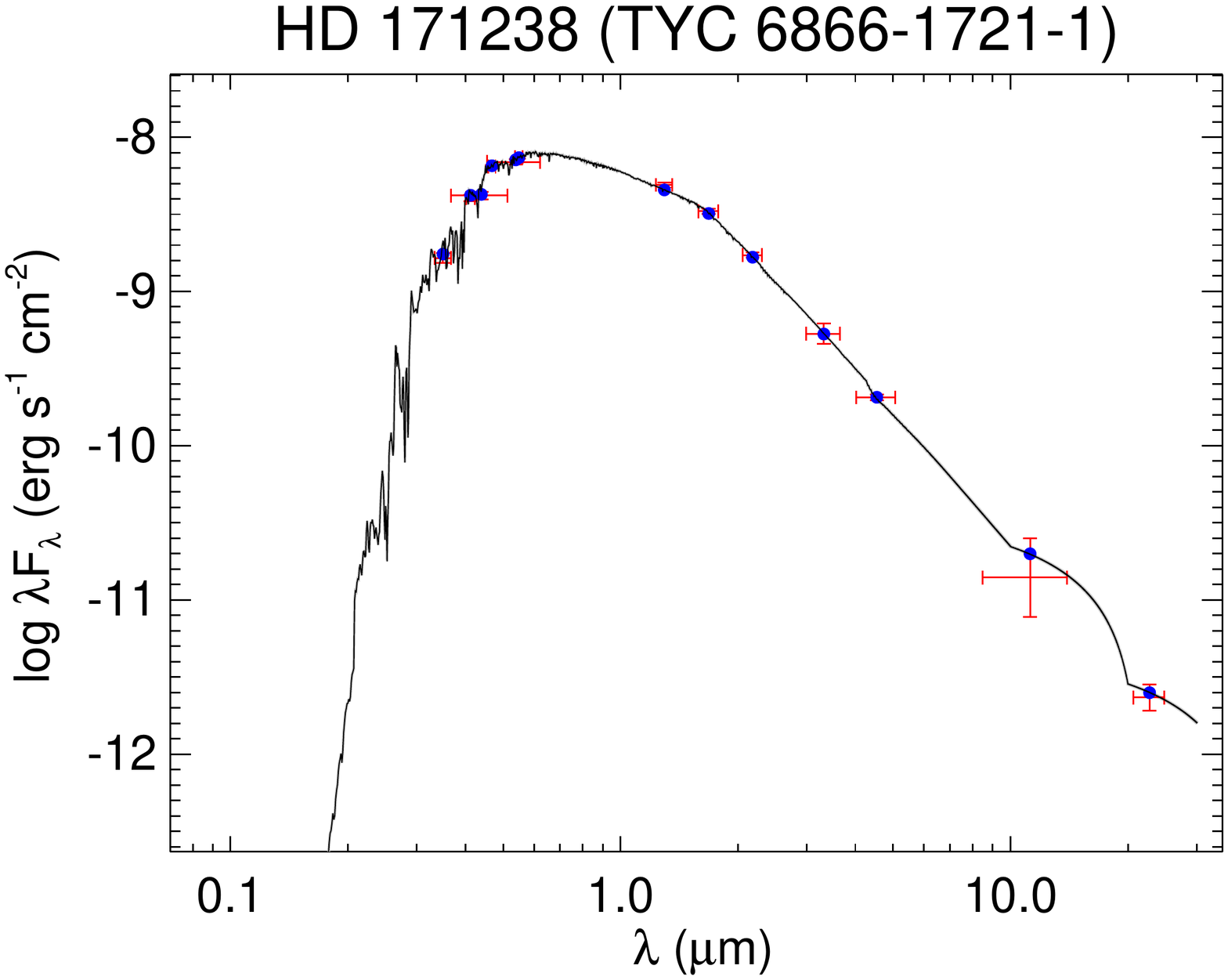}
  \caption{All labels, lines, symbols, and colors as in Figure \ref{fig:seds}.}
  \label{fig:seds_62}
\end{figure}

\begin{figure}[H]
  \centering
  \includegraphics[trim=60 60 60 60,clip,width=0.49\linewidth]{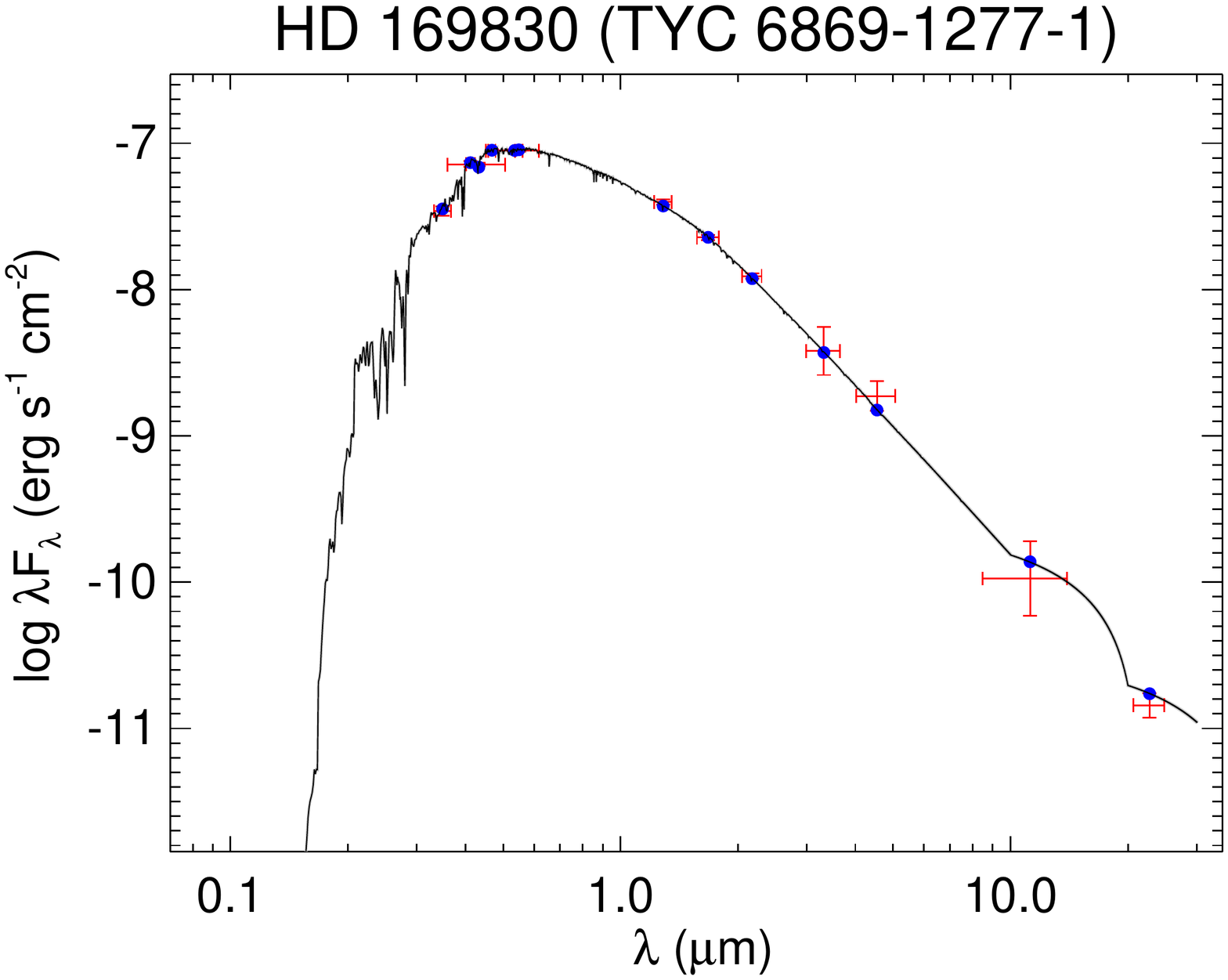}
  \includegraphics[trim=60 60 60 60,clip,width=0.49\linewidth]{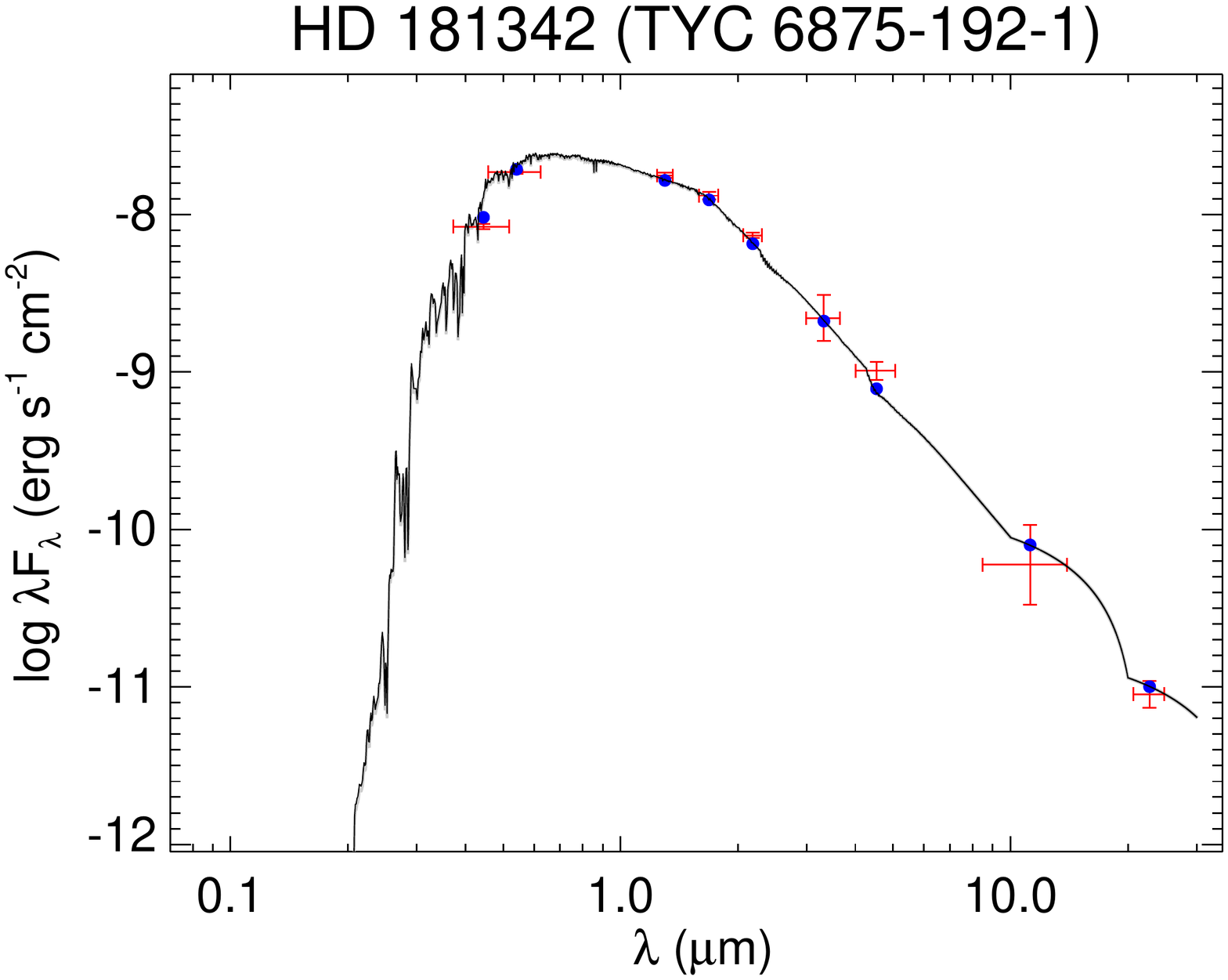}
  \includegraphics[trim=60 60 60 60,clip,width=0.49\linewidth]{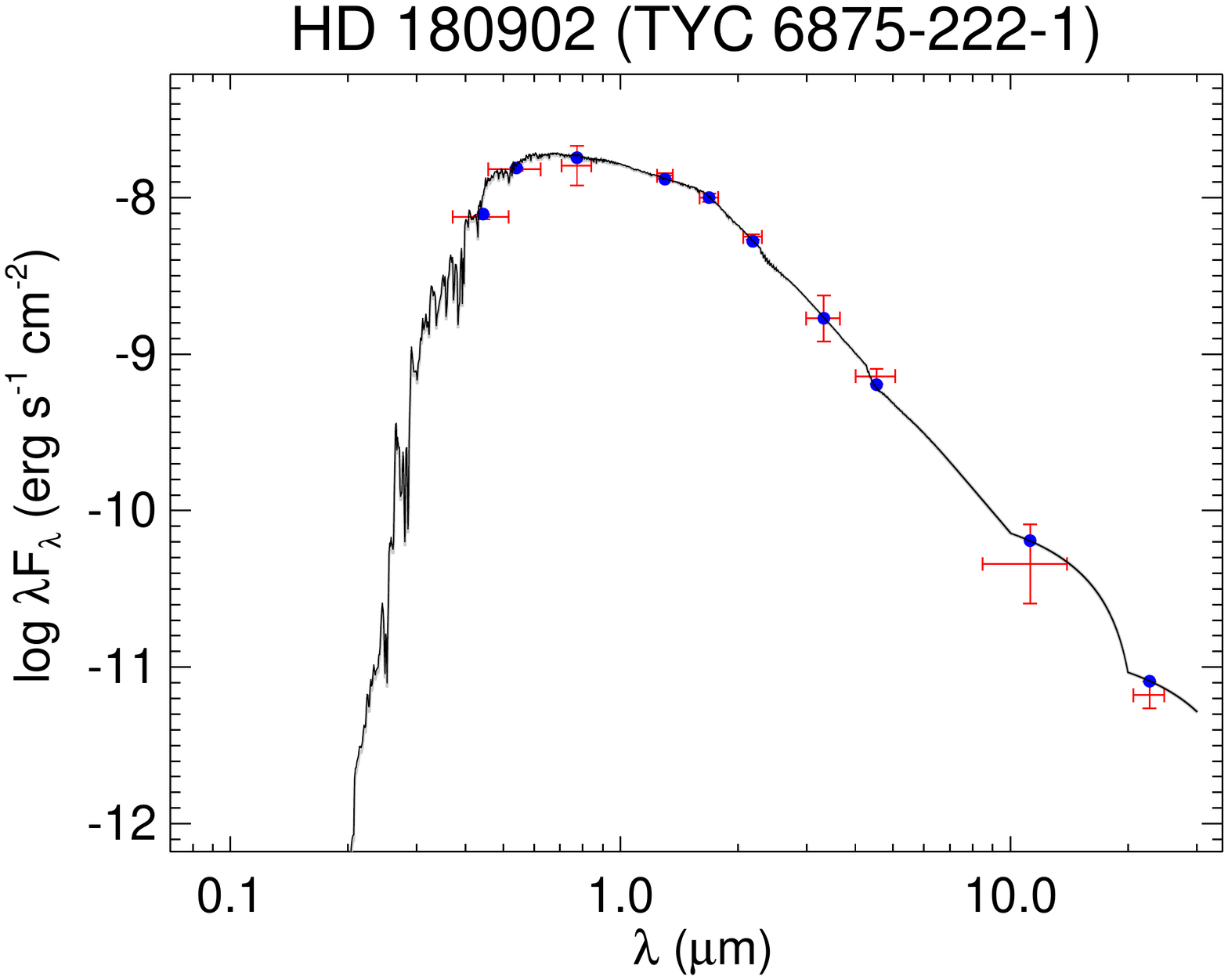}
  \includegraphics[trim=60 60 60 60,clip,width=0.49\linewidth]{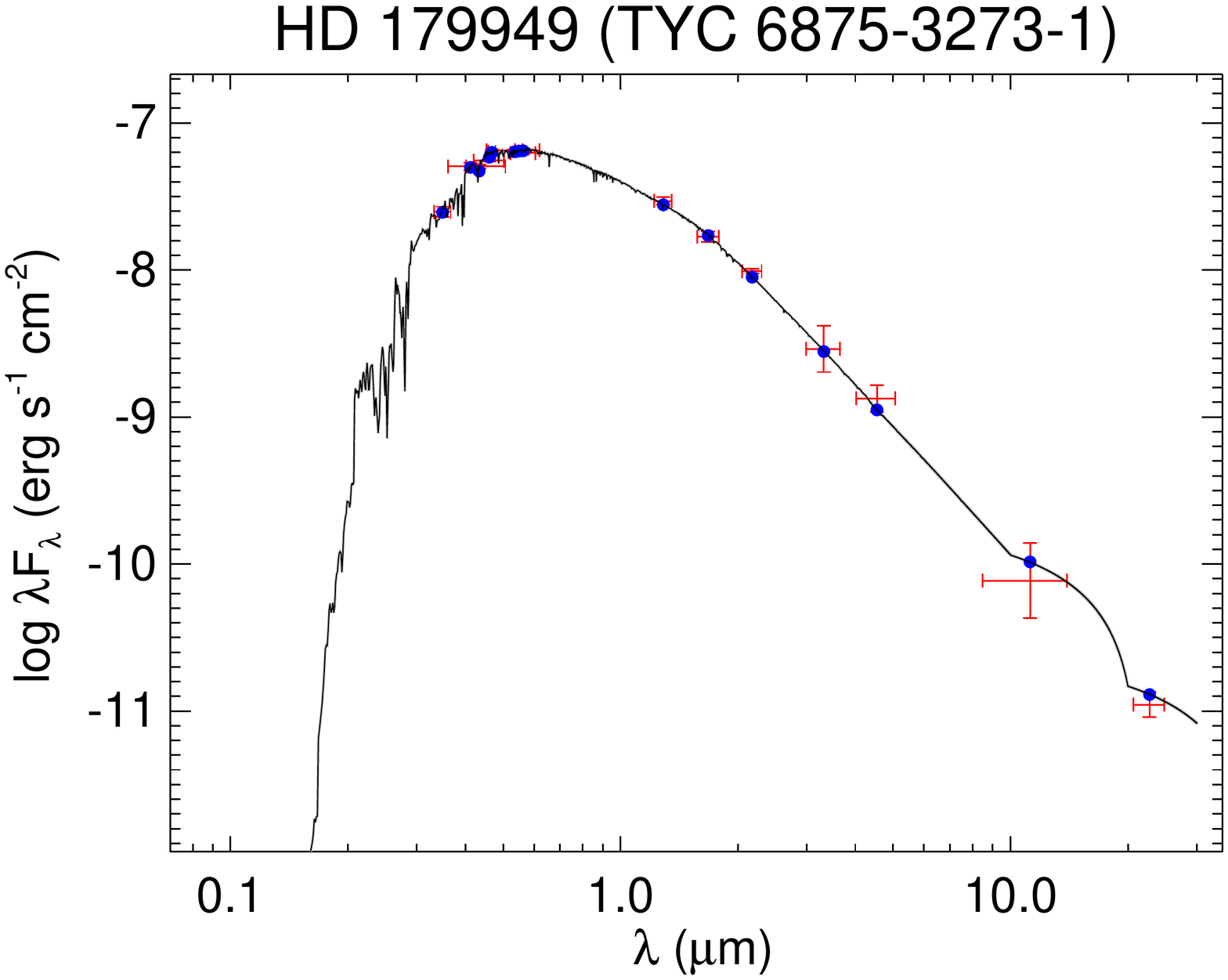}
  \includegraphics[trim=60 60 60 60,clip,width=0.49\linewidth]{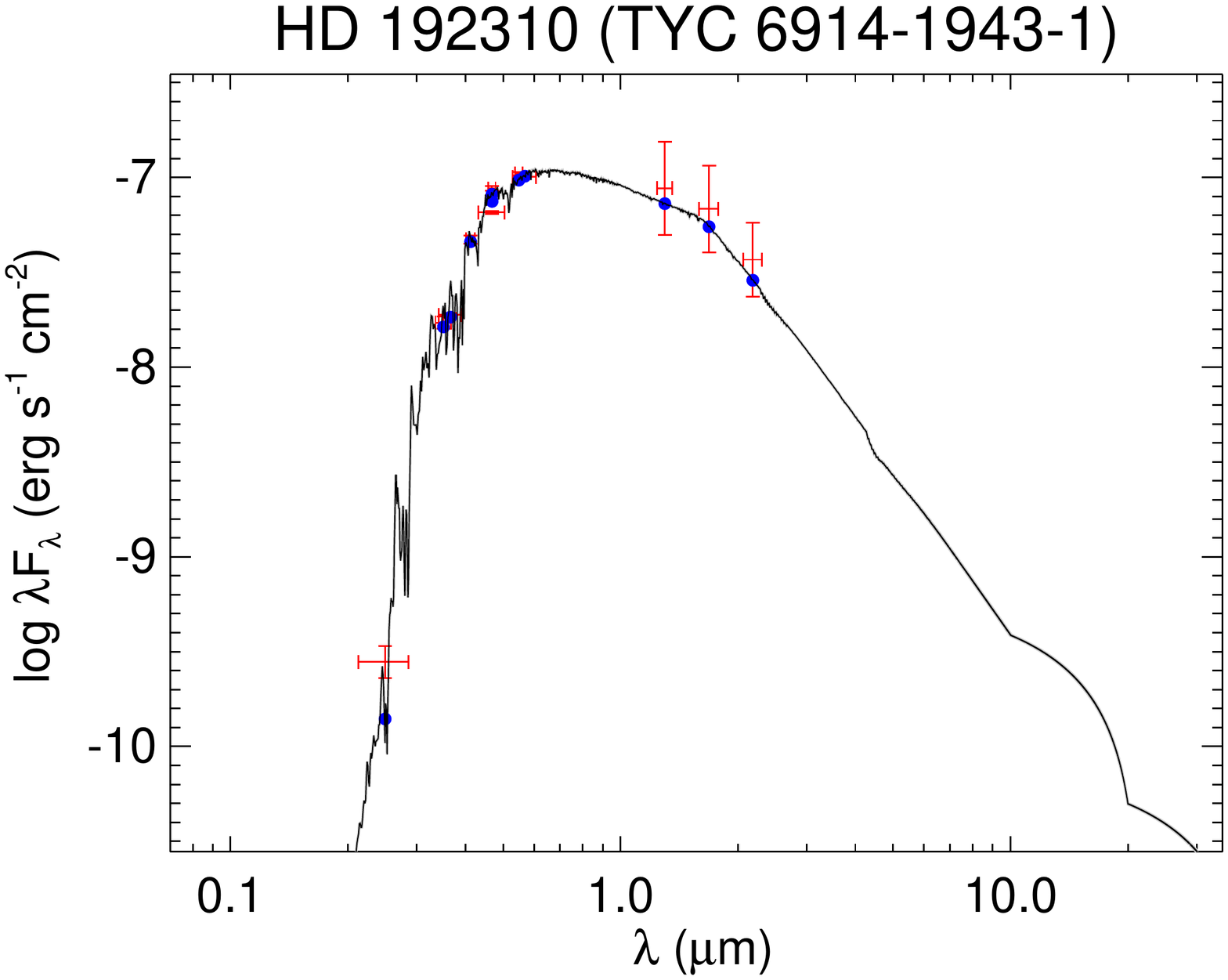}
  \includegraphics[trim=60 60 60 60,clip,width=0.49\linewidth]{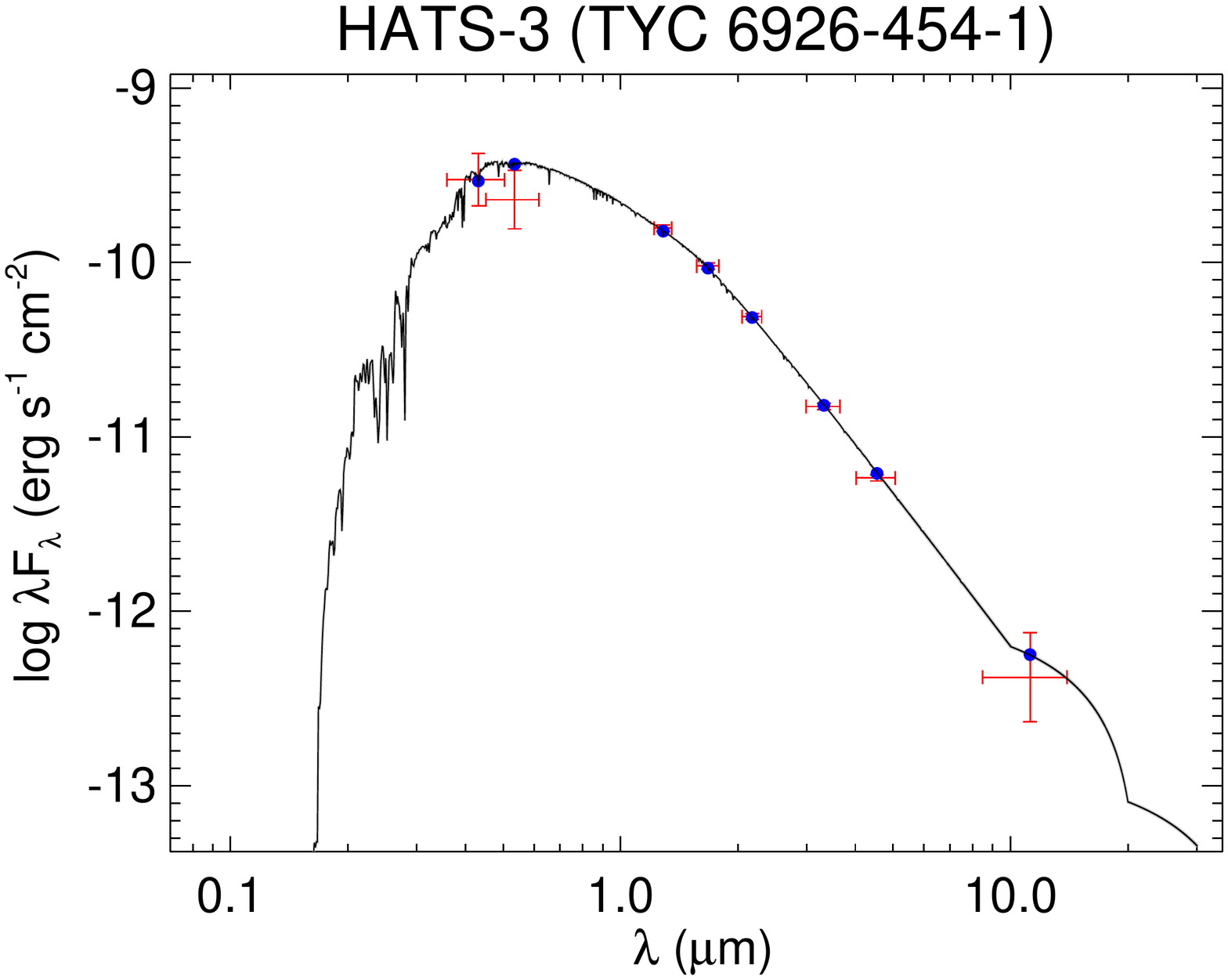}
  \caption{All labels, lines, symbols, and colors as in Figure \ref{fig:seds}.}
  \label{fig:seds_63}
\end{figure}

\begin{figure}[H]
  \centering
  \includegraphics[trim=60 60 60 60,clip,width=0.49\linewidth]{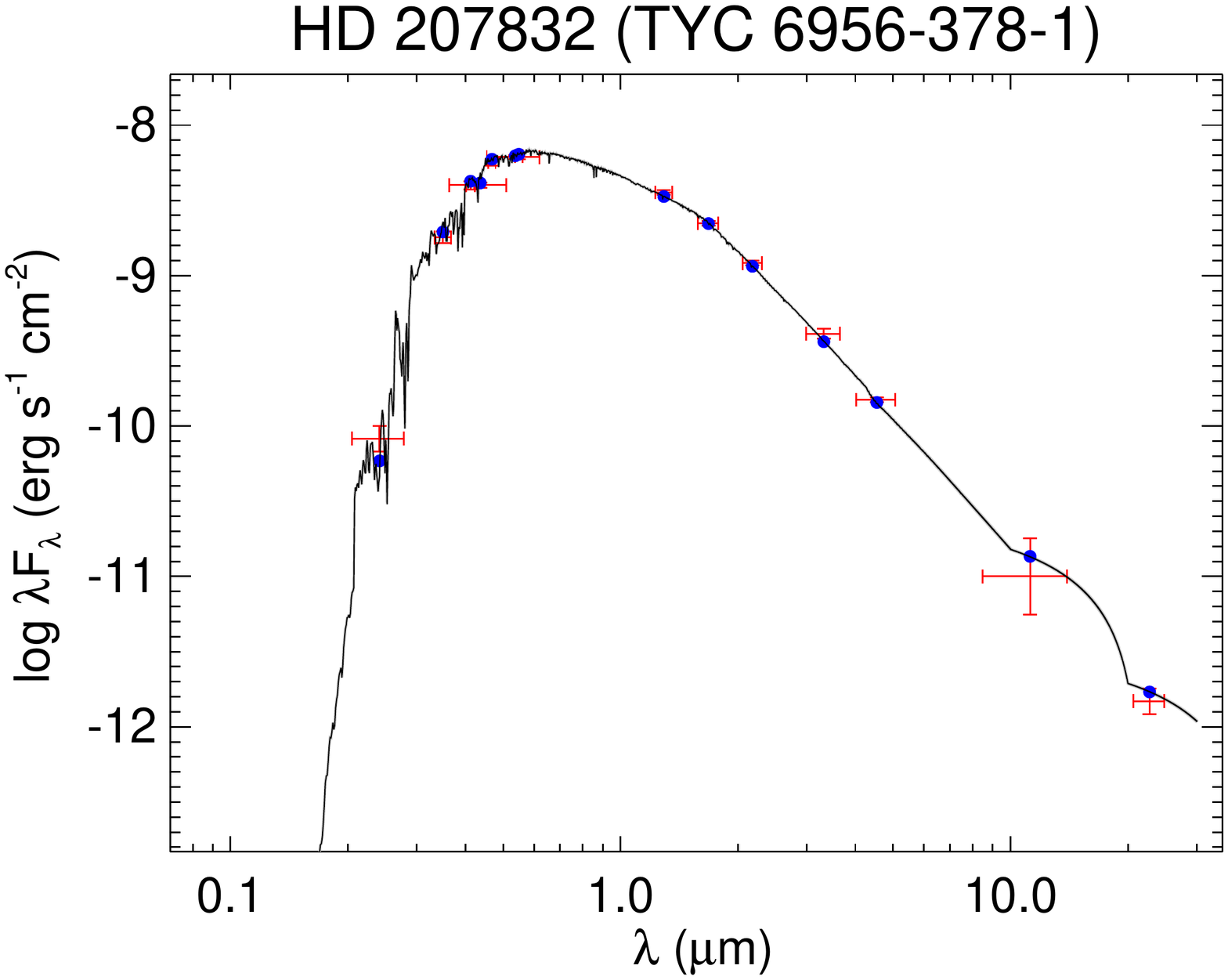}
  \includegraphics[trim=60 60 60 60,clip,width=0.49\linewidth]{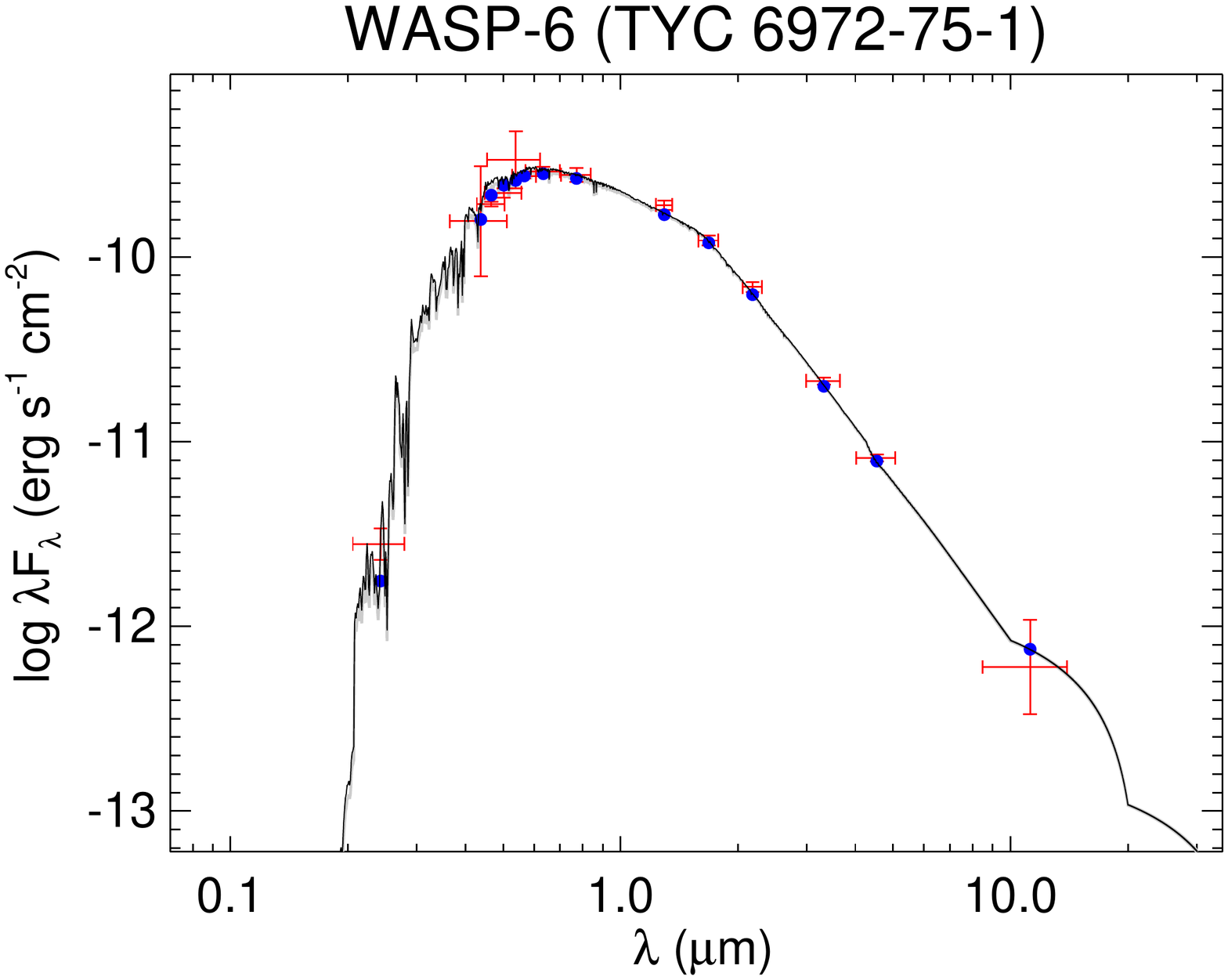}
  \includegraphics[trim=60 60 60 60,clip,width=0.49\linewidth]{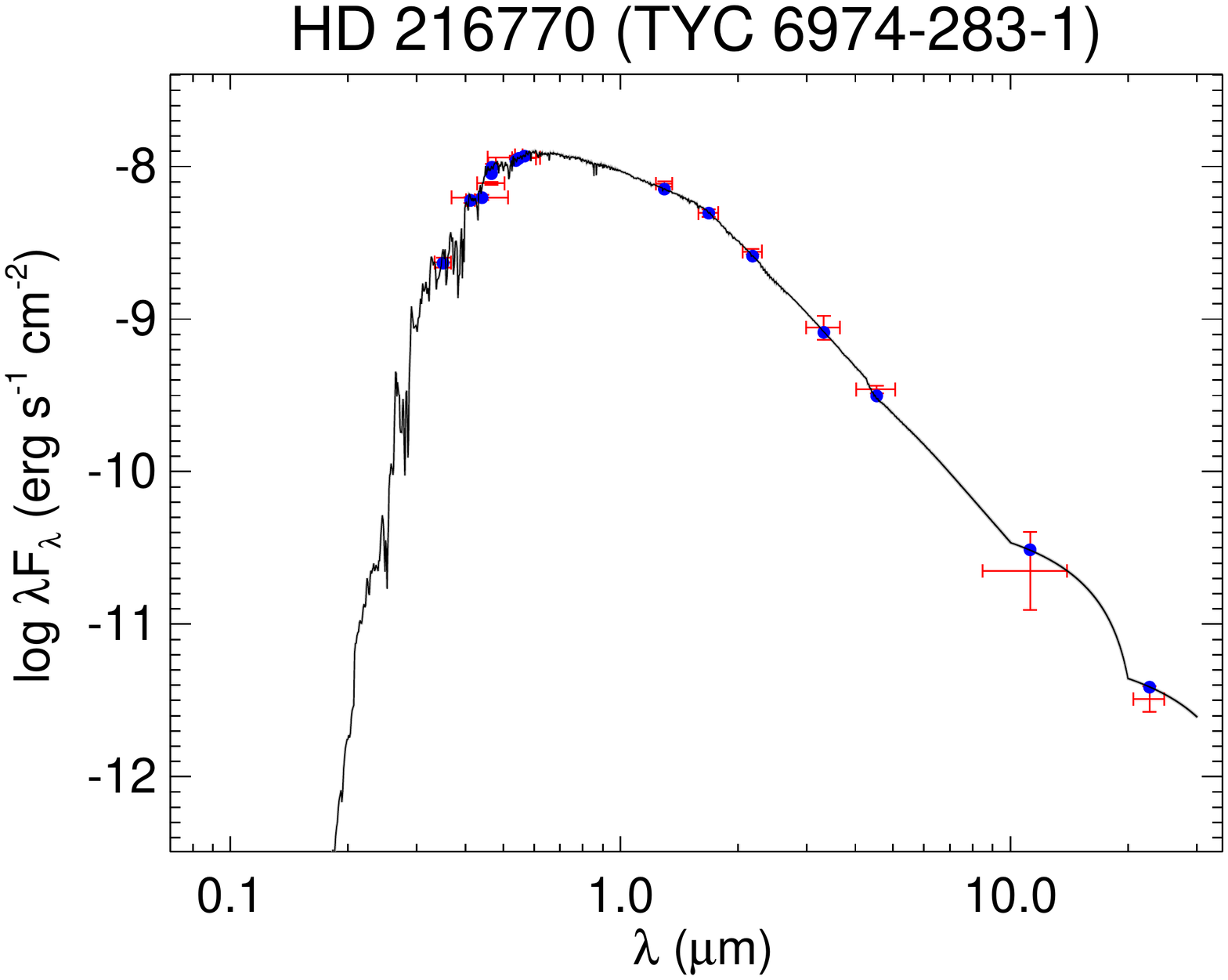}
  \includegraphics[trim=60 60 60 60,clip,width=0.49\linewidth]{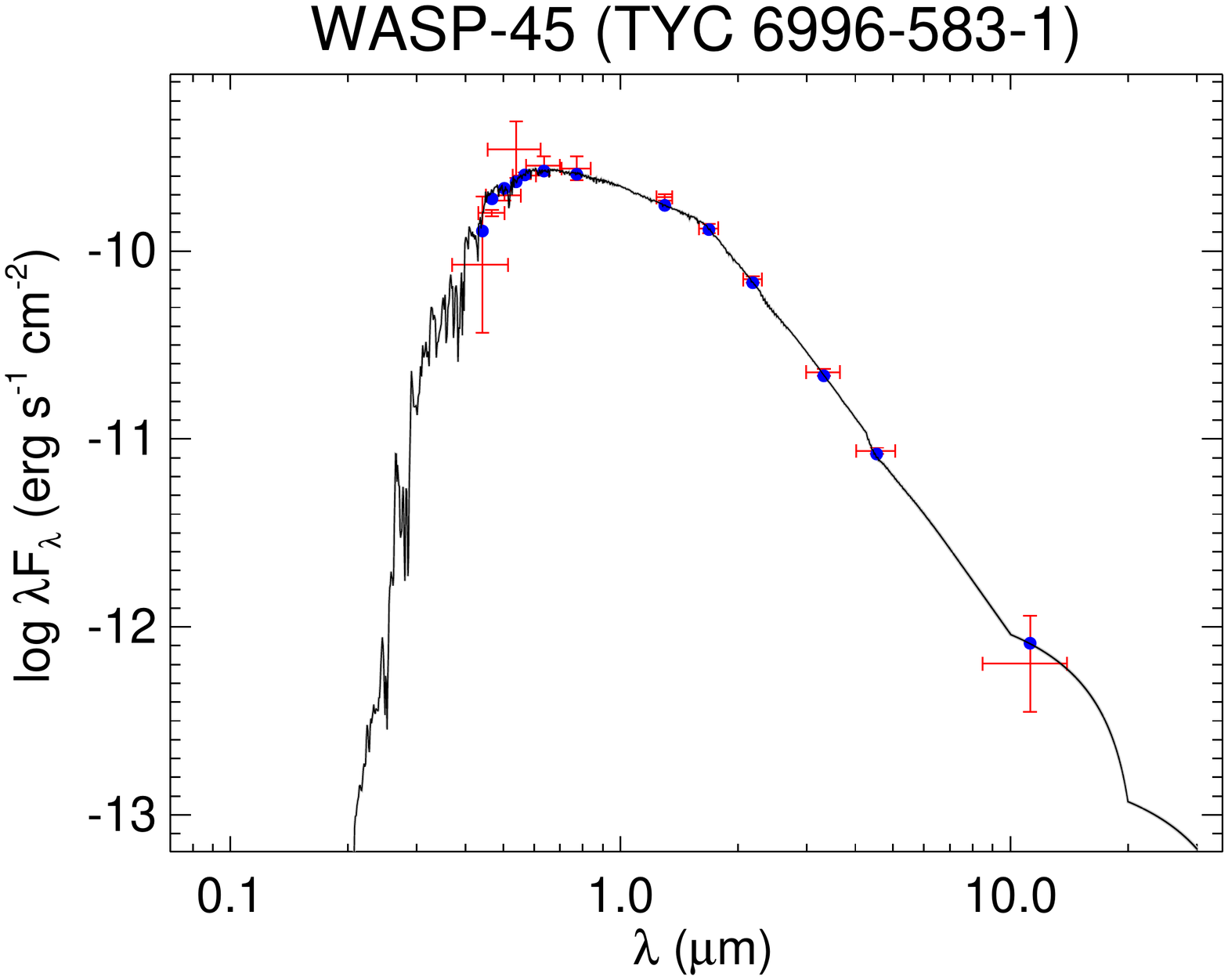}
  \includegraphics[trim=60 60 60 60,clip,width=0.49\linewidth]{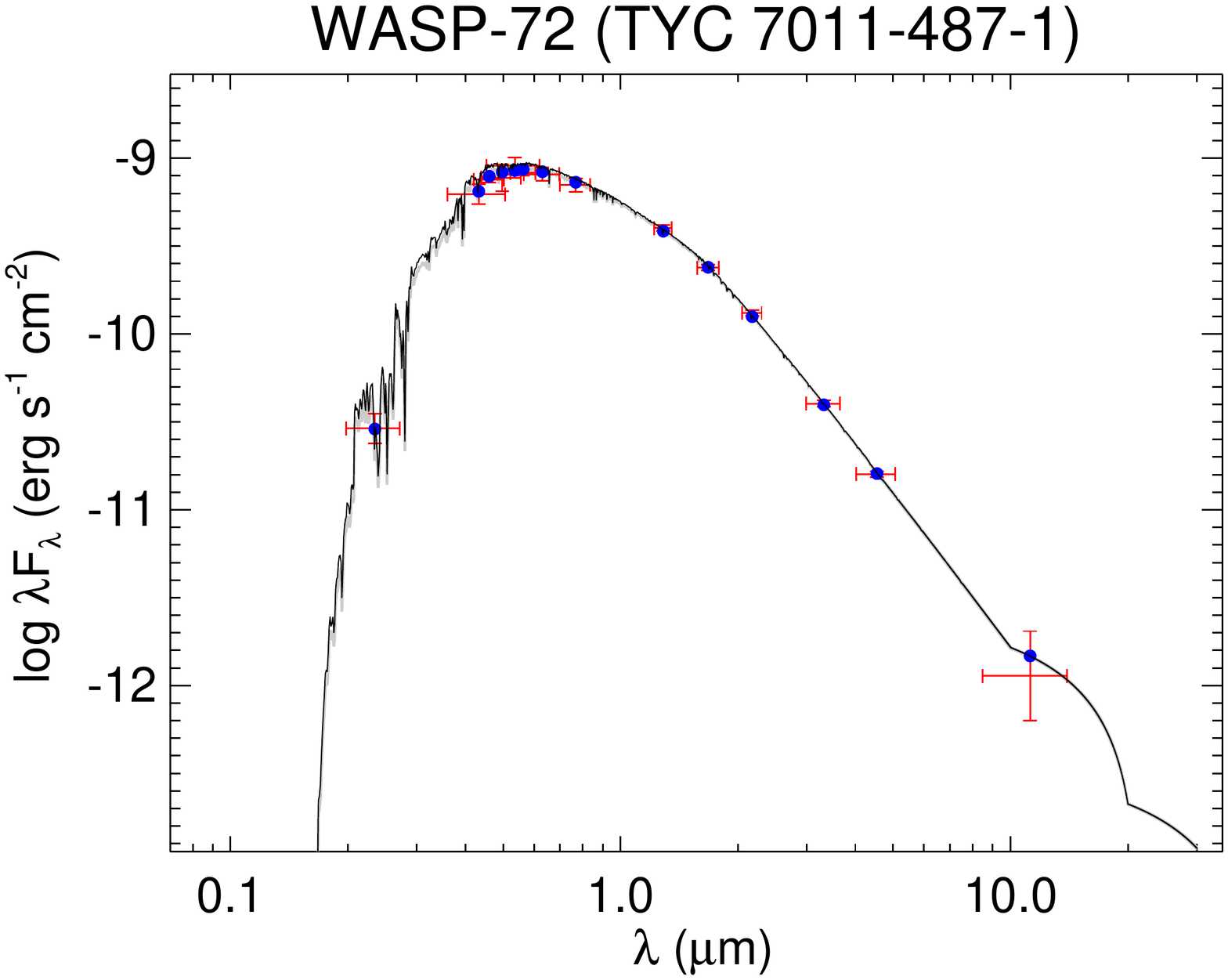}
  \includegraphics[trim=60 60 60 60,clip,width=0.49\linewidth]{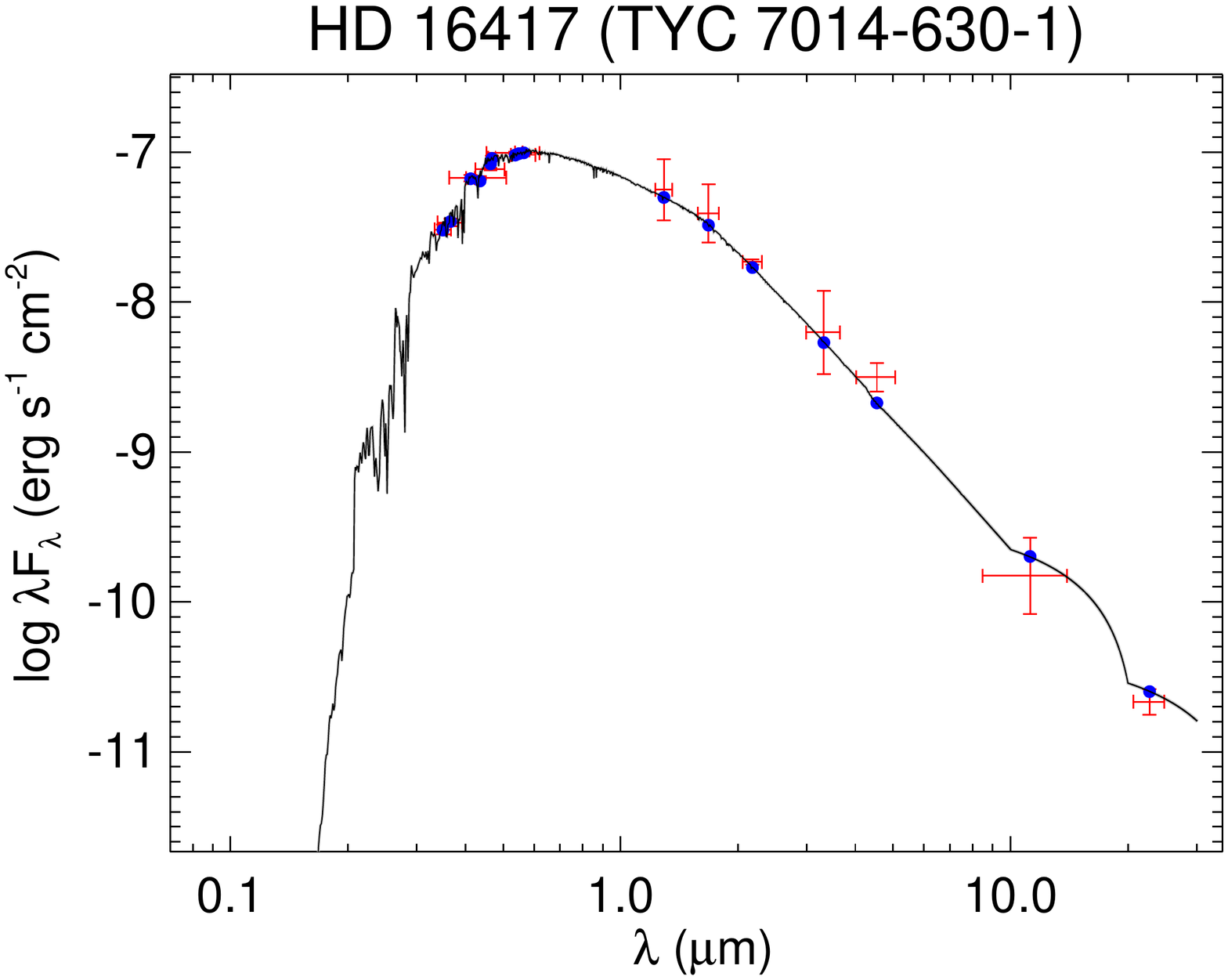}
  \caption{All labels, lines, symbols, and colors as in Figure \ref{fig:seds}.}
  \label{fig:seds_64}
\end{figure}

\begin{figure}[H]
  \centering
  \includegraphics[trim=60 60 60 60,clip,width=0.49\linewidth]{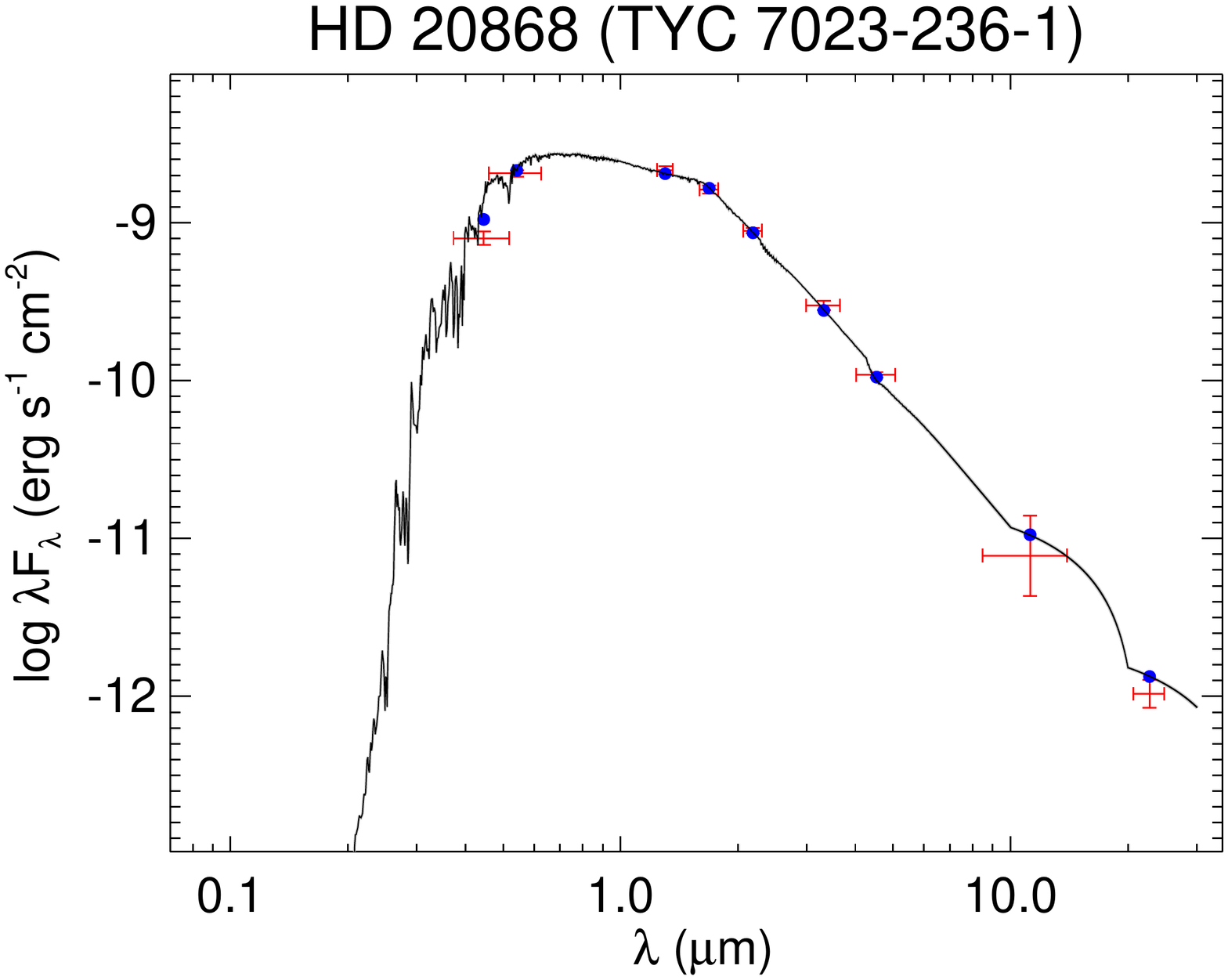}
  \includegraphics[trim=60 60 60 60,clip,width=0.49\linewidth]{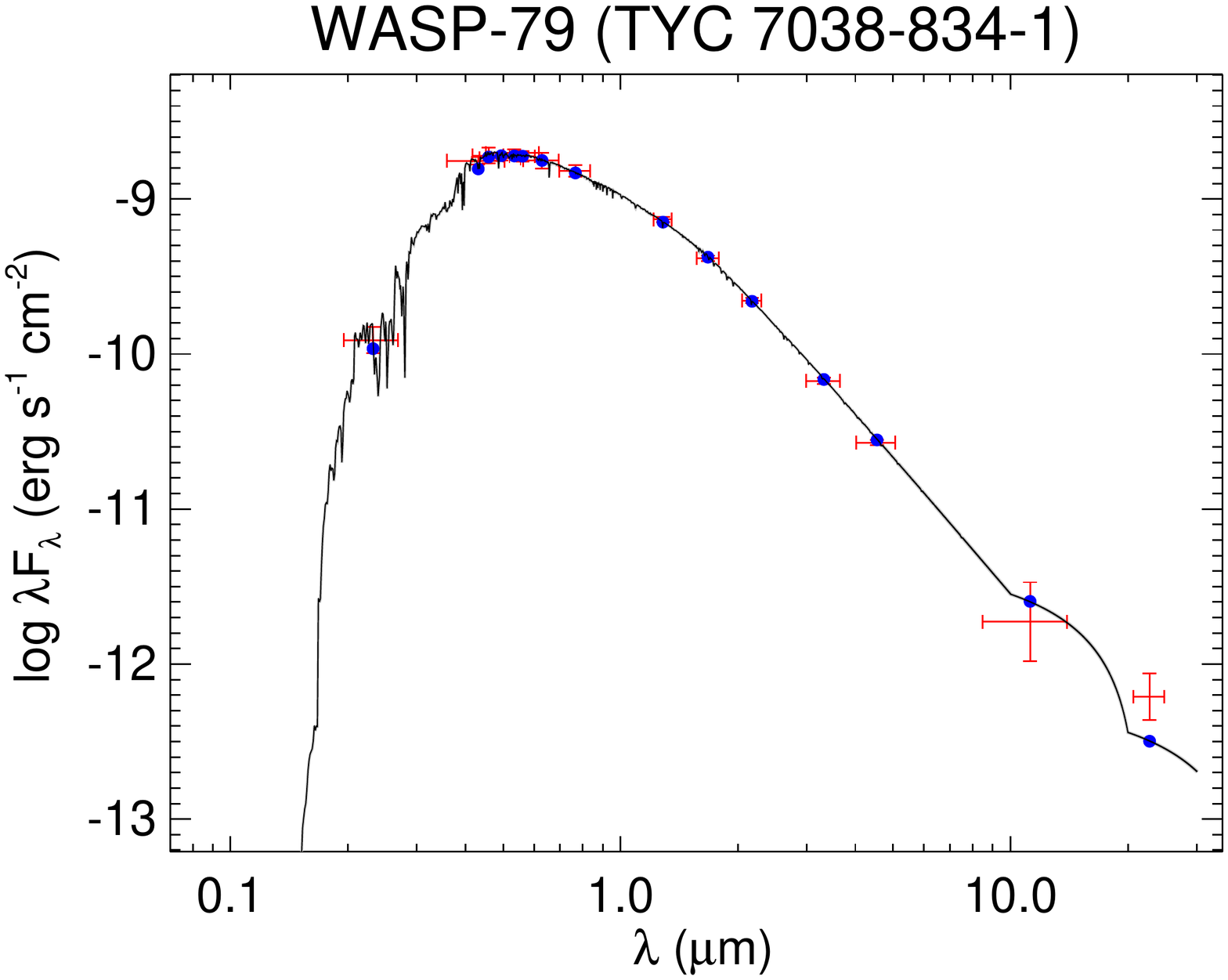}
  \includegraphics[trim=60 60 60 60,clip,width=0.49\linewidth]{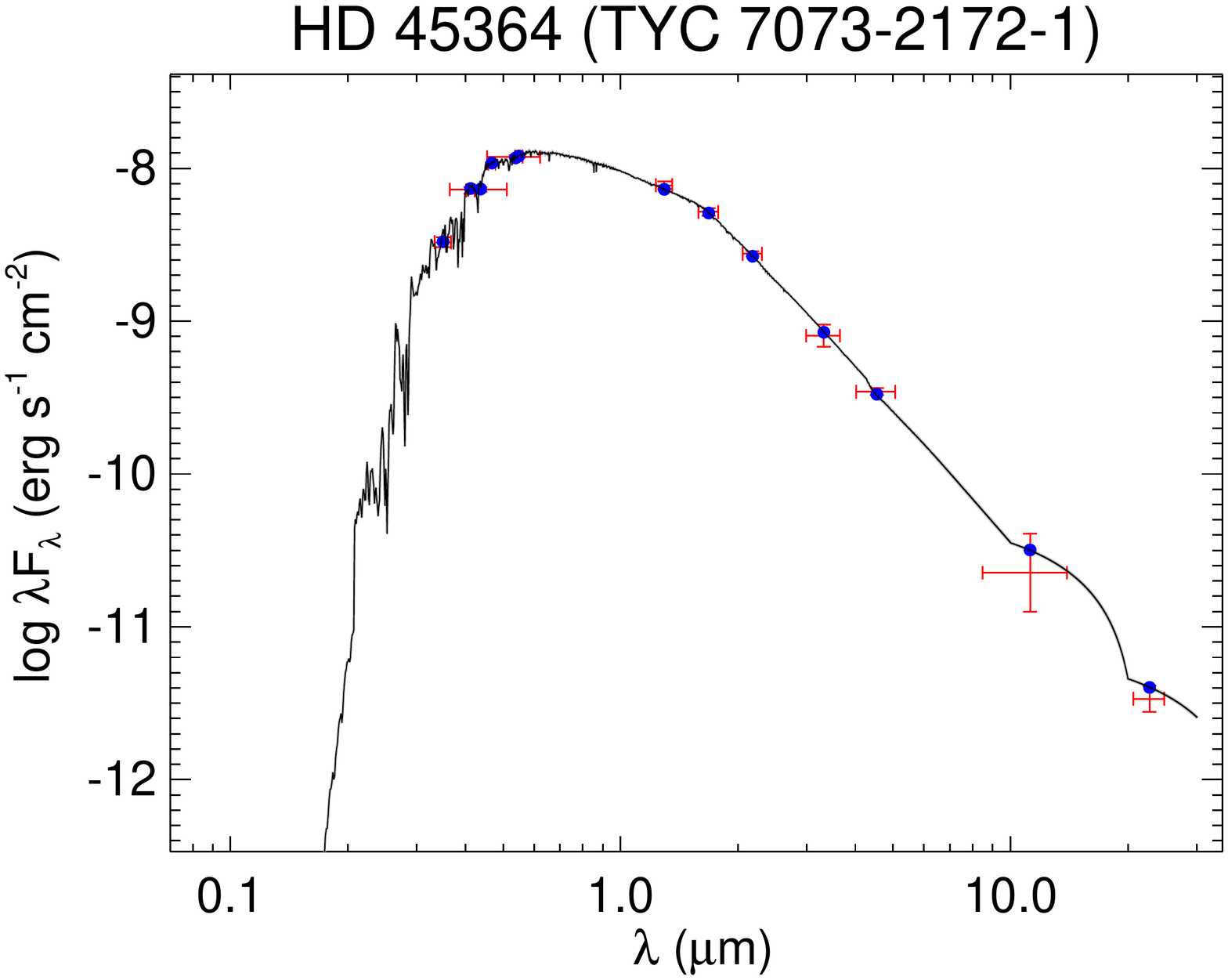}
  \includegraphics[trim=60 60 60 60,clip,width=0.49\linewidth]{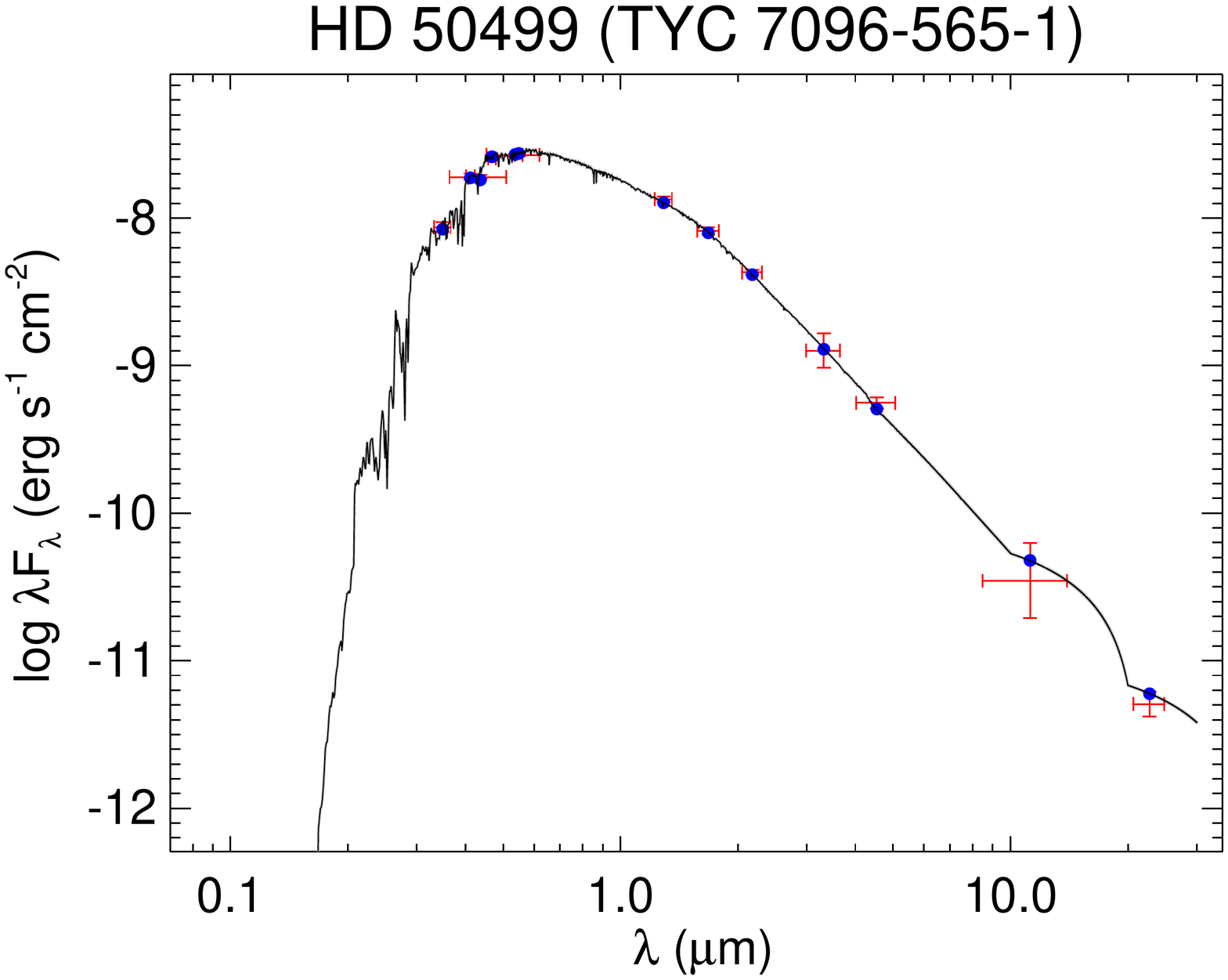}
  \includegraphics[trim=60 60 60 60,clip,width=0.49\linewidth]{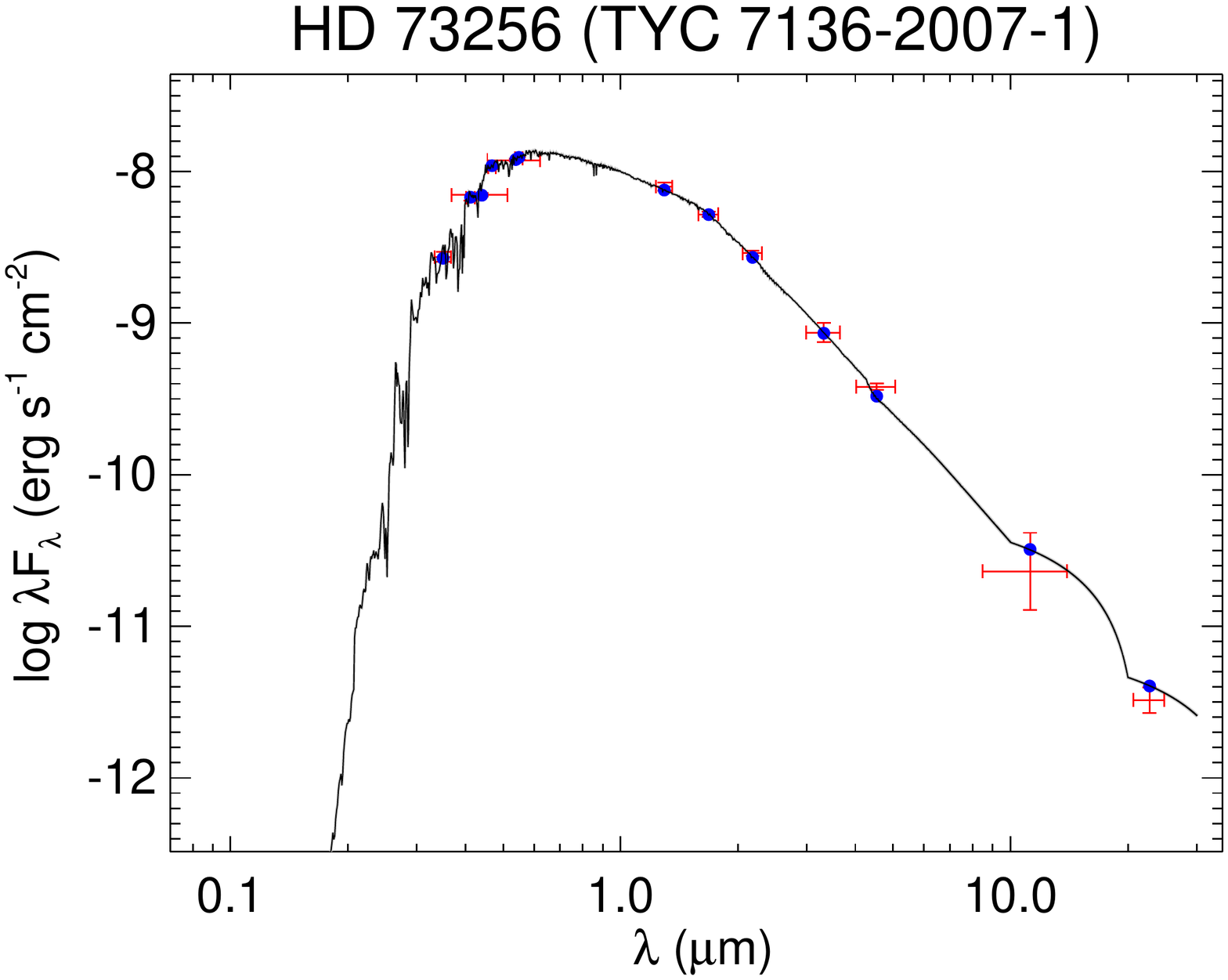}
  \includegraphics[trim=60 60 60 60,clip,width=0.49\linewidth]{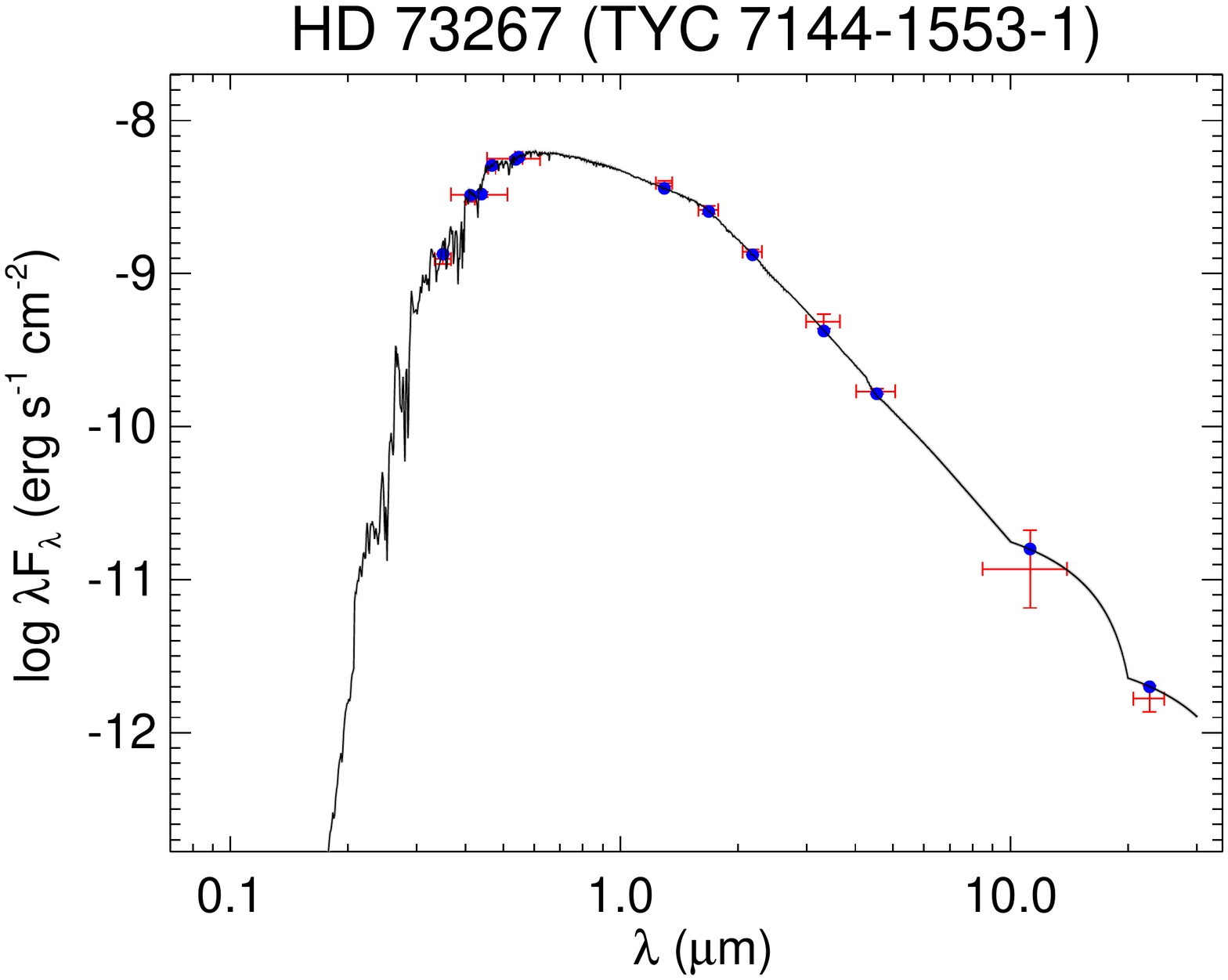}
  \caption{All labels, lines, symbols, and colors as in Figure \ref{fig:seds}.}
  \label{fig:seds_65}
\end{figure}

\begin{figure}[H]
  \centering
  \includegraphics[trim=60 60 60 60,clip,width=0.49\linewidth]{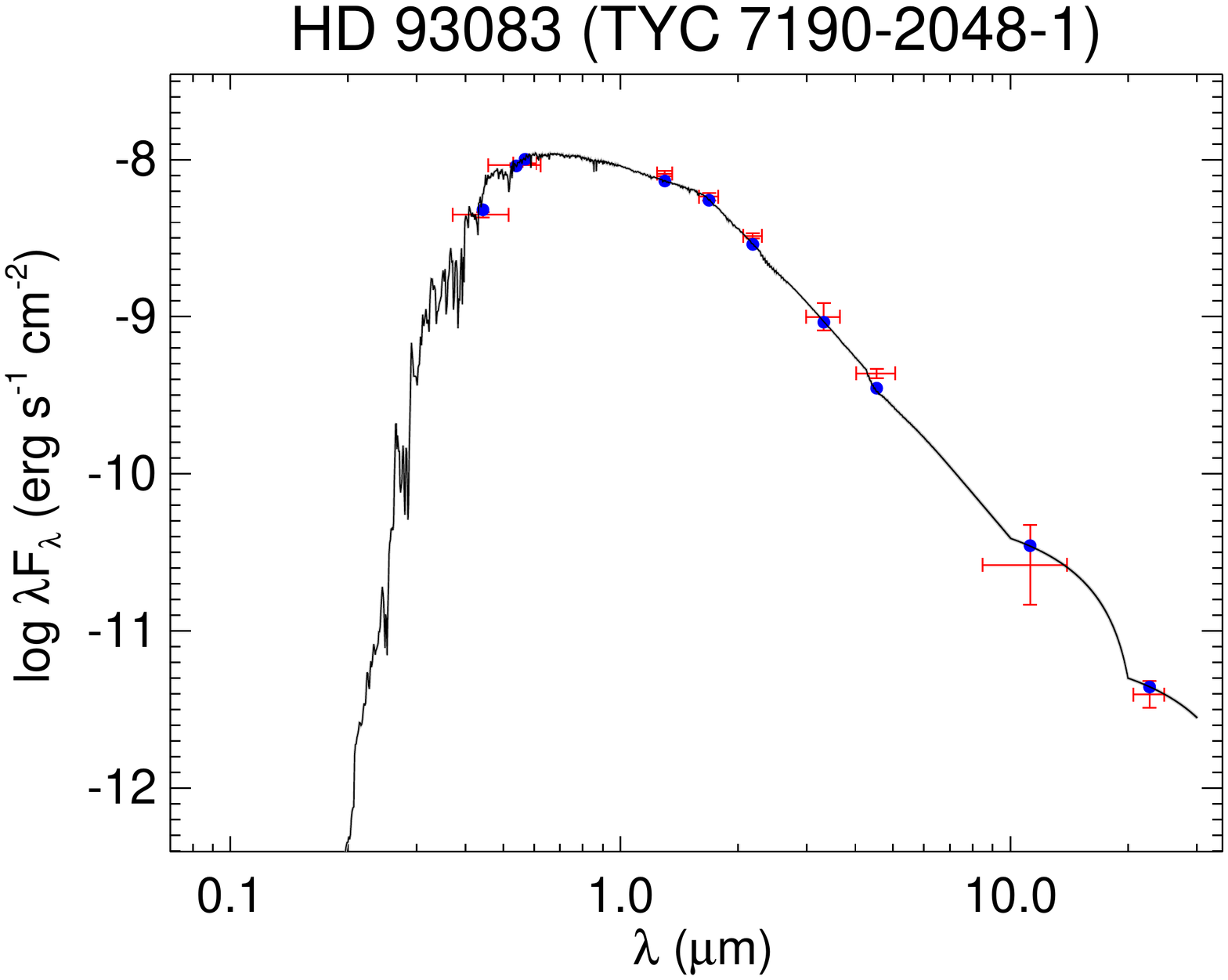}
  \includegraphics[trim=60 60 60 60,clip,width=0.49\linewidth]{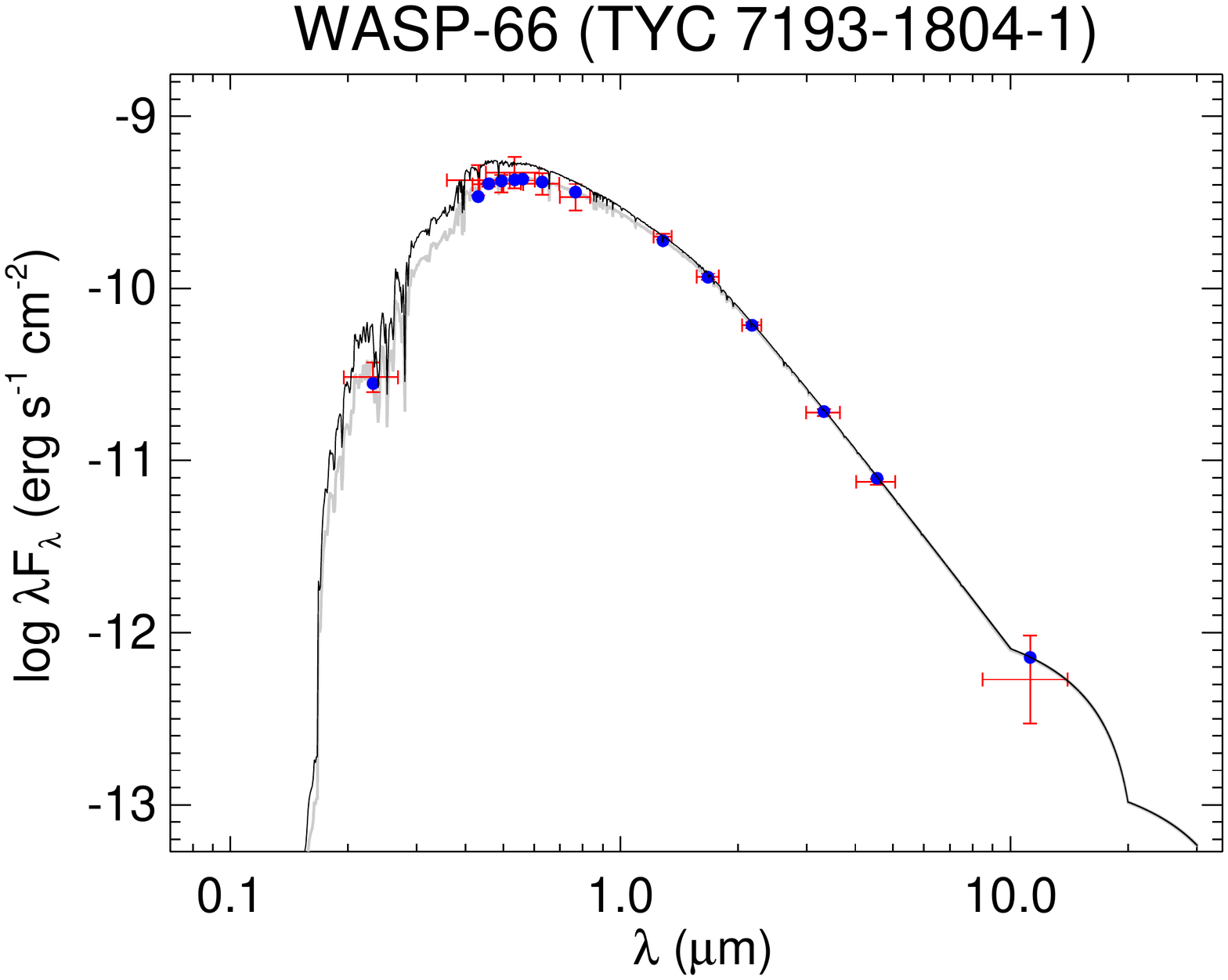}
  \includegraphics[trim=60 60 60 60,clip,width=0.49\linewidth]{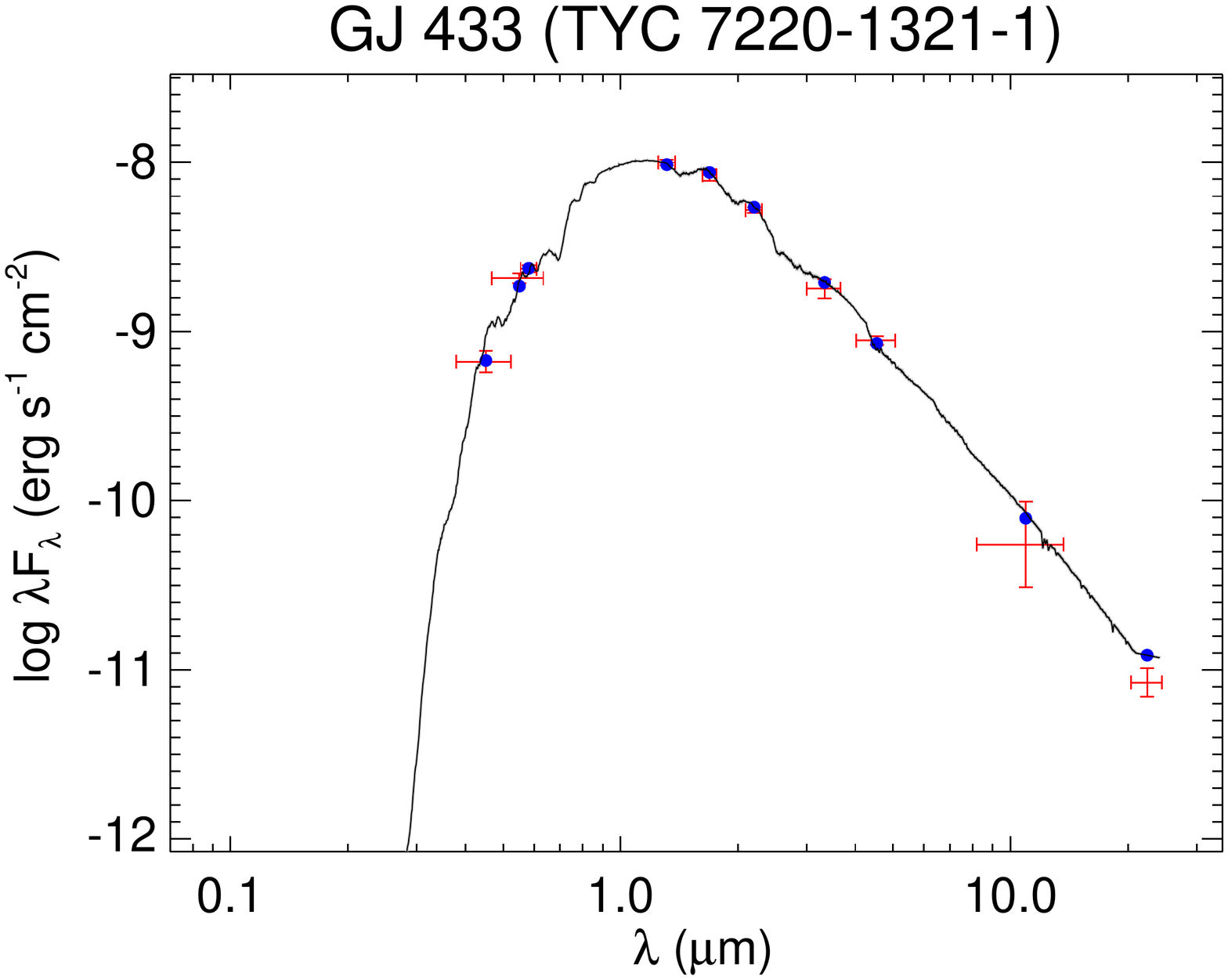}
  \includegraphics[trim=60 60 60 60,clip,width=0.49\linewidth]{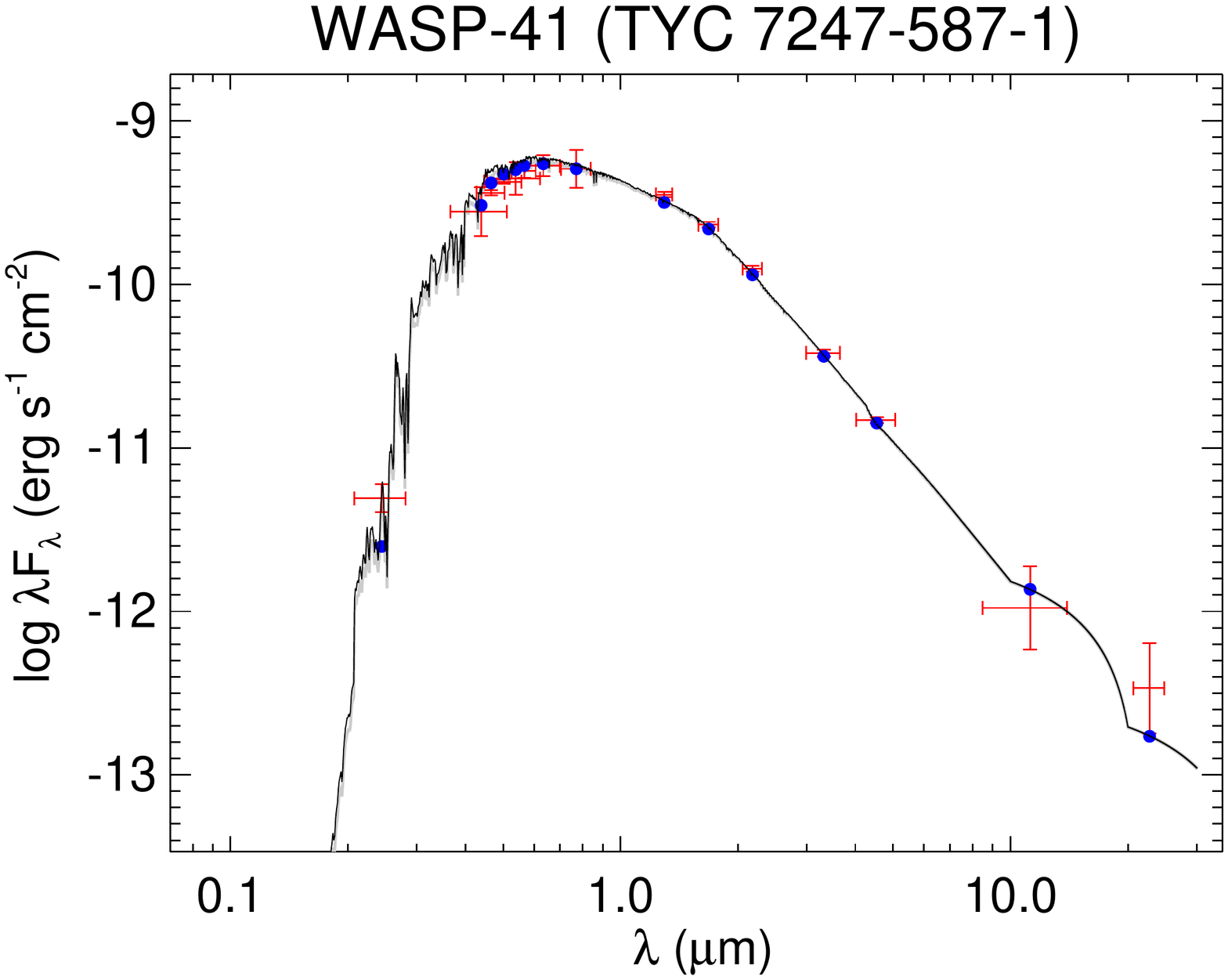}
  \includegraphics[trim=60 60 60 60,clip,width=0.49\linewidth]{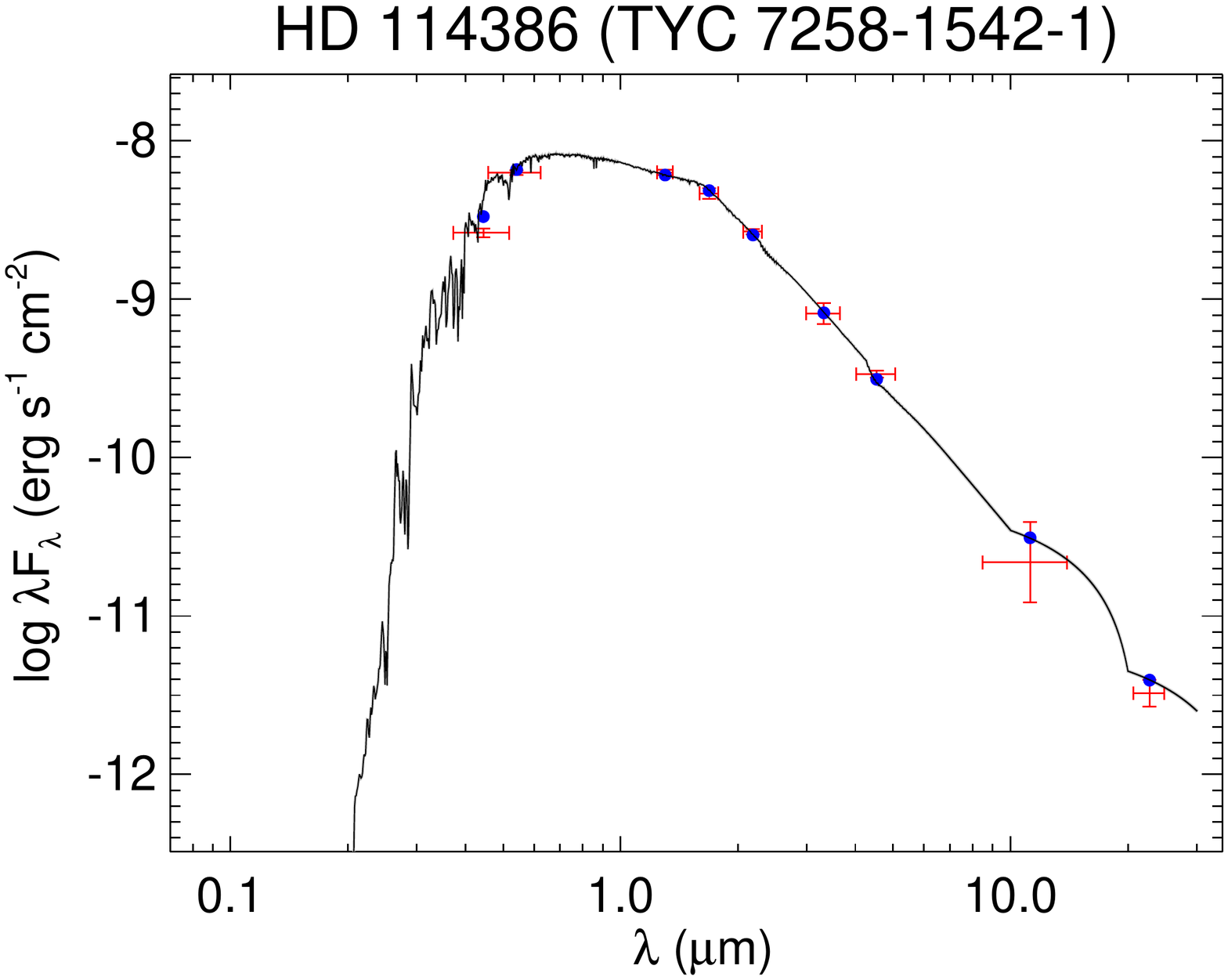}
  \includegraphics[trim=60 60 60 60,clip,width=0.49\linewidth]{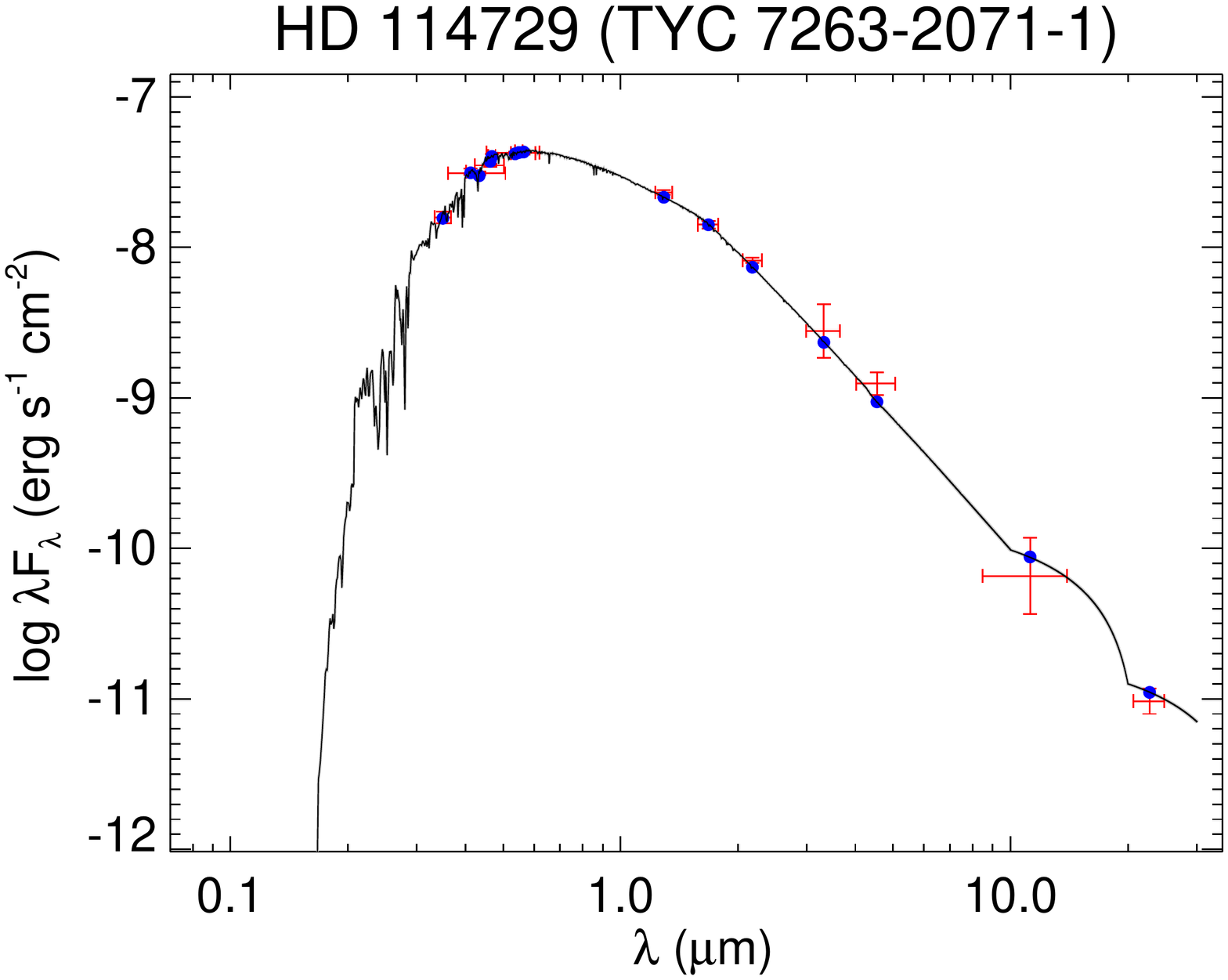}
  \caption{All labels, lines, symbols, and colors as in Figure \ref{fig:seds}.}
  \label{fig:seds_66}
\end{figure}

\begin{figure}[H]
  \centering
  \includegraphics[trim=60 60 60 60,clip,width=0.49\linewidth]{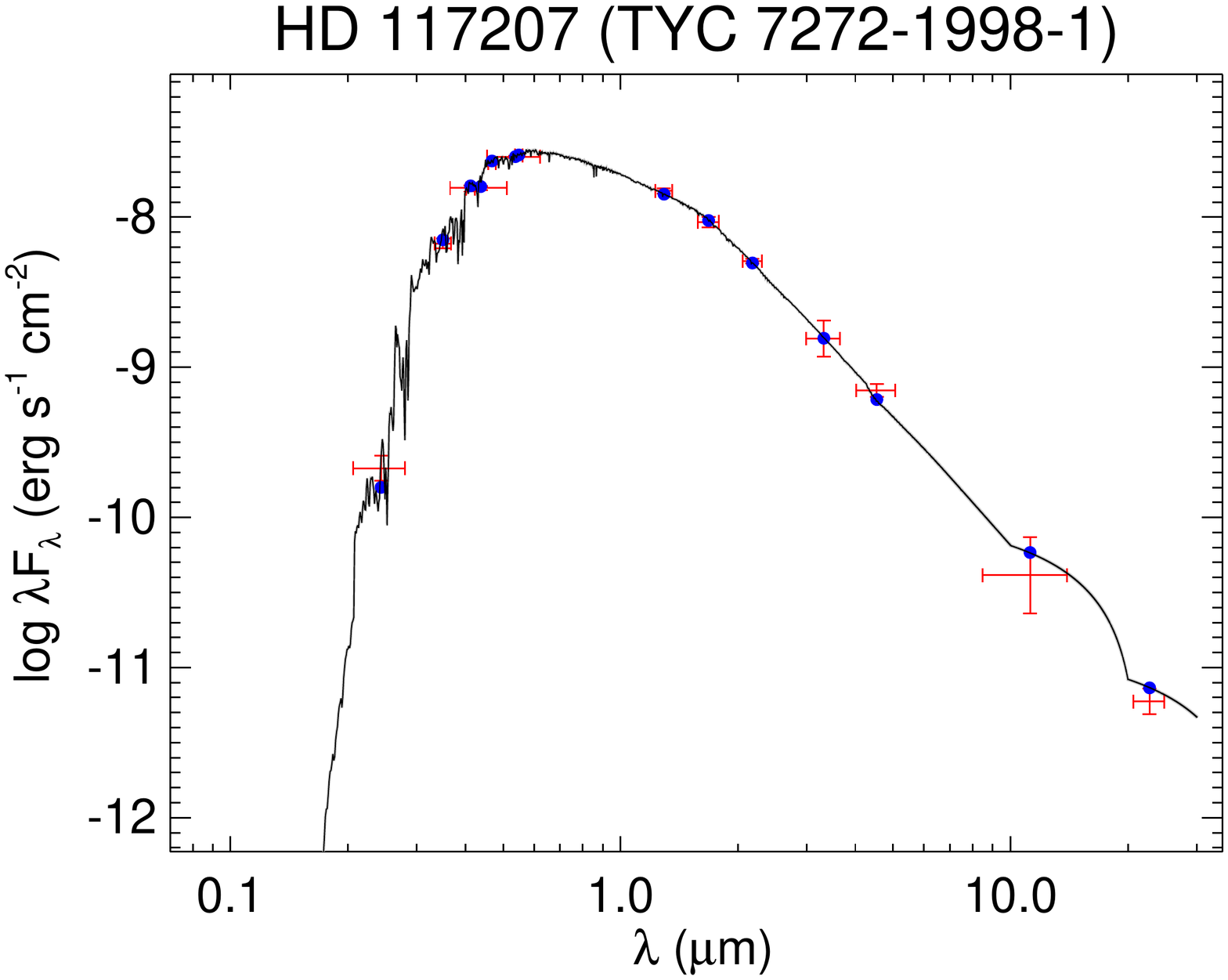}
  \includegraphics[trim=60 60 60 60,clip,width=0.49\linewidth]{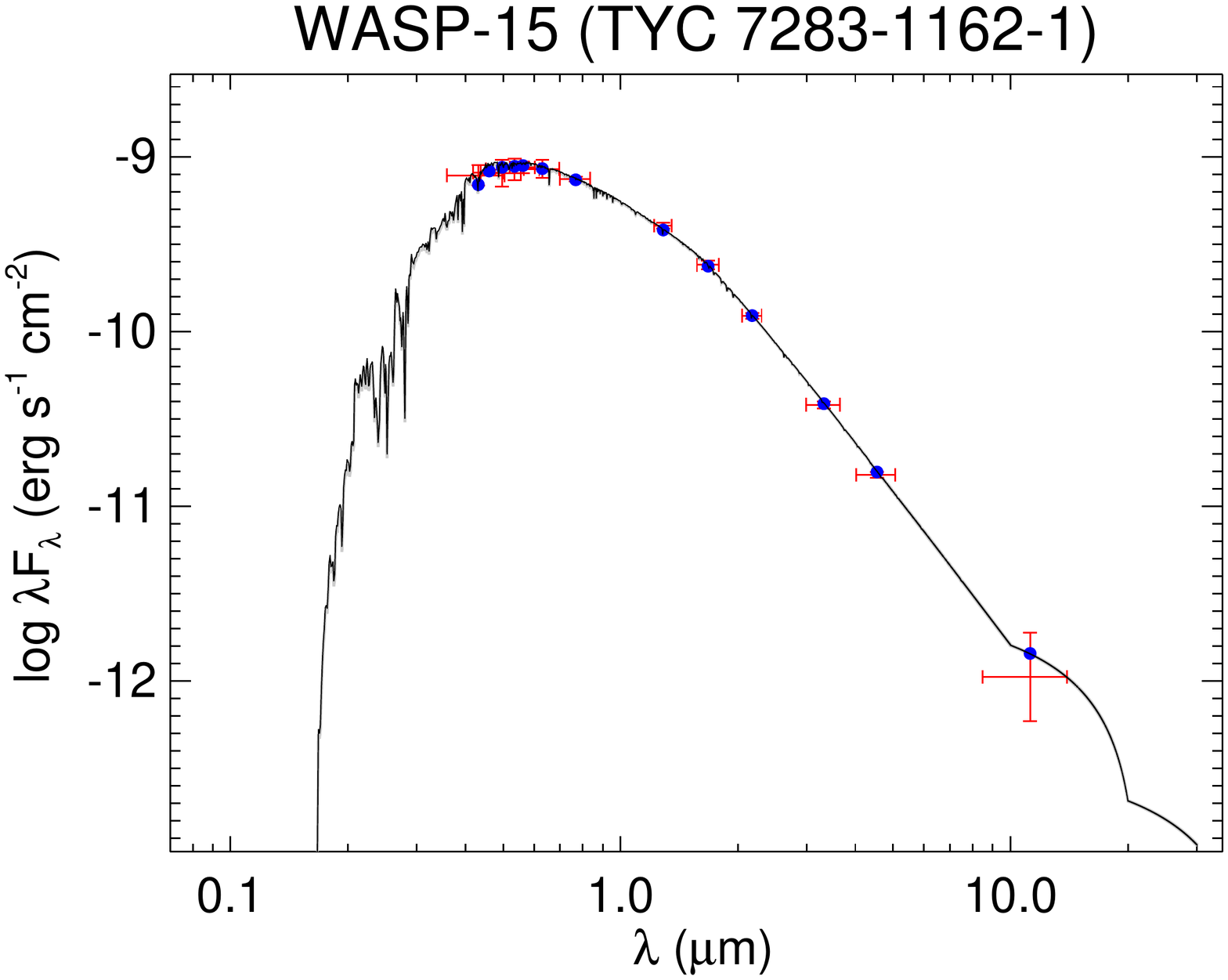}
  \includegraphics[trim=60 60 60 60,clip,width=0.49\linewidth]{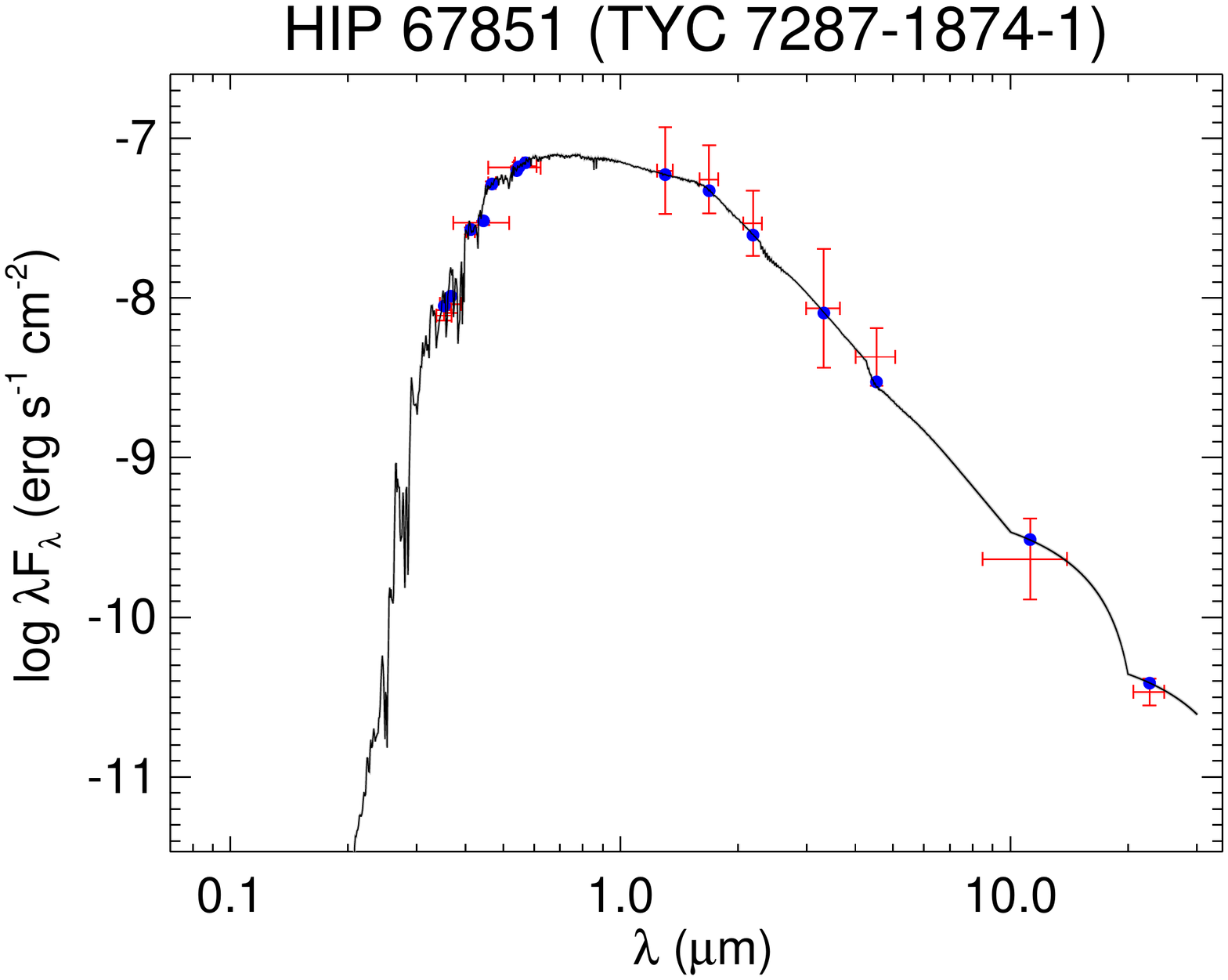}
  \includegraphics[trim=60 60 60 60,clip,width=0.49\linewidth]{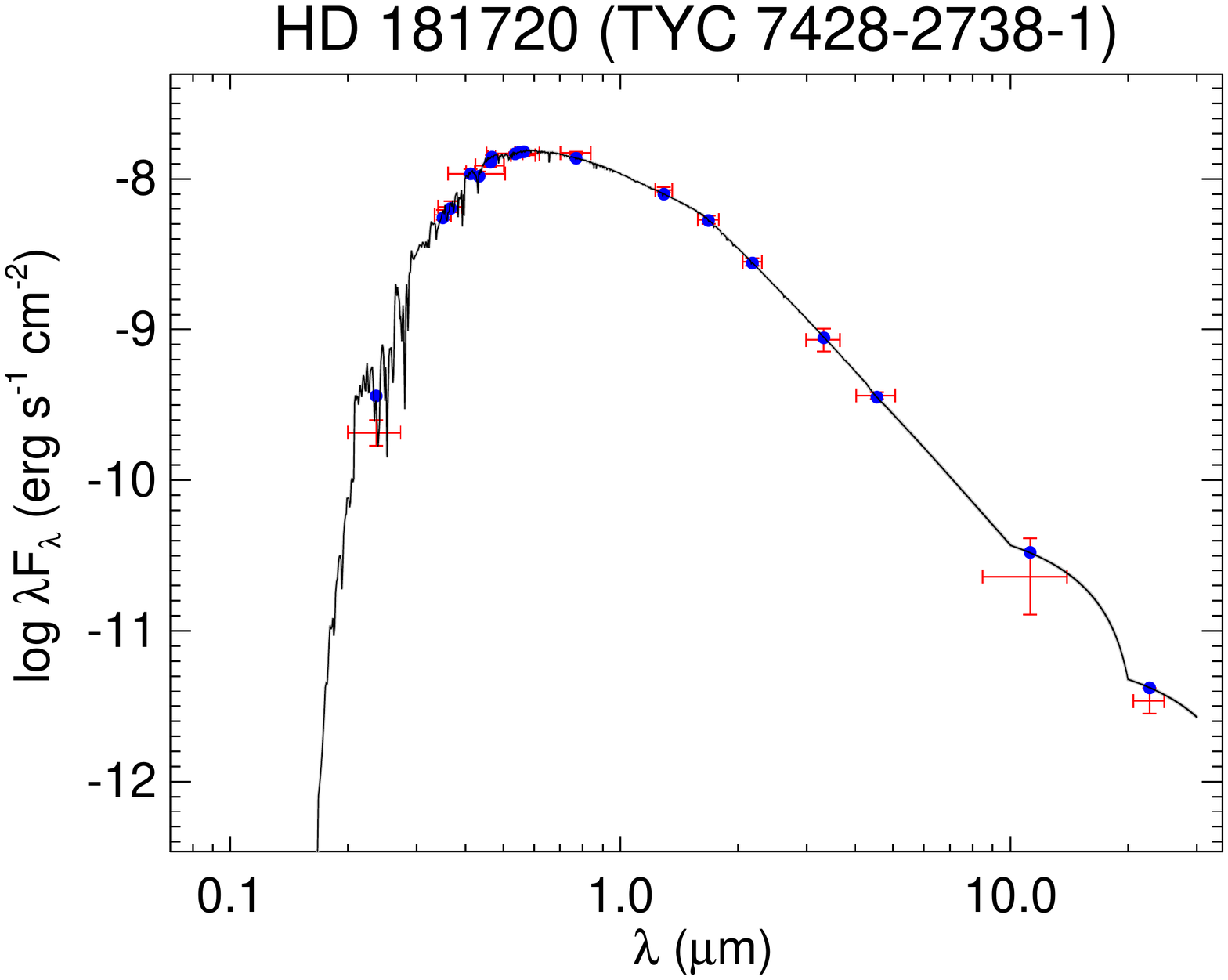}
  \includegraphics[trim=60 60 60 60,clip,width=0.49\linewidth]{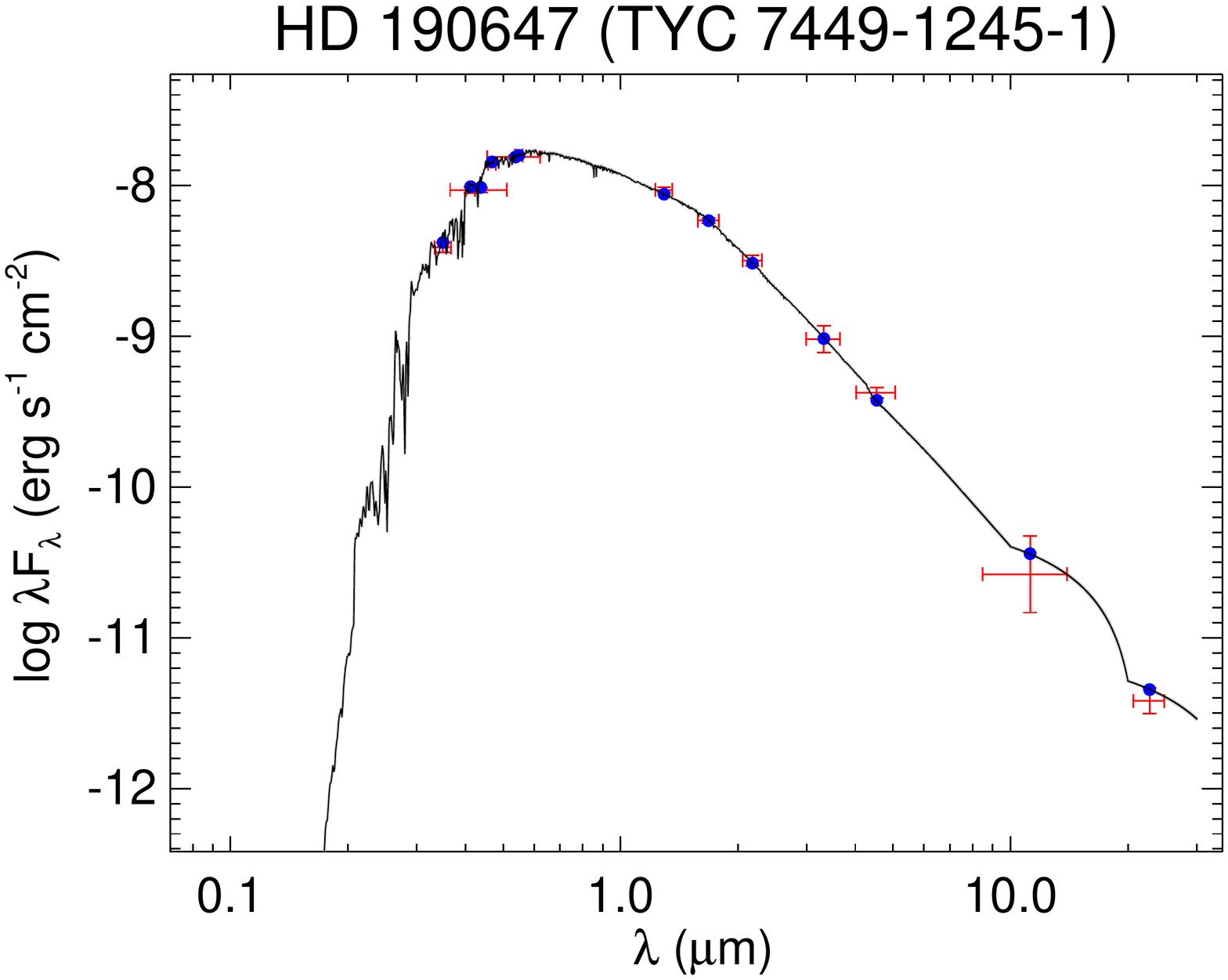}
  \includegraphics[trim=60 60 60 60,clip,width=0.49\linewidth]{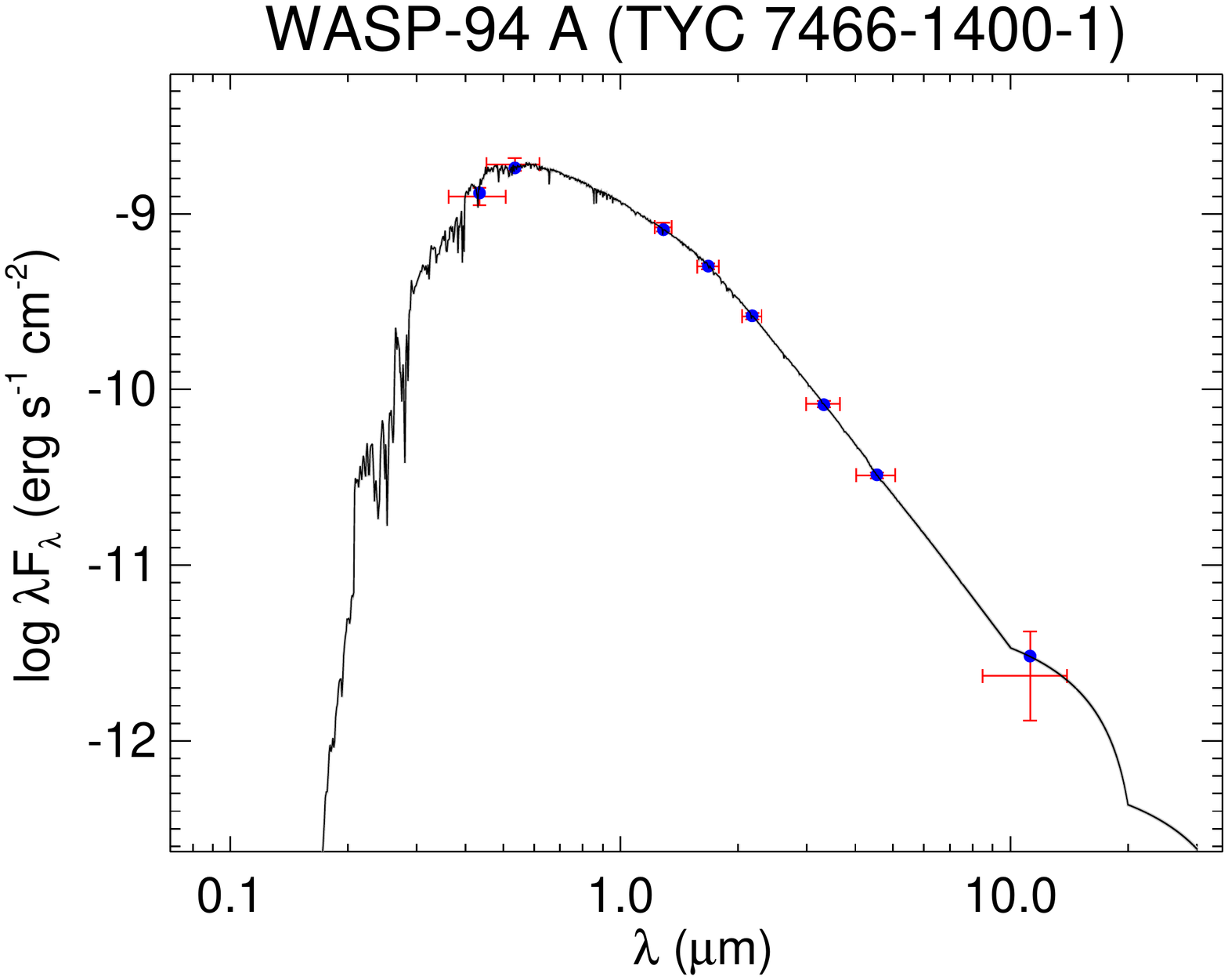}
  \caption{All labels, lines, symbols, and colors as in Figure \ref{fig:seds}.}
  \label{fig:seds_67}
\end{figure}

\begin{figure}[H]
  \centering
  \includegraphics[trim=60 60 60 60,clip,width=0.49\linewidth]{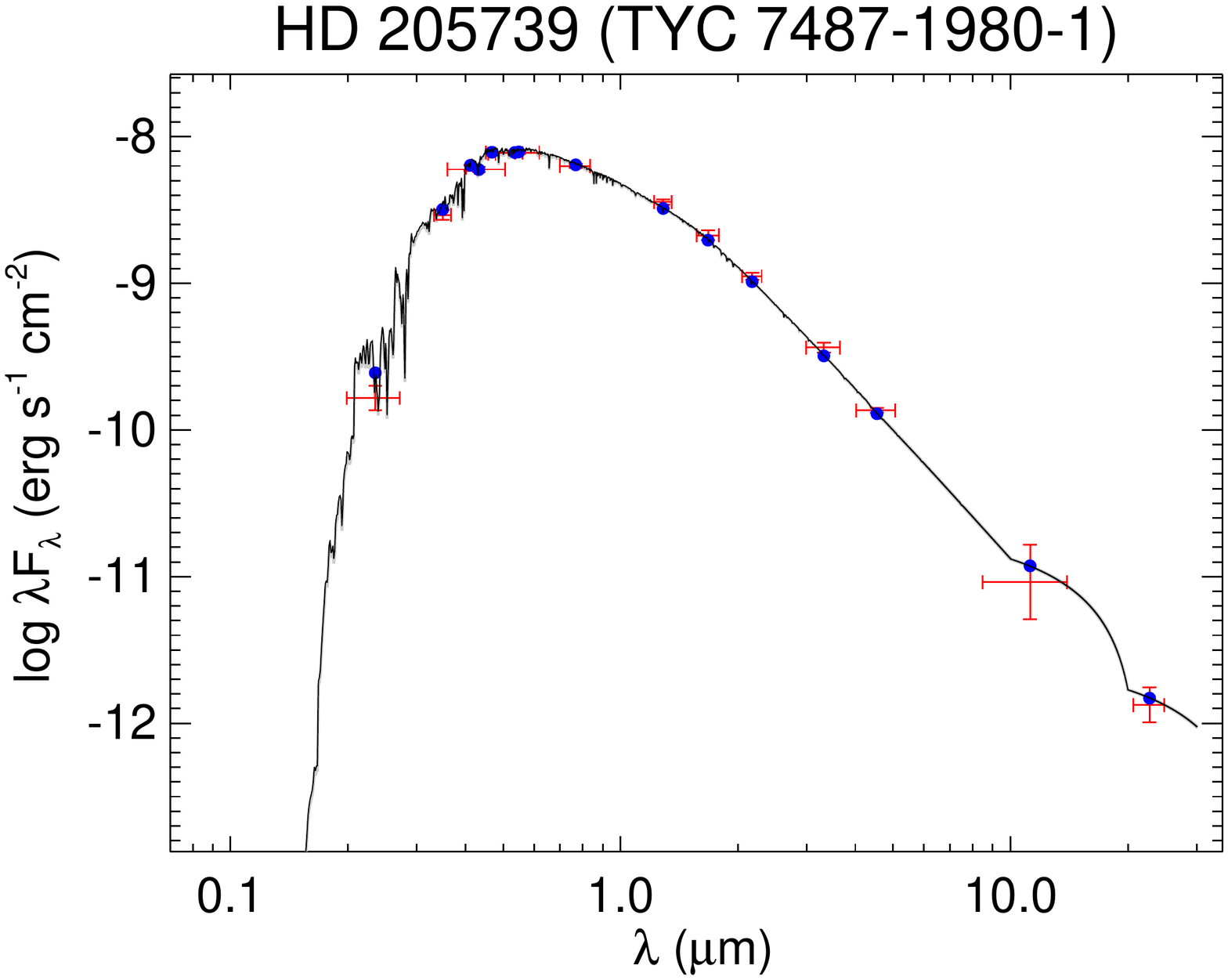}
  \includegraphics[trim=60 60 60 60,clip,width=0.49\linewidth]{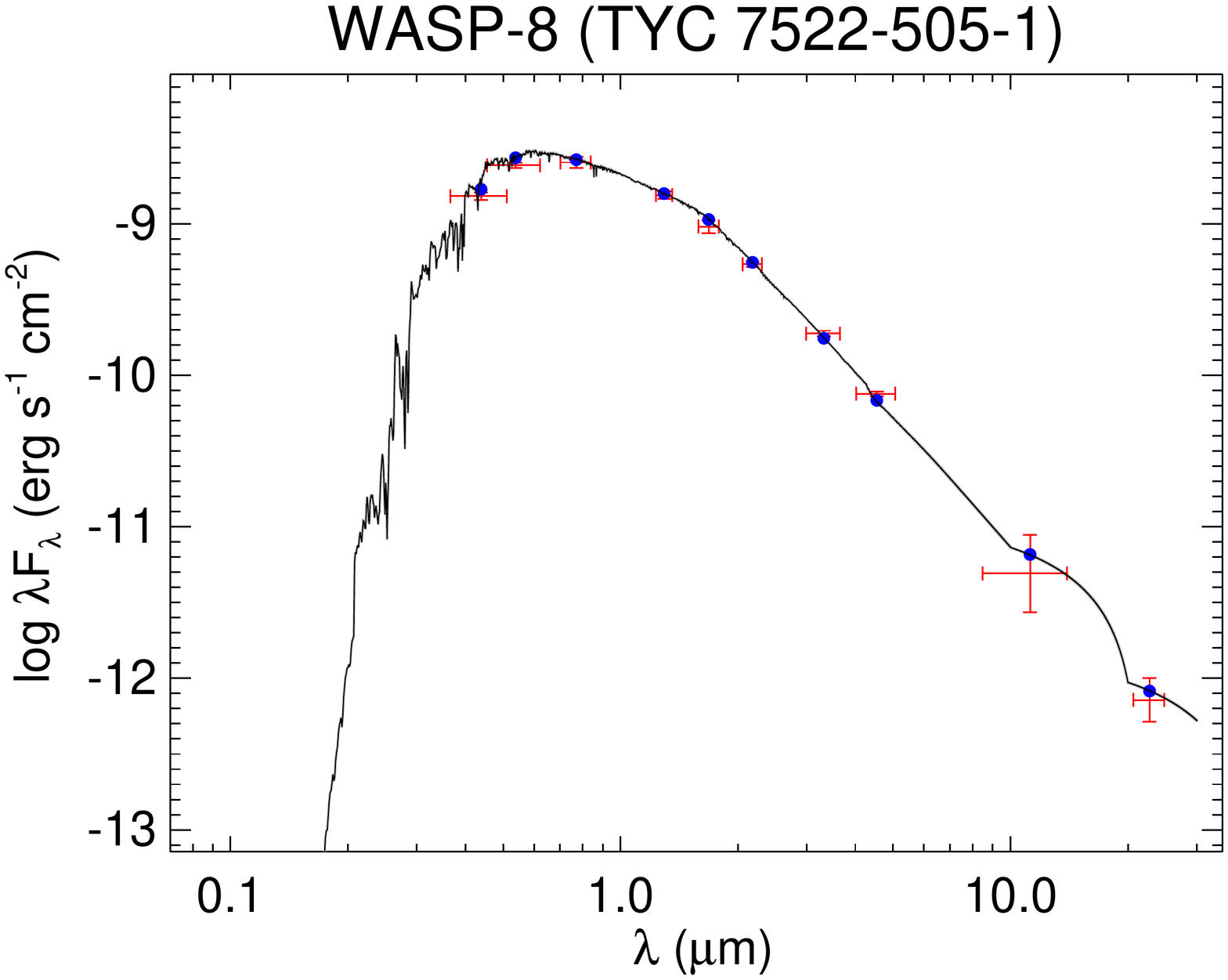}
  \includegraphics[trim=60 60 60 60,clip,width=0.49\linewidth]{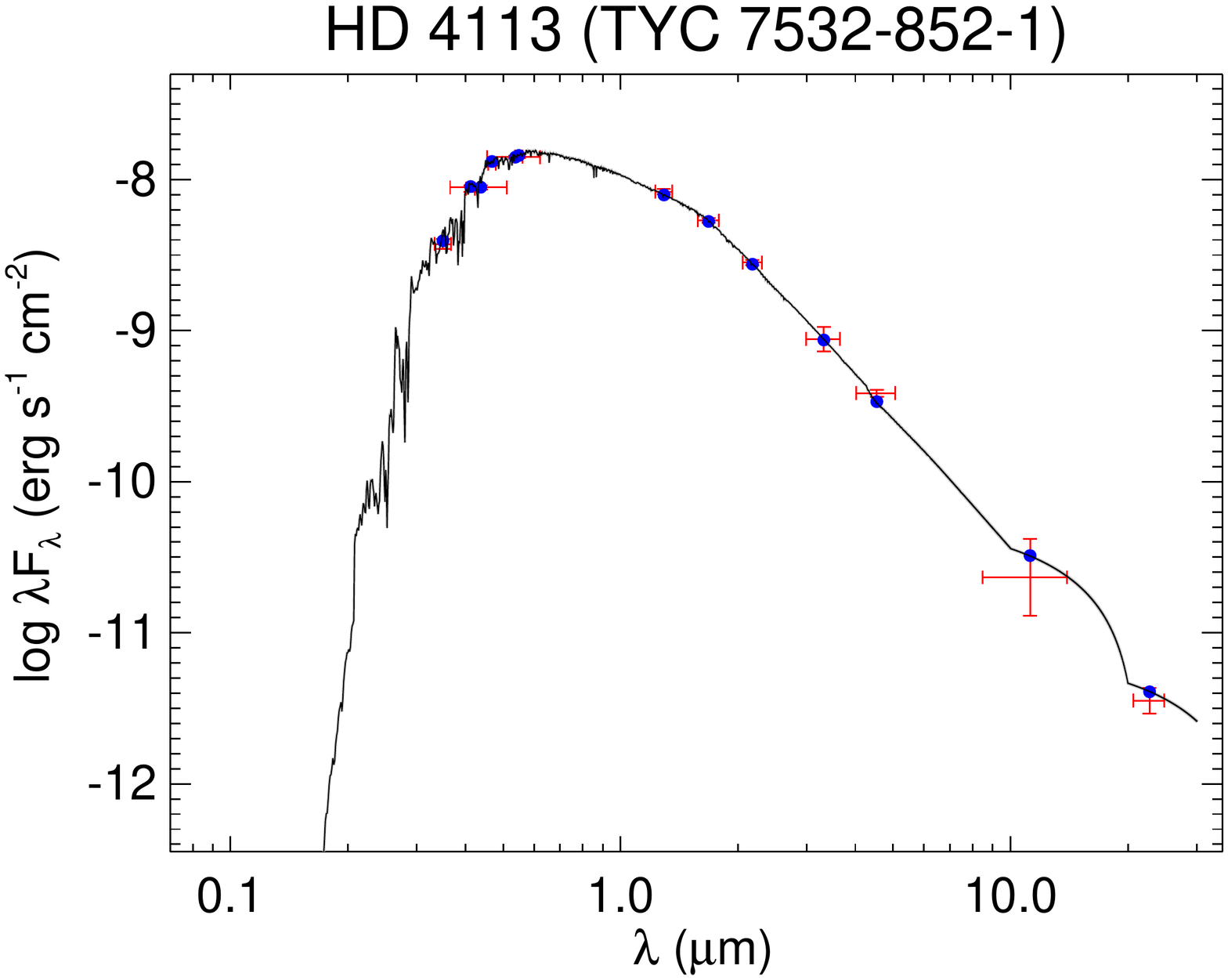}
  \includegraphics[trim=60 60 60 60,clip,width=0.49\linewidth]{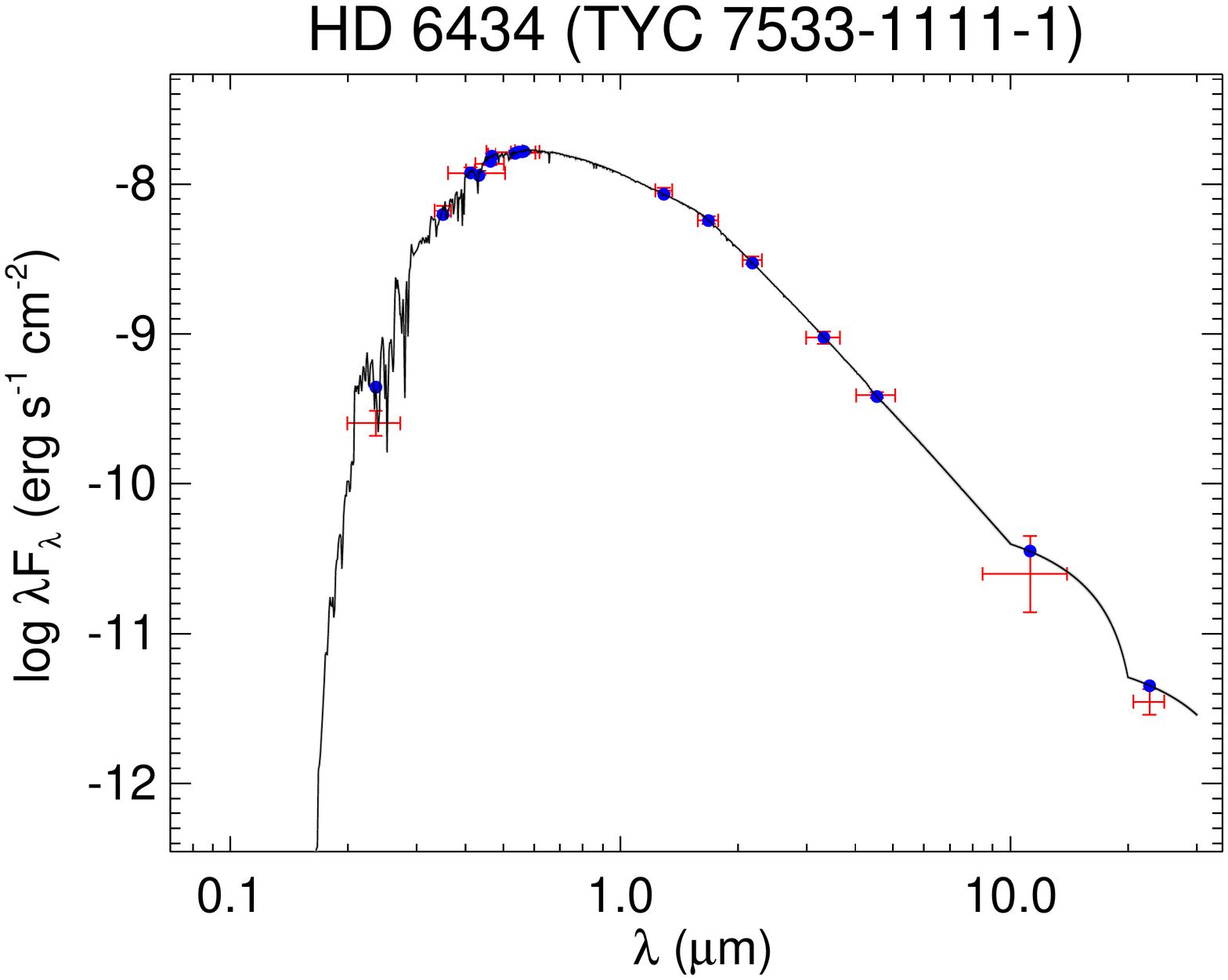}
  \includegraphics[trim=60 60 60 60,clip,width=0.49\linewidth]{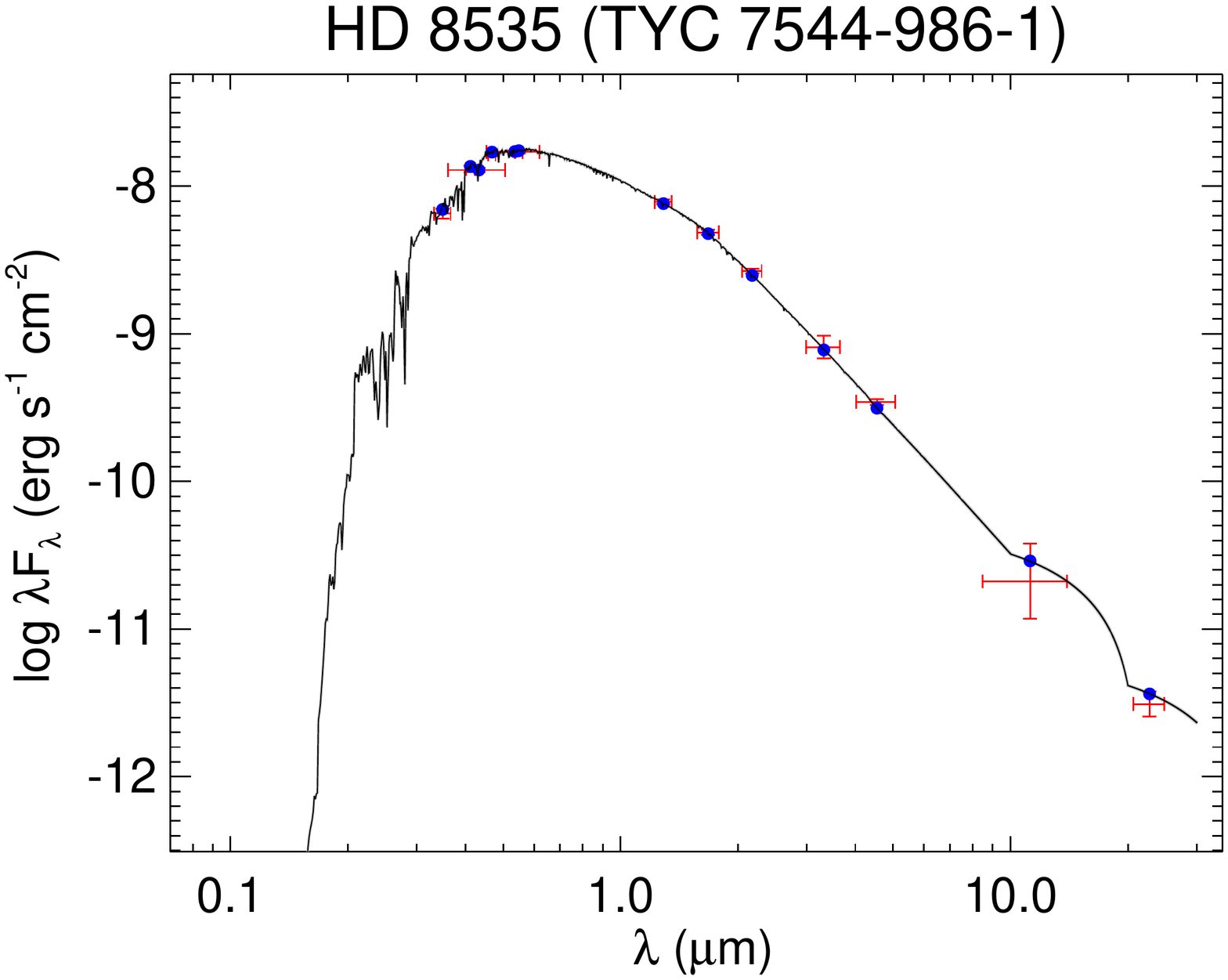}
  \includegraphics[trim=60 60 60 60,clip,width=0.49\linewidth]{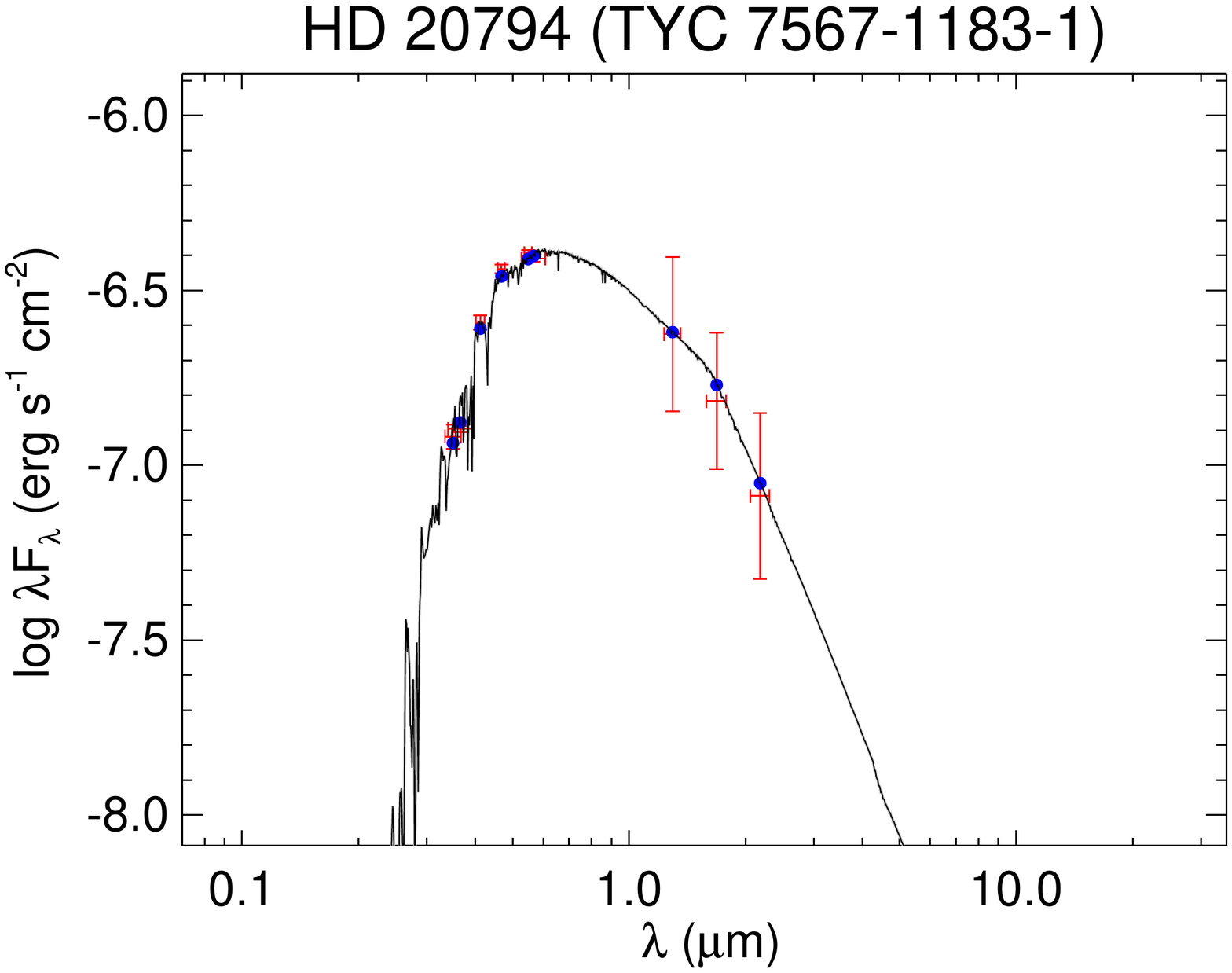}
  \caption{All labels, lines, symbols, and colors as in Figure \ref{fig:seds}.}
  \label{fig:seds_68}
\end{figure}

\begin{figure}[H]
  \centering
  \includegraphics[trim=60 60 60 60,clip,width=0.49\linewidth]{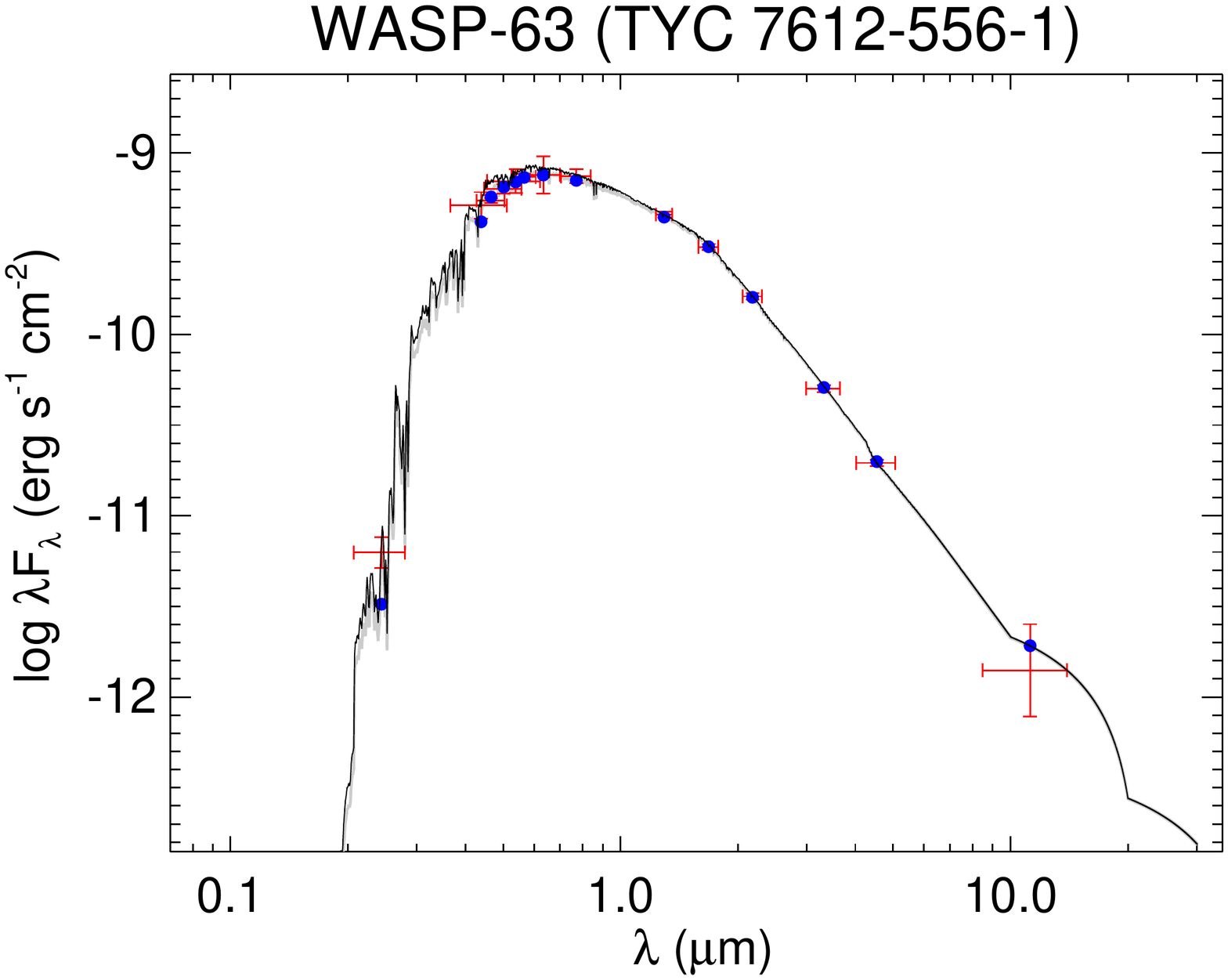}
  \includegraphics[trim=60 60 60 60,clip,width=0.49\linewidth]{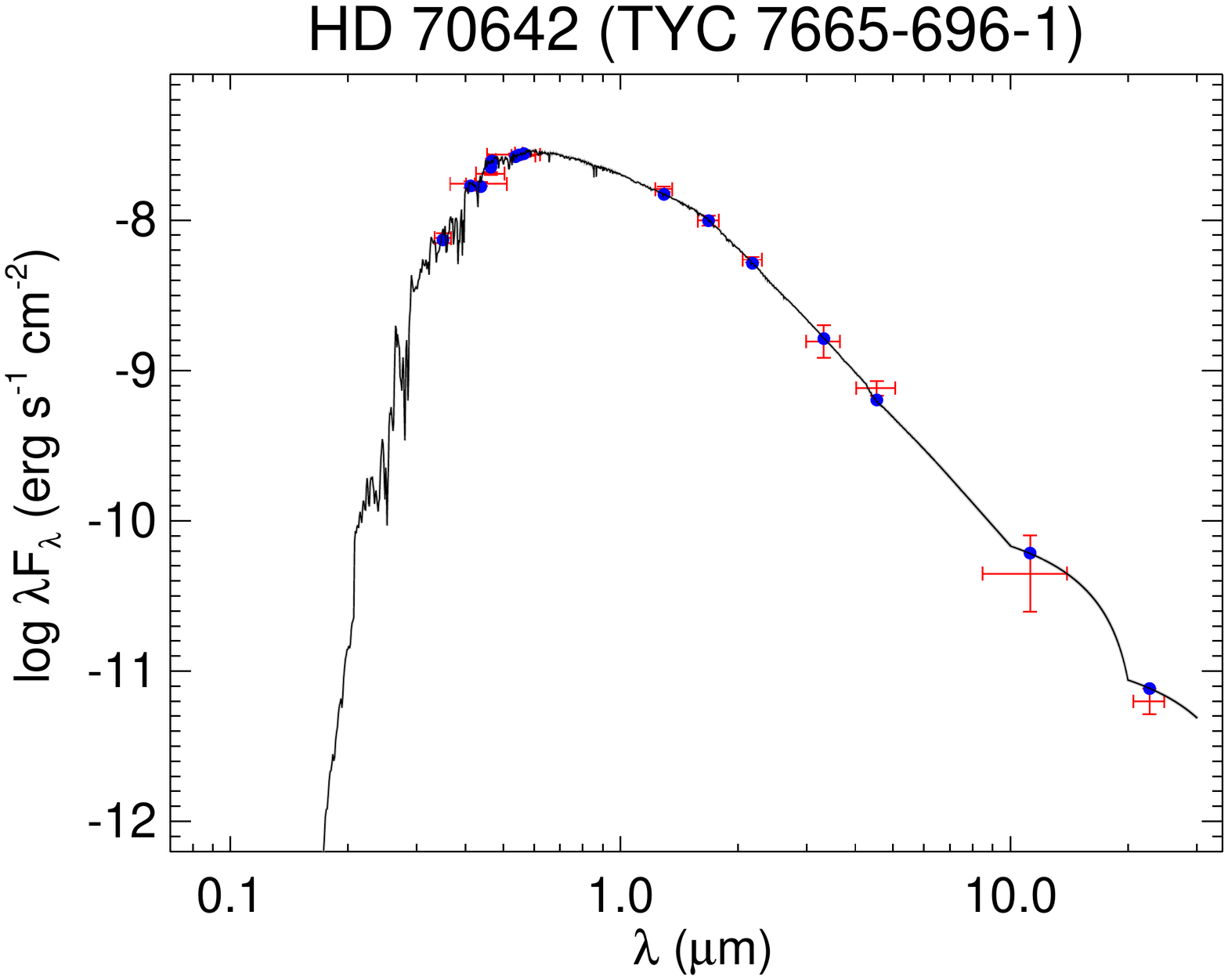}
  \includegraphics[trim=60 60 60 60,clip,width=0.49\linewidth]{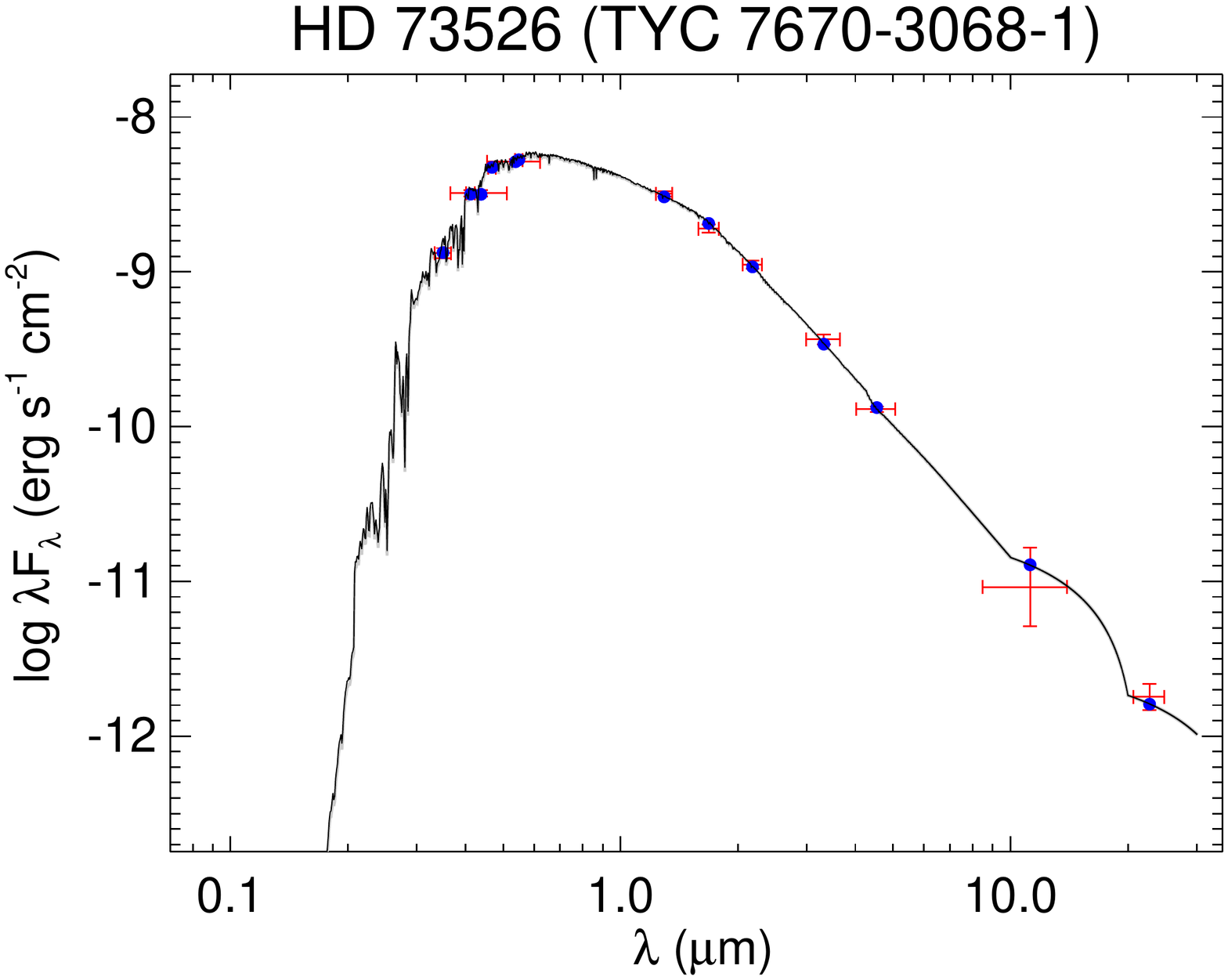}
  \includegraphics[trim=60 60 60 60,clip,width=0.49\linewidth]{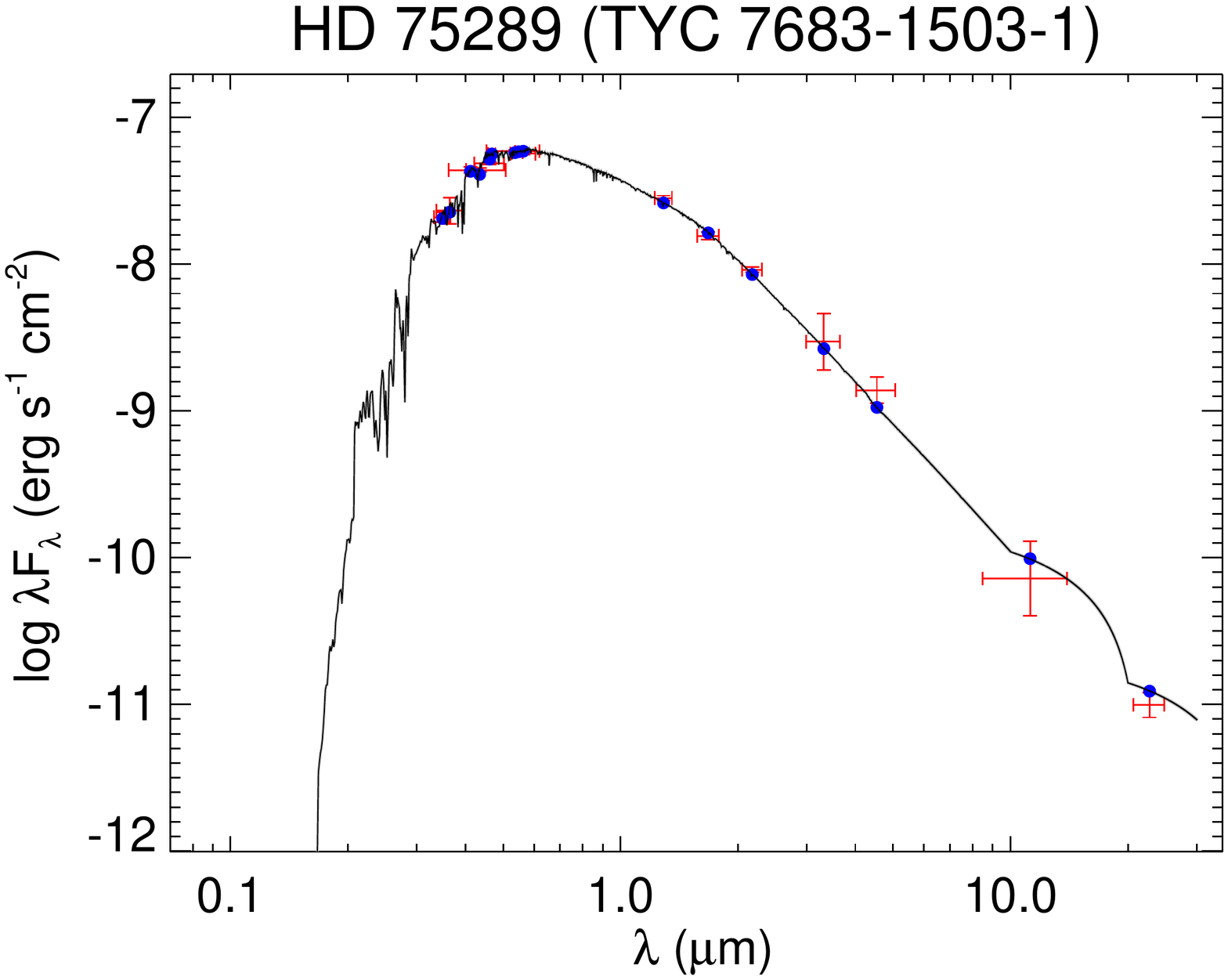}
  \includegraphics[trim=60 60 60 60,clip,width=0.49\linewidth]{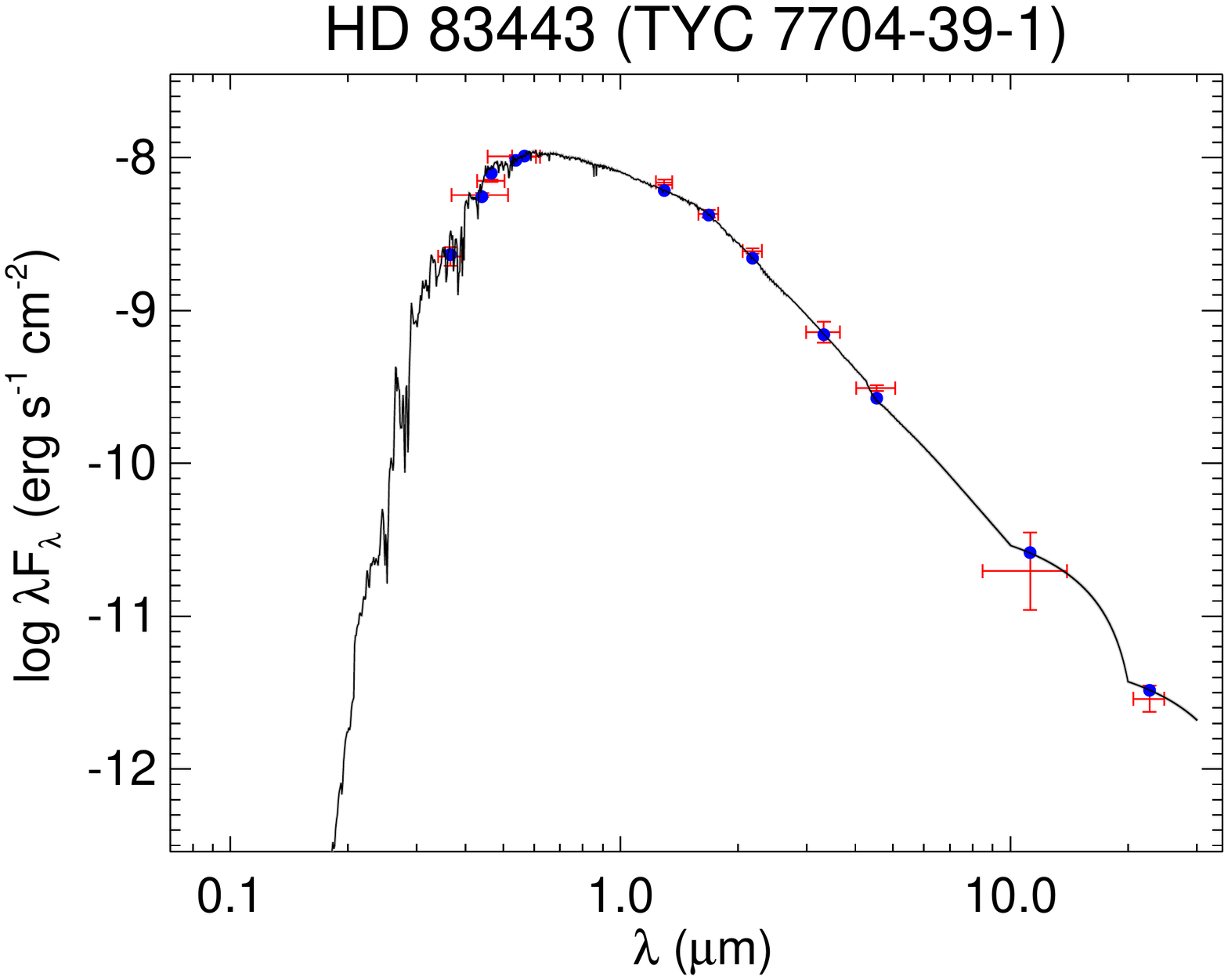}
  \includegraphics[trim=60 60 60 60,clip,width=0.49\linewidth]{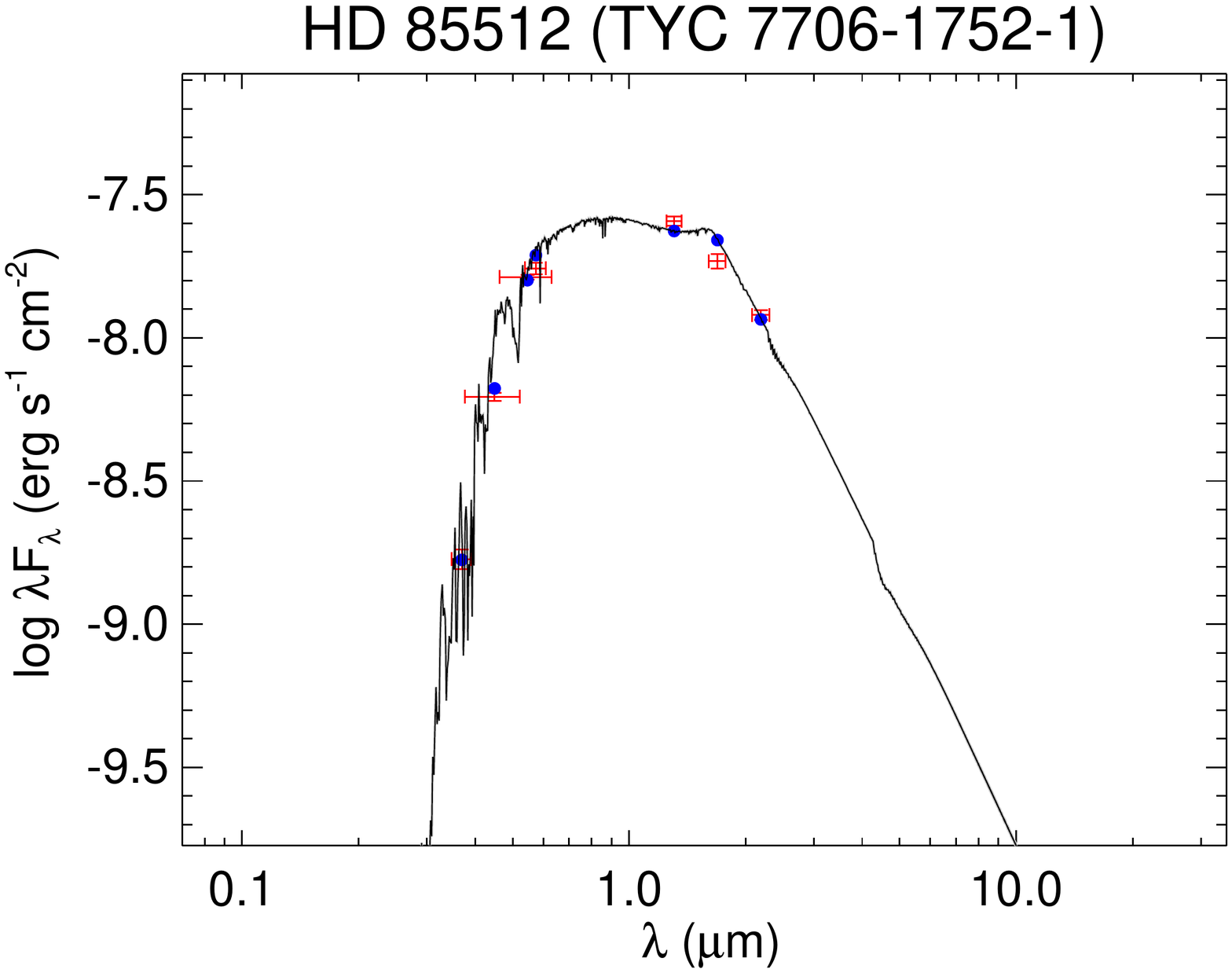}
  \caption{All labels, lines, symbols, and colors as in Figure \ref{fig:seds}.}
  \label{fig:seds_69}
\end{figure}

\begin{figure}[H]
  \centering
  \includegraphics[trim=60 60 60 60,clip,width=0.49\linewidth]{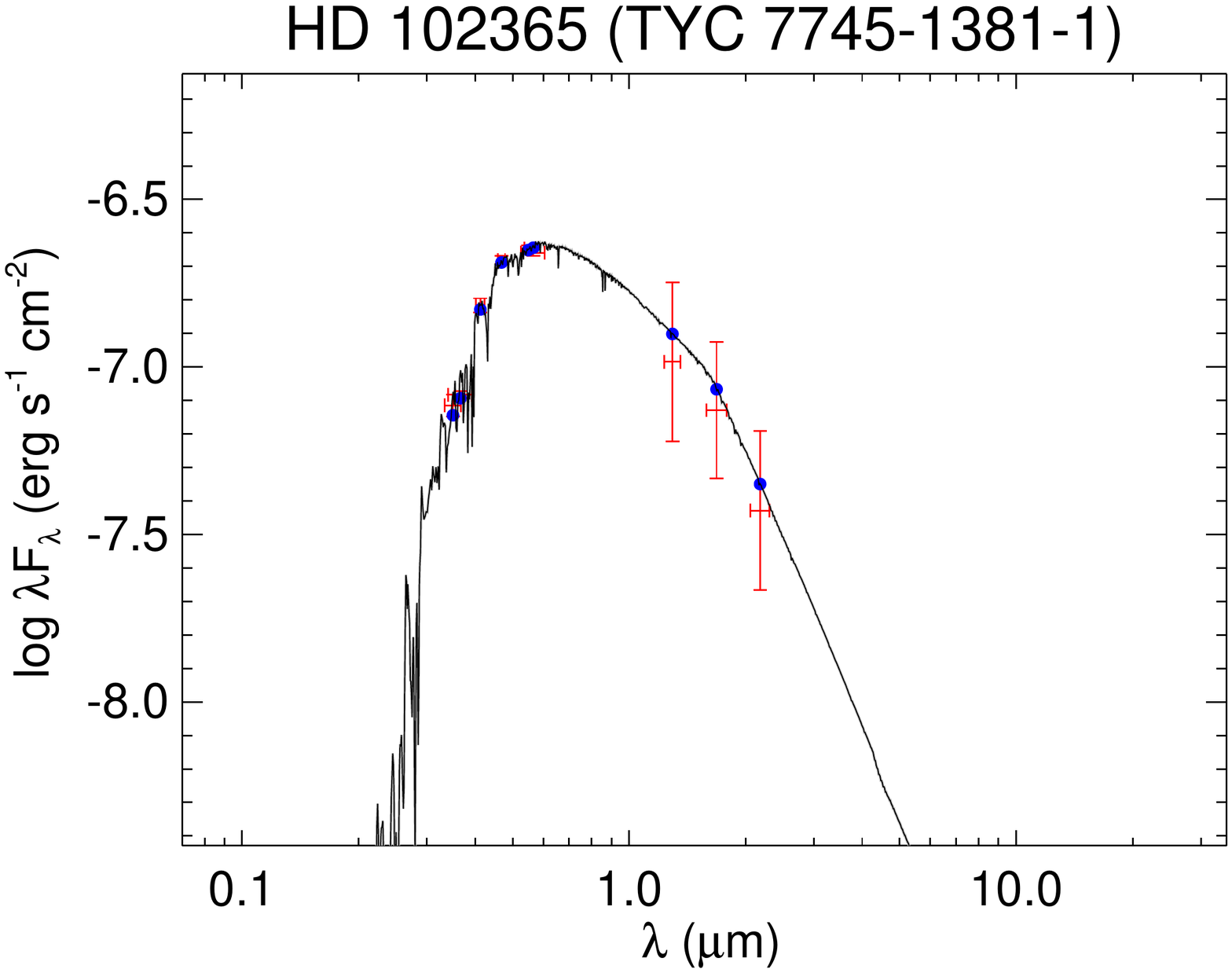}
  \includegraphics[trim=60 60 60 60,clip,width=0.49\linewidth]{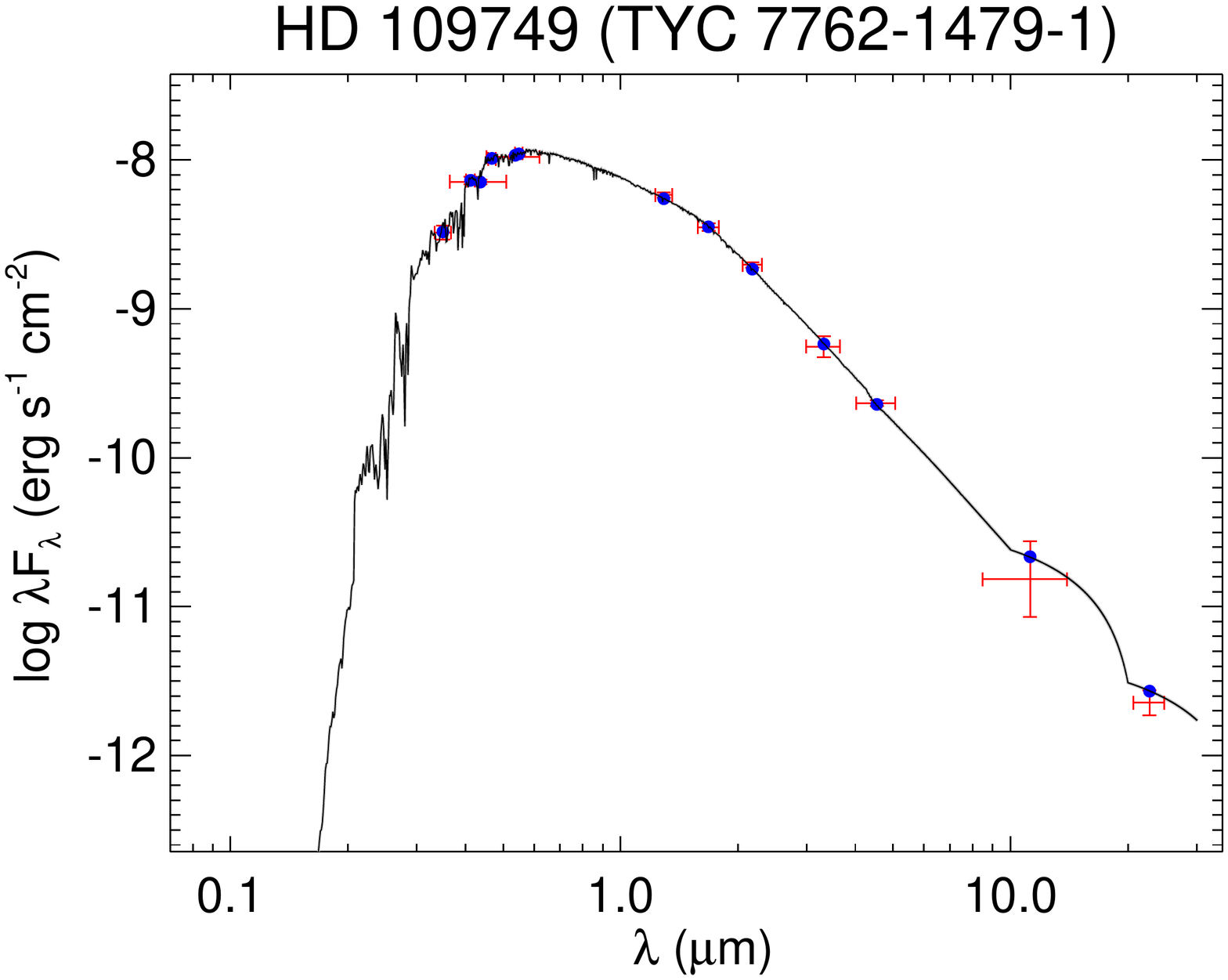}
  \includegraphics[trim=60 60 60 60,clip,width=0.49\linewidth]{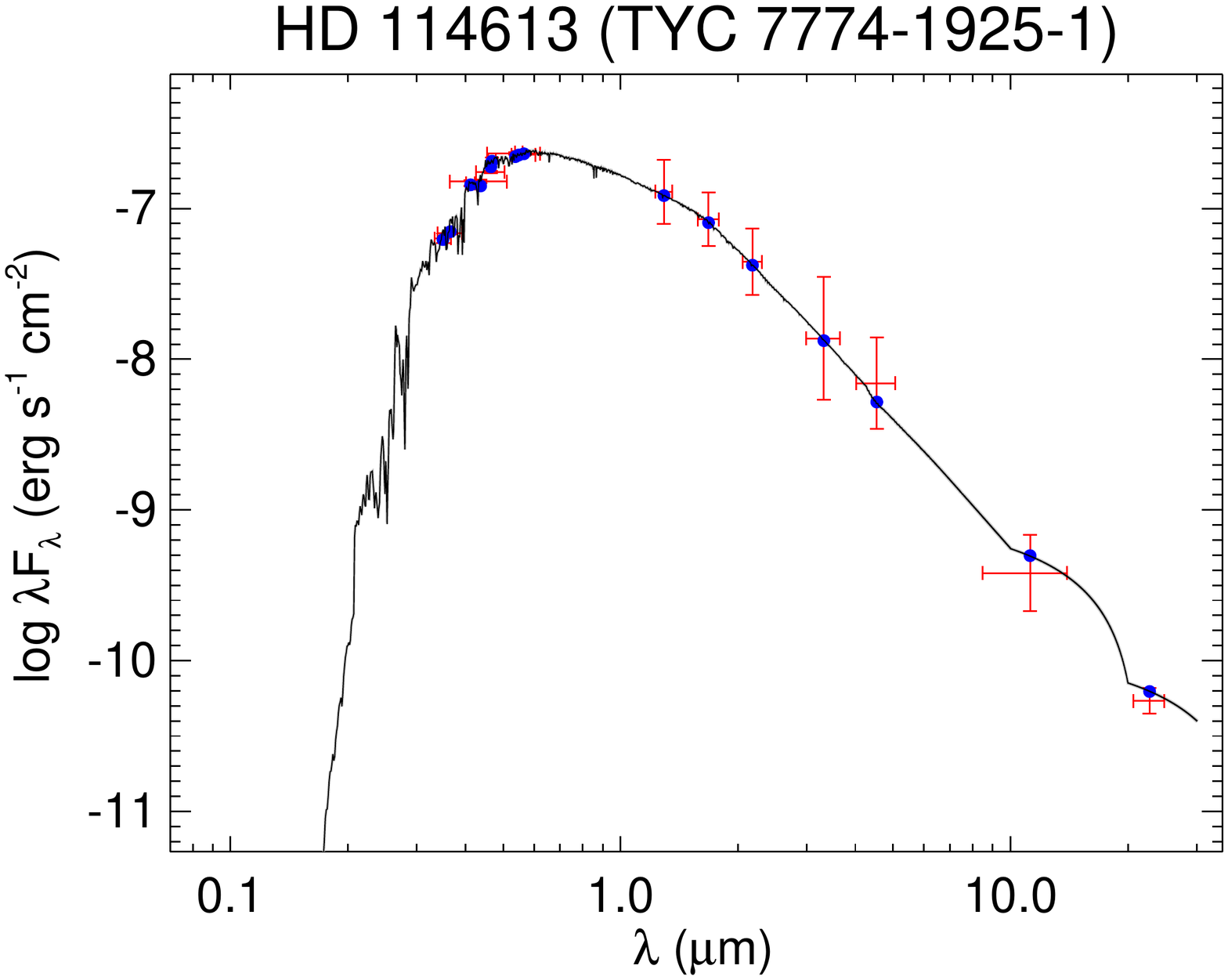}
  \includegraphics[trim=60 60 60 60,clip,width=0.49\linewidth]{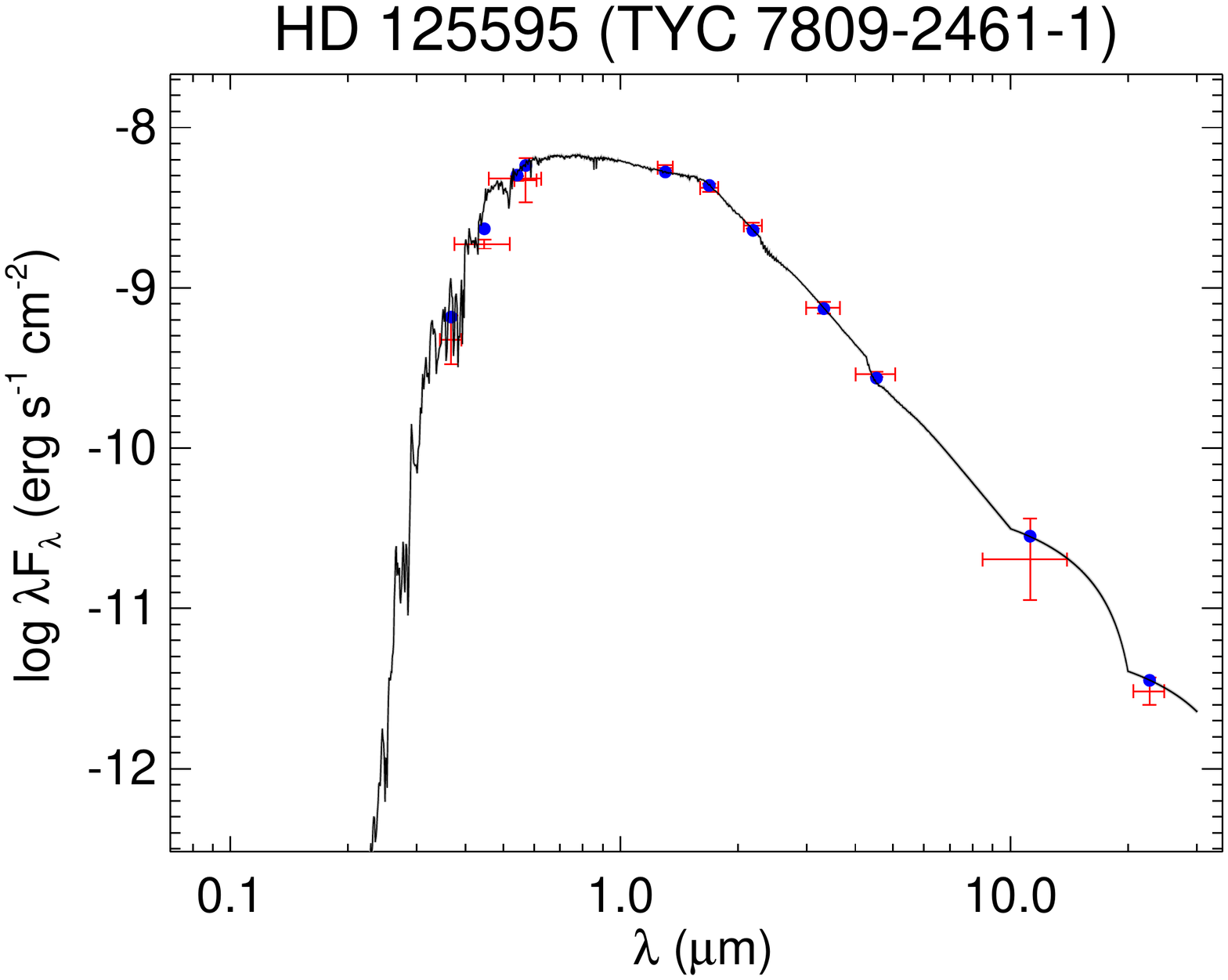}
  \includegraphics[trim=60 60 60 60,clip,width=0.49\linewidth]{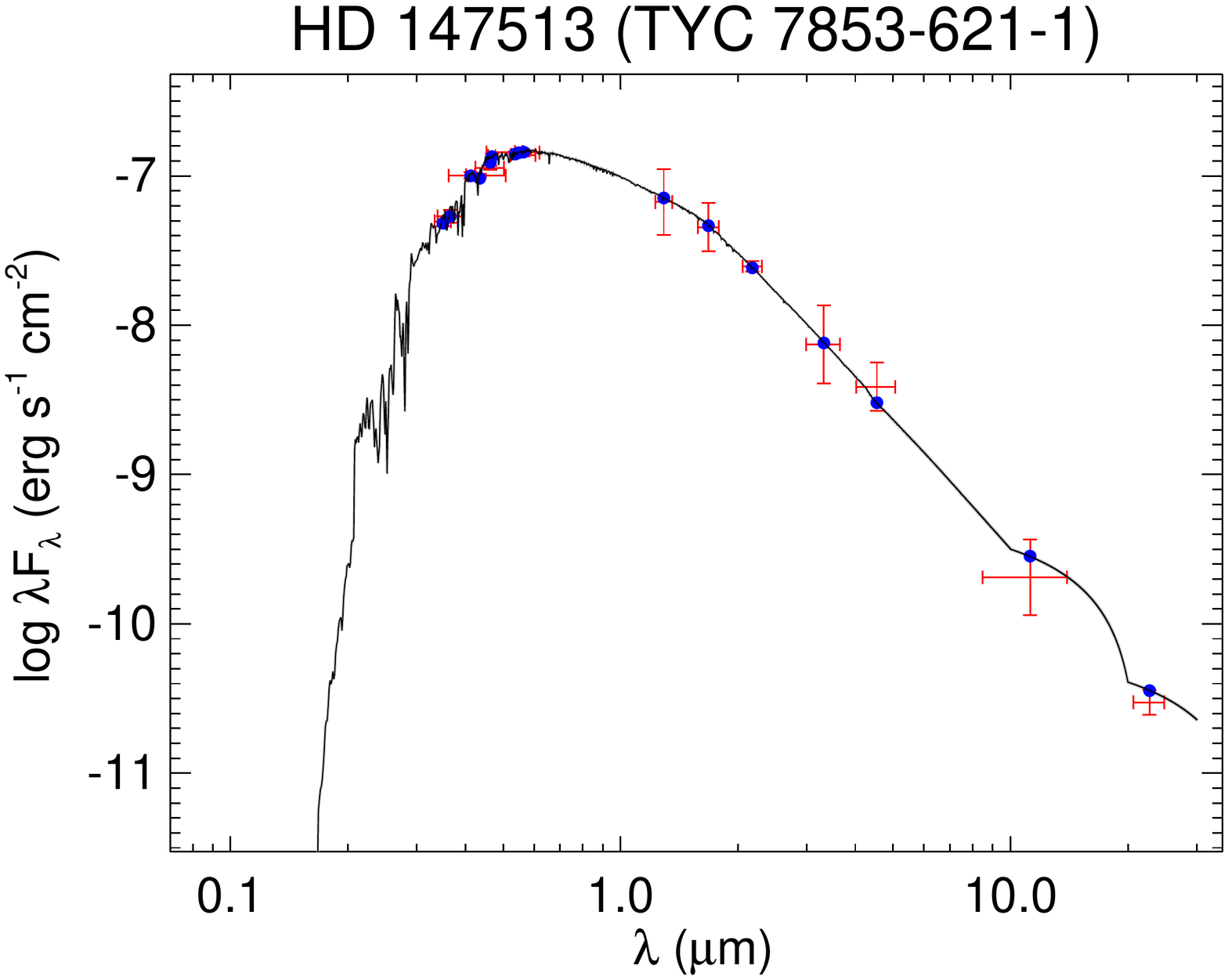}
  \includegraphics[trim=60 60 60 60,clip,width=0.49\linewidth]{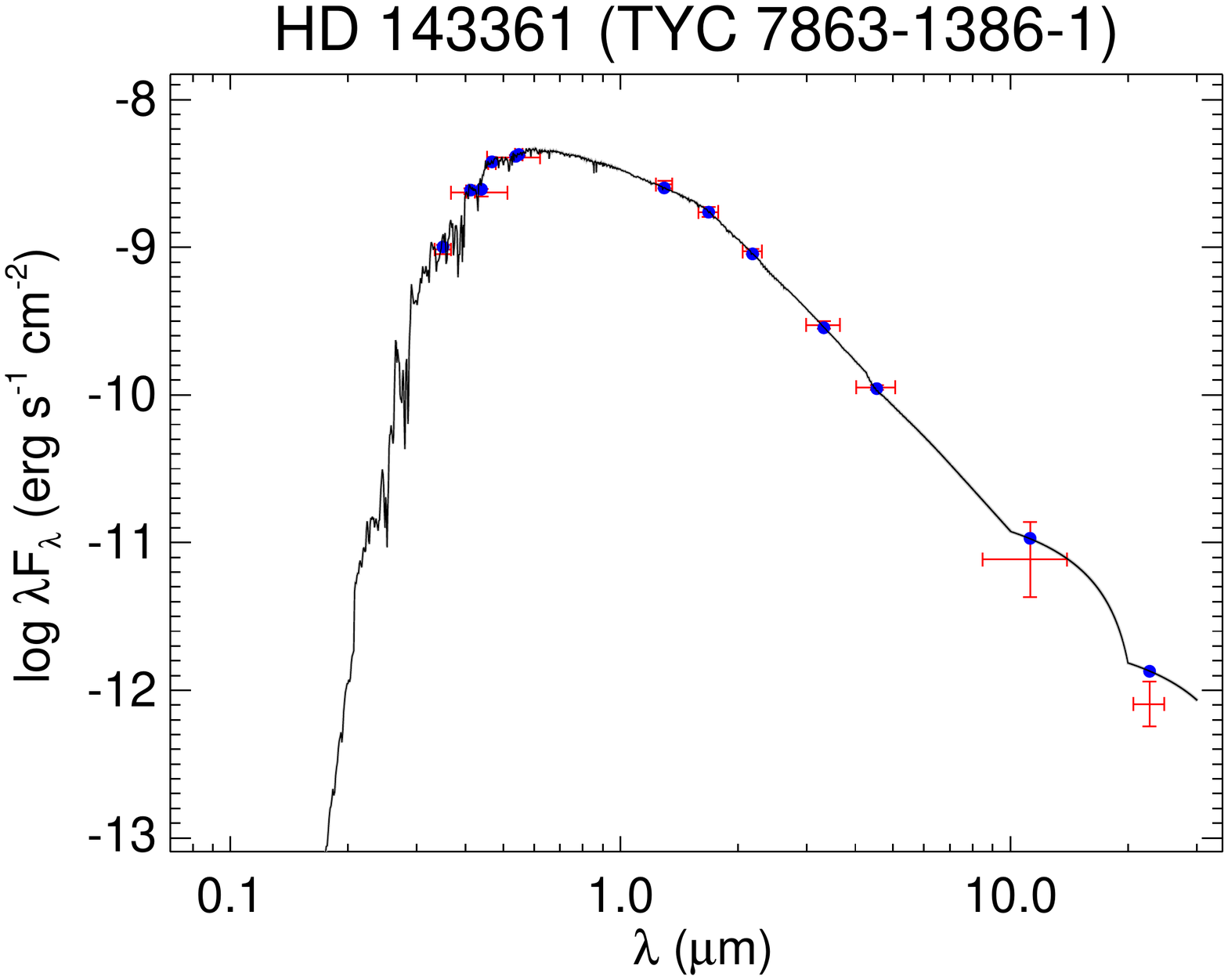}
  \caption{All labels, lines, symbols, and colors as in Figure \ref{fig:seds}.}
  \label{fig:seds_70}
\end{figure}

\begin{figure}[H]
  \centering
  \includegraphics[trim=60 60 60 60,clip,width=0.49\linewidth]{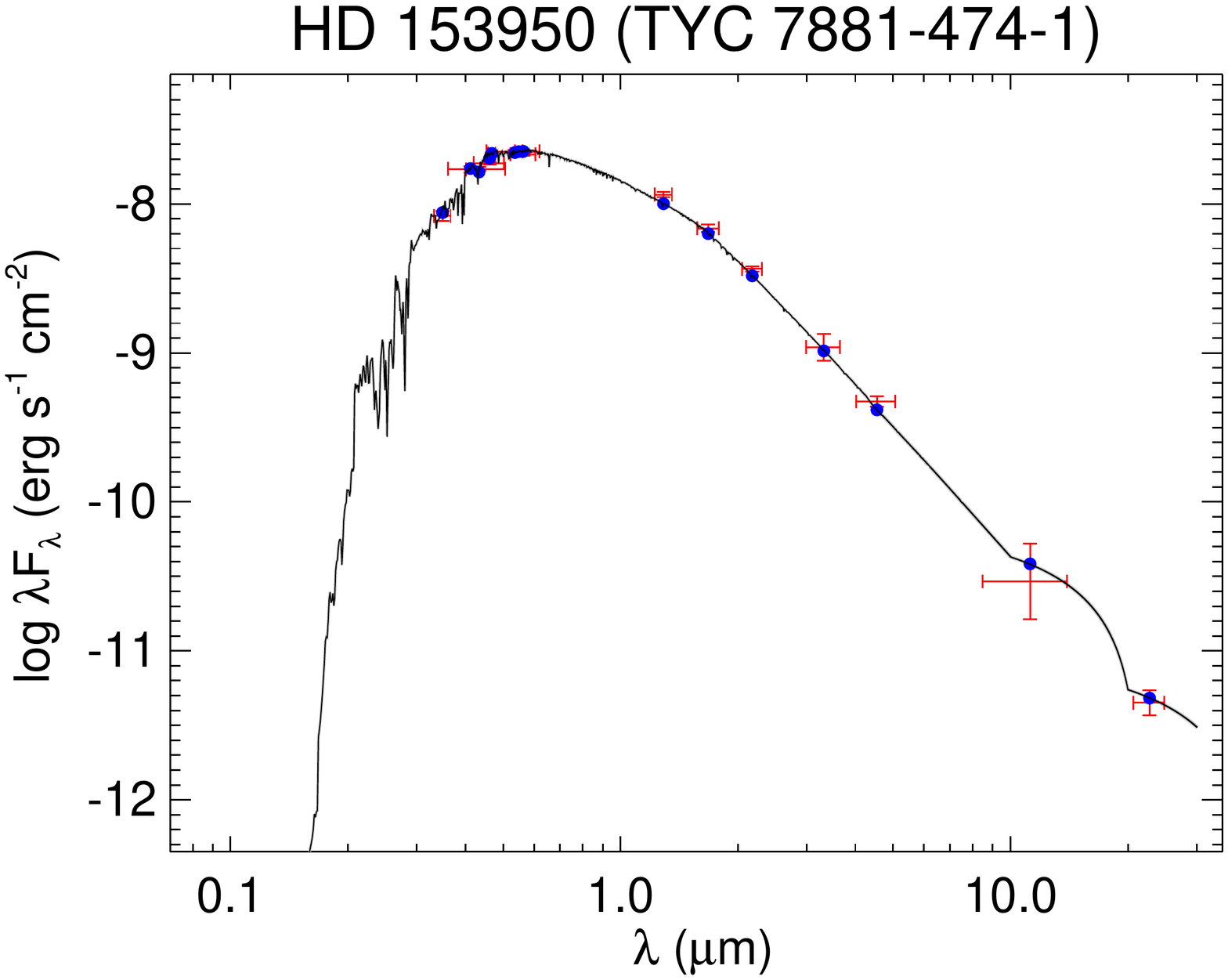}
  \includegraphics[trim=60 60 60 60,clip,width=0.49\linewidth]{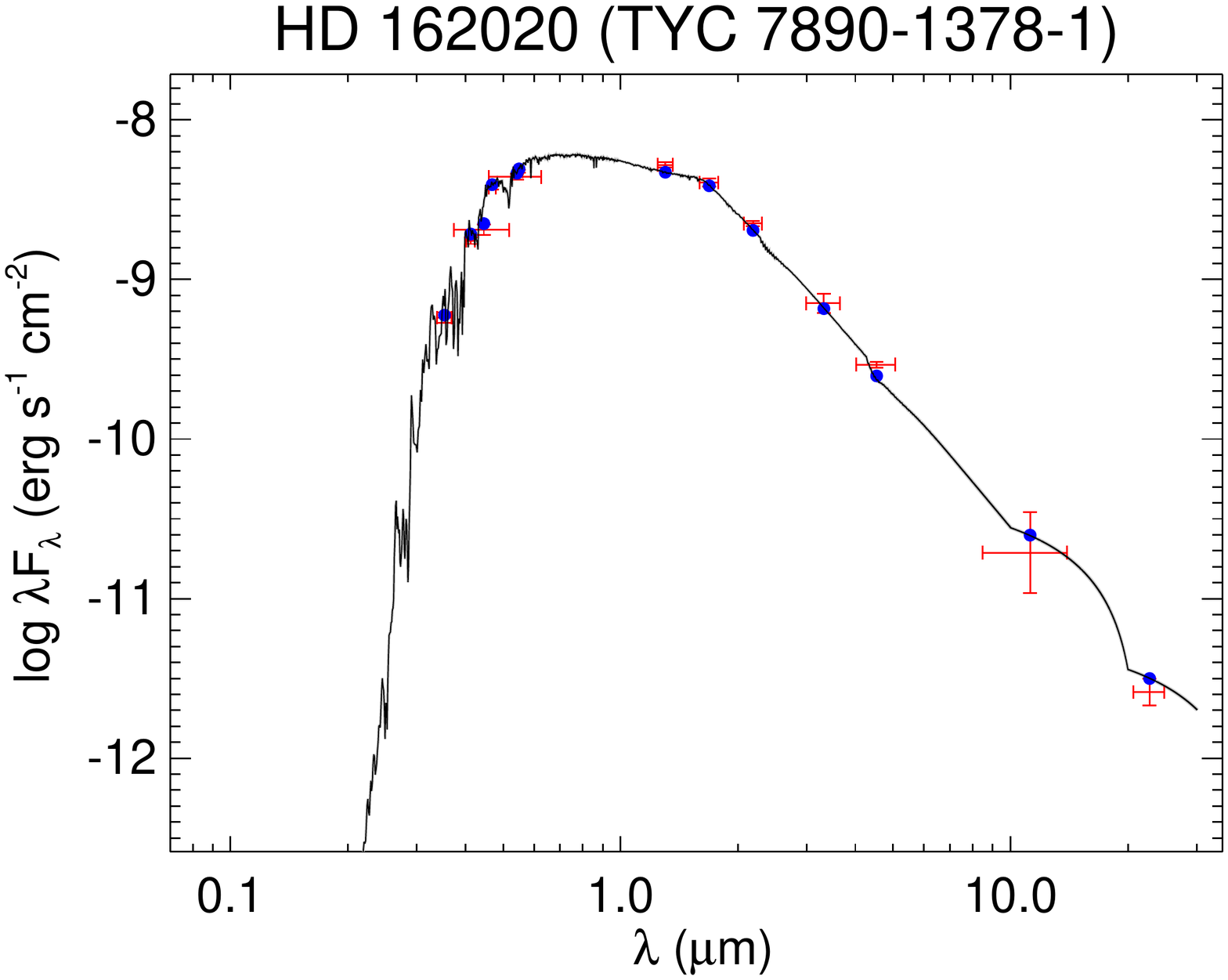}
  \includegraphics[trim=60 60 60 60,clip,width=0.49\linewidth]{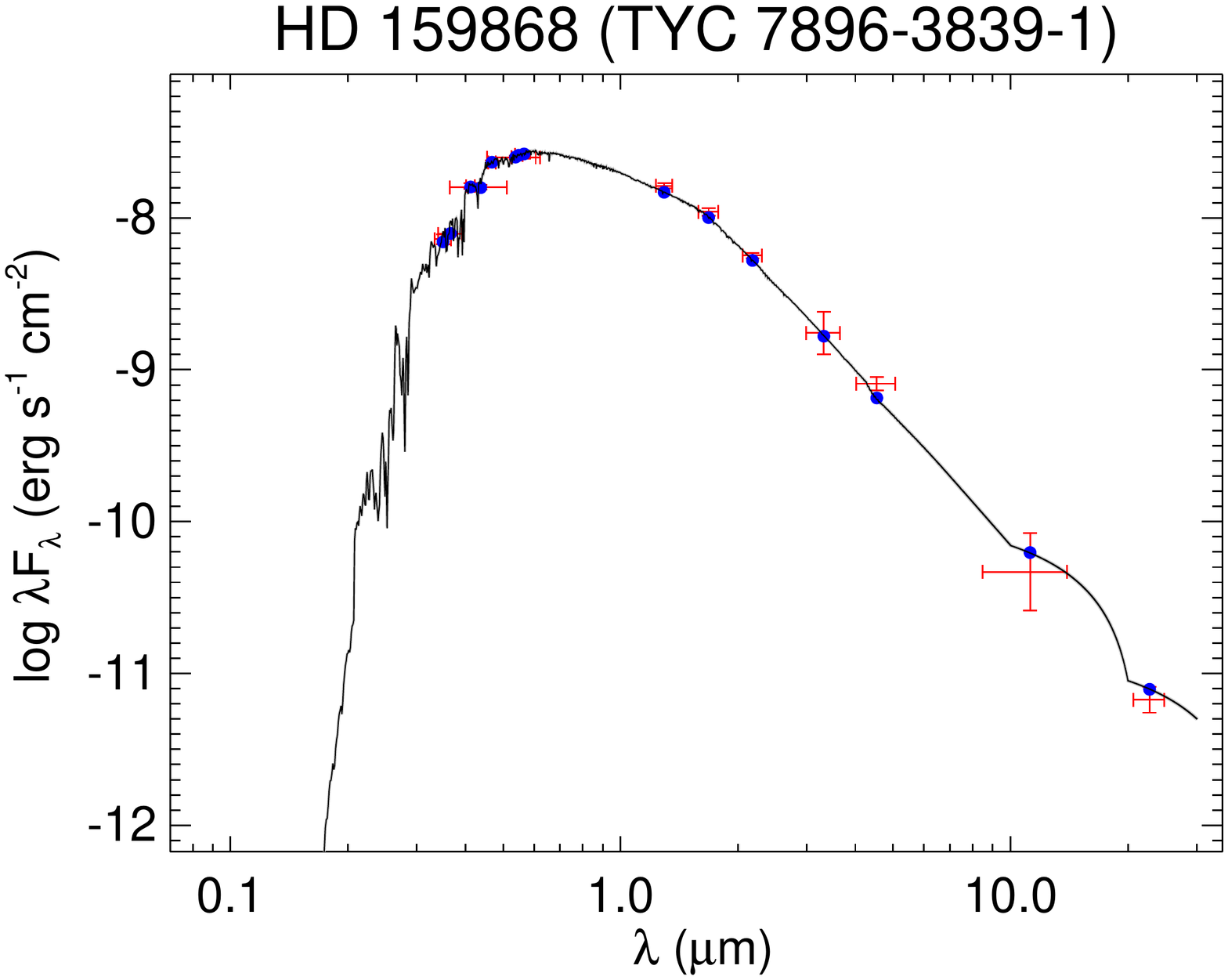}
  \includegraphics[trim=60 60 60 60,clip,width=0.49\linewidth]{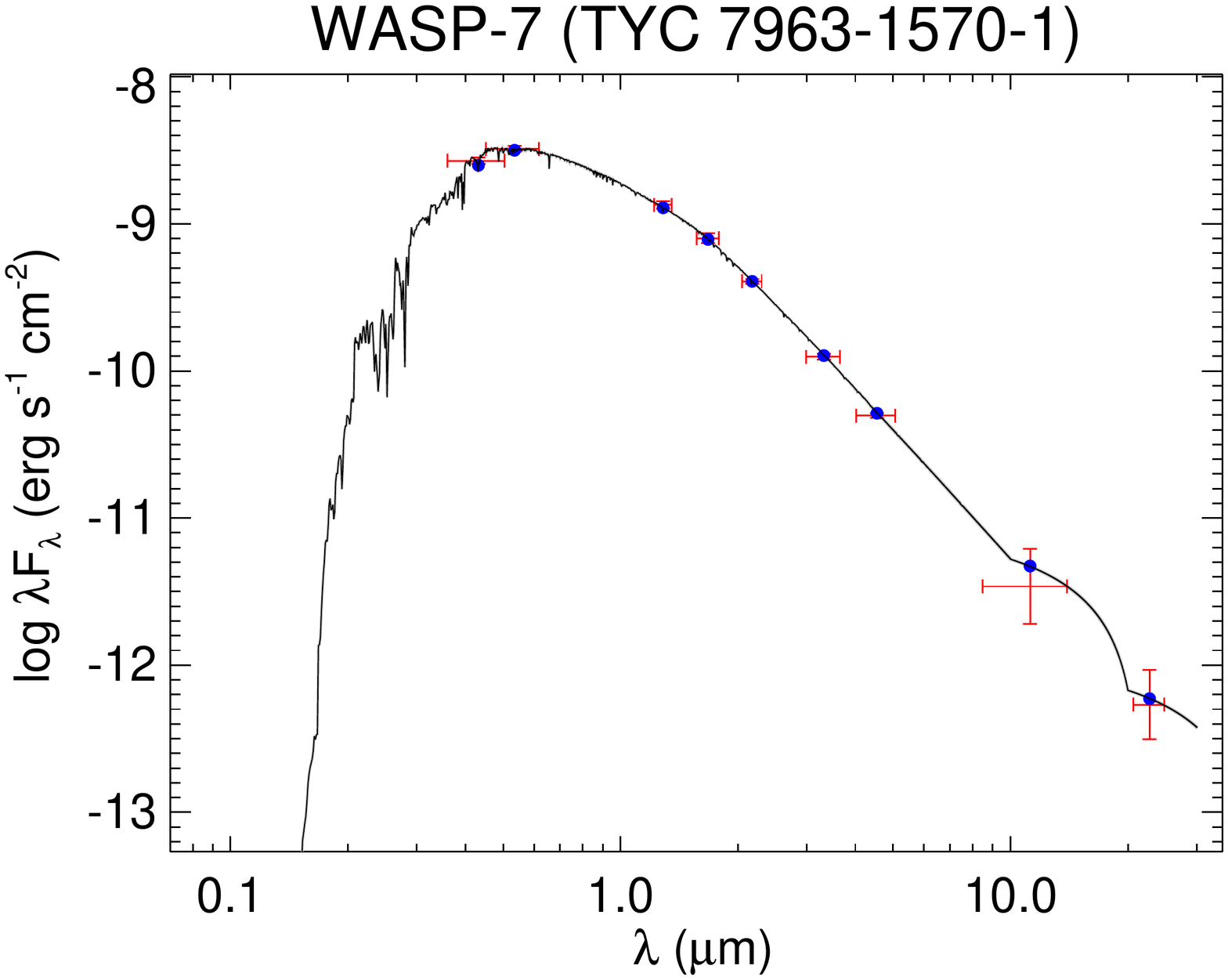}
  \includegraphics[trim=60 60 60 60,clip,width=0.49\linewidth]{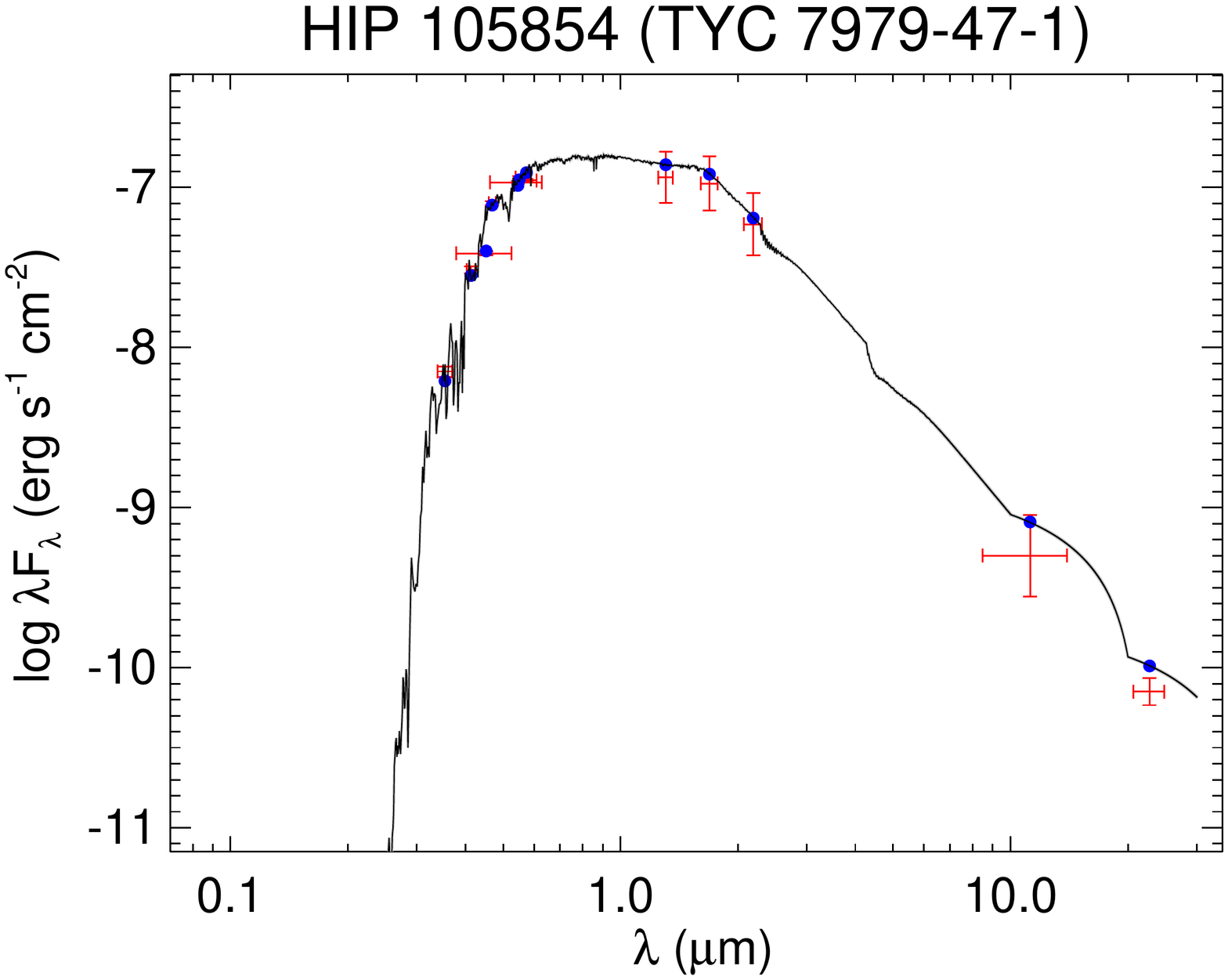}
  \includegraphics[trim=60 60 60 60,clip,width=0.49\linewidth]{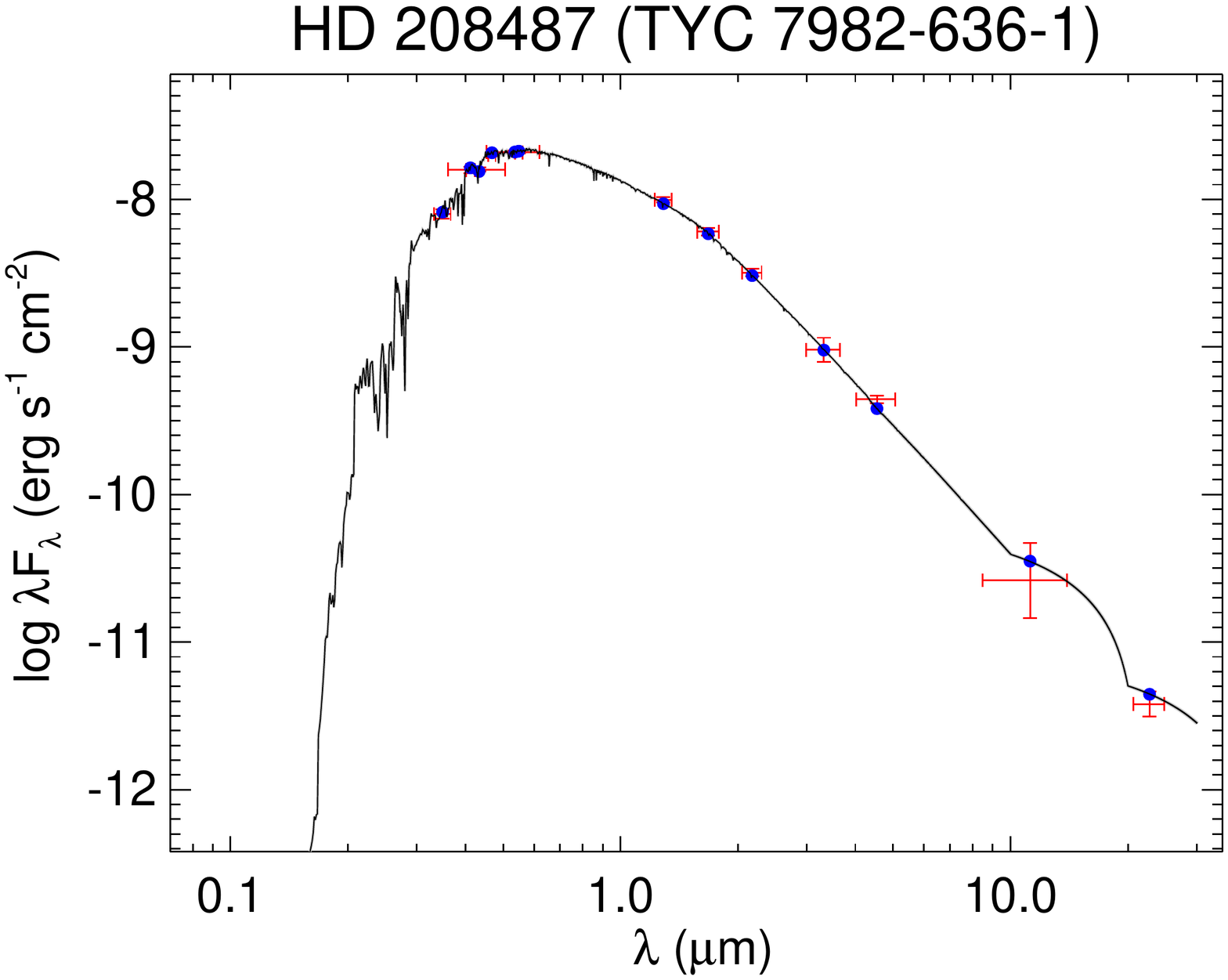}
  \caption{All labels, lines, symbols, and colors as in Figure \ref{fig:seds}.}
  \label{fig:seds_71}
\end{figure}

\begin{figure}[H]
  \centering
  \includegraphics[trim=60 60 60 60,clip,width=0.49\linewidth]{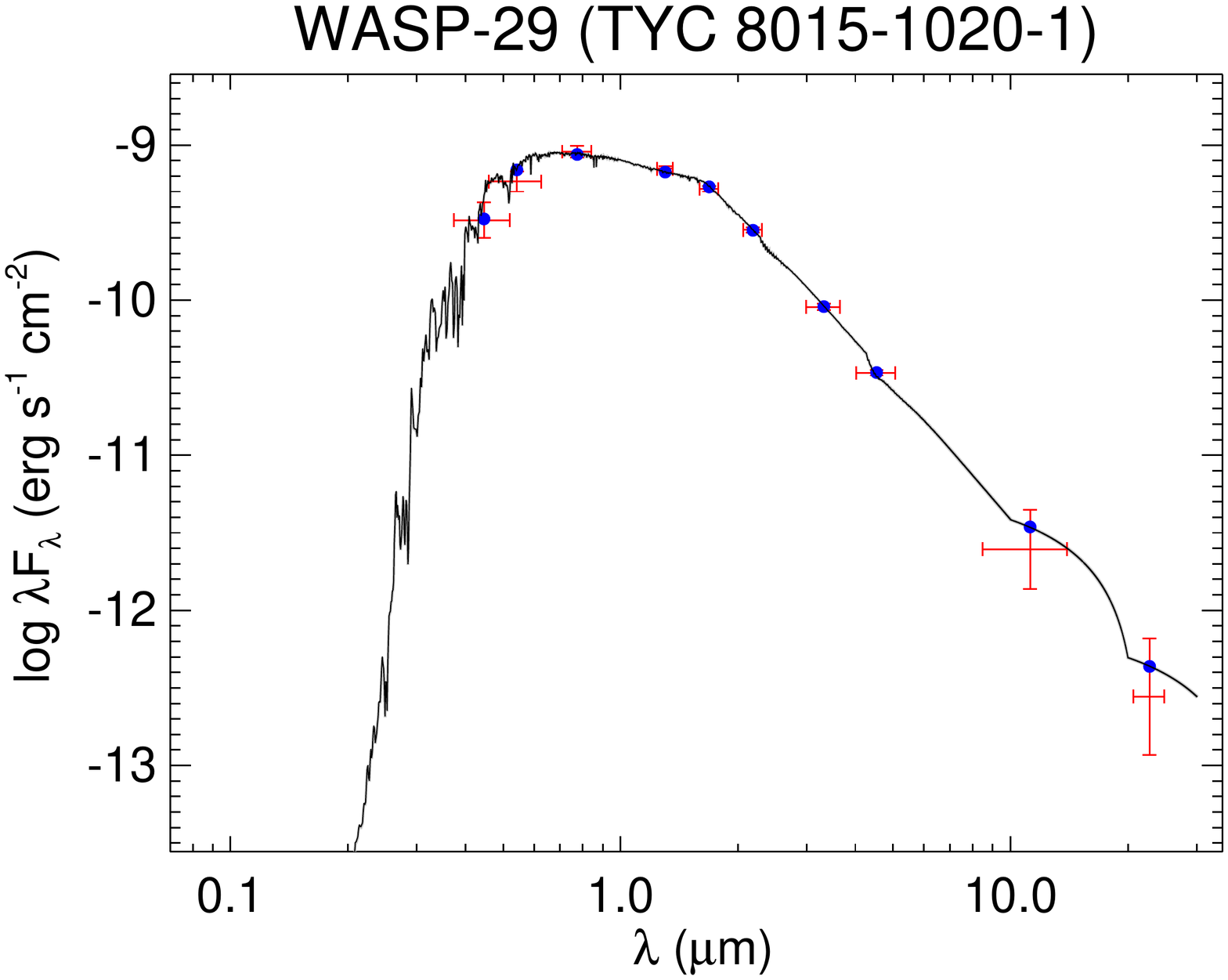}
  \includegraphics[trim=60 60 60 60,clip,width=0.49\linewidth]{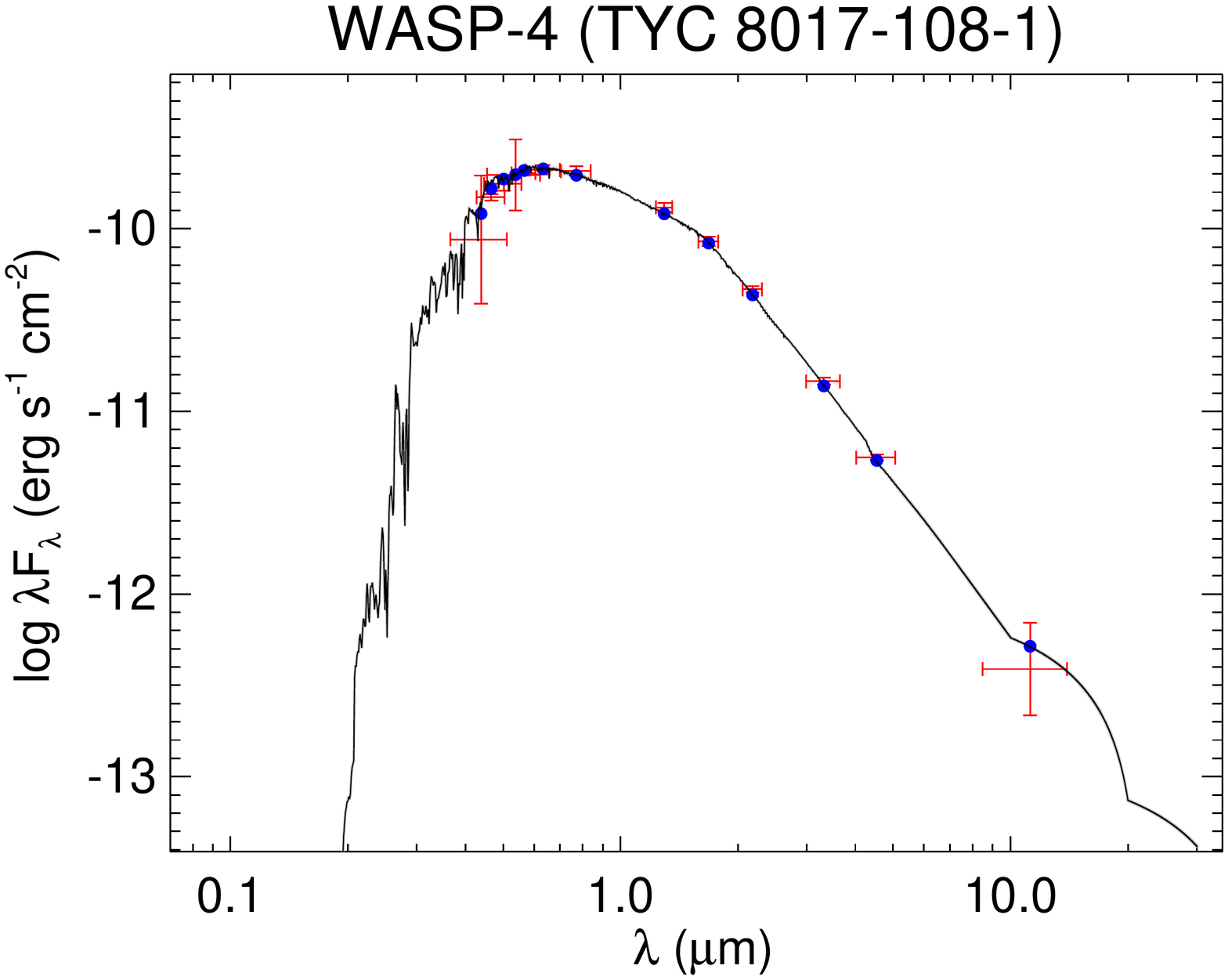}
  \includegraphics[trim=60 60 60 60,clip,width=0.49\linewidth]{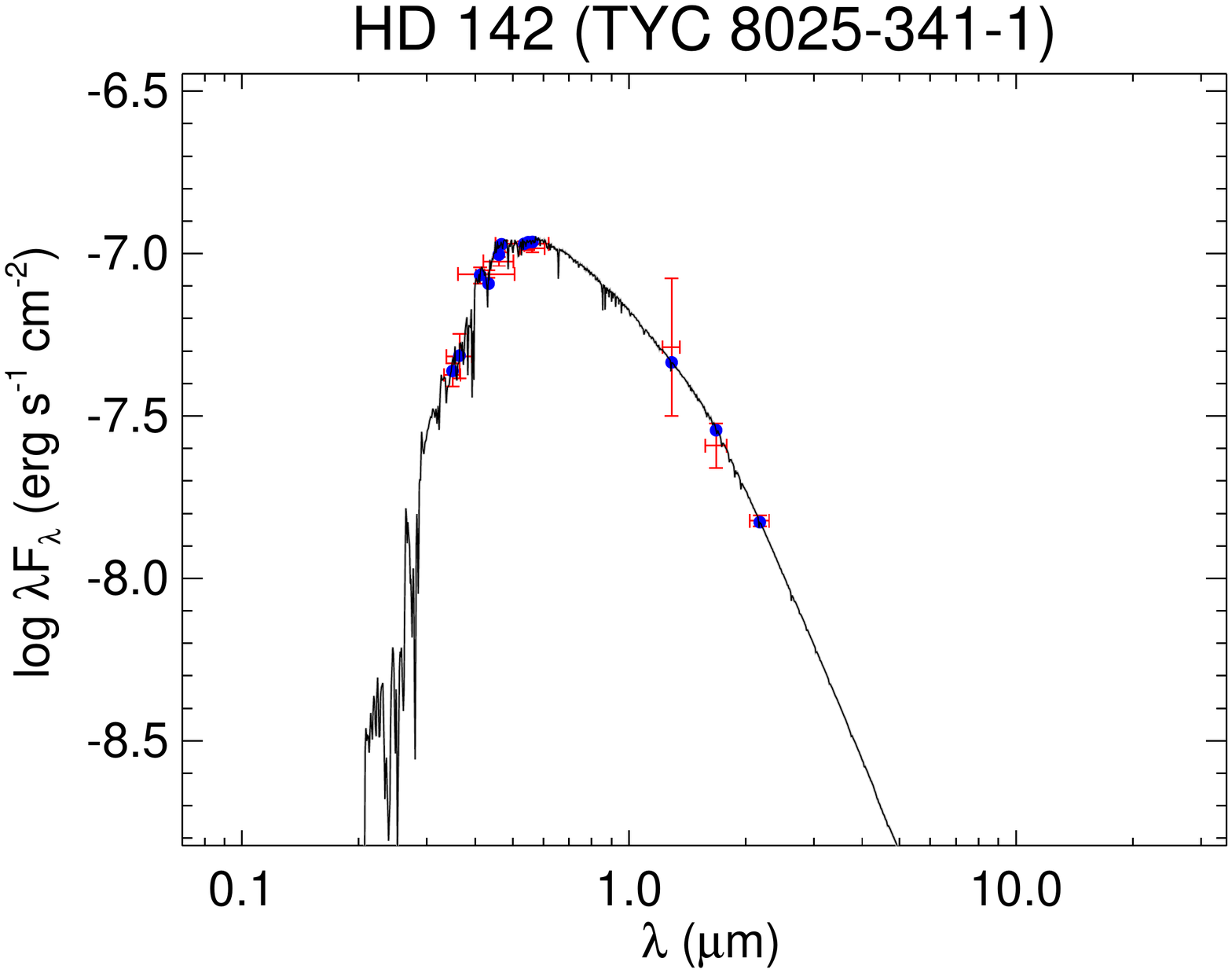}
  \includegraphics[trim=60 60 60 60,clip,width=0.49\linewidth]{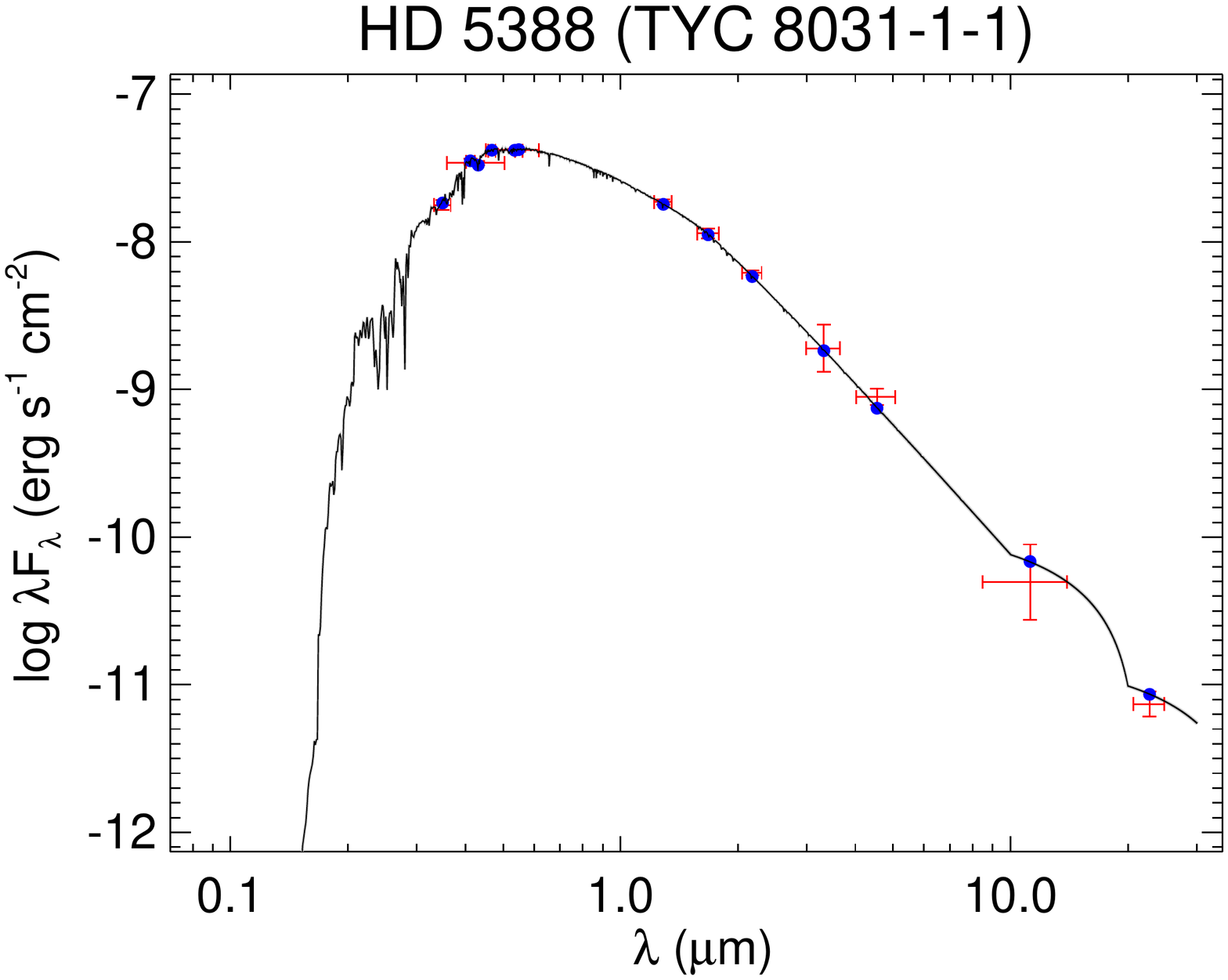}
  \includegraphics[trim=60 60 60 60,clip,width=0.49\linewidth]{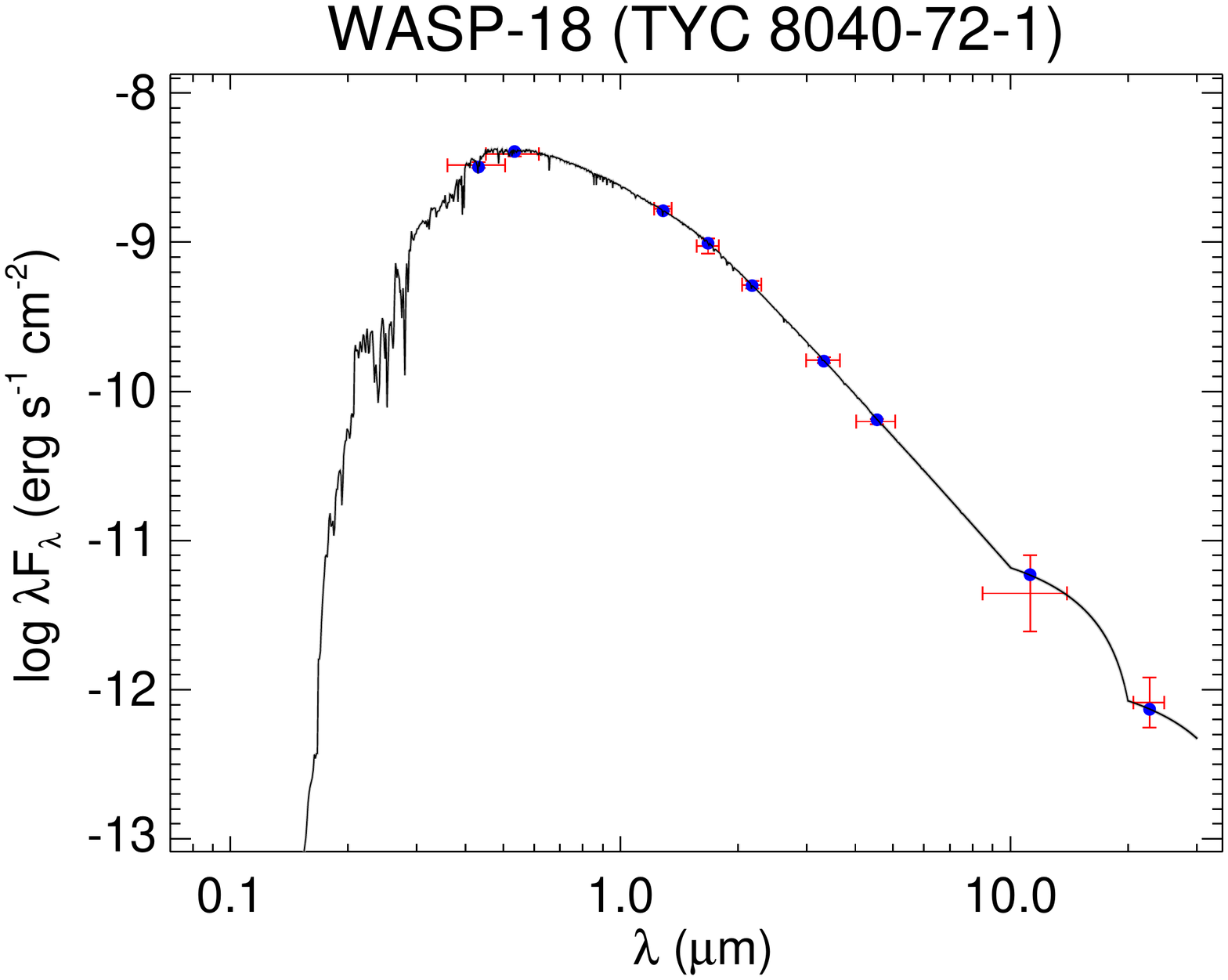}
  \includegraphics[trim=60 60 60 60,clip,width=0.49\linewidth]{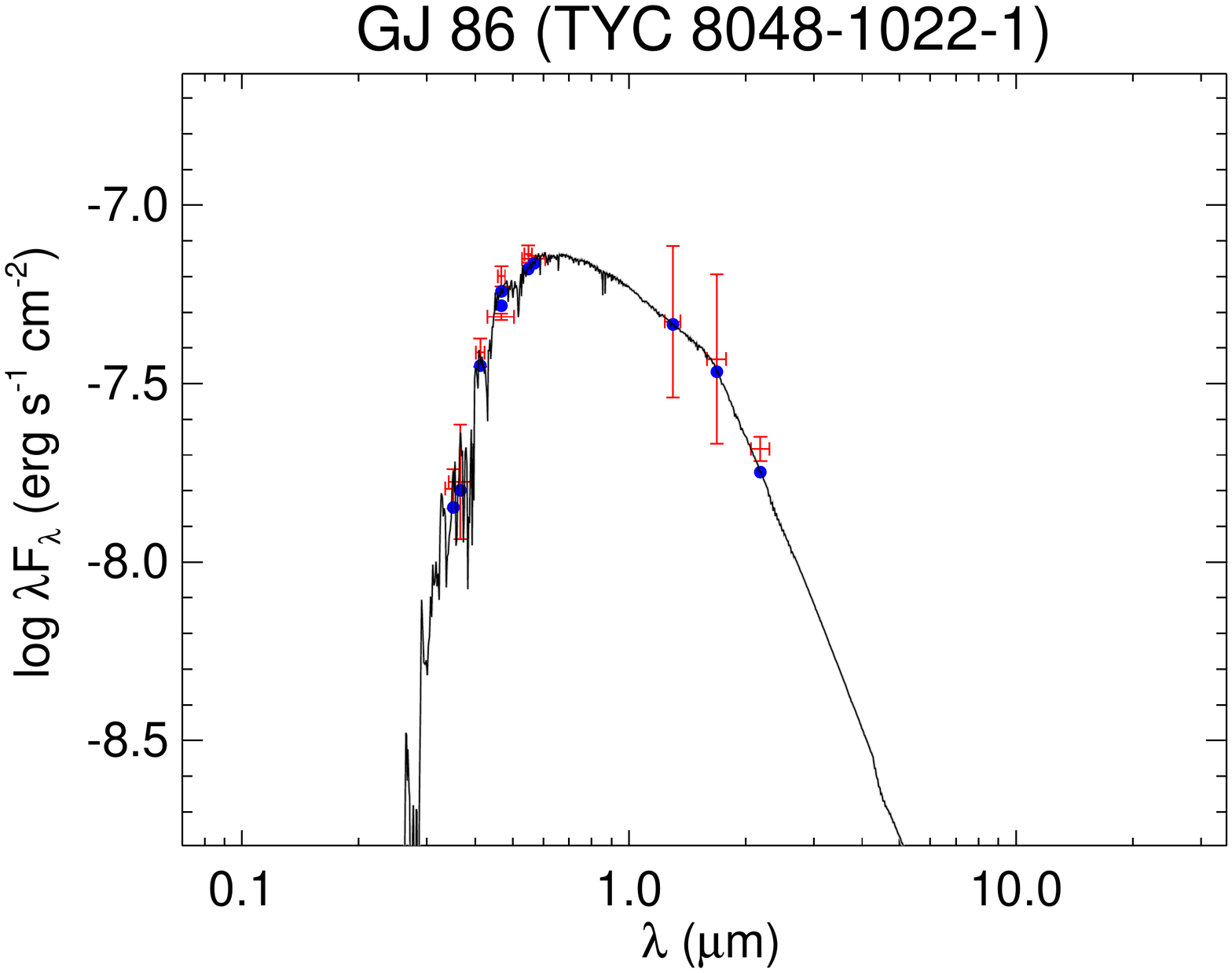}
  \caption{All labels, lines, symbols, and colors as in Figure \ref{fig:seds}.}
  \label{fig:seds_72}
\end{figure}

\begin{figure}[H]
  \centering
  \includegraphics[trim=60 60 60 60,clip,width=0.49\linewidth]{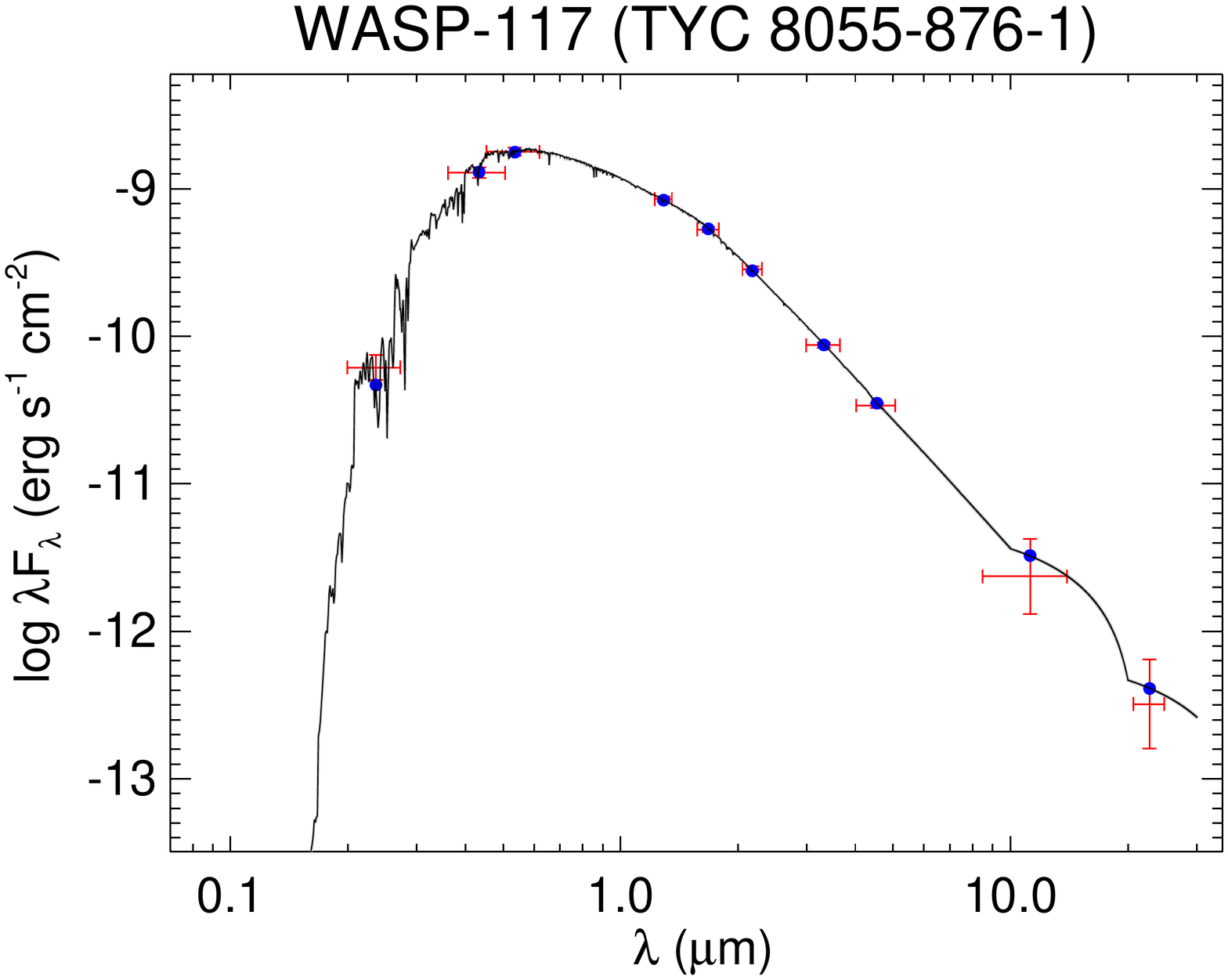}
  \includegraphics[trim=60 60 60 60,clip,width=0.49\linewidth]{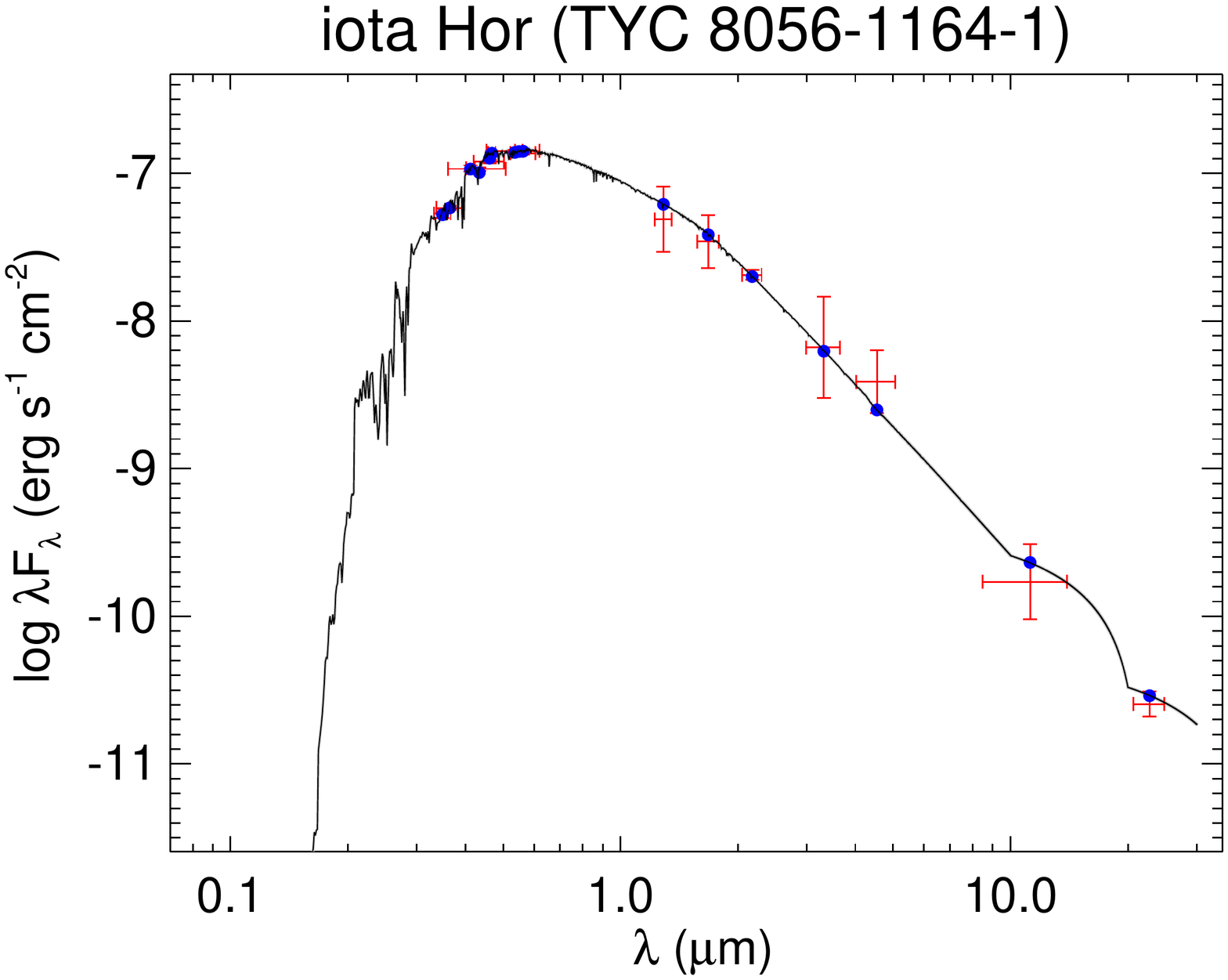}
  \includegraphics[trim=60 60 60 60,clip,width=0.49\linewidth]{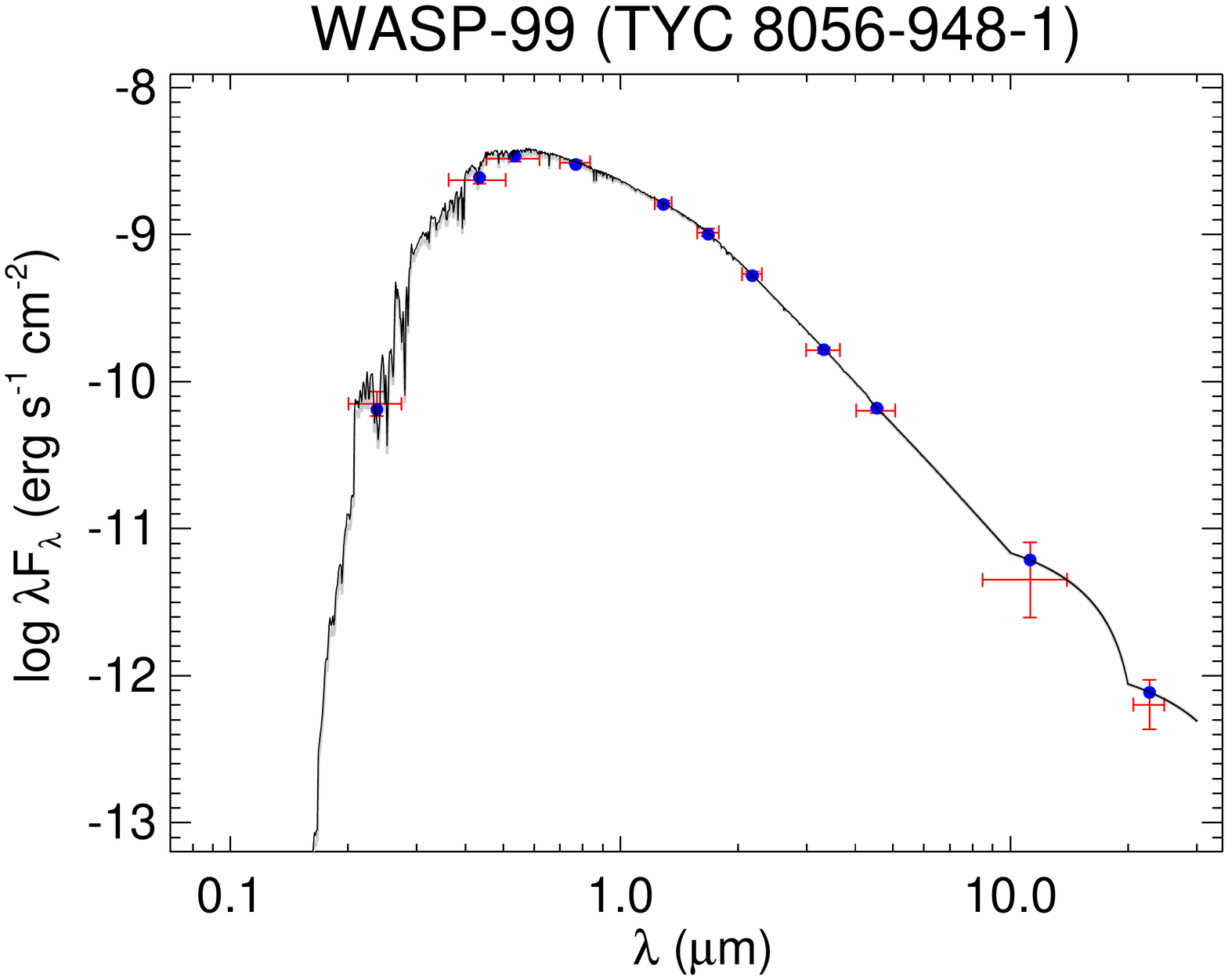}
  \includegraphics[trim=60 60 60 60,clip,width=0.49\linewidth]{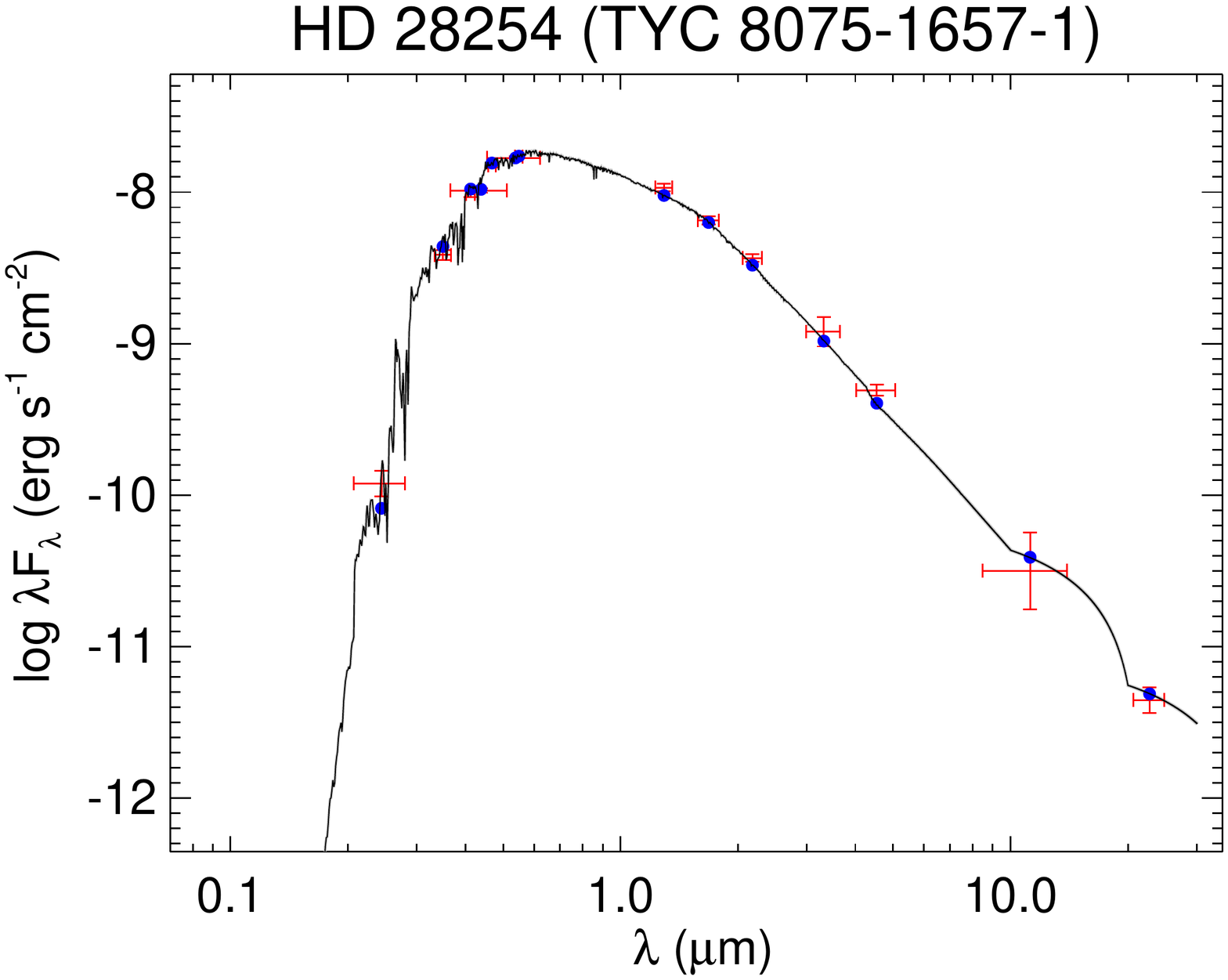}
  \includegraphics[trim=60 60 60 60,clip,width=0.49\linewidth]{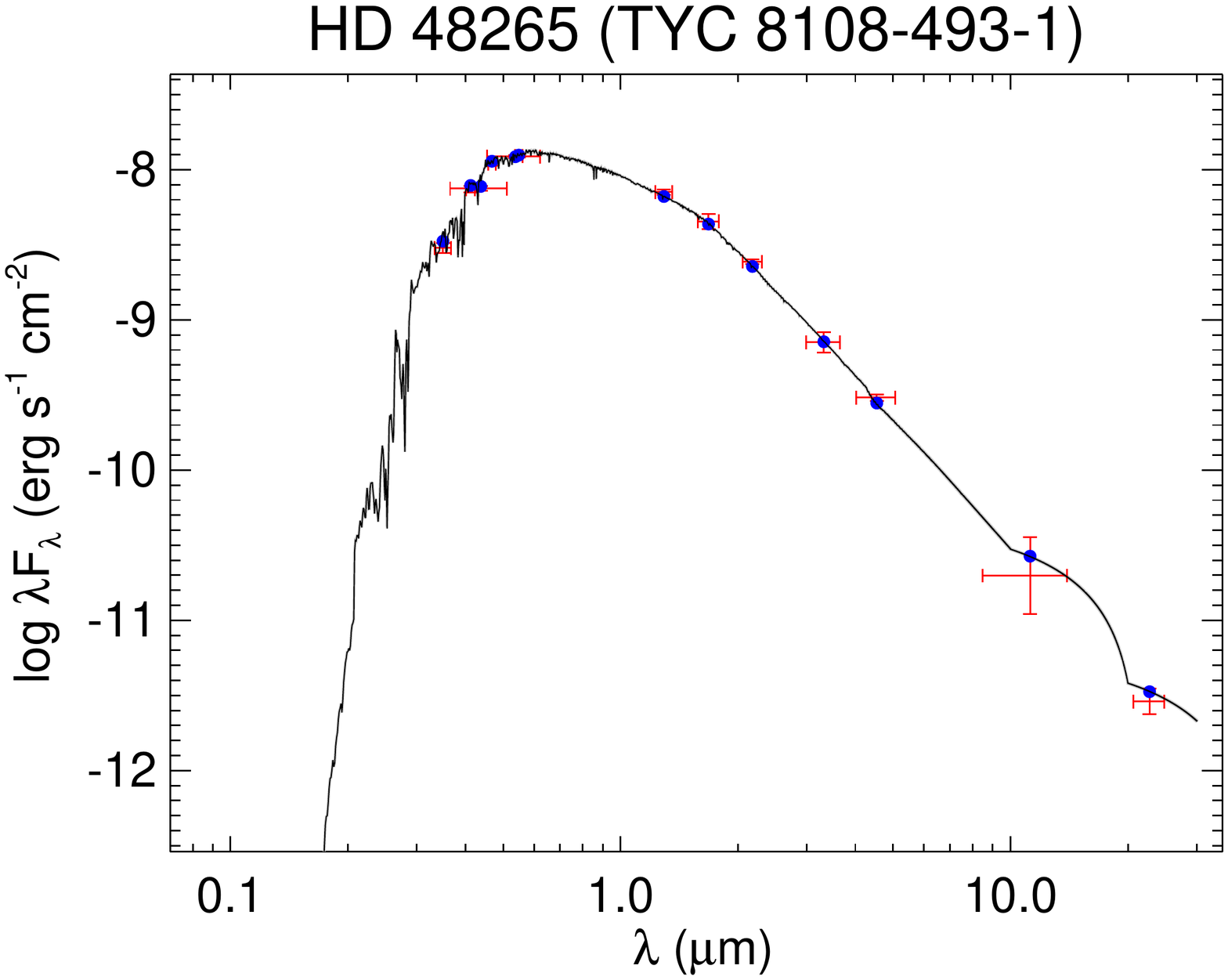}
  \includegraphics[trim=60 60 60 60,clip,width=0.49\linewidth]{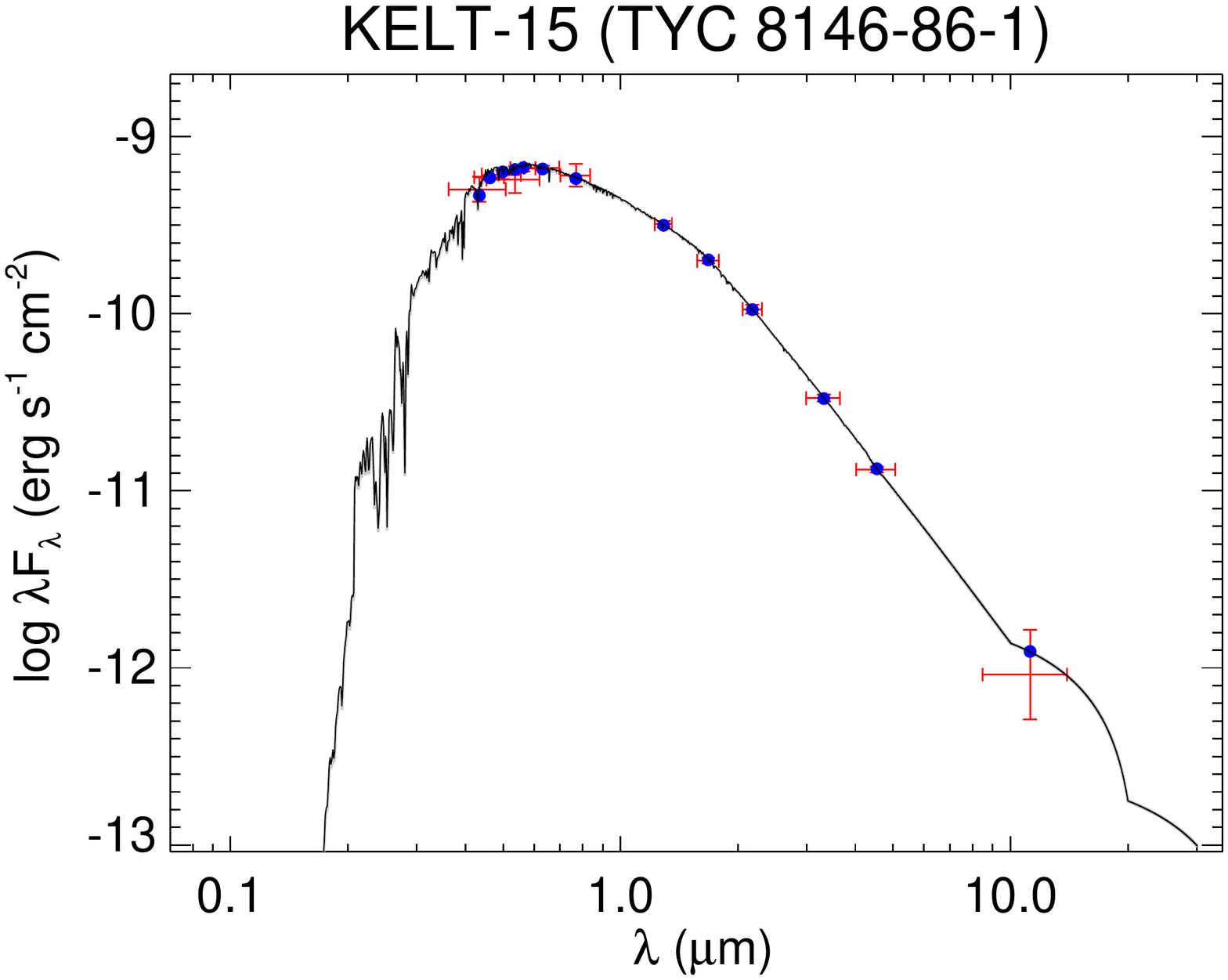}
  \caption{All labels, lines, symbols, and colors as in Figure \ref{fig:seds}.}
  \label{fig:seds_73}
\end{figure}

\begin{figure}[H]
  \centering
  \includegraphics[trim=60 60 60 60,clip,width=0.49\linewidth]{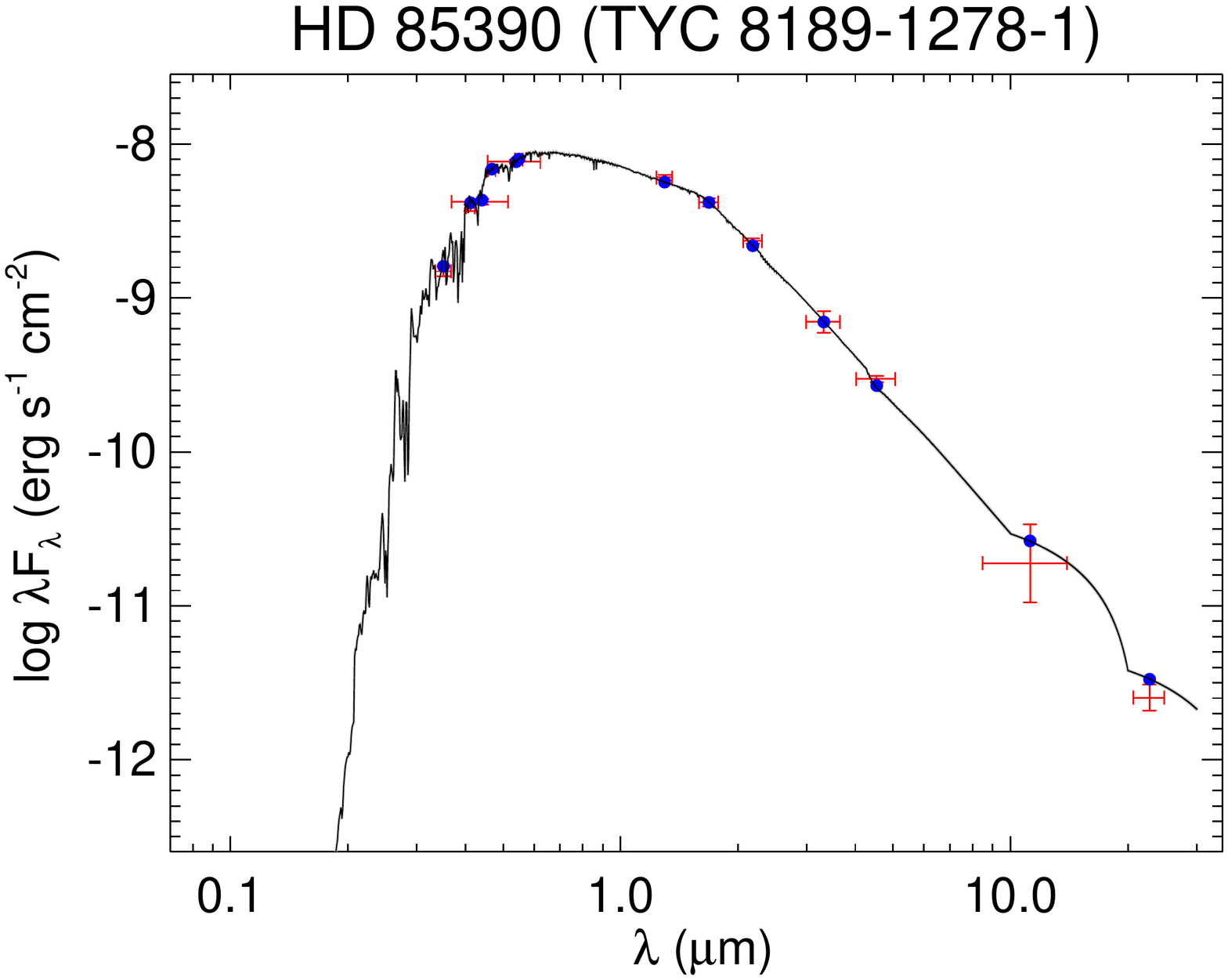}
  \includegraphics[trim=60 60 60 60,clip,width=0.49\linewidth]{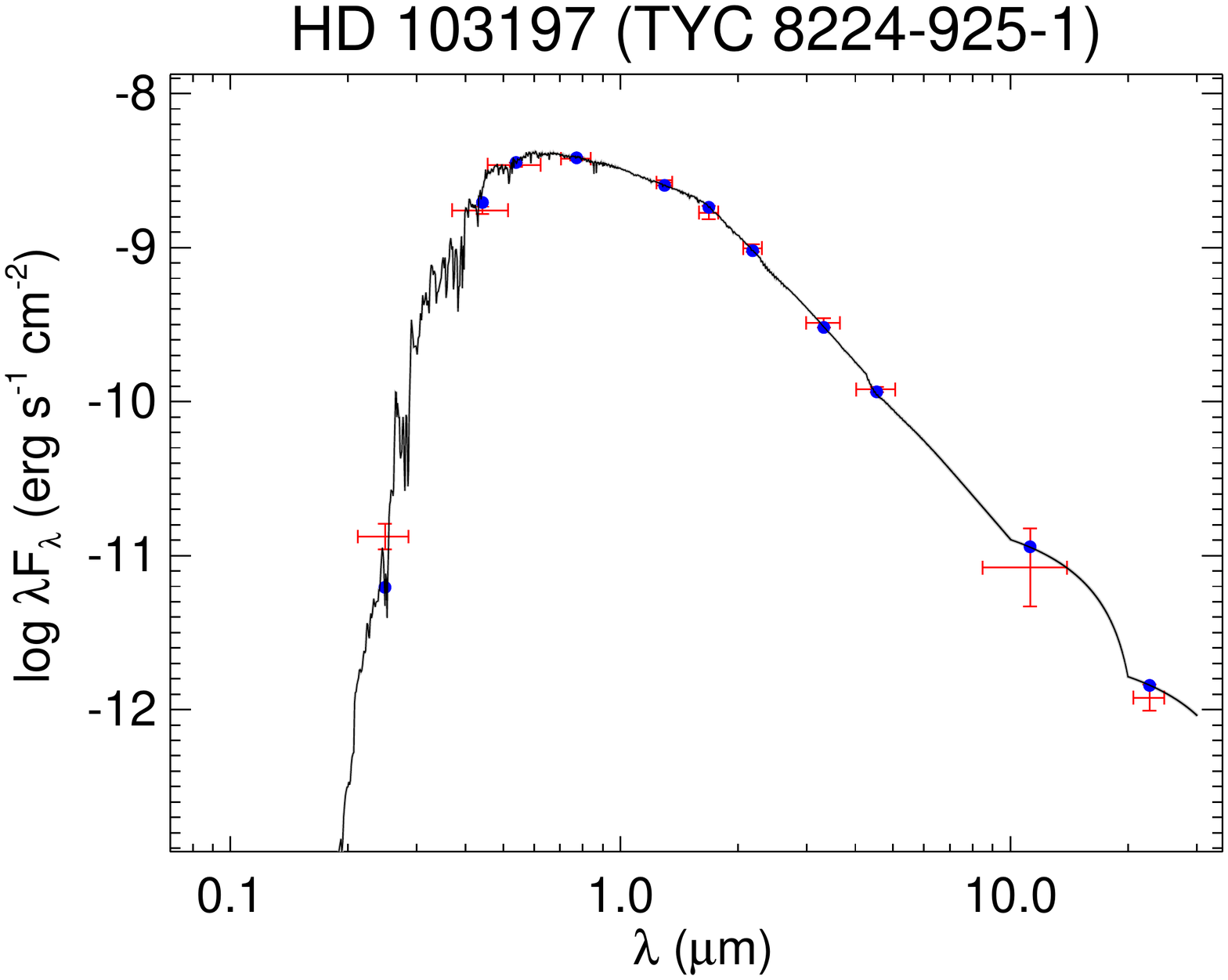}
  \includegraphics[trim=60 60 60 60,clip,width=0.49\linewidth]{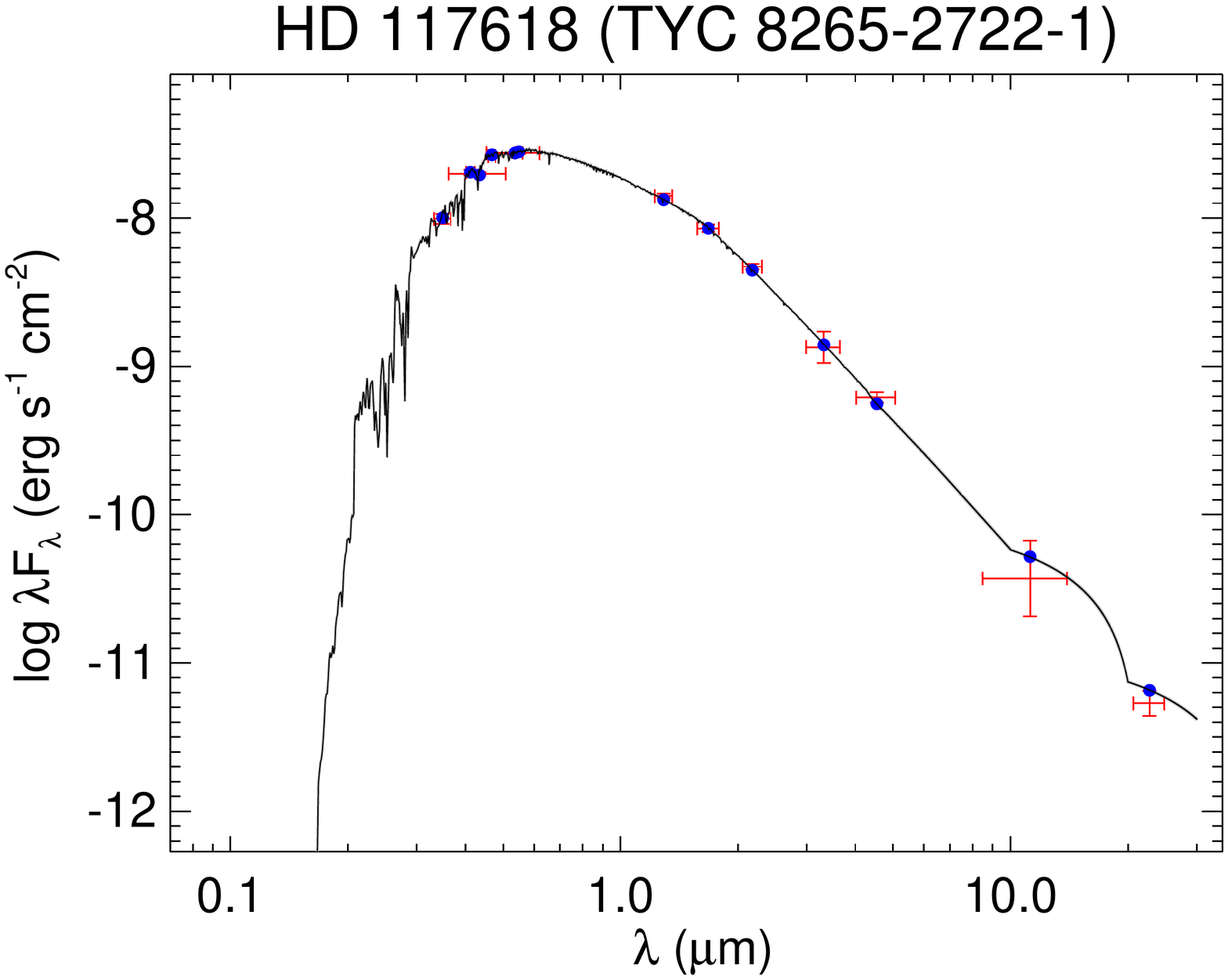}
  \includegraphics[trim=60 60 60 60,clip,width=0.49\linewidth]{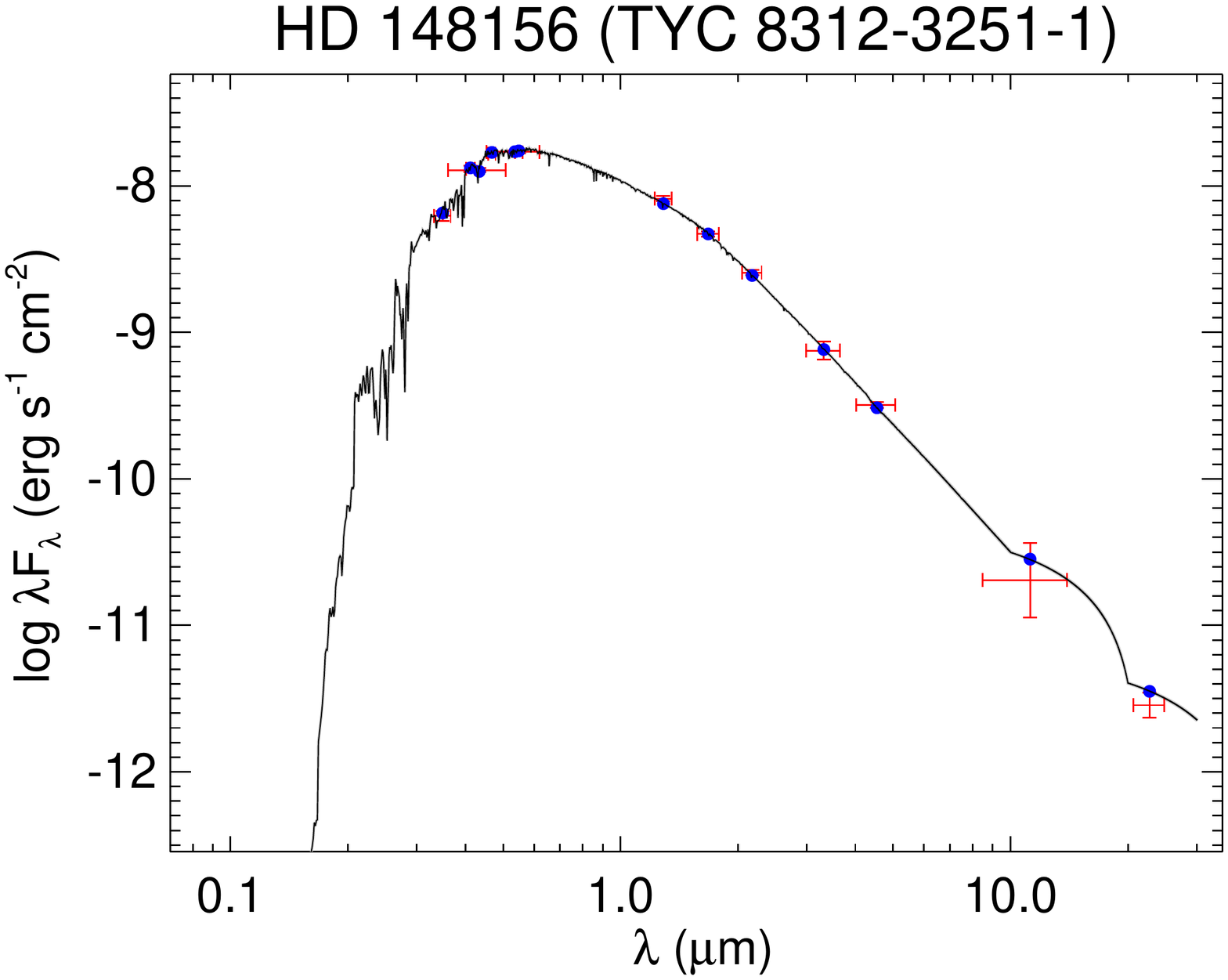}
  \includegraphics[trim=60 60 60 60,clip,width=0.49\linewidth]{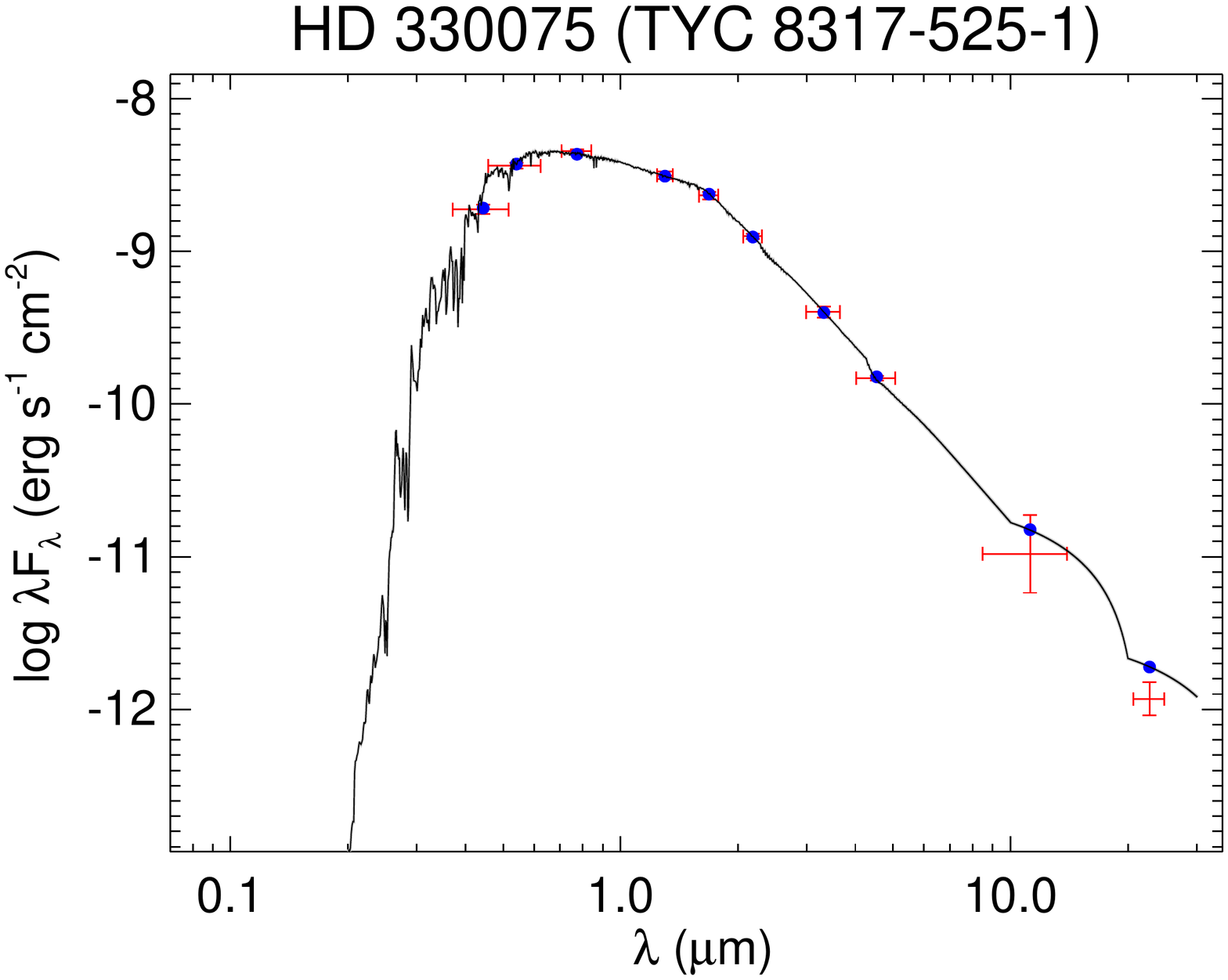}
  \includegraphics[trim=60 60 60 60,clip,width=0.49\linewidth]{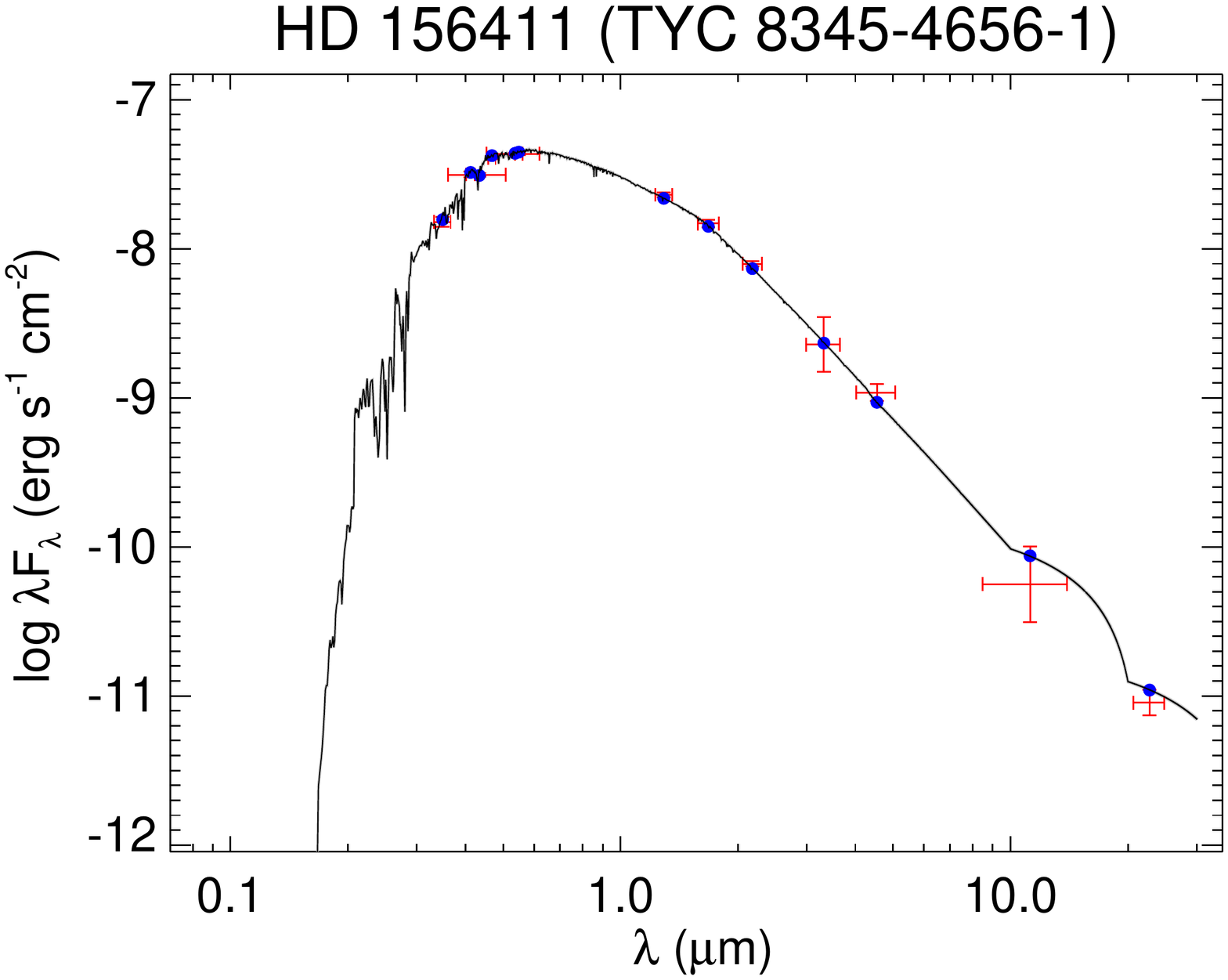}
  \caption{All labels, lines, symbols, and colors as in Figure \ref{fig:seds}.}
  \label{fig:seds_74}
\end{figure}

\begin{figure}[H]
  \centering
  \includegraphics[trim=60 60 60 60,clip,width=0.49\linewidth]{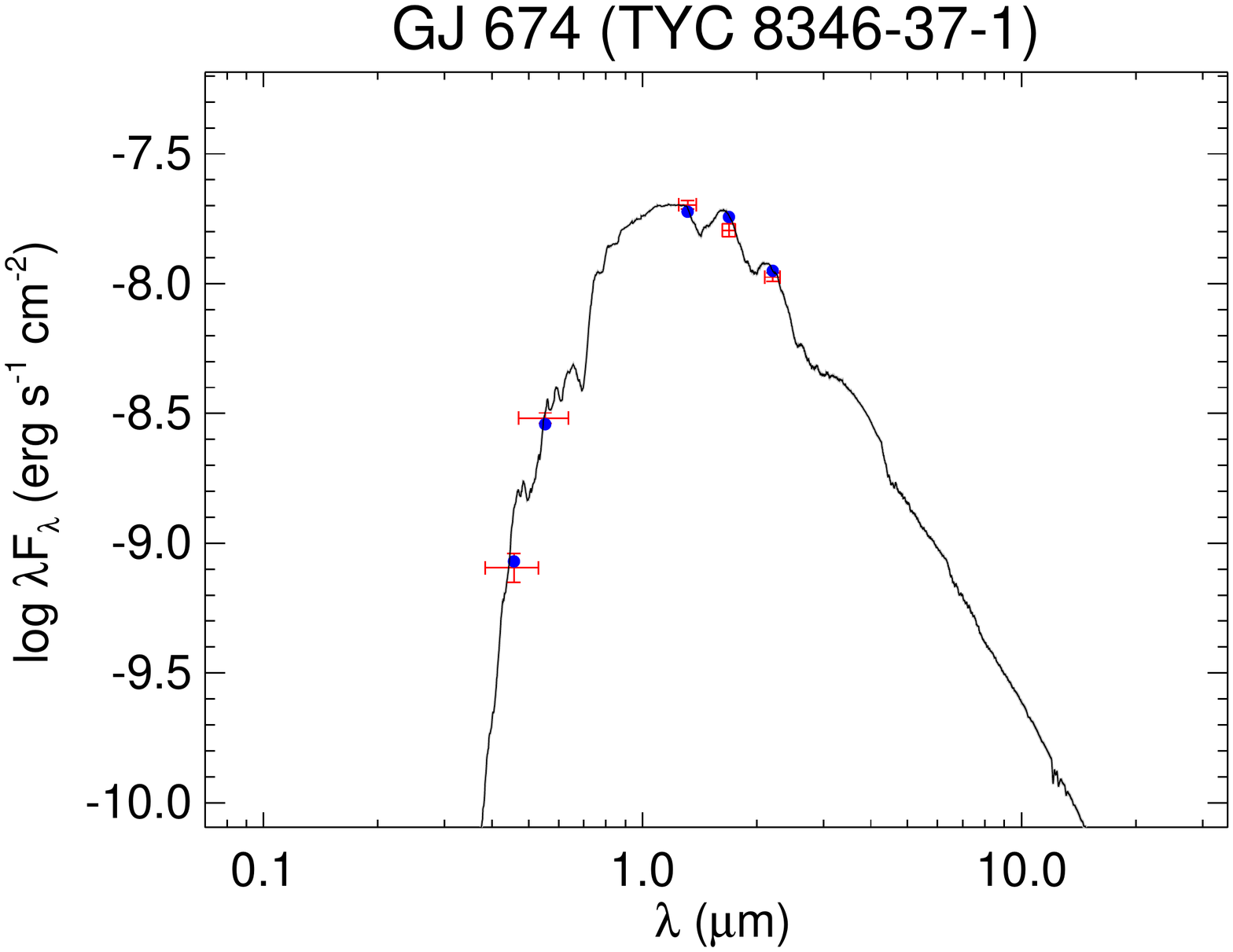}
  \includegraphics[trim=60 60 60 60,clip,width=0.49\linewidth]{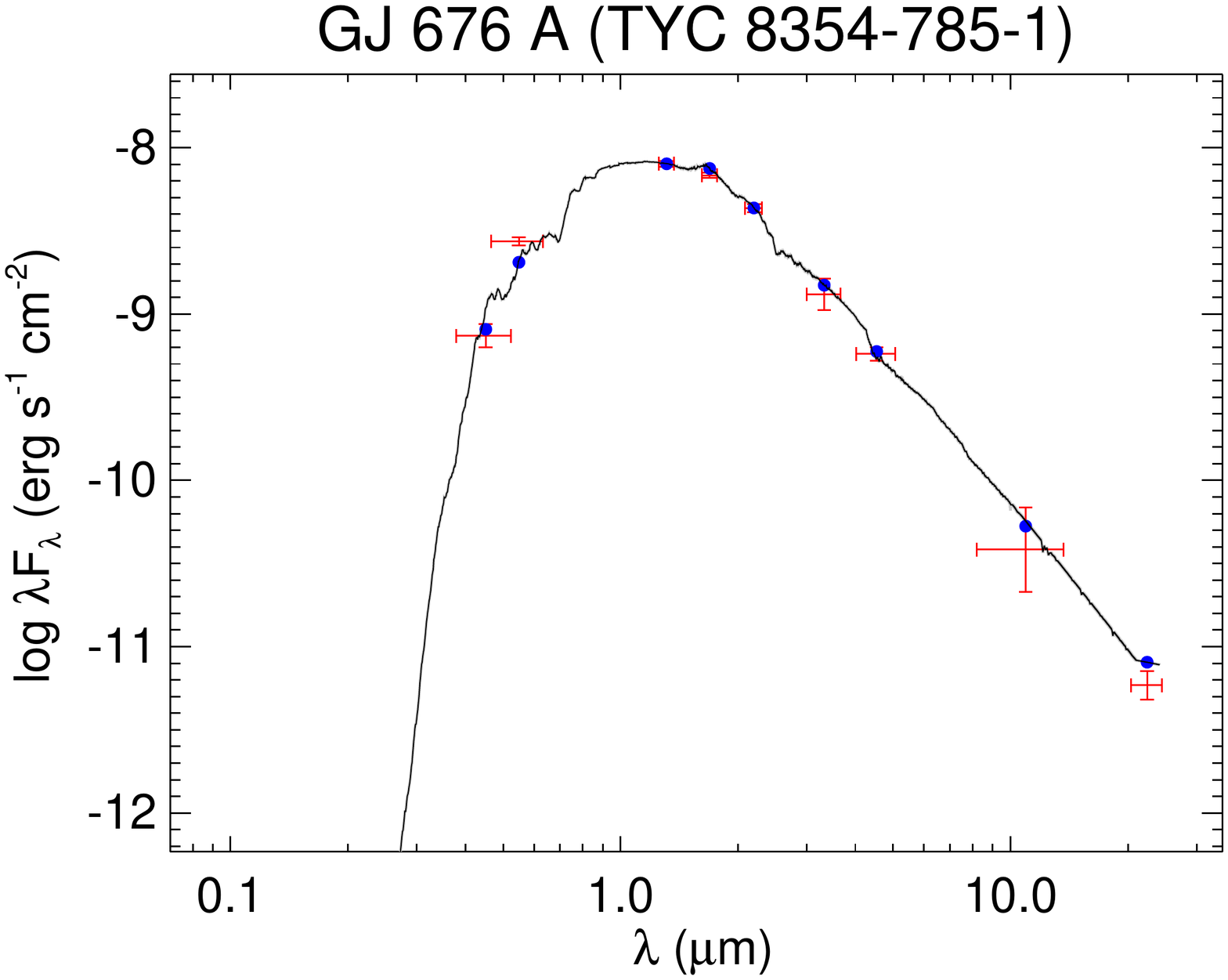}
  \includegraphics[trim=60 60 60 60,clip,width=0.49\linewidth]{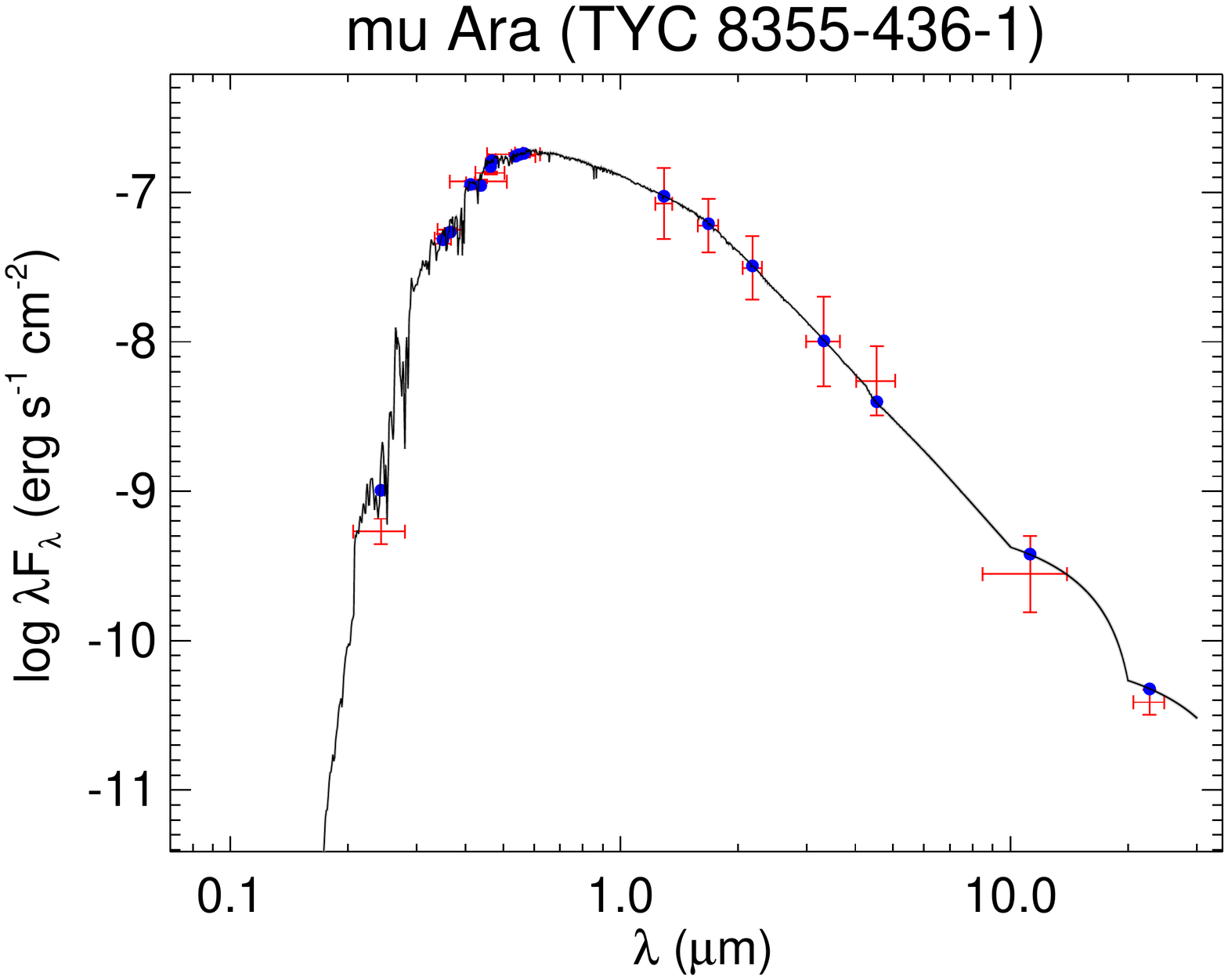}
  \includegraphics[trim=60 60 60 60,clip,width=0.49\linewidth]{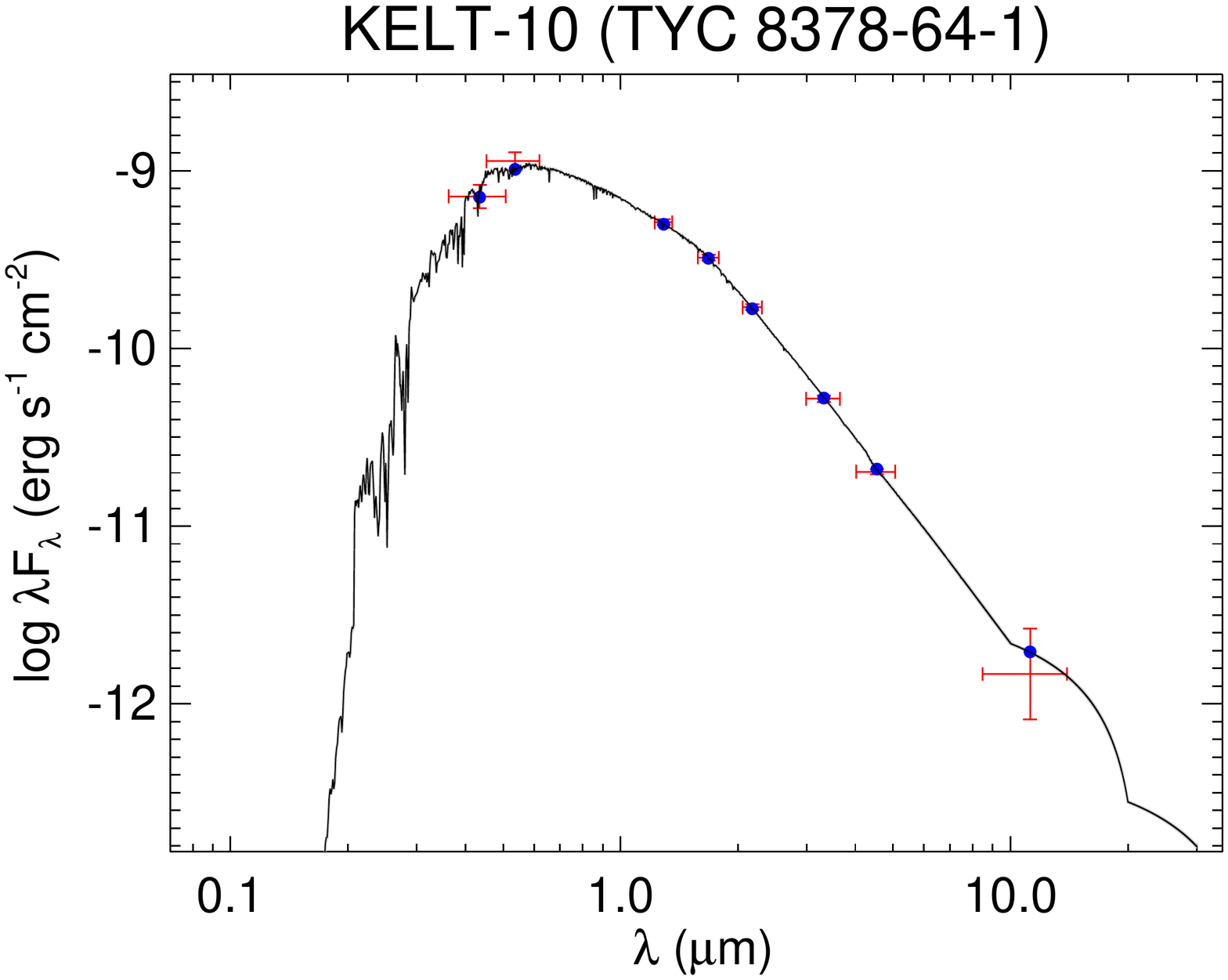}
  \includegraphics[trim=60 60 60 60,clip,width=0.49\linewidth]{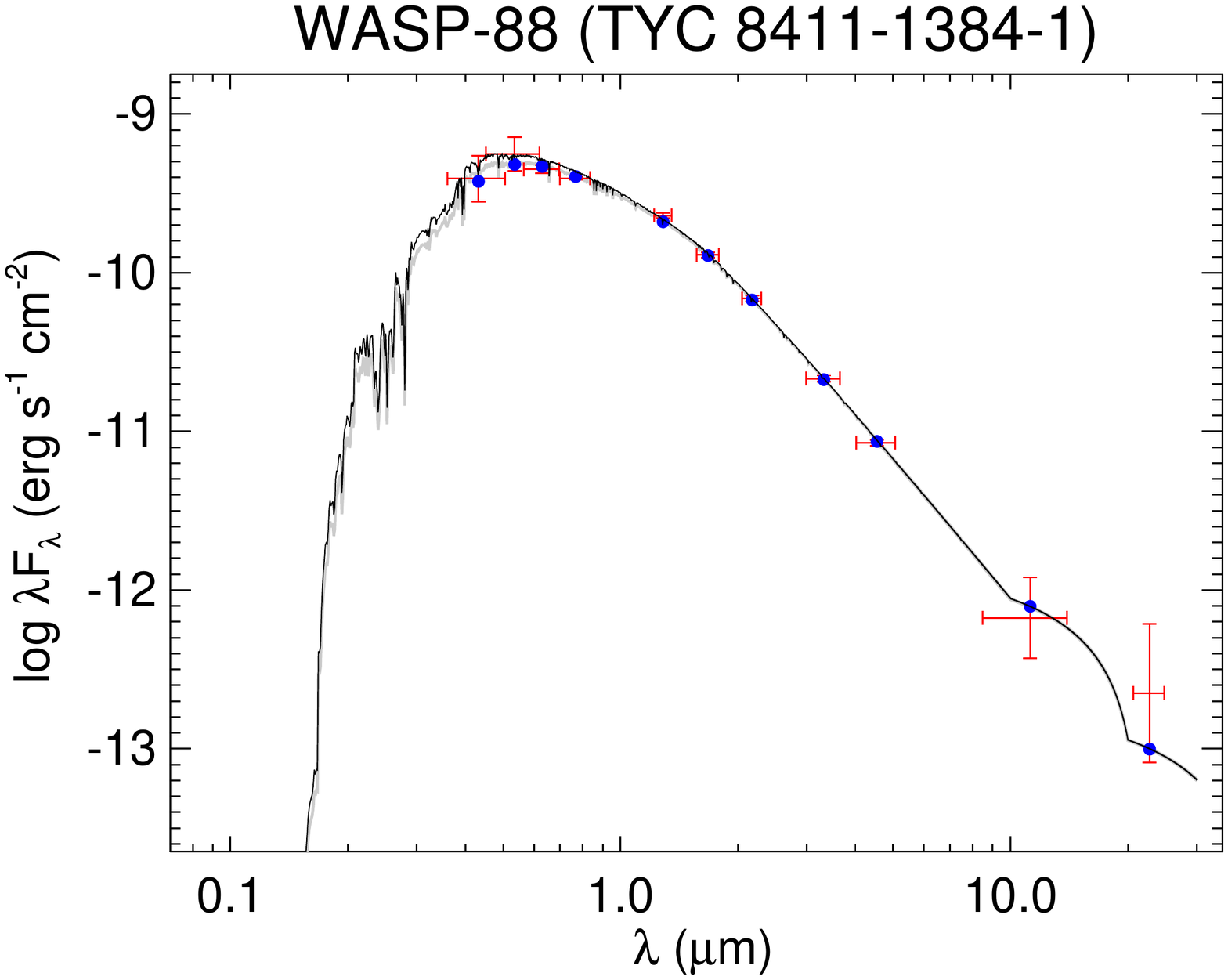}
  \includegraphics[trim=60 60 60 60,clip,width=0.49\linewidth]{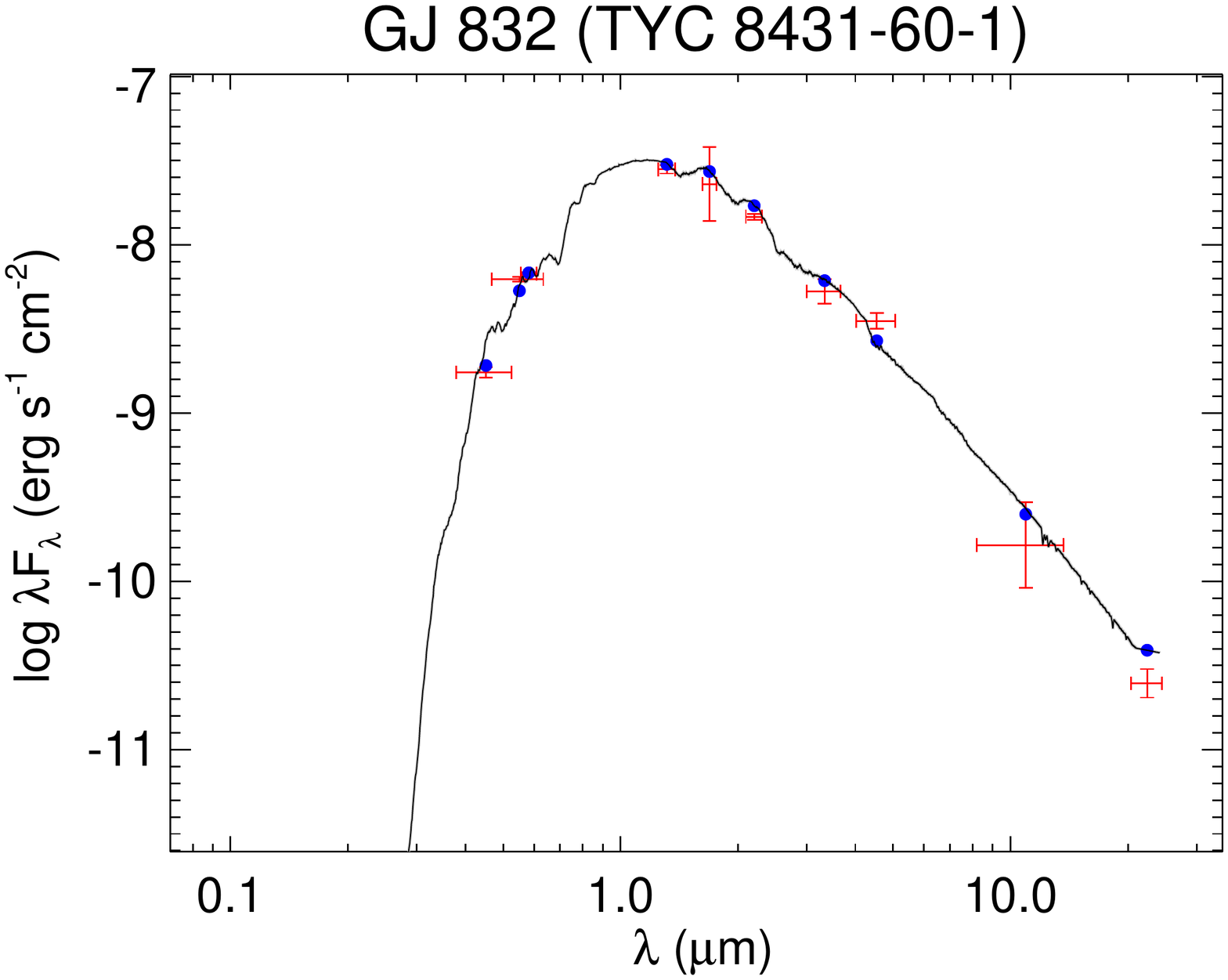}
  \caption{All labels, lines, symbols, and colors as in Figure \ref{fig:seds}.}
  \label{fig:seds_75}
\end{figure}

\begin{figure}[H]
  \centering
  \includegraphics[trim=60 60 60 60,clip,width=0.49\linewidth]{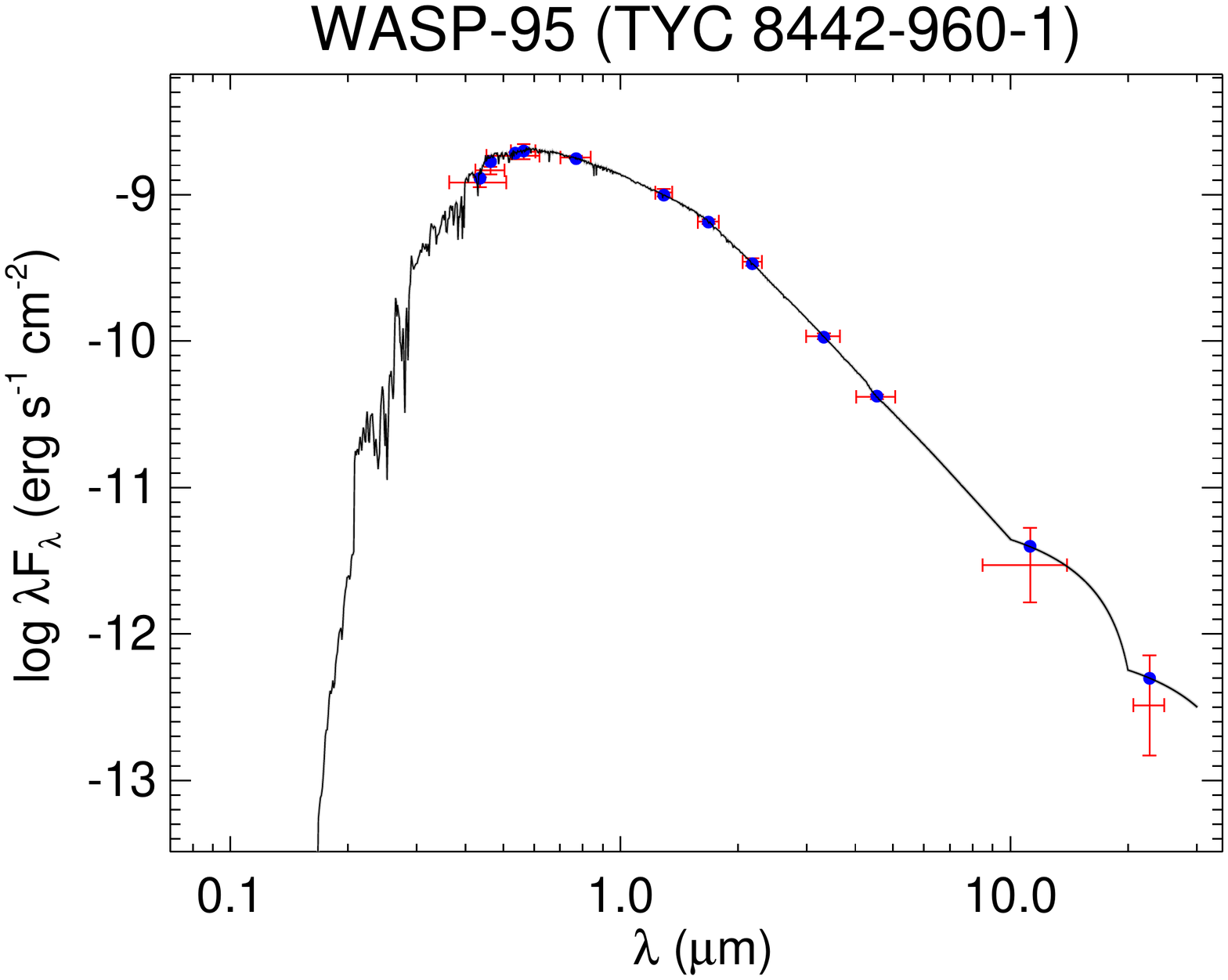}
  \includegraphics[trim=60 60 60 60,clip,width=0.49\linewidth]{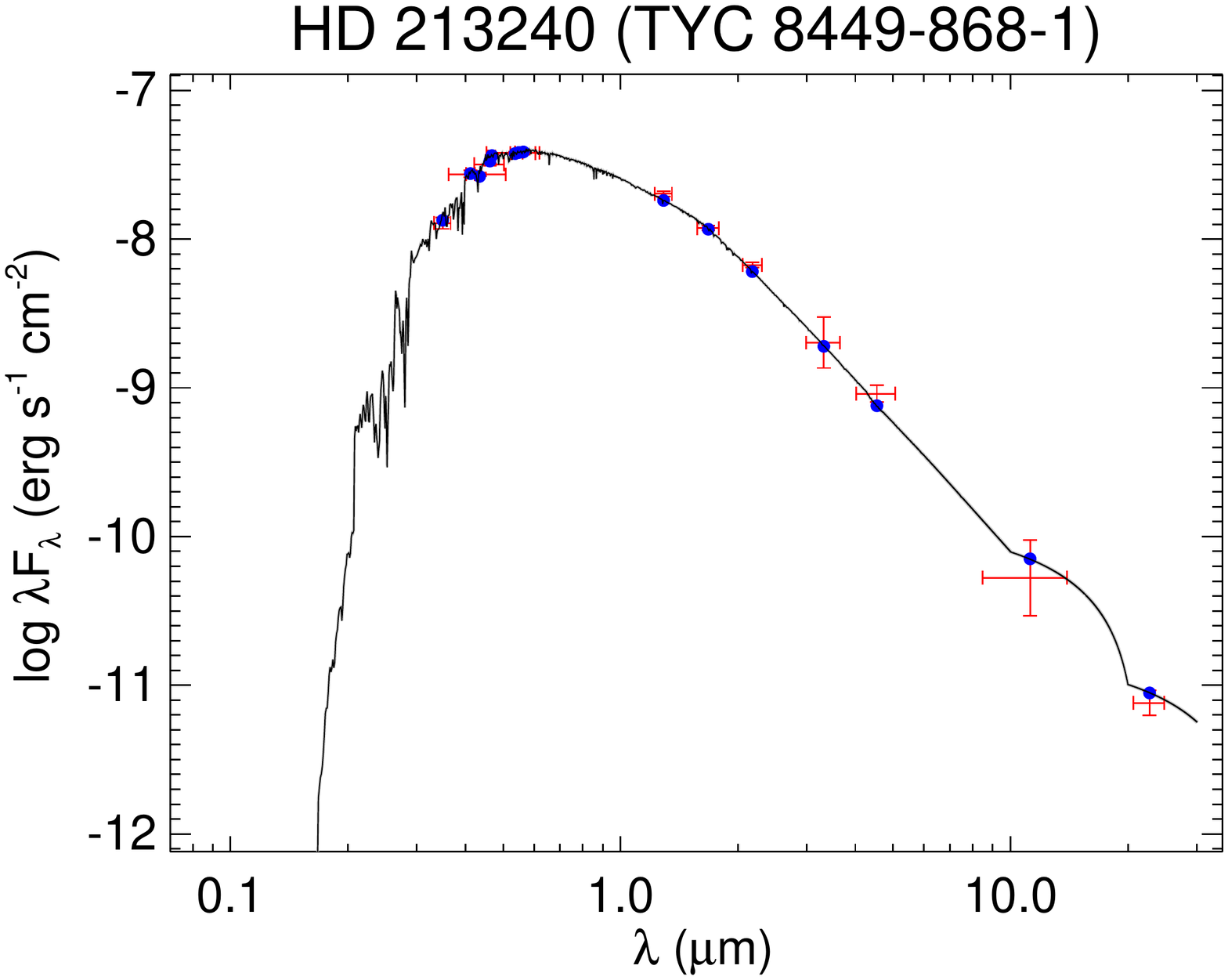}
  \includegraphics[trim=60 60 60 60,clip,width=0.49\linewidth]{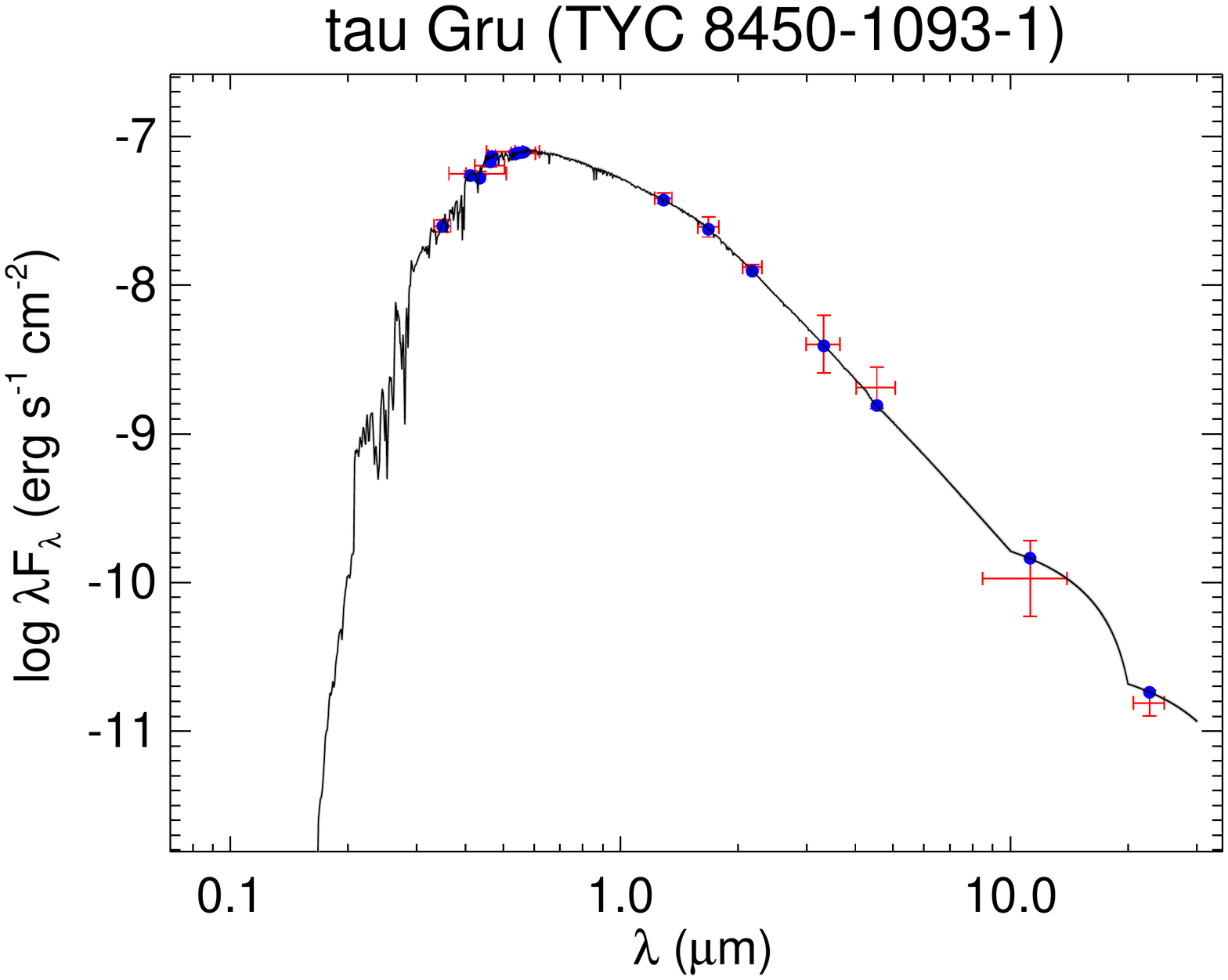}
  \includegraphics[trim=60 60 60 60,clip,width=0.49\linewidth]{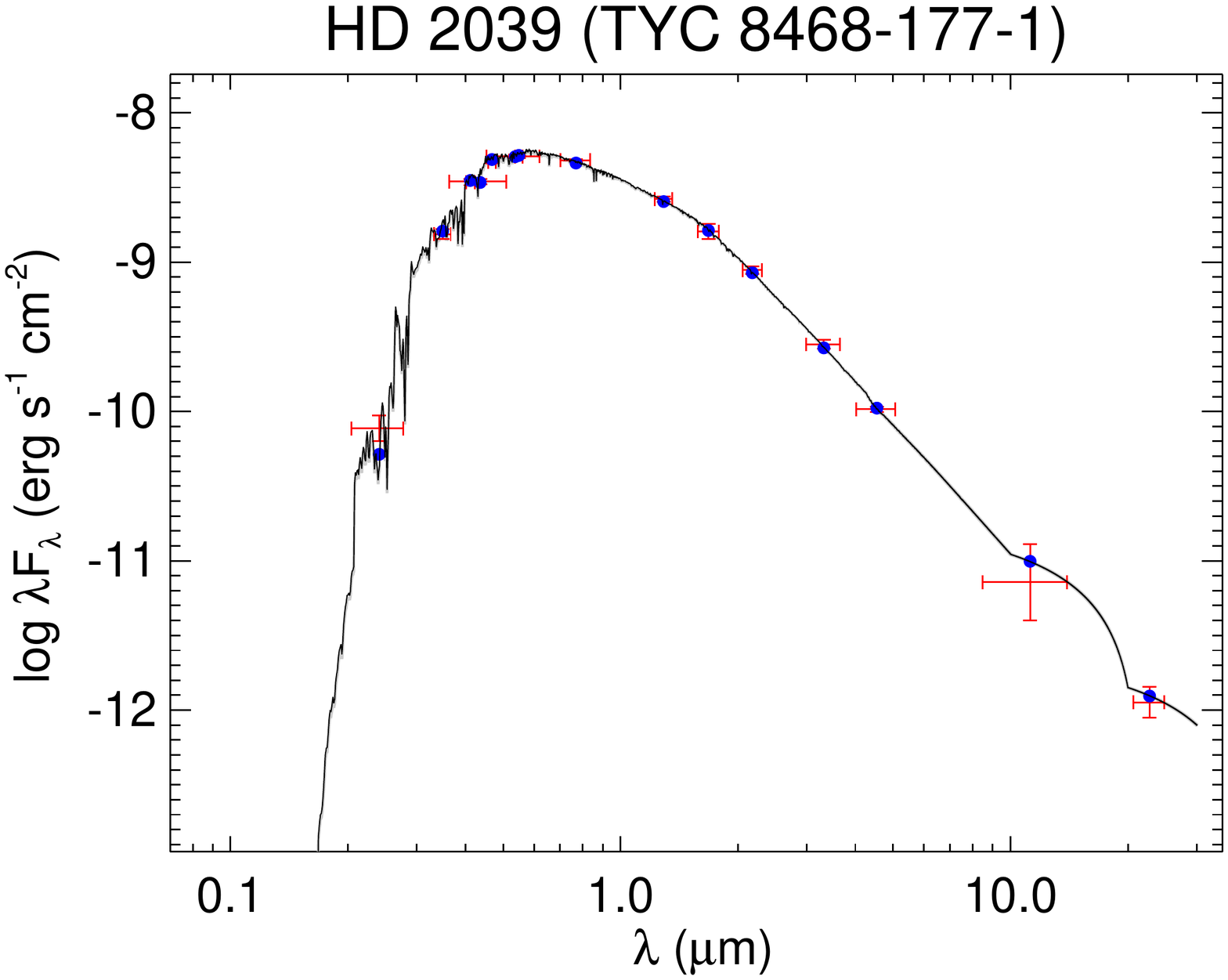}
  \includegraphics[trim=60 60 60 60,clip,width=0.49\linewidth]{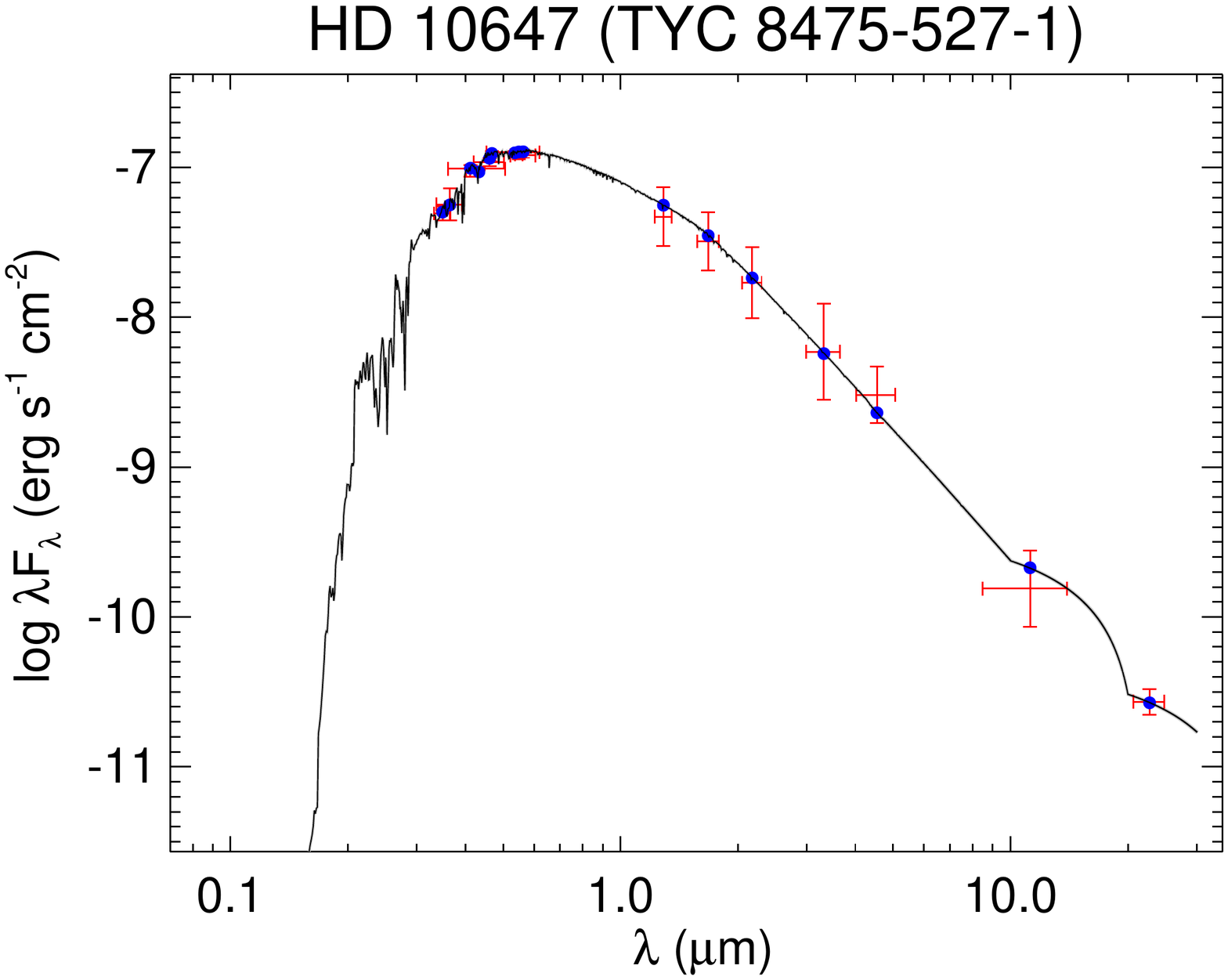}
  \includegraphics[trim=60 60 60 60,clip,width=0.49\linewidth]{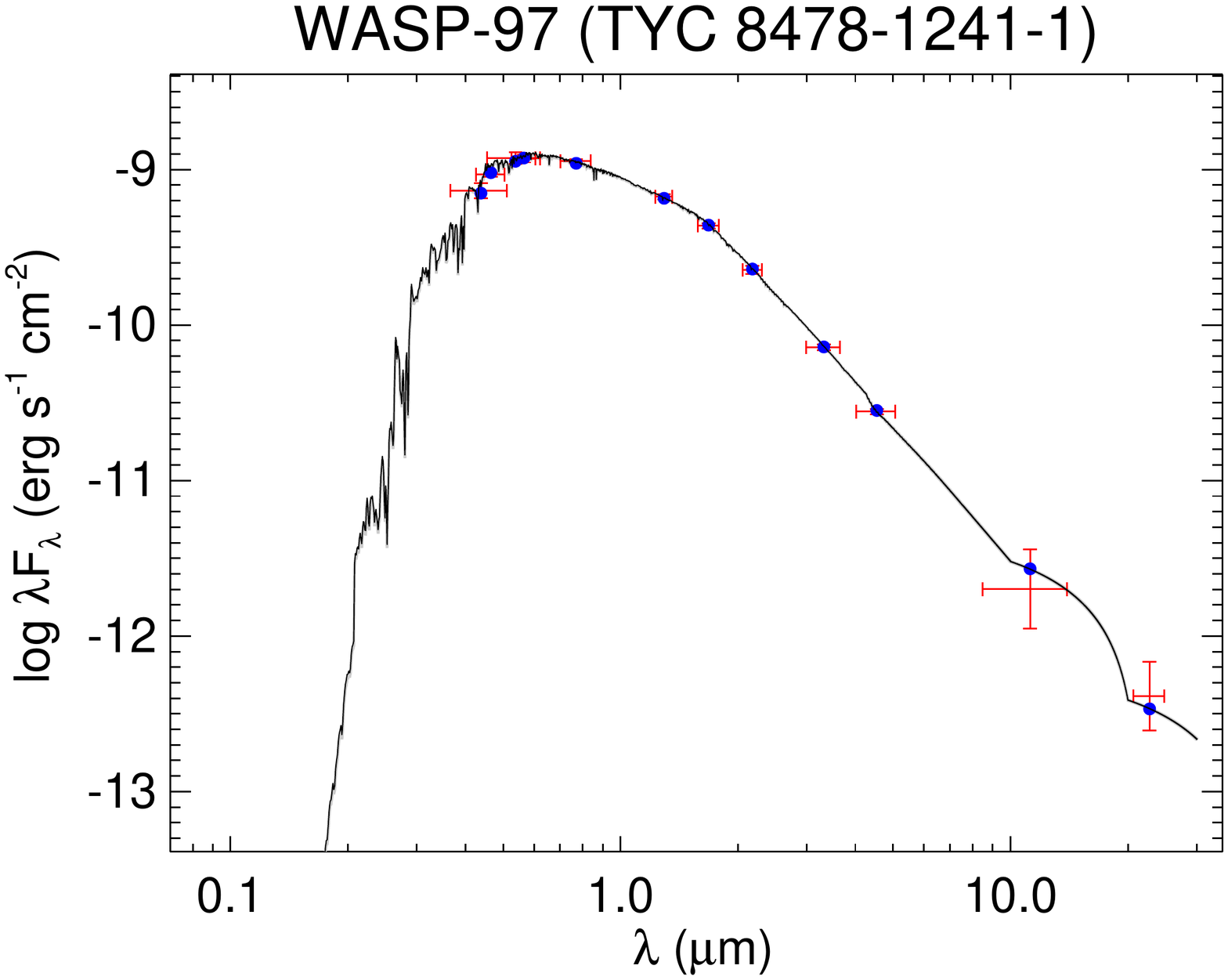}
  \caption{All labels, lines, symbols, and colors as in Figure \ref{fig:seds}.}
  \label{fig:seds_76}
\end{figure}

\begin{figure}[H]
  \centering
  \includegraphics[trim=60 60 60 60,clip,width=0.49\linewidth]{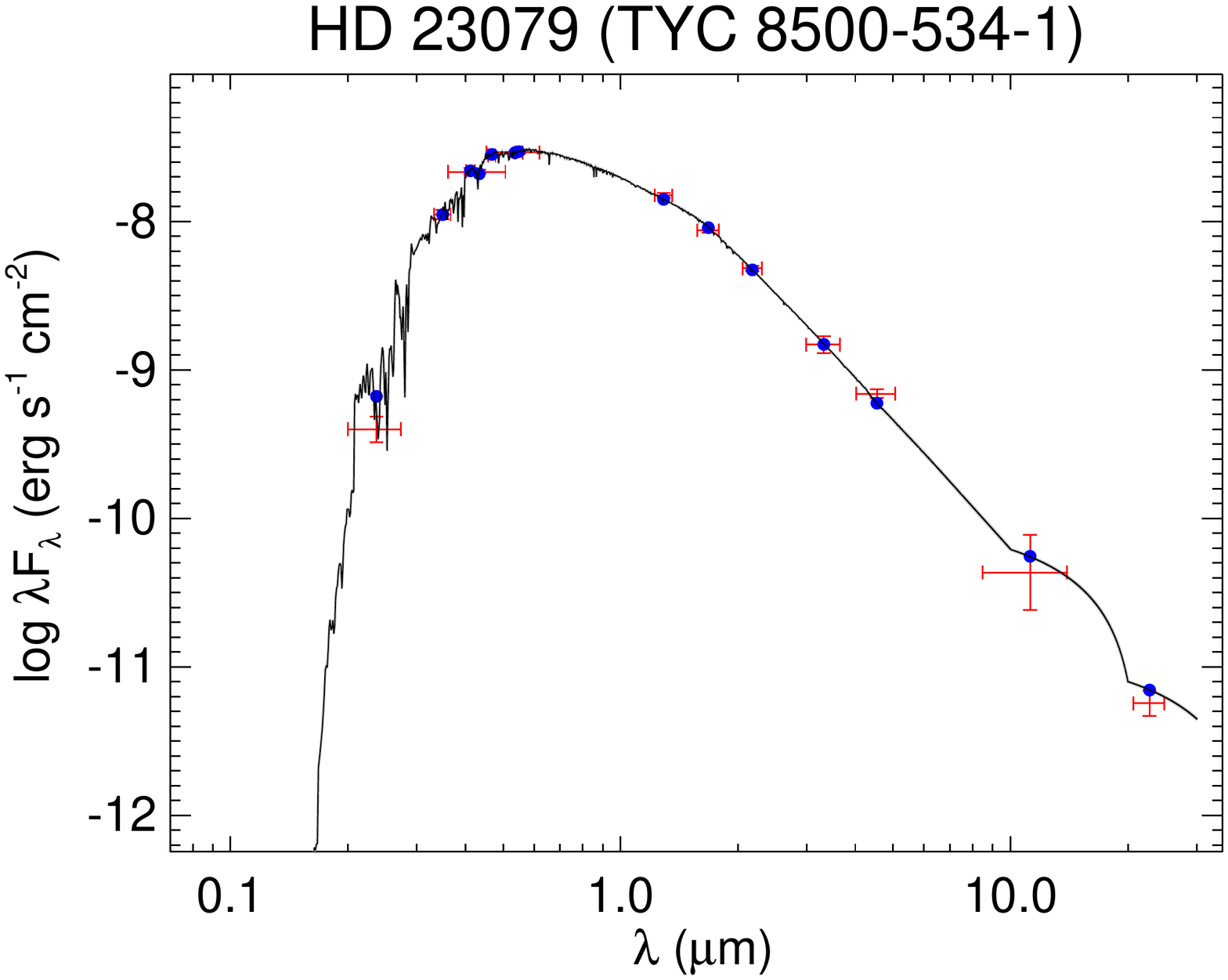}
  \includegraphics[trim=60 60 60 60,clip,width=0.49\linewidth]{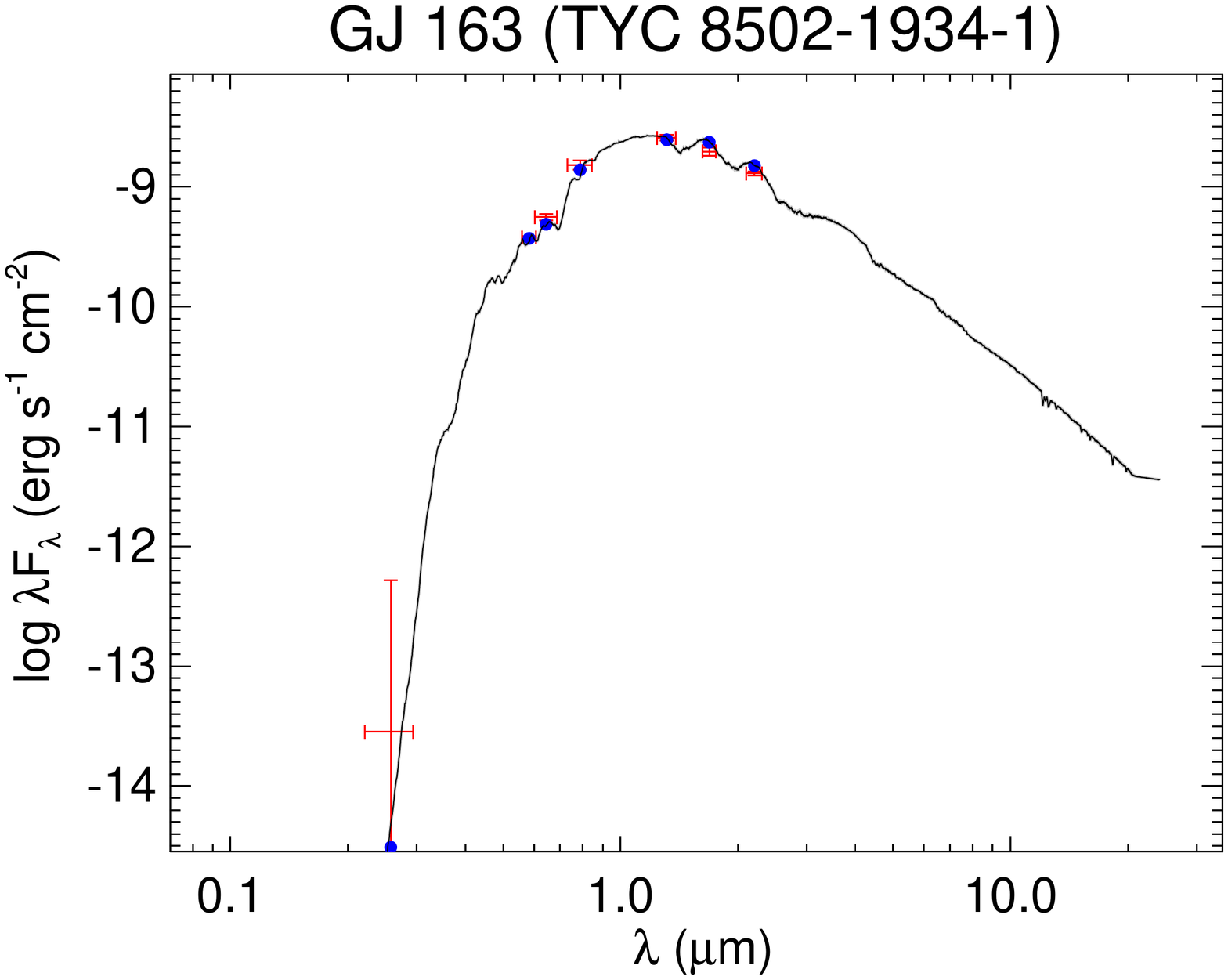}
  \includegraphics[trim=60 60 60 60,clip,width=0.49\linewidth]{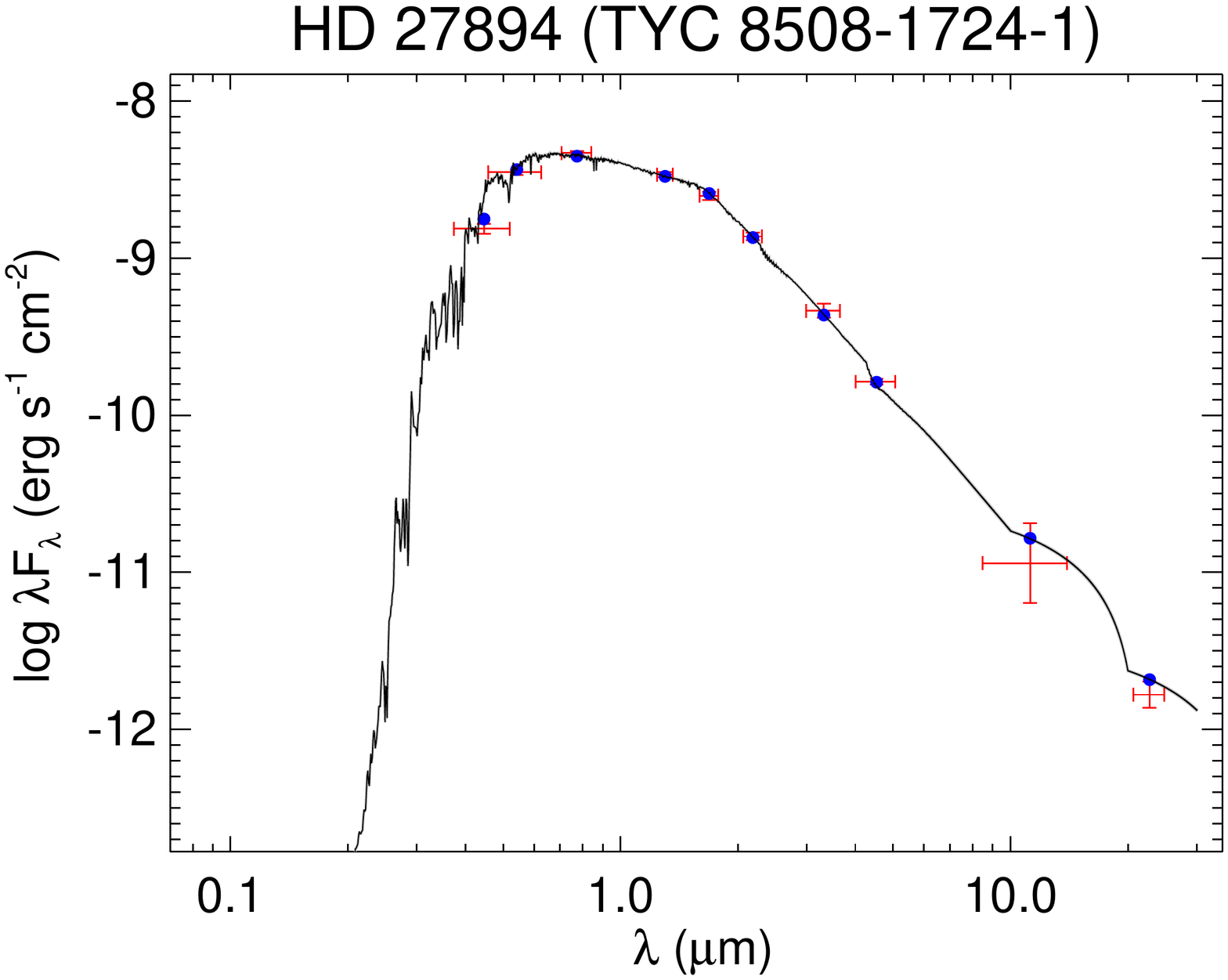}
  \includegraphics[trim=60 60 60 60,clip,width=0.49\linewidth]{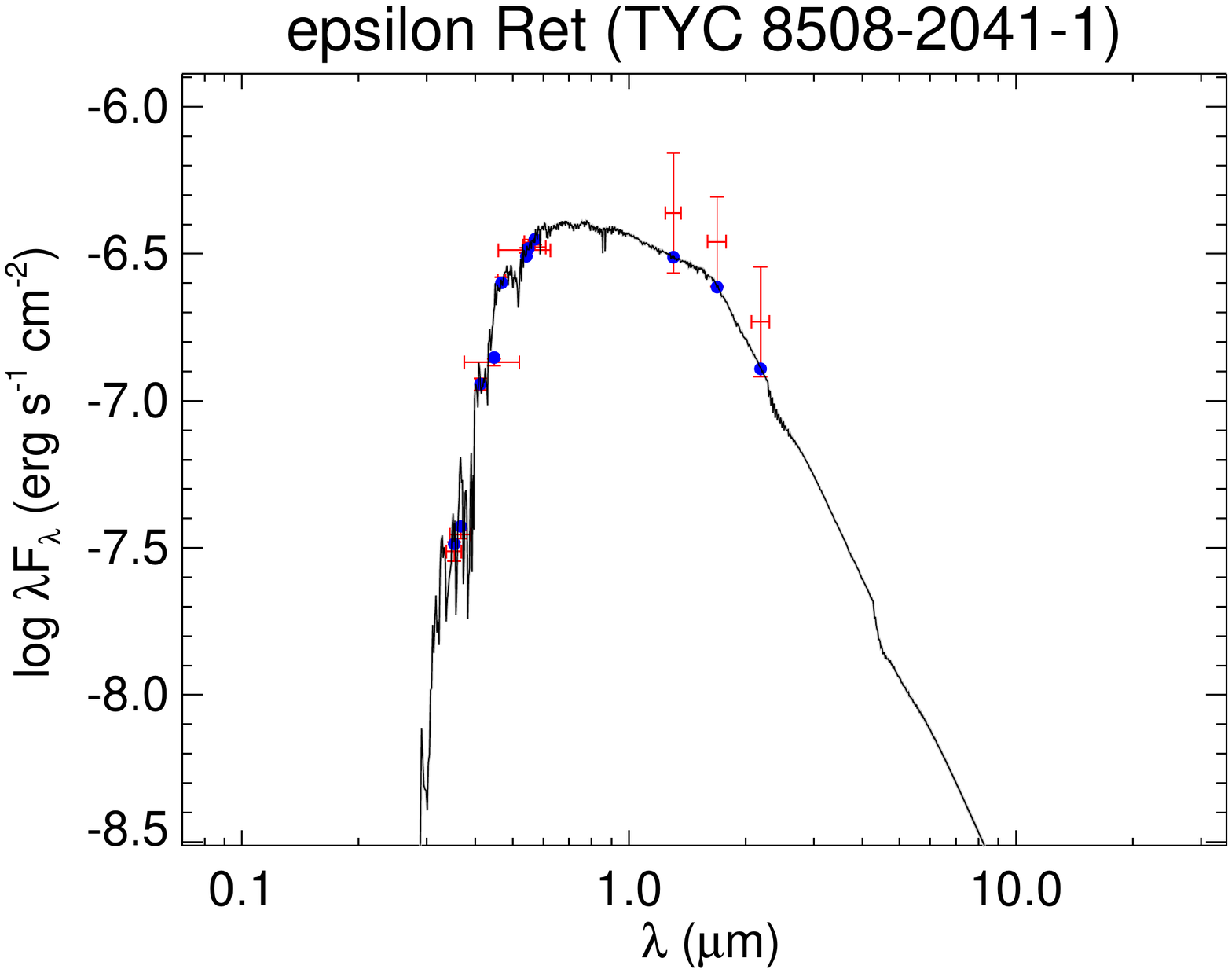}
  \includegraphics[trim=60 60 60 60,clip,width=0.49\linewidth]{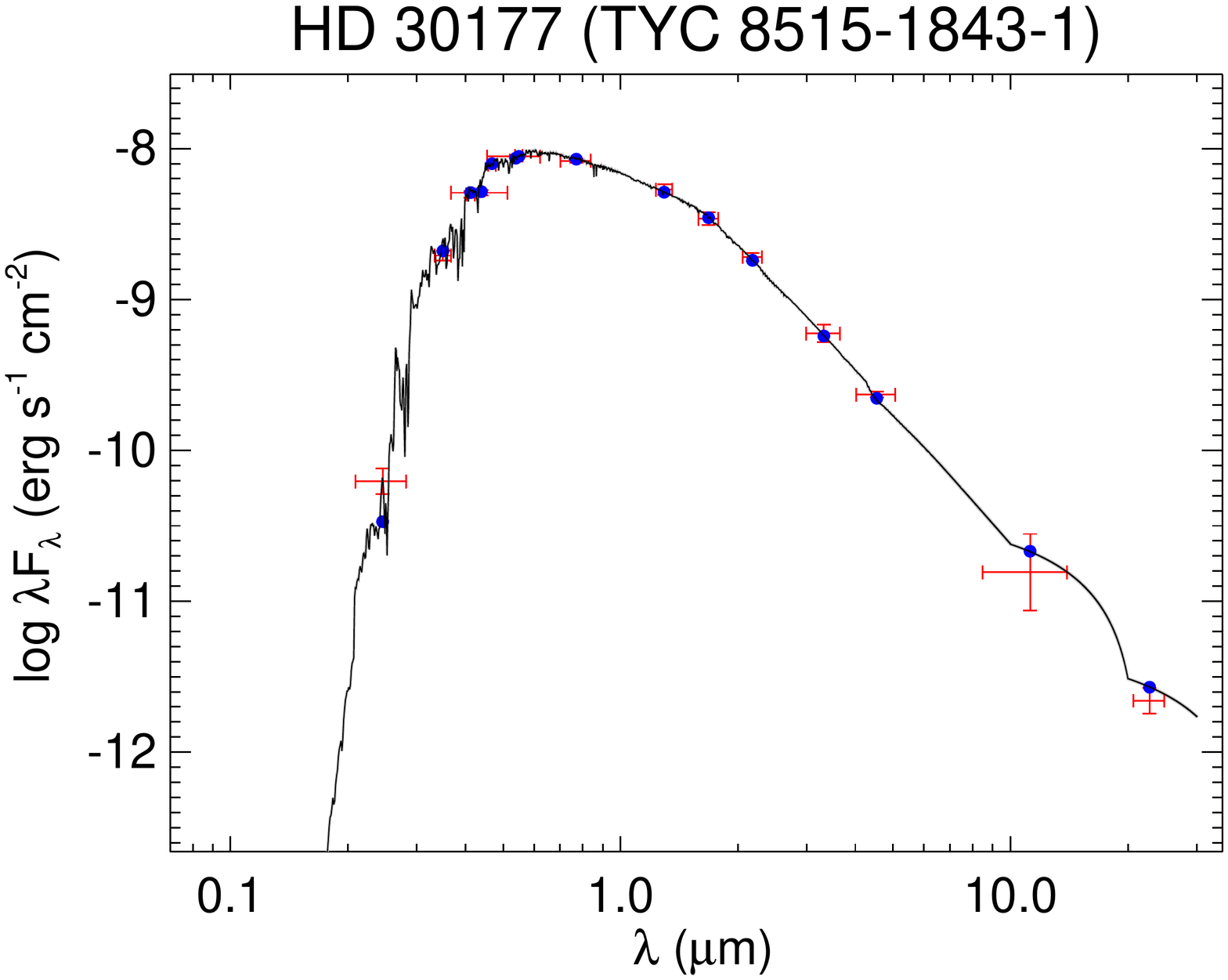}
  \includegraphics[trim=60 60 60 60,clip,width=0.49\linewidth]{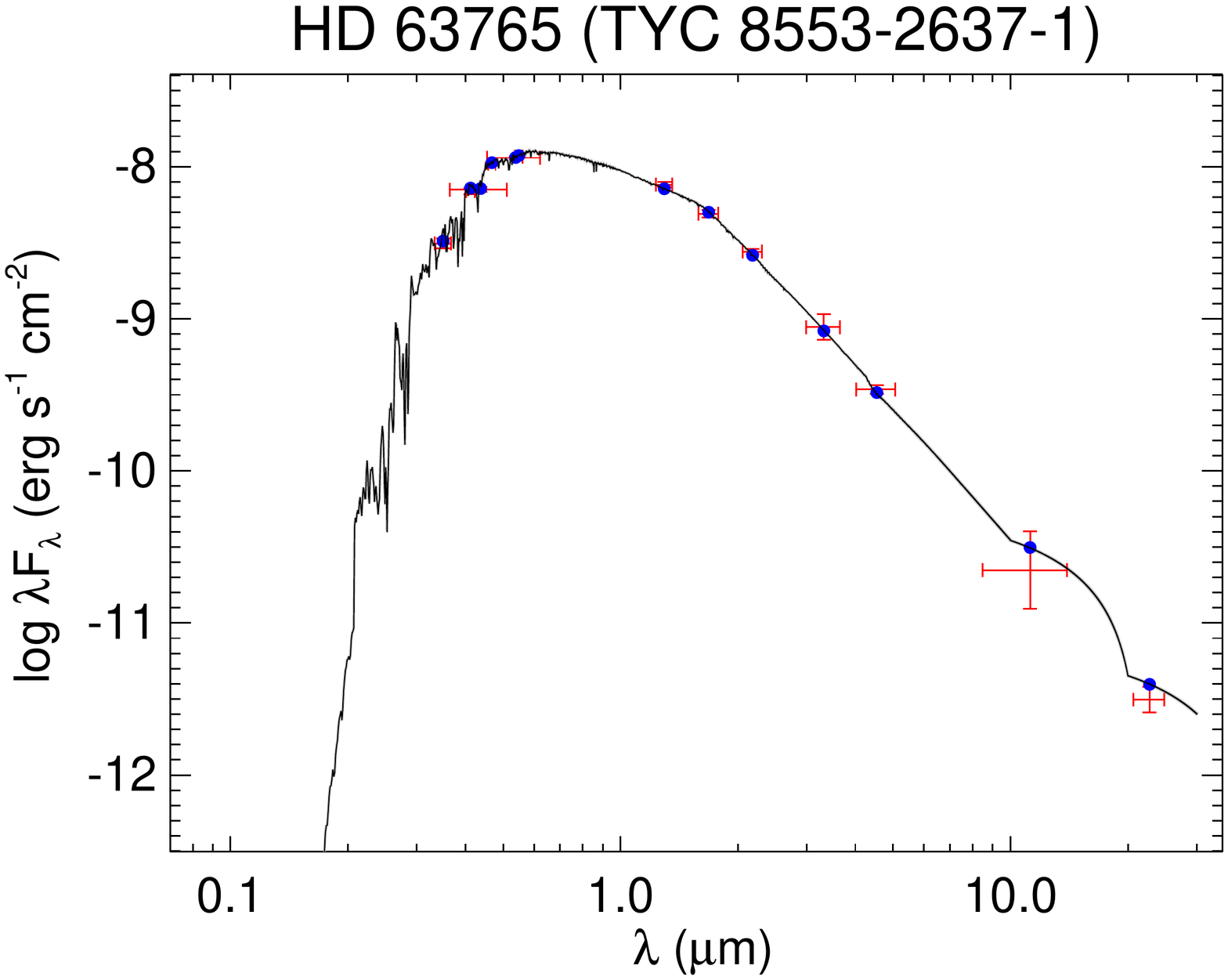}
  \caption{All labels, lines, symbols, and colors as in Figure \ref{fig:seds}.}
  \label{fig:seds_77}
\end{figure}

\begin{figure}[H]
  \centering
  \includegraphics[trim=60 60 60 60,clip,width=0.49\linewidth]{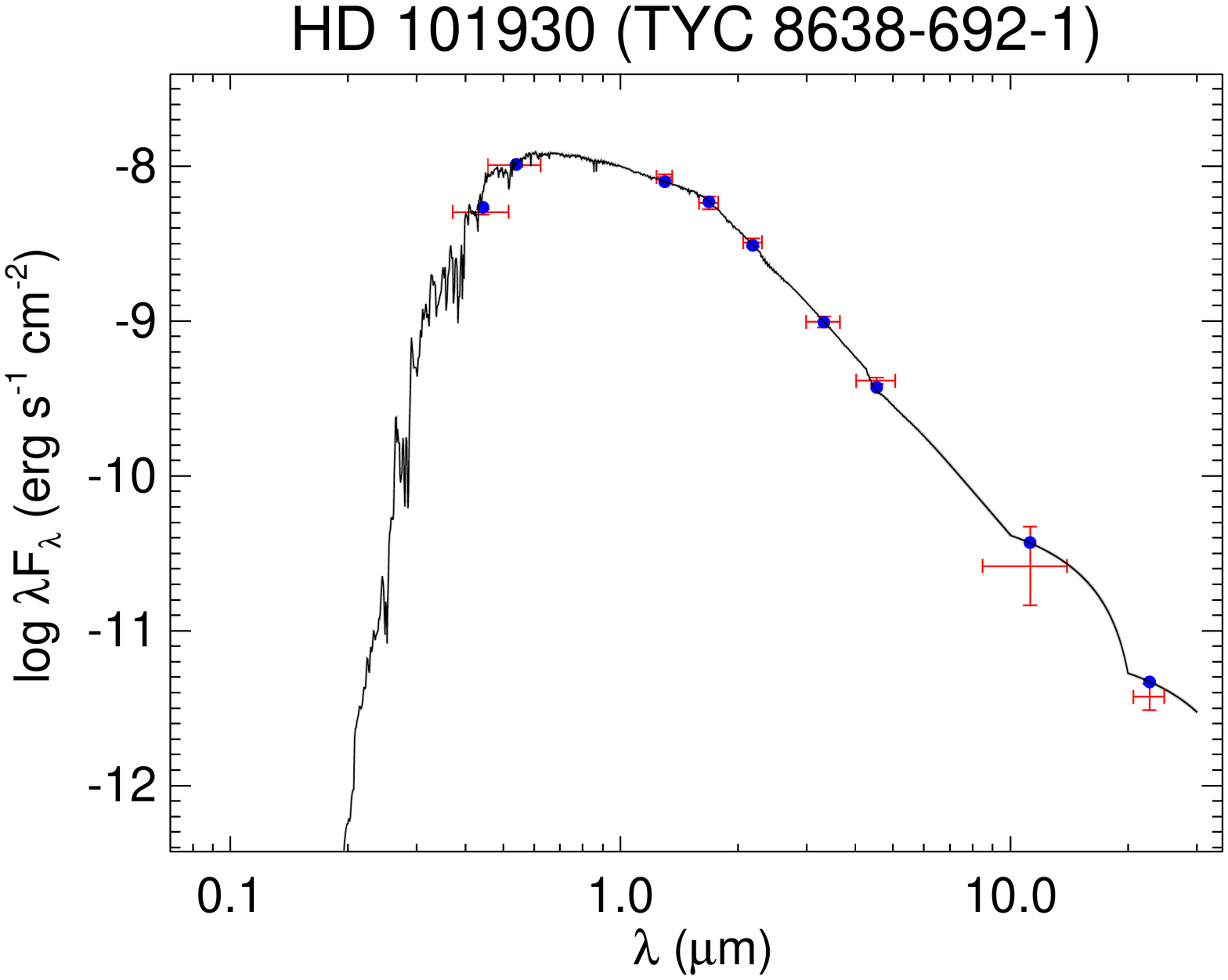}
  \includegraphics[trim=60 60 60 60,clip,width=0.49\linewidth]{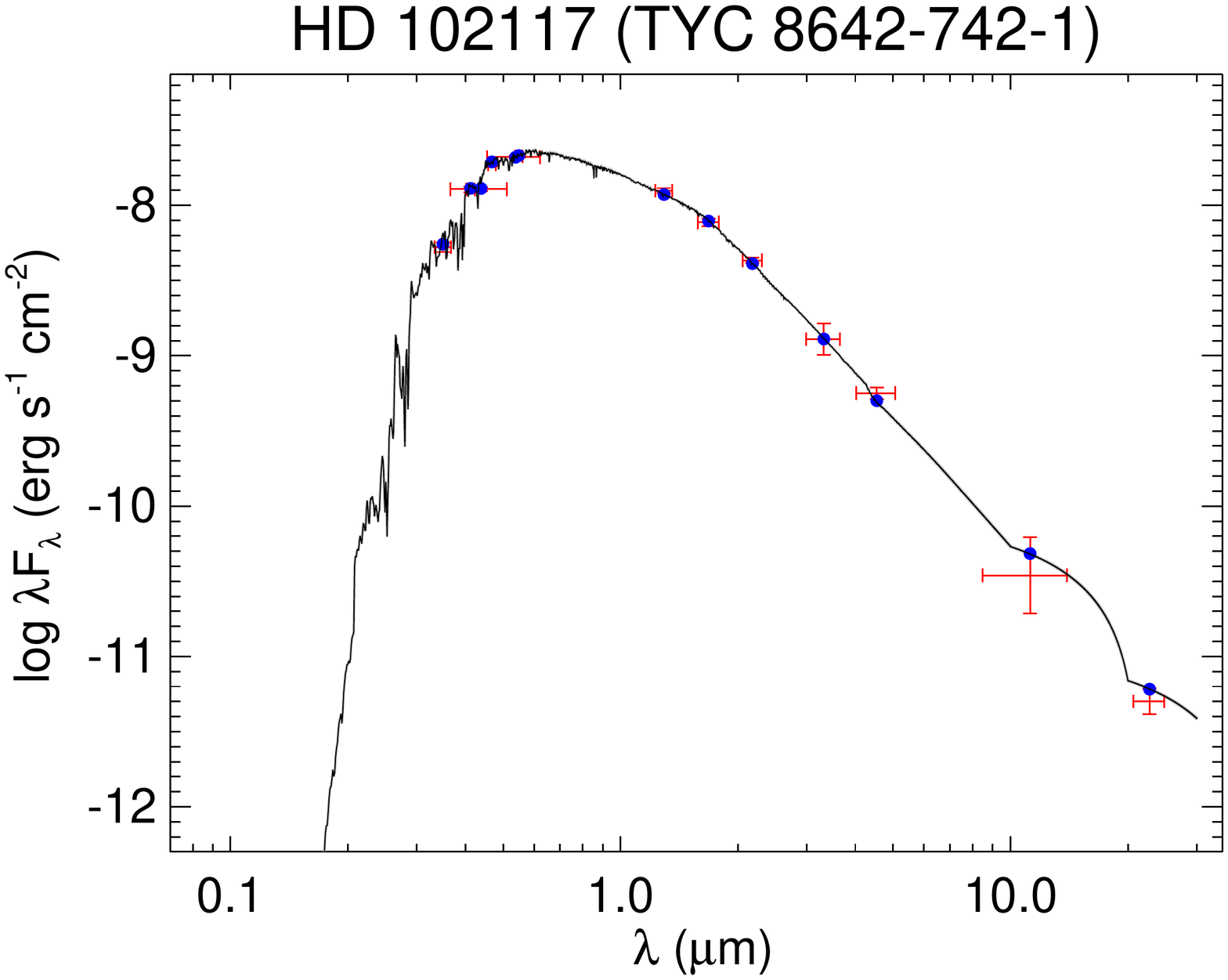}
  \includegraphics[trim=60 60 60 60,clip,width=0.49\linewidth]{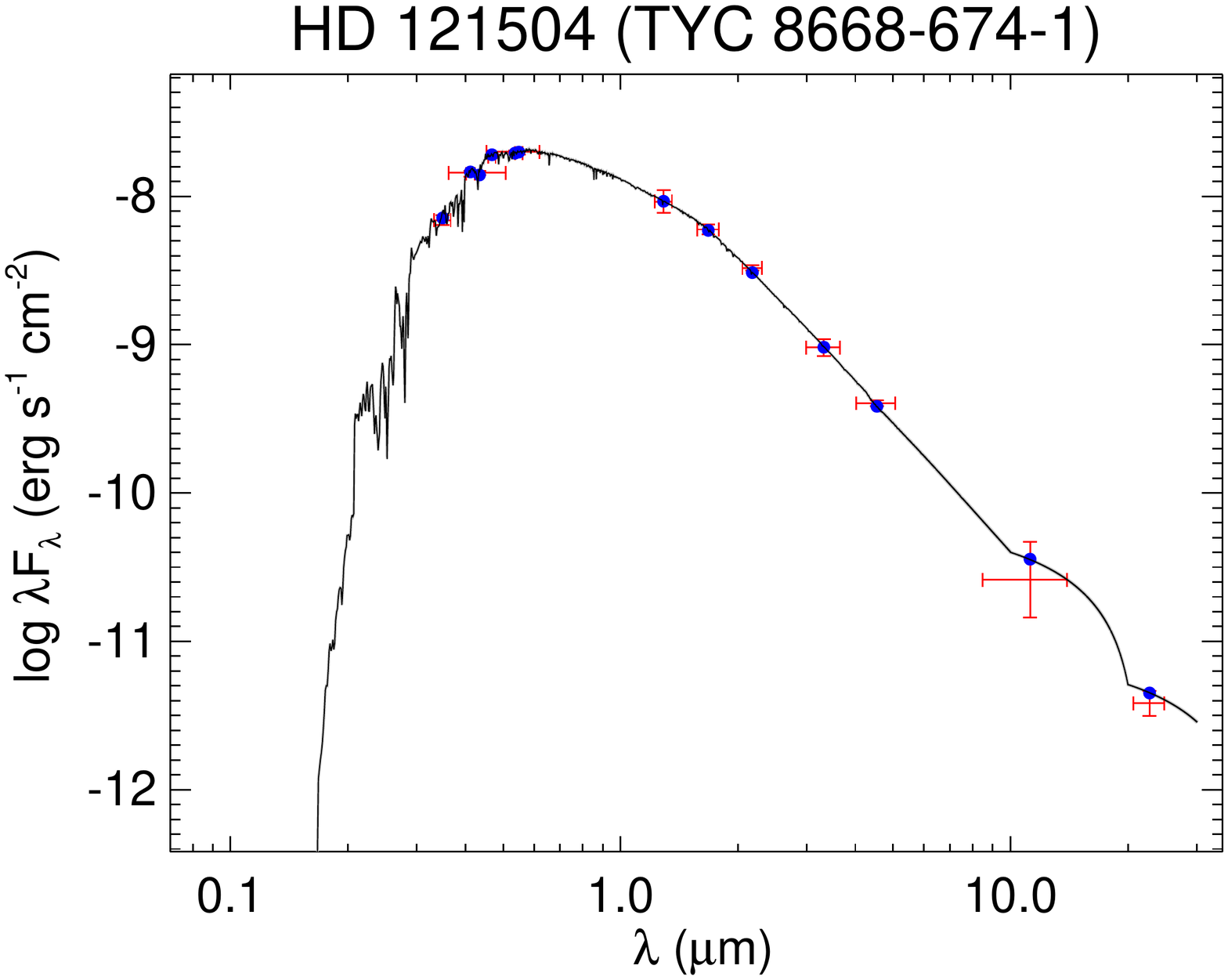}
  \includegraphics[trim=60 60 60 60,clip,width=0.49\linewidth]{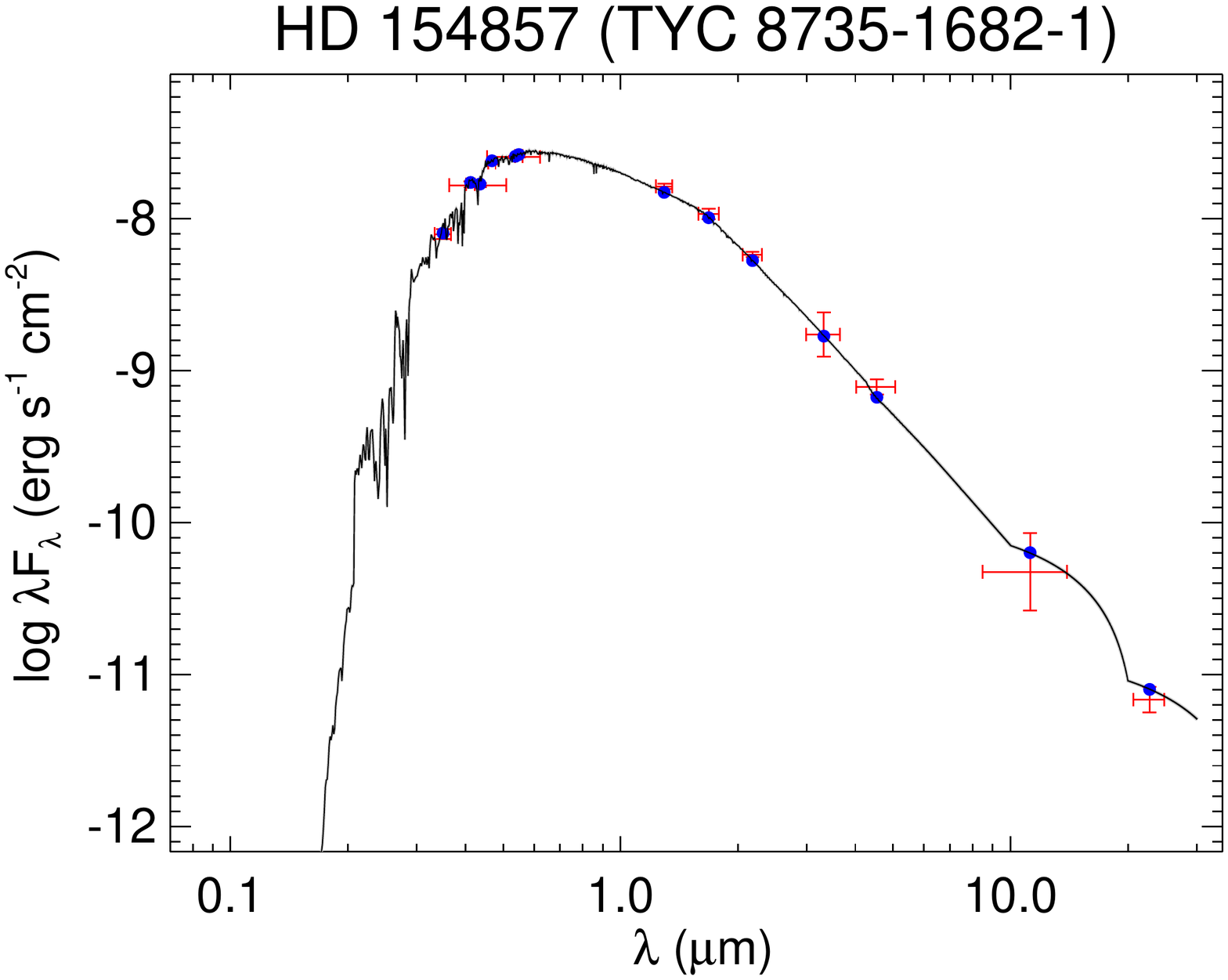}
  \includegraphics[trim=60 60 60 60,clip,width=0.49\linewidth]{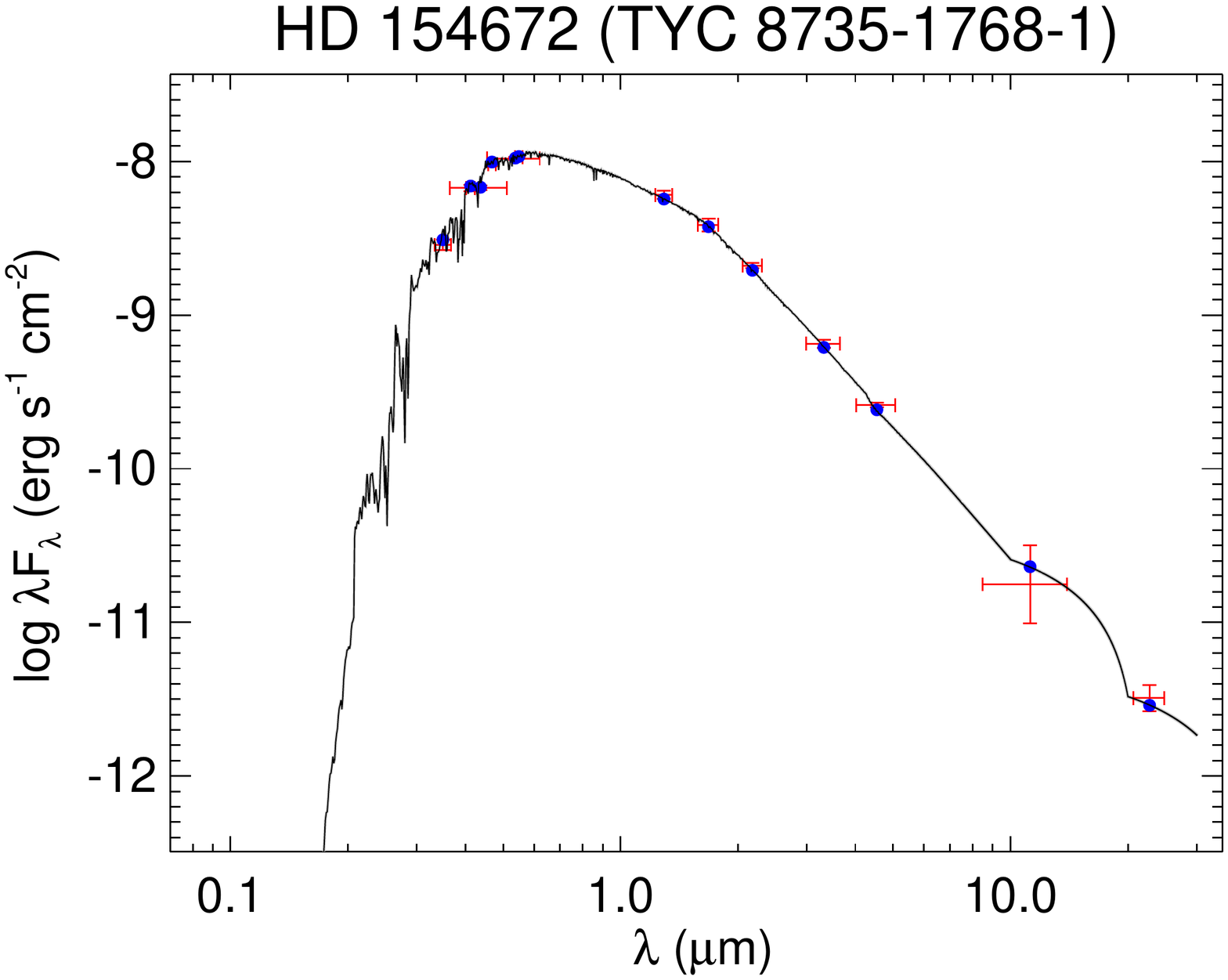}
  \includegraphics[trim=60 60 60 60,clip,width=0.49\linewidth]{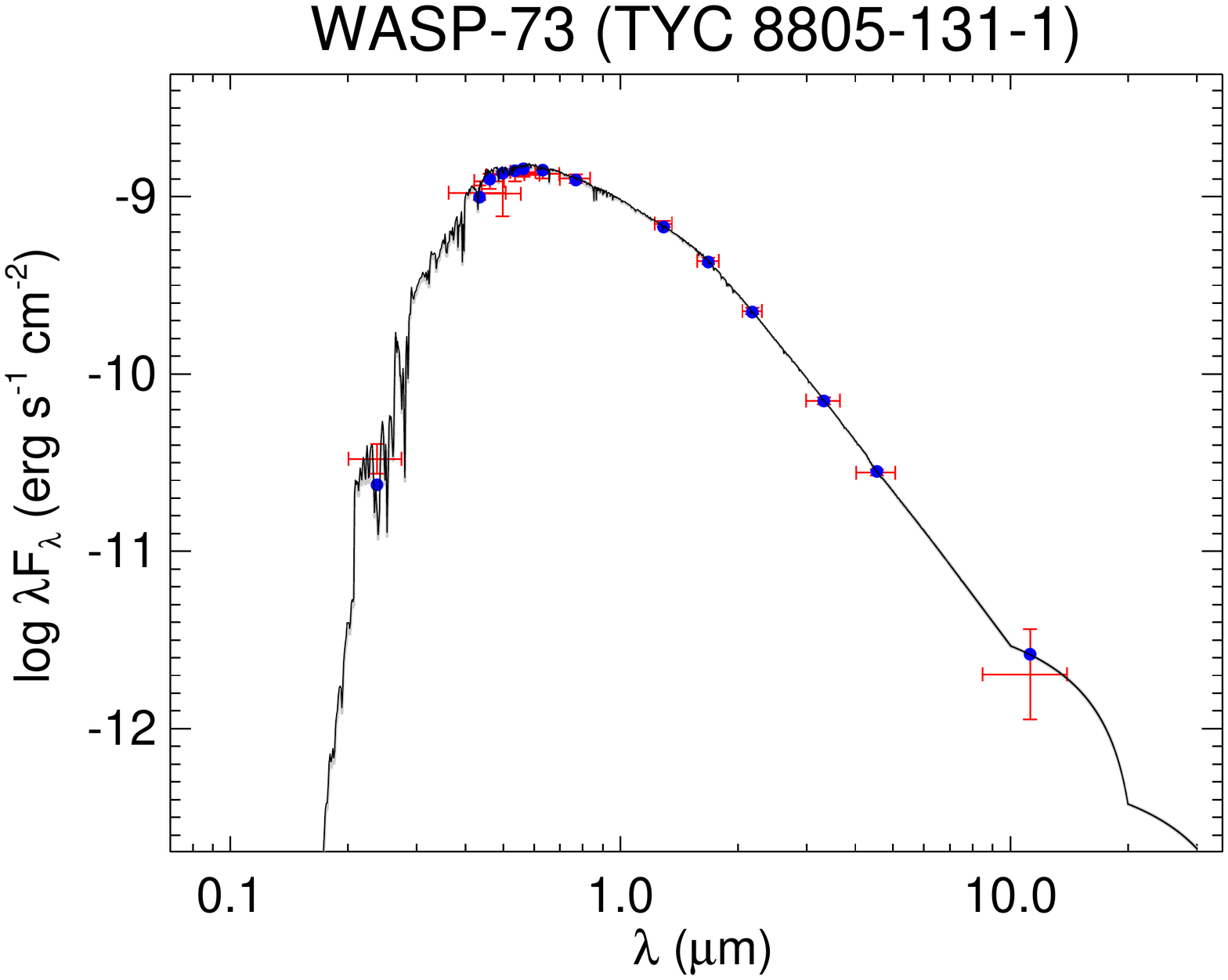}
  \caption{All labels, lines, symbols, and colors as in Figure \ref{fig:seds}.}
  \label{fig:seds_78}
\end{figure}

\begin{figure}[H]
  \centering
  \includegraphics[trim=60 60 60 60,clip,width=0.49\linewidth]{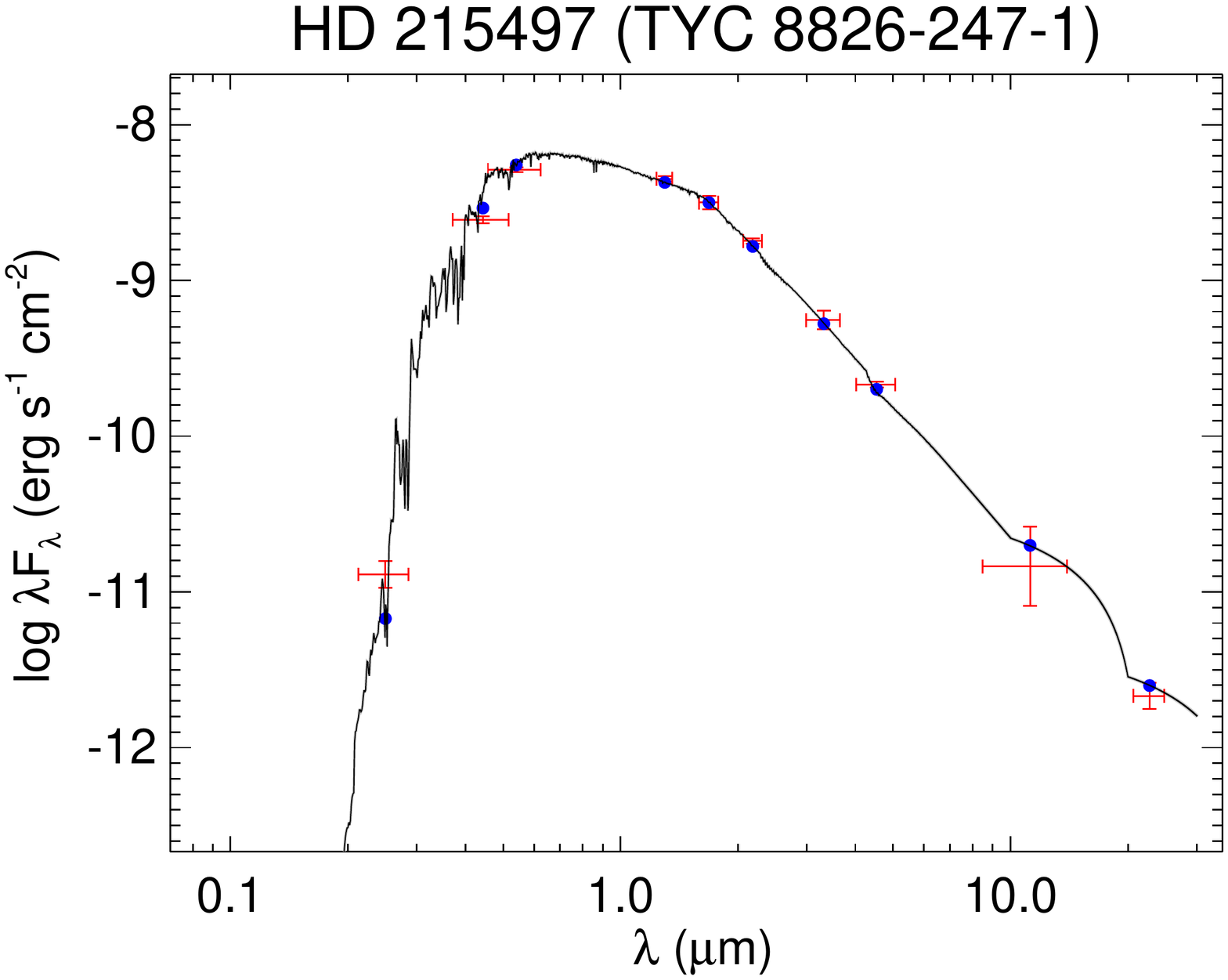}
  \includegraphics[trim=60 60 60 60,clip,width=0.49\linewidth]{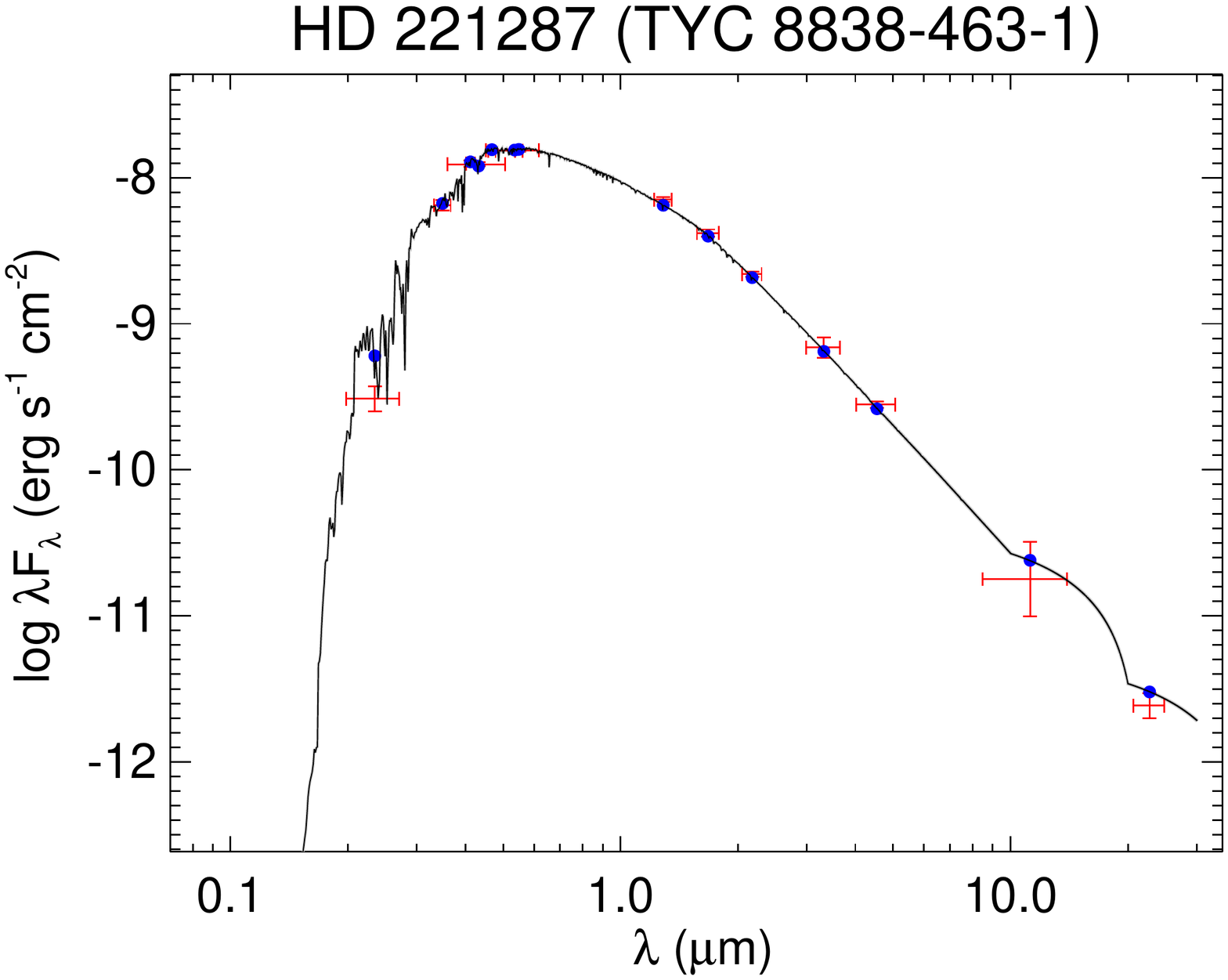}
  \includegraphics[trim=60 60 60 60,clip,width=0.49\linewidth]{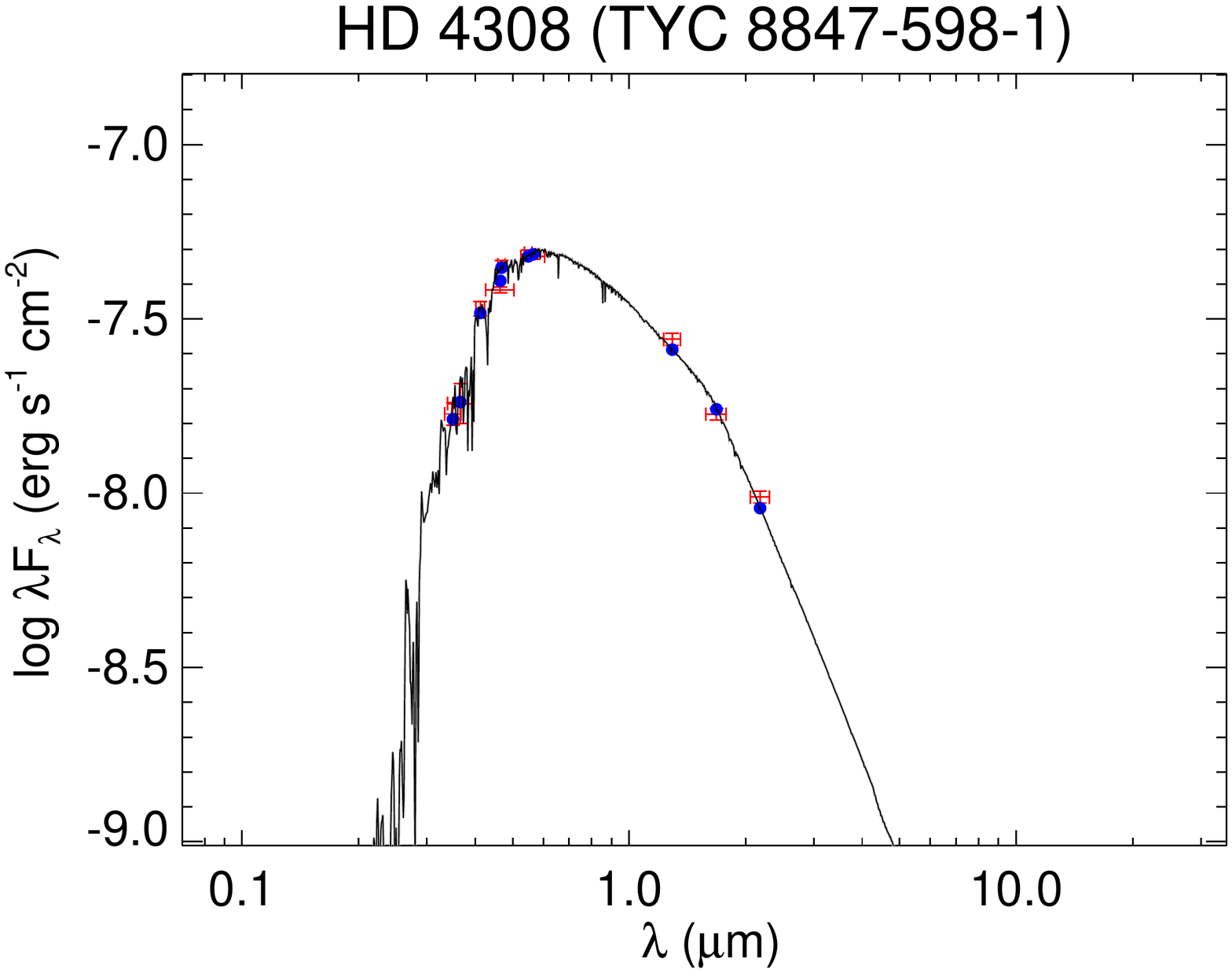}
  \includegraphics[trim=60 60 60 60,clip,width=0.49\linewidth]{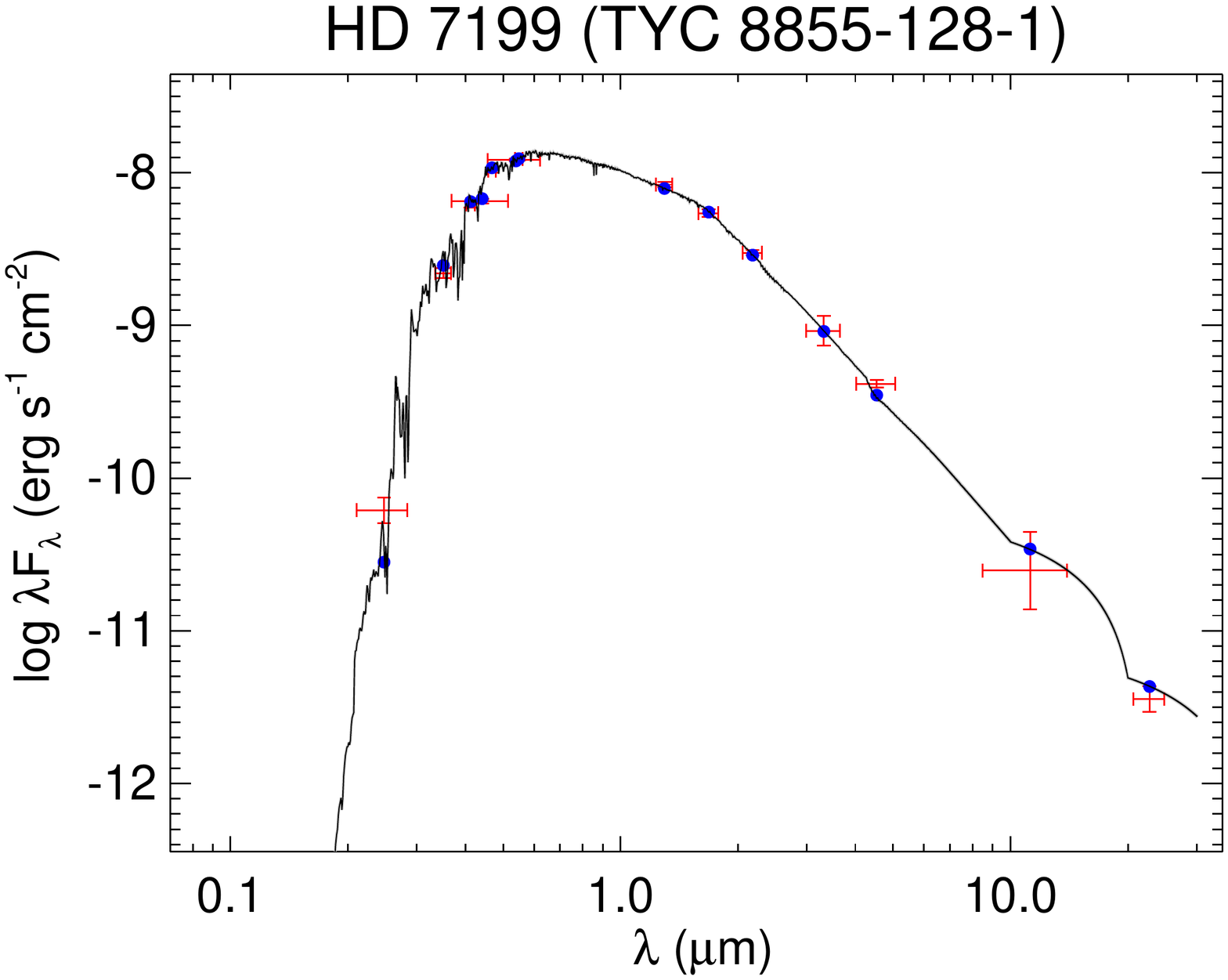}
  \includegraphics[trim=60 60 60 60,clip,width=0.49\linewidth]{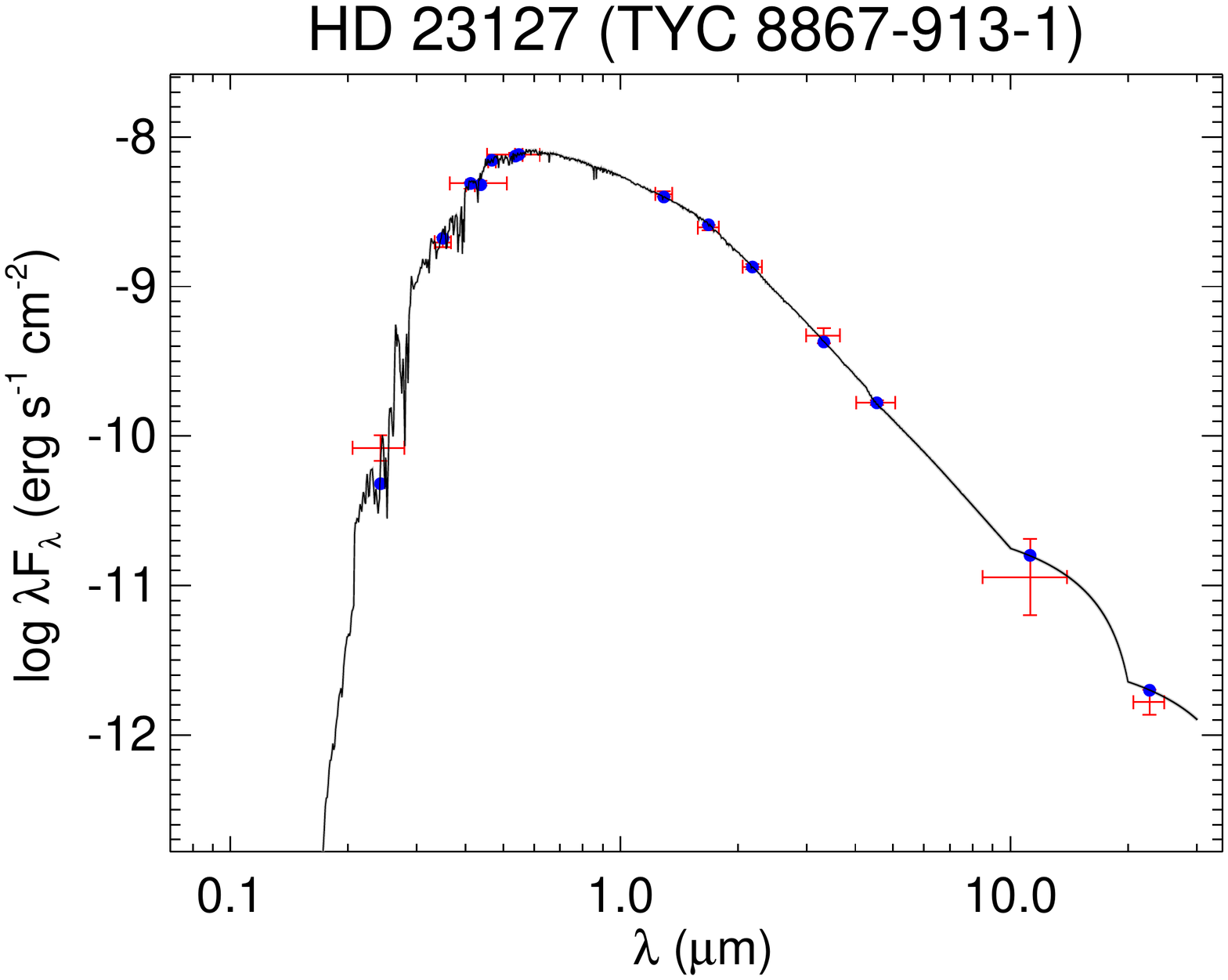}
  \includegraphics[trim=60 60 60 60,clip,width=0.49\linewidth]{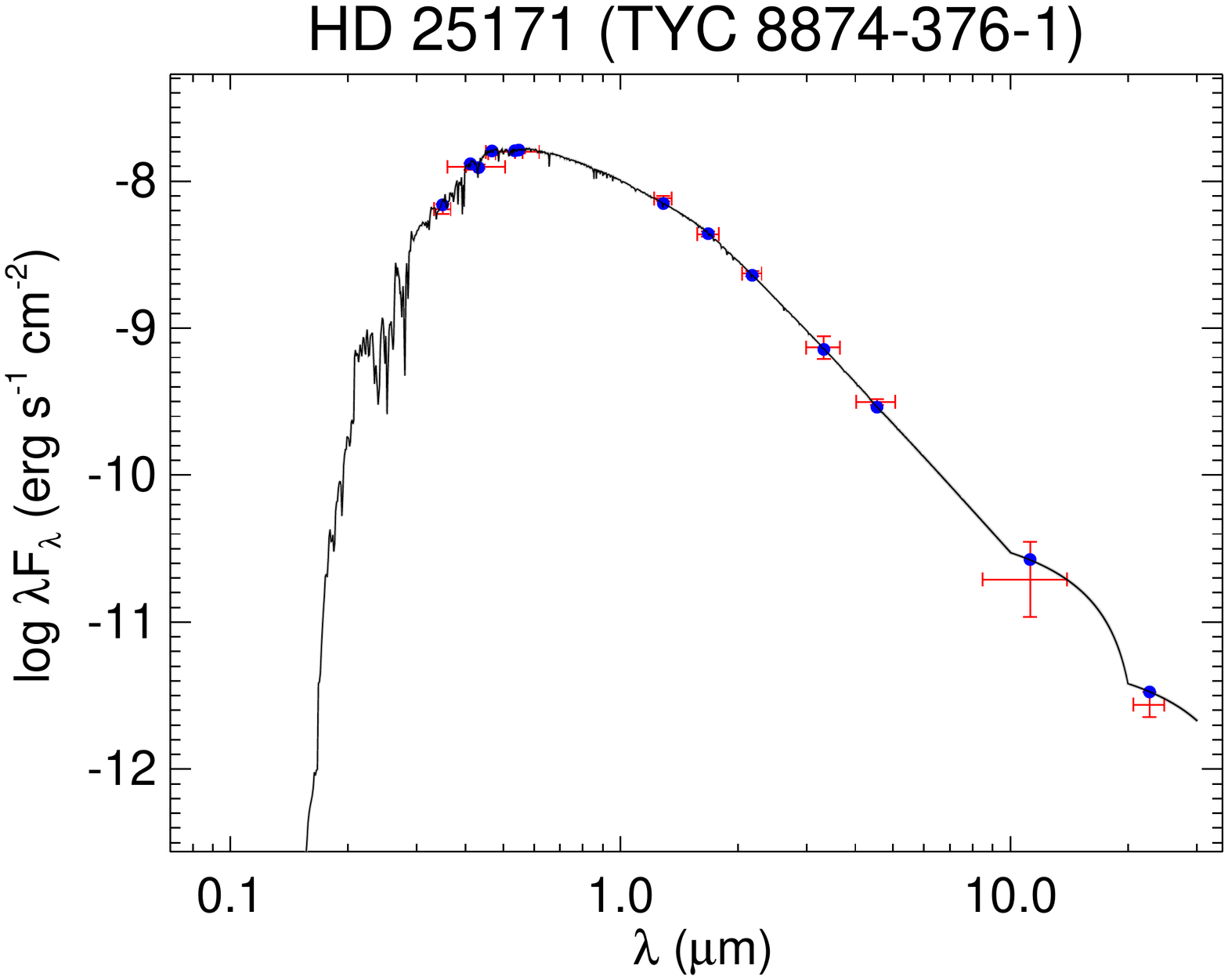}
  \caption{All labels, lines, symbols, and colors as in Figure \ref{fig:seds}.}
  \label{fig:seds_79}
\end{figure}

\begin{figure}[H]
  \centering
  \includegraphics[trim=60 60 60 60,clip,width=0.49\linewidth]{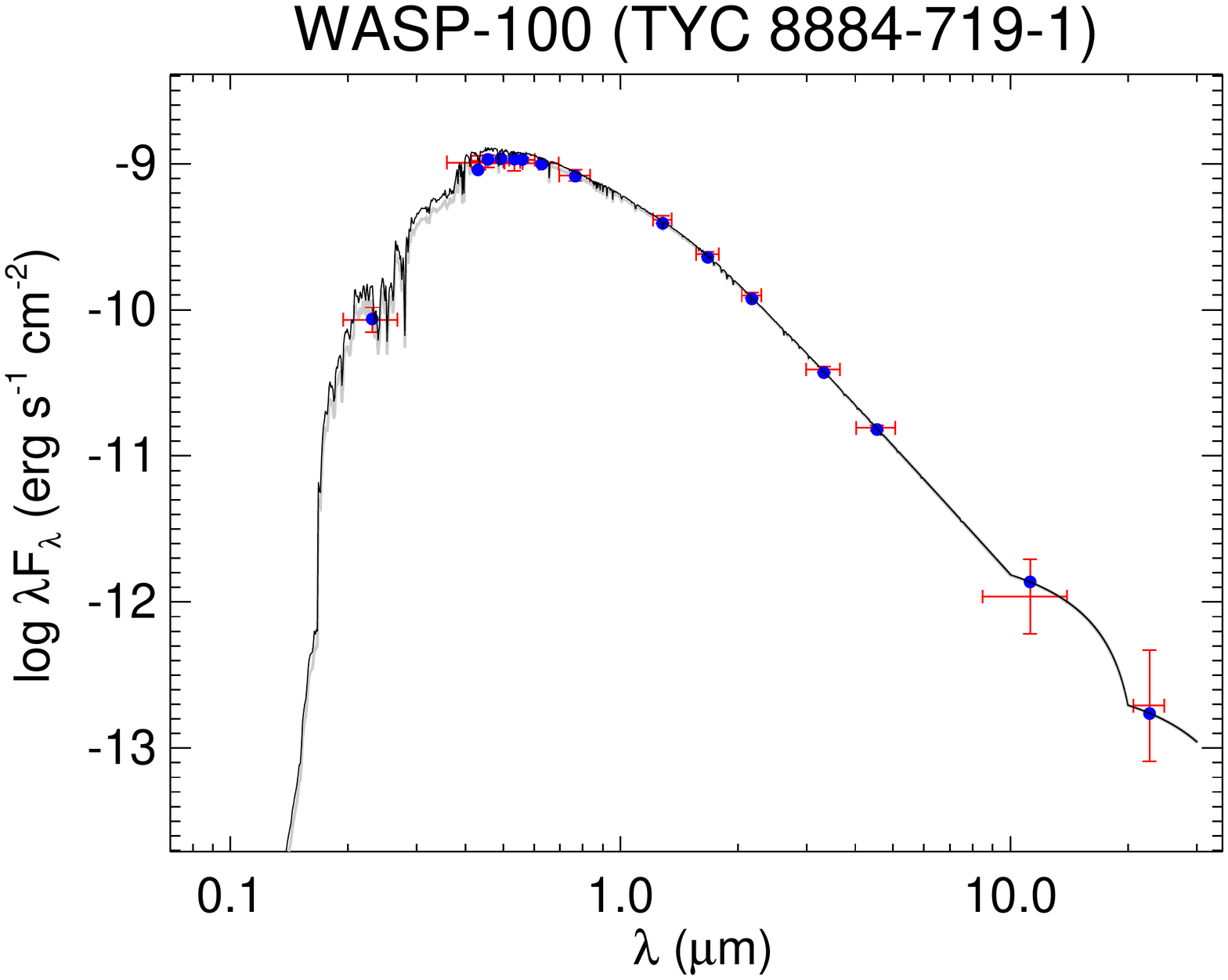}
  \includegraphics[trim=60 60 60 60,clip,width=0.49\linewidth]{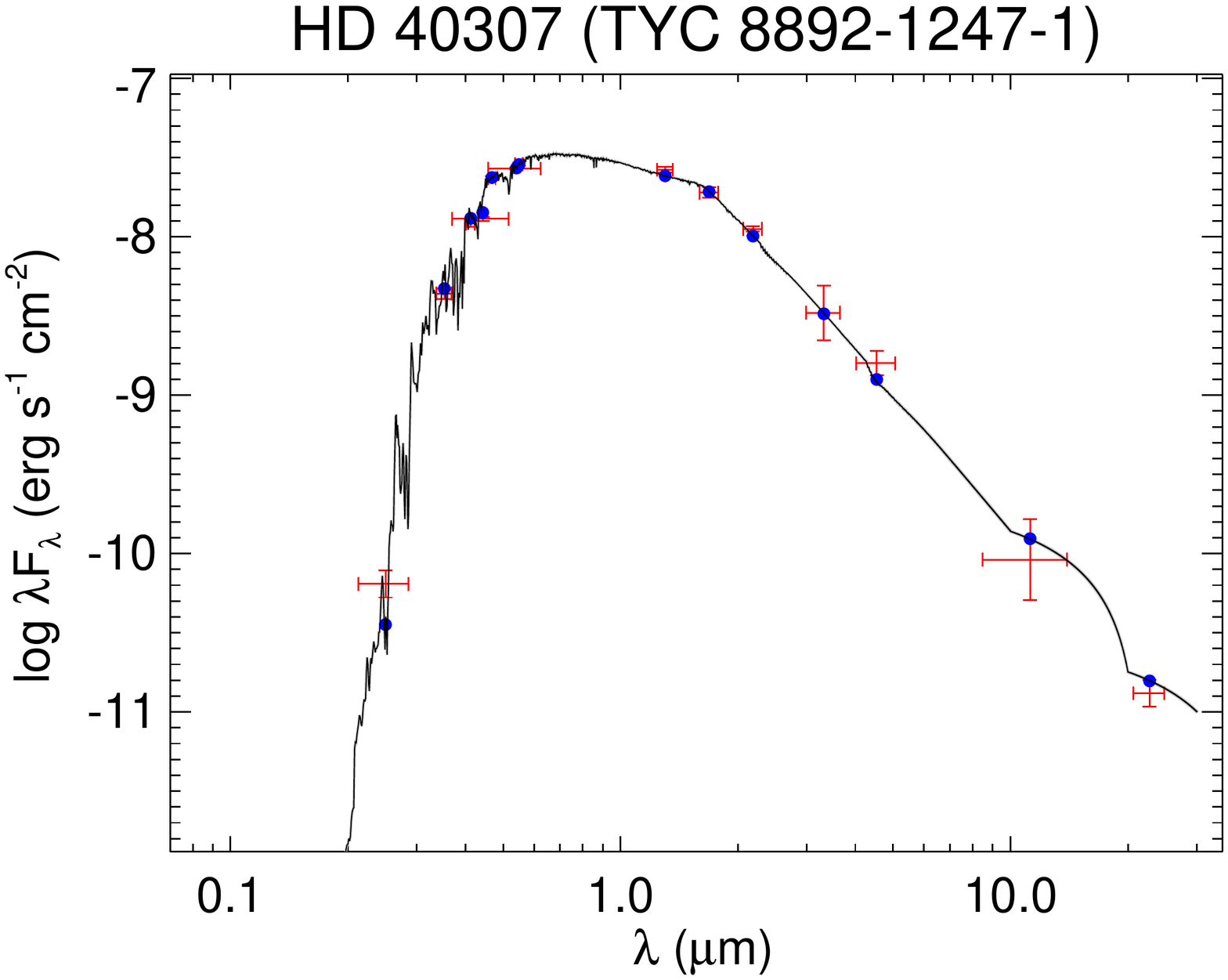}
  \includegraphics[trim=60 60 60 60,clip,width=0.49\linewidth]{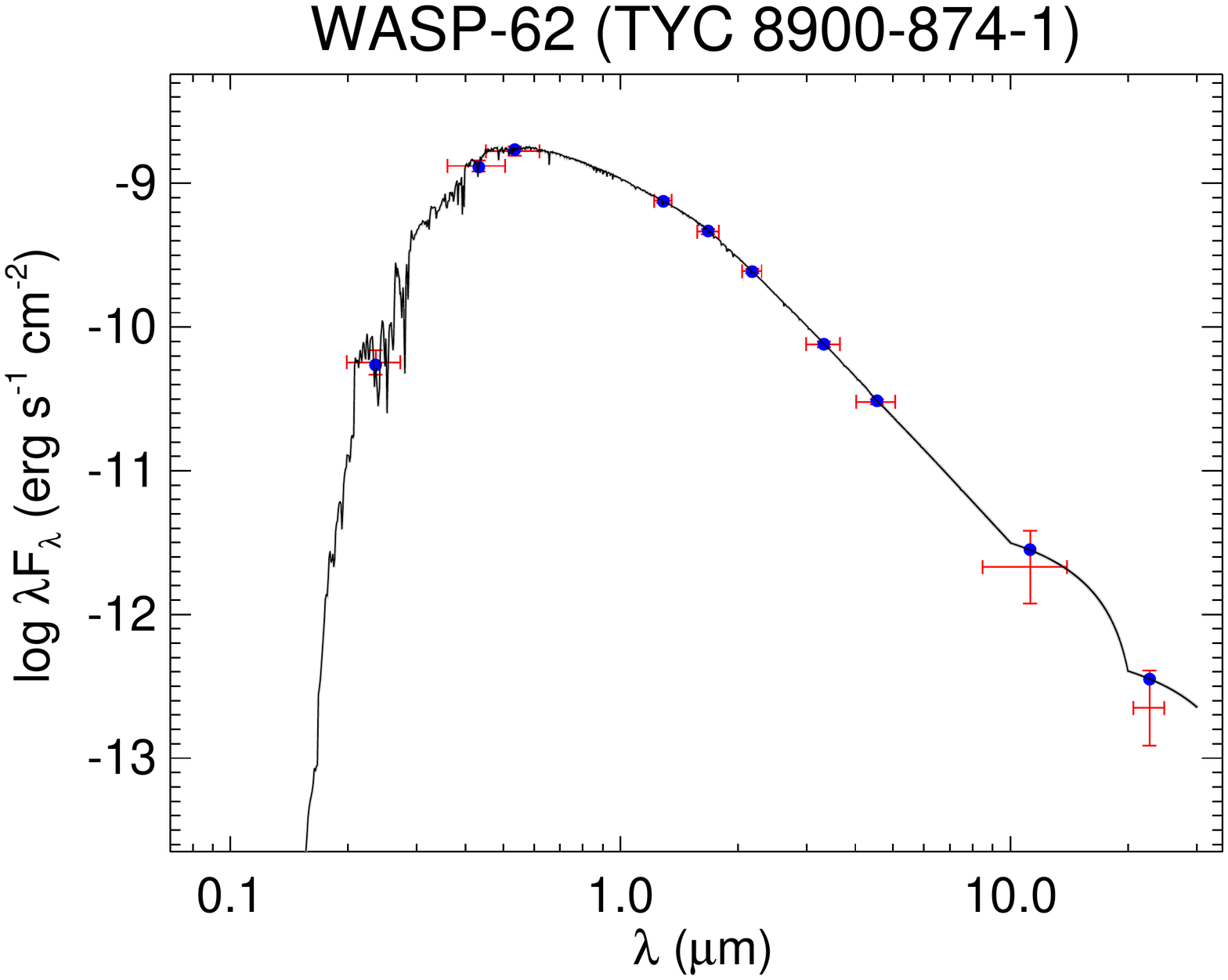}
  \includegraphics[trim=60 60 60 60,clip,width=0.49\linewidth]{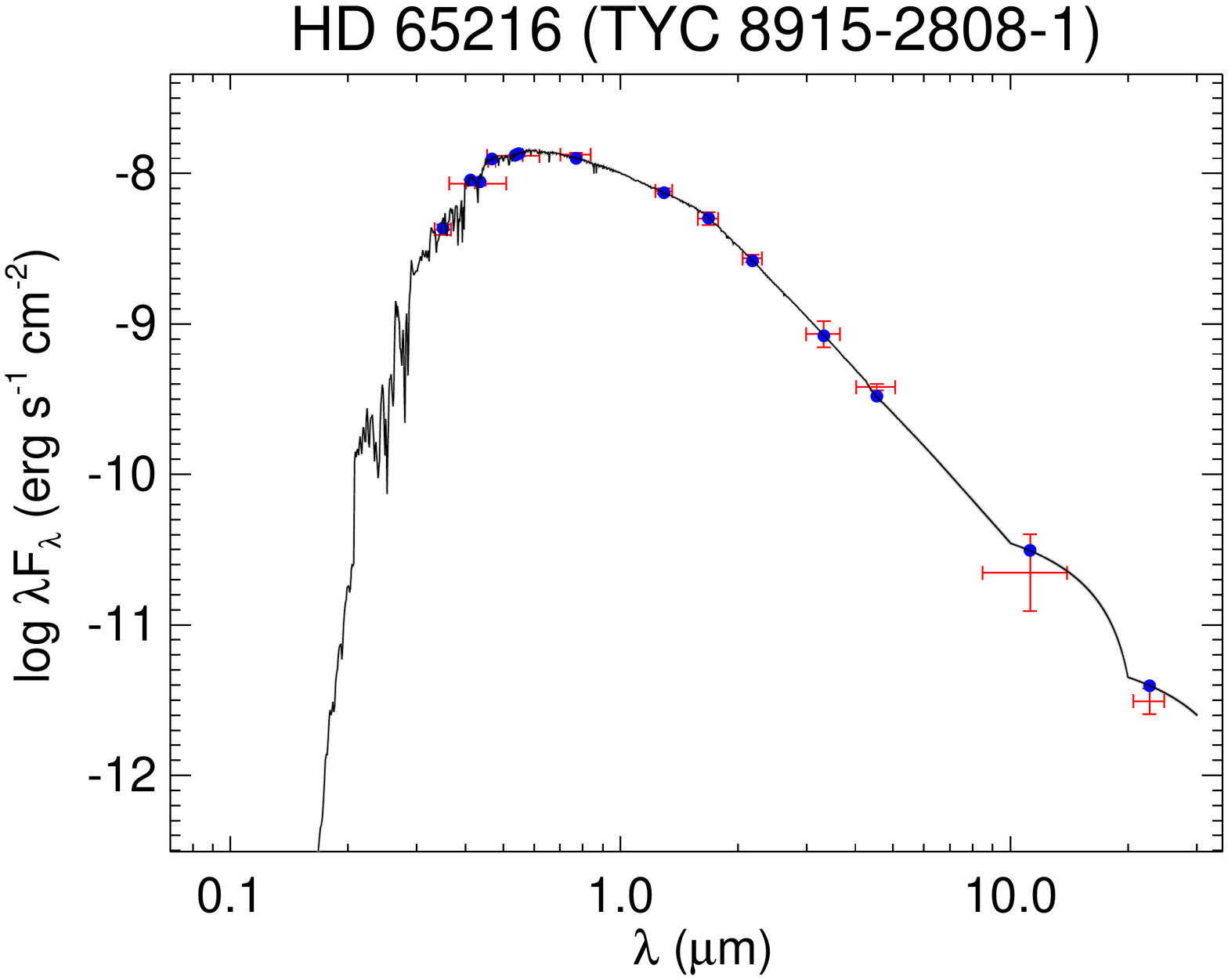}
  \includegraphics[trim=60 60 60 60,clip,width=0.49\linewidth]{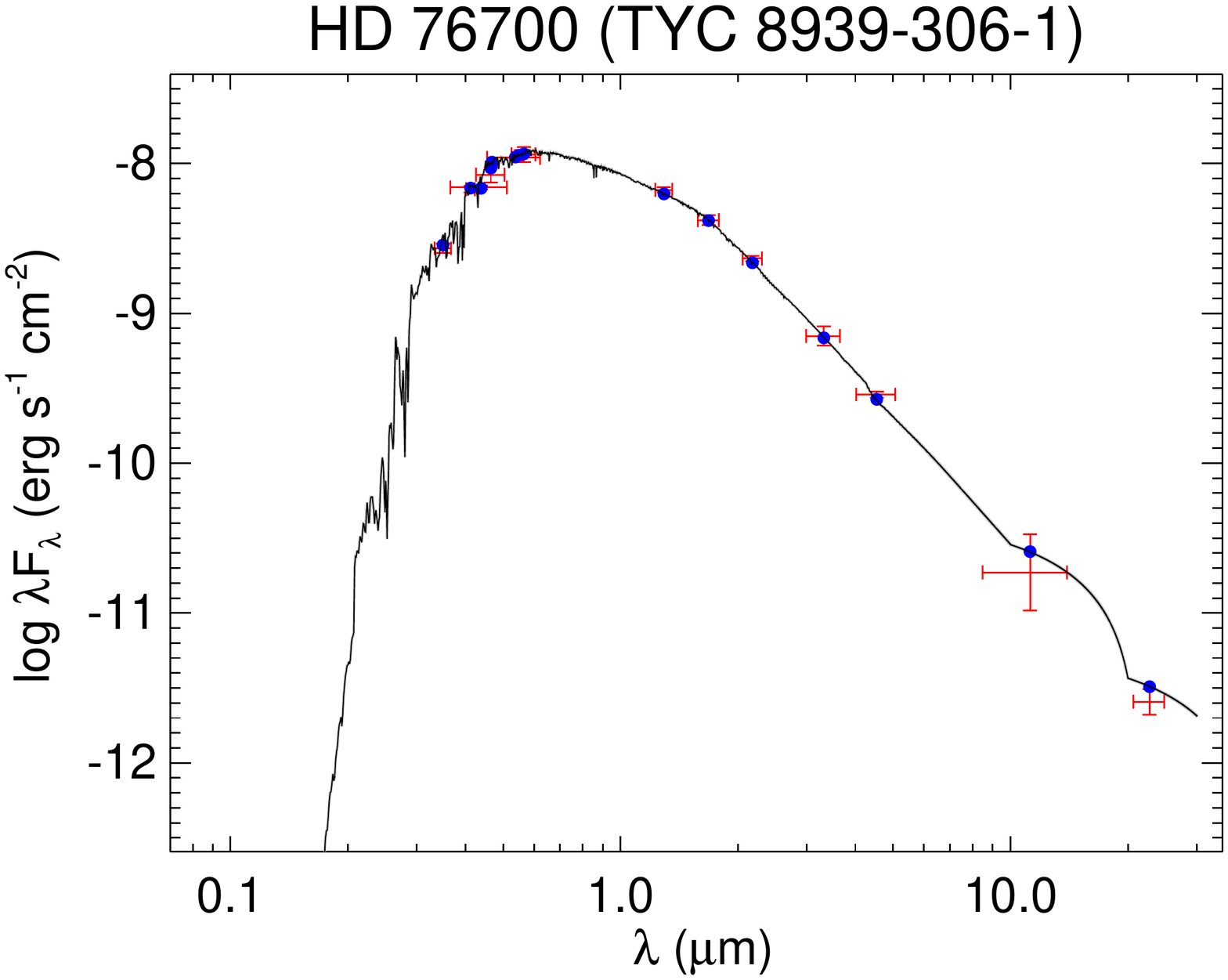}
  \includegraphics[trim=60 60 60 60,clip,width=0.49\linewidth]{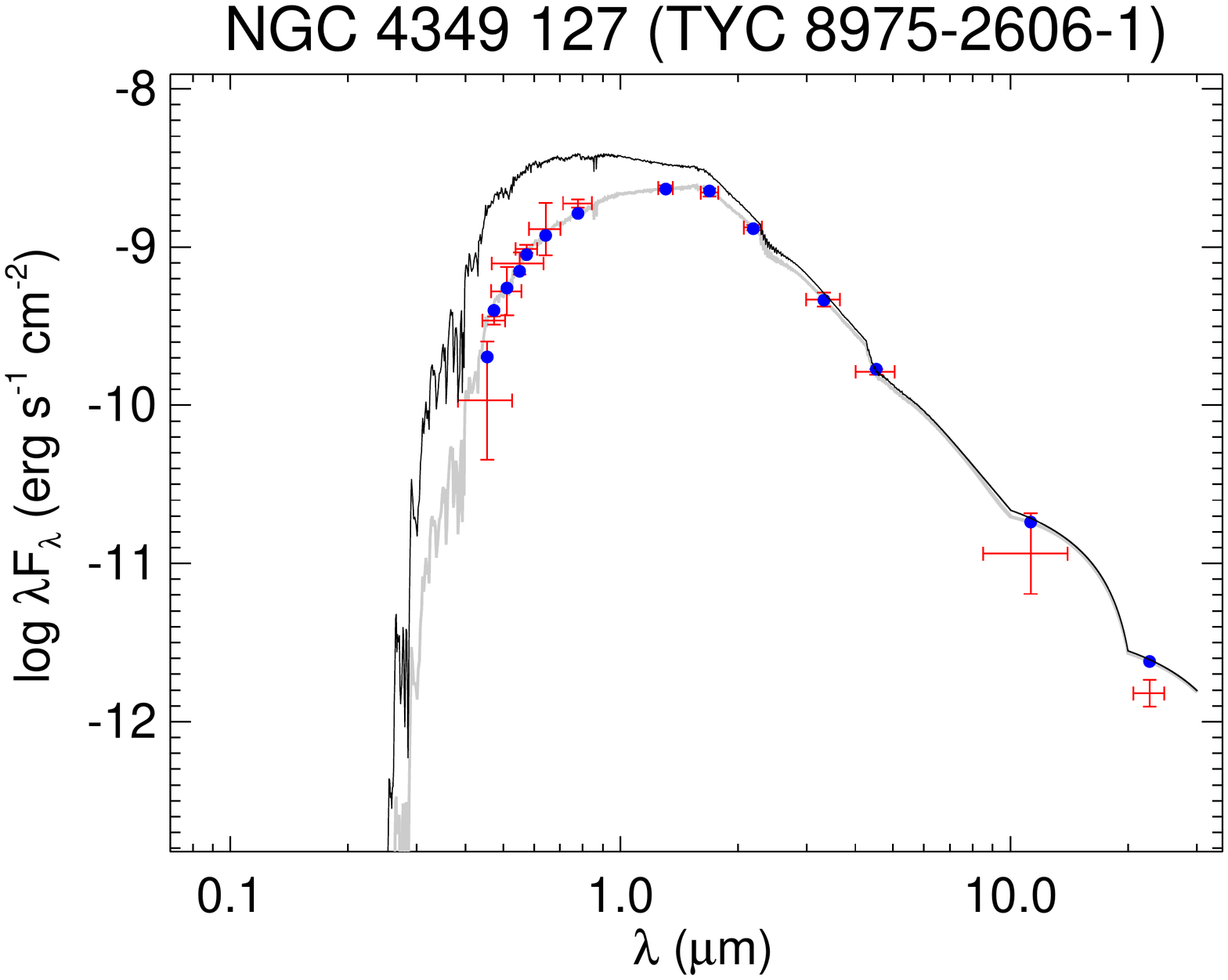}
  \caption{All labels, lines, symbols, and colors as in Figure \ref{fig:seds}.}
  \label{fig:seds_80}
\end{figure}

\begin{figure}[H]
  \centering
  \includegraphics[trim=60 60 60 60,clip,width=0.49\linewidth]{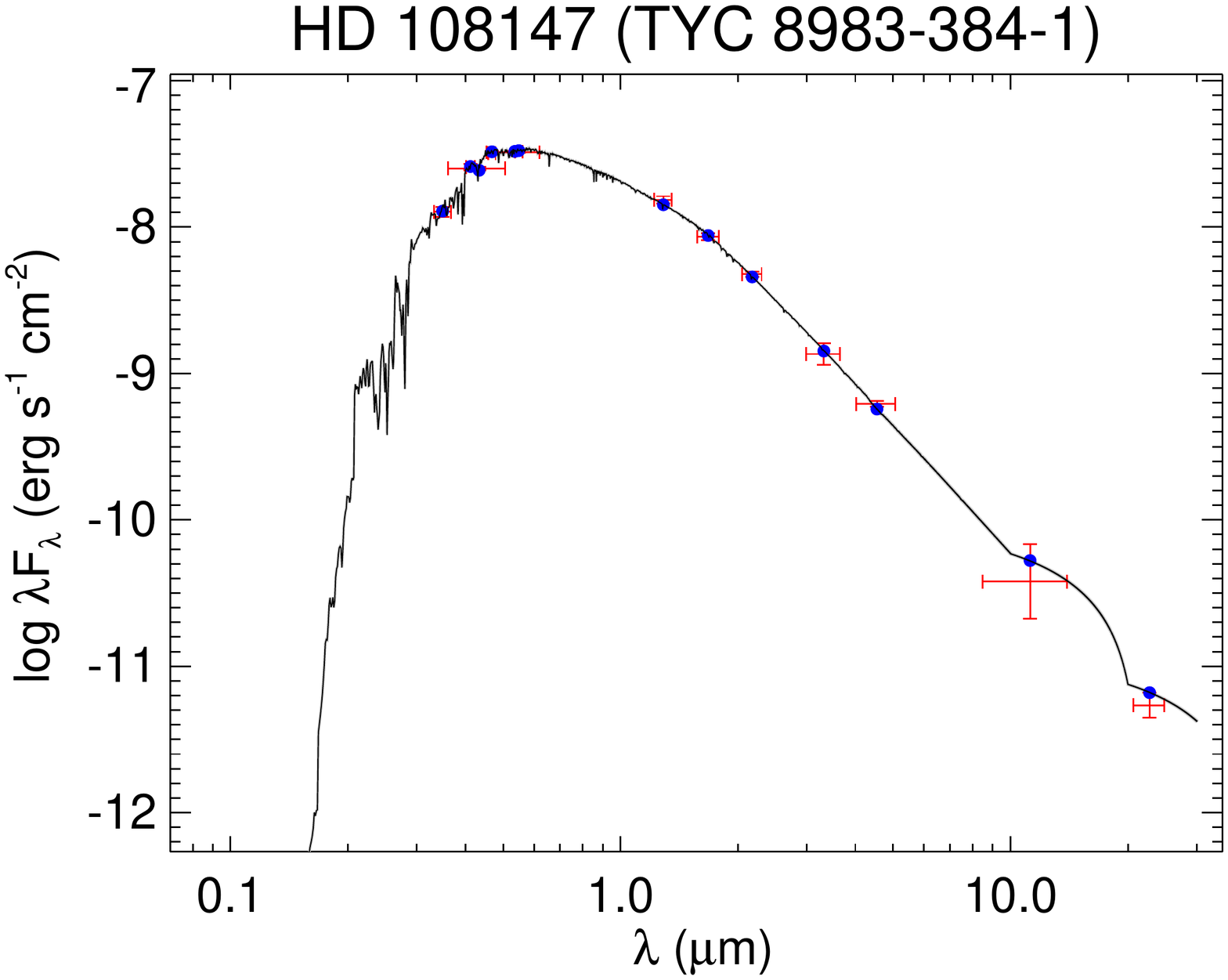}
  \includegraphics[trim=60 60 60 60,clip,width=0.49\linewidth]{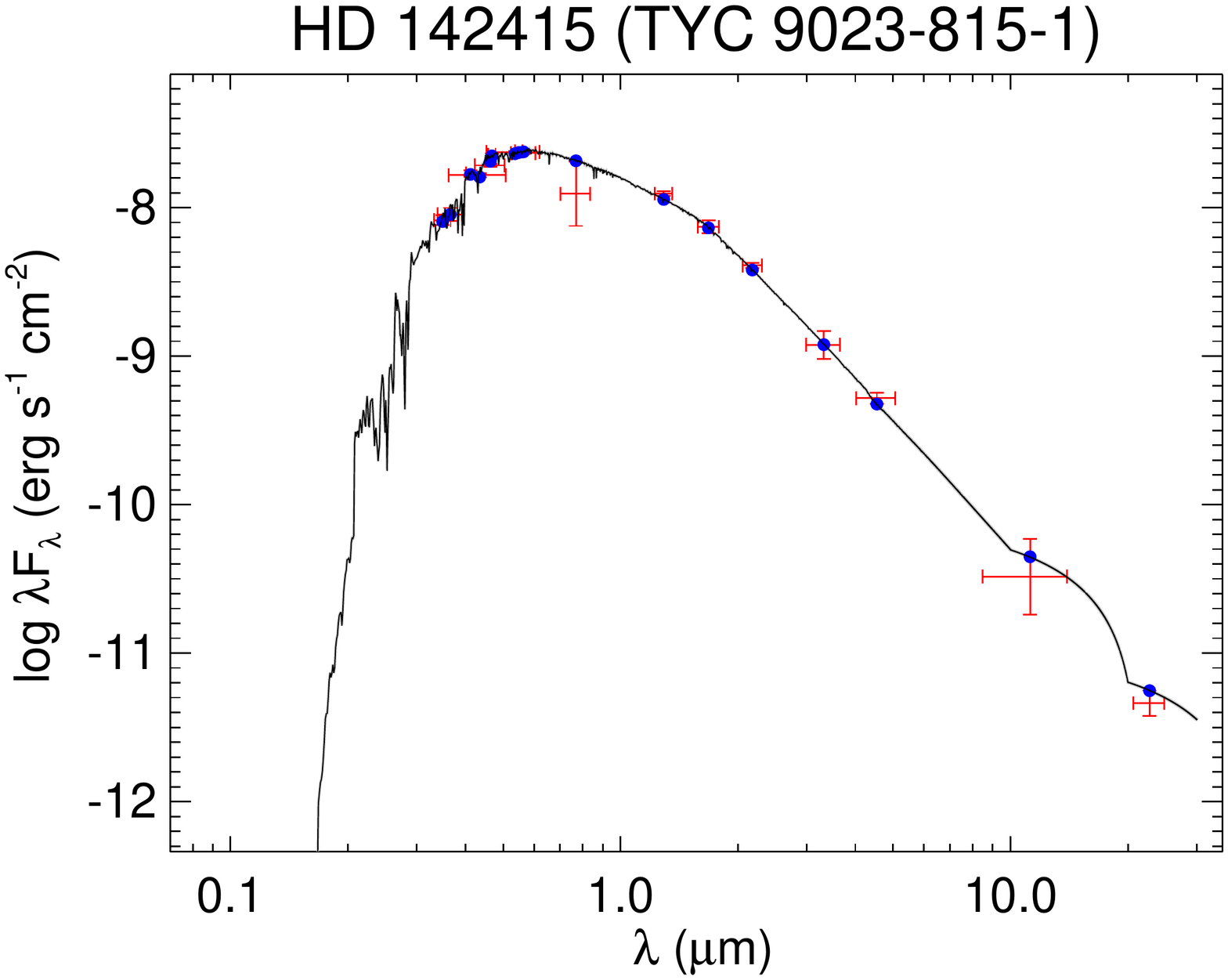}
  \includegraphics[trim=60 60 60 60,clip,width=0.49\linewidth]{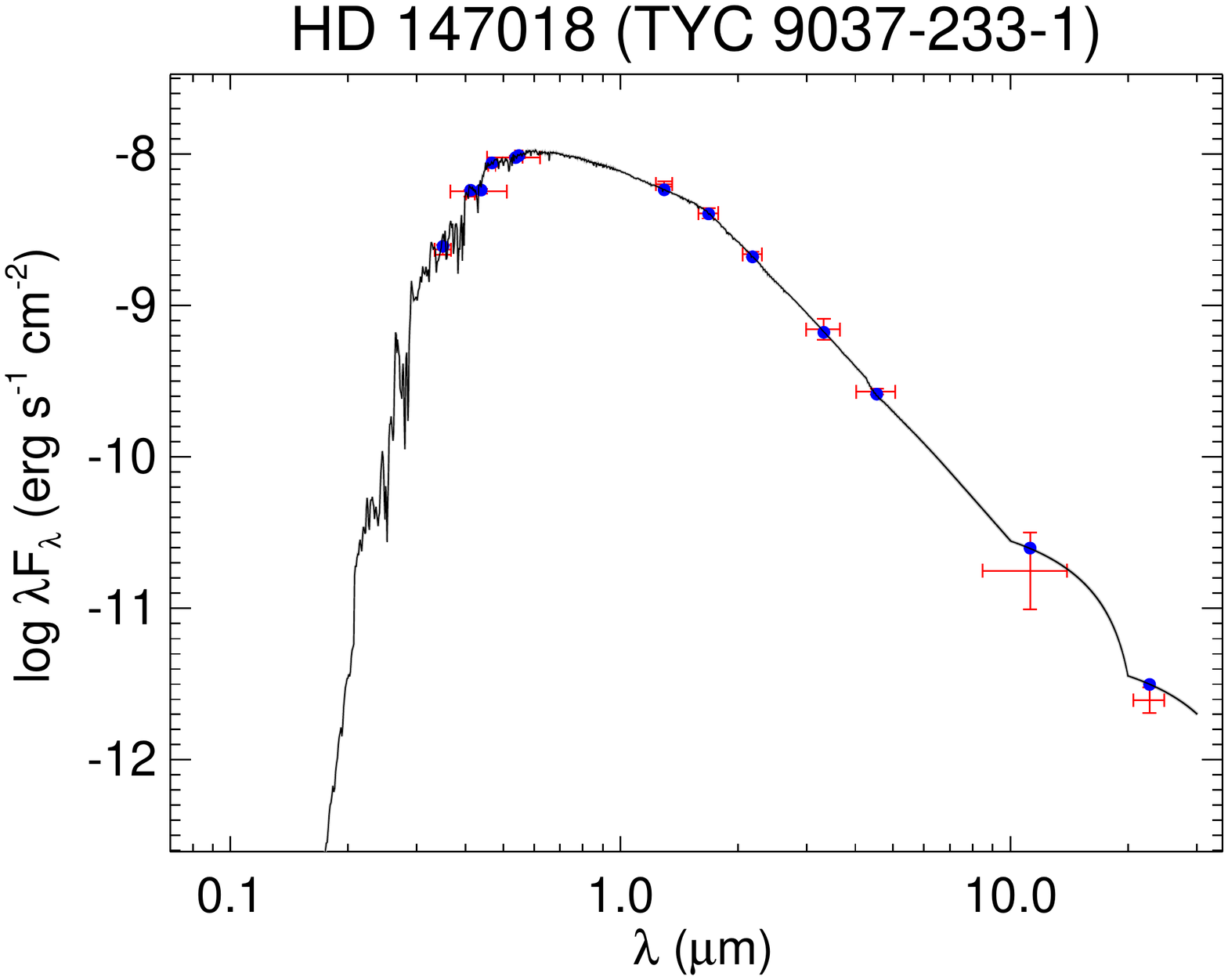}
  \includegraphics[trim=60 60 60 60,clip,width=0.49\linewidth]{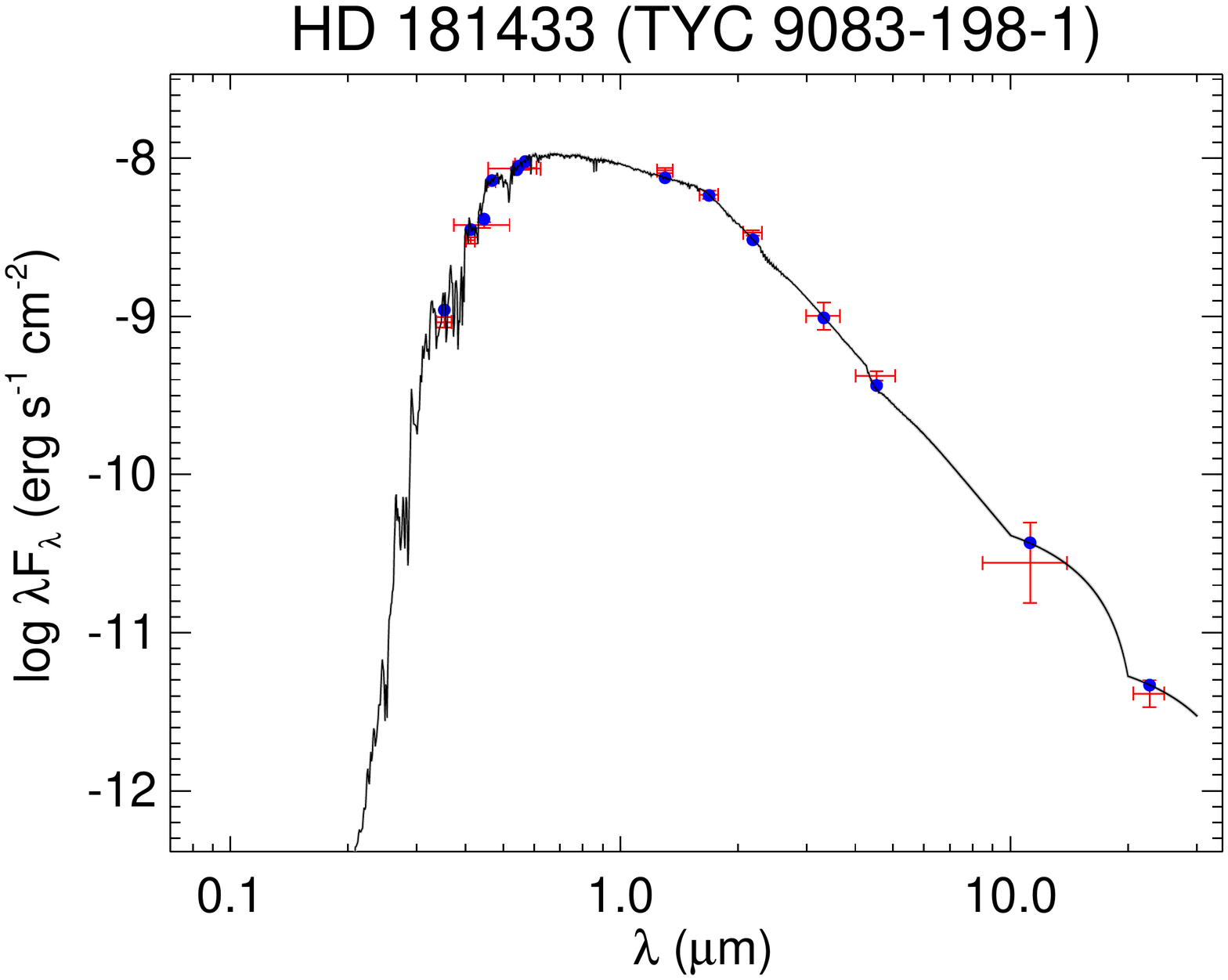}
  \includegraphics[trim=60 60 60 60,clip,width=0.49\linewidth]{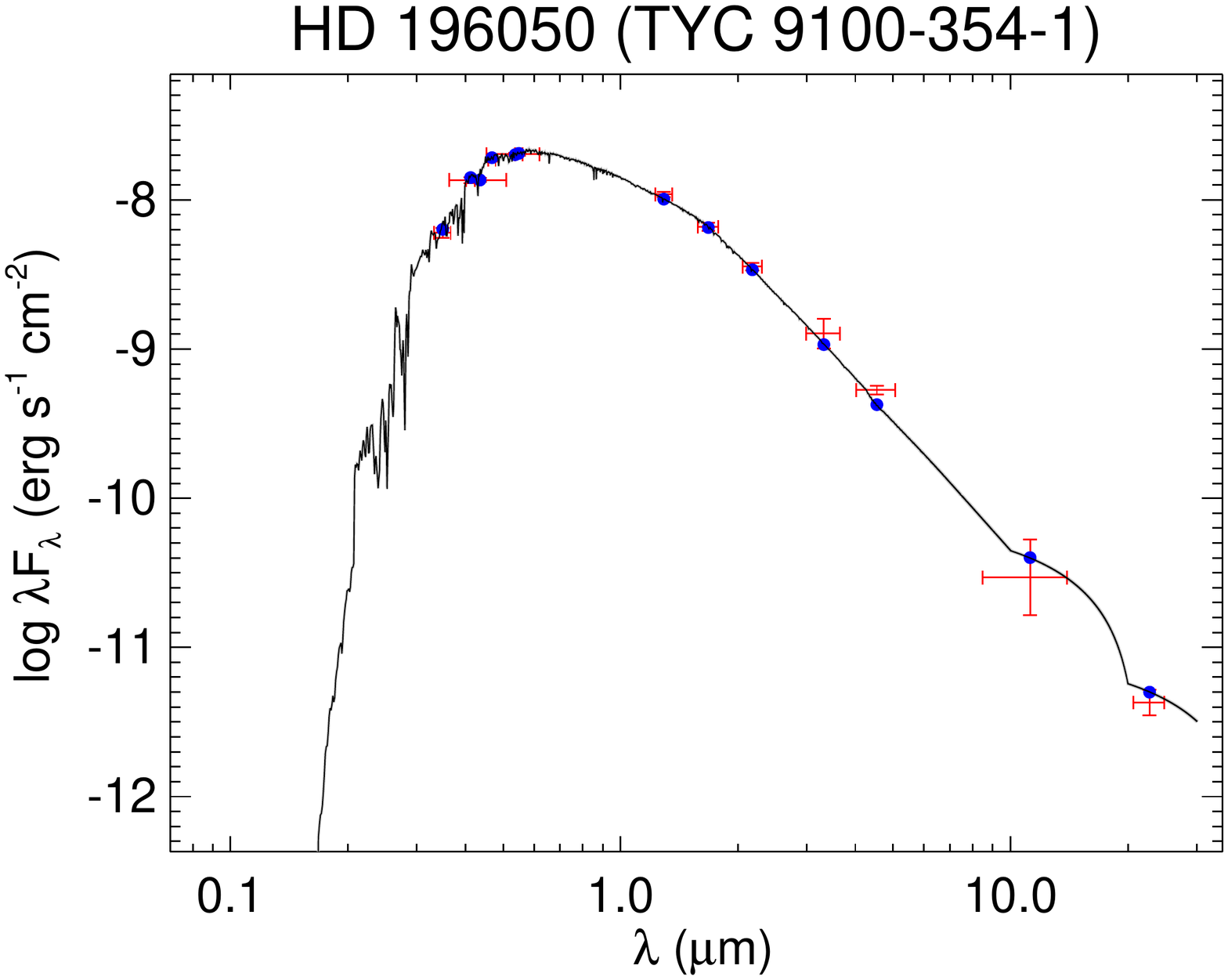}
  \includegraphics[trim=60 60 60 60,clip,width=0.49\linewidth]{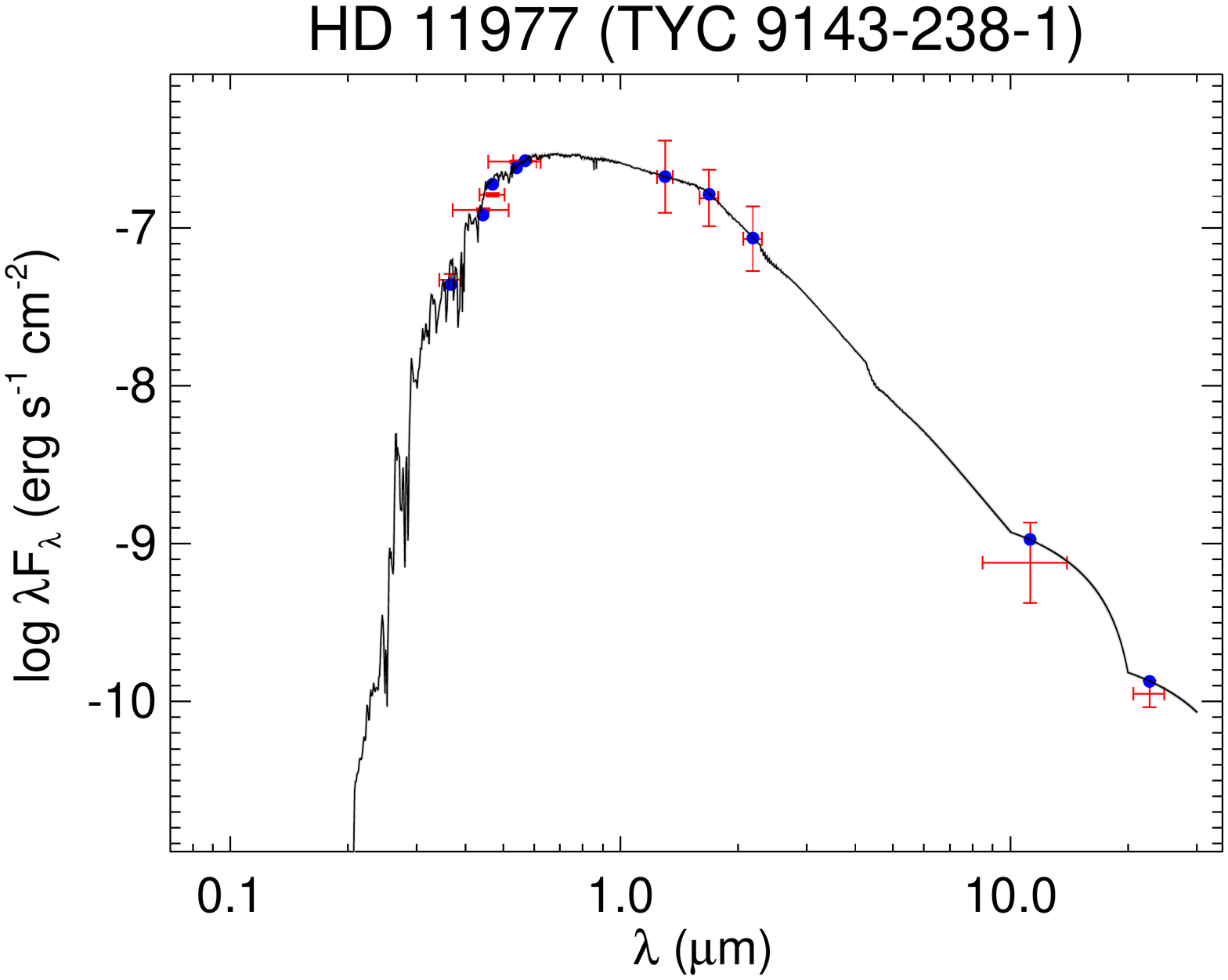}
  \caption{All labels, lines, symbols, and colors as in Figure \ref{fig:seds}.}
  \label{fig:seds_81}
\end{figure}

\begin{figure}[H]
  \centering
  \includegraphics[trim=60 60 60 60,clip,width=0.49\linewidth]{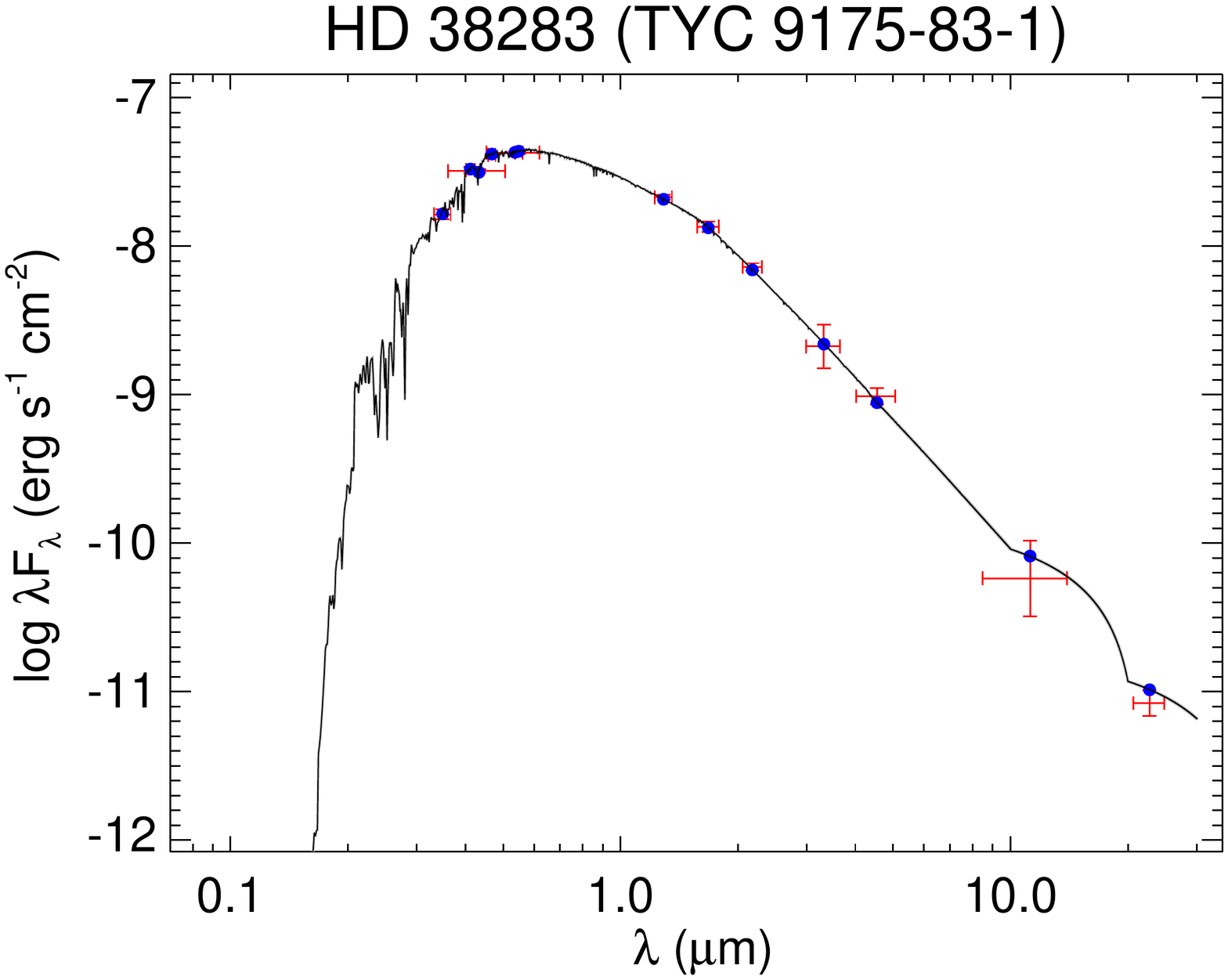}
  \includegraphics[trim=60 60 60 60,clip,width=0.49\linewidth]{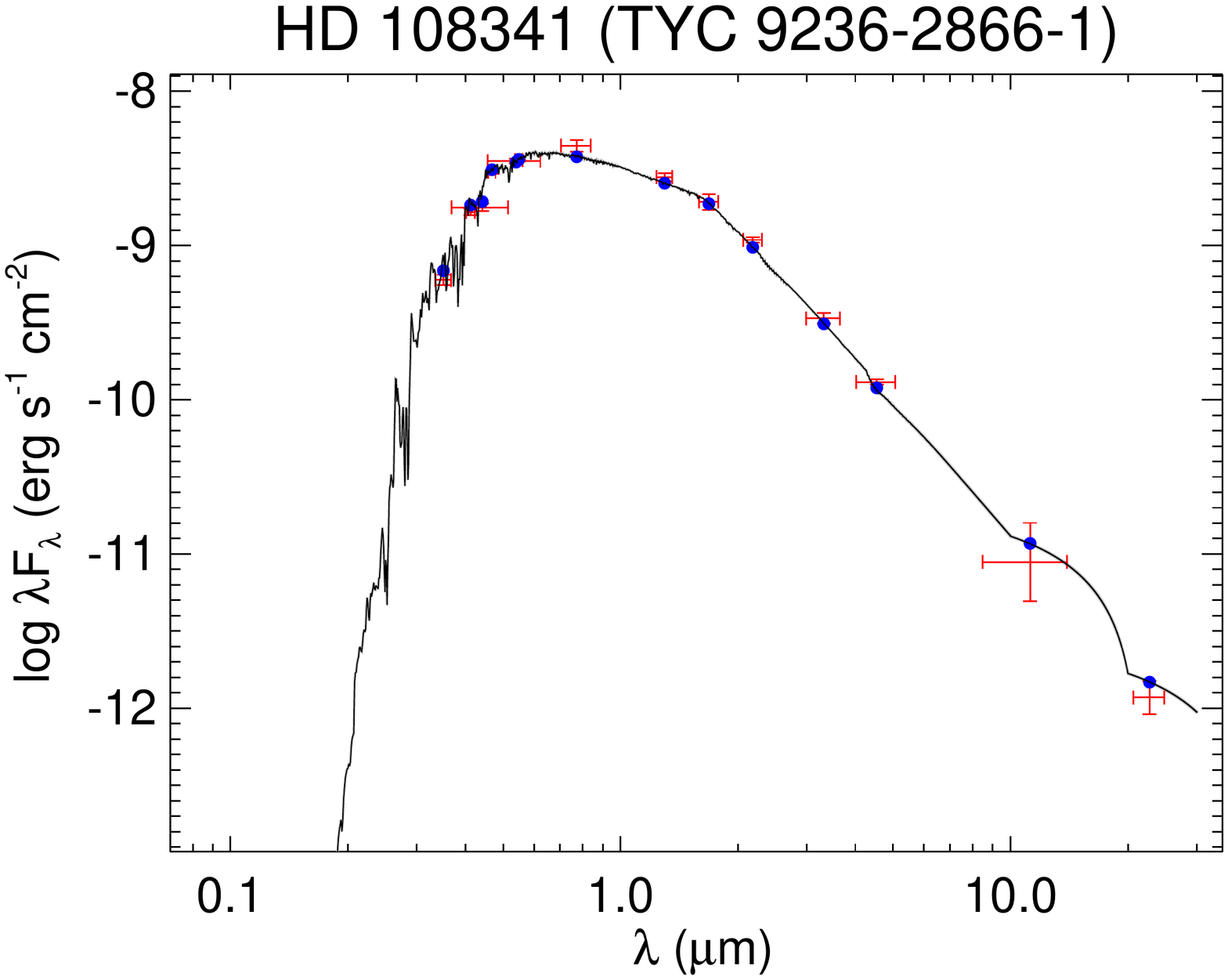}
  \includegraphics[trim=60 60 60 60,clip,width=0.49\linewidth]{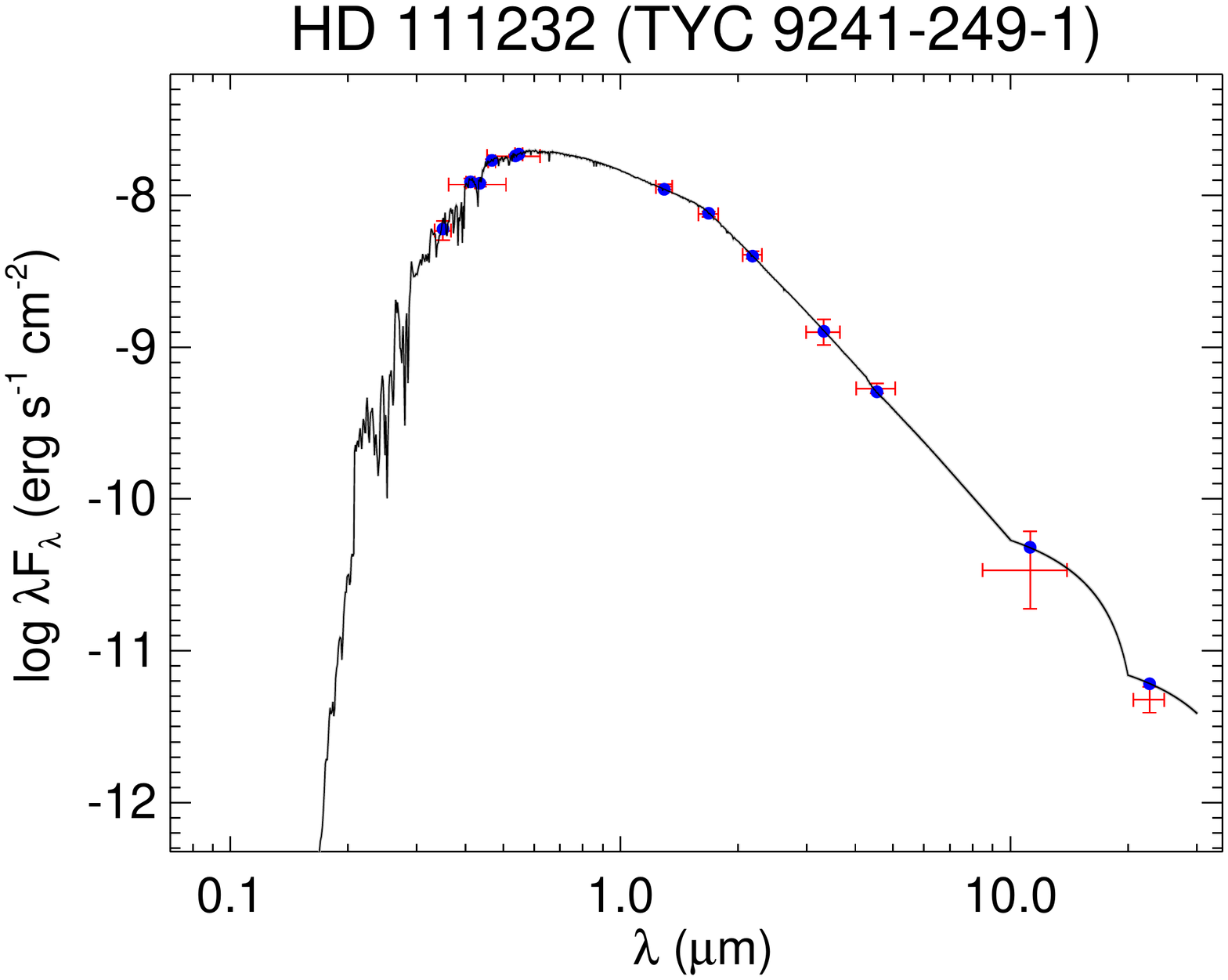}
  \includegraphics[trim=60 60 60 60,clip,width=0.49\linewidth]{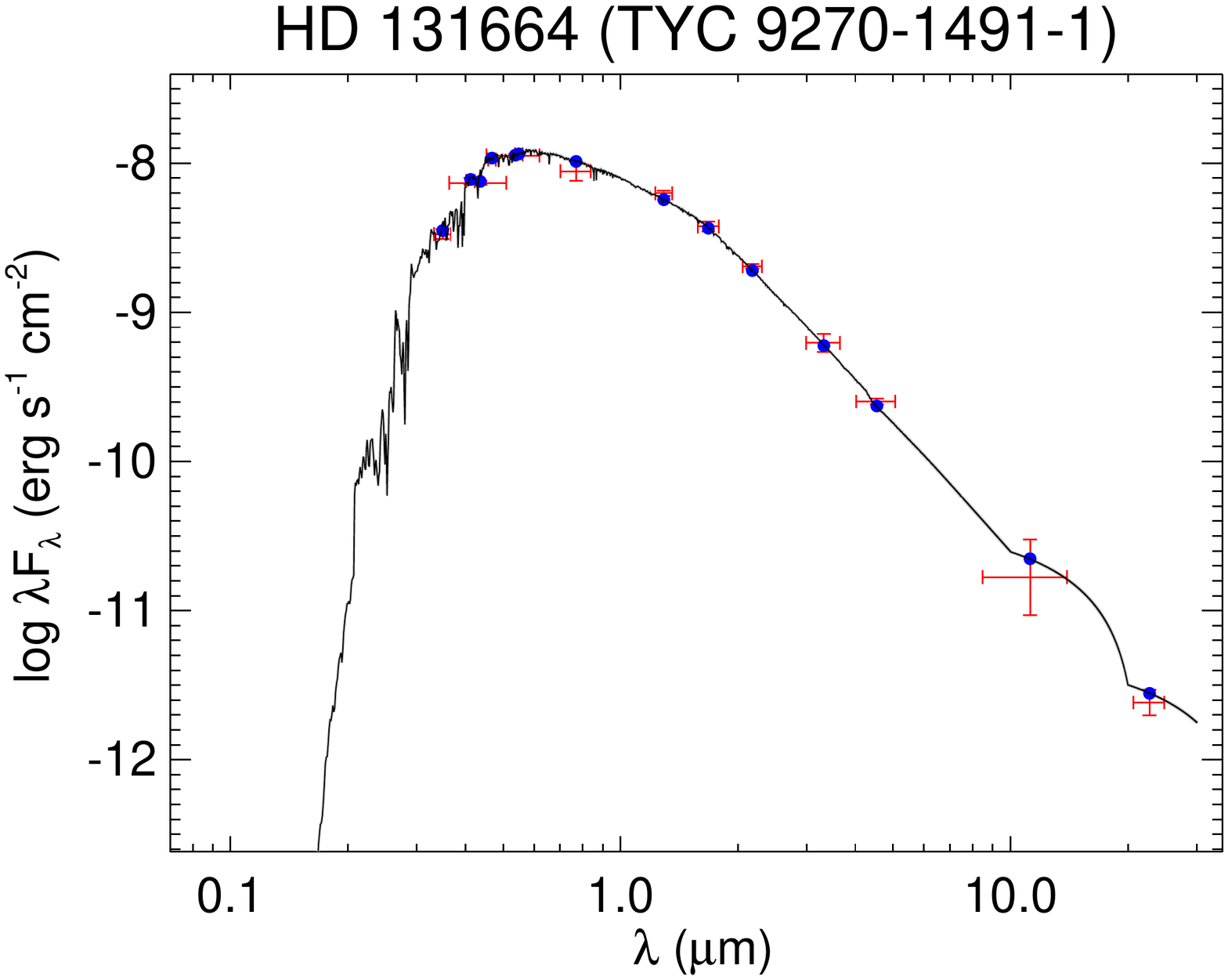}
  \includegraphics[trim=60 60 60 60,clip,width=0.49\linewidth]{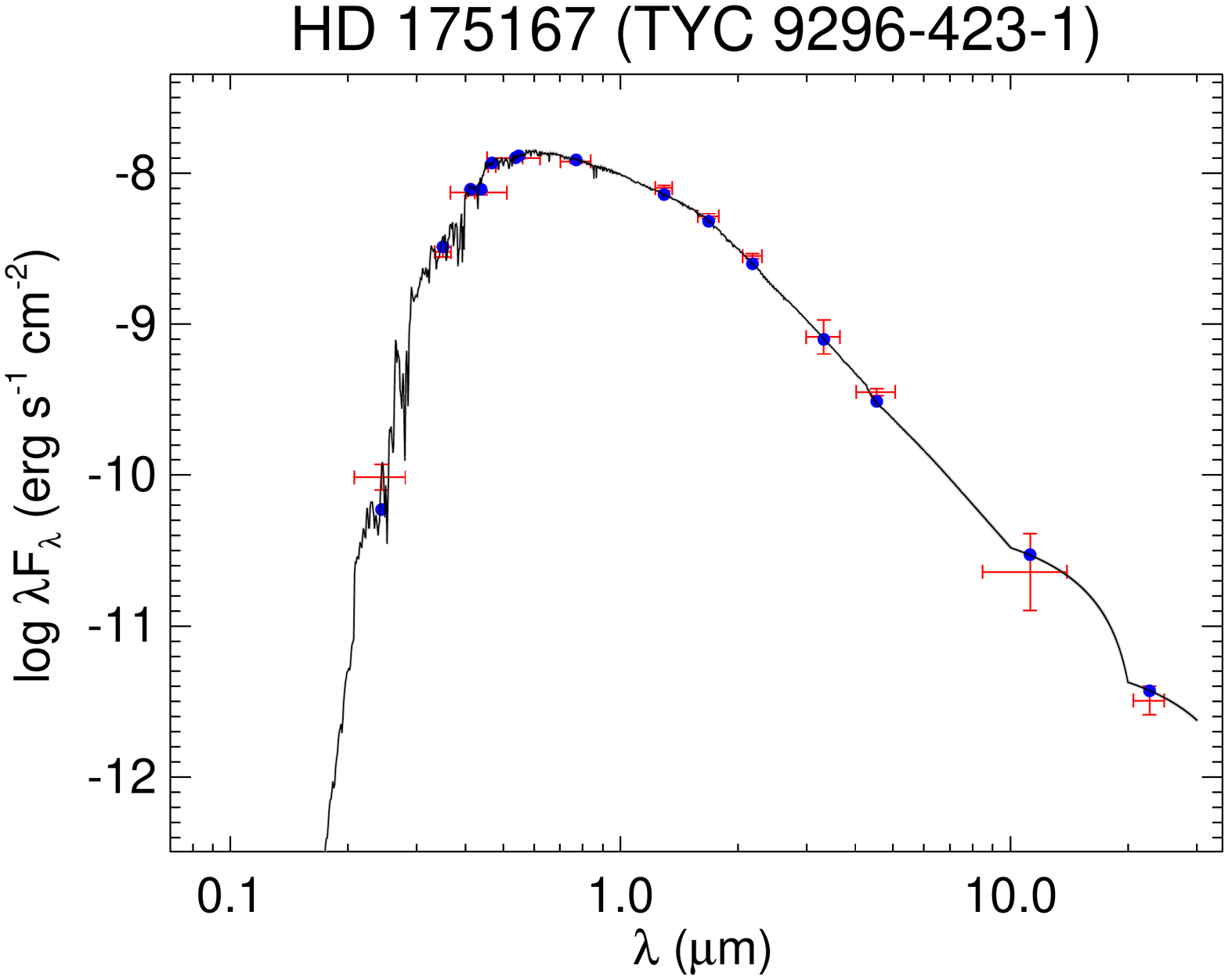}
  \includegraphics[trim=60 60 60 60,clip,width=0.49\linewidth]{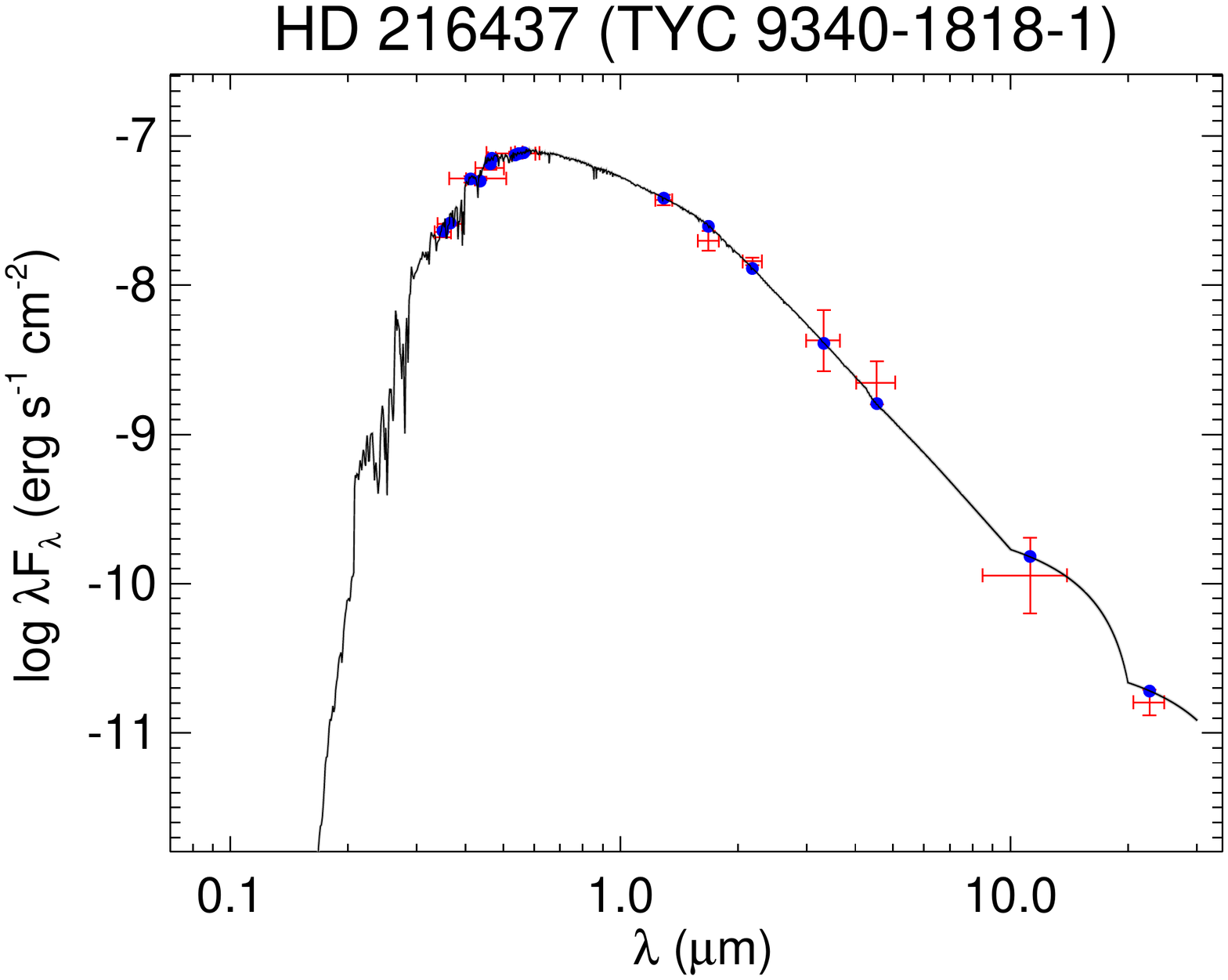}
  \caption{All labels, lines, symbols, and colors as in Figure \ref{fig:seds}.}
  \label{fig:seds_82}
\end{figure}

\begin{figure}[H]
  \centering
  \includegraphics[trim=60 60 60 60,clip,width=0.49\linewidth]{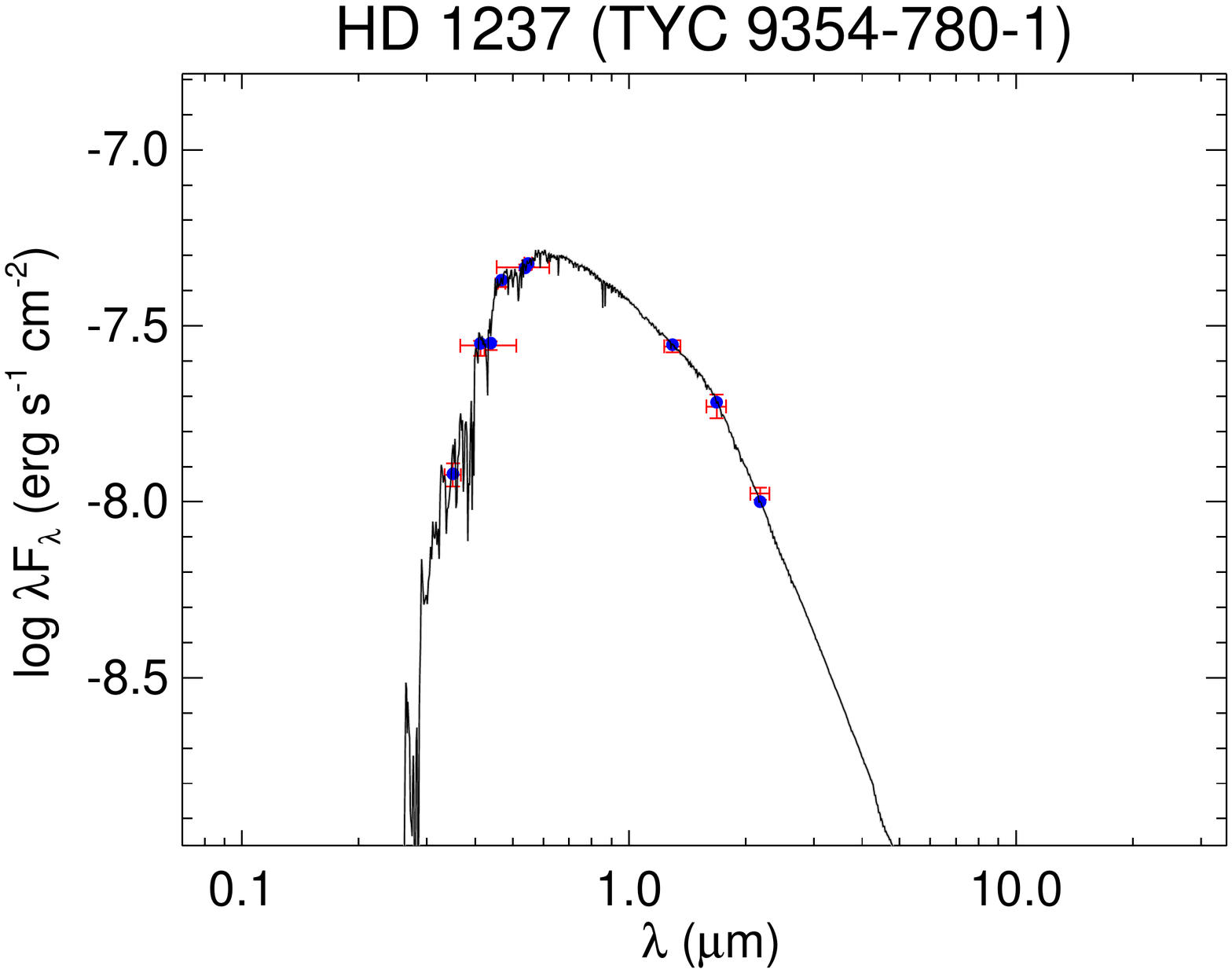}
  \includegraphics[trim=60 60 60 60,clip,width=0.49\linewidth]{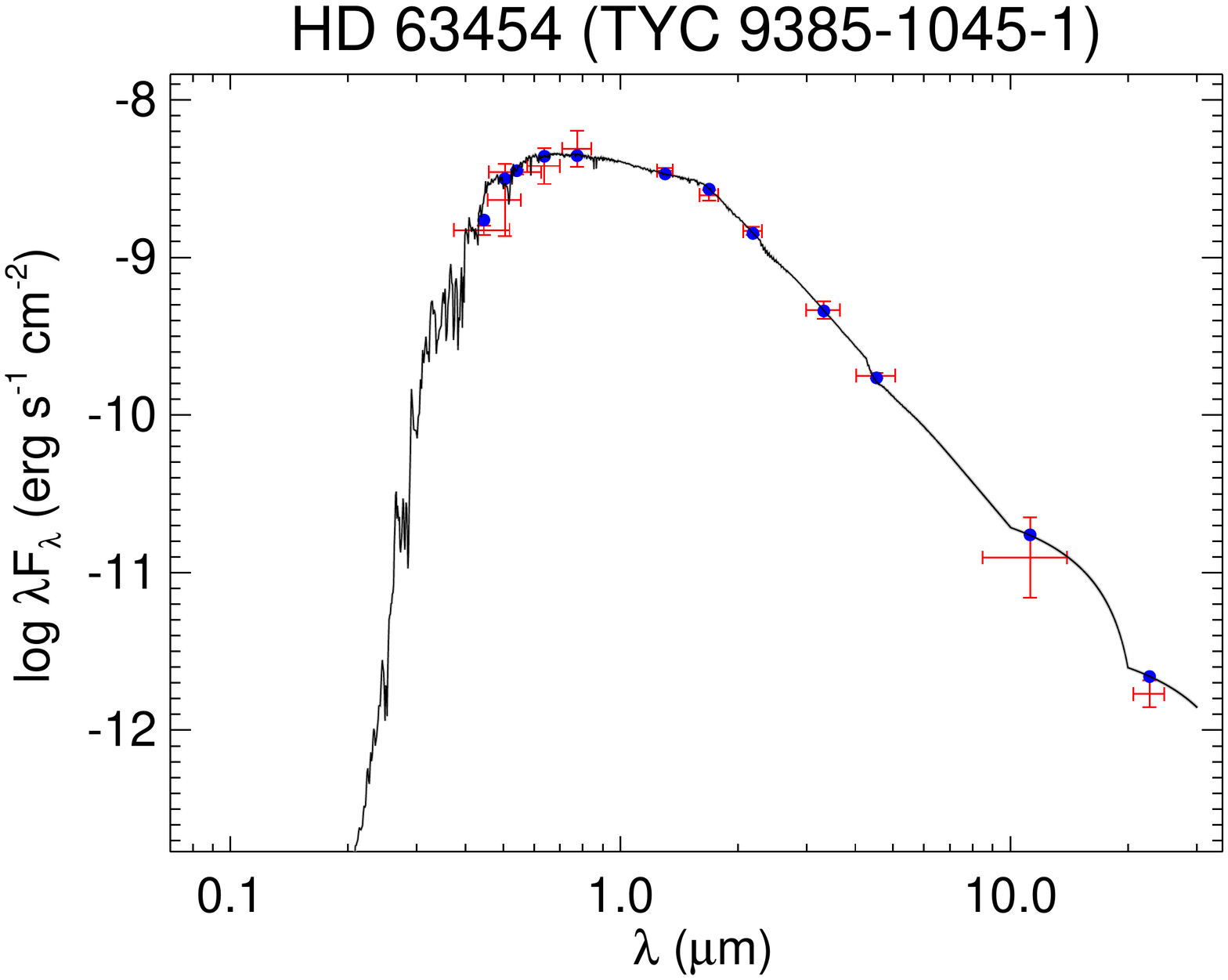}
  \includegraphics[trim=60 60 60 60,clip,width=0.49\linewidth]{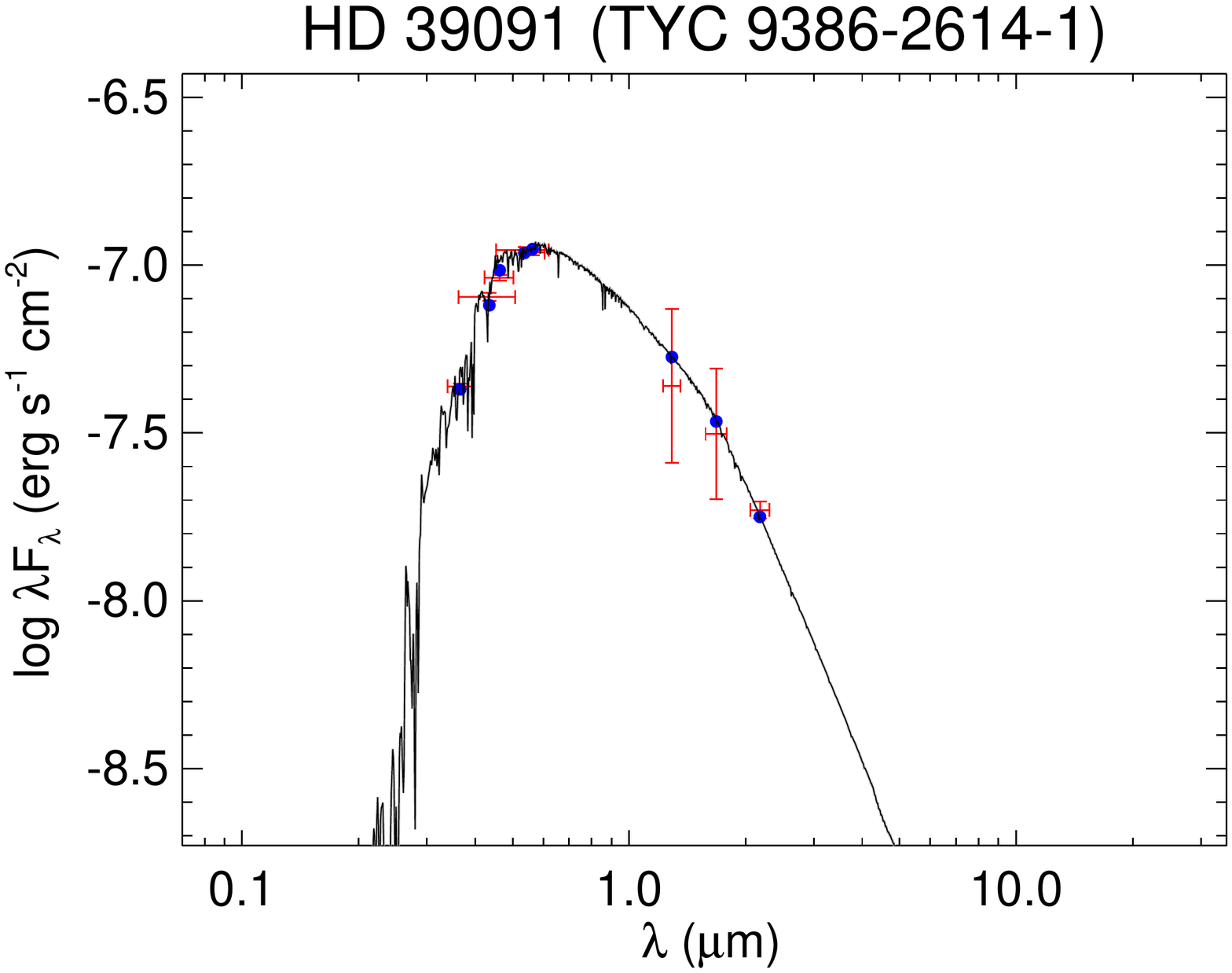}
  \includegraphics[trim=60 60 60 60,clip,width=0.49\linewidth]{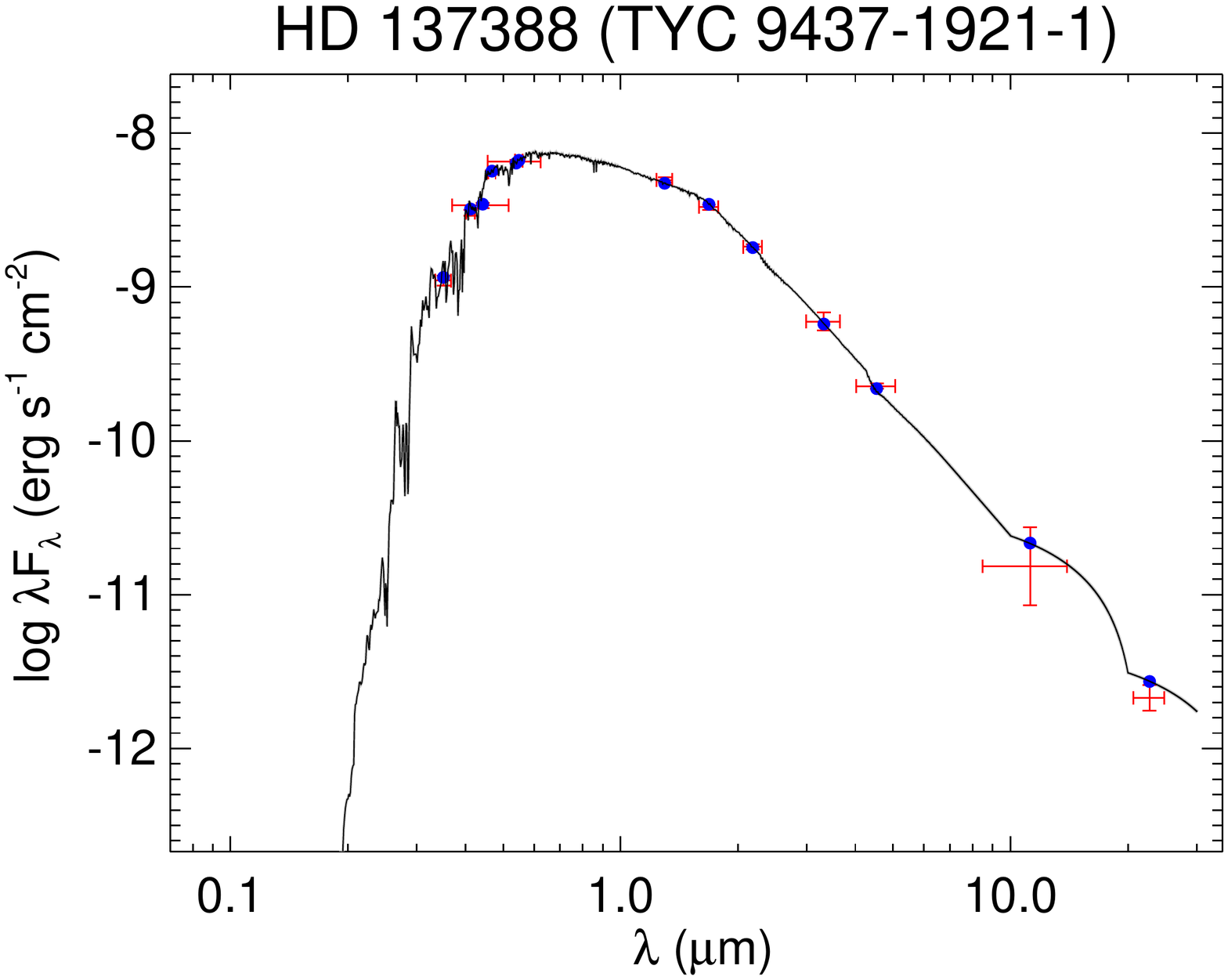}
  \includegraphics[trim=60 60 60 60,clip,width=0.49\linewidth]{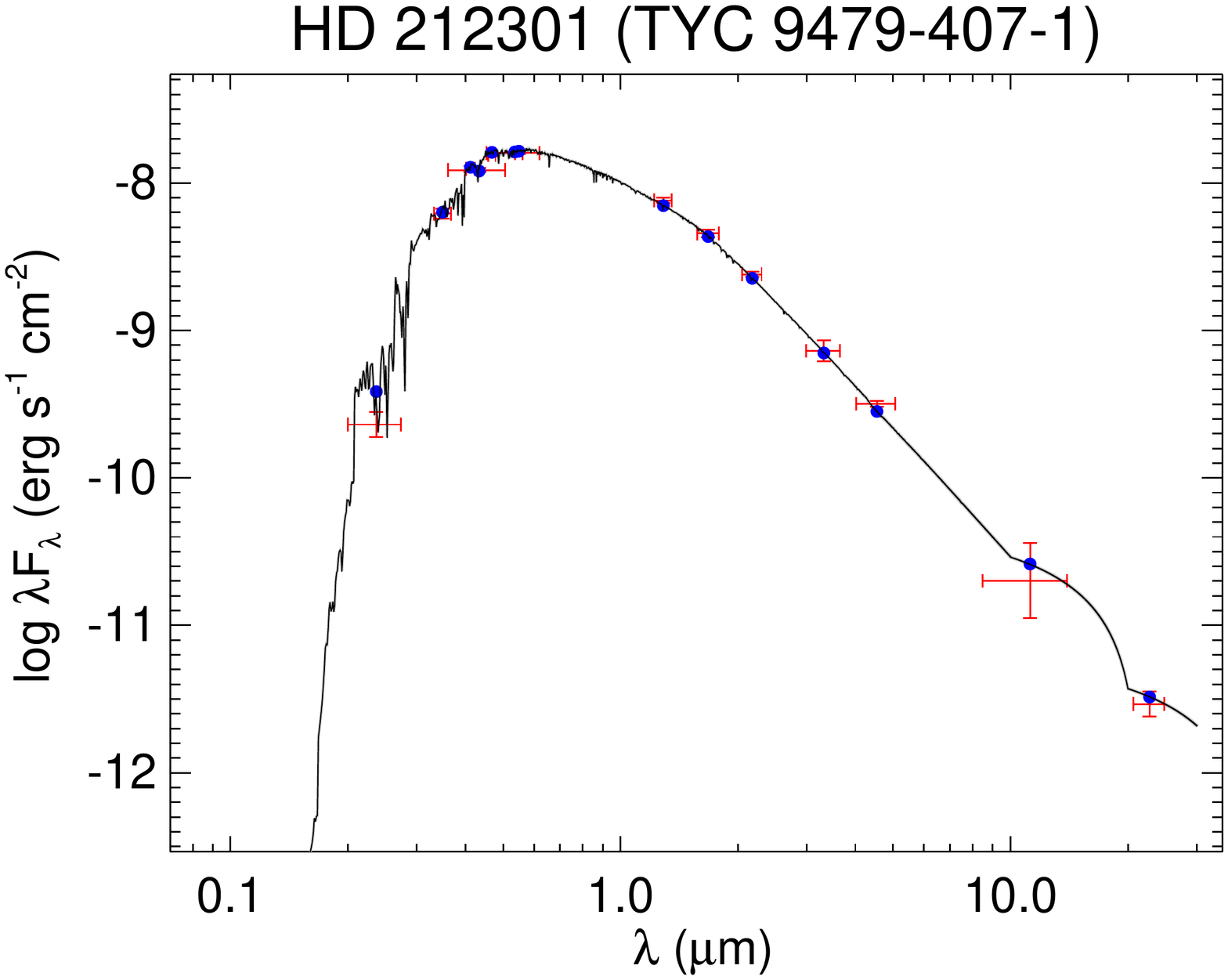}
  \includegraphics[trim=60 60 60 60,clip,width=0.49\linewidth]{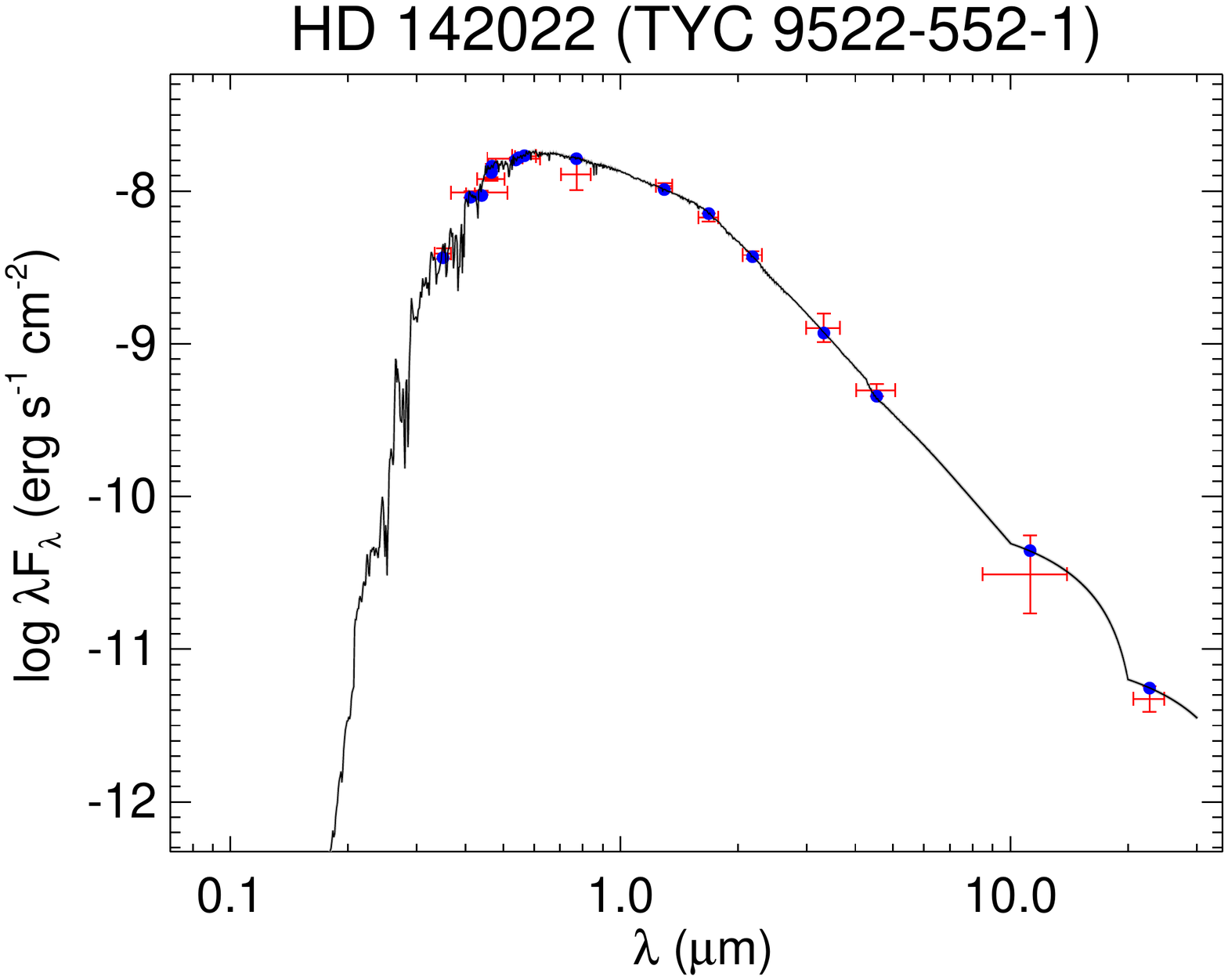}
  \caption{All labels, lines, symbols, and colors as in Figure \ref{fig:seds}.}
  \label{fig:seds_83}
\end{figure}


\end{document}